\documentclass[sn-mathphys,Numbered]{sn-jnl}

\makeatletter
\AtBeginDocument{
  \setlength{\oddsidemargin}{\evensidemargin}
}
\renewcommand\normalsize{%
   \@setfontsize\normalsize{9bp}{11bp}%
   \abovedisplayskip 10\p@ \@plus2\p@ \@minus1\p@
   \abovedisplayshortskip \z@ \@plus3\p@
   \belowdisplayshortskip 3\p@ \@plus3\p@ \@minus3\p@
   \belowdisplayskip \abovedisplayskip
   \let\@listi\@listI}
\normalsize
\DeclareMathSizes{9}{9}{6.5}{4.5}
\def\Titlefont{\reset@font\fontsize{15bp}{19bp}\selectfont\titraggedcenter}
\def\SubTitlefont{\reset@font\fontsize{12bp}{14bp}\selectfont\titraggedcenter}
\def\Authorfont{\reset@font\fontsize{11bp}{13bp}\selectfont\boldmath\titraggedcenter}
\def\addressfont{\reset@font\fontsize{10bp}{12bp}\selectfont\titraggedcenter}
\def\abstractheadfont{\reset@font\fontsize{8.5bp}{10.5bp}\bfseries\selectfont\titraggedcenter}
\def\abstractfont{\reset@font\fontsize{8.5bp}{10.5bp}\selectfont\leftskip=24pt\rightskip=24pt\parfillskip=0pt plus 1fil}
\def\keywordfont{\reset@font\fontsize{7.5bp}{9bp}\selectfont\leftskip=24pt\rightskip=24pt plus0.5fill}
\def\sectionfont{\reset@font\fontfamily{\rmdefault}\fontsize{12bp}{14bp}\bfseries\selectfont\raggedright\boldmath}
\def\subsectionfont{\reset@font\fontfamily{\rmdefault}\fontsize{10.5bp}{12bp}\bfseries\selectfont\raggedright\boldmath}
\def\subsubsectionfont{\reset@font\fontsize{9.5bp}{11bp}\bfseries\selectfont\raggedright\boldmath}
\def\paragraphfont{\reset@font\fontsize{9bp}{11bp}\bfseries\itshape\selectfont\raggedright}
\def\tablecaptionfont{\reset@font\fontsize{7.5bp}{9bp}\selectfont}
\def\tablebodyfont{\reset@font\fontsize{7.5bp}{9bp}\selectfont}
\def\tablecolheadfont{\reset@font\fontsize{7.5bp}{9bp}\selectfont\bfseries\boldmath}
\def\figurecaptionfont{\reset@font\fontfamily{\rmdefault}\fontsize{7.5bp}{9bp}\selectfont}
\makeatother

\usepackage{graphicx}
\usepackage{multirow}
\usepackage{amsmath,amssymb,amsfonts}
\usepackage{amsthm}
\usepackage{mathrsfs}
\usepackage[title]{appendix}
\usepackage{xcolor}
\usepackage{textcomp}
\usepackage{manyfoot}
\usepackage{booktabs}
\usepackage{algorithm}
\usepackage{algorithmicx}
\usepackage{algpseudocode}
\usepackage{listings}
\usepackage{subcaption}
\usepackage{url}
\usepackage{adjustbox}
\usepackage{morefloats}
\usepackage{adjustbox}
\usepackage{wrapfig}

\graphicspath{{figs/}}

\theoremstyle{thmstyleone}

\theoremstyle{thmstyletwo}

\theoremstyle{thmstylethree}

\newcommand{\pderfoper}[1]{{\frac{\partial}{\partial #1}}}
\newcommand{\pdernoper}[2]{{\frac{\partial^{#2}}{\partial #1^{#2}}}}
\newcommand{\pderf}[2]{{\frac{\partial #1}{\partial #2}}}
\newcommand{\pdern}[3]{{\frac{\partial^{#3} #1}{\partial #2^{#3}}}}

\newcommand{\Dx}[0]{{\Delta x}}
\newcommand{\Dt}[0]{{\Delta t}}
\newcommand{\Dtn}[1]{{{\Delta t}^{#1}}}
\newcommand{\intrefdom}[1]{{\int\limits_{0}^{1} d{#1}}}
\newcommand{\uphi}[0]{\smash[b]{\underline{\varphi}}}
\newcommand{\ophi}[0]{\smash[b]{\overline{\varphi}}}
\newcommand{\uvecphi}[0]{\smash[b]{\underline{\boldsymbol{\varphi}}}}
\newcommand{\ovecphi}[0]{\smash[b]{\overline{\boldsymbol{\varphi}}}}

\newcommand{\cfl}[0]{\mathrm{CFL}}

\newcommand{\done}{{\color{green!70!black}\checkmark}}
\newcommand{\fail}{{\color{red!70!black}$\times$}}

\begin{document}

\title[Stiff source terms inside ADER-DG method]{Stiff source terms inside ADER-DG method}

\author*[1]{\fnm{Ivan S.} \sur{Popov}}\email{diphosgen@mail.ru, popovis@omsu.ru}
\author[1]{\fnm{Anna P.} \sur{Popova}}\email{popovaap@omsu.ru}
\affil[1]{\orgdiv{Department of Theoretical Physics}, \orgname{Dostoevsky Omsk State University},
\orgaddress{\street{Mira prospekt}, \city{Omsk}, \postcode{644077}, \country{Russia}}}

\abstract{%
ADER-DG numerical method with the LST-DG predictor, when used to solve hyperbolic equations with stiff source terms, exhibits properties characteristic of IMEX computational schemes. With increasing stiffness, characterized by the parameter $z = \nu\Dtn{n} < 0$ for a linear advection-reaction equation with a decay source term, the boundary value of the Courant number $\mathrm{CFL}_{\rm max}$ of the ADER-DG numerical method increases. In the region of strong stiffness $|z| \gg 1$, a linear increase in the Courant number $\mathrm{CFL}_{\rm max}(z) \sim |z|$ occurs. In the region of small source terms $|z| \ll 1$, the results $\mathrm{CFL}_{\rm max}(z)$ transform into known results for the problem $\mathrm{CFL}_{\rm max}(0)$ without source terms. In the transition region $|z| \sim 1$, a range of insignificant decrease in the Courant number $\mathrm{CFL}_{\rm max}(z)$ is discovered. It is found that, in the linear case, instability is violated precisely when one of the matrix eigenvalues reaches $\lambda = -1$, regardless of the phase $\theta$ of the numerical solution. In this case, unlike the case of a linear advection equation, where this argument is confirmed only with a small accuracy, a small unstable eigenvalues with phase $\arg\lambda_{k} \in [-\pi/4, +\pi/4]$ did not arise even for the parameter $|z| > 10^{-5}$--$10^{-6}$. A rigorous stability theory of the ADER-DG numerical method for stiff problems for arbitrary degree $N$ of basis polynomials is constructed. The stability condition is simplified to solving a polynomial equation for the Courant number $\mathrm{CFL}$, with coefficients in the form of polynomials in the parameter $z$, instead of the original problem of finding the maximum value of the spectral radius of the evolution operator matrix, which depends on the continuous phase $\theta\in[0, 2\pi]$. Highly accurate dependencies of the Courant number $\mathrm{CFL}_{\rm max}(N, z)$ of the ADER-DG method on the parameter $z$ are obtained. A rigorous direct proof of the approximation of the ADER-DG method is presented. Approximation orders $p = N+1$ for arbitrary degrees $N$ are rigorously derived. A set of numerical experiments is carried out to apply the ADER-DG method to solving both a linear advection-reaction equation. The results obtained in these calculations confirm the theoretical results well. A numerical experiments to solving the system of Euler equations for a two-component reacting medium, with weak and strong stiff cases, is carried out. These results demonstrate that the obtained theoretical results are also correct in the nonlinear system with stiff source terms, and that the ADER-DG method with the LST-DG predictor is well applicable to solving such problems.
}

\keywords{%
arbitrarily high order methods, ADER-DG method, LST-DG predictor, hyperbolic equations, advection-reaction equation, stiffness, stiff source terms, convergence analysis, stability analysis, approximation analysis
}

\pacs[MSC Classification]{65M12, 35L40, 76M10, 65M60, 35Q31, 76N30, 76V05, 76L05}

\maketitle

\section*{Introduction}
\addcontentsline{toc}{section}{Introduction}
\label{intro}

This work is devoted to the construction of a rigorous theory and the study of the applicability of the ADER-DG numerical method for solving hyperbolic equations with stiff source terms. Finite element ADER-DG numerical methods are numerical methods of arbitrarily high order and are among the most highly accurate modern numerical methods for solving partial differential equations (PDE). The ADER paradigm is developed by Titarev and Toro~\cite{ader_init_1, ader_init_2} for use in finite-volume methods, and the main innovation was the use of a solution to the generalized Riemann problem based on calculating the evolution of a local space-time solution on the interfaces of finite-volume cells using the Cauchy-Kovalevskaya procedure, based on expanding the solution into high-order Taylor series in coordinates and time (see also~\cite{ader_init_3}), and was significantly developed in subsequent works~\cite{ader_init_4, ader_init_5, ader_init_6}. Dumbser \textit{et al}~\cite{ader_stiff_1, ader_stiff_2} significantly modified the ADER paradigm for finite-volume methods based on the use of a discrete space-time solution obtained in a separate stage of the local space-time DG-predictor (LST-DG predictor), and showed higher solution accuracy than the original version, especially for stiff problems. Finite-element ADER-DG methods are fundamentally linear methods for solving PDE systems, so such methods require limiters to preserve the monotonicity of the numerical solution, on the well-known Godunov theorem. Currently, finite-element ADER-DG methods are based on the development of the Multi-dimensional Optimal Order Detection (MOOD) paradigm~\cite{mood_par_1, mood_par_2, mood_par_3, mood_par_4}, which resulted in the development~\cite{ader_dg_dev_1, ader_dg_dev_2} of finite-element ADER-DG methods with \textit{a posteriori} correction of the solution in subcells by a finite-volume limiter, which can be the finite-volume ADER-WENO method~\cite{ader_weno_lstdg_ideal, ader_weno_lstdg_diss}. This approach made it possible to preserve the monotonicity of the numerical solution and the subgrid resolution characteristic of DG methods~\cite{ader_dg_dev_1}. However, the further development of purely finite-volume methods of the ADER family, carried out in the works~\cite{ader_weno_new_1, ader_weno_new_2, ader_weno_new_3, ader_weno_new_4, ader_weno_new_5, ader_weno_new_6, ader_weno_new_7}, is not directly related to the ADER-DG method with a posteriori correction of the solution.

The current modern state and formulations of finite-element ADER-DG numerical methods with the LST-DG predictor are presented in~\cite{ader_rev_2024, ader_dg_simple_mod_2025, ader_dg_axioms, exahype, fron_phys, ader_dg_simple_mod_2016, ader_dg_mod_1, ader_dg_mod_2, ader_dg_mod_3}. The DG methods were created in the work~\cite{lasl_rep_dg_1973} for numerical solution of the neutron transport equations. Cockburn, Shu \textit{et al} in a series of works~\cite{Cockburn_base_1, Cockburn_base_2, Cockburn_base_3, Cockburn_base_4, Cockburn_base_5} created an accurate and thoroughly developed mathematical basis of DG methods, which stimulated their further development and use for solving a wide class of problems. The ADER-DG numerical methods with the LST-DG predictor are widely used in solving problems of classical ideal and dissipative hydrodynamics and magnetohydrodynamics~\cite{ader_dg_ideal_flows, ader_dg_diss_flows, ader_dg_semiexpl, PNPM_DG_2010}, relativistic hydrodynamics and magnetohydrodynamics~\cite{ader_dg_grmhd, ader_dg_PNPM, PNPM_DG_2008, PNPM_DG_2009}, which are characterized by essentially stiff terms. In recent years, significant interest~\cite{ader_dg_gr_prd, ader_dg_gr_z4_2024, ader_dg_gr_z4_2026} has been associated with the development and use of ADER-DG numerical methods for solving Einstein equations of general relativity in the formulations of first-order hyperbolic systems~\cite{rel_hyd_book_2013, rel_hyd_book_2007, rel_hyd_book_2008, rel_hyd_book_2009, rel_hyd_book_2010, rel_hyd_book_2012, rel_hyd_book_2021}. The ADER-DG numerical methods are also used to solve the equations of motion of deformable media~\cite{ader_dg_axioms, fron_phys, ader_dg_hyperelastic}, seismic wave propagation~\cite{ader_dg_axioms, exahype, ader_dg_seiemic, ader_dg_seiemic_underwater}, the equations of shallow water theory~\cite{ADER_GRP_LST_DG_1, ader_dg_wb_shwater_2018, ader_dg_wb_shwater_2022}, simulating blood flow~\cite{ader_eno_fv_blood_2022}, simulating compressible barotropic two‐fluid flows~\cite{ader_dg_barotropic} and nonlinear dispersive systems~\cite{ader_dg_dispersive}. A special place is occupied by problems associated with the study of the reacting flows, which are characterized by abnormally high stiffness, for which the finite element ADER-DG methods and finite volume ADER-WENO methods were developed and studied in the works~\cite{ader_stiff_1, ader_stiff_2, ader_stiff_3, ader_stiff_4}. Application of finite element ADER-DG method for solving a wide range of other problems are also presented in~\cite{ader_dg_axioms, exahype}. Finite-element ADER-DG numerical method with the LST-DG predictor are developed for structured Cartesian meshes~\cite{ader_dg_dev_1, ader_dg_dev_2, ader_dg_grmhd, ader_dg_gr_prd, ader_dg_gr_z4_2024}, using the adaptive mesh refinement paradigm~\cite{ader_weno_lstdg_ideal, ader_weno_lstdg_diss, ader_dg_ideal_flows, ader_dg_diss_flows, ader_dg_PNPM}, unstructured triangular and tetrahedral meshes~\cite{ader_dg_simple_mod_2025, ader_dg_simple_mod_2016, PNPM_DG_2009, PNPM_DG_2010, ader_dg_axioms, ader_dg_wb_shwater_2018} and moving meshes~\cite{ader_dg_ale} (see also~\cite{ader_rev_2024}). A detailed discussion of the features of the mathematical and algorithmic formulations of numerical methods and their software implementation is presented in works~\cite{ader_dg_axioms, exahype, fron_phys, ader_dg_hpc_impl_1, ader_dg_hpc_impl_2, ader_dg_hpc_impl_3, ader_dg_hpc_impl_4, ader_eff_blas}. The implementation of the ADER-DG numerical method on a GPU is presented in~\cite{ader_dg_impl_gpu}. Dumbser \textit{et al} developed well-balanced ADER-DG method for the Euler equations in general relativity problems~\cite{ader_dg_gr_z4_2024} and for the multilayer shallow water equations~\cite{ader_dg_wb_shwater_2022}. Based on the ADER paradigm, a SPH numerical method was created~\cite{ader_weno_sph}, which demonstrated a very high quality of modeling hydrodynamic and magnetohydrodynamic flows with discontinuities, compared to classical SPH methods. Zhao \textit{et al} developed path-conservative discontinuous Galerkin (DG) method to solve shallow water equations~\cite{ader_dg_wb_pc_shwater_2024}. In~\cite{dg_entropy} Gaburro \textit{et al} developed entropy preserving ADER-DG schemes (see also~\cite{dg_entropy_add}). The numerical methods of the ADER family are significantly interconnected with the deferred correction (DeC) numerical methods, which was shown by Han Veiga \textit{et al}~\cite{dec_vs_ader_2021}. DeC methods have a long history and their application goes back to~\cite{dec_src_1968}, and are effectively used to solve partial differential equations~\cite{dec_abgrall_2017, dec_abgrall_2019} and ordinary differential systems of equations~\cite{dec_dutt_2000, dec_dutt_2000, dec_minion_2003, dec_shu_2008}. DeC methods, like the methods of the ADER family, allow one to obtain an arbitrarily high order and are characterized by a high accuracy of the numerical solution. The current state of research on DeC methods~\cite{dec_vs_ader_2021} shows that they compete with methods of the ADER family. The relationship between the ADER paradigm and the DeC paradigm are discussed in~\cite{dec_vs_ader_2021, dec_vs_ader_2023}.

An important feature of the ADER-DG method, like other DG methods, is a very strict limitation on the time step and on the permissible maximum Courant number $\mathrm{CFL}_{\rm max}$, especially for high degrees $N$ of the basis polynomials used in the DG representation of the solution. ADER-DG methods are characterized by very small values of the Courant number $\mathrm{CFL}_{\rm max}$, which is used in particular for subcell limiting of the solution with finite-volume limiters, which do not have such a strict limitation on the time step and can reliably compute solutions even for smaller coordinate grid steps~\cite{ader_dg_dev_1, ader_dg_dev_2, ader_dg_ideal_flows, ader_dg_diss_flows, ader_dg_PNPM}. There are several different estimates of the maximum Courant number $\mathrm{CFL}_{\rm max}(N)$ for the ADER-DG method with the LST-DG predictor. In many works (for example~\cite{ader_dg_dev_1, ader_dg_dev_2, ader_weno_lstdg_ideal, ader_dg_grmhd, ader_stiff_3, ader_stiff_4}) the widely used following well-known estimate for Courant number $\mathrm{CFL}$ is employed:
\begin{equation}\label{eq:cfl_cs}
\mathrm{CFL}_{\rm CS}(N) = \frac{1}{2N+1},\qquad
\mathrm{CFL} = \mathrm{C}\cdot\mathrm{CFL}_{\rm CS}(N),
\end{equation}
obtained by Cockburn and Shu~\cite{Cockburn_base_1, Cockburn_base_2, Cockburn_base_3, Cockburn_base_4, Cockburn_base_5} for classical DG methods, and hereinafter referred to as the Cockburn-Shu formula, where $\mathrm{C} \leqslant 1$ is the ``effective Courant number''. This estimate, in particular, forms the basis for a certain rule for choosing the number of subcells $N_{s} = 2N+1$ on each cell for a posteriori subgrid constraint of a solution by a finite-volume limiter~\cite{ader_dg_dev_1, ader_dg_dev_2, ader_weno_lstdg_ideal}, for which the Courant number is usually expected to be approximately one. In this case, other estimates of the admissible maximum Courant number $\mathrm{CFL}_{\max}(N)$ can be expressed in terms of the maximum effective Courant number $\mathrm{C}_{\max}(N)$ as follows:
\begin{equation}\label{eq:cfl_cs_and_cfl_max}
\mathrm{CFL}_{\max}(N) = \mathrm{C}_{\max}(N)\cdot\mathrm{CFL}_{\rm CS}(N),\qquad
\mathrm{C}_{\max}(N) = \frac{\mathrm{CFL}_{\max}(N)}{\mathrm{CFL}_{\rm CS}(N)},
\end{equation}
where now, in case $\mathrm{CFL}_{\max}(N) < \mathrm{CFL}_{\rm CS}(N)$, the constraint $\mathrm{C}_{\max}(N) < 1$ holds instead $\mathrm{C}_{\max}(N) = 1$. In~\cite{PNPM_DG_2008} the estimates of the maximum Courant number $\mathrm{CFL}_{\rm max}(N)$ for the ADER-DG method with the LST-DG predictor were first calculated for degrees $N = 1, \ldots, 4$ of basis polynomials. In a subsequent work~\cite{ader_dg_PNPM} estimates of the maximum Courant number $\mathrm{CFL}_{\rm max}(N)$ for the ADER-DG method with the LST-DG predictor were calculated for basis polynomials for degrees $N = 1, \ldots, 6$ of basis polynomials, which were in good agreement with the previous estimates~\cite{PNPM_DG_2008} in the intersection region $N \leqslant 4$.

A new study~\cite{ader_dg_stab} has recently been published in which estimates of the maximum Courant number have been calculated for degrees $N = 1, \ldots, 9$ of basis polynomials, and here differences with existing results~\cite{PNPM_DG_2008, ader_dg_PNPM} have begun to arise, which are described in detail in~\cite{ader_dg_stab}. It should be noted that estimates~\cite{ader_dg_stab, ader_dg_PNPM, PNPM_DG_2008} for large degrees $N \gg 1$ of the basis polynomials differ significantly from the widely used estimate (\ref{eq:cfl_cs}), which is significantly, sometimes by more than twofold, overestimated. Therefore, using estimates (\ref{eq:cfl_cs}) requires introducing an additional empirical factor $\mathrm{C}(N) \leqslant 1$, with $\mathrm{C}(N) \ll 1$ for large degrees $N$ of the basis polynomials. The calculations in~\cite{ader_dg_stab, ader_dg_PNPM, PNPM_DG_2008} were based on the Von Neumann stability criterion (spectral stability criterion) and were technically implemented by calculating the maximum spectral radius of the amplification matrix (the matrix of the evolution operator for a single time step $\Dt$) for a continuous set of phase $\theta\in[0, 2\pi]$. Clearly, such calculations require an estimate of the robustness of the eigenvalues and an extremely detailed discretization of the continuous set of phase $\theta\in[0, 2\pi]$. Moreover, the evolution operator matrix itself for a single time step, when directly calculated, even in the one-dimensional case, requires multiplying $(N+1)\times(N+1)^{2}$ and $(N+1)^{2}\times(N+1)$ matrices to obtain a $(N+1)\times(N+1)$ matrix, which also affects the final calculation error for large degrees $N$ of the basis polynomials, when using standard machine arithmetic to represent real numbers with single- or double-precision floating-point numbers. Another new study~\cite{ader_dg_stab_rig} carried out a detailed analysis and constructed a rigorous stability theory for the ADER-DG numerical method with the LST-DG predictor (but with a different choice of functional basis, which did not affect the quantitative values of the maximum Courant numbers $\mathrm{CFL}_{\rm max}(N)$ in the intersection region by degree $N$ of the results). This analysis is based on the empirical argument that stability loss occurs when one of the eigenvalues $\{\lambda_{k}\}$ of the evolution operator matrix reaches $\lambda = -1$. A qualitative theory was also constructed that agrees well with the rigorous theory, leading to a dependence $\mathrm{CFL}_{\rm max}(N) \propto 1/(N+1)^{2}$ rather than (\ref{eq:cfl_cs}). Moreover, in the latter study, the calculation of the maximum Courant number is based on the solution of two polynomial equations, which is a fairly simple and computationally stable problem, so in this case the discrepancy with the result (\ref{eq:cfl_cs}) cannot be explained by an error or other inaccuracy of the calculations. Therefore, it can currently be considered established that the estimate (\ref{eq:cfl_cs}) for the ADER-DG numerical method with the LST-DG predictor is overestimated (except for cases $N = 0$ and $1$), especially significantly for high degrees $N$ of the basis polynomials, and the actual dependence of the maximum Courant number $\mathrm{CFL}_{\rm max}(N)$ on the degree $N$ of the basis polynomials is closer to the asymptotic form $\mathrm{CFL}_{\rm max}(N) \propto 1/(N+1)^{2}$. However, this estimate is well applicable to hyperbolic equations without source terms and explains the use of the effective Courant number $\mathrm{C}_{\max}(N) \sim 0.1$--$0.3 \leqslant \mathrm{C}_{\max}(N)$ (\ref{eq:cfl_cs_and_cfl_max}) in calculations. However, for example, in~\cite{ader_stiff_4}, a study was carried out on the applicability of the ADER-DG numerical method with the LST-DG predictor to the modeling of one-dimensional and multidimensional detonation waves forming and propagating in a reacting medium --- this problem is characterized by anomalously high stiffness. To obtain correct results, an effective Courant number $\mathrm{C}_{\max}(N) = 0.90$ was used, including for high degrees $N \geqslant 5$ (up to $N = 9$) of the basis polynomials. This suggests that the maximum Courant number $\mathrm{CFL}_{\max}(N)$ of the ADER-DG numerical method with the LST-DG predictor when solving a problem with stiff source terms is higher (and possibly significantly) than when solving problems without stiff source terms. Therefore, the present work is devoted to solving this particular problem --- studying the stability, as well as approximation and convergence, of the ADER-DG numerical method with the LST-DG predictor for solving hyperbolic equations with stiff source terms. In many ways, the theory and study are structured similarly to the study presented in~\cite{ader_dg_stab_rig}, but for a different problem.

The article is divided into four main Sections. Section~\ref{sec:descr} ``General description of the numerical method'' presents a detailed description of the ADER-DG numerical method with the LST-DG predictor for the linear advection-reaction equation, for which a direct rigorous approximation theory and stability theory are constructed. This Section introduces all the basic concepts and notations, as well as the main properties of the ADER-DG numerical method with the LST-DG predictor, which are used in the subsequent Sections for research. Section~\ref{sec:approx_anal} ``Approximation analysis'' presents the direct rigorous approximation theory of the ADER-DG numerical method with the LST-DG predictor. Subsection~\ref{sec:approx_anal:local_sol} ``Discrete space-time solution'' is devoted to the derivation of the order and approximation conditions for the local discrete space-time solution obtained at the LST-DG predictor stage. Subsection~\ref{sec:approx_anal:final_sol} ``Corrected solution at time steps'' is devoted to the derivation of the approximation order for the adjusted solution determined at time steps. This Section also introduces and rigorously proves properties of the matrices that arise during a detailed analysis of the numerical method and are used in the next, main Section of the article. Section~\ref{sec:stab_anal} ``Stability analysis'' is the key Section of this article and is divided into four Subsections. Subsection~\ref{sec:stab_anal:qual_cons} ``Qualitative considerations'' is devoted to the derivation and justification of a new empirical estimate for the maximum Courant number $\mathrm{CFL}_{\rm max}(N, z) \propto |z|\mathrm{CFL}_{\rm max}(N, 0)$ of the ADER-DG numerical method with the LST-DG predictor, where $z = \nu\Dt < 0$ is a dimensionless real parameter characterizing the stiffness of the problem, $\nu < 0$ is the decay constant. Subsection~\ref{sec:stab_anal:statment_quant_cons} ``Statement of the problem of quantitative considerations'' is devoted to a detailed derivation of the matrices required for a detailed stability analysis of the ADER-DG numerical method with the LST-DG predictor, as well as the derivation and proof of the relations required for further research. Subsection~\ref{sec:stab_anal:empir_cons} ``Empirical considerations'' is devoted to the justification of the empirical assumption that stability is violated when one of the some eigenvalue $\lambda_{k}$ of the matrix of the evolution operator for a single time step $\Dt$ reaches the value of $\lambda = -1$. Subsection~\ref{sec:stab_anal:quant_cons} ``Rigorous quantitative considerations'' is devoted to a detailed derivation of equations whose solutions determine the maximum Courant numbers $\mathrm{CFL}_{\rm max}(N, z)$ and are finite polynomial equations. This Subsection presents the main results, including the obtained dependencies of the maximum Courant number $\mathrm{CFL}_{\rm max}(N, z)$ on the parameter $z$ for degrees $N = 1, \ldots, 12$ of the basis polynomials. Section~\ref{sec:app} ``Applications'' contains problem statements and results of numerical experiments that confirm previously developed theoretical results. Subsection~\ref{sec:app:lin_adv} ``Linear advection-reaction equation'' is devoted to numerical experiments for a linear advection-reaction equation. Subsection~\ref{sec:app:euler} ``ZND-detonation waves'' is devoted to numerical experiments for the system of Euler equations for a two-component reacting medium with a monomolecular reaction with energy release and with divergent initiation, which was chosen as a good example of a nonlinear problem with stiff terms for which ADER-DG numerical method with the LST-DG predictor are widely applicable.

\section{General description of the numerical method}
\label{sec:descr}

This Section presents a detailed description of the ADER-DG numerical method with the LST-DG predictor for the linear advection-reaction equation, for which a direct rigorous approximation theory and stability theory are constructed. The description of the numerical method is presented in a form convenient for subsequent analysis of approximation and stability, similar to the description presented in~\cite{ader_dg_stab_rig}, where approximation and stability were studied for a linear advection equation without source terms. The advection-reaction equation is chosen in the following standard form:
\begin{equation}\label{eq:adv_eq_src}
\pderf{u}{t} + a\pderf{u}{x} = \nu u,\quad
(x, t)\in\Omega\times\mathcal{T},\quad
\Omega\subset\mathcal{R},\quad
\mathcal{T} = \left[t_{0}, t_{f}\right]\subset\mathcal{R},
\end{equation}
where $a\in\mathcal{R}$ is the advection velocity, $\nu u$ is the decay term (or sink term), $\nu \leqslant 0$, $\nu\in\mathcal{R}$ is the decay constant, $u: \Omega\times\mathcal{T}\rightarrow\mathcal{R}$ is the desired function. The work studies in detail the case $\nu \leqslant 0$, which corresponds precisely to the classical definition of stiff terms. It can be assumed that this advection-reaction equation (\ref{eq:adv_eq_src}) is a generalization to the case of partial differential equations of the classical Dahlquist's test equation $\dot{u} = \mu u$, where $\mu\in\mathcal{C}$, $\Re[\mu] \leqslant 0$, from the theory of linear stability of numerical methods for solving systems of ordinary differential equations~\cite{Hairer_book_1, Hairer_book_2, Dekker_Verwer_1984}. The advection-reaction equation (\ref{eq:adv_eq_src}) has a well-known exact analytical solution $u(x, t) = \exp(\nu (t - t_{0}))u_{0}(x - a(t - t_{0}))$, where $u_{0}(x)$ is the solution at time $t_{0}$, which is a simple shift of the solution function in the direction of advection at velocity $a$ and a simple exponential decay of the solution function with decrement $|\nu|$. It should be noted that numerical DG methods, which include the ADER-DG numerical method, are linear numerical methods of arbitrarily high order, and therefore they are very sensitive to the smoothness of the solution to the problem, even in the case of a linear advection-reaction equation (\ref{eq:adv_eq_src}). Therefore, in the further transformations, it is assumed that the smoothness of the initial condition $u_{0}(x)$, determined at time $t_{0}$, is sufficient to achieve the expected orders of approximation and convergence of the ADER-DG numerical method. Condition $u_{0}(x) \in \mathcal{C}_{N}(\Omega)$, where $N$ is the degree of the polynomials used in the DG representation of the solution, is usually sufficient for this. Otherwise, a decrease in the order of the ADER-DG numerical method will be observed.

The coordinate domain $\Omega$ is discretized by a one-dimensional mesh represented by non-overlapping coordinate discretization domains $\Omega_{i}$, with nodes $x_{i}$ and a coordinate discretization step $\Dx = x_{i+1} - x_{i} \geqslant 0$, which is assumed to be constant for simplicity, but without loss of generality. The time domain $\mathcal{T}$ is also discretized by a one-dimensional mesh with time steps $t^{n}$ and non-overlapping time discretization domains $\mathcal{T}^{n}$, with a discretization time step $\Dtn{n} = t^{n+1} - t^{n} \geqslant 0$. The resulting coordinate-time mesh $\{\Xi_{i}^{n}\}$ is defined as follows:
\begin{equation}\label{eq:discr_domains}
\begin{split}
&\Omega_{i} = \left[x_{i}, x_{i+1}\right],\quad
\Omega = \bigcup\limits_{i} \Omega_{i},\quad
\mathcal{T}^{n} = \left[t^{n}, t^{n+1}\right],\quad
\mathcal{T} = \bigcup\limits_{n} \mathcal{T}^{n},\quad
\Xi_{i}^{n} = \mathcal{T}^{n} \times \Omega_{i},\quad
\Xi = \Omega\times\mathcal{T} = \bigcup\limits_{n, i} \Xi_{i}^{n},
\end{split}
\end{equation}
where $\Xi_{i}^{n}$ is the coordinate-time discretization domain. The ADER-DG numerical method with the LST-DG predictor is constructed on a reference coordinate-time element $\omega^{2} = [0, 1]^{2}$ using the following reversible mapping:
\begin{equation}\label{eq:mapping}
t(\tau) = t^{n} + \tau\Dtn{n},\quad
x(\xi) = x_{i} + \xi\Dx,
\end{equation}
where $\tau\in\omega$ and $\xi\in\omega$ are the rescaled coordinate $x$ and time $t$. The advection-reaction equation (\ref{eq:adv_eq_src}) in the reference coordinate-time element $\omega^{2}$ is presented in the following form:
\begin{equation}\label{eq:adv_eq_mapped}
\pderf{u}{\tau} + \frac{a\Dtn{n}}{\Dx}\pderf{u}{\xi} = \nu\Dtn{n} u.
\end{equation}
The ADER-DG numerical method is formulated using the weak form of the equation (\ref{eq:adv_eq_mapped}):
\begin{equation}\label{eq:adv_eq_weak}
\intrefdom{\tau}\intrefdom{\xi}\,\varphi_{k}(\xi)\left[\pderf{u}{\tau} + \frac{a\Dtn{n}}{\Dx}\pderf{u}{\xi} -  \nu\Dtn{n}u\right] = 0,
\end{equation}
for which a finite set $\{\varphi_{k}\}$ of basis functions $\varphi_{k}: \omega\rightarrow\mathcal{R}$ is introduced. In this work, a nodal polynomial basis $\{\varphi_{k}(\xi)\}$ is chosen, consisting of Lagrange interpolation polynomials $\varphi_{k}\in\mathcal{P}_{N}(\omega)$ with nodal points $\{\xi_{p}\}$ at the roots of shifted Legendre polynomials $\tilde{P}_{N+1}: \omega\rightarrow\mathcal{R}$. The basis functions $\{\varphi_{k}(\xi)\}$ are represented in the following form:
\begin{equation}\label{eq:phi_def}
\varphi_{p}(\xi) = \sum\limits_{k = 0}^{N} \varphi_{p, k} \xi^{k},\quad
\varphi_{p}(\xi) = \prod\limits_{k = 0}^{N}\frac{\xi - \xi_{k}}{\xi_{p} - \xi_{k}},\quad
\tilde{P}_{N+1}(\xi_{p}) = 0,\quad
\varphi_{p}(\xi_{q}) = \delta_{pq},
\end{equation}
where the last expression is used to calculate the polynomial coefficients $\{\varphi_{p, k}\}$, which is technically equivalent to calculating the inverse matrix of the Vandermonde matrix $\{\xi_{q}^{p}\}_{p,q}$. It is well known that Legendre polynomials $\tilde{P}_{N+1}(\xi)$ do not have multiple roots $\xi_{p}$, therefore the corresponding Vandermonde matrix is non-singular, and hence the interpolation polynomials that make up the basis $\{\varphi_{k}\}$ are unique. The chosen functional basis is orthogonal in the sense of the $\mathcal{L}_{2}(\omega)$-scalar product. The choice of this functional basis $\{\varphi_{k}\}$ (\ref{eq:phi_def}) allows for the convenient use of the Gauss-Legendre (GL) quadrature rule, in the following form:
\begin{equation}\label{eq:gl_rule}
\intrefdom{\xi}\,f(\xi) \approx \sum\limits_{p = 0}^{N} w_{p} f(\xi_{p}),\quad
\sum\limits_{p = 0}^{N} w_{p} = 1,\quad
w_{p} = \intrefdom{\xi}\,\varphi_{p}(\xi) = \intrefdom{\xi}\,\varphi_{p}^{2}(\xi) > 0,
\end{equation}
where $\{w_{p}\}$ is the weight of the GL quadrature rule. It should be noted that the GL quadrature rule is satisfied exactly for all functions $\mathcal{P}_{2N+1}(\omega)$. The chosen functional basis $\{\varphi_{k}\}$ (\ref{eq:phi_def}) allows for the use of the point-wise evaluation for the expansion coefficients of functions $f: \omega\rightarrow\mathcal{R}$ in this basis:
\begin{equation}\label{eq:pw_eval}
f(\xi) = \sum\limits_{p = 0}^{N} f_{p}\varphi_{p}(\xi) \mapsto \sum\limits_{p = 0}^{N} f(\xi_{p})\varphi_{p}(\xi),
\end{equation}
which is justified by the following chain of transformations:
\begin{equation}
f_{p} = \frac{1}{w_{p}} \intrefdom{\xi}\,f(\xi)\varphi_{p}(\xi) \approx
\frac{1}{w_{p}} \sum\limits_{q = 0}^{N} w_{q} f(\xi_{q})\varphi_{p}(\xi_{q}) =
\frac{1}{w_{p}} \sum\limits_{q = 0}^{N} w_{q} f(\xi_{q})\delta_{pq} = f(\xi_{p}),
\end{equation}
where it is used the orthogonality properties of functions $\{\varphi_{k}\}$, the nodal property of the interpolation polynomial (\ref{eq:phi_def}), and the GL quadrature rule (\ref{eq:gl_rule}). It should be noted that the representation (\ref{eq:pw_eval}) is exact for all functions $\mathcal{P}_{N}(\omega)$.

By introducing the DG representations of the solution at the current $t^{n}$ and next $t^{n+1}$ time steps in the following form:
\begin{equation}\label{eq:u_n_def}
u^{n}(\xi) = \sum\limits_{k = 0}^{N} \hat{u}^{n}_{k} \varphi_{k}(\xi),\quad
u^{n+1}(\xi) = \sum\limits_{k = 0}^{N} \hat{u}^{n+1}_{k} \varphi_{k}(\xi),
\end{equation}
and performing integration by parts in the flux term in the equation (\ref{eq:adv_eq_weak}), the following form of the completely single-step discrete finite element scheme of the ADER-DG numerical method is obtained:
\begin{equation}\label{eq:ader_dg_corr_temp_1}
\begin{split}
&\sum\limits_{l = 0}^{N} \intrefdom{\xi}\,\varphi_{k}(\xi)\varphi_{l}(\xi)\left[\hat{u}^{n+1}_{l} - \hat{u}^{n}_{l}\right] -
\frac{a\Dtn{n}}{\Dx}\intrefdom{\tau}\intrefdom{\xi}\,\varphi_{k}'(\xi) q(\tau, \xi)\\ & +
\frac{\Dtn{n}}{\Dx}\intrefdom{\tau}\,\Big[
  \varphi_{k}(1) F_{\rm RP}\left(q(\tau, 1),\, q^{+}(\tau, 0)\right) -
  \varphi_{k}(0) F_{\rm RP}\left(q^{-}(\tau, 1),\, q(\tau, 0)\right)
\Big]\\ & = \nu\Dtn{n}\intrefdom{\tau}\intrefdom{\xi}\,\varphi_{k}(\xi) q(\tau, \xi),
\end{split}
\end{equation}
where $q: \omega^{2}\rightarrow\mathcal{R}$ is the local discrete space-time solution used in ADER methods to calculate local and flux terms, and is determined for each coordinate-time element $\Xi_{i}^{n}$ separately, $F_{\rm RP}$ is the solver of the classical Riemann problem, which, in the case of the advection equation (\ref{eq:adv_eq_src}), is expressed in exact closed finite form:
\begin{equation}\label{eq:f_rp}
F_{\rm RP}\left(u_{L},\, u_{R}\right) = \frac{a+|a|}{2} u_{L} + \frac{a-|a|}{2} u_{R} = a \Big[\theta(a)u_{L} + \theta(-a)u_{R}\Big],
\end{equation}
where $\theta$ is the Heaviside $\theta$-function.

The local discrete space-time solution $q$ is a numerical solution of the equation (\ref{eq:adv_eq_mapped}) in coordinate-time discretization domain $\Xi_{i}^{n}$:
\begin{equation}\label{eq:lstdg_pred_eq_src}
\pderf{q}{\tau} + \frac{a\Dtn{n}}{\Dx}\pderf{q}{\xi} = \nu\Dtn{n}q,\quad q(0, \xi) = u^{n}(\xi),
\end{equation}
and is chosen in the form of an expansion over a functional basis $\{\Theta_{\mathrm{p}}(\tau, \xi)\, |\, \Theta_{\mathrm{p}}(\tau, \xi)\in\mathcal{P}_{N}(\omega)\times\mathcal{P}_{N}(\omega)\}$, chosen in the form of a tensor product of the functional basis $\{\varphi_{k}\}$:
\begin{equation}\label{eq:local_sol_def}
q(\tau, \xi) = \sum\limits_{\mathrm{p}} \hat{q}_{\mathrm{p}} \Theta_{\mathrm{p}}(\tau, \xi)
             = \sum\limits_{p_{0} = 0}^{N}\sum\limits_{p_{1} = 0}^{N} \hat{q}_{p_{0}p_{1}} \varphi_{p_{0}}(\tau)\varphi_{p_{1}}(\xi),
\end{equation}
where $\mathrm{p}$ is a multi-index for which projections $(p_{0}, p_{1})$ are defined:
\begin{equation}\label{eq:local_sol_idx}
\mathrm{p} = (p_{0},\, p_{1}),\quad p_{0} = \pi_{0}(\mathrm{p}),\quad p_{1} = \pi_{1}(\mathrm{p}),\quad
\mathrm{idx}(\mathrm{p}) = p_{0}\cdot(N+1) + p_{1},\quad 0 \leqslant p_{0},\, p_{1} \leqslant N.
\end{equation}
Choosing basis functions $\{\Theta_{\mathrm{p}}(\tau, \xi)\}$ in the form of a tensor product will allow us to conveniently represent the matrices of the linear transformations as Kronecker products of the matrices of the original basis $\{\varphi_{k}\}$. The equation (\ref{eq:lstdg_pred_eq_src}) for obtaining the local discrete space-time solution $q$ is considered in weak form:
\begin{equation}\label{eq:lstdg_pred_eq_weak}
\intrefdom{\tau}\intrefdom{\xi}\,\Theta_{\mathrm{p}}(\tau, \xi)\left[\pderf{q}{\tau} + \frac{a\Dtn{n}}{\Dx}\pderf{q}{\xi} - \nu\Dtn{n}q\right] = 0,\quad q(0, \xi) = u^{n}(\xi),
\end{equation}
which is the LST-DG predictor. It should be noted that the initial conditions in the problem (\ref{eq:lstdg_pred_eq_weak}) are also considered in weak form, so a discontinuity is allowed for the solution $q(\tau, \xi)$ at the initial time $\tau = 0$. Substituting the chosen solution representation (\ref{eq:local_sol_def}) into the equation (\ref{eq:lstdg_pred_eq_weak}) and calculating the terms included in the equation (\ref{eq:lstdg_pred_eq_weak}) leads to the following expression:
\begin{equation}\label{eq:lstdg_pred_eq_tmp_1}
\begin{split}
\sum\limits_{q_{0} = 0}^{N}\sum\limits_{q_{1} = 0}^{N} &\left[\vphantom{\intrefdom{\tau}}
  \varphi_{p_{0}}(1)\varphi_{q_{0}}(1) -
  \intrefdom{\tau}\,\varphi_{p_{0}}'(\tau)\varphi_{q_{0}}(\tau)
\right]\intrefdom{\xi}\,\varphi_{p_{1}}(\xi)\varphi_{q_{1}}(\xi)\cdot\hat{q}_{q_{0}q_{1}}\\
&+\frac{a\Dtn{n}}{\Dx}\,\sum\limits_{q_{0} = 0}^{N}\sum\limits_{q_{1} = 0}^{N}
  \intrefdom{\tau}\,\varphi_{p_{0}}(\tau)\varphi_{q_{0}}(\tau)\,\intrefdom{\xi}\,\varphi_{p_{1}}(\xi)\varphi_{q_{1}}'(\xi)\cdot\hat{q}_{q_{0}q_{1}} \\
&-\nu\Dtn{n}\,\sum\limits_{q_{0} = 0}^{N}\sum\limits_{q_{1} = 0}^{N}
  \intrefdom{\tau}\,\varphi_{p_{0}}(\tau)\varphi_{q_{0}}(\tau)\,\intrefdom{\xi}\,\varphi_{p_{1}}(\xi)\varphi_{q_{1}}(\xi)\cdot\hat{q}_{q_{0}q_{1}} \\ & =
\sum\limits_{q_{1} = 0}^{N} \varphi_{p_{0}}(0) \intrefdom{\xi}\,\varphi_{p_{1}}(\xi)\varphi_{q_{1}}(\xi)\cdot\hat{u}^{n}_{q_{1}},
\end{split}
\end{equation}
where integration by parts over time is performed in the time term to include the initial condition in weak form in the system of equations. To simplify the notation, the following matrix and vector elements are introduced:
\begin{equation}\label{eq:matrices_def}
\begin{split}
&\varrho_{pq} = \intrefdom{\xi}\, \varphi_{p}(\xi)\varphi_{q}'(\xi),\quad
m_{pq} = \intrefdom{\xi}\, \varphi_{p}(\xi)\varphi_{q}(\xi) = w_{p}\delta_{pq},\\
&\kappa_{pq} = \varphi_{p}(1)\varphi_{q}(1) - \intrefdom{\xi}\, \varphi_{p}'(\xi)\varphi_{q}(\xi)
             = \varphi_{p}(0)\varphi_{q}(0) + \intrefdom{\xi}\, \varphi_{p}(\xi)\varphi_{q}'(\xi),
\end{split}
\end{equation}
from which the following matrices and vectors are assembled:
\begin{equation}
\begin{split}
&\uphi_{p} = \varphi_{p}(0),\quad \ophi_{p} = \varphi_{p}(1),\quad
\uvecphi = \{\uphi_{p}\}_{p},\quad \ovecphi = \{\ophi_{p}\}_{p},\\
&\varrho = \{\varrho_{pq}\}_{p,q},\quad m = \{m_{pq}\}_{p,q} = \mathrm{diag}(w_{0}, \ldots, w_{N}),\\
&\kappa = \{\kappa_{pq}\}_{p,q} = \ovecphi\otimes\ovecphi^{T} - \varrho^{T} = \uvecphi\otimes\uvecphi^{T} + \varrho,
\end{split}
\end{equation}
where the $\mathcal{L}_{2}(\omega)$-orthogonality of the functional basis $\{\varphi_{k}\}$ is taken into account through the diagonality of the mass matrix $m$; matrix $\varrho$ is the stiffness matrix of the functional basis $\{\varphi_{k}\}$. The following two properties of matrix $\kappa$ can be verified by direct substitution and were derived in~\cite{ader_improving_2024, ader_dg_ivp_ode_sinum}:
\begin{equation}\label{eq:gl_rule:int_prop}
\begin{split}
\sum_{q = 0}^{N} \kappa_{pq} = \varphi_{p}(0),\qquad
\sum_{p = 0}^{N} \kappa_{pq} = \varphi_{q}(1),
\end{split}
\end{equation}
which can be rewritten in matrix form: $\kappa \mathbf{1} = \uvecphi$, $\kappa^{-1} \uvecphi = \mathbf{1}$, $\mathbf{1}^{T} \kappa = \ovecphi^{T}$, $\ovecphi^{T} \kappa^{-1} = \mathbf{1}^{T}$. The matrices included in the system of algebraic equations (\ref{eq:lstdg_pred_eq_tmp_1}) of the LST-DG predictor are conveniently expressed through the matrices and vectors introduced above as follows:
\begin{equation}
\begin{split}
&\mathrm{K}^{(\tau)}_{\mathrm{p}\mathrm{q}} = \mathrm{K}^{(\tau)}_{p_{0}p_{1},q_{0}q_{1}} = \kappa_{p_{0}q_{0}}m_{p_{1}q_{1}} = \kappa_{p_{0}q_{0}}w_{p_{1}}\delta_{p_{1}q_{1}},\\
&\mathrm{K}^{(\xi)}_{\mathrm{p}\mathrm{q}} = \mathrm{K}^{(\xi)}_{p_{0}p_{1},q_{0}q_{1}} = m_{p_{0}q_{0}}\varrho_{p_{1}q_{1}} = w_{p_{0}}\delta_{p_{0}q_{0}}\varrho_{p_{1}q_{1}},\\
&\mathrm{M}^{(\xi)}_{\mathrm{p}\mathrm{q}} = \mathrm{M}^{(\xi)}_{p_{0}p_{1},q_{0}q_{1}} = m_{p_{0}q_{0}}m_{p_{1}q_{1}} = w_{p_{0}}w_{p_{1}}\delta_{p_{0}q_{0}}\delta_{p_{1}q_{1}},\\
&\mathrm{F}^{(0)}_{\mathrm{p},k} = \mathrm{F}^{(0)}_{p_{0}p_{1},k} = \uphi_{p}m_{p_{1}q_{1}} = \uphi_{p}w_{p_{1}}\delta_{p_{1}q_{1}},
\end{split}
\end{equation}
where $\mathrm{p}$ and $\mathrm{q}$ are multi-indices (\ref{eq:local_sol_idx}). The resulting matrices can be expressed through the Kronecker products as follows:
\begin{equation}\label{eq:matrices_K_def}
\begin{split}
&\mathrm{K}^{(\tau)} = \left\{\mathrm{K}^{(\tau)}_{\mathrm{p}\mathrm{q}}\right\}_{\mathrm{p},\mathrm{q}} = \kappa \otimes m,\quad
 \mathrm{K}^{(\xi)} = \left\{\mathrm{K}^{(\xi)}_{\mathrm{p}\mathrm{q}}\right\}_{\mathrm{p},\mathrm{q}} = m \otimes \varrho,\\
&\mathrm{M}^{(\xi)} = \left\{\mathrm{M}^{(\xi)}_{\mathrm{p}\mathrm{q}}\right\}_{\mathrm{p},\mathrm{q}} = m \otimes m,\quad
 \mathrm{F}^{(0)} = \left\{\mathrm{F}^{(0)}_{\mathrm{p},k}\right\}_{\mathrm{p},k} = \uvecphi \otimes m,
\end{split}
\end{equation}
where $\otimes$ is the symbol of the Kronecker product.

\section{Approximation analysis}
\label{sec:approx_anal}

This Section is devoted to a rigorous direct analysis of the approximation of the ADER-DG numerical method with the LST-DG predictor. The Subsection~\ref{sec:approx_anal:local_sol} ``Discrete space-time solution'' presents a rigorous direct analysis of the approximation of a local discrete space-time solution $q(\tau, \xi)$ (\ref{eq:local_sol_def}). This Subsection~\ref{sec:approx_anal:final_sol} ``Corrected solution at time steps'' presents a rigorous direct analysis of the approximation of the numerical solution $u^{n+1}(\xi)$ at time steps obtained at the corrector stage (\ref{eq:ader_dg_corr}).

\subsection{Discrete space-time solution}
\label{sec:approx_anal:local_sol}

The vectors of the coefficients of the DG representations (\ref{eq:u_n_def}) and the coefficients of the local discrete space-times solution (\ref{eq:local_sol_def}) can be expressed as follows:
\begin{equation}\label{eq:vectors_u_q_def}
\begin{split}
&\hat{\mathbf{q}} = \left\{\hat{q}_{\mathrm{p}}\right\}_{\mathrm{p}},\quad \hat{\mathbf{u}}^{n} = \left\{\hat{u}^{n}_{k}\right\}_{k}.
\end{split}
\end{equation}
As a result of introducing matrices (\ref{eq:matrices_K_def}) and vectors (\ref{eq:vectors_u_q_def}), the system of equations (\ref{eq:lstdg_pred_eq_tmp_1}) of the LST-DG predictor takes the classical form of a system of linear algebraic equations:
\begin{equation}\label{eq:lstdg_pred_eq_tmp_2}
\left(\mathrm{K}^{(\tau)} + \frac{a\Dtn{n}}{\Dx}\,\mathrm{K}^{(\xi)} - \nu\Dtn{n}\mathrm{M}^{(\xi)}\right)\hat{\mathbf{q}} = \mathrm{F}^{(0)}\hat{\mathbf{u}}^{n},
\end{equation}
which leads to the following form:
\begin{equation}\label{eq:lstdg_pred_eq_tmp_3}
\left[I \otimes I + \frac{a\Dtn{n}}{\Dx}\,\left(\mathrm{K}^{(\tau)}\right)^{-1}\mathrm{K}^{(\xi)} -
\nu\Dtn{n}\,\left(\mathrm{K}^{(\tau)}\right)^{-1}\mathrm{M}^{(\xi)}\right]\hat{\mathbf{q}} =
\left[\left(\mathrm{K}^{(\tau)}\right)^{-1}\mathrm{F}^{(0)}\right]\hat{\mathbf{u}}^{n}.
\end{equation}
which requires the invertibility of matrix $\mathrm{K}^{(\tau)}$.

For the matrix $\mathrm{K}^{(\tau)} = \kappa \otimes m$ (\ref{eq:matrices_K_def}) to be invertible, both matrix $\kappa$ and matrix $m$ must be invertible. Matrix $m$ (\ref{eq:matrices_def}) is the mass matrix (or Gram matrix, a detailed discussion can be found on page~96 in~\cite{StrangFix_FEM_1973}) of a functional basis $\{\varphi_{p}\}$ consisting of linearly independent functions $\varphi_{p}$ (\ref{eq:phi_def}), so matrix $m$ is invertible. The invertibility of matrix $\kappa$ (\ref{eq:matrices_def}) is somewhat more difficult to rigorously prove, and can be done based on the following considerations: it is necessary to rigorously prove that the kernel of the linear operator defined by matrix $\kappa$ in $\mathcal{R}^{N+1}$ is trivial, which requires solving the system of linear algebraic equations $\kappa\mathbf{g} = \mathbf{0}$ for the vector $\mathbf{g} = \{g_{p}\}_{p}$ of unknown quantities $\{g_{p}\}$, and the invertibility of matrix $\kappa$ occurs if and only if result $\mathbf{g} = \mathbf{0}$ holds. In this case, it is sufficient to prove consequent $\mathbf{g} = \mathbf{0}$ from antecedent $\kappa\mathbf{g} = \mathbf{0}$, and the converse is obvious. The original expression can be represented in the following form:
\begin{equation}\label{eq:lstdg_pred_K_tau_inv_proof_1}
\begin{split}
\sum\limits_{q = 0}^{N} \kappa_{pq} g_{q} = 
\sum\limits_{q = 0}^{N} \Bigg[\varphi_{p}(0)\varphi_{q}(0) + \intrefdom{\xi}\, \varphi_{p}(\xi)\varphi_{q}'(\xi)\Bigg]g_{q} =
\varphi_{p}(0)\Bigg[\sum\limits_{q = 0}^{N}g_{q}\varphi_{q}(0)\Bigg] + \intrefdom{\xi}\, \varphi_{p}(\xi)\Bigg[\sum\limits_{q = 0}^{N}g_{q}\varphi_{q}'(\xi)\Bigg] = 0,
\end{split}
\end{equation}
which can be left-multiplied by the transposed vector $\mathbf{h} = \{h_{p}\}_{p}$: $\mathbf{h}^{T}\kappa\mathbf{c} = 0$, leading to the following expression: 
\begin{equation}\label{eq:lstdg_pred_K_tau_inv_proof_2}
\begin{split}
\sum\limits_{p = 0}^{N} \sum\limits_{q = 0}^{N} h_{p} \kappa_{pq} c_{q} & = 
\Bigg[\sum\limits_{p = 0}^{N} h_{p}\varphi_{p}(0)\Bigg]\Bigg[\sum\limits_{q = 0}^{N}c_{q}\varphi_{q}(0)\Bigg] +
\intrefdom{\xi}\, \Bigg[\sum\limits_{p = 0}^{N} h_{p}\varphi_{p}(\xi)\Bigg]\Bigg[\sum\limits_{q = 0}^{N}g_{q}\varphi_{q}'(\xi)\Bigg]\\ & =
\Bigg\{h(\xi) = \sum\limits_{p = 0}^{N} h_{p}\varphi_{p}(\xi),\ g(\xi) = \sum\limits_{p = 0}^{N} g_{p}\varphi_{p}(\xi)\Bigg\} =
h(0)g(0) + \intrefdom{\xi}\, h(\xi) g'(\xi) = 0,
\end{split}
\end{equation}
where two functions $g$, $h \in \mathcal{P}_{N}(\omega)$ have been separately identified. The resulting expression (\ref{eq:lstdg_pred_K_tau_inv_proof_2}) holds for all functions $h \in \mathcal{P}_{N}(\omega)$. Condition $\mathbf{g} = \mathbf{0}$ is equivalent to condition $g(\xi) \equiv 0$ due to the linear independence of the basis functions $\{\varphi_{p}\}_{p}$ (\ref{eq:phi_def}). Since $g \in \mathcal{P}_{N}(\omega)$, then $g' \in \mathcal{P}_{N-1}(\omega)$, so one can choose a convenient concrete function $h(\xi) = \xi g'(\xi)$ for which $h(0)g(0) = 0$ holds, and resulting expression (\ref{eq:lstdg_pred_K_tau_inv_proof_2}) takes the following form:
\begin{equation}\label{eq:lstdg_pred_K_tau_inv_proof_4}
h(0)g(0) + \intrefdom{\xi}\, h(\xi) g'(\xi) = \intrefdom{\xi}\, \xi \Big(g'(\xi)\Big)^{2} = 0,
\end{equation}
from which it immediately follows $g(\xi) \equiv C$. To determine the constant $C$, another function $f \in \mathcal{P}_{N}(\omega)$, $f \not\equiv h$, is chosen, for which the resulting expression (\ref{eq:lstdg_pred_K_tau_inv_proof_2}) takes the following form:
\begin{equation}\label{eq:lstdg_pred_K_tau_inv_proof_5}
f(0)g(0) + \intrefdom{\xi}\, f(\xi) g'(\xi) = f(0) \cdot C = 0.
\end{equation}
Function $f \in \mathcal{P}_{N}(\omega)$ is not specified, so it can always be chosen with condition $f(0) \neq 0$, from which result $C = 0$ (and $g(\xi) \equiv 0$) immediately follows. This proves $\mathbf{g} = \mathbf{0}$ and is a proof of the invertibility of matrix $\kappa$, which proves the expression (\ref{eq:lstdg_pred_eq_tmp_3}) for the system of linear algebraic equations of the LST-DG predictor.

The matrices arising in this form of the system of equations (\ref{eq:lstdg_pred_eq_tmp_3}) are calculated as follows:
\begin{equation}\label{eq:Kinv_mult_KMF}
\begin{split}
&\left(\mathrm{K}^{(\tau)}\right)^{-1}\mathrm{F}^{(0)} = 
 \left[\kappa \otimes m\right]^{-1} \left[\uvecphi \otimes m\right] =
 \left[\kappa^{-1} \uvecphi\right] \otimes \left[m^{-1} m\right] = \mathbf{1} \otimes I,\\
&\left(\mathrm{K}^{(\tau)}\right)^{-1}\mathrm{K}^{(\xi)} = 
 \left[\kappa \otimes m\right]^{-1} \left[m \otimes \varrho\right] =
 \left[\kappa^{-1} m\right] \otimes \left[m^{-1} \varrho\right] = \chi \otimes \mathrm{D},\\
&\left(\mathrm{K}^{(\tau)}\right)^{-1}\mathrm{M}^{(\xi)} = 
 \left[\kappa \otimes m\right]^{-1} \left[m \otimes m\right] =
 \left[\kappa^{-1} m\right] \otimes \left[m^{-1} m\right] = \chi \otimes I,
\end{split}
\end{equation}
where the notations for the two matrices $\chi$ and $\mathrm{D}$ are introduced:
\begin{equation}\label{eq:chi_and_tilde_ae_def}
\chi = \kappa^{-1} m,\quad \mathrm{D} = m^{-1} \varrho.
\end{equation}
This leads to the following explicit expression for the solution of the LST-DG predictor system of equations:
\begin{equation}\label{eq:lstdg_pred_eq_sol_temp_1}
\begin{split}
\hat{\mathbf{q}}
& = \left(I \otimes I + \frac{a\Dtn{n}}{\Dx} \cdot \chi \otimes \mathrm{D} - \nu\Dtn{n}\cdot \chi \otimes I\right)^{-1}\Big(\mathbf{1} \otimes \hat{\mathbf{u}}^{n}\Big)\\
& = \left(\Big(I - \nu\Dtn{n}\chi\Big) \otimes I + \frac{a\Dtn{n}}{\Dx} \cdot \chi \otimes \mathrm{D}\right)^{-1}\Big(\mathbf{1} \otimes \hat{\mathbf{u}}^{n}\Big),
\end{split}
\end{equation}
that is then used to obtain the numerical solution (\ref{eq:u_n_def}) obtained by an ADER-DG numerical method with LST-DG predictor in a closed form convenient for analyzing the approximation of the method. The resulting vector $\hat{\mathbf{q}}$ of expansion coefficients of the local discrete space-time solution $q$ (\ref{eq:local_sol_def}) in the original multi-index notation (\ref{eq:local_sol_idx}) is written in the following form:
\begin{equation}\label{eq:lstdg_pred_eq_sol_temp_1_ind_form}
\hat{q}_{\mathrm{p}} = 
  \sum\limits_{\mathrm{q}}
  \left[\left(\Big(I - \nu\Dtn{n}\chi\Big) \otimes I +
  \frac{a\Dtn{n}}{\Dx} \cdot \chi \otimes \mathrm{D}\right)^{-1}\right]_{\mathrm{p}\mathrm{q}}\cdot\hat{u}^{n}_{\pi_{1}(\mathrm{q})} =
  \sum\limits_{k = 0}^{N} \sigma_{\mathrm{p}, k} \hat{u}^{n}_{k},
\end{equation}
where the following matrix $\sigma = \{\sigma_{\mathrm{p}, k}\}_{\mathrm{p}, k}$ is introduced:
\begin{equation}\label{eq:sigma_matrix_def}
\sigma_{\mathrm{p}, k} \equiv \sigma_{p_{0}p_{1}, k} = 
  \sum\limits_{l = 0}^{N} \left[\left(\Big(I - \nu\Dtn{n}\chi\Big) \otimes I + \frac{a\Dtn{n}}{\Dx} \cdot \chi \otimes \mathrm{D}\right)^{-1}\right]_{p_{0}p_{1},lk},
\end{equation}
whose elements can be calculated in explicit closed form. The explicit calculation of $\sigma_{\mathrm{p}, k}$ (\ref{eq:chi_and_tilde_ae_def}) is based on the properties of matrix $\mathrm{D}$, for which the meaning of $D_{pq}$ is clarified. Despite the fact that this property seems obvious (see, for example, the Section~8.1 in the book~\cite{Funaro_book_1992} or Section~1.9 in the Zanotti's lectures on the ADER-DG method~\cite{Zanotti_lects_2016}), a rigorous proof was given by Jackson in~\cite{Jackson_2017}, as it was hidden by other properties of the ADER-DG method. In the notation of this paper, this property is derived similarly as follows. A direct calculation using the GL quadrature rule (\ref{eq:gl_rule}) shows
\begin{equation}
D_{pq} = \left[m^{-1} \varrho\right]_{pq} = \frac{\varrho_{pq}}{w_{p}} = \frac{1}{w_{p}} \intrefdom{\xi}\, \varphi_{p}(\xi)\varphi_{q}'(\xi) = \varphi_{q}'(\tau_{p}),
\end{equation}
that is exactly satisfied because the quadrature rule (\ref{eq:gl_rule}) exactly integrates polynomials in $\mathcal{P}_{2N+1}(\omega)$, and $(\varphi_{p}\cdot\varphi_{q}')\in\mathcal{P}_{2N-1}(\omega)\subset\mathcal{P}_{2N+1}(\omega)$. The arbitrary function $f\in\mathcal{P}_{N}(\omega)$ is considered, for which $f'\in\mathcal{P}_{N-1}(\omega)$ follow, which allowed the use of the following representation:
\begin{equation}\label{eq:tilde_ae_deriv_prop}
\begin{split}
&f(\tau) = \sum\limits_{p = 0}^{N} f^{(0)}_{p}\varphi_{p}(\tau),\quad f^{(0)}_{p} = f(\tau_{p}),\\
&f'(\tau) = \sum\limits_{p = 0}^{N} f^{(1)}_{p}\varphi_{p}(\tau),\quad
f^{(1)}_{p} = f'(\tau_{p}) = \sum\limits_{q = 0}^{N} f_{q}\varphi_{q}'(\tau_{p}) = \sum\limits_{q = 0}^{N} D_{pq} f_{q},
\end{split}
\end{equation}
which are exact, since the functional basis $\{\varphi_{p}\}$ (\ref{eq:phi_def}) allows for an exact representation of any function in $\mathcal{P}_{N}(\omega)$. The last expression shows that the matrix $\mathrm{D}$ (\ref{eq:chi_and_tilde_ae_def}) is a matrix of linear transformation of the vector of the expansion coefficients $\{f^{(0)}_{p}\}_{p}$ of the function $f$ into the vector of the expansion coefficients $\{f^{(1)}_{p}\}_{p}$ of the derivative $f'$ of the function $f$. For any function $f\in\mathcal{P}_{N}(\omega)$, the property $f^{(N+1)} \equiv 0$ is known, so $\mathrm{D}^{N+1} = 0$ holds --- the matrix $\mathrm{D}$ (\ref{eq:chi_and_tilde_ae_def}) is nilpotent with degree of nilpotency $N+1$. Using the property of the nilpotency of matrix $\mathrm{D}$, the inverse matrix in the solution of the LST-DG predictor system of equations (\ref{eq:lstdg_pred_eq_sol_temp_1}) (and (\ref{eq:lstdg_pred_eq_sol_temp_1_ind_form})) can be exactly represented in the form of a binomial theorem expansion:
\begin{equation}\label{eq:local_sol_inv_matrix}
\begin{split}
&\left[\Big(I - \nu\Dtn{n}\chi\Big) \otimes I + \frac{a\Dtn{n}}{\Dx} \cdot \chi \otimes \mathrm{D}\right]^{-1}\\ & =
 \Big[\Big(I - \nu\Dtn{n}\chi\Big) \otimes I\Big]^{-1}
 \left[I \otimes I + \frac{a\Dtn{n}}{\Dx} \cdot \Big[\Big(I - \nu\Dtn{n}\chi\Big) \otimes I\Big]^{-1}\Big(\chi \otimes \mathrm{D}\Big)\right]^{-1}\\ & = 
 \Big[\Big(I - \nu\Dtn{n}\chi\Big) \otimes I\Big]^{-1}
 \sum\limits_{s = 0}^{N} (-1)^{s} \left(\frac{a\Dtn{n}}{\Dx}\right)^{s} \Big[\Big(I - \nu\Dtn{n}\chi\Big) \otimes I\Big]^{-s}\left(\chi^{s} \otimes \mathrm{D}^{s}\right)\\ & = 
 \left[\Big(I - \nu\Dtn{n}\chi\Big)^{-1} \otimes I\right]
 \sum\limits_{s = 0}^{N} (-1)^{s} \left(\frac{a\Dtn{n}}{\Dx}\right)^{s} \left[\Big(I - \nu\Dtn{n}\chi\Big)^{-s}\chi^{s}\right] \otimes \mathrm{D}^{s},
\end{split}
\end{equation}
where only the finite part of the sum, containing $N+1$ terms, remains. Using expression (\ref{eq:local_sol_inv_matrix}), the expression (\ref{eq:sigma_matrix_def}) for the elements $\{\sigma_{\mathrm{p}, k}\}$ of the matrix $\sigma$ is rewritten as follows:
\begin{equation}\label{eq:sigma_matrix_temp_1}
\begin{split}
\sigma_{p_{0}p_{1}, k} & = 
\sum\limits_{l = 0}^{N} \left[
  \left[\Big(I - \nu\Dtn{n}\chi\Big)^{-1} \otimes I\right]
  \sum\limits_{s = 0}^{N} (-1)^{s} \left(\frac{a\Dtn{n}}{\Dx}\right)^{s} \left[\Big(I - \nu\Dtn{n}\chi\Big)^{-s}\chi^{s}\right] \otimes \mathrm{D}^{s}
\right]_{p_{0}p_{1},lk}\\ & = 
\sum\limits_{l = 0}^{N} \sum\limits_{s = 0}^{N} (-1)^{s} \left[\left(\frac{a\Dtn{n}}{\Dx}\right)^{s}
\left[\Big(I - \nu\Dtn{n}\chi\Big)^{-(s+1)}\chi^{s}\right] \otimes \mathrm{D}^{s}\right]_{p_{0}p_{1},lk}\\ & = 
\sum\limits_{s = 0}^{N} (-1)^{s} \left(\frac{a\Dtn{n}}{\Dx}\right)^{s}
\left(\sum\limits_{l = 0}^{N}\left[\Big(I - \nu\Dtn{n}\chi\Big)^{-(s+1)}\chi^{s}\right]_{p_{0}l}\right) \left[\mathrm{D}^{s}\right]_{p_{1}k}\\ & = 
\sum\limits_{s = 0}^{N} (-1)^{s} \left(\frac{a\Dtn{n}}{\Dx}\right)^{s}
\left(\sum\limits_{l = 0}^{N}\sum\limits_{i = 0}^{N}\left[\Big(I - \nu\Dtn{n}\chi\Big)^{-(s+1)}\right]_{p_{0}i}\left[\chi^{s}\right]_{il}\right)
\left[\mathrm{D}^{s}\right]_{p_{1}k}\\ & = 
\sum\limits_{s = 0}^{N} (-1)^{s} \left(\frac{a\Dtn{n}}{\Dx}\right)^{s}
\left\{\sum\limits_{i = 0}^{N}\left(\left[\Big(I - \nu\Dtn{n}\chi\Big)^{-(s+1)}\right]_{p_{0}i}\left[\sum\limits_{l = 0}^{N}\left[\chi^{s}\right]_{il}\right]\right)\right\}
\left[\mathrm{D}^{s}\right]_{p_{1}k}.
\end{split}
\end{equation}
An explicit calculation of this sums in (\ref{eq:local_sol_inv_matrix}) and (\ref{eq:sigma_matrix_temp_1}) is carried out using the property of matrix elements $\{\chi_{pq}\}$ proved by Han Veiga \textit{et al.} in~\cite{ader_improving_2024} (which is also interpreted as the simplifying condition $C(N)$~\cite{Butcher_book_2016, Hairer_book_1, Hairer_book_2, Dekker_Verwer_1984} in the case of using the ADER-DG method with LST-DG-predictor as Runge-Kutta method to solve the initial value problem for an ODE system~\cite{ader_improving_2024, ader_dg_ivp_ode_sinum} and~\cite{ader_dg_ivp_ode, ader_dg_ivp_dae}):
\begin{equation}\label{eq:c_prop_ode}
\sum\limits_{q = 0}^{N} \chi_{pq} \tau_{q}^{r} = \frac{\tau_{p}^{r+1}}{r+1},\quad 0 \leqslant p \leqslant N,\quad 0 \leqslant r \leqslant N-1.
\end{equation}
In the notation of this work, the property (\ref{eq:c_prop_ode}) is derived similarly as follows:\\[-2mm]
\begin{equation}
\begin{split}
\sum\limits_{q = 0}^{N} \left[\chi^{-1}\right]_{pq}\frac{\tau_{q}^{r+1}}{r+1} & =
\sum\limits_{q = 0}^{N} \frac{\kappa_{pq}\tau_{q}^{r+1}}{w_{p} (r+1)} = \sum\limits_{q = 0}^{N} \left[
	\smash[b]{\underline{\varphi}}_{p}\smash[b]{\underline{\varphi}}_{q} + \int\limits_{0}^{1} \varphi_{p}(\tau)\frac{d\varphi_{q}(\tau)}{d\tau}d\tau
\right] \frac{\tau_{q}^{r+1}}{w_{p} (r+1)}\\ & =
\sum\limits_{q = 0}^{N} \frac{\tau_{q}^{r+1}}{w_{p} (r+1)}\smash[b]{\underline{\varphi}}_{p}\smash[b]{\underline{\varphi}}_{q} +
\sum\limits_{q = 0}^{N} \frac{\tau_{q}^{r+1}}{w_{p} (r+1)}\int\limits_{0}^{1} \varphi_{p}(\tau)\frac{d\varphi_{q}(\tau)}{d\tau}d\tau\\ & =
\frac{\smash[b]{\underline{\varphi}}_{p}}{w_{p} (r+1)} \sum\limits_{q = 0}^{N} \tau_{q}^{r+1}\smash[b]{\underline{\varphi}}_{q} +
\int\limits_{0}^{1} \frac{\varphi_{p}(\tau)}{w_{p} (r+1)} \frac{d}{d\tau}\left[\sum\limits_{q = 0}^{N}\tau_{q}^{r+1}\varphi_{q}(\tau)\right]d\tau\\ & =
\frac{\smash[b]{\underline{\varphi}}_{p}}{w_{p} (r+1)}\cdot(0)^{r+1} +
\int\limits_{0}^{1} \frac{\varphi_{p}(\tau)}{w_{p} (r+1)} \frac{d\left(\tau^{r+1}\right)}{d\tau}d\tau =
\frac{1}{w_{p}}\int\limits_{0}^{1} \tau^{r}\varphi_{p}(\tau) d\tau = \tau_{p}^{r},
\end{split}
\end{equation}
which yields the original expression (\ref{eq:c_prop_ode}) after using the definition (\ref{eq:chi_and_tilde_ae_def}) of the matrix $\chi$. The calculation of the sum in the expression (\ref{eq:sigma_matrix_temp_1}) is based on the following expression:\\[-2mm]
\begin{equation}\label{eq:chi_int_prop}
\begin{split}
\sum\limits_{l = 0}^{N}\left[\chi^{s}\right]_{il} & =
\sum\limits_{l = 0}^{N} \left[\sum\limits_{i_{1} = 0}^{N}\ldots\sum\limits_{i_{s} = 0}^{N}\sum\limits_{i_{s} = 0}^{N} \chi_{i i_{1}}\ldots\chi_{i_{s-1}l}\chi_{i_{s}l}\right]\\ & = 
\sum\limits_{i_{1} = 0}^{N}\ldots\sum\limits_{i_{s-1} = 0}^{N}\sum\limits_{i_{s} = 0}^{N}
\chi_{i i_{1}}\ldots\chi_{i_{s-1}i_{s}}\left[\sum\limits_{l = 0}^{N}\chi_{i_{s}l}\right]\\ & =
\sum\limits_{i_{1} = 0}^{N}\ldots\sum\limits_{i_{s-1} = 0}^{N}\sum\limits_{i_{s} = 0}^{N} \chi_{i i_{1}}\ldots\chi_{i_{s-1}i_{s}}\tau_{i_{s}} =
\ldots = \frac{\tau_{i}^{s}}{s!},
\end{split}
\end{equation}
which is strictly satisfied for $0 \leqslant s \leqslant N$, leading to the expression for the elements $\{\sigma_{\mathrm{p}, k}\}$ of the matrix $\sigma$ (\ref{eq:sigma_matrix_temp_1}):
\begin{equation}
\begin{split}
\sigma_{p_{0}p_{1}, k} & = 
\sum\limits_{s = 0}^{N} \frac{(-1)^{s}}{s!} \left(\frac{a\Dtn{n}}{\Dx}\right)^{s}
\left\{\sum\limits_{i = 0}^{N}\left(\left[\Big(I - \nu\Dtn{n}\chi\Big)^{-(s+1)}\right]_{p_{0}i}\tau_{i}^{s}\right)\right\}
\left[\mathrm{D}^{s}\right]_{p_{1}k}.
\end{split}
\end{equation}
To obtain a local discrete space-time solution $q$ (\ref{eq:local_sol_def}), the property of matrix $\mathrm{D}$ (\ref{eq:tilde_ae_deriv_prop}) as a differentiation matrix of a function represented by a representation in the form of an expansion in basis functions $\{\varphi_{p}\}$ is used:
\begin{equation}
\frac{d^{s}u^{n}(\xi)}{d\xi^{s}} = \sum\limits_{k = 0}^{N} \hat{u}^{n}_{k} \varphi_{k}^{(s)}(\xi) = \sum\limits_{k = 0}^{N} \hat{u}^{s, n}_{k} \varphi_{k}(\xi),\quad
\hat{u}^{s, n}_{k} = \sum\limits_{l = 0}^{N} [\mathrm{D}^{s}]_{kl} \hat{u}^{n}_{l},
\end{equation}
where the set $\{\hat{u}^{s, n}_{k}\}$ corresponds to the representation of derivatives of $u^{n}(\xi)$ in the form of an expansion in basis functions $\{\varphi_{p}\}$, which, when using expression (\ref{eq:lstdg_pred_eq_sol_temp_1_ind_form}), leads to the following expression for the expansion coefficients $\hat{q}_{p_{0}p_{1}}$ of a local discrete space-time solution $q$:\\[-2mm]
\begin{equation}\label{eq:local_sol_coeffs}
\hat{q}_{p_{0}p_{1}} = 
  \sum\limits_{k = 0}^{N} \sigma_{p_{0}p_{1}, k} \hat{u}^{n}_{k} =
  \sum\limits_{s = 0}^{N} \frac{(-1)^{s}}{s!} \left(\frac{a\Dtn{n}}{\Dx}\right)^{s}
  \left\{\sum\limits_{i = 0}^{N}\left(\left[\Big(I - \nu\Dtn{n}\chi\Big)^{-(s+1)}\right]_{p_{0}i}\tau_{i}^{s}\right)\right\} \, \hat{u}^{s, n}_{p_{1}}.
\end{equation}
To calculate the result inner sum, the binomial theorem infinite expansion is used:
\begin{equation}
\begin{split}
\Big(I - \nu\Dtn{n}\chi\Big)^{-(s+1)} = 
\sum_{j = 0}^{\infty} \binom{s+j}{s} (\nu\Dtn{n})^{j} \chi^{j},
\end{split}
\end{equation}
which in this case does not terminate at a finite number of terms. Substituting this expansion into the inner sum in the formula (\ref{eq:local_sol_coeffs}) yields the following expression:
\begin{equation}
\begin{split}
&\sum\limits_{i = 0}^{N}\left(\left[\Big(I - \nu\Dtn{n}\chi\Big)^{-(s+1)}\right]_{p_{0}i}\tau_{i}^{s}\right) =
 \sum\limits_{i = 0}^{N}\left[\sum_{j = 0}^{\infty} \binom{s+j}{s} (\nu\Dtn{n})^{j} \left[\chi^{j}\right]_{p_{0}i}\tau_{i}^{s}\right]\\ & =
 \sum_{j = 0}^{\infty} \binom{s+j}{s} (\nu\Dtn{n})^{j} \left[\sum\limits_{i = 0}^{N}\left[\chi^{j}\right]_{p_{0}i}\tau_{i}^{s}\right] =
 \sum_{j = 0}^{N-s} \binom{s+j}{s} (\nu\Dtn{n})^{j} \left[\sum\limits_{i = 0}^{N}\left[\chi^{j}\right]_{p_{0}i}\tau_{i}^{s}\right]\\ & +
 \sum_{j = N-s+1}^{\infty} \binom{s+j}{s} (\nu\Dtn{n})^{j} \left[\sum\limits_{i = 0}^{N}\left[\chi^{j}\right]_{p_{0}i}\tau_{i}^{s}\right] =
 \sum_{j = 0}^{N-s} \binom{s+j}{s} (\nu\Dtn{n})^{j} \frac{s!}{(s+j)!}\tau_{p_{0}}^{s+j} + R_{s, p_{0}}^{\chi}(\nu\Dtn{n})\\ & =
 \sum_{j = 0}^{N-s} \frac{(s+j)!}{s!j!} (\nu\Dtn{n})^{j} \frac{s!}{(s+j)!}\tau_{p_{0}}^{s+j} + R_{s, p_{0}}^{\chi}(\nu\Dtn{n}) =
 \sum_{j = 0}^{N-s}  \frac{(\nu\Dtn{n})^{j}}{j!}\tau_{p_{0}}^{s+j} + R_{s, p_{0}}^{\chi}(\nu\Dtn{n})\\ & =
 \tau_{p_{0}}^{s} \sum_{j = 0}^{N-s} \frac{(\nu\Dtn{n}\tau_{p_{0}})^{j}}{j!} + R_{s, p_{0}}^{\chi}(\nu\Dtn{n}) =
 \tau_{p_{0}}^{s} \hat{\mathfrak{T}}_{N-s}\left[\exp\Big(\nu\Dtn{n}\tau_{p_{0}}\Big)\right] + R_{s, p_{0}}^{\chi}(\nu\Dtn{n}),
\end{split}
\end{equation}
where the leading and remainder terms are explicitly identified, and the following relation is used:
\begin{equation}
\sum\limits_{q = 0}^{N} \left[\chi^{j}\right]_{pq} \tau_{q}^{s} = \frac{s!}{(s+j)!}\tau_{p}^{s+j}, \quad 0 \leqslant s+j \leqslant N.
\end{equation}
which is a direct consequence of relation (\ref{eq:chi_int_prop}), and where $\hat{\mathfrak{T}}_{l}[f]$ is the Taylor expansion of the function $f$ up to the $l$-th term of the expansion. The expression for the remainder term $R_{s, p_{0}}^{\chi}(\nu\Dtn{n})$ has the following form and quantitative estimate:
\begin{equation}
R_{s, p_{0}}^{\chi}(\nu\Dtn{n}) = 
\sum_{j = N-s+1}^{\infty} \binom{s+j}{s} (\nu\Dtn{n})^{j} \left[\sum\limits_{i = 0}^{N}\left[\chi^{j}\right]_{p_{0}i}\tau_{i}^{s}\right] =
O\left((\nu\Dtn{n})^{N-s+1}\right),
\end{equation}
where the expression is clearly a convergent series due to the factorial growth of the denominators of the terms in the series. Substituting the resulting representation for the inner sum into the formula (\ref{eq:local_sol_coeffs}) yields the following expression for the coefficients of the representation of the local discrete space-time solution:
\begin{equation}
\hat{q}_{p_{0}p_{1}} = 
  \sum\limits_{k = 0}^{N} \sigma_{p_{0}p_{1}, k} \hat{u}^{n}_{k} =
  \sum\limits_{s = 0}^{N} (-1)^{s} \left(\frac{a\Dtn{n}}{\Dx}\right)^{s} \frac{\tau_{p_{0}}^{s}}{s!}
  \hat{\mathfrak{T}}_{N-s}\left[\exp\Big(\nu\Dtn{n}\tau_{p_{0}}\Big)\right] \, \hat{u}^{s, n}_{p_{1}} + R_{p_{0}p_{1}}^{\hat{q}}(\nu\Dtn{n}),
\end{equation}
where the remainder term $R_{p_{0}p_{1}}^{\hat{q}}(\nu\Dtn{n})$ is represented in the following form:
\begin{equation}
R_{p_{0}p_{1}}^{\hat{q}}(\nu\Dtn{n}) =
\sum\limits_{s = 0}^{N} \frac{(-1)^{s}}{s!} \left(\frac{a\Dtn{n}}{\Dx}\right)^{s} R_{s, p_{0}}^{\chi}(\nu\Dtn{n}) \, \hat{u}^{s, n}_{p_{1}}.
\end{equation}
The coordinate and time dependence of the local discrete space-time solution $q(\tau, \xi)$ (\ref{eq:local_sol_def}) itself is expressed as follows:
\begin{equation}\label{eq:lstdg_pred_eq_sol_temp_2}
\begin{split}
& q(\tau, \xi) = \sum\limits_{p_{0} = 0}^{N}\sum\limits_{p_{1} = 0}^{N}\sum\limits_{s = 0}^{N}
    (-1)^{s} \left(\frac{a\Dtn{n}}{\Dx}\right)^{s} \frac{\tau_{p_{0}}^{s}}{s!}
    \hat{\mathfrak{T}}_{N-s}\left[\exp\Big(\nu\Dtn{n}\tau_{p_{0}}\Big)\right] \,
    \hat{u}^{s, n}_{k} \varphi_{p_{0}}(\tau)\varphi_{p_{1}}(\xi)\\
& + \sum\limits_{p_{0} = 0}^{N}\sum\limits_{p_{1} = 0}^{N}\sum\limits_{s = 0}^{N}
    \frac{(-1)^{s}}{s!} \left(\frac{a\Dtn{n}}{\Dx}\right)^{s} R_{s, p_{0}}^{\chi}(\nu\Dtn{n}) \,
    \hat{u}^{s, n}_{k} \varphi_{p_{0}}(\tau)\varphi_{p_{1}}(\xi)\\
& = \sum\limits_{s = 0}^{N} \frac{(-1)^{s}}{s!} \left(\frac{a\Dtn{n}}{\Dx}\right)^{s}
    \left(\sum\limits_{p_{0} = 0}^{N} \tau_{p_{0}}^{s}
    \hat{\mathfrak{T}}_{N-s}\left[\exp\Big(\nu\Dtn{n}\tau_{p_{0}}\Big)\right]\varphi_{p_{0}}(\tau)\right)
    \left(\sum\limits_{p_{1} = 0}^{N} \hat{u}^{s, n}_{k}\varphi_{p_{1}}(\xi)\right)\\
& + \sum\limits_{s = 0}^{N} \frac{(-1)^{s}}{s!} \left(\frac{a\Dtn{n}}{\Dx}\right)^{s}
    \left(\sum\limits_{p_{0} = 0}^{N} R_{s, p_{0}}^{\chi}(\nu\Dtn{n})\varphi_{p_{0}}(\tau)\right) \,
    \left(\sum\limits_{p_{1} = 0}^{N} \hat{u}^{s, n}_{k}\varphi_{p_{1}}(\xi)\right)\\
& = \sum\limits_{s = 0}^{N} \frac{(-1)^{s}}{s!} \left(\frac{a\Dtn{n}}{\Dx}\right)^{s} \tau^{s}\,
    \hat{\mathfrak{T}}_{N-s}\left[\exp\Big(\nu\Dtn{n}\tau\Big)\right]\,\frac{d^{s}u^{n}(\xi)}{d\xi^{s}}\\
& + \sum\limits_{s = 0}^{N} \frac{(-1)^{s}}{s!} \left(\frac{a\Dtn{n}}{\Dx}\right)^{s}
    \left(\sum\limits_{p_{0} = 0}^{N} R_{s, p_{0}}^{\chi}(\nu\Dtn{n})\varphi_{p_{0}}(\tau)\right)\frac{d^{s}u^{n}(\xi)}{d\xi^{s}}
  = \sum\limits_{s = 0}^{N} \frac{(-1)^{s}}{s!} \left(\frac{a\Dtn{n}}{\Dx}\right)^{s}\\
& \cdot \left(\frac{t(\tau) - t^{n}}{\Dtn{n}}\right)^{s}
    \hat{\mathfrak{T}}_{N-s}\left[\exp\left(\nu\Dtn{n}\left(\frac{t(\tau) - t^{n}}{\Dtn{n}}\right)\right)\right] \left(\Dx\right)^{s}
    \left.\left[\frac{d^{s}}{dx^{s}}u^{n}(x)\right]\right|_{x(\xi)} + R^{q}(\tau, \xi)\\
& = \sum\limits_{s = 0}^{N} \frac{(-1)^{s}}{s!} \left[a\left(t(\tau) - t^{n}\right)\right]^{s}
    \hat{\mathfrak{T}}_{N-s}\Big[\exp\left(\nu\left(t(\tau) - t^{n}\right)\right)\Big]
    \left.\left[\frac{d^{s}}{dx^{s}}u^{n}(x)\right]\right|_{x(\xi)} + R^{q}(\tau, \xi),
\end{split}
\end{equation}
where the following notation is used:
\begin{equation}\label{eq:mapping_derx}
\frac{1}{\left(\Dx\right)^{s}}\frac{d^{s}u^{n}(\xi)}{d\xi^{s}} = \left.\left[\frac{d^{s}}{dx^{s}}u^{n}(x)\right]\right|_{x(\xi)},
\end{equation}
explicitly follows from the mapping (\ref{eq:mapping}) used, and the expression for the remainder term $R^{q}(\tau, \xi)$ has the following form and quantitative estimate:
\begin{equation}
\begin{split}
&R^{q}(\tau, \xi) = 
 \sum\limits_{s = 0}^{N} \frac{(-1)^{s}}{s!} \left(\frac{a\Dtn{n}}{\Dx}\right)^{s}
 \left(\sum\limits_{p_{0} = 0}^{N} R_{s, p_{0}}^{\chi}(\nu\Dtn{n})\varphi_{p_{0}}(\tau)\right)\frac{d^{s}u^{n}(\xi)}{d\xi^{s}} =
 \sum\limits_{s = 0}^{N} \frac{(-1)^{s}}{s!} \left(\frac{a\Dtn{n}}{\Dx}\right)^{s}\\ & \cdot
 \left[\sum\limits_{p_{0} = 0}^{N}
 \left(\sum_{j = N-s+1}^{\infty} \binom{s+j}{s} (\nu\Dtn{n})^{j} \left[\sum\limits_{i = 0}^{N}\left[\chi^{j}\right]_{p_{0}i}\tau_{i}^{s}\right]\right)
 \varphi_{p_{0}}(\tau)\right]\frac{d^{s}u^{n}(\xi)}{d\xi^{s}} = O\left((\nu\Dtn{n})^{N+1}\right).
\end{split}
\end{equation}
In the general case, when function $u(x(\xi), t^{n})$ is not exactly representable by a polynomial representation ($u(x(\xi), t^{n})|_{\Omega_{i}}\not\in\mathcal{P}_{N}(\omega)$), but are sufficiently smooth $u(x(\xi), t^{n})|_{\Omega_{i}}\in\mathcal{C}_{N+1}(\omega)$, the well-known quantitative estimates hold:
\begin{equation}\label{eq:errs_ests}
\begin{split}
&u^{n}(\xi) = \left.u(x(\xi), t^{n})\right|_{\Omega_{i}} + R^{(0)}_{u}(\xi),\quad
\frac{d^{s}}{dx^{s}}u^{n}(x) = \left.\pdernoper{x}{s}u(x(\xi), t^{n})\right|_{\Omega_{i}} + R^{(s)}_{u}(\xi),\\
&R^{(0)}_{u}(\xi) = \left[\frac{u^{(N+1)}(\xi_{0})}{(N+1)!}\hat{\omega}(\xi)\right]\Dx^{N+1} + o\left(\Dx^{N+1}\right) = O\left(\Dx^{N+1}\right),\\
&R^{(s)}_{u}(\xi) = \left[\frac{u^{(N+1)}(\xi_{s}(\xi))}{(N+1)!}\frac{d^{s}\hat{\omega}(\xi)}{d\xi^{s}}\right]\Dx^{N+1-s} + o\left(\Dx^{N+1-s}\right) = O\left(\Dx^{N+1-s}\right),
\end{split}
\end{equation}
where $R^{(0)}_{u}(\xi)$ and $R^{(s)}_{u}(\xi)$ are remainder terms, $\xi_{0}\in[0, 1]$ is a constant intermediate point (point from the mean value theorem), $\{\xi_{s}(\xi)\in\omega\}$ is a set of the intermediate points, $\hat{\omega}:\, \omega \rightarrow \mathcal{R}$ is a polynomial with leading coefficient~$1$ and roots at the roots $\{\xi_{p}\}$ of the shifted Legendre polynomial $\tilde{P}_{N+1}(\xi)$. It is interesting to quantitatively estimate the error and calculate the order of approximation of the local discrete space-time solution (\ref{eq:lstdg_pred_eq_sol_temp_2}) not for the original solution function $u(x, t)$, but for the pre-exponential factor $u(x, t)\exp(-\nu t)$. For this purpose, it is convenient to introduce a new function $v(x, t) = u(x, t)\exp(-\nu t)$, which is a pre-exponential factor and satisfies the classical linear advection equation:
\begin{equation}
u(x, t) = v(x, t)\exp(\nu t), \quad\mapsto\quad \pderf{u}{t} + a\pderf{u}{x} = \nu u, \quad\Rightarrow\quad \pderf{v}{t} + a\pderf{v}{x} = 0.
\end{equation}
This advection equation has a well-known exact analytical solution $v(x, t) = u(x - a(t - t_{n}), t_{n})$, where $u(x, t^{n})$ is the solution of advection-reaction equation (\ref{eq:adv_eq_src}) at time $t^{n}$. To calculate the derivatives of the new function with respect to $t$, the following transformations are performed:
\begin{equation}
\pderfoper{x}(au) = -\pderf{u}{t} + \nu u = -\exp(\nu t)\pderf{v}{t} = -\exp(\nu t)\pderfoper{t}\Big[u \exp(-\nu t)\Big],
\end{equation}
from which the expression for derivative ${\partial v}/{\partial t}$ follows:
\begin{equation}
\exp(-\nu t)\,\pderfoper{x}(au) = -\pderfoper{t}\Big[u \exp(-\nu t)\Big] \equiv -\pderf{v}{t}.
\end{equation}
Repeated transformation yielded the following relation for derivative ${\partial^{2} v}/{\partial t^{2}}$:
\begin{equation}
\begin{split}
\exp(-\nu t)\,\pdernoper{x}{2}(au) & = -\pderfoper{t}\left[\pderf{u}{x}\,\exp(-\nu t)\right]\\ & =
-\pderfoper{t}\left[\exp(\nu t)\,\left(-\pderfoper{t}\Big[u \exp(-\nu t)\Big]\right)\,\exp(-\nu t)\right] = \pdernoper{t}{2}\Big[u \exp(-\nu t)\Big] \equiv \pdern{v}{t}{2}.
\end{split}
\end{equation}
Next, the following general relation for derivative ${\partial^{s} v}/{\partial t^{s}}$ is shown by induction:
\begin{equation}
\exp(-\nu t)\,\pdernoper{x}{s}(au) = (-1)^{s} \pdernoper{t}{s}\Big[u \exp(-\nu t)\Big] \equiv (-1)^{s} \pdern{v}{t}{s}.
\end{equation}
Substituting the resulting relations into expression (\ref{eq:lstdg_pred_eq_sol_temp_2}), expressed as a pre-exponential factor $q(\tau, \xi)\exp(-\nu t(\tau))$, leads to the following expression:
\begin{equation}\label{eq:lstdg_pred_eq_sol_temp_3}
\begin{split}
q(\tau, \xi)\exp(-\nu t(\tau)) &= \sum\limits_{s = 0}^{N} \frac{(-1)^{s}}{s!} \left[a\left(t(\tau) - t^{n}\right)\right]^{s}\,\exp(-\nu t(\tau))\\
&\cdot\hat{\mathfrak{T}}_{N-s}\Big[\exp\left(\nu\left(t(\tau) - t^{n}\right)\right)\Big]\left.\left[\frac{d^{s}}{dx^{s}}u^{n}(x)\right]\right|_{x(\xi)}\hspace{-3mm} +
R^{q}(\tau, \xi)\exp(-\nu t(\tau)).
\end{split}
\end{equation}
The calculation of the factor included in this expression is performed as follows:
\begin{equation}\label{eq:lstdg_pred_eq_sol_temp_4}
\begin{split}
&\exp(-\nu t(\tau))\,\hat{\mathfrak{T}}_{N-s}\Big[\exp\left(\nu\left(t(\tau) - t^{n}\right)\right)\Big]\\ & =
 \exp(-\nu t^{n})\exp\left(-\nu\left(t(\tau) - t^{n}\right)\right)\, \hat{\mathfrak{T}}_{N-s}\Big[\exp\left(\nu\left(t(\tau) - t^{n}\right)\right)\Big]\\ & =
 \exp(-\nu t^{n})\exp\left(-\nu\left(t(\tau) - t^{n}\right)\right)
 \left[\exp\left(\nu\left(t(\tau) - t^{n}\right)\right) - \sum_{j = N-s+1}^{\infty} \frac{\big(\nu\left(t(\tau) - t^{n}\right)\big)^{j}}{j!}\right]\\ & =
 \exp(-\nu t^{n})\left\{1 - \exp\left(-\nu\left(t(\tau) - t^{n}\right)\right)
 \left[\sum_{j = N-s+1}^{\infty} \frac{\big(\nu\left(t(\tau) - t^{n}\right)\big)^{j}}{j!}\right]\right\}\\ & =
 \exp(-\nu t^{n})\left\{1 -
 \left[\sum_{j = 0}^{\infty} \frac{(-1)^{j}\big(\nu\left(t(\tau) - t^{n}\right)\big)^{j}}{j!} \vphantom{\sum_{j = N-s+1}^{\infty}}\right]
 \left[\sum_{j = N-s+1}^{\infty} \frac{\big(\nu\left(t(\tau) - t^{n}\right)\big)^{j}}{j!}\right]\right\}\\ & =
 \exp(-\nu t^{n}) + R^{\mathfrak{T}}_{s}(\tau),
\end{split}
\end{equation}
where $R^{\mathfrak{T}}_{s}(\tau)$ is the remainder term for which the following expression and quantitative estimate hold:
\begin{equation}\label{eq:lstdg_pred_eq_sol_temp_5}
R^{\mathfrak{T}}_{s}(\tau) = -\exp(-\nu t^{n})
 \left[\sum_{j = 0}^{\infty} \frac{(-1)^{j}\big(\nu\left(t(\tau) - t^{n}\right)\big)^{j}}{j!} \vphantom{\sum_{j = N-s+1}^{\infty}}\right]
 \left[\sum_{j = N-s+1}^{\infty} \frac{\big(\nu\left(t(\tau) - t^{n}\right)\big)^{j}}{j!}\right] =
O\left((t(\tau) - t^{n})^{N-s+1}\right),
\end{equation}
where an inequality $t(\tau) - t^{n} \leqslant \Dtn{n}$ valid for single step $\tau\in\omega$ in the argument $t$ is used. Substituting (\ref{eq:lstdg_pred_eq_sol_temp_4}) and (\ref{eq:lstdg_pred_eq_sol_temp_5}) into the original expression (\ref{eq:lstdg_pred_eq_sol_temp_3}) leads to the following expression to quantitatively estimate the error and calculate the order of approximation of the local discrete space-time solution $q(\tau, \xi)$ (\ref{eq:local_sol_def}):
\begin{equation}\label{eq:local_sol_app_order}
\begin{split}
q(\tau, \xi)\exp(-\nu t(\tau)) &= 
  \sum\limits_{s = 0}^{N} \frac{(-1)^{s}}{s!} \left[a\left(t(\tau) - t^{n}\right)\right]^{s}\,\exp(-\nu t^{n})
  \left.\left[\frac{d^{s}}{dx^{s}}u^{n}(x)\right]\right|_{x(\xi)}\\ & + 
  \sum\limits_{s = 0}^{N} \frac{(-1)^{s}}{s!} \left[a\left(t(\tau) - t^{n}\right)\right]^{s}\,R^{\mathfrak{T}}_{s}(\tau)
  \left.\left[\frac{d^{s}}{dx^{s}}u^{n}(x)\right]\right|_{x(\xi)}\hspace{-3mm} + R^{q}(\tau, \xi)\exp(-\nu t(\tau))\\ & =
  \sum\limits_{s = 0}^{N} \frac{(-1)^{s}}{s!} \left[a\left(t(\tau) - t^{n}\right)\right]^{s}
  \left.\left[\frac{d^{s}}{dx^{s}}u^{n}(x)\exp(-\nu t^{n})\right]\right|_{x(\xi)}\hspace{-3mm} + O\left((\nu\Dtn{n})^{N+1}\right)\\ & =
  u^{n}(x + a(t(\tau) - t^{n}), t^{n}))\exp(-\nu t(\tau)) + O\left((\nu\Dtn{n})^{N+1}\right).
\end{split}
\end{equation}
The obtained approximation estimate showed that the local discrete space-time solution $q(\tau, \xi)$ (\ref{eq:local_sol_def}) has an approximation order $p_{\rm L} = N+1$, however, of course, this estimate takes place only locally within element $\Xi_{i}^{n}$ (\ref{eq:discr_domains}) only in the space-time domain of influence $\{\smash[b]{\tilde{\Xi}}_{i}^{n}\subseteq\Xi_{i}^{n}\, |\, x\in\Omega_{i},\, t - t_{n}\leqslant \theta(a)a(x - x_{i}) - \theta(-a)a(x_{i+1} - x)\}$ of the discretization domain $\Omega_{i}$.

\subsection{Corrected solution at time steps}
\label{sec:approx_anal:final_sol}

Substitution of the diagonal matrix $m$ (\ref{eq:matrices_def}) for the basis functions $\{\varphi\}$ (\ref{eq:phi_def}) into the expression for the completely one-step ADER-DG scheme (\ref{eq:ader_dg_corr_temp_1}) (which, with respect to the LST-DG predictor step, is a corrector step) leads to the following expression:
\begin{equation}\label{eq:ader_dg_corr_temp_2}
\begin{split}
\left(\hat{u}^{n+1}_{k} - \hat{u}^{n}_{k}\right) & -
\frac{1}{w_{k}}\frac{a\Dtn{n}}{\Dx}\intrefdom{\tau}\intrefdom{\xi}\,\varphi_{k}'(\xi) q(\tau, \xi) +
\frac{1}{w_{k}}\frac{\Dtn{n}}{\Dx}\intrefdom{\tau}\,\Big[
  \ophi_{k} F_{\rm RP}\left(q(\tau, 1),\, q^{+}(\tau, 0)\right) \\ & -
  \uphi_{k} F_{\rm RP}\left(q^{-}(\tau, 1),\, q(\tau, 0)\right)
\Big] = \frac{\nu\Dtn{n}}{w_{k}}\intrefdom{\tau}\intrefdom{\xi}\,\varphi_{k}(\xi) q(\tau, \xi).
\end{split}
\end{equation}
The local term of the resulting expression (\ref{eq:ader_dg_corr_temp_2}), after detailed calculation and use of notations (\ref{eq:matrices_def}) and (\ref{eq:chi_and_tilde_ae_def}), is reduced to the form
\begin{equation}\label{eq:ader_dg_corr_local_term_temp_1}
\begin{split}
\frac{1}{w_{k}}\frac{a\Dtn{n}}{\Dx}&\intrefdom{\tau}\intrefdom{\xi}\,\varphi_{k}'(\xi) q(\tau, \xi)
    = \frac{1}{w_{k}}\frac{a\Dtn{n}}{\Dx}\sum\limits_{p_{0} = 0}^{N}\sum\limits_{p_{1} = 0}^{N}
      \left[\intrefdom{\tau}\, \varphi_{p_{0}}(\tau)\right] \left[\intrefdom{\xi}\, \varphi_{k}'(\xi)\varphi_{p_{1}}(\xi)\right]\,\hat{q}_{p_{0}p_{1}}\\
  & = \frac{1}{w_{k}}\frac{a\Dtn{n}}{\Dx}\sum\limits_{p_{0} = 0}^{N}\sum\limits_{p_{1} = 0}^{N} w_{p_{0}} \varrho_{p_{1}k} \,\hat{q}_{p_{0}p_{1}}
    = \frac{1}{w_{k}}\sum\limits_{p_{0} = 0}^{N}\sum\limits_{p_{1} = 0}^{N} w_{p_{0}} \varrho_{p_{1}k}
      \sum\limits_{s = 0}^{N} \frac{(-1)^{s}}{s!} \left(\frac{a\Dtn{n}}{\Dx}\right)^{s}\\
  & \cdot \left\{\sum\limits_{i = 0}^{N}\left(\left[\Big(I - \nu\Dtn{n}\chi\Big)^{-(s+1)}\right]_{p_{0}i}\tau_{i}^{s}\right)\right\} \, \hat{u}^{s, n}_{p_{1}}
    = \frac{1}{w_{k}}\sum\limits_{s = 0}^{N} \frac{(-1)^{s}}{s!}\left(\frac{a\Dtn{n}}{\Dx}\right)^{s+1}\\
  & \cdot \left[\sum\limits_{p_{0} = 0}^{N}w_{p_{0}}
      \left\{\sum\limits_{i = 0}^{N}\left(\left[\Big(I - \nu\Dtn{n}\chi\Big)^{-(s+1)}\right]_{p_{0}i}\tau_{i}^{s}\right)\right\}\right]
      \left[\sum\limits_{p_{1} = 0}^{N} \varrho_{p_{1}k} \hat{u}^{s, n}_{k}\right].
\end{split}
\end{equation}
Explicitly calculating the first inner sum in the resulting expression (\ref{eq:ader_dg_corr_local_term_temp_1}) is accomplished in two steps. The first stage uses the expressions (\ref{eq:c_prop_ode}) and (\ref{eq:chi_int_prop}) and trivial differentiation to the parameter $\nu\Dtn{n}$ yields the following expression:
\begin{equation}\label{eq:ader_dg_corr_local_term_temp_1_temp_1}
\frac{d^{s}}{d(\nu\Dtn{n})^{s}}\left[\Big(I - \nu\Dtn{n}\chi\Big)^{-1}\mathbf{1}\right]_{p_{0}} =
\left[\Big(I - \nu\Dtn{n}\chi\Big)^{-(s+1)}s!\,\chi^{s}\mathbf{1}\right]_{p_{0}} =
\sum\limits_{i = 0}^{N}\left[\Big(I - \nu\Dtn{n}\chi\Big)^{-(s+1)}\right]_{p_{0}i}\tau_{i}^{s},
\end{equation}
which allows the inner sum to be rewritten in the following form:
\begin{equation}\label{eq:ader_dg_corr_local_term_temp_1_temp_2}
\begin{split}
\sum\limits_{p_{0} = 0}^{N}w_{p_{0}}\left\{\sum\limits_{i = 0}^{N}\left(\left[\Big(I - \nu\Dtn{n}\chi\Big)^{-(s+1)}\right]_{p_{0}i}\tau_{i}^{s}\right)\right\} =
\frac{d^{s}}{d(\nu\Dtn{n})^{s}}\left\{\sum\limits_{p_{0} = 0}^{N}w_{p_{0}}\left[\Big(I - \nu\Dtn{n}\chi\Big)^{-1}\mathbf{1}\right]_{p_{0}}\right\},
\end{split}
\end{equation}
which is now free of high powers of the inverse matrix $(I - \nu\Dtn{n}\chi)^{-1}$. In the second stage, the semantics of expression $(I - \nu\Dtn{n}\chi)^{-1}\mathbf{1}$ must be clarified. To do this, it is necessary to consider the application of the ADER-DG numerical method with the LST-DG predictor to solve the initial value problem for an ordinary differential equation:
\begin{equation}\label{eq:dahlquist_test_eq}
\frac{dy}{dt} = c y,\quad y(t_{0}) = y^{0},\quad c\in\mathcal{C},\quad \Re[c] < 0,\quad t\in\mathcal{T},
\end{equation}
which is the Dahlquist's test equation~\cite{Hairer_book_1, Hairer_book_2, Dekker_Verwer_1984}, where $y^{0}$ is the initial value of the solution, which is set at the point $t_{0}$. The condition $\Re[c] < 0$ that $c$ belongs to the left complex half-plane is necessary for the classical formulation of the Dahlquist's test equation to study the decay region of the solution, but is not used in any subsequent results --- this is correct for any $c\in\mathcal{C}$. Below, a brief description of the ADER-DG method with the LST-DG predictor for solving problem (\ref{eq:dahlquist_test_eq}), adapted for this work and necessary for further use, is presented (for a full description and detailed study, see works~\cite{ader_dg_ivp_ode, ader_dg_ivp_dae, ader_improving_2024, ader_proofs_2025, ader_dg_ivp_ode_sinum}). The original ordinary differential equation (\ref{eq:dahlquist_test_eq}) is rewritten in integral form for the time-discretization domain $\mathcal{T}^{n}$:
\begin{equation}
y^{n+1} = y^{n} + c \int\limits_{t^{n}}^{t^{n+1}} dt\, y(t) = y^{n} + c\Dtn{n}\intrefdom{\tau}\, y(t(\tau)),
\end{equation}
in which the local discrete time solution $r^{n}: \omega\rightarrow\mathcal{C}$ in the time-discretization domain $\mathcal{T}^{n}$ is then used in the integrand, leading to the following formula for the numerical solution $y^{n+1}$ at the next time step $t^{n+1}$ in terms of the numerical solution $y^{n}$ at the previous time step $t^{n}$:
\begin{equation}\label{eq:dahlquist_test_eq_int}
y^{n+1} = y^{n} + c\Dtn{n}\intrefdom{\tau}\, r^{n}(\tau),
\end{equation}
where $\tau\in\omega$ is the previously used mapping variable (\ref{eq:mapping}). To obtain the local discrete time solution $r^{n}$ the same equation (\ref{eq:dahlquist_test_eq}) is used, defined in the discretization domain with the initial condition $y^{n}$ at the previous time step $t^{n}$:
\begin{equation}\label{eq:dahlquist_test_local}
\frac{dr^{n}}{d\tau} = c\Dtn{n}r^{n},\qquad r(0) = y^{n},\qquad \tau\in\omega.
\end{equation}
The representation of the local time solution $r^{n}$ is chosen as an expansion over a set of basis functions $\{\varphi_{p}\}_{p}$ (\ref{eq:phi_def}):
\begin{equation}
r^{n}(\tau) = \sum\limits_{p = 0}^{N} \hat{r}_{p}^{n} \varphi_{p}(\tau),
\end{equation}
where $\{\hat{r}_{p}^{n}\}_{p}$ is the set of expansion coefficients. The equation (\ref{eq:dahlquist_test_local}) for the local discrete time solution $r^{n}$ is rewritten in weak form:
\begin{equation}\label{eq:dahlquist_test_local_int}
\intrefdom{\tau}\,\varphi_{p}(\tau)\left[\frac{dr^{n}(\tau)}{d\tau} - c\Dtn{n}r^{n}(\tau)\right] = 0,\qquad r(0) = y^{n},
\end{equation}
while it could have been initially chosen in this form for use in a broader class of problems, in this case, only a problem (\ref{eq:dahlquist_test_eq}) for which the weak (\ref{eq:dahlquist_test_local_int}) and source (\ref{eq:dahlquist_test_local}) forms are equivalent is of interest. The initial condition in problem (\ref{eq:dahlquist_test_local_int}) is now also set in a weak form: it has been shown in~\cite{ader_dg_ivp_ode, ader_dg_ivp_ode_sinum} that the initial condition in this case is not strictly satisfied --- there are discontinuities on the right side of the local solution at the nodes (also see~\cite{ader_dg_imp_ode}, where an improved local solution is constructed that is free of this drawback). Direct evaluation of the expression (\ref{eq:dahlquist_test_local}) with integration of the first term by parts and using the initial condition $r(0) = y^{n}$ in weak form leads to the following system of linear algebraic equations of local time-DG predictor for the expansion coefficients $\{\hat{r}_{p}^{n}\}$ of the local discrete time solution $r^{n}$:
\begin{equation}
\hat{r}_{p}^{n} - c\Dtn{n} \sum\limits_{q = 0}^{N} \chi_{pq} \hat{r}_{q}^{n} = y^{n},
\end{equation}
the solution of which can be represented as follows:
\begin{equation}
\hat{r}_{p}^{n} = \left[\Big(I - c\Dtn{n}\chi\Big)^{-1}\mathbf{1}\right]_{p} y^{n}.
\end{equation}
Substituting the obtained local discrete time solution $r^{n}$ back into the original expression (\ref{eq:dahlquist_test_eq_int}) allows us to obtain the numerical solution $y^{n+1}$ at the next time step $t^{n+1}$ in terms of the numerical solution $y^{n}$ at the previous time step $t^{n}$:
\begin{equation}
\begin{split}
y^{n+1} & = y^{n} + c\Dtn{n} \sum\limits_{p = 0}^{N} w_{p} \left[\sum\limits_{q = 0}^{N} \hat{r}_{q}^{n} \varphi_{q}(\tau_{p})\right]
          = y^{n} + c\Dtn{n} \sum\limits_{p = 0}^{N} w_{p} \hat{r}_{p}^{n}\\
        & = y^{n} + c\Dtn{n} \sum\limits_{p = 0}^{N} w_{p} \left[\Big(I - c\Dtn{n}\chi\Big)^{-1}\mathbf{1}\right]_{p} y^{n}\\
        & = \left\{1 + c\Dtn{n} \sum\limits_{p = 0}^{N} w_{p} \left[\Big(I - c\Dtn{n}\chi\Big)^{-1}\mathbf{1}\right]_{p}\right\} y^{n}
          \equiv R_{\rm RK}(c\Dtn{n}) y^{n},
\end{split}
\end{equation}
where the GL quadrature formula is used, and the stability function $R_{\rm RK}(c\Dtn{n})$ is introduced:
\begin{equation}\label{eq:stab_func_def}
R_{\rm RK}(c\Dtn{n}) = 1 + \nu\Dtn{n}\sum\limits_{p_{0} = 0}^{N}w_{p_{0}}\left[\Big(I - c\Dtn{n}\chi\Big)^{-1}\mathbf{1}\right]_{p_{0}},
\end{equation}
which defines the numerical approximation of the evolution operator $R_{\rm RK}$ for the problem (\ref{eq:dahlquist_test_eq}) for one time step $\Dtn{n}$, and the subscript ${\rm RK}$ is associated with the fact that this mathematical apparatus is similar to the study of the stability of Runge-Kutta numerical methods and similar single-step numerical methods for solving the initial value problem for a system of ordinary differential equations~\cite{Hairer_book_1, Hairer_book_2, Dekker_Verwer_1984, Butcher_book_2016}. Using the resulting expression for the evolution operator $R_{\rm RK}(c\Dtn{n})$ (\ref{eq:stab_func_def}), the original desired expression (\ref{eq:ader_dg_corr_local_term_temp_1_temp_2}) can be represented in the formula as follows:\\[-2mm]
\begin{equation}\label{eq:exp_int_by_stab_func}
\sum\limits_{p_{0} = 0}^{N}w_{p_{0}}\left[\Big(I - \nu\Dtn{n}\chi\Big)^{-1}\mathbf{1}\right]_{p_{0}} = \frac{R_{\rm RK}(\nu\Dtn{n}) - 1}{\nu\Dtn{n}},
\end{equation}
which leads to the final relation:\\[-2mm]
\begin{equation}\label{eq:exp_int_diff_by_stab_func}
\sum\limits_{p_{0} = 0}^{N}\left\{w_{p_{0}}\sum\limits_{i = 0}^{N}\left[\Big(I - \nu\Dtn{n}\chi\Big)^{-(s+1)}\right]_{p_{0}i}\tau_{i}^{s}\right\} =
\frac{d^{s}}{d(\nu\Dtn{n})^{s}}\left[\frac{R_{\rm RK}(\nu\Dtn{n}) - 1}{\nu\Dtn{n}}\right],
\end{equation}
necessary for calculating the first inner sum in the expressions (\ref{eq:ader_dg_corr_local_term_temp_1_temp_2}) and (\ref{eq:ader_dg_corr_local_term_temp_1}). {\"O}ffner \textit{et al}~\cite{ader_proofs_2025} found that the evolution operator of the ADER-DG numerical method is a $[N, N+1]$-Pad\'{e} approximation of the exact evolution operator $\exp(\nu\Dtn{n})$ of the Dahlquist's test equation (\ref{eq:dahlquist_test_eq}):\\[-2mm]
\begin{equation}\label{eq:stab_func_ader_dg}
R_{\rm RK}(\nu\Dtn{n}) = [N, N+1]_{\exp(\nu\Dtn{n})} =
\frac{\displaystyle\sum_{k=0}^{N-1} \left[\frac{(2N - 1 - k)! (N-1)!}{(2N - 1)! (N - 1 - k)! k!}\right] (\nu\Dtn{n})^k}
{\displaystyle\sum_{k=0}^{N} \left[\frac{(2N - 1 - k)! N!}{(2N - 1)! (N - k)! k!}\right] (-\nu\Dtn{n})^k},
\end{equation}
as a result, the ADER-DG numerical method with the time-DG predictor to solve the initial value problem for an ordinary differential equation is $A$-stable and $L$-stable (the latter is noted in~\cite{ader_dg_ivp_ode_sinum}). The resulting expression has clear and transparent semantics: it is an exponential integrator~\cite{exp_int_rev, exp_int_cox}, whose operation is well represented as follows:
\begin{equation}
\frac{dy}{dt} = c y, \quad\Rightarrow\quad
\frac{dy(t)}{dt} = c \left[\lim\limits_{\Dtn{n} \rightarrow 0_{+}} \frac{\exp(c\Dtn{n})-1}{c\Dtn{n}}\right] y(t)
                 = c \left[\lim\limits_{\Dtn{n} \rightarrow 0_{+}} \frac{R_{\rm RK}(\nu\Dtn{n})-1}{c\Dtn{n}}\right] y(t),
\end{equation}
and the approximation order $2N+1$~\cite{ader_dg_ivp_ode} of the ADER-DG numerical method with the time-DG predictor to solve the initial value problem for an ordinary differential equation is determined by the Pad\'{e} approximation error estimates of the original entire function $(\exp(c\Dtn{n}) - 1)/(c\Dtn{n})$~\cite{exp_int_rev} (see also Chapter~8 in the book~\cite{Varga_2000}):\\[-2mm]
\begin{equation}\label{eq:exp_pade_approx}
\begin{split}
&R_{\rm RK}(c\Dtn{n}) = \exp(c\Dtn{n}) + O\left((c\Dtn{n})^{2N+2}\right),\\
&\frac{R_{\rm RK}(c\Dtn{n})-1}{c\Dtn{n}} = \frac{\exp(c\Dtn{n})-1}{c\Dtn{n}} + O\left((c\Dtn{n})^{2N+1}\right).
\end{split}
\end{equation}
The second inner sum in the expression (\ref{eq:ader_dg_corr_local_term_temp_1_temp_2}) is conveniently transformed using integration by parts as follows:\\[-2mm]
\begin{equation}\label{eq:ader_dg_corr_local_term_temp_2}
\begin{split}
\sum\limits_{p_{1} = 0}^{N} \varrho_{p_{1}k} \hat{u}^{s, n}_{k}
  & = \sum\limits_{p_{1} = 0}^{N} w_{p_{1}} \varphi_{k}'(\xi_{p_{1}}) \hat{u}^{s, n}_{k}
    = \intrefdom{\xi}\,\varphi_{k}'(\xi)\,\frac{d^{s}u^{n}(\xi)}{d\xi^{s}}
    = \left.\left[\varphi_{k}(\xi)\,\frac{d^{s}u^{n}(\xi)}{d\xi^{s}}\right]\right|_{0}^{1}\\
  & - \intrefdom{\xi}\,\varphi_{k}(\xi)\,\frac{d^{s+1}u^{n}(\xi)}{d\xi^{s+1}}
    = \ophi_{k}\,\frac{d^{s}u^{n}}{d\xi^{s}}(1) - \uphi_{k}\,\frac{d^{s}u^{n}}{d\xi^{s}}(0) - w_{k}\,\hat{u}^{s+1, n}_{k},
\end{split}
\end{equation}
which is necessary to isolate the terms that will subsequently be used to ``compensate'' for the flux terms. As a result of substituting expressions (\ref{eq:exp_int_diff_by_stab_func}) and (\ref{eq:ader_dg_corr_local_term_temp_2}) into expression (\ref{eq:ader_dg_corr_local_term_temp_1}), the local term of the resulting expression (\ref{eq:ader_dg_corr_temp_2}) takes the following form:\\[-2mm]
\begin{equation}\label{eq:ader_dg_corr_local_term}
\begin{split}
\frac{1}{w_{k}}\frac{a\Dtn{n}}{\Dx}\intrefdom{\tau}\intrefdom{\xi}\,\varphi_{k}'(\xi) q(\tau, \xi) =
\sum\limits_{s = 0}^{N} & \frac{(-1)^{s}}{s!}\left(\frac{a\Dtn{n}}{\Dx}\right)^{s+1}
\left\{\frac{d^{s}}{d(\nu\Dtn{n})^{s}}\Bigg[\frac{R_{\rm RK}(\nu\Dtn{n}) - 1}{\nu\Dtn{n}}\Bigg]\right\}\\ & \cdot
\Bigg[\frac{\ophi_{k}}{w_{k}}\,\frac{d^{s}u^{n}}{d\xi^{s}}(1) - \frac{\uphi_{k}}{w_{k}}\,\frac{d^{s}u^{n}}{d\xi^{s}}(0) - \hat{u}^{s+1, n}_{k}\Bigg],
\end{split}
\end{equation}
where it is necessary to take into account that, due to additional differentiation, the zeroth term $\hat{u}^{N+1, n}_{k} = 0$ is taken into account. Similarly, but without any integration by parts, the source term in the resulting expression (\ref{eq:ader_dg_corr_local_term_temp_1}) is transformed:
\begin{equation}\label{eq:ader_dg_corr_source_term}
\begin{split}
&\frac{\nu\Dtn{n}}{w_{k}}\intrefdom{\tau}\intrefdom{\xi}\,\varphi_{k}(\xi) q(\tau, \xi)
  = \frac{\nu\Dtn{n}}{w_{k}}\sum\limits_{p_{0} = 0}^{N}\sum\limits_{p_{1} = 0}^{N}
    \left[\intrefdom{\tau}\, \varphi_{p_{0}}(\tau)\right] \left[\intrefdom{\xi}\, \varphi_{k}(\xi)\varphi_{p_{1}}(\xi)\right]\,\hat{q}_{p_{0}p_{1}}\\
& = \frac{\nu\Dtn{n}}{w_{k}}\sum\limits_{p_{0} = 0}^{N}\sum\limits_{p_{1} = 0}^{N} w_{p_{0}} m_{kp_{1}} \,\hat{q}_{p_{0}p_{1}}
  = \frac{\nu\Dtn{n}}{w_{k}}\sum\limits_{p_{0} = 0}^{N}\sum\limits_{p_{1} = 0}^{N} w_{p_{0}} w_{p_{1}}\, \delta_{kp_{1}} \,\hat{q}_{p_{0}p_{1}}
  = \frac{\nu\Dtn{n}}{w_{k}}\sum\limits_{p_{0} = 0}^{N} w_{p_{0}} w_{k} \,\hat{q}_{p_{0}k}\\
& = \nu\Dtn{n}\sum\limits_{p_{0} = 0}^{N} w_{p_{0}} \hat{q}_{p_{0}k}
  = \nu\Dtn{n}\sum\limits_{p_{0} = 0}^{N} w_{p_{0}}
    \sum\limits_{s = 0}^{N} \frac{(-1)^{s}}{s!} \left(\frac{a\Dtn{n}}{\Dx}\right)^{s}
    \left\{\sum\limits_{i = 0}^{N}\left(\left[\Big(I - \nu\Dtn{n}\chi\Big)^{-(s+1)}\right]_{p_{0}i}\tau_{i}^{s}\right)\right\} \, \hat{u}^{s, n}_{k}\\
& = \nu\Dtn{n}\sum\limits_{s = 0}^{N} \frac{(-1)^{s}}{s!}\left(\frac{a\Dtn{n}}{\Dx}\right)^{s} \left[\sum\limits_{p_{0} = 0}^{N}w_{p_{0}}
    \left\{\sum\limits_{i = 0}^{N}\left(\left[\Big(I - \nu\Dtn{n}\chi\Big)^{-(s+1)}\right]_{p_{0}i}\tau_{i}^{s}\right)\right\}\right] \hat{u}^{s, n}_{k}\\
& = \nu\Dtn{n}\sum\limits_{s = 0}^{N} \frac{(-1)^{s}}{s!}\left(\frac{a\Dtn{n}}{\Dx}\right)^{s}
    \left\{\frac{d^{s}}{d(\nu\Dtn{n})^{s}}\left[\frac{R_{\rm RK}(\nu\Dtn{n}) - 1}{\nu\Dtn{n}}\right]\right\}\hat{u}^{s, n}_{k}.
\end{split}
\end{equation}
The union of the local parts of the local term defined by expression (\ref{eq:ader_dg_corr_local_term}) and the source term defined by expression (\ref{eq:ader_dg_corr_source_term}) results in the following expression:
\begin{equation}
\begin{split}
&\sum\limits_{s = 0}^{N} \frac{(-1)^{s}}{s!}\left(\frac{a\Dtn{n}}{\Dx}\right)^{s+1}
 \left\{\frac{d^{s}}{d(\nu\Dtn{n})^{s}}\Bigg[\frac{R_{\rm RK}(\nu\Dtn{n}) - 1}{\nu\Dtn{n}}\Bigg]\right\}\hat{u}^{s+1, n}_{k}\\ & -
 \nu\Dtn{n}\sum\limits_{s = 0}^{N} \frac{(-1)^{s}}{s!}\left(\frac{a\Dtn{n}}{\Dx}\right)^{s}
 \left\{\frac{d^{s}}{d(\nu\Dtn{n})^{s}}\left[\frac{R_{\rm RK}(\nu\Dtn{n}) - 1}{\nu\Dtn{n}}\right]\right\}\hat{u}^{s, n}_{k}\\ & =
 \sum\limits_{s = 0}^{N} \frac{(-1)^{s}}{s!}\left(\frac{a\Dtn{n}}{\Dx}\right)^{s}
 \left\{\frac{d^{s}}{d(\nu\Dtn{n})^{s}}\left[\frac{R_{\rm RK}(\nu\Dtn{n}) - 1}{\nu\Dtn{n}}\right]\right\}
 \left[\frac{a\Dtn{n}}{\Dx}\hat{u}^{s+1, n}_{k} - \nu\Dtn{n}\hat{u}^{s, n}_{k}\right],
\end{split}
\end{equation}
regrouping the summations and explicit evaluation in which yields the following result:
\begin{equation}\label{eq:ader_dg_corr_local_part_of_terms}
\begin{split}
&\sum\limits_{k = 0}^{N}\left[\sum\limits_{s = 0}^{N} \frac{(-1)^{s}}{s!}\left(\frac{a\Dtn{n}}{\Dx}\right)^{s+1}
 \left\{\frac{d^{s}}{d(\nu\Dtn{n})^{s}}\Bigg[\frac{R_{\rm RK}(\nu\Dtn{n}) - 1}{\nu\Dtn{n}}\Bigg]\right\}\hat{u}^{s+1, n}_{k}
 \rule{0pt}{0.75cm}\right.\\ & \left.\rule{0pt}{0.75cm} -
 \nu\Dtn{n}\sum\limits_{s = 0}^{N} \frac{(-1)^{s}}{s!}\left(\frac{a\Dtn{n}}{\Dx}\right)^{s}
 \left\{\frac{d^{s}}{d(\nu\Dtn{n})^{s}}\left[\frac{R_{\rm RK}(\nu\Dtn{n}) - 1}{\nu\Dtn{n}}\right]\right\}\hat{u}^{s, n}_{k}\right]\varphi_{k}(\xi)\\ & =
 \sum\limits_{s = 0}^{N} \frac{(-1)^{s}}{s!}\left(\frac{a\Dtn{n}}{\Dx}\right)^{s}
 \left\{\frac{d^{s}}{d(\nu\Dtn{n})^{s}}\left[\frac{R_{\rm RK}(\nu\Dtn{n}) - 1}{\nu\Dtn{n}}\right]\right\}
 \left[\frac{a\Dtn{n}}{\Dx}\frac{d^{s+1}u^{n}(\xi)}{d\xi^{s+1}} - \nu\Dtn{n}\frac{d^{s}u^{n}(\xi)}{d\xi^{s}}\right]\\ & =
 \sum\limits_{s = 0}^{N} \frac{(-1)^{s}}{s!}\left(a\Dtn{n}\right)^{s+1}
 \left\{\frac{d^{s}}{d(\nu\Dtn{n})^{s}}\left[\frac{R_{\rm RK}(\nu\Dtn{n}) - 1}{\nu\Dtn{n}}\right]\right\}
 \left\{\left.\left[\frac{d^{s+1}}{dx^{s+1}}u^{n}(x)\right]\right|_{x(\xi)} -
 \frac{\nu}{a}\left.\left[\frac{d^{s}}{dx^{s}}u^{n}(x)\right]\right|_{x(\xi)}\right\},
\end{split}
\end{equation}
where the mapping (\ref{eq:mapping_derx}) is used. The first two terms in the expression (\ref{eq:ader_dg_corr_local_term}) take into account the values of the local discrete space-time solution $q$ at the boundary points of discretization domain $\Omega_{i}$, which is evident from the factors $\ophi_{k}/w_{k}$ and $\uphi_{k}/w_{k}$ included in these terms. To substantiate this conclusion, the delta-function $\delta$ in polynomial space $\mathcal{P}_{N}(\omega)$ is considered, represented as an expansion in basis functions $\{\varphi_{p}\}$ (\ref{eq:phi_def}):\\[-3mm]
\begin{equation}\label{eq:delta_func_varphi_coeffs}
\delta_{\mathcal{P}_{N}}(\tau) = \sum\limits_{k = 0}^{N} \iota_{k}\varphi_{k}(\tau),\quad
\iota_{k} = \frac{1}{w_{k}} \intrefdom{\tau}\, \delta(\tau)\varphi_{k}(\tau) = \frac{\uphi_{k}}{w_{k}},
\end{equation}
where $\{\iota_{k}\}$ is the set of the expansion coefficients. From the resulting representation, it is clear that $\delta_{\mathcal{P}_{N}}$ accurately represents the sifting (filtering) property of the delta-function $\delta$ when acting on any function $f\in\mathcal{P}_{N}(\omega)$:\\[-2mm]
\begin{equation}\label{eq:delta_func_varphi_action}
\begin{split}
f(\tau) = \sum\limits_{k = 0}^{N} f(\tau_{k})\varphi_{k}(\tau),\ \Rightarrow\ 
\intrefdom{\tau}\, f(\tau)\delta_{\mathcal{P}_{N}}(\tau)
  = \sum\limits_{k = 0}^{N}\sum\limits_{l = 0}^{N} f(\tau_{k}) \frac{\uphi_{l}}{w_{l}} \intrefdom{\tau}\, \varphi_{k}(\tau)\varphi_{l}(\tau)
  = \sum\limits_{k = 0}^{N} f(\tau_{k})\uphi_{k} = f(0).
\end{split}
\end{equation}
It should be noted that despite the ``flux form'' of the boundary terms in the local term (\ref{eq:ader_dg_corr_local_term}), these are local contributions to the final numerical solution $\{u_{k}^{n+1}\}$ (\ref{eq:u_n_def}) --- only the local discrete space-time solution $q$ at this coordinate-time discretization domain $\Xi_{i}^{n}$ is used there. The flux term in the expression (\ref{eq:ader_dg_corr_temp_2}) contains a local discrete space-time solution $q$ from two nearest coordinate-time discretization domains $\Xi_{i-1}^{n}$ and $\Xi_{i+1}^{n}$ (this is the case in the general form of recording the formulation of the numerical method, and in each specific case it contains a local discrete space-time solution from only one nearest cell --- left or right, depending on the sign of the advection velocity $a$ (\ref{eq:adv_eq_src}), in accordance with the upwind property, which is determined by the $\theta$-functions in the exact Riemann solver (\ref{eq:f_rp})) --- $q^{-}$ and $q^{+}$, which are selected in the following form:
\begin{equation}
\begin{split}
&q^{+}(\tau, \xi) = \sum\limits_{p_{0} = 0}^{N}\sum\limits_{p_{1} = 0}^{N} \hat{q}^{+}_{p_{0}p_{1}} \varphi_{p_{0}}(\tau)\varphi_{p_{1}}(\xi),\quad
\hat{q}^{+}_{p_{0}p_{1}} = 
  \sum\limits_{s = 0}^{N} (-1)^{s} \left(\frac{a\Dtn{n}}{\Dx}\right)^{s} \frac{\tau_{p_{0}}^{s}}{s!} \,
  \sum\limits_{l = 0}^{N} [\mathrm{D}^{s}]_{p_{1}l} \hat{u}^{+,n}_{l},\\
&q^{-}(\tau, \xi) = \sum\limits_{p_{0} = 0}^{N}\sum\limits_{p_{1} = 0}^{N} \hat{q}^{-}_{p_{0}p_{1}} \varphi_{p_{0}}(\tau)\varphi_{p_{1}}(\xi),\quad
\hat{q}^{-}_{p_{0}p_{1}} = 
  \sum\limits_{s = 0}^{N} (-1)^{s} \left(\frac{a\Dtn{n}}{\Dx}\right)^{s} \frac{\tau_{p_{0}}^{s}}{s!} \,
  \sum\limits_{l = 0}^{N} [\mathrm{D}^{s}]_{p_{1}l} \hat{u}^{-,n}_{l},\\
\end{split}
\end{equation}
where upper index ``$+$'' corresponds to the right cell, upper index ``$-$'' corresponds to the left cell, the last expressions for the expansion coefficients $\{\hat{q}^{+}_{p_{0}p_{1}}\}$ and $\{\hat{q}^{-}_{p_{0}p_{1}}\}$ correspond to formula (\ref{eq:local_sol_coeffs}). After substituting the Riemann solver (\ref{eq:f_rp}) expression into the flux term of the expression (\ref{eq:ader_dg_corr_temp_2}) and rearranging the components of the formula, the following expression is obtained:
\begin{equation}
\begin{split}
&\frac{1}{w_{k}}\frac{\Dtn{n}}{\Dx}\intrefdom{\tau}\,\Big[\ophi_{k} F_{\rm RP}\left(q(\tau, 1),\, q^{+}(\tau, 0)\right) -
                                                          \uphi_{k} F_{\rm RP}\left(q^{-}(\tau, 1),\, q(\tau, 0)\right)\Big]\\ & =
\frac{1}{w_{k}}\frac{a\Dtn{n}}{\Dx}\intrefdom{\tau}\,\Bigg\{\ophi_{k}\Big[\theta(a)q(\tau, 1) + \theta(-a)q^{+}(\tau, 0)\Big] -
                                                            \uphi_{k}\Big[\theta(a)q^{-}(\tau, 1) + \theta(-a)q(\tau, 0)\Big]\Bigg\}\\ & =
\frac{1}{w_{k}}\frac{a\Dtn{n}}{\Dx}\sum\limits_{p = 0}^{N} w_{p}\Bigg\{\theta( a)\Big[\ophi_{k}q(\tau_{p}, 1) - \uphi_{k}q^{-}(\tau_{p}, 1)\Big] +
                                                                       \theta(-a)\Big[\ophi_{k}q^{+}(\tau_{p}, 0) - \uphi_{k}q(\tau_{p}, 0)\Big]\Bigg\},
\end{split}
\end{equation}
where the terms are explicitly separated in accordance with the sign of the advection velocity $a$. A detailed calculation and transformation of the first term, nonzero under condition $a > 0$, leads to the following expression:
\begin{equation}
\begin{split}
&\frac{1}{w_{k}}\frac{a\Dtn{n}}{\Dx}\sum\limits_{p = 0}^{N} w_{p} \Big[\ophi_{k}q(\tau_{p}, 1) - \uphi_{k}q^{-}(\tau_{p}, 1)\Big] =
\frac{1}{w_{k}} \sum\limits_{p = 0}^{N} w_{p} \sum\limits_{p_{0} = 0}^{N}\sum\limits_{p_{1} = 0}^{N}\sum\limits_{s = 0}^{N}
  \frac{(-1)^{s}}{s!} \varphi_{p_{0}}(\tau_{p}) \ophi_{p_{1}}\\ & \cdot \left(\frac{a\Dtn{n}}{\Dx}\right)^{s+1}
  \left\{\sum\limits_{i = 0}^{N}\left(\left[\Big(I - \nu\Dtn{n}\chi\Big)^{-(s+1)}\right]_{p_{0}i}\tau_{i}^{s}\right)\right\}
  \sum\limits_{l = 0}^{N} [\mathrm{D}^{s}]_{p_{1}l} \Big[\ophi_{k}\hat{u}^{n}_{l} - \uphi_{k}\hat{u}^{-,n}_{l}\Big]\\ & =
\frac{1}{w_{k}} \sum\limits_{s = 0}^{N} (-1)^{s} \left(\frac{a\Dtn{n}}{\Dx}\right)^{s+1}
  \left[\sum\limits_{p_{0} = 0}^{N} w_{p_{0}} \left\{\sum\limits_{i = 0}^{N}
  \left(\left[\Big(I - \nu\Dtn{n}\chi\Big)^{-(s+1)}\right]_{p_{0}i}\tau_{i}^{s}\right)\right\}\right]\\ & \cdot
  \left[\sum\limits_{p_{1} = 0}^{N} \ophi_{p_{1}} \sum\limits_{l = 0}^{N} [\mathrm{D}^{s}]_{p_{1}l}
        \Big(\ophi_{k}\hat{u}^{n}_{l} - \uphi_{k}\hat{u}^{-,n}_{l}\Big)\right] =
\frac{1}{w_{k}} \sum\limits_{s = 0}^{N} \frac{(-1)^{s}}{s!} \left(\frac{a\Dtn{n}}{\Dx}\right)^{s+1}\\ & \cdot
  \left\{\frac{d^{s}}{d(\nu\Dtn{n})^{s}}\left[\frac{R_{\rm RK}(\nu\Dtn{n}) - 1}{\nu\Dtn{n}}\right]\right\}
  \left[\sum\limits_{p_{1} = 0}^{N} \ophi_{p_{1}} \left(\ophi_{k}\frac{d^{s}u^{n}}{d\xi^{s}}(\xi_{p_{1}}) -
                                                        \uphi_{k}\frac{d^{s}u^{-,n}}{d\xi^{s}}(\xi_{p_{1}})\right)\right]\\ & =
  \frac{1}{w_{k}} \sum\limits_{s = 0}^{N} \frac{(-1)^{s}}{s!} \left(\frac{a\Dtn{n}}{\Dx}\right)^{s+1}
    \left\{\frac{d^{s}}{d(\nu\Dtn{n})^{s}}\left[\frac{R_{\rm RK}(\nu\Dtn{n}) - 1}{\nu\Dtn{n}}\right]\right\}\\ & \cdot
    \left[\ophi_{k}\left(\sum\limits_{p_{1} = 0}^{N} \ophi_{p_{1}}\frac{d^{s}u^{n}}{d\xi^{s}}(\xi_{p_{1}})\right) -
          \uphi_{k}\left(\sum\limits_{p_{1} = 0}^{N} \ophi_{p_{1}}\frac{d^{s}u^{-,n}}{d\xi^{s}}(\xi_{p_{1}})\right)\right]\\ & =
  \sum\limits_{s = 0}^{N} \frac{(-1)^{s}}{s!} \left(\frac{a\Dtn{n}}{\Dx}\right)^{s+1}
  \left\{\frac{d^{s}}{d(\nu\Dtn{n})^{s}}\left[\frac{R_{\rm RK}(\nu\Dtn{n}) - 1}{\nu\Dtn{n}}\right]\right\}
  \left[\frac{\ophi_{k}}{w_{k}}\frac{d^{s}u^{n}}{d\xi^{s}}(1) - \frac{\uphi_{k}}{w_{k}}\frac{d^{s}u^{-,n}}{d\xi^{s}}(1)\right],
\end{split}
\end{equation}
where expressions (\ref{eq:exp_int_by_stab_func}) and (\ref{eq:tilde_ae_deriv_prop}) are used. A similar calculation and transformation of the second term, different from zero under condition $a < 0$, leads to the following expression:
\begin{equation}
\begin{split}
&\frac{1}{w_{k}}\frac{a\Dtn{n}}{\Dx}\sum\limits_{p = 0}^{N} w_{p} \Big[\ophi_{k}q^{+}(\tau_{p}, 0) - \uphi_{k}q(\tau_{p}, 0)\Big] =
\frac{1}{w_{k}} \sum\limits_{p = 0}^{N} w_{p} \sum\limits_{p_{0} = 0}^{N}\sum\limits_{p_{1} = 0}^{N}\sum\limits_{s = 0}^{N}
  \frac{(-1)^{s}}{s!} \varphi_{p_{0}}(\tau_{p}) \uphi_{p_{1}}\\ & \cdot \left(\frac{a\Dtn{n}}{\Dx}\right)^{s+1}
  \left\{\sum\limits_{i = 0}^{N}\left(\left[\Big(I - \nu\Dtn{n}\chi\Big)^{-(s+1)}\right]_{p_{0}i}\tau_{i}^{s}\right)\right\}
  \sum\limits_{l = 0}^{N} [\mathrm{D}^{s}]_{p_{1}l} \Big[\ophi_{k}\hat{u}^{+,n}_{l} - \uphi_{k}\hat{u}^{n}_{l}\Big]\\ & =
\frac{1}{w_{k}} \sum\limits_{s = 0}^{N} (-1)^{s} \left(\frac{a\Dtn{n}}{\Dx}\right)^{s+1}
  \left[\sum\limits_{p_{0} = 0}^{N} w_{p_{0}} \left\{\sum\limits_{i = 0}^{N}
  \left(\left[\Big(I - \nu\Dtn{n}\chi\Big)^{-(s+1)}\right]_{p_{0}i}\tau_{i}^{s}\right)\right\}\right]\\ & \cdot
  \left[\sum\limits_{p_{1} = 0}^{N} \uphi_{p_{1}} \sum\limits_{l = 0}^{N} [\mathrm{D}^{s}]_{p_{1}l}
        \Big(\ophi_{k}\hat{u}^{+,n}_{l} - \uphi_{k}\hat{u}^{n}_{l}\Big)\right] =
\frac{1}{w_{k}} \sum\limits_{s = 0}^{N} \frac{(-1)^{s}}{s!} \left(\frac{a\Dtn{n}}{\Dx}\right)^{s+1}\\ & \cdot
  \left\{\frac{d^{s}}{d(\nu\Dtn{n})^{s}}\left[\frac{R_{\rm RK}(\nu\Dtn{n}) - 1}{\nu\Dtn{n}}\right]\right\}
  \left[\sum\limits_{p_{1} = 0}^{N} \uphi_{p_{1}} \left(\ophi_{k}\frac{d^{s}u^{+,n}}{d\xi^{s}}(\xi_{p_{1}}) -
                                                        \uphi_{k}\frac{d^{s}u^{n}}{d\xi^{s}}(\xi_{p_{1}})\right)\right]\\ & =
  \frac{1}{w_{k}} \sum\limits_{s = 0}^{N} \frac{(-1)^{s}}{s!} \left(\frac{a\Dtn{n}}{\Dx}\right)^{s+1}
    \left\{\frac{d^{s}}{d(\nu\Dtn{n})^{s}}\left[\frac{R_{\rm RK}(\nu\Dtn{n}) - 1}{\nu\Dtn{n}}\right]\right\}\\ & \cdot
    \left[\ophi_{k}\left(\sum\limits_{p_{1} = 0}^{N} \uphi_{p_{1}}\frac{d^{s}u^{+,n}}{d\xi^{s}}(\xi_{p_{1}})\right) -
          \uphi_{k}\left(\sum\limits_{p_{1} = 0}^{N} \uphi_{p_{1}}\frac{d^{s}u^{n}}{d\xi^{s}}(\xi_{p_{1}})\right)\right]\\ & =
  \sum\limits_{s = 0}^{N} \frac{(-1)^{s}}{s!} \left(\frac{a\Dtn{n}}{\Dx}\right)^{s+1}
  \left\{\frac{d^{s}}{d(\nu\Dtn{n})^{s}}\left[\frac{R_{\rm RK}(\nu\Dtn{n}) - 1}{\nu\Dtn{n}}\right]\right\}
  \left[\frac{\ophi_{k}}{w_{k}}\frac{d^{s}u^{+,n}}{d\xi^{s}}(0) - \frac{\uphi_{k}}{w_{k}}\frac{d^{s}u^{n}}{d\xi^{s}}(0)\right].
\end{split}
\end{equation}
The final expression for the flux term of the expression (\ref{eq:ader_dg_corr_temp_2}) takes the following form:
\begin{equation}\label{eq:ader_dg_corr_flux_term}
\begin{split}
\frac{1}{w_{k}}\frac{a\Dtn{n}}{\Dx}&\intrefdom{\tau}\,\Big[\ophi_{k} F_{\rm RP}\left(q(\tau, 1),\, q^{+}(\tau, 0)\right) -
                                                           \uphi_{k} F_{\rm RP}\left(q^{-}(\tau, 1),\, q(\tau, 0)\right)\Big]\\ & =
\sum\limits_{s = 0}^{N} \frac{(-1)^{s}}{s!} \left(\frac{a\Dtn{n}}{\Dx}\right)^{s+1}
\left\{\frac{d^{s}}{d(\nu\Dtn{n})^{s}}\left[\frac{R_{\rm RK}(\nu\Dtn{n}) - 1}{\nu\Dtn{n}}\right]\right\}\\ & \cdot
\Bigg\{
\theta( a)\left[\frac{\ophi_{k}}{w_{k}}\frac{d^{s}u^{n}}{d\xi^{s}}(1) - \frac{\uphi_{k}}{w_{k}}\frac{d^{s}u^{-,n}}{d\xi^{s}}(1)\right] +
\theta(-a)\left[\frac{\ophi_{k}}{w_{k}}\frac{d^{s}u^{+,n}}{d\xi^{s}}(0) - \frac{\uphi_{k}}{w_{k}}\frac{d^{s}u^{n}}{d\xi^{s}}(0)\right]
\Bigg\}.
\end{split}
\end{equation}
Substituting the resulting expressions for the local term (\ref{eq:ader_dg_corr_local_term}) and source term (\ref{eq:ader_dg_corr_source_term}), expression (\ref{eq:ader_dg_corr_local_part_of_terms}) and the expression for the flux term (\ref{eq:ader_dg_corr_flux_term}) into the original expression (\ref{eq:ader_dg_corr_temp_2}) for the ADER-DG numerical method leads to the following expression:
\begin{equation}
\begin{split}
\hat{u}^{n+1}_{k} - \hat{u}^{n}_{k}
  & = \sum\limits_{s = 0}^{N} \frac{(-1)^{s}}{s!}\left(\frac{a\Dtn{n}}{\Dx}\right)^{s+1}
      \left\{\frac{d^{s}}{d(\nu\Dtn{n})^{s}}\Bigg[\frac{R_{\rm RK}(\nu\Dtn{n}) - 1}{\nu\Dtn{n}}\Bigg]\right\}\\ & \cdot
      \Bigg\{
        \nu\Dtn{n}\frac{d^{s}u^{n}(\xi)}{d\xi^{s}} -
        \Bigg[\frac{\ophi_{k}}{w_{k}}\,\frac{d^{s}u^{n}}{d\xi^{s}}(1) - \frac{\uphi_{k}}{w_{k}}\,\frac{d^{s}u^{n}}{d\xi^{s}}(0) - \hat{u}^{s+1, n}_{k}\Bigg]\\ & -
        \frac{\theta( a)}{w_{k}}\Bigg[\ophi_{k}\frac{d^{s}u^{n}}{d\xi^{s}}(1) - \uphi_{k}\frac{d^{s}u^{-,n}}{d\xi^{s}}(1)\Bigg] -
        \frac{\theta(-a)}{w_{k}}\Bigg[\ophi_{k}\frac{d^{s}u^{+,n}}{d\xi^{s}}(0) - \uphi_{k}\frac{d^{s}u^{n}}{d\xi^{s}}(0)\Bigg]
      \Bigg\},
\end{split}
\end{equation}
which, after convolution with a set of basis functions $\{\varphi_{k}\}_{k}$ (\ref{eq:phi_def}), takes the following form:
\begin{equation}\label{eq:ader_dg_corr}
\begin{split}
&u^{n+1}(\xi) - u^{n}(\xi) = \sum\limits_{k = 0}^{N} \left(\hat{u}^{n+1}_{k} - \hat{u}^{n}_{k}\right)\varphi_{k}(\xi)\\ & =
 \sum\limits_{s = 0}^{N} \frac{(-1)^{s}}{s!}\left(a\Dtn{n}\right)^{s+1}
 \left\{\frac{d^{s}}{d(\nu\Dtn{n})^{s}}\left[\frac{R_{\rm RK}(\nu\Dtn{n}) - 1}{\nu\Dtn{n}}\right]\right\}
 \left\{\frac{\nu}{a}\pdernoper{x}{s}u^{n}(x) - \pdernoper{x}{s+1}u^{n}(x)\right\}\\
   & - \theta( a)\sum\limits_{k = 0}^{N}\left[\frac{\uphi_{k}}{w_{k}} \sum\limits_{s = 0}^{N} \frac{(-1)^{s}}{s!}
       \left\{\frac{d^{s}}{d(\nu\Dtn{n})^{s}}\left[\frac{R_{\rm RK}(\nu\Dtn{n}) - 1}{\nu\Dtn{n}}\right]\right\}
       \left(\frac{a\Dtn{n}}{\Dx}\right)^{s+1}\left(\frac{d^{s}u^{n}}{d\xi^{s}}(0) - \frac{d^{s}u^{-,n}}{d\xi^{s}}(1)\right)\right]\varphi_{k}(\xi)\\
   & - \theta(-a)\sum\limits_{k = 0}^{N}\left[\frac{\ophi_{k}}{w_{k}} \sum\limits_{s = 0}^{N} \frac{(-1)^{s}}{s!}
       \left\{\frac{d^{s}}{d(\nu\Dtn{n})^{s}}\left[\frac{R_{\rm RK}(\nu\Dtn{n}) - 1}{\nu\Dtn{n}}\right]\right\}
       \left(\frac{a\Dtn{n}}{\Dx}\right)^{s+1}\left(\frac{d^{s}u^{+,n}}{d\xi^{s}}(0) - \frac{d^{s}u^{n}}{d\xi^{s}}(1)\right)\right]\varphi_{k}(\xi)\\ & =
\sum\limits_{s = 0}^{N} \frac{(-1)^{s}}{s!}\left(a\Dtn{n}\right)^{s+1}
\left\{\frac{d^{s}}{d(\nu\Dtn{n})^{s}}\left[\frac{R_{\rm RK}(\nu\Dtn{n}) - 1}{\nu\Dtn{n}}\right]\right\}
\left\{\frac{\nu}{a}\pdernoper{x}{s}u^{n}(x) - \pdernoper{x}{s+1}u^{n}(x)\right\}\\
   & - \theta( a)\delta_{\mathcal{P}_{N}}(\xi)\left[\frac{\uphi_{k}}{w_{k}} \sum\limits_{s = 0}^{N} \frac{(-1)^{s}}{s!}
       \left\{\frac{d^{s}}{d(\nu\Dtn{n})^{s}}\left[\frac{R_{\rm RK}(\nu\Dtn{n}) - 1}{\nu\Dtn{n}}\right]\right\}
       \left(\frac{a\Dtn{n}}{\Dx}\right)^{s+1}\left(\frac{d^{s}u^{n}}{d\xi^{s}}(0) - \frac{d^{s}u^{-,n}}{d\xi^{s}}(1)\right)\right]\\
   & - \theta(-a)\delta_{\mathcal{P}_{N}}(\xi-1)\left[\frac{\ophi_{k}}{w_{k}} \sum\limits_{s = 0}^{N} \frac{(-1)^{s}}{s!}
       \left\{\frac{d^{s}}{d(\nu\Dtn{n})^{s}}\left[\frac{R_{\rm RK}(\nu\Dtn{n}) - 1}{\nu\Dtn{n}}\right]\right\}
       \left(\frac{a\Dtn{n}}{\Dx}\right)^{s+1}\left(\frac{d^{s}u^{+,n}}{d\xi^{s}}(0) - \frac{d^{s}u^{n}}{d\xi^{s}}(1)\right)\right],
\end{split}
\end{equation}
where the expression of an exponential integrator (\ref{eq:exp_int_by_stab_func}) and definition of delta-functions (\ref{eq:delta_func_varphi_coeffs}) are used. The first term on the right-hand side of the resulting expression (\ref{eq:ader_dg_corr}) is clearly local in nature and is associated with local terms inherent to DG methods as such and source terms, while the second and third terms are associated with discontinuities in the numerical solution at the interfaces $\{x_{i}\}\times\mathcal{T}^{n}$ and $\{x_{i+1}\}\times\mathcal{T}^{n}$ of the space-time discretization domains $\Xi_{i}^{n}$ and $\Xi_{i \pm 1}^{n}$. First, the semantics and quantitative estimates of the second term on the right-hand side of the resulting expression (\ref{eq:ader_dg_corr}), which makes a nonzero contribution in the case $a > 0$, are studied. The internal sum of this term is rewritten in the following form, explicitly taking into account the discontinuity of the numerical solution at the interface $\{x_{i}\}\times\mathcal{T}^{n}$ of the space-time discretization domains $\Xi_{i}^{n}$ and $\Xi_{i-1}^{n}$:
\begin{equation}\label{eq:ader_dg_corr_total_flux_a_plus}
\begin{split}
&\sum\limits_{s = 0}^{N} \frac{(-1)^{s}}{s!}\left\{\frac{d^{s}}{d(\nu\Dtn{n})^{s}}\left[\frac{R_{\rm RK}(\nu\Dtn{n}) - 1}{\nu\Dtn{n}}\right]\right\}
 \left(\frac{a\Dtn{n}}{\Dx}\right)^{s+1}\left(\frac{d^{s}u^{n}}{d\xi^{s}}(0) - \frac{d^{s}u^{-,n}}{d\xi^{s}}(1)\right)\\ & =
 \frac{1}{\Dx}\sum\limits_{s = 0}^{N} \frac{(-1)^{s}}{s!}\left\{\frac{d^{s}}{d(\nu\Dtn{n})^{s}}\left[\frac{R_{\rm RK}(\nu\Dtn{n}) - 1}{\nu\Dtn{n}}\right]\right\}
 \left(a\Dtn{n}\right)^{s+1}\left(\pdern{u^{n}}{x}{s}(x_{i}^{+}) - \pdern{u^{-,n}}{x}{s}(x_{i}^{-})\right)\\ & =
 \frac{1}{\Dx}\sum\limits_{s = 0}^{N} \frac{(-1)^{s}}{s!}\left\{\frac{d^{s}}{d(\nu\Dtn{n})^{s}}\left[\frac{R_{\rm RK}(\nu\Dtn{n}) - 1}{\nu\Dtn{n}}\right]\right\}
 \left(a\Dtn{n}\right)^{s+1}\left[\!\!\left[\pdern{u^{n}}{x}{s}\right]\!\!\right](x_{i}, t_{n}),
\end{split}
\end{equation}
where $[\![f]\!](x_{i}, t) = f(x_{i}^{+}, t) - f(x_{i}^{-}, t)$ is the operator determining the magnitude of the discontinuity of the function $f = f(x, t)$ at point $x_{i}$, and the mapping (\ref{eq:mapping_derx}) is used. To clarify the meaning of the last expression obtained, two auxiliary problems are studied, which together constitute a certain analogue of the generalized Riemann problem (in the linear case of the advection equation, even with a reactive source term, its complete solution presents no problems):
\begin{equation}\label{eq:aux_probl_w_eqs}
\pderf{w_{\pm}}{t} + a\pderf{w_{\pm}}{x} = \nu w_{\pm}, \quad w_{+}(x, t^{n}) = u^{n}(x) \equiv u^{n}_{+}(x), \quad w_{-}(x, t^{n}) = u^{-,n}(x) \equiv u^{n}_{-}(x).
\end{equation}
The solution to these problems is written in the well-known form:
\begin{equation}\label{eq:aux_probl_w_ex_sols}
w_{\pm}(x, t) = \exp(\nu(t - t^{n}))\, u^{n}_{\pm}(x - a(t - t^{n})),
\end{equation}
as a result, the expression for the discontinuity of the solution $[\![w]\!](x_{i}, t)$ at the interface $\{x_{i}\}\times\mathcal{T}^{n}$ of the space-time discretization domains $\Xi_{i}^{n}$ and $\Xi_{i-1}^{n}$ takes the following form:
\begin{equation}
[\![w]\!](x_{i}, t) = w_{+}(x_{i}, t) - w_{-}(x_{i}, t) = \exp(\nu(t - t^{n}))\left[u^{n}_{+}(x - a(t - t^{n})) - u^{n}_{-}(x - a(t - t^{n}))\right].
\end{equation}
Since the functions $u^{n}_{-} \in \mathcal{P}_{N}(\Omega_{i-1})$ and $u^{n}_{+} \in \mathcal{P}_{N}(\Omega_{i})$, the pre-exponential factors of the solutions can be exactly represented in the following form:\\[-2mm]
\begin{equation}
u^{n}_{\pm}(x - a(t - t^{n})) = \exp(\nu(t - t^{n}))\sum\limits_{s = 0}^{N} \frac{(-1)^{s}}{s!}\left[a(t - t^{n})\right]^{s}\pdern{u^{n}_{\pm}}{x}{s}(x, t^{n}).
\end{equation}
Then the exact expression for the total flux $F_{i}^{\rm ex}$ through the interface $x_{i}$ over the entire time interval $\mathcal{T}^{n}$ is determined by the following expression:\\[-2mm]
\begin{equation}\label{eq:aux_probl_w_flux_ex}
\begin{split}
F_{i}^{\rm ex} & = \int\limits_{t^{n}}^{t^{n+1}} dt\, [\![a w]\!](x_{i}, t) 
                 = a \int\limits_{t^{n}}^{t^{n+1}} dt\, [\![w]\!](x_{i}, t)
                 = a \int\limits_{0}^{\Dtn{n}} d\zeta\, [\![w]\!](x_{i}, t^{n} + \zeta)\\
               & = a \sum\limits_{s = 0}^{N} \frac{(-1)^{s} a^{s}}{s!}\left(\int\limits_{0}^{\Dtn{n}} d\zeta\, \zeta^{s}\exp(\nu\zeta)\right)
                   \left[\!\!\left[\pdern{u^{n}_{\pm}}{x}{s}\right]\!\!\right](x, t^{n})\\
               & = \sum\limits_{s = 0}^{N} \frac{(-1)^{s} a^{s+1}}{s!}\left\{\pdernoper{\nu}{s}\left[\frac{\exp(\nu\Dtn{n}) - 1}{\nu}\right]\right\}
                   \left[\!\!\left[\pdern{u^{n}_{\pm}}{x}{s}\right]\!\!\right](x, t^{n})\\
               & = \sum\limits_{s = 0}^{N} \frac{(-1)^{s} }{s!}\left(a\Dtn{n}\right)^{s+1}
                   \left\{\frac{d^{s}}{d(\nu\Dtn{n})^{s}}\left[\frac{\exp(\nu\Dtn{n}) - 1}{\nu\Dtn{n}}\right]\right\}
                   \left[\!\!\left[\pdern{u^{n}_{\pm}}{x}{s}\right]\!\!\right](x, t^{n}).
\end{split}
\end{equation}
Next, to construct an approximate numerical total flux $F_{i}^{\rm num}$ through the interface $x_{i}$ over the entire time interval $\mathcal{T}^{n}$, the following approximate solution $\tilde{w}^{n}_{\pm}$ is used to solve the auxiliary problems (\ref{eq:aux_probl_w_eqs}):
\begin{equation}
\begin{split}
\tilde{w}^{n}_{\pm}(x - a(t - t^{n})) = 
  \sum\limits_{s = 0}^{N} (-1)^{s}\frac{s+1}{s!}
  \left\{\frac{d^{s}}{d(\nu\Dtn{n})^{s}}\left[\frac{R_{\rm RK}(\nu\Dtn{n}) - 1}{\nu\Dtn{n}}\right]\right\}
  \left[a(t - t^{n})\right]^{s}\pdern{u^{n}_{\pm}}{x}{s}(x, t^{n}),
\end{split}
\end{equation}
where the factor $s+1$ is needed for further reduction during integration to obtain the numerical total flux $F_{i}^{\rm num}$, and an exponential integrator approximation is used in the following form derived earlier (\ref{eq:exp_pade_approx}):
\begin{equation}
\frac{\exp(\nu\Dtn{n}) - 1}{\nu\Dtn{n}} = \frac{R_{\rm RK}(\nu\Dtn{n}) - 1}{\nu\Dtn{n}} + O\left((\nu\Dtn{n})^{2N+1}\right),
\end{equation}
which guarantees an order of approximation $2N+1$ significantly higher than the order of approximation $N+1$ (\ref{eq:errs_ests}) provided by the DG representation of the numerical solution $u^{n}_{\pm}$ (\ref{eq:u_n_def}). The expression for the magnitude of the discontinuity of the approximate solution $\tilde{w}^{n}_{\pm}$ then takes the following form:
\begin{equation}
\begin{split}
[\![\tilde{w}]\!](x_{i}, t) & = \tilde{w}_{+}(x_{i}, t) - \tilde{w}_{-}(x_{i}, t) \\ & =
  \sum\limits_{s = 0}^{N} (-1)^{s}\frac{s+1}{s!}
  \left\{\frac{d^{s}}{d(\nu\Dtn{n})^{s}}\left[\frac{R_{\rm RK}(\nu\Dtn{n}) - 1}{\nu\Dtn{n}}\right]\right\}
  \left[a(t - t^{n})\right]^{s}\left[\!\!\left[\pdern{u^{n}_{\pm}}{x}{s}\right]\!\!\right](x, t^{n}),
\end{split}
\end{equation}
which results in the following expression for the approximate numerical total flux $F_{i}^{\rm num}$ through the interface $x_{i}$ over the entire time interval $\mathcal{T}^{n}$:\\[-2mm]
\begin{equation}\label{eq:aux_probl_w_flux_num}
\begin{split}
F_{i}^{\rm num} & = \int\limits_{t^{n}}^{t^{n+1}} dt\, [\![a \tilde{w}]\!](x_{i}, t) 
                  = a \int\limits_{t^{n}}^{t^{n+1}} dt\, [\![\tilde{w}]\!](x_{i}, t)
                  = a \int\limits_{0}^{\Dtn{n}} d\zeta\, [\![\tilde{w}]\!](x_{i}, t^{n} + \zeta)\\
                & = a \sum\limits_{s = 0}^{N} (-1)^{s}\frac{s+1}{s!} a^{s}
                    \left\{\frac{d^{s}}{d(\nu\Dtn{n})^{s}}\left[\frac{R_{\rm RK}(\nu\Dtn{n}) - 1}{\nu\Dtn{n}}\right]\right\}
                    \left(\int\limits_{0}^{\Dtn{n}} d\zeta\, \zeta^{s}\right) \left[\!\!\left[\pdern{u^{n}_{\pm}}{x}{s}\right]\!\!\right](x, t^{n})\\
                & = \sum\limits_{s = 0}^{N} (-1)^{s}\frac{s+1}{s!}
                    \left\{\frac{d^{s}}{d(\nu\Dtn{n})^{s}}\left[\frac{R_{\rm RK}(\nu\Dtn{n}) - 1}{\nu\Dtn{n}}\right]\right\}
                    \frac{(a\Dtn{n})^{s+1}}{s+1} \left[\!\!\left[\pdern{u^{n}_{\pm}}{x}{s}\right]\!\!\right](x, t^{n})\\
                & = \sum\limits_{s = 0}^{N} \frac{(-1)^{s}}{s!}\left(a\Dtn{n}\right)^{s+1}
                    \left\{\frac{d^{s}}{d(\nu\Dtn{n})^{s}}\left[\frac{R_{\rm RK}(\nu\Dtn{n}) - 1}{\nu\Dtn{n}}\right]\right\}
                    \left[\!\!\left[\pdern{u^{n}_{\pm}}{x}{s}\right]\!\!\right](x, t^{n}),
\end{split}
\end{equation}
which, up to a normalization dimensional factor $(\Dx)^{-1}$, coincides with the term (\ref{eq:ader_dg_corr_total_flux_a_plus}) in the expression (\ref{eq:ader_dg_corr}), while providing a quantitative estimate of the error in the total numerical flux:
\begin{equation}
F_{i}^{\rm num} = F_{i}^{\rm ex} + O\left((\nu\Dtn{n})^{N+1}\right).
\end{equation}
By comparing expressions (\ref{eq:aux_probl_w_flux_ex}) and (\ref{eq:aux_probl_w_flux_num}), and taking into account the previous error estimate for $F_{i}^{\rm num}$, the following relation holds for the second term in expression (\ref{eq:ader_dg_corr}):
\begin{equation}\label{eq:ader_dg_corr_flux_errs_a_plus}
\begin{split}
&\sum\limits_{s = 0}^{N} \frac{(-1)^{s}}{s!}\left\{\frac{d^{s}}{d(\nu\Dtn{n})^{s}}\left[\frac{R_{\rm RK}(\nu\Dtn{n}) - 1}{\nu\Dtn{n}}\right]\right\}
\left(a\Dtn{n}\right)^{s+1}\\ & \cdot\left[\!\!\left[\pdern{u^{n}}{x}{s}\right]\!\!\right](x_{i}, t_{n}) =
\frac{1}{\Dx}\int\limits_{t^{n}}^{t^{n+1}} dt\, [\![a u]\!](x_{i}, t) + O\left((\nu\Dtn{n})^{N+1}\right),
\end{split}
\end{equation}
where the constraint $\Dtn{n} \propto \Dx$ is used. Secondly, the semantics and quantitative estimates of the third term on the right-hand side of the resulting expression (\ref{eq:ader_dg_corr}), which makes a nonzero contribution in the case $a < 0$, are studied. The internal sum of this term is rewritten in the following form, explicitly taking into account the discontinuity of the numerical solution at the interface $\{x_{i+1}\}\times\mathcal{T}^{n}$ of the space-time discretization domains $\Xi_{i+1}^{n}$ and $\Xi_{i}^{n}$:
\begin{equation}
\begin{split}
&\sum\limits_{s = 0}^{N} \frac{(-1)^{s}}{s!}\left\{\frac{d^{s}}{d(\nu\Dtn{n})^{s}}\left[\frac{R_{\rm RK}(\nu\Dtn{n}) - 1}{\nu\Dtn{n}}\right]\right\}
 \left(\frac{a\Dtn{n}}{\Dx}\right)^{s+1}\left(\frac{d^{s}u^{+,n}}{d\xi^{s}}(0) - \frac{d^{s}u^{n}}{d\xi^{s}}(1)\right)\\ & =
 \frac{1}{\Dx}\sum\limits_{s = 0}^{N} \frac{(-1)^{s}}{s!}\left\{\frac{d^{s}}{d(\nu\Dtn{n})^{s}}\left[\frac{R_{\rm RK}(\nu\Dtn{n}) - 1}{\nu\Dtn{n}}\right]\right\}
 \left(a\Dtn{n}\right)^{s+1}\left(\pdern{u^{+,n}}{x}{s}(x_{i+1}^{+}) - \pdern{u^{n}}{x}{s}(x_{i+1}^{-})\right)\\ & =
 \frac{1}{\Dx}\sum\limits_{s = 0}^{N} \frac{(-1)^{s}}{s!}\left\{\frac{d^{s}}{d(\nu\Dtn{n})^{s}}\left[\frac{R_{\rm RK}(\nu\Dtn{n}) - 1}{\nu\Dtn{n}}\right]\right\}
 \left(a\Dtn{n}\right)^{s+1}\left[\!\!\left[\pdern{u^{n}}{x}{s}\right]\!\!\right](x_{i+1}, t_{n}).
\end{split}
\end{equation}
Calculations and transformations completely analogous to expressions (\ref{eq:aux_probl_w_flux_ex}) and (\ref{eq:aux_probl_w_flux_num}) lead to the following expression for the third term in expression (\ref{eq:ader_dg_corr})
\begin{equation}\label{eq:ader_dg_corr_flux_errs_a_minus}
\begin{split}
&\sum\limits_{s = 0}^{N} \frac{(-1)^{s}}{s!}\left\{\frac{d^{s}}{d(\nu\Dtn{n})^{s}}\left[\frac{R_{\rm RK}(\nu\Dtn{n}) - 1}{\nu\Dtn{n}}\right]\right\}
\left(a\Dtn{n}\right)^{s+1}\\ & \cdot\left[\!\!\left[\pdern{u^{n}}{x}{s}\right]\!\!\right](x_{i+1}, t_{n}) =
\frac{1}{\Dx}\int\limits_{t^{n}}^{t^{n+1}} dt\, [\![a u]\!](x_{i+1}, t) + O\left((\nu\Dtn{n})^{N+1}\right).
\end{split}
\end{equation}
where the constraint $\Dtn{n} \propto \Dx$ is used. To estimate the total approximation error of the numerical solution given by expression (\ref{eq:ader_dg_corr}), it is necessary to a Taylor series expansion of the increment $u(x, t^{n+1}) - u(x, t^{n})$ of the exact solution $u(x, t)$ at a new time step $t^{n}$. This is done using expressions for the time derivatives of the exact solution $u(x, t)$, expressed in terms of the coordinate derivatives of the original linear advection-reaction equation (\ref{eq:adv_eq_src}):
\begin{equation}
\pderf{u}{t} = \left(\nu - a\pderfoper{x}\right) u,\quad
\pdern{u}{t}{s} = \left(\nu - a\pderfoper{x}\right)^{s} u,
\end{equation}
which immediately yields the following Taylor series expansion:
\begin{equation}
\begin{split}
u(x, t^{n+1}) = \sum\limits_{j = 0}^{\infty} \frac{(\Dtn{n})^{j}}{j!} \pdern{u}{t}{j}(x, t^{n})
              = \sum\limits_{j = 0}^{\infty} \frac{(\Dtn{n})^{j}}{j!} \left(\nu - a\pderfoper{x}\right)^{j} u(x, t^{n})
              \equiv \exp\left(\nu - a\pderfoper{x}\right) u(x, t^{n}),
\end{split}
\end{equation}
where the final expression represents the solution expressed in terms of the evolution operator of the original equation (\ref{eq:adv_eq_src}). Using this expression, the Taylor series expansion of the increment $u(x, t^{n+1}) - u(x, t^{n})$ of the exact solution $u(x, t)$ at a new time step $t^{n}$ is written as follows:
\begin{equation}
\begin{split}
u(x, t^{n+1}) - u(x, t^{n}) & = 
\sum\limits_{j = 1}^{\infty} \frac{(\Dtn{n})^{j}}{j!} \left(\nu - a\pderfoper{x}\right)^{j} u(x, t^{n})\\ & =
\Dtn{n}\left(\nu - a\pderfoper{x}\right) \sum\limits_{j = 0}^{\infty} \frac{(\Dtn{n})^{j}}{(j+1)!} \left(\nu - a\pderfoper{x}\right)^{j} u(x, t^{n})\\ & =
\Dtn{n}\left\{\frac{\displaystyle\exp\left[\Dtn{n}\left(\nu - a\pderfoper{x}\right)\right] - 1}{\displaystyle\Dtn{n}\left(\nu - a\pderfoper{x}\right)}\right\}
\left(\nu - a\pderfoper{x}\right) u(x, t^{n}),
\end{split}
\end{equation}
where the operator construct in the denominator is used in the sense of an inverse operator. The remaining first term in expression (\ref{eq:adv_eq_src}), which is entirely local in nature, is rewritten as follows:
\begin{equation}
\begin{split}
&\sum\limits_{s = 0}^{N} \frac{(-1)^{s}}{s!}\left(a\Dtn{n}\right)^{s+1}
 \left\{\frac{d^{s}}{d(\nu\Dtn{n})^{s}}\left[\frac{R_{\rm RK}(\nu\Dtn{n}) - 1}{\nu\Dtn{n}}\right]\right\}
 \left\{\frac{\nu}{a}\pdernoper{x}{s}u^{n}(x) - \pdernoper{x}{s+1}u^{n}(x)\right\}\\ & =
 \sum\limits_{s = 0}^{N} \frac{(-1)^{s}}{s!}\left(a\Dtn{n}\right)^{s+1}
 \left\{\frac{d^{s}}{d(\nu\Dtn{n})^{s}}\left[\frac{R_{\rm RK}(\nu\Dtn{n}) - 1}{\nu\Dtn{n}}\right]\right\}
 \frac{1}{a}\pdernoper{x}{s}\left\{\nu - \pderfoper{x}\right\} u^{n}(x)\\ & =
  \sum\limits_{s = 0}^{N} \frac{(-1)^{s}}{s!}\left(a\Dtn{n}\right)^{s+1}
  \left\{\frac{d^{s}}{d(\nu\Dtn{n})^{s}}\left[\frac{R_{\rm RK}(\nu\Dtn{n}) - 1}{\nu\Dtn{n}}\right]\right\}
  \frac{1}{a}\pdernoper{x}{s}\left\{\nu - \pderfoper{x}\right\} u(x, t^{n}) + O\left((\nu\Dtn{n})^{N+1}\right),
\end{split}
\end{equation}
where the constraint $\Dtn{n} \propto \Dx$ and the estimates (\ref{eq:errs_ests}) are used. The expression allows us to calculate the following difference between the increment $u(x, t^{n+1}) - u(x, t^{n})$ of the exact solution $u(x, t)$ and the local increment of the numerical solution $u^{n}(x)$:
\begin{equation}
\begin{split}
u(x, t^{n+1}) - u(x, t^{n}) & -
\sum\limits_{s = 0}^{N} \frac{(-1)^{s}}{s!}\left(a\Dtn{n}\right)^{s+1}
\left\{\frac{d^{s}}{d(\nu\Dtn{n})^{s}}\left[\frac{R_{\rm RK}(\nu\Dtn{n}) - 1}{\nu\Dtn{n}}\right]\right\}
\frac{1}{a}\pdernoper{x}{s}\left\{\nu - \pderfoper{x}\right\} u^{n}(x)\\ & =
\Dtn{n}\left[\frac{\displaystyle\exp\left[\Dtn{n}\left(\nu - a\pderfoper{x}\right)\right] - 1}{\displaystyle\Dtn{n}\left(\nu - a\pderfoper{x}\right)} -
\sum\limits_{s = 0}^{N} \frac{(-1)^{s}}{s!}\left(a\Dtn{n}\pderfoper{x}\right)^{s}\right.\\ & \left.\cdot
\vphantom{\frac{\displaystyle\exp\left[\Dtn{n}\left(\nu - a\pderfoper{x}\right)\right] - 1}{\displaystyle\Dtn{n}\left(\nu - a\pderfoper{x}\right)}}
\left\{\frac{d^{s}}{d(\nu\Dtn{n})^{s}}\left[\frac{R_{\rm RK}(\nu\Dtn{n}) - 1}{\nu\Dtn{n}}\right]\right\}\right]
\left(\nu - a\pderfoper{x}\right) u(x, t^{n}) + O\left((\nu\Dtn{n})^{N+1}\right),
\end{split}
\end{equation}
in which the main construction for estimating the approximation error is enclosed in large square brackets:
\begin{equation}
\begin{split}
&\frac{\displaystyle\exp\left[\Dtn{n}\left(\nu - a\pderfoper{x}\right)\right] - 1}{\displaystyle\Dtn{n}\left(\nu - a\pderfoper{x}\right)} -
 \sum\limits_{s = 0}^{N} \frac{(-1)^{s}}{s!}\left(a\Dtn{n}\pderfoper{x}\right)^{s}
 \left\{\frac{d^{s}}{d(\nu\Dtn{n})^{s}}\left[\frac{R_{\rm RK}(\nu\Dtn{n}) - 1}{\nu\Dtn{n}}\right]\right\}\\ & =
 \left[
  \frac{\displaystyle\exp\left[\Dtn{n}\left(\nu - a\pderfoper{x}\right)\right] - 1}{\displaystyle\Dtn{n}\left(\nu - a\pderfoper{x}\right)} -
  \sum\limits_{s = 0}^{N} \frac{(-1)^{s}}{s!}\left(a\Dtn{n}\pderfoper{x}\right)^{s}
  \left\{\frac{d^{s}}{d(\nu\Dtn{n})^{s}}\left[\frac{\exp(\nu\Dtn{n}) - 1}{\nu\Dtn{n}}\right]\right\}
 \right]\\ & + 
 \left[
  \vphantom{\frac{\displaystyle\exp\left[\Dtn{n}\left(\nu - a\pderfoper{x}\right)\right] - 1}{\displaystyle\Dtn{n}\left(\nu - a\pderfoper{x}\right)}}
  \sum\limits_{s = 0}^{N} \frac{(-1)^{s}}{s!}\left(a\Dtn{n}\pderfoper{x}\right)^{s}
  \left\{\frac{d^{s}}{d(\nu\Dtn{n})^{s}}\left[\frac{\exp(\nu\Dtn{n}) - 1}{\nu\Dtn{n}} - \frac{R_{\rm RK}(\nu\Dtn{n}) - 1}{\nu\Dtn{n}}\right]\right\}
 \right].
\end{split}
\end{equation}
This operator construction, when acting on a sufficiently smooth function $u(\,\cdot\,, t^{n})\in\mathcal{C}_{N}(\Omega_{i})\subset\mathcal{C}_{\infty}(\Omega_{i})$, has a quantitative estimate $O((\Dtn{n})^{N+1})$: the first term has a quantitative estimate $O((\Dtn{n})^{N+1})$, and the second term has a quantitative estimate $O((\Dtn{n})^{2N+1})$. As a result, the first term in expression (\ref{eq:ader_dg_corr}), which is entirely local in nature, has the following quantitative estimate:
\begin{equation}
\begin{split}
u(x, t^{n+1}) - u(x, t^{n}) & -
\sum\limits_{s = 0}^{N} \frac{(-1)^{s}}{s!}\left(a\Dtn{n}\right)^{s+1}
\left\{\frac{d^{s}}{d(\nu\Dtn{n})^{s}}\left[\frac{R_{\rm RK}(\nu\Dtn{n}) - 1}{\nu\Dtn{n}}\right]\right\}\\ & \cdot
\frac{1}{a}\pdernoper{x}{s}\left\{\nu - \pderfoper{x}\right\} u(x, t^{n}) = O\left((\nu\Dtn{n})^{N+1}\right),
\end{split}
\end{equation}
which, when combined with the estimates (\ref{eq:ader_dg_corr_flux_errs_a_plus}) for the terms at the interface $\{x_{i}\}\times\mathcal{T}^{n}$ of the space-time discretization domains $\Xi_{i}^{n}$ and $\Xi_{i-1}^{n}$ in the case $a > 0$ and (\ref{eq:ader_dg_corr_flux_errs_a_plus}) for the terms at the interface $\{x_{i+1}\}\times\mathcal{T}^{n}$ of the space-time discretization domains $\Xi_{i+1}^{n}$ and $\Xi_{i}^{n}$  in the case $a < 0$, yields the final expression for estimating the error of the ADER-DG numerical method with LST-DG predictor:
\begin{equation}\label{eq:errs_ests_final}
\begin{split}
&u^{n+1}(\xi) - u^{n}(\xi) =
  u(x(\xi), t^{n+1}) - u(x(\xi), t^{n}) + O\left((\nu\Dtn{n})^{N+1}\right)\\ & - \Bigg[
  \theta( a) \frac{\delta_{\mathcal{P}_{N}}(\xi)}{\Dx}\int\limits_{t_{n}}^{t_{n+1}}dt\,[\![a u]\!](x_{i}, t) -
  \theta(-a) \frac{\delta_{\mathcal{P}_{N}}(\xi-1)}{\Dx}\int\limits_{t_{n}}^{t_{n+1}}dt\,[\![a u]\!](x_{i+1}, t) + O\left(\Dx^{N+1}\right)\Bigg]\\ & =
  u(x(\xi), t^{n+1}) - u(x(\xi), t^{n}) + O\left(\Dx^{N+1}\right),
\end{split}
\end{equation}
The expression (\ref{eq:errs_ests_final}) shows that the approximation order of the ADER-DG numerical method with LST-DG predictor is $p = N+1$.

\section{Stability analysis}
\label{sec:stab_anal}

This Section is devoted to the stability analysis of the ADER-DG numerical method with the LST-DG predictor. The Subsection~\ref{sec:stab_anal:qual_cons} ``Qualitative considerations'' presents a qualitative analysis of the ADER-DG numerical method's stability and obtains an asymptotic estimate of the boundary value of Courant number $\mathrm{CFL}_{\rm max}^{\rm est}(z)$ for high degree $N \gg 1$ of basis polynomials. The Subsection~\ref{sec:stab_anal:statment_quant_cons} ``Statement of the problem of quantitative considerations'' presents a detailed derivation of the relations necessary for a rigorous stability analysis of the ADER-DG numerical method with the LST-DG predictor within the framework of the Von Neumann criterion (spectral stability criterion). The Subsection~\ref{sec:stab_anal:empir_cons} ``Empirical considerations'' presents a comprehensive set of empirical results from analyzing the spectrum of the evolution operator matrix for a single time step. Based on this analysis, several assumptions are made and substantiated regarding the localization of the eigenvalue boundary upon reaching which the stability of the ADER-DG numerical method with the LST-DG predictor is lost. The Subsection~\ref{sec:stab_anal:quant_cons} ``Rigorous quantitative considerations'', based on the obtained empirical results, significantly simplifies the general stability problem of the ADER-DG numerical method with the LST-DG predictor. This results in a set of polynomials whose roots determine the boundary value of Courant number $\mathrm{CFL}_{\rm max}(N)$.

\subsection{Qualitative considerations}
\label{sec:stab_anal:qual_cons}

Before developing a quantitative stability analysis for the ADER-DG numerical method with the LST-DG predictor, it is necessary to clarify the main expected qualitative stability properties of the numerical method to generally understand the expected features of the stability properties. The rigorous formulation of the ADER-DG numerical method with the LST-DG predictor contains an inverse matrix in expression (\ref{eq:lstdg_pred_eq_sol_temp_1}) for the local discrete space-time solution $q(\tau, \xi)$ (\ref{eq:local_sol_def}), which enters the right-hand side of the numerical solution increment (\ref{eq:ader_dg_corr}) linearly and uniformly. Explicitly calculating this inverse matrix leads to matrix factors $(I - \nu\Dtn{n}\cdot\chi)^{-s}$ (\ref{eq:local_sol_coeffs}), which can be interpreted in terms of IMEX numerical methods. To establish an analogy, the following auxiliary problem for a linear advection-reaction equation is considered:
\begin{equation}\label{eq:qcs:adv_eq}
\pderf{w}{t} + a\pderf{w}{x} = \nu w,\quad
(x, t)\in\Omega\times\mathcal{T},\quad
\Omega\subset\mathcal{R},\quad
\mathcal{T} = \left[t_{0}, t_{f}\right]\subset\mathcal{R},\quad
\nu \leqslant 0,
\end{equation}
which is identical to the original problem (\ref{eq:adv_eq_src}) for a linear advection-reaction equation, except for the notation to avoid confusion. The simplest classical finite-difference IMEX scheme for the problem (\ref{eq:qcs:adv_eq}) can be formulated as follows:
\begin{equation}\label{eq:qcs:imex_fd_method}
\frac{w^{n+1}_{l} - w^{n}_{l}}{\Dtn{n}} + a\Bigg[\theta( a)\frac{w^{n}_{l} - w^{n}_{l-1}}{\Dx} + \theta(-a)\frac{w^{n}_{l+1} - w^{n}_{l}}{\Dx}\Bigg] = \nu\Dtn{n}w^{n+1}_{l},
\end{equation}
where the flux term on the left-hand side is discretized by an explicit upwind finite-difference scheme, and the source term on the right-hand side is discretized by an implicit scheme. This yields a typical classical IMEX scheme, which well demonstrates some of the basic stability properties of this class of numerical methods. A trivial analysis of the Von Neumann criterion of stability (spectral criterion) for the numerical method (\ref{eq:qcs:imex_fd_method}) leads to the following results:
\begin{equation}\label{eq:qcs:imex_fd_stab_cond}
|\mathrm{CFL}| \leqslant \mathrm{CFL}_{\rm max}^{\rm IMEX}(z),\quad
\mathrm{CFL} = \frac{a\Dtn{n}}{\Dx},\quad
\mathrm{CFL}_{\rm max}^{\rm IMEX}(z) = 1 + \frac{z}{2},\quad
z = |\nu|\Dtn{n} \geqslant 0,
\end{equation}
from which it follows that with increasing stiffness of the problem, characterized by $|\nu|$ or $z$, the stability region of the numerical method expands. This allows the use of IMEX numerical methods in the case of essentially stiff systems of equations, where the characteristic time scales $\sim 1/|\nu|$ of the reaction processes are significantly smaller than the characteristic time scales $\sim \Dx/a$ of the advection (of course, this does not guarantee any high-precision resolution of the reaction processes, but only stable calculations). In this case, the following two known asymptotics of the dependency $\mathrm{CFL}_{\rm max}^{\rm IMEX}(z)$ hold:
\begin{equation}\label{eq:qcs:imex_fd_stab_asymp}
\mathrm{CFL}_{\rm max}^{\rm IMEX}(0) = 1,\qquad
\mathrm{CFL}_{\rm max}^{\rm IMEX}(z \gg 1) \approx \frac{z}{2}\cdot\mathrm{CFL}_{\rm max}^{\rm IMEX}(0),
\end{equation}
the first of which corresponds to the case of a linear advection equation (formally corresponding to case $\nu = 0$), and the second of which corresponds to the case of an anomalously strong stiffness of the equation. The analogy between the ADER-DG numerical method with the LST-DG predictor and the IMEX numerical method becomes clear when comparing expressions (\ref{eq:local_sol_coeffs}) and (\ref{eq:ader_dg_corr}), taking into account the semantics of the inverse matrix $(I - \nu\Dtn{n}\cdot\chi)^{-1}$, with expression (\ref{eq:qcs:imex_fd_method}), which represents the explicit form of locally implicit numerical scheme (\ref{eq:qcs:imex_fd_method}):
\begin{equation}\label{eq:qcs:imex_fd_method_expl}
w^{n+1}_{l} =
\frac{1}{1 - \nu\Dtn{n}}\Bigg\{w^{n}_{l} - \frac{a\Dtn{n}}{\Dx}\Big[\theta( a)(w^{n}_{l} - w^{n}_{l-1}) + \theta(-a)(w^{n}_{l+1} - w^{n}_{l})\Big]\Bigg\}.
\end{equation}
Conceptually, the denominator $(1 - \nu\Dtn{n})$ of this expression for the finite-difference IMEX numerical method is equivalent, from the standpoint of the basic stability semantics, to the inverse matrix in expressions (\ref{eq:local_sol_coeffs}) and (\ref{eq:ader_dg_corr}). From the expression (\ref{eq:qcs:imex_fd_method_expl}), substituting the $k$-harmonic $w^{n}_{l} = \exp(ikx_{l}) = \exp(ik\Dx\cdot l)$ and extracting the amplification factor $R(z, \theta)$ (or the evolution operator for this $k$-harmonic) as $w^{n+1}_{l} = R(z, \theta)w^{n}_{l}$, where $\theta = k\Dx$, yields the following expression:
\begin{equation}\label{eq:imex_spectr_compr}
R(z, \theta) = \frac{1 - |\mathrm{CFL}|(1 - \exp(i\theta))}{1 - z}.
\end{equation}
This shows that with increasing parameter $|z|$ in $z < 0$, the spectrum of the evolution operator $R(z, \theta)$ of the finite-difference IMEX numerical method ``compresses''. Qualitatively similar behavior is expected from the ADER-DG numerical method with the LST-DG predictor. In~\cite{ader_dg_stab_rig} it was qualitatively established that the linear advection equation (formally corresponding to case $\nu = 0$) satisfies the following approximate asymptotic relationship for the boundary value of Courant number $\mathrm{CFL}_{\rm max}^{\rm est}$:
\begin{equation}
\begin{split}
\mathrm{CFL}_{\rm max}^{\rm est}(N, 0) = \frac{a}{a^{\rm eff}(N)} \propto \frac{1}{(N+1)^{2}},
\end{split}
\end{equation}
where parameter $a^{\rm eff}(N)$ determines the maximum effective ``advection velocity'' of the harmonics of the numerical solution:
\begin{equation}
\begin{split}\label{eq:a_eff}
a^{\rm eff}(N) = a \cdot \max\limits_{k}\left[\left|\frac{\ophi_{k}}{w_{k}}\right|\right]
               = a \cdot \max\limits_{k}\left[\left|\frac{\uphi_{k}}{w_{k}}\right|\right]
                \propto (N+1)^{2},
\end{split}
\end{equation}
and quantitative estimates were obtained based on an analysis of the asymptotic behavior of the roots and derivatives of the shifted Legendre polynomials $\tilde{P}$ in the case $N \gg 1$. In this case, it is expected that for the advection-reaction equation studied in this work, approximately the following asymptotic behavior will hold:
\begin{equation}\label{eq:cfls_asymp_est}
\begin{split}
&\mathrm{CFL}_{\rm max}^{\rm est}(N, z) \approx \mathrm{CFL}_{\rm max}^{\rm est}(0) \propto \frac{1}{(N+1)^{2}},\hspace{1.4cm} |z| \ll 1,\ z < 0,\\
&\mathrm{CFL}_{\rm max}^{\rm est}(N, z) \approx \frac{z}{2}\cdot\mathrm{CFL}_{\rm max}^{\rm est}(0) \propto \frac{z}{2(N+1)^{2}},\qquad |z| \gg 1,\ z < 0,
\end{split}
\end{equation}
the first of which corresponds to the case of a linear advection equation (formally corresponding to case $\nu = 0$ in (\ref{eq:adv_eq_src})), and the second of which corresponds to the case of anomalously strong stiffness of the equation (\ref{eq:adv_eq_src}). It should be noted immediately that this estimate $a^{\rm eff}$ overestimates (by $21$--$47$\%) the original exact expression (\ref{eq:a_eff}), which is, in general, only an estimate and should not be expected to provide even satisfactory quantitative correspondence to the exact values $\mathrm{CFL}_{\rm max}^{\rm est}$. However, the resulting expression (\ref{eq:cfls_asymp_est}) correctly expresses the asymptotic behavior for large polynomial degrees $N \gg 1$, which may be important for preliminary estimates of computational complexity and computational costs of the ADER-DG numerical method with LST-DG predictor.

\subsection{Statement of the problem of quantitative considerations}
\label{sec:stab_anal:statment_quant_cons}

A rigorous quantitative analysis of stability of the ADER-DG numerical method with LST-DG predictor in this work is carried out on the basis of the use of the Von Neumann criterion  of stability (spectral criterion), which has proven itself well for studying the stability of linear numerical methods for linear equations, and is based on the use of the previously obtained expression (\ref{eq:ader_dg_corr}), which is rewritten in the following form:
\begin{equation}\label{eq:evol_oper_prop}
u(x, t^{n}) = \exp(ikx),\qquad u(x, t^{n+1}) = R\exp(ikx) = R \cdot u(x, t^{n}),
\end{equation}
where the following notations are introduced:
\begin{equation}\label{eq:evol_oper_def}
R = R(\mathrm{CFL}, z, \theta),\quad \mathrm{CFL} = \frac{a\Dtn{n}}{\Dx},\quad z = |\nu|\Dtn{n},\quad \theta = k\Dx,
\end{equation}
where $R(\mathrm{CFL}, z, \theta)$ is the evolution operator (or the transition operator) for one time step $\Dtn{n}$, and the dependence on $a\Dtn{n}/\Dx$, $z$ and $k\Dx$ obviously follows from considerations of similarity and the dimensionality of the problem: $\mathrm{CFL}$ and $\theta$ are conveniently chosen three unique dimensionless parameters of the problem. It is important to note that in the case of a linear advection-reaction equation (\ref{eq:adv_eq_src}), the dependence has a much more complex structure, in contrast to the case of a linear advection equation~\cite{ader_dg_stab_rig} (which is formally an advection-reaction equation (\ref{eq:adv_eq_src}) in case $\nu = 0$) --- an explicit dependence on parameter $z$ arises. The use of the Von Neumann criterion of stability (spectral criterion) presupposes consideration of the evolution over one time step $\Dtn{n}$ of one characteristic harmonic of the numerical solution, which is chosen in the form of a plane wave $\exp(ikx)$. Representing the solution (\ref{eq:u_n_def}) as an expansion in basis polynomials $\{\varphi_{k}\}_{k}$ is a linear operation, therefore, the transition to the solution $\hat{u}^{\pm}(\xi)$ in neighboring cells $\Omega_{i \pm 1}\equiv\Omega_{\pm}$ in the form of multiplication by the ``shift exponential factor'' $\exp(\pm ik\Dx)$ has a similar form for individual expansion coefficients $\hat{u}_{k}$ of the numerical solution (\ref{eq:u_n_def}). In this case, the solution in neighboring coordinate cells $\hat{u}^{\pm}(\xi)$ is represented in the following form:
\begin{equation}\label{eq:r_matrix_def}
\hat{u}^{\pm, n}_{k} = \exp(\pm ik\Dx)\cdot\hat{u}^{n}_{k},\quad
\hat{u}^{n+1}_{k} = \sum\limits_{l = 0}^{N} R_{kl} \hat{u}^{n}_{l},\quad
\mathrm{R}(\mathrm{CFL}, z, \theta) = \{R_{kl}(\mathrm{CFL}, z, \theta)\}_{k,l},
\end{equation}
and the evolution operator $R(\mathrm{CFL}, z, \theta)$ is conveniently represented in the form of a matrix $\{R_{kl}(\mathrm{CFL}, z, \theta)\}_{k,l}$ defining a linear operator (the ADER-DG numerical method ((\ref{eq:ader_dg_corr_temp_1}) and (\ref{eq:errs_ests_final})) and the original problem (\ref{eq:adv_eq_src}) are linear) in the finite-dimensional representation space of the numerical solution (\ref{eq:u_n_def}). The local discrete space-time solution $q(\tau, \xi)$ (\ref{eq:local_sol_def}) obtained as a result of solving the system of algebraic equations of the LST-DG predictor (\ref{eq:lstdg_pred_eq_weak}) (see the expression (\ref{eq:lstdg_pred_eq_sol_temp_1_ind_form})) also depends linearly on the numerical solution $\hat{u}(\xi)$ (\ref{eq:u_n_def}), therefore it also has the following expression for the expansion coefficients $\{\hat{q}_{p_{0}p_{1}}\}_{p_{0}p_{1}}$ of the solution in neighboring coordinate cells $\Omega_{\pm}$:
\begin{equation}\label{eq:q_exp_shift}
\hat{q}^{\pm}_{p_{0}p_{1}} = \exp(\pm ik\Dx)\hat{q}_{p_{0}p_{1}}.
\end{equation}
Using the expression (\ref{eq:ader_dg_corr_temp_1}) for the ADER-DG numerical method and the exact Riemann solver (\ref{eq:f_rp}) for the problem (\ref{eq:adv_eq_src}) leads to the following expression for the coefficients of the numerical solution at the new time step:
\begin{equation}
\begin{split}
\hat{u}^{n+1}_{k} & =
  \hat{u}^{n}_{k} + \frac{\mathrm{CFL}}{w_{k}}\Bigg\{\intrefdom{\tau}\intrefdom{\xi}\,\varphi_{k}'(\xi) q(\tau, \xi) -
  \intrefdom{\tau}\,\Bigg[\ophi_{k} \Big(\theta(a)q(\tau, 1) + \theta(-a)q^{+}(\tau, 0)\Big)\\ & -
  \uphi_{k} \Big(\theta(a)q^{-}(\tau, 1) + \theta(-a)q(\tau, 0)\Big)\Bigg]\Bigg\} + \frac{z}{w_{k}}\intrefdom{\tau}\intrefdom{\xi}\,\varphi_{k}(\xi) q(\tau, \xi)\\ & =
  \hat{u}^{n}_{k} + \frac{\mathrm{CFL}}{w_{k}}\Bigg\{\intrefdom{\tau}\intrefdom{\xi}\,\varphi_{k}'(\xi) q(\tau, \xi) -
  \intrefdom{\tau}\,\Bigg[\theta(a) \Big(\ophi_{k} - \uphi_{k}\exp(-i\theta)\Big)q(\tau, 1)\\ & +
  \theta(-a)\Big(\ophi_{k}\exp(i\theta) - \uphi_{k}\Big)q(\tau, 0)\Bigg]\Bigg\} + \frac{z}{w_{k}}\intrefdom{\tau}\intrefdom{\xi}\,\varphi_{k}(\xi) q(\tau, \xi),
\end{split}
\end{equation}
where the properties (\ref{eq:phi_def}) and (\ref{eq:gl_rule}) of the chosen functional basis $\{\varphi_{k}\}_{k}$ and the expression (\ref{eq:q_exp_shift}) are used. Substituting the representation of the local discrete space-time solution $q(\tau, \xi)$ in the form (\ref{eq:local_sol_def}) of an expansion in basis functions $\{\varphi_{p_{0}}\varphi_{p_{1}}\}_{p_{0}p_{1}}$ leads to the following expression:
\begin{equation}\label{eq:r_matrix_deriv_temp_1}
\begin{split}
\hat{u}^{n+1}_{k} & = \hat{u}^{n}_{k} + \frac{\mathrm{CFL}}{w_{k}}\Bigg\{
  \sum\limits_{p_{0} = 0}^{N}\sum\limits_{p_{1} = 0}^{N} \intrefdom{\tau}\,\varphi_{p_{0}}(\tau)
  \intrefdom{\xi}\,\varphi_{k}'(\xi)\varphi_{p_{1}}(\xi)\,\hat{q}_{p_{0}p_{1}}\\ & - \sum\limits_{p_{0} = 0}^{N}\sum\limits_{p_{1} = 0}^{N}
  \intrefdom{\tau}\,\varphi_{p_{0}}(\tau)\Bigg[
  \theta(a) \ophi_{p_{1}}\Big(\ophi_{k} - \uphi_{k}\exp(-i\theta)\Big)\,\hat{q}_{p_{0}p_{1}}\\ & +
  \theta(-a)\uphi_{p_{1}}\Big(\ophi_{k}\exp(i\theta) - \uphi_{k}\Big)\,\hat{q}_{p_{0}p_{1}}\Bigg]
\Bigg\}\\ & + \frac{z}{w_{k}}\sum\limits_{p_{0} = 0}^{N}\sum\limits_{p_{1} = 0}^{N} \intrefdom{\tau}\,\varphi_{p_{0}}(\tau)
\intrefdom{\xi}\,\varphi_{k}(\xi)\varphi_{p_{1}}(\xi)\,\hat{q}_{p_{0}p_{1}}\\ & =
\hat{u}^{n}_{k} + \frac{\mathrm{CFL}}{w_{k}}\sum\limits_{p_{0} = 0}^{N}\sum\limits_{p_{1} = 0}^{N} w_{p_{0}} \Bigg\{
  \varrho_{p_{1}k} - \Bigg[\theta(a) \ophi_{p_{1}}\Big(\ophi_{k} - \uphi_{k}\exp(-i\theta)\Big)\\ & +
  \theta(-a)\uphi_{p_{1}}\Big(\ophi_{k}\exp(i\theta) - \uphi_{k}\Big)\Bigg] +
  \frac{z}{\mathrm{CFL}} m_{kp_{1}}
\Bigg\}\hat{q}_{p_{0}p_{1}},
\end{split}
\end{equation}
where the definition (\ref{eq:matrices_def}) of matrix $\varrho$ is used. Substituting expression (\ref{eq:lstdg_pred_eq_sol_temp_1_ind_form}) into (\ref{eq:r_matrix_deriv_temp_1}) yields the following expression:
\begin{equation}\label{eq:r_matrix_deriv_temp_2}
\begin{split}
\hat{u}^{n+1}_{k} & =
\sum\limits_{l = 0}^{N} \left[
  \vphantom{\intrefdom{\tau}}\right.
  \delta_{kl} + \frac{\mathrm{CFL}}{w_{k}}\sum\limits_{p_{0} = 0}^{N}\sum\limits_{p_{1} = 0}^{N} w_{p_{0}} \Bigg\{
    \varrho_{p_{1}k} - \Bigg[
      \theta(a) \ophi_{p_{1}}\Big(\ophi_{k} - \uphi_{k}\exp(-i\theta)\Big)\\ & +
      \theta(-a)\uphi_{p_{1}}\Big(\ophi_{k}\exp(i\theta) - \uphi_{k}\Big)
    \Bigg] + \frac{z}{\mathrm{CFL}} m_{kp_{1}}
  \Bigg\}\sigma_{p_{0}p_{1}, l}
  \left.\vphantom{\intrefdom{\tau}}
\right] \hat{u}^{n}_{l}\\ & =
\sum\limits_{l = 0}^{N} \left[
  \vphantom{\intrefdom{\tau}}\right.
  \delta_{kl} + 
    \frac{\mathrm{CFL}}{w_{k}}\sum\limits_{p_{0} = 0}^{N}\sum\limits_{p_{1} = 0}^{N} w_{p_{0}} \Bigg\{
      \varrho_{p_{1}k} - \Bigg[
        \theta(a) \ophi_{p_{1}}\Big(\ophi_{k} - \uphi_{k}\exp(-i\theta)\Big)\\ & +
        \theta(-a)\uphi_{p_{1}}\Big(\ophi_{k}\exp(i\theta) - \uphi_{k}\Big)
      \Bigg] + \frac{z}{\mathrm{CFL}} m_{kp_{1}}
    \Bigg\}
  \Bigg[
    \sum\limits_{s = 0}^{N} \frac{(-1)^{s}}{s!} \mathrm{CFL}^{s}\\ & \cdot
    \left\{
      \sum\limits_{i = 0}^{N}\left(\left[(I - z\chi)^{-(s+1)}\right]_{p_{0}i}
      \tau_{i}^{s}\right)
    \right\}
    \left[\mathrm{D}^{s}\right]_{p_{1}l}
  \Bigg]
  \left.\vphantom{\intrefdom{\tau}}
\right] \hat{u}^{n}_{l}.
\end{split}
\end{equation}
The last expression (\ref{eq:r_matrix_deriv_temp_2}) allowed us to explicitly obtain an expression for the elements of matrix $\mathrm{R}$ (\ref{eq:r_matrix_def}) of the evolution operator $R(\mathrm{CFL}, \theta)$:
\begin{equation}\label{eq:r_matrix_deriv_temp_2_add}
\begin{split}
R_{kl}(\mathrm{CFL}, \theta) & = \delta_{kl} + 
\frac{\mathrm{CFL}}{w_{k}}\sum\limits_{p_{0} = 0}^{N}\sum\limits_{p_{1} = 0}^{N} w_{p_{0}} \Bigg\{
  \varrho_{p_{1}k} - \Bigg[
    \theta(a) \ophi_{p_{1}}\Big(\ophi_{k} - \uphi_{k}\exp(-i\theta)\Big)\\ & +
    \theta(-a)\uphi_{p_{1}}\Big(\ophi_{k}\exp(i\theta) - \uphi_{k}\Big)
  \Bigg] + \frac{z}{\mathrm{CFL}} m_{kp_{1}}
\Bigg\}
\Bigg[
\sum\limits_{s = 0}^{N} \frac{(-1)^{s}}{s!} \mathrm{CFL}^{s}\\ & \cdot
\left\{
  \sum\limits_{i = 0}^{N}\left(\left[(I - z\chi)^{-(s+1)}\right]_{p_{0}i}
  \tau_{i}^{s}\right)
\right\}
\left[\mathrm{D}^{s}\right]_{p_{1}l}
\Bigg]\\ & = \delta_{kl} +
\frac{\mathrm{CFL}}{w_{k}}\sum\limits_{p_{0} = 0}^{N}\sum\limits_{p_{1} = 0}^{N} w_{p_{0}} \Bigg\{
  w_{p_{1}}D_{p_{1}k} - \Bigg[
    \theta(a) \ophi_{p_{1}}\Big(\ophi_{k} - \uphi_{k}\exp(-i\theta)\Big)\\ & +
    \theta(-a)\uphi_{p_{1}}\Big(\ophi_{k}\exp(i\theta) - \uphi_{k}\Big)
  \Bigg] + \frac{z}{\mathrm{CFL}} w_{k}\delta_{kp_{1}}
\Bigg\} \Bigg[
  \sum\limits_{s = 0}^{N} \frac{(-1)^{s}}{s!} \mathrm{CFL}^{s}\\ & \cdot
  \left\{
    \sum\limits_{i = 0}^{N}\left(\left[(I - z\chi)^{-(s+1)}\right]_{p_{0}i}
    \tau_{i}^{s}\right)
  \right\}
  \left[\mathrm{D}^{s}\right]_{p_{1}l}
\Bigg],
\end{split}
\end{equation}
The last expression (\ref{eq:r_matrix_deriv_temp_2}) allowed us to explicitly obtain an expression for the elements of matrix $\mathrm{R}$ (\ref{eq:r_matrix_def}) of the evolution operator $R(\mathrm{CFL}, \theta)$:
\begin{equation}\label{eq:r_matrix_elems_expr}
\begin{split}
R_{kl}(\mathrm{CFL}, \theta) & = \delta_{kl} +
\mathrm{CFL}\sum\limits_{p_{1} = 0}^{N} \Bigg[\frac{1}{w_{k}}\Bigg\{
  w_{p_{1}}D_{p_{1}k} - \Bigg[
    \theta(a) \ophi_{p_{1}}\Big(\ophi_{k} - \uphi_{k}\exp(-i\theta)\Big)\\ & +
    \theta(-a)\uphi_{p_{1}}\Big(\ophi_{k}\exp(i\theta) - \uphi_{k}\Big)
  \Bigg]\Bigg\} + \frac{z}{\mathrm{CFL}}\delta_{kp_{1}}
\Bigg] \Bigg[
  \sum\limits_{s = 0}^{N} \frac{(-1)^{s}}{s!} \mathrm{CFL}^{s}\\ & \cdot
  \Bigg[\sum\limits_{p_{0} = 0}^{N} w_{p_{0}} \left\{
    \sum\limits_{i = 0}^{N}\left(\left[(I - z\chi)^{-(s+1)}\right]_{p_{0}i}
    \tau_{i}^{s}\right)
  \right\}\Bigg]
  \left[\mathrm{D}^{s}\right]_{p_{1}l}
\Bigg]\\ & = \delta_{kl} +
\sum\limits_{p_{1} = 0}^{N} \Bigg[\frac{\mathrm{CFL}}{w_{k}}\Bigg\{
  w_{p_{1}}D_{p_{1}k} - \Bigg[
    \theta(a) \ophi_{p_{1}}\Big(\ophi_{k} - \uphi_{k}\exp(-i\theta)\Big)\\ & +
    \theta(-a)\uphi_{p_{1}}\Big(\ophi_{k}\exp(i\theta) - \uphi_{k}\Big)
  \Bigg]\Bigg\} + z\delta_{kp_{1}}
\Bigg]\\ & \cdot\Bigg[
  \sum\limits_{s = 0}^{N} \frac{(-1)^{s}}{s!} \mathrm{CFL}^{s}
  \frac{d^{s}}{dz^{s}}\left[\frac{R_{\rm RK}(z) - 1}{z}\right]
  \left[\mathrm{D}^{s}\right]_{p_{1}l}
\Bigg],
\end{split}
\end{equation}
where expressions (\ref{eq:ader_dg_corr_local_term_temp_1_temp_1}) and (\ref{eq:exp_int_diff_by_stab_func}) are used. The expression (\ref{eq:r_matrix_elems_expr}) is conveniently rewritten in pure matrix form:
\begin{equation}\label{eq:r_matrix_in_matrix_form}
\begin{split}
&\mathrm{R}(\mathrm{CFL}, z, \theta) = I + \mathrm{A}(\mathrm{CFL}, z, \theta)\mathrm{B}(\mathrm{CFL}, z),\\
&\mathrm{A}(\mathrm{CFL}, z, \theta) = \{A_{kl}(\mathrm{CFL}, z, \theta)\}_{k,l},\quad
 \mathrm{B}(\mathrm{CFL}, z) = \{B_{kl}(\mathrm{CFL}, z)\}_{k,l},
\end{split}
\end{equation}
where $I = \{\delta_{kl}\}_{kl}$ is the identity matrix, and the matrices $\mathrm{A}(\theta)$ and $\mathrm{B}(\mathrm{CFL})$ are introduced with following elements:
\begin{equation}\label{eq:a_and_b_matrices_def}
\begin{split}
&A_{kl}(\mathrm{CFL}, z, \theta) = \frac{\mathrm{CFL}}{w_{k}}\Bigg\{
 w_{l}D_{lk} - \Bigg[\theta(a) \ophi_{l}\Big(\ophi_{k} - \uphi_{k}\exp(-i\theta)\Big) +
                     \theta(-a)\uphi_{l}\Big(\ophi_{k}\exp(i\theta) - \uphi_{k}\Big)\Bigg]\Bigg\} +
 z\delta_{kp_{1}},\\
&B_{kl}(\mathrm{CFL}, z) = \sum\limits_{s = 0}^{N} \frac{(-1)^{s}}{s!} \mathrm{CFL}^{s}
 \frac{d^{s}}{dz^{s}}\left[\frac{R_{\rm RK}(z) - 1}{z}\right]\left[\mathrm{D}^{s}\right]_{kl}.
\end{split}
\end{equation}

The expressions (\ref{eq:r_matrix_elems_expr}), (\ref{eq:r_matrix_in_matrix_form}) and (\ref{eq:a_and_b_matrices_def}) obtained above for the matrix elements $R_{kl}$ the matrix $\mathrm{R}$ (\ref{eq:r_matrix_def}) of the evolution operator $R(\mathrm{CFL}, \theta)$ can be obtained by using pure matrix operations, which is of interest from the perspective of developing a high-performance software implementation of the ADER-DG numerical method with the LST-DG predictor using the BLAS interface~\cite{ader_eff_blas}. Below is presented the derivation of the matrix $\mathrm{R}$ of the evolution operator $R(\mathrm{CFL}, \theta)$, for which the expression (\ref{eq:lstdg_pred_eq_sol_temp_1}) for the vector of expansion coefficients of the local discrete space-time solution is used. The following vector and matrix notations are introduced:\\[-2mm]
\begin{equation}
\boldsymbol{\Theta}(\tau, \xi) = \{\Theta_{\mathrm{p}}(\tau, \xi)\}_{\mathrm{p}},\quad
\boldsymbol{\varphi}(\xi) = \{\varphi_{k}(\xi)\}_{k},\quad
\boldsymbol{\Theta}(\tau, \xi) = \boldsymbol{\varphi}(\tau)\otimes\boldsymbol{\varphi}(\xi),
\end{equation}
using which the local discrete space-time solution $q(\tau, \xi)$ (\ref{eq:local_sol_def}) and solutions $u^{n}(\xi)$ and $u^{n+1}(\xi)$ (\ref{eq:u_n_def}) are expressed in convolution forms:\\[-2mm]
\begin{equation}\label{eq:matrix_notation_base_def}
\begin{split}
q(\tau, \xi) = \boldsymbol{\Theta}(\tau, \xi)^{T}\hat{\mathbf{q}} = \hat{\mathbf{q}}^{T}\boldsymbol{\Theta}(\tau, \xi),\quad
u^{n}(\xi) = \boldsymbol{\varphi}(\xi)^{T}\hat{\mathbf{u}}^{n} = [\hat{\mathbf{u}}^{n}]^{T}\boldsymbol{\varphi}(\xi),\quad
u^{n+1}(\xi) = \boldsymbol{\varphi}(\xi)^{T}\hat{\mathbf{u}}^{n+1} = [\hat{\mathbf{u}}^{n+1}]^{T}\boldsymbol{\varphi}(\xi).
\end{split}
\end{equation}
Substituting the introduced matrix form of the solutions $q(\tau, \xi)$ (\ref{eq:local_sol_def}), $u^{n}(\xi)$ and $u^{n+1}(\xi)$ (\ref{eq:u_n_def}) into the expression for the ADER-DG numerical method (\ref{eq:ader_dg_corr_temp_1}) leads to the following expression
\begin{equation}
\begin{split}
\hat{\mathbf{u}}^{n+1} & =
  \hat{\mathbf{u}}^{n} + \mathrm{CFL} \cdot m^{-1}\Bigg\{\intrefdom{\tau}\intrefdom{\xi}\,\boldsymbol{\varphi}'(\xi) q(\tau, \xi)
  - \intrefdom{\tau}\,\Bigg[\theta(a) \Big(\ovecphi - \uvecphi\exp(-i\theta)\Big)q(\tau, 1)\\
  & + \theta(-a)\Big(\ovecphi\exp(i\theta) - \uvecphi\Big)q(\tau, 0)\Bigg]\Bigg\} + z \cdot m^{-1} \intrefdom{\tau}\intrefdom{\xi}\,\boldsymbol{\varphi}(\xi) q(\tau, \xi) \\ & =
  \hat{\mathbf{u}}^{n} + \Bigg[\mathrm{CFL} \cdot m^{-1}\Bigg\{\intrefdom{\tau}\intrefdom{\xi}\,\boldsymbol{\varphi}'(\xi)
  [\boldsymbol{\varphi}(\tau)\otimes\boldsymbol{\varphi}(\xi)]^{T}\\ 
  & - \intrefdom{\tau}\,\Bigg[\theta(a) \Big(\ovecphi - \uvecphi\exp(-i\theta)\Big)[\boldsymbol{\varphi}(\tau)\otimes\ovecphi]^{T}
  + \theta(-a)\Big(\ovecphi\exp(i\theta) - \uvecphi\Big)[\boldsymbol{\varphi}(\tau)\otimes\uvecphi]^{T}\Bigg]\Bigg\}\\
  & + z \cdot m^{-1}\intrefdom{\tau}\intrefdom{\xi}\,\boldsymbol{\varphi}(\xi) [\boldsymbol{\varphi}(\tau)\otimes\boldsymbol{\varphi}(\xi)]^{T}\Bigg]
  \Bigg[\Big(I - z\chi\Big) \otimes I + \mathrm{CFL} \cdot \chi \otimes \mathrm{D}\Bigg]^{-1}\Big(\mathbf{1} \otimes \hat{\mathbf{u}}^{n}\Big),
\end{split}
\end{equation}
where expressions (\ref{eq:matrices_K_def}) are used to represent the transformation matrices in the form of Kronecker products of the main matrices of the functional basis $\{\varphi_{k}\}_{k}$. Explicit transformations were implemented in the resulting expression, leading to the following expression:
\begin{equation}
\begin{split}
\hat{\mathbf{u}}^{n+1} =
 \Bigg(I & + \Bigg[m^{-1}\Bigg\{\intrefdom{\tau}\intrefdom{\xi}\,\boldsymbol{\varphi}'(\xi) [\boldsymbol{\varphi}(\tau)\otimes\boldsymbol{\varphi}(\xi)]^{T} -
 \intrefdom{\tau}\,\Bigg[\theta(a) \Big(\ovecphi - \uvecphi\exp(-i\theta)\Big)[\boldsymbol{\varphi}(\tau)\otimes\ovecphi]^{T}\\ & +
 \theta(-a)\Big(\ovecphi\exp(i\theta) - \uvecphi\Big)[\boldsymbol{\varphi}(\tau)\otimes\uvecphi]^{T}\Bigg]\Bigg\} +
 \frac{z}{\mathrm{CFL}} \cdot m^{-1}\intrefdom{\tau}\intrefdom{\xi}\,\boldsymbol{\varphi}(\xi)
 [\boldsymbol{\varphi}(\tau)\otimes\boldsymbol{\varphi}(\xi)]^{T}\Bigg]\\ & \cdot \Bigg[
 \sum\limits_{s = 0}^{N} \frac{(-1)^{s}}{s!} \mathrm{CFL}^{s+1}
 \left(\Big(I - z\chi\Big)^{-(s+1)}\boldsymbol{\tau}_{s}\right)\otimes\mathrm{D}^{s}
 \Bigg]\Bigg)\hat{\mathbf{u}}^{n},
\end{split}
\end{equation}
where the following property of the Kronecker product is used:
\begin{equation}
(\mathrm{G}\otimes\mathrm{H})(\mathbf{a}\otimes\mathbf{b}) = (\mathrm{G}\mathbf{a})\otimes(\mathrm{H}\mathbf{b}) =
(\mathrm{G}\mathbf{a})\otimes(\mathrm{H}I\mathbf{b}) = ((\mathrm{G}\mathbf{a})\otimes(\mathrm{H}I))(1\otimes\mathbf{b}) =
((\mathrm{G}\mathbf{a})\otimes\mathrm{H})\mathbf{b},
\end{equation}
which holds for any matrices $\mathrm{G}$, $\mathrm{H} \in \mathcal{R}^{n \times n}$ and vectors $\mathbf{a}$, $\mathbf{b} \in \mathcal{R}^{n}$, and the following notation for property (\ref{eq:chi_int_prop}) is introduced:
\begin{equation}
\chi\boldsymbol{\tau}_{s} = \frac{\boldsymbol{\tau}_{s+1}}{s+1},\quad
\chi\mathbf{1} = \boldsymbol{\tau},\quad \boldsymbol{\tau}_{s} = \{\tau_{p}^{s}\}_{p},
\end{equation}
as well as the nilpotency $\mathrm{D}^{N+1} = 0$ of the differentiation matrix $\mathrm{D}$ is used (see also (\ref{eq:local_sol_inv_matrix})). From the last expression obtained and the definition (\ref{eq:r_matrix_def}), the expression for the matrix $\mathrm{R}(\mathrm{CFL}, z, \theta)$ follows:
\begin{equation}\label{eq:r_matrix_deriv_temp_3}
\begin{split}
\mathrm{R}(\mathrm{CFL}, z, \theta) = 
I & + \Bigg[\mathrm{CFL} \cdot m^{-1}\Bigg\{\intrefdom{\tau}\intrefdom{\xi}\,\boldsymbol{\varphi}'(\xi) [\boldsymbol{\varphi}(\tau)\otimes\boldsymbol{\varphi}(\xi)]^{T}\\ & -
\intrefdom{\tau}\,\Bigg[\theta(a) \Big(\ovecphi - \uvecphi\exp(-i\theta)\Big)[\boldsymbol{\varphi}(\tau)\otimes\ovecphi]^{T} +
\theta(-a)\Big(\ovecphi\exp(i\theta) - \uvecphi\Big)[\boldsymbol{\varphi}(\tau)\otimes\uvecphi]^{T}\Bigg]\Bigg\}\\ & +
z \cdot m^{-1}\intrefdom{\tau}\intrefdom{\xi}\,\boldsymbol{\varphi}(\xi)
[\boldsymbol{\varphi}(\tau)\otimes\boldsymbol{\varphi}(\xi)]^{T}\Bigg] \Bigg[
\sum\limits_{s = 0}^{N} \frac{(-1)^{s}}{s!} \mathrm{CFL}^{s}
\left(\Big(I - z\chi\Big)^{-(s+1)}\boldsymbol{\tau}_{s}\right)\otimes\mathrm{D}^{s}
\Bigg].
\end{split}
\end{equation}
The first term in the first brackets of the last expression (\ref{eq:r_matrix_deriv_temp_3}), that corresponds to the classical local term of DG methods, is expressed in the following form
\begin{equation}\label{eq:r_matrix_deriv_temp_4_1}
\begin{split}
\intrefdom{\tau}\intrefdom{\xi}&\,\boldsymbol{\varphi}'(\xi) [\boldsymbol{\varphi}(\tau)\otimes\boldsymbol{\varphi}(\xi)]^{T} =
\intrefdom{\xi}\intrefdom{\xi}\,\boldsymbol{\varphi}'(\xi)\otimes\boldsymbol{\varphi}^{T}(\tau)\otimes\boldsymbol{\varphi}^{T}(\xi)\\ & =
\intrefdom{\xi}\,\boldsymbol{\varphi}'(\xi)\otimes\left[\intrefdom{\tau}\,\boldsymbol{\varphi}^{T}(\tau)\right]\otimes\boldsymbol{\varphi}^{T}(\xi) =
\intrefdom{\xi}\,\boldsymbol{\varphi}'(\xi)\otimes\mathbf{w}^{T}\otimes\boldsymbol{\varphi}^{T}(\xi)\\ & \equiv
\mathbf{w}^{T} \otimes \intrefdom{\xi}\,(\boldsymbol{\varphi}'(\xi) \otimes \boldsymbol{\varphi}^{T}(\xi)) = \mathbf{w}^{T} \otimes \varrho^{T},
\end{split}
\end{equation}
where the properties $\mathbf{a}^{T}\otimes\mathbf{b} = \mathbf{b}\otimes\mathbf{a}^{T}$, $\mathbf{a}(\mathbf{b}\otimes\mathbf{c})^{T} = \mathbf{a}\otimes\mathbf{b}^{T}\otimes\mathbf{c}^{T}$ of the Kronecker product are used. The last terms in the first brackets of the expression (\ref{eq:r_matrix_deriv_temp_3}), that corresponds to the source term, are transformed similarly (\ref{eq:r_matrix_deriv_temp_4_1}):\\[-2mm]
\begin{equation}\label{eq:r_matrix_deriv_temp_4_2}
\begin{split}
\intrefdom{\tau}\intrefdom{\xi}&\,\boldsymbol{\varphi}(\xi) [\boldsymbol{\varphi}(\tau)\otimes\boldsymbol{\varphi}(\xi)]^{T} =
\intrefdom{\xi}\intrefdom{\xi}\,\boldsymbol{\varphi}(\xi)\otimes\boldsymbol{\varphi}^{T}(\tau)\otimes\boldsymbol{\varphi}^{T}(\xi)\\ & =
\intrefdom{\xi}\,\boldsymbol{\varphi}(\xi)\otimes\left[\intrefdom{\tau}\,\boldsymbol{\varphi}^{T}(\tau)\right]\otimes\boldsymbol{\varphi}^{T}(\xi) =
\intrefdom{\xi}\,\boldsymbol{\varphi}(\xi)\otimes\mathbf{w}^{T}\otimes\boldsymbol{\varphi}^{T}(\xi)\\ & \equiv
\mathbf{w}^{T} \otimes \intrefdom{\xi}\,(\boldsymbol{\varphi}(\xi) \otimes \boldsymbol{\varphi}^{T}(\xi)) = \mathbf{w}^{T} \otimes m.
\end{split}
\end{equation}
The second terms in the first brackets of the expression (\ref{eq:r_matrix_deriv_temp_3}), that corresponds to the flux terms, are transformed similarly (\ref{eq:r_matrix_deriv_temp_4_1}) and (\ref{eq:r_matrix_deriv_temp_4_2}):
\begin{equation}\label{eq:r_matrix_deriv_temp_4_3}
\begin{split}
&\intrefdom{\tau}\,\ovecphi\,[\boldsymbol{\varphi}(\tau)\otimes\ovecphi]^{T} \equiv
 \left[\intrefdom{\tau}\,\boldsymbol{\varphi}(\tau)^{T}\right] \otimes \ovecphi \otimes \ovecphi^{T} =
 \mathbf{w}^{T} \otimes \ovecphi \otimes \ovecphi^{T},\\
&\intrefdom{\tau}\,\uvecphi\,[\boldsymbol{\varphi}(\tau)\otimes\ovecphi]^{T} =
 \mathbf{w}^{T} \otimes \uvecphi \otimes \ovecphi^{T},\quad
 \intrefdom{\tau}\,\ovecphi\,[\boldsymbol{\varphi}(\tau)\otimes\uvecphi]^{T} =
 \mathbf{w}^{T} \otimes \ovecphi \otimes \uvecphi^{T},\quad
 \intrefdom{\tau}\,\uvecphi\,[\boldsymbol{\varphi}(\tau)\otimes\uvecphi]^{T} =
 \mathbf{w}^{T} \otimes \uvecphi \otimes \uvecphi^{T}.
\end{split}
\end{equation}
Full expansion of the brackets in the expression (\ref{eq:r_matrix_deriv_temp_3}) required the following transformation:
\begin{equation}\label{eq:r_matrix_deriv_temp_5_1}
\Big(I - z\chi\Big)^{-(s+1)}\boldsymbol{\tau}_{s}
= \Big(I - z\chi\Big)^{-(s+1)}s!\,\chi^{s}\mathbf{1}
= \frac{d^{s}}{dz^{s}}\left[\Big(I - z\chi\Big)^{-1}\mathbf{1}\right],
\end{equation}
which is a reformulation of expression (\ref{eq:ader_dg_corr_local_term_temp_1_temp_1}) in matrix terms, and the following transformation:
\begin{equation}\label{eq:r_matrix_deriv_temp_5_2}
\mathbf{w}^{T}\Big(I - z\chi\Big)^{-1}\mathbf{1} = \frac{R_{\rm RK}(z) - 1}{z},
\end{equation}
which is a reformulation of exponential integrator (\ref{eq:exp_int_by_stab_func}) in matrix terms. Substituting the resulting expressions (\ref{eq:r_matrix_deriv_temp_4_1}), (\ref{eq:r_matrix_deriv_temp_4_2}), (\ref{eq:r_matrix_deriv_temp_4_3}), (\ref{eq:r_matrix_deriv_temp_5_1}) and (\ref{eq:r_matrix_deriv_temp_5_2}) into the expression (\ref{eq:r_matrix_deriv_temp_3}) yielded the following expression:
\begin{equation}\label{eq:r_matrix_elems_expr_dup}
\begin{split}
\mathrm{R}(\mathrm{CFL}, z, \theta) & = I + \Bigg[\mathrm{CFL} \cdot m^{-1}\Bigg(\varrho^{T} - \Big[\theta(a) \Big(\ovecphi - \uvecphi\exp(-i\theta)\Big)\otimes\ovecphi^{T} +
\theta(-a)\Big(\ovecphi\exp(i\theta) - \uvecphi\Big)\otimes\uvecphi^{T}\Big]\Bigg)\\ & \hphantom{= I +\Bigg[} + z I\Bigg]
\Bigg[\sum\limits_{s = 0}^{N} \frac{(-1)^{s}}{s!} \mathrm{CFL}^{s} \frac{d^{s}}{dz^{s}}\left[\mathbf{w}^{T}\Big(I - z\chi\Big)^{-1}\mathbf{1}\right]\mathrm{D}^{s}\Bigg]\\ & =
I + \Bigg[\mathrm{CFL} \cdot m^{-1}\Bigg(\mathrm{D}^{T}m - \Big[\theta(a) \Big(\ovecphi - \uvecphi\exp(-i\theta)\Big)\otimes\ovecphi^{T} +
\theta(-a)\Big(\ovecphi\exp(i\theta) - \uvecphi\Big)\otimes\uvecphi^{T}\Big]\Bigg)\\ & \hphantom{= I +\Bigg[} + z I\Bigg]
\Bigg[\sum\limits_{s = 0}^{N} \frac{(-1)^{s}}{s!} \mathrm{CFL}^{s} \frac{d^{s}}{dz^{s}}\left[\frac{R_{\rm RK}(z) - 1}{z}\right]\mathrm{D}^{s}\Bigg],
\end{split}
\end{equation}
which, taking into account the definition of matrix $\mathrm{R}(\mathrm{CFL}, \theta) = I + \mathrm{A}(\theta)\mathrm{B}(\mathrm{CFL})$ (\ref{eq:r_matrix_in_matrix_form}) in its representation form, allowed us to determine expressions for matrices $\mathrm{A}(\theta)$ and $\mathrm{B}(\mathrm{CFL})$:
\begin{equation}\label{eq:a_and_b_matrices_def_dup}
\begin{split}
&\mathrm{A}(\mathrm{CFL}, z, \theta) = \mathrm{CFL} \cdot m^{-1}\Bigg(\mathrm{D}^{T}m - \Big[\theta(a) \Big(\ovecphi - \uvecphi\exp(-i\theta)\Big)\otimes\ovecphi^{T}
 + \theta(-a)\Big(\ovecphi\exp(i\theta) - \uvecphi\Big)\otimes\uvecphi^{T}\Big]\Bigg) + z I,\\
&\mathrm{B}(\mathrm{CFL}, z) = \sum\limits_{s = 0}^{N} \frac{(-1)^{s}}{s!} \mathrm{CFL}^{s} \frac{d^{s}}{dz^{s}}\left[\frac{R_{\rm RK}(z) - 1}{z}\right]\mathrm{D}^{s},
\end{split}
\end{equation}
which is of course consistent with expression (\ref{eq:a_and_b_matrices_def}).

It should be noted that in the case of initial conditions $u(x, t_{0})$ of the problem (\ref{eq:adv_eq_src}) that are explicitly independent of the coordinate $x$: $u(x, t_{0}) \equiv \mathrm{const}$, the exact solution $u(x, t) \equiv \mathrm{const}\cdot\exp(\nu t)$ will also be independent of the coordinate $x$. In the work~\cite{ader_dg_stab}, this property was noted for the ADER-DG numerical method with LST-DG predictor for the linear advection equation (the case of equation (\ref{eq:adv_eq_src}) at $\nu = 0$), and it was noted that the result is independent of the Courant number $\mathrm{CFL}$ (for a linear uniform problem (\ref{eq:adv_eq_src}) with $\nu = 0$, it is clear that this is reduced to the correct reproduction of a ``unit solution'' $u(\xi) \equiv 1$, sometimes called ``constant-preserving property''). In the work~\cite{ader_dg_stab_rig}, this property of the ADER-DG numerical method with LST-DG predictor was rigorously proven, including its independence from the Courant number $\mathrm{CFL}$. Clearly, in the general case $\nu \neq 0$ of a linear advection-reaction equation (\ref{eq:adv_eq_src}), there is no conserved constant solution $u(x, t)$ --- there is dissipation of the solution, characterized by exponential decay $u(x, t) \propto \exp(\nu t)$. However, in the case of an initial condition $u(x, t_{0})$ of the problem (\ref{eq:adv_eq_src}) that is explicitly independent of the coordinate $x$: $u(x, t_{0}) \equiv \mathrm{const}$, the partial differential equation (\ref{eq:adv_eq_src}) at the initial moment $t_{0}$ of time ``transforms'' into an ordinary differential equation $\dot{u} = \nu u$ with an elementary solution $u(x, t) = \mathrm{const}\cdot\exp(\nu t)$. Therefore, it is interesting to rigorously prove that, in the case of an initial condition $u(x, t_{0}) \equiv \mathrm{const}$ that is explicitly independent of the coordinate $x$ (it is clear that this is reduced case of a ``unit solution'' $u(\xi) \equiv 1$), the ADER-DG numerical method with LST-DG predictor allows one to obtain a numerical solution $u(x, t)$ of an ordinary differential equation $\dot{u} = \nu u$, independent of the Courant number $\mathrm{CFL}$. In this work, this ``decay property'' of the ADER-DG numerical method with LST-DG predictor is rigorously proven from the obtained expression for the matrix $\mathrm{R}(\mathrm{CFL}, z, \theta)$ of the evolution operator $R$ for one time step $\Dtn{n}$. For a constant unit solution $u^{n}(\xi) \equiv 1$, the wave vector $k = 0$ is chosen, as a result of which the phase $\theta = 0$ is chosen. The expansion coefficients $\mathbf{u}^{n}$ of the numerical solution at the previous time step are calculated trivially:
\begin{equation}
u^{n}(\xi) \equiv 1,\ \Rightarrow\ 
\hat{u}_{n}^{k} = \frac{1}{w_{k}}\intrefdom{\xi}\, u^{n}(\xi)\varphi_{k}(\xi) \equiv u^{n}(\xi_{k}) = 1,\ \Rightarrow\ 
\hat{\mathbf{u}}^{n} = \mathbf{1}.
\end{equation}
The expansion coefficients of the numerical solution $\mathbf{u}^{n+1}$ at the new time step are determined by the following expression:
\begin{equation}
\hat{\mathbf{u}}^{n+1} = \mathrm{R}(\mathrm{CFL}, z, 0)\hat{\mathbf{u}}^{n} =
\mathrm{R}(\mathrm{CFL}, z, 0)\mathbf{1} = I\mathbf{1} + \mathrm{A}(\mathrm{CFL}, z, 0)(\mathrm{B}(\mathrm{CFL}, z)\mathbf{1}).
\end{equation}
Then, this ``decay property'' reduces to requirement $\mathrm{A}(\mathrm{CFL}, z, 0)(\mathrm{B}(\mathrm{CFL}, z)\mathbf{1}) = ([\exp](\nu t) - 1)\mathbf{1}$, where $[\exp](\nu t)$ is some approximation of the $\exp(\nu t)$. For matrix $\mathrm{B}(\mathrm{CFL})$, an obvious relation holds
\begin{equation}
\mathrm{B}(\mathrm{CFL}, z)\mathbf{1} =
\sum\limits_{s = 0}^{N} \frac{(-1)^{s}}{s!} \mathrm{CFL}^{s} \frac{d^{s}}{dz^{s}}\left[\frac{R_{\rm RK}(z) - 1}{z}\right]\, (\mathrm{D}^{s}\mathbf{1}) =
\left[\frac{R_{\rm RK}(z) - 1}{z}\right]\mathbf{1},
\end{equation}
since the derivative of the constant is zero: $\mathrm{D}^{s} \mathbf{1} = 0$ if $s > 0$. For the action of matrix $\mathrm{A}(\mathrm{CFL}, z, 0)$ on this vector $\mathrm{B}(\mathrm{CFL}, z)\mathbf{1}$, the following chain of relations is obtained:
\begin{equation}
\begin{split}
&\mathrm{A}(\mathrm{CFL}, z, 0)(\mathrm{B}(\mathrm{CFL}, z)\mathbf{1})\\ & = \Bigg[\frac{R_{\rm RK}(z) - 1}{z}\Bigg]\Bigg[\mathrm{CFL} \cdot m^{-1}\Big[\mathrm{D}^{T}m\mathbf{1} - 
\Big[\ovecphi - \uvecphi\Big]\otimes\Big[\theta(a) \ovecphi^{T} + \theta(-a)\otimes\uvecphi^{T}\Big]\Big] + zI\Bigg]\mathbf{1}\\ & =
\Bigg[\frac{R_{\rm RK}(z) - 1}{z}\Bigg]\Bigg[\mathrm{CFL} \cdot m^{-1}\Big[\mathrm{D}^{T}m\mathbf{1} - \Big[\ovecphi - \uvecphi\Big]\otimes
\Big[\theta(a) (\ovecphi^{T}\mathbf{1}) + \theta(-a)(\uvecphi^{T}\mathbf{1})\Big]\Big] + z\mathbf{1}\Bigg]\\ & =
\Bigg[\frac{R_{\rm RK}(z) - 1}{z}\Bigg]\Bigg[\mathrm{CFL} \cdot m^{-1}\Big[\mathrm{D}^{T}m\mathbf{1} -
\Big[\ovecphi - \uvecphi\Big]\Big[\theta(a) + \theta(-a)\Big]\Big] + z\mathbf{1}\Bigg]\\ & =
\Bigg[\frac{R_{\rm RK}(z) - 1}{z}\Bigg]\Bigg[\mathrm{CFL} \cdot m^{-1}\Big[\mathrm{D}^{T}m\mathbf{1} - \Big[\ovecphi - \uvecphi\Big]\Big] + z\mathbf{1}\Bigg].
\end{split}
\end{equation}
Direct calculation of the first term in the brackets of the last expression obtained is accomplished by the following chain of transformations the $p$-th component of the resulting vector:
\begin{equation}
\begin{split}
\big[\mathrm{D}^{T} m \mathbf{1}\big]_{p} & = \sum\limits_{q = 0}^{N} w_{q} D_{qp} = \sum\limits_{q = 0}^{N} w_{q} \frac{{\varrho}_{qp}}{w_{q}} =
\sum\limits_{q = 0}^{N} {\varrho}_{qp} = \sum\limits_{q = 0}^{N} \intrefdom{\xi}\, \varphi_{q}(\xi) \varphi_{p}'(\xi)\\ & =
\intrefdom{\xi}\, \left[\sum\limits_{q = 0}^{N} \varphi_{q}(\xi)\right] \varphi_{p}'(\xi) =
\intrefdom{\xi}\, \varphi_{p}'(\xi) = \ophi_{p} - \uphi_{p},
\end{split}
\end{equation}
which led to the following expression and the chain of implications
\begin{equation}\label{eq:cont_presiv_prop}
\mathrm{D}^{T} m \mathbf{1} = \ovecphi - \uvecphi,\quad\Rightarrow\quad
\mathrm{A}(\mathrm{CFL}, z, 0)(\mathrm{B}(\mathrm{CFL}, z)\mathbf{1}) = (R_{\rm RK}(z) - 1)\mathbf{1},\quad\Rightarrow\quad
\mathrm{R}(\mathrm{CFL}, 0)\mathbf{1} = R_{\rm RK}(z)\mathbf{1},
\end{equation}
which it turns out that some approximation $[\exp](\nu t)$ of the exponential $\exp(\nu t)$ for the ADER-DG numerical method with LST-DG predictor is the Pad\'{e} approximation $R_{\rm RK}(z)$ (\ref{eq:stab_func_ader_dg}). Based on this, solutions are written for the expansion coefficients $\{\hat{\mathbf{u}}^{n+1}\}$ of the DG representation (\ref{eq:u_n_def}) of the solution at a new time step $t^{n+1}$:
\begin{equation}
\hat{\mathbf{u}}^{n+1} = R_{\rm RK}(\lambda\Dtn{n})\mathbf{1} = \exp(\lambda\Dtn{n})\mathbf{1} + O\left((\Dtn{n})^{2N+2}\right),
\end{equation}
which then yields the numerical solution $u^{n+1}(\xi)$ at a new time step $t^{n+1}$:
\begin{equation}
u^{n+1}(\xi) = \boldsymbol{\varphi}(\xi)^{T}\hat{\mathbf{u}}^{n+1} = R_{\rm RK}(\lambda\Dtn{n})\Big[\boldsymbol{\varphi}(\xi)^{T}\mathbf{1}\Big] 
             = R_{\rm RK}(\lambda\Dtn{n}) = \exp(\lambda\Dtn{n}) + O\left((\Dtn{n})^{2N+2}\right),
\end{equation}
which directly confirms~\cite{ader_dg_ivp_ode, ader_dg_ivp_ode_sinum} the order of approximation of $2N+1$ for the ADER-DG numerical method with LST-DG predictor for solving a system of ordinary differential equations, and also confirms the correctness of the reproduction of the coordinate-independent numerical solution, independent of the Courant number $\mathrm{CFL}$. This results rigorously proves the ``decay property'' of the ADER-DG method with LST-DG predictor.

It is clear that the Von Neumann stability criterion (spectral criterion) consists of the requirement on the Courant number $\mathrm{CFL}$ that the spectral radius $\rho[\mathrm{R}(\mathrm{CFL}, \theta)]$ of the matrix $\mathrm{R}(\mathrm{CFL}, \theta)$ (\ref{eq:r_matrix_in_matrix_form}) of the evolution operator $R$ for one time step $\Dtn{n}$ be no greater than one for any value of phase $\theta$:
\begin{equation}\label{eq:cfl_max_base_def}
\mathrm{CFL}_{\rm max} = \sup\Big\{\mathrm{CFL}\,\Big|\, \mathrm{CFL} \in \mathcal{R}_{+},\, \forall \theta \in [0, 2\pi] \subset \mathcal{R},\,
\rho[\mathrm{R}(\mathrm{CFL}, z, \theta)] \leqslant 1\Big\},
\end{equation}
where the supremum is chosen from considerations of the presence of an upper limit of the time step $\Dtn{n}$ for a fixed coordinate step $\Dx$ ($\mathrm{CFL}_{\rm max} \in \mathcal{R}_{+}$ for explicit methods, which include the ADER-DG numerical method with the LST-DG predictor) and the continuity of the set of values of the time step $\Dtn{n}$.

\subsection{Empirical considerations}
\label{sec:stab_anal:empir_cons}

\begin{figure}[h!]
\centering
\includegraphics[width=0.028125\textwidth]{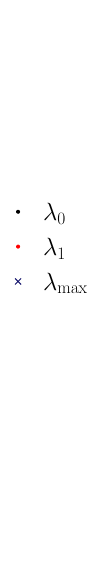}
\includegraphics[width=0.15\textwidth]{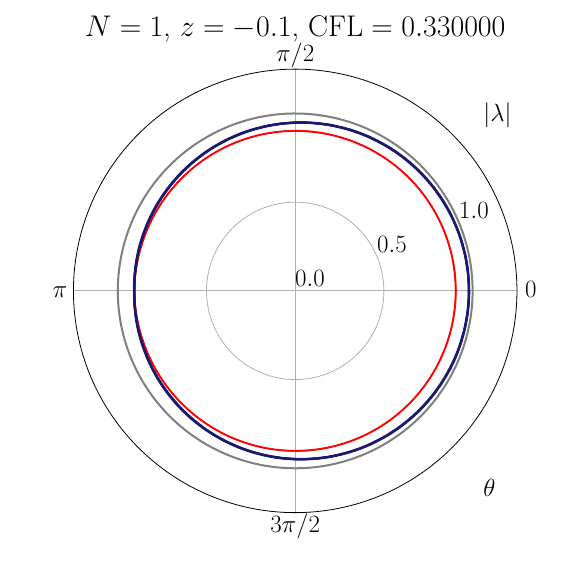}
\includegraphics[width=0.15\textwidth]{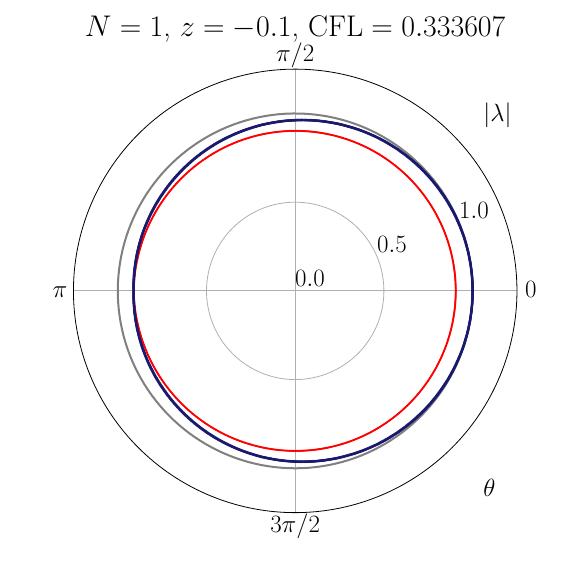}
\includegraphics[width=0.15\textwidth]{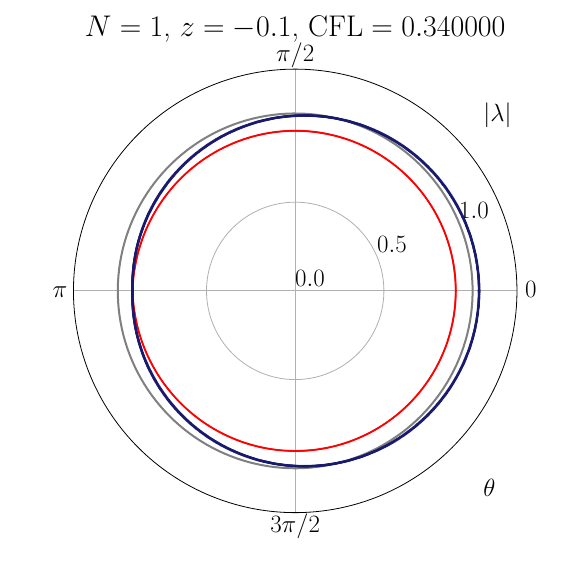}
\includegraphics[width=0.15\textwidth]{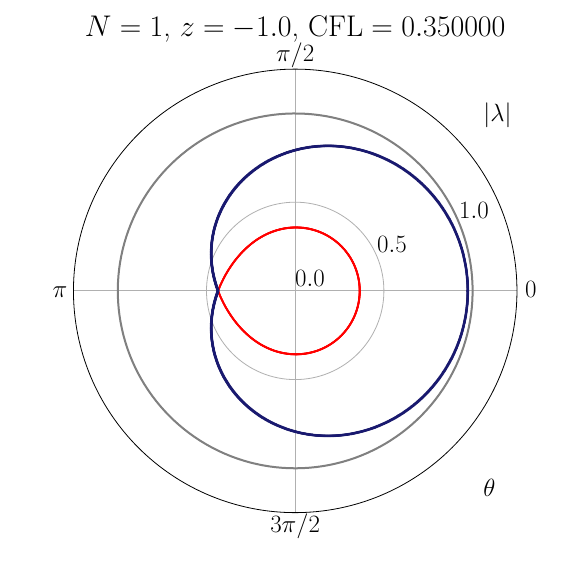}
\includegraphics[width=0.15\textwidth]{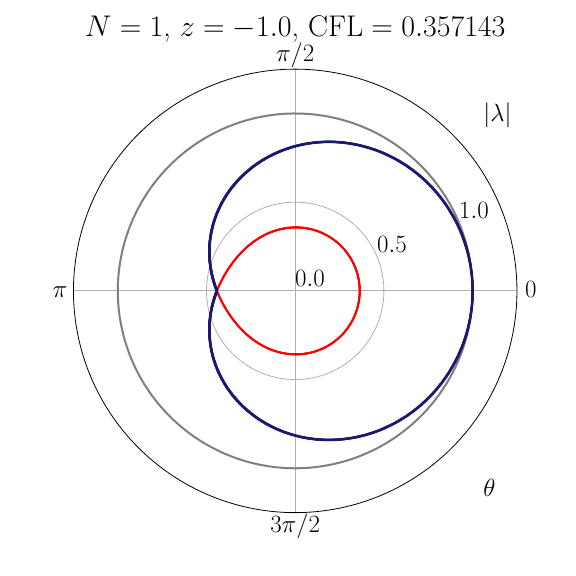}
\includegraphics[width=0.15\textwidth]{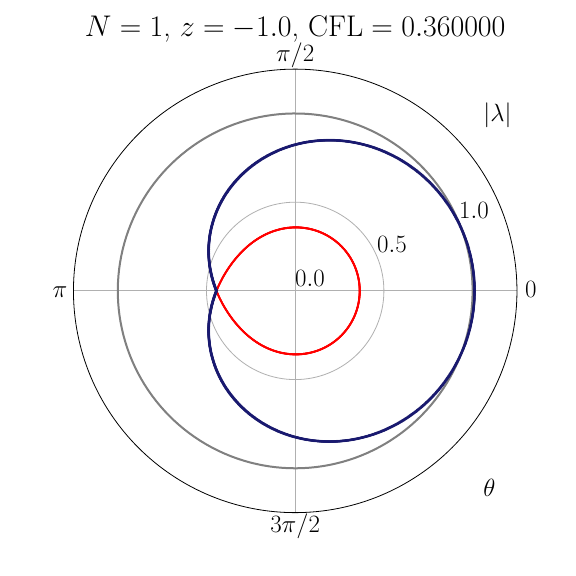}\\
\includegraphics[width=0.028125\textwidth]{rhos_on_theta_legend_N_1.png}
\includegraphics[width=0.15\textwidth]{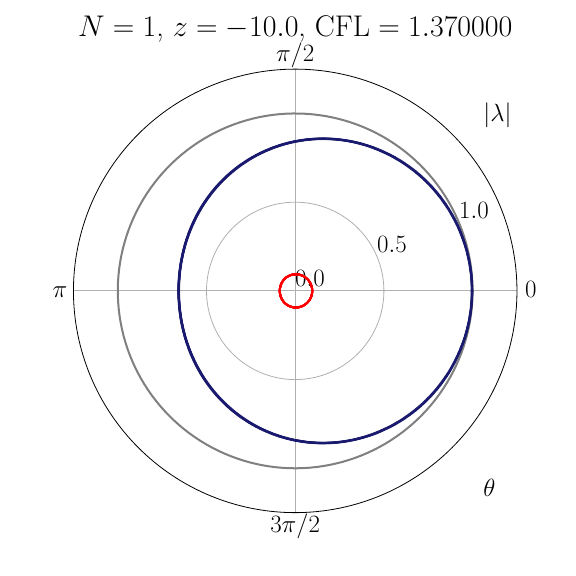}
\includegraphics[width=0.15\textwidth]{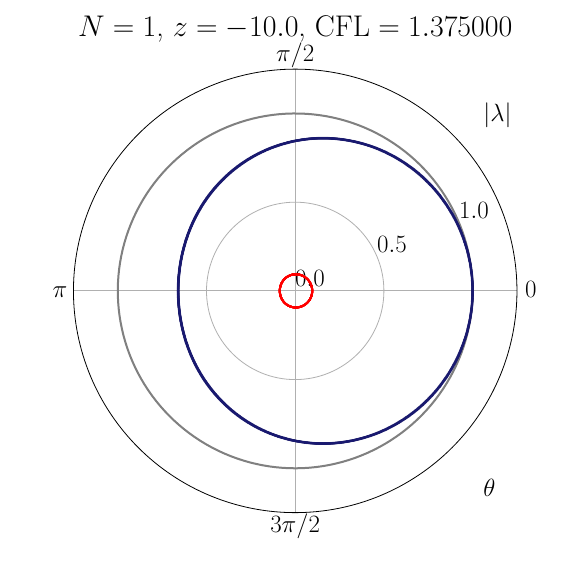}
\includegraphics[width=0.15\textwidth]{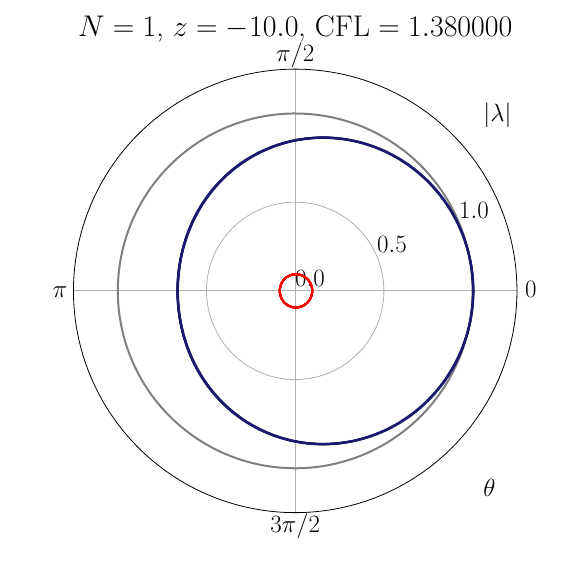}
\includegraphics[width=0.15\textwidth]{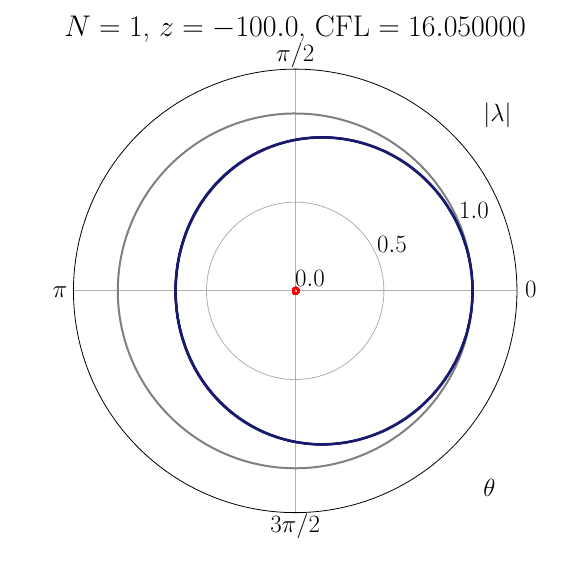}
\includegraphics[width=0.15\textwidth]{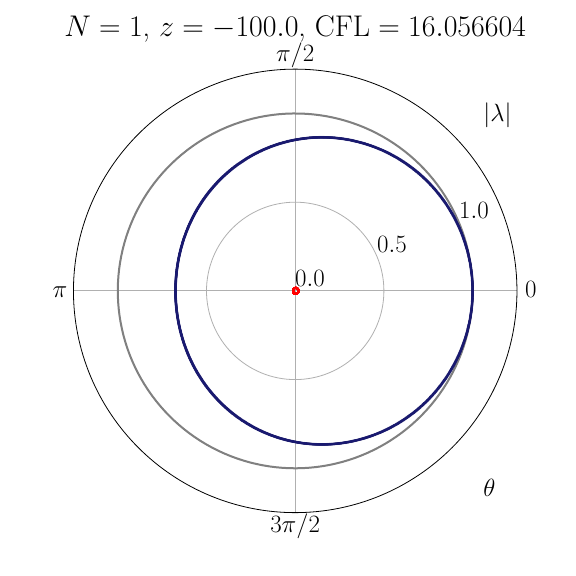}
\includegraphics[width=0.15\textwidth]{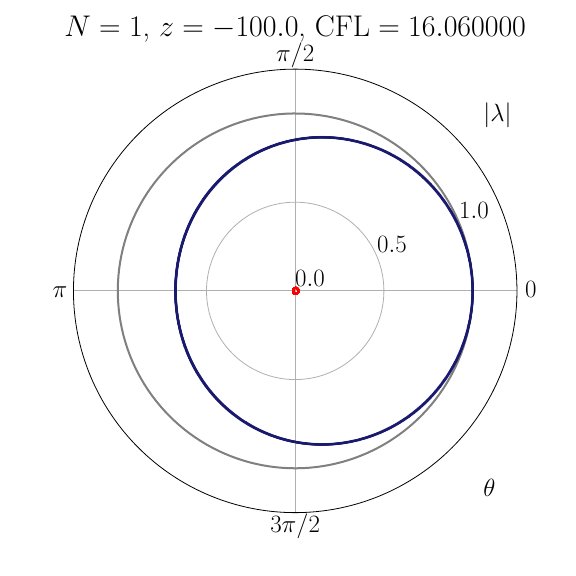}\\
\includegraphics[width=0.028125\textwidth]{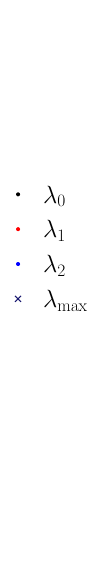}
\includegraphics[width=0.15\textwidth]{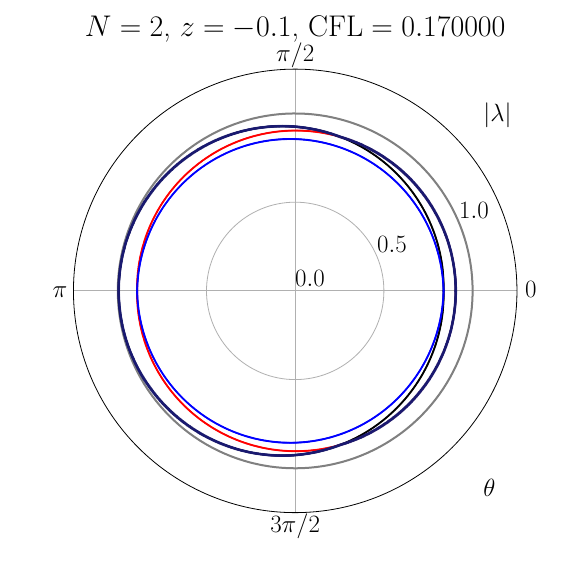}
\includegraphics[width=0.15\textwidth]{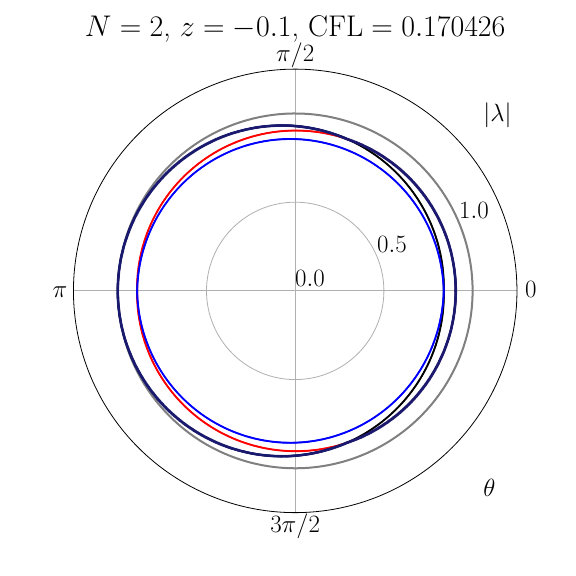}
\includegraphics[width=0.15\textwidth]{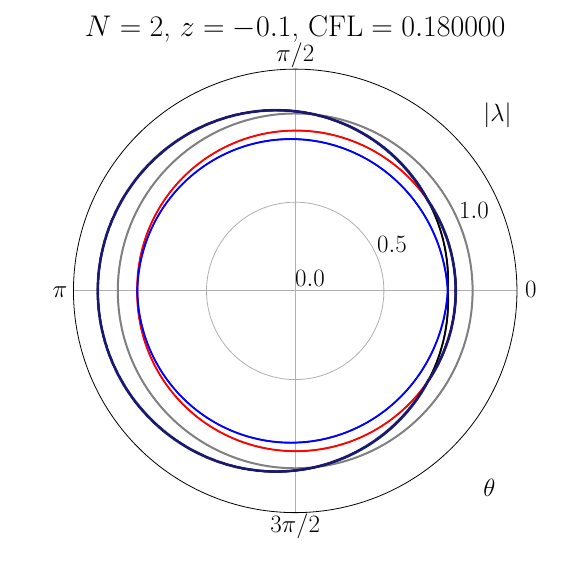}
\includegraphics[width=0.15\textwidth]{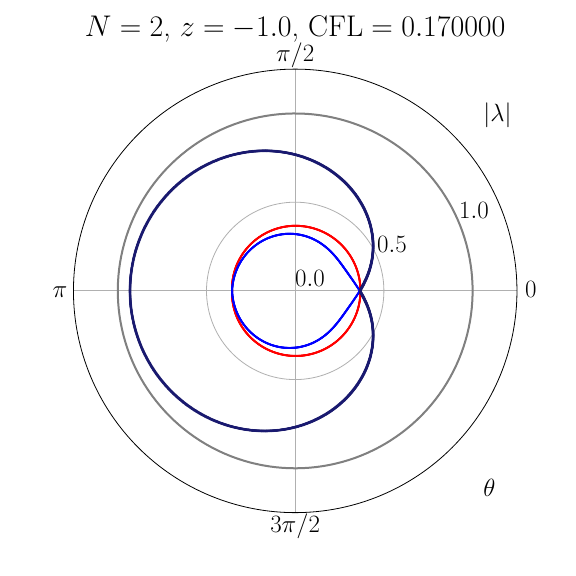}
\includegraphics[width=0.15\textwidth]{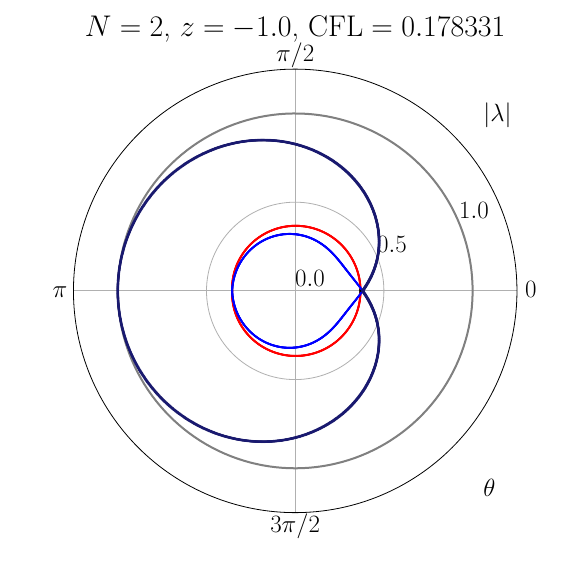}
\includegraphics[width=0.15\textwidth]{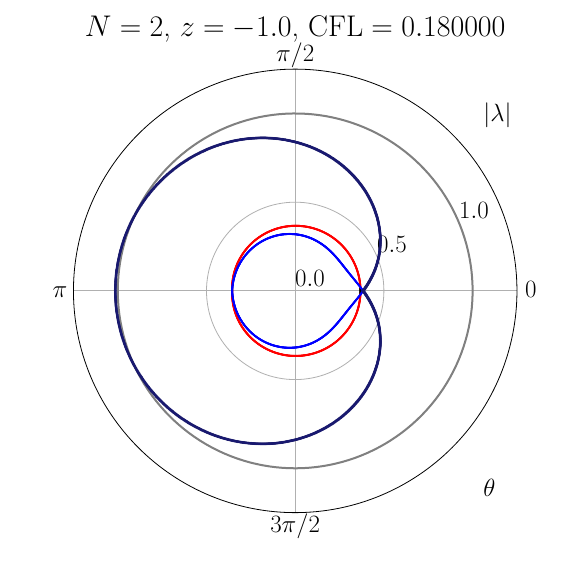}\\
\includegraphics[width=0.028125\textwidth]{rhos_on_theta_legend_N_2.png}
\includegraphics[width=0.15\textwidth]{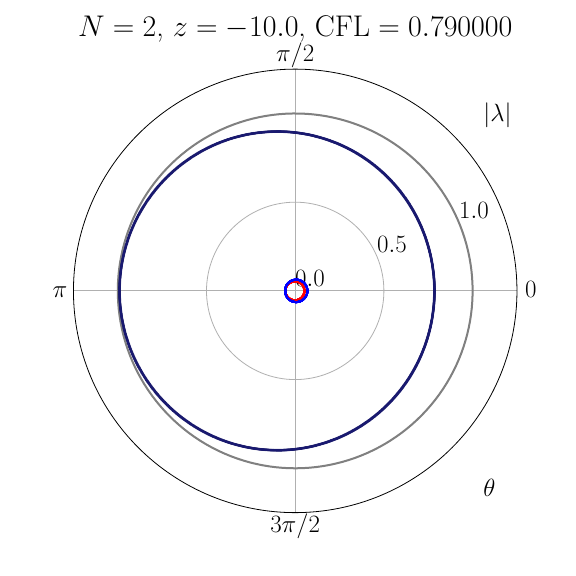}
\includegraphics[width=0.15\textwidth]{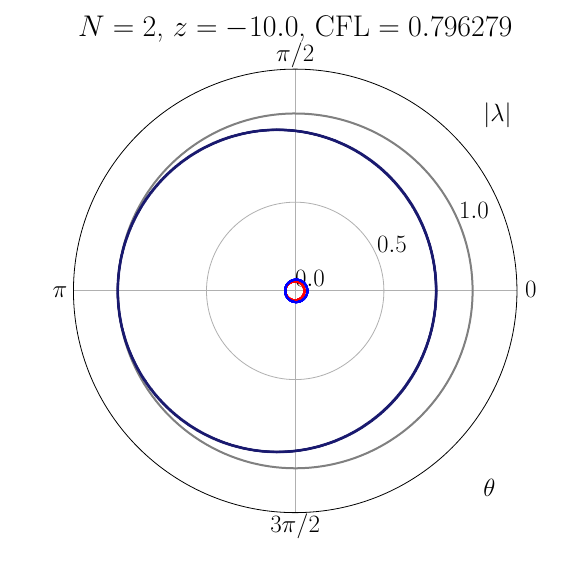}
\includegraphics[width=0.15\textwidth]{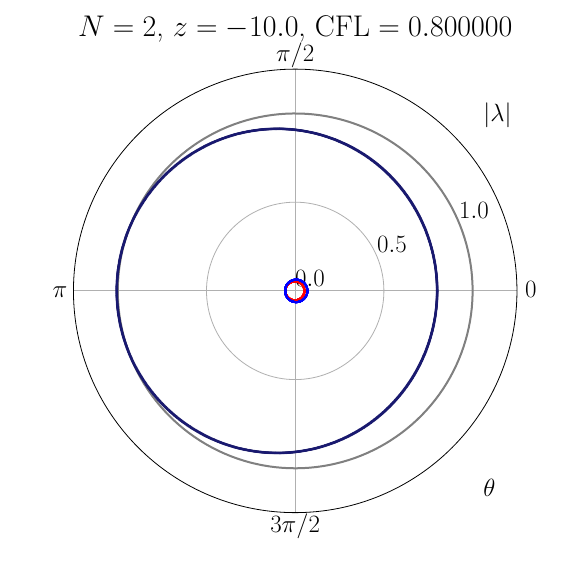}
\includegraphics[width=0.15\textwidth]{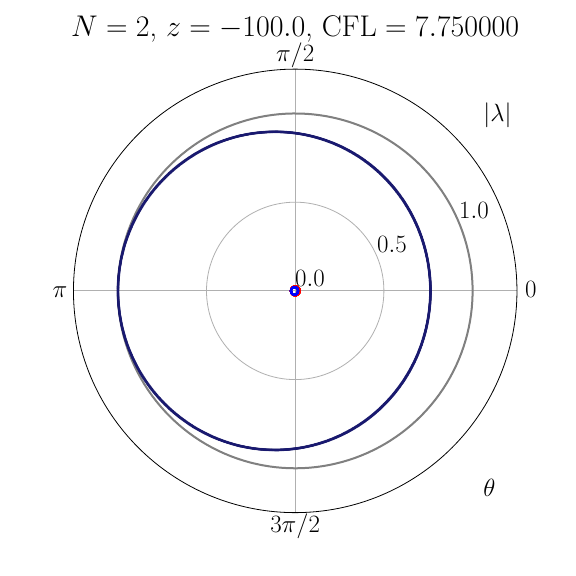}
\includegraphics[width=0.15\textwidth]{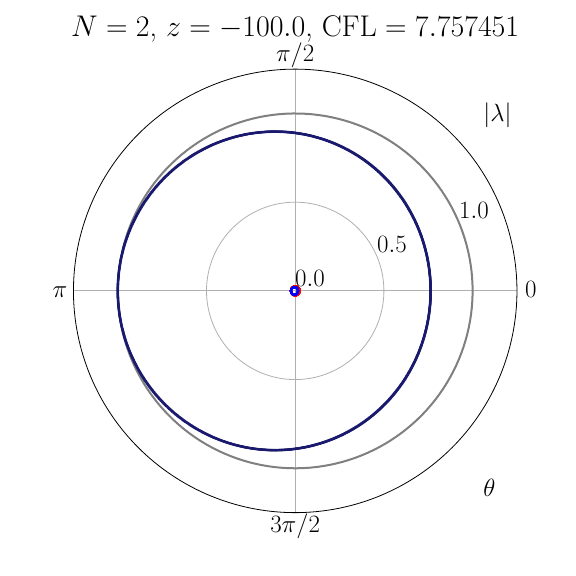}
\includegraphics[width=0.15\textwidth]{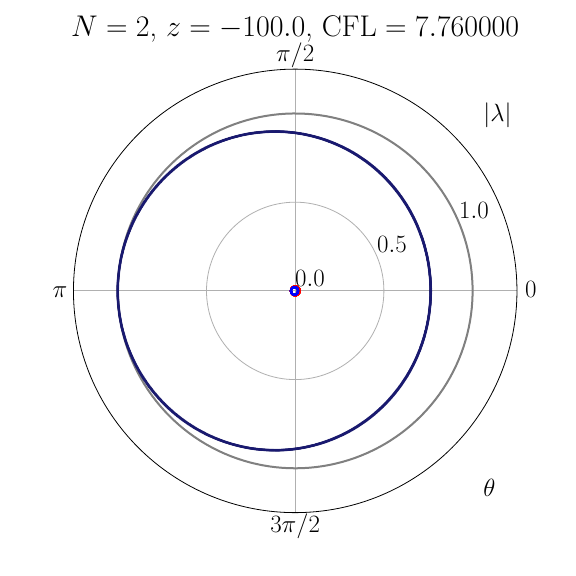}\\
\includegraphics[width=0.028125\textwidth]{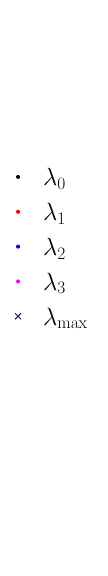}
\includegraphics[width=0.15\textwidth]{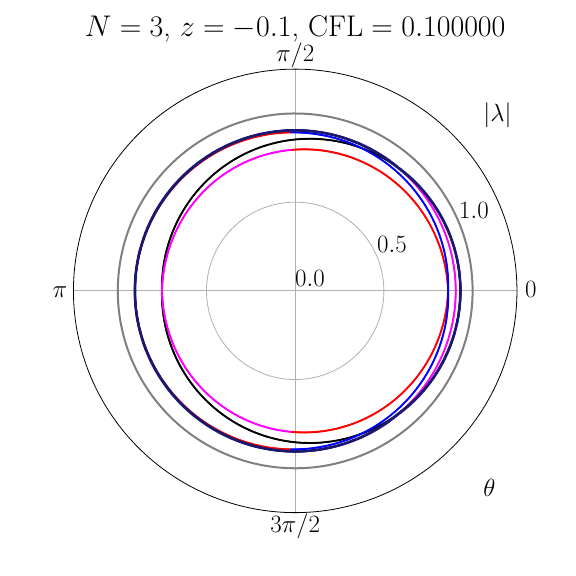}
\includegraphics[width=0.15\textwidth]{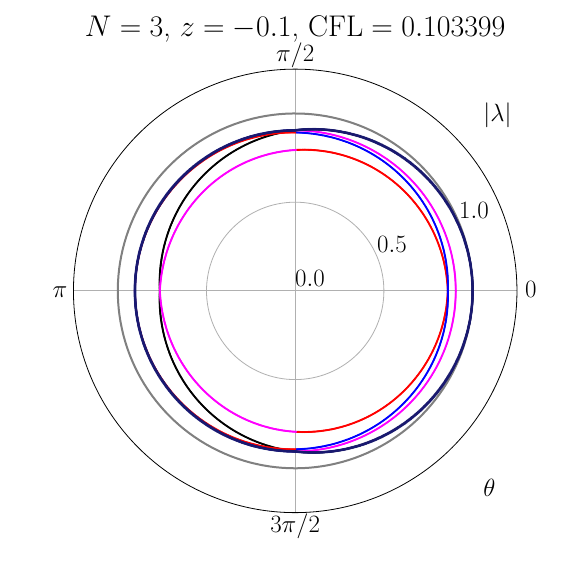}
\includegraphics[width=0.15\textwidth]{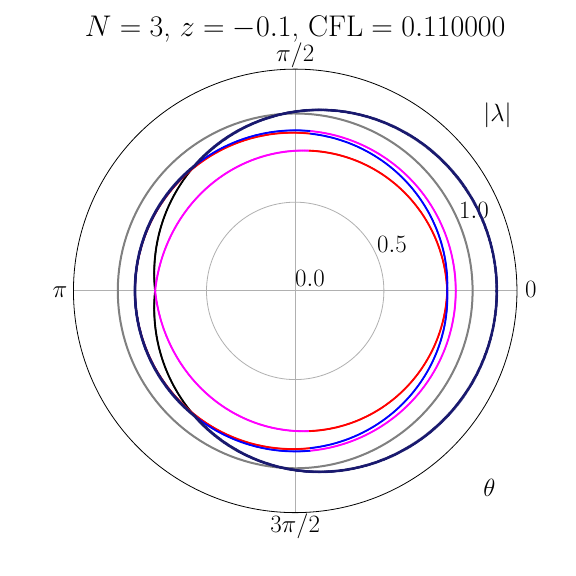}
\includegraphics[width=0.15\textwidth]{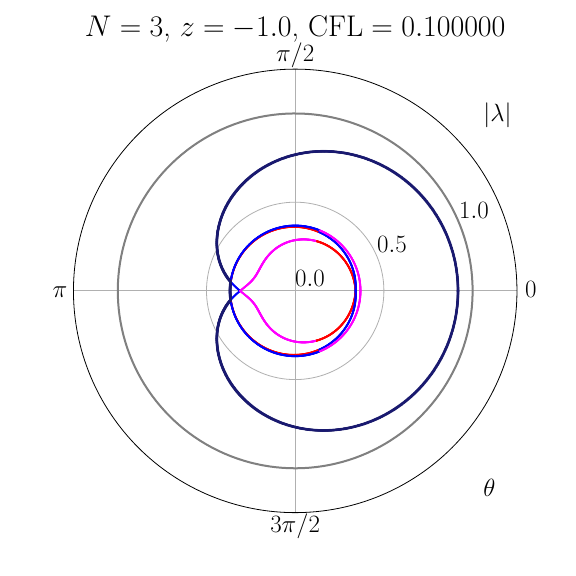}
\includegraphics[width=0.15\textwidth]{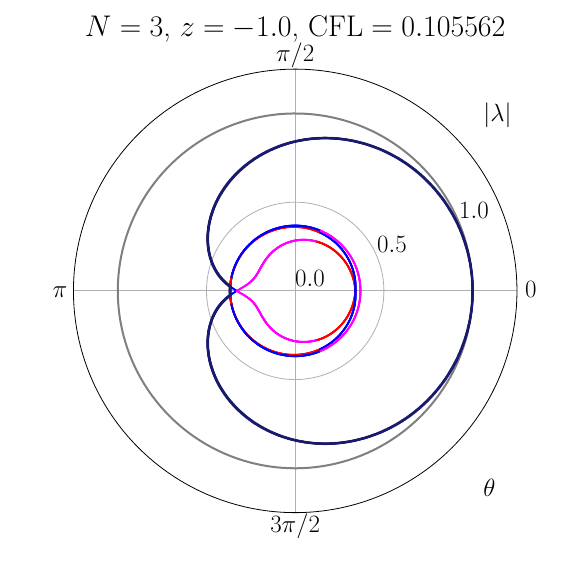}
\includegraphics[width=0.15\textwidth]{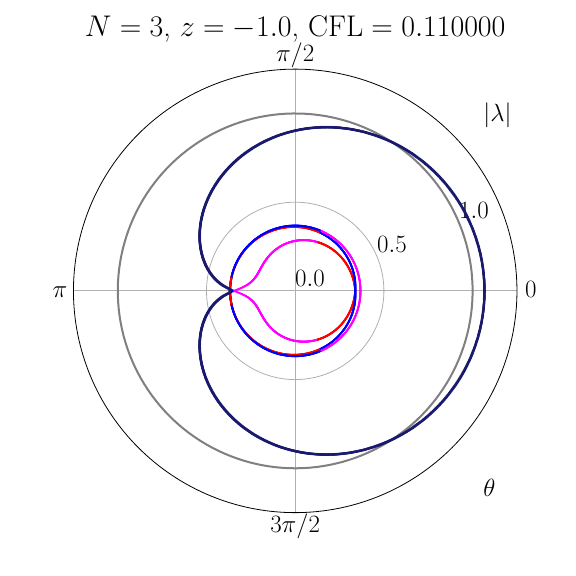}\\
\includegraphics[width=0.028125\textwidth]{rhos_on_theta_legend_N_3.png}
\includegraphics[width=0.15\textwidth]{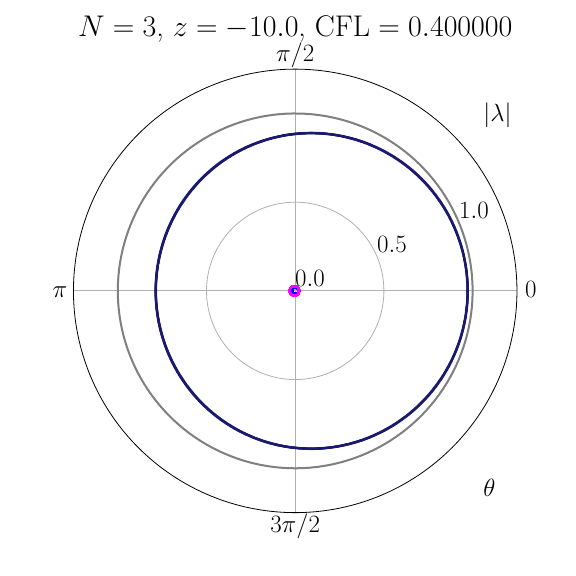}
\includegraphics[width=0.15\textwidth]{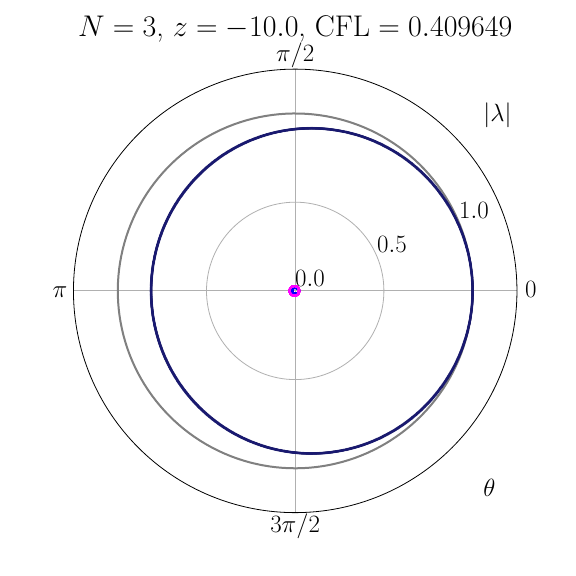}
\includegraphics[width=0.15\textwidth]{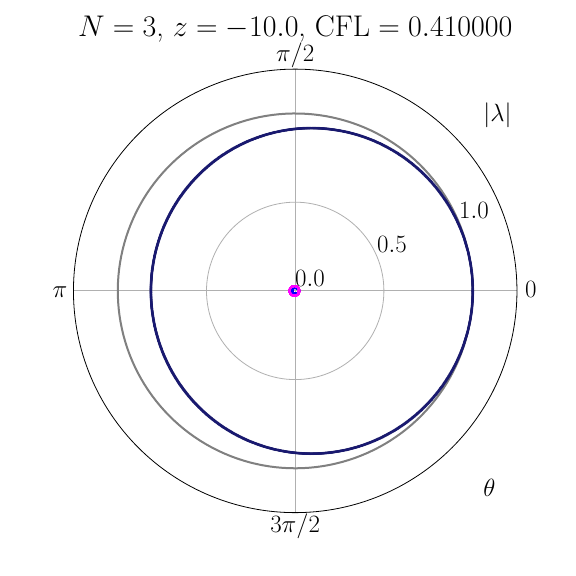}
\includegraphics[width=0.15\textwidth]{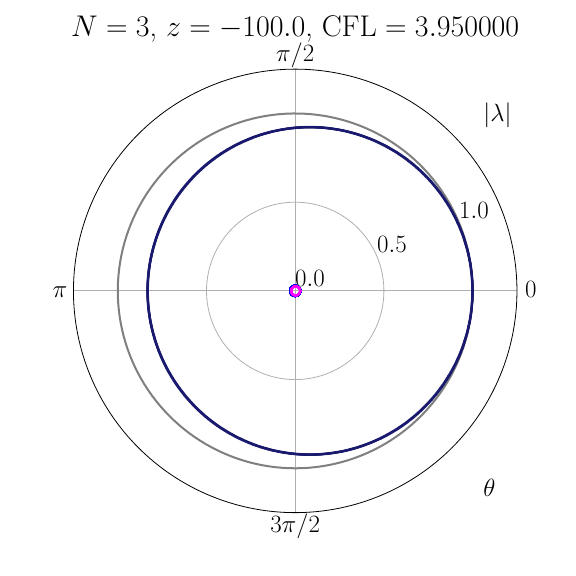}
\includegraphics[width=0.15\textwidth]{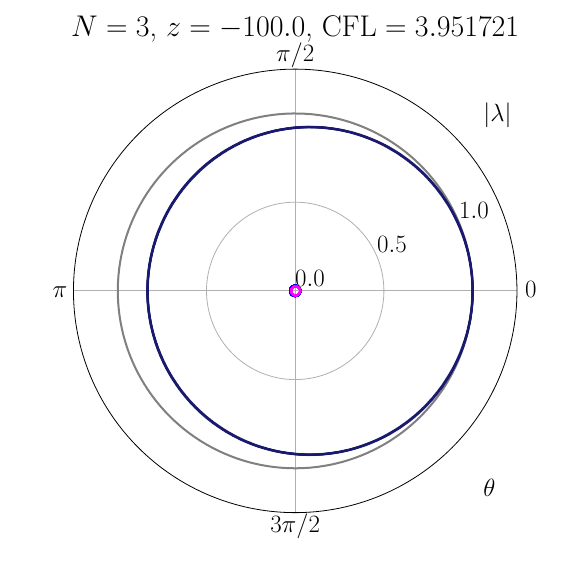}
\includegraphics[width=0.15\textwidth]{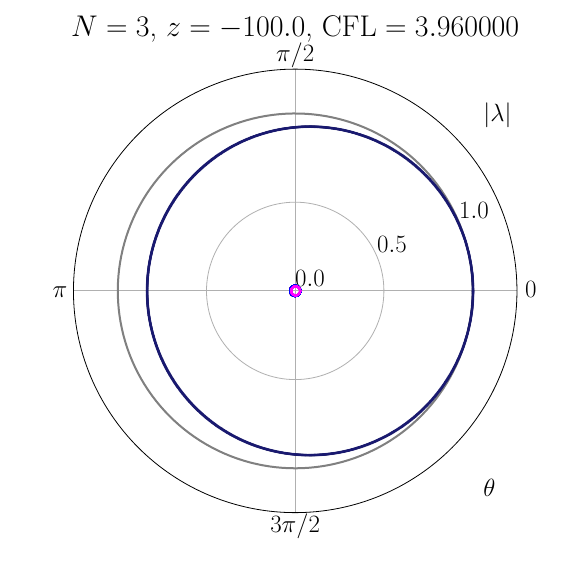}
\caption{%
The dependence of the absolute values $|\lambda_{k}|$ of the spectrum of the matrix $\mathrm{R}(\mathrm{CFL}, z, \theta)$ (\ref{eq:r_matrix_elems_expr_dup}) (eigenvalues $\lambda_{k} = \lambda_{k}(\mathrm{CFL}, z, \theta)$, $k = 0, \ldots, N$) of the evolution operator $R$ (\ref{eq:evol_oper_prop}) for a single time step $\Dtn{n}$ on phase $\theta = k\Dx$ for several selected values of the Courant number $\mathrm{CFL}$ for $z = -0.1$, $-1.0$, $-10.0$, $-100.0$ and for polynomial degrees $N = 1, \ldots, 3$ --- the polar plot with phase $\theta$ as angle and absolute value $|\lambda|$ as radius. The range of phase $\theta\in[0, 2\pi)$ is sampled on a uniform grid of $1000$ nodes. Legends for each row of the graphs are located on the left. The gray circle of unit radius defines the stability boundary. $\lambda_{\rm max}$ is the eigenvalue $\lambda_{k}$ with the largest absolute value $|\lambda_{k}|$. The Courant numbers $\mathrm{CFL}$ are taken both at the boundary value of the Courant number $\mathrm{CFL}_{\rm max}$ and at two values $\mathrm{CFL}$ near it --- in the guaranteed stability region $\mathrm{CFL} < \mathrm{CFL}_{\rm max}$ and in the guaranteed instability region $\mathrm{CFL} > \mathrm{CFL}_{\rm max}$, by rounding off at the second significant decimal place (the values $\mathrm{CFL}_{\rm max}$ are calculated further in this work and presented in Figure~\ref{fig:cfls_on_z}). \textit{Note}: the phase $\theta$ is not the phase $\phi_{k} \equiv \arg \lambda_{k}$ of eigenvalue $\lambda_{k}$; the dependence of eigenvalues $\lambda_{k}$ of the matrix $\mathrm{R}(\mathrm{CFL}, z, \theta)$ (\ref{eq:r_matrix_elems_expr_dup}) for these values $z$ and these polynomial degrees $N$ is presented in Figure~\ref{fig:spectrum_exact_eigvals_degrees_1_3}.
}
\label{fig:rhos_on_theta_degrees_1_3}
\end{figure}
\begin{figure}[h!]
\centering
\includegraphics[width=0.028125\textwidth]{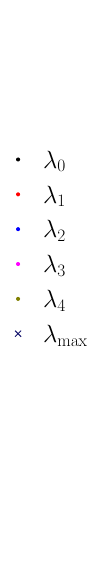}
\includegraphics[width=0.15\textwidth]{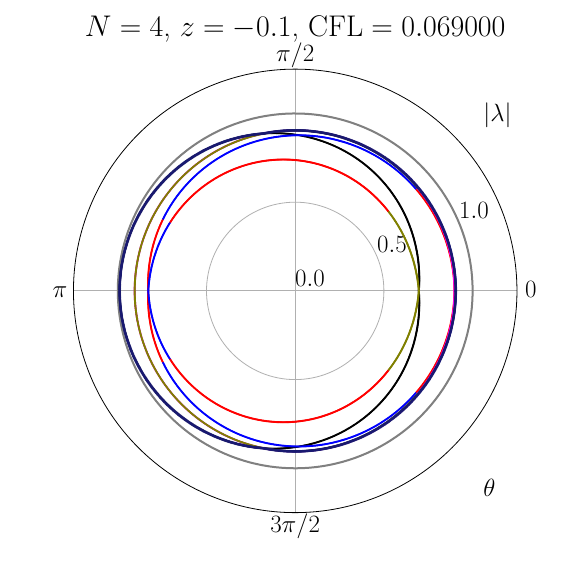}
\includegraphics[width=0.15\textwidth]{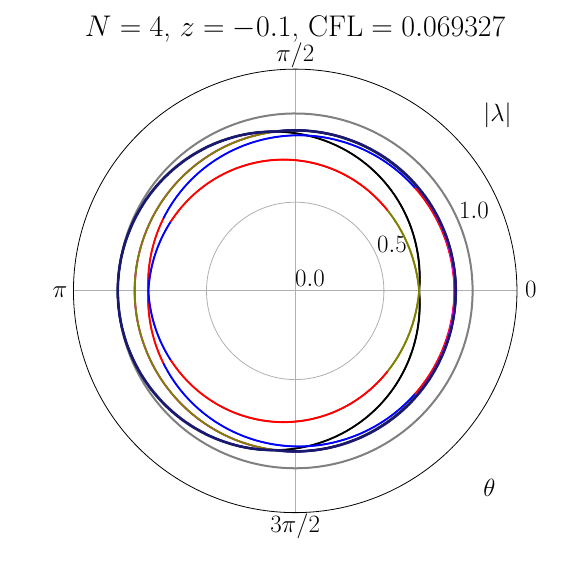}
\includegraphics[width=0.15\textwidth]{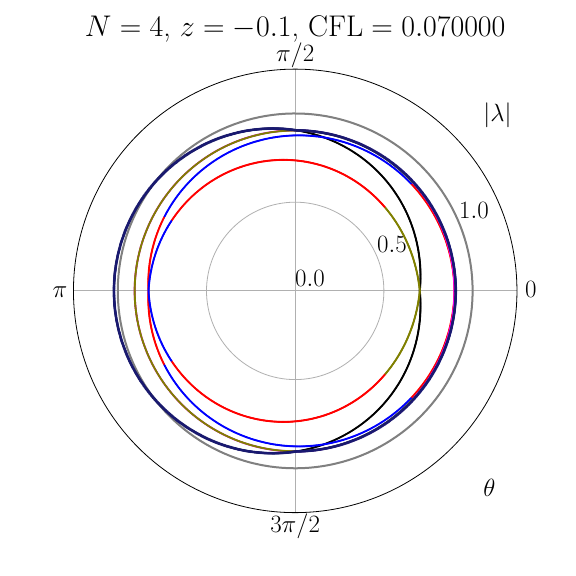}
\includegraphics[width=0.15\textwidth]{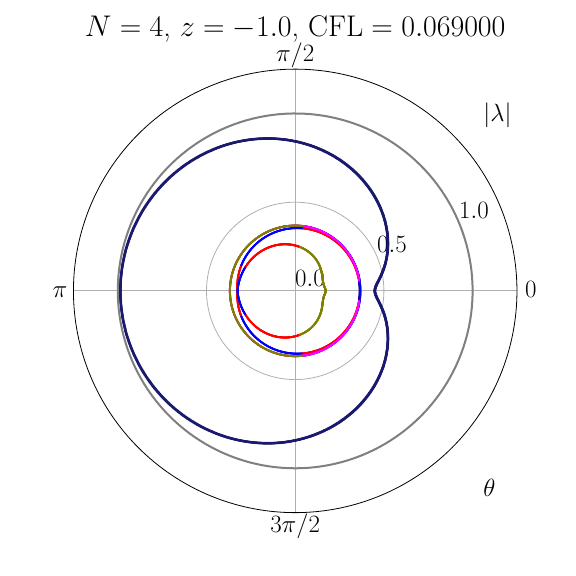}
\includegraphics[width=0.15\textwidth]{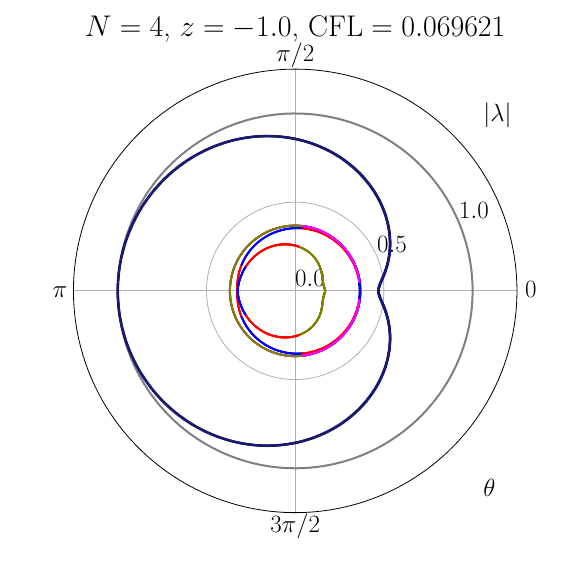}
\includegraphics[width=0.15\textwidth]{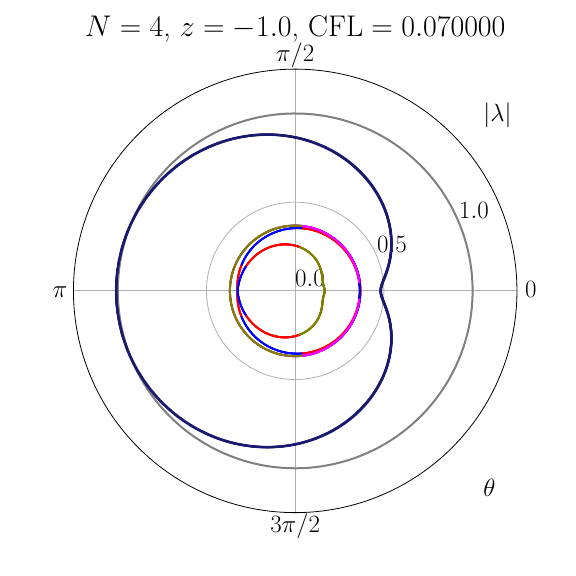}\\
\includegraphics[width=0.028125\textwidth]{rhos_on_theta_legend_N_4.png}
\includegraphics[width=0.15\textwidth]{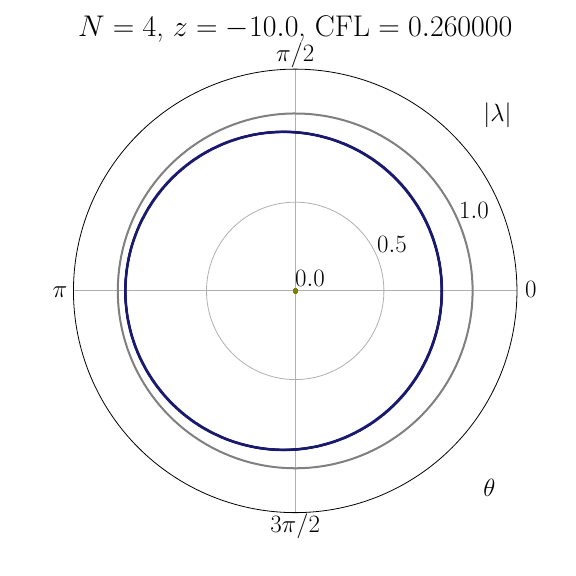}
\includegraphics[width=0.15\textwidth]{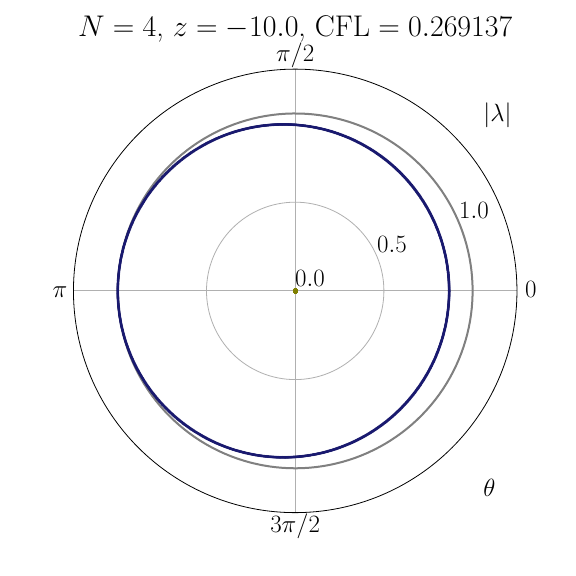}
\includegraphics[width=0.15\textwidth]{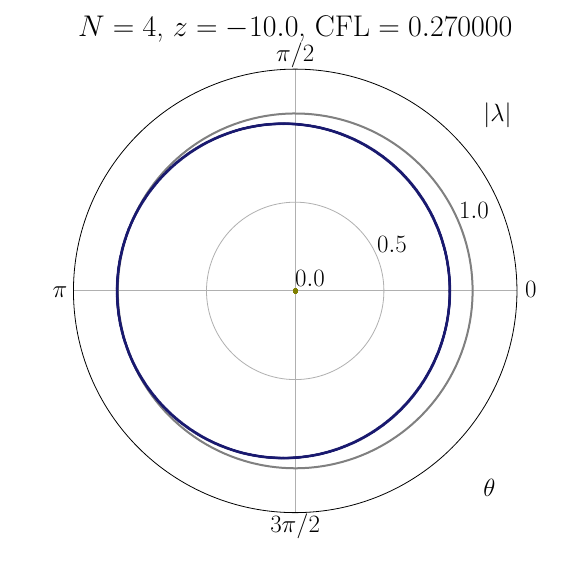}
\includegraphics[width=0.15\textwidth]{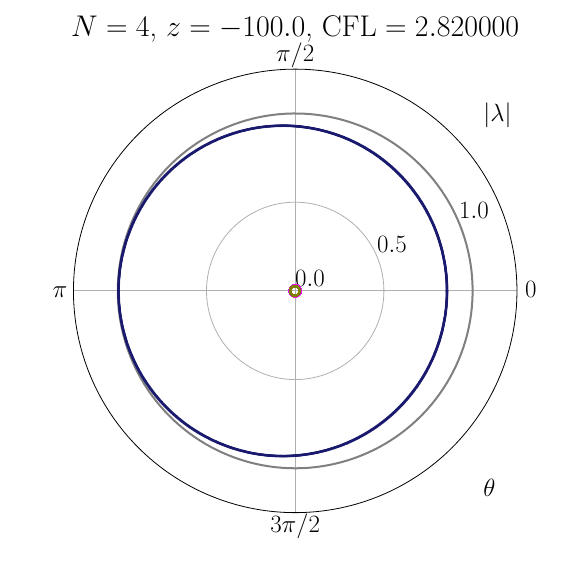}
\includegraphics[width=0.15\textwidth]{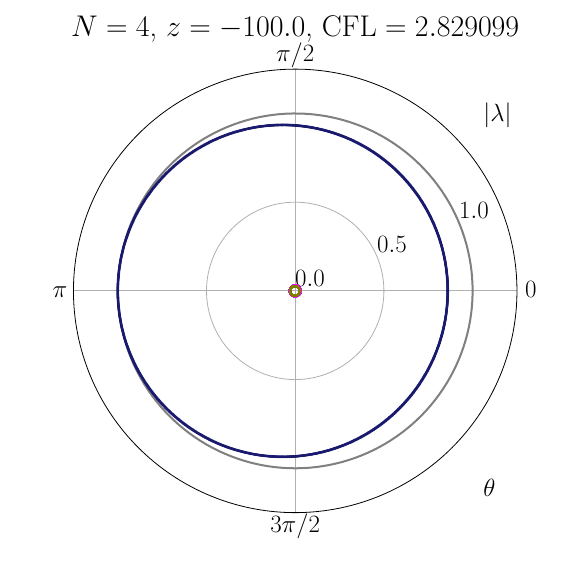}
\includegraphics[width=0.15\textwidth]{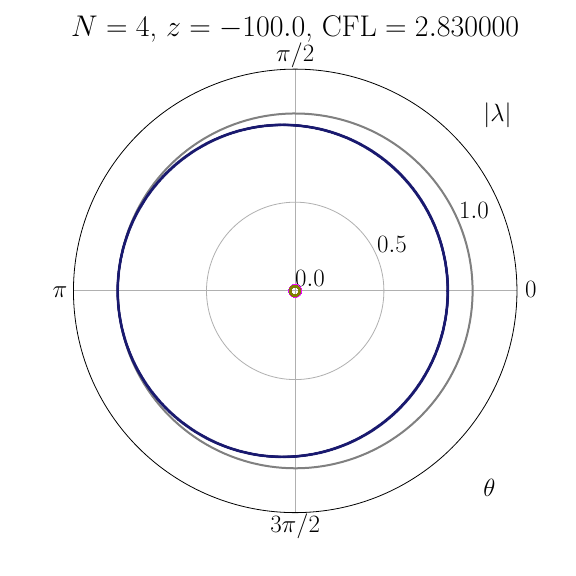}\\
\includegraphics[width=0.028125\textwidth]{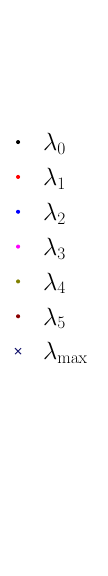}
\includegraphics[width=0.15\textwidth]{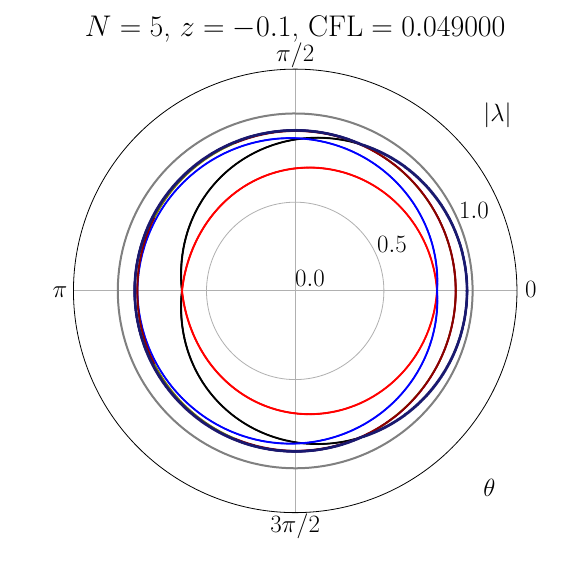}
\includegraphics[width=0.15\textwidth]{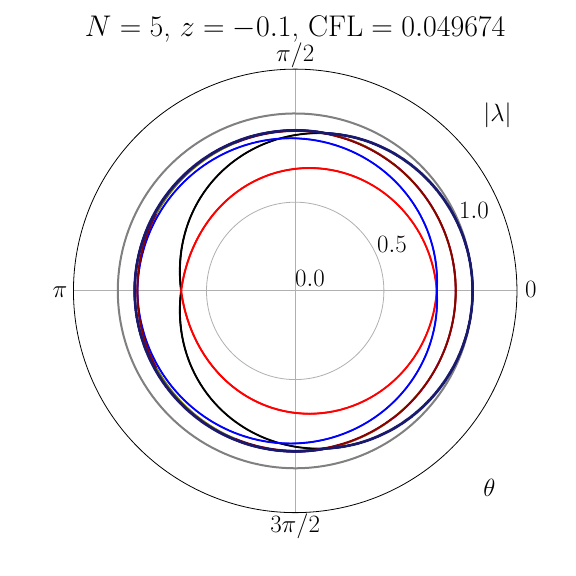}
\includegraphics[width=0.15\textwidth]{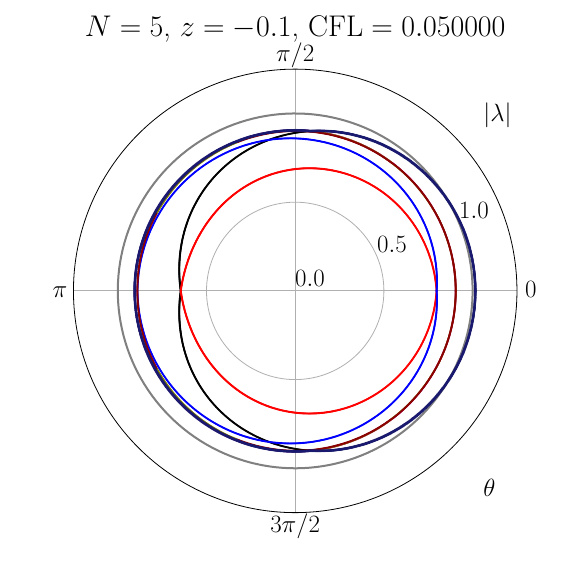}
\includegraphics[width=0.15\textwidth]{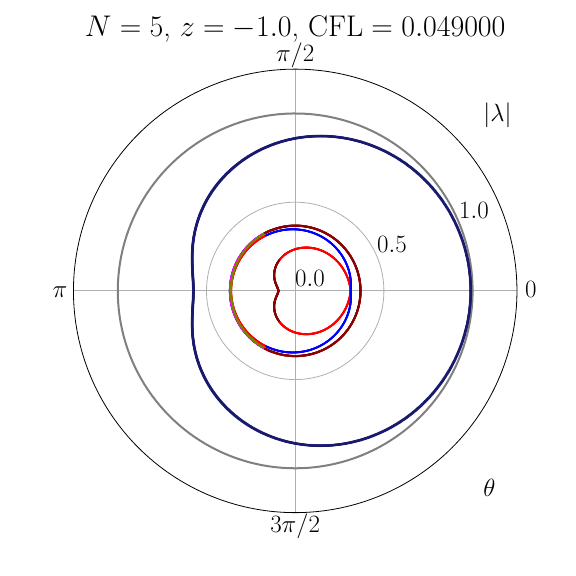}
\includegraphics[width=0.15\textwidth]{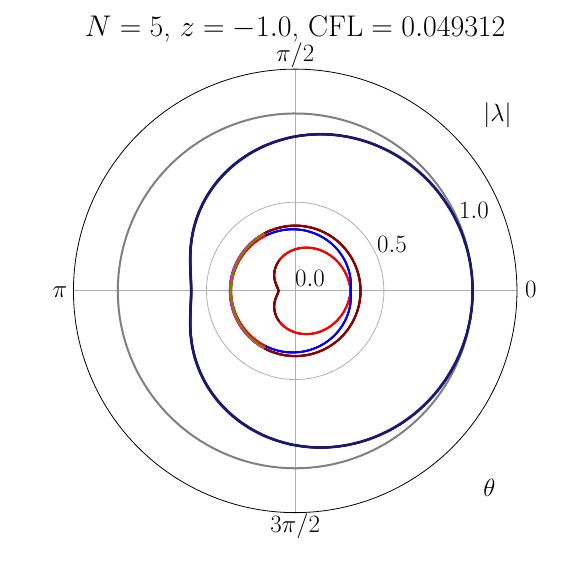}
\includegraphics[width=0.15\textwidth]{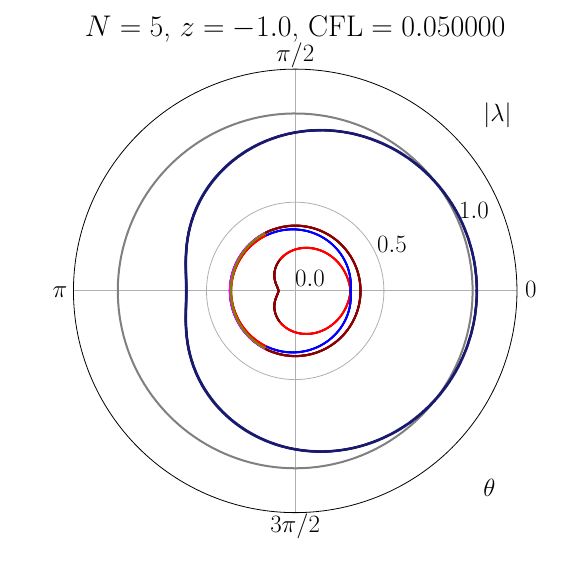}\\
\includegraphics[width=0.028125\textwidth]{rhos_on_theta_legend_N_5.png}
\includegraphics[width=0.15\textwidth]{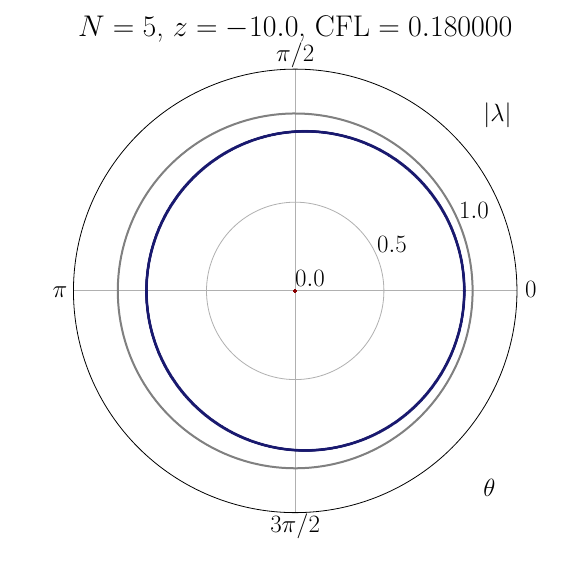}
\includegraphics[width=0.15\textwidth]{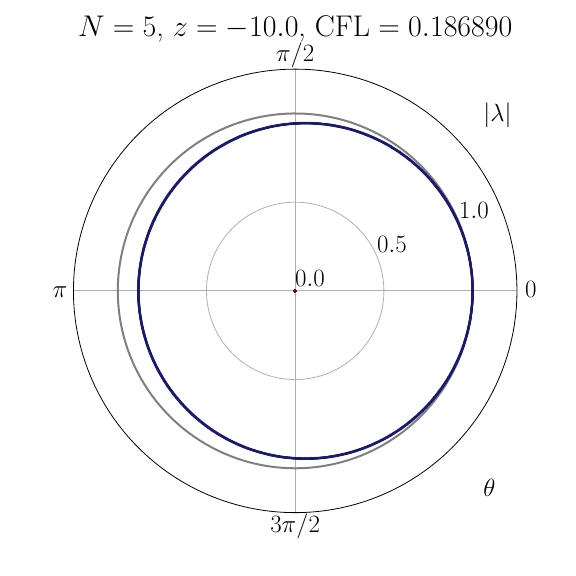}
\includegraphics[width=0.15\textwidth]{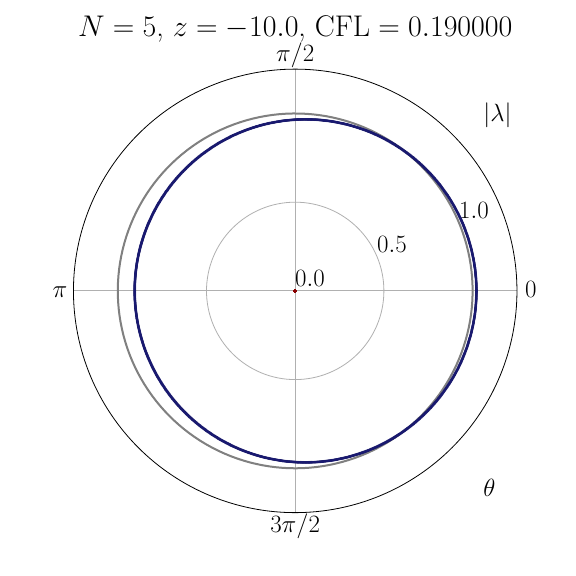}
\includegraphics[width=0.15\textwidth]{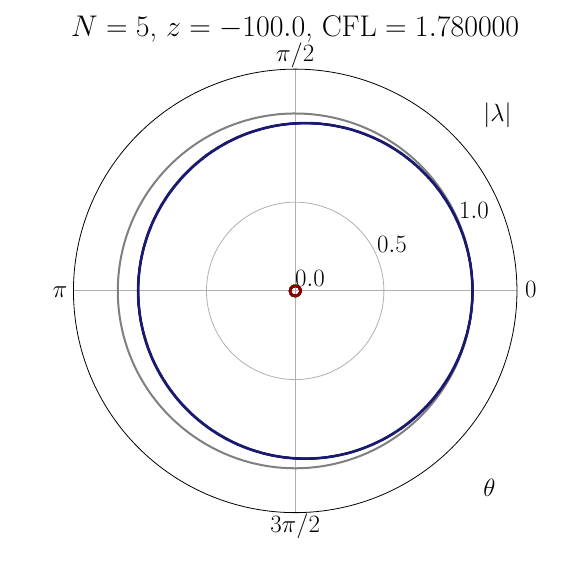}
\includegraphics[width=0.15\textwidth]{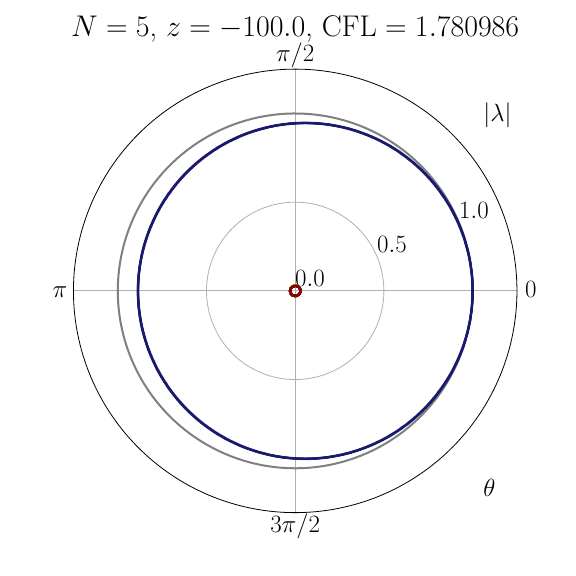}
\includegraphics[width=0.15\textwidth]{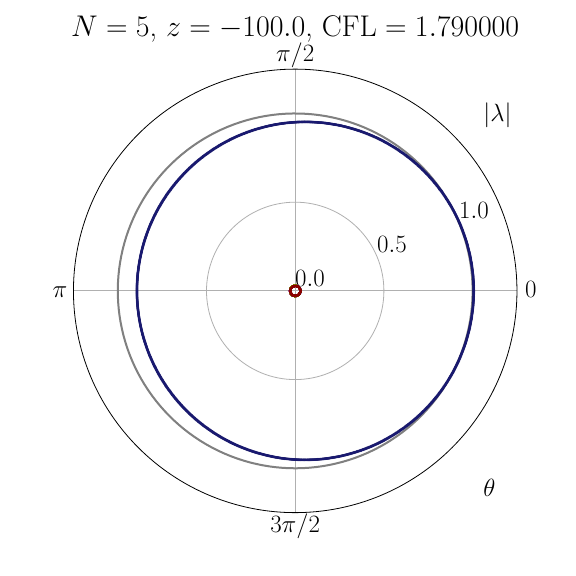}\\
\includegraphics[width=0.028125\textwidth]{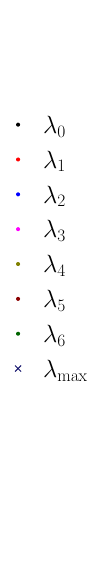}
\includegraphics[width=0.15\textwidth]{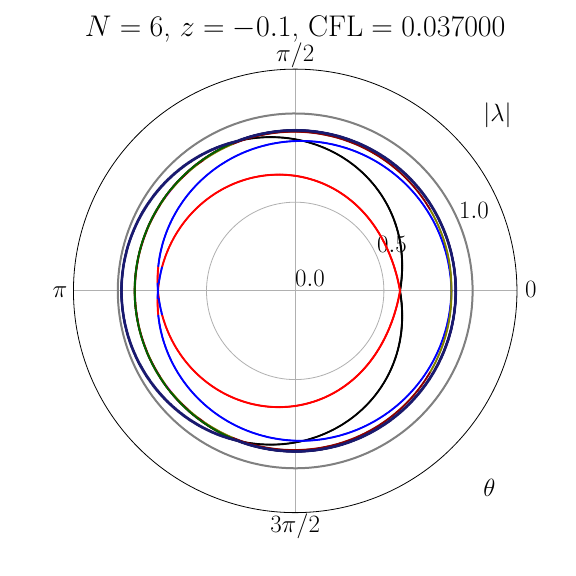}
\includegraphics[width=0.15\textwidth]{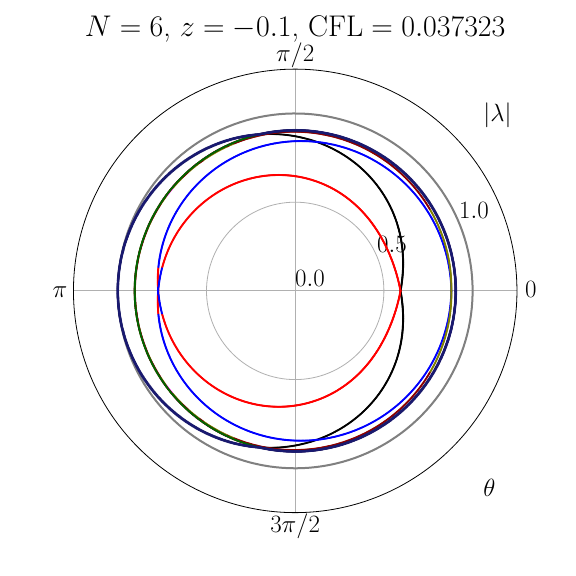}
\includegraphics[width=0.15\textwidth]{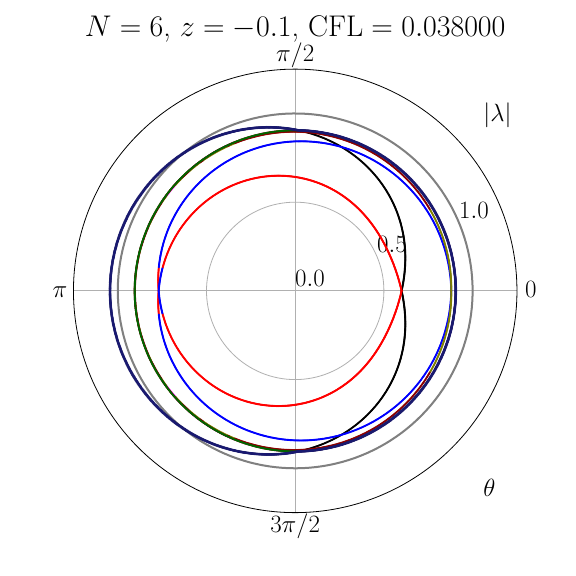}
\includegraphics[width=0.15\textwidth]{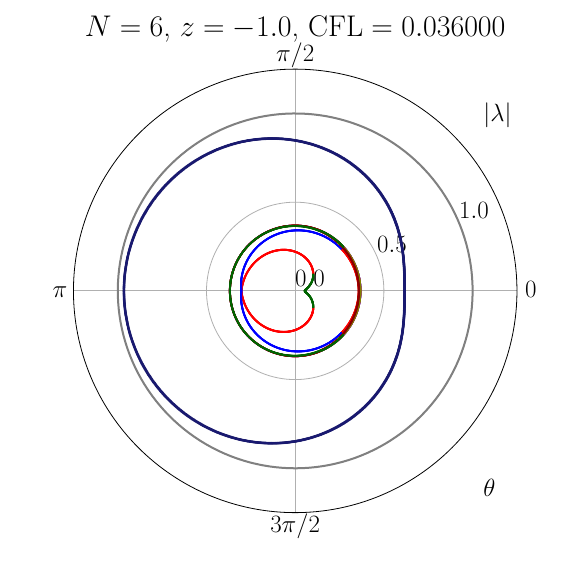}
\includegraphics[width=0.15\textwidth]{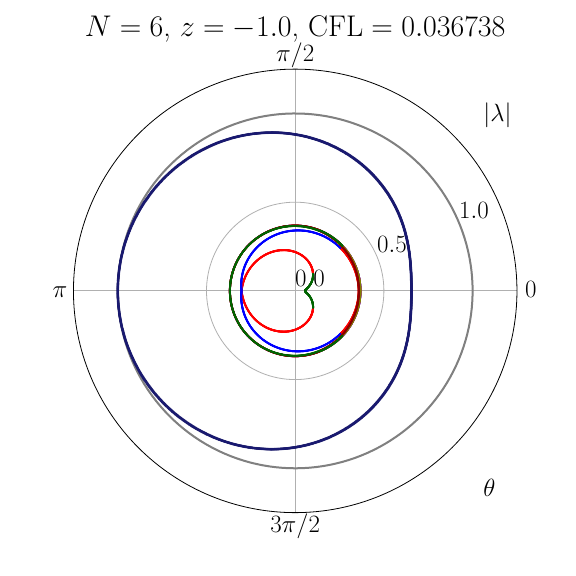}
\includegraphics[width=0.15\textwidth]{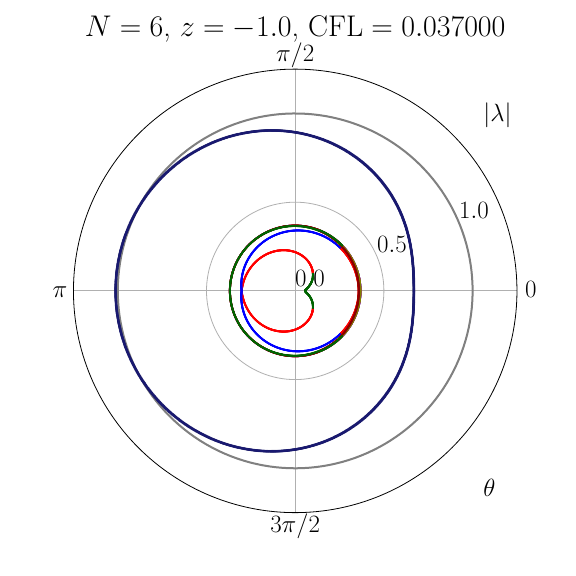}\\
\includegraphics[width=0.028125\textwidth]{rhos_on_theta_legend_N_6.png}
\includegraphics[width=0.15\textwidth]{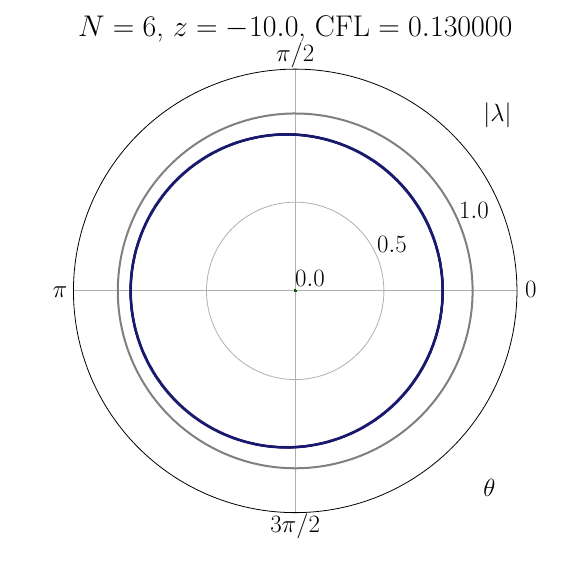}
\includegraphics[width=0.15\textwidth]{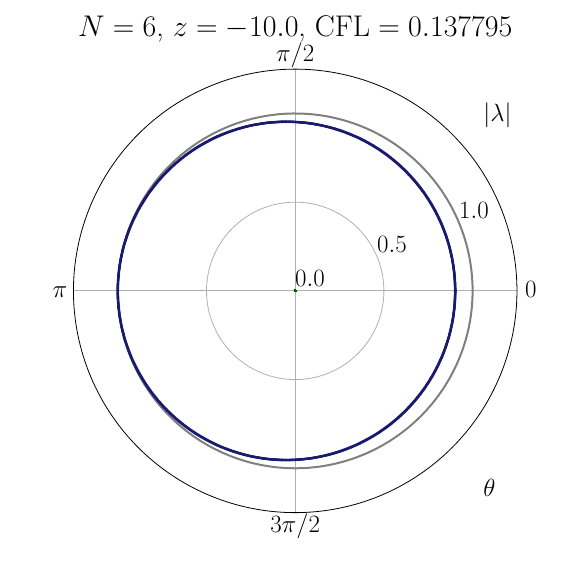}
\includegraphics[width=0.15\textwidth]{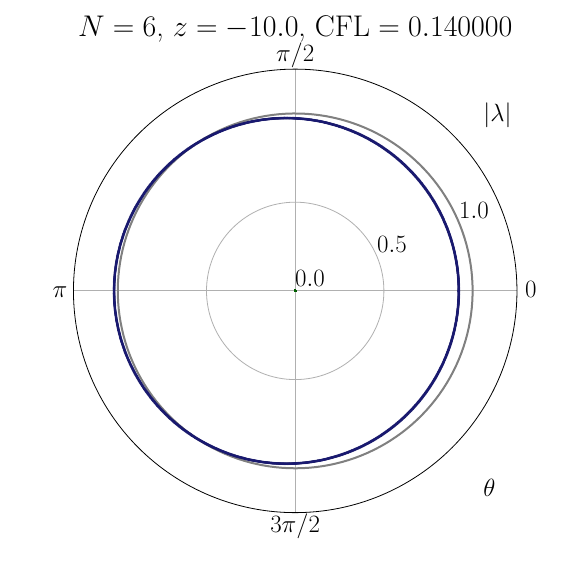}
\includegraphics[width=0.15\textwidth]{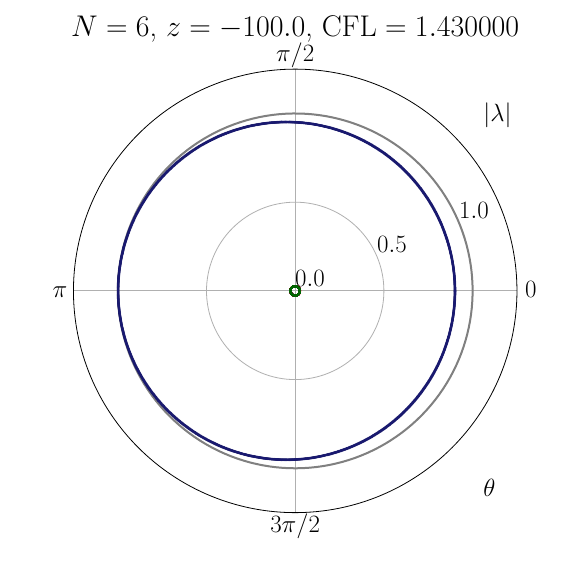}
\includegraphics[width=0.15\textwidth]{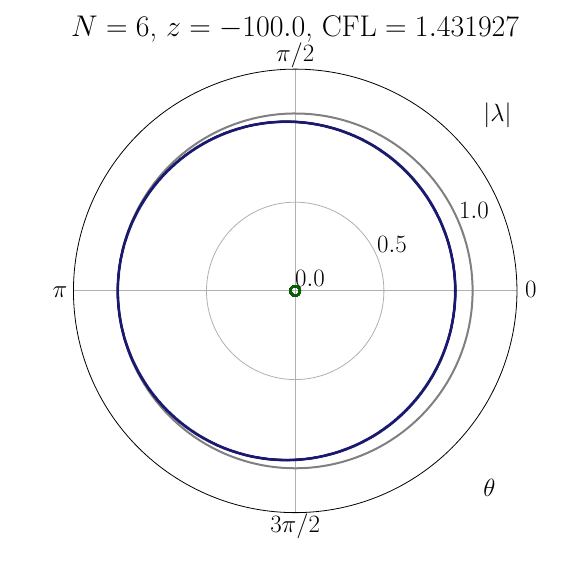}
\includegraphics[width=0.15\textwidth]{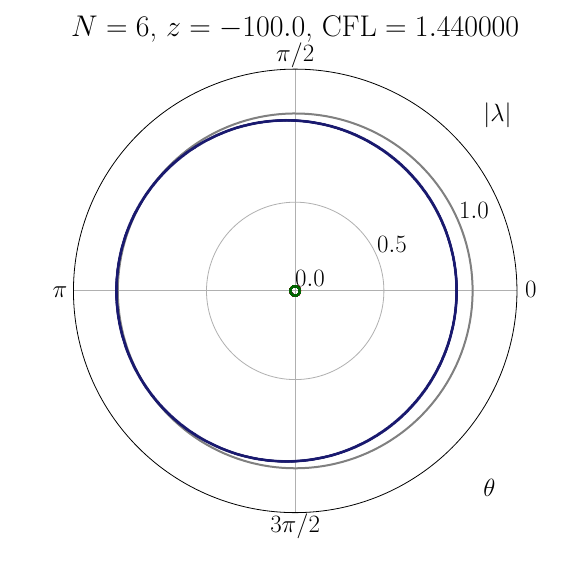}
\caption{%
The dependence of the absolute values $|\lambda_{k}|$ of the spectrum of the matrix $\mathrm{R}(\mathrm{CFL}, z, \theta)$ (\ref{eq:r_matrix_elems_expr_dup}) (eigenvalues $\lambda_{k} = \lambda_{k}(\mathrm{CFL}, z, \theta)$, $k = 0, \ldots, N$) of the evolution operator $R$ (\ref{eq:evol_oper_prop}) for a single time step $\Dtn{n}$ on phase $\theta = k\Dx$ for several selected values of the Courant number $\mathrm{CFL}$ for $z = -0.1$, $-1.0$, $-10.0$, $-100.0$ and for polynomial degrees $N = 4, \ldots, 6$ --- the polar plot with phase $\theta$ as angle and absolute value $|\lambda|$ as radius. The range of phase $\theta\in[0, 2\pi)$ is sampled on a uniform grid of $1000$ nodes. Legends for each row of the graphs are located on the left. The gray circle of unit radius defines the stability boundary. $\lambda_{\rm max}$ is the eigenvalue $\lambda_{k}$ with the largest absolute value $|\lambda_{k}|$. The Courant numbers $\mathrm{CFL}$ are taken both at the boundary value of the Courant number $\mathrm{CFL}_{\rm max}$ and at two values $\mathrm{CFL}$ near it --- in the guaranteed stability region $\mathrm{CFL} < \mathrm{CFL}_{\rm max}$ and in the guaranteed instability region $\mathrm{CFL} > \mathrm{CFL}_{\rm max}$, by rounding off at the second significant decimal place (the values $\mathrm{CFL}_{\rm max}$ are calculated further in this work and presented in Figure~\ref{fig:cfls_on_z}). \textit{Note}: the phase $\theta$ is not the phase $\phi_{k} \equiv \arg \lambda_{k}$ of eigenvalue $\lambda_{k}$; the dependence of eigenvalues $\lambda_{k}$ of the matrix $\mathrm{R}(\mathrm{CFL}, z, \theta)$ (\ref{eq:r_matrix_elems_expr_dup}) for these values $z$ and these polynomial degrees $N$ is presented in Figure~\ref{fig:spectrum_exact_eigvals_degrees_4_6}.
}
\label{fig:rhos_on_theta_degrees_4_6}
\end{figure}
\begin{figure}[h!]
\centering
\includegraphics[width=0.028125\textwidth]{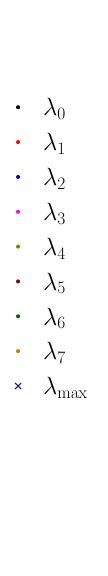}
\includegraphics[width=0.15\textwidth]{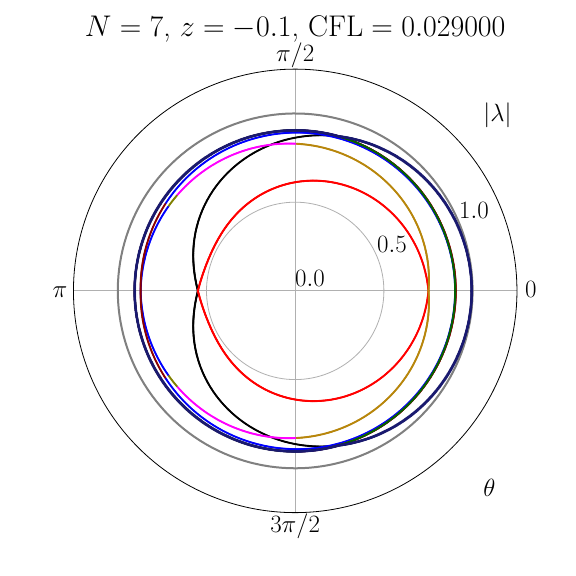}
\includegraphics[width=0.15\textwidth]{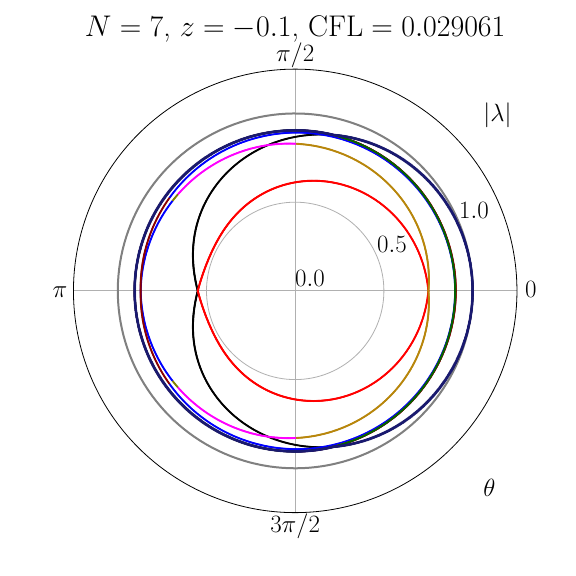}
\includegraphics[width=0.15\textwidth]{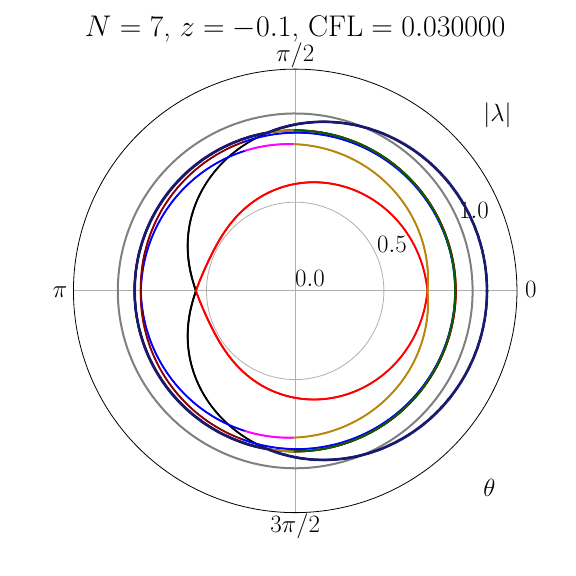}
\includegraphics[width=0.15\textwidth]{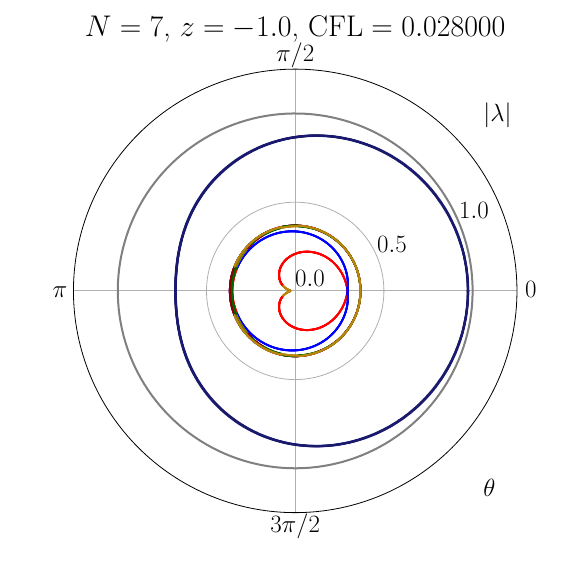}
\includegraphics[width=0.15\textwidth]{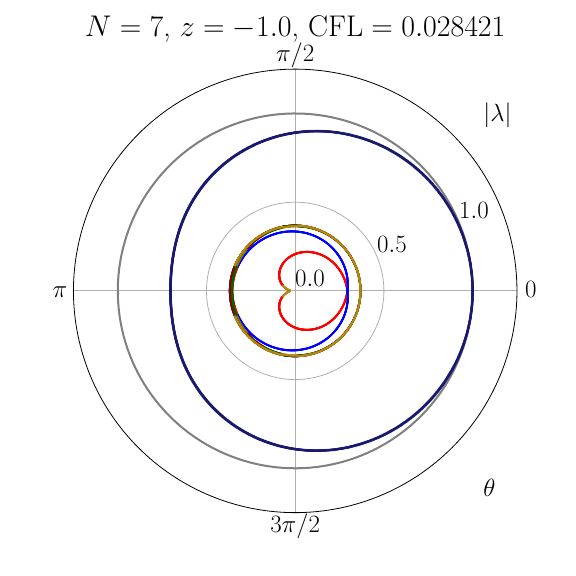}
\includegraphics[width=0.15\textwidth]{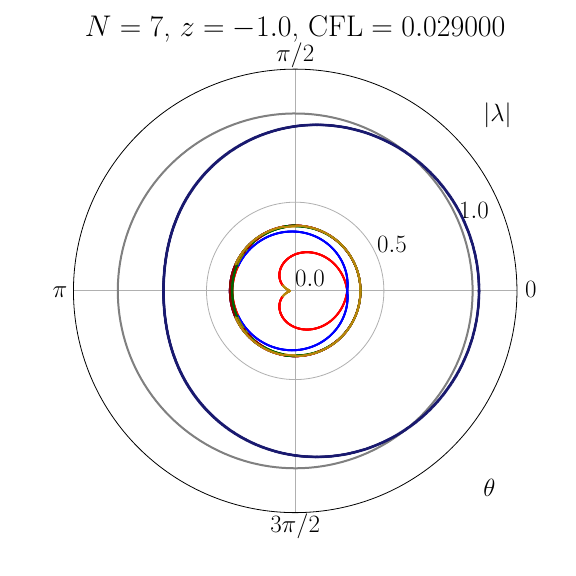}\\
\includegraphics[width=0.028125\textwidth]{rhos_on_theta_legend_N_7.png}
\includegraphics[width=0.15\textwidth]{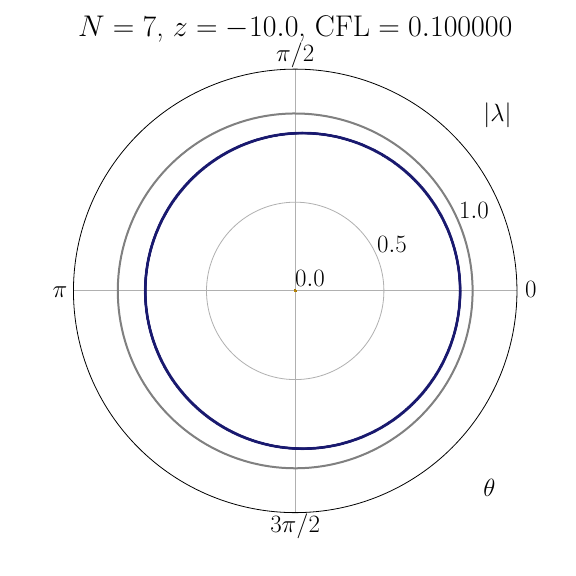}
\includegraphics[width=0.15\textwidth]{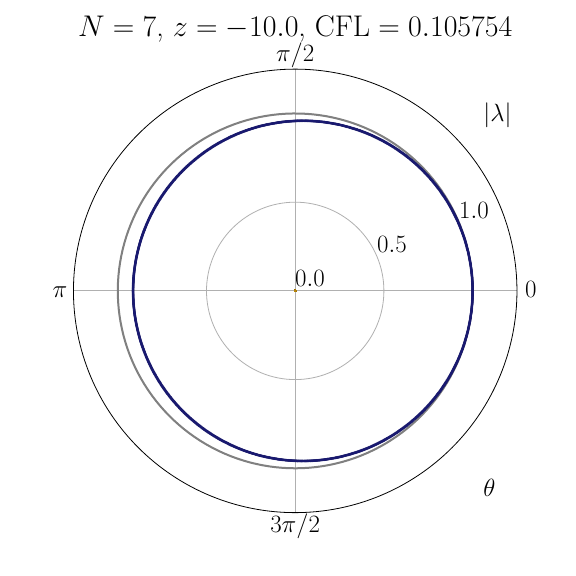}
\includegraphics[width=0.15\textwidth]{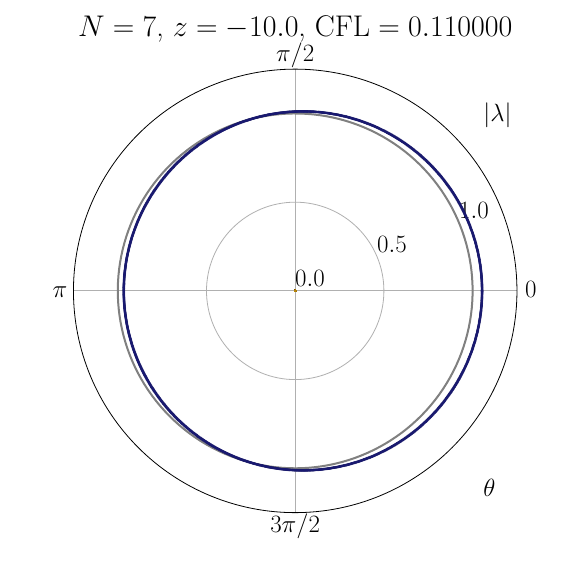}
\includegraphics[width=0.15\textwidth]{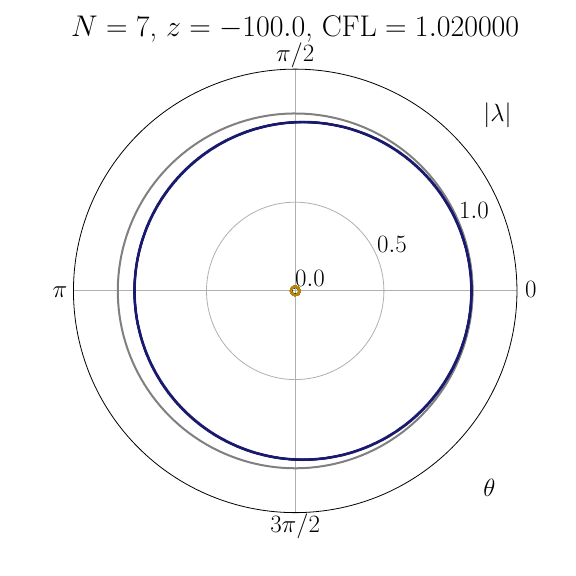}
\includegraphics[width=0.15\textwidth]{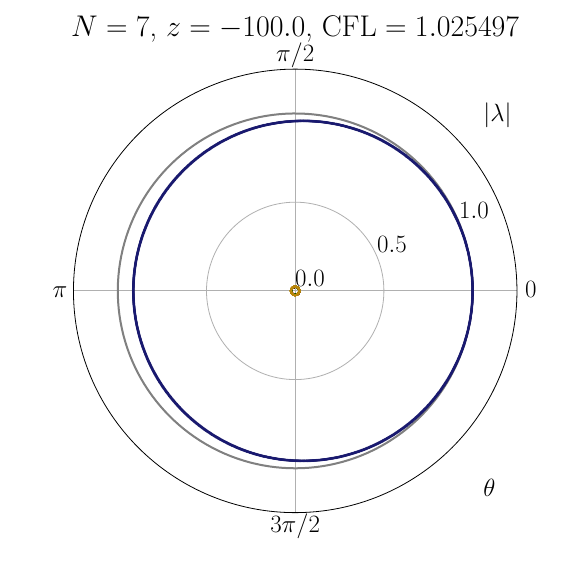}
\includegraphics[width=0.15\textwidth]{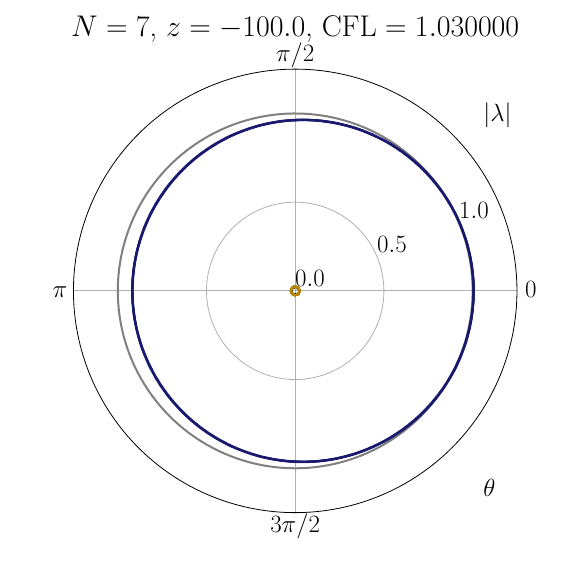}\\
\includegraphics[width=0.028125\textwidth]{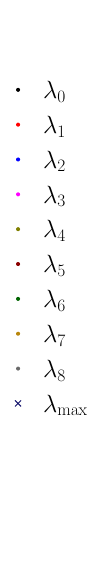}
\includegraphics[width=0.15\textwidth]{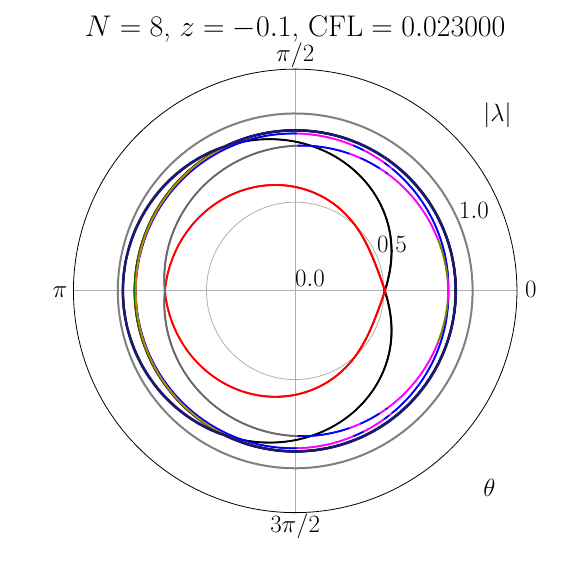}
\includegraphics[width=0.15\textwidth]{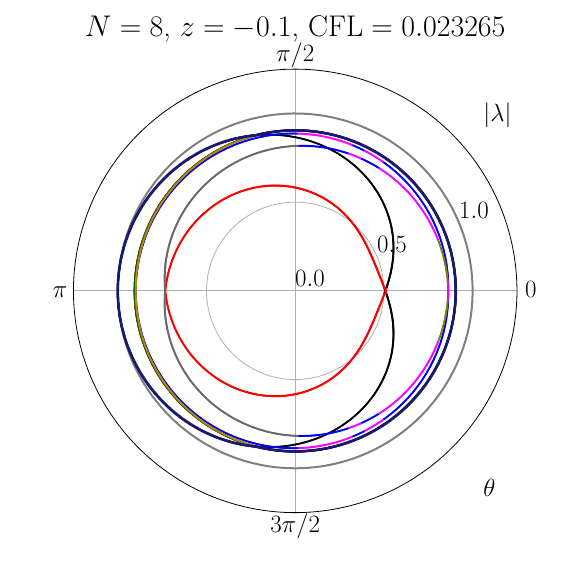}
\includegraphics[width=0.15\textwidth]{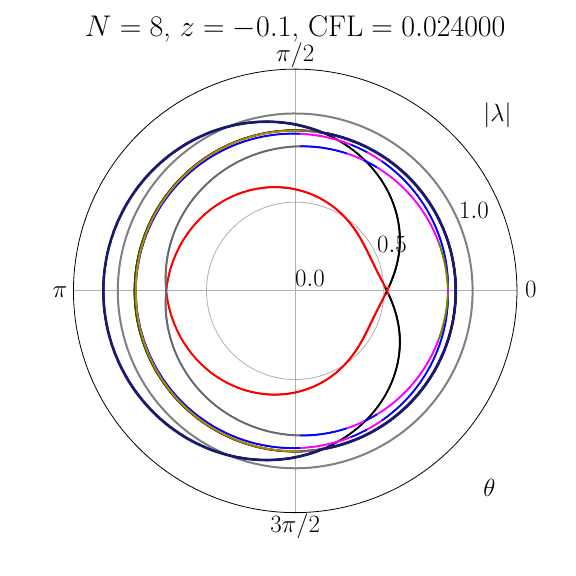}
\includegraphics[width=0.15\textwidth]{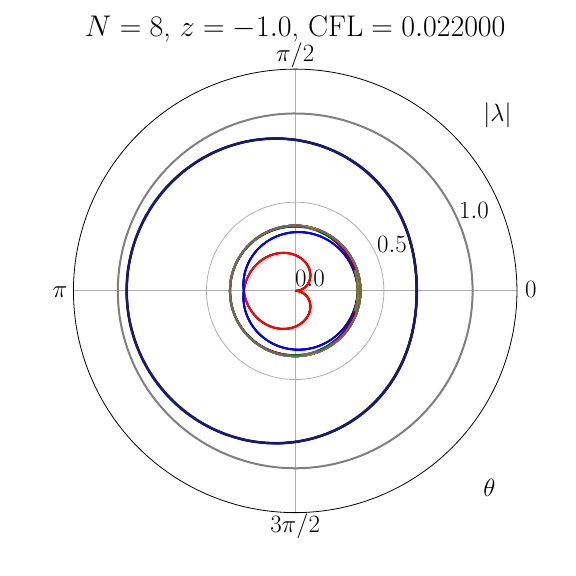}
\includegraphics[width=0.15\textwidth]{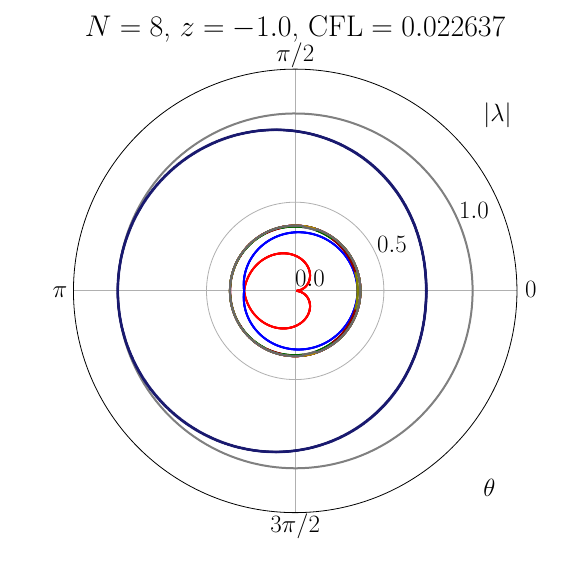}
\includegraphics[width=0.15\textwidth]{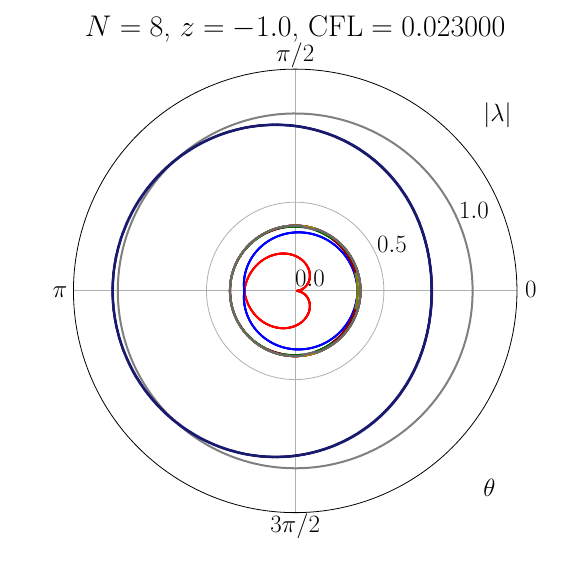}\\
\includegraphics[width=0.028125\textwidth]{rhos_on_theta_legend_N_8.png}
\includegraphics[width=0.15\textwidth]{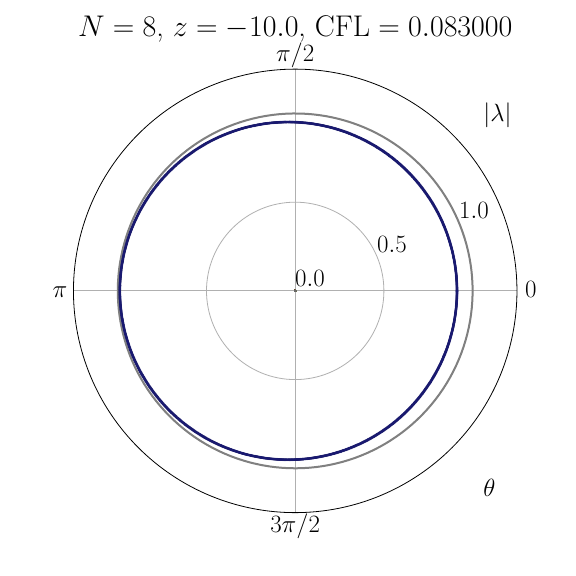}
\includegraphics[width=0.15\textwidth]{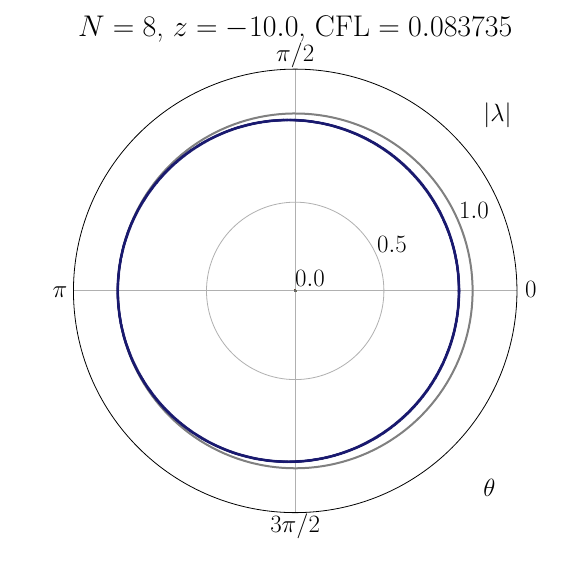}
\includegraphics[width=0.15\textwidth]{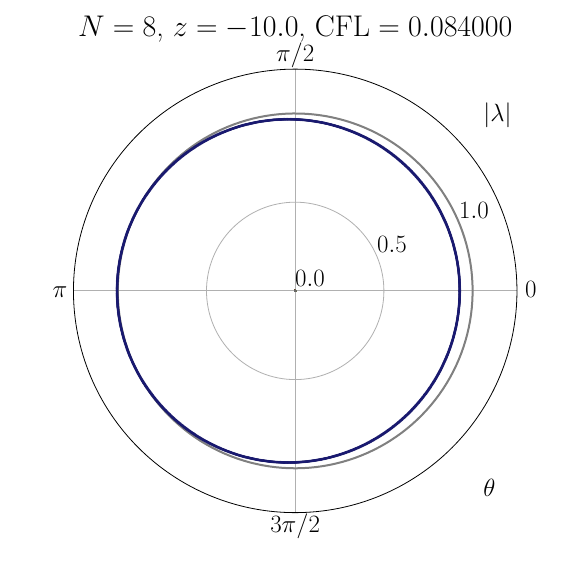}
\includegraphics[width=0.15\textwidth]{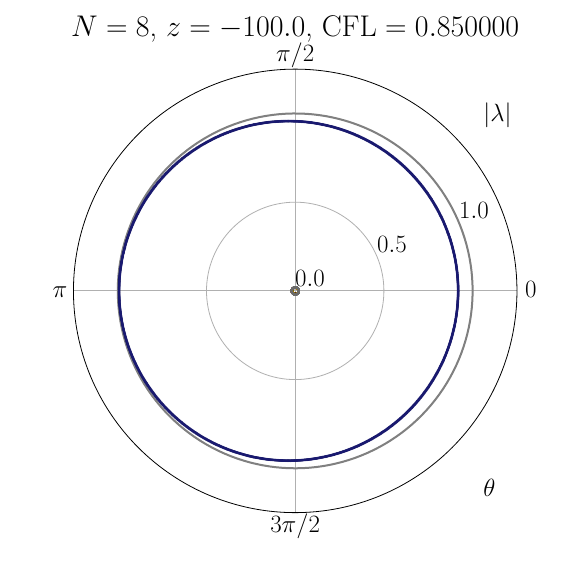}
\includegraphics[width=0.15\textwidth]{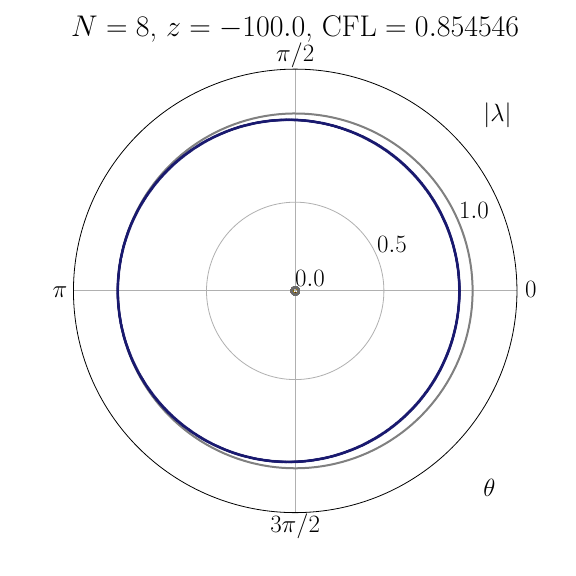}
\includegraphics[width=0.15\textwidth]{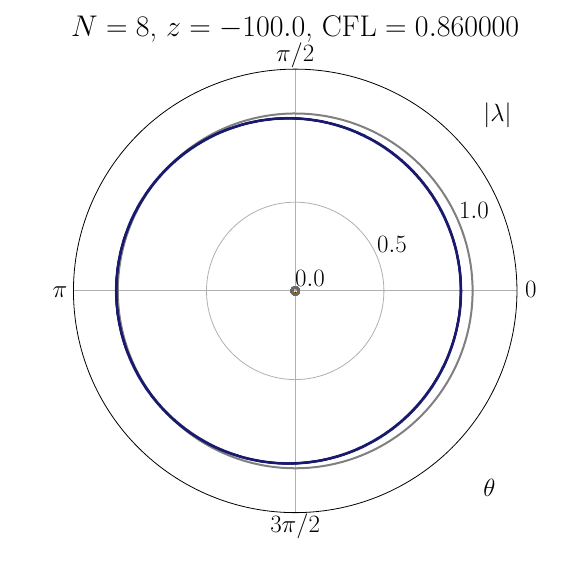}\\
\includegraphics[width=0.028125\textwidth]{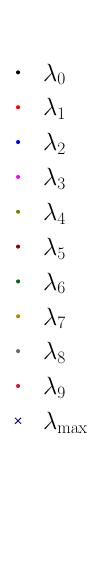}
\includegraphics[width=0.15\textwidth]{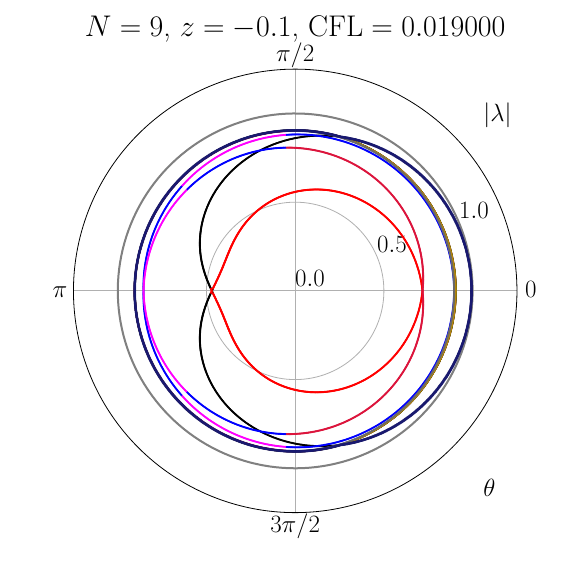}
\includegraphics[width=0.15\textwidth]{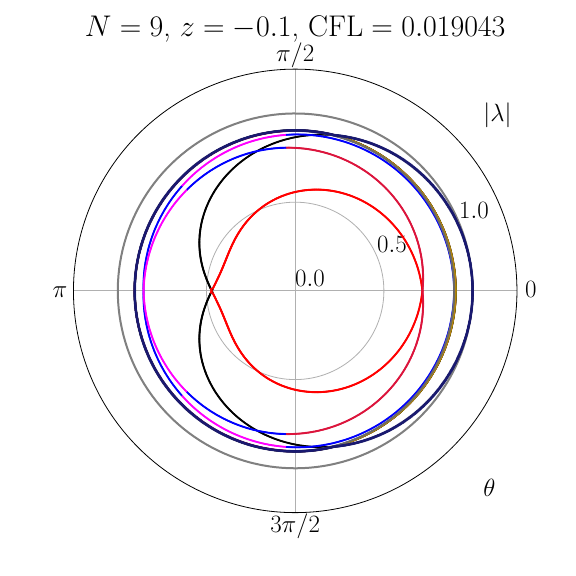}
\includegraphics[width=0.15\textwidth]{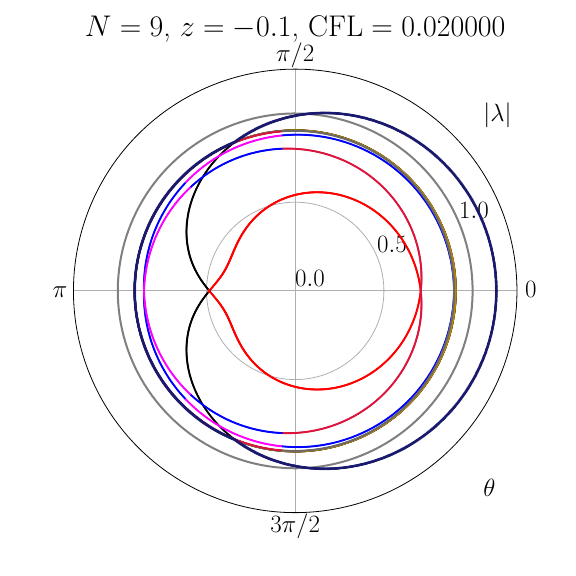}
\includegraphics[width=0.15\textwidth]{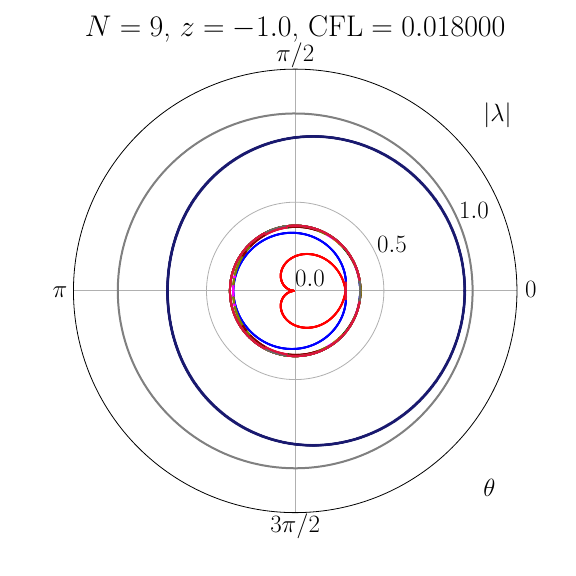}
\includegraphics[width=0.15\textwidth]{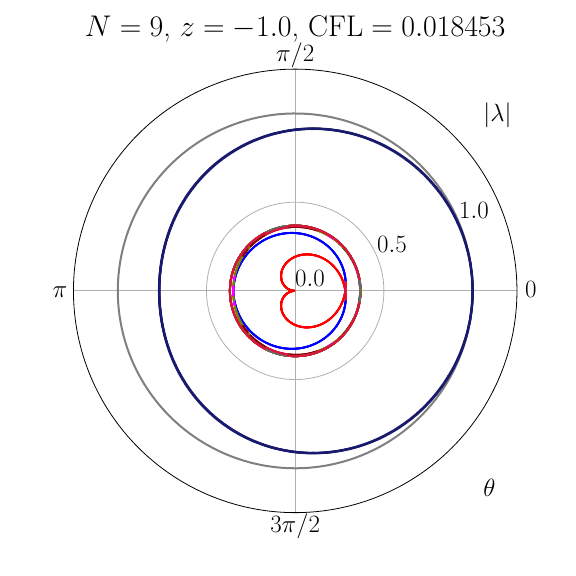}
\includegraphics[width=0.15\textwidth]{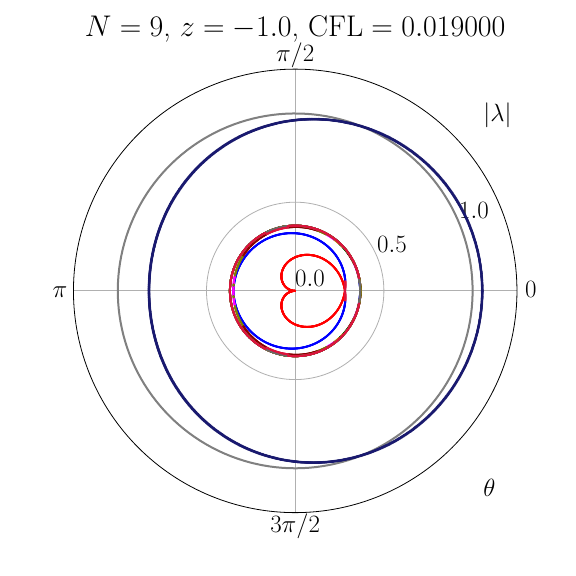}\\
\includegraphics[width=0.028125\textwidth]{rhos_on_theta_legend_N_9.png}
\includegraphics[width=0.15\textwidth]{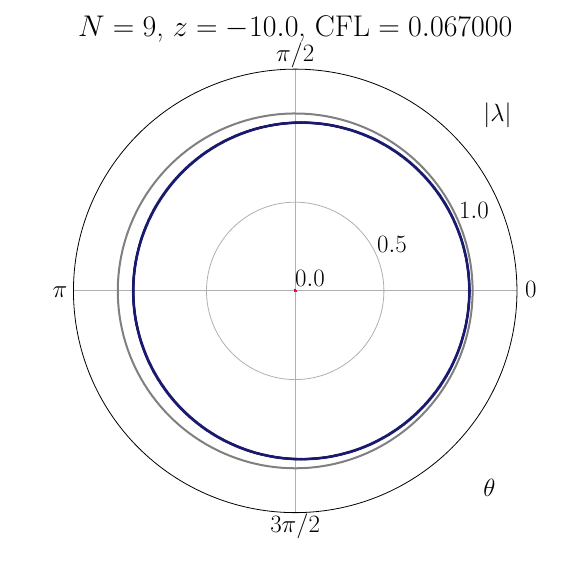}
\includegraphics[width=0.15\textwidth]{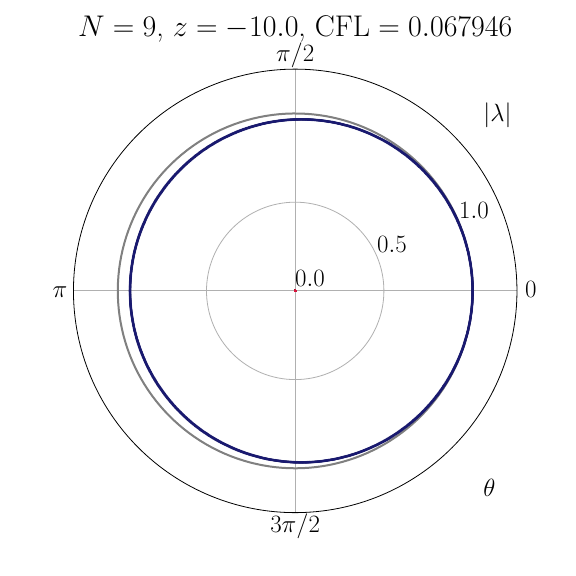}
\includegraphics[width=0.15\textwidth]{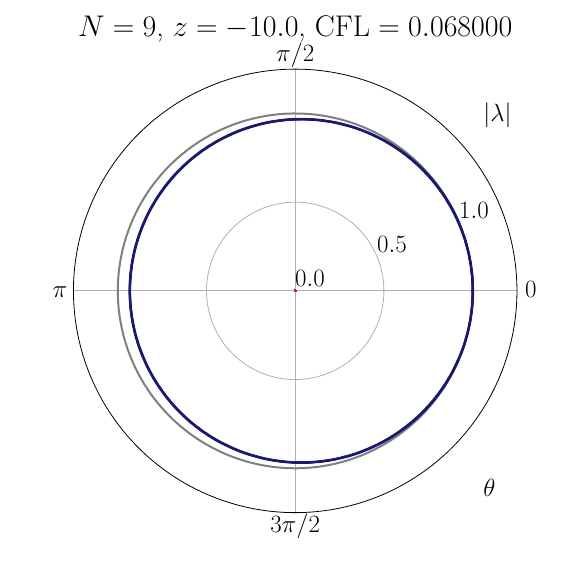}
\includegraphics[width=0.15\textwidth]{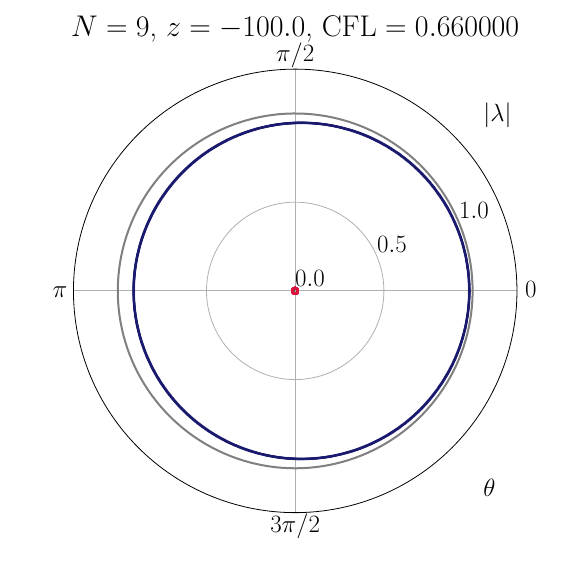}
\includegraphics[width=0.15\textwidth]{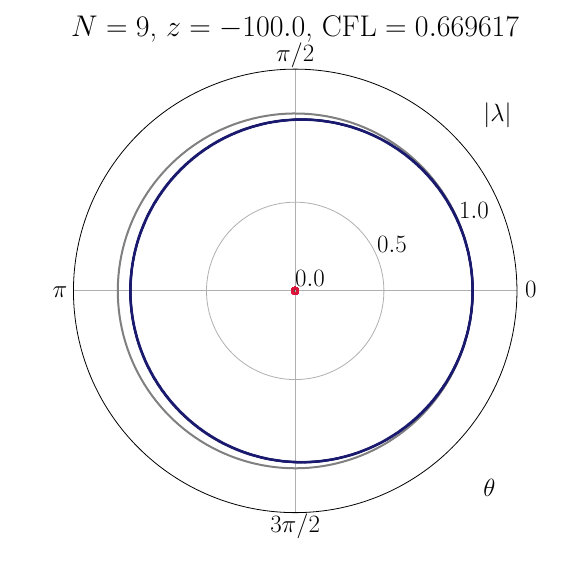}
\includegraphics[width=0.15\textwidth]{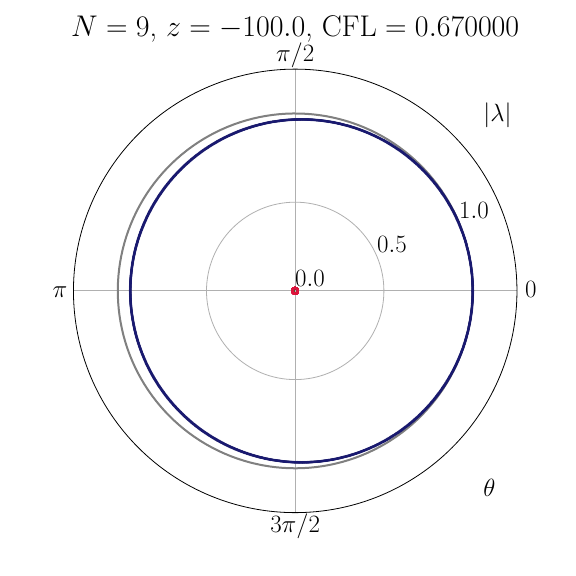}
\caption{%
The dependence of the absolute values $|\lambda_{k}|$ of the spectrum of the matrix $\mathrm{R}(\mathrm{CFL}, z, \theta)$ (\ref{eq:r_matrix_elems_expr_dup}) (eigenvalues $\lambda_{k} = \lambda_{k}(\mathrm{CFL}, z, \theta)$, $k = 0, \ldots, N$) of the evolution operator $R$ (\ref{eq:evol_oper_prop}) for a single time step $\Dtn{n}$ on phase $\theta = k\Dx$ for several selected values of the Courant number $\mathrm{CFL}$ for $z = -0.1$, $-1.0$, $-10.0$, $-100.0$ and for polynomial degrees $N = 7, \ldots, 9$ --- the polar plot with phase $\theta$ as angle and absolute value $|\lambda|$ as radius. The range of phase $\theta\in[0, 2\pi)$ is sampled on a uniform grid of $1000$ nodes. Legends for each row of the graphs are located on the left. The gray circle of unit radius defines the stability boundary. $\lambda_{\rm max}$ is the eigenvalue $\lambda_{k}$ with the largest absolute value $|\lambda_{k}|$. The Courant numbers $\mathrm{CFL}$ are taken both at the boundary value of the Courant number $\mathrm{CFL}_{\rm max}$ and at two values $\mathrm{CFL}$ near it --- in the guaranteed stability region $\mathrm{CFL} < \mathrm{CFL}_{\rm max}$ and in the guaranteed instability region $\mathrm{CFL} > \mathrm{CFL}_{\rm max}$, by rounding off at the second significant decimal place (the values $\mathrm{CFL}_{\rm max}$ are calculated further in this work and presented in Figure~\ref{fig:cfls_on_z}). \textit{Note}: the phase $\theta$ is not the phase $\phi_{k} \equiv \arg \lambda_{k}$ of eigenvalue $\lambda_{k}$; the dependence of eigenvalues $\lambda_{k}$ of the matrix $\mathrm{R}(\mathrm{CFL}, z, \theta)$ (\ref{eq:r_matrix_elems_expr_dup}) for these values $z$ and these polynomial degrees $N$ is presented in Figure~\ref{fig:spectrum_exact_eigvals_degrees_7_9}.
}
\label{fig:rhos_on_theta_degrees_7_9}
\end{figure}
\begin{figure}[h!]
\centering
\includegraphics[width=0.028125\textwidth]{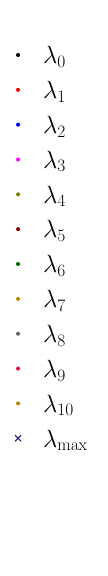}
\includegraphics[width=0.15\textwidth]{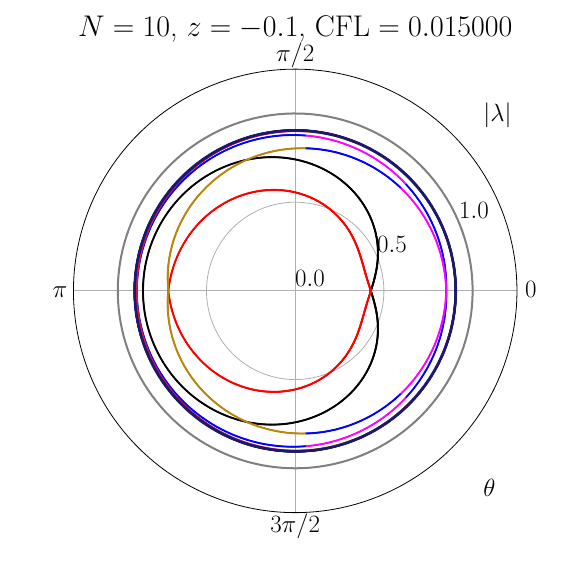}
\includegraphics[width=0.15\textwidth]{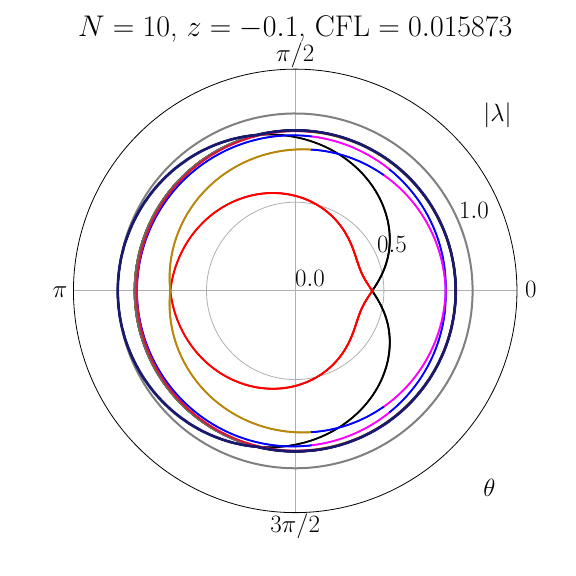}
\includegraphics[width=0.15\textwidth]{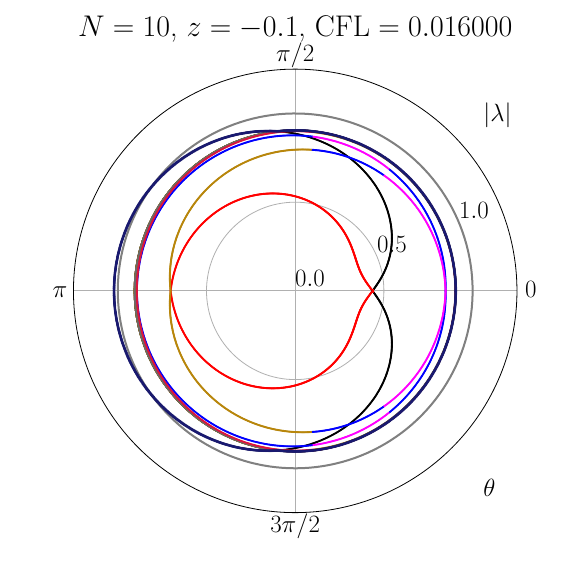}
\includegraphics[width=0.15\textwidth]{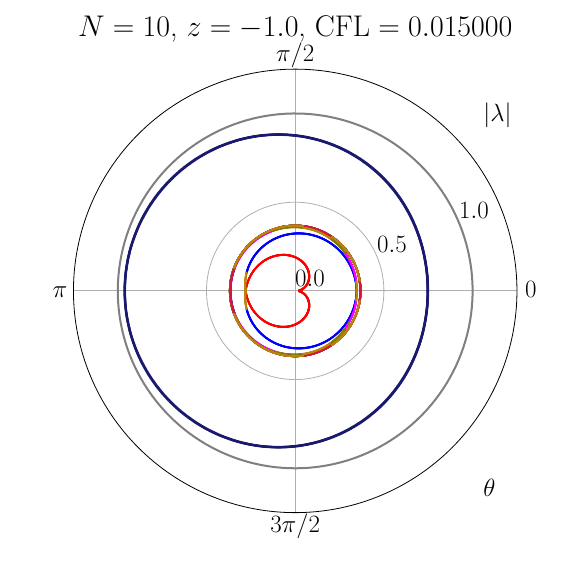}
\includegraphics[width=0.15\textwidth]{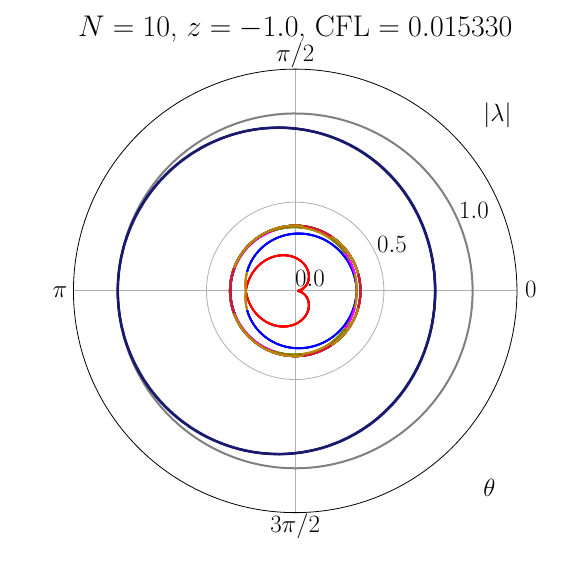}
\includegraphics[width=0.15\textwidth]{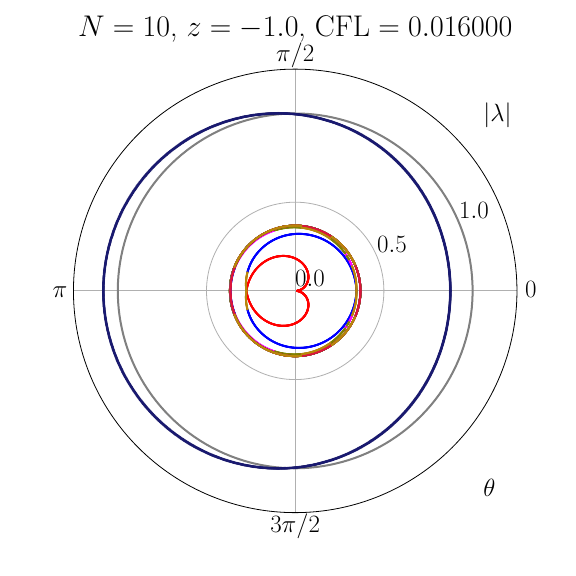}\\
\includegraphics[width=0.028125\textwidth]{rhos_on_theta_legend_N_10.png}
\includegraphics[width=0.15\textwidth]{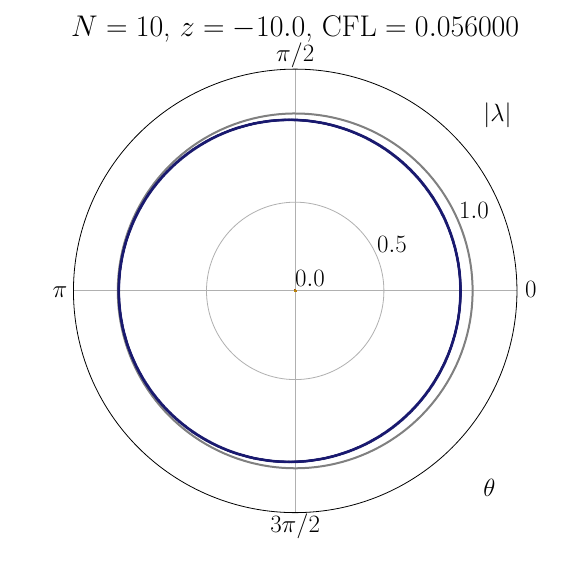}
\includegraphics[width=0.15\textwidth]{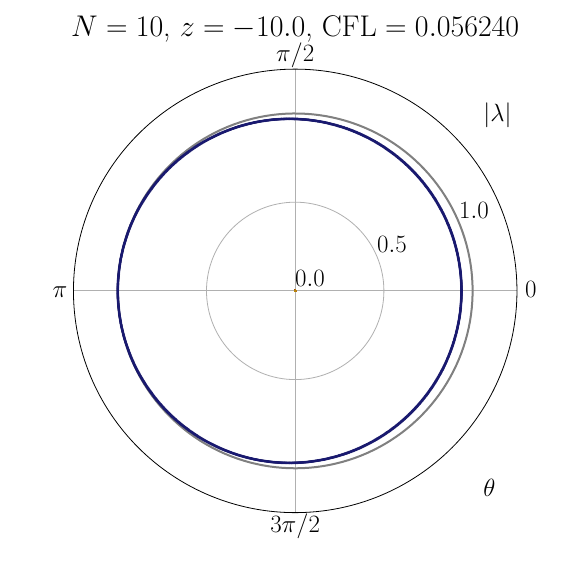}
\includegraphics[width=0.15\textwidth]{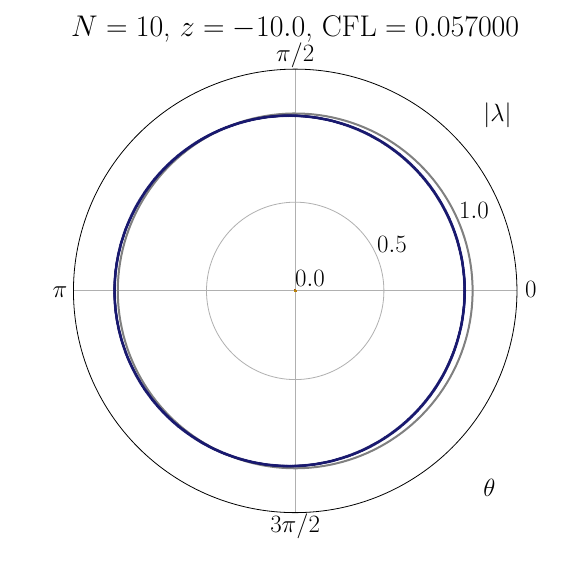}
\includegraphics[width=0.15\textwidth]{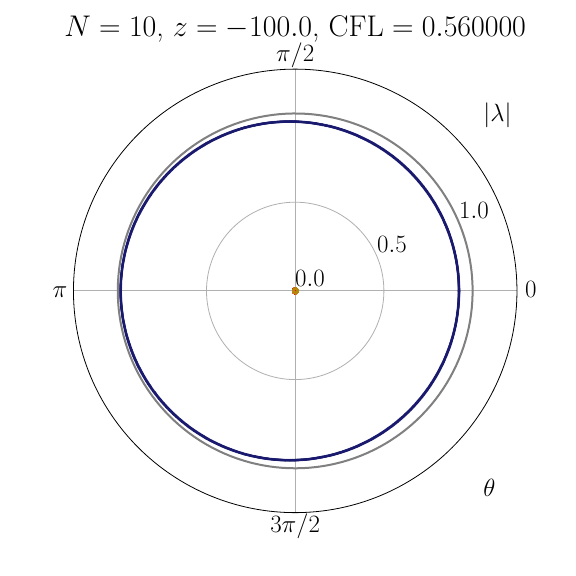}
\includegraphics[width=0.15\textwidth]{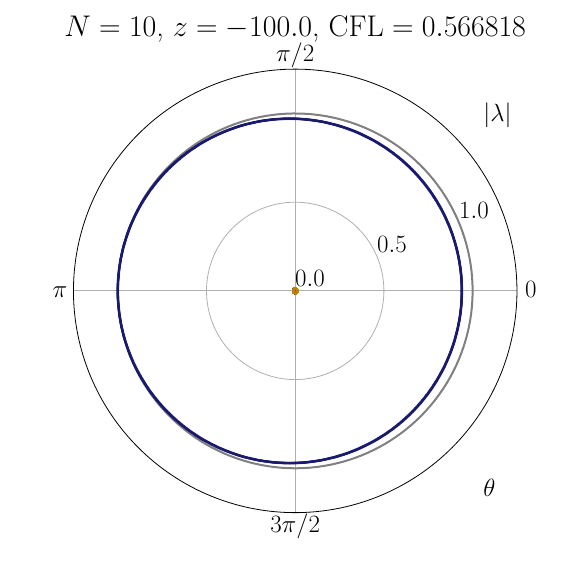}
\includegraphics[width=0.15\textwidth]{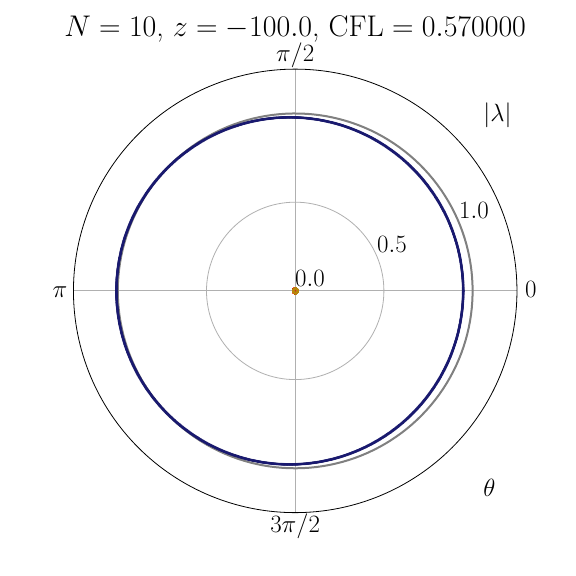}\\
\includegraphics[width=0.028125\textwidth]{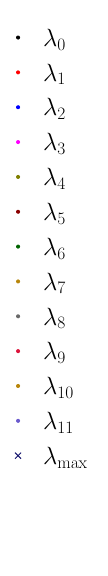}
\includegraphics[width=0.15\textwidth]{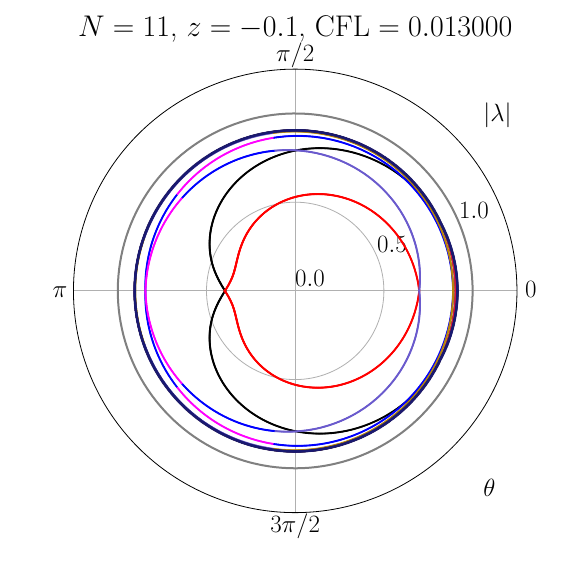}
\includegraphics[width=0.15\textwidth]{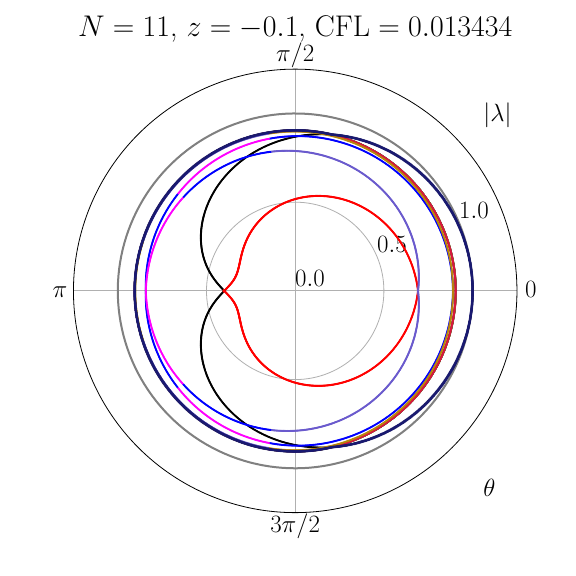}
\includegraphics[width=0.15\textwidth]{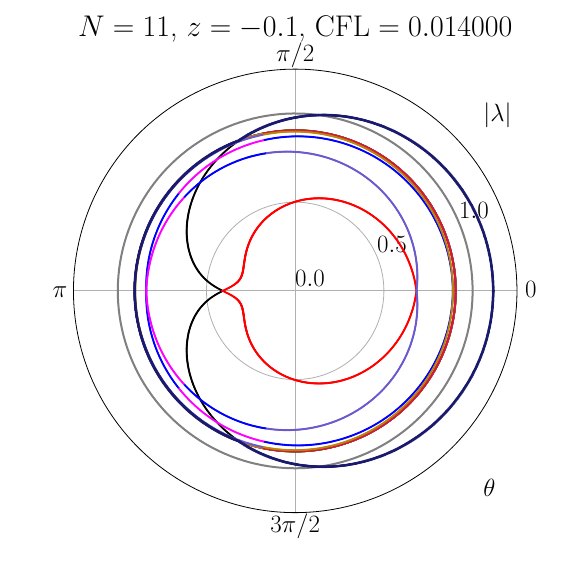}
\includegraphics[width=0.15\textwidth]{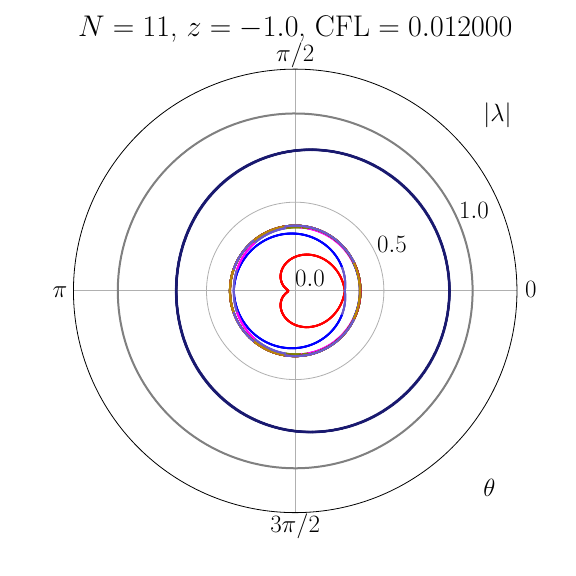}
\includegraphics[width=0.15\textwidth]{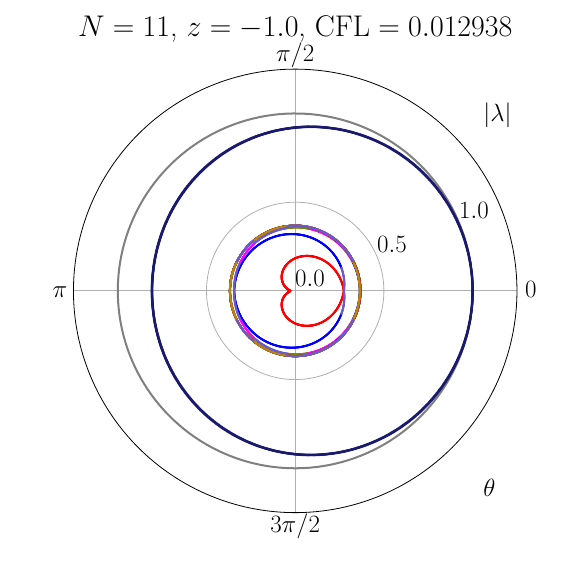}
\includegraphics[width=0.15\textwidth]{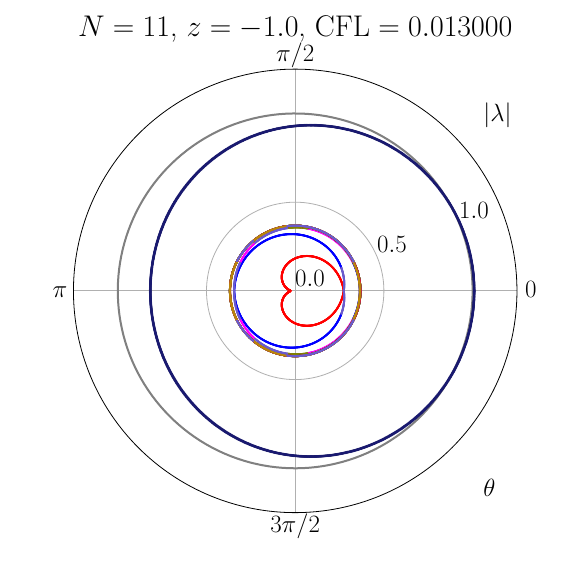}\\
\includegraphics[width=0.028125\textwidth]{rhos_on_theta_legend_N_11.png}
\includegraphics[width=0.15\textwidth]{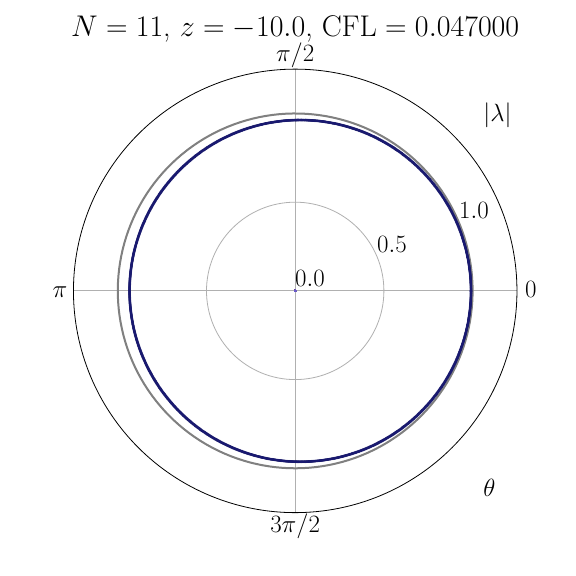}
\includegraphics[width=0.15\textwidth]{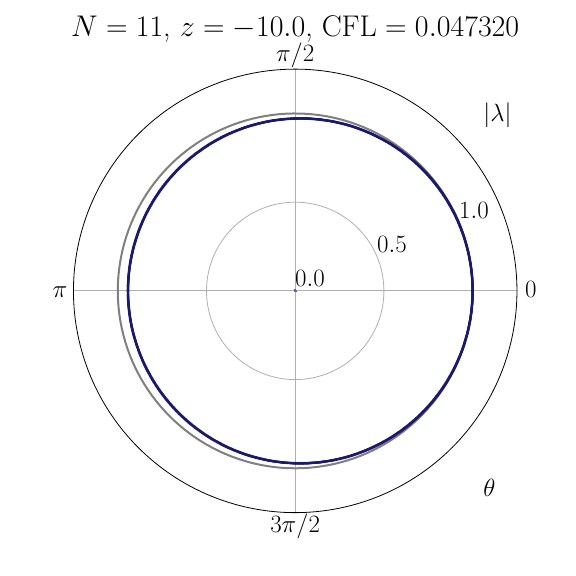}
\includegraphics[width=0.15\textwidth]{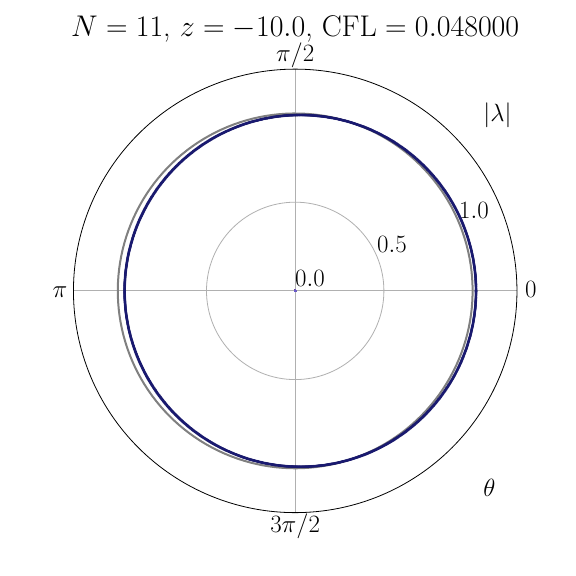}
\includegraphics[width=0.15\textwidth]{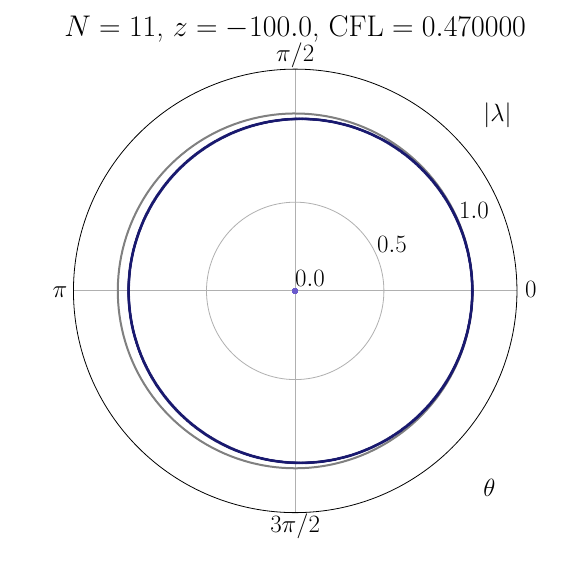}
\includegraphics[width=0.15\textwidth]{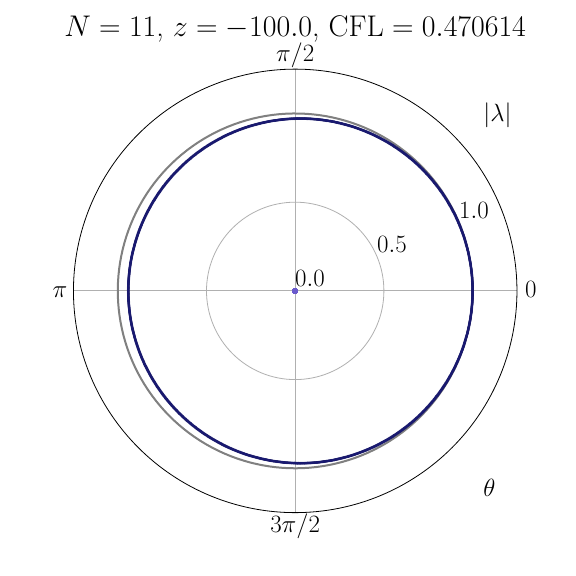}
\includegraphics[width=0.15\textwidth]{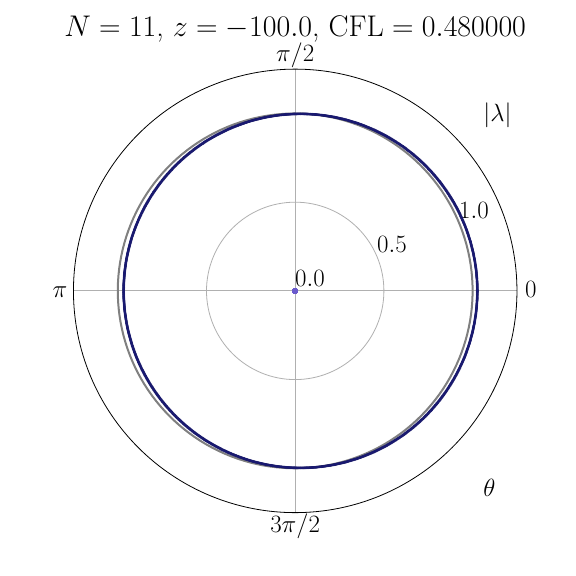}\\
\includegraphics[width=0.028125\textwidth]{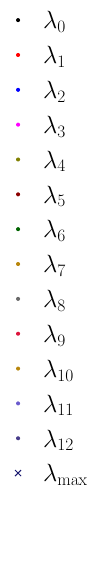}
\includegraphics[width=0.15\textwidth]{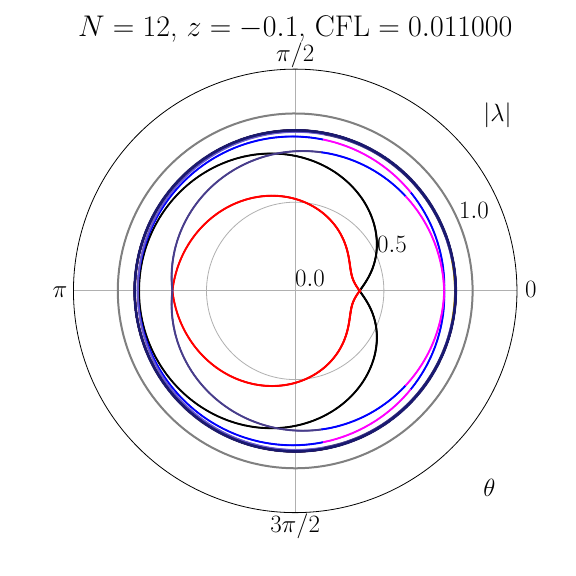}
\includegraphics[width=0.15\textwidth]{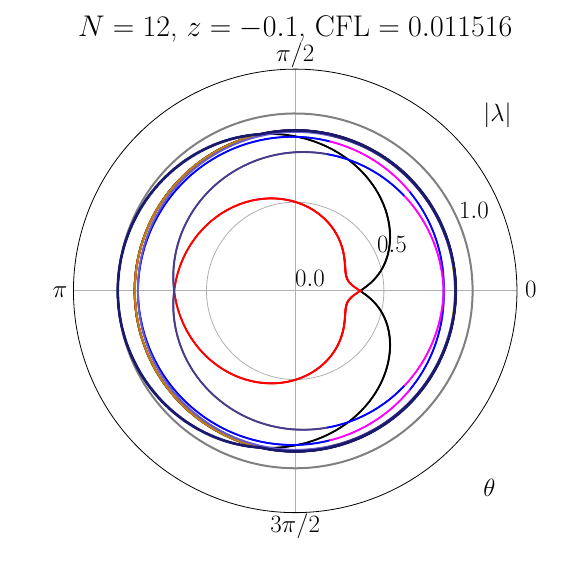}
\includegraphics[width=0.15\textwidth]{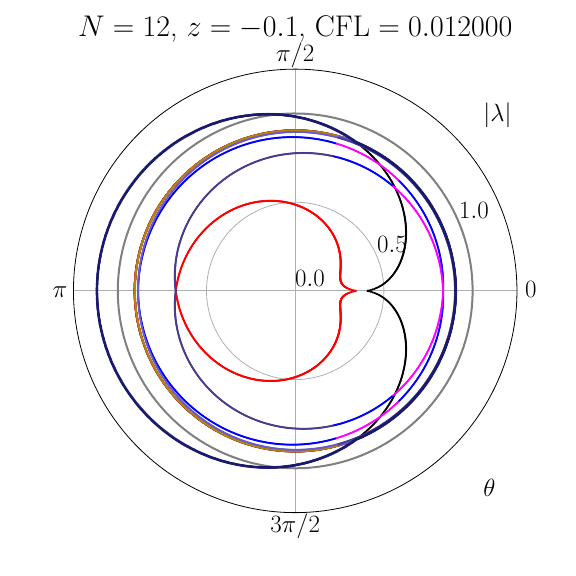}
\includegraphics[width=0.15\textwidth]{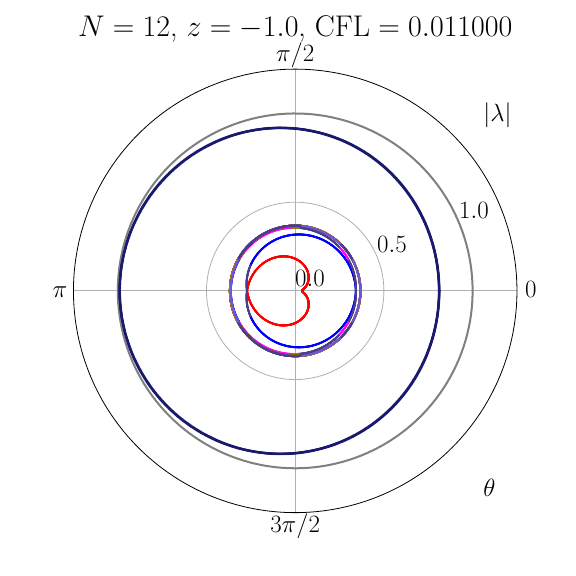}
\includegraphics[width=0.15\textwidth]{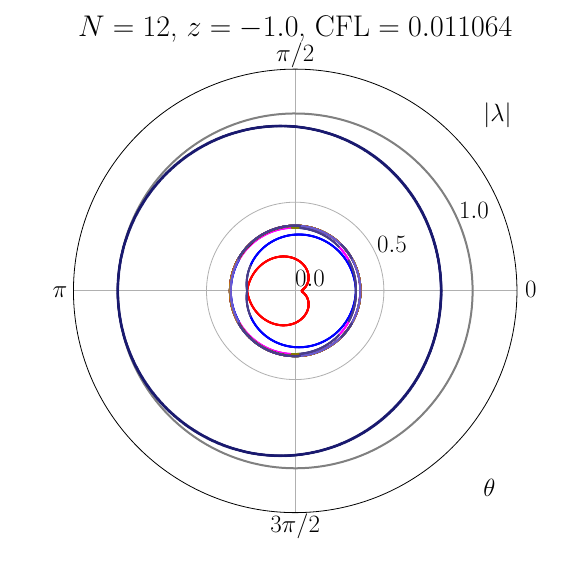}
\includegraphics[width=0.15\textwidth]{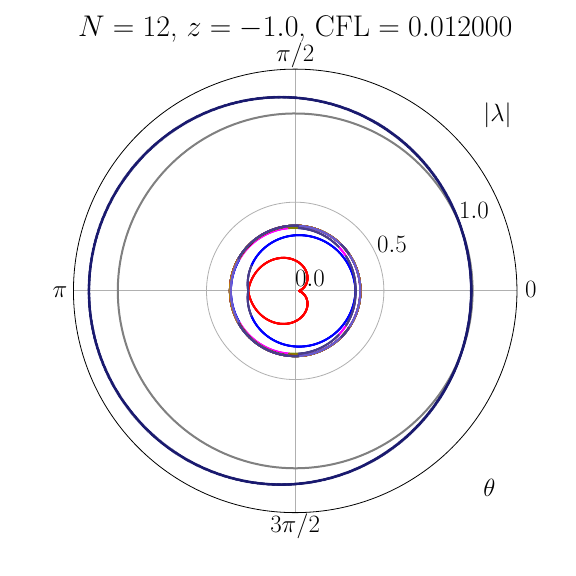}\\
\includegraphics[width=0.028125\textwidth]{rhos_on_theta_legend_N_12.png}
\includegraphics[width=0.15\textwidth]{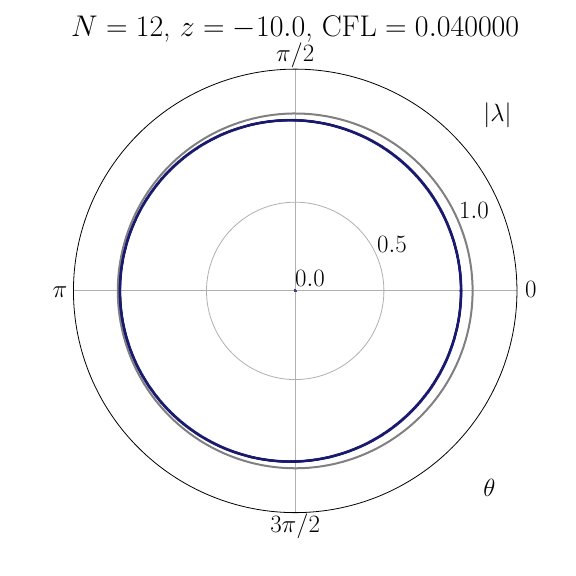}
\includegraphics[width=0.15\textwidth]{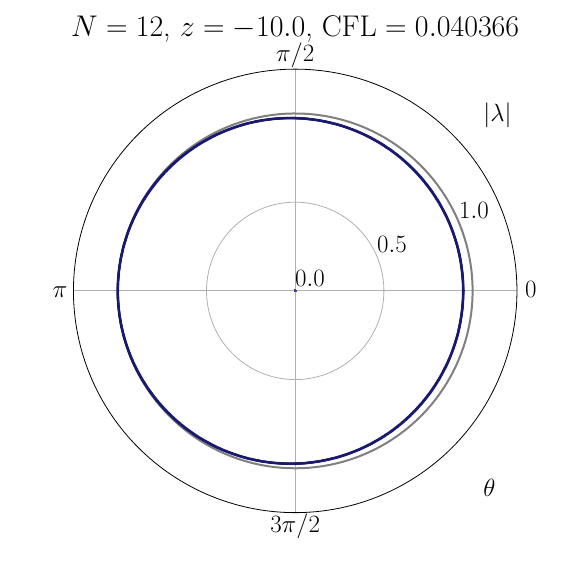}
\includegraphics[width=0.15\textwidth]{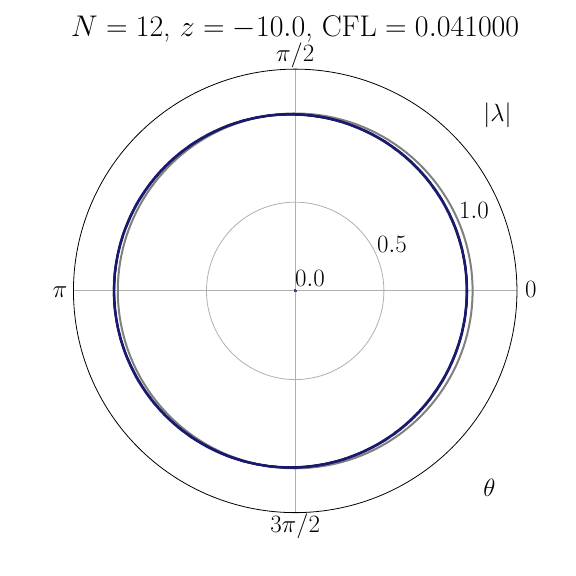}
\includegraphics[width=0.15\textwidth]{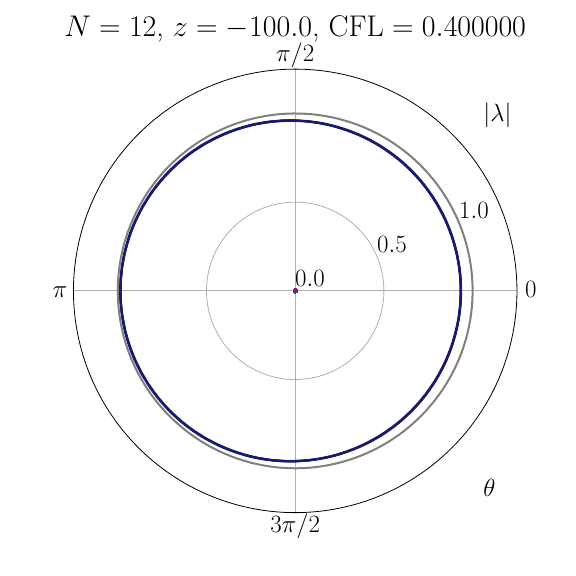}
\includegraphics[width=0.15\textwidth]{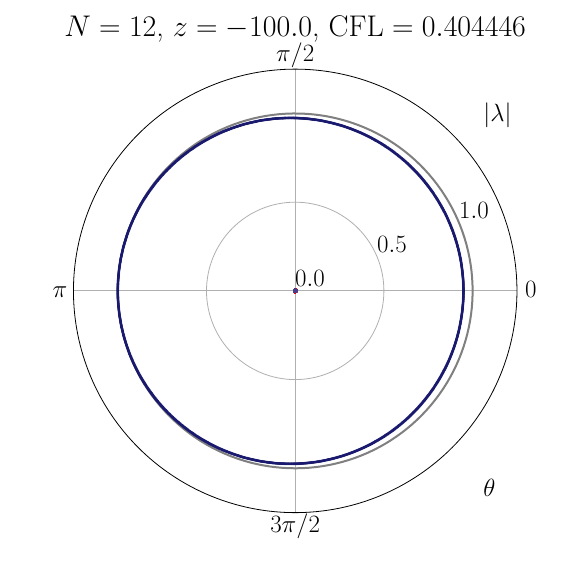}
\includegraphics[width=0.15\textwidth]{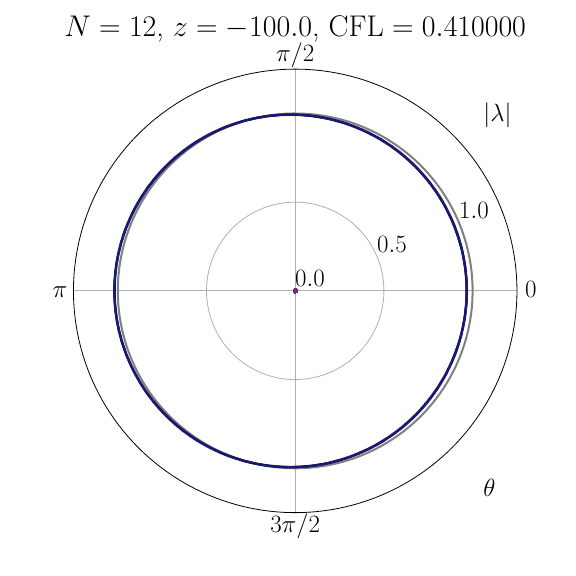}
\caption{%
The dependence of the absolute values $|\lambda_{k}|$ of the spectrum of the matrix $\mathrm{R}(\mathrm{CFL}, z, \theta)$ (\ref{eq:r_matrix_elems_expr_dup}) (eigenvalues $\lambda_{k} = \lambda_{k}(\mathrm{CFL}, z, \theta)$, $k = 0, \ldots, N$) of the evolution operator $R$ (\ref{eq:evol_oper_prop}) for a single time step $\Dtn{n}$ on phase $\theta = k\Dx$ for several selected values of the Courant number $\mathrm{CFL}$ for $z = -0.1$, $-1.0$, $-10.0$, $-100.0$ and for polynomial degrees $N = 10, \ldots, 12$ --- the polar plot with phase $\theta$ as angle and absolute value $|\lambda|$ as radius. The range of phase $\theta\in[0, 2\pi)$ is sampled on a uniform grid of $1000$ nodes. Legends for each row of the graphs are located on the left. The gray circle of unit radius defines the stability boundary. $\lambda_{\rm max}$ is the eigenvalue $\lambda_{k}$ with the largest absolute value $|\lambda_{k}|$. The Courant numbers $\mathrm{CFL}$ are taken both at the boundary value of the Courant number $\mathrm{CFL}_{\rm max}$ and at two values $\mathrm{CFL}$ near it --- in the guaranteed stability region $\mathrm{CFL} < \mathrm{CFL}_{\rm max}$ and in the guaranteed instability region $\mathrm{CFL} > \mathrm{CFL}_{\rm max}$, by rounding off at the second significant decimal place (the values $\mathrm{CFL}_{\rm max}$ are calculated further in this work and presented in Figure~\ref{fig:cfls_on_z}). \textit{Note}: the phase $\theta$ is not the phase $\phi_{k} \equiv \arg \lambda_{k}$ of eigenvalue $\lambda_{k}$; the dependence of eigenvalues $\lambda_{k}$ of the matrix $\mathrm{R}(\mathrm{CFL}, z, \theta)$ (\ref{eq:r_matrix_elems_expr_dup}) for these values $z$ and these polynomial degrees $N$ is presented in Figure~\ref{fig:spectrum_exact_eigvals_degrees_10_12}.
}
\label{fig:rhos_on_theta_degrees_10_12}
\end{figure}

\begin{figure}[h!]
\centering
\includegraphics[width=0.028125\textwidth]{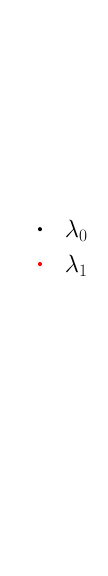}
\includegraphics[width=0.15\textwidth]{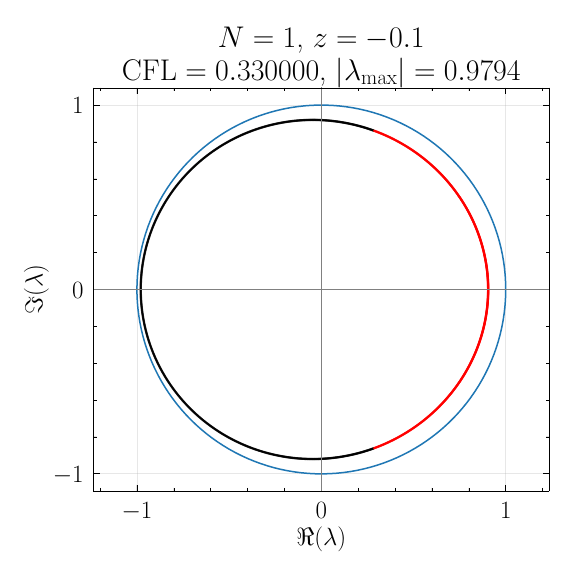}
\includegraphics[width=0.15\textwidth]{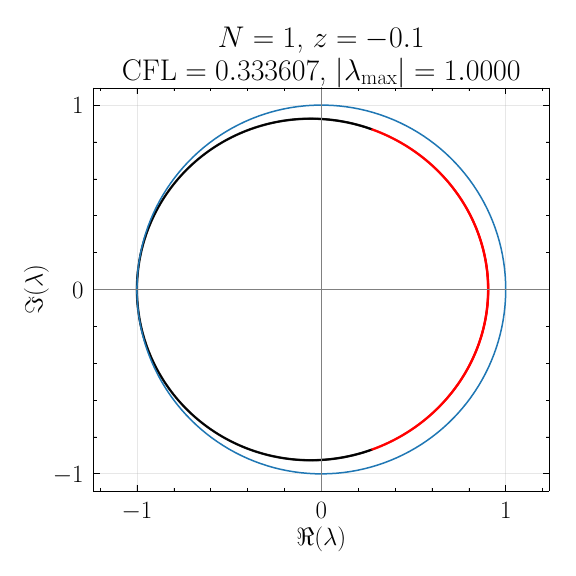}
\includegraphics[width=0.15\textwidth]{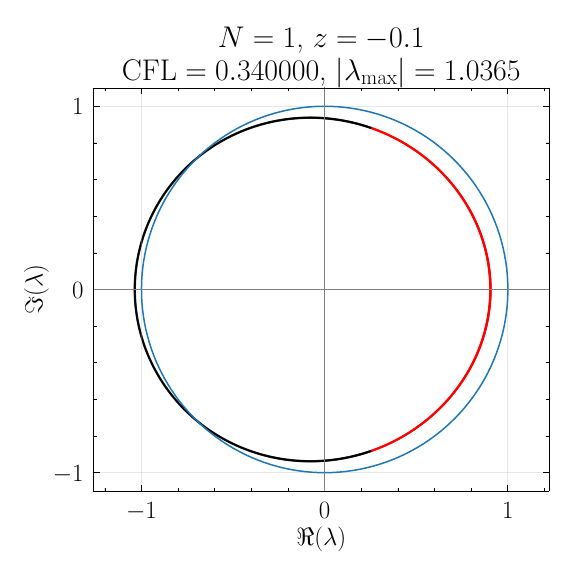}
\includegraphics[width=0.15\textwidth]{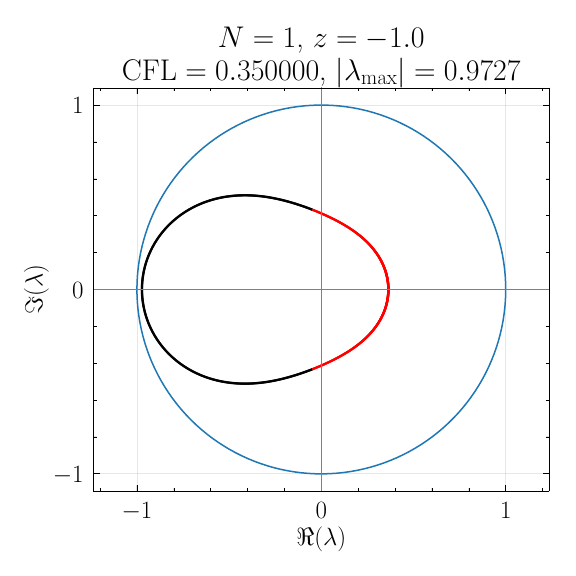}
\includegraphics[width=0.15\textwidth]{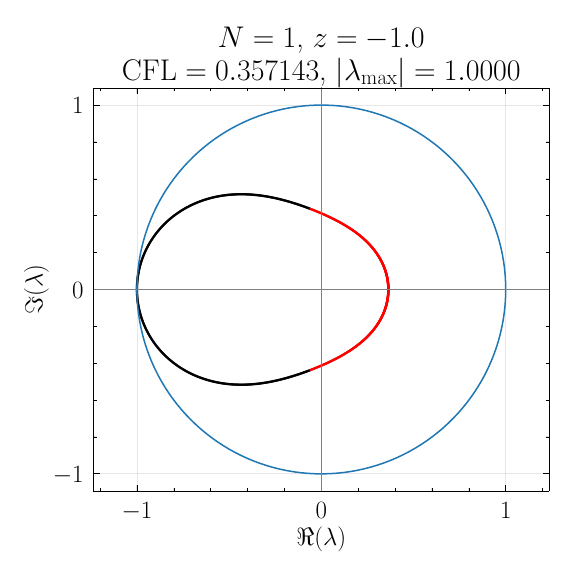}
\includegraphics[width=0.15\textwidth]{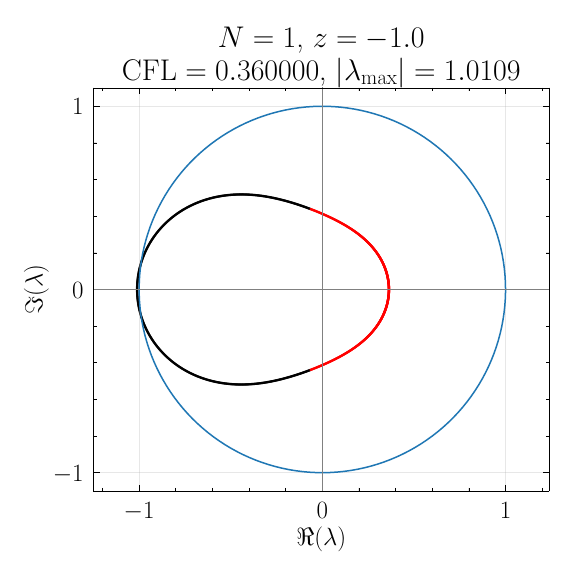}\\
\includegraphics[width=0.028125\textwidth]{spectrum_exact_eigvals_legend_N_1.png}
\includegraphics[width=0.15\textwidth]{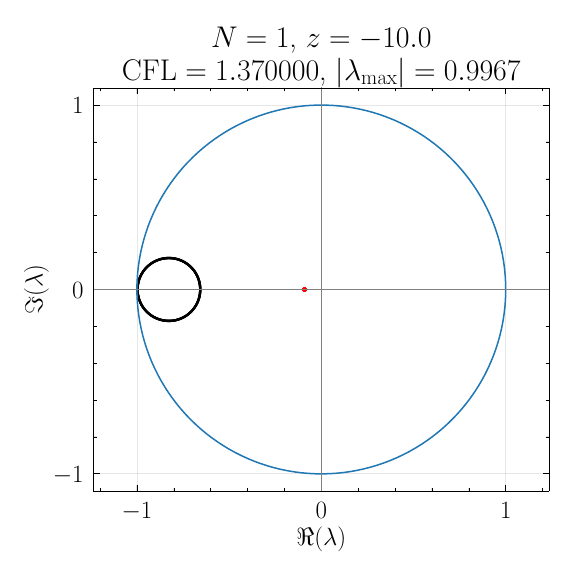}
\includegraphics[width=0.15\textwidth]{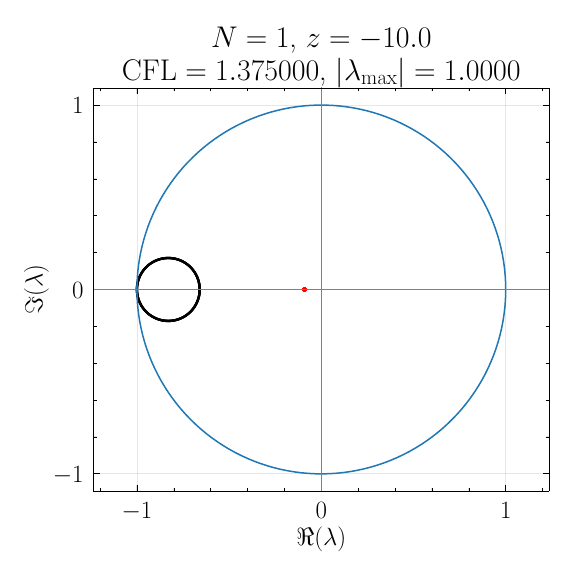}
\includegraphics[width=0.15\textwidth]{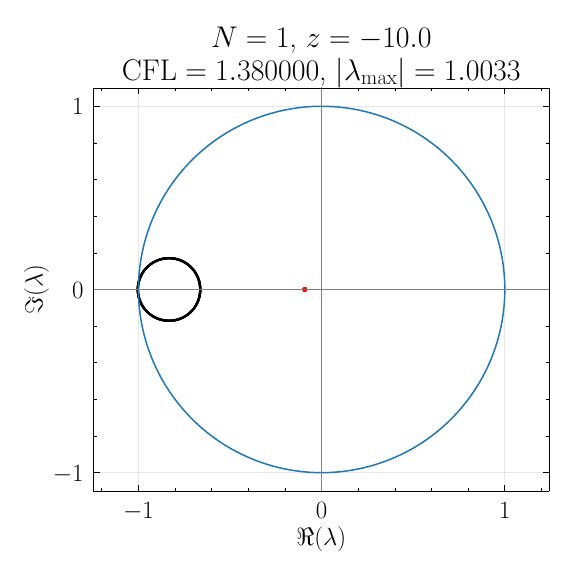}
\includegraphics[width=0.15\textwidth]{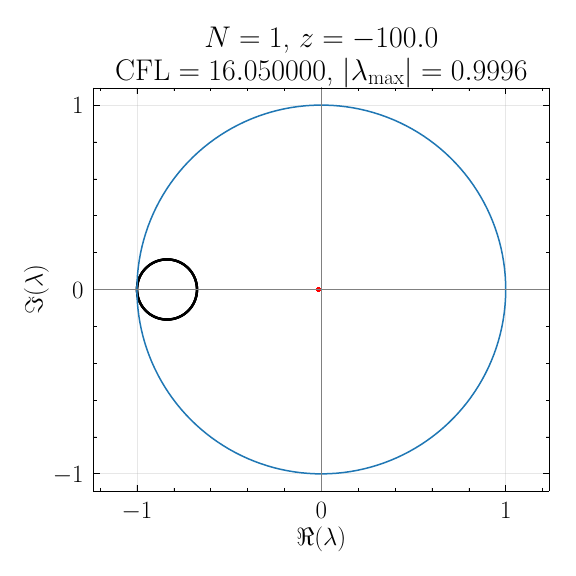}
\includegraphics[width=0.15\textwidth]{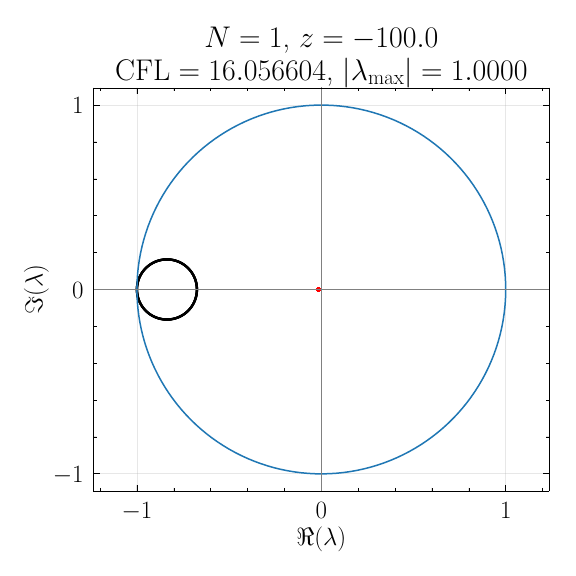}
\includegraphics[width=0.15\textwidth]{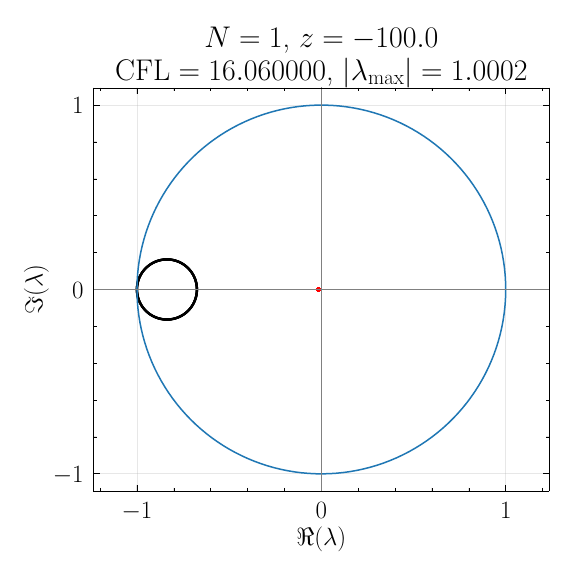}\\
\includegraphics[width=0.028125\textwidth]{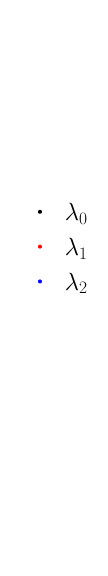}
\includegraphics[width=0.15\textwidth]{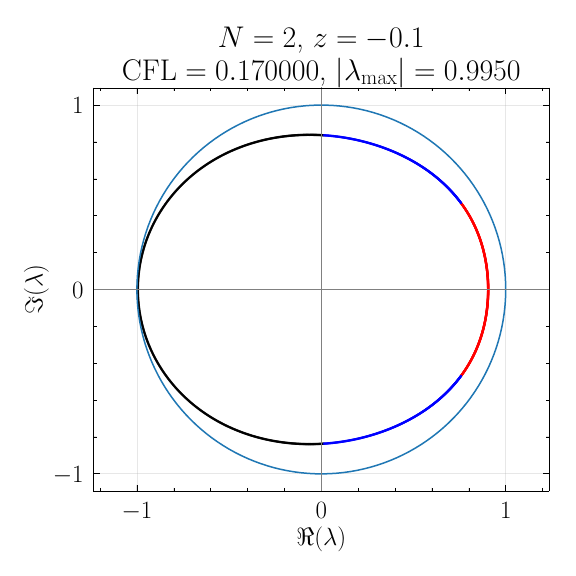}
\includegraphics[width=0.15\textwidth]{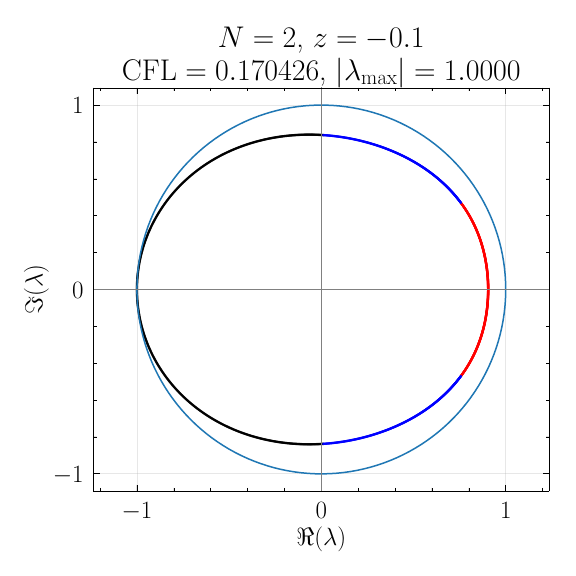}
\includegraphics[width=0.15\textwidth]{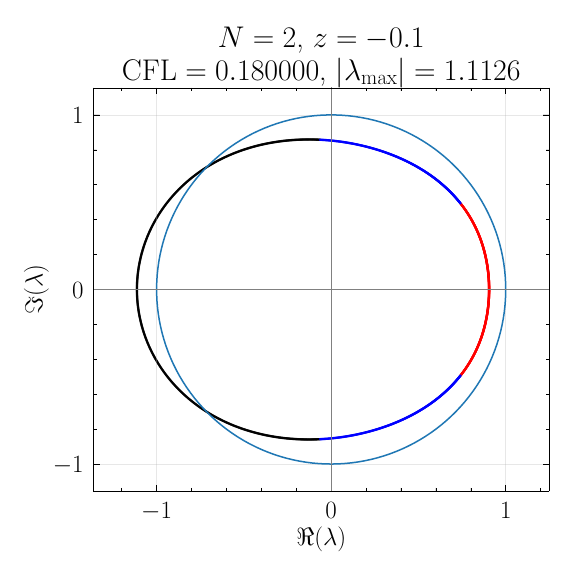}
\includegraphics[width=0.15\textwidth]{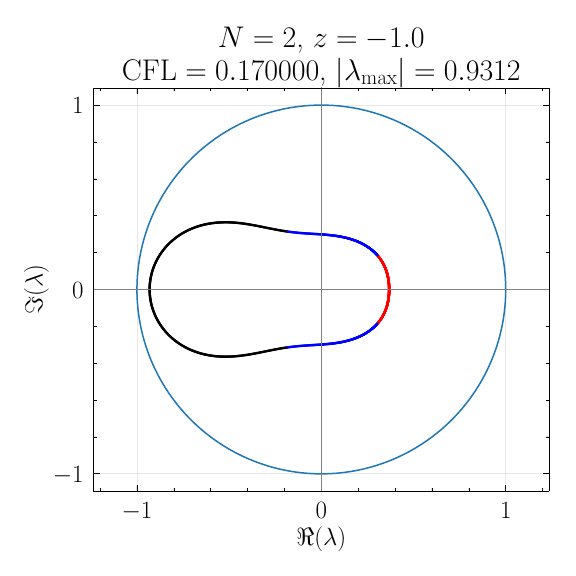}
\includegraphics[width=0.15\textwidth]{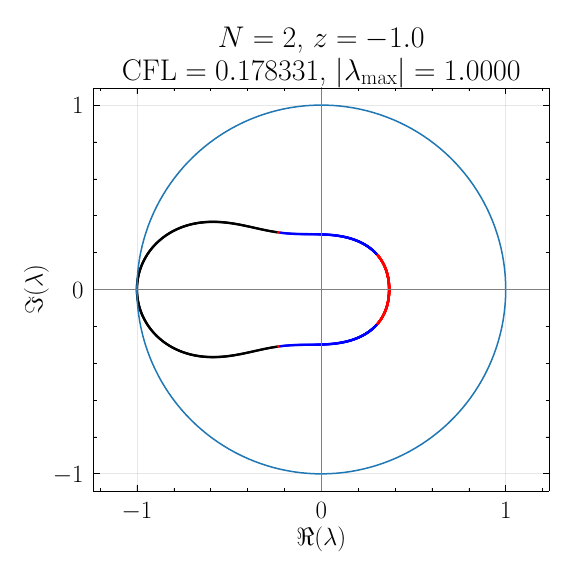}
\includegraphics[width=0.15\textwidth]{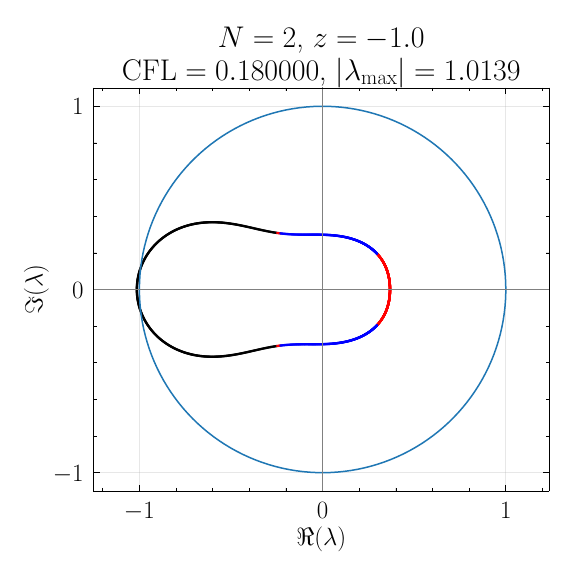}\\
\includegraphics[width=0.028125\textwidth]{spectrum_exact_eigvals_legend_N_2.png}
\includegraphics[width=0.15\textwidth]{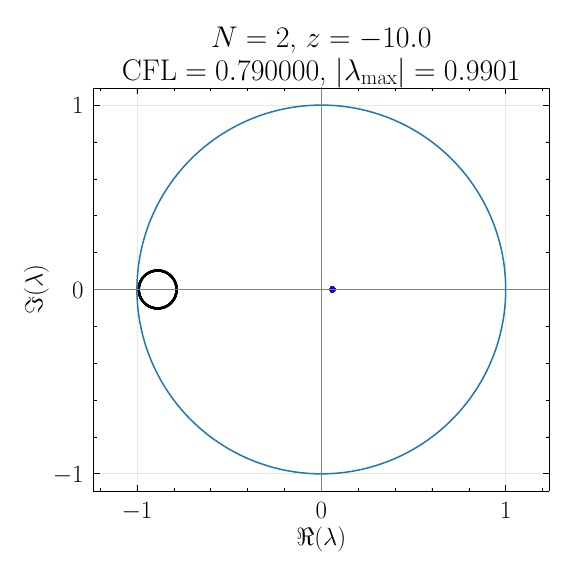}
\includegraphics[width=0.15\textwidth]{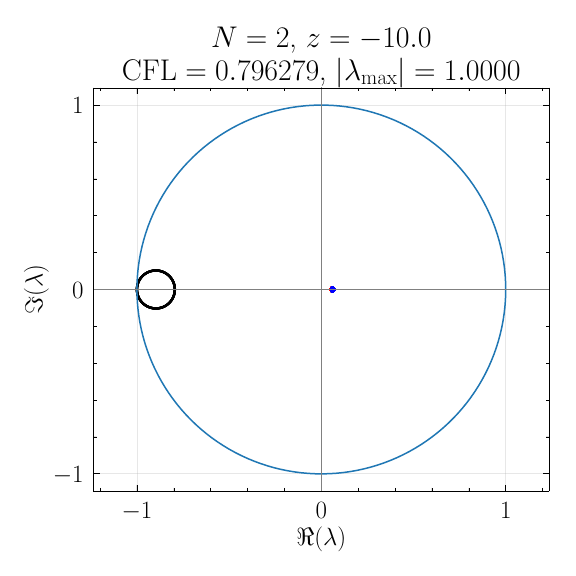}
\includegraphics[width=0.15\textwidth]{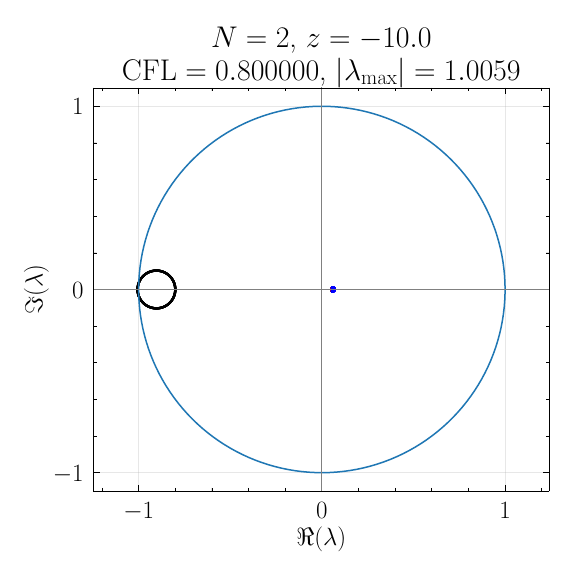}
\includegraphics[width=0.15\textwidth]{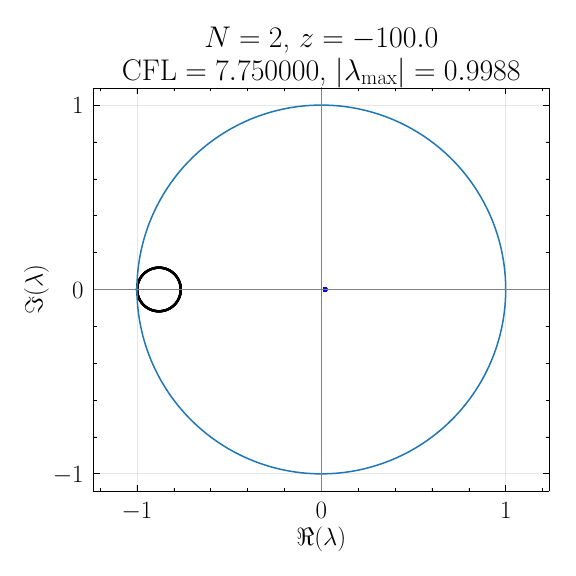}
\includegraphics[width=0.15\textwidth]{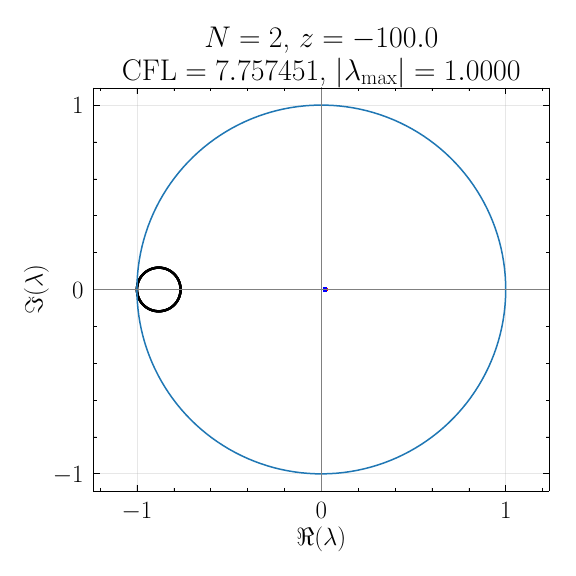}
\includegraphics[width=0.15\textwidth]{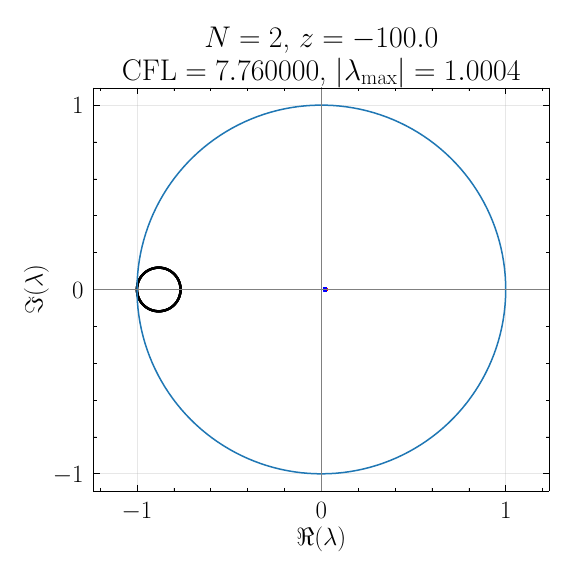}\\
\includegraphics[width=0.028125\textwidth]{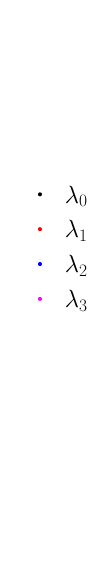}
\includegraphics[width=0.15\textwidth]{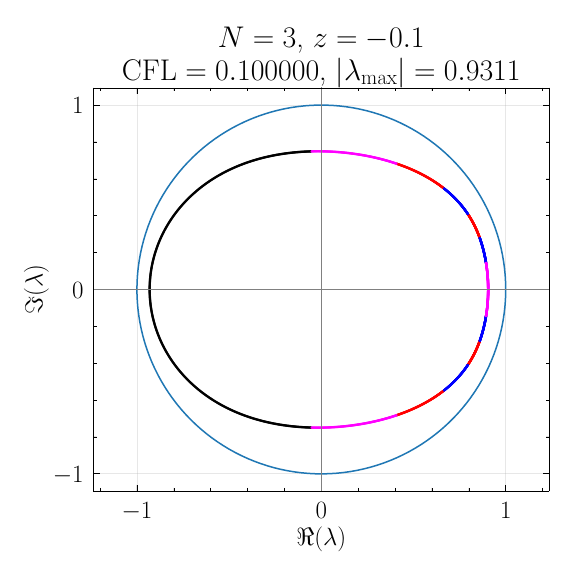}
\includegraphics[width=0.15\textwidth]{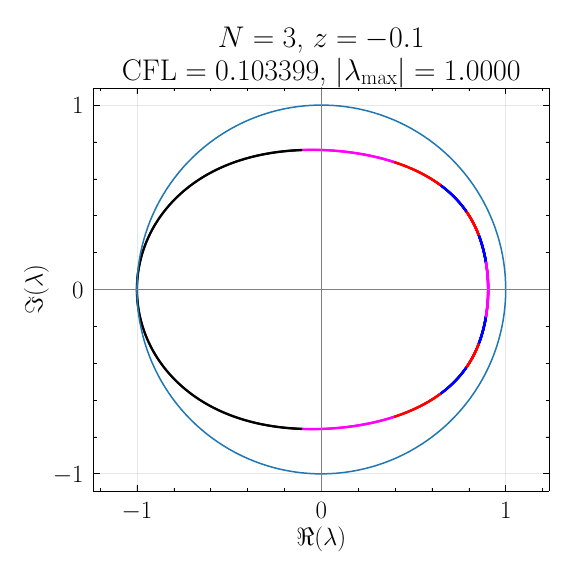}
\includegraphics[width=0.15\textwidth]{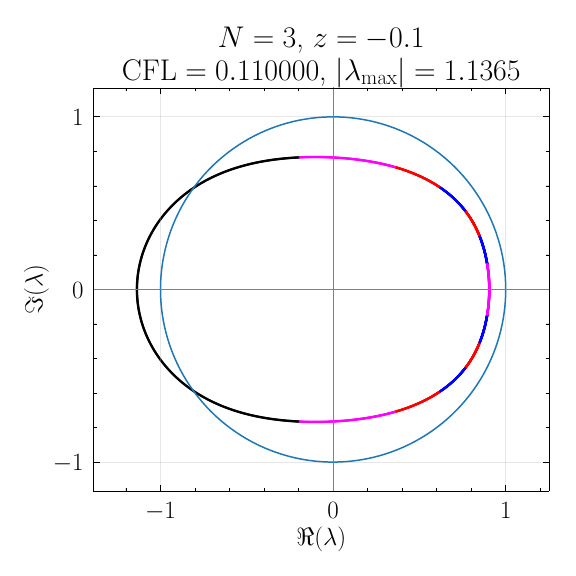}
\includegraphics[width=0.15\textwidth]{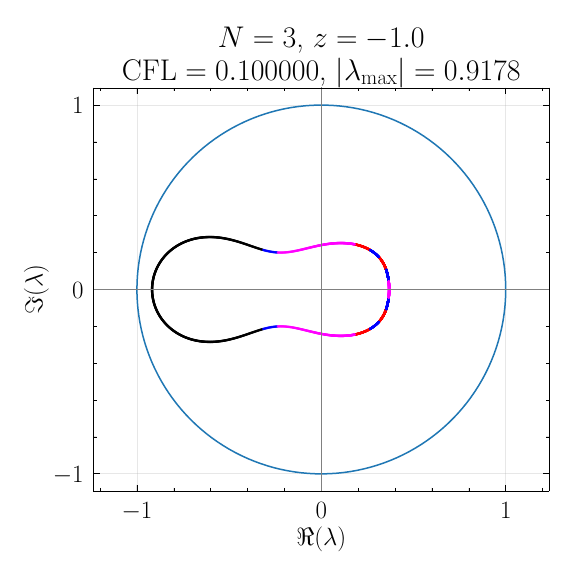}
\includegraphics[width=0.15\textwidth]{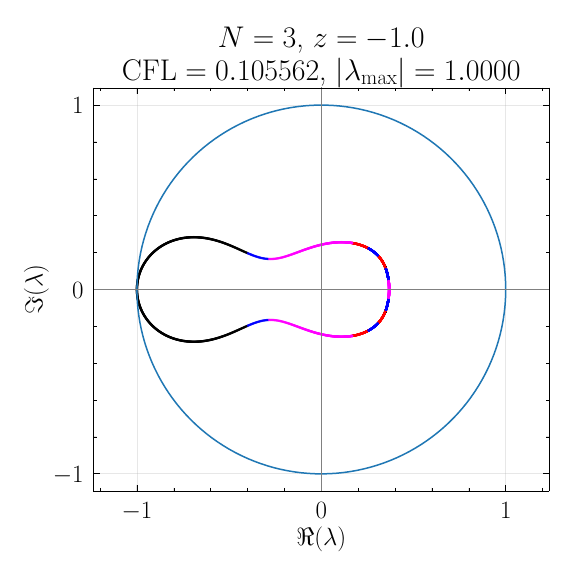}
\includegraphics[width=0.15\textwidth]{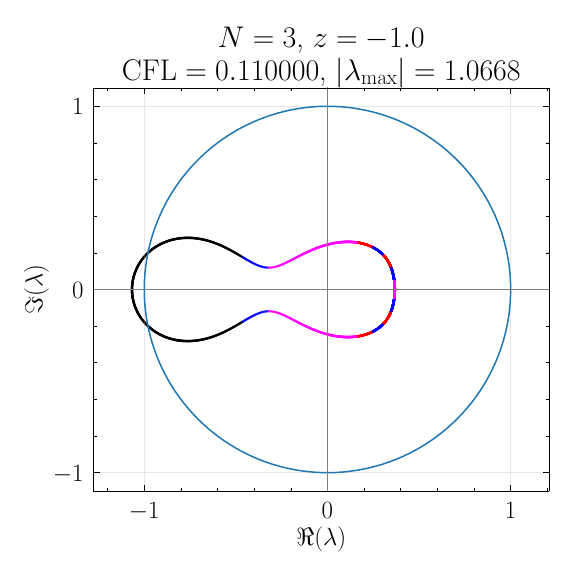}\\
\includegraphics[width=0.028125\textwidth]{spectrum_exact_eigvals_legend_N_3.png}
\includegraphics[width=0.15\textwidth]{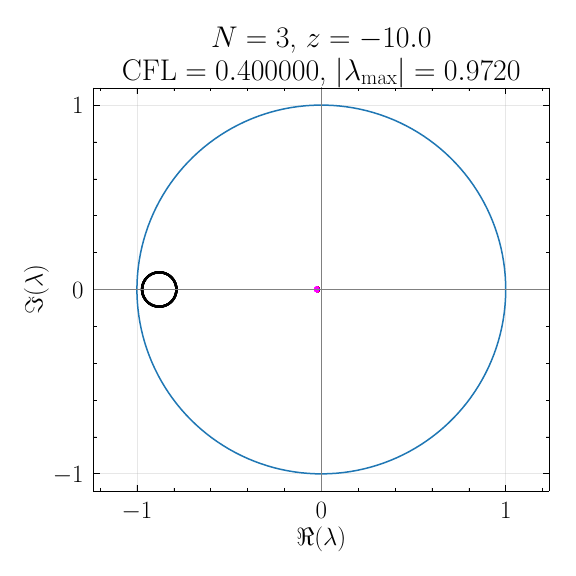}
\includegraphics[width=0.15\textwidth]{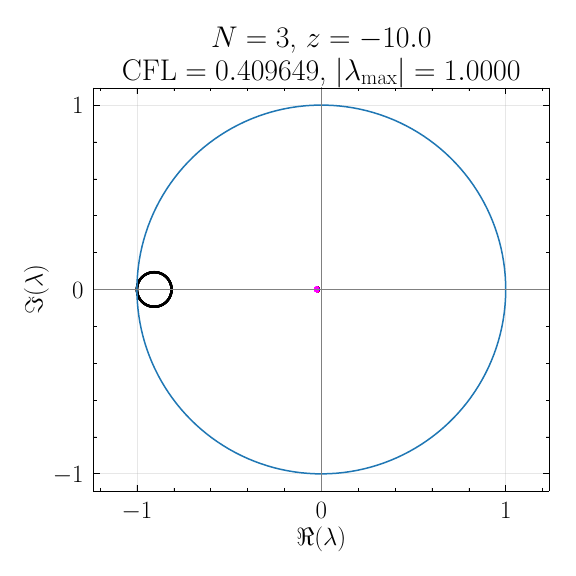}
\includegraphics[width=0.15\textwidth]{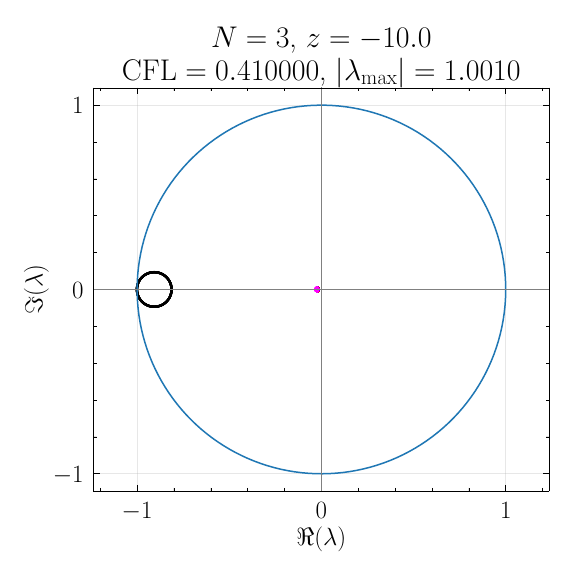}
\includegraphics[width=0.15\textwidth]{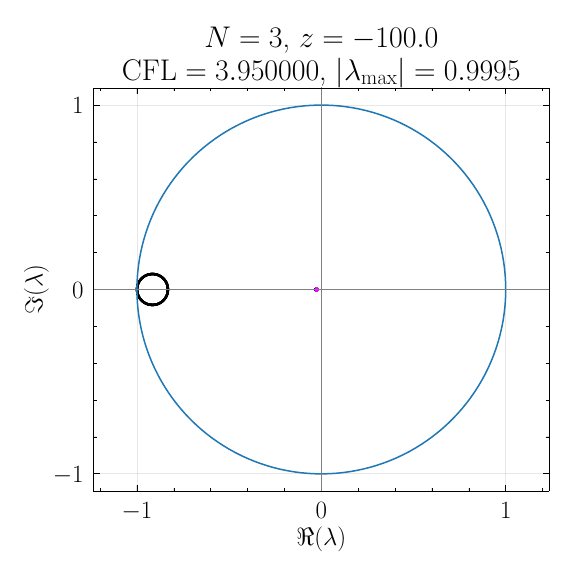}
\includegraphics[width=0.15\textwidth]{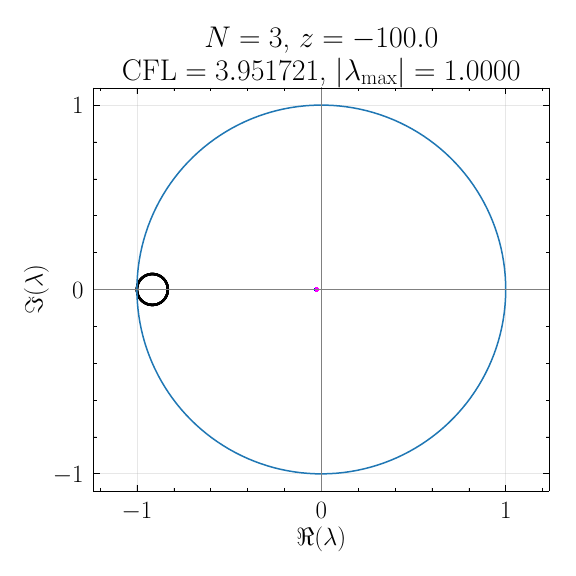}
\includegraphics[width=0.15\textwidth]{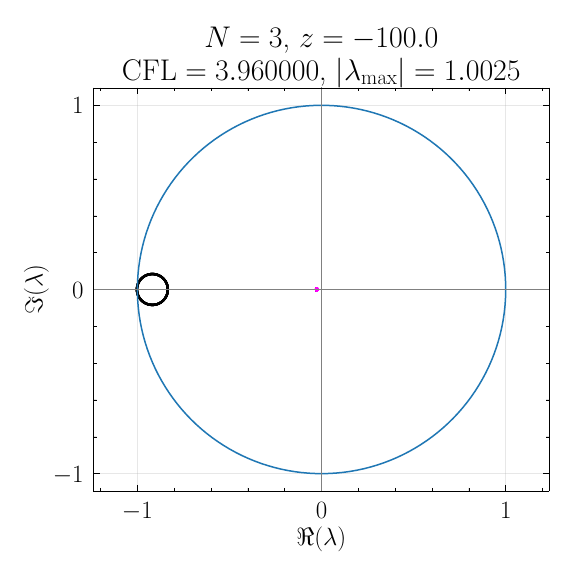}
\caption{%
Spectrum of the matrix $\mathrm{R}(\mathrm{CFL}, z, \theta)$ (\ref{eq:r_matrix_elems_expr_dup}) (eigenvalues $\lambda_{k} = \lambda_{k}(\mathrm{CFL}, z, \theta)$, $k = 0, \ldots, N$) of the evolution operator $R$ (\ref{eq:evol_oper_prop}) for a single time step $\Dtn{n}$ for several selected values of the Courant number $\mathrm{CFL}$ for $z = -0.1$, $-1.0$, $-10.0$, $-100.0$ and for polynomial degrees $N = 1, \ldots, 3$. The range of phase $\theta\in[0, 2\pi)$ is sampled on a uniform grid of $1000$ nodes. The gray circle of unit radius defines the stability boundary. Legends for each row of the graphs are located on the left. The gray circle of unit radius defines the stability boundary. The Courant numbers $\mathrm{CFL}$ are taken both at the boundary value of the Courant number $\mathrm{CFL}_{\rm max}$ and at two values $\mathrm{CFL}$ near it --- in the guaranteed stability region $\mathrm{CFL} < \mathrm{CFL}_{\rm max}$ and in the guaranteed instability region $\mathrm{CFL} > \mathrm{CFL}_{\rm max}$, by rounding off at the second significant decimal place (the values $\mathrm{CFL}_{\rm max}$ are calculated further in this work and presented in Figure~\ref{fig:cfls_on_z}). \textit{Note}: the phase $\arg \lambda_{k}$ of eigenvalue $\lambda_{k}$ is not the phase $\theta$; the dependence of the absolute value $|\lambda_{k}|$ on phase $\theta$ for these polynomial degrees $N$ is presented in Figure~\ref{fig:rhos_on_theta_degrees_1_3}.
}
\label{fig:spectrum_exact_eigvals_degrees_1_3}
\end{figure}
\begin{figure}[h!]
\centering
\includegraphics[width=0.028125\textwidth]{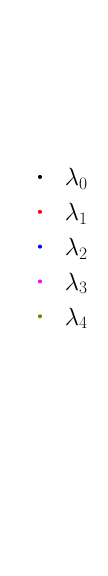}
\includegraphics[width=0.15\textwidth]{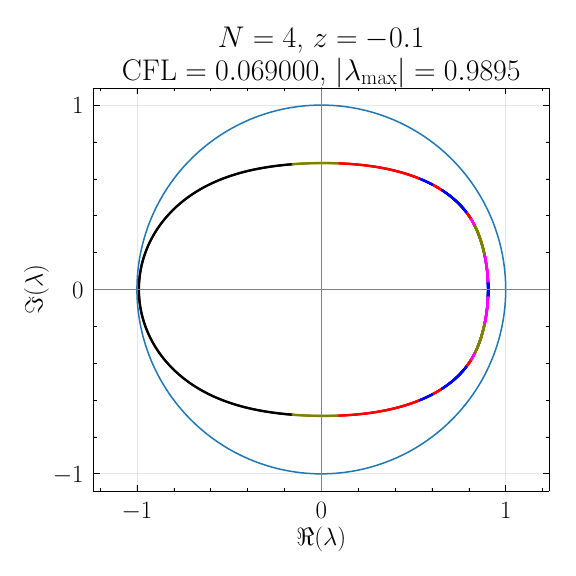}
\includegraphics[width=0.15\textwidth]{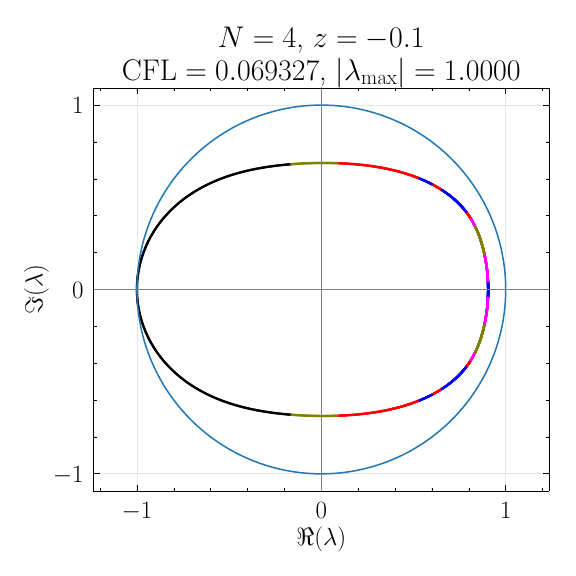}
\includegraphics[width=0.15\textwidth]{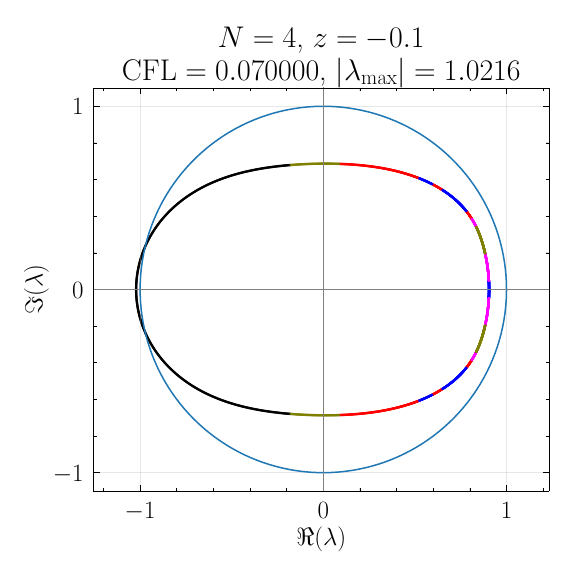}
\includegraphics[width=0.15\textwidth]{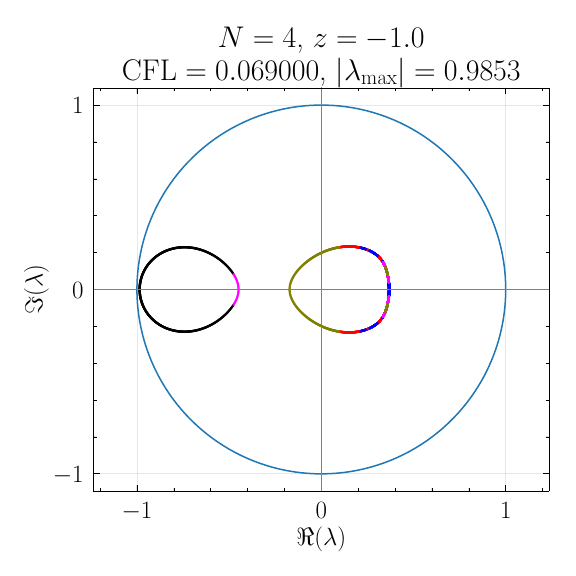}
\includegraphics[width=0.15\textwidth]{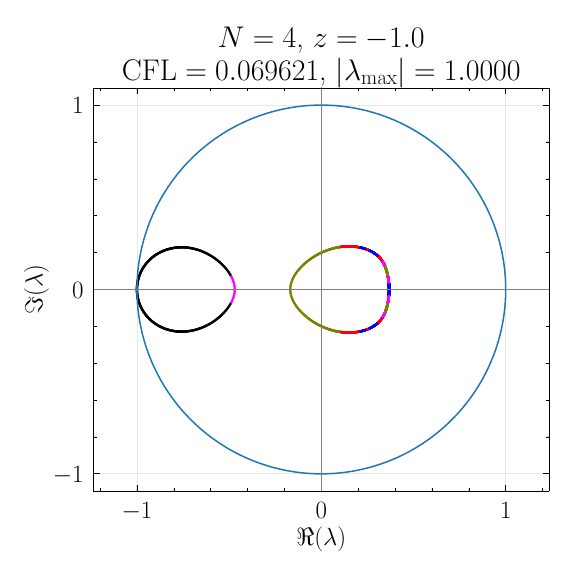}
\includegraphics[width=0.15\textwidth]{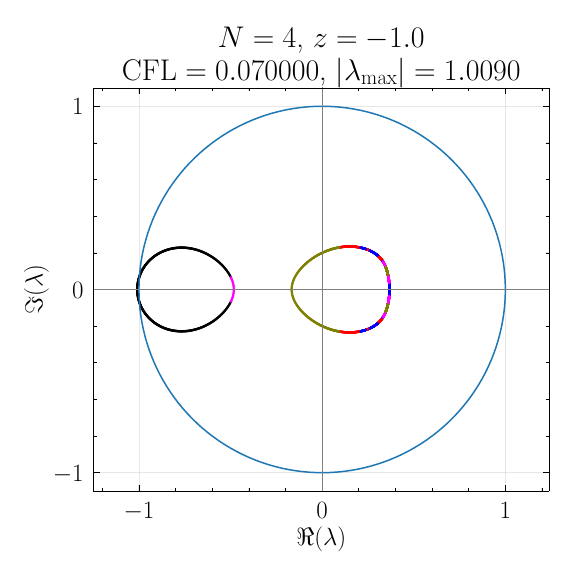}\\
\includegraphics[width=0.028125\textwidth]{spectrum_exact_eigvals_legend_N_4.png}
\includegraphics[width=0.15\textwidth]{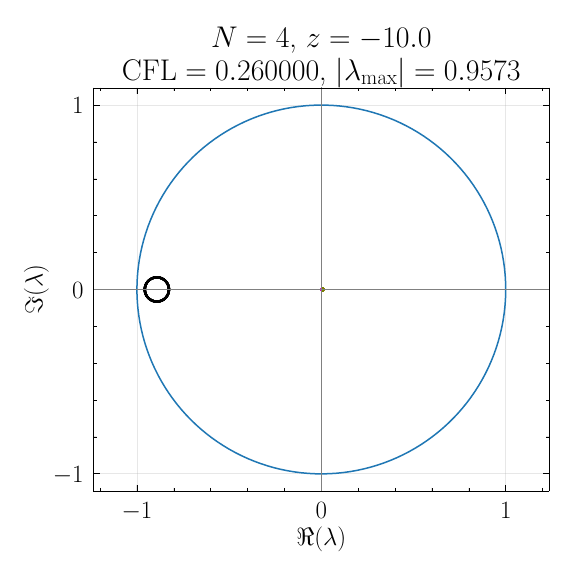}
\includegraphics[width=0.15\textwidth]{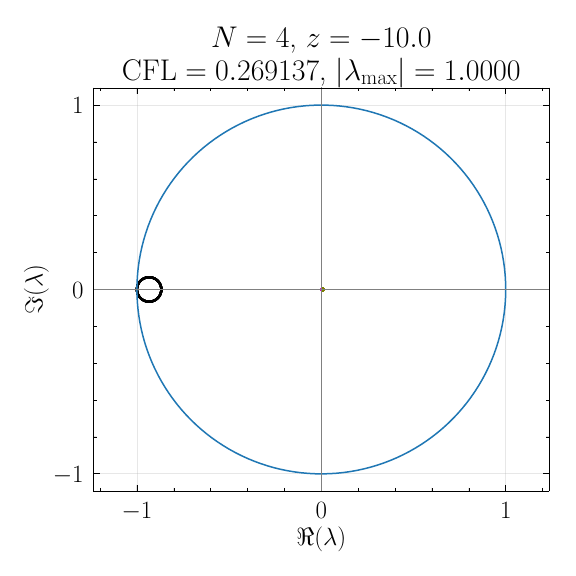}
\includegraphics[width=0.15\textwidth]{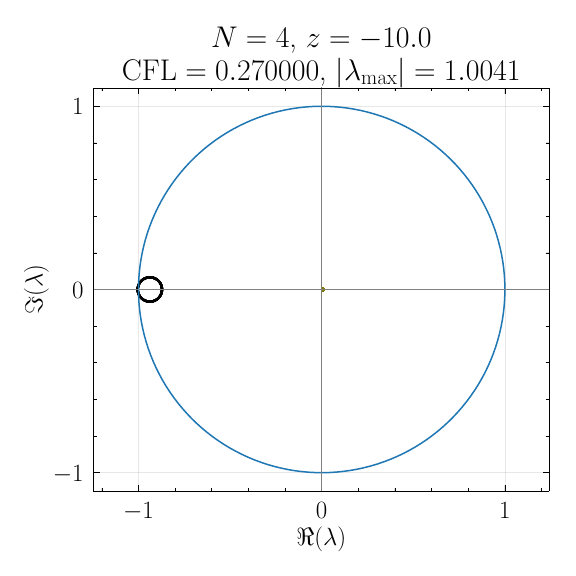}
\includegraphics[width=0.15\textwidth]{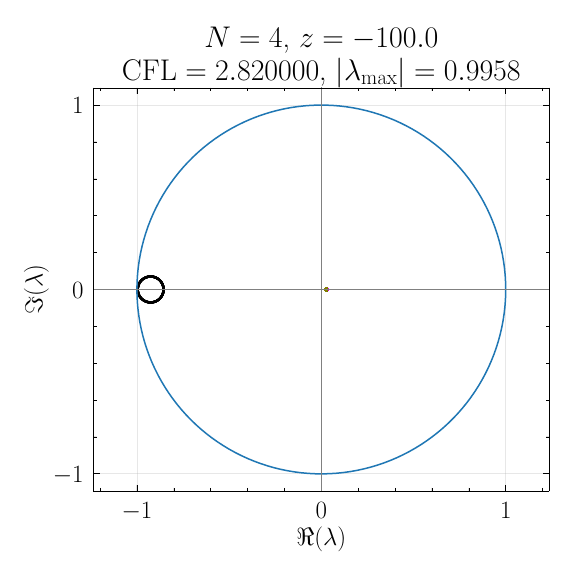}
\includegraphics[width=0.15\textwidth]{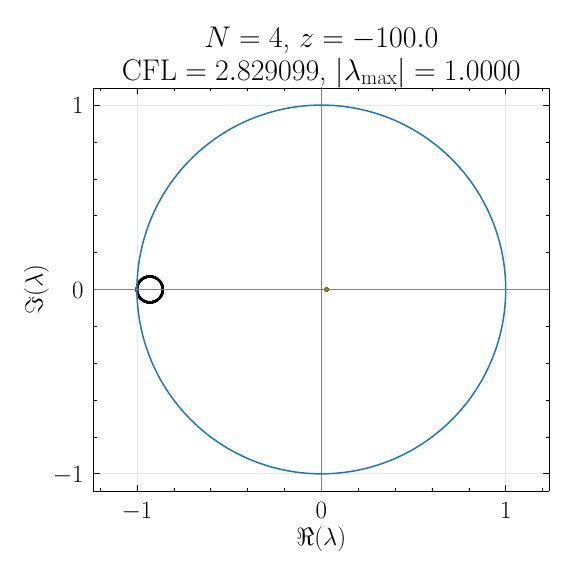}
\includegraphics[width=0.15\textwidth]{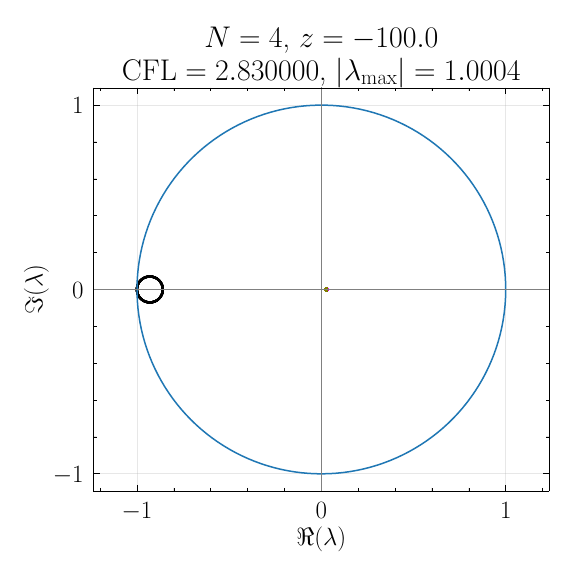}\\
\includegraphics[width=0.028125\textwidth]{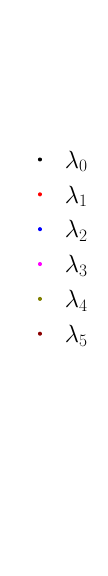}
\includegraphics[width=0.15\textwidth]{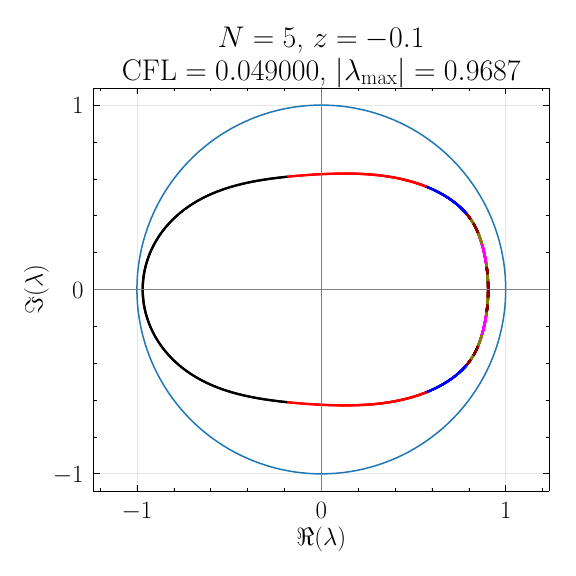}
\includegraphics[width=0.15\textwidth]{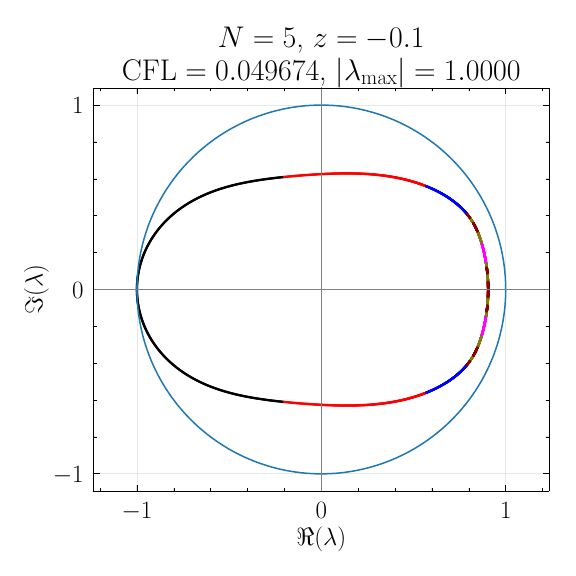}
\includegraphics[width=0.15\textwidth]{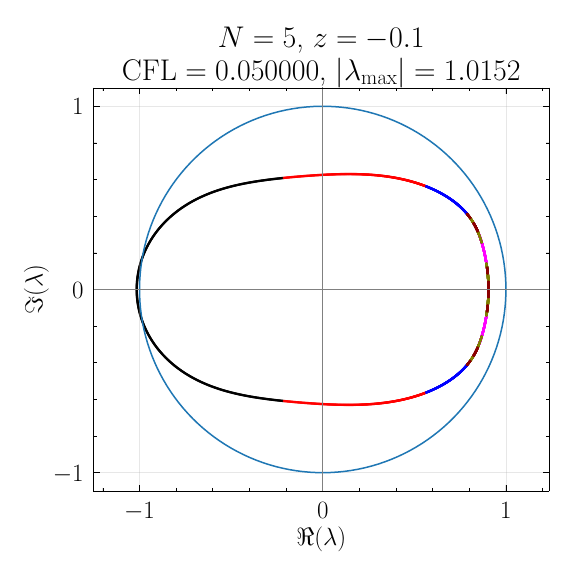}
\includegraphics[width=0.15\textwidth]{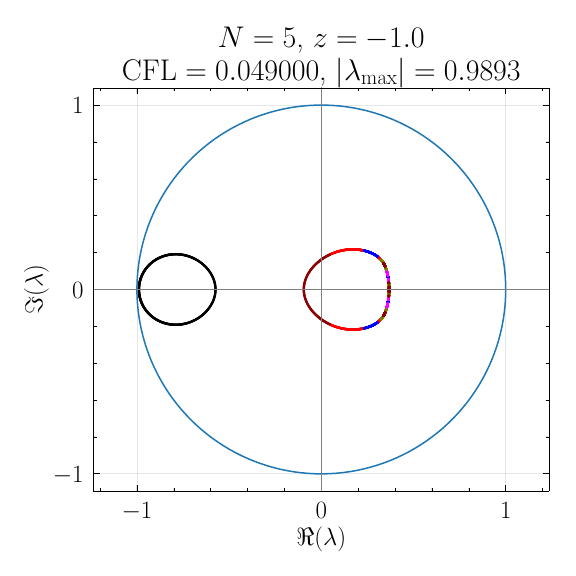}
\includegraphics[width=0.15\textwidth]{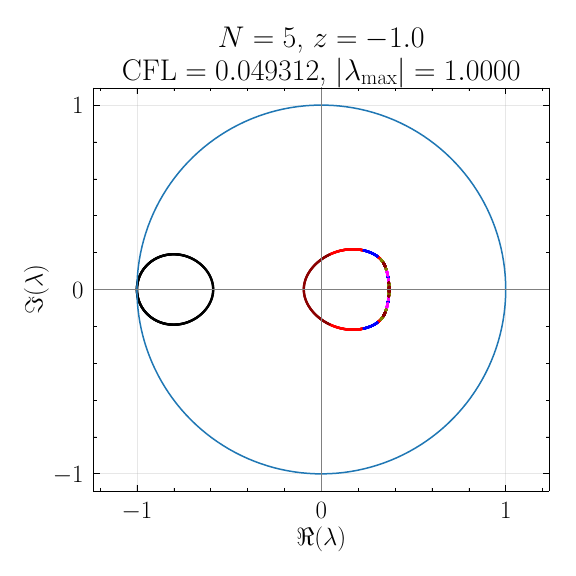}
\includegraphics[width=0.15\textwidth]{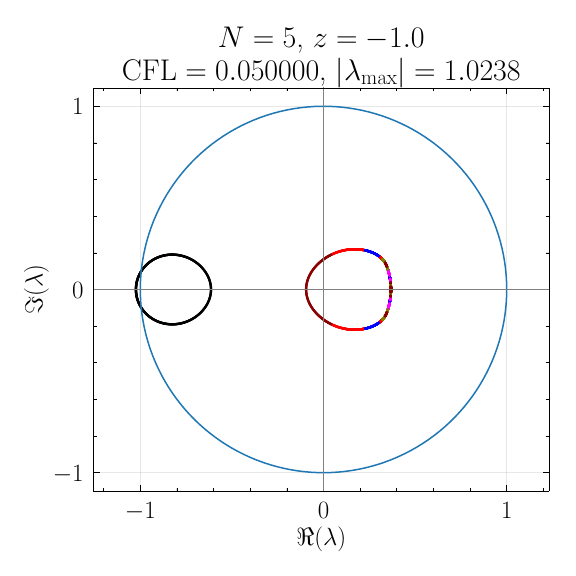}\\
\includegraphics[width=0.028125\textwidth]{spectrum_exact_eigvals_legend_N_5.png}
\includegraphics[width=0.15\textwidth]{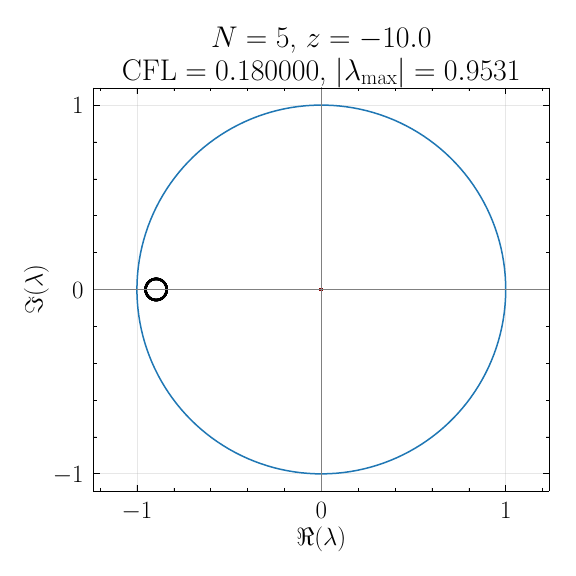}
\includegraphics[width=0.15\textwidth]{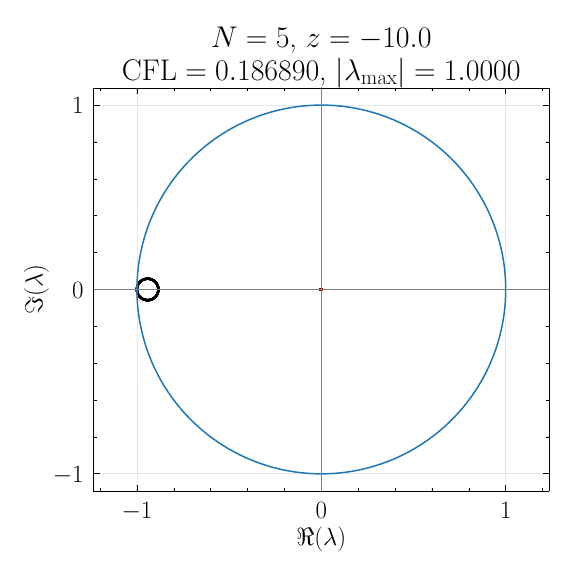}
\includegraphics[width=0.15\textwidth]{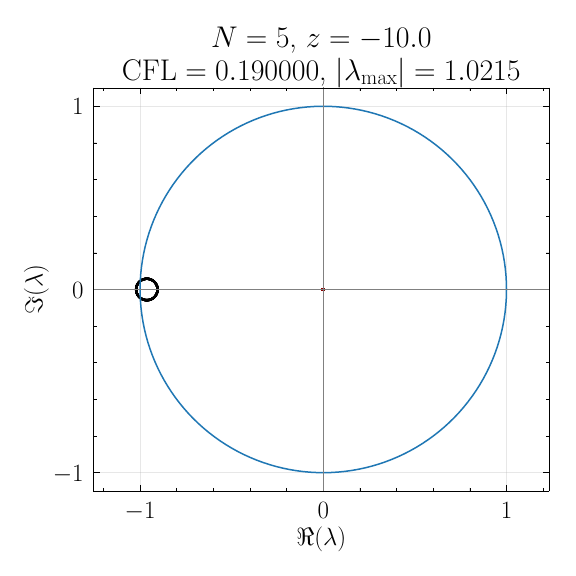}
\includegraphics[width=0.15\textwidth]{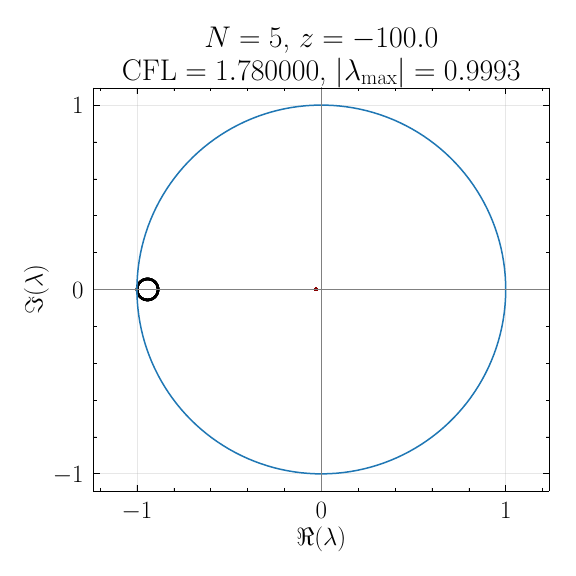}
\includegraphics[width=0.15\textwidth]{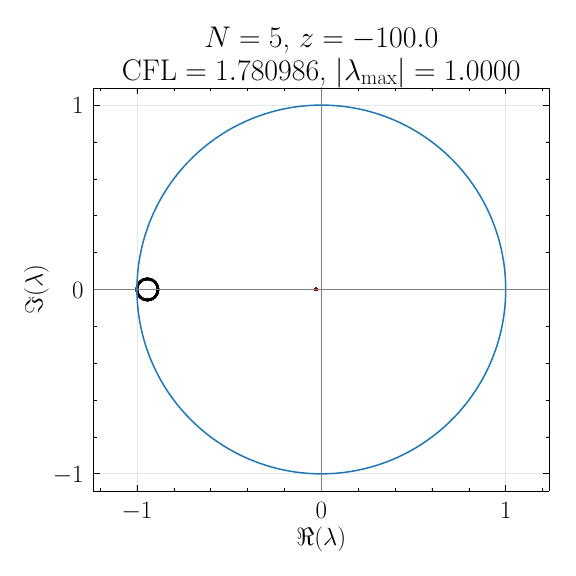}
\includegraphics[width=0.15\textwidth]{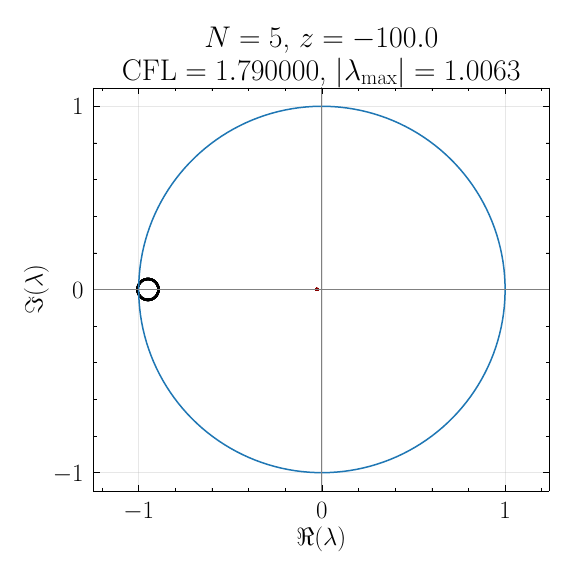}\\
\includegraphics[width=0.028125\textwidth]{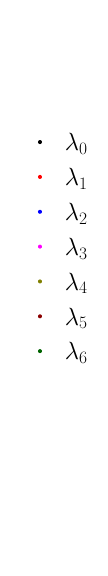}
\includegraphics[width=0.15\textwidth]{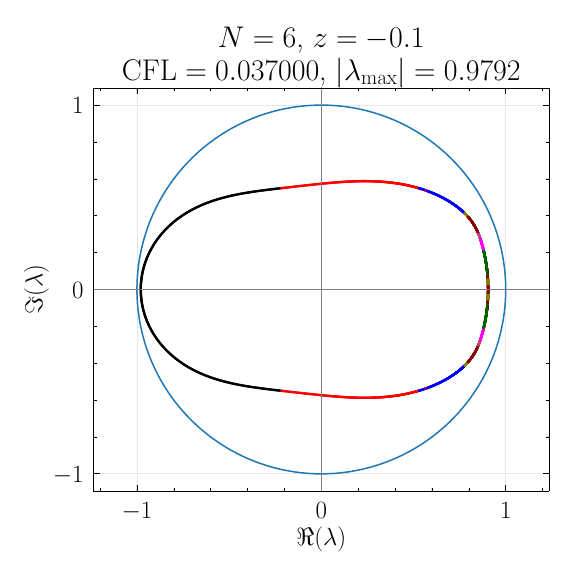}
\includegraphics[width=0.15\textwidth]{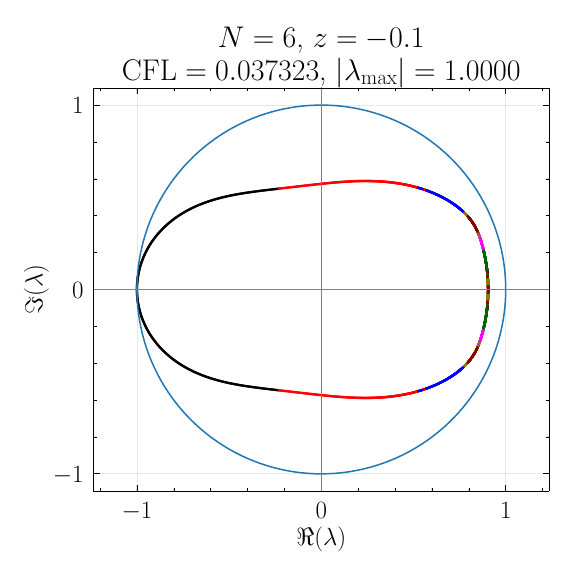}
\includegraphics[width=0.15\textwidth]{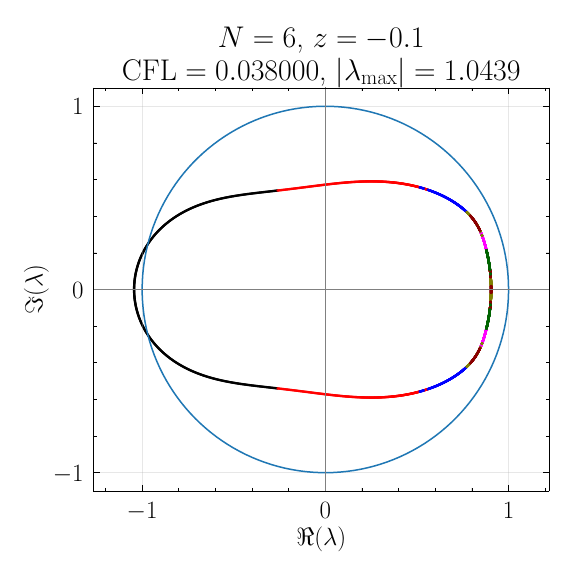}
\includegraphics[width=0.15\textwidth]{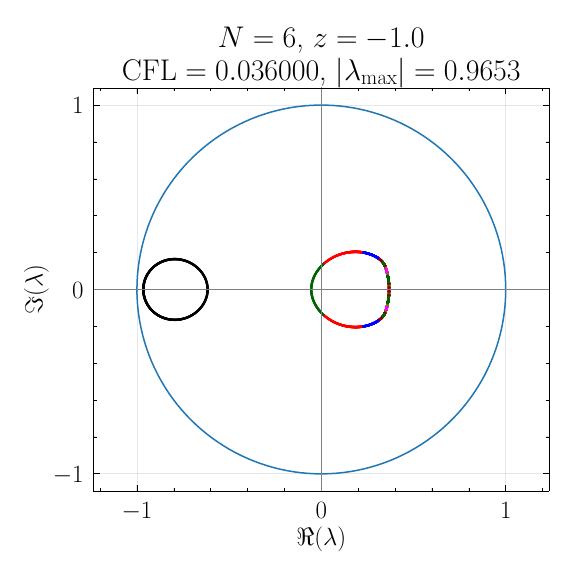}
\includegraphics[width=0.15\textwidth]{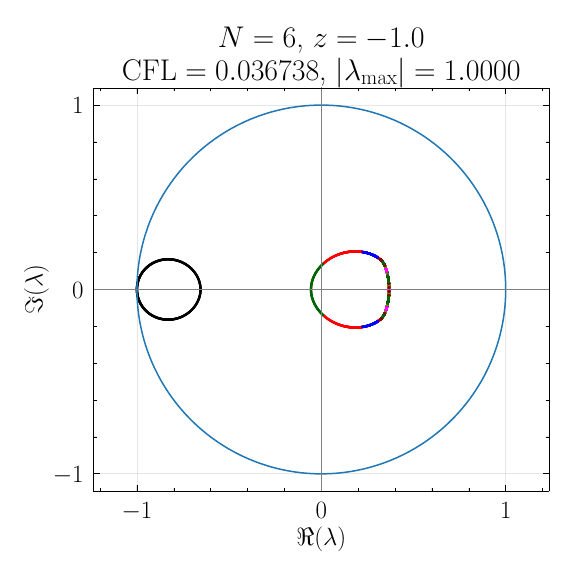}
\includegraphics[width=0.15\textwidth]{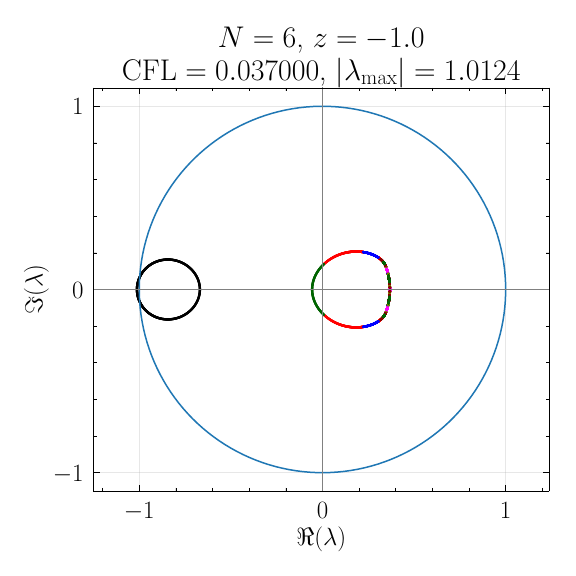}\\
\includegraphics[width=0.028125\textwidth]{spectrum_exact_eigvals_legend_N_6.png}
\includegraphics[width=0.15\textwidth]{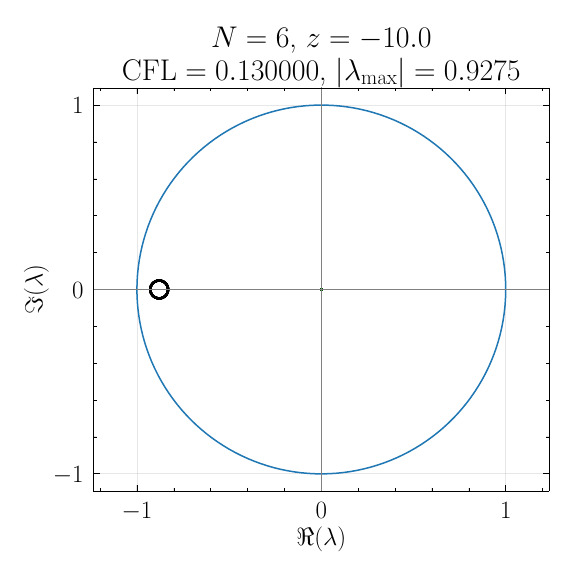}
\includegraphics[width=0.15\textwidth]{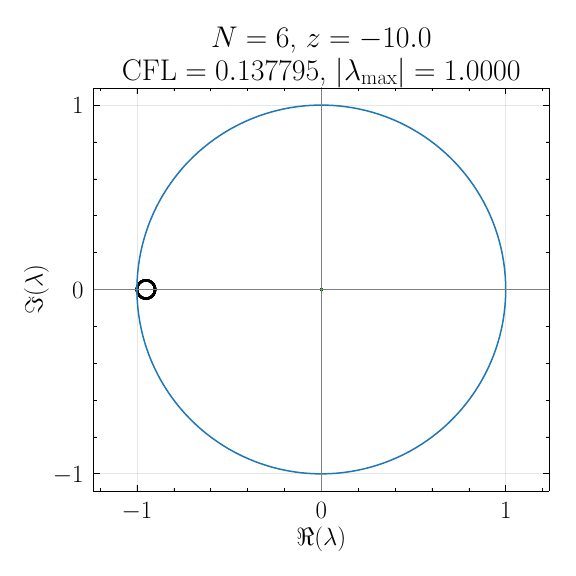}
\includegraphics[width=0.15\textwidth]{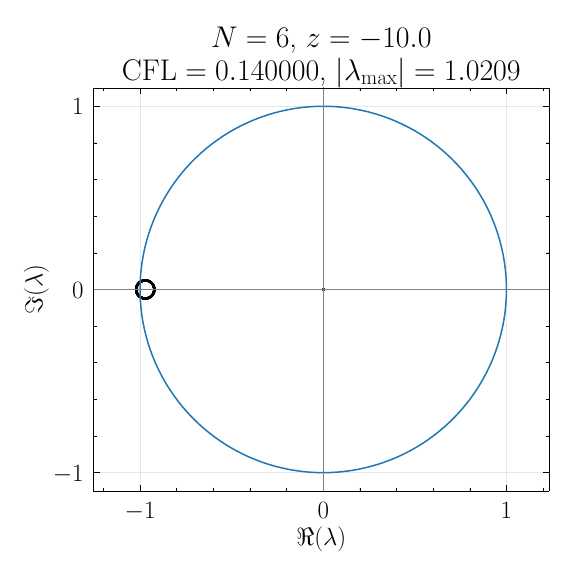}
\includegraphics[width=0.15\textwidth]{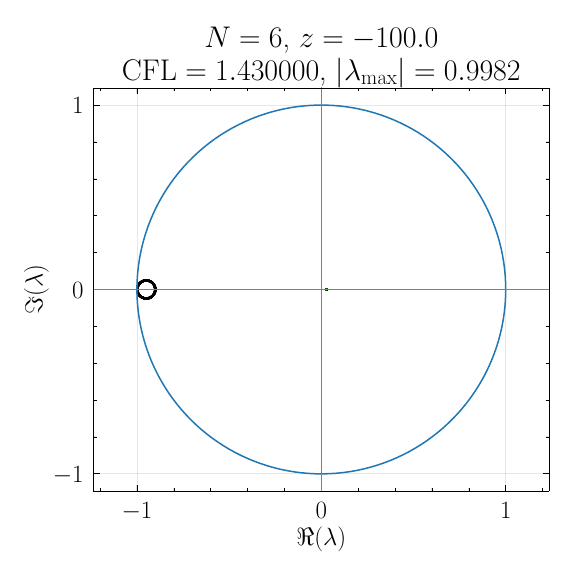}
\includegraphics[width=0.15\textwidth]{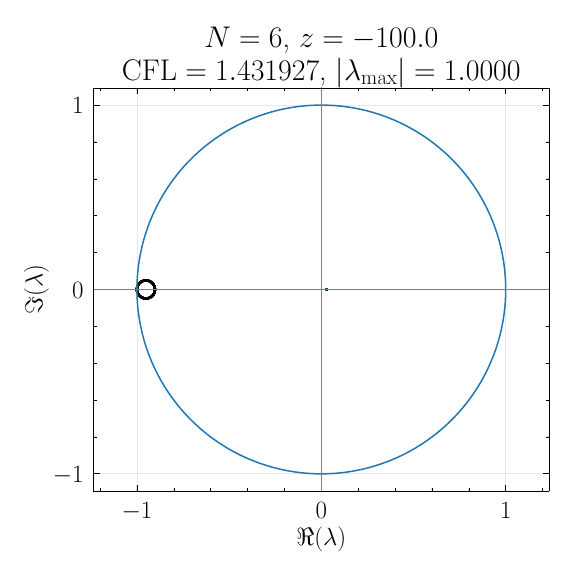}
\includegraphics[width=0.15\textwidth]{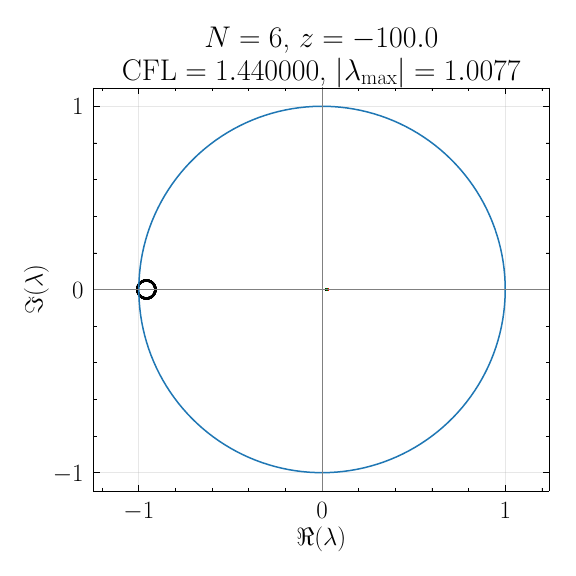}
\caption{%
Spectrum of the matrix $\mathrm{R}(\mathrm{CFL}, z, \theta)$ (\ref{eq:r_matrix_elems_expr_dup}) (eigenvalues $\lambda_{k} = \lambda_{k}(\mathrm{CFL}, z, \theta)$, $k = 0, \ldots, N$) of the evolution operator $R$ (\ref{eq:evol_oper_prop}) for a single time step $\Dtn{n}$ for several selected values of the Courant number $\mathrm{CFL}$ for $z = -0.1$, $-1.0$, $-10.0$, $-100.0$ and for polynomial degrees $N = 4, \ldots, 6$. The range of phase $\theta\in[0, 2\pi)$ is sampled on a uniform grid of $1000$ nodes. The gray circle of unit radius defines the stability boundary. Legends for each row of the graphs are located on the left. The gray circle of unit radius defines the stability boundary. The Courant numbers $\mathrm{CFL}$ are taken both at the boundary value of the Courant number $\mathrm{CFL}_{\rm max}$ and at two values $\mathrm{CFL}$ near it --- in the guaranteed stability region $\mathrm{CFL} < \mathrm{CFL}_{\rm max}$ and in the guaranteed instability region $\mathrm{CFL} > \mathrm{CFL}_{\rm max}$, by rounding off at the second significant decimal place (the values $\mathrm{CFL}_{\rm max}$ are calculated further in this work and presented in Figure~\ref{fig:cfls_on_z}). \textit{Note}: the phase $\arg \lambda_{k}$ of eigenvalue $\lambda_{k}$ is not the phase $\theta$; the dependence of the absolute value $|\lambda_{k}|$ on phase $\theta$ for these polynomial degrees $N$ is presented in Figure~\ref{fig:rhos_on_theta_degrees_4_6}.
}
\label{fig:spectrum_exact_eigvals_degrees_4_6}
\end{figure}
\begin{figure}[h!]
\centering
\includegraphics[width=0.028125\textwidth]{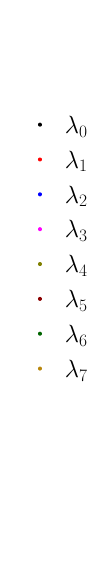}
\includegraphics[width=0.15\textwidth]{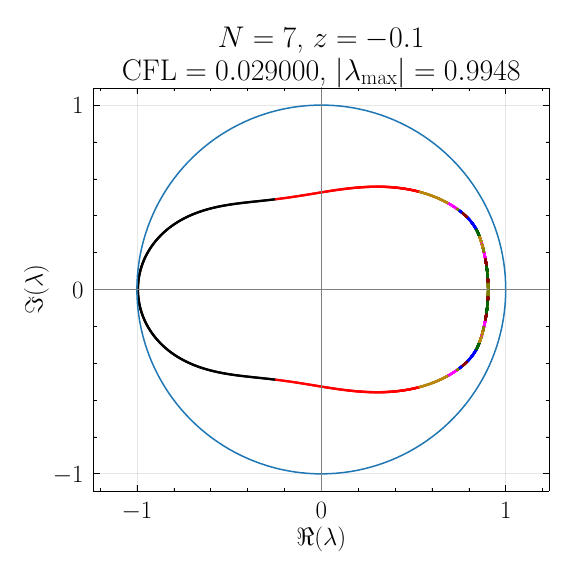}
\includegraphics[width=0.15\textwidth]{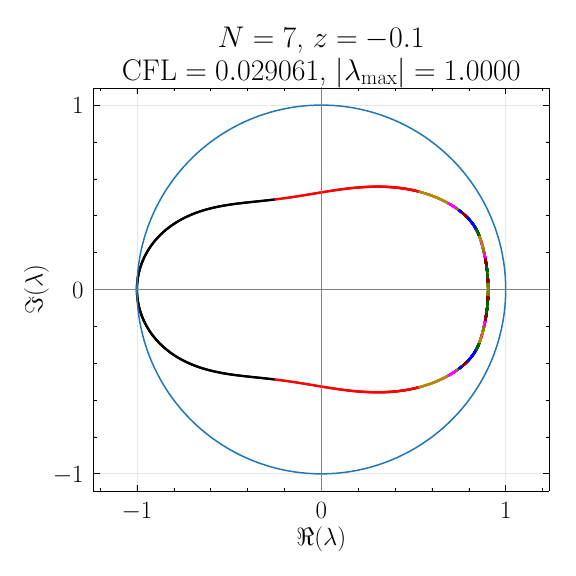}
\includegraphics[width=0.15\textwidth]{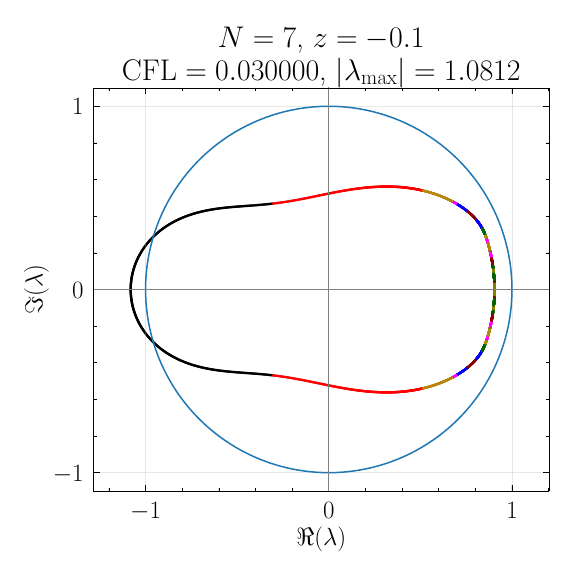}
\includegraphics[width=0.15\textwidth]{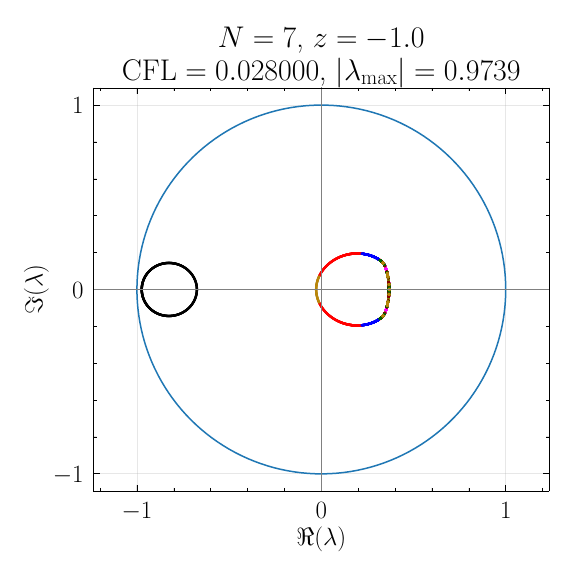}
\includegraphics[width=0.15\textwidth]{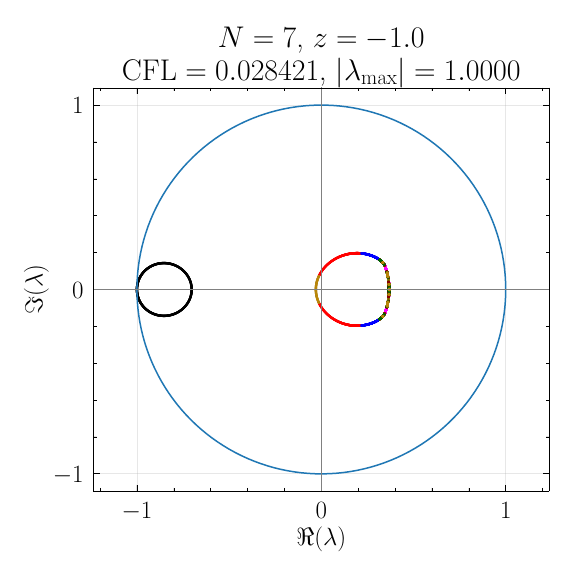}
\includegraphics[width=0.15\textwidth]{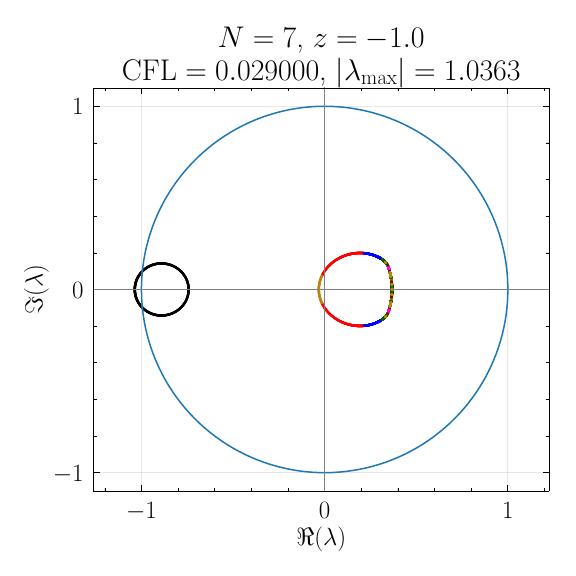}\\
\includegraphics[width=0.028125\textwidth]{spectrum_exact_eigvals_legend_N_7.png}
\includegraphics[width=0.15\textwidth]{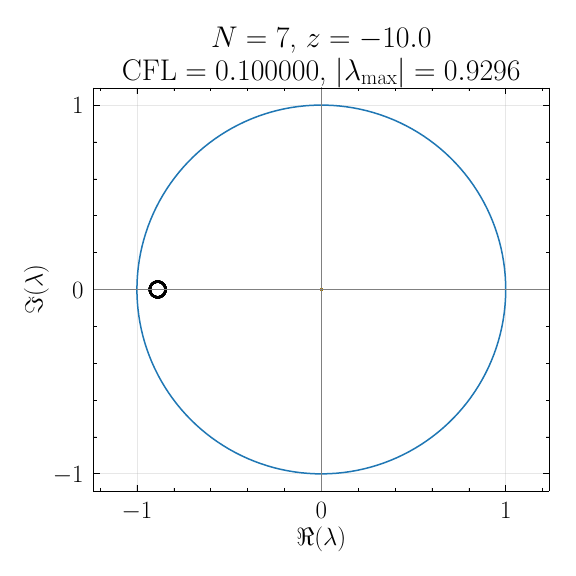}
\includegraphics[width=0.15\textwidth]{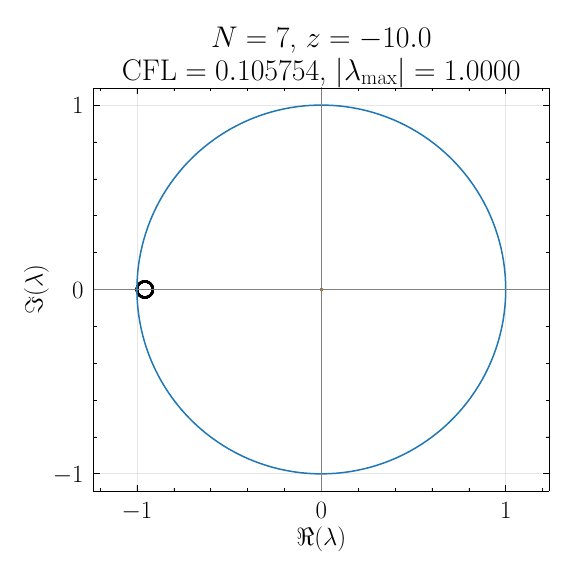}
\includegraphics[width=0.15\textwidth]{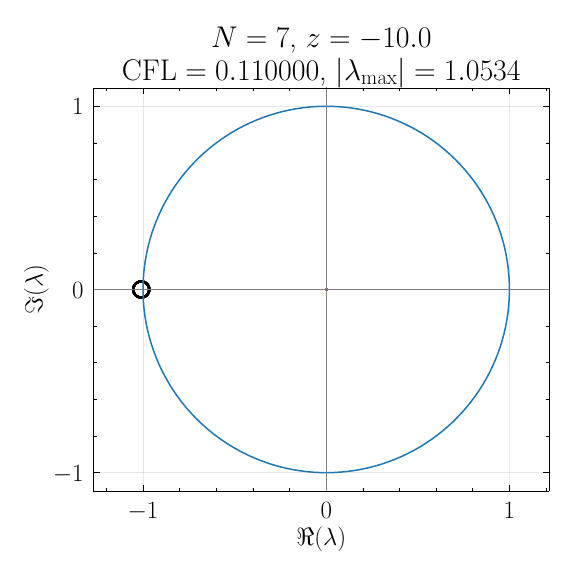}
\includegraphics[width=0.15\textwidth]{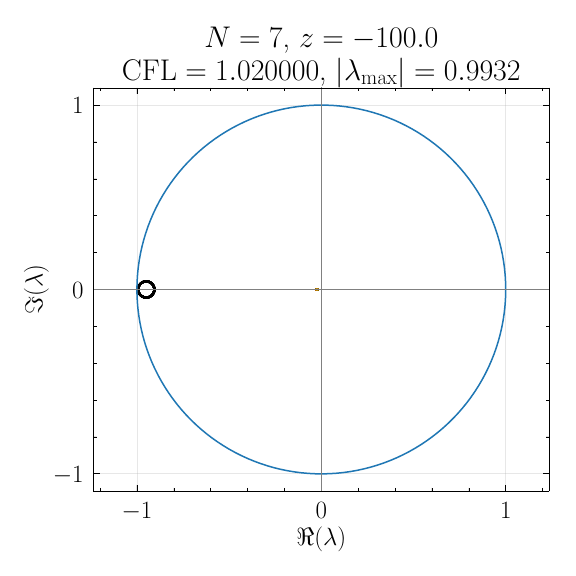}
\includegraphics[width=0.15\textwidth]{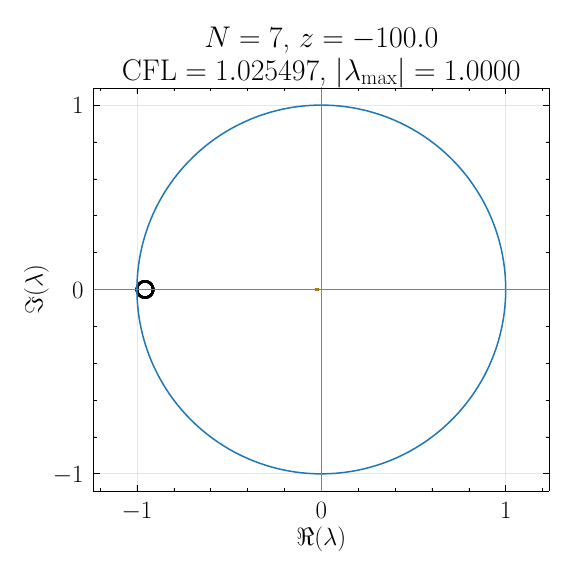}
\includegraphics[width=0.15\textwidth]{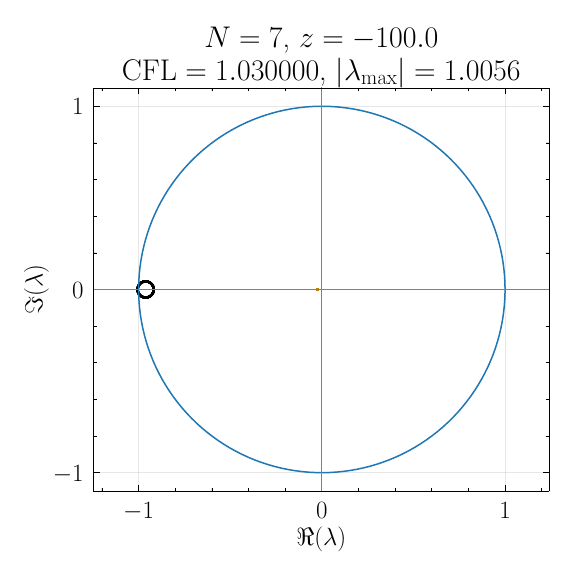}\\
\includegraphics[width=0.028125\textwidth]{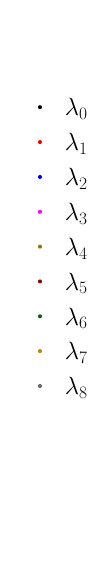}
\includegraphics[width=0.15\textwidth]{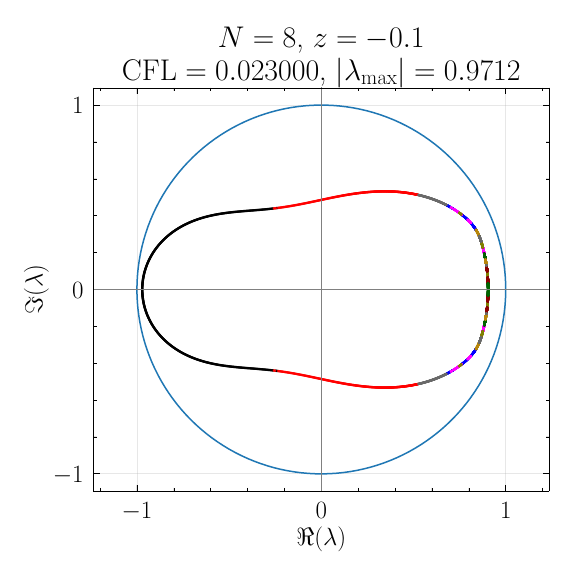}
\includegraphics[width=0.15\textwidth]{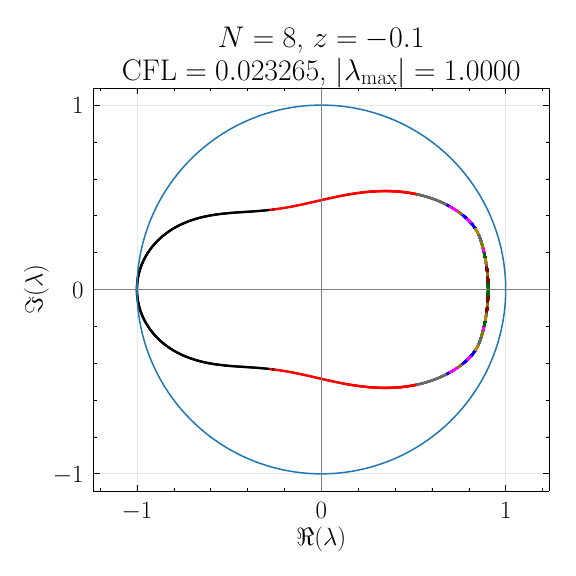}
\includegraphics[width=0.15\textwidth]{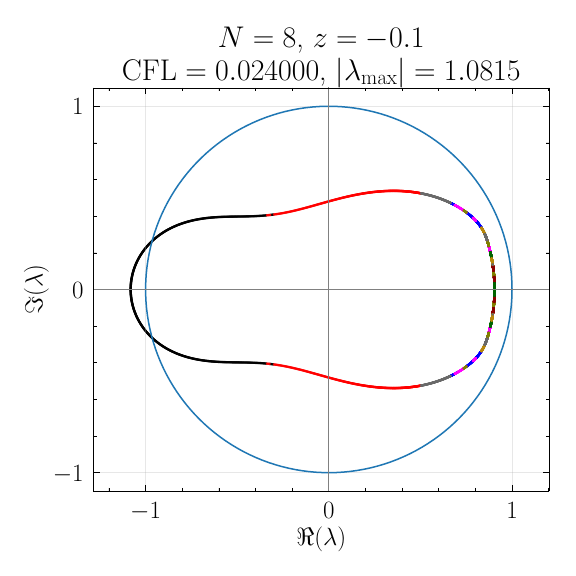}
\includegraphics[width=0.15\textwidth]{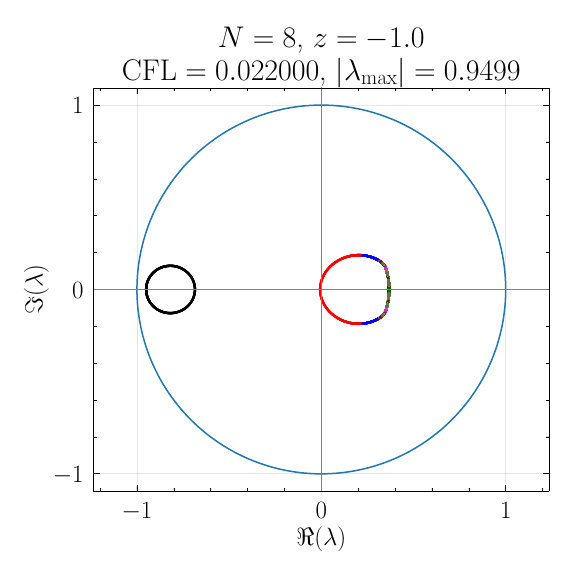}
\includegraphics[width=0.15\textwidth]{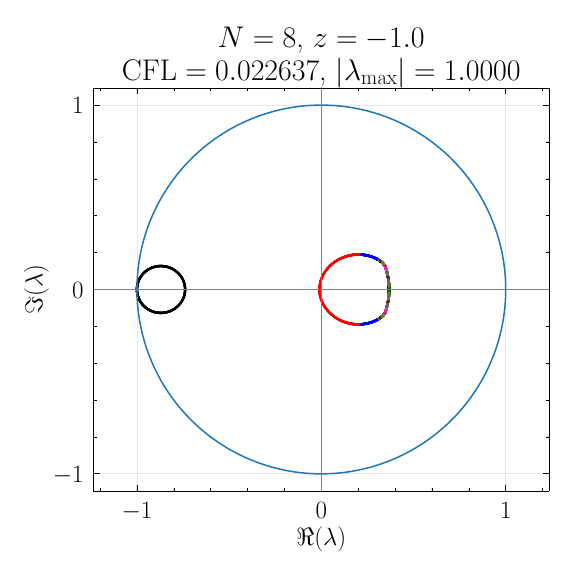}
\includegraphics[width=0.15\textwidth]{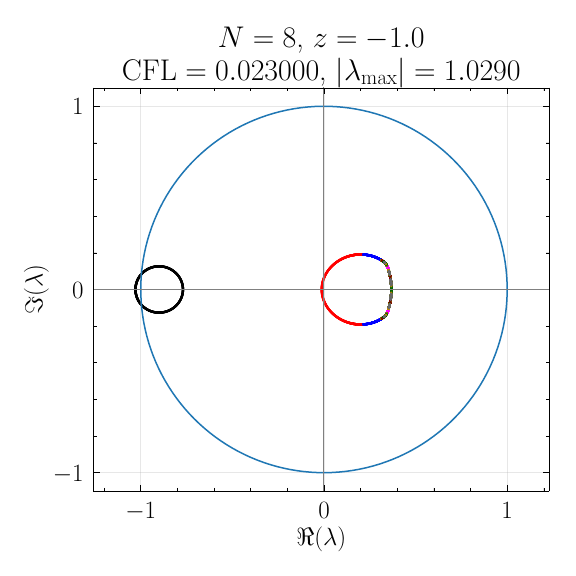}\\
\includegraphics[width=0.028125\textwidth]{spectrum_exact_eigvals_legend_N_8.png}
\includegraphics[width=0.15\textwidth]{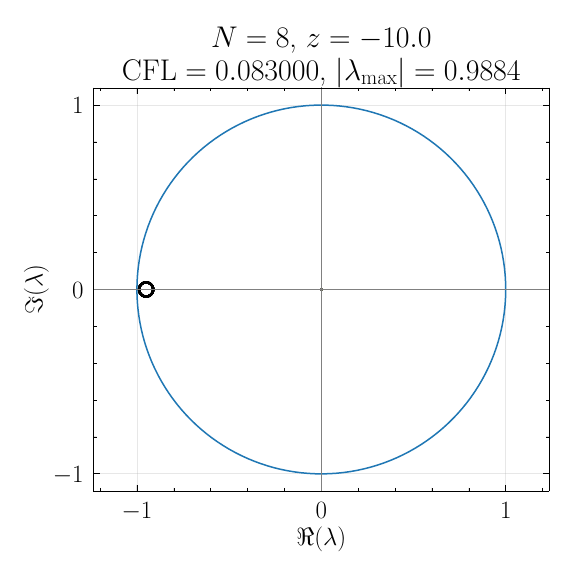}
\includegraphics[width=0.15\textwidth]{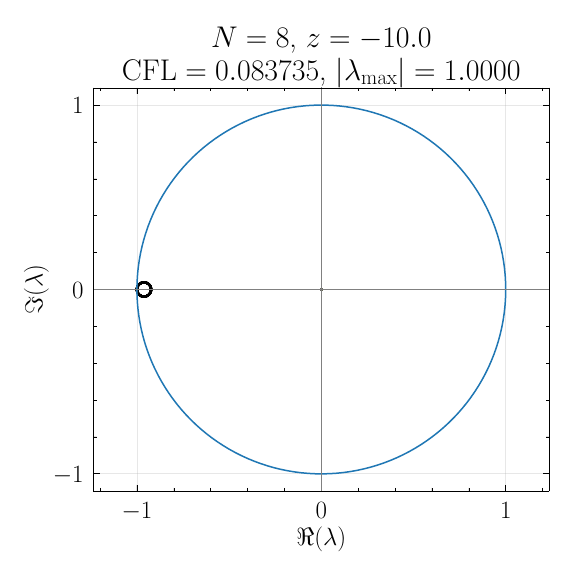}
\includegraphics[width=0.15\textwidth]{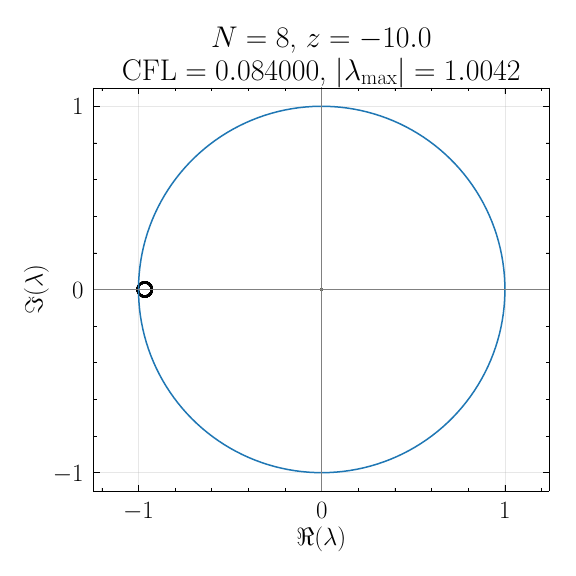}
\includegraphics[width=0.15\textwidth]{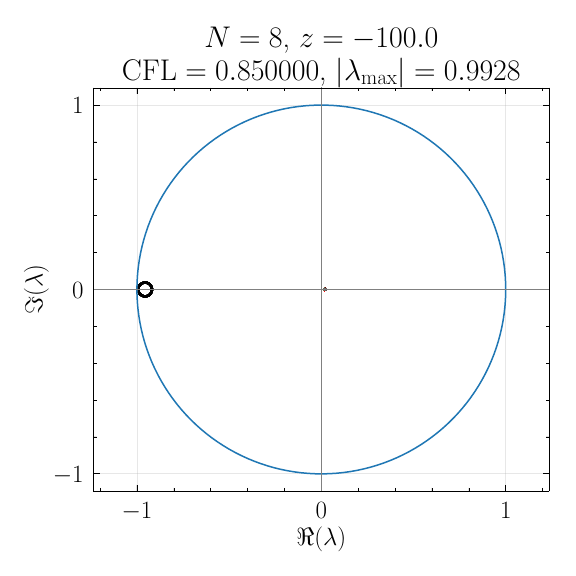}
\includegraphics[width=0.15\textwidth]{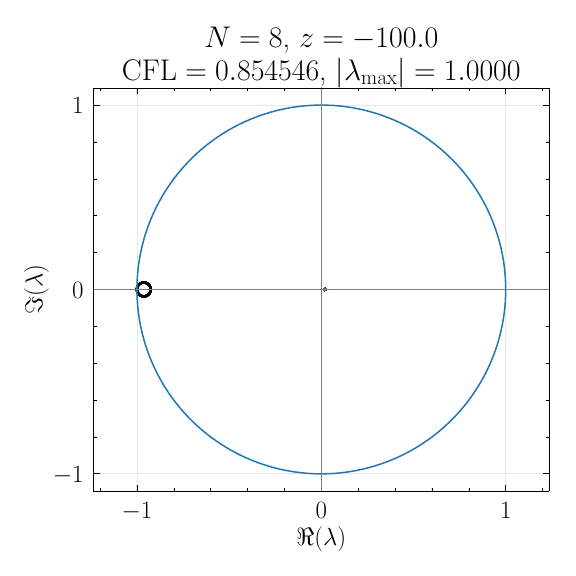}
\includegraphics[width=0.15\textwidth]{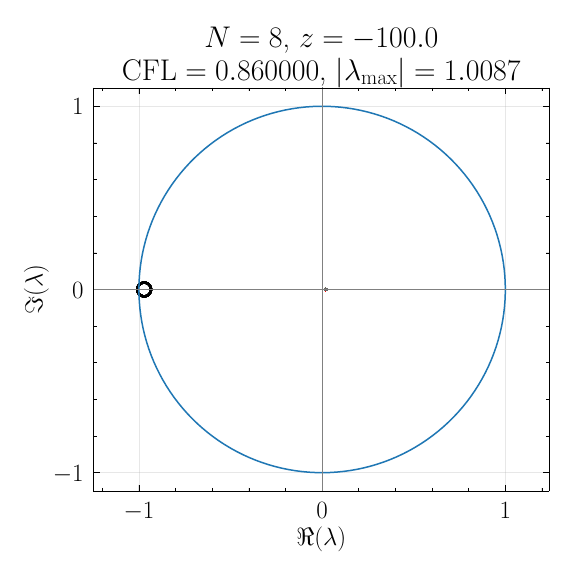}\\
\includegraphics[width=0.028125\textwidth]{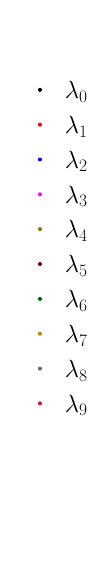}
\includegraphics[width=0.15\textwidth]{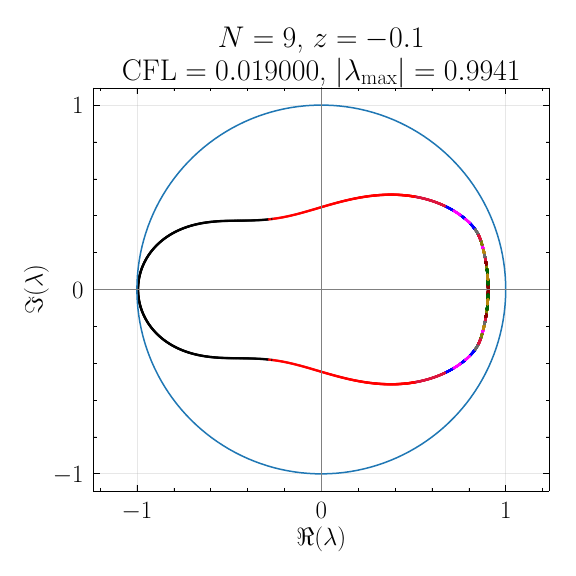}
\includegraphics[width=0.15\textwidth]{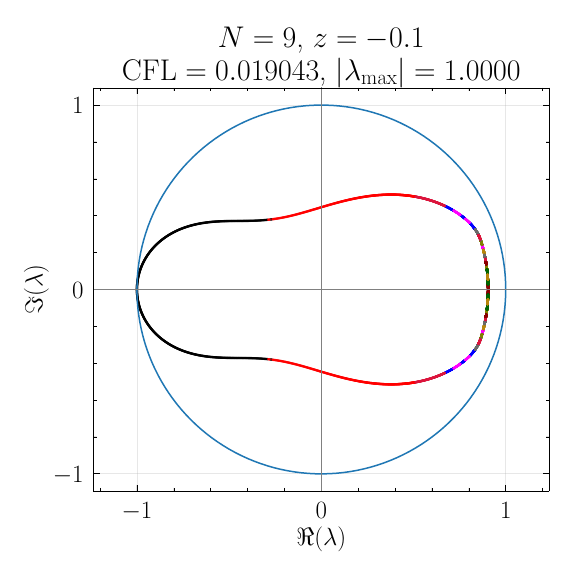}
\includegraphics[width=0.15\textwidth]{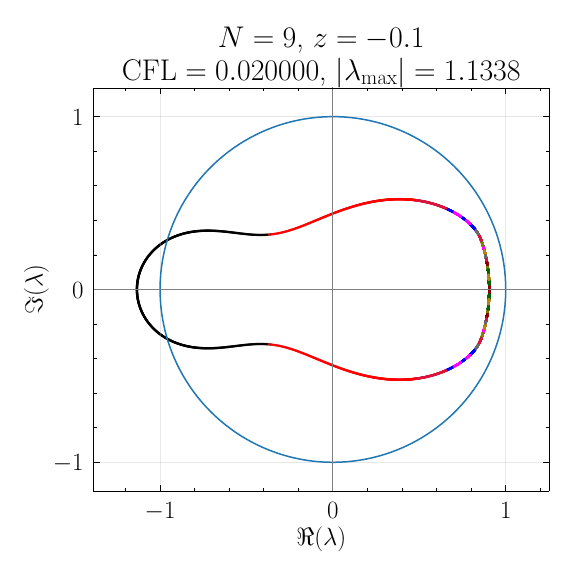}
\includegraphics[width=0.15\textwidth]{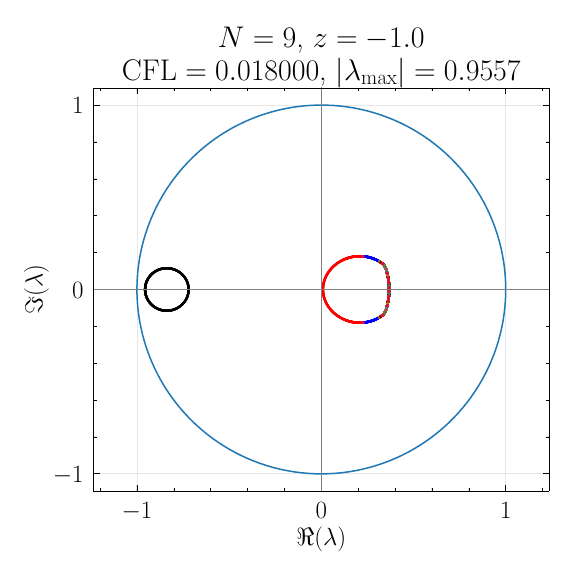}
\includegraphics[width=0.15\textwidth]{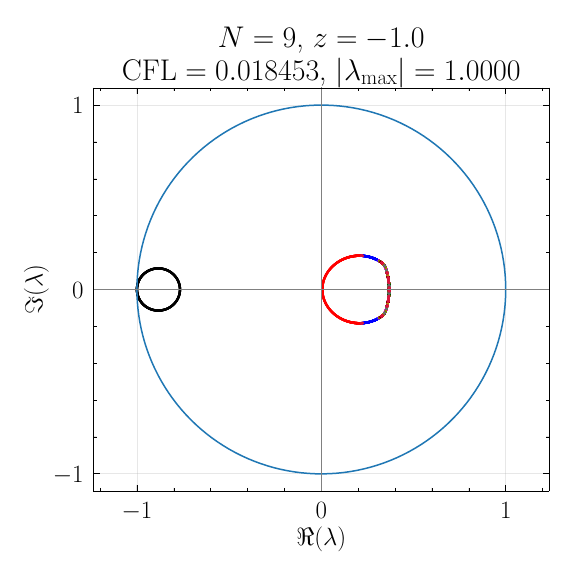}
\includegraphics[width=0.15\textwidth]{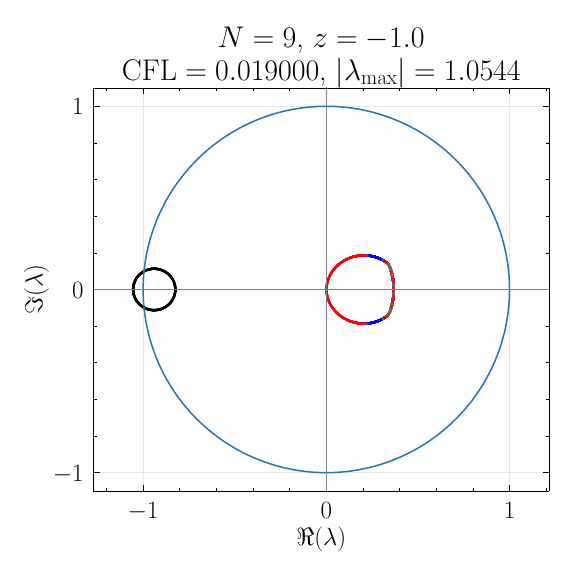}\\
\includegraphics[width=0.028125\textwidth]{spectrum_exact_eigvals_legend_N_9.png}
\includegraphics[width=0.15\textwidth]{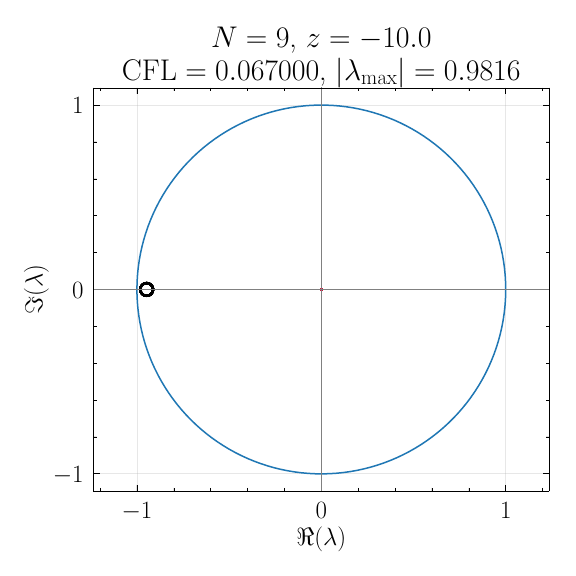}
\includegraphics[width=0.15\textwidth]{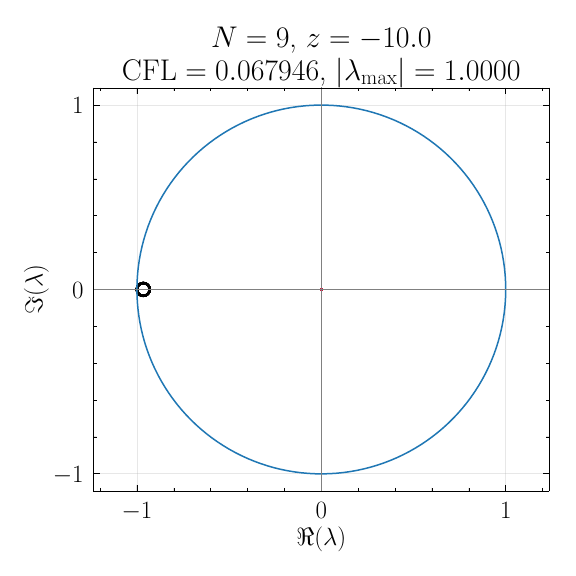}
\includegraphics[width=0.15\textwidth]{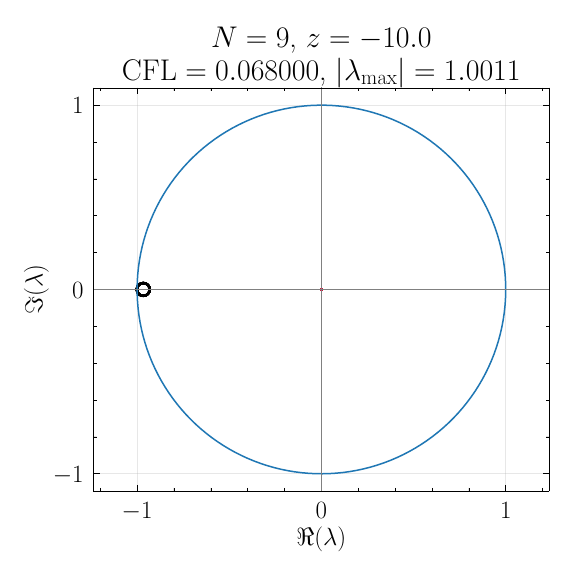}
\includegraphics[width=0.15\textwidth]{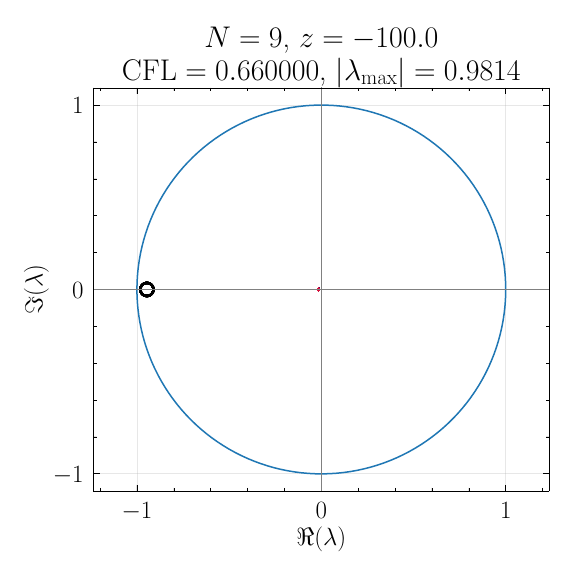}
\includegraphics[width=0.15\textwidth]{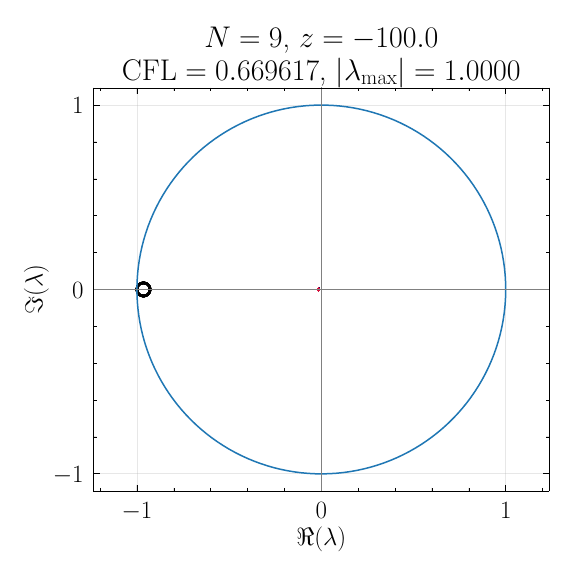}
\includegraphics[width=0.15\textwidth]{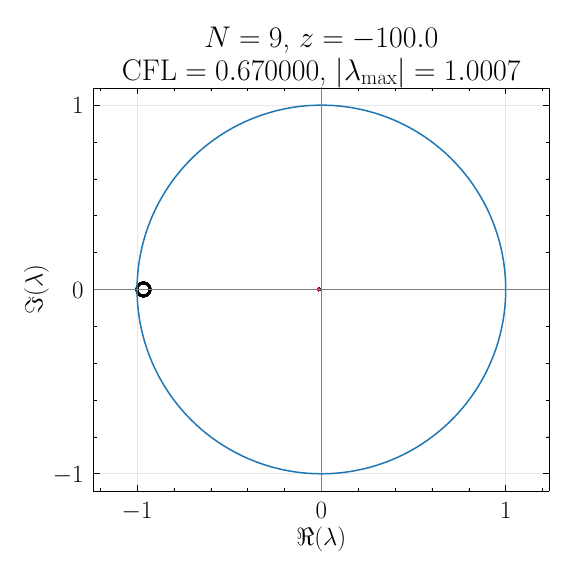}
\caption{%
Spectrum of the matrix $\mathrm{R}(\mathrm{CFL}, z, \theta)$ (\ref{eq:r_matrix_elems_expr_dup}) (eigenvalues $\lambda_{k} = \lambda_{k}(\mathrm{CFL}, z, \theta)$, $k = 0, \ldots, N$) of the evolution operator $R$ (\ref{eq:evol_oper_prop}) for a single time step $\Dtn{n}$ for several selected values of the Courant number $\mathrm{CFL}$ for $z = -0.1$, $-1.0$, $-10.0$, $-100.0$ and for polynomial degrees $N = 7, \ldots, 9$. The range of phase $\theta\in[0, 2\pi)$ is sampled on a uniform grid of $1000$ nodes. The gray circle of unit radius defines the stability boundary. Legends for each row of the graphs are located on the left. The gray circle of unit radius defines the stability boundary. The Courant numbers $\mathrm{CFL}$ are taken both at the boundary value of the Courant number $\mathrm{CFL}_{\rm max}$ and at two values $\mathrm{CFL}$ near it --- in the guaranteed stability region $\mathrm{CFL} < \mathrm{CFL}_{\rm max}$ and in the guaranteed instability region $\mathrm{CFL} > \mathrm{CFL}_{\rm max}$, by rounding off at the second significant decimal place (the values $\mathrm{CFL}_{\rm max}$ are calculated further in this work and presented in Figure~\ref{fig:cfls_on_z}). \textit{Note}: the phase $\arg \lambda_{k}$ of eigenvalue $\lambda_{k}$ is not the phase $\theta$; the dependence of the absolute value $|\lambda_{k}|$ on phase $\theta$ for these polynomial degrees $N$ is presented in Figure~\ref{fig:rhos_on_theta_degrees_7_9}.
}
\label{fig:spectrum_exact_eigvals_degrees_7_9}
\end{figure}
\begin{figure}[h!]
\centering
\includegraphics[width=0.028125\textwidth]{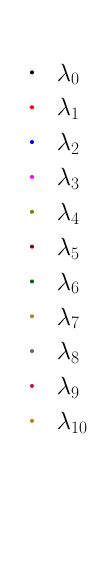}
\includegraphics[width=0.15\textwidth]{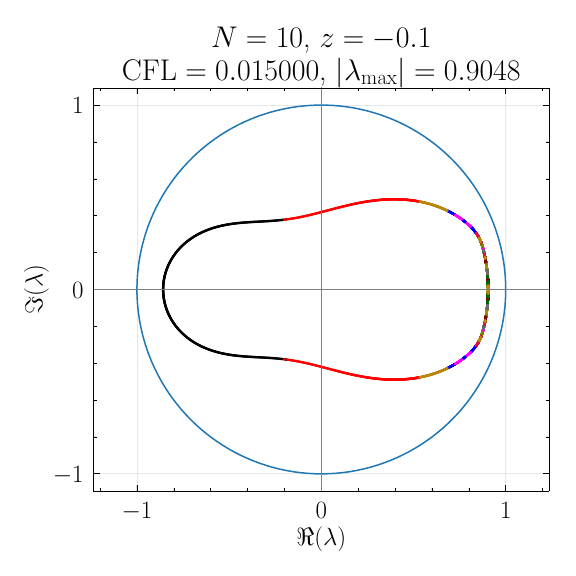}
\includegraphics[width=0.15\textwidth]{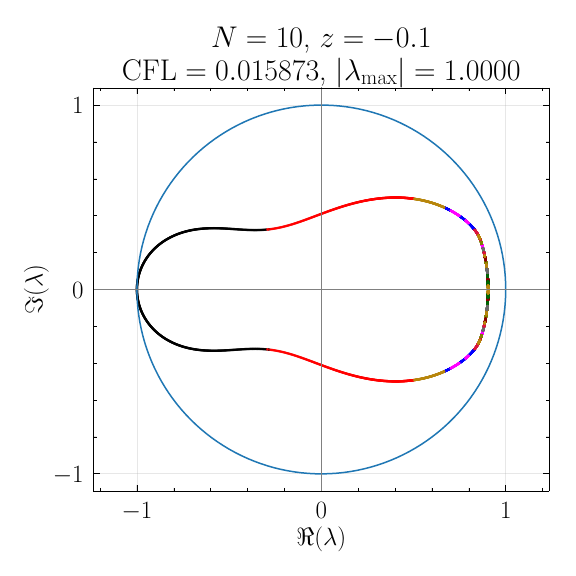}
\includegraphics[width=0.15\textwidth]{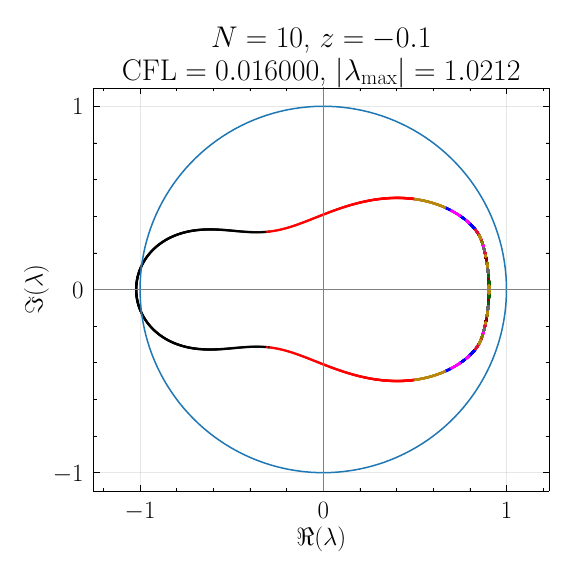}
\includegraphics[width=0.15\textwidth]{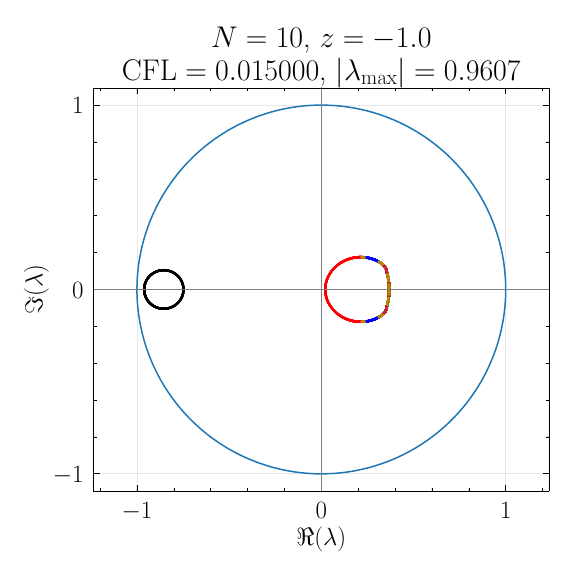}
\includegraphics[width=0.15\textwidth]{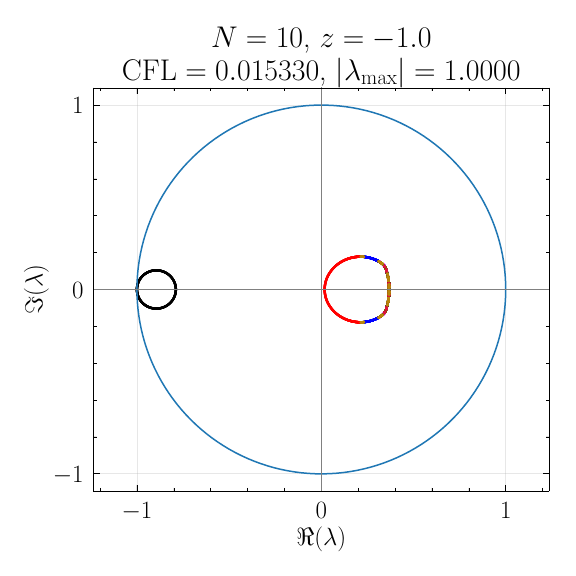}
\includegraphics[width=0.15\textwidth]{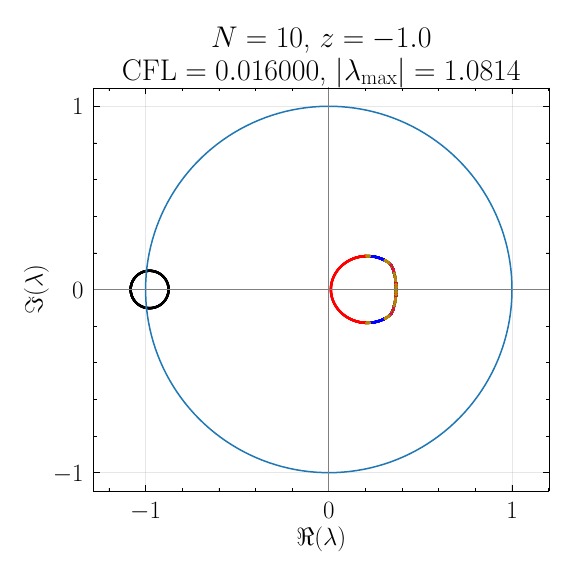}\\
\includegraphics[width=0.028125\textwidth]{spectrum_exact_eigvals_legend_N_10.png}
\includegraphics[width=0.15\textwidth]{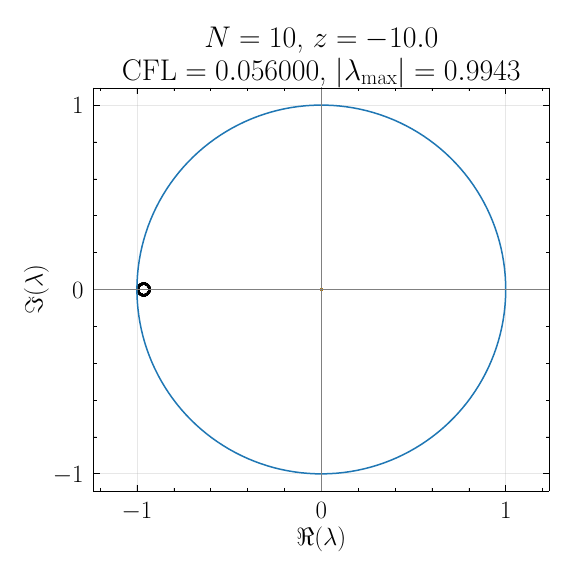}
\includegraphics[width=0.15\textwidth]{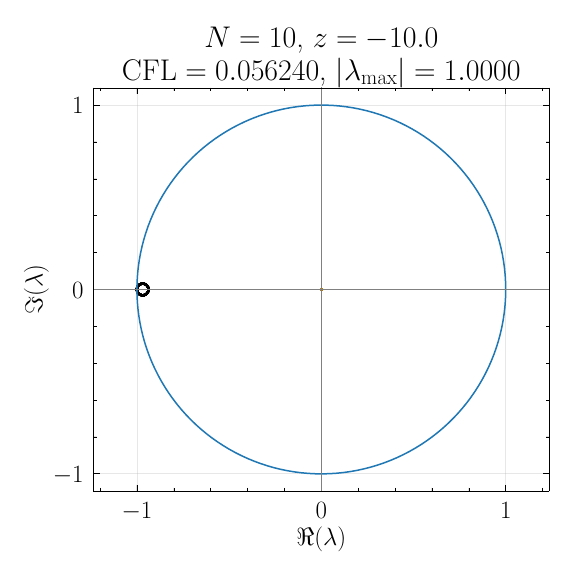}
\includegraphics[width=0.15\textwidth]{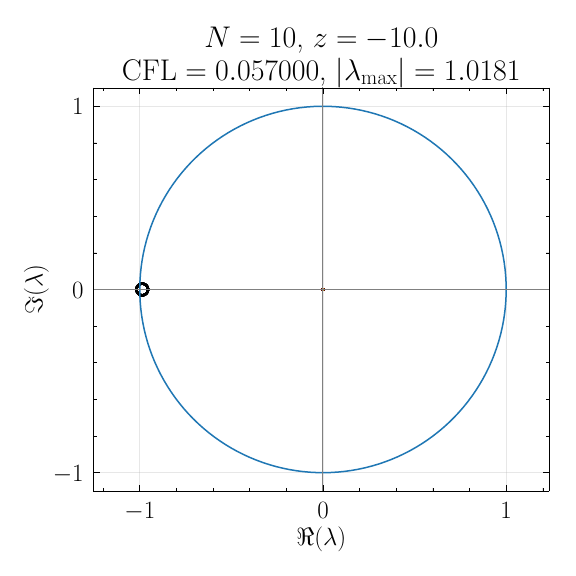}
\includegraphics[width=0.15\textwidth]{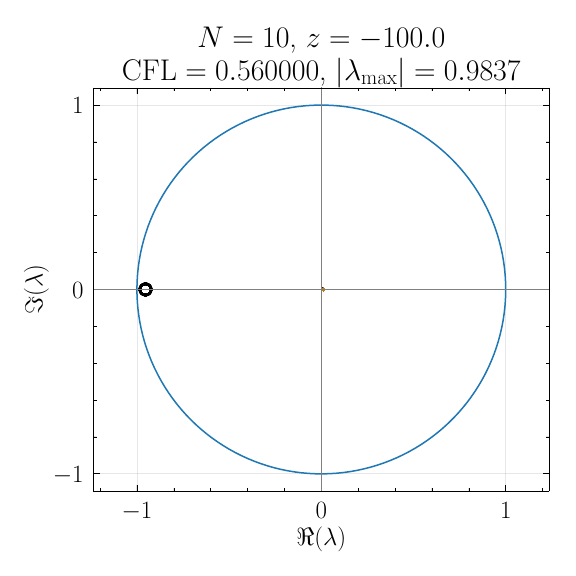}
\includegraphics[width=0.15\textwidth]{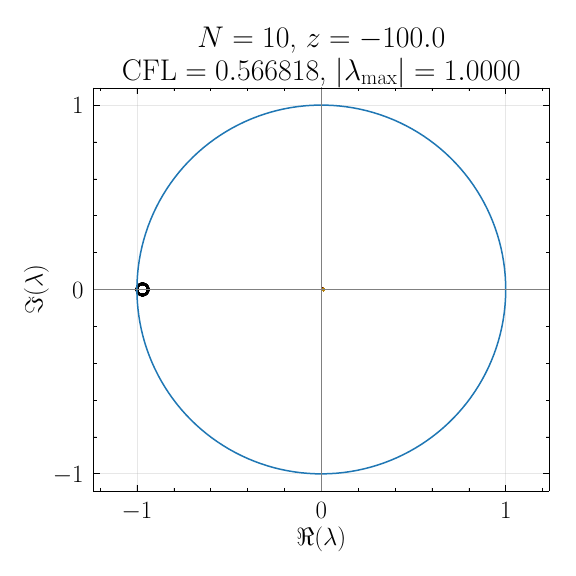}
\includegraphics[width=0.15\textwidth]{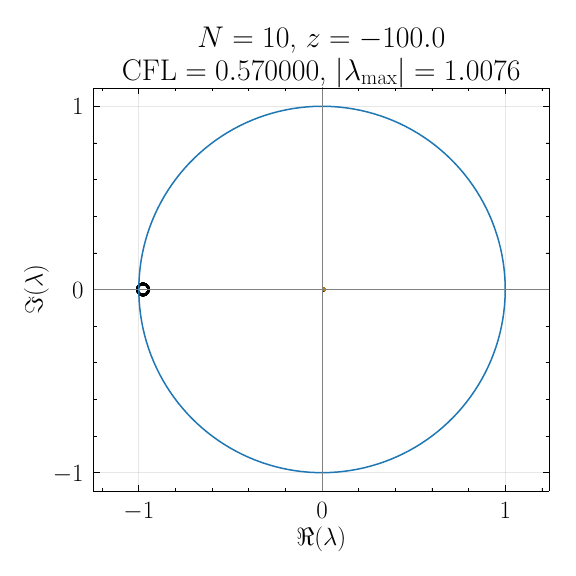}\\
\includegraphics[width=0.028125\textwidth]{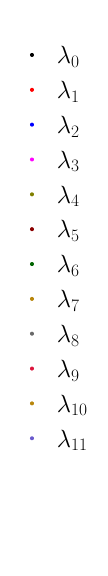}
\includegraphics[width=0.15\textwidth]{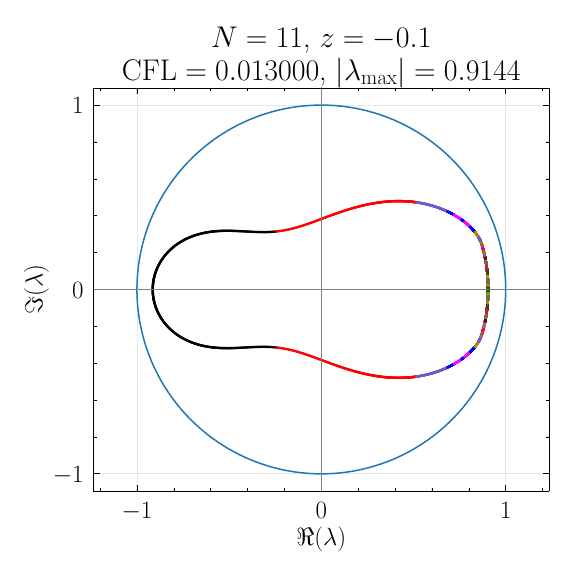}
\includegraphics[width=0.15\textwidth]{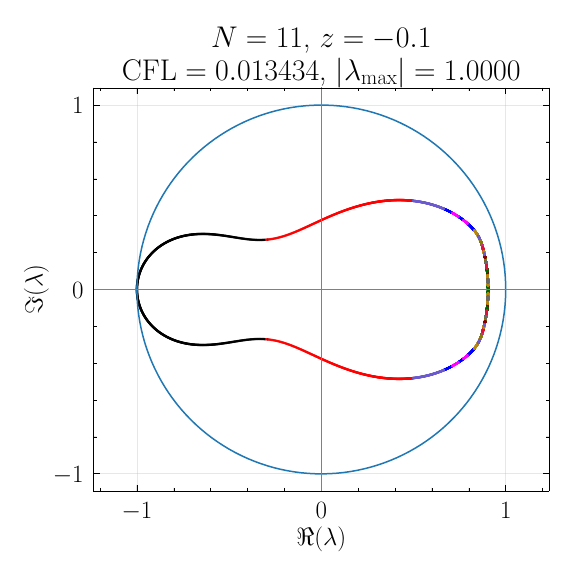}
\includegraphics[width=0.15\textwidth]{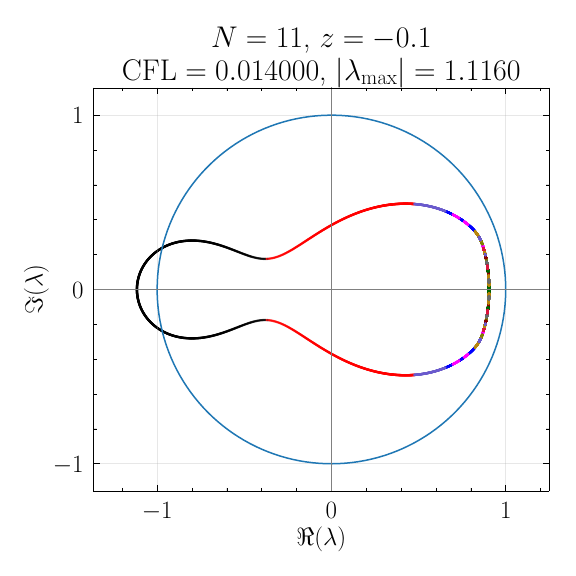}
\includegraphics[width=0.15\textwidth]{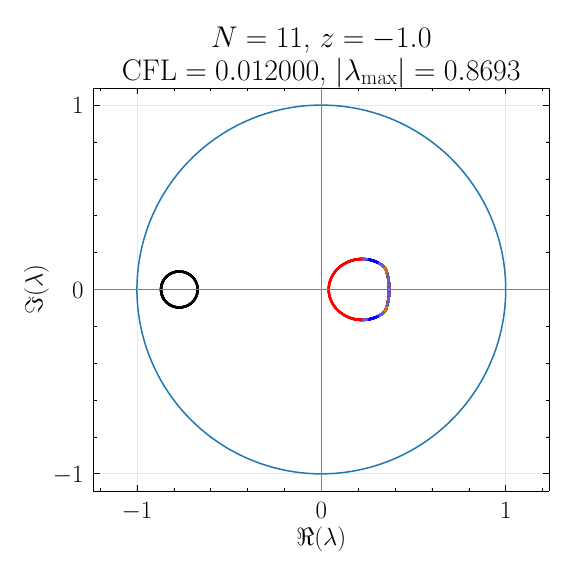}
\includegraphics[width=0.15\textwidth]{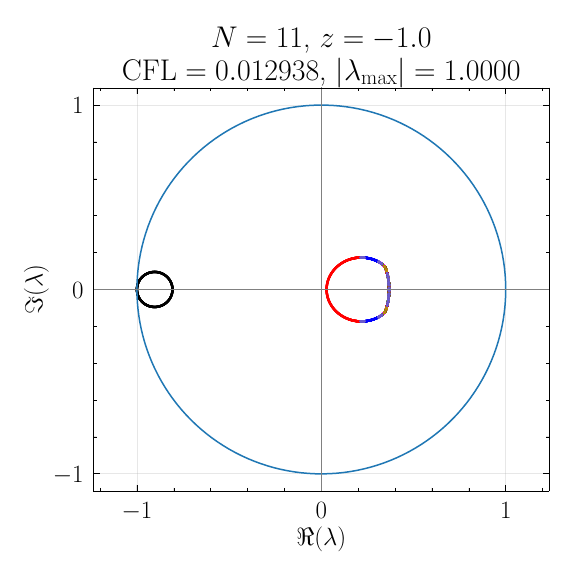}
\includegraphics[width=0.15\textwidth]{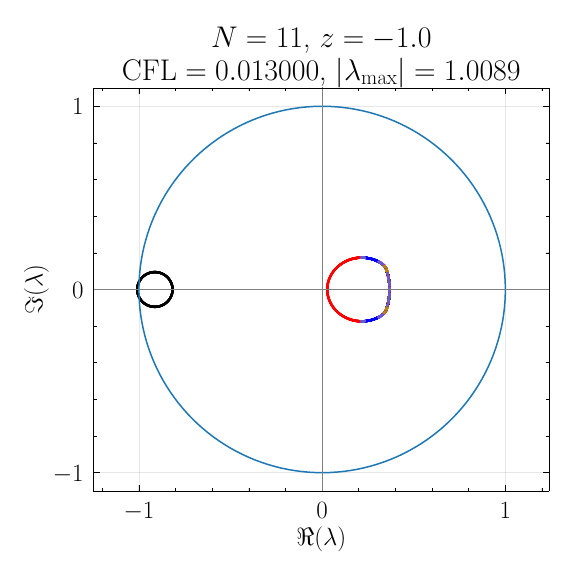}\\
\includegraphics[width=0.028125\textwidth]{spectrum_exact_eigvals_legend_N_11.png}
\includegraphics[width=0.15\textwidth]{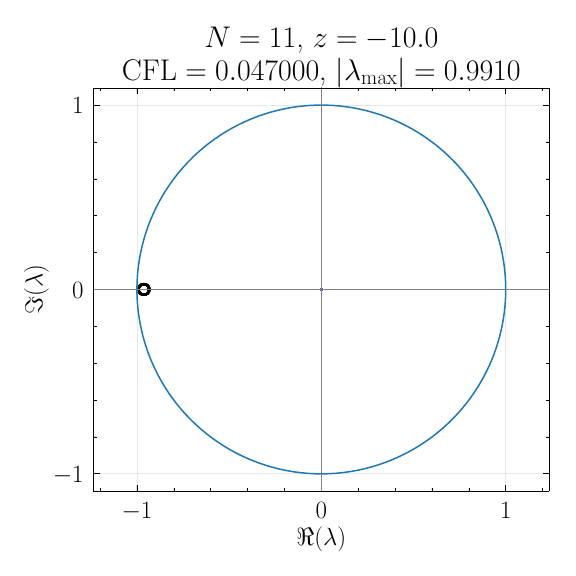}
\includegraphics[width=0.15\textwidth]{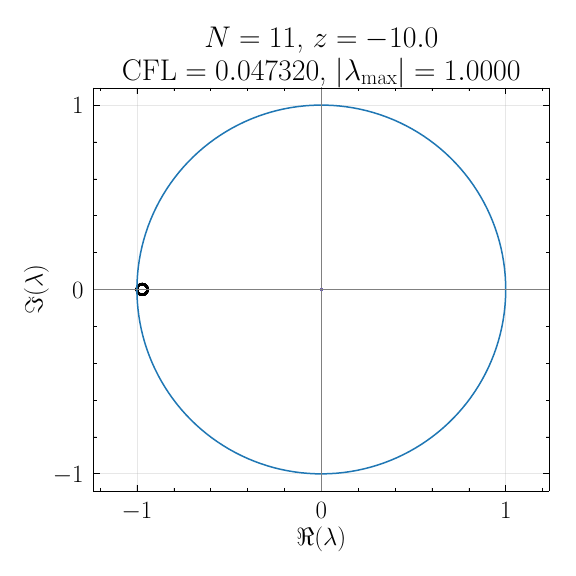}
\includegraphics[width=0.15\textwidth]{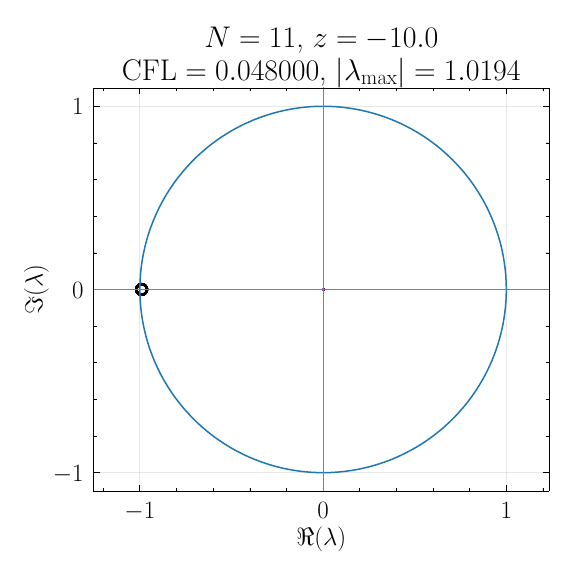}
\includegraphics[width=0.15\textwidth]{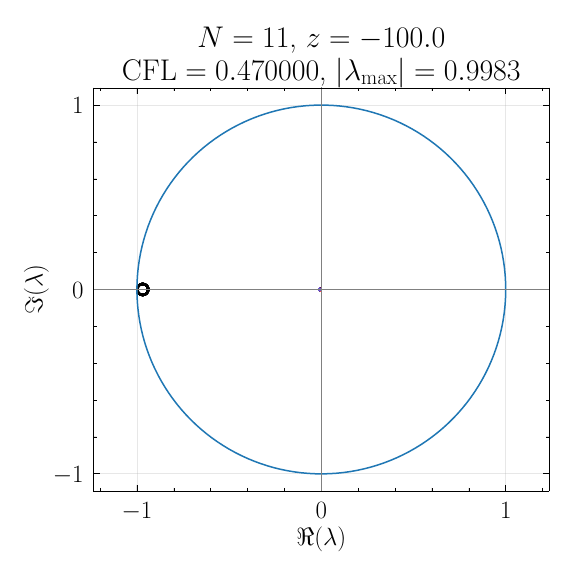}
\includegraphics[width=0.15\textwidth]{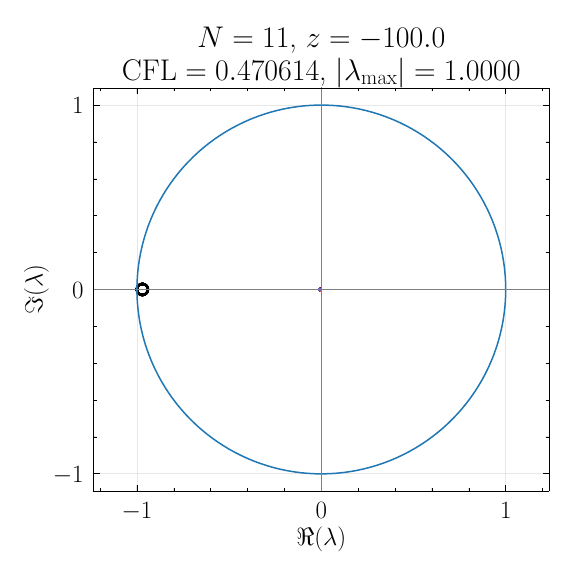}
\includegraphics[width=0.15\textwidth]{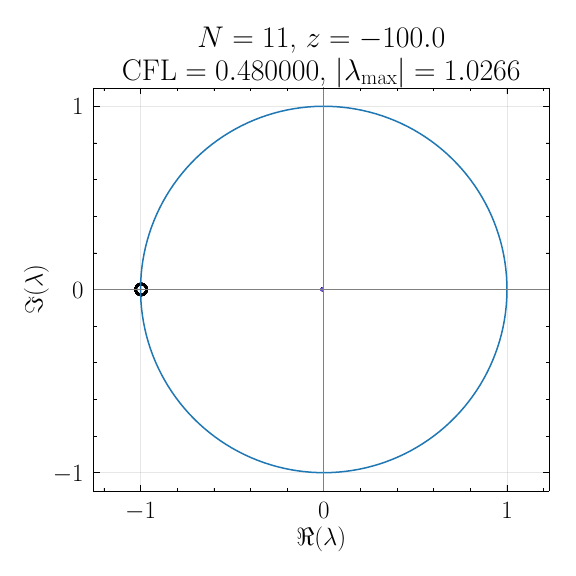}\\
\includegraphics[width=0.028125\textwidth]{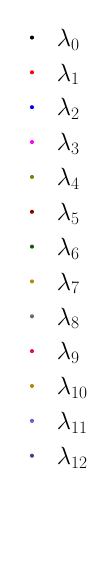}
\includegraphics[width=0.15\textwidth]{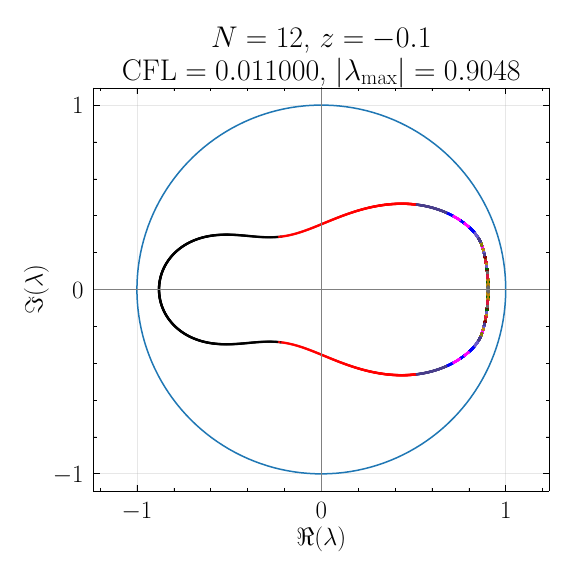}
\includegraphics[width=0.15\textwidth]{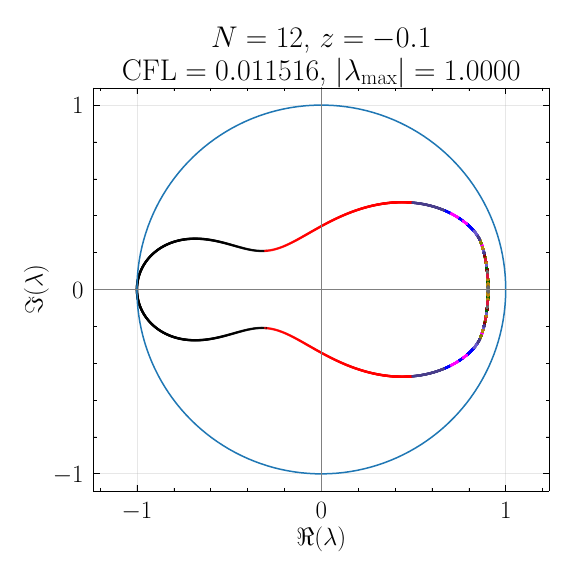}
\includegraphics[width=0.15\textwidth]{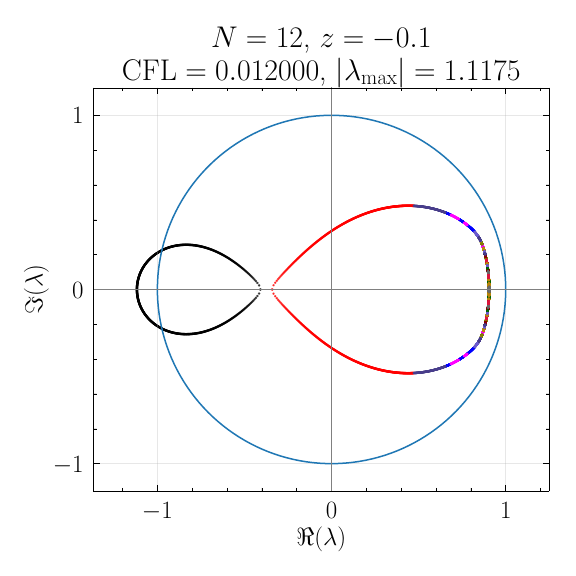}
\includegraphics[width=0.15\textwidth]{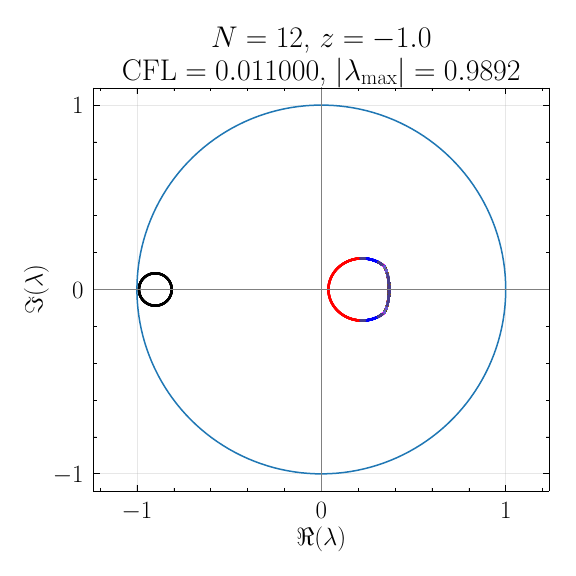}
\includegraphics[width=0.15\textwidth]{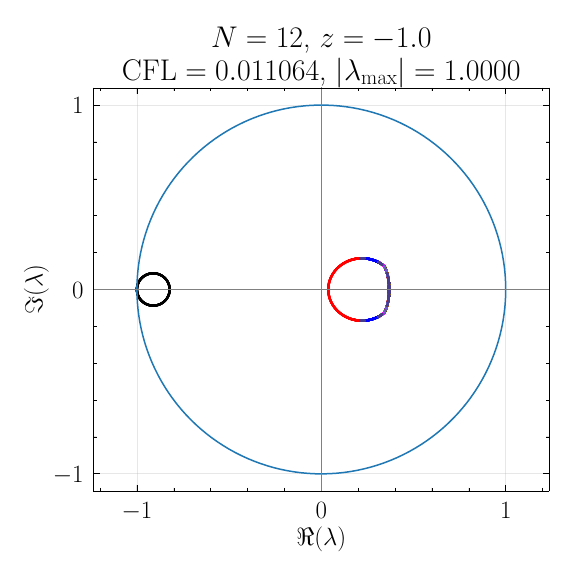}
\includegraphics[width=0.15\textwidth]{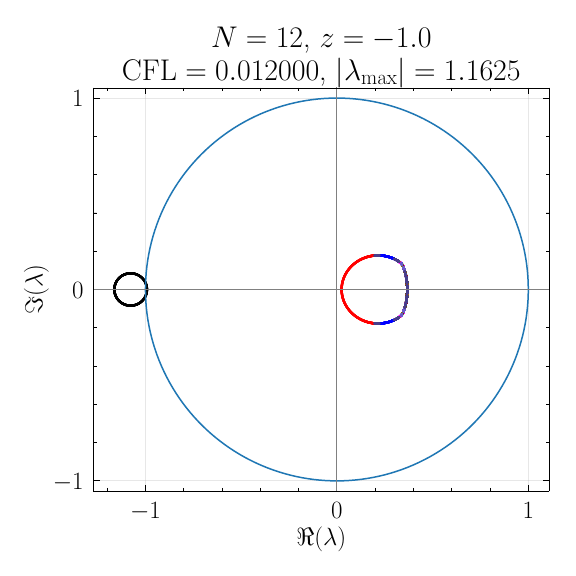}\\
\includegraphics[width=0.028125\textwidth]{spectrum_exact_eigvals_legend_N_12.png}
\includegraphics[width=0.15\textwidth]{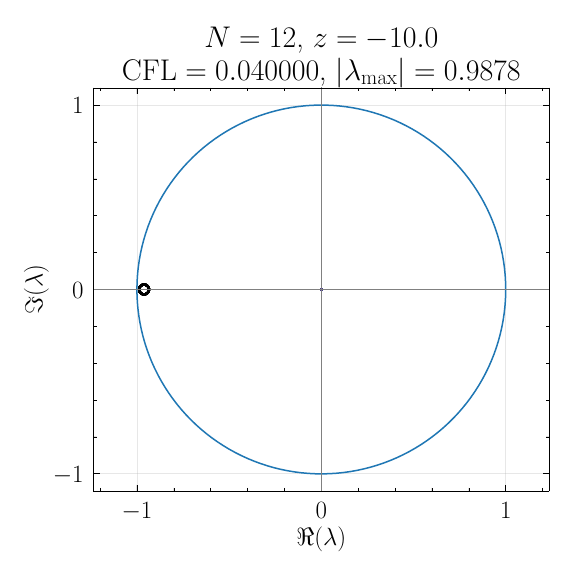}
\includegraphics[width=0.15\textwidth]{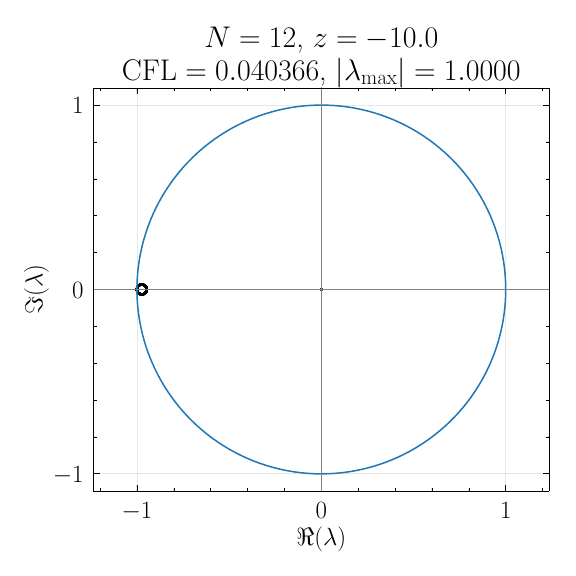}
\includegraphics[width=0.15\textwidth]{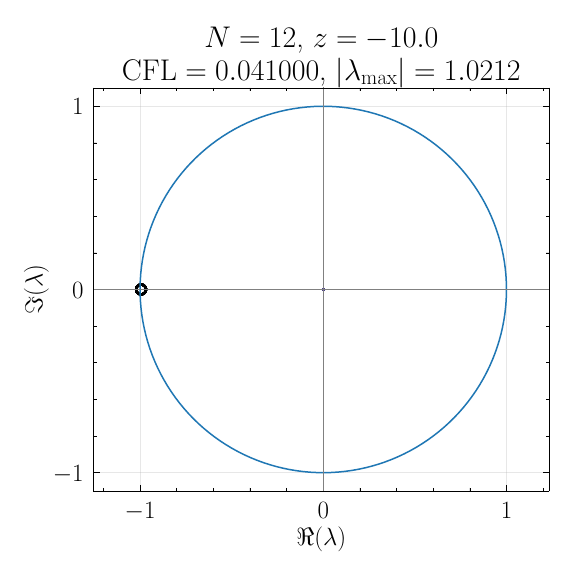}
\includegraphics[width=0.15\textwidth]{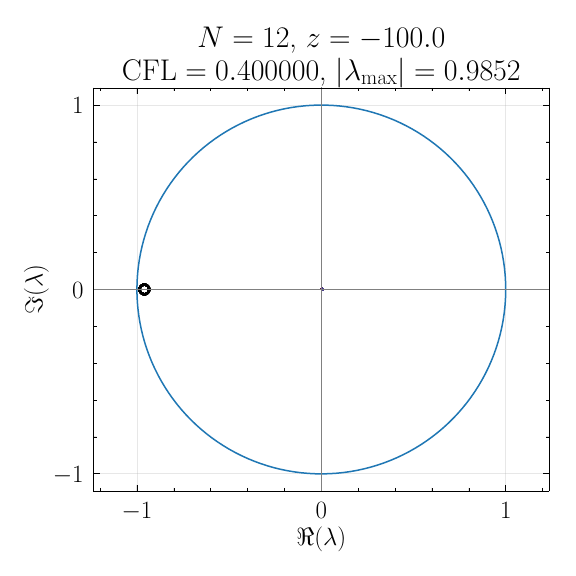}
\includegraphics[width=0.15\textwidth]{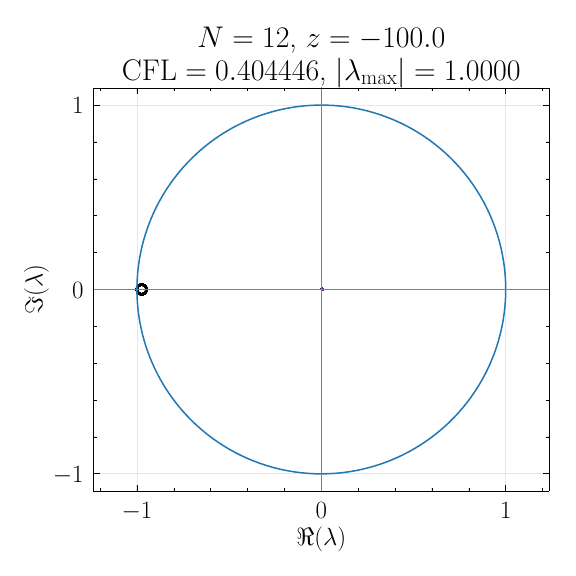}
\includegraphics[width=0.15\textwidth]{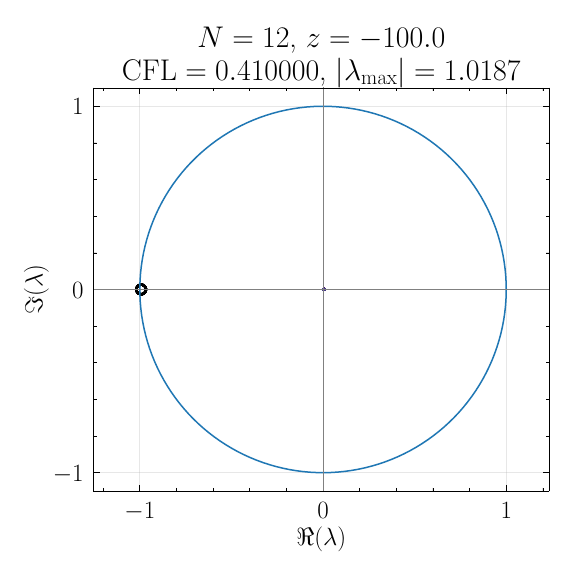}
\caption{%
Spectrum of the matrix $\mathrm{R}(\mathrm{CFL}, z, \theta)$ (\ref{eq:r_matrix_elems_expr_dup}) (eigenvalues $\lambda_{k} = \lambda_{k}(\mathrm{CFL}, z, \theta)$, $k = 0, \ldots, N$) of the evolution operator $R$ (\ref{eq:evol_oper_prop}) for a single time step $\Dtn{n}$ for several selected values of the Courant number $\mathrm{CFL}$ for $z = -0.1$, $-1.0$, $-10.0$, $-100.0$ and for polynomial degrees $N = 10, \ldots, 12$. The range of phase $\theta\in[0, 2\pi)$ is sampled on a uniform grid of $1000$ nodes. The gray circle of unit radius defines the stability boundary. Legends for each row of the graphs are located on the left. The gray circle of unit radius defines the stability boundary. The Courant numbers $\mathrm{CFL}$ are taken both at the boundary value of the Courant number $\mathrm{CFL}_{\rm max}$ and at two values $\mathrm{CFL}$ near it --- in the guaranteed stability region $\mathrm{CFL} < \mathrm{CFL}_{\rm max}$ and in the guaranteed instability region $\mathrm{CFL} > \mathrm{CFL}_{\rm max}$, by rounding off at the second significant decimal place (the values $\mathrm{CFL}_{\rm max}$ are calculated further in this work and presented in Figure~\ref{fig:cfls_on_z}). \textit{Note}: the phase $\arg \lambda_{k}$ of eigenvalue $\lambda_{k}$ is not the phase $\theta$; the dependence of the absolute value $|\lambda_{k}|$ on phase $\theta$ for these polynomial degrees $N$ is presented in Figure~\ref{fig:rhos_on_theta_degrees_10_12}.
}
\label{fig:spectrum_exact_eigvals_degrees_10_12}
\end{figure}

In general, it is impossible to perform a rigorous analytical analysis of the matrix spectrum $\rho[\mathrm{R}(\mathrm{CFL}, z, \theta)]$ (\ref{eq:r_matrix_elems_expr_dup}) to determine the stability region $[0, \mathrm{CFL}_{\rm max}]$ for an arbitrary degree $N$ of polynomials $\{\varphi_{p}\}$ (\ref{eq:phi_def}). This was impossible even for a linear advection equation~\cite{ader_dg_stab_rig}, and even more so for the full advection-reaction equation (\ref{eq:adv_eq_src}), for which the matrix $\rho[\mathrm{R}(\mathrm{CFL}, z, \theta)]$ (\ref{eq:r_matrix_elems_expr_dup}) has a significantly more complex form. In this work, a rigorous study combines both a rigorous derivation of algebraic equations (more precisely, polynomials on the $\mathrm{CFL}_{\rm max}$) for the boundary value of the Courant number $\mathrm{CFL}_{\rm max}$ and the derivation and substantiation of some purely empirical considerations necessary for obtaining these algebraic equations. The equation for the boundary of the stability region is quite complex for at least two reasons: it is precisely an inequality $\rho[\mathrm{R}(\mathrm{CFL}, z, \theta)] \leqslant 1$, and not an equation in the strict sense, since the problem contains an additional parameter --- the phase $\theta$, and the solution requires that the inequality be satisfied for any phase $\theta \in [0, 2\pi]$; and the eigenvalues themselves are complex numbers $\lambda_{k}$ --- $\lambda_{k} \in \mathcal{C}$. Therefore, before solving the problem of determining the stability region $[0, \mathrm{CFL}_{\rm max}]$, graphical representations of the dependencies of the eigenvalues were constructed, which are presented in Figures~\ref{fig:rhos_on_theta_degrees_1_3},~\ref{fig:rhos_on_theta_degrees_4_6},~\ref{fig:rhos_on_theta_degrees_7_9},~\ref{fig:rhos_on_theta_degrees_10_12}, where dependence of the absolute values $|\lambda_{k}|$ of the spectrum of the matrix $\mathrm{R}(\mathrm{CFL}, z, \theta)$ (\ref{eq:r_matrix_elems_expr_dup}) (eigenvalues $\lambda_{k} = \lambda_{k}(\mathrm{CFL}, z, \theta)$, $k = 0, \ldots, N$) of the evolution operator $R$ (\ref{eq:evol_oper_prop}) for a single time step $\Dtn{n}$ on phase $\theta = k\Dx$ for several selected values of the Courant number $\mathrm{CFL}$ for $z = -0.1$, $-1.0$, $-10.0$, $-100.0$ are presented, and in Figures~\ref{fig:spectrum_exact_eigvals_degrees_1_3},~\ref{fig:spectrum_exact_eigvals_degrees_4_6},~\ref{fig:spectrum_exact_eigvals_degrees_7_9},~\ref{fig:spectrum_exact_eigvals_degrees_10_12}, where the spectrum of the matrix $\mathrm{R}(\mathrm{CFL}, z, \theta)$ (\ref{eq:r_matrix_elems_expr_dup}) (eigenvalues $\lambda_{k} = \lambda_{k}(\mathrm{CFL}, z, \theta)$, $k = 0, \ldots, N$) of the evolution operator $R$ (\ref{eq:evol_oper_prop}) for a single time step $\Dtn{n}$ (also for several selected values of the Courant number $\mathrm{CFL}$ for $z = -0.1$, $-1.0$, $-10.0$, $-100.0$) are presented. The calculations to obtain these results are performed using a developed program in the \texttt{python} programming language. Floating-point calculations are performed using the \texttt{mpmath} module~\cite{mpmath_man} (with \texttt{gmp2} module), which enables calculations with arbitrarily high precision representation of real numbers as floating-point numbers. The presented results are calculated using \texttt{mpmath.mp.dps = 2000} (for comparison, for double-precision floating-point numbers \texttt{double}, \texttt{mpmath.mp.dps = 15}). The phase $\theta \in [0, 2\pi)$ is discretized using a uniform grid with $1000$ nodes. The calculation of the eigenvalues is additionally controlled by verification the control error $\|(\mathrm{R} - \lambda_{k}I)\mathbf{r}_{k}\|_{\infty}$, where $\mathbf{r}_{k}$ is the right eigenvector, and the control error did not exceed $\sim 10^{-1996}$--$10^{-2000}$ for degrees $N \leqslant 4$ and up to $\sim 10^{-1993}$--$10^{-1998}$ for degree $N = 12$. It is important to note that the results presented in the Figures~\ref{fig:rhos_on_theta_degrees_1_3},~\ref{fig:rhos_on_theta_degrees_4_6},~\ref{fig:rhos_on_theta_degrees_7_9},~\ref{fig:rhos_on_theta_degrees_10_12},~\ref{fig:spectrum_exact_eigvals_degrees_1_3},~\ref{fig:spectrum_exact_eigvals_degrees_4_6},~\ref{fig:spectrum_exact_eigvals_degrees_7_9} and~\ref{fig:spectrum_exact_eigvals_degrees_10_12} include the Courant numbers $\mathrm{CFL}_{\rm max}$ (\ref{eq:cfl_max_base_def}) obtained in this work, which will be presented later in the text in Figure~\ref{fig:cfls_on_z}.

Figures~\ref{fig:rhos_on_theta_degrees_1_3},~\ref{fig:rhos_on_theta_degrees_4_6},~\ref{fig:rhos_on_theta_degrees_7_9} and~\ref{fig:rhos_on_theta_degrees_10_12} demonstrate the dependence of the absolute value $|\lambda_{k}|$ of each eigenvalue $\lambda_{k}$ of the matrix $\mathrm{R}(\mathrm{CFL}, z, \theta)$ (\ref{eq:r_matrix_in_matrix_form}) of the evolution operator $R$ on the phase $\theta$ for $z = -0.1$, $-1.0$, $-10.0$, $-100.0$ and for several selected values of the Courant number $\mathrm{CFL}$ for polynomial degrees $N = 1, \ldots, 12$: for polynomial degrees $N = 1, \ldots, 3$ (Figure~\ref{fig:rhos_on_theta_degrees_1_3}), for $N = 4, \ldots, 6$ (Figure~\ref{fig:rhos_on_theta_degrees_4_6}), for $N = 7, \ldots, 9$ (Figure~\ref{fig:rhos_on_theta_degrees_7_9}) and for $N = 10, \ldots, 12$ (Figure~\ref{fig:rhos_on_theta_degrees_10_12}). The Courant numbers $\mathrm{CFL}$ are taken both at the boundary value of the Courant number $\mathrm{CFL}_{\rm max}$ and at two values $\mathrm{CFL}$ near it --- in the guaranteed stability region $\mathrm{CFL} < \mathrm{CFL}_{\rm max}$ and in the guaranteed instability region $\mathrm{CFL} > \mathrm{CFL}_{\rm max}$, by rounding off at the second significant decimal place (the values $\mathrm{CFL}_{\rm max}$ are calculated further in this work and presented in Figure~\ref{fig:cfls_on_z}). The presented dependencies $|\lambda_{k}| = |\lambda_{k}|(\theta)$ clearly demonstrate the complex structure of the phase $\theta$ dependence, especially for high-degree polynomials $N \geqslant 4$ (for low-degree polynomials, the boundary values of the Courant $\mathrm{CFL}_{\rm max}$ (\ref{eq:cfl_max_base_def}) number are simply calculated precisely analytically), which requires a very high degree of precision in resolving the intersection points of the curves of the maximum eigenvalue versus phase $\theta$ with the unit circle. Therefore, these results ``seem to directly suggest'' that it is desirable to eliminate the phase $\theta$ dependence in the form of a complex exponent $\exp(\pm i\theta)$ within the framework of a rigorous analysis, for example, by isolating the complex exponent $\exp(\pm i\theta)$ as a common factor on one side of the equation and then taking the absolute value of both sides of the equation (and using the property $|\exp(\pm i\theta)| = 1$ for $\theta\in[0, 2\pi]\subset\mathcal{R}$). It is necessary to emphasize the most important difference between the case of the linear advection-reaction equation (\ref{eq:adv_eq_src}) studied in this work and the linear advection equation studied similarly in~\cite{ader_dg_stab_rig}: with increasing value $|z|$ in the region $z < 0$, a significant partial compression of the matrix spectrum $\rho[\mathrm{R}(\mathrm{CFL}, z, \theta)]$ (\ref{eq:r_matrix_elems_expr_dup}) occurs. All eigenvalues $\lambda_{k}$, except one, decrease significantly in absolute value $|\lambda_{k}|$ with increasing value $|z|$ in the region $z < 0$. This behavior of the matrix spectrum $\rho[\mathrm{R}(\mathrm{CFL}, z, \theta)]$ is consistent with the characteristic qualitative features of the spectrum compression (\ref{eq:imex_spectr_compr}) of the matrix of classical IMEX schemes, which is presented in detail in Subsection~\ref{sec:stab_anal:qual_cons} ``Qualitative considerations''. Moreover, the decrease in the characteristic absolute value of the eigenvalues $\lambda_{k}$ of the compressed part of the spectrum $\rho[\mathrm{R}(\mathrm{CFL}, z, \theta)]$ for high degrees of polynomials $N \geqslant 4$ does not occur entirely monotonically with increasing value $|z|$, which is clearly seen from the comparison of the practical ``points'' into which the eigenvalues merge on a given scale for the case $z = -10.0$ and the fairly clearly distinguishable ``circles'' into which the eigenvalues merge on a given scale for the case $z = -100.0$. It is also interesting to note that with an increase in the degree $N$ of polynomials, the phase $\theta$ for which the largest eigenvalue in absolute value is achieved alternates between the values $0$ and $\pi$: $\theta = 0$ for odd degrees $N$ of polynomials, $\theta = \pi$ for even degrees $N$ of polynomials.

Figures~\ref{fig:spectrum_exact_eigvals_degrees_1_3},~\ref{fig:spectrum_exact_eigvals_degrees_4_6},~\ref{fig:spectrum_exact_eigvals_degrees_7_9} and~\ref{fig:spectrum_exact_eigvals_degrees_10_12} demonstrate the set of eigenvalues $\lambda_{k}$ of the matrix $\mathrm{R}(\mathrm{CFL}, z, \theta)$ (\ref{eq:r_matrix_in_matrix_form}) of the evolution operator $R$ for all phase values $\theta\in[0, 2\pi]$ in the complex plane for $z = -0.1$, $-1.0$, $-10.0$, $-100.0$ and for polynomial degrees $N = 1, \ldots, 12$: for polynomial degrees $N = 1, \ldots, 3$ (Figure~\ref{fig:spectrum_exact_eigvals_degrees_1_3}), for $N = 4, \ldots, 6$ (Figure~\ref{fig:spectrum_exact_eigvals_degrees_4_6}), for $N = 7, \ldots, 9$ (Figure~\ref{fig:spectrum_exact_eigvals_degrees_7_9}) and for $N = 10, \ldots, 12$ (Figure~\ref{fig:spectrum_exact_eigvals_degrees_10_12}). The presented dependencies $\lambda_{k} = \lambda_{k}(\mathrm{CFL}, z, \theta)$ also clearly demonstrate the complex structure of the dependency, especially for high polynomial degrees $N$. However, in this case, the phase $\theta$ dependence is not evident and does not ``obscure the overall picture'' of the studied dependency. The most important conclusion that clearly follows from the presented dependencies is that the intersection of the unit circle by the dependency $|\lambda_{k}| = |\lambda_{k}|(\mathrm{CFL}, z, \theta)$ with increasing Courant number occurs in the vicinity of the point $\lambda = -1$, therefore, it is expected that it is precisely upon reaching this eigenvalue $\lambda_{k} = -1$ with increasing Courant number $\mathrm{CFL}$ that the stability of the numerical method is lost. In general, this result could be considered expected, since a significant loss of stability by high-order numerical methods usually occurs when oscillations of the initially sufficiently smooth solution occur already at the first step of time integration, and for such oscillations to occur, it is necessary for the eigenvalue $\lambda$ to have a negative real part (which leads to a characteristic factor $(-1)^{n}$ for one or more harmonics $u^{n+1}_{k}$ of the numerical solution $u^{n+1}(\xi)$); and the location on the real axis may have symmetry reasons. It is important to note the following: the phase of an eigenvalue $\arg \lambda_{k}$ and the phase $\theta$ are different quantities --- the $\arg \lambda_{k}$ is the result of a calculation, and the phase $\theta$ is an independent argument in the problem of calculating the eigenvalues of the matrix $\mathrm{R}(\mathrm{CFL}, z, \theta)$ (\ref{eq:r_matrix_in_matrix_form}) of the evolution operator $R$. Therefore, the loss of stability in the vicinity of an eigenvalue $\lambda = -1$ does not clearly demonstrate anything about the phase $\theta$ for which this loss of stability occurs (of course, the phase $\theta$ can be calculated separately). The obtained results showed that it is better to analyze the stability boundary $\mathrm{CFL}_{\rm max}$ (\ref{eq:cfl_max_base_def}) as follows: first, transform the original problem $\rho[\mathrm{R}(\mathrm{CFL}, z, \theta)] \leqslant 1$ in order to exclude the phase $\theta$ included in the analyzed expression in the form $\exp(\pm i\theta)$, and then try to simplify this expression as much as possible under the assumption of searching for the boundary value of the Courant number $\mathrm{CFL}_{\rm max}$ for which one of the eigenvalues $\lambda_{k}$ of the matrix (for any phase $\theta$) reaches the value $\lambda = -1$. It is important to emphasize the significant difference in the results for the advection-reaction equation (\ref{eq:adv_eq_src}) from the linear advection equation studied similarly in~\cite{ader_dg_stab_rig}, which are revealed in Figures~\ref{fig:spectrum_exact_eigvals_degrees_1_3},~\ref{fig:spectrum_exact_eigvals_degrees_4_6},~\ref{fig:spectrum_exact_eigvals_degrees_7_9} and~\ref{fig:spectrum_exact_eigvals_degrees_10_12}. In the case of a linear advection equation, the following interesting property took place: although the stability $\rho[\mathrm{R}(\mathrm{CFL}, z, \theta)] \leqslant 1$ of the ADER-DG numerical method with LST-DG predictor is lost when one of the eigenvalues reaches $\lambda = -1$, for sufficiently large polynomial degrees $N$, even at lower Courant numbers $\mathrm{CFL} \ll \mathrm{CFL}_{\rm max}$ (\ref{eq:cfl_max_base_def}), some eigenvalues $\lambda_{k}$ of the matrix $\mathrm{R}(\mathrm{CFL}, z, \theta)$ (\ref{eq:r_matrix_in_matrix_form}) of the evolution operator $R$ with absolute values $|\lambda_{k}|$ greater than one remain in the vicinity of point $\lambda = +1$, approximately in the region of the eigenvalue phase $\arg\lambda_{k} \in [-\pi/4, +\pi/4]$; however, they exceed unity only by $\sim 10^{-5}$--$10^{-9}$, and the Courant numbers $\mathrm{CFL}$ for which the condition is achieved in this region are unusually small --- $1.5$--$4.0$ times smaller than the boundary value $\mathrm{CFL}_{\rm max}$. However, in the case of the advection-reaction equation (\ref{eq:adv_eq_src}), this property no longer holds --- the stability is lost strictly when one of the eigenvalues reaches $\lambda = -1$. Clearly, this property is likely not violated for any arbitrarily small parameter $z < 0$, but rather, there exists a range of $z$ in which this property is smoothly violated. However, this range is limited by $|z| \leqslant 10^{-5}$--$10^{-6}$ and is of no significant interest.

It is also necessary to note some interesting features of the distribution of eigenvalues $\{\lambda_{k}(\mathrm{CFL}, z, \theta)\}$ for all available values of the phase $\theta\in[0, 2\pi]$ on the complex plane. It is evident that for small deviations of the Courant number $\mathrm{CFL}$ from the boundary value of the Courant number $\mathrm{CFL}_{\rm max}(N, z)$, at small values of the parameter $z$, the set $\{\lambda_{k}(\mathrm{CFL}, z, \theta)\}$ of eigenvalues is closed curves. As the parameter $z$ increases in absolute value, this form of the set $\{\lambda_{k}(\mathrm{CFL}, z, \theta)\}$ of eigenvalues is violated: there are now two closed curves. One of them contains exactly one eigenvalue and is located in the neighborhood of point $\lambda = -1$, and its characteristic radius decreases with increasing $z$. The second curve contains all the remaining eigenvalues and is located closer to point $\lambda = 0$; and its characteristic radius decreases with increasing $z$, but significantly faster than for the first curve. The value of the parameter $z$, which separates these two modes of the distribution of eigenvalues $\{\lambda_{k}(\mathrm{CFL}, z, \theta)\}$ on the complex plane, depends significantly on the polynomial degrees $N$: from the results presented in Figures~\ref{fig:spectrum_exact_eigvals_degrees_1_3},~\ref{fig:spectrum_exact_eigvals_degrees_4_6},~\ref{fig:spectrum_exact_eigvals_degrees_7_9} and~\ref{fig:spectrum_exact_eigvals_degrees_10_12}, it is $z = -10.0$ for polynomial degrees $N = 1, \ldots, 3$, and $z = -1.0$ for polynomial degrees $N = 4, \ldots, 12$.

\subsection{Rigorous quantitative considerations}
\label{sec:stab_anal:quant_cons}

The eigenvalues of the matrix $\mathrm{R}(\mathrm{CFL}, z, \theta)$ (\ref{eq:r_matrix_in_matrix_form}) are determined by the roots of the characteristic equation, which, up to the sign on the left-hand side, has the form $\det[R(\mathrm{CFL}, z, \theta) - \lambda I] = 0$. This equation will then be subjected to a series of transformations that make the equation free of the phase factor $\exp(\pm i\theta)$, which contains the entire phase $\theta$ dependence of the characteristic equation. An equation will then be obtained for the boundary value of Courant number $\mathrm{CFL}_{\rm max}$, which determines the stability boundary (\ref{eq:cfl_max_base_def}) of the ADER-DG numerical method with the LST-DG predictor. For the last transformation, it is also shown that the boundary value of Courant number $\mathrm{CFL}_{\rm max}$ indeed corresponds to one of the eigenvalues $\lambda_{k}$ reaching the point $\lambda = -1$.

A detailed derivation depends significantly on the sign of the advection velocity $a$, so the derivation is first made for case $a > 0$. In this case, the expression for matrices $\mathrm{A}(\mathrm{CFL}, z, \theta)$ and $\mathrm{B}(\mathrm{CFL}, z)$ (\ref{eq:a_and_b_matrices_def_dup}), which define the matrix $\mathrm{R}(\mathrm{CFL}, z, \theta) = I + \mathrm{A}(\mathrm{CFL}, z, \theta)\mathrm{B}(\mathrm{CFL}, z)$ (\ref{eq:r_matrix_in_matrix_form}) of the evolution operator $R$, takes the following form:
\begin{equation}\label{eq:f_derivd_temp_1_a_plus}
\begin{split}
&\mathrm{A}(\mathrm{CFL}, z, \theta) = \mathrm{CFL} \cdot m^{-1}\Big[\mathrm{D}^{T}m - \Big(\ovecphi - \uvecphi\exp(-i\theta)\Big)\otimes\ovecphi^{T}\Big] + z I,\\
&\mathrm{B}(\mathrm{CFL}, z) = \sum\limits_{s = 0}^{N} \frac{(-1)^{s}}{s!} \mathrm{CFL}^{s} \frac{d^{s}}{dz^{s}}\left[\frac{R_{\rm RK}(z) - 1}{z}\right]\mathrm{D}^{s}.
\end{split}
\end{equation}
To obtain an expression for the characteristic equation $\det[R(\mathrm{CFL}, z, \theta) - \lambda I] = 0$ in a form suitable for quantitative estimates, an integration-by-parts identity is introduced for the chosen polynomial $f\in\mathcal{P}_{N}$ representation in the chosen functional basis $\{\varphi_{p}\}_{p}$:
\begin{equation}\label{eq:ibp_prop}
m\mathrm{D} + \mathrm{D}^{T}m = \ovecphi\otimes\ovecphi^{T} - \uvecphi\otimes\uvecphi^{T},\quad
\varrho + \varrho^{T} = \ovecphi\otimes\ovecphi^{T} - \uvecphi\otimes\uvecphi^{T},
\end{equation}
which is proved by simple integration-by-parts of the element $\varrho_{pq}$ of matrix $\varrho$ (\ref{eq:matrices_def}):
\begin{equation}
\begin{split}
&\intrefdom{\xi}\,\varphi_{p}(\xi)\varphi_{q}'(\xi) = \varphi_{p}(1)\varphi_{q}(1) - \varphi_{p}(0)\varphi_{q}(0) - \intrefdom{\xi}\,\varphi_{p}'(\xi)\varphi_{q}(\xi),\\
&\intrefdom{\xi}\,\varphi_{p}(\xi)\varphi_{q}'(\xi) + \intrefdom{\xi}\,\varphi_{p}'(\xi)\varphi_{q}(\xi) = \varphi_{p}(1)\varphi_{q}(1) - \varphi_{p}(0)\varphi_{q}(0).
\end{split}
\end{equation}
The resulting integration-by-parts identity (\ref{eq:ibp_prop}) is used to express the matrix $\mathrm{D}^{T}m = \ovecphi\otimes\ovecphi^{T} - \uvecphi\otimes\uvecphi^{T} - m\mathrm{D}$, leading to the following chain of transformations for matrix $\mathrm{A}(\mathrm{CFL}, z, \theta)$:
\begin{equation}
\begin{split}
\mathrm{A}(\mathrm{CFL}, z, \theta) & = \mathrm{CFL} \cdot m^{-1}\Big[\mathrm{D}^{T}m - \Big(\ovecphi - \uvecphi\exp(-i\theta)\Big)\otimes\ovecphi^{T}\Big] + z I \\ & =
z I + \mathrm{CFL} \cdot m^{-1}\Big[\mathrm{D}^{T}m - \ovecphi\otimes\ovecphi^{T} + \exp(-i\theta)\uvecphi\otimes\ovecphi^{T}\Big] \\ & =
z I + \mathrm{CFL} \cdot m^{-1}\Big[\ovecphi\otimes\ovecphi^{T} - \uvecphi\otimes\uvecphi^{T} - m\mathrm{D} - 
\ovecphi\otimes\ovecphi^{T} + \exp(-i\theta)\uvecphi\otimes\ovecphi^{T}\Big]\\ & =
z I + \mathrm{CFL} \cdot m^{-1}\Big[\exp(-i\theta)\uvecphi\otimes\ovecphi^{T} - \uvecphi\otimes\uvecphi^{T} - m\mathrm{D}\Big]\\ & =
z I - \mathrm{CFL} \cdot \mathrm{D} + \mathrm{CFL} \cdot m^{-1}\Big[\uvecphi\otimes\Big(\exp(-i\theta)\ovecphi^{T} - \uvecphi^{T}\Big)\Big],
\end{split}
\end{equation}
where the following matrix $\mathrm{Z}^{+}(\mathrm{CFL}, z)$ is introduced:
\begin{equation}\label{eq:z_plus_matrix_def}
\mathrm{Z}^{+}(\mathrm{CFL}, z) = z I - \mathrm{CFL} \cdot \mathrm{D},
\end{equation}
using which, the expression for matrix $\mathrm{A}(\mathrm{CFL}, z, \theta)$ takes the following form:
\begin{equation}\label{eq:f_derivd_temp_1_1_a_plus}
\mathrm{A}(\mathrm{CFL}, z, \theta) = \mathrm{Z}^{+}(\mathrm{CFL}, z) + \mathrm{CFL} \cdot m^{-1}\Big[\uvecphi\otimes\Big(\exp(-i\theta)\ovecphi^{T} - \uvecphi^{T}\Big)\Big].
\end{equation}
Treating the expression (\ref{eq:f_derivd_temp_1_a_plus}) for the matrix $\mathrm{B}(\mathrm{CFL}, z)$ as a Taylor series expansion and taking into account the nilpotency of matrix $\mathrm{D}$: $\mathrm{D}^{N+1} = 0$, the following relation is obtained:
\begin{equation}\label{eq:f_derivd_temp_1_2_a_plus}
\begin{split}
\mathrm{B}(\mathrm{CFL}, z) & = \sum\limits_{s = 0}^{N} \frac{(-1)^{s}}{s!} \mathrm{CFL}^{s} \frac{d^{s}}{dz^{s}}\left[\frac{R_{\rm RK}(z) - 1}{z}\right]\mathrm{D}^{s} =
\sum\limits_{s = 0}^{N} \left\{\frac{d^{s}}{dz^{s}}\left[\frac{R_{\rm RK}(z) - 1}{z}\right]\right\} \frac{(-\mathrm{CFL}\cdot\mathrm{D})^{s}}{s!}\\ & =
\sum\limits_{s = 0}^{N} \left\{\frac{d^{s}}{dz^{s}}\Big[(R_{\rm RK}(z I) - I)z^{-1} I\Big]\right\} \frac{(-\mathrm{CFL}\cdot\mathrm{D})^{s}}{s!} =
\Big[R_{\rm RK}\Big(z I - \mathrm{CFL} \cdot \mathrm{D}\Big) - I\Big]\\ & \cdot \Big[z I - \mathrm{CFL} \cdot \mathrm{D}\Big]^{-1} =
\Big[R_{\rm RK}(\mathrm{Z}^{+}(\mathrm{CFL}, z)) - I\Big]\Big[\mathrm{Z}^{+}(\mathrm{CFL}, z)\Big]^{-1},
\end{split}
\end{equation}
where it is noted that, in case $z \neq 0$, matrix $\mathrm{Z}^{+}(\mathrm{CFL}, z)$ is invertible, being the sum of a diagonal matrix $z I$ and a nilpotent matrix $\mathrm{CFL} \cdot \mathrm{D}$ with all zero eigenvalues, and $R_{\rm RK}(z) I = R_{\rm RK}(z I)$, since all powers of the identity matrix $I$ are equal to the identity matrix $I$. Taking into account the previously obtained expression (\ref{eq:f_derivd_temp_1_1_a_plus}) for the matrix $\mathrm{A}(\mathrm{CFL}, z, \theta)$ and expression (\ref{eq:f_derivd_temp_1_2_a_plus}) for the matrix $\mathrm{B}(\mathrm{CFL}, z)$, the expression for matrix $\mathrm{R}(\mathrm{CFL}, z, \theta)$ is rewritten in the following form:
\begin{equation}
\begin{split}
\mathrm{R}(\mathrm{CFL}, z, \theta) & = I + \mathrm{A}(\mathrm{CFL}, z, \theta) \mathrm{B}(\mathrm{CFL}, z) =
I + \Big[\mathrm{Z}^{+}(\mathrm{CFL}, z) + \mathrm{CFL} \cdot m^{-1}\Big[\uvecphi\otimes\Big(\exp(-i\theta)\ovecphi^{T} - \uvecphi^{T}\Big)\Big]\Big]\\ & \cdot
\mathrm{B}(\mathrm{CFL}, z) = I + \mathrm{Z}^{+}(\mathrm{CFL}, z)\mathrm{B}(\mathrm{CFL}, z) +
\mathrm{CFL} \cdot m^{-1}\Big[\uvecphi\otimes\Big(\exp(-i\theta)\ovecphi^{T} - \uvecphi^{T}\Big)\Big]\mathrm{B}(\mathrm{CFL}, z)\\ & =
R_{\rm RK}(\mathrm{Z}^{+}(\mathrm{CFL}, z)) + \mathrm{CFL} \cdot m^{-1}\Big[\uvecphi\otimes
\Big(\exp(-i\theta)\ovecphi^{T}\mathrm{B}(\mathrm{CFL}, z) - \uvecphi^{T}\mathrm{B}(\mathrm{CFL}, z)\Big)\Big],
\end{split}
\end{equation}
where a consequence $\mathrm{Z}^{+}(\mathrm{CFL}, z)\mathrm{B}(\mathrm{CFL}, z) = R_{\rm RK}(\mathrm{Z}^{+}(\mathrm{CFL}, z)) - I$ of relation (\ref{eq:f_derivd_temp_1_2_a_plus}) is explicitly taken into account. The expression for the characteristic equation in a form suitable for quantitative estimates is obtained by applying a well-known relation $\det(\mathrm{K}+\mathbf{a}\otimes\mathbf{b}^{T}) = \det{K}\cdot(1 + \mathbf{b}^{T}\mathrm{K}^{-1}\mathbf{a})$ within the expression for the characteristic equation $\det[R(\mathrm{CFL}, z, \theta) - \lambda I] = 0$:\\[-2mm]
\begin{equation}\label{eq:f_derivd_temp_2_a_plus}
\begin{split}
\det\Big[\mathrm{R}(\mathrm{CFL}, z, \theta) - \lambda I\Big] & = 
\det\Big[\Big(R_{\rm RK}(\mathrm{Z}^{+}(\mathrm{CFL}, z)) - \lambda I\Big) + \mathrm{CFL} \cdot m^{-1}\Big[\uvecphi\otimes
\Big(\exp(-i\theta)\ovecphi^{T}\mathrm{B}(\mathrm{CFL}, z)\\ & - \uvecphi^{T}\mathrm{B}(\mathrm{CFL}, z)\Big)\Big]\Big] =
\det\Big[R_{\rm RK}(\mathrm{Z}^{+}(\mathrm{CFL}, z)) - \lambda I\Big]\cdot\Big\{
1 + \mathrm{CFL}\cdot\Big(\exp(-i\theta)\\ & \cdot\ovecphi^{T}\mathrm{B}(\mathrm{CFL}, z) - \uvecphi^{T}\mathrm{B}(\mathrm{CFL}, z)\Big)
\Big[R_{\rm RK}(\mathrm{Z}^{+}(\mathrm{CFL}, z)) - \lambda I\Big]^{-1} \Big(m^{-1}\uvecphi\Big)\Big\}\\ & =
\det\Big[R_{\rm RK}(\mathrm{Z}^{+}(\mathrm{CFL}, z)) - \lambda I\Big]\cdot\Big\{
1 + \mathrm{CFL}\cdot\exp(-i\theta)\ovecphi^{T}\mathrm{B}(\mathrm{CFL}, z)\\ & \cdot
\Big[R_{\rm RK}(\mathrm{Z}^{+}(\mathrm{CFL}, z)) - \lambda I\Big]^{-1} \Big(m^{-1}\uvecphi\Big) -
\mathrm{CFL}\cdot\uvecphi^{T}\mathrm{B}(\mathrm{CFL}, z)\\ & \cdot
\Big[R_{\rm RK}(\mathrm{Z}^{+}(\mathrm{CFL}, z)) - \lambda I\Big]^{-1} \Big(m^{-1}\uvecphi\Big)\Big\}.
\end{split}
\end{equation}
After introducing the following notation:\\[-2mm]
\begin{equation}\label{eq:upsilons_def_plus}
\begin{split}
&\Upsilon^{+}_{\mathrm{A}}(\mathrm{CFL}, z, \lambda) = \ovecphi^{T}\mathrm{B}(\mathrm{CFL}, z)\Big[R_{\rm RK}(\mathrm{Z}^{+}(\mathrm{CFL}, z)) - \lambda I\Big]^{-1} m^{-1}\uvecphi,\\
&\Upsilon^{+}_{\mathrm{B}}(\mathrm{CFL}, z, \lambda) = \uvecphi^{T}\mathrm{B}(\mathrm{CFL}, z)\Big[R_{\rm RK}(\mathrm{Z}^{+}(\mathrm{CFL}, z)) - \lambda I\Big]^{-1} m^{-1}\uvecphi,
\end{split}
\end{equation}
the equation (\ref{eq:f_derivd_temp_2_a_plus}) is transformed to the following form:\\[-2mm]
\begin{equation}\label{eq:upsilon_equation_plus_def_theta}
\begin{split}
\det\Big[\mathrm{R}(\mathrm{CFL}, z, \theta) - \lambda I\Big] = 
\det\Big[R_{\rm RK}(\mathrm{Z}^{+}(\mathrm{CFL}, z)) - \lambda I\Big]\cdot
\Big[&\mathrm{CFL}\cdot\exp(-i\theta)\Upsilon^{+}_{\mathrm{A}}(\mathrm{CFL}, \lambda)\\ & + 1 - \mathrm{CFL}\cdot\Upsilon^{+}_{\mathrm{B}}(\mathrm{CFL}, \lambda)\Big] = 0,
\end{split}
\end{equation}
where the first factor is the characteristic equation for matrix $R_{\rm RK}(\mathrm{Z}^{+}(\mathrm{CFL}, z))$. Since matrix $\mathrm{D}$ has only zero eigenvalues, it can be represented in the following form:
\begin{equation}
\mathrm{D} = \mathrm{S}\mathrm{D}_{\triangle}\mathrm{S}^{-1},\quad
\mathrm{D}_{\triangle} = \mathrm{S}^{-1}\mathrm{D}\mathrm{S},
\end{equation}
where $\mathrm{D}_{\triangle} \in SUT_{N+1}(\mathcal{R})$ is a strictly upper triangular matrix, and $\mathrm{S}$ is a transition matrix. Therefore, matrix $\mathrm{Z}^{+}(\mathrm{CFL}, z)$ (\ref{eq:z_plus_matrix_def}) can be represented as follows:
\begin{equation}
\mathrm{Z}^{+}(\mathrm{CFL}, z) = z I - \mathrm{CFL} \cdot \mathrm{D} =
z \cdot \mathrm{S}I\mathrm{S}^{-1} - \mathrm{CFL} \cdot \mathrm{S}\mathrm{D}_{\triangle}\mathrm{S}^{-1} =
\mathrm{S}(zI - \mathrm{D}_{\triangle})\mathrm{S}^{-1},
\end{equation}
which, taking into account elementary property $R_{\rm RK}(zI) = R_{\rm RK}(z)I$, allows the expression for the characteristic equation $\det[R_{\rm RK}(\mathrm{Z}^{+}(\mathrm{CFL}, z)) - \lambda I]$ of the eigenvalue problem for matrix $R_{\rm RK}(\mathrm{Z}^{+}(\mathrm{CFL}, z))$ to be rewritten as follows:\\[-2mm]
\begin{equation}
\begin{split}
&\det\Big[R_{\rm RK}(\mathrm{Z}^{+}(\mathrm{CFL}, z)) - \lambda I\Big] = 
 \det\Big[R_{\rm RK}(\mathrm{S}(zI - \mathrm{D}_{\triangle})\mathrm{S}^{-1}) - \lambda \mathrm{S}I\mathrm{S}^{-1}\Big]\\ & = 
 \det(S)\det\Big[R_{\rm RK}(zI - \mathrm{D}_{\triangle}) - \lambda I\Big]\det(S^{-1}) =
 \det\Big[R_{\rm RK}(z)I - \lambda I\Big] = (R_{\rm RK}(z) - \lambda)^{N+1},
\end{split}
\end{equation}
from which it follows that the matrix $R_{\rm RK}(\mathrm{Z}^{+}(\mathrm{CFL}, z))$ has only the eigenvalue $\lambda = R_{\rm RK}(z)$ with algebraic multiplicity $N+1$. It does not follow from this result and the previous expression (\ref{eq:upsilon_equation_plus_def_theta}) that the matrix $\mathrm{R}(\mathrm{CFL}, z, \theta)$ is also characterized by the same set of eigenvalues --- the expressions for functions $\Upsilon^{+}_{\mathrm{A}}(\mathrm{CFL}, z, \lambda)$ and $\Upsilon^{+}_{\mathrm{B}}(\mathrm{CFL}, z, \lambda)$ (\ref{eq:upsilons_def_plus}) in the case $\theta \neq 0$ (and also equivalent case $\theta \neq 2\pi$) contain singular terms in expression $[R_{\rm RK}(\mathrm{Z}^{+}(\mathrm{CFL}, z)) - \lambda I]^{-1}$, because these matrices are non-invertible in the case $\lambda = R_{\rm RK}(z)$. In the case $\theta \neq 0$ (and also equivalent case $\theta \neq 2\pi$) indirectly considered in the end of Subsection~\ref{sec:stab_anal:statment_quant_cons} ``Statement of the problem of quantitative considerations'', it clearly follows from the expression $\mathrm{R}(\mathrm{CFL}, 0)\mathbf{1} = R_{\rm RK}(z)\mathbf{1}$ (\ref{eq:cont_presiv_prop}) that each $\varphi_{k}$-harmonic evolves by multiplying by $R_{\rm RK}(z)$ for each time step $\Dtn{n}$, and the result does not depend on the $\varphi_{k}$-harmonic number and on the Courant number $\mathrm{CFL}$, so this result strictly shows that in case $\theta = 0$ (or also equivalent case $\theta = 2\pi$) the matrix $\mathrm{R}$ has only one eigenvalue $R_{\rm RK}(z)$ with multiplicity $N+1$. To clarify this problem in the case $\theta \in (0, 2\pi)$, the representation of matrix $R_{\rm RK}(\mathrm{Z}^{+}(\mathrm{CFL}, z))$ as a Taylor series expansion is considered:\\[-2mm]
\begin{equation}
\begin{split}
R_{\rm RK}(\mathrm{Z}^{+}(\mathrm{CFL}, z)) = R_{\rm RK}(z) I + \sum\limits_{s = 1}^{N} \frac{d^{s}R_{\rm RK}(z)}{dz^{s}} (-\mathrm{CFL}\cdot\mathrm{D})^{s},
\end{split}
\end{equation}
where the nilpotency $\mathrm{D}^{N+1} = 0$ of matrix $\mathrm{D}$ is also taken into account. Also considering the following relation:\\[-2mm]
\begin{equation}
\left[\sum\limits_{s = 1}^{N} \frac{d^{s}R_{\rm RK}(z)}{dz^{s}} (-\mathrm{CFL}\cdot\mathrm{D})^{s}\right]^{N+1} = 0,
\end{equation}
the following expression for the inverse matrix $[R_{\rm RK}(\mathrm{Z}^{+}(\mathrm{CFL}, z)) - \lambda I]^{-1}$ is obtained:\\[-2mm]
\begin{equation}
\begin{split}
\Big[R_{\rm RK}(\mathrm{Z}^{+}(\mathrm{CFL}, z)) - \lambda I\Big]^{-1} & = 
\left[(R_{\rm RK}(z) - \lambda) I + \sum\limits_{s = 1}^{N} \frac{d^{s}R_{\rm RK}(z)}{dz^{s}} (-\mathrm{CFL}\cdot\mathrm{D})^{s}\right]^{-1}\\ & =
\sum\limits_{j = 0}^{N} \frac{(-1)^{j}}{(R_{\rm RK}(z) - \lambda)^{j+1}} \left[\sum\limits_{s = 1}^{N} \frac{d^{s}R_{\rm RK}(z)}{dz^{s}} (-\mathrm{CFL}\cdot\mathrm{D})^{s}\right]^{j},
\end{split}
\end{equation}
from which the following asymptotic behavior in the vicinity of point $\lambda = R_{\rm RK}(z)$ follows:
\begin{equation}
\Big[R_{\rm RK}(\mathrm{Z}^{+}(\mathrm{CFL}, z)) - \lambda I\Big]^{-1} \sim \frac{1}{(R_{\rm RK}(z) - \lambda)^{N+1}} \quad\mathrm{as}\quad \lambda \rightarrow R_{\rm RK}(z).
\end{equation}
From the obtained asymptotic estimate and the fact that the characteristic polynomial is indeed a polynomial, the following form of the left side of the characteristic equation $\det[\mathrm{R}(\mathrm{CFL}, z, \theta) - \lambda I] = 0$ follows:\\[-2mm]
\begin{equation}\label{eq:R_eigenval_prop_a_plus}
\det\Big[\mathrm{R}(\mathrm{CFL}, z, \theta) - \lambda I\Big] = \mathfrak{h}^{+}(\lambda; \mathrm{CFL}, z, \theta),
\end{equation}
where $\mathfrak{h}^{+} \in \mathcal{P}_{N}(\mathcal{C})$ is a polynomial of degree $N+1$ with no roots at the point $\lambda = R_{\rm RK}(z)$: $\mathfrak{h}^{+}(R_{\rm RK}(z); \mathrm{CFL}, z, \theta) \neq 0$. Therefore, the matrix $\mathrm{R}(\mathrm{CFL}, z, \theta)$ has no an eigenvalue $\lambda = R_{\rm RK}(z)$ in its spectrum. Consequently, only the second factor remains in the characteristic equation (\ref{eq:upsilon_equation_plus_def_theta}), leading to the following equation:\\[-2mm]
\begin{equation}\label{eq:upsilon_equation_plus_def_theta_final}
\mathrm{CFL}\cdot\exp(-i\theta)\Upsilon^{+}_{\mathrm{B}}(\mathrm{CFL}, z, \lambda) = 1 - \mathrm{CFL}\cdot\Upsilon^{+}_{\mathrm{A}}(\mathrm{CFL}, z, \lambda).
\end{equation}
Now, to eliminate the need to account for the phase $\theta \in [0, 2\pi]$, it is necessary to take the absolute value of both sides of the resulting equation:
\begin{equation}\label{eq:upsilon_equation_plus_def_theta_final_abs}
|\mathrm{CFL}\cdot\Upsilon^{+}_{\mathrm{B}}(\mathrm{CFL}, z, \lambda)|^{2} = |1 - \mathrm{CFL}\cdot\Upsilon^{+}_{\mathrm{A}}(\mathrm{CFL}, z, \lambda)|^{2},
\end{equation}
which leads to a phase-independent relationship involving the Courant number $\mathrm{CFL}$. The functions $\Upsilon^{+}_{\mathrm{A}}(\mathrm{CFL}, \lambda)$ and $\Upsilon^{+}_{\mathrm{B}}(\mathrm{CFL}, \lambda)$ (\ref{eq:upsilons_def_plus}) are conveniently defined by introducing the following polynomial representation:
\begin{equation}
\Upsilon^{+}_{\mathrm{A}}(\mathrm{CFL}, z, \lambda) = \ovecphi^{T}\mathbf{h}^{+}(\mathrm{CFL}, z, \lambda),\quad
\Upsilon^{+}_{\mathrm{B}}(\mathrm{CFL}, z, \lambda) = \uvecphi^{T}\mathbf{h}^{+}(\mathrm{CFL}, z, \lambda),
\end{equation}
where the polynomial coefficients $\mathbf{h}^{+} = \{h^{+}_{p}\}_{p}$ are defined as follows:\\[-2mm]
\begin{equation}\label{eq:h_plus_coeffs_def}
h^{+}_{p}(\mathrm{CFL}, z, \lambda) = \Big\{\mathrm{B}(\mathrm{CFL}, z)\Big[R_{\rm RK}(\mathrm{Z}^{+}(\mathrm{CFL}, z)) - \lambda I\Big]^{-1} m^{-1}\uvecphi\Big\}_{p},
\end{equation}
and determine the following polynomial on $\xi\in\omega$:\\[-2mm]
\begin{equation}\label{eq:h_plus_def}
H^{+}(\xi; \mathrm{CFL}, z, \lambda) = \sum\limits_{p = 0}^{N} h^{+}_{p}(\mathrm{CFL}, z, \lambda) \varphi_{p}(\xi).
\end{equation}
Using this polynomial $H^{+}(\xi;\, \mathrm{CFL}, \lambda)$, the functions $\Upsilon^{+}_{\mathrm{A}}(\mathrm{CFL}, \lambda)$ and $\Upsilon^{+}_{\mathrm{B}}(\mathrm{CFL}, \lambda)$ (\ref{eq:upsilons_def_plus}) are defined as follows:
\begin{equation}\label{eq:upsilons_def_by_h_plus}
\Upsilon^{+}_{\mathrm{A}}(\mathrm{CFL}, z, \lambda) = H^{+}(1; \mathrm{CFL}, z, \lambda),\quad
\Upsilon^{+}_{\mathrm{B}}(\mathrm{CFL}, z, \lambda) = H^{+}(0; \mathrm{CFL, z}, \lambda),
\end{equation}
and the equation (\ref{eq:upsilon_equation_plus_def_theta_final}) takes the following form:
\begin{equation}\label{eq:h_equation_plus_def}
\mathrm{CFL}\cdot\exp(-i\theta)H^{+}(1; \mathrm{CFL}, z, \lambda) = 1 - \mathrm{CFL} \cdot H^{+}(0; \mathrm{CFL, z}, \lambda),
\end{equation}
and the equation (\ref{eq:upsilon_equation_plus_def_theta_final_abs}) takes the following form:
\begin{equation}\label{eq:h_abs_equation_plus_def}
|\mathrm{CFL} \cdot H^{+}(1; \mathrm{CFL}, z, \lambda)|^{2} = |1 - \mathrm{CFL} \cdot H^{+}(0; \mathrm{CFL}, z, \lambda)|^{2}.
\end{equation}
The resulting equation must have a solution $\mathrm{CFL}$ for some eigenvalue $\lambda$ at the stability boundary $|\lambda| = 1$:
\begin{equation}\label{eq:f_def_plus}
F^{+}(\phi; \mathrm{CFL}, z) = |1 - \mathrm{CFL} \cdot H^{+}(0; \mathrm{CFL}, z, \exp(i\phi))|^{2} - |\mathrm{CFL} \cdot H^{+}(1; \mathrm{CFL}, z, \exp(i\phi))|^{2} = 0,
\end{equation}
where definition $\lambda = \exp(i\phi)$, based on the argument $\phi$ of the eigenvalue $\lambda$, is used. In the general case, the boundary value of the Courant number $\mathrm{CFL}_{\rm max}$ can be defined as follows:
\begin{equation}\label{eq:cfl_by_h_src_a_plus}
\mathrm{CFL}_{\rm max} = \inf\left\{\mathrm{CFL}\, \left|
\begin{split}
&\mathrm{CFL}\in\mathcal{R}_{+},\, \exists\, \lambda \in \mathcal{C},\, |\lambda| \geqslant 1:\\
&\left[|\mathrm{CFL} \cdot H^{+}(1; \mathrm{CFL}, z, \lambda)|^{2} = |1 - \mathrm{CFL} \cdot H^{+}(0; \mathrm{CFL}, z, \lambda)|^{2}\right]
\end{split}
\right. \right\},
\end{equation}
or, the boundary value of the Courant number $\mathrm{CFL}_{\rm max}$ is the infimum of the set of Courant numbers for which the equation has a solution for some complex number $|\lambda| \geqslant 1$.

The simplification of the characteristic equation $\det[R(\mathrm{CFL}, \theta) - \lambda I] = 0$ in the case $a < 0$ is carried out similarly to the case $a > 0$, with only technical differences. First, it should be noted that in the case $a < 0$, taking into account conditions $\Dtn{n} > 0$ and $\Dx > 0$, the Courant number $\mathrm{CFL}$ introduced in this work (\ref{eq:evol_oper_def}) will be negative: $\mathrm{CFL} = -|\mathrm{CFL}| < 0$. In this case, the expression for matrices $\mathrm{A}(\mathrm{CFL}, z, \theta)$ and $\mathrm{B}(\mathrm{CFL}, z)$ (\ref{eq:a_and_b_matrices_def_dup}), which define the matrix $\mathrm{R}(\mathrm{CFL}, z, \theta) = I + \mathrm{A}(\mathrm{CFL}, z, \theta)\mathrm{B}(\mathrm{CFL}, z)$ (\ref{eq:r_matrix_in_matrix_form}) of the evolution operator $R$, takes the following form:\\[-2mm]
\begin{equation}\label{eq:f_derivd_temp_1_a_minus}
\begin{split}
&\mathrm{A}(\mathrm{CFL}, z, \theta) = -|\mathrm{CFL}| \cdot m^{-1}\Big[\mathrm{D}^{T}m - \Big(\ovecphi\exp(i\theta) - \uvecphi\Big)\otimes\uvecphi^{T}\Big] + z I,\\
&\mathrm{B}(\mathrm{CFL}, z) = \sum\limits_{s = 0}^{N} \frac{(-1)^{s}}{s!} (-|\mathrm{CFL}|)^{s} \frac{d^{s}}{dz^{s}}\left[\frac{R_{\rm RK}(z) - 1}{z}\right]\mathrm{D}^{s}.
\end{split}
\end{equation}
The resulting integration-by-parts identity (\ref{eq:ibp_prop}) is used to express the matrix $\mathrm{D}^{T}m = \ovecphi\otimes\ovecphi^{T} - \uvecphi\otimes\uvecphi^{T} - m\mathrm{D}$, leading to the following chain of transformations for matrix $\mathrm{A}(\mathrm{CFL}, z, \theta)$:\\[-2mm]
\begin{equation}
\begin{split}
\mathrm{A}(\mathrm{CFL}, z, \theta) & = -|\mathrm{CFL}| \cdot m^{-1}\Big[\mathrm{D}^{T}m - \Big(\ovecphi\exp(i\theta) - \uvecphi\Big)\otimes\uvecphi^{T}\Big] + z I \\ & =
z I - |\mathrm{CFL}| \cdot m^{-1}\Big[\mathrm{D}^{T}m - \exp(i\theta)\ovecphi\otimes\uvecphi^{T} + \uvecphi\otimes\uvecphi^{T}\Big] \\ & =
z I - |\mathrm{CFL}| \cdot m^{-1}\Big[\ovecphi\otimes\ovecphi^{T} - \uvecphi\otimes\uvecphi^{T} - m\mathrm{D} - 
\exp(i\theta)\ovecphi\otimes\uvecphi^{T} + \uvecphi\otimes\uvecphi^{T}\Big]\\ & =
z I - |\mathrm{CFL}| \cdot m^{-1}\Big[\ovecphi\otimes\ovecphi^{T} - \exp(i\theta)\ovecphi\otimes\uvecphi^{T} - m\mathrm{D}\Big]\\ & =
z I + |\mathrm{CFL}| \cdot \mathrm{D} + |\mathrm{CFL}| \cdot m^{-1}\Big[\ovecphi\otimes\Big(\exp(i\theta)\uvecphi^{T} - \ovecphi^{T}\Big)\Big],
\end{split}
\end{equation}
where the following matrix $\mathrm{Z}^{+}(\mathrm{CFL}, z)$ is introduced:\\[-2mm]
\begin{equation}\label{eq:z_minus_matrix_def}
\mathrm{Z}^{-}(\mathrm{CFL}, z) = z I + |\mathrm{CFL}| \cdot \mathrm{D},
\end{equation}
using which, the expression for matrix $\mathrm{A}(\mathrm{CFL}, z, \theta)$ takes the following form:\\[-2mm]
\begin{equation}\label{eq:f_derivd_temp_1_1_a_minus}
\mathrm{A}(\mathrm{CFL}, z, \theta) = \mathrm{Z}^{-}(\mathrm{CFL}, z) + |\mathrm{CFL}| \cdot m^{-1}\Big[\ovecphi\otimes\Big(\exp(i\theta)\uvecphi^{T} - \ovecphi^{T}\Big)\Big].
\end{equation}
Treating the expression (\ref{eq:f_derivd_temp_1_a_minus}) for the matrix $\mathrm{B}(\mathrm{CFL}, z)$ as a Taylor series expansion and taking into account the nilpotency of matrix $\mathrm{D}$: $\mathrm{D}^{N+1} = 0$, the following relation is obtained:\\[-2mm]
\begin{equation}\label{eq:f_derivd_temp_1_2_a_minus}
\begin{split}
\mathrm{B}(\mathrm{CFL}, z) & = \sum\limits_{s = 0}^{N} \frac{(-1)^{s}}{s!} (-|\mathrm{CFL}|)^{s} \frac{d^{s}}{dz^{s}}\left[\frac{R_{\rm RK}(z) - 1}{z}\right]\mathrm{D}^{s} =
\sum\limits_{s = 0}^{N} \left\{\frac{d^{s}}{dz^{s}}\left[\frac{R_{\rm RK}(z) - 1}{z}\right]\right\} \frac{(|\mathrm{CFL}|\cdot\mathrm{D})^{s}}{s!}\\ & =
\sum\limits_{s = 0}^{N} \left\{\frac{d^{s}}{dz^{s}}\Big[(R_{\rm RK}(z I) - I)z^{-1} I\Big]\right\} \frac{(|\mathrm{CFL}|\cdot\mathrm{D})^{s}}{s!} =
\Big[R_{\rm RK}\Big(z I + |\mathrm{CFL}| \cdot \mathrm{D}\Big) - I\Big]\\ & \cdot \Big[z I + |\mathrm{CFL}| \cdot \mathrm{D}\Big]^{-1} =
\Big[R_{\rm RK}(\mathrm{Z}^{-}(\mathrm{CFL}, z)) - I\Big]\Big[\mathrm{Z}^{-}(\mathrm{CFL}, z)\Big]^{-1},
\end{split}
\end{equation}
where it is noted that, in case $z \neq 0$, matrix $\mathrm{Z}^{-}(\mathrm{CFL}, z)$ is invertible similar matrix $\mathrm{Z}^{+}(\mathrm{CFL}, z)$ (\ref{eq:z_plus_matrix_def}). Taking into account the previously obtained expression (\ref{eq:f_derivd_temp_1_1_a_plus}) for the matrix $\mathrm{A}(\mathrm{CFL}, z, \theta)$ and expression (\ref{eq:f_derivd_temp_1_2_a_plus}) for the matrix $\mathrm{B}(\mathrm{CFL}, z)$, the expression for matrix $\mathrm{R}(\mathrm{CFL}, z, \theta)$ is rewritten in the following form:\\[-2mm]
\begin{equation}
\begin{split}
R(\mathrm{CFL}, z, \theta) & = I + \mathrm{A}(\mathrm{CFL}, z, \theta) \mathrm{B}(\mathrm{CFL}, z) =
I + \Big[\mathrm{Z}^{-}(\mathrm{CFL}, z) - |\mathrm{CFL}| \cdot m^{-1}\Big[\ovecphi\otimes\Big(\exp(i\theta)\uvecphi^{T} - \ovecphi^{T}\Big)\Big]\Big]\\ & \cdot
\mathrm{B}(\mathrm{CFL}, z) = I + \mathrm{Z}^{-}(\mathrm{CFL}, z)\mathrm{B}(\mathrm{CFL}, z) -
|\mathrm{CFL}| \cdot m^{-1}\Big[\uvecphi\otimes\Big(\exp(-i\theta)\ovecphi^{T} - \uvecphi^{T}\Big)\Big]\mathrm{B}(\mathrm{CFL}, z)\\ & =
R_{\rm RK}(\mathrm{Z}^{-}(\mathrm{CFL}, z)) - |\mathrm{CFL}| \cdot m^{-1}\Big[\uvecphi\otimes
\Big(\exp(-i\theta)\ovecphi^{T}\mathrm{B}(\mathrm{CFL}, z) - \uvecphi^{T}\mathrm{B}(\mathrm{CFL}, z)\Big)\Big],
\end{split}
\end{equation}
where a consequence $\mathrm{Z}^{-}(\mathrm{CFL}, z)\mathrm{B}(\mathrm{CFL}, z) = R_{\rm RK}(\mathrm{Z}^{-}(\mathrm{CFL}, z)) - I$ of relation (\ref{eq:f_derivd_temp_1_2_a_minus}) is explicitly taken into account. The expression for the characteristic equation in a form suitable for quantitative estimates is obtained by applying a well-known relation $\det(\mathrm{K}+\mathbf{a}\otimes\mathbf{b}^{T}) = \det{K}\cdot(1 + \mathbf{b}^{T}\mathrm{K}^{-1}\mathbf{a})$, similar derivation (\ref{eq:f_derivd_temp_2_a_plus}), within the expression for the characteristic equation $\det[R(\mathrm{CFL}, z, \theta) - \lambda I] = 0$:\\[-2mm]
\begin{equation}\label{eq:f_derivd_temp_2_a_minus}
\begin{split}
\det\Big[R(\mathrm{CFL}, z, \theta) - \lambda I\Big] & = 
\det\Big[\Big(R_{\rm RK}(\mathrm{Z}^{-}(\mathrm{CFL}, z)) - \lambda I\Big) + |\mathrm{CFL}| \cdot m^{-1}\Big[\ovecphi\otimes
\Big(\exp(i\theta)\uvecphi^{T}\mathrm{B}(\mathrm{CFL}, z)\\ & - \ovecphi^{T}\mathrm{B}(\mathrm{CFL}, z)\Big)\Big]\Big] =
\det\Big[R_{\rm RK}(\mathrm{Z}^{-}(\mathrm{CFL}, z)) - \lambda I\Big]\cdot\Big\{
1 + |\mathrm{CFL}|\cdot\Big(\exp(i\theta)\\ & \cdot\uvecphi^{T}\mathrm{B}(\mathrm{CFL}, z) - \ovecphi^{T}\mathrm{B}(\mathrm{CFL}, z)\Big)
\Big[R_{\rm RK}(\mathrm{Z}^{-}(\mathrm{CFL}, z)) - \lambda I\Big]^{-1} \Big(m^{-1}\ovecphi\Big)\Big\}\\ & =
\det\Big[R_{\rm RK}(\mathrm{Z}^{-}(\mathrm{CFL}, z)) - \lambda I\Big]\cdot\Big\{
1 + |\mathrm{CFL}|\cdot\exp(i\theta)\uvecphi^{T}\mathrm{B}(\mathrm{CFL}, z)\\ & \cdot
\Big[R_{\rm RK}(\mathrm{Z}^{-}(\mathrm{CFL}, z)) - \lambda I\Big]^{-1} \Big(m^{-1}\ovecphi\Big) -
|\mathrm{CFL}|\cdot\ovecphi^{T}\mathrm{B}(\mathrm{CFL}, z)\\ & \cdot
\Big[R_{\rm RK}(\mathrm{Z}^{-}(\mathrm{CFL}, z)) - \lambda I\Big]^{-1} \Big(m^{-1}\ovecphi\Big)\Big\}.
\end{split}
\end{equation}
After introducing the following notation:
\begin{equation}\label{eq:upsilons_def_minus}
\begin{split}
&\Upsilon^{-}_{\mathrm{A}}(\mathrm{CFL}, z, \lambda) = \uvecphi^{T}\mathrm{B}(\mathrm{CFL}, z)\Big[R_{\rm RK}(\mathrm{Z}^{-}(\mathrm{CFL}, z)) - \lambda I\Big]^{-1} m^{-1}\ovecphi,\\
&\Upsilon^{-}_{\mathrm{B}}(\mathrm{CFL}, z, \lambda) = \ovecphi^{T}\mathrm{B}(\mathrm{CFL}, z)\Big[R_{\rm RK}(\mathrm{Z}^{-}(\mathrm{CFL}, z)) - \lambda I\Big]^{-1} m^{-1}\ovecphi,
\end{split}
\end{equation}
the equation (\ref{eq:f_derivd_temp_2_a_minus}) is transformed to the following form:
\begin{equation}\label{eq:upsilon_equation_minus_def_theta}
\begin{split}
\det\Big[R(\mathrm{CFL}, z, \theta) - \lambda I\Big] = 
\det\Big[R_{\rm RK}(\mathrm{Z}^{-}(\mathrm{CFL}, z)) - \lambda I\Big]\cdot
\Big[&|\mathrm{CFL}|\cdot\exp(i\theta)\Upsilon^{-}_{\mathrm{A}}(\mathrm{CFL}, \lambda)\\ & + 1 - |\mathrm{CFL}|\cdot\Upsilon^{-}_{\mathrm{B}}(\mathrm{CFL}, \lambda)\Big] = 0.
\end{split}
\end{equation}
As it is done previously for $a > 0$ (\ref{eq:R_eigenval_prop_a_plus}), the following asymptotic property can also be demonstrated:
\begin{equation}\label{eq:R_eigenval_prop_a_minus}
\Big[R_{\rm RK}(\mathrm{Z}^{-}(\mathrm{CFL}, z)) - \lambda I\Big]^{-1} \sim \frac{1}{(R_{\rm RK}(z) - \lambda)^{N+1}} \quad\mathrm{as}\quad \lambda \rightarrow R_{\rm RK}(z),
\end{equation}
and from this the obtained asymptotic estimate and the fact that the characteristic polynomial is indeed a polynomial, the following form of the left side of the characteristic equation $\det[\mathrm{R}(\mathrm{CFL}, z, \theta) - \lambda I] = 0$ follows:
\begin{equation}
\det\Big[R(\mathrm{CFL}, z, \theta) - \lambda I\Big] = \mathfrak{h}^{-}(\lambda; \mathrm{CFL}, z, \theta),
\end{equation}
where $\mathfrak{h}^{-} \in \mathcal{P}_{N}(\mathcal{C})$ is a polynomial of degree $N+1$ with no roots at the point $\lambda = R_{\rm RK}(z)$: $\mathfrak{h}^{-}(R_{\rm RK}(z); \mathrm{CFL}, z, \theta) \neq 0$. Therefore, the matrix $\mathrm{R}(\mathrm{CFL}, z, \theta)$ in the case $a < 0$ has no an eigenvalue $\lambda = R_{\rm RK}(z)$ in the case $\theta \in (0, 2\pi)$. Consequently, only the second factor remains in the characteristic equation (\ref{eq:upsilon_equation_minus_def_theta}), leading to the following equation:
\begin{equation}\label{eq:upsilon_equation_minus_def_theta_final}
|\mathrm{CFL}|\cdot\exp(-i\theta)\Upsilon^{-}_{\mathrm{A}}(\mathrm{CFL}, z, \lambda) = 1 - |\mathrm{CFL}|\cdot\Upsilon^{-}_{\mathrm{B}}(\mathrm{CFL}, z, \lambda),
\end{equation}
and to eliminate the need to account for the phase $\theta \in [0, 2\pi]$, it is necessary to take the absolute value of both sides of the resulting equation:
\begin{equation}\label{eq:upsilon_equation_minus_def_theta_final_abs}
|\mathrm{CFL}\cdot\Upsilon^{-}_{\mathrm{A}}(\mathrm{CFL}, z, \lambda)|^{2} = |1 - |\mathrm{CFL}|\cdot\Upsilon^{-}_{\mathrm{B}}(\mathrm{CFL}, z, \lambda)|^{2},
\end{equation}
which leads to a phase-independent relationship involving the Courant number $\mathrm{CFL}$. The functions $\Upsilon^{+}_{\mathrm{A}}(\mathrm{CFL}, \lambda)$ and $\Upsilon^{+}_{\mathrm{B}}(\mathrm{CFL}, \lambda)$ (\ref{eq:upsilons_def_minus}) are conveniently defined by introducing the following polynomial representation:
\begin{equation}
\Upsilon^{-}_{\mathrm{A}}(\mathrm{CFL}, z, \lambda) = \uvecphi^{T}\mathbf{h}^{-}(\mathrm{CFL}, z, \lambda),\quad
\Upsilon^{-}_{\mathrm{B}}(\mathrm{CFL}, z, \lambda) = \ovecphi^{T}\mathbf{h}^{-}(\mathrm{CFL}, z, \lambda),
\end{equation}
where the polynomial coefficients $\mathbf{h}^{+} = \{h^{+}_{p}\}_{p}$ are defined as follows:\\[-2mm]
\begin{equation}\label{eq:h_minus_coeffs_def}
h^{-}_{p}(\mathrm{CFL}, z, \lambda) = \Big\{\mathrm{B}(\mathrm{CFL}, z)\Big[R_{\rm RK}(\mathrm{Z}^{-}(\mathrm{CFL}, z)) - \lambda I\Big]^{-1} m^{-1}\ovecphi\Big\}_{p},
\end{equation}
and determine the following polynomial on $\xi\in\omega$:\\[-2mm]
\begin{equation}\label{eq:h_minus_def}
H^{-}(\xi; \mathrm{CFL}, z, \lambda) = \sum\limits_{p = 0}^{N} h^{-}_{p}(\mathrm{CFL}, z, \lambda) \varphi_{p}(\xi).
\end{equation}
Using this polynomial $H^{-}(\xi;\, \mathrm{CFL}, \lambda)$, the functions $\Upsilon^{-}_{\mathrm{A}}(\mathrm{CFL}, \lambda)$ and $\Upsilon^{-}_{\mathrm{B}}(\mathrm{CFL}, \lambda)$ (\ref{eq:upsilons_def_minus}) are defined as follows:\\[-2mm]
\begin{equation}\label{eq:upsilons_def_by_h_minus}
\Upsilon^{-}_{\mathrm{A}}(\mathrm{CFL}, z, \lambda) = H^{-}(0; \mathrm{CFL}, z, \lambda),\quad
\Upsilon^{-}_{\mathrm{B}}(\mathrm{CFL}, z, \lambda) = H^{-}(1; \mathrm{CFL, z}, \lambda),
\end{equation}
and the equation (\ref{eq:upsilon_equation_minus_def_theta_final}) takes the following form:\\[-2mm]
\begin{equation}\label{eq:h_equation_minus_def}
|\mathrm{CFL}|\cdot\exp(i\theta)H^{-}(0; \mathrm{CFL}, z, \lambda) = 1 - |\mathrm{CFL}| \cdot H^{-}(1; \mathrm{CFL, z}, \lambda),
\end{equation}
and the equation (\ref{eq:upsilon_equation_minus_def_theta_final_abs}) takes the following form:\\[-2mm]
\begin{equation}\label{eq:h_abs_equation_minus_def}
|\mathrm{CFL} \cdot H^{-}(0; \mathrm{CFL}, z, \lambda)|^{2} = |1 - |\mathrm{CFL}| \cdot H^{-}(1; \mathrm{CFL}, z, \lambda)|^{2}.
\end{equation}
The resulting equation must have a solution $\mathrm{CFL}$ for some eigenvalue $\lambda$ at the stability boundary $|\lambda| = 1$:\\[-2mm]
\begin{equation}\label{eq:f_def_minus}
F^{-}(\phi; \mathrm{CFL}, z) = |1 - |\mathrm{CFL}| \cdot H^{-}(1; \mathrm{CFL}, z, \lambda)|^{2} - |\mathrm{CFL} \cdot H^{-}(0; \mathrm{CFL}, z, \lambda)|^{2} = 0,
\end{equation}
where definition $\lambda = \exp(i\phi)$, based on the argument $\phi$ of the eigenvalue $\lambda$, is used. In the general case, combining (\ref{eq:f_def_minus}) with the case $a > 0$ represented by formula (\ref{eq:cfl_by_h_src_a_plus}), the boundary value of the Courant number $\mathrm{CFL}_{\rm max}$ can be defined as follows:
\begin{equation}\label{eq:cfl_by_h_src}
\mathrm{CFL}_{\rm max} = \inf\left\{\mathrm{CFL}\, \left|
\begin{split}
&\mathrm{CFL}\in\mathcal{R}_{+},\, \exists\, \lambda \in \mathcal{C},\, |\lambda| \geqslant 1:\\
&a > 0\,\wedge\,\left[|\mathrm{CFL} \cdot H^{+}(1; \mathrm{CFL}, z, \lambda)|^{2} = |1 - \mathrm{CFL} \cdot H^{+}(0; \mathrm{CFL}, z, \lambda)|^{2}\right]\ \vee\\
&a < 0\,\wedge\,\left[|\mathrm{CFL} \cdot H^{-}(0; \mathrm{CFL}, z, \lambda)|^{2} = |1 - |\mathrm{CFL}| \cdot H^{-}(1; \mathrm{CFL}, z, \lambda)|^{2}\right]
\end{split}
\right. \right\},
\end{equation}
or, the boundary value of the Courant number $\mathrm{CFL}_{\rm max}$ is the infimum of the set of Courant numbers for which the equation has a solution for some complex number $|\lambda| \geqslant 1$. Instead of using two functions $F^{+}(\phi; \mathrm{CFL}, z)$ (\ref{eq:f_def_plus}) and $F^{-}(\phi; \mathrm{CFL}, z)$ (\ref{eq:f_def_minus}), it is convenient to introduce a single function $F(\phi; \mathrm{CFL}, z)$:\\[-2mm]
\begin{equation}\label{eq:f_def}
F(\phi; \mathrm{CFL}, z) = \theta(a) F^{+}(\phi; \mathrm{CFL}, z) + \theta(-a) F^{-}(\phi; \mathrm{CFL}, z) = 0,
\end{equation}
which defines the equation for the boundary value of the Courant number $\mathrm{CFL}_{\rm max}$. It is important to note that, up to the choice of sign for the Courant number $\mathrm{CFL} \leftrightarrow -|\mathrm{CFL}|$, the expressions for cases $F^{+}(\phi; \mathrm{CFL}, z)$ (\ref{eq:f_def_plus}) and $F^{-}(\phi; \mathrm{CFL}, z)$ (\ref{eq:f_def_minus}) yield the same function $F(\phi; \mathrm{CFL}, z)$ (\ref{eq:f_def}) values for cases $a > 0$ and $a < 0$, respectively.

\begin{figure}[h!]
\centering
\includegraphics[width=0.240\textwidth]{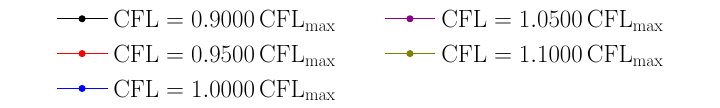}
\includegraphics[width=0.240\textwidth]{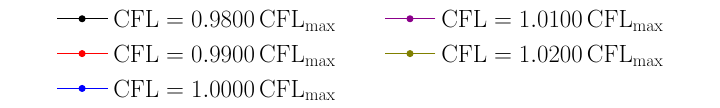}
\includegraphics[width=0.240\textwidth]{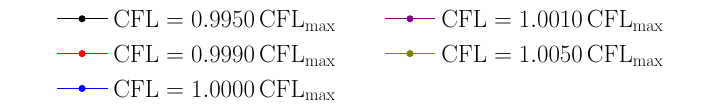}
\includegraphics[width=0.240\textwidth]{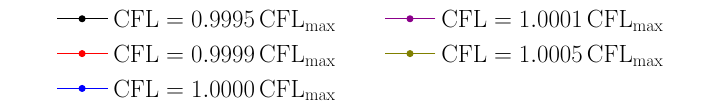}\\[1mm]
\includegraphics[width=0.240\textwidth]{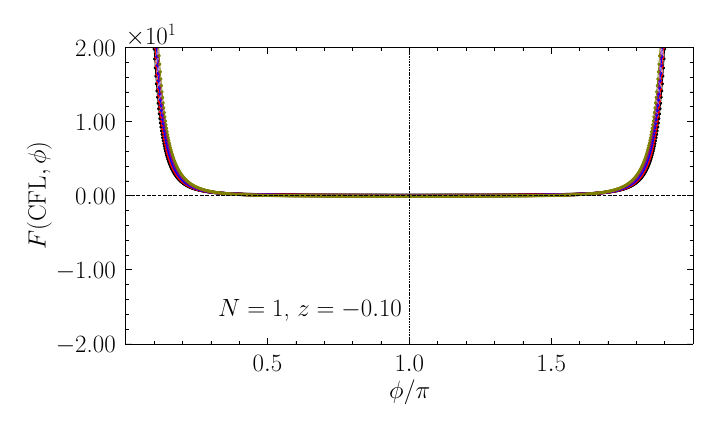}
\includegraphics[width=0.240\textwidth]{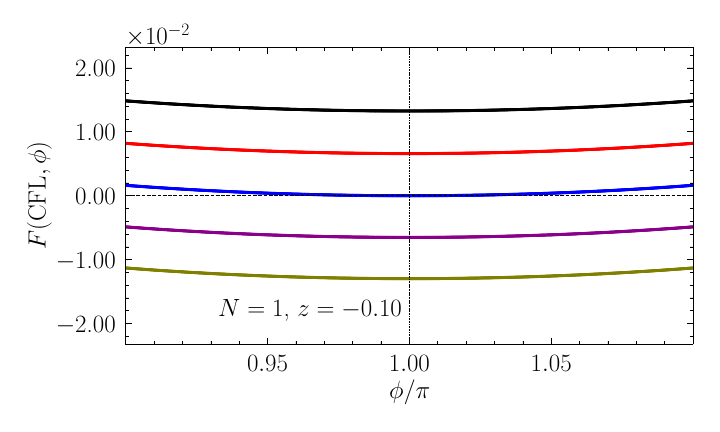}
\includegraphics[width=0.240\textwidth]{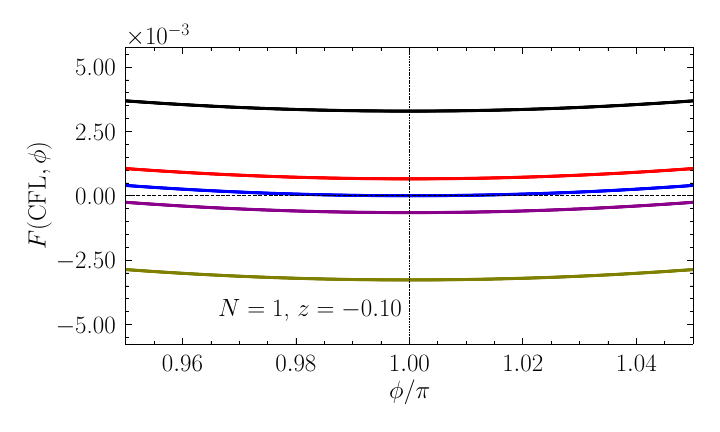}
\includegraphics[width=0.240\textwidth]{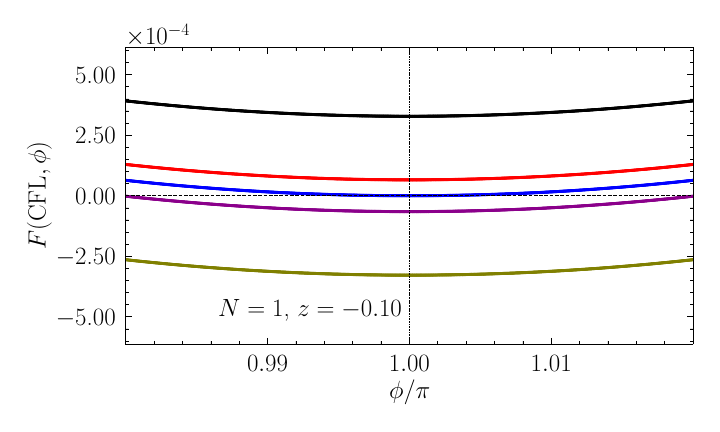}\\[1mm]
\includegraphics[width=0.240\textwidth]{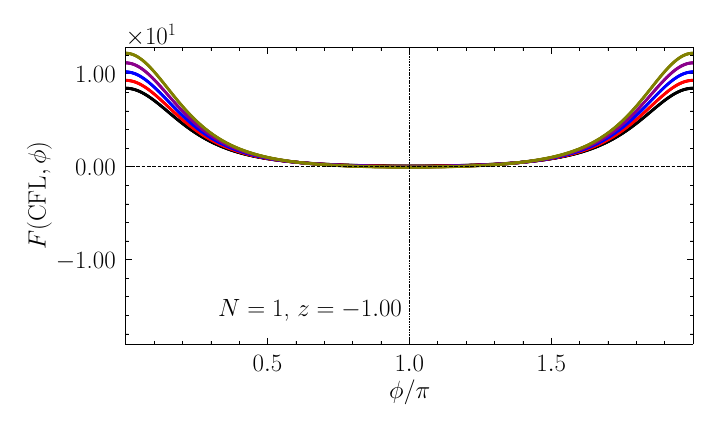}
\includegraphics[width=0.240\textwidth]{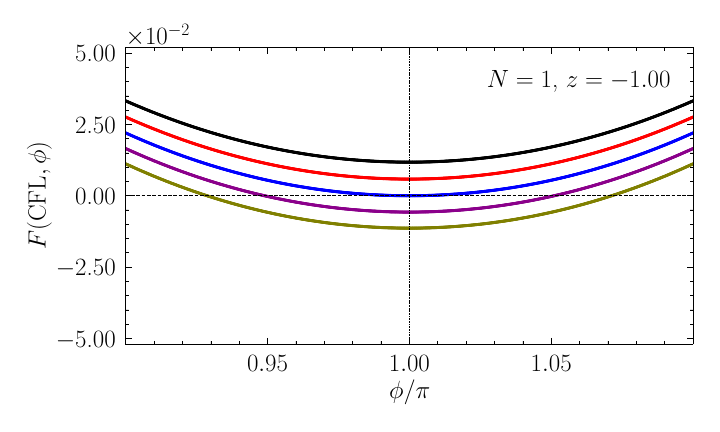}
\includegraphics[width=0.240\textwidth]{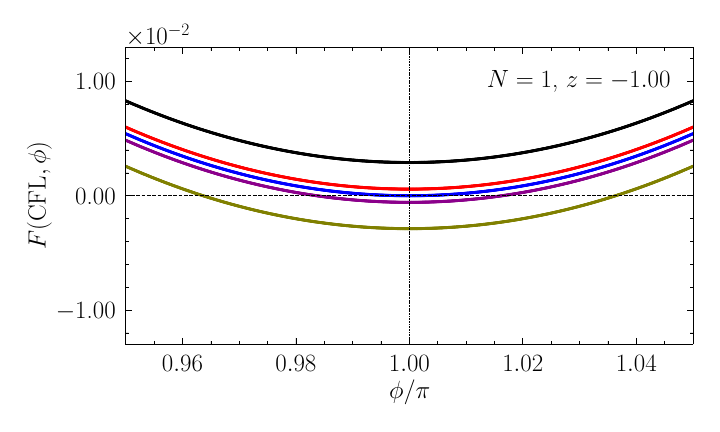}
\includegraphics[width=0.240\textwidth]{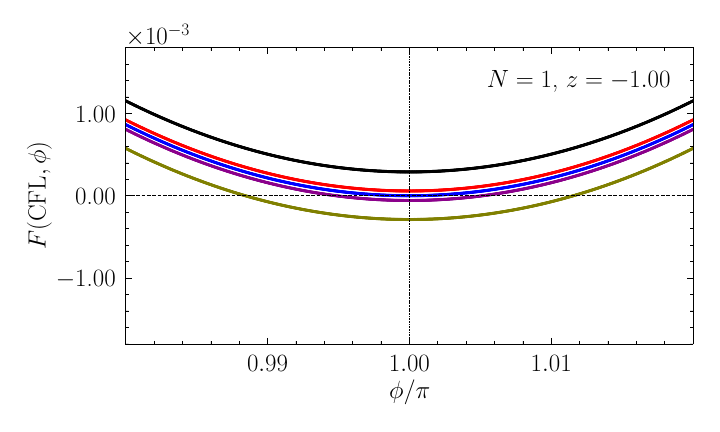}\\[1mm]
\includegraphics[width=0.240\textwidth]{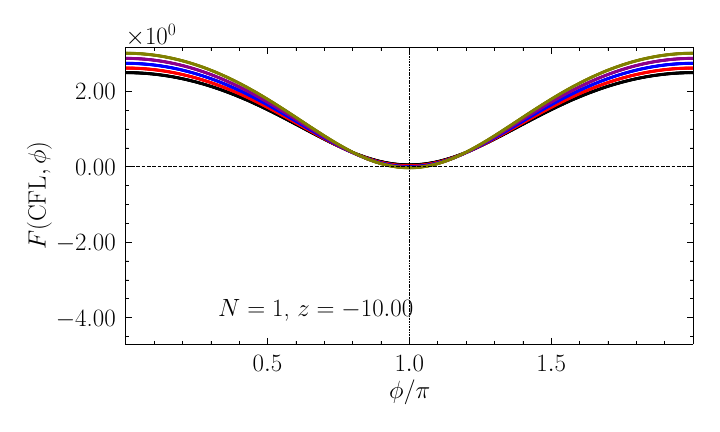}
\includegraphics[width=0.240\textwidth]{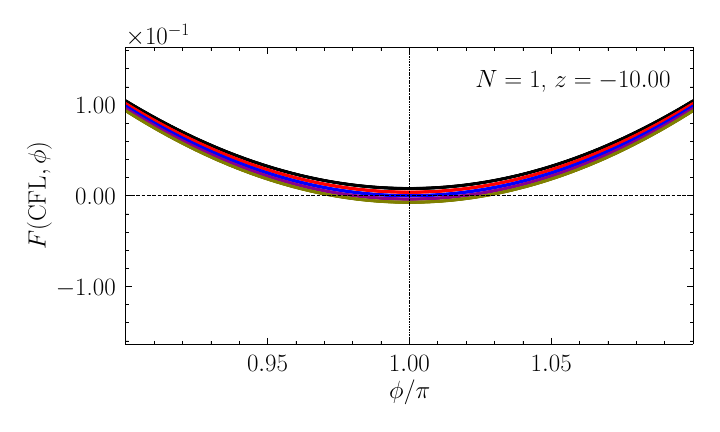}
\includegraphics[width=0.240\textwidth]{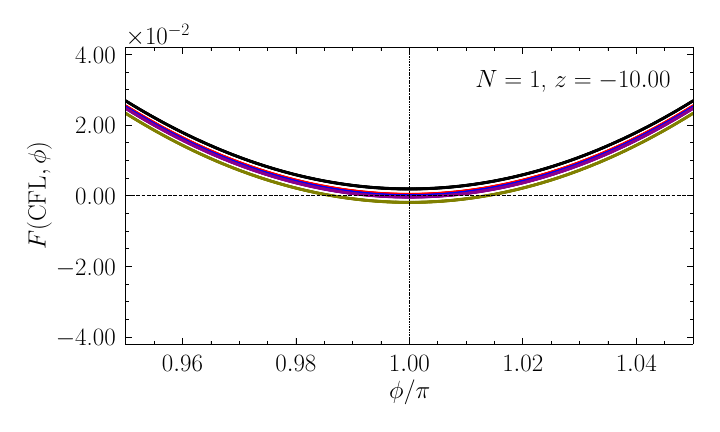}
\includegraphics[width=0.240\textwidth]{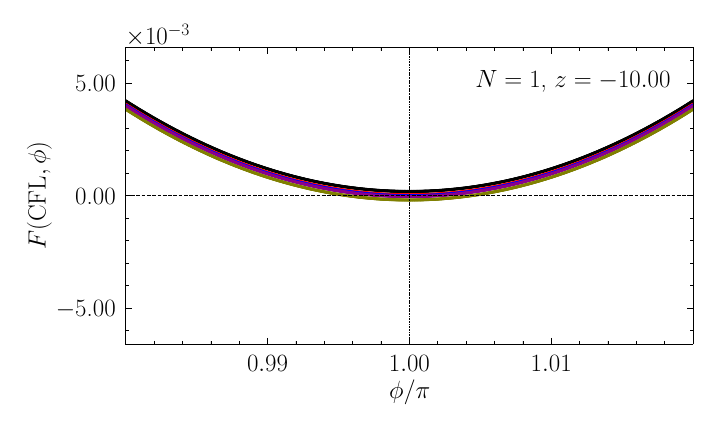}\\[1mm]
\includegraphics[width=0.240\textwidth]{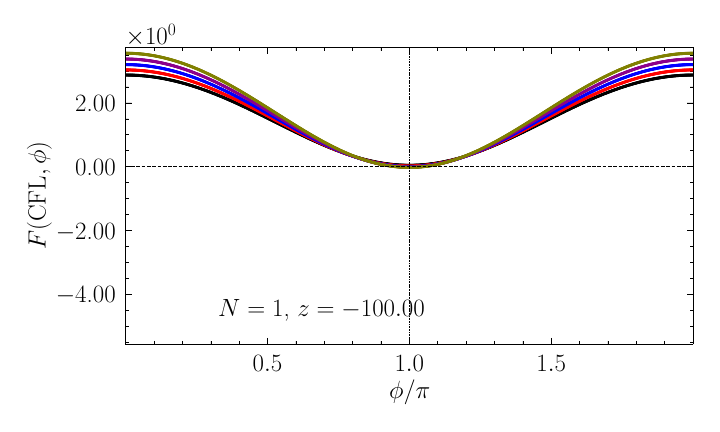}
\includegraphics[width=0.240\textwidth]{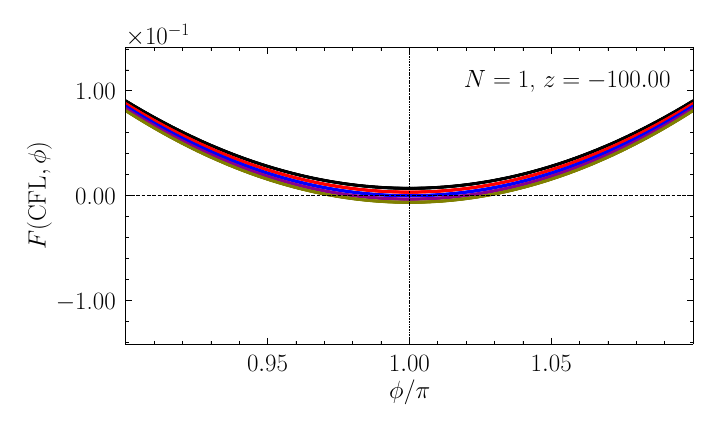}
\includegraphics[width=0.240\textwidth]{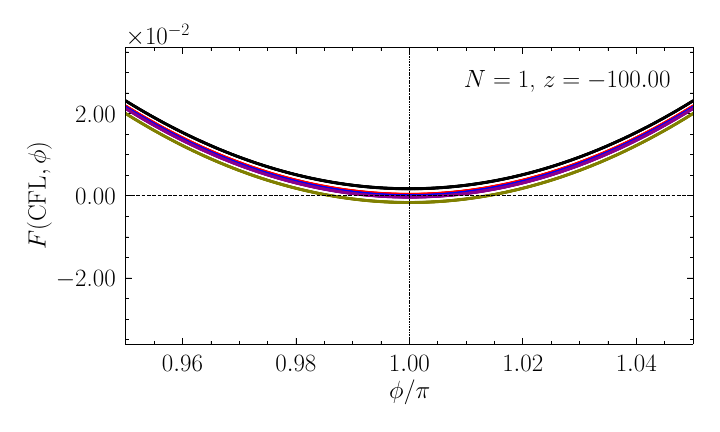}
\includegraphics[width=0.240\textwidth]{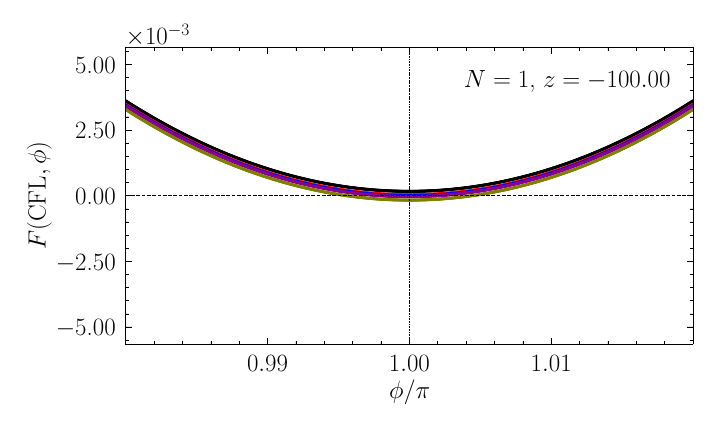}\\[1mm]
\includegraphics[width=0.240\textwidth]{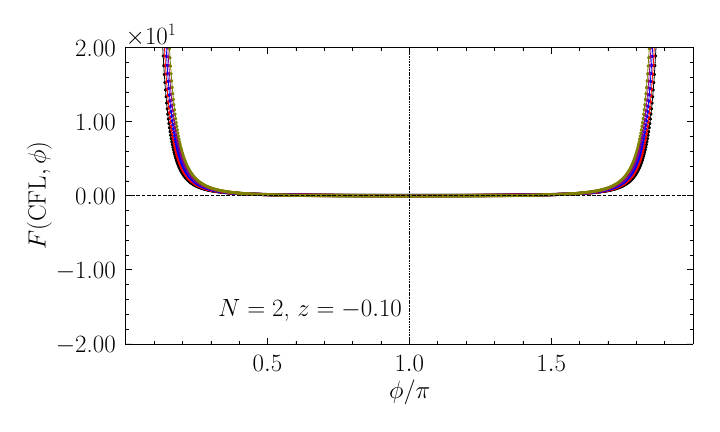}
\includegraphics[width=0.240\textwidth]{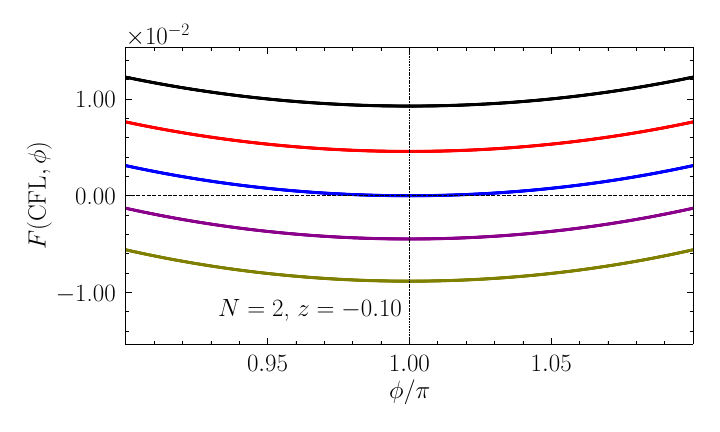}
\includegraphics[width=0.240\textwidth]{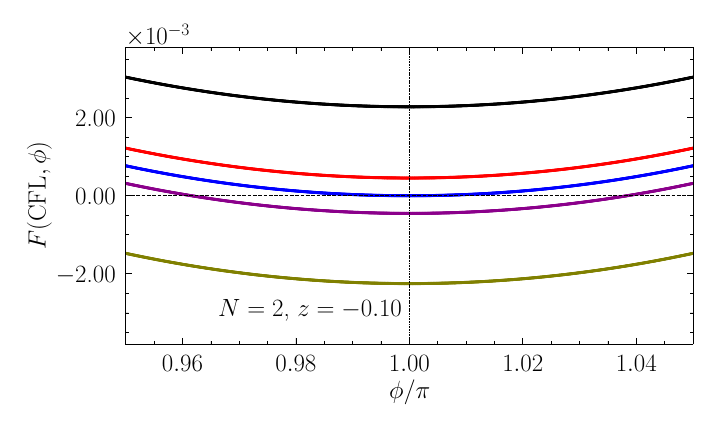}
\includegraphics[width=0.240\textwidth]{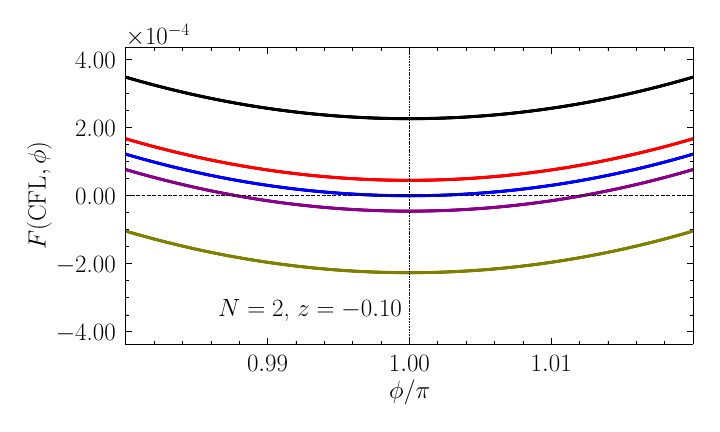}\\[1mm]
\includegraphics[width=0.240\textwidth]{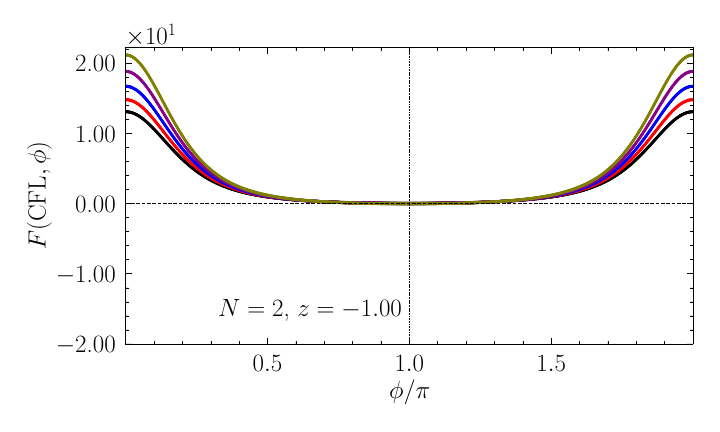}
\includegraphics[width=0.240\textwidth]{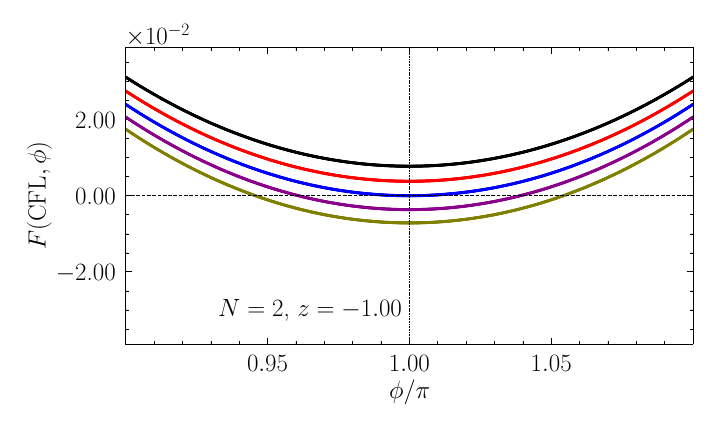}
\includegraphics[width=0.240\textwidth]{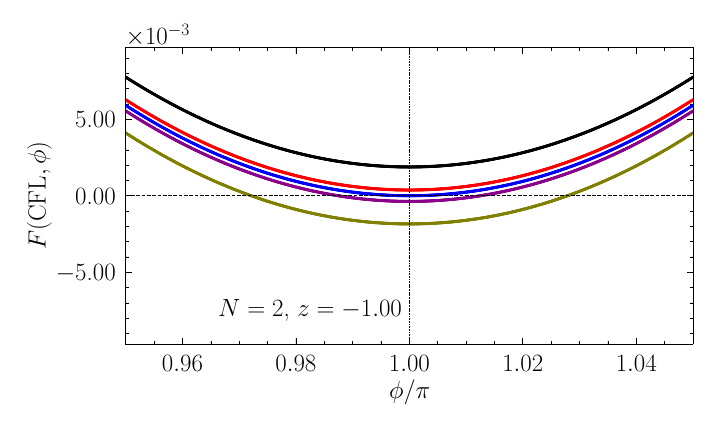}
\includegraphics[width=0.240\textwidth]{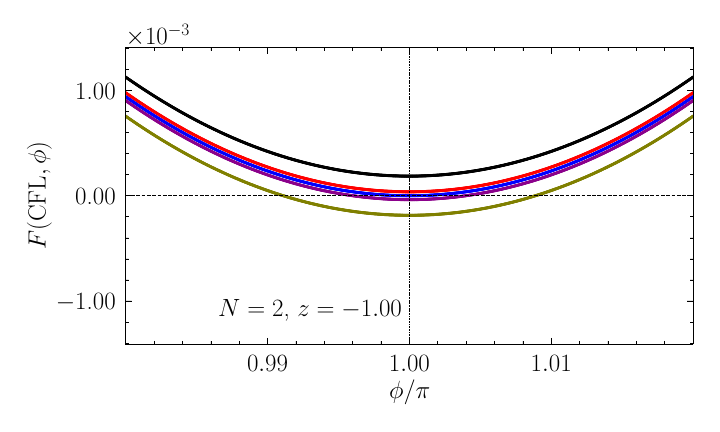}\\[1mm]
\includegraphics[width=0.240\textwidth]{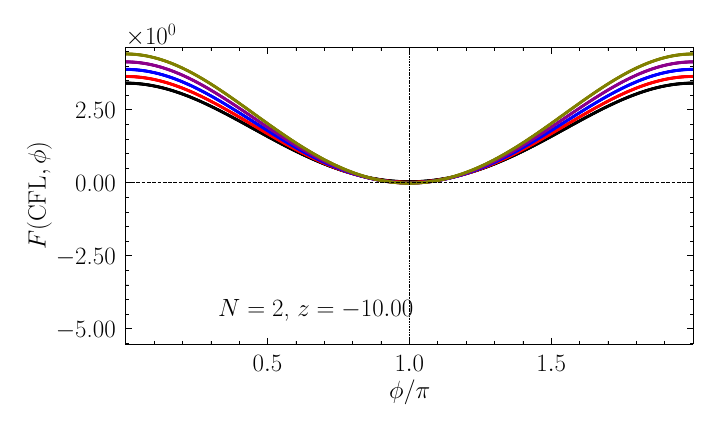}
\includegraphics[width=0.240\textwidth]{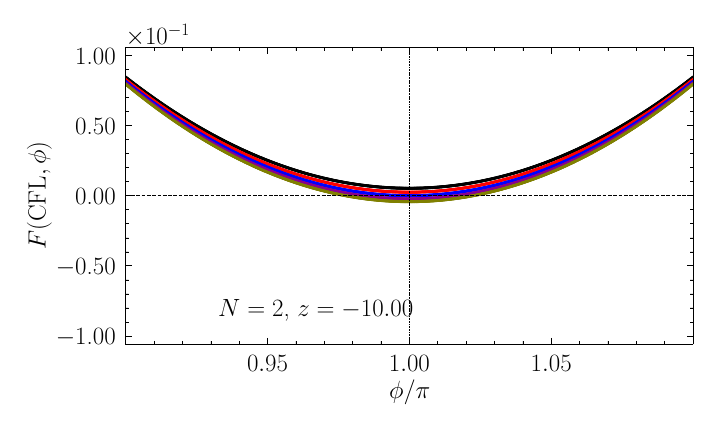}
\includegraphics[width=0.240\textwidth]{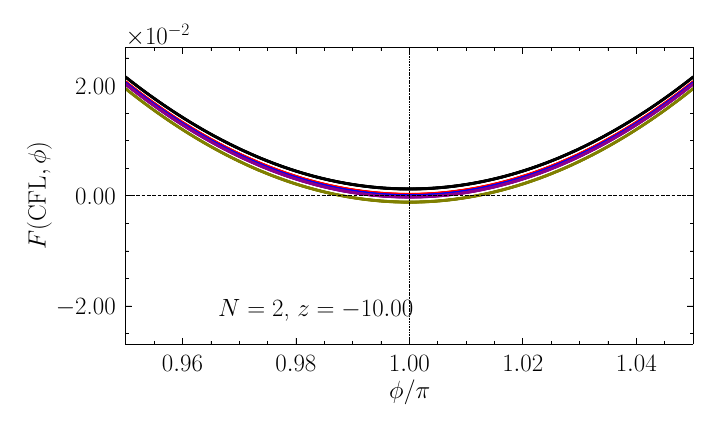}
\includegraphics[width=0.240\textwidth]{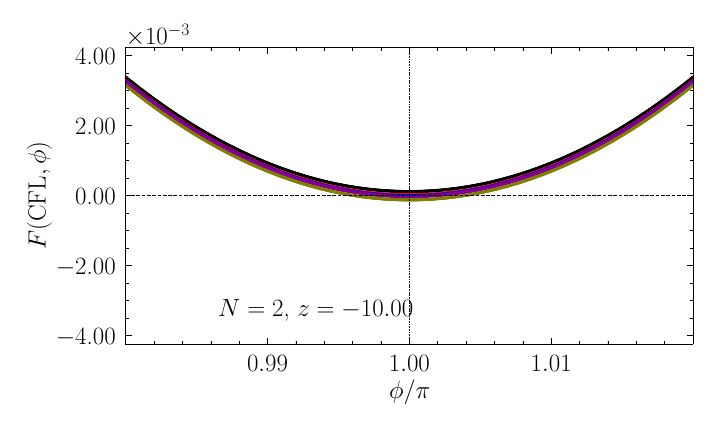}\\[1mm]
\includegraphics[width=0.240\textwidth]{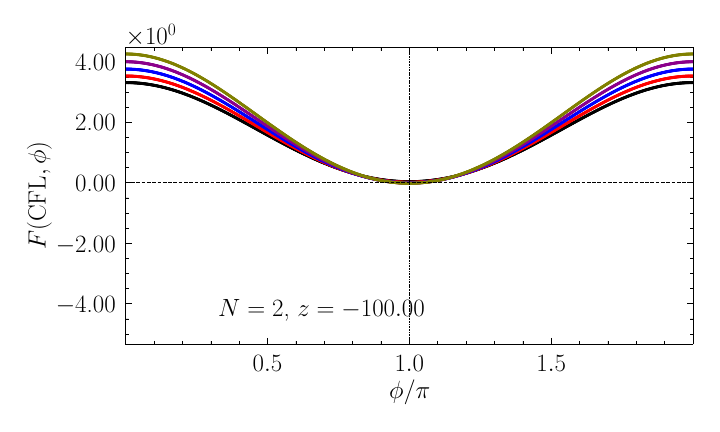}
\includegraphics[width=0.240\textwidth]{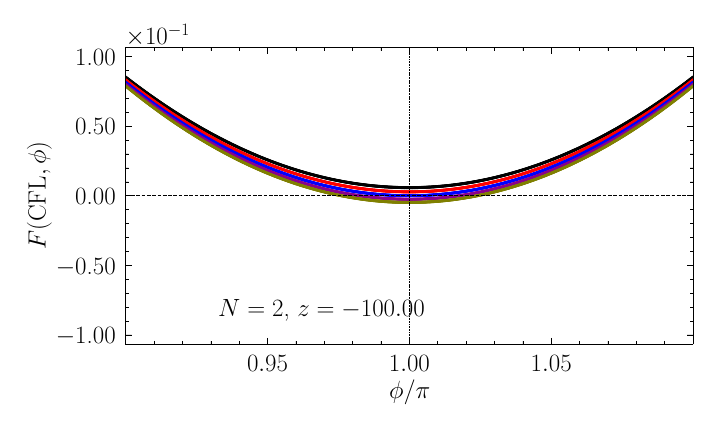}
\includegraphics[width=0.240\textwidth]{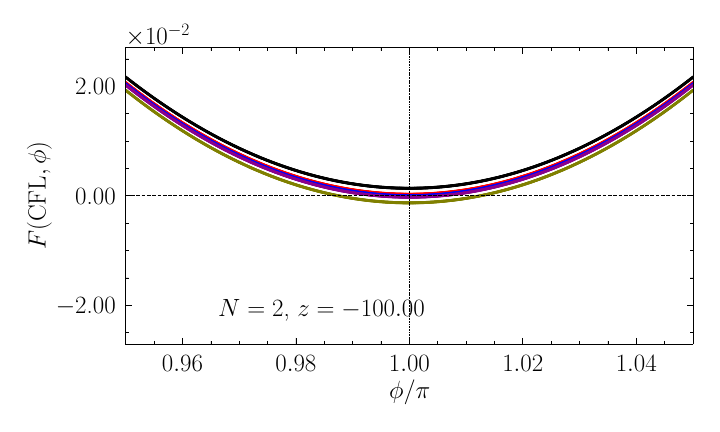}
\includegraphics[width=0.240\textwidth]{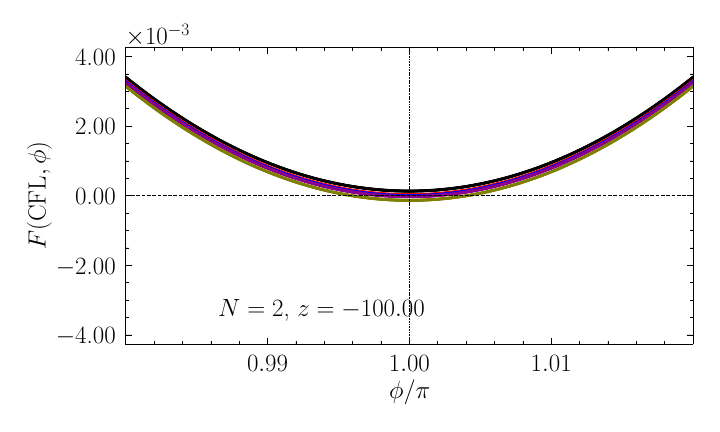}
\caption{%
The dependence of function $F(\phi; \mathrm{CFL}, z)$ (\ref{eq:f_def}) on argument $\phi\in[0, 2\pi]$ for several values of the Courant number $\mathrm{CFL}$, selected in the vicinity of the stability boundary $\mathrm{CFL}_{\rm max}(z)$, defined further in Figure~\ref{fig:cfls_on_z}, for $z = -0.1$, $-1.0$, $-10.0$, $-100.0$ and for polynomial degrees $N = 1$ and $2$. The graph legends are distributed among columns and located at the top of the columns. From left to right, the graphs represent increasingly narrower neighborhoods of the stability boundary $\mathrm{CFL}_{\rm max}$, from $\mathrm{CFL} \in [0.90\, \mathrm{CFL}_{\rm max},\, 1.10\, \mathrm{CFL}_{\rm max}]$ in the left column to $\mathrm{CFL} \in [0.9995\, \mathrm{CFL}_{\rm max},\, 1.0005\, \mathrm{CFL}_{\rm max}]$ in the right column; the range of phase $\phi$ variation also narrows when moving from the left to the right columns of the graphs to ensure a correct and clear presentation of the root $\phi = \pi$ of the equation $F(\pi; \mathrm{CFL}, z) = 0$. The horizontal dotted line indicates $F = 0$, the vertical dotted line indicates $\phi = \pi$ ($\lambda = \exp(\pm i\phi) = -1$). \textit{Note}: the phase $\phi$ of eigenvalue $\lambda_{k} = |\lambda_{k}|\exp(i\phi)$ is not the phase $\theta$.
}
\label{fig:dep_f_on_phi_diff_cfls_degrees_1_2}
\end{figure}
\begin{figure}[h!]
\centering
\includegraphics[width=0.240\textwidth]{dep_f_on_phi_diff_cfls_legend_Cidx_0.png}
\includegraphics[width=0.240\textwidth]{dep_f_on_phi_diff_cfls_legend_Cidx_1.png}
\includegraphics[width=0.240\textwidth]{dep_f_on_phi_diff_cfls_legend_Cidx_2.png}
\includegraphics[width=0.240\textwidth]{dep_f_on_phi_diff_cfls_legend_Cidx_3.png}\\[1mm]
\includegraphics[width=0.240\textwidth]{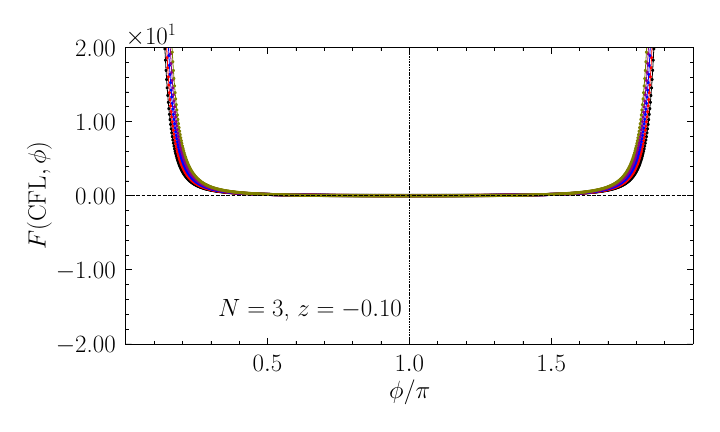}
\includegraphics[width=0.240\textwidth]{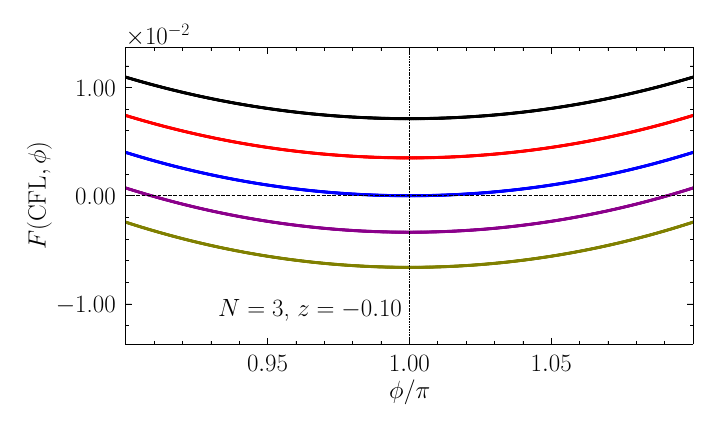}
\includegraphics[width=0.240\textwidth]{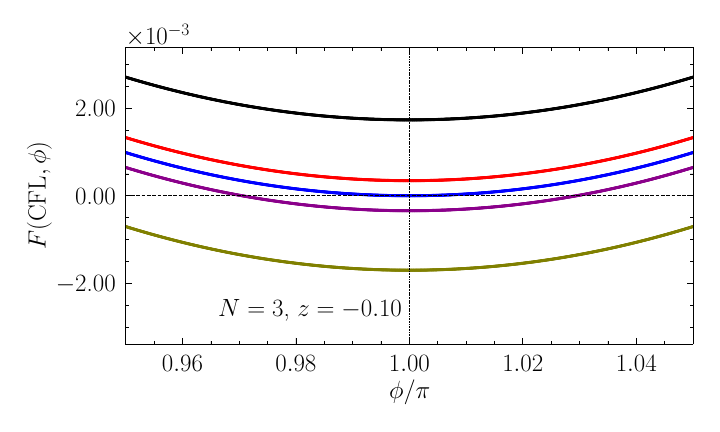}
\includegraphics[width=0.240\textwidth]{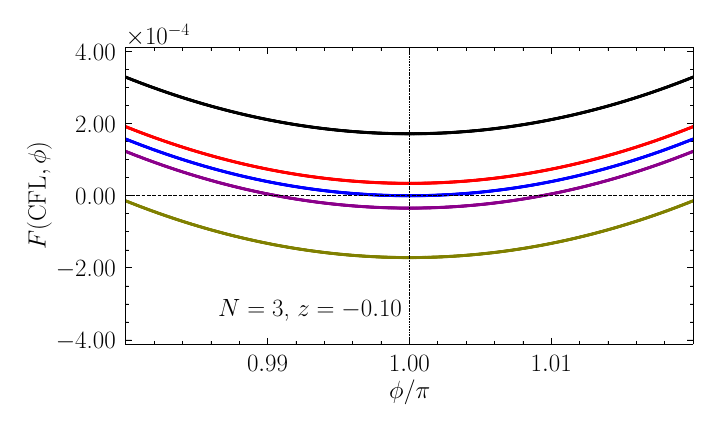}\\[1mm]
\includegraphics[width=0.240\textwidth]{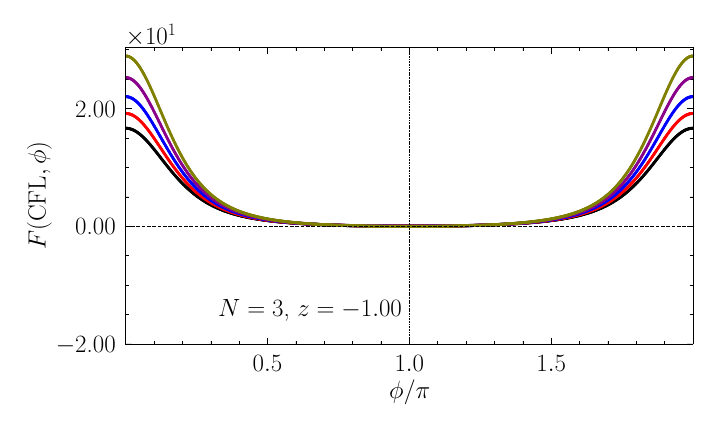}
\includegraphics[width=0.240\textwidth]{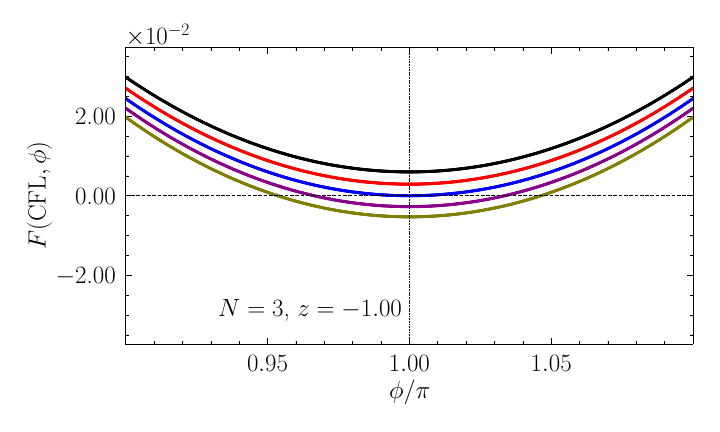}
\includegraphics[width=0.240\textwidth]{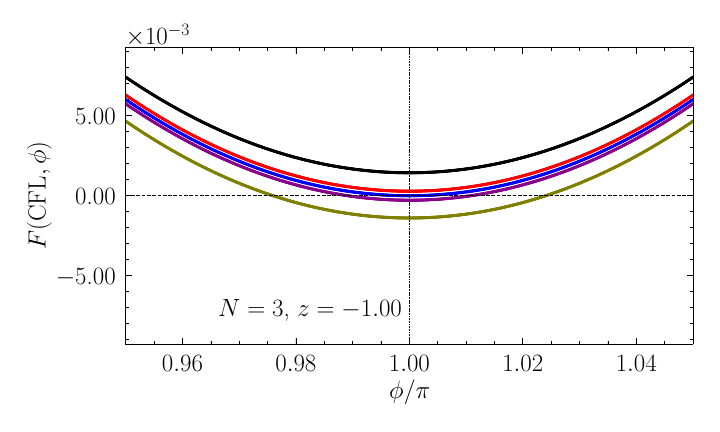}
\includegraphics[width=0.240\textwidth]{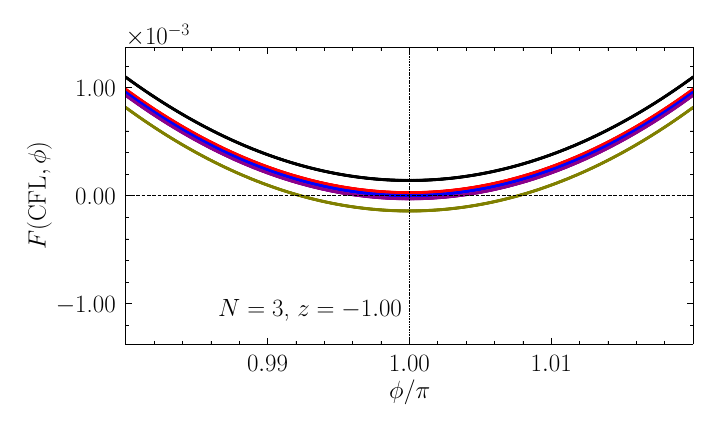}\\[1mm]
\includegraphics[width=0.240\textwidth]{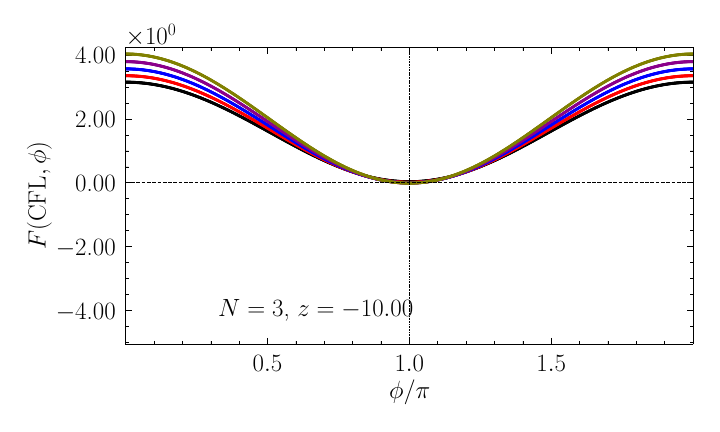}
\includegraphics[width=0.240\textwidth]{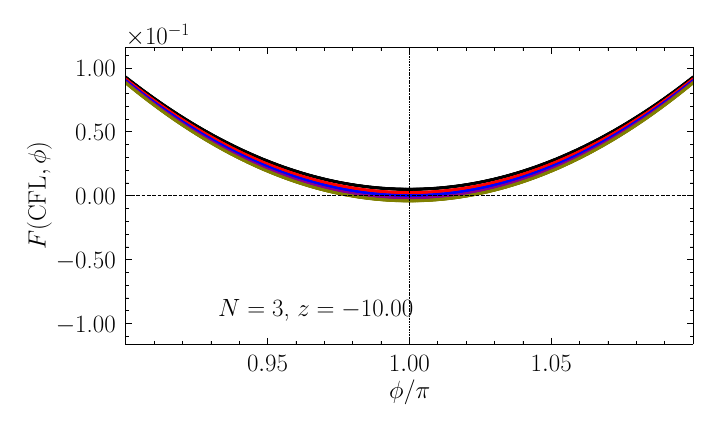}
\includegraphics[width=0.240\textwidth]{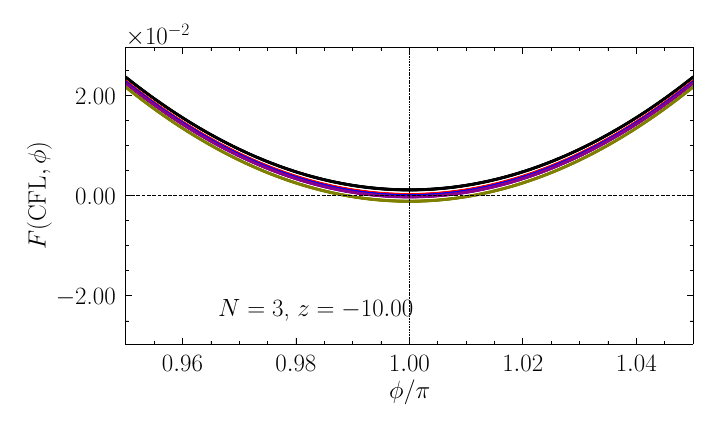}
\includegraphics[width=0.240\textwidth]{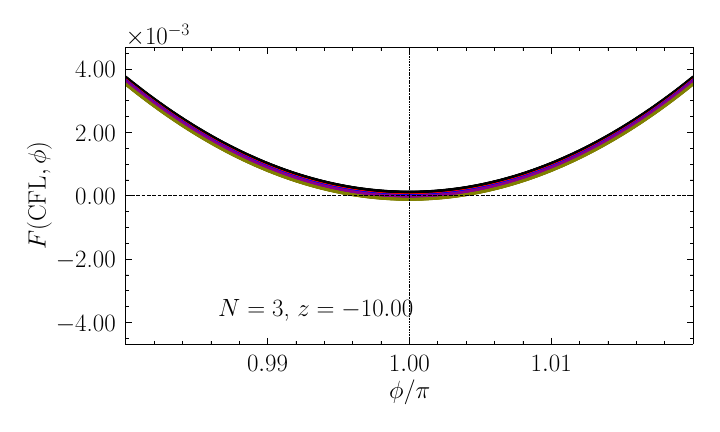}\\[1mm]
\includegraphics[width=0.240\textwidth]{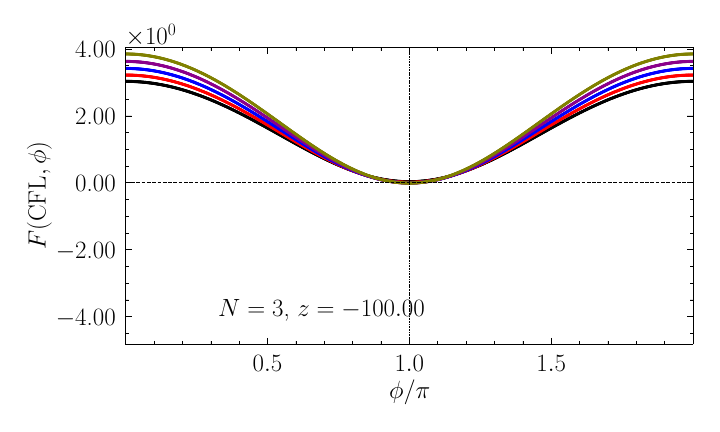}
\includegraphics[width=0.240\textwidth]{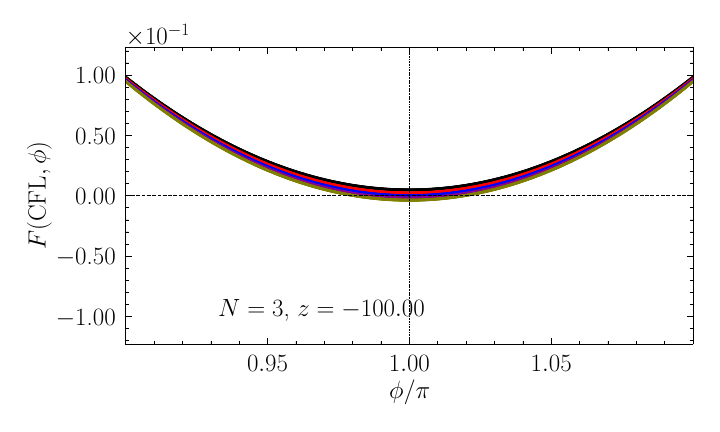}
\includegraphics[width=0.240\textwidth]{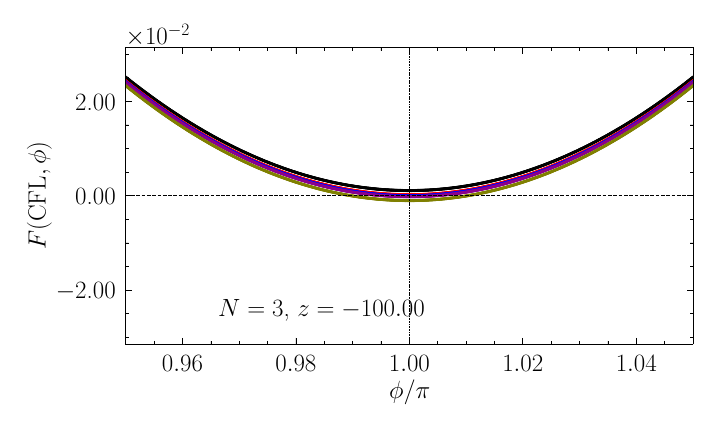}
\includegraphics[width=0.240\textwidth]{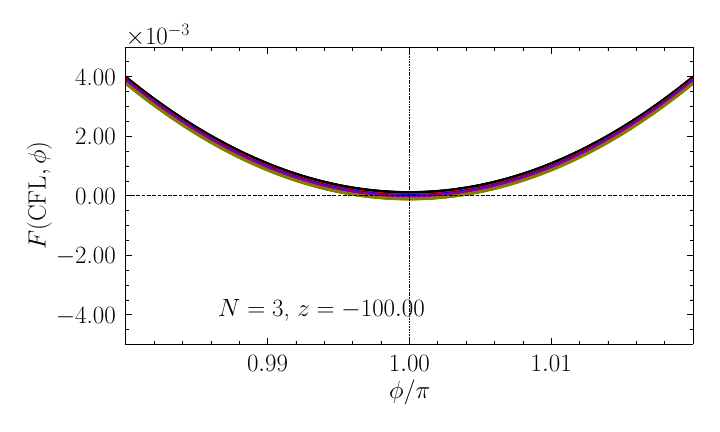}\\[1mm]
\includegraphics[width=0.240\textwidth]{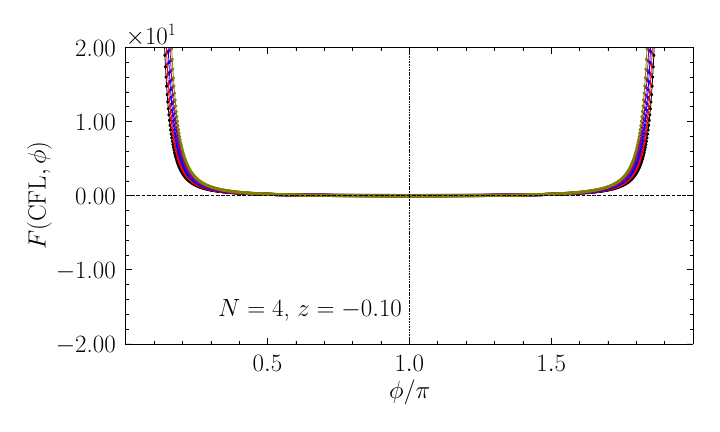}
\includegraphics[width=0.240\textwidth]{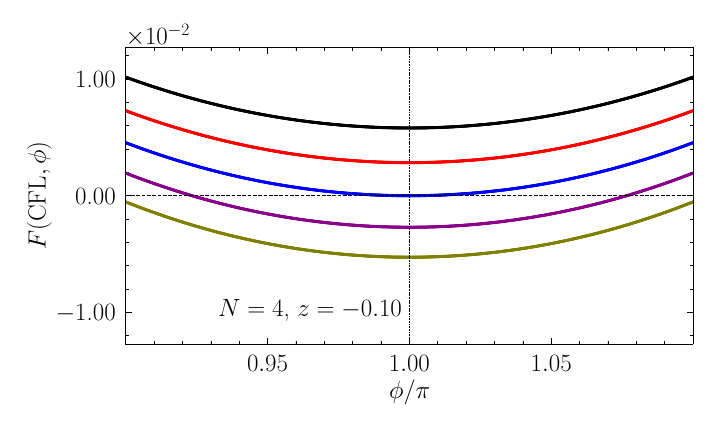}
\includegraphics[width=0.240\textwidth]{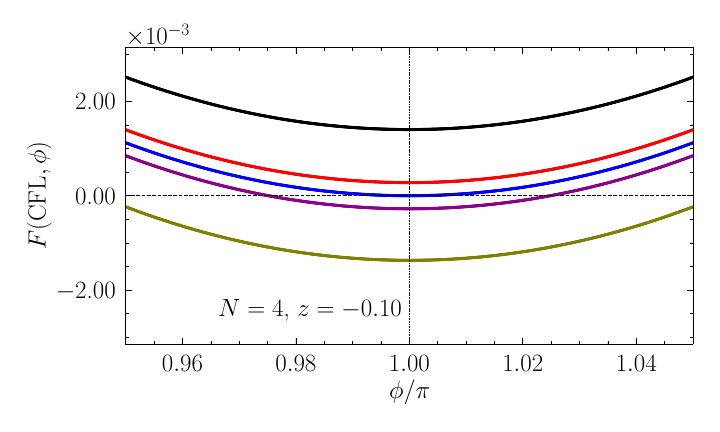}
\includegraphics[width=0.240\textwidth]{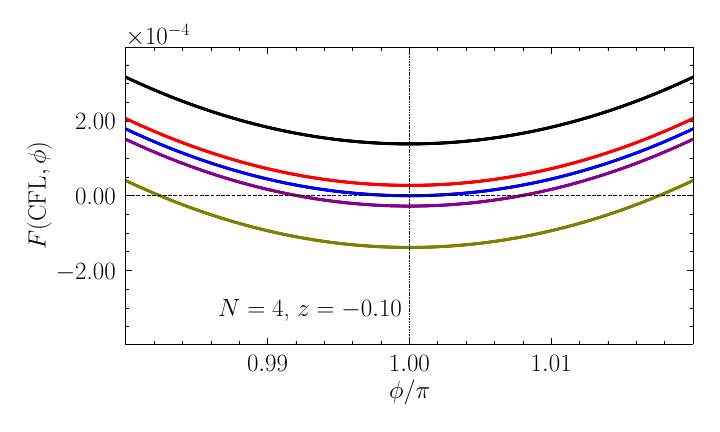}\\[1mm]
\includegraphics[width=0.240\textwidth]{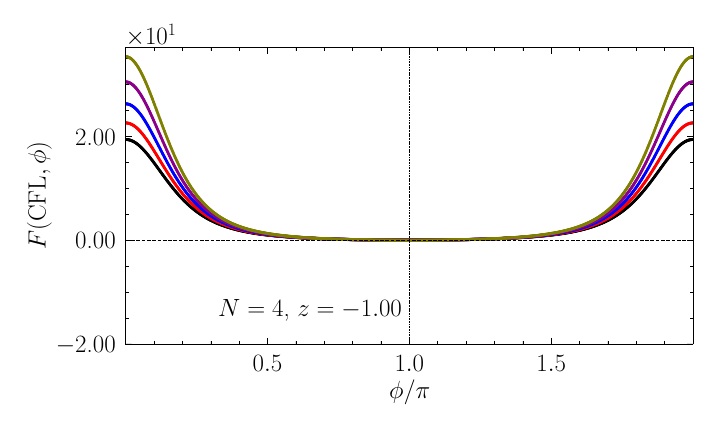}
\includegraphics[width=0.240\textwidth]{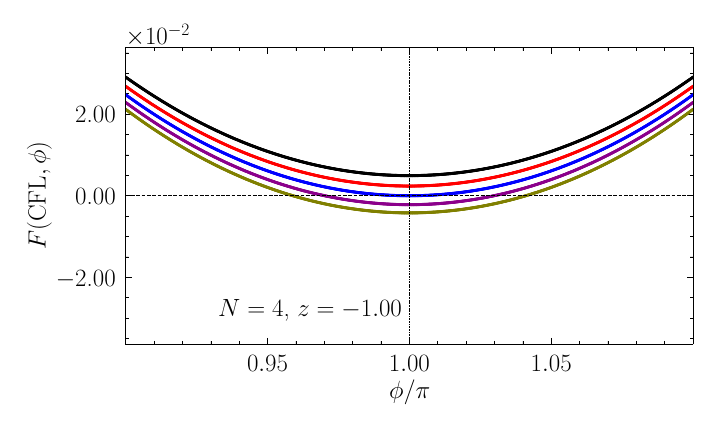}
\includegraphics[width=0.240\textwidth]{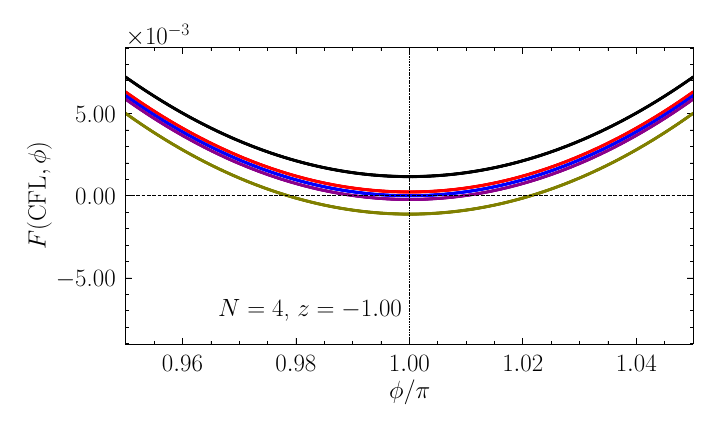}
\includegraphics[width=0.240\textwidth]{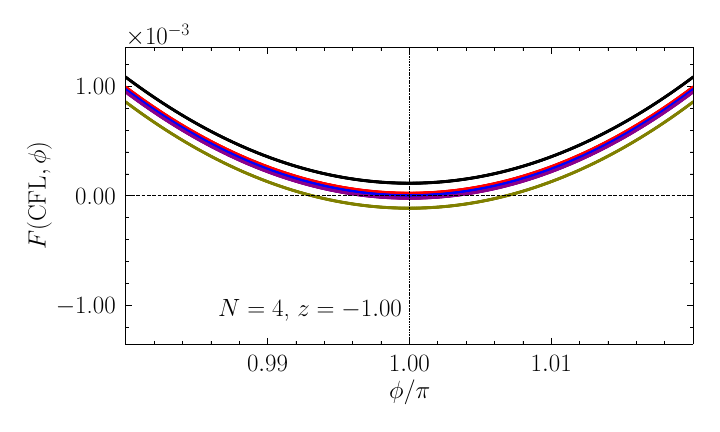}\\[1mm]
\includegraphics[width=0.240\textwidth]{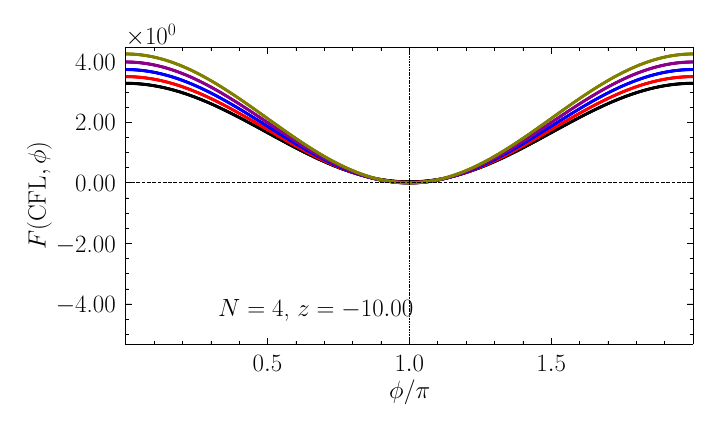}
\includegraphics[width=0.240\textwidth]{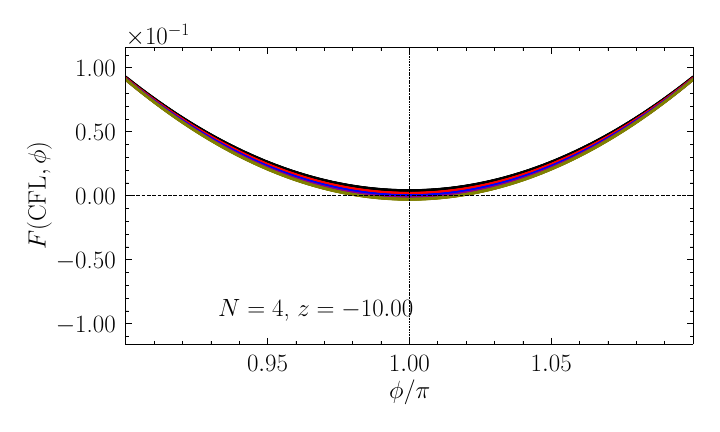}
\includegraphics[width=0.240\textwidth]{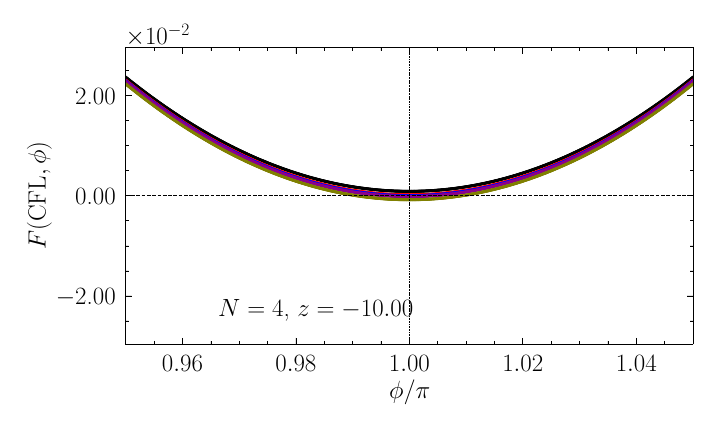}
\includegraphics[width=0.240\textwidth]{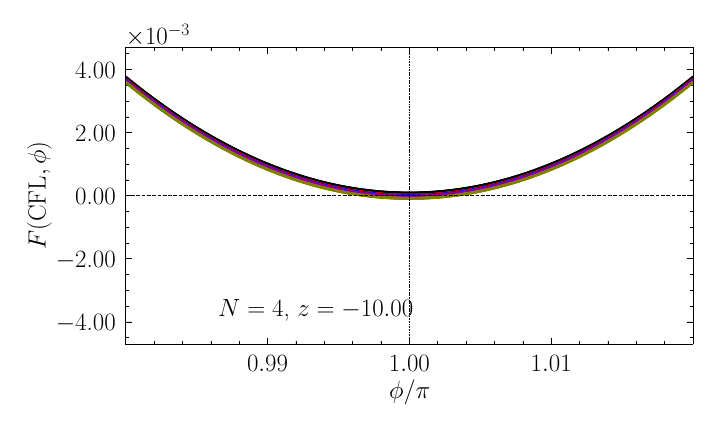}\\[1mm]
\includegraphics[width=0.240\textwidth]{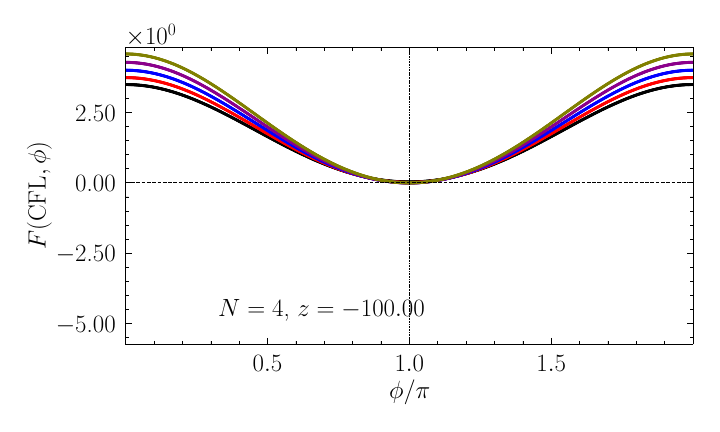}
\includegraphics[width=0.240\textwidth]{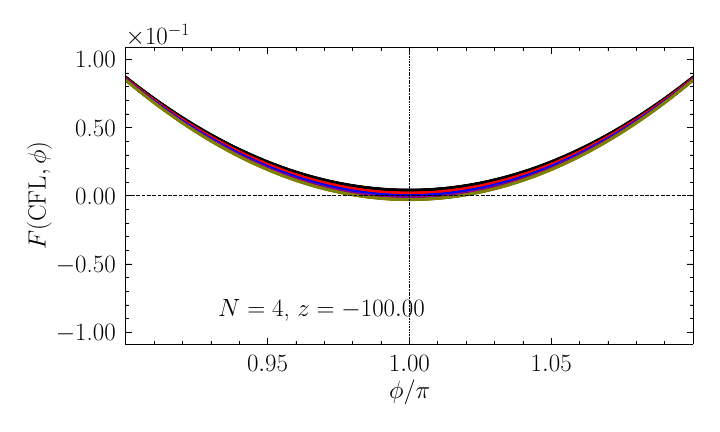}
\includegraphics[width=0.240\textwidth]{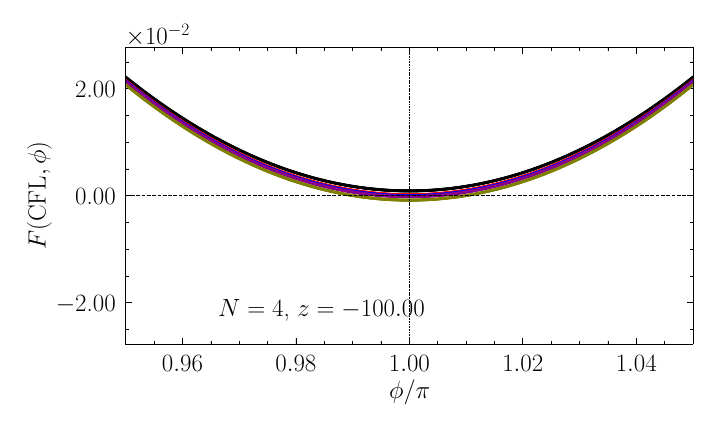}
\includegraphics[width=0.240\textwidth]{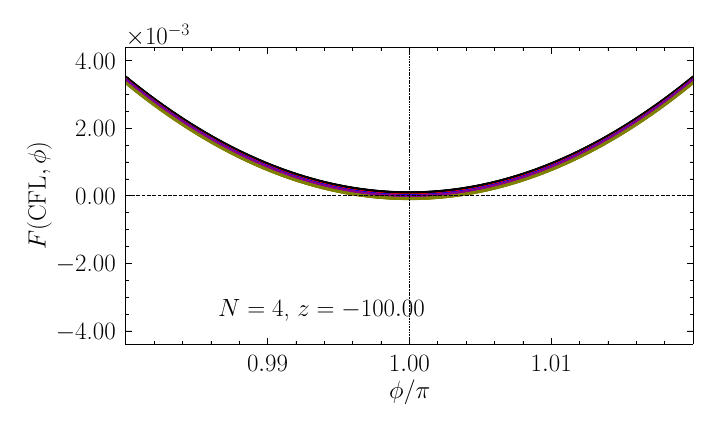}
\caption{%
The dependence of function $F(\phi; \mathrm{CFL}, z)$ (\ref{eq:f_def}) on argument $\phi\in[0, 2\pi]$ for several values of the Courant number $\mathrm{CFL}$, selected in the vicinity of the stability boundary $\mathrm{CFL}_{\rm max}(z)$, defined further in Figure~\ref{fig:cfls_on_z}, for $z = -0.1$, $-1.0$, $-10.0$, $-100.0$ and for polynomial degrees $N = 3$ and $4$. The graph legends are distributed among columns and located at the top of the columns. From left to right, the graphs represent increasingly narrower neighborhoods of the stability boundary $\mathrm{CFL}_{\rm max}$, from $\mathrm{CFL} \in [0.90\, \mathrm{CFL}_{\rm max},\, 1.10\, \mathrm{CFL}_{\rm max}]$ in the left column to $\mathrm{CFL} \in [0.9995\, \mathrm{CFL}_{\rm max},\, 1.0005\, \mathrm{CFL}_{\rm max}]$ in the right column; the range of phase $\phi$ variation also narrows when moving from the left to the right columns of the graphs to ensure a correct and clear presentation of the root $\phi = \pi$ of the equation $F(\pi; \mathrm{CFL}, z) = 0$. The horizontal dotted line indicates $F = 0$, the vertical dotted line indicates $\phi = \pi$ ($\lambda = \exp(\pm i\phi) = -1$). \textit{Note}: the phase $\phi$ of eigenvalue $\lambda_{k} = |\lambda_{k}|\exp(i\phi)$ is not the phase $\theta$.
}
\label{fig:dep_f_on_phi_diff_cfls_degrees_3_4}
\end{figure}
\begin{figure}[h!]
\centering
\includegraphics[width=0.240\textwidth]{dep_f_on_phi_diff_cfls_legend_Cidx_0.png}
\includegraphics[width=0.240\textwidth]{dep_f_on_phi_diff_cfls_legend_Cidx_1.png}
\includegraphics[width=0.240\textwidth]{dep_f_on_phi_diff_cfls_legend_Cidx_2.png}
\includegraphics[width=0.240\textwidth]{dep_f_on_phi_diff_cfls_legend_Cidx_3.png}\\[1mm]
\includegraphics[width=0.240\textwidth]{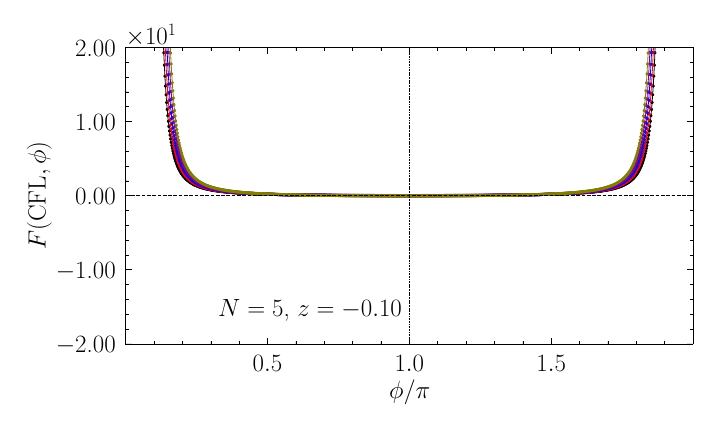}
\includegraphics[width=0.240\textwidth]{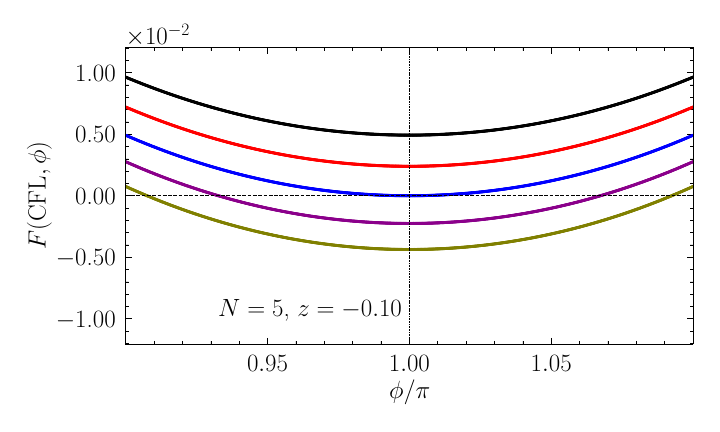}
\includegraphics[width=0.240\textwidth]{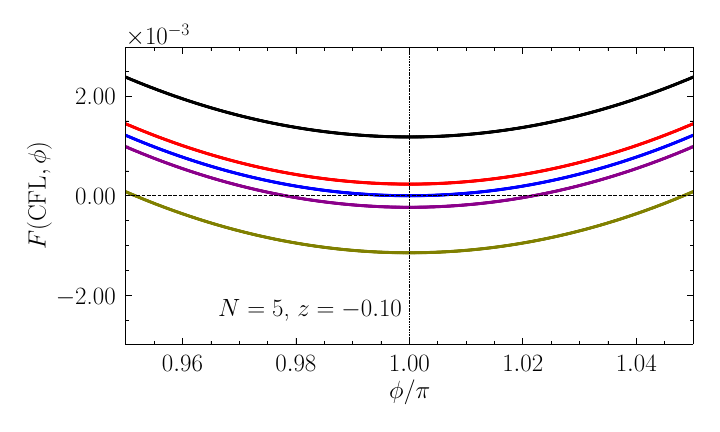}
\includegraphics[width=0.240\textwidth]{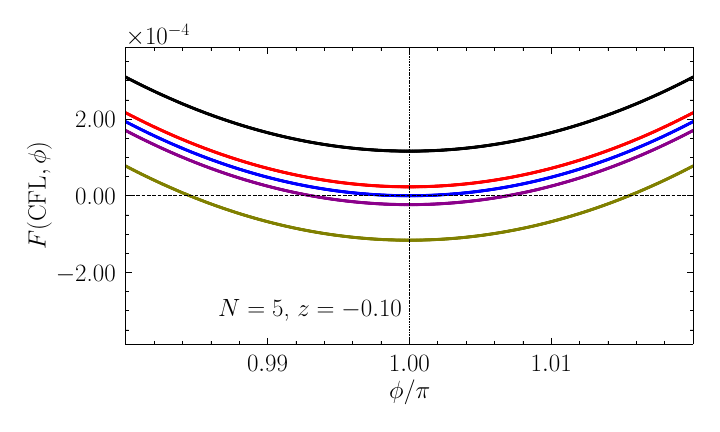}\\[1mm]
\includegraphics[width=0.240\textwidth]{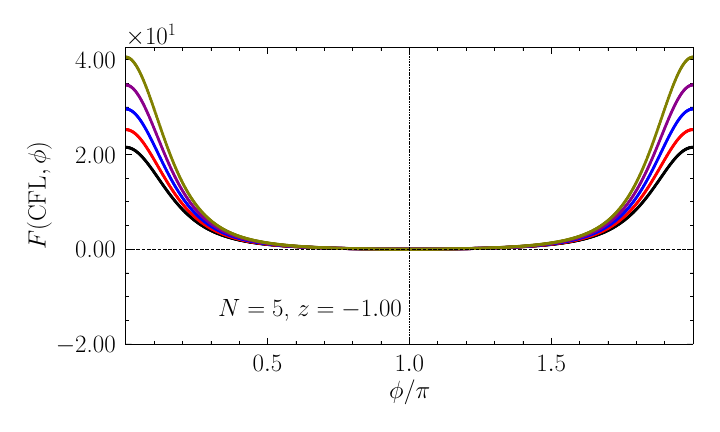}
\includegraphics[width=0.240\textwidth]{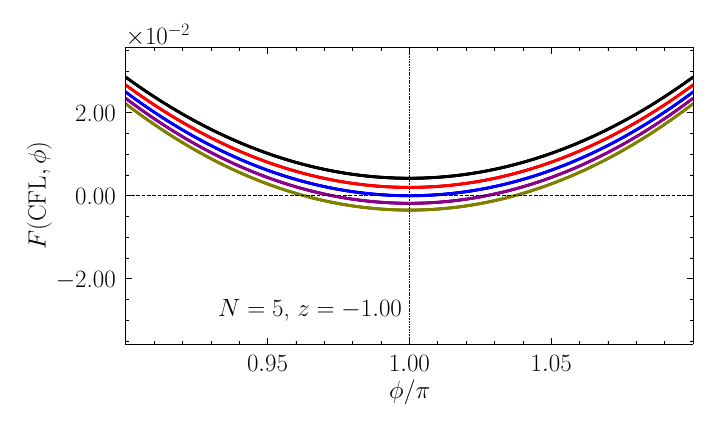}
\includegraphics[width=0.240\textwidth]{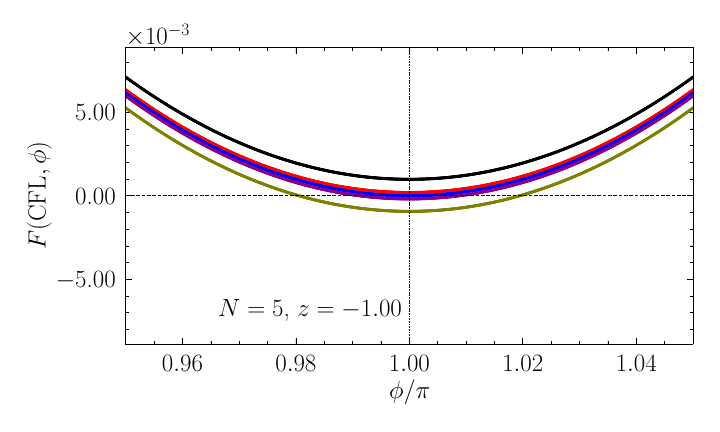}
\includegraphics[width=0.240\textwidth]{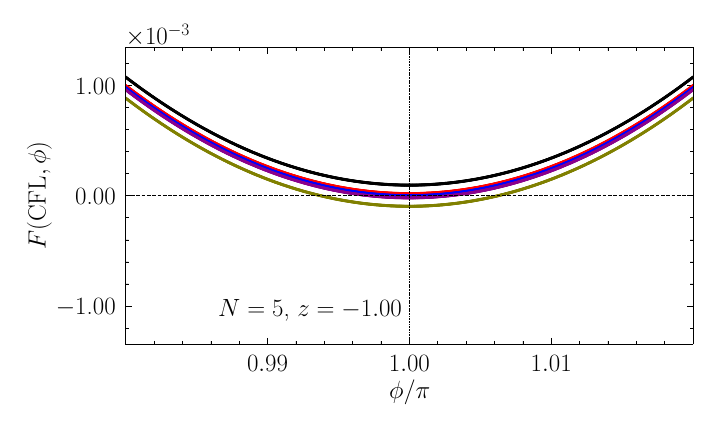}\\[1mm]
\includegraphics[width=0.240\textwidth]{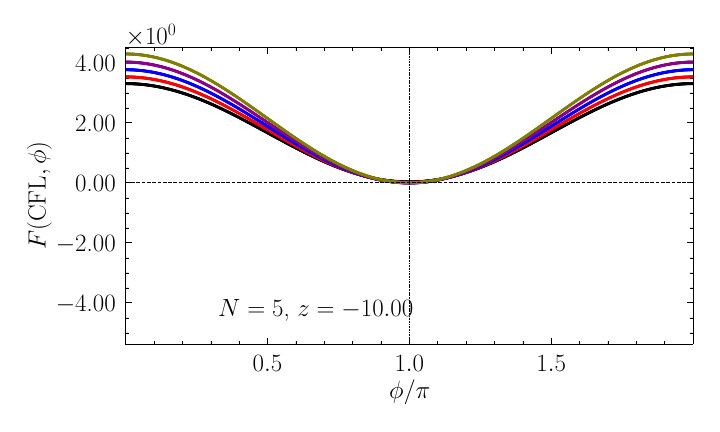}
\includegraphics[width=0.240\textwidth]{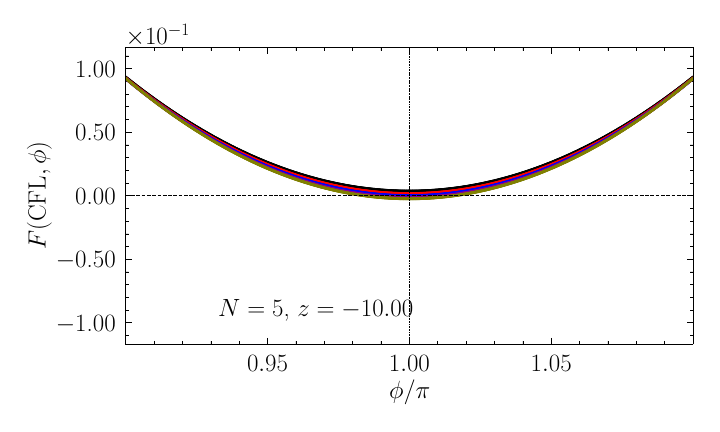}
\includegraphics[width=0.240\textwidth]{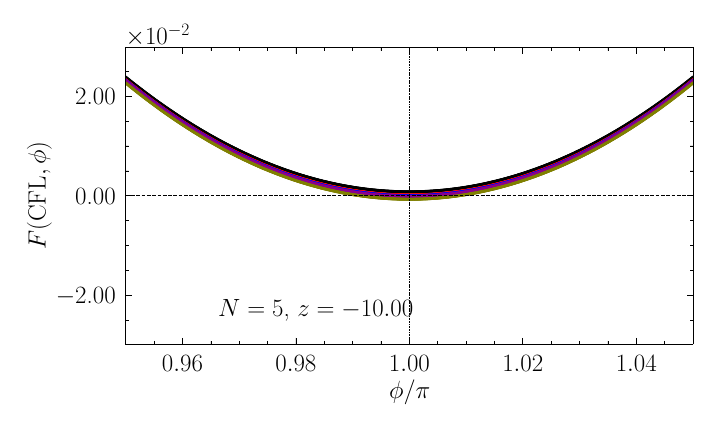}
\includegraphics[width=0.240\textwidth]{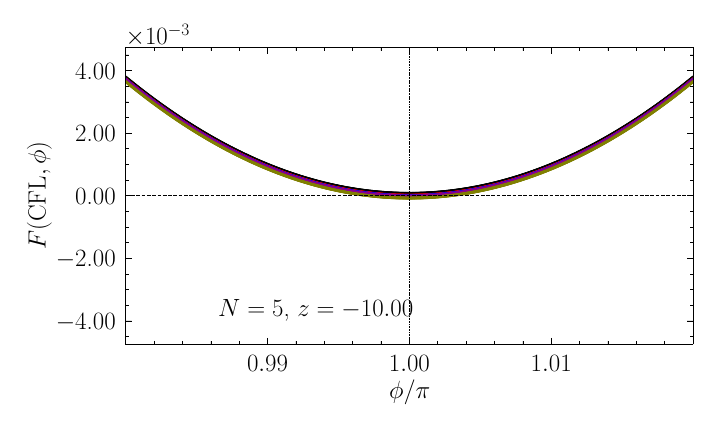}\\[1mm]
\includegraphics[width=0.240\textwidth]{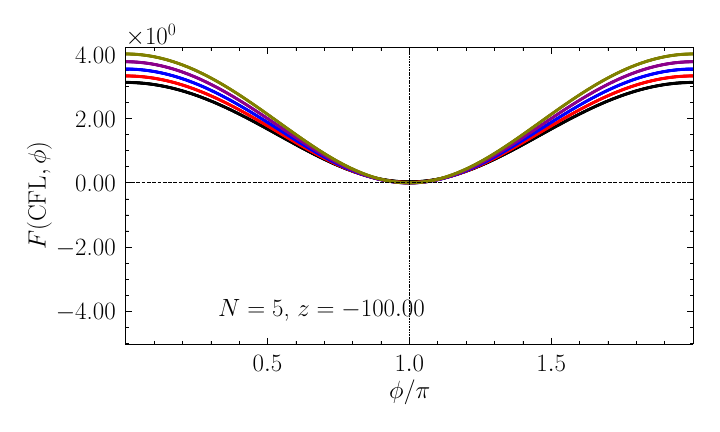}
\includegraphics[width=0.240\textwidth]{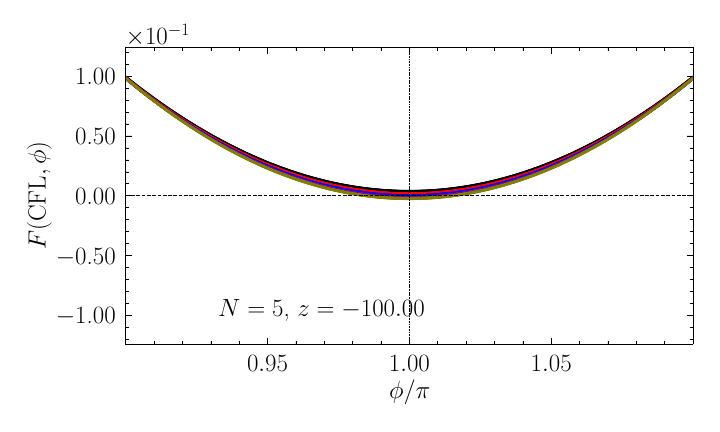}
\includegraphics[width=0.240\textwidth]{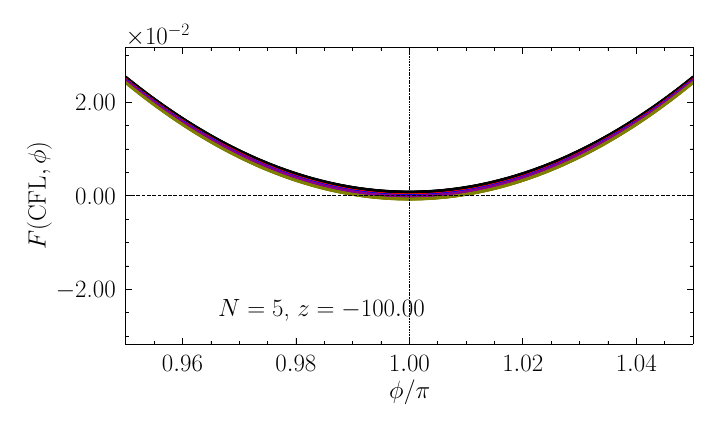}
\includegraphics[width=0.240\textwidth]{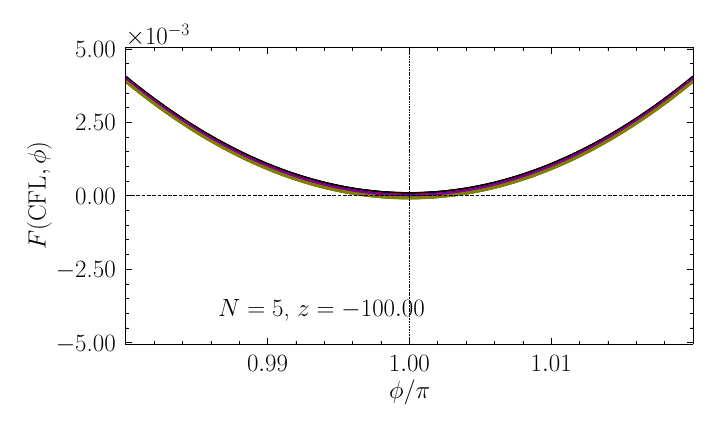}\\[1mm]
\includegraphics[width=0.240\textwidth]{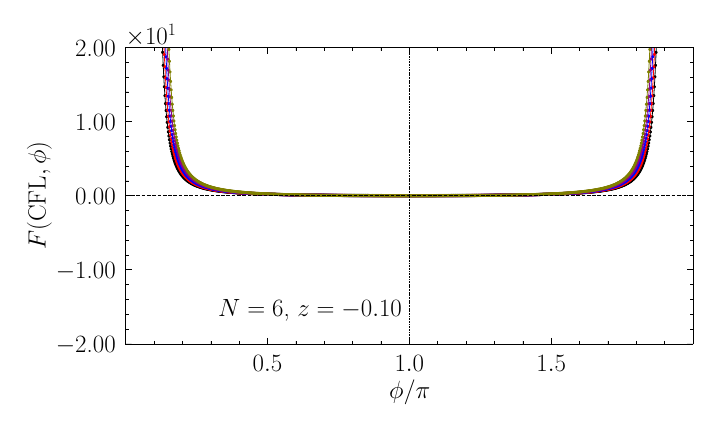}
\includegraphics[width=0.240\textwidth]{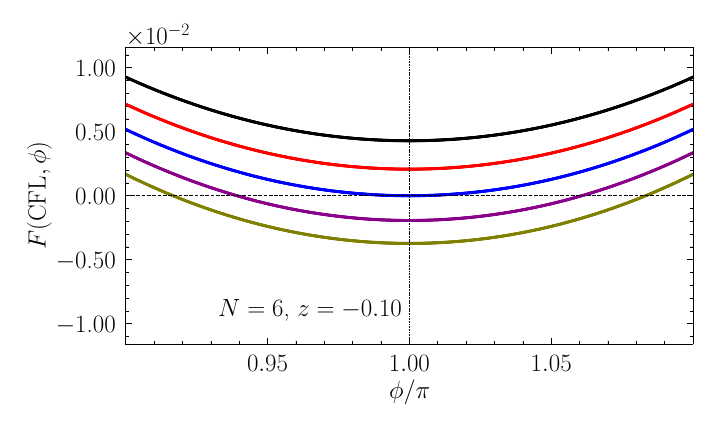}
\includegraphics[width=0.240\textwidth]{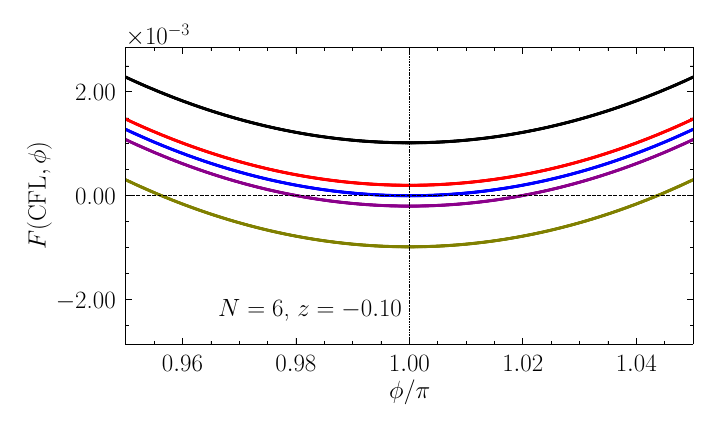}
\includegraphics[width=0.240\textwidth]{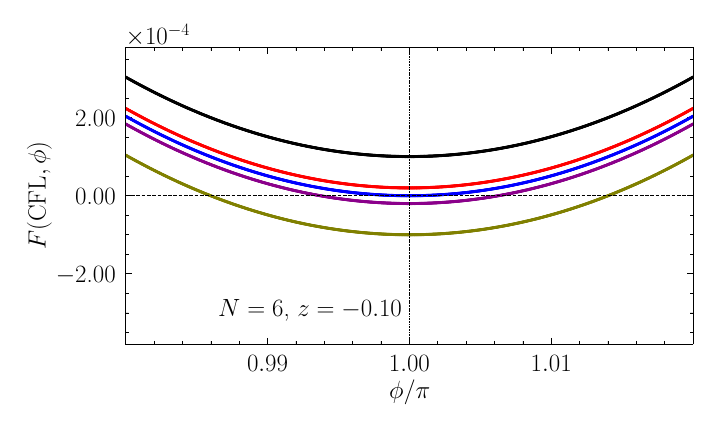}\\[1mm]
\includegraphics[width=0.240\textwidth]{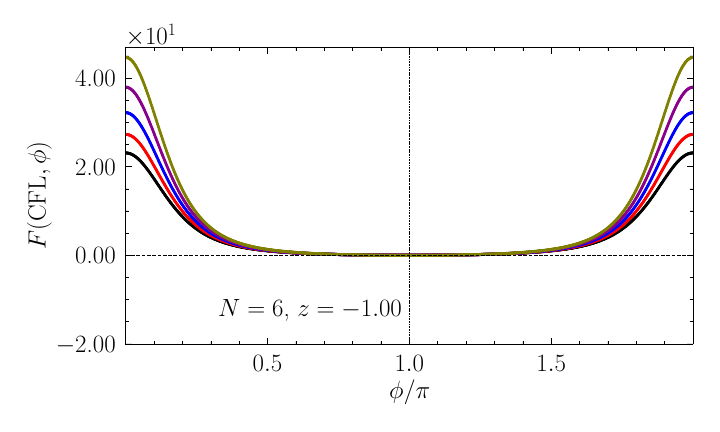}
\includegraphics[width=0.240\textwidth]{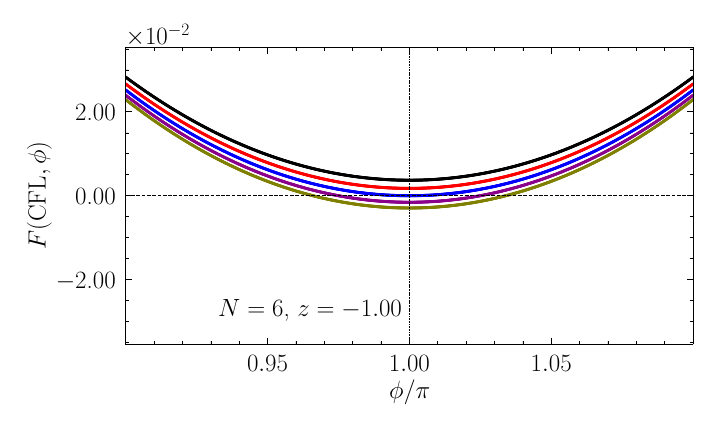}
\includegraphics[width=0.240\textwidth]{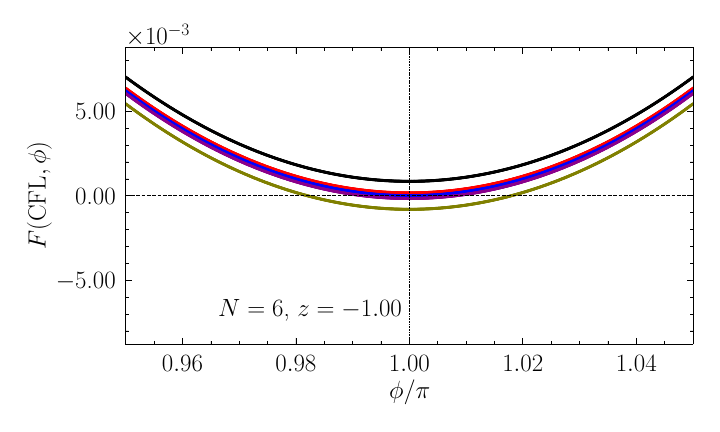}
\includegraphics[width=0.240\textwidth]{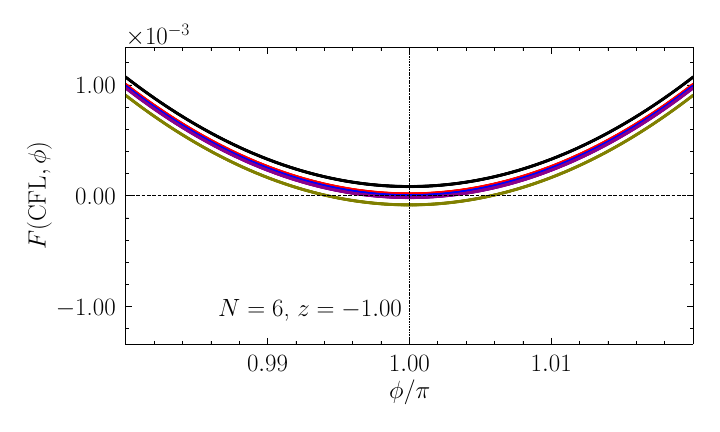}\\[1mm]
\includegraphics[width=0.240\textwidth]{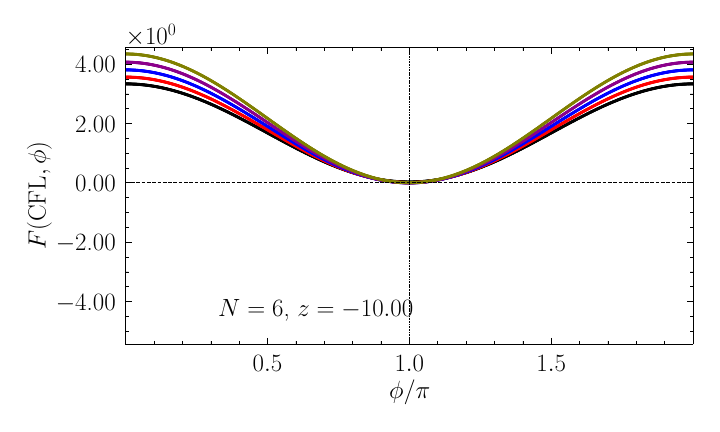}
\includegraphics[width=0.240\textwidth]{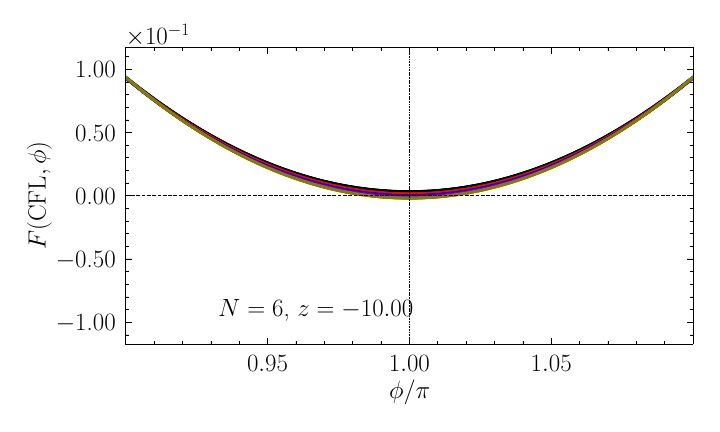}
\includegraphics[width=0.240\textwidth]{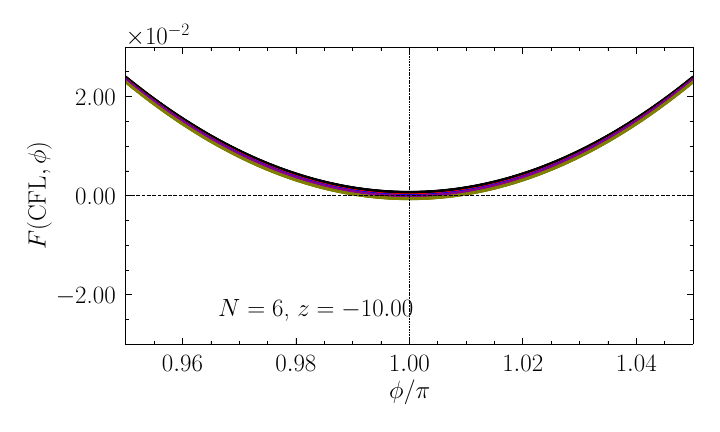}
\includegraphics[width=0.240\textwidth]{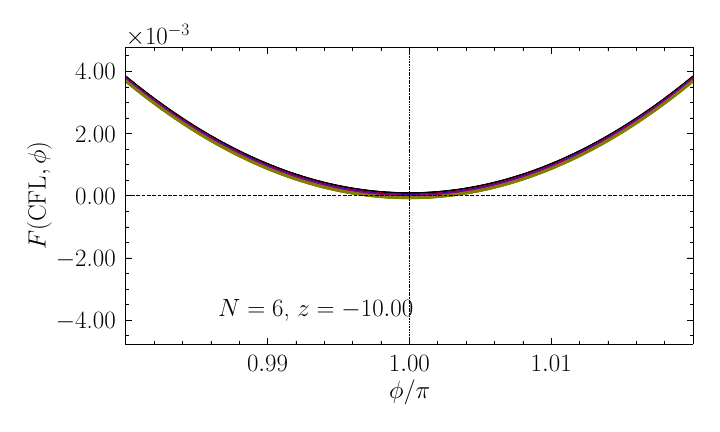}\\[1mm]
\includegraphics[width=0.240\textwidth]{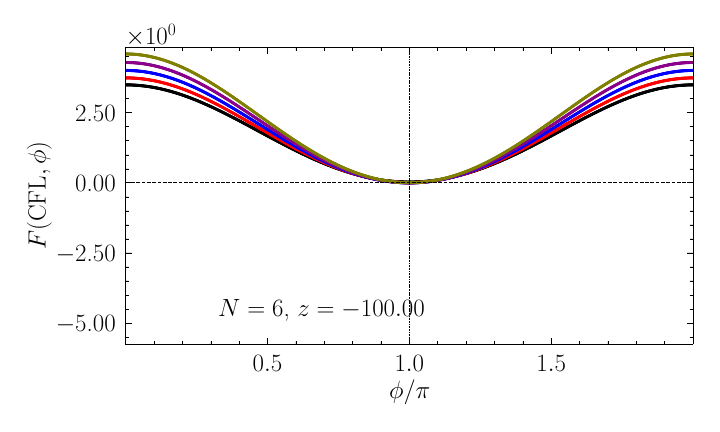}
\includegraphics[width=0.240\textwidth]{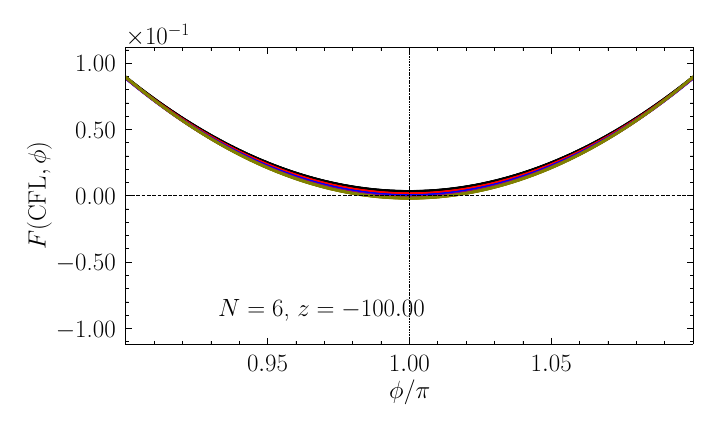}
\includegraphics[width=0.240\textwidth]{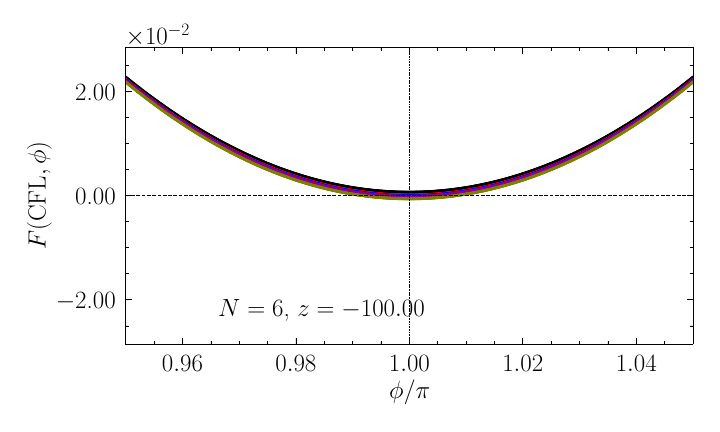}
\includegraphics[width=0.240\textwidth]{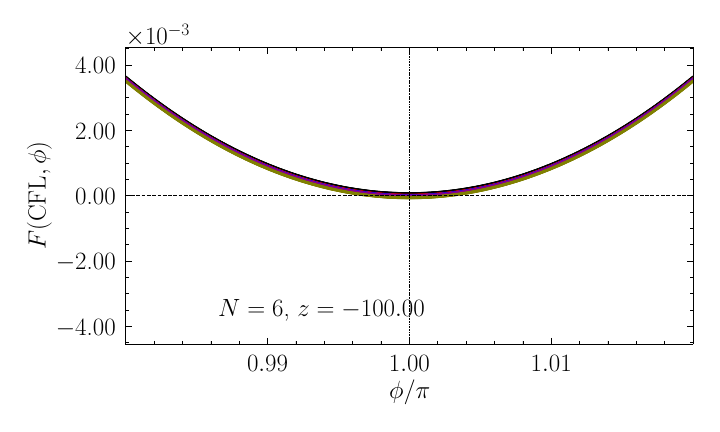}
\caption{%
The dependence of function $F(\phi; \mathrm{CFL}, z)$ (\ref{eq:f_def}) on argument $\phi\in[0, 2\pi]$ for several values of the Courant number $\mathrm{CFL}$, selected in the vicinity of the stability boundary $\mathrm{CFL}_{\rm max}(z)$, defined further in Figure~\ref{fig:cfls_on_z}, for $z = -0.1$, $-1.0$, $-10.0$, $-100.0$ and for polynomial degrees $N = 5$ and $6$. The graph legends are distributed among columns and located at the top of the columns. From left to right, the graphs represent increasingly narrower neighborhoods of the stability boundary $\mathrm{CFL}_{\rm max}$, from $\mathrm{CFL} \in [0.90\, \mathrm{CFL}_{\rm max},\, 1.10\, \mathrm{CFL}_{\rm max}]$ in the left column to $\mathrm{CFL} \in [0.9995\, \mathrm{CFL}_{\rm max},\, 1.0005\, \mathrm{CFL}_{\rm max}]$ in the right column; the range of phase $\phi$ variation also narrows when moving from the left to the right columns of the graphs to ensure a correct and clear presentation of the root $\phi = \pi$ of the equation $F(\pi; \mathrm{CFL}, z) = 0$. The horizontal dotted line indicates $F = 0$, the vertical dotted line indicates $\phi = \pi$ ($\lambda = \exp(\pm i\phi) = -1$). \textit{Note}: the phase $\phi$ of eigenvalue $\lambda_{k} = |\lambda_{k}|\exp(i\phi)$ is not the phase $\theta$.
}
\label{fig:dep_f_on_phi_diff_cfls_degrees_5_6}
\end{figure}
\begin{figure}[h!]
\centering
\includegraphics[width=0.240\textwidth]{dep_f_on_phi_diff_cfls_legend_Cidx_0.png}
\includegraphics[width=0.240\textwidth]{dep_f_on_phi_diff_cfls_legend_Cidx_1.png}
\includegraphics[width=0.240\textwidth]{dep_f_on_phi_diff_cfls_legend_Cidx_2.png}
\includegraphics[width=0.240\textwidth]{dep_f_on_phi_diff_cfls_legend_Cidx_3.png}\\[1mm]
\includegraphics[width=0.240\textwidth]{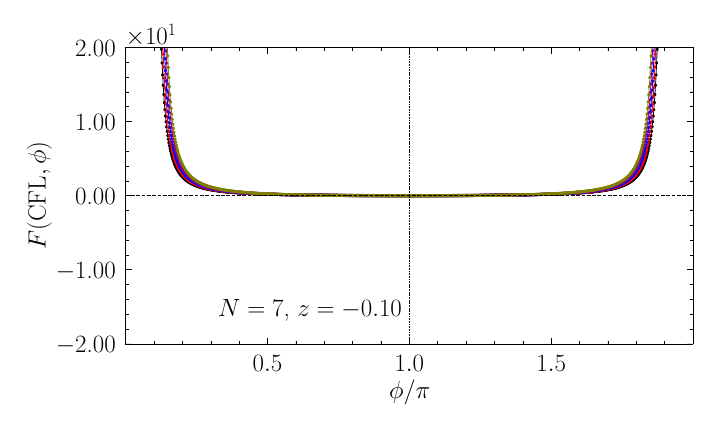}
\includegraphics[width=0.240\textwidth]{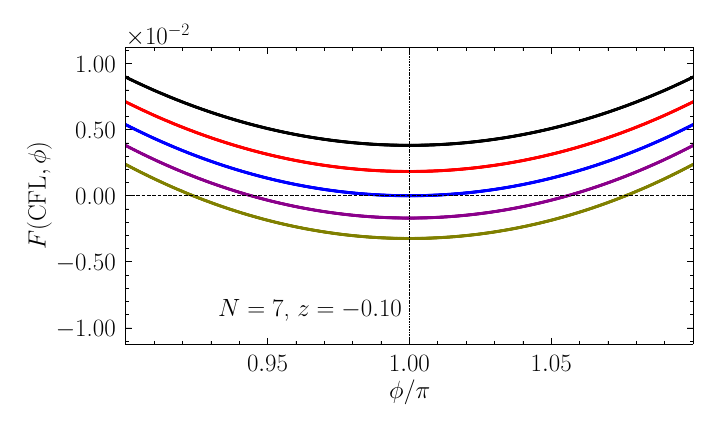}
\includegraphics[width=0.240\textwidth]{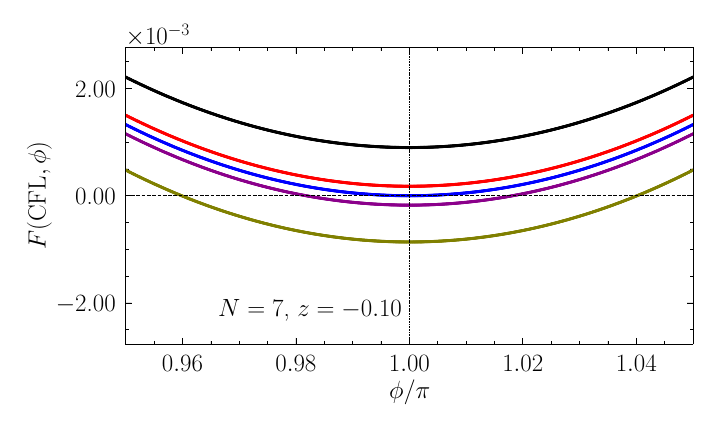}
\includegraphics[width=0.240\textwidth]{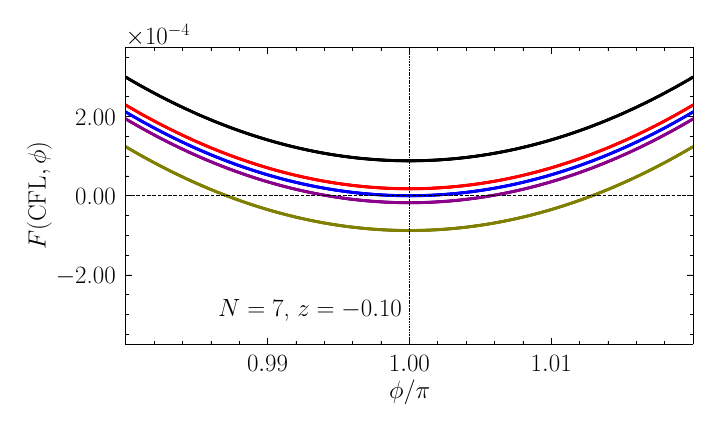}\\[1mm]
\includegraphics[width=0.240\textwidth]{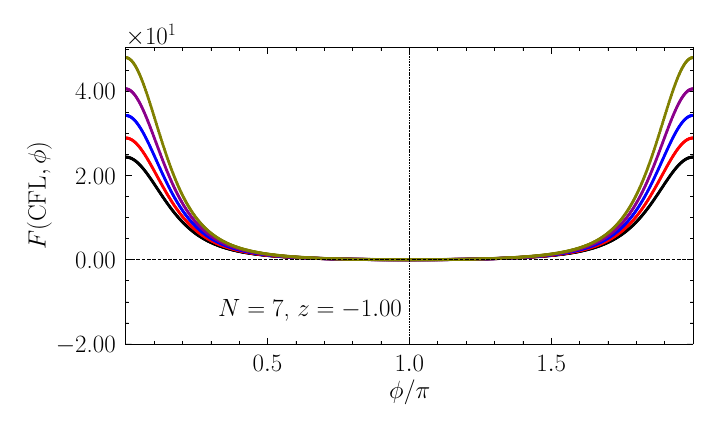}
\includegraphics[width=0.240\textwidth]{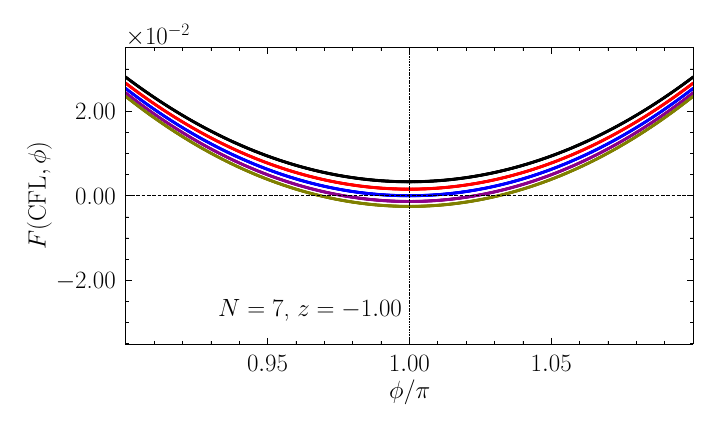}
\includegraphics[width=0.240\textwidth]{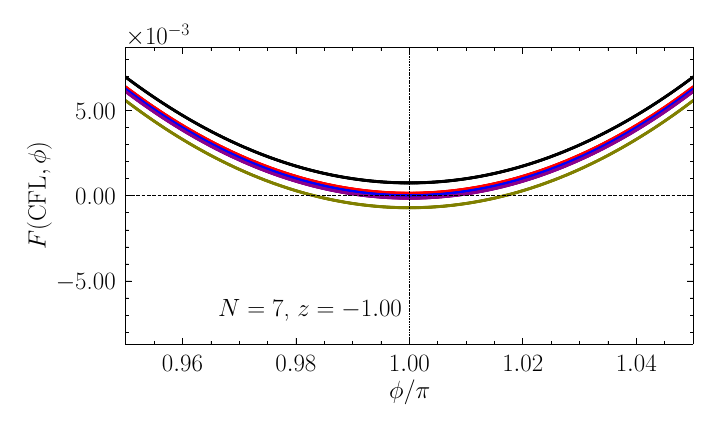}
\includegraphics[width=0.240\textwidth]{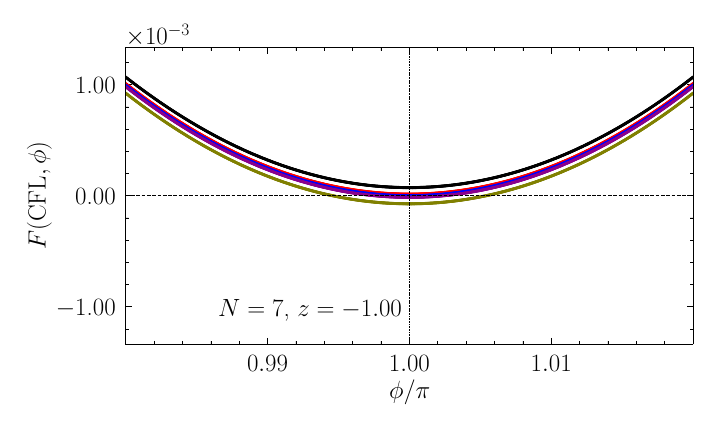}\\[1mm]
\includegraphics[width=0.240\textwidth]{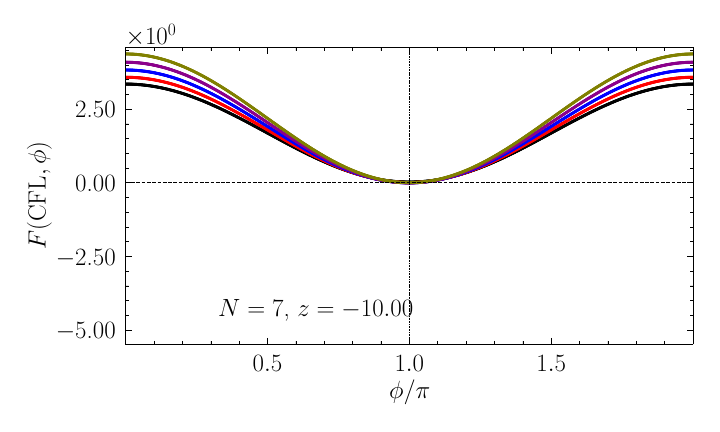}
\includegraphics[width=0.240\textwidth]{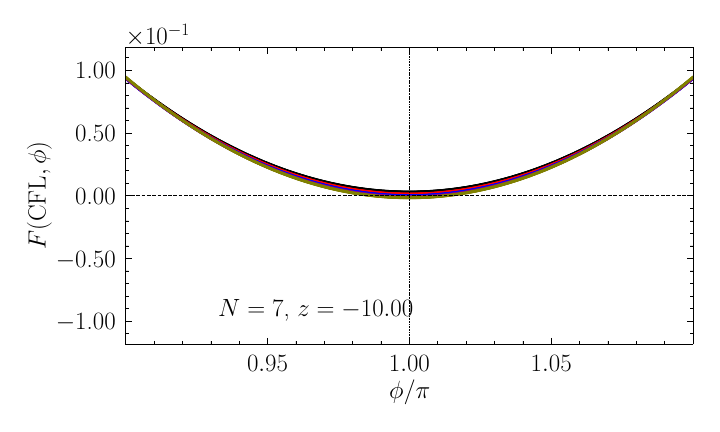}
\includegraphics[width=0.240\textwidth]{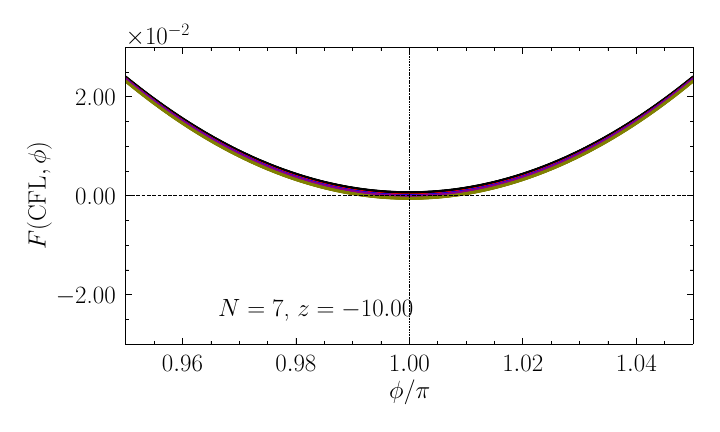}
\includegraphics[width=0.240\textwidth]{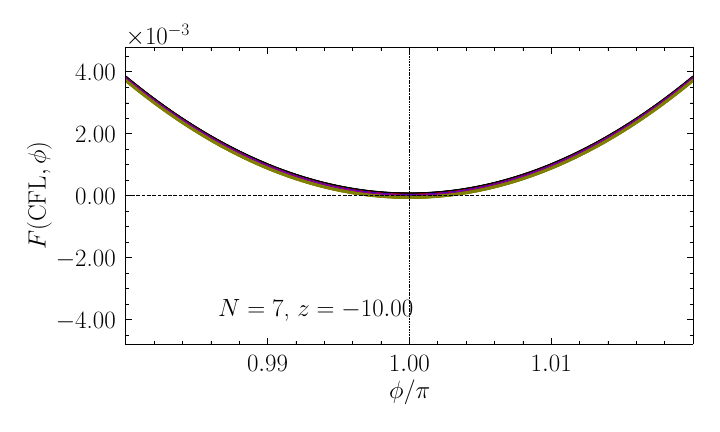}\\[1mm]
\includegraphics[width=0.240\textwidth]{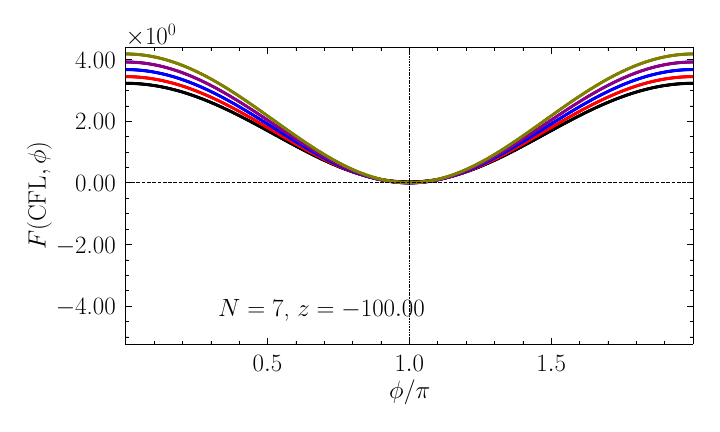}
\includegraphics[width=0.240\textwidth]{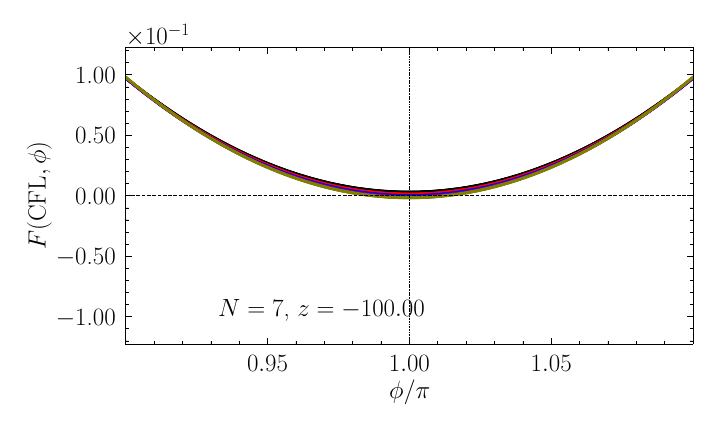}
\includegraphics[width=0.240\textwidth]{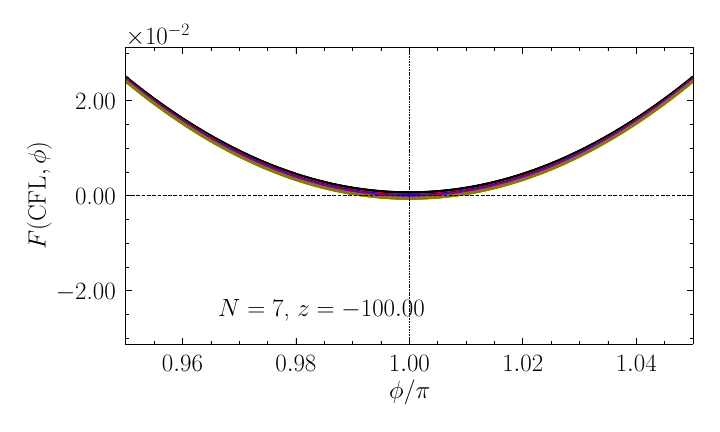}
\includegraphics[width=0.240\textwidth]{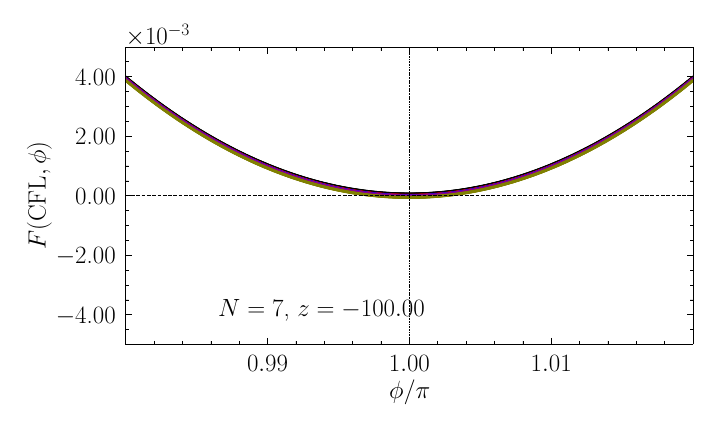}\\[1mm]
\includegraphics[width=0.240\textwidth]{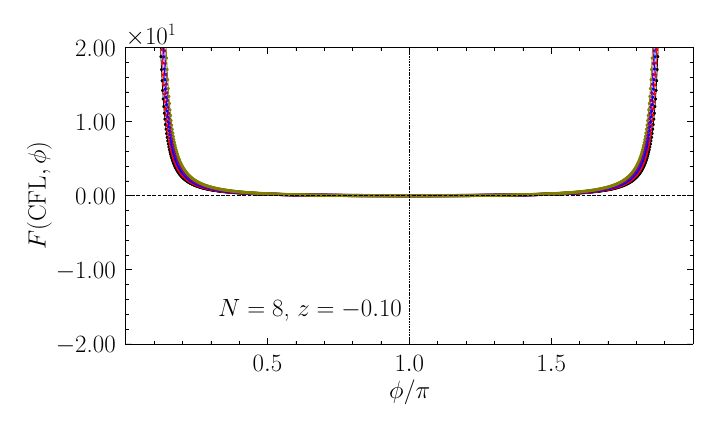}
\includegraphics[width=0.240\textwidth]{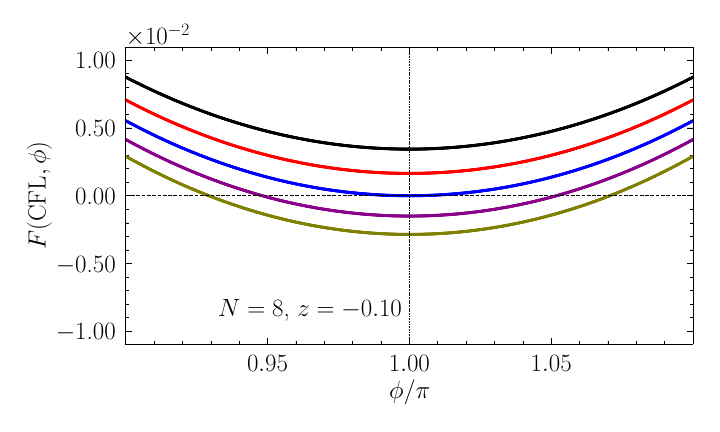}
\includegraphics[width=0.240\textwidth]{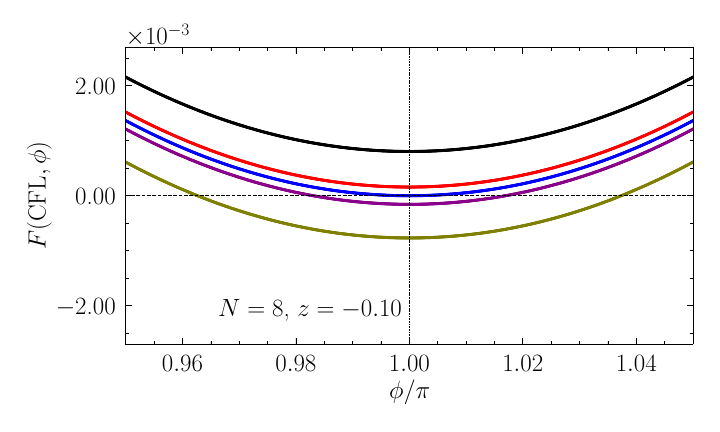}
\includegraphics[width=0.240\textwidth]{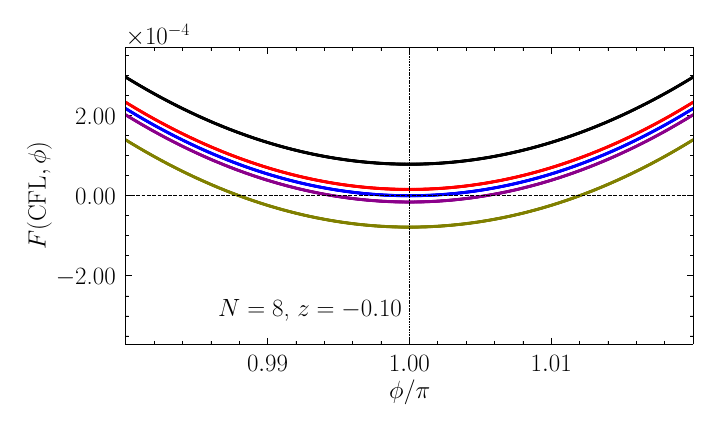}\\[1mm]
\includegraphics[width=0.240\textwidth]{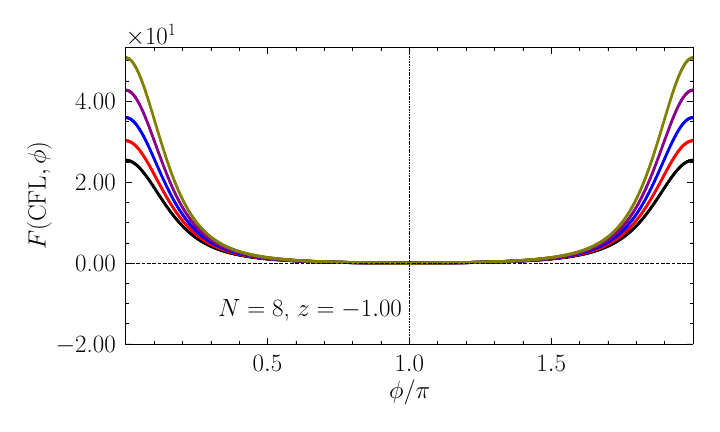}
\includegraphics[width=0.240\textwidth]{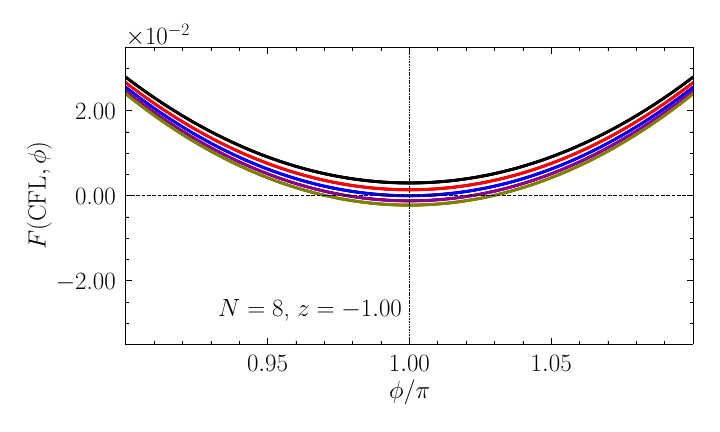}
\includegraphics[width=0.240\textwidth]{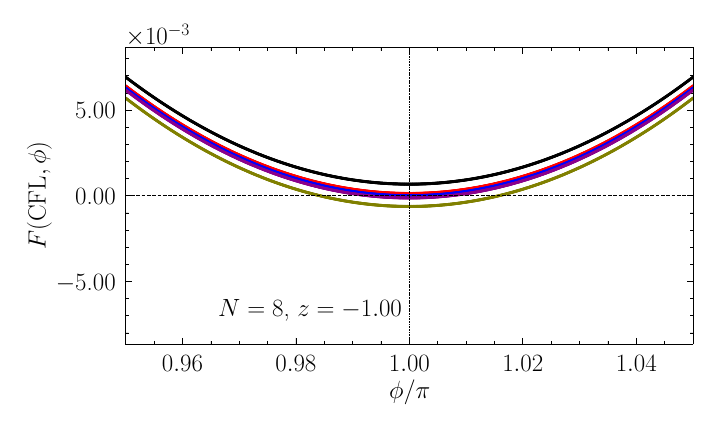}
\includegraphics[width=0.240\textwidth]{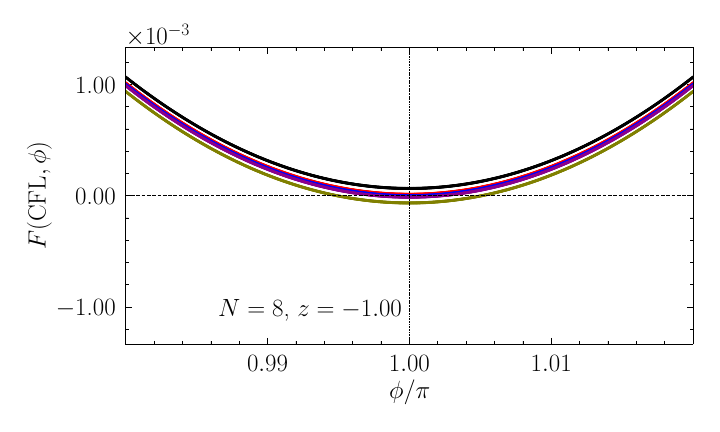}\\[1mm]
\includegraphics[width=0.240\textwidth]{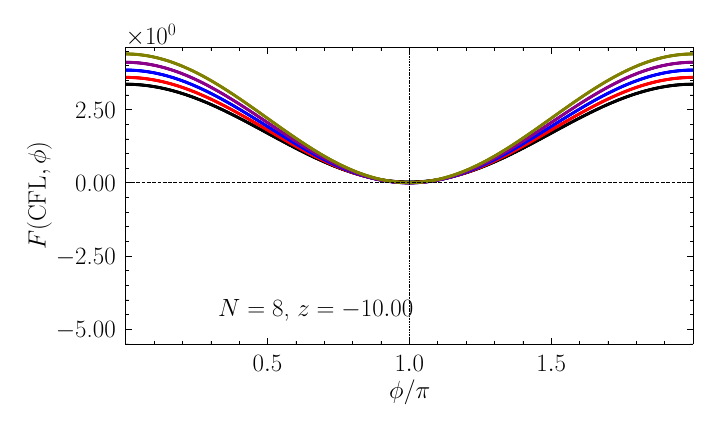}
\includegraphics[width=0.240\textwidth]{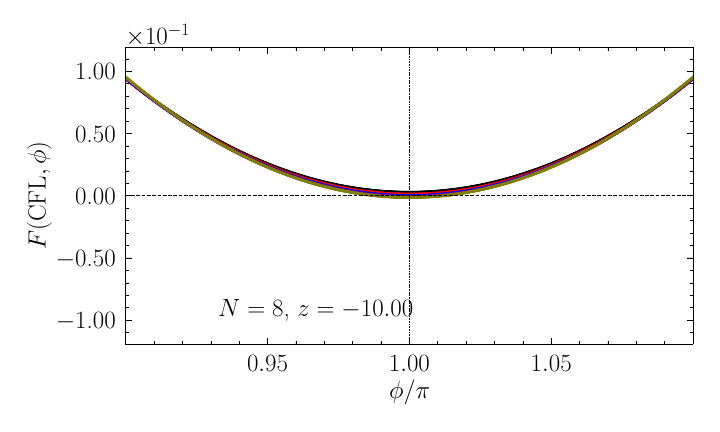}
\includegraphics[width=0.240\textwidth]{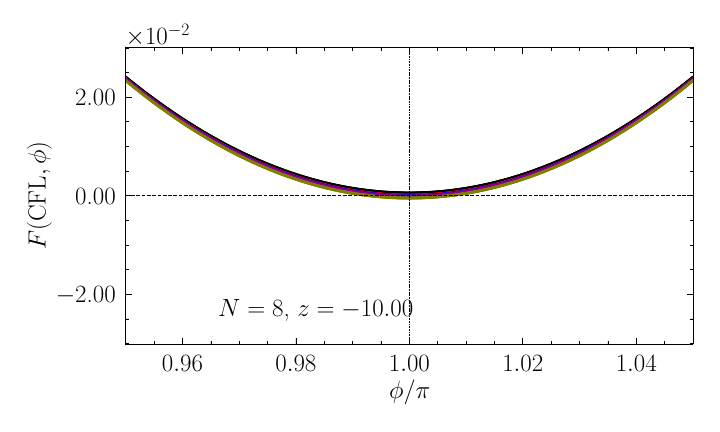}
\includegraphics[width=0.240\textwidth]{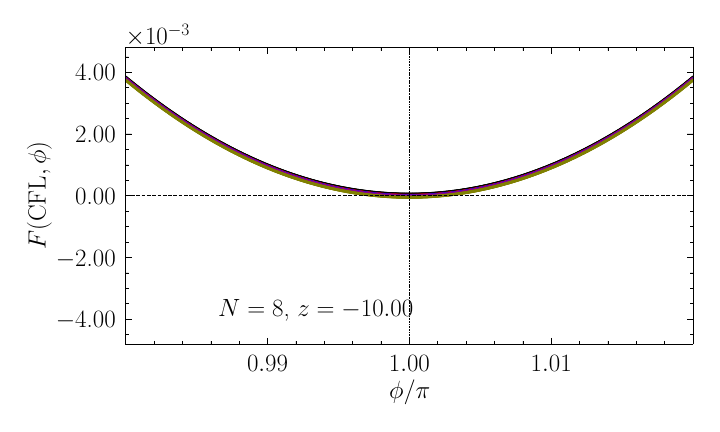}\\[1mm]
\includegraphics[width=0.240\textwidth]{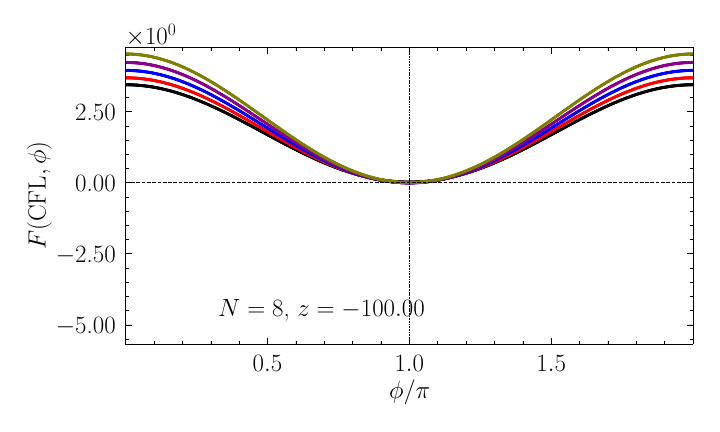}
\includegraphics[width=0.240\textwidth]{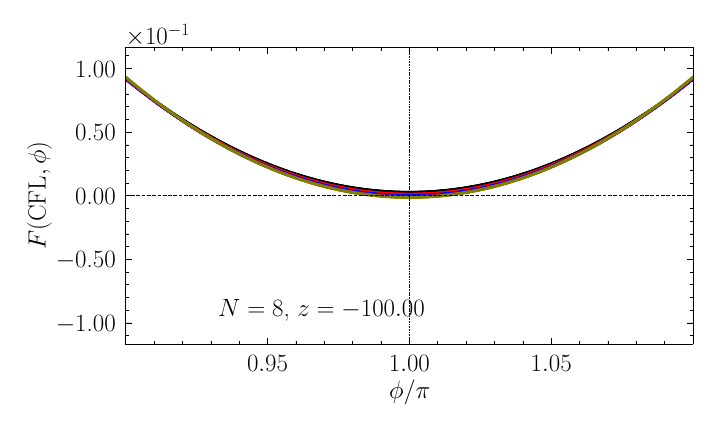}
\includegraphics[width=0.240\textwidth]{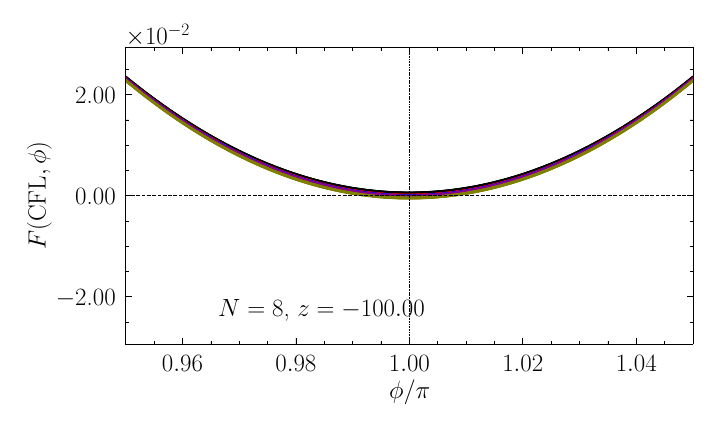}
\includegraphics[width=0.240\textwidth]{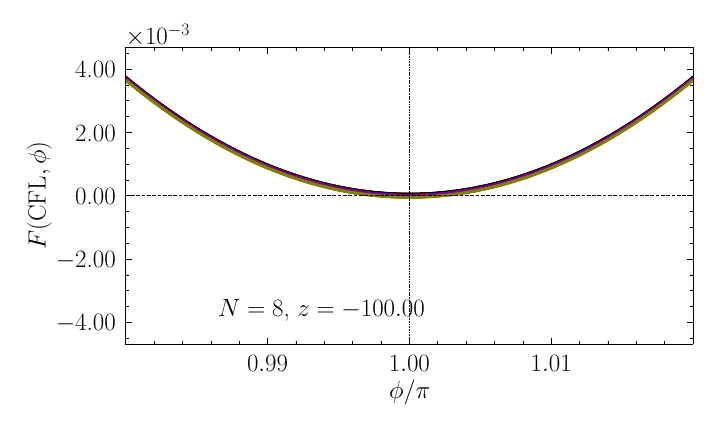}
\caption{%
The dependence of function $F(\phi; \mathrm{CFL}, z)$ (\ref{eq:f_def}) on argument $\phi\in[0, 2\pi]$ for several values of the Courant number $\mathrm{CFL}$, selected in the vicinity of the stability boundary $\mathrm{CFL}_{\rm max}(z)$, defined further in Figure~\ref{fig:cfls_on_z}, for $z = -0.1$, $-1.0$, $-10.0$, $-100.0$ and for polynomial degrees $N = 7$ and $8$. The graph legends are distributed among columns and located at the top of the columns. From left to right, the graphs represent increasingly narrower neighborhoods of the stability boundary $\mathrm{CFL}_{\rm max}$, from $\mathrm{CFL} \in [0.90\, \mathrm{CFL}_{\rm max},\, 1.10\, \mathrm{CFL}_{\rm max}]$ in the left column to $\mathrm{CFL} \in [0.9995\, \mathrm{CFL}_{\rm max},\, 1.0005\, \mathrm{CFL}_{\rm max}]$ in the right column; the range of phase $\phi$ variation also narrows when moving from the left to the right columns of the graphs to ensure a correct and clear presentation of the root $\phi = \pi$ of the equation $F(\pi; \mathrm{CFL}, z) = 0$. The horizontal dotted line indicates $F = 0$, the vertical dotted line indicates $\phi = \pi$ ($\lambda = \exp(\pm i\phi) = -1$). \textit{Note}: the phase $\phi$ of eigenvalue $\lambda_{k} = |\lambda_{k}|\exp(i\phi)$ is not the phase $\theta$.
}
\label{fig:dep_f_on_phi_diff_cfls_degrees_7_8}
\end{figure}
\begin{figure}[h!]
\centering
\includegraphics[width=0.240\textwidth]{dep_f_on_phi_diff_cfls_legend_Cidx_0.png}
\includegraphics[width=0.240\textwidth]{dep_f_on_phi_diff_cfls_legend_Cidx_1.png}
\includegraphics[width=0.240\textwidth]{dep_f_on_phi_diff_cfls_legend_Cidx_2.png}
\includegraphics[width=0.240\textwidth]{dep_f_on_phi_diff_cfls_legend_Cidx_3.png}\\[1mm]
\includegraphics[width=0.240\textwidth]{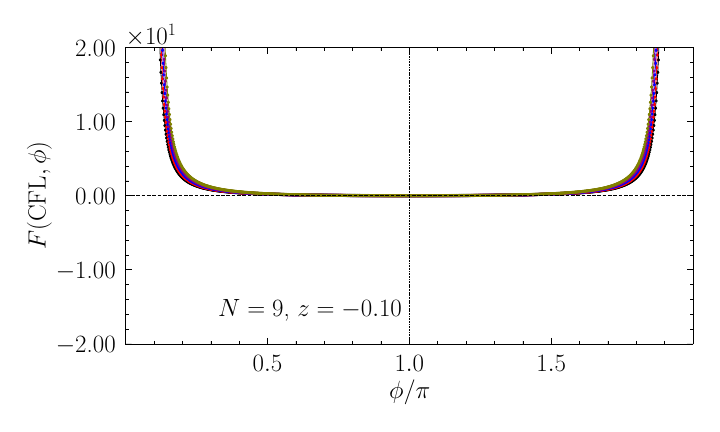}
\includegraphics[width=0.240\textwidth]{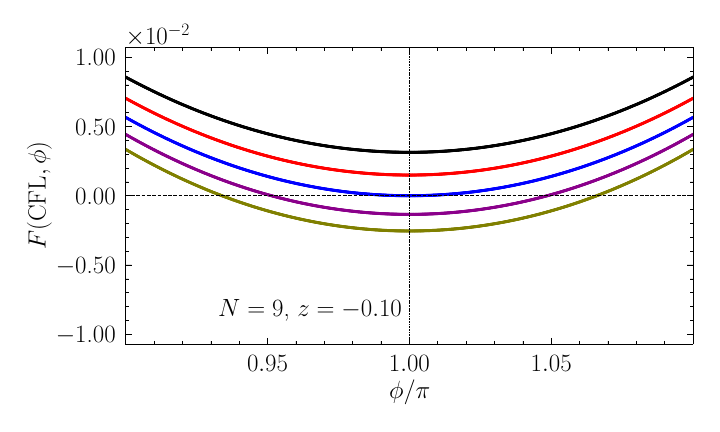}
\includegraphics[width=0.240\textwidth]{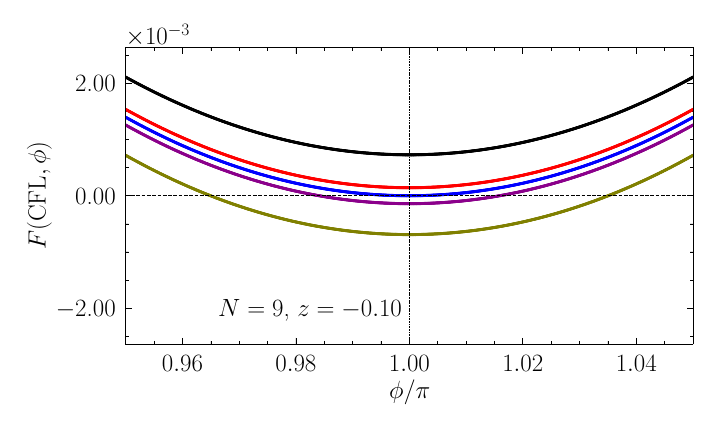}
\includegraphics[width=0.240\textwidth]{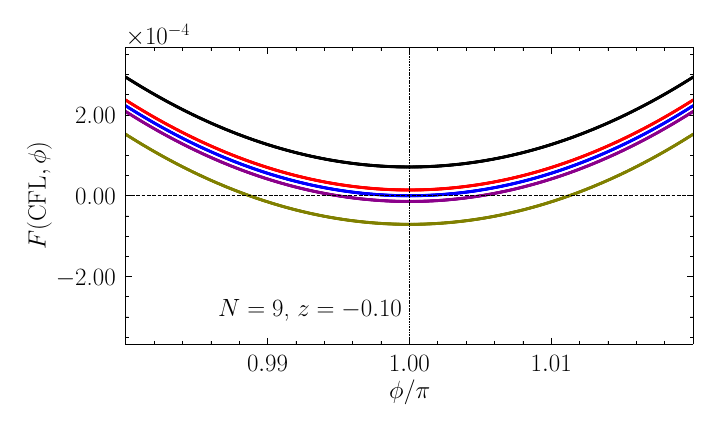}\\[1mm]
\includegraphics[width=0.240\textwidth]{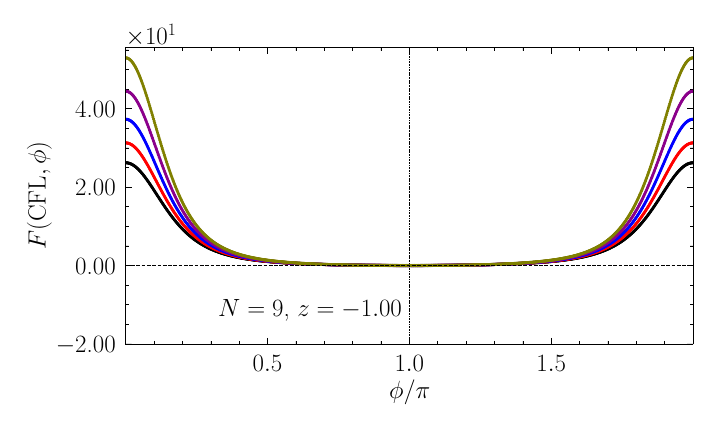}
\includegraphics[width=0.240\textwidth]{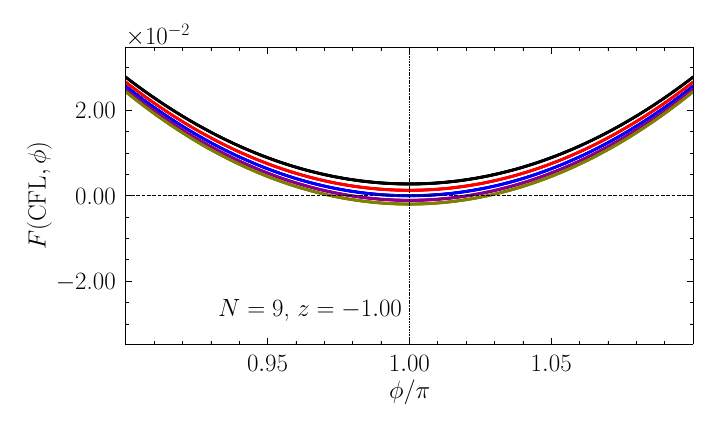}
\includegraphics[width=0.240\textwidth]{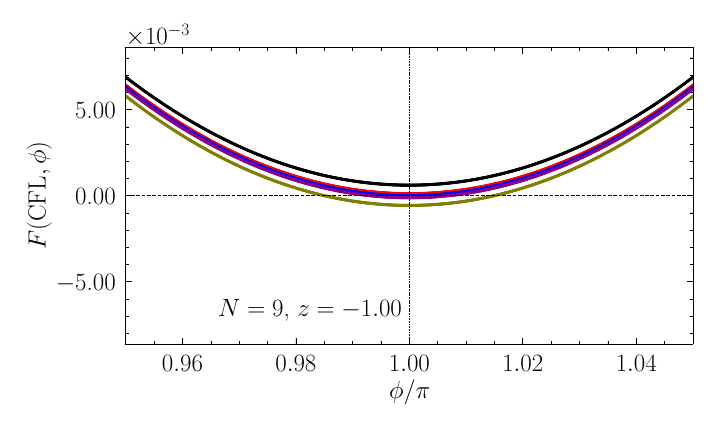}
\includegraphics[width=0.240\textwidth]{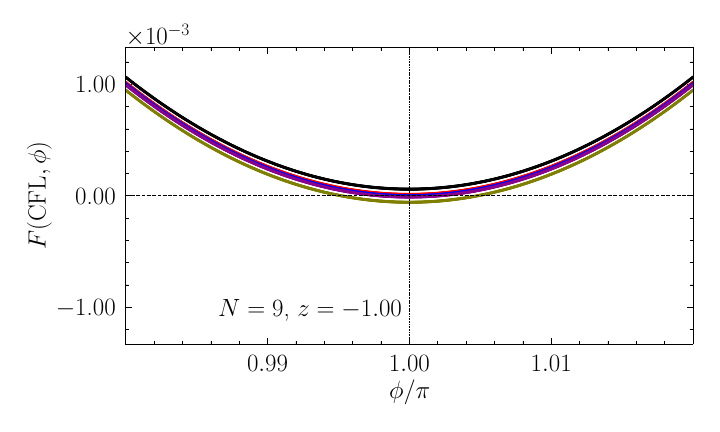}\\[1mm]
\includegraphics[width=0.240\textwidth]{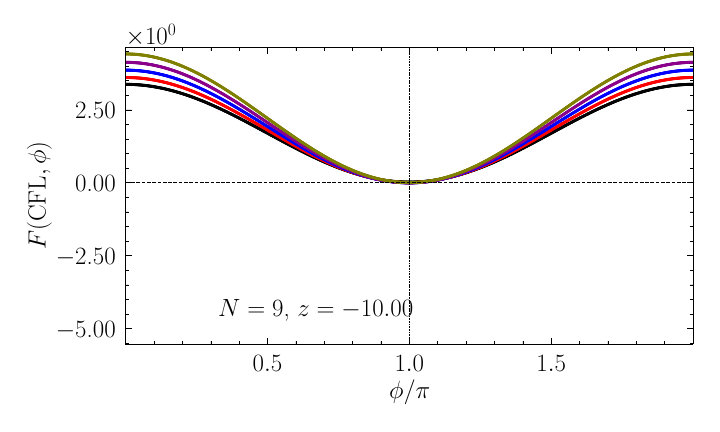}
\includegraphics[width=0.240\textwidth]{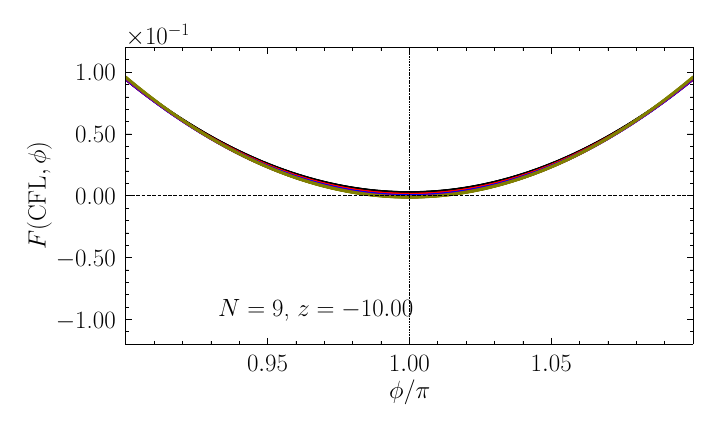}
\includegraphics[width=0.240\textwidth]{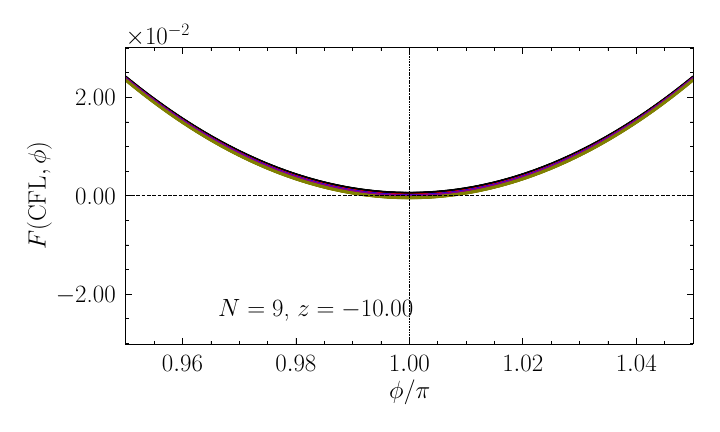}
\includegraphics[width=0.240\textwidth]{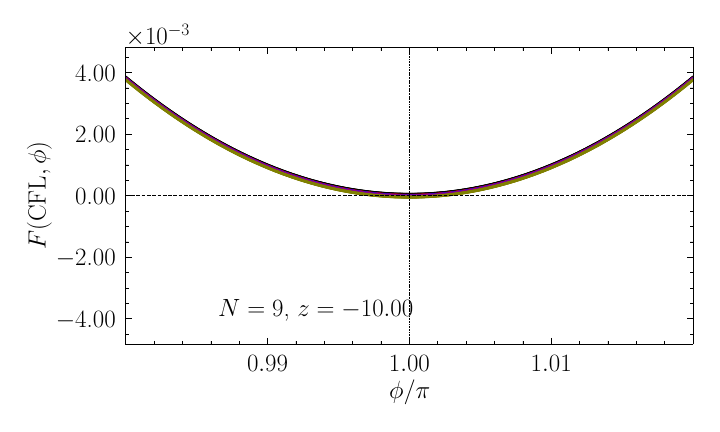}\\[1mm]
\includegraphics[width=0.240\textwidth]{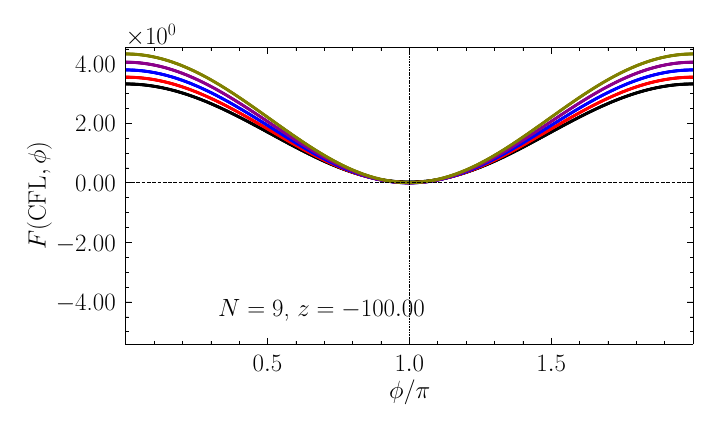}
\includegraphics[width=0.240\textwidth]{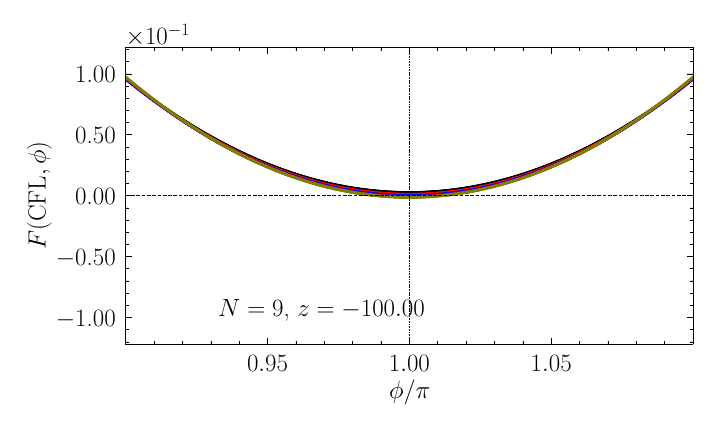}
\includegraphics[width=0.240\textwidth]{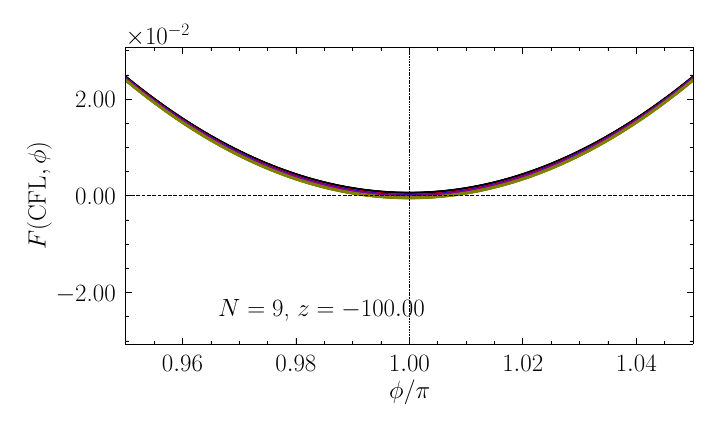}
\includegraphics[width=0.240\textwidth]{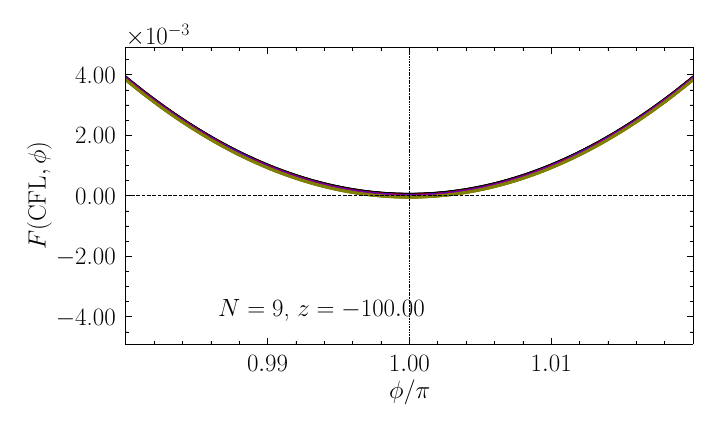}\\[1mm]
\includegraphics[width=0.240\textwidth]{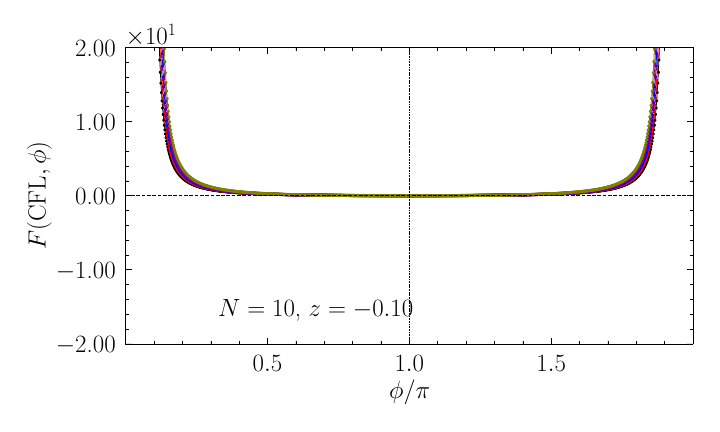}
\includegraphics[width=0.240\textwidth]{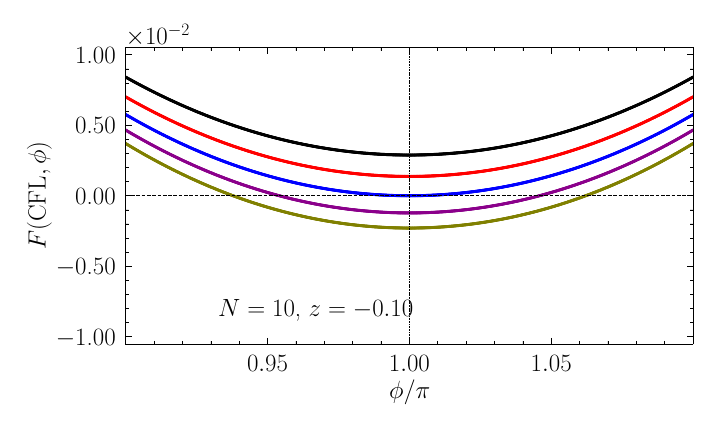}
\includegraphics[width=0.240\textwidth]{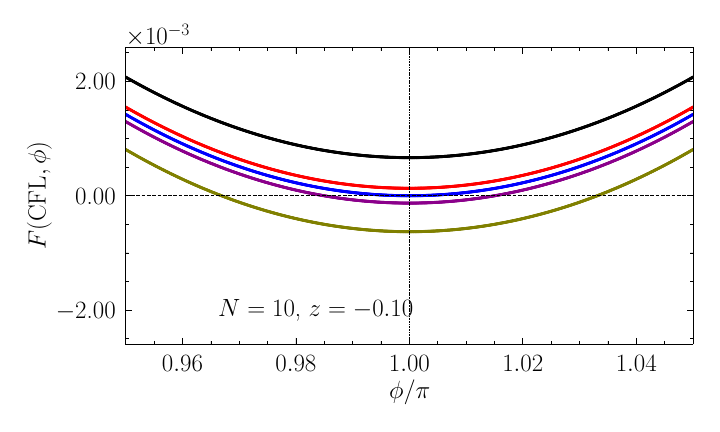}
\includegraphics[width=0.240\textwidth]{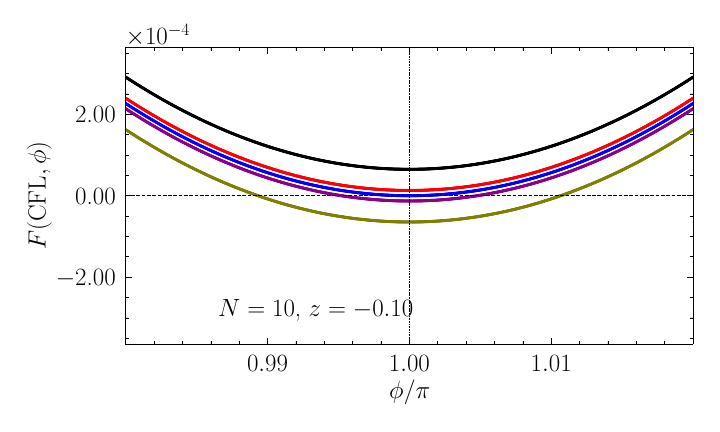}\\[1mm]
\includegraphics[width=0.240\textwidth]{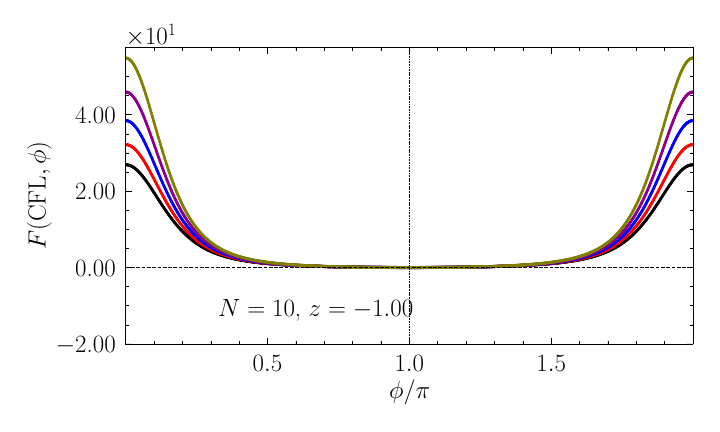}
\includegraphics[width=0.240\textwidth]{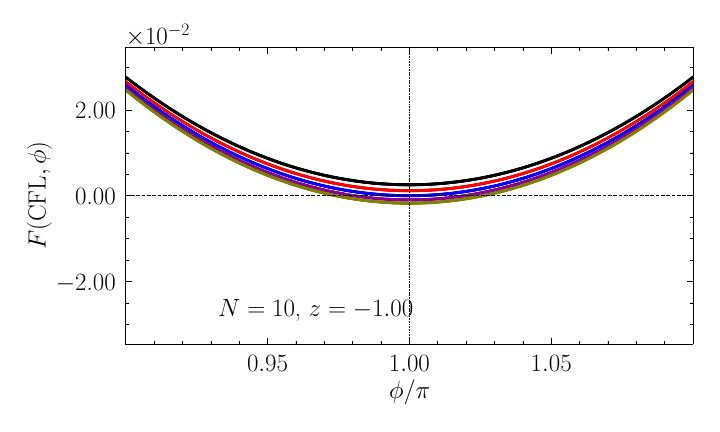}
\includegraphics[width=0.240\textwidth]{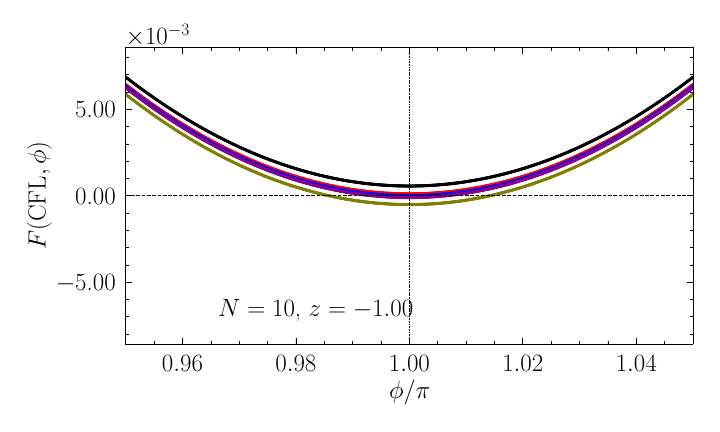}
\includegraphics[width=0.240\textwidth]{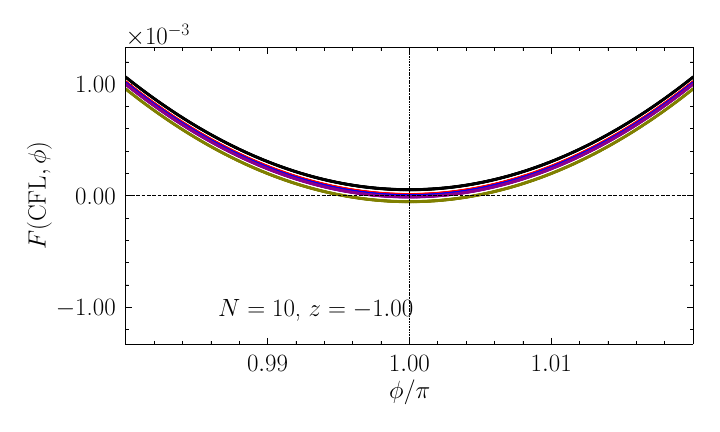}\\[1mm]
\includegraphics[width=0.240\textwidth]{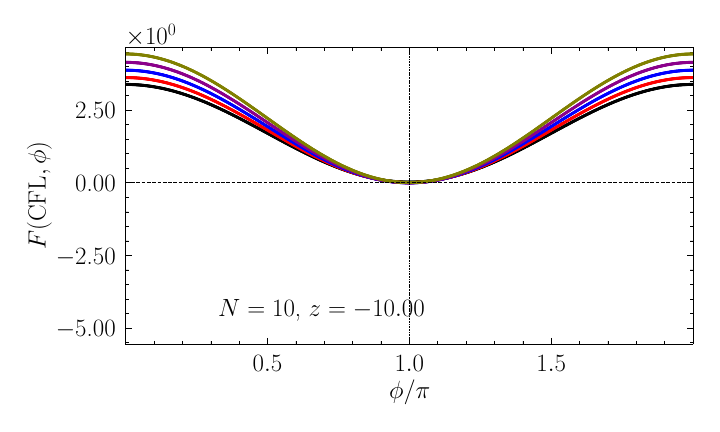}
\includegraphics[width=0.240\textwidth]{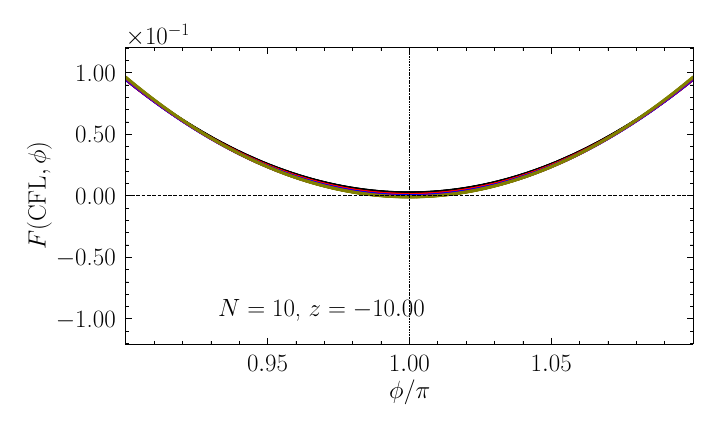}
\includegraphics[width=0.240\textwidth]{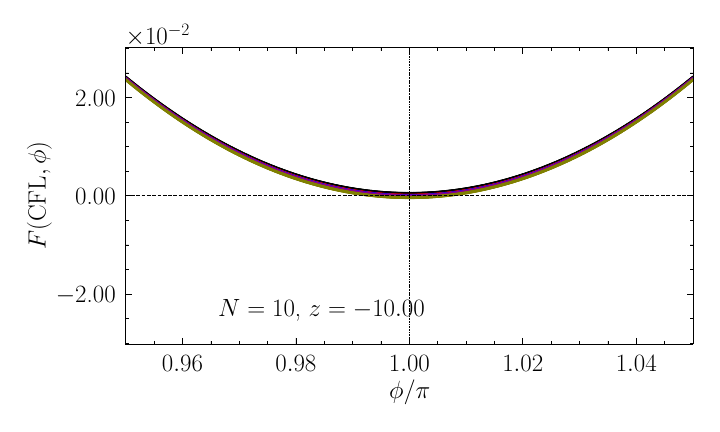}
\includegraphics[width=0.240\textwidth]{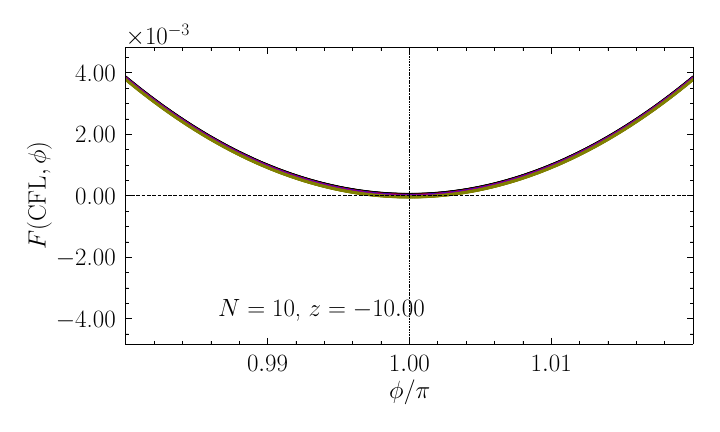}\\[1mm]
\includegraphics[width=0.240\textwidth]{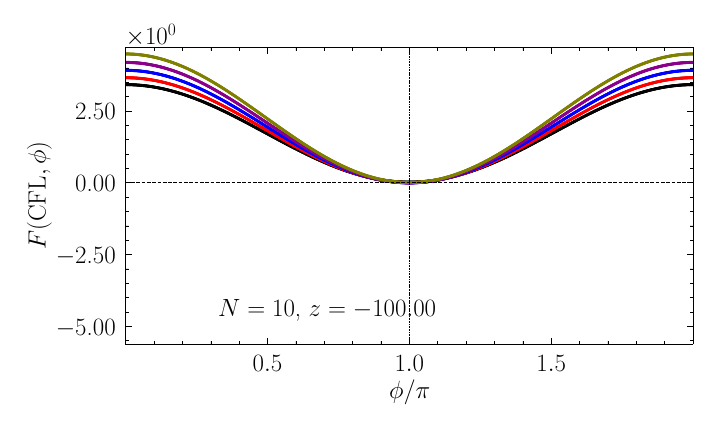}
\includegraphics[width=0.240\textwidth]{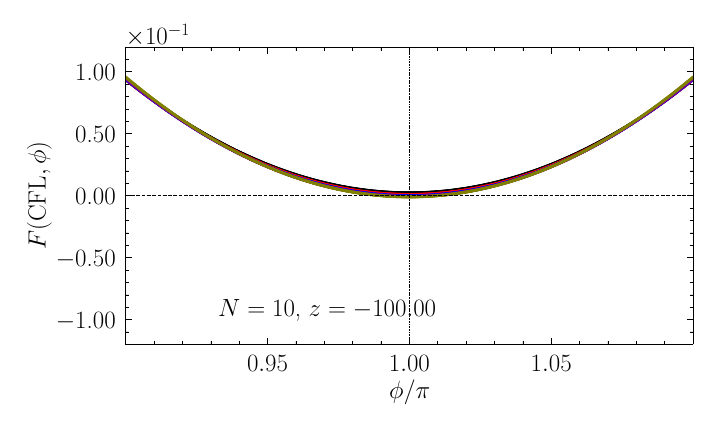}
\includegraphics[width=0.240\textwidth]{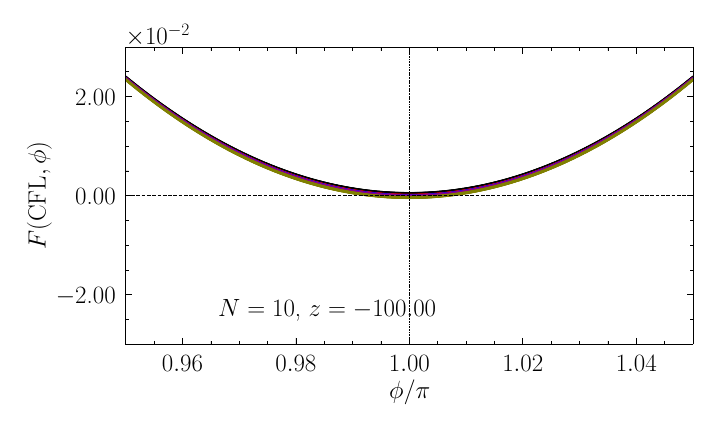}
\includegraphics[width=0.240\textwidth]{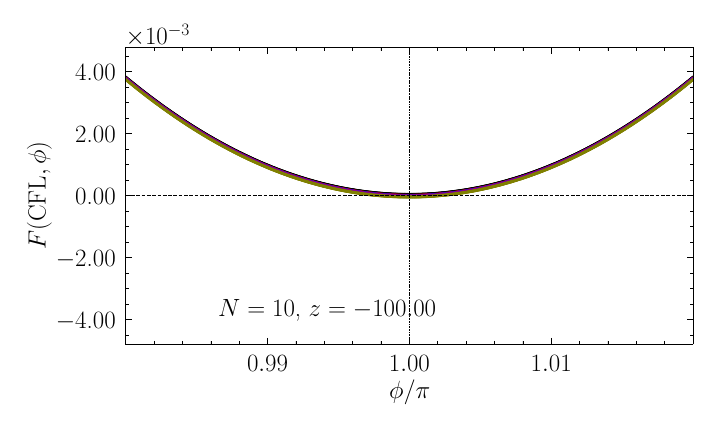}
\caption{%
The dependence of function $F(\phi; \mathrm{CFL}, z)$ (\ref{eq:f_def}) on argument $\phi\in[0, 2\pi]$ for several values of the Courant number $\mathrm{CFL}$, selected in the vicinity of the stability boundary $\mathrm{CFL}_{\rm max}(z)$, defined further in Figure~\ref{fig:cfls_on_z}, for $z = -0.1$, $-1.0$, $-10.0$, $-100.0$ and for polynomial degrees $N = 9$ and $10$. The graph legends are distributed among columns and located at the top of the columns. From left to right, the graphs represent increasingly narrower neighborhoods of the stability boundary $\mathrm{CFL}_{\rm max}$, from $\mathrm{CFL} \in [0.90\, \mathrm{CFL}_{\rm max},\, 1.10\, \mathrm{CFL}_{\rm max}]$ in the left column to $\mathrm{CFL} \in [0.9995\, \mathrm{CFL}_{\rm max},\, 1.0005\, \mathrm{CFL}_{\rm max}]$ in the right column; the range of phase $\phi$ variation also narrows when moving from the left to the right columns of the graphs to ensure a correct and clear presentation of the root $\phi = \pi$ of the equation $F(\pi; \mathrm{CFL}, z) = 0$. The horizontal dotted line indicates $F = 0$, the vertical dotted line indicates $\phi = \pi$ ($\lambda = \exp(\pm i\phi) = -1$). \textit{Note}: the phase $\phi$ of eigenvalue $\lambda_{k} = |\lambda_{k}|\exp(i\phi)$ is not the phase $\theta$.
}
\label{fig:dep_f_on_phi_diff_cfls_degrees_9_10}
\end{figure}
\begin{figure}[h!]
\centering
\includegraphics[width=0.240\textwidth]{dep_f_on_phi_diff_cfls_legend_Cidx_0.png}
\includegraphics[width=0.240\textwidth]{dep_f_on_phi_diff_cfls_legend_Cidx_1.png}
\includegraphics[width=0.240\textwidth]{dep_f_on_phi_diff_cfls_legend_Cidx_2.png}
\includegraphics[width=0.240\textwidth]{dep_f_on_phi_diff_cfls_legend_Cidx_3.png}\\[1mm]
\includegraphics[width=0.240\textwidth]{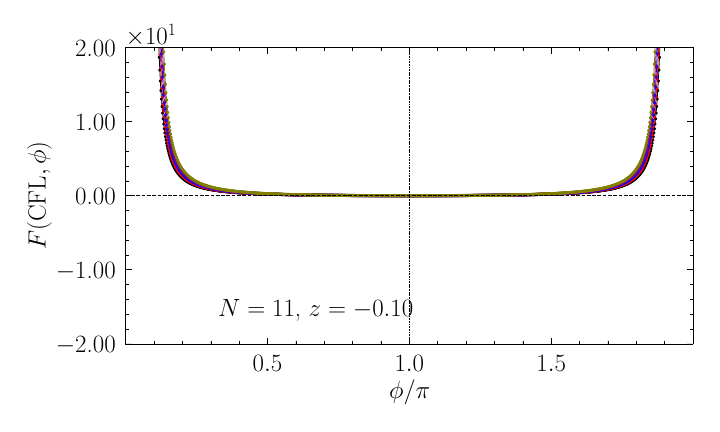}
\includegraphics[width=0.240\textwidth]{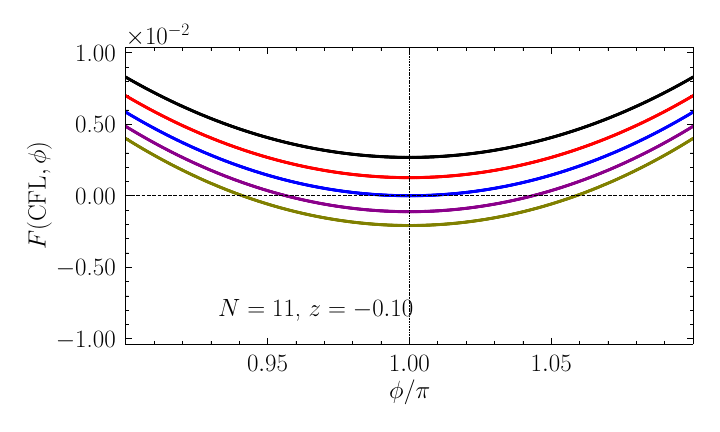}
\includegraphics[width=0.240\textwidth]{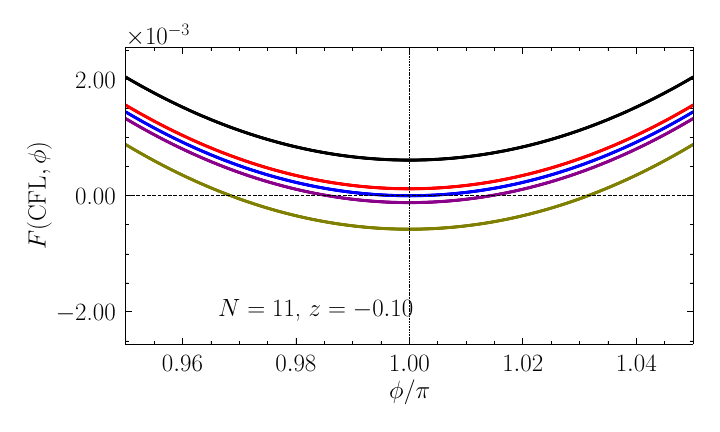}
\includegraphics[width=0.240\textwidth]{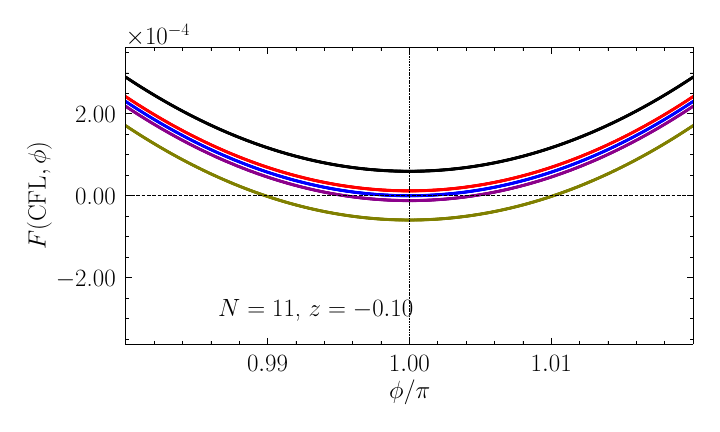}\\[1mm]
\includegraphics[width=0.240\textwidth]{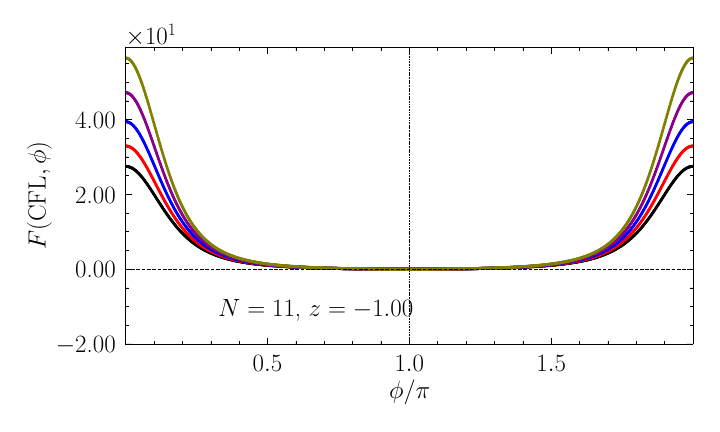}
\includegraphics[width=0.240\textwidth]{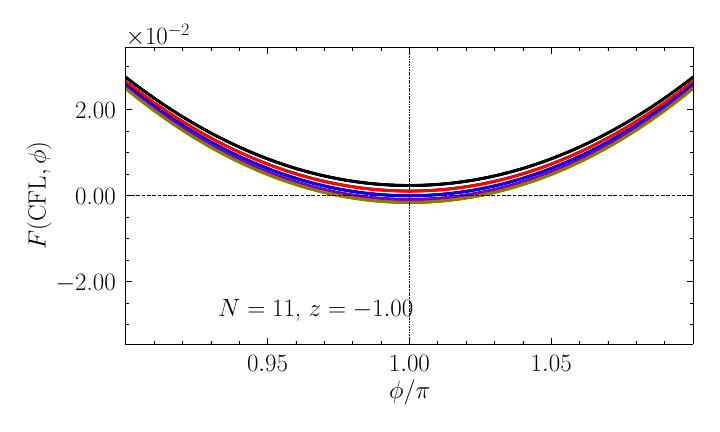}
\includegraphics[width=0.240\textwidth]{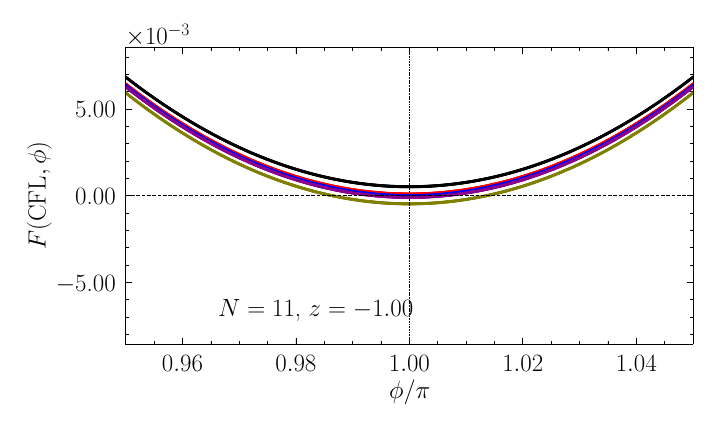}
\includegraphics[width=0.240\textwidth]{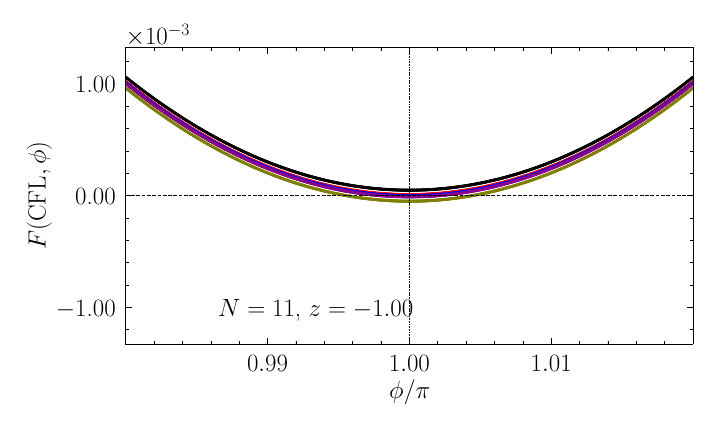}\\[1mm]
\includegraphics[width=0.240\textwidth]{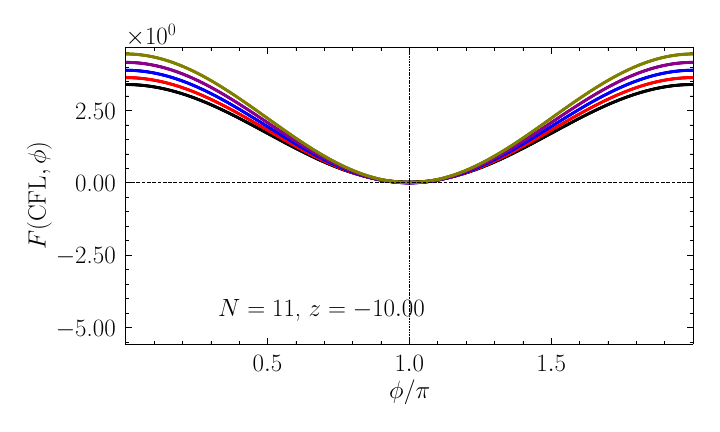}
\includegraphics[width=0.240\textwidth]{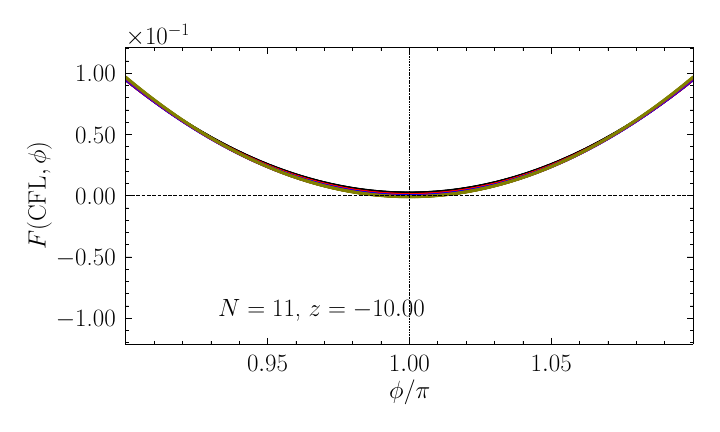}
\includegraphics[width=0.240\textwidth]{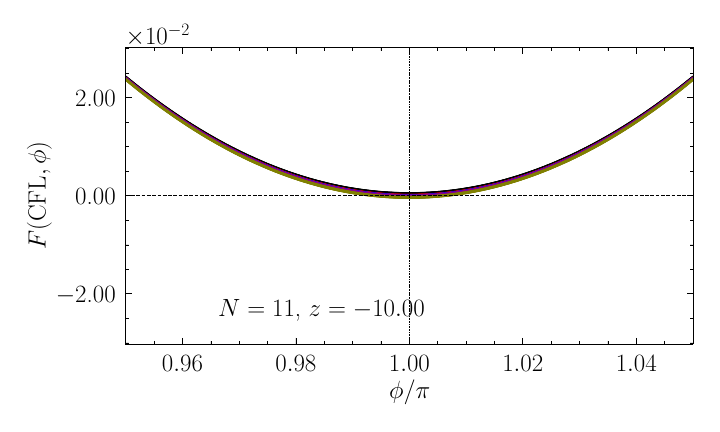}
\includegraphics[width=0.240\textwidth]{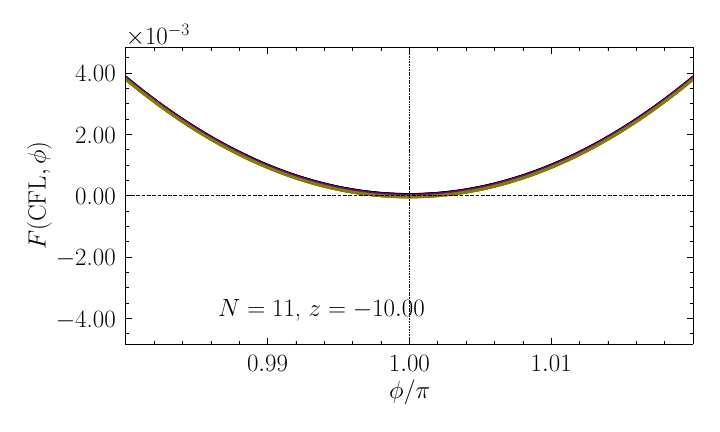}\\[1mm]
\includegraphics[width=0.240\textwidth]{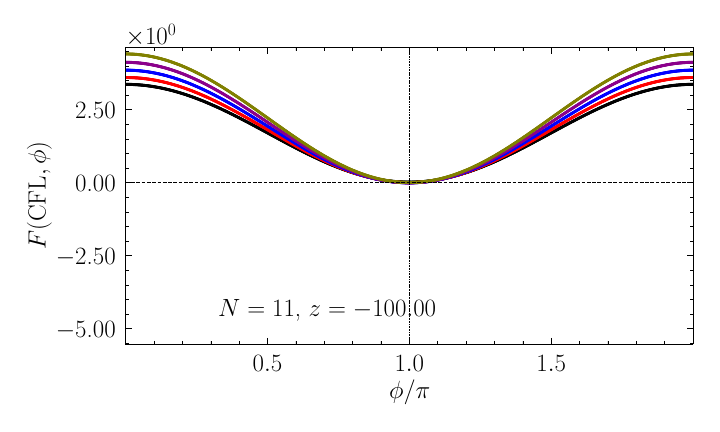}
\includegraphics[width=0.240\textwidth]{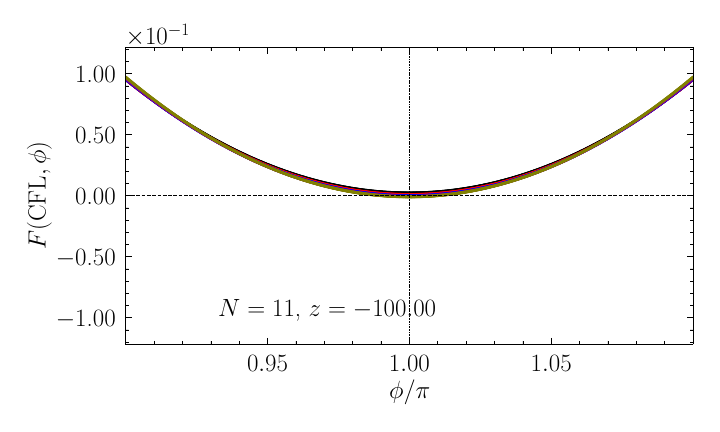}
\includegraphics[width=0.240\textwidth]{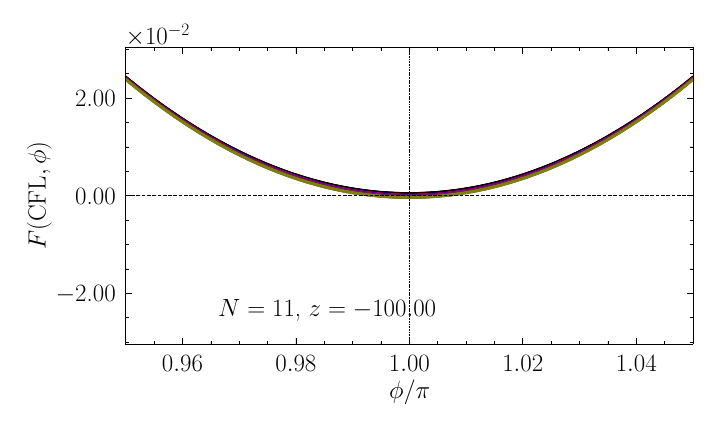}
\includegraphics[width=0.240\textwidth]{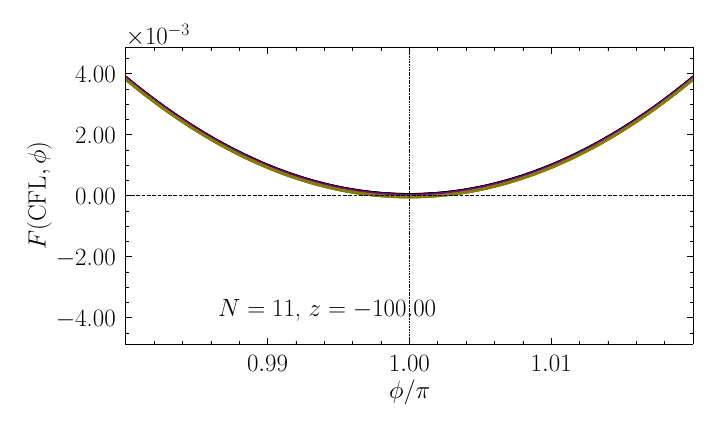}\\[1mm]
\includegraphics[width=0.240\textwidth]{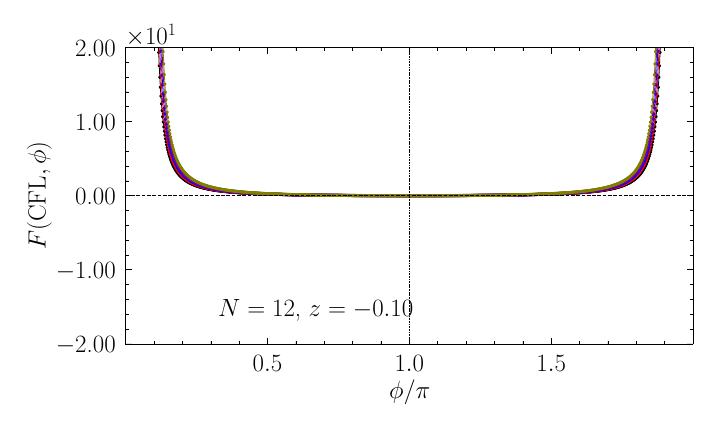}
\includegraphics[width=0.240\textwidth]{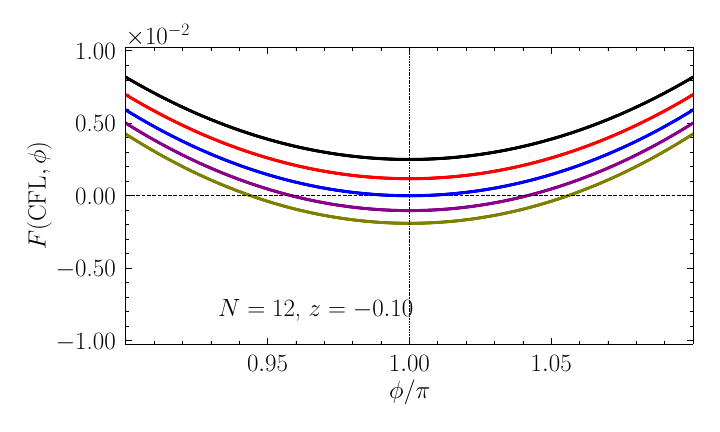}
\includegraphics[width=0.240\textwidth]{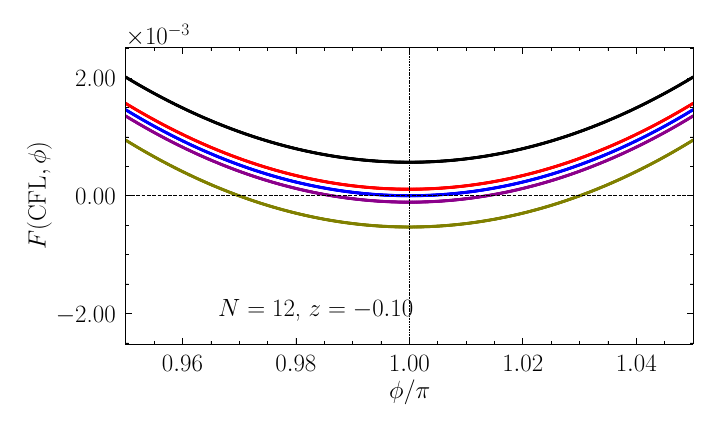}
\includegraphics[width=0.240\textwidth]{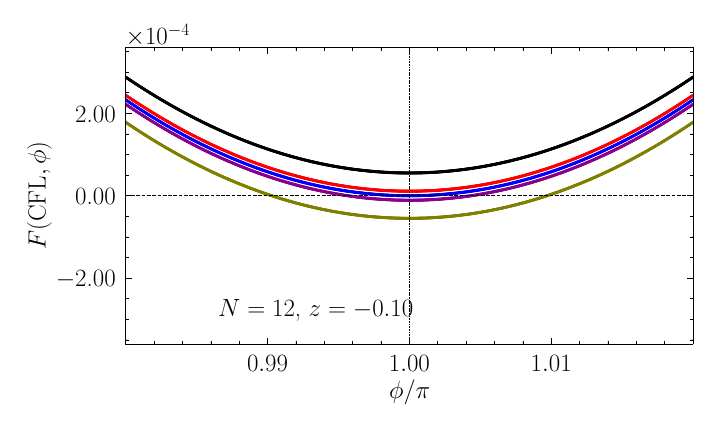}\\[1mm]
\includegraphics[width=0.240\textwidth]{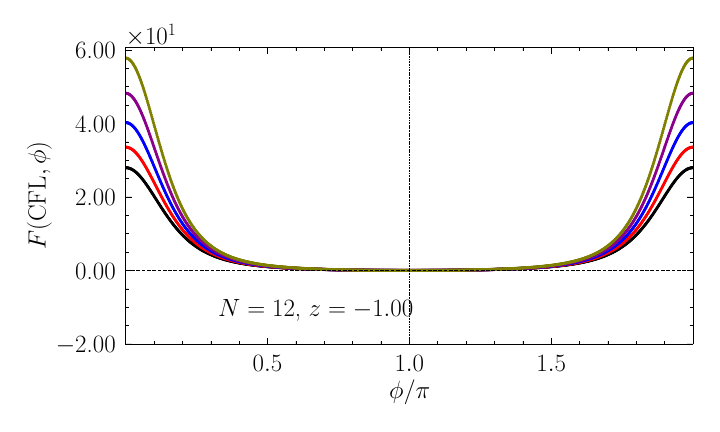}
\includegraphics[width=0.240\textwidth]{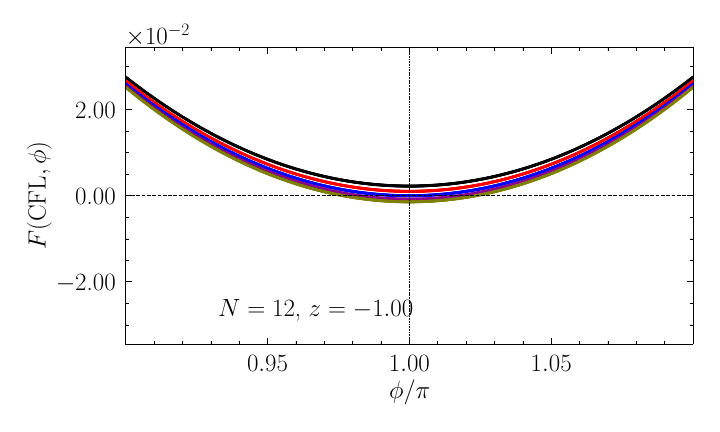}
\includegraphics[width=0.240\textwidth]{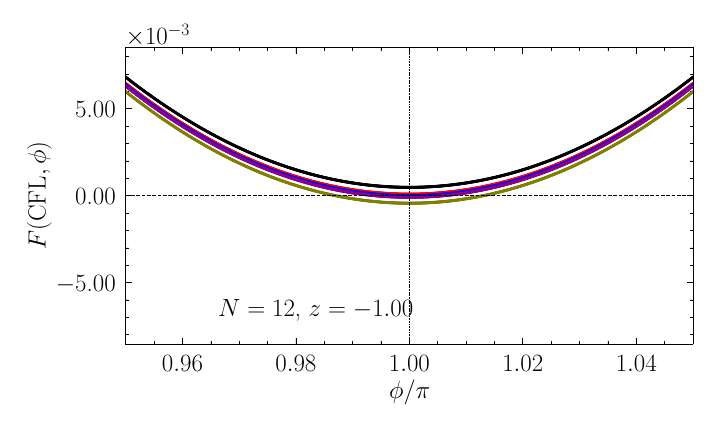}
\includegraphics[width=0.240\textwidth]{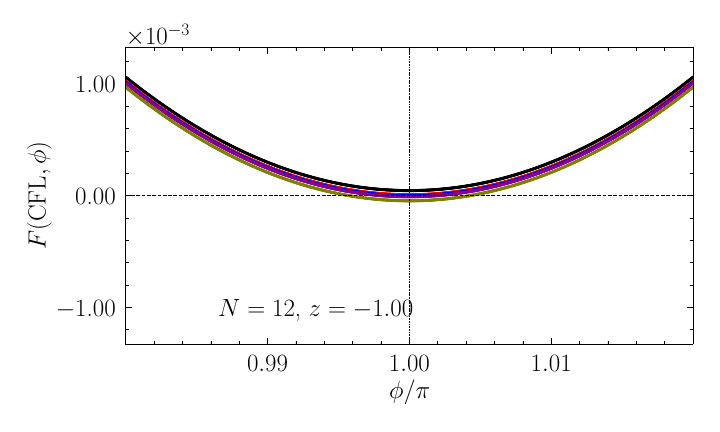}\\[1mm]
\includegraphics[width=0.240\textwidth]{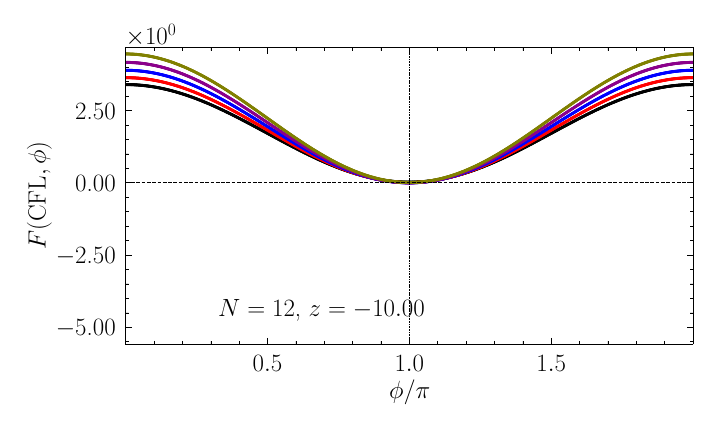}
\includegraphics[width=0.240\textwidth]{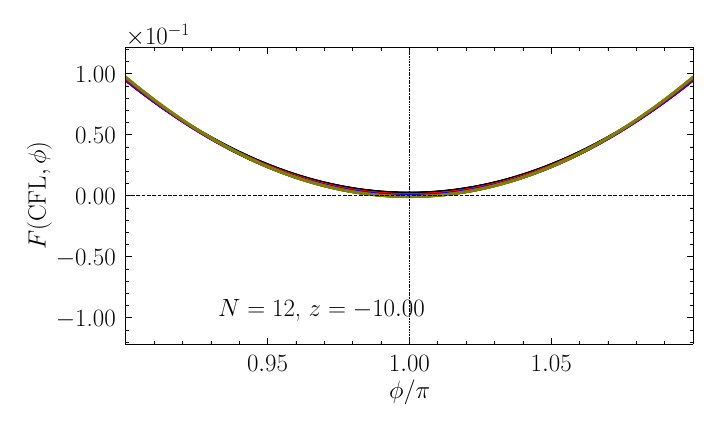}
\includegraphics[width=0.240\textwidth]{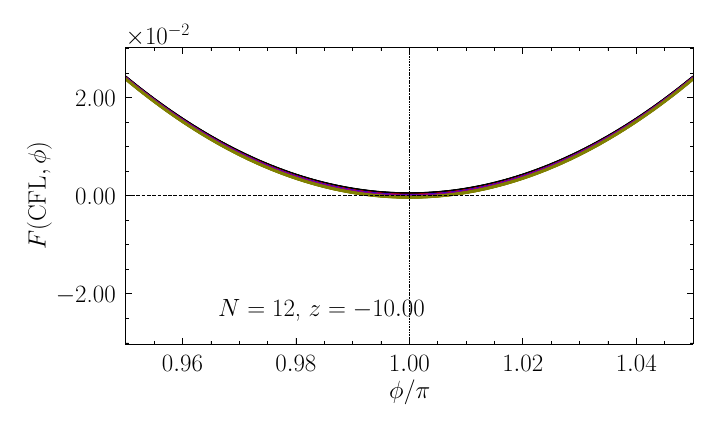}
\includegraphics[width=0.240\textwidth]{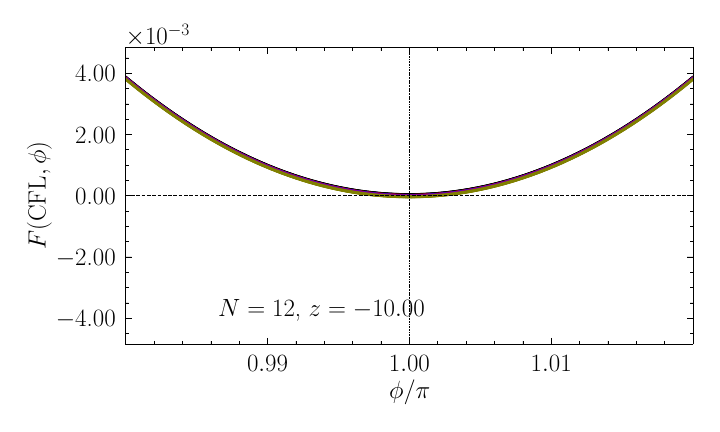}\\[1mm]
\includegraphics[width=0.240\textwidth]{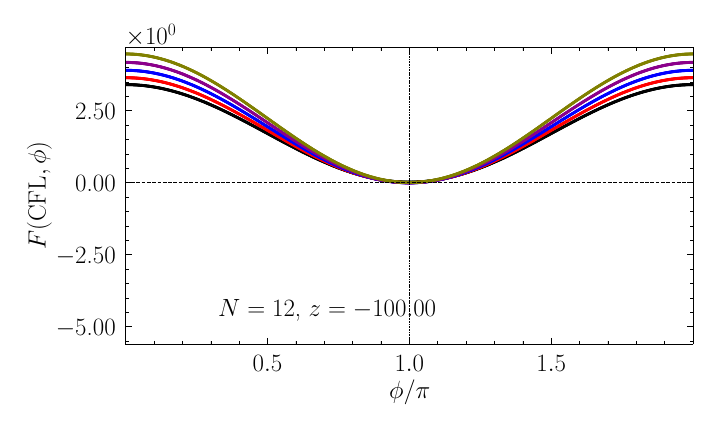}
\includegraphics[width=0.240\textwidth]{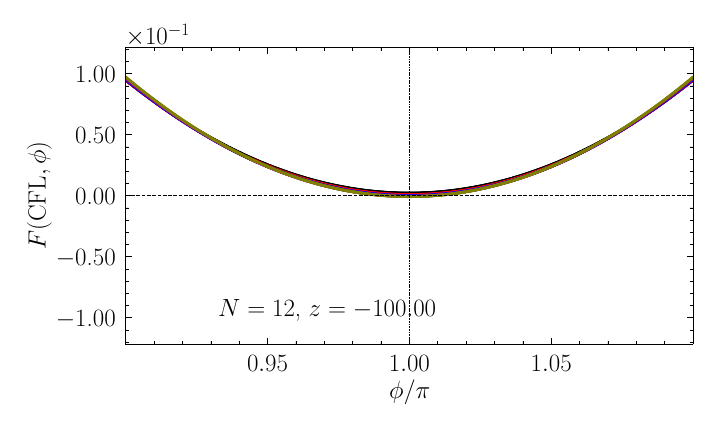}
\includegraphics[width=0.240\textwidth]{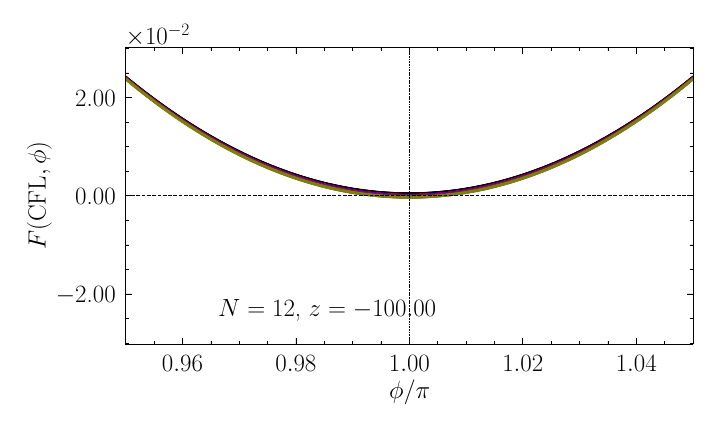}
\includegraphics[width=0.240\textwidth]{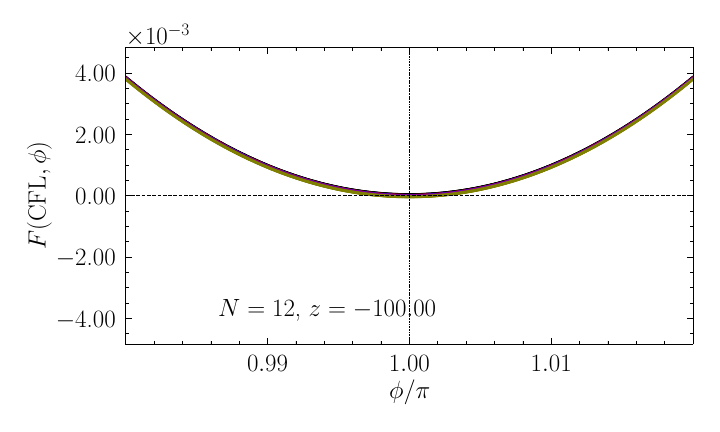}
\caption{%
The dependence of function $F(\phi; \mathrm{CFL}, z)$ (\ref{eq:f_def}) on argument $\phi\in[0, 2\pi]$ for several values of the Courant number $\mathrm{CFL}$, selected in the vicinity of the stability boundary $\mathrm{CFL}_{\rm max}(z)$, defined further in Figure~\ref{fig:cfls_on_z}, for $z = -0.1$, $-1.0$, $-10.0$, $-100.0$ and for polynomial degrees $N = 11$ and $12$. The graph legends are distributed among columns and located at the top of the columns. From left to right, the graphs represent increasingly narrower neighborhoods of the stability boundary $\mathrm{CFL}_{\rm max}$, from $\mathrm{CFL} \in [0.90\, \mathrm{CFL}_{\rm max},\, 1.10\, \mathrm{CFL}_{\rm max}]$ in the left column to $\mathrm{CFL} \in [0.9995\, \mathrm{CFL}_{\rm max},\, 1.0005\, \mathrm{CFL}_{\rm max}]$ in the right column; the range of phase $\phi$ variation also narrows when moving from the left to the right columns of the graphs to ensure a correct and clear presentation of the root $\phi = \pi$ of the equation $F(\pi; \mathrm{CFL}, z) = 0$. The horizontal dotted line indicates $F = 0$, the vertical dotted line indicates $\phi = \pi$ ($\lambda = \exp(\pm i\phi) = -1$). \textit{Note}: the phase $\phi$ of eigenvalue $\lambda_{k} = |\lambda_{k}|\exp(i\phi)$ is not the phase $\theta$.
}
\label{fig:dep_f_on_phi_diff_cfls_degrees_11_12}
\end{figure}

\begin{figure}[h!]
\centering
\captionsetup[subfigure]{%
	position=bottom,
	font+=smaller,
	textfont=normalfont,
	singlelinecheck=off,
	justification=raggedright
}
\begin{subfigure}{0.32\textwidth}
\includegraphics[width=\textwidth]{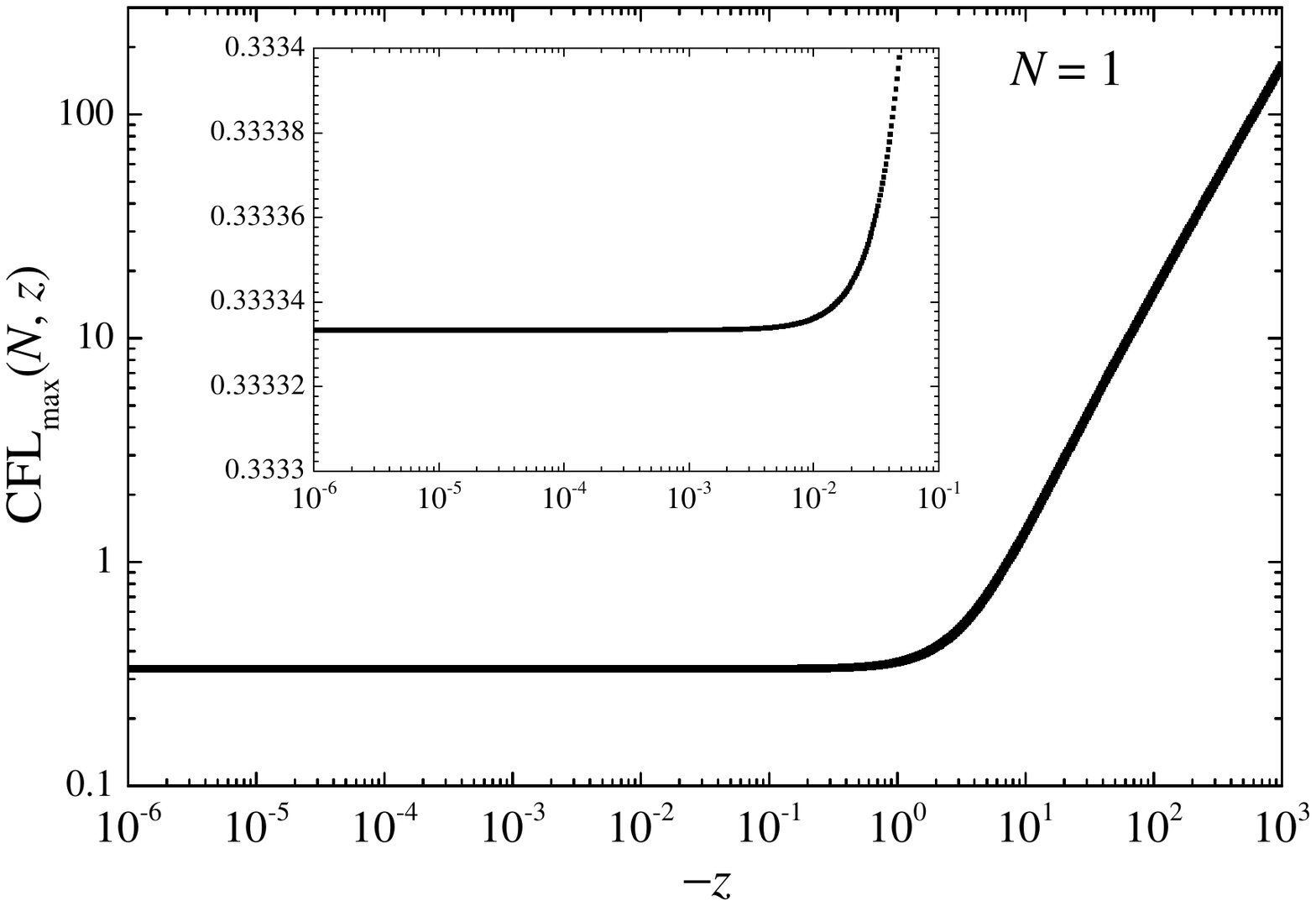}
\vspace{-8mm}\caption{\label{fig:cfls_on_z_degree_1}}
\end{subfigure}
\begin{subfigure}{0.32\textwidth}
\includegraphics[width=\textwidth]{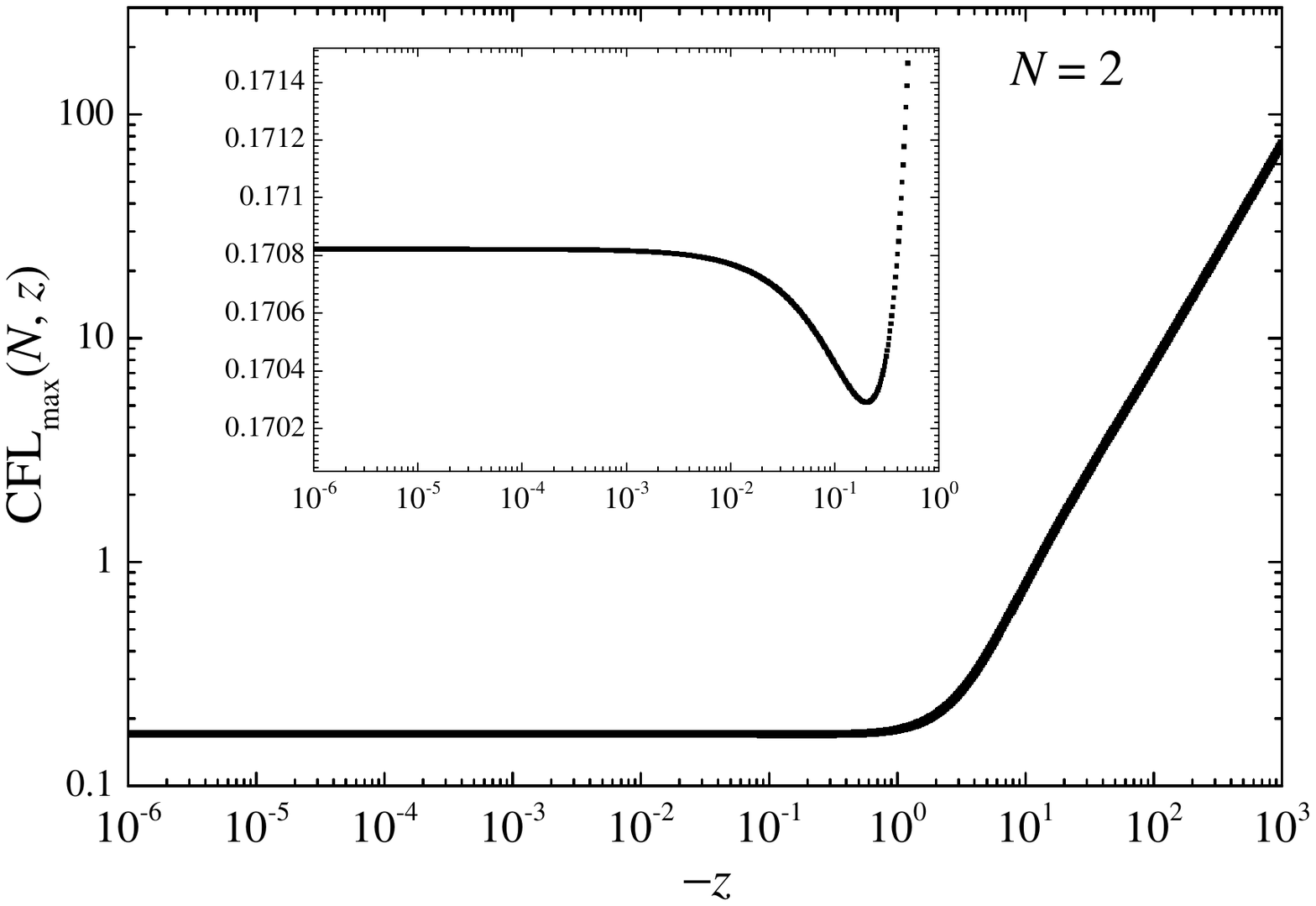}
\vspace{-8mm}\caption{\label{fig:cfls_on_z_degree_2}}
\end{subfigure}
\begin{subfigure}{0.32\textwidth}
\includegraphics[width=\textwidth]{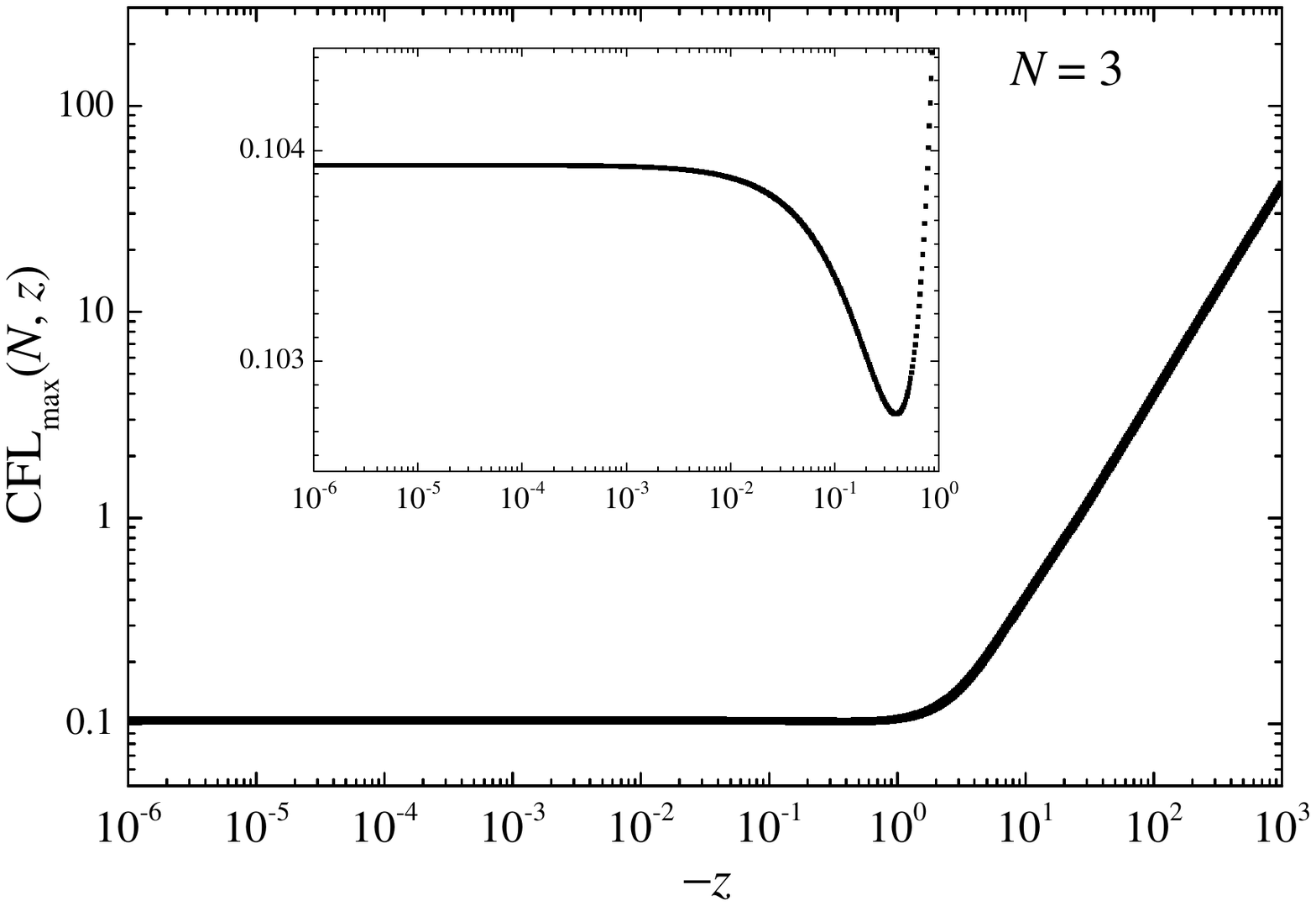}
\vspace{-8mm}\caption{\label{fig:cfls_on_z_degree_3}}
\end{subfigure}\\
\begin{subfigure}{0.32\textwidth}
\includegraphics[width=\textwidth]{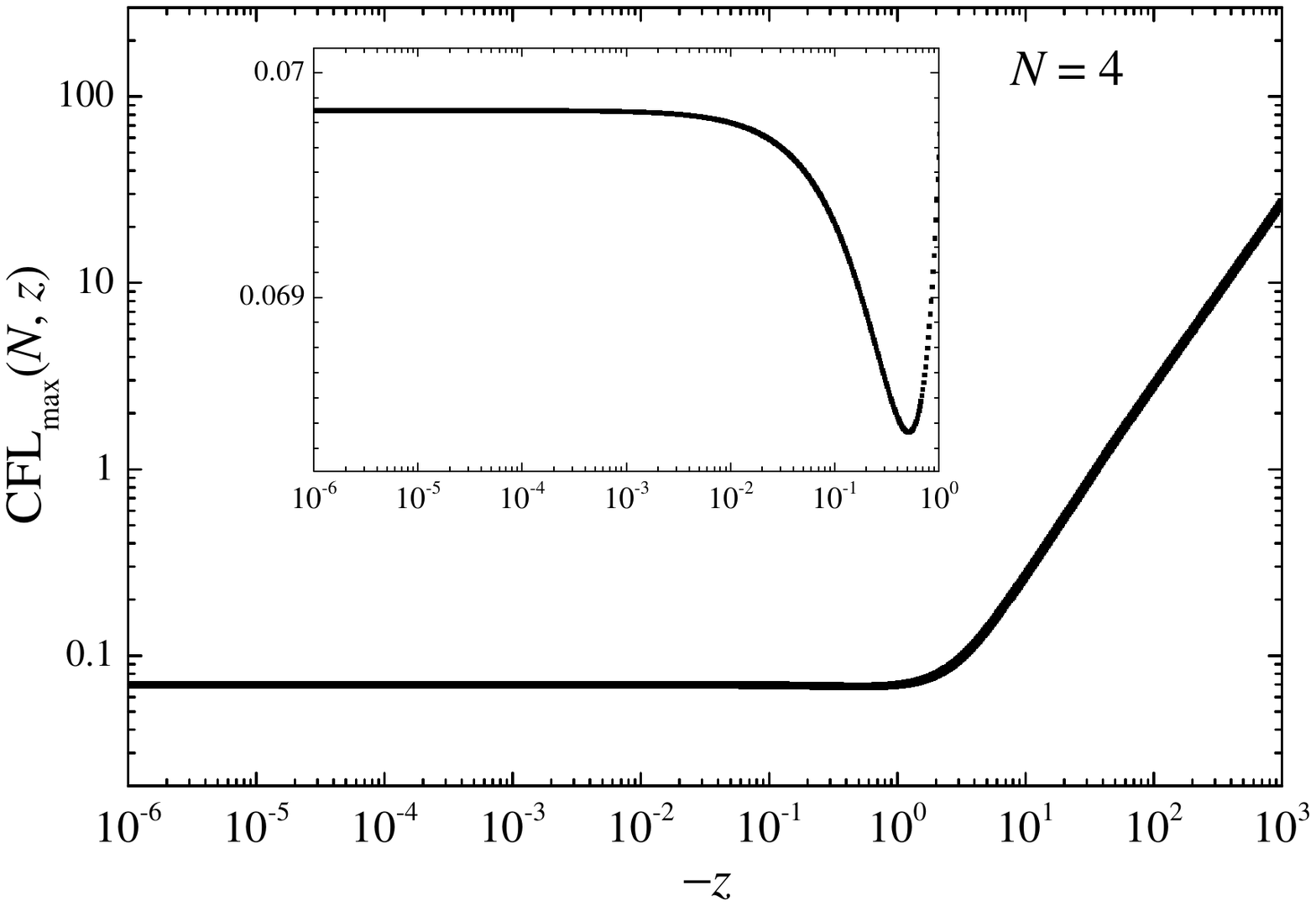}
\vspace{-8mm}\caption{\label{fig:cfls_on_z_degree_4}}
\end{subfigure}
\begin{subfigure}{0.32\textwidth}
\includegraphics[width=\textwidth]{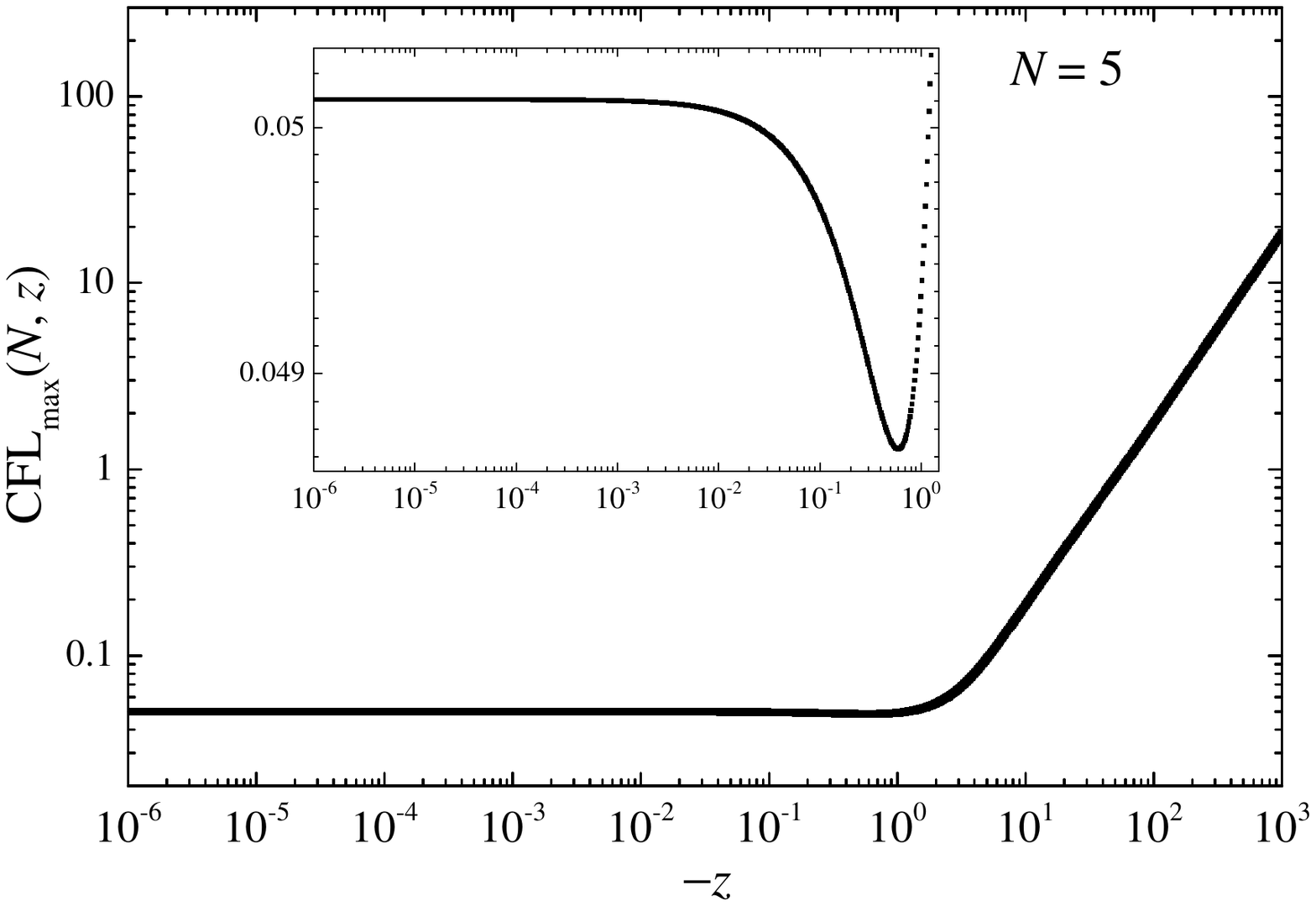}
\vspace{-8mm}\caption{\label{fig:cfls_on_z_degree_5}}
\end{subfigure}
\begin{subfigure}{0.32\textwidth}
\includegraphics[width=\textwidth]{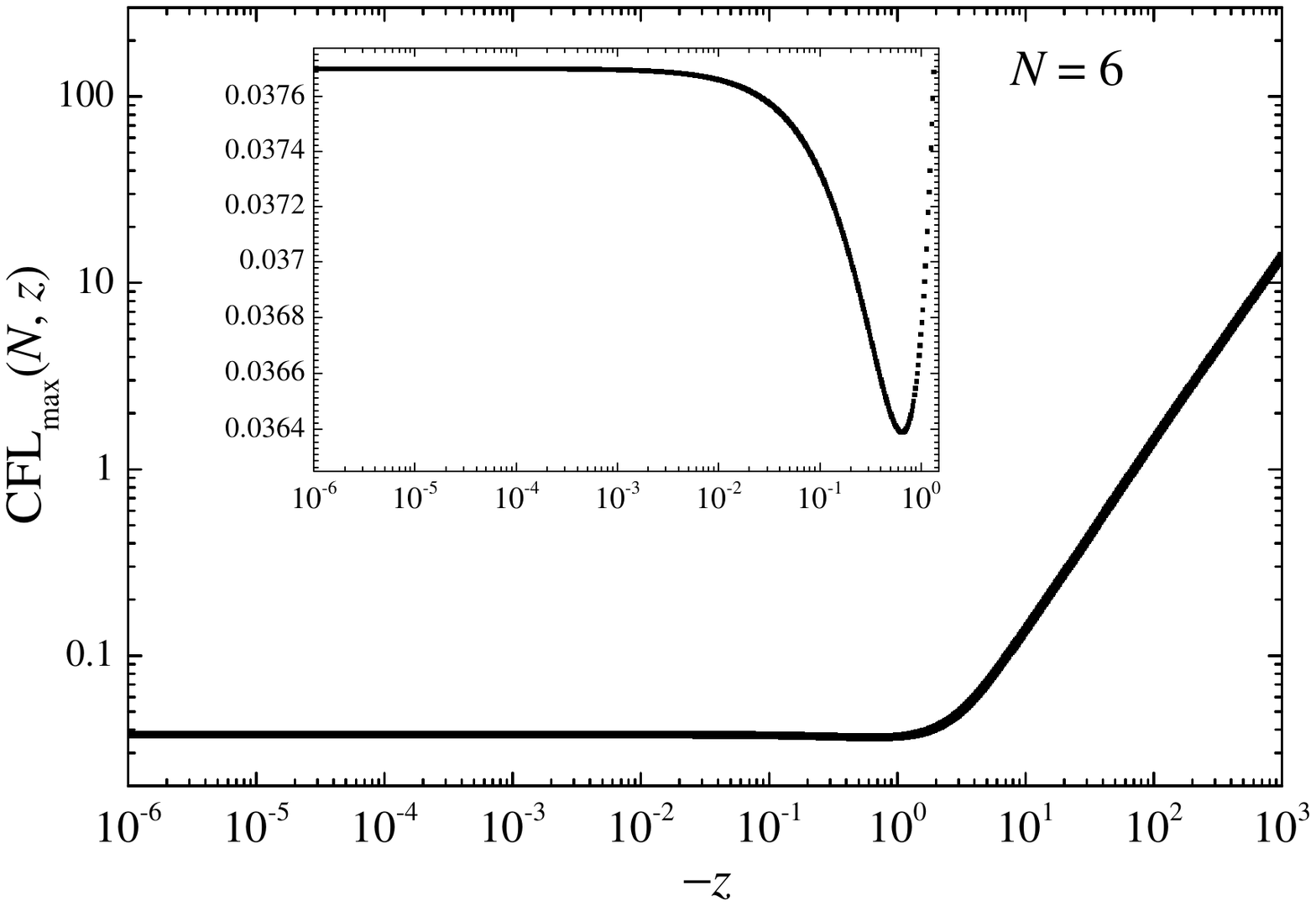}
\vspace{-8mm}\caption{\label{fig:cfls_on_z_degree_6}}
\end{subfigure}\\
\begin{subfigure}{0.32\textwidth}
\includegraphics[width=\textwidth]{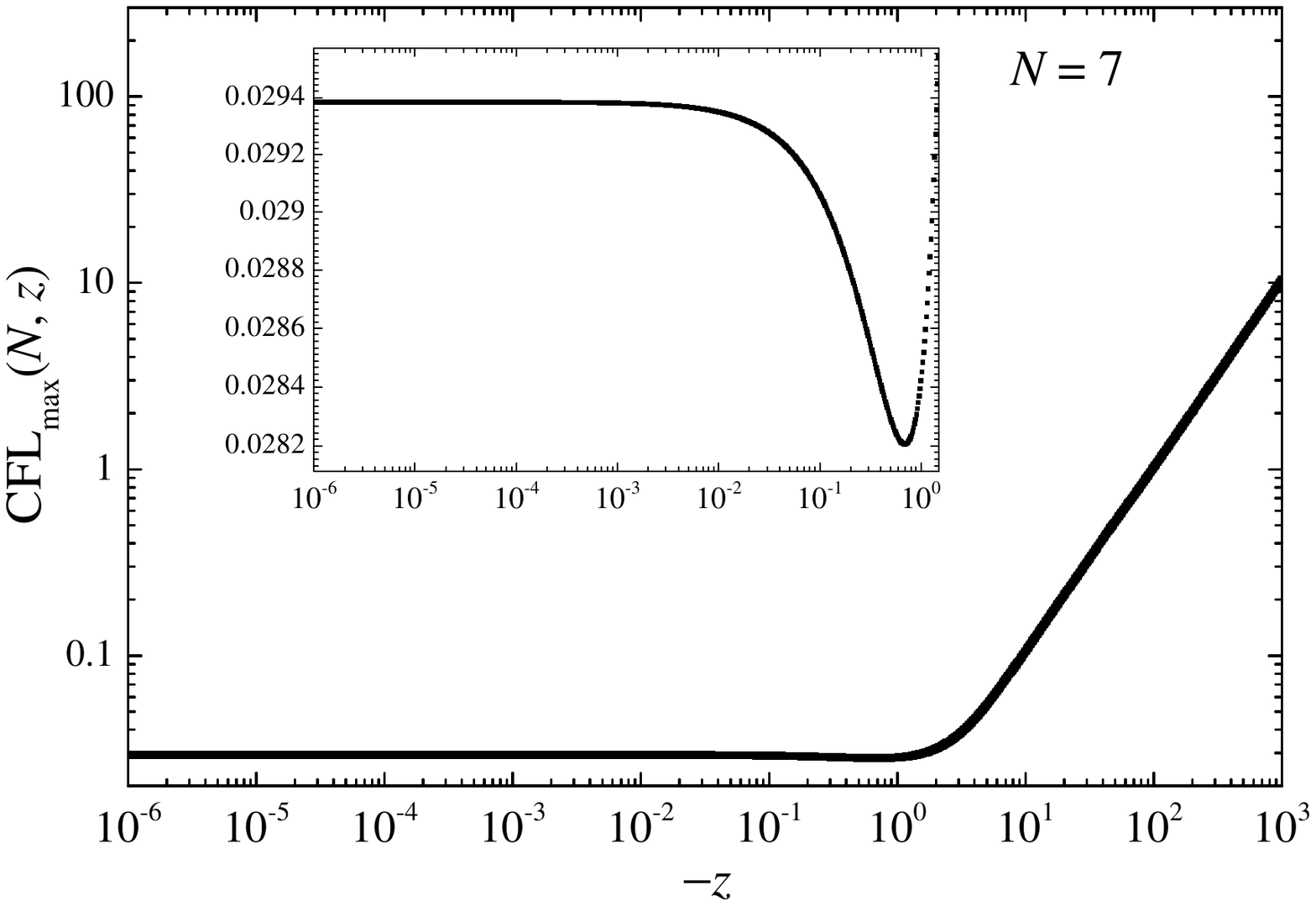}
\vspace{-8mm}\caption{\label{fig:cfls_on_z_degree_7}}
\end{subfigure}
\begin{subfigure}{0.32\textwidth}
\includegraphics[width=\textwidth]{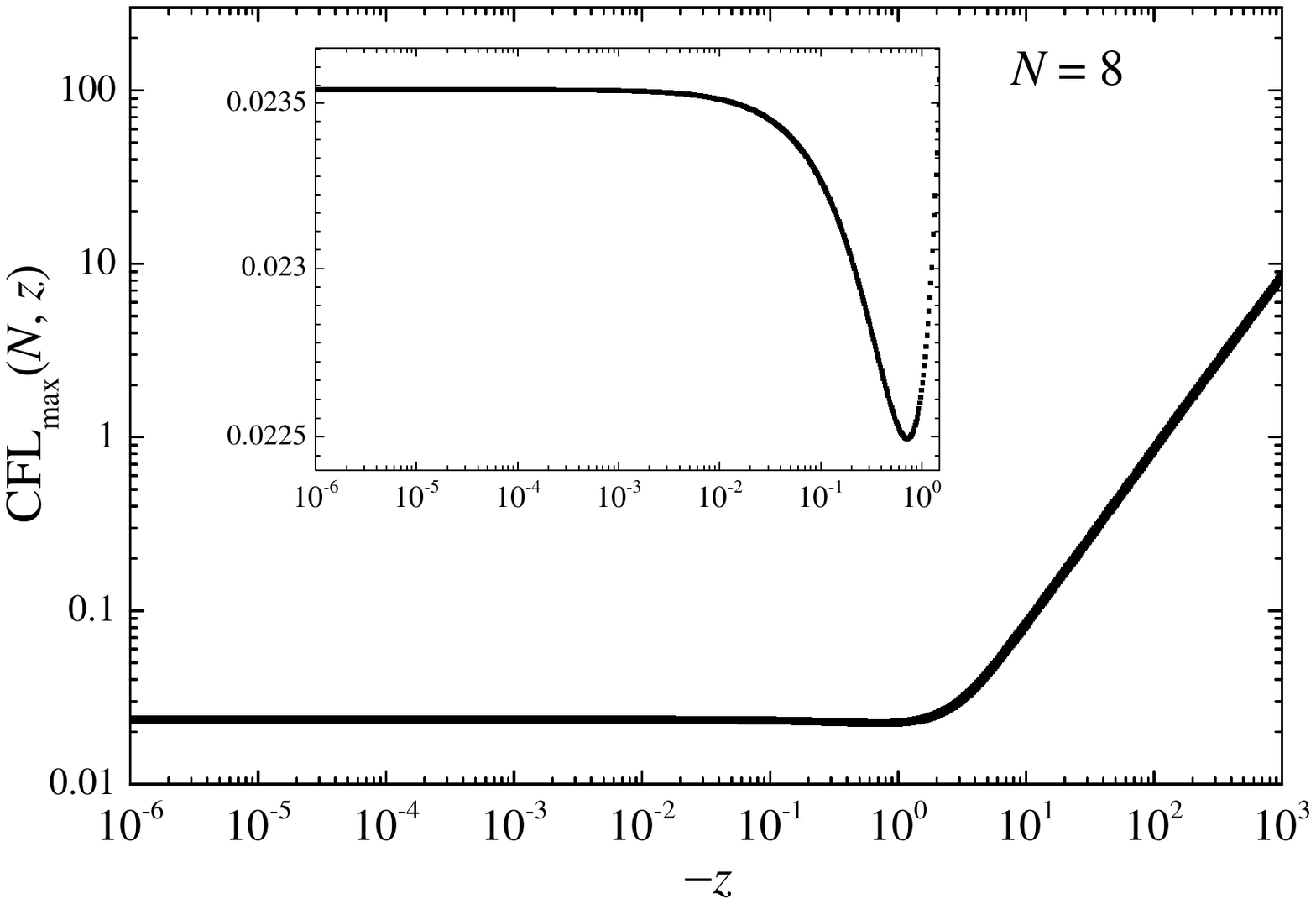}
\vspace{-8mm}\caption{\label{fig:cfls_on_z_degree_8}}
\end{subfigure}
\begin{subfigure}{0.32\textwidth}
\includegraphics[width=\textwidth]{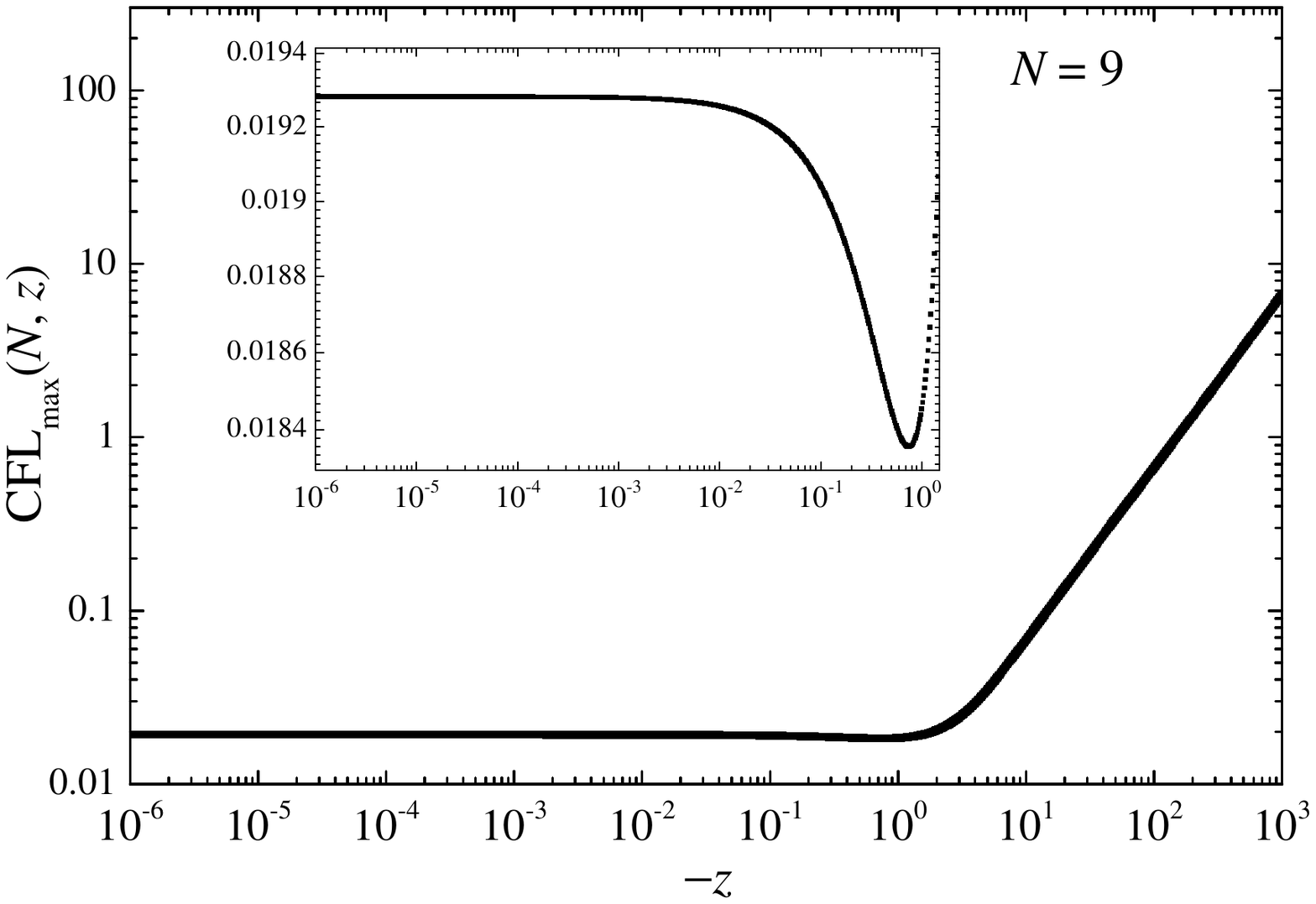}
\vspace{-8mm}\caption{\label{fig:cfls_on_z_degree_9}}
\end{subfigure}\\
\begin{subfigure}{0.32\textwidth}
\includegraphics[width=\textwidth]{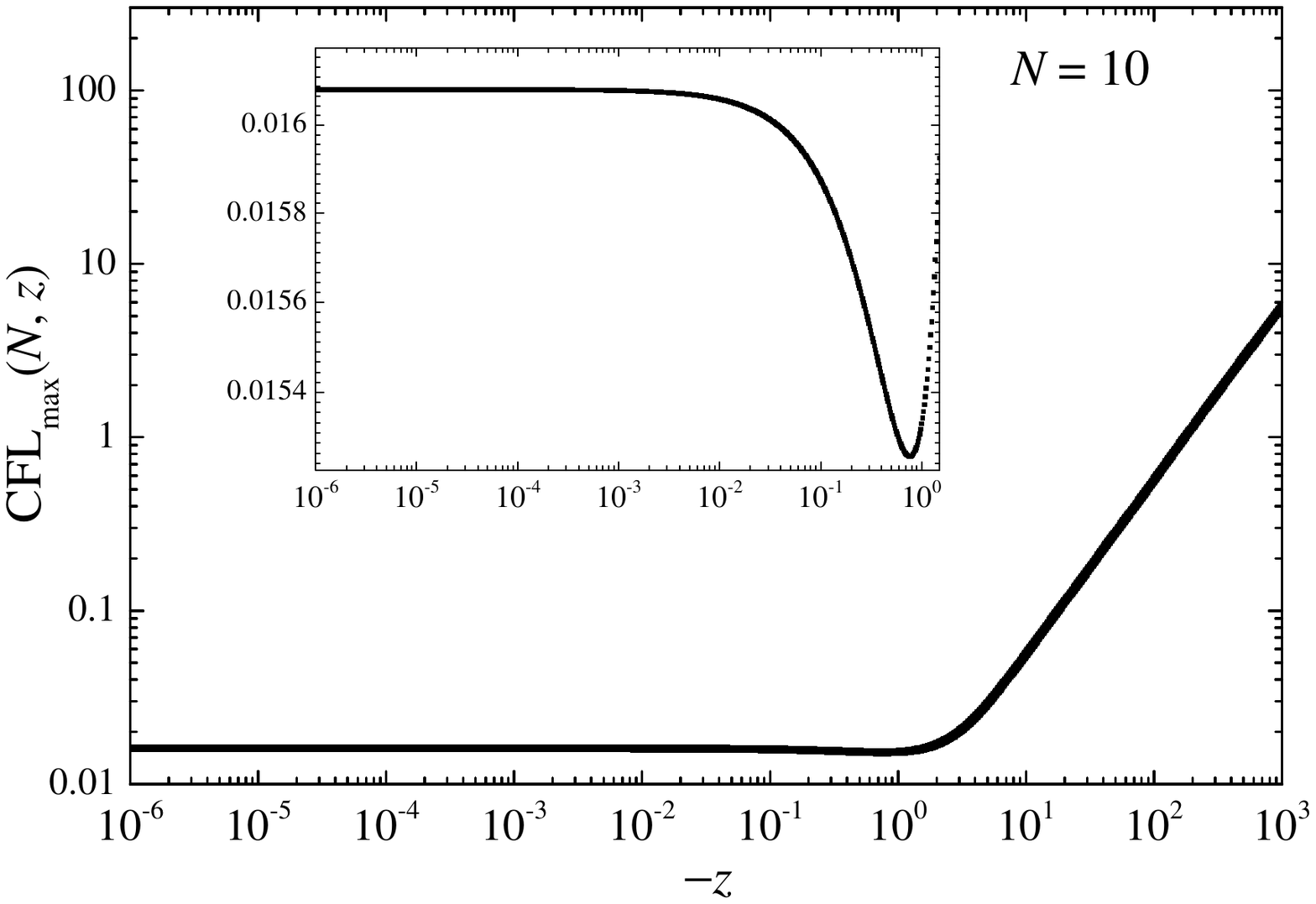}
\vspace{-8mm}\caption{\label{fig:cfls_on_z_degree_10}}
\end{subfigure}
\begin{subfigure}{0.32\textwidth}
\includegraphics[width=\textwidth]{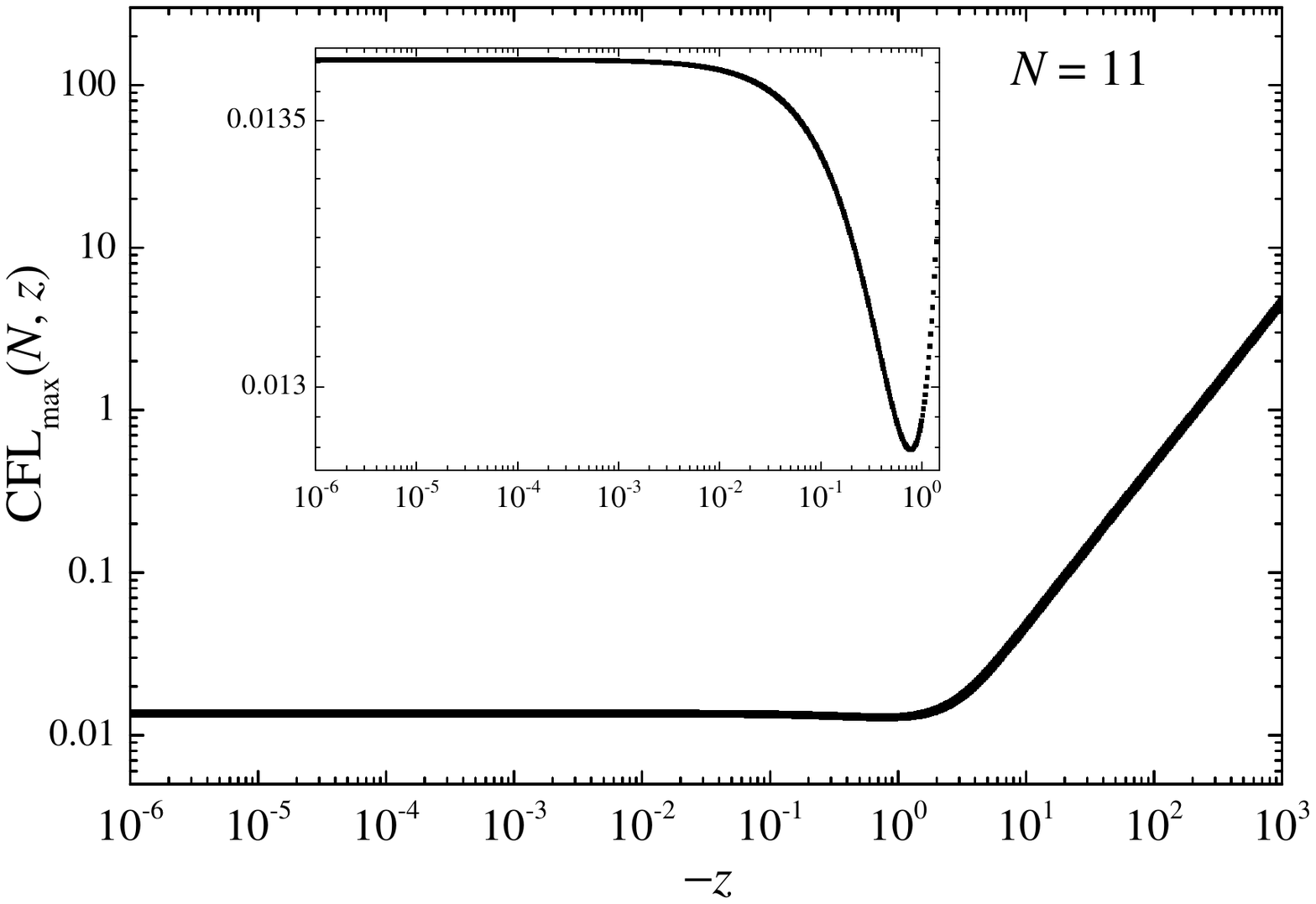}
\vspace{-8mm}\caption{\label{fig:cfls_on_z_degree_11}}
\end{subfigure}
\begin{subfigure}{0.32\textwidth}
\includegraphics[width=\textwidth]{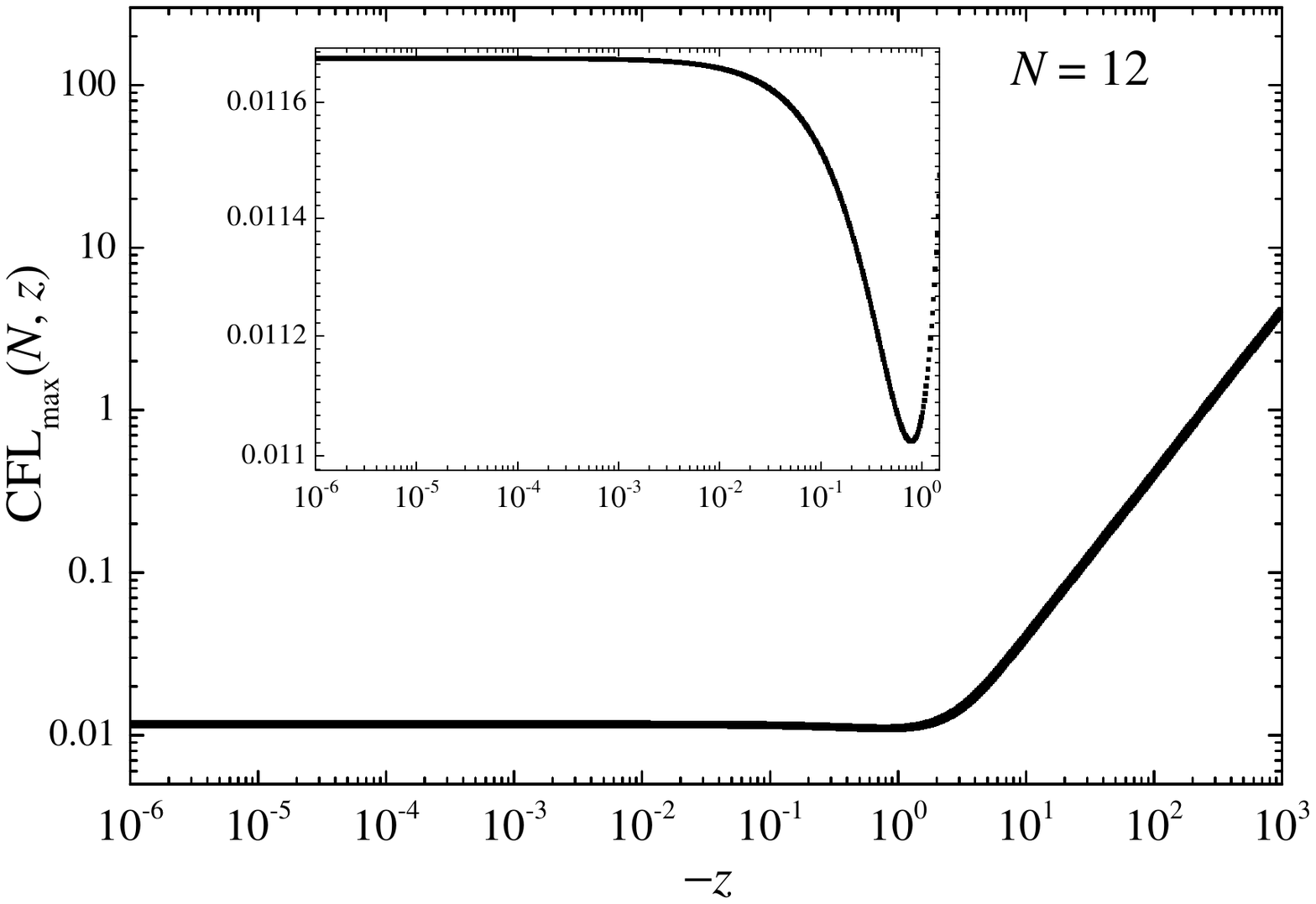}
\vspace{-8mm}\caption{\label{fig:cfls_on_z_degree_12}}
\end{subfigure}
\caption{%
Dependence of the boundary value of the Courant number $\mathrm{CFL}_{\rm max}(N, z)$ on the parameter $z$ for polynomial degrees $N = 1, \ldots, 12$. The insets show zoomed regions where a plateau $\mathrm{CFL}_{\rm max}(N, z) \approx \mathrm{CFL}_{\rm max}(N, 0)$ and a local decrease in the dependence $\mathrm{CFL}_{\rm max}(N, z)$ with increasing parameter $z$ occur.
}
\label{fig:cfls_on_z}
\end{figure}

Figures~\ref{fig:dep_f_on_phi_diff_cfls_degrees_1_2},~\ref{fig:dep_f_on_phi_diff_cfls_degrees_3_4},~\ref{fig:dep_f_on_phi_diff_cfls_degrees_5_6},~\ref{fig:dep_f_on_phi_diff_cfls_degrees_7_8},~\ref{fig:dep_f_on_phi_diff_cfls_degrees_9_10} and~\ref{fig:dep_f_on_phi_diff_cfls_degrees_11_12} show the dependence of function $F(\phi; \mathrm{CFL}, z)$ (\ref{eq:f_def}) on argument $\phi\in[0, 2\pi]$ for several values of the Courant number $\mathrm{CFL}$, selected in the vicinity of the stability boundary $\mathrm{CFL}_{\rm max}(z)$, defined further in Figure~\ref{fig:cfls_on_z}, for $z = -0.1$, $-1.0$, $-10.0$, $-100.0$ and for polynomial degrees $N = 1, \ldots, 12$: Figure~\ref{fig:dep_f_on_phi_diff_cfls_degrees_1_2} shows the results for polynomial degrees $N = 1$ and $2$, Figure~\ref{fig:dep_f_on_phi_diff_cfls_degrees_3_4} shows the results for polynomial degrees $N = 3$ and $4$, Figure~\ref{fig:dep_f_on_phi_diff_cfls_degrees_5_6} --- for polynomial degrees $N = 5$ and $6$, Figure~\ref{fig:dep_f_on_phi_diff_cfls_degrees_7_8} --- for polynomial degrees $N = 7$ and $8$, Figure~\ref{fig:dep_f_on_phi_diff_cfls_degrees_9_10} --- for polynomial degrees $N = 9$ and $10$, Figure~\ref{fig:dep_f_on_phi_diff_cfls_degrees_11_12} --- for polynomial degrees $N = 11$ and $12$. The calculations to obtain these results are performed using a developed program in the \texttt{python} programming language. Floating-point calculations are performed using the \texttt{mpmath} module~\cite{mpmath_man} (with \texttt{gmp2} module), which enables calculations with arbitrarily high precision representation of real numbers as floating-point numbers. The presented results are calculated using \texttt{mpmath.mp.dps = 2000}. The phase $\phi$, which determines the eigenvalue $\lambda_{k} = |\lambda_{k}|\exp(i\phi)$, is discretized using a uniform grid with $1000$ nodes on each graph (taking into account the narrowing of the phase range). It is important to note the following: the phase $\phi$ is the argument of the limit eigenvalue $\lambda = \exp(i\phi)$ and it is not the phase $\theta$.  From left to right, the graphs in these Figures represent increasingly narrower neighborhoods of the stability boundary $\mathrm{CFL}_{\rm max}$, from $\mathrm{CFL} \in [0.90\, \mathrm{CFL}_{\rm max},\, 1.10\, \mathrm{CFL}_{\rm max}]$ in the left column to $\mathrm{CFL} \in [0.9995\, \mathrm{CFL}_{\rm max},\, 1.0005\, \mathrm{CFL}_{\rm max}]$ in the right column; the range of phase $\phi$ variation also narrows when moving from the left to the right columns of the graphs to ensure a correct and clear presentation of the root $\phi = \pi$ of the equation $F(\pi; \mathrm{CFL}, z) = 0$. It is necessary to note two significant differences in the behavior of the function dependence in the case of a linear advection-reaction equation (\ref{eq:adv_eq_src}) from the case of a pure linear advection equation (with $\nu = 0$), which was studied in~\cite{ader_dg_stab_rig}. Firstly, there is no divergence $F(\phi; \mathrm{CFL}, z) \rightarrow \infty$ (\ref{eq:f_def}) of the function at the boundary points of the phase $\phi = 0$ and $\phi = 2\pi$. Furthermore, as $|z|$ increases, the function values $F(0; \mathrm{CFL}, z)$ and $F(2\pi; \mathrm{CFL}, z)$ decrease significantly. Secondly, there are no ``false'' roots of the equation $F(\phi; \mathrm{CFL}, z) = 0$ in the neighborhood of the boundary points of the phase $\phi = 0$ and $\phi = 2\pi$, which were associated for linear advection equation~\cite{ader_dg_stab_rig} with phenomenon of ``false'' violation of the stability condition for sufficiently large polynomial degrees $N$ even at lower Courant numbers $\mathrm{CFL} \ll \mathrm{CFL}_{\rm max}$ (\ref{eq:cfl_max_base_def}): some eigenvalues $\lambda_{k}$ of the matrix $\mathrm{R}(\mathrm{CFL}, z, \theta)$ (\ref{eq:r_matrix_in_matrix_form}) of the evolution operator $R$ with absolute values $|\lambda_{k}|$ greater than one remain in the vicinity of point $\lambda = +1$, approximately in the region of the eigenvalue phase $\arg\lambda_{k} \in [-\pi/4, +\pi/4]$. This is noted above in this work in Subsection~\ref{sec:stab_anal:empir_cons} ``Empirical considerations'' and is also confirmed by the behavior of function $F(\phi; \mathrm{CFL}, z)$ (\ref{eq:f_def}). The obtained results $F(\pi; \mathrm{CFL}_{\rm max}, z) = 0$ well confirm the results presented above in Subsection~\ref{sec:stab_anal:empir_cons} ``Empirical considerations'' regarding the choice of eigenvalue $\lambda = -1$ (and $\phi = \pi$) as the eigenvalue upon reaching which the stability of the ADER-DG numerical method with LST-DG predictor is violated.

Substituting the empirically determined eigenvalue of $\lambda = -1$ into the equation (\ref{eq:f_def}) yields the following equation:
\begin{equation}
F(\pi; \mathrm{CFL}, z) = |1 - \mathrm{CFL} \cdot H^{+}(0; \mathrm{CFL}, z, -1)|^{2} - |\mathrm{CFL} \cdot H^{+}(1; \mathrm{CFL}, z, -1)|^{2} = 0,
\end{equation}
the solution $\mathrm{CFL}$ to which determines the boundary value of the Courant number $\mathrm{CFL}_{\rm max}$. To simplify the notation, two new functions $\mathfrak{A}(\mathrm{CFL}, z) = H^{+}(1; \mathrm{CFL}, z, -1)$ and $\mathfrak{B}(\mathrm{CFL}, z) = H^{+}(0; \mathrm{CFL}, z, -1)$ are introduced, allowing the last equation to be rewritten in the following form:
\begin{equation}\label{eq:cfl_eq_plynomial_src}
\mathrm{CFL} \cdot \Big[\mathfrak{B}(\mathrm{CFL}, z) \pm \mathfrak{A}(\mathrm{CFL}, z)\Big] = 1.
\end{equation}
The functions $\mathfrak{A}(\mathrm{CFL}, z)$ and $\mathfrak{B}(\mathrm{CFL}, z)$ can be expressed using relations (\ref{eq:h_plus_def}) as follows:
\begin{equation}
\begin{split}
&\mathfrak{A}(\mathrm{CFL}, z) = \ovecphi^{T}\mathrm{B}(\mathrm{CFL}, z)\Big[R_{\rm RK}(\mathrm{Z}^{+}(\mathrm{CFL}, z)) + I\Big]^{-1} m^{-1}\uvecphi,\\
&\mathfrak{B}(\mathrm{CFL}, z) = \uvecphi^{T}\mathrm{B}(\mathrm{CFL}, z)\Big[R_{\rm RK}(\mathrm{Z}^{+}(\mathrm{CFL}, z)) + I\Big]^{-1} m^{-1}\uvecphi,
\end{split}
\end{equation}
which, taking into account the following expression:
\begin{equation}
\mathrm{B}(\mathrm{CFL}, z)\Big[R_{\rm RK}(\mathrm{Z}^{+}(\mathrm{CFL}, z)) + I\Big]^{-1} = 
\Big[R_{\rm RK}(\mathrm{Z}^{+}(\mathrm{CFL}, z)) + I\Big]^{-1}\Big[\mathrm{Z}^{+}(\mathrm{CFL}, z)\Big]^{-1}\Big[R_{\rm RK}(\mathrm{Z}^{+}(\mathrm{CFL}, z)) - I\Big],
\end{equation}
which is a consequence of expression (\ref{eq:f_derivd_temp_1_2_a_plus}), and by introducing a new function $\mathfrak{L}(z)$:
\begin{equation}
\mathfrak{L}(z) = \frac{R_{\rm RK}(z) - 1}{z(R_{\rm RK}(z) + 1)},
\end{equation}
and its Taylor series expansion for matrix argument $\mathrm{Z}^{+}(\mathrm{CFL}, z)$:
\begin{equation}
\mathfrak{L}(\mathrm{Z}^{+}(\mathrm{CFL}, z)) = 
\sum\limits_{s = 0}^{N} \frac{1}{s!}\frac{d^{s}\mathfrak{L}(z)}{dz^{s}}\Big[\mathrm{Z}^{+}(\mathrm{CFL}, z) - z I\Big]^{s} =
\sum\limits_{s = 0}^{N} \frac{1}{s!}\frac{d^{s}\mathfrak{L}(z)}{dz^{s}}(-\mathrm{CFL})^{s} \mathrm{D}^{s},
\end{equation}
taking into account the definition of matrix (\ref{eq:z_plus_matrix_def}) and the nilpotency property of matrix $\mathrm{D}$: $\mathrm{D}^{N+1} = 0$, can be represented in the following form, convenient for calculations:
\begin{equation}\label{eq:A_and_B_exp_def_init}
\mathfrak{A}(\mathrm{CFL}, z) = \sum\limits_{s = 0}^{N} \frac{1}{s!}\frac{d^{s}\mathfrak{L}(z)}{dz^{s}}(-\mathrm{CFL})^{s}\, \Big[\ovecphi^{T}\mathrm{D}^{s}m^{-1}\uvecphi\Big],\quad
\mathfrak{B}(\mathrm{CFL}, z) = \sum\limits_{s = 0}^{N} \frac{1}{s!}\frac{d^{s}\mathfrak{L}(z)}{dz^{s}}(-\mathrm{CFL})^{s}\, \Big[\uvecphi^{T}\mathrm{D}^{s}m^{-1}\uvecphi\Big].
\end{equation}
The terms in square brackets within the resulting expressions are evidently values of the delta function $\delta_{\mathcal{P}_{N}}(\xi)$ (\ref{eq:delta_func_varphi_action}) and its derivatives $\delta_{\mathcal{P}_{N}}^{(s)}(\xi)$ (accounting for the properties (\ref{eq:tilde_ae_deriv_prop}) of the differentiation matrix $\mathrm{D}$ and its powers), written in matrix notation (\ref{eq:matrix_notation_base_def}):
\begin{equation}
\delta_{\mathcal{P}_{N}}(\xi) = \boldsymbol{\varphi}^{T}(\xi) \Big[m^{-1}\uvecphi\Big],\qquad
\delta_{\mathcal{P}_{N}}^{(s)}(\xi) = \boldsymbol{\varphi}^{T}(\xi) \Big[\mathrm{D}^{s}m^{-1}\uvecphi\Big],
\end{equation}
evaluated at points $\xi = 0$ and $\xi = 1$:
\begin{equation}
\ovecphi^{T}\mathrm{D}^{s}m^{-1}\uvecphi = \delta_{\mathcal{P}_{N}}^{(s)}(1),\qquad
\uvecphi^{T}\mathrm{D}^{s}m^{-1}\uvecphi = \delta_{\mathcal{P}_{N}}^{(s)}(0),
\end{equation}
whereby the expressions (\ref{eq:A_and_B_exp_def_init}) for the functions $\mathfrak{A}(\mathrm{CFL}, z)$ and $\mathfrak{B}(\mathrm{CFL}, z)$ take the following form:
\begin{equation}\label{eq:A_and_B_exp_prefinal}
\begin{split}
\mathfrak{A}(\mathrm{CFL}, z) & =
\sum\limits_{s = 0}^{N} \frac{1}{s!}
\left\{\frac{d^{s}}{dz^{s}}\left[\frac{R_{\rm RK}(z) - 1}{z(R_{\rm RK}(z) + 1)}\right]\right\}
(-\mathrm{CFL})^{s}\delta_{\mathcal{P}_{N}}^{(s)}(1),\\
\mathfrak{B}(\mathrm{CFL}, z) & =
\sum\limits_{s = 0}^{N} \frac{1}{s!}
\left\{\frac{d^{s}}{dz^{s}}\left[\frac{R_{\rm RK}(z) - 1}{z(R_{\rm RK}(z) + 1)}\right]\right\}
(-\mathrm{CFL})^{s}\delta_{\mathcal{P}_{N}}^{(s)}(0).
\end{split}
\end{equation}

\begin{table}[h!]
\centering
\caption{%
Expressions for the numerator $\mathfrak{A}^{(n)}(\mathrm{CFL}, z)$ and denominator $\mathfrak{A}^{(d)}(\mathrm{CFL}, z)$ of function $\mathfrak{A}(\mathrm{CFL}, z)$, and the numerator $\mathfrak{B}^{(n)}(\mathrm{CFL}, z)$ and denominator $\mathfrak{B}^{(d)}(\mathrm{CFL}, z)$ of function $\mathfrak{B}(\mathrm{CFL}, z)$, calculated symbolically from expression (\ref{eq:A_and_B_exp_prefinal}), for polynomial degrees $N = 1, \ldots, 6$.
}
\label{tab:A_and_B_functions_1}
\begin{tabular}{@{}|p{0.175cm}|p{1.325cm}|p{12.5cm}|@{}}
\toprule
$N$ & formula & expression \\ 
\midrule
$1$ & $\mathfrak{A}^{(n)}$ & $\cfl \big(6 z^{2} - 72 z\big) + 2 z^{3} - 16 z^{2} + 48 z - 144$ \\
\cmidrule{2-3}
    & $\mathfrak{B}^{(n)}$  & $\cfl \big(6 z^{2} - 72 z\big) - 4 z^{3} + 32 z^{2} - 96 z + 288$ \\
\cmidrule{2-3}
    & $\mathfrak{A}^{(d)}$, $\mathfrak{B}^{(d)}$ & $z^{4} - 4 z^{3} + 28 z^{2} - 48 z + 144$ \\
\midrule
$2$ & $\mathfrak{A}^{(n)}$ & $\cfl^{2} \big(- 60 z^{6} + 1080 z^{5} - 32400 z^{4} + 334080 z^{3} - 1166400 z^{2} - 1296000 z + 4320000\big)$ \\
    &                & $+ \cfl \big(- 24 z^{7} + 576 z^{6} - 9504 z^{5} + 104256 z^{4} - 449280 z^{3} + 1036800 z^{2} - 3456000 z\big)$ \\
    &                & $- 3 z^{8} + 90 z^{7} - 1116 z^{6} + 8928 z^{5} - 48816 z^{4} + 158112 z^{3} - 639360 z^{2} + 777600 z - 2592000$ \\
    & $\mathfrak{B}^{(n)}$  & $\cfl^{2} \big(- 60 z^{6} + 1080 z^{5} - 32400 z^{4} + 334080 z^{3} - 1166400 z^{2} - 1296000 z + 4320000\big)$ \\
    &                & $+ \cfl \big(36 z^{7} - 864 z^{6} + 14256 z^{5} - 156384 z^{4} + 673920 z^{3} - 1555200 z^{2} + 5184000 z\big) - 9 z^{8}$ \\
    &                & $+ 270 z^{7} - 3348 z^{6} + 26784 z^{5} - 146448 z^{4} + 474336 z^{3} - 1918080 z^{2} + 2332800 z - 7776000$ \\
    & $\mathfrak{A}^{(d)}$, $\mathfrak{B}^{(d)}$ & $z^{9} - 36 z^{8} + 468 z^{7} - 2952 z^{6} + 14256 z^{5} - 65664 z^{4}$ \\
    &                & $+ 148608 z^{3} - 570240 z^{2} + 518400 z - 1728000$ \\
\midrule
$3$ & $\mathfrak{A}^{(n)}$ & $\cfl^{3} \big(840 z^{12} - 67200 z^{11} + 806400 z^{10} - 14918400 z^{9} + 398160000 z^{8}$ \\
    &                & $- 6123801600 z^{7} + 67036032000 z^{6} - 532901376000 z^{5} + 2447698176000 z^{4}$ \\
    &                & $- 5215795200000 z^{3} - 31863767040000 z^{2} + 111523184640000 z\big) + \cfl^{2} \big(360 z^{13} - 25920 z^{12}$ \\
    &                & $+ 518400 z^{11} - 10022400 z^{10} + 144720000 z^{9} - 2136844800 z^{8} + 24126336000 z^{7} - 203793408000 z^{6}$ \\
    &                & $+ 1171438848000 z^{5} - 4750617600000 z^{4} + 5922201600000 z^{3} - 16826019840000 z^{2}$ \\
    &                & $- 34139750400000 z + 119489126400000\big) + \cfl \big(60 z^{14} - 3840 z^{13} + 102240 z^{12} - 2102400 z^{11}$ \\
    &                & $+ 29102400 z^{10} - 363801600 z^{9} + 3396384000 z^{8} - 29590272000 z^{7} + 175153536000 z^{6}$ \\
    &                & $- 895656960000 z^{5} + 2577899520000 z^{4} - 10323210240000 z^{3} + 11379916800000 z^{2}$ \\
    &                & $- 39829708800000 z\big) + 4 z^{15} - 224 z^{14} + 7008 z^{13} - 149952 z^{12} + 2312640 z^{11} - 27843840 z^{10}$ \\
    &                & $+ 253221120 z^{9} - 1924646400 z^{8} + 10917504000 z^{7} - 58749235200 z^{6} + 196798464000 z^{5}$ \\
    &                & $- 860875776000 z^{4} + 1562222592000 z^{3} - 5988003840000 z^{2} + 4551966720000 z - 15931883520000$ \\
\cmidrule{2-3}
    & $\mathfrak{B}^{(n)}$  & $\cfl^{3} \big(840 z^{12} - 67200 z^{11} + 806400 z^{10} - 14918400 z^{9} + 398160000 z^{8} - 6123801600 z^{7}$ \\
    &                & $+ 67036032000 z^{6} - 532901376000 z^{5} + 2447698176000 z^{4} - 5215795200000 z^{3}$ \\
    &                & $- 31863767040000 z^{2} + 111523184640000 z\big) + \cfl^{2} \big(- 480 z^{13} + 34560 z^{12} - 691200 z^{11}$ \\
    &                & $+ 13363200 z^{10} - 192960000 z^{9} + 2849126400 z^{8} - 32168448000 z^{7} + 271724544000 z^{6}$ \\
    &                & $- 1561918464000 z^{5} + 6334156800000 z^{4} - 7896268800000 z^{3} + 22434693120000 z^{2}$ \\
    &                & $+ 45519667200000 z - 159318835200000\big) + \cfl \big(120 z^{14} - 7680 z^{13} + 204480 z^{12} - 4204800 z^{11}$ \\
    &                & $+ 58204800 z^{10} - 727603200 z^{9} + 6792768000 z^{8} - 59180544000 z^{7} + 350307072000 z^{6}$ \\
    &                & $- 1791313920000 z^{5} + 5155799040000 z^{4} - 20646420480000 z^{3} + 22759833600000 z^{2}$ \\
    &                & $- 79659417600000 z\big) - 16 z^{15} + 896 z^{14} - 28032 z^{13} + 599808 z^{12} - 9250560 z^{11} + 111375360 z^{10}$ \\
    &                & $- 1012884480 z^{9} + 7698585600 z^{8} - 43670016000 z^{7} + 234996940800 z^{6} - 787193856000 z^{5}$ \\
    &                & $+ 3443503104000 z^{4} - 6248890368000 z^{3} + 23952015360000 z^{2} - 18207866880000 z$ \\
    &                & $+ 63727534080000$ \\
\cmidrule{2-3}
    & $\mathfrak{A}^{(d)}$, $\mathfrak{B}^{(d)}$ & $z^{16} - 48 z^{15} + 1584 z^{14} - 33312 z^{13} + 550176 z^{12} - 6618240 z^{11} + 64990080 z^{10} - 467112960 z^{9}$ \\
    &                & $+ 2903904000 z^{8} - 12643430400 z^{7} + 61349529600 z^{6} - 160330752000 z^{5} + 668874240000 z^{4}$ \\
    &                & $- 970776576000 z^{3} + 3657830400000 z^{2} - 2275983360000 z + 7965941760000$ \\
\midrule
$4$ & $\mathfrak{A}^{(n)}$ & $\cfl^{4} \big(- 15120 z^{20} + 1512000 z^{19} - 158760000 z^{18} + 7901712000 z^{17} - 225275904000 z^{16}$ \\
    &                & $+ 3844256371200 z^{15} - 50897185920000 z^{14} + 822874671360000 z^{13} - 14479674048000000 z^{12}$ \\
    &                & $+ 197843267543040000 z^{11} - 2110137175222272000 z^{10} + 18582897728102400000 z^{9}$ \\
    &                & $- 134241424644096000000 z^{8} + 700869684333772800000 z^{7} - 2900789725962240000000 z^{6}$ \\
    &                & $+ 215832348694609920000 z^{5} + 43240973599014912000000 z^{4} - 424875898774978560000000 z^{3}$ \\
    &                & $+ 1483894930049925120000000 z^{2} + 442534346003054592000000 z$ \\
    &                & $- 1593123645610996531200000\big) + \cfl^{3} \big(- 6720 z^{21} + 739200 z^{20} - 62496000 z^{19}$ \\
    &                & $+ 3161088000 z^{18} - 95268096000 z^{17} + 1930753843200 z^{16} - 30484726272000 z^{15}$ \\
    &                & $+ 445984358400000 z^{14} - 6022008115200000 z^{13} + 74369923891200000 z^{12}$ \\
    &                & $- 831008637591552000 z^{11} + 7820897892433920000 z^{10} - 62448044514508800000 z^{9}$ \\
    &                & $+ 394907101824614400000 z^{8} - 2177088059355955200000 z^{7} + 7148999637278392320000 z^{6}$ \\
    &                & $- 22134907562832691200000 z^{5} - 40758601336750080000000 z^{4} + 227901285772296192000000 z^{3}$ \\
    &                & $- 786727726227652608000000 z^{2} + 2832219814419549388800000 z\big) + \cfl^{2} \big(- 1260 z^{22}$ \\
    &                & $+ 151200 z^{21} - 10886400 z^{20} + 522396000 z^{19} - 16344720000 z^{18} + 362352009600 z^{17}$ \\
    &                & $- 6422576832000 z^{16} + 96284619648000 z^{15} - 1244170368000000 z^{14} + 14283487330560000 z^{13}$ \\
    &                & $- 148906885994496000 z^{12} + 1345792801489920000 z^{11} - 11184719078615040000 z^{10}$ \\
    &                & $+ 75381371167334400000 z^{9} - 469964714358988800000 z^{8} + 2072153951238389760000 z^{7}$ \\
    &                & $- 9472678295180083200000 z^{6} + 19639667755524096000000 z^{5} - 72619325386997760000000 z^{4}$ \\
    &                & $+ 15577156947050496000000 z^{3} - 18029177059383705600000 z^{2} - 368778621669212160000000 z$ \\
    &                & $+ 1327603038009163776000000\big) + \cfl \big(- 120 z^{23} + 15600 z^{22} - 1022400 z^{21} + 44654400 z^{20}$ \\
    &                & $- 1384272000 z^{19} + 32047027200 z^{18} - 600638976000 z^{17} + 9448576128000 z^{16}$ \\
    &                & $- 125091993600000 z^{15} + 1455561031680000 z^{14} - 14703330926592000 z^{13}$ \\
    &                & $+ 130481760245760000 z^{12} - 1046233947709440000 z^{11} + 7031074604359680000 z^{10}$ \\
    &                & $- 45190002390220800000 z^{9} + 217682791445790720000 z^{8} - 1114304519739801600000 z^{7}$ \\
    &                & $+ 3401157612109824000000 z^{6} - 14778055729859788800000 z^{5} + 25130341280120832000000 z^{4}$ \\
    &                & $- 97716578694148915200000 z^{3} + 70243546984611840000000 z^{2} - 252876769144602624000000 z\big)$ \\
    &                & $- 5 z^{24} + 700 z^{23} - 44700 z^{22} + 1799400 z^{21} - 52879200 z^{20} + 1219228800 z^{19} - 23013432000 z^{18}$ \\
    &                & $+ 366867504000 z^{17} - 4991258880000 z^{16} + 58833451008000 z^{15} - 607342796928000 z^{14}$ \\
    &                & $+ 5407374309120000 z^{13} - 43521525573120000 z^{12} + 290826969108480000 z^{11}$ \\
    &                & $- 1866700344176640000 z^{10} + 9080139984814080000 z^{9} - 48100820098867200000 z^{8}$ \\
    &                & $+ 161445969248256000000 z^{7} - 740171706391756800000 z^{6} + 1606254089674752000000 z^{5}$ \\
    &                & $- 6627367848561868800000 z^{4} + 8331355440414720000000 z^{3} - 31804717106921472000000 z^{2}$ \\
    &                & $+ 17560886746152960000000 z - 63219192286150656000000$ \\
\bottomrule
\end{tabular}
\end{table}
\begin{table}[h!]
\centering
\caption{Table~\ref{tab:A_and_B_functions_1} continued.}
\label{tab:A_and_B_functions_2}
\begin{tabular}{@{}|p{0.175cm}|p{1.325cm}|p{12.5cm}|@{}}
\toprule
$N$ & formula & expression \\ 
\midrule
    & $\mathfrak{B}^{(n)}$  & $\cfl^{4} \big(- 15120 z^{20} + 1512000 z^{19} - 158760000 z^{18} + 7901712000 z^{17} - 225275904000 z^{16}$ \\
    &                & $+ 3844256371200 z^{15} - 50897185920000 z^{14} + 822874671360000 z^{13} - 14479674048000000 z^{12}$ \\
    &                & $+ 197843267543040000 z^{11} - 2110137175222272000 z^{10} + 18582897728102400000 z^{9}$ \\
    &                & $- 134241424644096000000 z^{8} + 700869684333772800000 z^{7} - 2900789725962240000000 z^{6}$ \\
    &                & $+ 215832348694609920000 z^{5} + 43240973599014912000000 z^{4} - 424875898774978560000000 z^{3}$ \\
    &                & $+ 1483894930049925120000000 z^{2} + 442534346003054592000000 z$ \\
    &                & $- 1593123645610996531200000\big) + \cfl^{3} \big(8400 z^{21} - 924000 z^{20} + 78120000 z^{19} - 3951360000 z^{18}$ \\
    &                & $+ 119085120000 z^{17} - 2413442304000 z^{16} + 38105907840000 z^{15} - 557480448000000 z^{14}$ \\
    &                & $+ 7527510144000000 z^{13} - 92962404864000000 z^{12} + 1038760796989440000 z^{11}$ \\
    &                & $- 9776122365542400000 z^{10} + 78060055643136000000 z^{9} - 493633877280768000000 z^{8}$ \\
    &                & $+ 2721360074194944000000 z^{7} - 8936249546597990400000 z^{6} + 27668634453540864000000 z^{5}$ \\
    &                & $+ 50948251670937600000000 z^{4} - 284876607215370240000000 z^{3}$ \\
    &                & $+ 983409657784565760000000 z^{2} - 3540274768024436736000000 z\big) + \cfl^{2} \big(- 2100 z^{22}$ \\
    &                & $+ 252000 z^{21} - 18144000 z^{20} + 870660000 z^{19} - 27241200000 z^{18} + 603920016000 z^{17}$ \\
    &                & $- 10704294720000 z^{16} + 160474366080000 z^{15} - 2073617280000000 z^{14}$ \\
    &                & $+ 23805812217600000 z^{13} - 248178143324160000 z^{12} + 2242988002483200000 z^{11}$ \\
    &                & $- 18641198464358400000 z^{10} + 125635618612224000000 z^{9} - 783274523931648000000 z^{8}$ \\
    &                & $+ 3453589918730649600000 z^{7} - 15787797158633472000000 z^{6} + 32732779592540160000000 z^{5}$ \\
    &                & $- 121032208978329600000000 z^{4} + 25961928245084160000000 z^{3}$ \\
    &                & $- 30048628432306176000000 z^{2} - 614631036115353600000000 z$ \\
    &                & $+ 2212671730015272960000000\big) + \cfl \big(300 z^{23} - 39000 z^{22} + 2556000 z^{21} - 111636000 z^{20}$ \\
    &                & $+ 3460680000 z^{19} - 80117568000 z^{18} + 1501597440000 z^{17} - 23621440320000 z^{16}$ \\
    &                & $+ 312729984000000 z^{15} - 3638902579200000 z^{14} + 36758327316480000 z^{13}$ \\
    &                & $- 326204400614400000 z^{12} + 2615584869273600000 z^{11} - 17577686510899200000 z^{10}$ \\
    &                & $+ 112975005975552000000 z^{9} - 544206978614476800000 z^{8} + 2785761299349504000000 z^{7}$ \\
    &                & $- 8502894030274560000000 z^{6} + 36945139324649472000000 z^{5} - 62825853200302080000000 z^{4}$ \\
    &                & $+ 244291446735372288000000 z^{3} - 175608867461529600000000 z^{2}$ \\
    &                & $+ 632191922861506560000000 z\big) - 25 z^{24} + 3500 z^{23} - 223500 z^{22} + 8997000 z^{21}$ \\
    &                & $- 264396000 z^{20} + 6096144000 z^{19} - 115067160000 z^{18} + 1834337520000 z^{17}$ \\
    &                & $- 24956294400000 z^{16} + 294167255040000 z^{15} - 3036713984640000 z^{14}$ \\
    &                & $+ 27036871545600000 z^{13} - 217607627865600000 z^{12} + 1454134845542400000 z^{11}$ \\
    &                & $- 9333501720883200000 z^{10} + 45400699924070400000 z^{9} - 240504100494336000000 z^{8}$ \\
    &                & $+ 807229846241280000000 z^{7} - 3700858531958784000000 z^{6} + 8031270448373760000000 z^{5}$ \\
    &                & $- 33136839242809344000000 z^{4} + 41656777202073600000000 z^{3}$ \\
    &                & $- 159023585534607360000000 z^{2} + 87804433730764800000000 z - 316095961430753280000000$ \\
\cmidrule{2-3}
    & $\mathfrak{A}^{(d)}$, $\mathfrak{B}^{(d)}$ & $z^{25} - 150 z^{24} + 9900 z^{23} - 394800 z^{22} + 11258400 z^{21} - 254635200 z^{20} + 4747608000 z^{19}$ \\
    &                & $- 74430720000 z^{18} + 1009181952000 z^{17} - 11833357824000 z^{16} + 120966742656000 z^{15}$ \\
    &                & $- 1097664099840000 z^{14} + 8486682785280000 z^{13} - 60764543400960000 z^{12}$ \\
    &                & $+ 347971600650240000 z^{11} - 2016518168125440000 z^{10} + 8225676168192000000 z^{9}$ \\
    &                & $- 40630638724300800000 z^{8} + 113723946749952000000 z^{7} - 500147889596006400000 z^{6}$ \\
    &                & $+ 910457435691417600000 z^{5} - 3675982602829824000000 z^{4} + 3917905067704320000000 z^{3}$ \\
    &                & $- 14829193252306944000000 z^{2} + 7024354698461184000000 z - 25287676914460262400000$ \\
\midrule
$5$ & $\mathfrak{A}^{(n)}$ & $\cfl^{5} \big(332640 z^{30} - 83825280 z^{29} + 5029516800 z^{28} - 274388083200 z^{27} + 19544702284800 z^{26}$ \\
    &                & $- 1021253445273600 z^{25} + 40783305591705600 z^{24} - 1306536651063705600 z^{23}$ \\
    &                & $+ 33725923025743872000 z^{22} - 726153567434330112000 z^{21} + 13279981358383939584000 z^{20}$ \\
    &                & $- 216507322744631525376000 z^{19} + 3299468417886246617088000 z^{18}$ \\
    &                & $- 48870626813761965391872000 z^{17} + 681993250323520392069120000 z^{16}$ \\
    &                & $- 8572335712949421425295360000 z^{15} + 91413239877713218110750720000 z^{14}$ \\
    &                & $- 839760875742003008687308800000 z^{13} + 6482393389617389064243118080000 z^{12}$ \\
    &                & $- 43776173457786164638635786240000 z^{11} + 214348133624186286113292288000000 z^{10}$ \\
    &                & $- 844924797657029414267073331200000 z^{9} - 789634598017870685880188928000000 z^{8}$ \\
    &                & $+ 25599356799049615084214235955200000 z^{7} - 268347194005614546609342775296000000 z^{6}$ \\
    &                & $+ 1500652350011654558937837364838400000 z^{5} - 5160622485160397615960294424576000000 z^{4}$ \\
    &                & $+ 16979379778873451056519169703936000000 z^{3} + 19898003172905449654985076768768000000 z^{2}$ \\
    &                & $- 72959344967319982068278614818816000000 z\big) + \cfl^{4} \big(151200 z^{31} - 36288000 z^{30}$ \\
    &                & $+ 2667168000 z^{29} - 154822752000 z^{28} + 8557799040000 z^{27} - 412836455808000 z^{26}$ \\
    &                & $+ 16662112459776000 z^{25} - 544958661542400000 z^{24} + 14741063827660800000 z^{23}$ \\
    &                & $- 333881381901312000000 z^{22} + 6427574016621649920000 z^{21} - 108285672843165081600000 z^{20}$ \\
    &                & $+ 1631841601029625036800000 z^{19} - 22831778914923692851200000 z^{18}$ \\
    &                & $+ 296418930589569392640000000 z^{17} - 3589398478560336209510400000 z^{16}$ \\
    &                & $+ 38952682588932496234905600000 z^{15} - 384497615910135797121024000000 z^{14}$ \\
    &                & $+ 3273470995570873461374976000000 z^{13} - 25079894259703673282887680000000 z^{12}$ \\
    &                & $+ 156061679590879386133856256000000 z^{11} - 897290609591952433241653248000000 z^{10}$ \\
    &                & $+ 3335895635409973891553034240000000 z^{9} - 11586553903281599170813624320000000 z^{8}$ \\
    &                & $- 17482635738264225383763148800000000 z^{7} + 230118978988281461292354502656000000 z^{6}$ \\
    &                & $- 1783368216129179609939123896320000000 z^{5} + 7758432987350856488815574384640000000 z^{4}$ \\
    &                & $- 12389412144372646015959155343360000000 z^{3} + 43973264992788538270921890201600000000 z^{2}$ \\
    &                & $+ 14896900771159160169507544104960000000 z - 54621969494250253954860995051520000000\big)$ \\
    &                & $+ \cfl^{3} \big(30240 z^{32} - 6894720 z^{31} + 586051200 z^{30} - 36375091200 z^{29} + 1833792307200 z^{28}$ \\
    &                & $- 80596635033600 z^{27} + 3062266112102400 z^{26} - 99118037317017600 z^{25}$ \\
    &                & $+ 2722544894232576000 z^{24} - 63537041360830464000 z^{23} + 1275946028732841984000 z^{22}$ \\
    &                & $- 22297721664370360320000 z^{21} + 344873903644617080832000 z^{20}$ \\
    &                & $- 4810521759049978675200000 z^{19} + 60927263077661059645440000 z^{18}$ \\
    &                & $- 712511059464418897428480000 z^{17} + 7559347826210484667023360000 z^{16}$ \\
    &                & $- 74669032396816049394155520000 z^{15} + 651911268092847078113280000000 z^{14}$ \\
\bottomrule
\end{tabular}
\end{table}
\begin{table}[h!]
\centering
\caption{Table~\ref{tab:A_and_B_functions_2} continued.}
\label{tab:A_and_B_functions_3}
\begin{tabular}{@{}|p{0.175cm}|p{1.325cm}|p{12.5cm}|@{}}
\toprule
$N$ & formula & expression \\ 
\midrule
    &                & $- 5311074980582186024750284800000 z^{13} + 36488303402367029878141747200000 z^{12}$ \\
    &                & $- 238508928354663398567942553600000 z^{11} + 1187797210700926548942677606400000 z^{10}$ \\
    &                & $- 6075234333280270661676171264000000 z^{9} + 17706005310517293527567695872000000 z^{8}$ \\
    &                & $- 65583044524143157783118531788800000 z^{7} - 6681845037850679751393411072000000 z^{6}$ \\
    &                & $+ 226483910276223572545439268864000000 z^{5} - 2147866146868822372072116191232000000 z^{4}$ \\
    &                & $+ 9039172834447946394340436213760000000 z^{3} - 11917520616927328135606035283968000000 z^{2}$ \\
    &                & $+ 43697575595400203163888796041216000000 z\big) + \cfl^{2} \big(3360 z^{33} - 725760 z^{32} + 68342400 z^{31}$ \\
    &                & $- 4507776000 z^{30} + 226988697600 z^{29} - 9526717670400 z^{28} + 339786974822400 z^{27}$ \\
    &                & $- 10522674576691200 z^{26} + 282104829167616000 z^{25} - 6587810457919488000 z^{24}$ \\
    &                & $+ 134034312434319360000 z^{23} - 2399750754741043200000 z^{22} + 38043288469317648384000 z^{21}$ \\
    &                & $- 540268359952399368192000 z^{20} + 6905548551014159155200000 z^{19}$ \\
    &                & $- 80360031638654254448640000 z^{18} + 846589651414359913267200000 z^{17}$ \\
    &                & $- 8251150121710495232163840000 z^{16} + 71912976492391630724136960000 z^{15}$ \\
    &                & $- 590953694522474049306624000000 z^{14} + 4187807435735004326186188800000 z^{13}$ \\
    &                & $- 28828677603025947318445670400000 z^{12} + 159113270780830378888475443200000 z^{11}$ \\
    &                & $- 908917189346684797546659840000000 z^{10} + 3596987116084482609405493248000000 z^{9} $ \\
    &                & $- 17128178273040350872912212787200000 z^{8} + 41588775707434524959959941120000000 z^{7}$ \\
    &                & $- 170660129513655570537652420608000000 z^{6} + 184409475031270959705973850112000000 z^{5}$ \\
    &                & $- 650461403420305416212069744640000000 z^{4} - 285292141082193078646726459392000000 z^{3}$ \\
    &                & $+ 1369311080985336944873925771264000000 z^{2} - 3310422393590924482112787578880000000 z$ \\
    &                & $+ 12138215443166723101080221122560000000\big) + \cfl \big(210 z^{34} - 42840 z^{33} + 4309200 z^{32}$ \\
    &                & $- 298216800 z^{31} + 15533834400 z^{30} - 653621270400 z^{29} + 22854717043200 z^{28}$ \\
    &                & $- 684558996710400 z^{27} + 17736476688000000 z^{26} - 403846302230784000 z^{25}$ \\
    &                & $+ 8091340448744448000 z^{24} - 144459134012390400000 z^{23} + 2294548056233975808000 z^{22}$ \\
    &                & $- 32911210952599891968000 z^{21} + 423247693020393185280000 z^{20}$ \\
    &                & $- 4974161115871598837760000 z^{19} + 52510511140559128903680000 z^{18}$ \\
    &                & $- 513448432172924738273280000 z^{17} + 4467240447027535465021440000 z^{16}$ \\
    &                & $- 36757537863744237054197760000 z^{15} + 261547109562639519940608000000 z^{14}$ \\
    &                & $- 1825292331993783212782387200000 z^{13} + 10382420029732362187230412800000 z^{12}$ \\
    &                & $- 61830772666136403231375360000000 z^{11} + 269228069248639068783142502400000 z^{10}$ \\
    &                & $- 1382820714894920173449589555200000 z^{9} + 4305600944807042405646729216000000 z^{8}$ \\
    &                & $- 19592420104121879821182566400000000 z^{7} + 40278611723066271828975550464000000 z^{6}$ \\
    &                & $- 167063571797235459036769419264000000 z^{5} + 201182637314820358936480186368000000 z^{4}$ \\
    &                & $- 778074657281503273163251777536000000 z^{3} + 413802799198865560264098447360000000 z^{2}$ \\
    &                & $- 1517276930395840387635027640320000000 z\big) + 6 z^{35} - 1152 z^{34} + 119520 z^{33} - 8509680 z^{32}$ \\
    &                & $+ 458788320 z^{31} - 19757831040 z^{30} + 701837740800 z^{29} - 21066239980800 z^{28}$ \\
    &                & $+ 543041428608000 z^{27} - 12205876386816000 z^{26} + 241485684217344000 z^{25}$ \\
    &                & $- 4253121893002752000 z^{24} + 66955636679565312000 z^{23} - 951895632117362688000 z^{22}$ \\
    &                & $+ 12194459595369676800000 z^{21} - 142701366335096217600000 z^{20}$ \\
    &                & $+ 1504759438336186368000000 z^{19} - 14685677756846695710720000 z^{18}$ \\
    &                & $+ 127688610609019091681280000 z^{17} - 1049470402600477936189440000 z^{16}$ \\
    &                & $+ 7464904982694026890444800000 z^{15} - 52152575863012607891865600000 z^{14}$ \\
    &                & $+ 297988039883430268659302400000 z^{13} - 1791289267455092726484172800000 z^{12}$ \\
    &                & $+ 7984051362373351629127680000000 z^{11} - 42042420200551658142302208000000 z^{10}$ \\
    &                & $+ 140549235029833808280865996800000 z^{9} - 663470958491956063449907200000000 z^{8}$ \\
    &                & $+ 1587243544581374302677368832000000 z^{7} - 6868792932907997184965738496000000 z^{6}$ \\
    &                & $+ 11021452119050370536409189580800000 z^{5} - 44541575785040924650466023833600000 z^{4}$ \\
    &                & $+ 42764508095089372737250000896000000 z^{3} - 163729765570460226009258000384000000 z^{2}$ \\
    &                & $+ 70937622719805524616702590976000000 z - 260104616639286923594576166912000000$ \\
\cmidrule{2-3}
    & $\mathfrak{B}^{(n)}$  & $\cfl^{5} \big(332640 z^{30} - 83825280 z^{29} + 5029516800 z^{28} - 274388083200 z^{27} + 19544702284800 z^{26}$ \\
    &                & $- 1021253445273600 z^{25} + 40783305591705600 z^{24} - 1306536651063705600 z^{23}$ \\
    &                & $+ 33725923025743872000 z^{22} - 726153567434330112000 z^{21} + 13279981358383939584000 z^{20}$ \\
    &                & $- 216507322744631525376000 z^{19} + 3299468417886246617088000 z^{18}$ \\
    &                & $- 48870626813761965391872000 z^{17} + 681993250323520392069120000 z^{16}$ \\
    &                & $- 8572335712949421425295360000 z^{15} + 91413239877713218110750720000 z^{14}$ \\
    &                & $- 839760875742003008687308800000 z^{13} + 6482393389617389064243118080000 z^{12}$ \\
    &                & $- 43776173457786164638635786240000 z^{11} + 214348133624186286113292288000000 z^{10}$ \\
    &                & $- 844924797657029414267073331200000 z^{9} - 789634598017870685880188928000000 z^{8}$ \\
    &                & $+ 25599356799049615084214235955200000 z^{7} - 268347194005614546609342775296000000 z^{6}$ \\
    &                & $+ 1500652350011654558937837364838400000 z^{5} - 5160622485160397615960294424576000000 z^{4}$ \\
    &                & $+ 16979379778873451056519169703936000000 z^{3} + 19898003172905449654985076768768000000 z^{2}$ \\
    &                & $- 72959344967319982068278614818816000000 z\big) + \cfl^{4} \big(- 181440 z^{31} + 43545600 z^{30}$ \\
    &                & $- 3200601600 z^{29} + 185787302400 z^{28} - 10269358848000 z^{27} + 495403746969600 z^{26}$ \\
    &                & $- 19994534951731200 z^{25} + 653950393850880000 z^{24} - 17689276593192960000 z^{23}$ \\
    &                & $+ 400657658281574400000 z^{22} - 7713088819945979904000 z^{21} + 129942807411798097920000 z^{20}$ \\
    &                & $- 1958209921235550044160000 z^{19} + 27398134697908431421440000 z^{18}$ \\
    &                & $- 355702716707483271168000000 z^{17} + 4307278174272403451412480000 z^{16}$ \\
    &                & $- 46743219106718995481886720000 z^{15} + 461397139092162956545228800000 z^{14}$ \\
    &                & $- 3928165194685048153649971200000 z^{13} + 30095873111644407939465216000000 z^{12}$ \\
    &                & $- 187274015509055263360627507200000 z^{11} + 1076748731510342919889983897600000 z^{10}$ \\
    &                & $- 4003074762491968669863641088000000 z^{9} + 13903864683937919004976349184000000 z^{8}$ \\
    &                & $+ 20979162885917070460515778560000000 z^{7} - 276142774785937753550825403187200000 z^{6}$ \\
    &                & $+ 2140041859355015531926948675584000000 z^{5} - 9310119584821027786578689261568000000 z^{4}$ \\
    &                & $+ 14867294573247175219150986412032000000 z^{3} - 52767917991346245925106268241920000000 z^{2}$ \\
    &                & $- 17876280925390992203409052925952000000 z + 65546363393100304745833194061824000000\big)$ \\
    &                & $+ \cfl^{3} \big(45360 z^{32} - 10342080 z^{31} + 879076800 z^{30} - 54562636800 z^{29} + 2750688460800 z^{28}$ \\
\bottomrule
\end{tabular}
\end{table}
\begin{table}[h!]
\centering
\caption{Table~\ref{tab:A_and_B_functions_3} continued.}
\label{tab:A_and_B_functions_4}
\begin{tabular}{@{}|p{0.175cm}|p{1.325cm}|p{12.5cm}|@{}}
\toprule
$N$ & formula & expression \\ 
\midrule
    &                & $- 120894952550400 z^{27} + 4593399168153600 z^{26} - 148677055975526400 z^{25}$ \\
    &                & $+ 4083817341348864000 z^{24} - 95305562041245696000 z^{23} + 1913919043099262976000 z^{22}$ \\
    &                & $- 33446582496555540480000 z^{21} + 517310855466925621248000 z^{20}$ \\
    &                & $- 7215782638574968012800000 z^{19} + 91390894616491589468160000 z^{18}$ \\
    &                & $- 1068766589196628346142720000 z^{17} + 11339021739315727000535040000 z^{16}$ \\
    &                & $- 112003548595224074091233280000 z^{15} + 977866902139270617169920000000 z^{14}$ \\
    &                & $- 7966612470873279037125427200000 z^{13} + 54732455103550544817212620800000 z^{12}$ \\
    &                & $- 357763392531995097851913830400000 z^{11} + 1781695816051389823414016409600000 z^{10}$ \\
    &                & $- 9112851499920405992514256896000000 z^{9} + 26559007965775940291351543808000000 z^{8}$ \\
    &                & $- 98374566786214736674677797683200000 z^{7} - 10022767556776019627090116608000000 z^{6}$ \\
    &                & $+ 339725865414335358818158903296000000 z^{5} - 3221799220303233558108174286848000000 z^{4}$ \\
    &                & $+ 13558759251671919591510654320640000000 z^{3} - 17876280925390992203409052925952000000 z^{2}$ \\
    &                & $+ 65546363393100304745833194061824000000 z\big) + \cfl^{2} \big(- 6720 z^{33} + 1451520 z^{32}$ \\
    &                & $- 136684800 z^{31} + 9015552000 z^{30} - 453977395200 z^{29} + 19053435340800 z^{28}$ \\
    &                & $- 679573949644800 z^{27} + 21045349153382400 z^{26} - 564209658335232000 z^{25}$ \\
    &                & $+ 13175620915838976000 z^{24} - 268068624868638720000 z^{23} + 4799501509482086400000 z^{22}$ \\
    &                & $- 76086576938635296768000 z^{21} + 1080536719904798736384000 z^{20}$ \\
    &                & $- 13811097102028318310400000 z^{19} + 160720063277308508897280000 z^{18}$ \\
    &                & $- 1693179302828719826534400000 z^{17} + 16502300243420990464327680000 z^{16}$ \\
    &                & $- 143825952984783261448273920000 z^{15} + 1181907389044948098613248000000 z^{14}$ \\
    &                & $- 8375614871470008652372377600000 z^{13} + 57657355206051894636891340800000 z^{12}$ \\
    &                & $- 318226541561660757776950886400000 z^{11} + 1817834378693369595093319680000000 z^{10}$ \\
    &                & $- 7193974232168965218810986496000000 z^{9} + 34256356546080701745824425574400000 z^{8}$ \\
    &                & $- 83177551414869049919919882240000000 z^{7} + 341320259027311141075304841216000000 z^{6}$ \\
    &                & $- 368818950062541919411947700224000000 z^{5} + 1300922806840610832424139489280000000 z^{4}$ \\
    &                & $+ 570584282164386157293452918784000000 z^{3} - 2738622161970673889747851542528000000 z^{2}$ \\
    &                & $+ 6620844787181848964225575157760000000 z - 24276430886333446202160442245120000000\big)$ \\
    &                & $+ \cfl \big(630 z^{34} - 128520 z^{33} + 12927600 z^{32} - 894650400 z^{31} + 46601503200 z^{30}$ \\
    &                & $- 1960863811200 z^{29} + 68564151129600 z^{28} - 2053676990131200 z^{27}$ \\
    &                & $+ 53209430064000000 z^{26} - 1211538906692352000 z^{25} + 24274021346233344000 z^{24}$ \\
    &                & $- 433377402037171200000 z^{23} + 6883644168701927424000 z^{22} - 98733632857799675904000 z^{21}$ \\
    &                & $+ 1269743079061179555840000 z^{20} - 14922483347614796513280000 z^{19}$ \\
    &                & $+ 157531533421677386711040000 z^{18} - 1540345296518774214819840000 z^{17}$ \\
    &                & $+ 13401721341082606395064320000 z^{16} - 110272613591232711162593280000 z^{15}$ \\
    &                & $+ 784641328687918559821824000000 z^{14} - 5475876995981349638347161600000 z^{13}$ \\
    &                & $+ 31147260089197086561691238400000 z^{12} - 185492317998409209694126080000000 z^{11}$ \\
    &                & $+ 807684207745917206349427507200000 z^{10} - 4148462144684760520348768665600000 z^{9}$ \\
    &                & $+ 12916802834421127216940187648000000 z^{8} - 58777260312365639463547699200000000 z^{7}$ \\
    &                & $+ 120835835169198815486926651392000000 z^{6} - 501190715391706377110308257792000000 z^{5}$ \\
    &                & $+ 603547911944461076809440559104000000 z^{4} - 2334223971844509819489755332608000000 z^{3}$ \\
    &                & $+ 1241408397596596680792295342080000000 z^{2} - 4551830791187521162905082920960000000 z\big)$ \\
    &                & $- 36 z^{35} + 6912 z^{34} - 717120 z^{33} + 51058080 z^{32} - 2752729920 z^{31} + 118546986240 z^{30}$ \\
    &                & $- 4211026444800 z^{29} + 126397439884800 z^{28} - 3258248571648000 z^{27} + 73235258320896000 z^{26}$ \\
    &                & $- 1448914105304064000 z^{25} + 25518731358016512000 z^{24} - 401733820077391872000 z^{23}$ \\
    &                & $+ 5711373792704176128000 z^{22} - 73166757572218060800000 z^{21}$ \\
    &                & $+ 856208198010577305600000 z^{20} - 9028556630017118208000000 z^{19}$ \\
    &                & $+ 88114066541080174264320000 z^{18} - 766131663654114550087680000 z^{17}$ \\
    &                & $+ 6296822415602867617136640000 z^{16} - 44789429896164161342668800000 z^{15}$ \\
    &                & $+ 312915455178075647351193600000 z^{14} - 1787928239300581611955814400000 z^{13}$ \\
    &                & $+ 10747735604730556358905036800000 z^{12} - 47904308174240109774766080000000 z^{11}$ \\
    &                & $+ 252254521203309948853813248000000 z^{10} - 843295410179002849685195980800000 z^{9}$ \\
    &                & $+ 3980825750951736380699443200000000 z^{8} - 9523461267488245816064212992000000 z^{7}$ \\
    &                & $+ 41212757597447983109794430976000000 z^{6} - 66128712714302223218455137484800000 z^{5}$ \\
    &                & $+ 267249454710245547902796143001600000 z^{4} - 256587048570536236423500005376000000 z^{3}$ \\
    &                & $+ 982378593422761356055548002304000000 z^{2} - 425625736318833147700215545856000000 z$ \\
    &                & $+ 1560627699835721541567457001472000000$ \\
\cmidrule{2-3}
    & $\mathfrak{A}^{(d)}$, $\mathfrak{B}^{(d)}$ & $z^{36} - 180 z^{35} + 18540 z^{34} - 1316160 z^{33} + 71571600 z^{32} - 3111013440 z^{31} + 111940591680 z^{30}$ \\
    &                & $- 3395558880000 z^{29} + 88489140768000 z^{28} - 2002141217664000 z^{27} + 39883502784384000 z^{26}$ \\
    &                & $- 703816862386176000 z^{25} + 11134067052681216000 z^{24} - 157996290085079040000 z^{23}$ \\
    &                & $+ 2036218024805529600000 z^{22} - 23656661249676779520000 z^{21}$ \\
    &                & $+ 252579976001911111680000 z^{20} - 2419651963041267548160000 z^{19}$ \\
    &                & $+ 21649076048778130391040000 z^{18} - 170520979618990109491200000 z^{17}$ \\
    &                & $+ 1290447454665883336704000000 z^{16} - 8250721994388540712550400000 z^{15}$ \\
    &                & $+ 53380923626281155467673600000 z^{14} - 270783200020939388013772800000 z^{13}$ \\
    &                & $+ 1522312184321115391682150400000 z^{12} - 5940858072394109964150374400000 z^{11}$ \\
    &                & $+ 29711561267505010021761024000000 z^{10} - 86289160914618786124922880000000 z^{9}$ \\
    &                & $+ 393068793462386061314359296000000 z^{8} - 816216186186879900070182912000000 z^{7}$ \\
    &                & $+ 3453929935318793236593298636800000 z^{6} - 4828878883658251678626349056000000 z^{5}$ \\
    &                & $+ 19274822167971200178377981952000000 z^{4} - 16225325551691055485102850048000000 z^{3}$ \\
    &                & $+ 61801716763466934325157560320000000 z^{2} - 23645874239935174872234196992000000 z$ \\
    &                & $+ 86701538879762307864858722304000000$ \\
\bottomrule
\end{tabular}
\end{table}

The only non-universal component of the obtained relations (\ref{eq:A_and_B_exp_prefinal}) is delta function $\delta_{\mathcal{P}_{N}}(\xi)$ (\ref{eq:delta_func_varphi_action}), the definition of which is essentially related to the choice of a functional basis $\{\varphi_{p}\}_{p}$ (\ref{eq:phi_def}) in space $\mathcal{P}_{N}(\omega)$. However, it is obvious that the delta function $\delta_{\mathcal{P}_{N}}(\xi)$ acting in space $\mathcal{P}_{N}(\omega)$ should not depend on the choice of a functional basis $\{\varphi_{p}\}_{p}$ (\ref{eq:phi_def}) --- only the specific values of the expansion coefficients $\{\ophi_{p}/w_{p}\}_{p}$ (\ref{eq:delta_func_varphi_coeffs}) in terms of the functional basis $\{\varphi_{p}\}_{p}$ (\ref{eq:phi_def}) chosen for representing the function depend on its choice. Therefore, below it is used a representation of the delta function $\delta_{\mathcal{P}_{N}}(\xi)$ in space $\mathcal{P}_{N}(\omega)$ convenient for analytical calculations, which is based on a functional basis $\{\tilde{L}_{k}\}_{k = 0}^{N}$, consisting of shifted Legendre polynomials $\tilde{L}_{k}$ (\ref{eq:phi_def}):
\begin{equation}\label{eq:delta_func_legendre_coeffs}
\delta_{\mathcal{P}_{N}}(\xi) = \sum\limits_{k = 0}^{N} (-1)^{k}(2k+1)\tilde{P}_{k}(\xi).
\end{equation}
It is easy to rigorously prove that this representation (\ref{eq:delta_func_legendre_coeffs}) of the delta function $\delta_{\mathcal{P}_{N}}(\xi)$ is identical to representation (\ref{eq:delta_func_varphi_coeffs}), with the differences arising only from the choice of functional basis. To do this, the expansion coefficients for the delta function $\delta_{\mathcal{P}_{N}}(\xi)$ defined by representation (\ref{eq:delta_func_varphi_coeffs}) is calculated in functional basis $\{\tilde{L}_{k}\}_{k}$:
\begin{equation}
\delta_{\mathcal{P}_{N}}(\xi) = \sum\limits_{k = 0}^{N} c_{k}\tilde{P}_{k}(\xi),\qquad
c_{k} = \frac{\displaystyle\intrefdom{\xi}\,f(\xi)\tilde{P}_{k}(\xi)}{\displaystyle\intrefdom{\xi}\,\tilde{P}_{k}^{2}(\xi)} =
(2k+1)\intrefdom{\xi}\,f(\xi)\tilde{P}_{k}(\xi),
\end{equation}
using the well-known $\mathcal{L}_{2}(\omega)$-norm $\|\tilde{L}_{k}\|_{\mathcal{L}_{2}(\omega)}^{-2} = 2k+1$ for shifted Legendre polynomials. Direct calculation, using property (\ref{eq:delta_func_varphi_action}), yields the following expression for the expansion coefficients:
\begin{equation}
\begin{split}
&\delta^{\tilde{P}}_{k} = (2k+1)\intrefdom{\xi}\,\delta_{\mathcal{P}_{N}}(\xi)\tilde{P}_{k}(\xi) = 
\Big\{\tilde{P}_{k}\in\mathcal{P}_{k}(\omega)\subseteq\mathcal{P}_{N}(\omega)\Big\} = (2k+1)\tilde{P}_{k}(0) = (-1)^{k}(2k+1),
\end{split}
\end{equation}
which concludes the proof. Using well-known expressions for the values of the derivatives of Legendre polynomials $\tilde{L}_{k}^{(m)}$ (recalculated taking into account the shift in the domain of the Legendre polynomials $[-1, +1] \rightarrow [0, 1]$):
\begin{equation}
\tilde{P}_{k}^{(s)}(0) = (-1)^{k-s}\frac{(k+s)!}{s!(k-s)!},\qquad
\tilde{P}_{k}^{(s)}(1) = \frac{(k+s)!}{s!(k-s)!},
\end{equation}
the following expressions for the derivatives of the delta functions at points $\xi = 0$ and $1$ are obtained:
\begin{equation}\label{eq:delta_func_in_0_and_1_legendre_coeffs}
\begin{split}
&\delta_{\mathcal{P}_{N}}^{(s)}(0) = \sum\limits_{k = s}^{N} (-1)^{k} (-1)^{k-s}\frac{(2k+1)(k+s)!}{s!(k-s)!} = (-1)^{s}\sum\limits_{k = s}^{N} \frac{(2k+1)(k+s)!}{s!(k-s)!},\\
&\delta_{\mathcal{P}_{N}}^{(s)}(1) = \sum\limits_{k = 0}^{N} (-1)^{k}(2k+1)\tilde{P}_{k}^{(s)}(1) = \sum\limits_{k = s}^{N} (-1)^{k} \frac{(2k+1)(k+s)!}{s!(k-s)!}.
\end{split}
\end{equation}
Substituting the expression (\ref{eq:delta_func_in_0_and_1_legendre_coeffs}) into the expression (\ref{eq:A_and_B_exp_prefinal}) yields the following expression:\\[-1mm]
\begin{equation}\label{eq:A_and_B_exp_final}
\begin{split}
\mathfrak{A}(\mathrm{CFL}, z) & =
\sum\limits_{s = 0}^{N} \frac{(-\mathrm{CFL})^{s}}{s!}
\Bigg\{\frac{d^{s}}{dz^{s}}\Bigg[\frac{R_{\rm RK}(z) - 1}{z(R_{\rm RK}(z) + 1)}\Bigg]\Bigg\}
\Bigg[\sum\limits_{k = s}^{N} (-1)^{k} \frac{(2k+1)(k+s)!}{s!(k-s)!}\Bigg] \equiv \frac{\mathfrak{A}^{(n)}(\mathrm{CFL}, z)}{\mathfrak{A}^{(d)}(z)},\\
\mathfrak{B}(\mathrm{CFL}, z) & =
\sum\limits_{s = 0}^{N} \frac{\mathrm{CFL}^{s}}{s!}
\Bigg\{\frac{d^{s}}{dz^{s}}\Bigg[\frac{R_{\rm RK}(z) - 1}{z(R_{\rm RK}(z) + 1)}\Bigg]\Bigg\}
\Bigg[\sum\limits_{k = s}^{N} \frac{(2k+1)(k+s)!}{s!(k-s)!}\Bigg] \equiv \frac{\mathfrak{B}^{(n)}(\mathrm{CFL}, z)}{\mathfrak{B}^{(d)}(z)}.
\end{split}
\end{equation}
The expressions (\ref{eq:A_and_B_exp_final}) of the functions $\mathfrak{A}(\mathrm{CFL}, z)$ and $\mathfrak{B}(\mathrm{CFL}, z)$ are computed exactly in symbolic form using the \texttt{sympy} module of the \texttt{python} programming language for polynomial degrees $N = 1, \ldots, 12$. Tables~\ref{tab:A_and_B_functions_1},~\ref{tab:A_and_B_functions_2},~\ref{tab:A_and_B_functions_3} and~\ref{tab:A_and_B_functions_4} present expressions for the numerators $\mathfrak{A}^{(n)}(\mathrm{CFL}, z)$, $\mathfrak{B}^{(n)}(\mathrm{CFL}, z)$ and denominators $\mathfrak{A}^{(d)}(z)$, $\mathfrak{B}^{(d)}(z)$ of the functions $\mathfrak{A}(\mathrm{CFL}, z)$ and $\mathfrak{B}(\mathrm{CFL}, z)$ (\ref{eq:A_and_B_exp_final}) for polynomial degrees $N = 1, \ldots, 5$; these expressions for polynomial degrees $N \geqslant 6$ are very large and are not presented here in their final form, but they can also be easily computed symbolically based on the original expressions (\ref{eq:A_and_B_exp_final}). It is evident from the structure of the expressions (\ref{eq:A_and_B_exp_final}) that the denominators $\mathfrak{A}^{(d)}(z)$ and $\mathfrak{B}^{(d)}(z)$ of functions $\mathfrak{A}(\mathrm{CFL}, z)$ and $\mathfrak{B}(\mathrm{CFL}, z)$ do not depend on the Courant number $\mathrm{CFL}$ but rather depend on $z$, and that they are identical: $\mathfrak{A}^{(d)}(z) \equiv \mathfrak{B}^{(d)}(z) \equiv \mathfrak{d}(z)$. Substituting the expression (\ref{eq:A_and_B_exp_final}) into the expression (\ref{eq:cfl_eq_plynomial_src}) yields the following equation:
\begin{equation}\label{eq:cfl_eq_plynomial_prefinal}
\begin{split}
\tilde{\Phi}_{\pm}(\mathrm{CFL}; N, z) =
\sum\limits_{s = 0}^{N} \frac{1}{s!}\Bigg\{\frac{d^{s}}{dz^{s}}\Bigg[\frac{R_{\rm RK}(z) - 1}{z(R_{\rm RK}(z) + 1)}\Bigg]\Bigg\}
\Bigg[\sum\limits_{k = s}^{N} \Big[(-1)^{k+s} \pm 1\Big]\frac{(2k+1)(k+s)!}{s!(k-s)!}\Bigg]
\mathrm{CFL}^{s+1} - 1 = 0,
\end{split}
\end{equation}
the minimum real solution $\mathrm{CFL}$ to which determines the boundary value of the Courant number $\mathrm{CFL}_{\rm max}$. For computational purposes, it is convenient to use the fully polynomial form of the expression (\ref{eq:cfl_eq_plynomial_prefinal}), obtained by multiplying $\tilde{\Phi}_{\pm}(\mathrm{CFL}; N, z)$ by the common denominator $\mathfrak{d}(z)$; and is written as follows:
\begin{equation}\label{eq:cfl_eq_plynomial_final}
\Phi_{\pm}(\mathrm{CFL}; N, z) \equiv \mathfrak{d}(z)\tilde{\Phi}_{\pm}(\mathrm{CFL}; N, z) =
\mathrm{CFL} \cdot \Big[\mathfrak{B}^{(n)}(\mathrm{CFL}, z) \pm \mathfrak{A}^{(n)}(\mathrm{CFL}, z)\Big] - \mathfrak{d}(z) = 0,
\end{equation}
which is a polynomial of degree $N+1$ for case $\Phi_{+}$ and a polynomial of degree $N$ for case $\Phi_{-}$: $\Phi_{+}(\mathrm{CFL}; N, z) \in \mathcal{P}_{N+1}(\mathcal{R}_{+})$, $\Phi_{-}(\mathrm{CFL}; N, z) \in \mathcal{P}_{N}(\mathcal{R}_{+})$, as can be seen from the form of the coefficients associated with powers of $\mathrm{CFL}$ in expression (\ref{eq:cfl_eq_plynomial_prefinal}). The boundary value of the Courant number $\mathrm{CFL}_{\rm max}$ is determined as follows:\\[-1mm]
\begin{equation}
\mathrm{CFL}_{\rm max}(N, z) =
\min\Big\{\mathrm{CFL}\,\Big|\,
\Phi_{-}(\mathrm{CFL}; N, z) = 0\,\wedge\,\mathrm{CFL}\in\mathcal{R}_{+}\,\vee\,
\Phi_{+}(\mathrm{CFL}; N, z) = 0\,\wedge\,\mathrm{CFL}\in\mathcal{R}_{+}
\Big\}.
\end{equation}
Exact expressions for the function $\Phi_{\pm}(\mathrm{CFL}; N, z)$ (\ref{eq:cfl_eq_plynomial_final}) are calculated symbolically using the expressions for functions $\mathfrak{A}^{(n)}(\mathrm{CFL}, z)$, $\mathfrak{B}^{(n)}(\mathrm{CFL}, z)$ and $\mathfrak{d}(z)$ (\ref{eq:A_and_B_exp_final}), in particular, those presented in Tables~\ref{tab:A_and_B_functions_1},~\ref{tab:A_and_B_functions_2},~\ref{tab:A_and_B_functions_3} and~\ref{tab:A_and_B_functions_4} for polynomial degrees $N = 1, \ldots, 5$. The expressions for the function $\Phi_{\pm}(\mathrm{CFL}; N, z)$ (\ref{eq:cfl_eq_plynomial_final}) for the function in cases $N = 1, 2, 3$ are as follows:
\begin{align}
\Phi_{+}(\mathrm{CFL}; 1, z) & =
\mathrm{CFL}^{2}\cdot\big(12 z^{4} - 168 z^{3} + 432 z^{2} - 1728 z\big)\\[-1mm] \nonumber &
+ \mathrm{CFL}\cdot\big(- 2 z^{5} + 20 z^{4} - 104 z^{3} + 432 z^{2} - 864 z + 1728\big)\\[-1mm] \nonumber &
-z^{6} + 6 z^{5} - 48 z^{4} + 152 z^{3} - 576 z^{2} + 864 z - 1728,\\[-1mm] \nonumber
\Phi_{-}(\mathrm{CFL}; 1, z) & = \mathrm{CFL}\cdot\big(36 - 6 z\big) - z^{2} + 2 z - 12,\\[-1mm] \nonumber
\Phi_{+}(\mathrm{CFL}; 2, z) & =
\mathrm{CFL}^{3}\cdot\big(- 120 z^{15} + 6480 z^{14} - 198720 z^{13} + 4366080 z^{12} - 64800000 z^{11} + 624049920 z^{10}\\[-1mm] \nonumber &
- 4045662720 z^{9} + 19532067840 z^{8} - 76359075840 z^{7} + 228296724480 z^{6} - 471636172800 z^{5}\\[-1mm] \nonumber &
+ 836075520000 z^{4} + 398131200000 z^{3} - 2239488000000 z^{2} + 8957952000000 z\\[-1mm] \nonumber &
- 14929920000000\big) + \mathrm{CFL}^{2}\cdot \big(12 z^{16} - 720 z^{15} + 20736 z^{14} - 393408 z^{13} + 5346432 z^{12}\\[-1mm] \nonumber &
- 51922944 z^{11} + 367842816 z^{10} - 2072770560 z^{9} + 9841056768 z^{8} - 37868580864 z^{7}\\[-1mm] \nonumber &
+ 124744458240 z^{6} - 353839104000 z^{5} + 758937600000 z^{4} - 1642291200000 z^{3}\\[-1mm] \nonumber &
+ 1791590400000 z^{2} - 2985984000000 z\big) + \mathrm{CFL}\cdot\big(- 12 z^{17} + 792 z^{16} - 23040 z^{15} + 400320 z^{14}\\[-1mm] \nonumber &
- 4803840 z^{13} + 43473024 z^{12} - 311192064 z^{11} + 1830159360 z^{10} - 9189780480 z^{9}\\[-1mm] \nonumber &
+ 39276472320 z^{8} - 144340475904 z^{7} + 474441007104 z^{6} - 1255685898240 z^{5}\\[-1mm] \nonumber &
+ 3266666496000 z^{4} - 5733089280000 z^{3} + 11943936000000 z^{2} - 10749542400000 z\\[-1mm] \nonumber &
+ 17915904000000\big) - z^{18} + 72 z^{17} - 2232 z^{16} + 39600 z^{15} - 460080 z^{14} + 3920832 z^{13}\\[-1mm] \nonumber &
- 27082944 z^{12} + 157469184 z^{11} - 772104960 z^{10} + 3324119040 z^{9} - 12525207552 z^{8}\\[-1mm] \nonumber &
+ 40453115904 z^{7} - 121955549184 z^{6} + 286833623040 z^{5} - 706185216000 z^{4}\\[-1mm] \nonumber &
+ 1104814080000 z^{3} - 2239488000000 z^{2} + 1791590400000 z - 2985984000000,\\[-1mm] \nonumber
\Phi_{-}(\mathrm{CFL}; 2, z) & =
\mathrm{CFL}^{2}\cdot\big(60 z^{7} - 1440 z^{6} + 23760 z^{5} - 260640 z^{4} + 1123200 z^{3} - 2592000 z^{2} + 8640000 z\big)\\[-1mm] \nonumber &
+ \mathrm{CFL}\cdot\big(- 6 z^{8} + 180 z^{7} - 2232 z^{6} + 17856 z^{5} - 97632 z^{4} + 316224 z^{3} - 1278720 z^{2}\\[-1mm] \nonumber &
+ 1555200 z - 5184000\big) - z^{9} + 36 z^{8} - 468 z^{7} + 2952 z^{6} - 14256 z^{5} + 65664 z^{4} - 148608 z^{3}\\[-1mm] \nonumber &
+ 570240 z^{2} - 518400 z + 1728000,\\[-1mm] \nonumber
\Phi_{+}(\mathrm{CFL}; 3, z) & =
\mathrm{CFL}^{4}\cdot\big(1680 z^{36} - 255360 z^{35} + 16732800 z^{34} - 749548800 z^{33} + 25662067200 z^{32}\\[-1mm] \nonumber &
- 727863413760 z^{31} + 17821229368320 z^{30} - 387965555712000 z^{29} + 7606656884736000 z^{28}\\[-1mm] \nonumber &
- 134922572273664000 z^{27} + 2159892993691238400 z^{26} - 31078412348910796800 z^{25}\\[-1mm] \nonumber &
+ 400173570114232320000 z^{24} - 4595302612751155200000 z^{23}\\[-1mm] \nonumber & + 46935945819587543040000 z^{22} - 425550423678430740480000 z^{21}\\[-1mm] \nonumber &
+ 3418617675457794539520000 z^{20} - 24292835770440351744000000 z^{19}\\[-1mm] \nonumber &
+ 152359060817785651200000000 z^{18} - 841785868284418326528000000 z^{17}\\[-1mm] \nonumber &
+ 4078059462318751167283200000 z^{16} - 17249151267709494755328000000 z^{15}\\[-1mm] \nonumber &
+ 62448828578904790794240000000 z^{14} - 189151896857673392455680000000 z^{13}\\[-1mm] \nonumber &
+ 423238269137221110988800000000 z^{12} - 484259918898436987944960000000 z^{11}\\[-1mm] \nonumber &
- 1994153298048168996372480000000 z^{10} + 14601189837007025602560000000000 z^{9}\\[-1mm] \nonumber &
- 65127367713162050666496000000000 z^{8} + 207766698590074621132800000000000 z^{7}\\[-1mm] \nonumber &
- 554267127114989964361728000000000 z^{6} + 1285567340305654178906112000000000 z^{5}\\[-1mm] \nonumber &
- 2222973201138625711964160000000000 z^{4} + 3987081725469065950003200000000000 z^{3}\\[-1mm] \nonumber &
- 3581977165127289195724800000000000 z^{2} + 5014768031178204874014720000000000 z\big)\\[-1mm] \nonumber &
+ \mathrm{CFL}^{3}\cdot\big(- 120 z^{37} + 17280 z^{36} - 1183680 z^{35} + 55785600 z^{34} - 2002003200 z^{33}\\[-1mm] \nonumber &
+ 58580029440 z^{32} - 1447386140160 z^{31} + 31057944514560 z^{30} - 588585056256000 z^{29}\\[-1mm] \nonumber &
+ 9982882188288000 z^{28} - 152569772942745600 z^{27} + 2110230493492838400 z^{26}\\[-1mm] \nonumber &
- 26424741923057664000 z^{25} + 299436123786485760000 z^{24} - 3063603311939911680000 z^{23}\\[-1mm] \nonumber &
+ 28255463078071173120000 z^{22} - 234429501123663298560000 z^{21}\\[-1mm] \nonumber &
+ 1748886306356194836480000 z^{20} - 11724416550426968064000000 z^{19}\\[-1mm] \nonumber &
+ 70746538375093616640000000 z^{18} - 384434082357094829260800000 z^{17}\\[-1mm] \nonumber &
+ 1887469717177543163904000000 z^{16} - 8359974075786443882496000000 z^{15}\\[-1mm] \nonumber &
+ 33572052916023444111360000000 z^{14} - 120965955870340310630400000000 z^{13}\\[-1mm] \nonumber &
+ 397134843304421558845440000000 z^{12} - 1140552126242616414044160000000 z^{11}\\[-1mm] \nonumber &
+ 3021408997160997460377600000000 z^{10} - 6476622156248616271872000000000 z^{9}\\[-1mm] \nonumber &
+ 13135453145420116525056000000000 z^{8} - 15984638626893637091328000000000 z^{7}\\[-1mm] \nonumber &
+ 16247351149153247821824000000000 z^{6} + 35961965403272556576768000000000 z^{5}\\[-1mm] \nonumber &
- 127057499054563708108800000000000 z^{4} + 336136705546748166144000000000000 z^{3}\\[-1mm] \nonumber &
- 627759773582257060577280000000000 z^{2} + 639638779487015927808000000000000 z\\[-1mm] \nonumber &
- 895494291281822298931200000000000\big) + \mathrm{CFL}^{2}\cdot\big(180 z^{38} - 24480 z^{37} + 1719360 z^{36}\\[-1mm] \nonumber &
- 83730240 z^{35} + 3121113600 z^{34} - 94460117760 z^{33} + 2397389736960 z^{32}\\[-1mm] \nonumber &
- 52266778398720 z^{31} + 994115970017280 z^{30} - 16707227277312000 z^{29}\\[-1mm] \nonumber &
+ 250243555556966400 z^{28} - 3365398257817190400 z^{27} + 40827343785504768000 z^{26}\\[-1mm] \nonumber &
- 448636925844209664000 z^{25} + 4473793168807034880000 z^{24} - 40564892602158612480000 z^{23}\\[-1mm] \nonumber &
+ 334465200695256514560000 z^{22} - 2511117496070252789760000 z^{21}\\[-1mm] \nonumber &
+ 17160448519376850124800000 z^{20} - 107015430324142669824000000 z^{19}\\[-1mm] \nonumber &
+ 608770837794513302323200000 z^{18} - 3173883876738376925184000000 z^{17}\\[-1mm] \nonumber &
+ 15127808265487182200832000000 z^{16} - 66434046160003129147392000000 z^{15}\\[-1mm] \nonumber &
+ 266463729930430867046400000000 z^{14} - 991999753255903203164160000000 z^{13}\\[-1mm] \nonumber &
+ 3343123970808237690716160000000 z^{12} - 10597720570081165128499200000000 z^{11}\\[-1mm] \nonumber &
+ 29690124363571758995865600000000 z^{10} - 80321651114321582751744000000000 z^{9}\\[-1mm] \nonumber &
+ 182503810418637822492672000000000 z^{8} - 421358183112310159048704000000000 z^{7}\\[-1mm] \nonumber &
+ 739178368663476234092544000000000 z^{6} - 1453024337646933529067520000000000 z^{5}\\[-1mm] \nonumber &
+ 1775976652024663611801600000000000 z^{4} - 2957872470284957940449280000000000 z^{3}\\[-1mm] \nonumber &
+ 1918916338461047783424000000000000 z^{2} - 2686482873845466896793600000000000 z\big)\\[-1mm] \nonumber &
+ \mathrm{CFL}\cdot\big(- 12 z^{39} + 1536 z^{38} - 108288 z^{37} + 5341824 z^{36} - 203565312 z^{35} + 6310462464 z^{34}\\[-1mm] \nonumber &
- 164233723392 z^{33} + 3664847586816 z^{32} - 71152085587968 z^{31} + 1214970242678784 z^{30}\\[-1mm] \nonumber &
- 18393287614709760 z^{29} + 248450059163566080 z^{28} - 3009070516506624000 z^{27}\\[-1mm] \nonumber &
+ 32817263071798886400 z^{26} - 323400955469384908800 z^{25} + 2889670537495019520000 z^{24}\\[-1mm] \nonumber &
- 23476590294934781952000 z^{23} + 173994742163412615168000 z^{22}\\[-1mm] \nonumber &
- 1179128443499878809600000 z^{21} + 7332117628984845926400000 z^{20}\\[-1mm] \nonumber &
- 41892779691490101166080000 z^{19} + 220853233490559447859200000 z^{18}\\[-1mm] \nonumber &
- 1073286872046497798553600000 z^{17} + 4839046781142596006707200000 z^{16}\\[-1mm] \nonumber &
- 20108818055560243131187200000 z^{15} + 78056620612183375478784000000 z^{14}\\[-1mm] \nonumber &
- 277489101062922873864192000000 z^{13} + 931688405793147086438400000000 z^{12}\\[-1mm] \nonumber &
- 2816667979711066889256960000000 z^{11} + 8202822555628779727749120000000 z^{10}\\[-1mm] \nonumber &
- 20780807334252307808256000000000 z^{9} + 52544171124907584572620800000000 z^{8}\\[-1mm] \nonumber &
- 108443713131041042674483200000000 z^{7} + 237894476179223441571840000000000 z^{6}\\[-1mm] \nonumber &
- 379587386485662076108800000000000 z^{5} + 720650988986942883889152000000000 z^{4}\\[-1mm] \nonumber &
- 799940089938047674613760000000000 z^{3} + 1308518188893438298030080000000000 z^{2}\\[-1mm] \nonumber &
- 767566535384419113369600000000000 z + 1074593149538186758717440000000000\big)\\[-1mm] \nonumber &
- z^{40} + 120 z^{39} - 8280 z^{38} + 402960 z^{37} - 15290160 z^{36} + 473926464 z^{35} - 12384239040 z^{34}\\[-1mm] \nonumber &
+ 278114031360 z^{33} - 5445864472320 z^{32} + 93891229470720 z^{31} - 1436474447437824 z^{30}\\[-1mm] \nonumber &
+ 19609123496140800 z^{29} - 239976674644377600 z^{28} + 2641902598228377600 z^{27}\\[-1mm] \nonumber &
- 26254645119553536000 z^{26} + 236152028132868096000 z^{25} - 1928992444800196608000 z^{24}\\[-1mm] \nonumber &
+ 14346764104655831040000 z^{23} - 97534606940350709760000 z^{22}\\[-1mm] \nonumber &
+ 607527517846593208320000 z^{21} - 3482870697435060633600000 z^{20}\\[-1mm] \nonumber &
+ 18391393738961230233600000 z^{19} - 89961428296817403494400000 z^{18}\\[-1mm] \nonumber &
+ 406439961686745808896000000 z^{17} - 1712350519246231437312000000 z^{16}\\[-1mm] \nonumber &
+ 6650551710735102325555200000 z^{15} - 24301979912423834910720000000 z^{14}\\[-1mm] \nonumber &
+ 81086703584584390410240000000 z^{13} - 258231683825243038679040000000 z^{12}\\[-1mm] \nonumber &
+ 734881491229627278950400000000 z^{11} - 2045572537398466078310400000000 z^{10}\\[-1mm] \nonumber &
+ 4885928568959785972531200000000 z^{9} - 11901623692792185225216000000000 z^{8}\\[-1mm] \nonumber &
+ 23164931408666636058624000000000 z^{7} - 49361145729292475301888000000000 z^{6}\\[-1mm] \nonumber &
+ 74197166002447948185600000000000 z^{5} - 138033613484876617482240000000000 z^{4}\\[-1mm] \nonumber &
+ 143983994647791544565760000000000 z^{3} - 233011269670270087987200000000000 z^{2}\\[-1mm] \nonumber &
+ 127927755897403185561600000000000 z - 179098858256364459786240000000000,\\[-1mm] \nonumber
\Phi_{-}(\mathrm{CFL}; 3, z) & =
\mathrm{CFL}^{3}\cdot\big(- 840 z^{21} + 80640 z^{20} - 3084480 z^{19} + 86728320 z^{18} - 1816819200 z^{17}\\[-1mm] \nonumber &
+ 32799513600 z^{16} - 505097510400 z^{15} + 6921776332800 z^{14} - 81031249152000 z^{13}\\[-1mm] \nonumber &
+ 809050346496000 z^{12} - 6639420275712000 z^{11} + 44786963128320000 z^{10}\\[-1mm] \nonumber &
- 241527005798400000 z^{9} + 1068425853788160000 z^{8} - 3635045251891200000 z^{7}\\[-1mm] \nonumber &
+ 11475270284083200000 z^{6} - 19670056132608000000 z^{5} + 43741213802496000000 z^{4}\\[-1mm] \nonumber &
+ 17779982008320000000 z^{3} - 93947130740736000000 z^{2} + 337246110351360000000 z\\[-1mm] \nonumber &
- 786907590819840000000\big) + \mathrm{CFL}^{2}\cdot\big(60 z^{22} - 5280 z^{21} + 224640 z^{20} - 6765120 z^{19}\\[-1mm] \nonumber &
+ 150917760 z^{18}- 2741472000 z^{17} + 40690598400 z^{16} - 519344179200 z^{15}\\[-1mm] \nonumber &
+ 5621166720000 z^{14} - 53342143488000 z^{13} + 428060132352000 z^{12} - 2992271984640000 z^{11}\\[-1mm] \nonumber &
+ 17213770598400000 z^{10} - 87226587832320000 z^{9} + 358152357888000000 z^{8}\\[-1mm] \nonumber &
- 1394889884467200000 z^{7} + 3935255585587200000 z^{6} - 12448913670144000000 z^{5}\\[-1mm] \nonumber &
+ 21811203735552000000 z^{4} - 58387166724096000000 z^{3} + 48178015764480000000 z^{2}\\[-1mm] \nonumber &
- 112415370117120000000 z\big) + \mathrm{CFL}\cdot\big(- 20 z^{23} + 1600 z^{22} - 72000 z^{21} + 2246400 z^{20}\\[-1mm] \nonumber &
- 53097600 z^{19} + 999344640 z^{18} - 15313996800 z^{17} + 195503270400 z^{16} - 2090128665600 z^{15}\\[-1mm] \nonumber &
+ 19054126080000 z^{14} - 147538990080000 z^{13} + 995457116160000 z^{12}\\[-1mm] \nonumber &
- 5732266752000000 z^{11} + 29645451878400000 z^{10} - 128115368755200000 z^{9}\\[-1mm] \nonumber &
+ 530896508928000000 z^{8} - 1693026717696000000 z^{7} + 5844285997056000000 z^{6}\\[-1mm] \nonumber &
- 13092663951360000000 z^{5} + 38815920783360000000 z^{4} - 54863944482816000000 z^{3}\\[-1mm] \nonumber &
+ 143004586475520000000 z^{2} - 96356031528960000000 z + 224830740234240000000\big)\\[-1mm] \nonumber &
- z^{24} + 72 z^{23} - 3240 z^{22} + 100080 z^{21} - 2405520 z^{20} + 45772992 z^{19} - 718853184 z^{18}\\[-1mm] \nonumber &
+ 9320866560 z^{17} - 102092797440 z^{16} + 937197250560 z^{15} - 7380937728000 z^{14}\\[-1mm] \nonumber &
+ 49137891840000 z^{13} - 288774291456000 z^{12} + 1442780190720000 z^{11}\\[-1mm] \nonumber &
- 6691567749120000 z^{10} + 25266540625920000 z^{9} - 96850440806400000 z^{8}\\[-1mm] \nonumber &
+ 271090918195200000 z^{7} - 888776346009600000 z^{6} + 1753083150336000000 z^{5}\\[-1mm] \nonumber &
- 5042070473932800000 z^{4} + 6289361377689600000 z^{3} - 16174048149504000000 z^{2}\\[-1mm] \nonumber &
+ 9635603152896000000 z - 22483074023424000000.
\end{align}
The expressions for the function $\Phi_{\pm}(\mathrm{CFL}; N, z)$ (\ref{eq:cfl_eq_plynomial_final}) for polynomial degrees $N \geqslant 4$ are very large and are not presented here in their final form, but they can also be computed symbolically based on the original expressions (\ref{eq:A_and_B_exp_final} and (\ref{eq:cfl_eq_plynomial_final}). Note that in the case of polynomial degree $N = 12$, the coefficients at the highest power of the Courant number $\mathrm{CFL}$ have the form of a polynomial in $z$ of degree $1170$ and the leading coefficient is $-64764752532480000$. The values of the coefficients at $-z \sim 10^{1}$--$10^{2}$ have significantly offsetting values, which can lead to the well-known problem of catastrophic cancellation. The calculations to obtain these results are performed using a developed program in the \texttt{python} programming language. Floating-point calculations are performed using the \texttt{mpmath} module~\cite{mpmath_man} (with \texttt{gmp2} module), which enables calculations with arbitrarily high precision representation of real numbers as floating-point numbers. The presented results are calculated using \texttt{mpmath.mp.dps = 5000}. The calculation of roots of $\Phi_{\pm}(\mathrm{CFL}; N, z) = 0$ (\ref{eq:cfl_eq_plynomial_final}) is carried out using function \texttt{mpmath.polyroots} (implements the Durand-Kerner method~\cite{DurandKerner_method_0, DurandKerner_method_1, DurandKerner_method_2, DurandKerner_method_3}) with parameters \texttt{extraprec=200} and \texttt{maxsteps=100000}. Note that this problem could have been solved by directly calculating the coefficients associated with powers of $\mathrm{CFL}$ in the expression (\ref{eq:cfl_eq_plynomial_prefinal}) using automatic differentiation~\cite{autodiff_1, autodiff_2} of the function $\mathfrak{L}(z)$ rather than by intensive calculations with arbitrary-precision floating-point arithmetic. This would have allowed for the significant increase in the powers of the parameter $z$ and the coefficients in the corresponding expressions to be addressed. However, in the present case, this approach seemed rather excessively complicated for the problem, and no significant increase in computational costs is observed when using arbitrary-precision floating-point arithmetic (at least up to polynomial degrees $N \leqslant 12$).

\begin{table*}[h!]
\centering
\small
\caption{%
Parameter $|z_{\rm min}|$ at which the minimum boundary value of the Courant number $\mathrm{CFL}_{\rm max}(N, z_{\rm min})$ is reached; the boundary value of the Courant number $\mathrm{CFL}_{\rm max}(N, 0)$ in the case of a linear advection equation ($\nu = 0$) and their ratio $\mathrm{CFL}_{\rm max}(N, z_{\rm min})/\mathrm{CFL}_{\rm max}(N, 0)$; the slope coefficient $r(N)$ of the linear dependence $\mathrm{CFL}_{\rm max}(N, z) \approx r(N)|z|$ in the region $|z| \gg 1$; ratios $2r(N)/\mathrm{CFL}_{\rm max}(N, 0)$ and $2r(N)/\mathrm{CFL}_{\rm max}(N, z_{\rm min})$ for comparison $r(N)$ with the expected qualitative estimate (\ref{eq:cfls_asymp_est}).
}
\label{tab:min_cfls_on_z}
\begin{tabular}{@{}|c|c|c|c|c|c|c|c|@{}}
\toprule
$N$ & $|z_{\rm min}|$ & $\mathrm{CFL}_{\rm max}(N, z_{\rm min})$ &
$\mathrm{CFL}_{\rm max}(N, 0)$ & $\cfrac{\mathrm{CFL}_{\rm max}(N, z_{\rm min})}{\mathrm{CFL}_{\rm max}(N, 0)}$ & $r(N)$ &
$\cfrac{2 r(N)}{\mathrm{CFL}_{\rm max}(N, z_{\rm min})}$ &
$\cfrac{2 r(N)}{\mathrm{CFL}_{\rm max}(N, 0)}$ \\ 
\midrule
$1$  & $0.000$ & $0.33333333$ & $0.33333333$ & $1.00000000$ & $0.16664$ & $0.99984$ & $0.99984$ \\
$2$  & $0.202$ & $0.17029099$ & $0.17082039$ & $0.99690083$ & $0.07389$ & $0.86781$ & $0.86512$ \\
$3$  & $0.385$ & $0.10274852$ & $0.10392896$ & $0.98864187$ & $0.04167$ & $0.81111$ & $0.80189$ \\
$4$  & $0.515$ & $0.06840530$ & $0.06983091$ & $0.97958490$ & $0.02688$ & $0.78590$ & $0.76986$ \\
$5$  & $0.595$ & $0.04869690$ & $0.05011556$ & $0.97169236$ & $0.01861$ & $0.76432$ & $0.74268$ \\
$6$  & $0.647$ & $0.03638936$ & $0.03770124$ & $0.96520337$ & $0.01385$ & $0.76121$ & $0.73472$ \\
$7$  & $0.688$ & $0.02820580$ & $0.02938384$ & $0.95990861$ & $0.01048$ & $0.74311$ & $0.71332$ \\
$8$  & $0.718$ & $0.02249531$ & $0.02354150$ & $0.95555967$ & $0.00846$ & $0.75216$ & $0.71873$ \\
$9$  & $0.748$ & $0.01835533$ & $0.01928191$ & $0.95194541$ & $0.00670$ & $0.73003$ & $0.69495$ \\
$10$ & $0.764$ & $0.01525958$ & $0.01608128$ & $0.94890351$ & $0.00571$ & $0.74838$ & $0.71014$ \\
$11$ & $0.780$ & $0.01288480$ & $0.01361576$ & $0.94631569$ & $0.00465$ & $0.72178$ & $0.68303$ \\
$12$ & $0.796$ & $0.01102360$ & $0.01167641$ & $0.94409172$ & $0.00411$ & $0.74567$ & $0.70398$ \\
\bottomrule
\end{tabular}
\end{table*}

To study the influence of stiff source terms in the advection-reaction equation (\ref{eq:adv_eq_src}) the range of the parameter $-z \in [10^{-6},\, 10^{3}]$ is specifically selected. This range of the parameter $z$ covers both cases $|z| \ll 1$ where the influence of the source terms stiffness can be practically neglected (at short times $t \ll 1/\nu$), and cases $|z| \gg 1$ of extremely stiff source terms, and of course covers intermediate cases $|z| \sim 1$. The obtained dependencies of the boundary values of the Courant number $\mathrm{CFL}_{\rm max}$ on the parameter $z$ for polynomial degrees $N = 1, \ldots, 12$ are presented in Figure~\ref{fig:cfls_on_z}. It is immediately worth noting that the result expected from the qualitative study in Subsection~\ref{sec:stab_anal:empir_cons} ``Empirical considerations'' a significant increase in the boundary values of the Courant number $\mathrm{CFL}_{\rm max}$ for the large values of $z$ --- is confirmed by the obtained results. The boundary values of the Courant number $\mathrm{CFL}_{\rm max}$ correspond to the values $\mathrm{CFL}_{\rm max}(N, 0)$ obtained in~\cite{ader_dg_stab_rig} for the case of a linear advection equation ($\nu = 0$) for the small $|z| \ll 1$. Between these regions $|z| \ll 1$ and $|z| \gg 1$ there is a small range $|z| \sim 1$ in which the boundary value of the Courant number $\mathrm{CFL}_{\rm max}(N, z)$ decreases slightly, reaching a minimum value $\mathrm{CFL}_{\rm max}(N, z_{\rm min})$ at the point $z_{\rm min}$. Table~\ref{tab:min_cfls_on_z} presents the calculated $z_{\rm min}(N)$ and the minimum values of the boundary value of the Courant number $\mathrm{CFL}_{\rm max}(N, z_{\rm min})$ at the point $z_{\rm min}$. For convenient comparison, the boundary values of the Courant number $\mathrm{CFL}_{\rm max}(N, 0)$ for the linear advection equation ($\nu = 0$) and the relative decrease $\mathrm{CFL}_{\rm max}(N, z_{\rm min})/\mathrm{CFL}_{\rm max}(N, 0)$ in the boundary value of the Courant number $\mathrm{CFL}_{\rm max}(N, z_{\rm min})$ at the point $z_{\rm min}$ are also presented. The presented results show that in the case of polynomial degree $N = 1$, there is no decrease in the boundary value of the Courant number $\mathrm{CFL}_{\rm max}(N, z)$ with an increase in the parameter $|z|$ in region $|z| \sim 1$ --- the boundary value of the Courant number $\mathrm{CFL}_{\rm max}(N, z)$ immediately begins to increase with an increase in the parameter $|z|$. With an increase of the polynomial degree $N$, the parameter $|z_{\rm min}|$ gradually increases, starting from a value $0.202$ for polynomial degree $N = 2$ up to a value $0.796$ for polynomial degree $N = 12$, while the growth rate of $|z_{\rm min}|$ significantly decreases with an increase in the polynomial degree $N$, and it can be expected that the parameter $|z_{\rm min}|$ will not exceed $1$ for all values of the polynomial degrees $N$. The relative decrease in the boundary value of the Courant number $\mathrm{CFL}_{\rm max}(N, z_{\rm min})$ at the point $z_{\rm min}$ increases with the polynomial degree $N$, however, it is very small and does not exceed $6$\% even in the case of a polynomial degree $N = 12$. From the insets presented in Figure~\ref{fig:cfls_on_z}, it is clear that the main region of decrease in the boundary value of the Courant number $\mathrm{CFL}_{\rm max}(N, z)$ at point $z_{\rm min}$ is concentrated in the close vicinity of $|z - z_{\rm min}| \ll 1$. A nearly linear increase in the boundary value of the Courant number $\mathrm{CFL}_{\rm max}(N, z) \propto -z$ is observed for the large $|z| \gg 1$, which becomes asymptotically linear in the region of very large $|z| \sim 10^{2}$--$10^{3}$. This asymptotic behavior could be expected from the asymptotic properties of equation $\Phi_{\pm}(\mathrm{CFL}; N, z) = 0$ (\ref{eq:cfl_eq_plynomial_final}), where the following is seen: the maximum power of the parameter $z$ in the coefficient at the highest power of the Courant number $\mathrm{CFL}$ (and for large Courant number $\mathrm{CFL} > 1$ or, especially, very large Courant number $\mathrm{CFL} \gg 1$, these coefficients provide the main contribution to the ``$\mathrm{CFL}$-dependent'' part of the expressions $\Phi_{\pm}(\mathrm{CFL}; N, z)$) is always $N+1$ less for $\Phi_{+}(\mathrm{CFL}; N, z)$ and $N$ less for $\Phi_{-}(\mathrm{CFL}; N, z)$ than the maximum power of the parameter $z$ in the coefficient without the Courant number $\mathrm{CFL}$ (free term), so equations $\Phi_{+}(\mathrm{CFL}; N, z) = 0$ and $\Phi_{-}(\mathrm{CFL}; N, z) = 0$ in the region of very large $|z| \gg 1$ take the following asymptotic forms:
\begin{equation}
\begin{split}
&\Phi_{+}(\mathrm{CFL}; N, z) \sim a_{+} z^{c_{+}} \mathrm{CFL}^{N+1} - b_{+} z^{c_{+}+N+1} = 0,\\
&\Phi_{-}(\mathrm{CFL}; N, z) \sim a_{-} z^{c_{-}} \mathrm{CFL}^{N} - b_{-} z^{c_{-}+N} = 0,
\end{split}
\end{equation}
where $a_{+}$, $a_{-}$, $b_{+}$, $b_{-}$ are the leading coefficients of the polynomial dependencies on the parameter $z$ of the coefficients at $\mathrm{CFL}^{N+1}$ (for $\Phi_{+}(\mathrm{CFL}; N, z)$), $\mathrm{CFL}^{N}$ (for $\Phi_{-}(\mathrm{CFL}; N, z)$), $\mathrm{CFL}^{0}$, and $c_{+}$ and $c_{-}$ are the highest powers of these polynomial dependencies on the parameter $z$. This asymptotic behavior immediately implies a linear dependence of the boundary value of the Courant number $\mathrm{CFL}_{\rm max}(N, z) \propto -z$ in $|z| \gg 1$. This result also unexpectedly confirms results $\mathrm{CFL}_{\rm max}^{\rm est}(N, z)$ (\ref{eq:cfls_asymp_est}), expected from the qualitative study in Subsection~\ref{sec:stab_anal:empir_cons} ``Empirical considerations''. To quantitatively characterize the linear growth of the boundary value of the Courant number $\mathrm{CFL}_{\rm max}(N, z)$ on the parameter $|z|$, the slope coefficient of the linear dependence $r(N)$ is calculated: $\mathrm{CFL}_{\rm max}(N, z) \approx r(N) |z|$. The qualitative estimates $\mathrm{CFL}_{\rm max}^{\rm est}(N, z)$, presented in expression (\ref{eq:cfls_asymp_est}), assumed $r = \mathrm{CFL}_{\rm max}^{\rm est}(0)/2$. Table~\ref{tab:min_cfls_on_z} presents the calculated slope coefficient $r(N)$ for polynomial degrees $N = 1, \ldots, 12$. It can be immediately concluded that as the polynomial degree $N$ increases, the ``rate of growth'' $r(N)$ of the boundary value of the Courant number $\mathrm{CFL}_{\rm max}(N, z)$ decreases with increasing parameter $|z|$. In the case of qualitative estimates (\ref{eq:cfls_asymp_est}), it is necessary to take into account the presence of a multiplier $\mathrm{CFL}_{\rm max}(N, 0)$ (which can be chosen as either $\mathrm{CFL}_{\rm max}(N, 0)$ or $\mathrm{CFL}_{\rm max}(N, z_{\rm min})$), which is equal $1$ in the case of the classical IMEX scheme (\ref{eq:qcs:imex_fd_method}). Therefore, for convenient comparison, Table~\ref{tab:min_cfls_on_z} also presents the ratios $2r(N)/\mathrm{CFL}_{\rm max}(N, 0)$ and $2r(N)/\mathrm{CFL}_{\rm max}(N, z_{\rm min})$, which quantitatively characterize the accuracy of the qualitative estimate (\ref{eq:cfls_asymp_est}). It is evident that in the case of polynomial degree $N = 1$, there is an almost perfect match with the initial qualitative estimate (it should be noted that the  ADER-DG numerical method with the LST-DG predictor in the case of polynomial degree $N = 1$ has nothing in common in terms of its internal structure with the classical finite-difference IMEX scheme (\ref{eq:qcs:imex_fd_method}) considered in Subsection~\ref{sec:stab_anal:empir_cons} ``Empirical considerations''). However, as the polynomial degree $N$ increases, differences with the estimates (\ref{eq:qcs:imex_fd_method}) arise, but they stop systematically increasing starting from polynomial degree $N = 8$ and do not exceed approximately $26$\%. Therefore, the qualitative estimates (\ref{eq:cfls_asymp_est}) can be considered sufficiently correct. From this conclusion it follows that the decrease in the growth rate of the boundary value of the Courant number $\mathrm{CFL}_{\rm max}(N, z)$ on the parameter $|z|$ is associated primarily with the decrease in the Courant number $\mathrm{CFL}_{\rm max}(N, 0) \propto 1/(N+1)^{2}$ with the increase in the polynomial degree $N$.

\section{Applications}
\label{sec:app}

\subsection{Linear advection-reaction equation}
\label{sec:app:lin_adv}

The study of the convergence of the numerical ADER-DG method with the LST-DG predictor based on the solution of the linear advection-reaction equation in the original formulation (\ref{eq:adv_eq_src}) is meaningless, since for the large times $t_{f} - t_{0}$ and large discretization time steps $\Dtn{n}$ the solution becomes very small: even for typical values $\nu = -10^{2}$ and $t_{f} - t_{0} = 10^{4}$, it turns out to $|u(x, t_{f})| = \exp(\nu (t_{f} - t_{0}))|u_{0}(x - a(t_{f} - t_{0}))| \sim \exp(-10^{6}) \approx 3.7 \cdot 10^{-434295}$, which is somewhat lower than the minimum positive normalized value $\mathtt{DBL\_MIN} \approx 2.2 \cdot 10^{-308}$ for representing a real number as a double-precision floating-point number type \texttt{double}. Usually~\cite{Hairer_book_2, Dekker_Verwer_1984}, in such real-world problems~\cite{ader_stiff_1, ader_stiff_2, ader_stiff_3, ader_stiff_4}, it is not necessary to obtain a numerical solution $u(x, t)$ of such a small magnitude, and such stiff relaxation processes compete (linearly additive or nonlinear) with other compensating source terms. Therefore, the problem chosen to study the stability and convergence of the ADER-DG numerical method with the LST-DG predictor was one in the form of an inhomogeneous linear advection-reaction equation of the following form:
\begin{equation}\label{eq:adv_eq_s_src}
\pderf{u}{t} + a\pderf{u}{x} = \nu u + s(t, x),\quad
(x, t)\in\Omega\times\mathcal{T},\quad
\Omega\subset\mathcal{R},\quad
\mathcal{T} = \left[t_{0}, t_{f}\right]\subset\mathcal{R},
\end{equation}
where $\nu < 0$ and all notations correspond to the original problem formulation (\ref{eq:adv_eq_src}), and $s: \mathcal{T}\times\Omega\rightarrow\mathcal{R}$ is an additional source term that generates the non-homogeneity of the equation. The choice of an additional source term in this study is made in the form $s(t, x) = -\nu u_{\rm init}(x - a t)$, where $u_{\rm init}(x) = u(x, 0)$ is a function of the initial condition, then the solution $u(x, t) = u_{\rm init}(x - a t)$ to the problem (\ref{eq:adv_eq_s_src}) is take the form of a simple advection of the initial solution with an advection velocity $a$.

The numerical ADER-DG method with the LST-DG predictor for the case of a linear inhomogeneous advection-reaction equation (\ref{eq:adv_eq_s_src}) is constructed similarly to the case of a homogeneous equation (\ref{eq:adv_eq_src}), as presented above in Subsection~\ref{sec:descr} ``General description of the numerical method''. By mapping (\ref{eq:mapping}) the arguments $(t, x)$ of the coordinate-time discretization domain $\Xi_{i}^{n}$ on the rescaled arguments $(\tau, \xi)$ reference domain $\omega^{2} = [0, 1]^{2}$ of the desired function $u(x, t)$, the original linear inhomogeneous advection-reaction equation (\ref{eq:adv_eq_s_src}) in a weak formulation is written in the following form (similar to expression (\ref{eq:adv_eq_weak}) for the case of a homogeneous equation (\ref{eq:adv_eq_src})):
\begin{equation}\label{eq:adv_eq_s_weak}
\intrefdom{\tau}\intrefdom{\xi}\,\varphi_{k}(\xi)\left[\pderf{u}{\tau} + \frac{a\Dtn{n}}{\Dx}\pderf{u}{\xi} - \nu\Dtn{n}u - \Dtn{n}s(t, x)\right] = 0,
\end{equation}
which, using the representation of the solution at the previous $t^{n}$ and next $t^{n+1} = t^{n} + \Dtn{n}$ time steps (\ref{eq:u_n_def}), the following formulation of the completely one-step ADER-DG scheme is obtained (similar to expression (\ref{eq:ader_dg_corr_temp_1}) for the case of a homogeneous equation (\ref{eq:adv_eq_src})):
\begin{equation}\label{eq:ader_dg_s_corr_temp_1}
\begin{split}
&\sum\limits_{l = 0}^{N} \intrefdom{\xi}\,\varphi_{k}(\xi)\varphi_{l}(\xi)\left[\hat{u}^{n+1}_{l} - \hat{u}^{n}_{l}\right] -
\frac{a\Dtn{n}}{\Dx}\intrefdom{\tau}\intrefdom{\xi}\,\varphi_{k}'(\xi) q(\tau, \xi) +
\frac{\Dtn{n}}{\Dx}\intrefdom{\tau}\,\Big[
  \varphi_{k}(1) F_{\rm RP}\left(q(\tau, 1),\, q^{+}(\tau, 0)\right)\\ & -
  \varphi_{k}(0) F_{\rm RP}\left(q^{-}(\tau, 1),\, q(\tau, 0)\right)
\Big] = \nu\Dtn{n}\intrefdom{\tau}\intrefdom{\xi}\,\varphi_{k}(\xi) q(\tau, \xi) + \Dtn{n}\intrefdom{\tau}\intrefdom{\xi}\,\varphi_{k}(\xi) s(t(\tau), x(\xi)),
\end{split}
\end{equation}
where $q: \omega^{2}\rightarrow\mathcal{R}$ is the local discrete space-time solution $q(\tau, \xi)$ (\ref{eq:local_sol_def}) used in ADER methods to calculate local and flux terms, and is determined for each coordinate-time element $\Xi_{i}^{n}$ separately, $F_{\rm RP}$ is the solver of the classical Riemann problem (\ref{eq:f_rp}), which, in the case of the advection equation (\ref{eq:adv_eq_s_src}). The system of equations of the LST-DG predictor for calculating the local discrete space-time solution $q(\tau, \xi)$ (\ref{eq:local_sol_def}) is obtained similarly to system of equations (\ref{eq:lstdg_pred_eq_tmp_1}):
\begin{equation}\label{eq:lstdg_s_pred_eq_tmp_1}
\begin{split}
&\sum\limits_{q_{0} = 0}^{N}\sum\limits_{q_{1} = 0}^{N}
  \left[\varphi_{p_{0}}(1)\varphi_{q_{0}}(1) - \intrefdom{\tau}\,\varphi_{p_{0}}'(\tau)\varphi_{q_{0}}(\tau)\right]
  \intrefdom{\xi}\,\varphi_{p_{1}}(\xi)\varphi_{q_{1}}(\xi)\cdot\hat{q}_{q_{0}q_{1}}\\
&+\frac{a\Dtn{n}}{\Dx}\,\sum\limits_{q_{0} = 0}^{N}\sum\limits_{q_{1} = 0}^{N}
  \intrefdom{\tau}\,\varphi_{p_{0}}(\tau)\varphi_{q_{0}}(\tau)\,\intrefdom{\xi}\,\varphi_{p_{1}}(\xi)\varphi_{q_{1}}'(\xi)\cdot\hat{q}_{q_{0}q_{1}} \\
&-\nu\Dtn{n}\,\sum\limits_{q_{0} = 0}^{N}\sum\limits_{q_{1} = 0}^{N}
  \intrefdom{\tau}\,\varphi_{p_{0}}(\tau)\varphi_{q_{0}}(\tau)\,\intrefdom{\xi}\,\varphi_{p_{1}}(\xi)\varphi_{q_{1}}(\xi)\cdot\hat{q}_{q_{0}q_{1}} =
\sum\limits_{q_{1} = 0}^{N} \varphi_{p_{0}}(0) \intrefdom{\xi}\,\varphi_{p_{1}}(\xi)\varphi_{q_{1}}(\xi)\cdot\hat{u}^{n}_{q_{1}} \\ & +
\Dtn{n}\,\intrefdom{\tau}\intrefdom{\xi}\,\varphi_{p_{0}}(\tau)\varphi_{p_{1}}(\xi) \cdot s(t(\tau), x(\xi)) =
\sum\limits_{q_{1} = 0}^{N} \varphi_{p_{0}}(0) \intrefdom{\xi}\,\varphi_{p_{1}}(\xi)\varphi_{q_{1}}(\xi)\cdot\hat{u}^{n}_{q_{1}} \\ & +
\Dtn{n}\,\sum\limits_{q_{0} = 0}^{N}\sum\limits_{q_{1} = 0}^{N}
\intrefdom{\tau}\intrefdom{\xi}\,\varphi_{p_{0}}(\tau)\varphi_{q_{0}}(\tau)\varphi_{p_{1}}(\xi)\varphi_{q_{1}}(\xi) \cdot s(t(\tau_{q_{0}}), x(\xi_{q_{1}})),
\end{split}
\end{equation}
where the point-wise evaluation estimate (\ref{eq:pw_eval}) is used:
\begin{equation}
s(t(\tau), x(\xi)) \mapsto
\sum\limits_{p_{0} = 0}^{N}\sum\limits_{p_{1} = 0}^{N} \hat{s}_{p_{0}p_{1}}\varphi_{p_{0}}(\tau)\varphi_{p_{1}}(\xi),\quad
\hat{s}_{p_{0}p_{1}} = s(t(\tau_{p_{0}}), x(\xi_{p_{1}})),\quad
\hat{\mathbf{s}} = \{\hat{s}_{\pi_{0}(\mathrm{p})\pi_{1}(\mathrm{p})}\}_{\mathrm{p}},
\end{equation}
where multi-index $\mathrm{p}$ and projections $\pi_{0}(\mathrm{p})$, $\pi_{1}(\mathrm{p})$ (\ref{eq:local_sol_idx}) are used. Then the expression (\ref{eq:lstdg_s_pred_eq_tmp_1}) can be rewritten using matrix notations (\ref{eq:matrices_def}):
\begin{equation}
\Big[\kappa \otimes m - z \cdot m \otimes m + \mathrm{CFL} \cdot m\otimes\varrho\Big]\hat{\mathbf{q}} =
(\uvecphi \otimes m)\hat{\mathbf{u}}^{n} + \Dtn{n} (m \otimes m) \hat{\mathbf{s}},
\end{equation}
which similar to expression (\ref{eq:lstdg_pred_eq_tmp_2}) for the case of a homogeneous equation (\ref{eq:adv_eq_src}) and which can be rewritten using notation (\ref{eq:chi_and_tilde_ae_def}) as follows:
\begin{equation}
\Big[(I - z \chi) \otimes I + \mathrm{CFL}\cdot\chi\otimes\mathrm{D}\Big]\hat{\mathbf{q}} =
\mathbf{1} \otimes \hat{\mathbf{u}}^{n} + \Dtn{n} (\chi \otimes I) \hat{\mathbf{s}}.
\end{equation}
The vector of the coefficient of the local discrete space-time solution $q(\tau, \xi)$ (\ref{eq:local_sol_def}) in this case is written as follows:
\begin{equation}\label{eq:lstdg_s_pred_eq_sol_temp_1}
\begin{split}
\hat{\mathbf{q}} =
\Big[(I - z \chi) \otimes I + \mathrm{CFL}\cdot\chi\otimes\mathrm{D}\Big]^{-1}(\mathbf{1} \otimes \hat{\mathbf{u}}^{n}) +
\Dtn{n}\Big\{\Big[(I - z \chi) \otimes I + \mathrm{CFL}\cdot\chi\otimes\mathrm{D}\Big]^{-1}(\chi \otimes I)\Big\}\,\hat{\mathbf{s}},
\end{split}
\end{equation}
which differs from the case (\ref{eq:lstdg_pred_eq_sol_temp_1}) of a homogeneous equation (\ref{eq:adv_eq_src}) by an additional term with $\hat{\mathbf{s}}$ on the right-hand side. The expression for a completely single-step ADER-DG scheme (\ref{eq:ader_dg_s_corr_temp_1}) in vector and matrix notation (\ref{eq:matrix_notation_base_def}) can be represented as follows:
\begin{equation}\label{eq:ader_dg_corr_s_sol}
\begin{split}
\hat{\mathbf{u}}^{n+1} & =
\hat{\mathbf{u}}^{n} + \mathrm{CFL} \cdot m^{-1}\Bigg\{\intrefdom{\tau}\intrefdom{\xi}\,
\Big[\boldsymbol{\varphi}'(\xi) (\boldsymbol{\varphi}(\tau)\otimes\boldsymbol{\varphi}(\xi))^{T}\Big]\hat{\mathbf{q}} -
\intrefdom{\tau}\,\Bigg[\theta(a) \Big\{\Big[\ovecphi(\boldsymbol{\varphi}(\tau)\otimes\ovecphi)^{T}\Big]\hat{\mathbf{q}}\\ & -
\Big[\uvecphi(\boldsymbol{\varphi}(\tau)\otimes\ovecphi)^{T}\Big]\hat{\mathbf{q}}^{-}\Big\} +
\theta(-a) \Big\{\Big[\ovecphi(\boldsymbol{\varphi}(\tau)\otimes\uvecphi)^{T}\Big]\hat{\mathbf{q}}^{+} -
\Big[\uvecphi(\boldsymbol{\varphi}(\tau)\otimes\uvecphi)^{T}\Big]\hat{\mathbf{q}}\Big\}\Bigg]\Bigg\}\\ & +
z \cdot m^{-1}\intrefdom{\tau}\intrefdom{\xi}\,\Big[\boldsymbol{\varphi}(\xi) (\boldsymbol{\varphi}(\tau)\otimes\boldsymbol{\varphi}(\xi))^{T}\Big]\hat{\mathbf{q}} +
\Dtn{n} \cdot m^{-1}\intrefdom{\tau}\intrefdom{\xi}\,\Big[\boldsymbol{\varphi}(\xi) (\boldsymbol{\varphi}(\tau)\otimes\boldsymbol{\varphi}(\xi))^{T}\Big]\,\hat{\mathbf{s}},
\end{split}
\end{equation}
convenient for vectorized implementation. These expressions clearly demonstrates that the difference in the formulation of the ADER-DG numerical method with LST-DG predictor for the case of an inhomogeneous linear advection-diffusion equation (\ref{eq:adv_eq_s_src}) from the case of a homogeneous equation (\ref{eq:adv_eq_src}) lies only in the additional term $\propto \Dtn{n}\hat{\mathbf{s}}$ on the right-hand side to (\ref{eq:ader_dg_corr_temp_1}) (including after substituting the expression (\ref{eq:lstdg_s_pred_eq_sol_temp_1}) and expanding the brackets), which obviously has no effect on the stability condition obtained for the homogeneous equation (\ref{eq:adv_eq_src}). All calculations, the results of which are presented in this Subsection, were performed using a software implementation of the ADER-DG numerical method with a LST-DG predictor, which was presented in~\cite{ader_stiff_3, ader_stiff_4, ader_eff_blas}, using flow and source terms corresponding to the linear advection-diffusion equation (\ref{eq:adv_eq_s_src}), and an exact Riemann solver (\ref{eq:f_rp}). The only fundamental difference in this case is the direct calculation of the vector of the coefficient of the local discrete space-time solution $\hat{\mathbf{q}}$ (\ref{eq:lstdg_s_pred_eq_sol_temp_1}) instead of the iterative Picard method in~\cite{ader_stiff_3, ader_stiff_4, ader_eff_blas}: at the beginning of each calculation, when the main parameters $\nu$, $\mathrm{CFL}$ are $\Dx$ are set, the matrices on the right-hand side of the expression (\ref{eq:lstdg_s_pred_eq_sol_temp_1}) are calculated once (for this set of parameters), where the inverse matrix is calculated using the standard \texttt{LAPACK} procedures: \texttt{dgetrf} (with \texttt{ipiv}) and \texttt{dgetri}. All other matrix-vector and matrix-matrix calculations in expressions (\ref{eq:lstdg_s_pred_eq_sol_temp_1}) and (\ref{eq:ader_dg_corr_s_sol}) are performed using the \texttt{gemm} functions of the \texttt{BLAS} interface, as in~\cite{ader_eff_blas}.

\begin{wraptable}{l}{0.55\textwidth}
\centering
\small
\caption{%
Quantitative comparison of the values $\mathrm{CFL}_{\rm max}(N, 0)$ and $\mathrm{CFL}_{\rm CS}(N)$ (\ref{eq:cfl_cs}) with the values $\mathrm{CFL} = 1.0 \cdot 10^{r}$ used in this work.
}
\label{tab:comp_cfl_10r_and_cfl_max}
\begin{tabular}{@{}|c|c|l|c|l|@{}}
\toprule
$N$ &
$\mathrm{CFL}_{\rm max}(N, 0)$ &
$\cfrac{1.0 \cdot 10^{r}}{\mathrm{CFL}_{\rm max}(N, 0)}$ & $\mathrm{CFL}_{\rm CS}(N)$ & $\cfrac{1.0 \cdot 10^{r}}{\mathrm{CFL}_{\rm CS}(N)}$ \\ 
\midrule
$1$ & $0.33333333$ & $3.00000000 \cdot 10^{r}$ & $0.33333333$ & $3.0 \cdot 10^{r}$ \\
$2$ & $0.17082039$ & $5.85410197 \cdot 10^{r}$ & $0.20000000$ & $5.0 \cdot 10^{r}$ \\
$3$ & $0.10392896$ & $9.62195698 \cdot 10^{r}$ & $0.14285714$ & $7.0 \cdot 10^{r}$ \\
$4$ & $0.06983091$ & $1.43203063 \cdot 10^{r+1}$ & $0.11111111$ & $9.0 \cdot 10^{r}$ \\
$5$ & $0.05011556$ & $1.99538839 \cdot 10^{r+1}$ & $0.09090909$ & $1.1 \cdot 10^{r+1}$ \\
$6$ & $0.03770124$ & $2.65243276 \cdot 10^{r+1}$ & $0.07692308$ & $1.3 \cdot 10^{r+1}$ \\
$7$ & $0.02938384$ & $3.40323170 \cdot 10^{r+1}$ & $0.06666667$ & $1.5 \cdot 10^{r+1}$ \\
$8$ & $0.02354150$ & $4.24781750 \cdot 10^{r+1}$ & $0.05882353$ & $1.7 \cdot 10^{r+1}$ \\
$9$ & $0.01928191$ & $5.18620712 \cdot 10^{r+1}$ & $0.05263158$ & $1.9 \cdot 10^{r+1}$ \\
$10$ & $0.01608128$ & $6.21841020 \cdot 10^{r+1}$ & $0.04761905$ & $2.1 \cdot 10^{r+1}$ \\
$11$ & $0.01361576$ & $7.34443255 \cdot 10^{r+1}$ & $0.04347826$ & $2.3 \cdot 10^{r+1}$ \\
$12$ & $0.01167641$ & $8.56427782 \cdot 10^{r+1}$ & $0.04000000$ & $2.5 \cdot 10^{r+1}$ \\
\bottomrule
\end{tabular}
\end{wraptable}

Numerical instability of the numerical method depends significantly on the spectral composition $u_{\rm init}(k)$ of the solution $u_{\rm init}(x)$. Therefore, trivial tests for sine wave advection may ``miss'' the instability boundary if the wave vector $k$ for the selected sine wave in the initial conditions $u_{\rm init}(x) \propto \sin(kx)$ and coordinate discretization steps $\Dx$ are chosen such that the corresponding phases $\theta = k\Dx$ (\ref{eq:r_matrix_def}) for the boundary values of the Courant number $\mathrm{CFL}_{\rm max}(N, z)$ do not characterized eigenvalues $\lambda_{k}$ on or outside the circle of stability $|\lambda_{k}| \leqslant 1$. The choice of the Gaussian profile $u_{\rm init}(x) \propto \exp(-x^{2})$ (or Gaussian bell), as was done in the work~\cite{ader_dg_stab}, also revealed certain limitations: despite the very narrow profile (and, accordingly, a very broad solution spectrum at the initial instant of time), the Gaussian profile exhibits a very rapid decay (also having the form of a Gaussian profile) of the spectrum $k$ with increasing distance from zero $k = 0$. Moreover, mathematically rigorously, the standard Gaussian profile is not a periodic function, although it can, of course, attain very small solution values at the solution domain $\Omega$ boundary at the initial and reference instants of time. In this work, a compromise variant is chosen, as was done in the work~\cite{ader_dg_stab_rig} --- the well-known Lorentzian profile $u_{\rm init}(x) \propto 1/[1+(x/\alpha)^{2}]$ (or Lorentzian bell), which is characterized by a significantly slower decay with increasing coordinate $x$, compared to a Gaussian profile, but is also characterized only by an exponential decay of the spectrum with increasing wave vector $k$: $u_{\rm init}(k) \propto \exp(-\alpha|k|)$. To achieve periodicity, the initial condition is chosen in the form of a superposition of an infinite number of Lorentzian profiles on the coordinate domain $[-L/2,\, +L/2]$ shifted on $L k$ with $k\in\mathcal{Z}$:\\[-2mm]
\begin{equation}\label{eq:lorentz_bell}
u_{\rm init}(x)
= u_{0} + \sum\limits_{k = -\infty}^{+\infty} \cfrac{A/\alpha}{1 + \left[\cfrac{x - L k}{\alpha}\right]^{2}}
= u_{0} + \frac{\pi\alpha A}{L}\cfrac{\sinh\left(\cfrac{2\pi\alpha}{L}\right)}{\cosh\left(\cfrac{2\pi\alpha}{L}\right) - \cos\left(\cfrac{2\pi x}{L}\right)},
\end{equation}
where the summation is carried out based on trivial considerations of the representation of the cotangent in the complex plane $\mathcal{C}$ and regarding the residue theory, $\alpha\in\mathcal{R}_{+}$ is the width of the Lorentz profile at half the level, $L\in\mathcal{R}_{+}$ is the coordinate period, $A\in\mathcal{R}_{+}$ is the intensity of the Lorentz profile (reduced height), and the wave vector $k$ has only a discrete (countable) set of values due to the periodicity of the function $u(x, 0)$ (therefore, of course, the exact expression for the spectrum $u_{\rm init}(k)$ contains an additional factor in the form of a sum of delta functions $\delta(k - k_{l})$). The spectrum of chosen $u_{\rm init}(x)$ is equal to $u_{\rm init}(k) \propto \pi A \exp(-\alpha|k|)$. In this study the following simulation parameters are chosen:\\[-2mm]
\begin{equation}
\begin{split}
\Omega = [-50.0, +50.0],\quad
L = 100.0,\quad
A = 1.0,\quad
\alpha = 5.0,\quad
u_{0} = 0.1,\quad
t_{0} = 0.0,\quad
t_{f} = 1.0 \cdot 10^{6},\quad
a = 1.0.
\end{split}
\end{equation}
Calculations are performed for degrees $N = 1, \ldots, 12$ of basis polynomials $\{\varphi_{p}\}_{p}$ (\ref{eq:phi_def}). The boundary conditions are periodic: $u(-50.0, t) = u(+50.0, t)$. The time $t_{f} - t_{0} \equiv t_{f} = 10^{6}$ is chosen so that the periodic coordinate dependence of the solution $u(x, t)$ transferred the solution domain $\Omega$ exactly $10^{4}$ times. Two values of the decay parameter $\nu$ are studied: $\nu = -10^{2}$ and $\nu = -10^{3}$, which correspond to situations of strong stiff problems. The stability condition of the ADER-DG numerical method with LST-DG predictor in the case of an inhomogeneous linear advection-diffusion equation (\ref{eq:adv_eq_s_src}) becomes significantly more complex than in the case of a homogeneous equation (\ref{eq:adv_eq_src}), which was studied in~\cite{ader_dg_stab_rig}. The constraint on the time step $\Dtn{n}$, expressed in the following form:\\[-2mm]
\begin{equation}\label{eq:stab_criterion}
a\,\frac{\Dtn{n}}{\Dx} \equiv \mathrm{CFL} \leqslant \,\mathrm{CFL}_{\max}(N, -\nu\Dtn{n}),
\end{equation}
is no longer an inequality resolved with respect to the time discretization step $\Dtn{n}$.\par
\begin{wrapfigure}{l}{0.4\textwidth}
\includegraphics[width=0.4\textwidth]{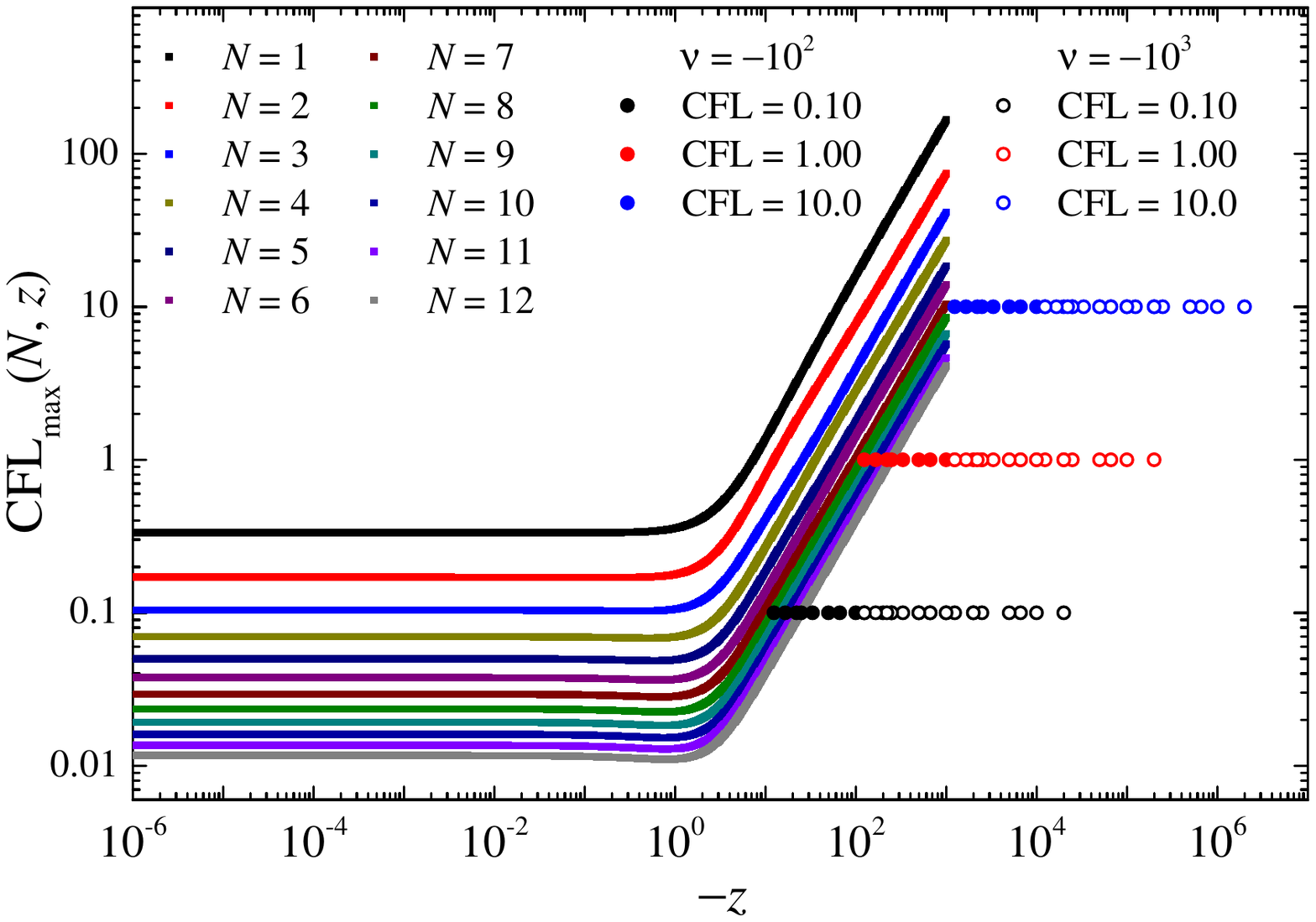}
\caption{%
The selected parameters $z = \nu\Dtn{n}$ for cases $\nu = -10^{2}$ and $-10^{3}$ for the selected Courant numbers $\mathrm{CFL} = 0.1$, $1.0$ and $10.0$. The data $\mathrm{CFL}_{\max}(N, z)$ correspond to the results presented in Figure~\ref{fig:cfls_on_z}. It is demonstrated that the selected calculation parameters $\mathrm{CFL}$ and $z$ are in the stability region $[0, \mathrm{CFL}_{\max}(N, z)]$ of the ADER-DG numerical method with LST-DG predictor.
}
\label{fig:cfls_on_z_def_for_tests}
\end{wrapfigure}
\noindent The study of convergence, especially at very high values of the Courant number $\mathrm{CFL} \gg \mathrm{CFL}_{\rm max}(N, 0)$, must be carried out under ``sufficiently controlled conditions'', which are defined by fixed simulation parameters, in particular, the decay constant $\nu$, the coordinate discretization step $\Dx$ (with decreasing numerical solution error, the empirical order of convergence is determined by the value of the Courant number $\mathrm{CFL}$ itself). Therefore, in this case, the parameter $z = -\nu\Dtn{n}$ remains unfixed, since it clearly depends on the time discretization step $\Dtn{n}$, which, for a fixed Courant number $\mathrm{CFL}$, depends on the coordinate discretization step $\Dx$. It is important to note that the stability criterion (\ref{eq:stab_criterion}) depends significantly on the parameter $z$. In this work, three values of the Courant number $\mathrm{CFL}$ are selected for the convergence study: $\mathrm{CFL} = 0.1$, $1.0$, $10.0$. The Courant number $\mathrm{CFL} = 0.1$ is in the stability region $[0, \mathrm{CFL}_{\rm max}(N, 0)]$ for the advection equation ($\nu = 0$) only for polynomial degrees $N = 1$, $2$, $3$ (see Table~\ref{tab:min_cfls_on_z}), and for polynomial degrees $N \geqslant 4$, this value of the Courant number $\mathrm{CFL} = 0.1$ already goes beyond the stability boundary $\mathrm{CFL}_{\rm max}(N, 0)$ of the ADER-DG numerical method with LST-DG predictor for the advection equation. The Courant number $\mathrm{CFL} = 1.0$ is within the stability region $[0, \mathrm{CFL}_{\rm max}(N, 0)]$ for the advection equation (without reaction) only for polynomial degree $N = 1$ --- this is a trivial case and is not investigated in this work; therefore, this value of the Courant number $\mathrm{CFL} = 1.0$ is already beyond the stability boundary $\mathrm{CFL}_{\rm max}(N, 0)$ of the ADER-DG numerical method with LST-DG predictor for the advection equation. The Courant number $\mathrm{CFL} = 10.0 \gg \mathrm{CFL}_{\rm max}(N, 0)$ is far beyond the stability region $[0, \mathrm{CFL}_{\rm max}(N, 0)]$ for all polynomial degrees $N$ for the advection equation. Therefore, the chosen values of the Courant numbers $\mathrm{CFL}$ allow us to clearly demonstrate the capabilities of stable computations using the ADER-DG numerical method with LST-DG predictor for solving significantly stiff problems similar (\ref{eq:adv_eq_src}). This is clearly demonstrated quantitatively in Table~\ref{tab:comp_cfl_10r_and_cfl_max}, where cases $\mathrm{CFL} = 0.1$, $1.0$, $10.0$ correspond to $r = -1$, $0$, $+1$, respectively. The number of coordinate discretization domains $\Omega_{i}$ chosen to be not very large in order to maintain very large values of the time discretization step $\Dtn{n}\propto\Dx$. For ease of presentation, Figure~\ref{fig:cfls_on_z_def_for_tests} presents the values of the $(z, \mathrm{CFL})$ pairs for $\nu = -10^{2}$ and $\nu = -10^{3}$ for which the calculations are performed, based on which the convergence of the ADER-DG numerical method with LST-DG predictor is studied in this work (for each polynomial degree $N$, only those pairs $(z, \mathrm{CFL})$ that are in the stability region (\ref{eq:stab_criterion}) are selected). Even larger coordinate discretization steps $\Dx$ are not investigated due to the anomalously large numerical values in the source terms on the right-hand side of the linear advection-reaction equation (\ref{eq:adv_eq_src}), the rounding errors of double-precision floating-point numbers type \texttt{double} in the calculation of which already significantly affect the accuracy of the numerical solution and do not allow the asymptotics of convergence to be correctly identified.

\begin{figure}[h!]
\captionsetup[subfigure]{%
	position=bottom,
	font+=smaller,
	textfont=normalfont,
	singlelinecheck=off,
	justification=raggedright
}
\centering
\begin{subfigure}{0.24\textwidth}
\includegraphics[width=\textwidth]{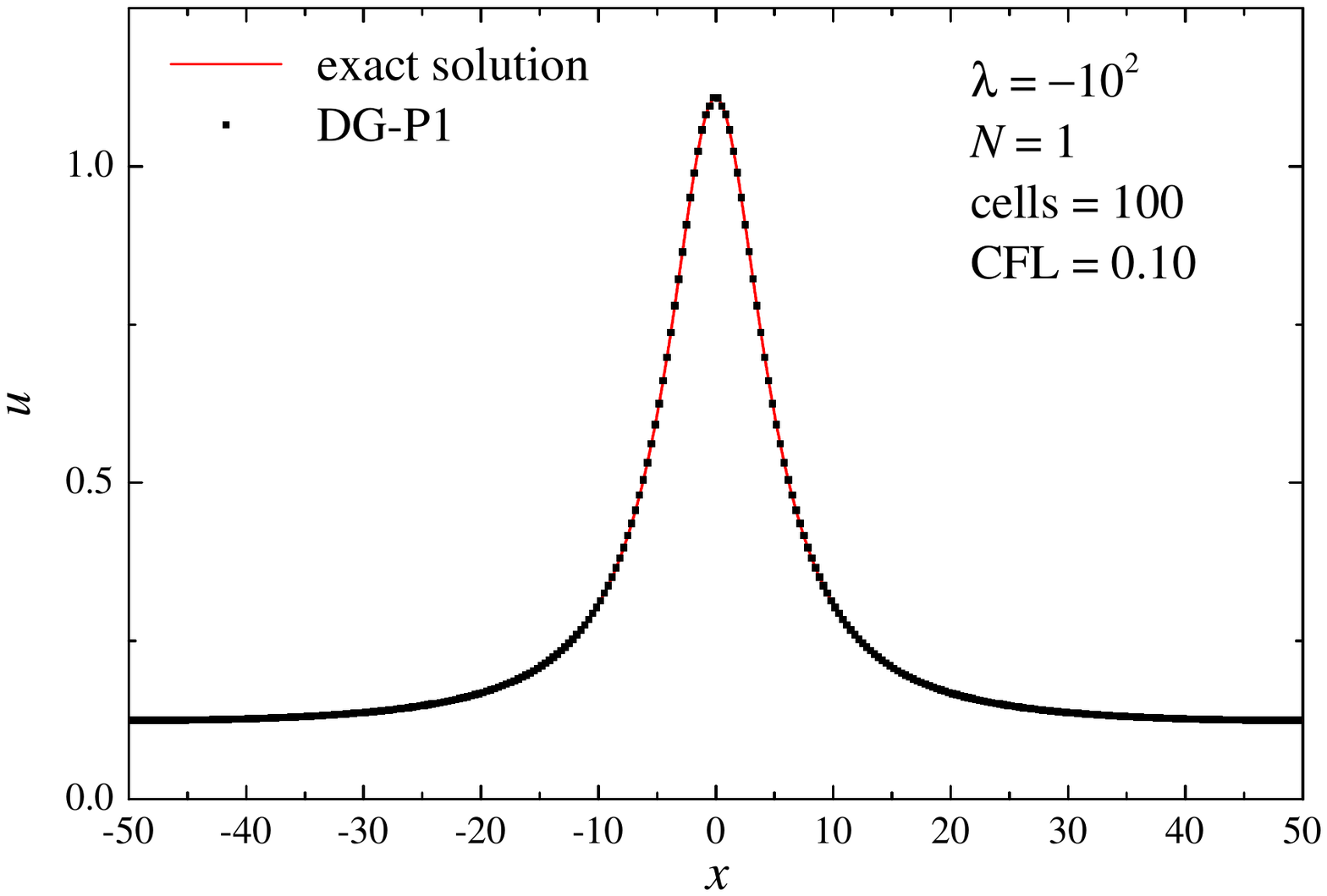}
\vspace{-8mm}\caption{\label{fig:test_adveq_stiff_z100_lrnz_1d_cfl_010:1}}
\end{subfigure}
\begin{subfigure}{0.24\textwidth}
\includegraphics[width=\textwidth]{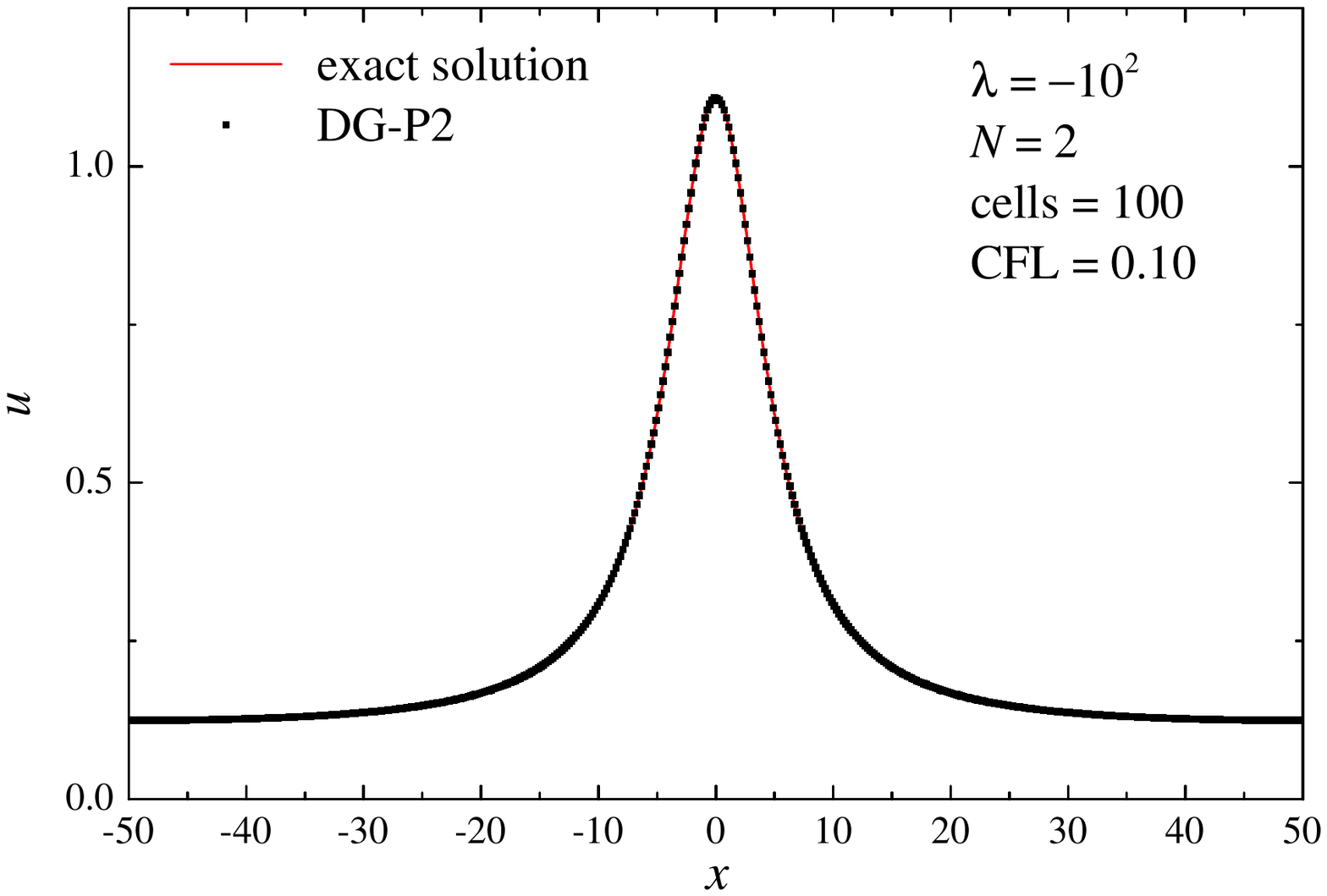}
\vspace{-8mm}\caption{\label{fig:test_adveq_stiff_z100_lrnz_1d_cfl_010:2}}
\end{subfigure}
\begin{subfigure}{0.24\textwidth}
\includegraphics[width=\textwidth]{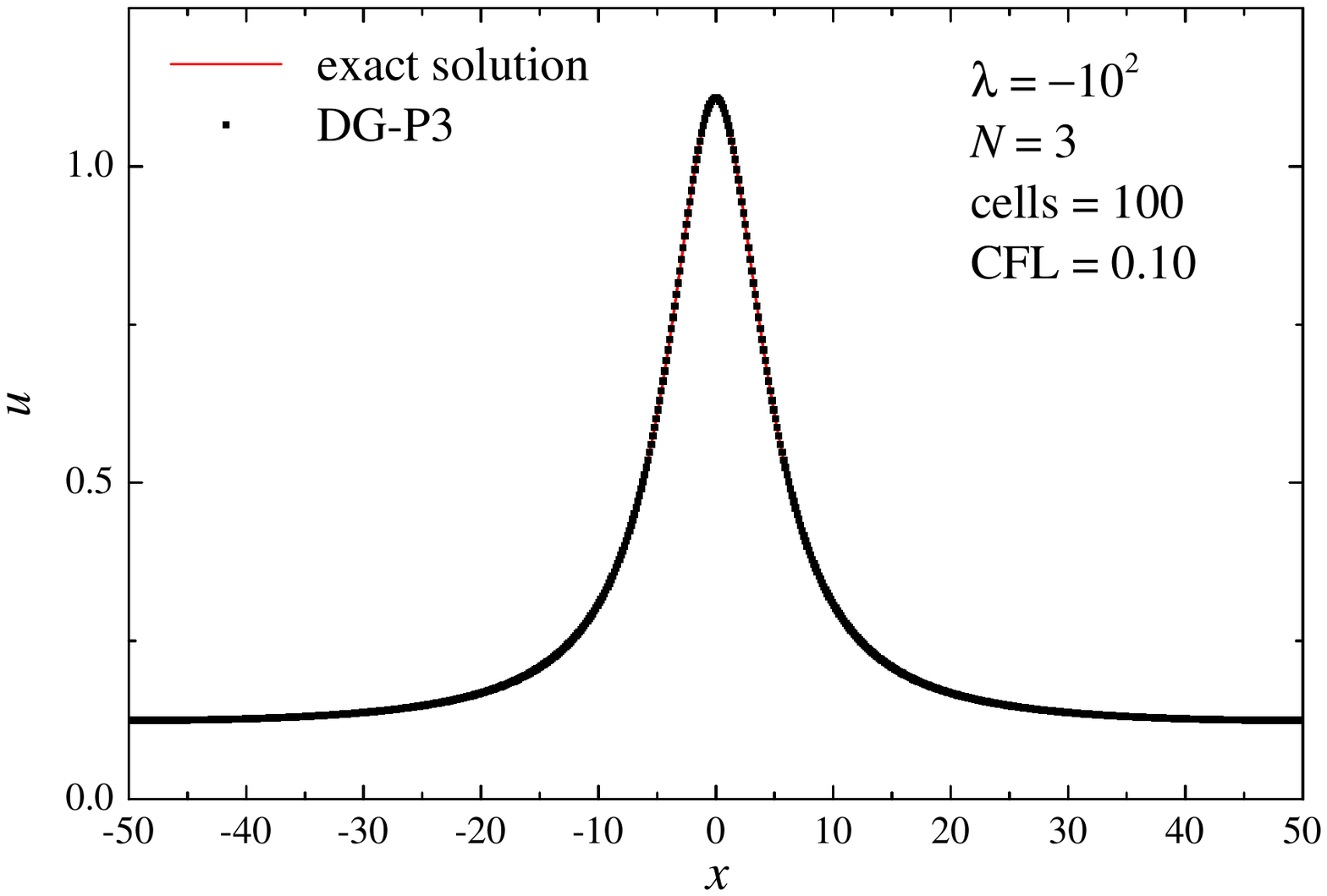}
\vspace{-8mm}\caption{\label{fig:test_adveq_stiff_z100_lrnz_1d_cfl_010:3}}
\end{subfigure}
\begin{subfigure}{0.24\textwidth}
\includegraphics[width=\textwidth]{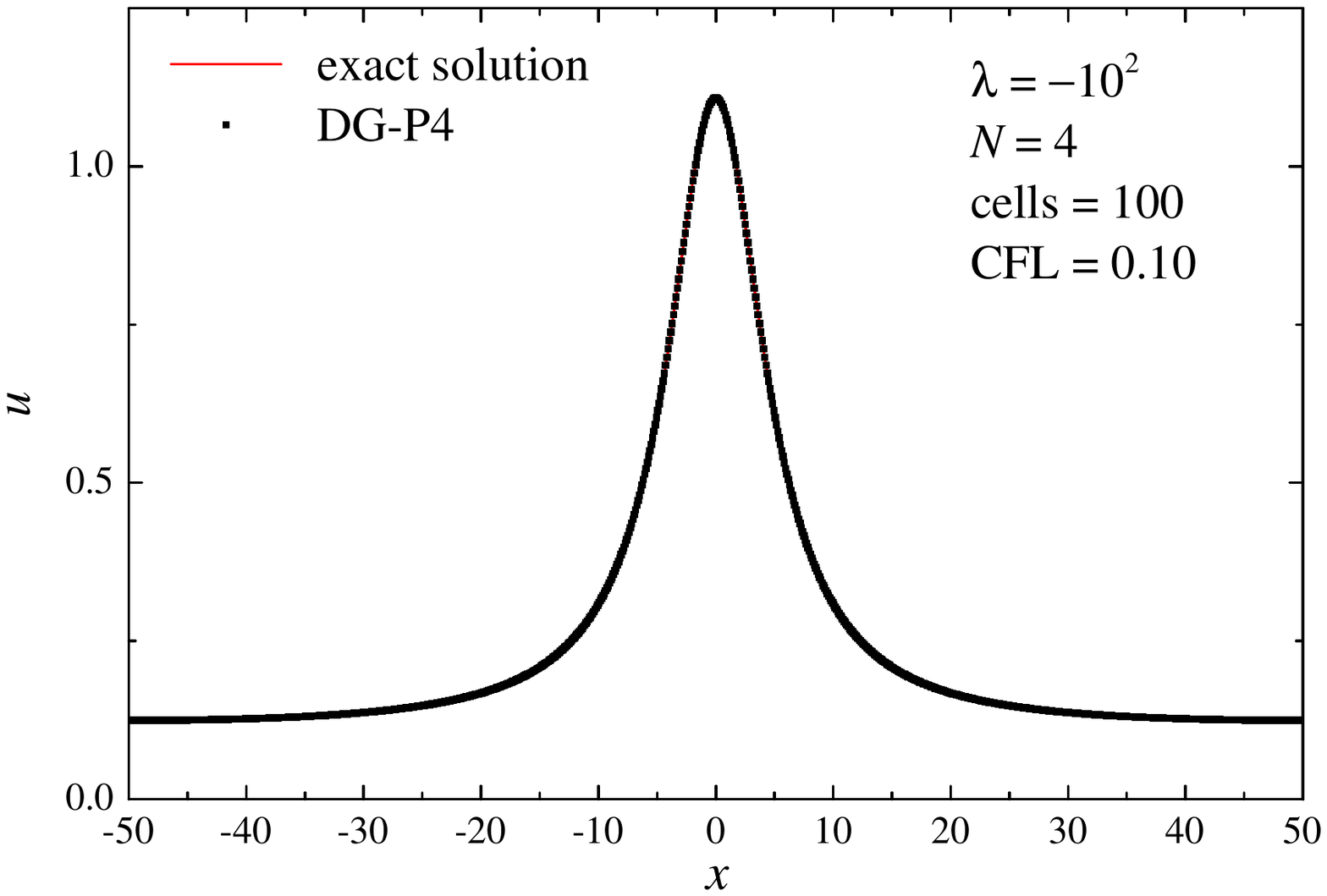}
\vspace{-8mm}\caption{\label{fig:test_adveq_stiff_z100_lrnz_1d_cfl_010:4}}
\end{subfigure}\\[-2.5mm]
\begin{subfigure}{0.24\textwidth}
\includegraphics[width=\textwidth]{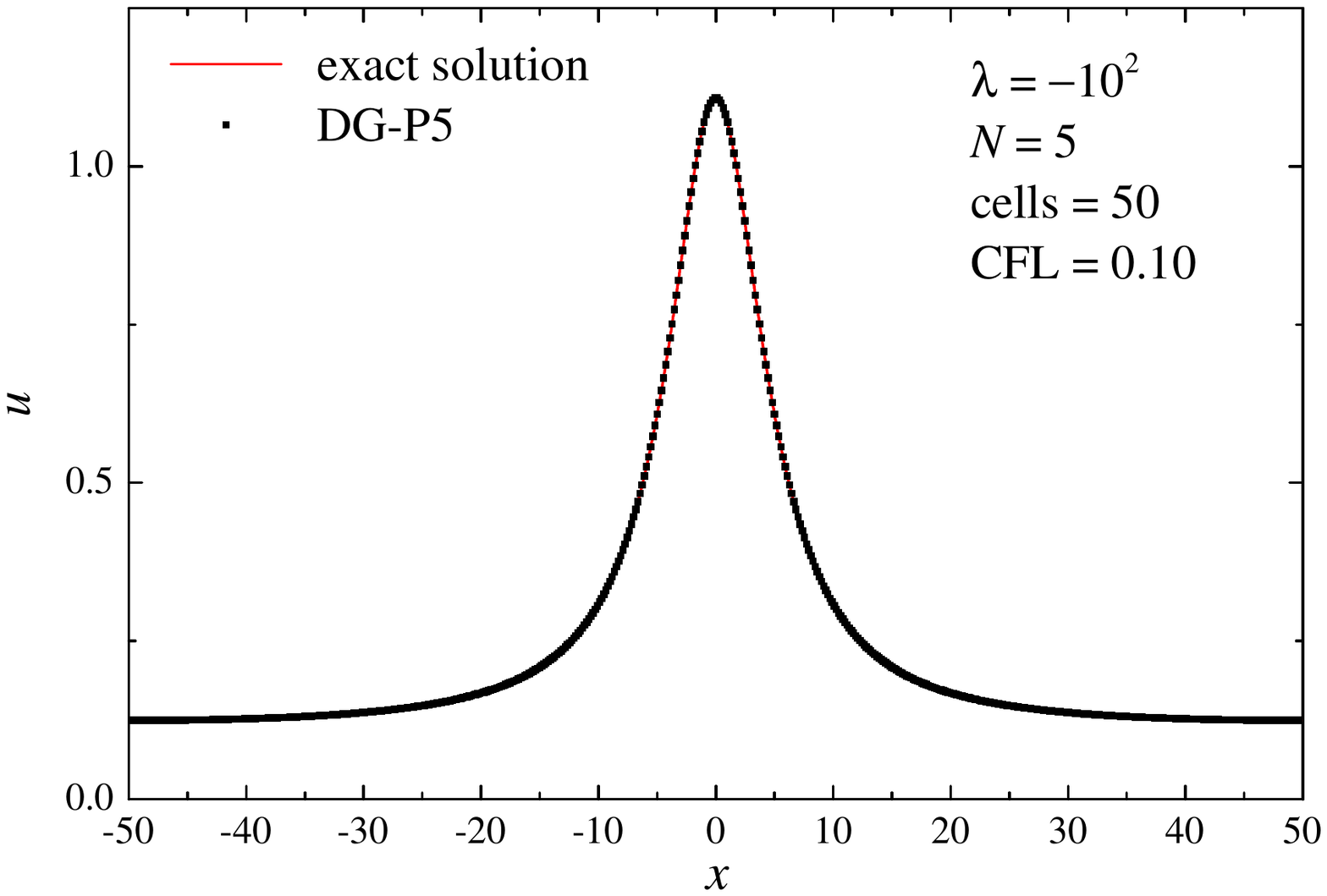}
\vspace{-8mm}\caption{\label{fig:test_adveq_stiff_z100_lrnz_1d_cfl_010:5}}
\end{subfigure}
\begin{subfigure}{0.24\textwidth}
\includegraphics[width=\textwidth]{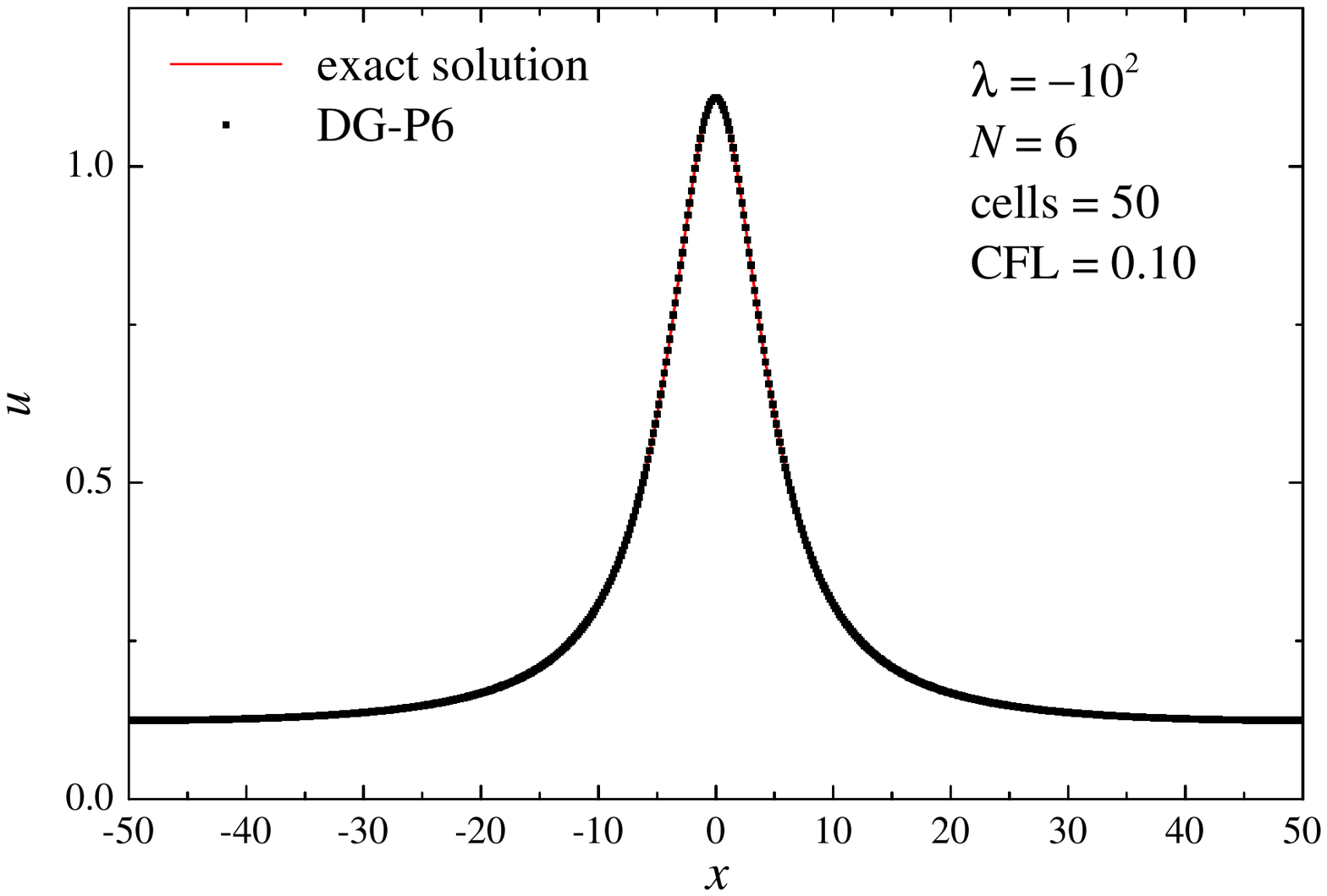}
\vspace{-8mm}\caption{\label{fig:test_adveq_stiff_z100_lrnz_1d_cfl_010:6}}
\end{subfigure}
\begin{subfigure}{0.24\textwidth}
\includegraphics[width=\textwidth]{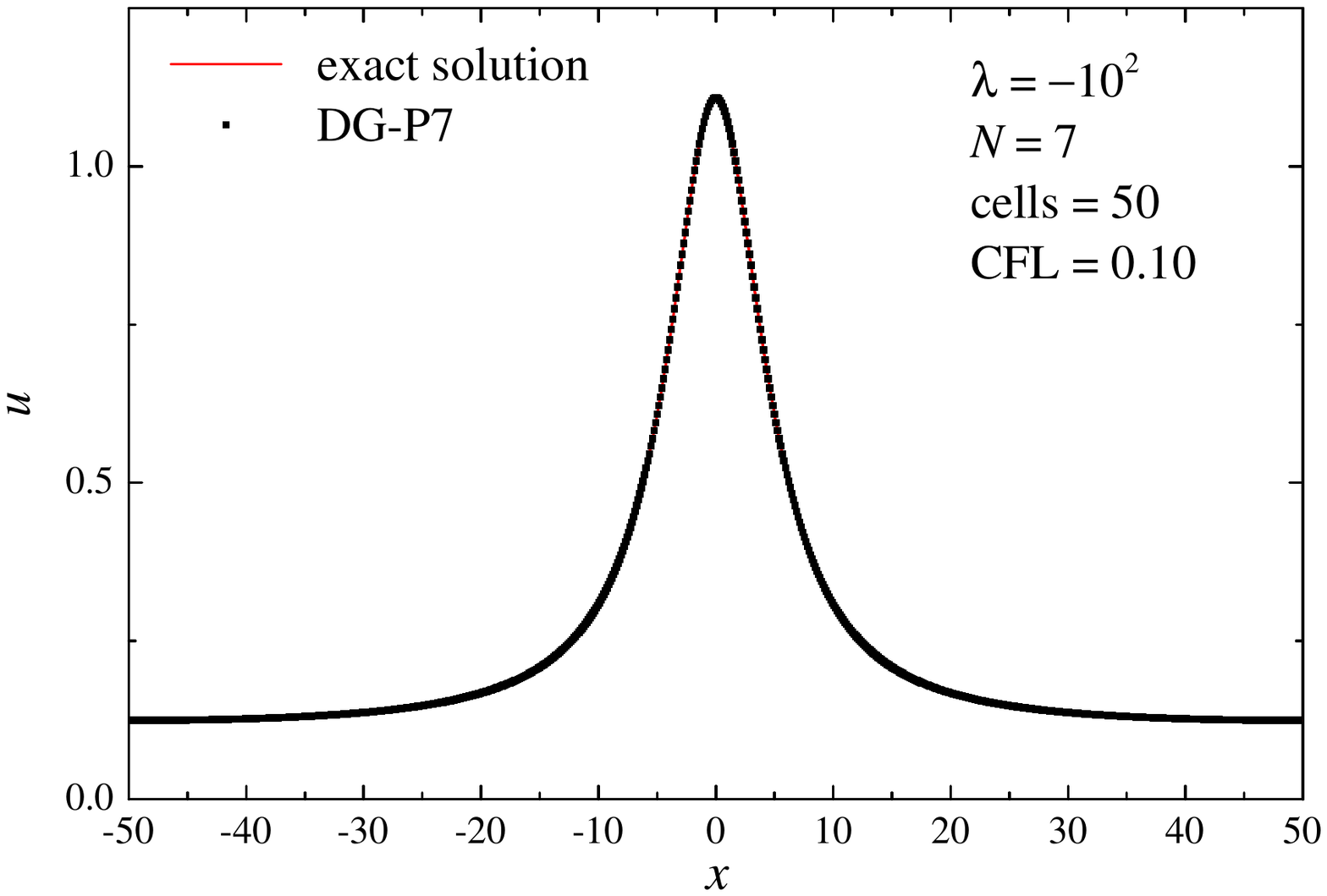}
\vspace{-8mm}\caption{\label{fig:test_adveq_stiff_z100_lrnz_1d_cfl_010:7}}
\end{subfigure}
\begin{subfigure}{0.24\textwidth}
\includegraphics[width=\textwidth]{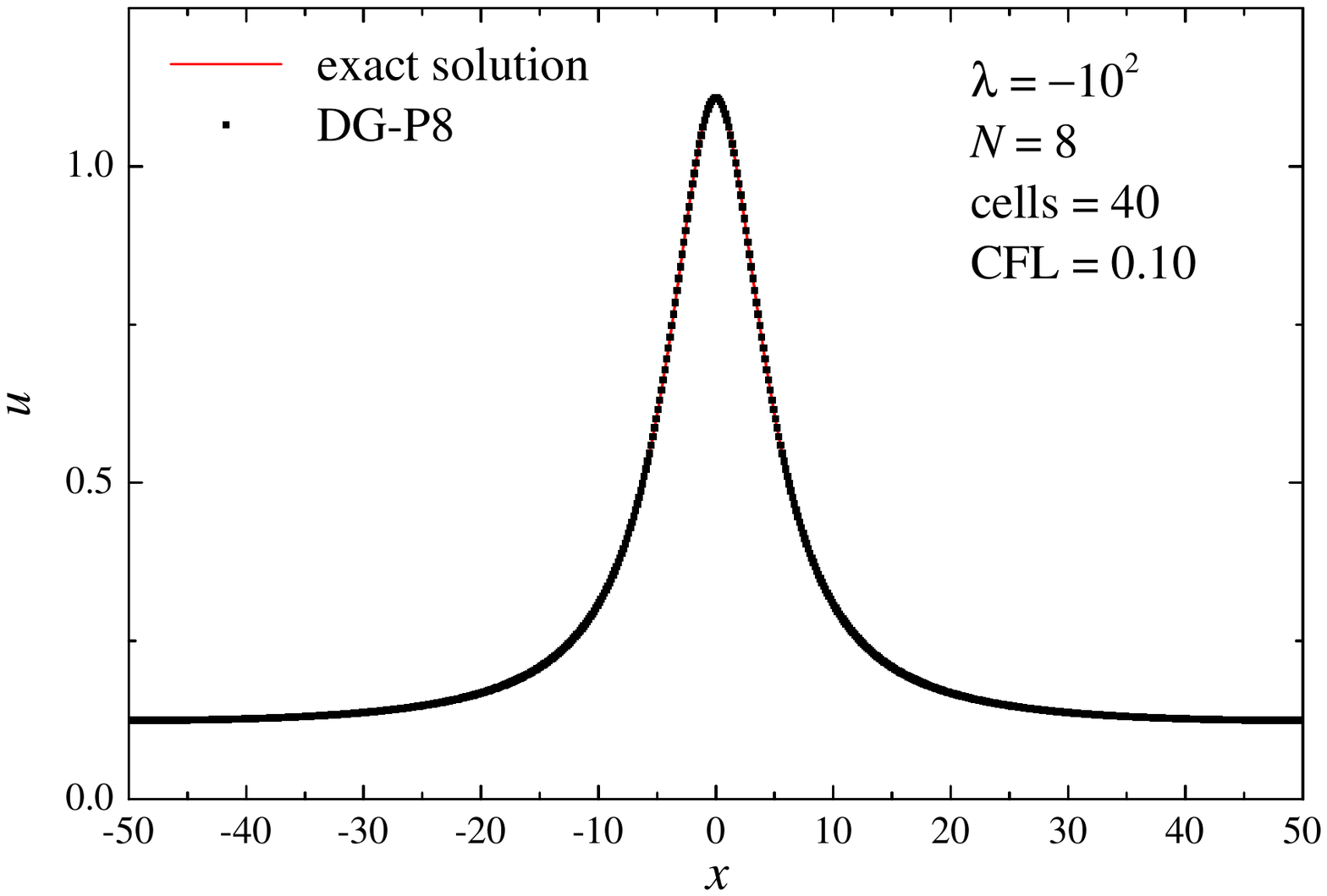}
\vspace{-8mm}\caption{\label{fig:test_adveq_stiff_z100_lrnz_1d_cfl_010:8}}
\end{subfigure}\\[-2.5mm]
\begin{subfigure}{0.24\textwidth}
\includegraphics[width=\textwidth]{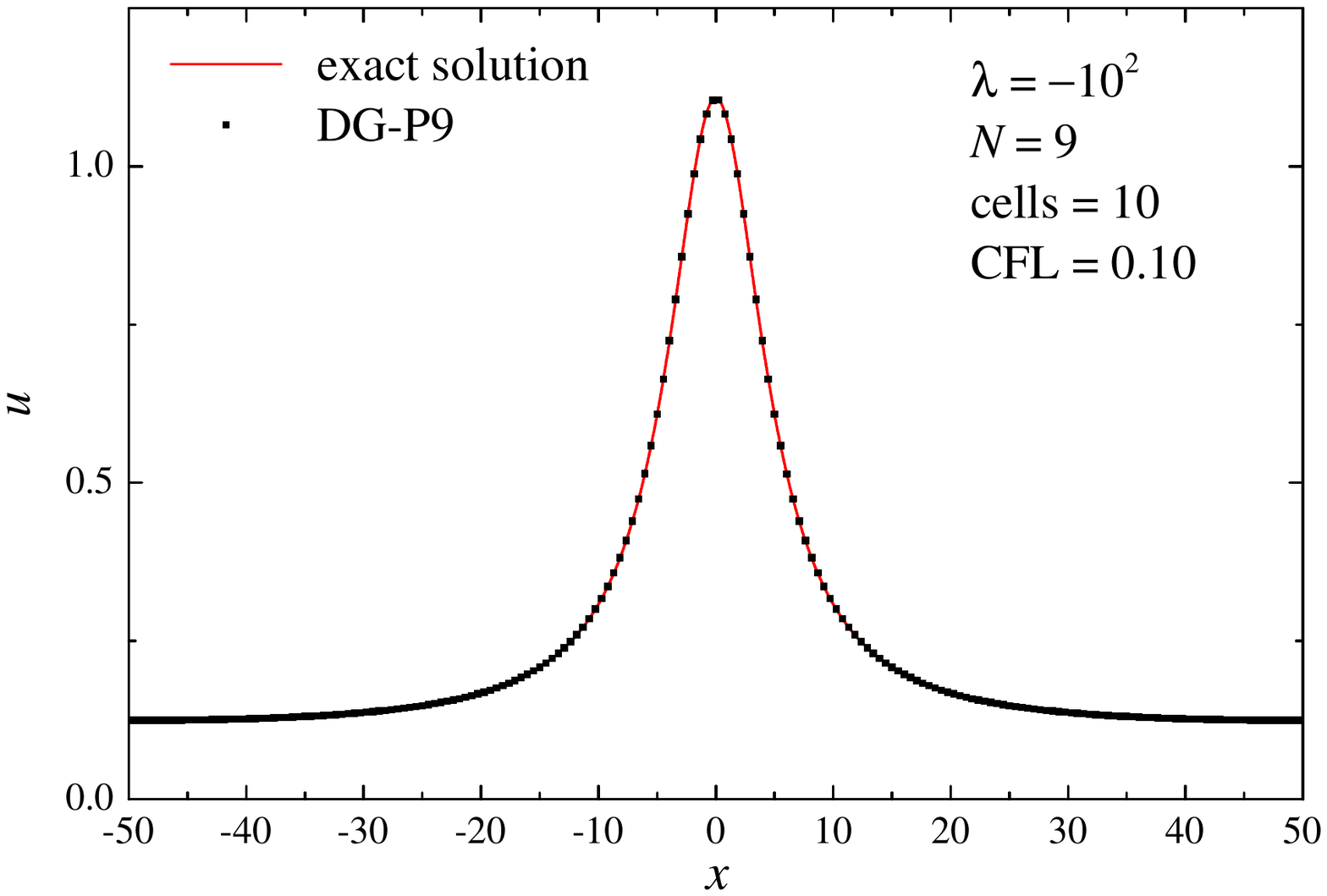}
\vspace{-8mm}\caption{\label{fig:test_adveq_stiff_z100_lrnz_1d_cfl_010:9}}
\end{subfigure}
\begin{subfigure}{0.24\textwidth}
\includegraphics[width=\textwidth]{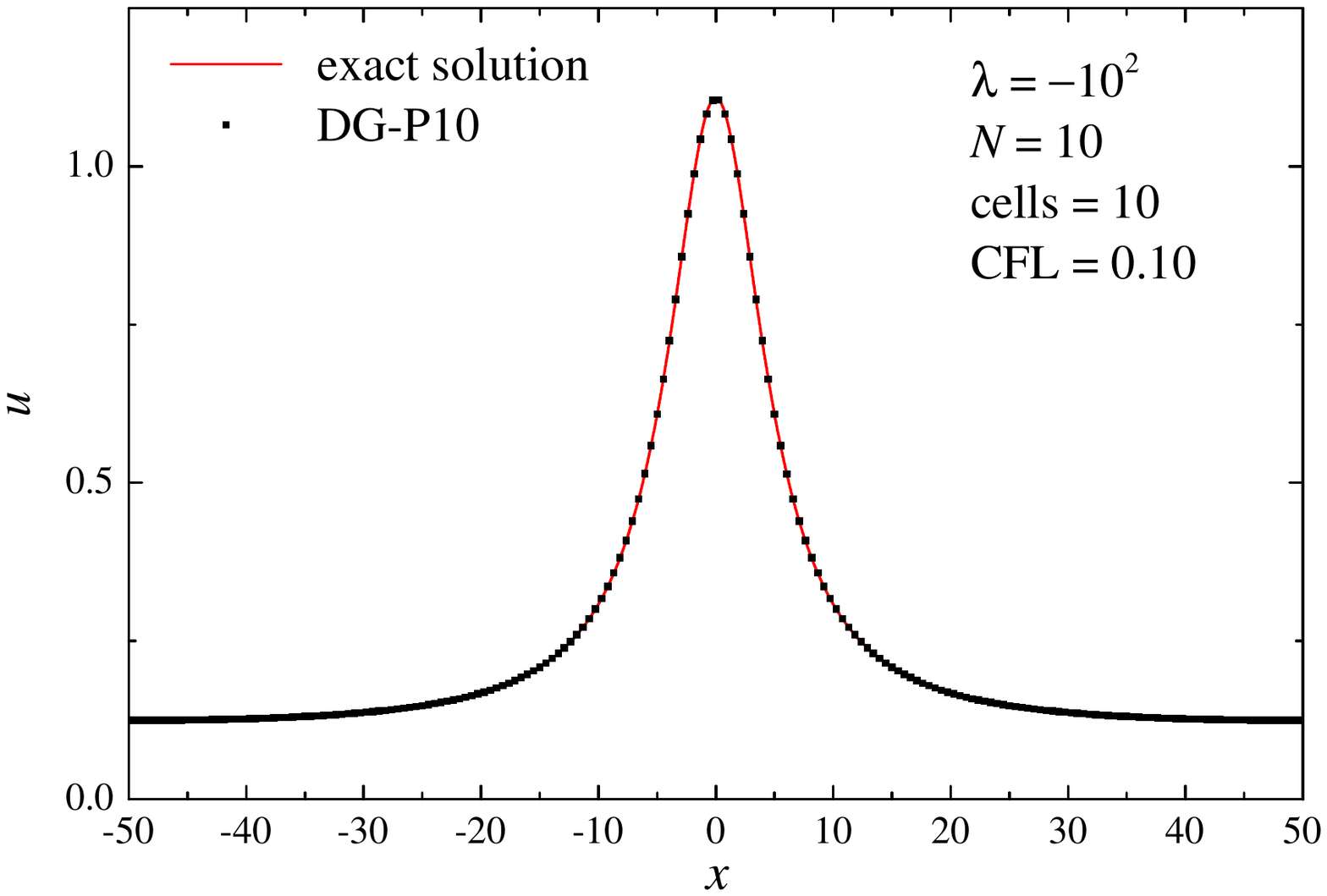}
\vspace{-8mm}\caption{\label{fig:test_adveq_stiff_z100_lrnz_1d_cfl_010:10}}
\end{subfigure}
\begin{subfigure}{0.24\textwidth}
\includegraphics[width=\textwidth]{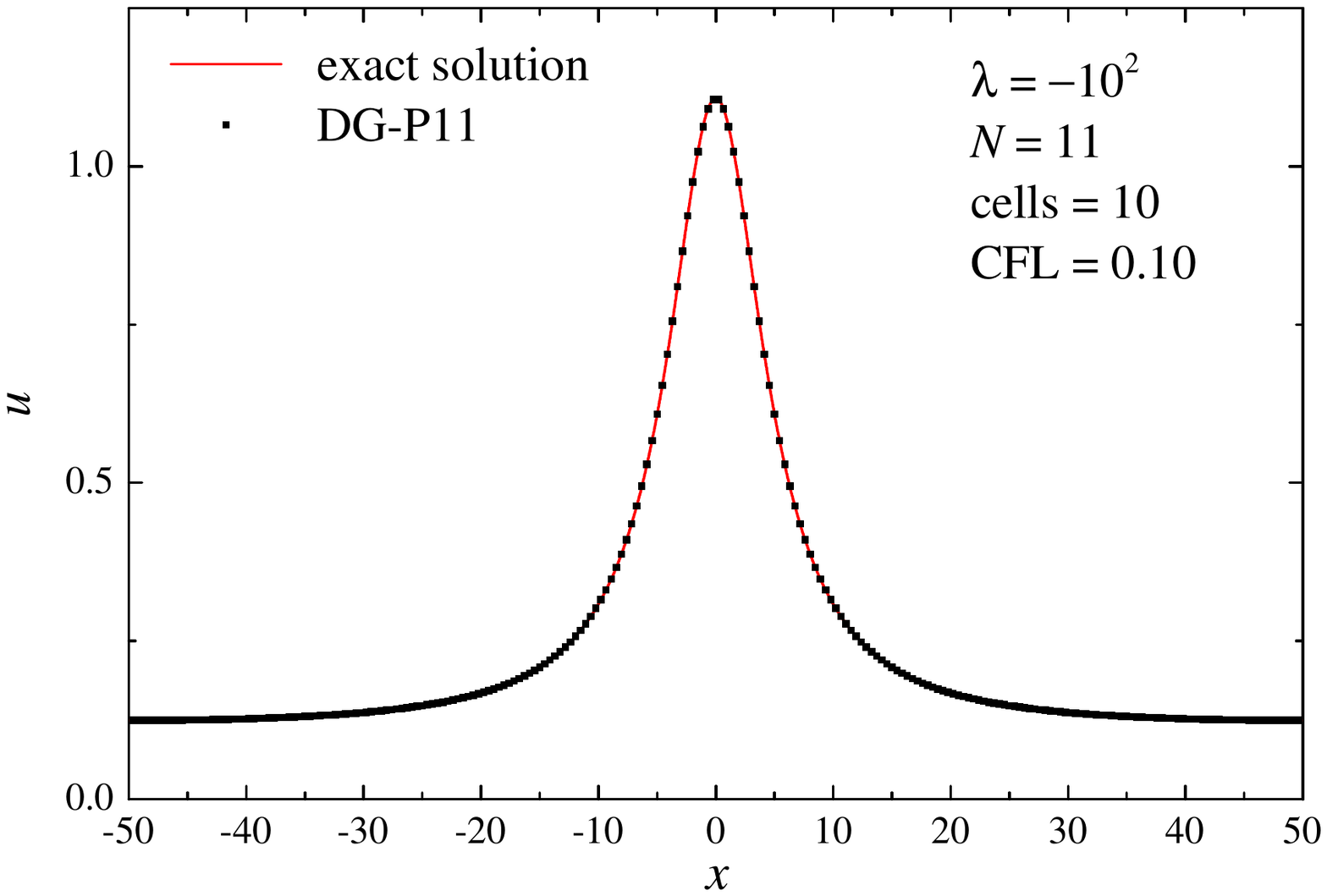}
\vspace{-8mm}\caption{\label{fig:test_adveq_stiff_z100_lrnz_1d_cfl_010:11}}
\end{subfigure}
\begin{subfigure}{0.24\textwidth}
\includegraphics[width=\textwidth]{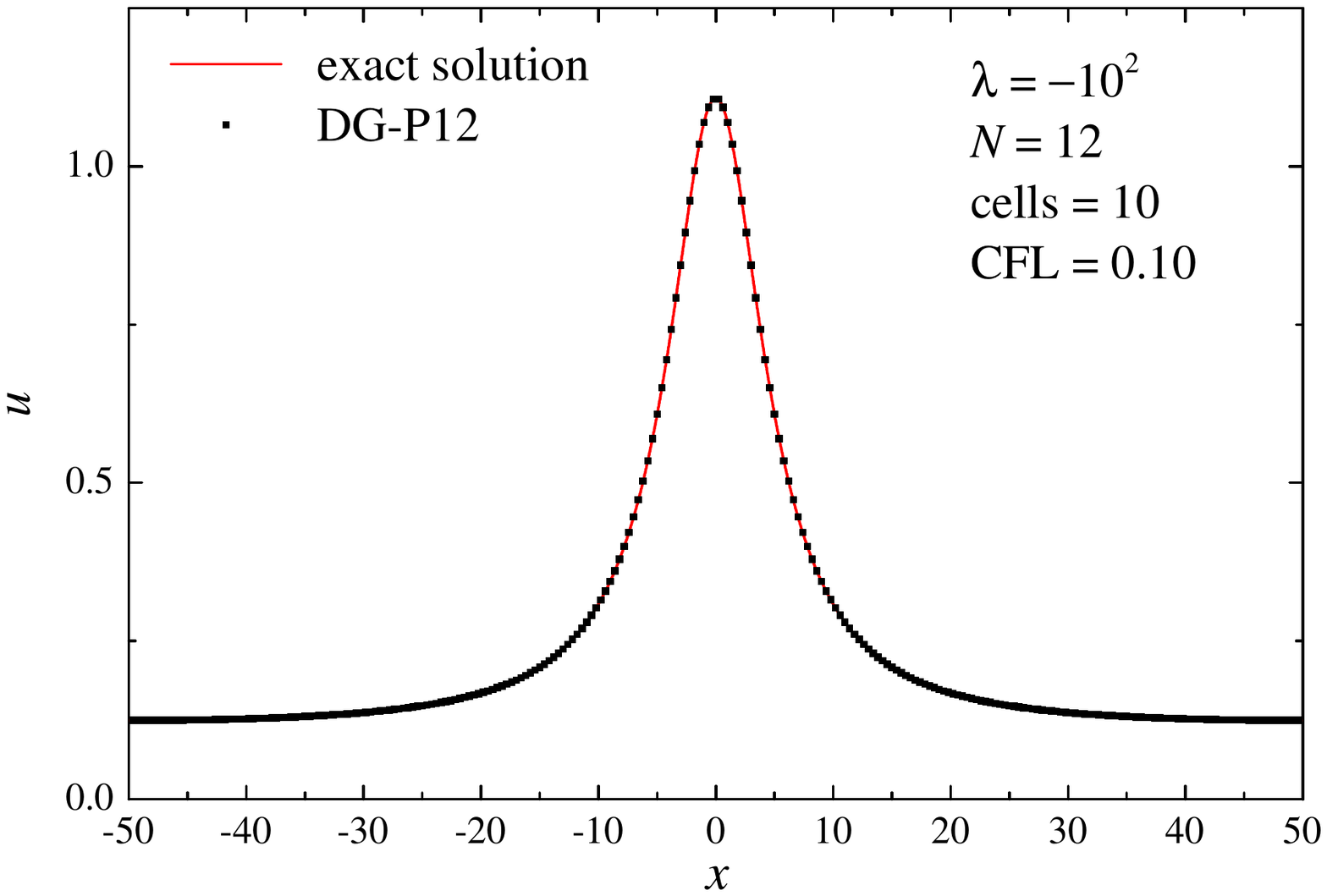}
\vspace{-8mm}\caption{\label{fig:test_adveq_stiff_z100_lrnz_1d_cfl_010:12}}
\end{subfigure}
\caption{%
Coordinate dependencies of the solution $u(x, t_{f})$ to the linear advection-reaction equation (\ref{eq:adv_eq_src}) at the final time $t_{f}$, obtained by the ADER-DG numerical method with the LST-DG predictor, for the degrees $N = 1, \ldots, 12$ of the basis polynomials for the Courant number $\mathrm{CFL} = 0.1$ and the decay parameter $\nu = -10^{2}$. The red line represents the exact analytical solution.
}
\label{fig:test_adveq_stiff_z100_lrnz_1d_cfl_010}
\end{figure}
\begin{figure}[h!]
\captionsetup[subfigure]{%
	position=bottom,
	font+=smaller,
	textfont=normalfont,
	singlelinecheck=off,
	justification=raggedright
}
\centering
\begin{subfigure}{0.24\textwidth}
\includegraphics[width=\textwidth]{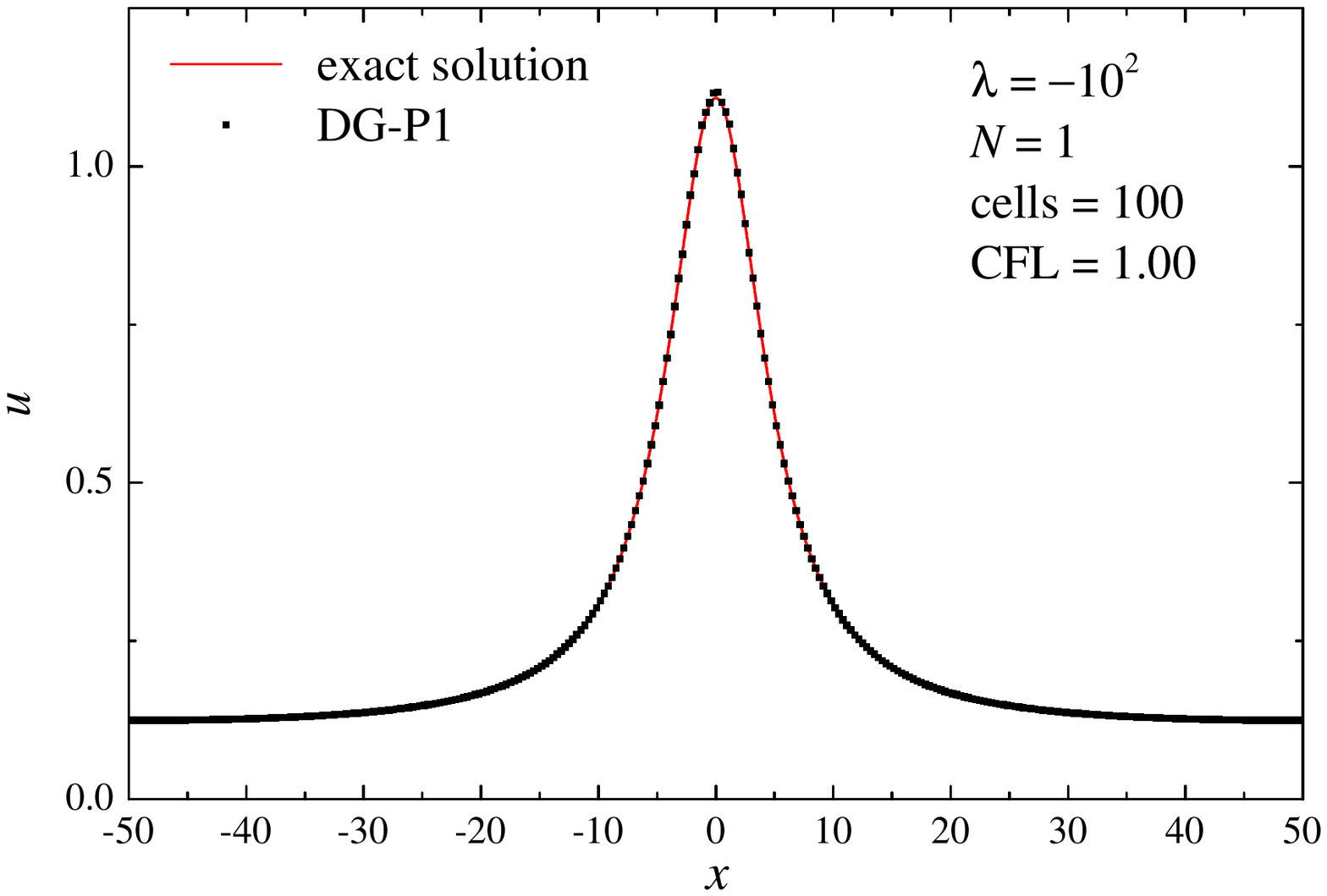}
\vspace{-8mm}\caption{\label{fig:test_adveq_stiff_z100_lrnz_1d_cfl_100:1}}
\end{subfigure}
\begin{subfigure}{0.24\textwidth}
\includegraphics[width=\textwidth]{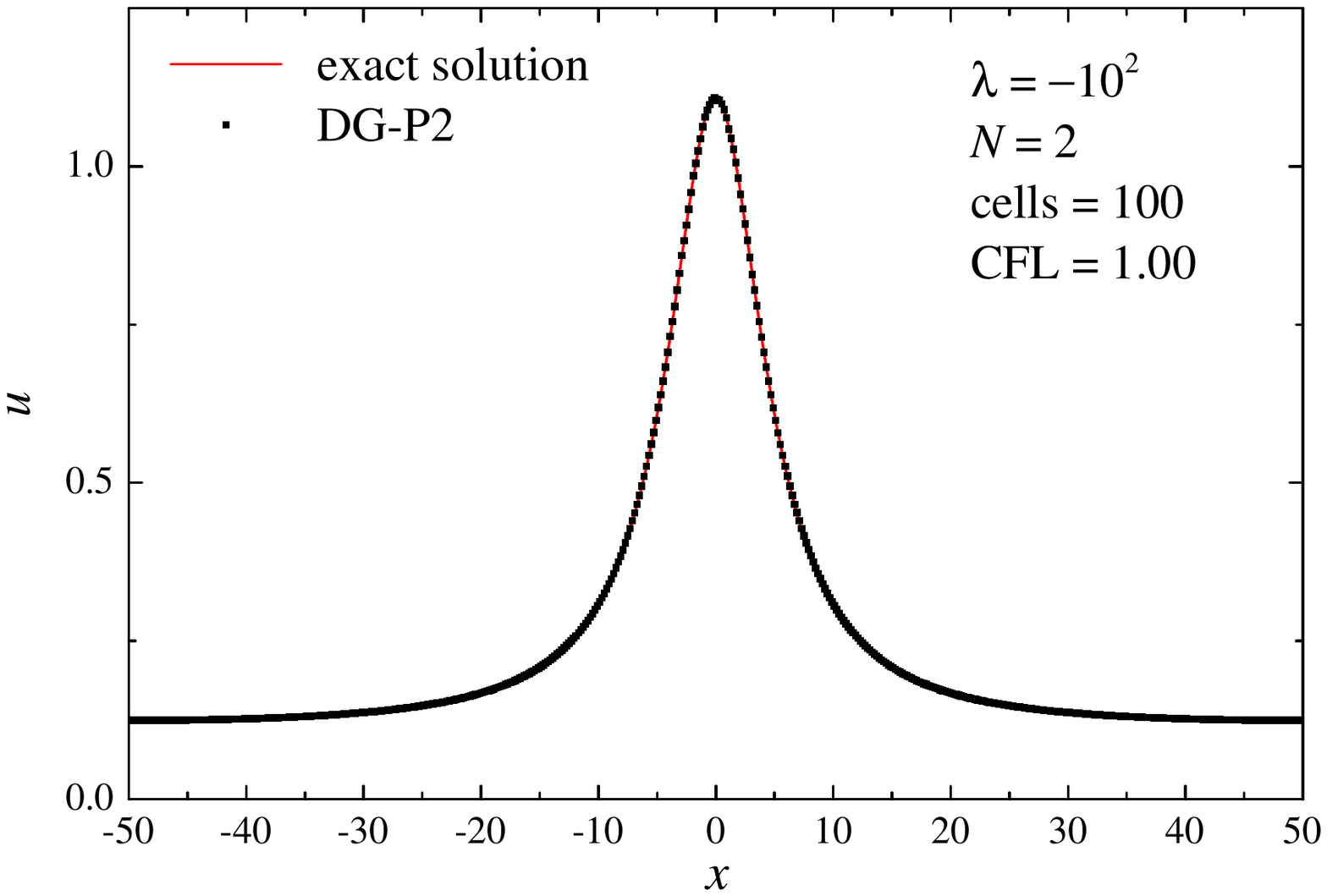}
\vspace{-8mm}\caption{\label{fig:test_adveq_stiff_z100_lrnz_1d_cfl_100:2}}
\end{subfigure}
\begin{subfigure}{0.24\textwidth}
\includegraphics[width=\textwidth]{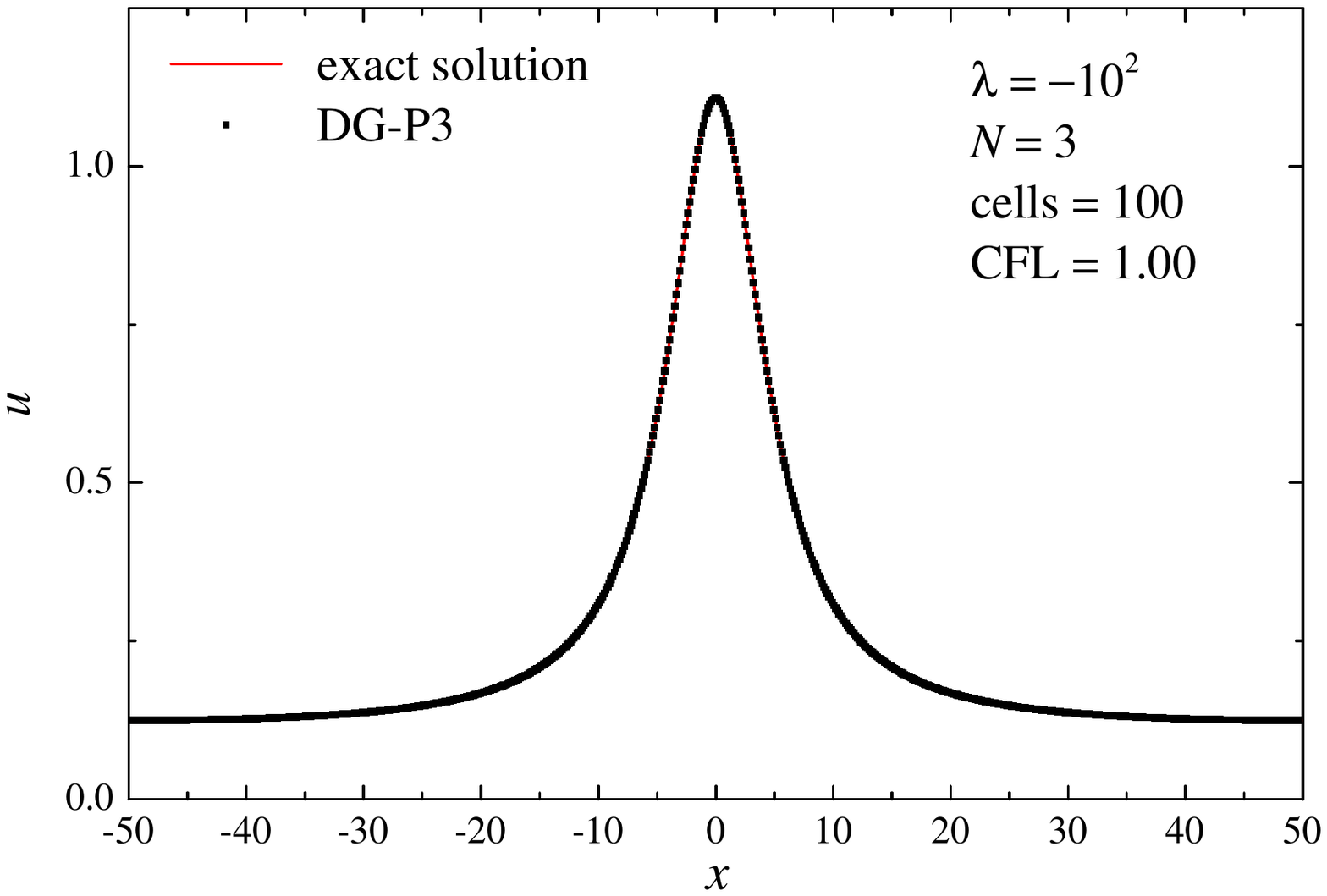}
\vspace{-8mm}\caption{\label{fig:test_adveq_stiff_z100_lrnz_1d_cfl_100:3}}
\end{subfigure}
\begin{subfigure}{0.24\textwidth}
\includegraphics[width=\textwidth]{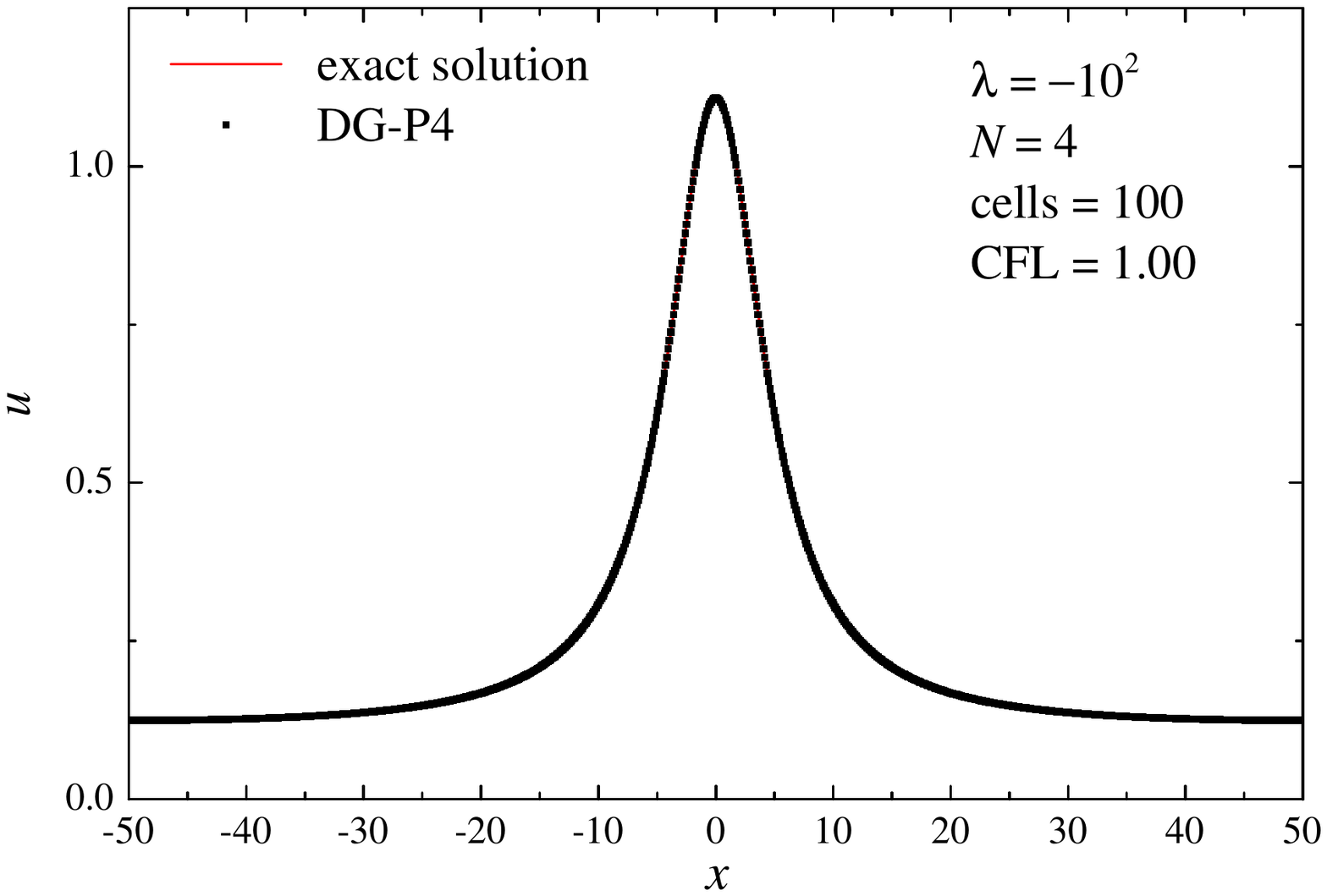}
\vspace{-8mm}\caption{\label{fig:test_adveq_stiff_z100_lrnz_1d_cfl_100:4}}
\end{subfigure}\\[-2.5mm]
\begin{subfigure}{0.24\textwidth}
\includegraphics[width=\textwidth]{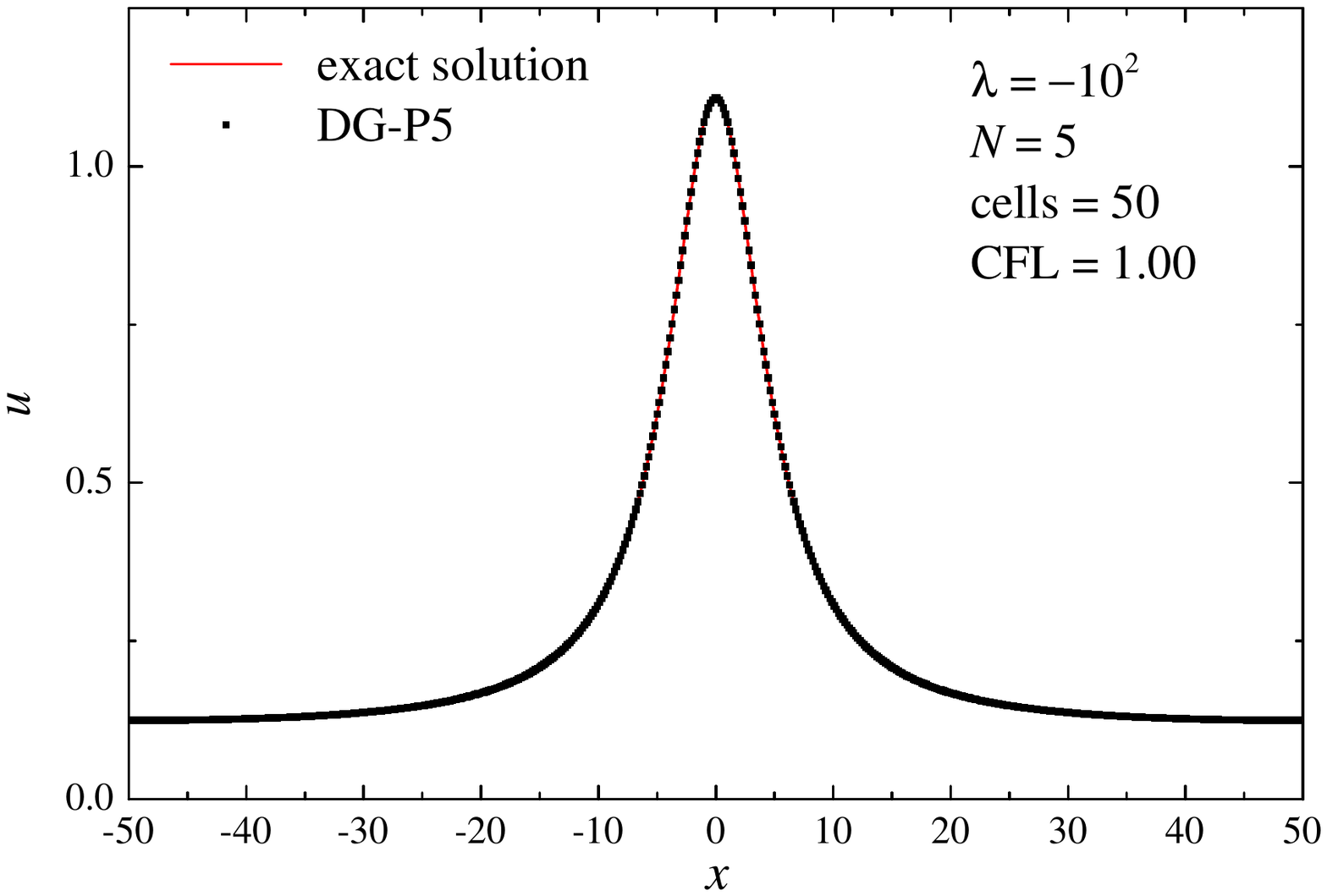}
\vspace{-8mm}\caption{\label{fig:test_adveq_stiff_z100_lrnz_1d_cfl_100:5}}
\end{subfigure}
\begin{subfigure}{0.24\textwidth}
\includegraphics[width=\textwidth]{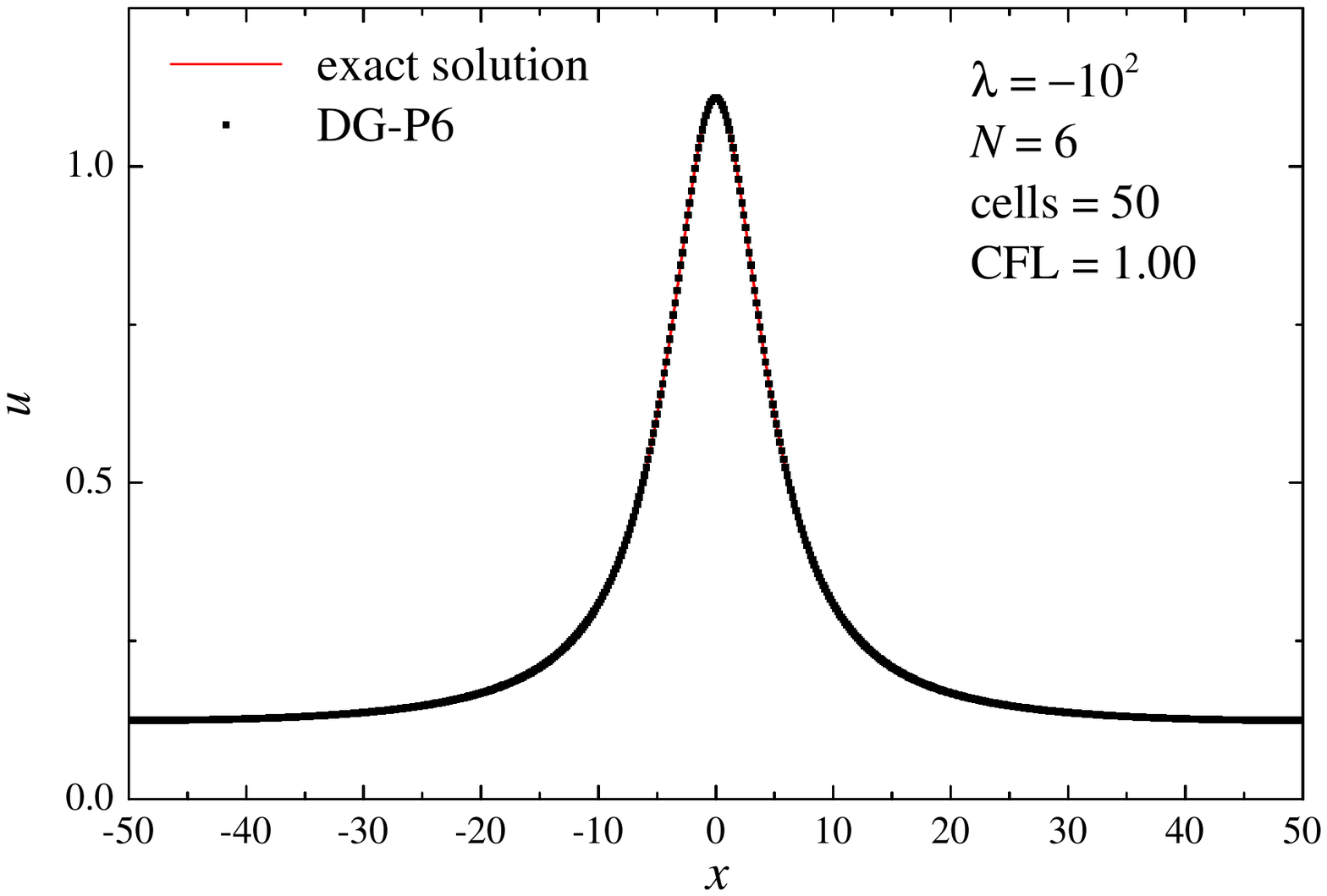}
\vspace{-8mm}\caption{\label{fig:test_adveq_stiff_z100_lrnz_1d_cfl_100:6}}
\end{subfigure}
\begin{subfigure}{0.24\textwidth}
\includegraphics[width=\textwidth]{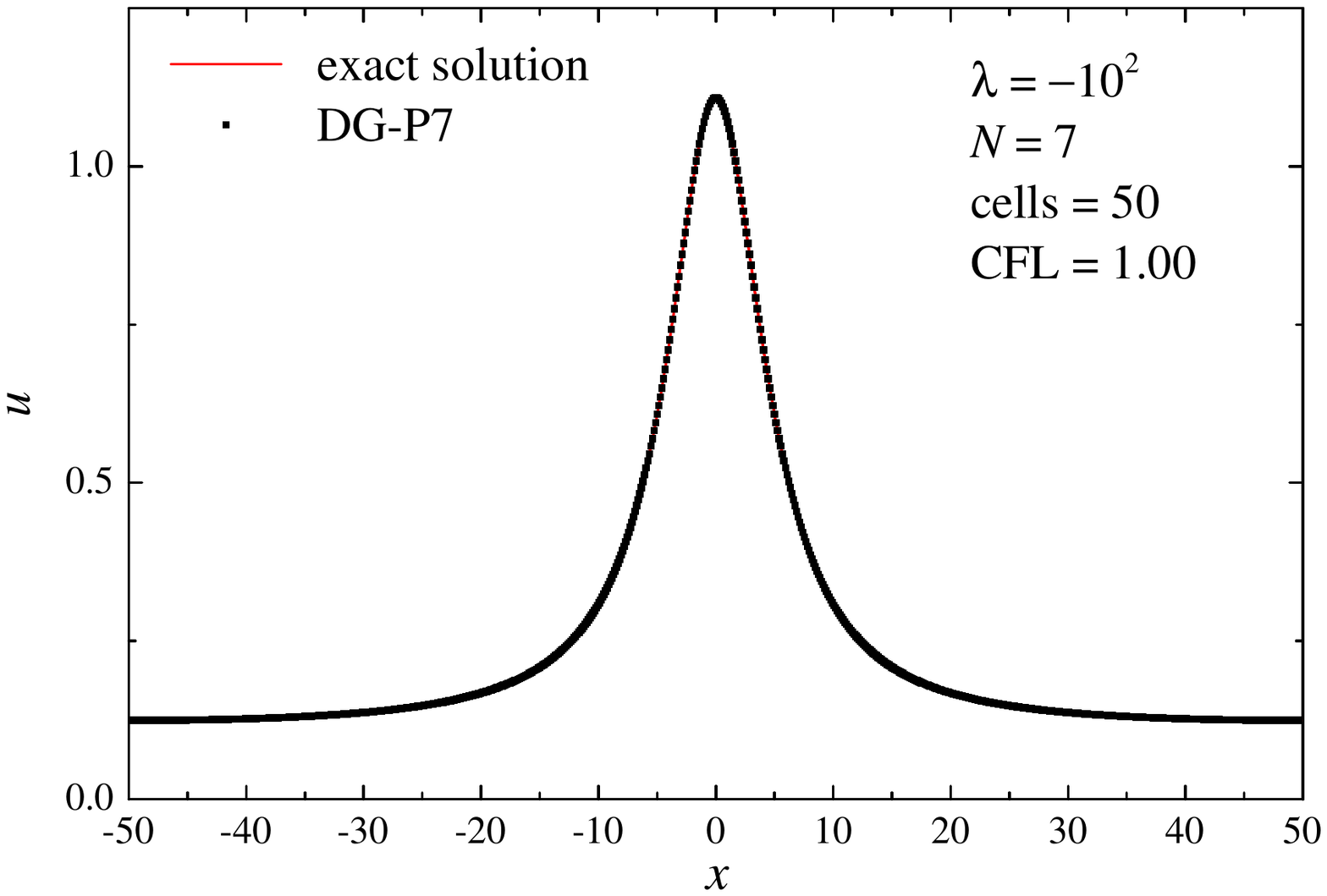}
\vspace{-8mm}\caption{\label{fig:test_adveq_stiff_z100_lrnz_1d_cfl_100:7}}
\end{subfigure}
\begin{subfigure}{0.24\textwidth}
\includegraphics[width=\textwidth]{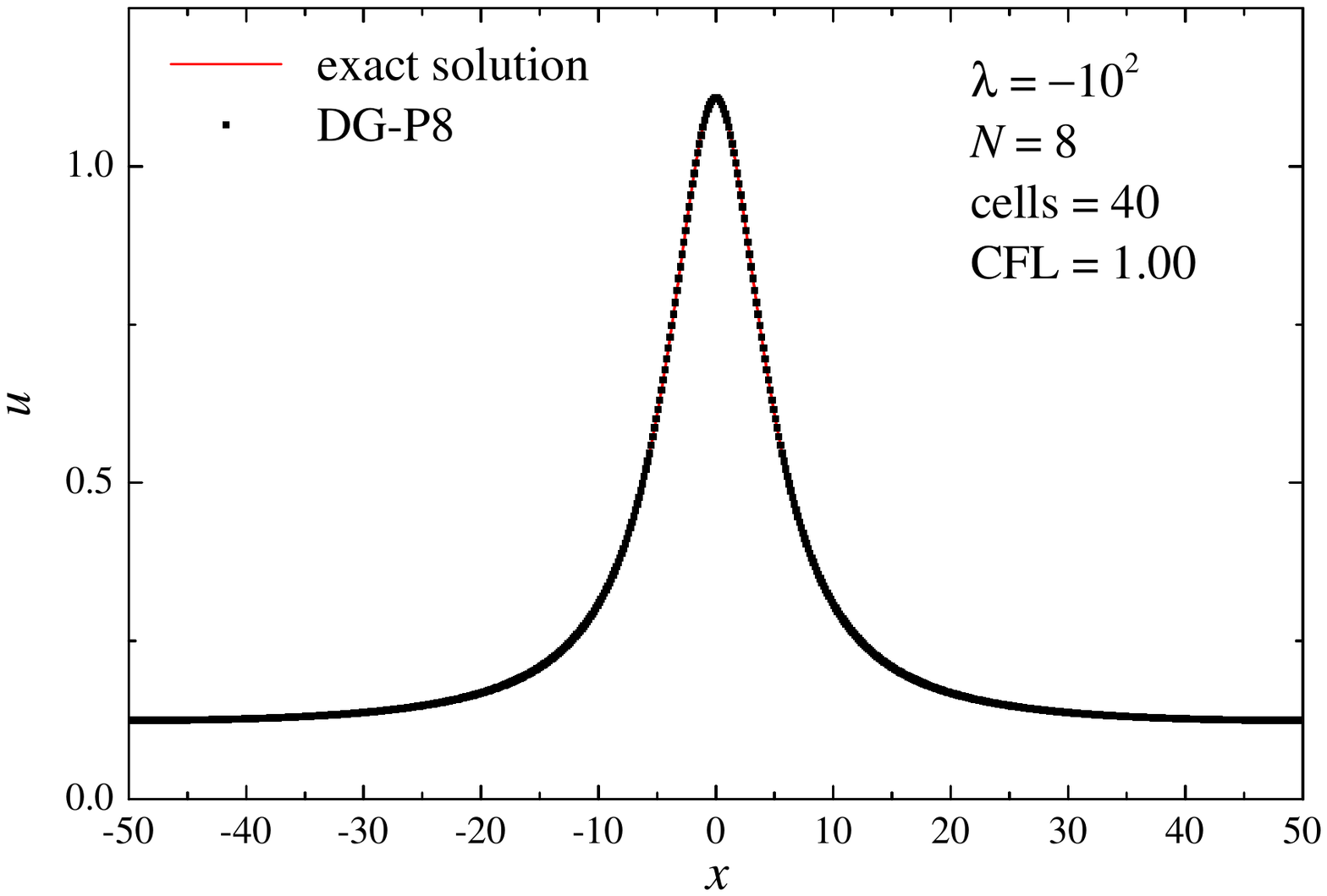}
\vspace{-8mm}\caption{\label{fig:test_adveq_stiff_z100_lrnz_1d_cfl_100:8}}
\end{subfigure}\\[-2.5mm]
\begin{subfigure}{0.24\textwidth}
\includegraphics[width=\textwidth]{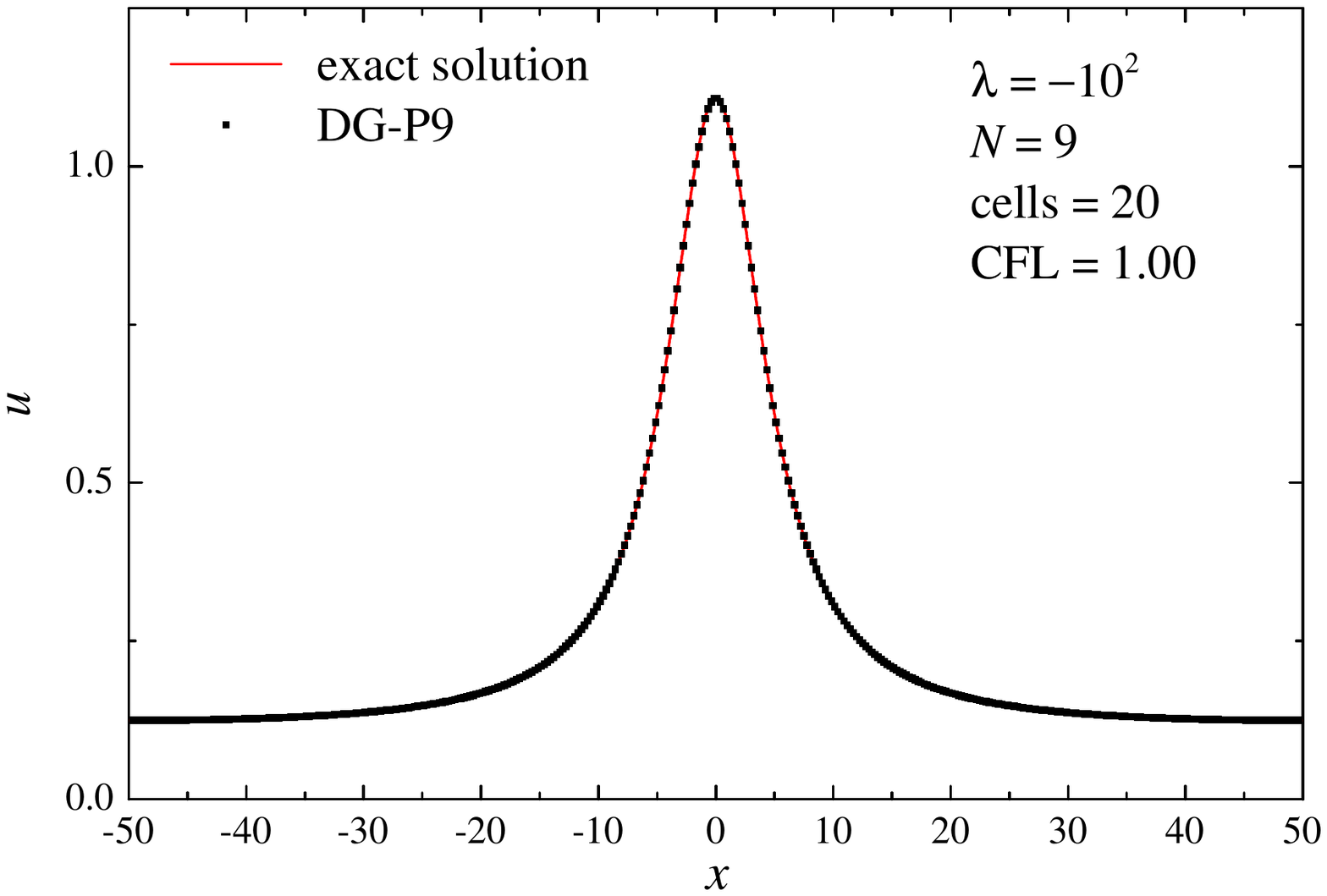}
\vspace{-8mm}\caption{\label{fig:test_adveq_stiff_z100_lrnz_1d_cfl_100:9}}
\end{subfigure}
\begin{subfigure}{0.24\textwidth}
\includegraphics[width=\textwidth]{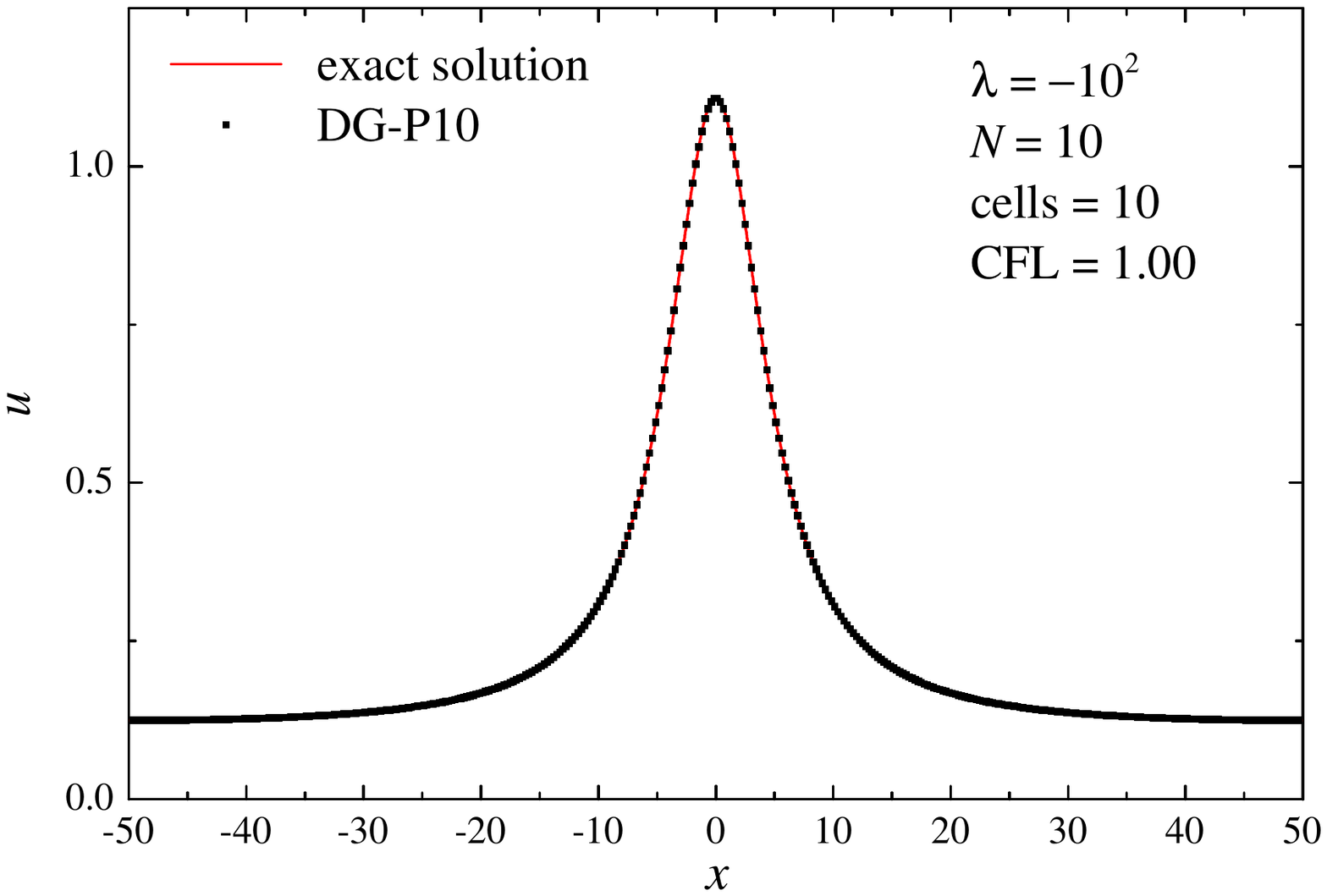}
\vspace{-8mm}\caption{\label{fig:test_adveq_stiff_z100_lrnz_1d_cfl_100:10}}
\end{subfigure}
\begin{subfigure}{0.24\textwidth}
\includegraphics[width=\textwidth]{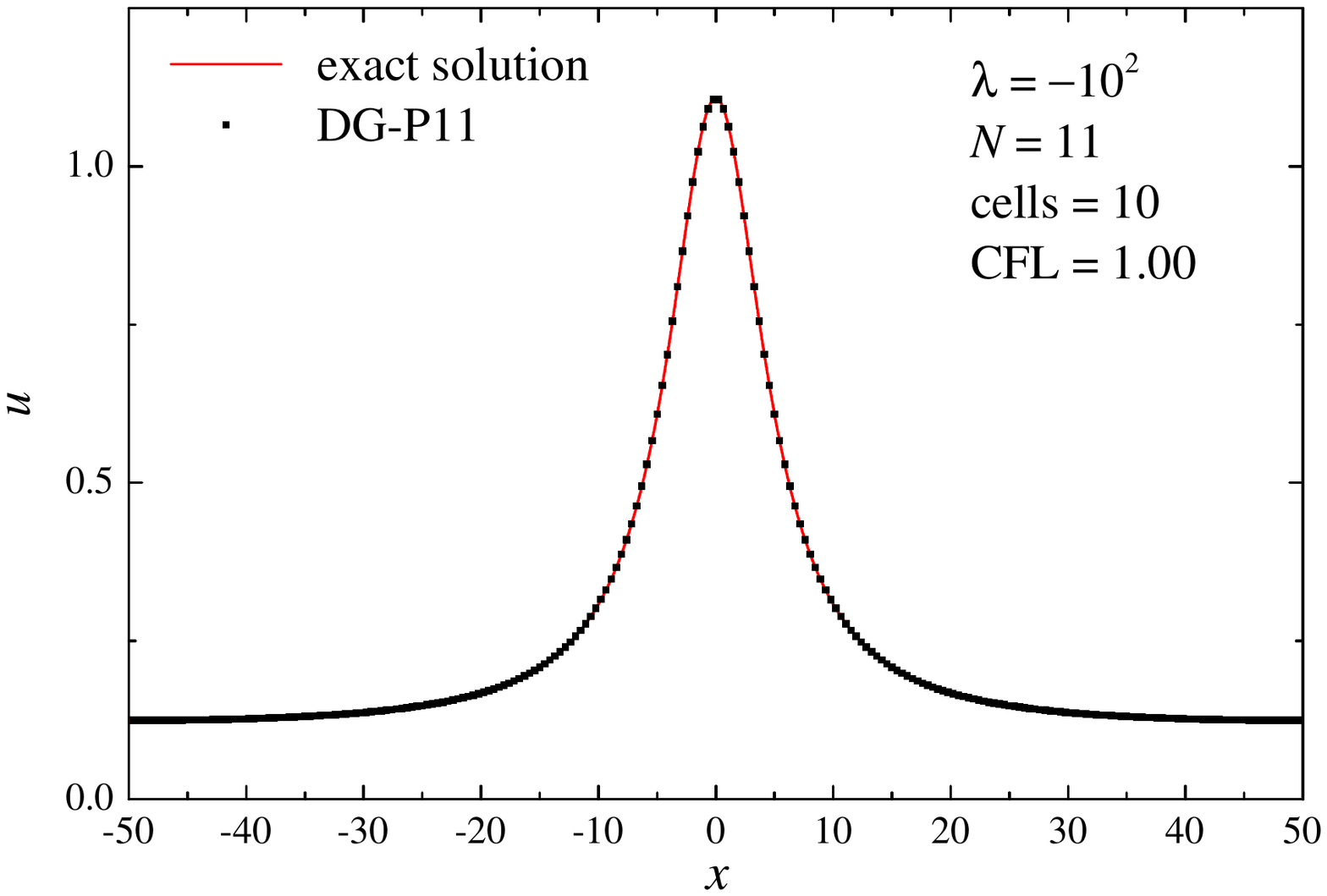}
\vspace{-8mm}\caption{\label{fig:test_adveq_stiff_z100_lrnz_1d_cfl_100:11}}
\end{subfigure}
\begin{subfigure}{0.24\textwidth}
\includegraphics[width=\textwidth]{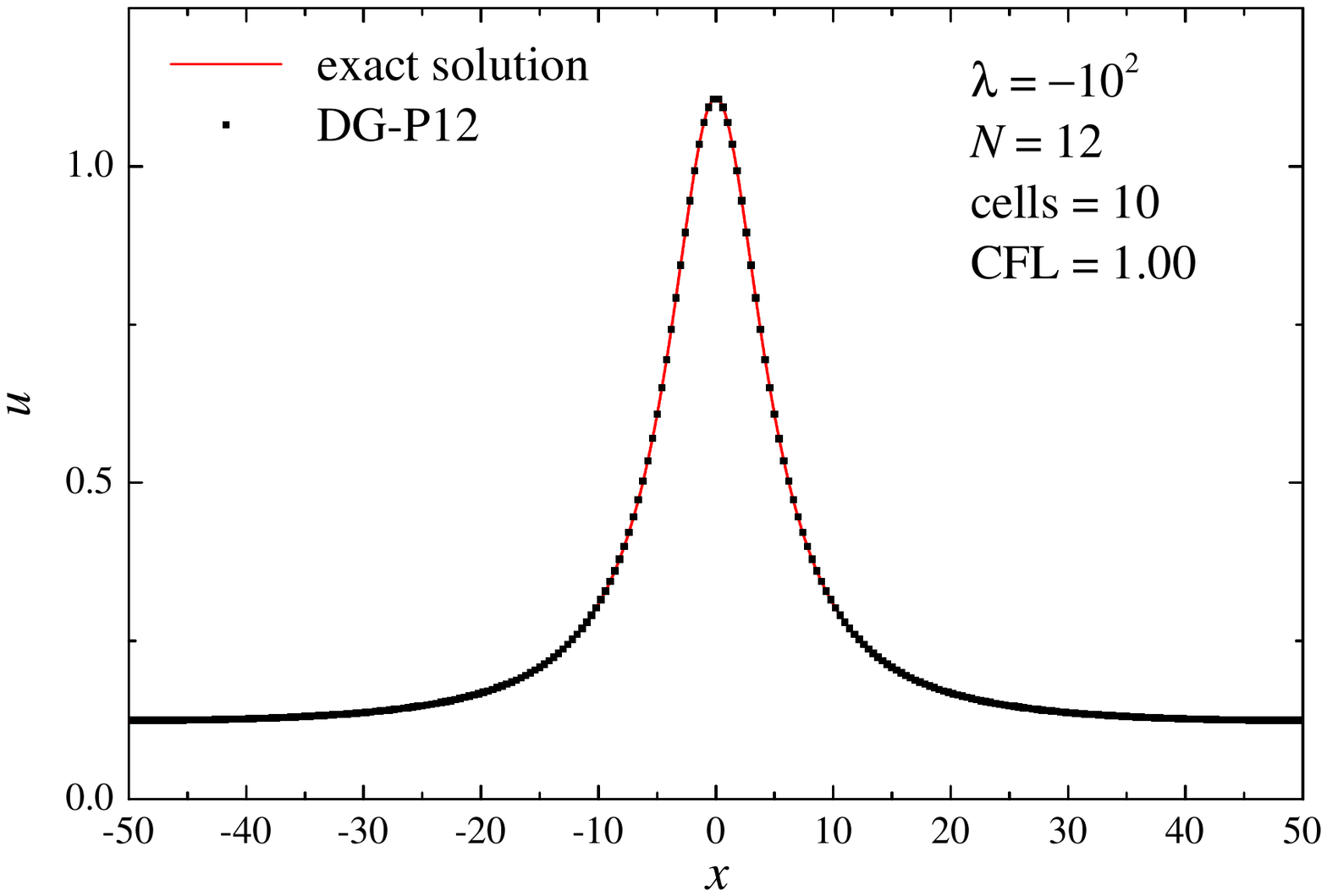}
\vspace{-8mm}\caption{\label{fig:test_adveq_stiff_z100_lrnz_1d_cfl_100:12}}
\end{subfigure}
\caption{%
Coordinate dependencies of the solution $u(x, t_{f})$ to the linear advection-reaction equation (\ref{eq:adv_eq_src}) at the final time $t_{f}$, obtained by the ADER-DG numerical method with the LST-DG predictor, for the degrees $N = 1, \ldots, 12$ of the basis polynomials for the Courant number $\mathrm{CFL} = 1.0$ and the decay parameter $\nu = -10^{2}$. The red line represents the exact analytical solution.
}
\label{fig:test_adveq_stiff_z100_lrnz_1d_cfl_100}
\end{figure}
\begin{figure}[h!]
\captionsetup[subfigure]{%
	position=bottom,
	font+=smaller,
	textfont=normalfont,
	singlelinecheck=off,
	justification=raggedright
}
\centering
\begin{subfigure}{0.24\textwidth}
\includegraphics[width=\textwidth]{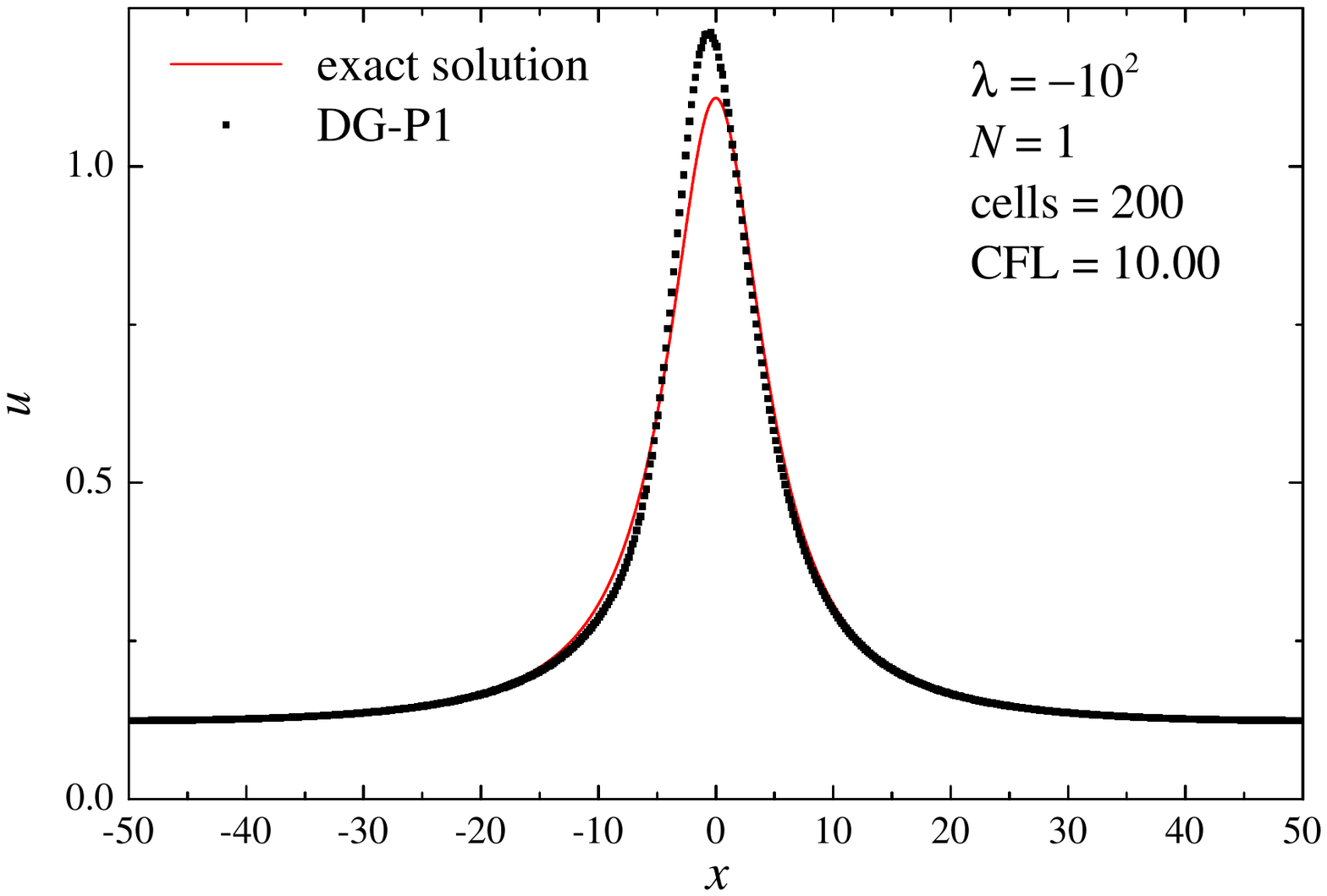}
\vspace{-8mm}\caption{\label{fig:test_adveq_stiff_z100_lrnz_1d_cfl_1000:1}}
\end{subfigure}
\begin{subfigure}{0.24\textwidth}
\includegraphics[width=\textwidth]{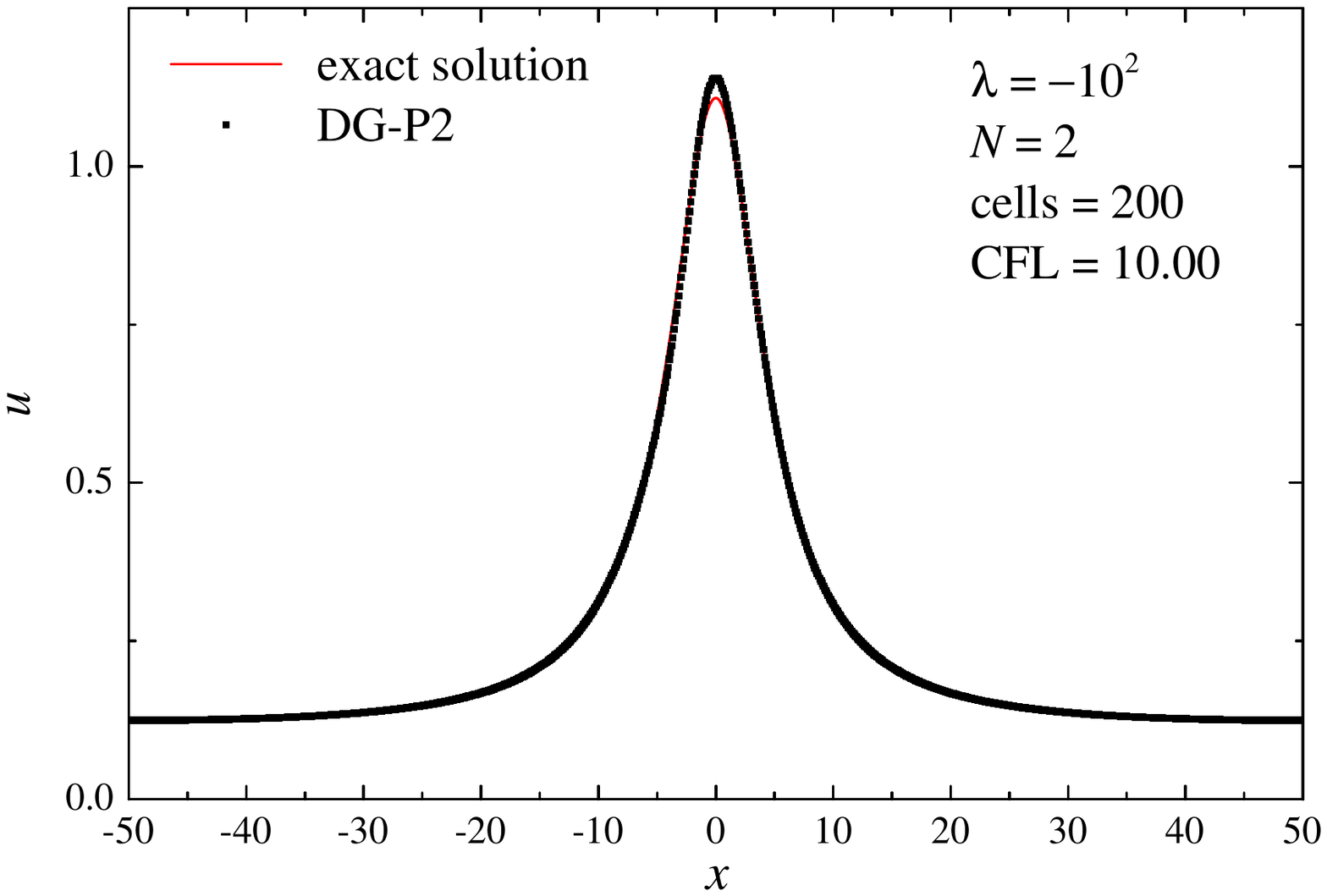}
\vspace{-8mm}\caption{\label{fig:test_adveq_stiff_z100_lrnz_1d_cfl_1000:2}}
\end{subfigure}
\begin{subfigure}{0.24\textwidth}
\includegraphics[width=\textwidth]{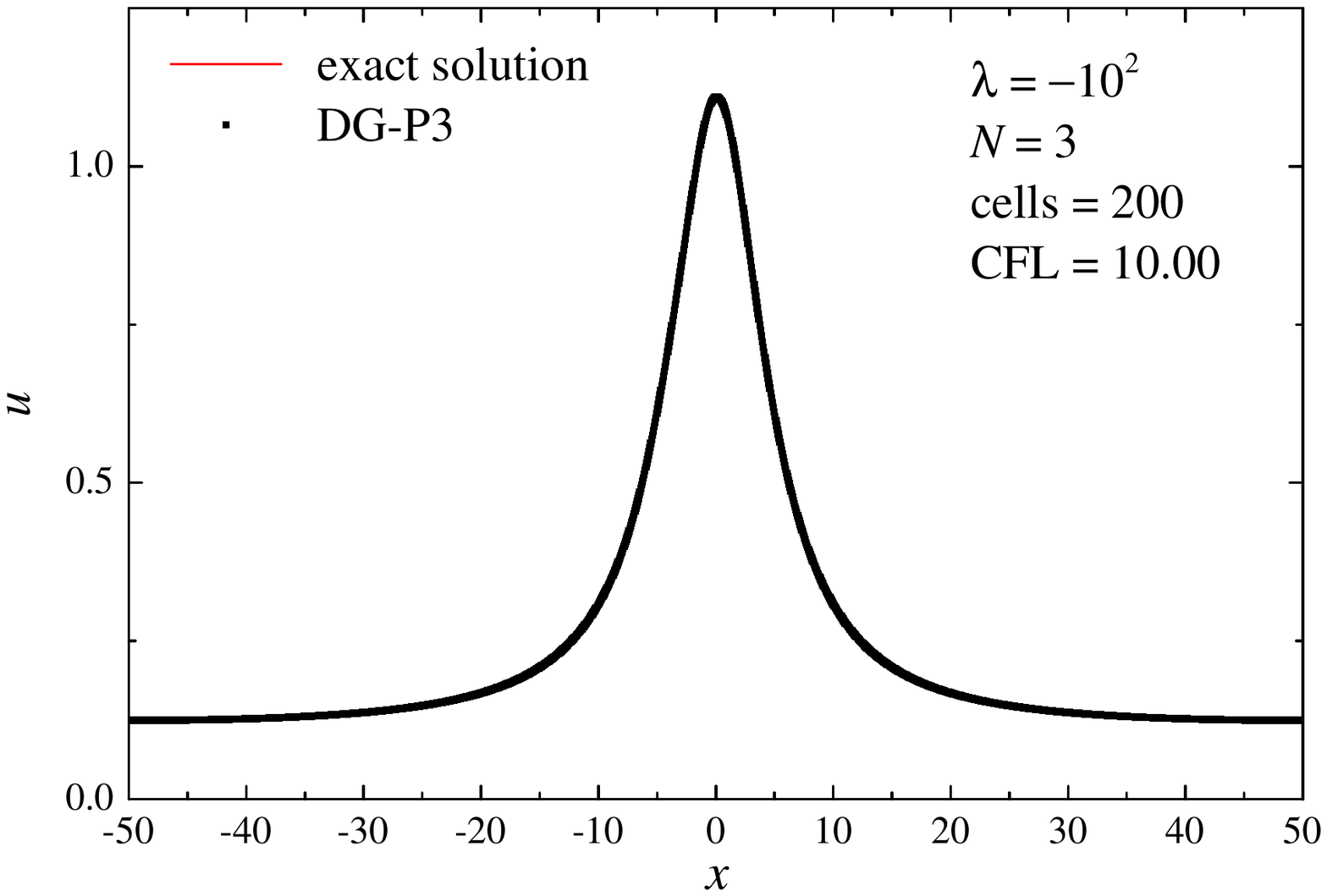}
\vspace{-8mm}\caption{\label{fig:test_adveq_stiff_z100_lrnz_1d_cfl_1000:3}}
\end{subfigure}
\begin{subfigure}{0.24\textwidth}
\includegraphics[width=\textwidth]{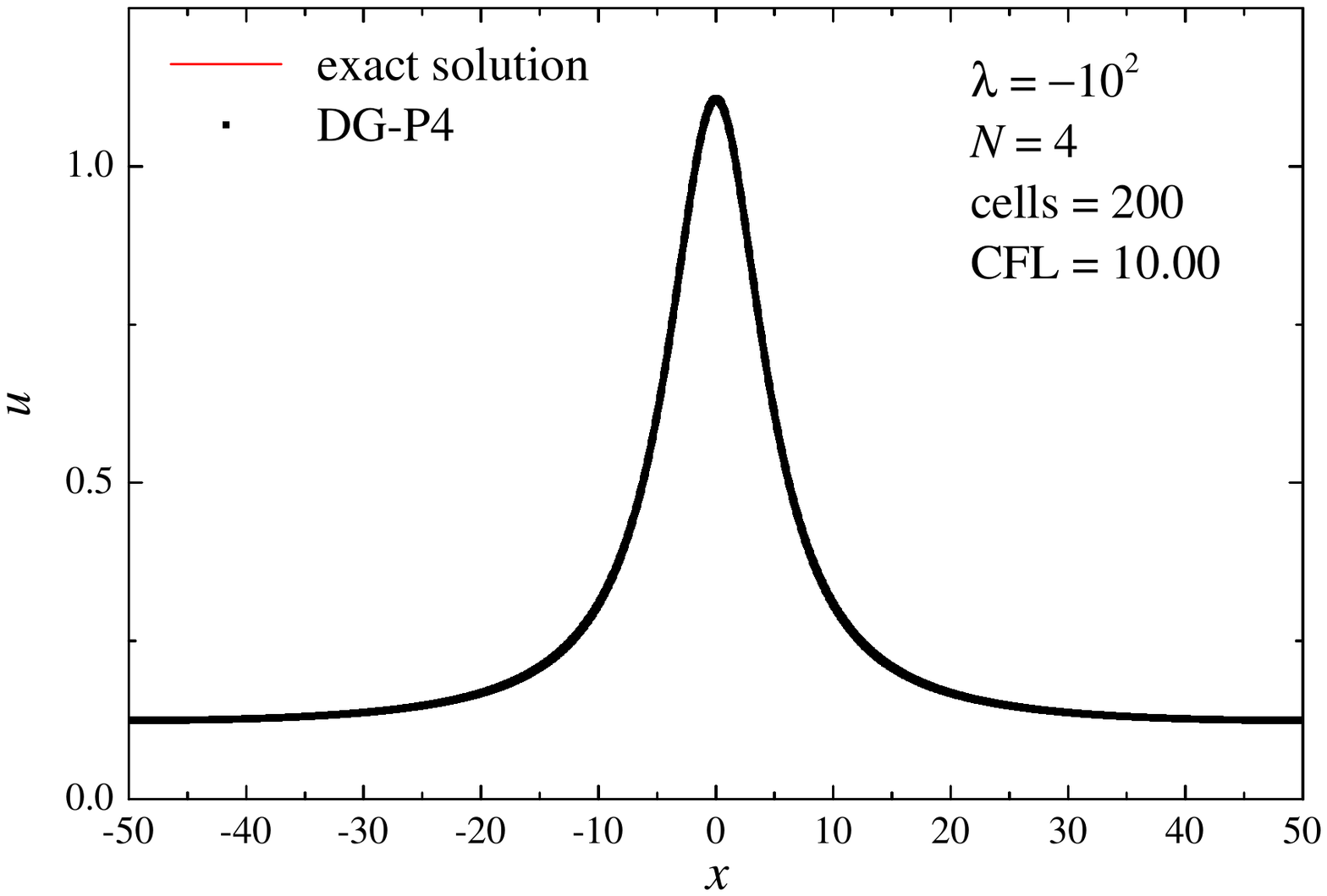}
\vspace{-8mm}\caption{\label{fig:test_adveq_stiff_z100_lrnz_1d_cfl_1000:4}}
\end{subfigure}\\[-2.5mm]
\begin{subfigure}{0.24\textwidth}
\includegraphics[width=\textwidth]{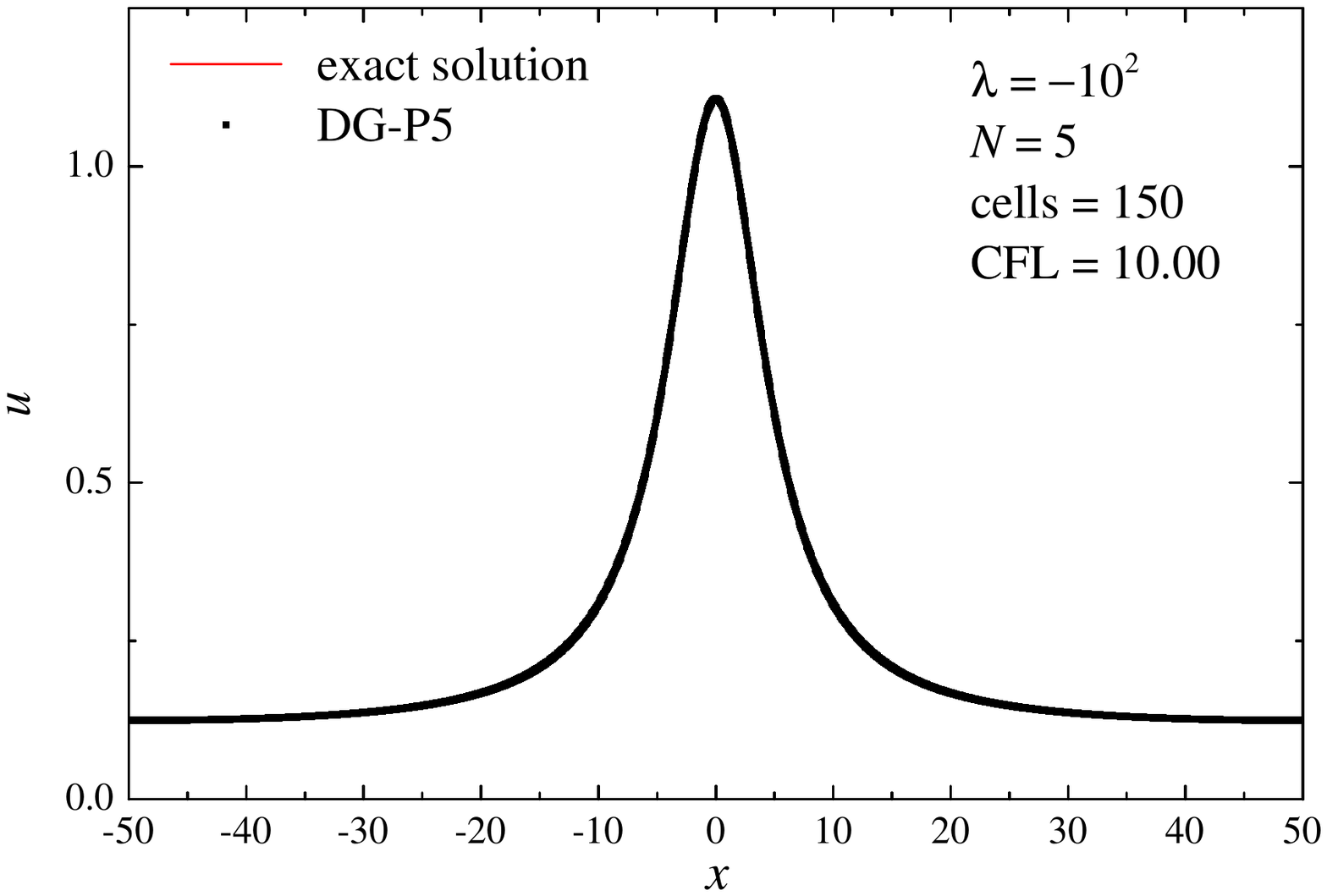}
\vspace{-8mm}\caption{\label{fig:test_adveq_stiff_z100_lrnz_1d_cfl_1000:5}}
\end{subfigure}
\begin{subfigure}{0.24\textwidth}
\includegraphics[width=\textwidth]{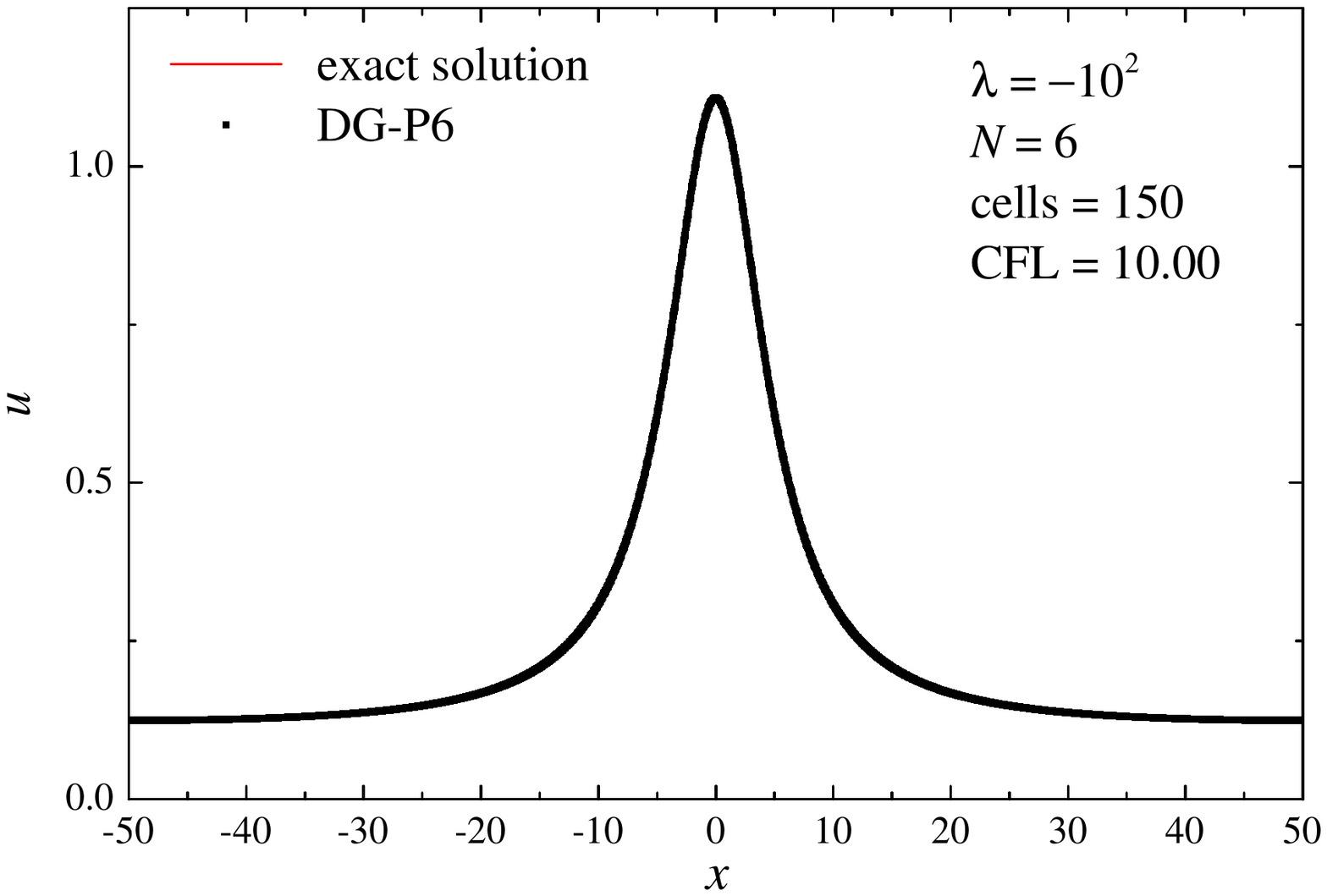}
\vspace{-8mm}\caption{\label{fig:test_adveq_stiff_z100_lrnz_1d_cfl_1000:6}}
\end{subfigure}
\begin{subfigure}{0.24\textwidth}
\includegraphics[width=\textwidth]{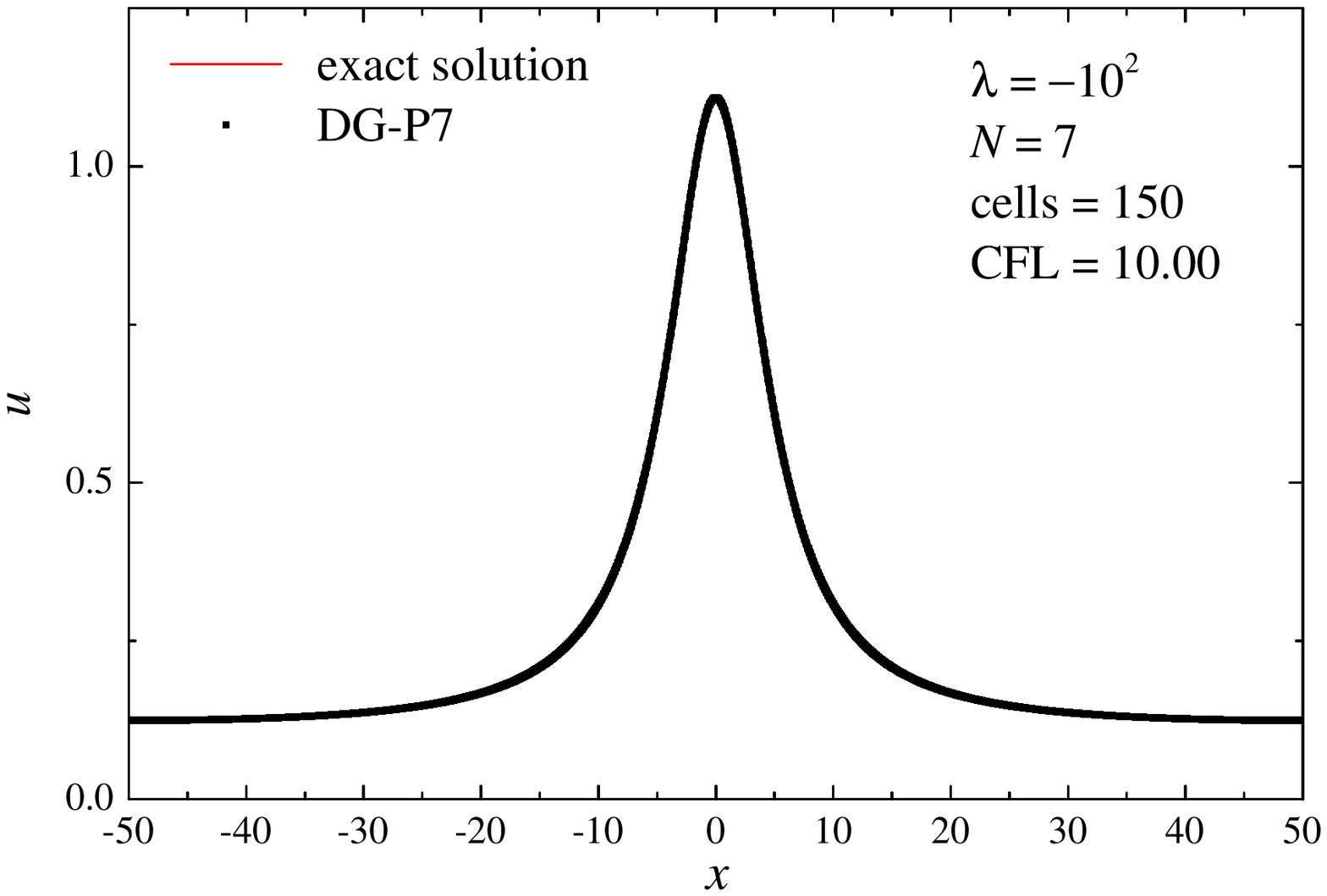}
\vspace{-8mm}\caption{\label{fig:test_adveq_stiff_z100_lrnz_1d_cfl_1000:7}}
\end{subfigure}
\begin{subfigure}{0.24\textwidth}
\includegraphics[width=\textwidth]{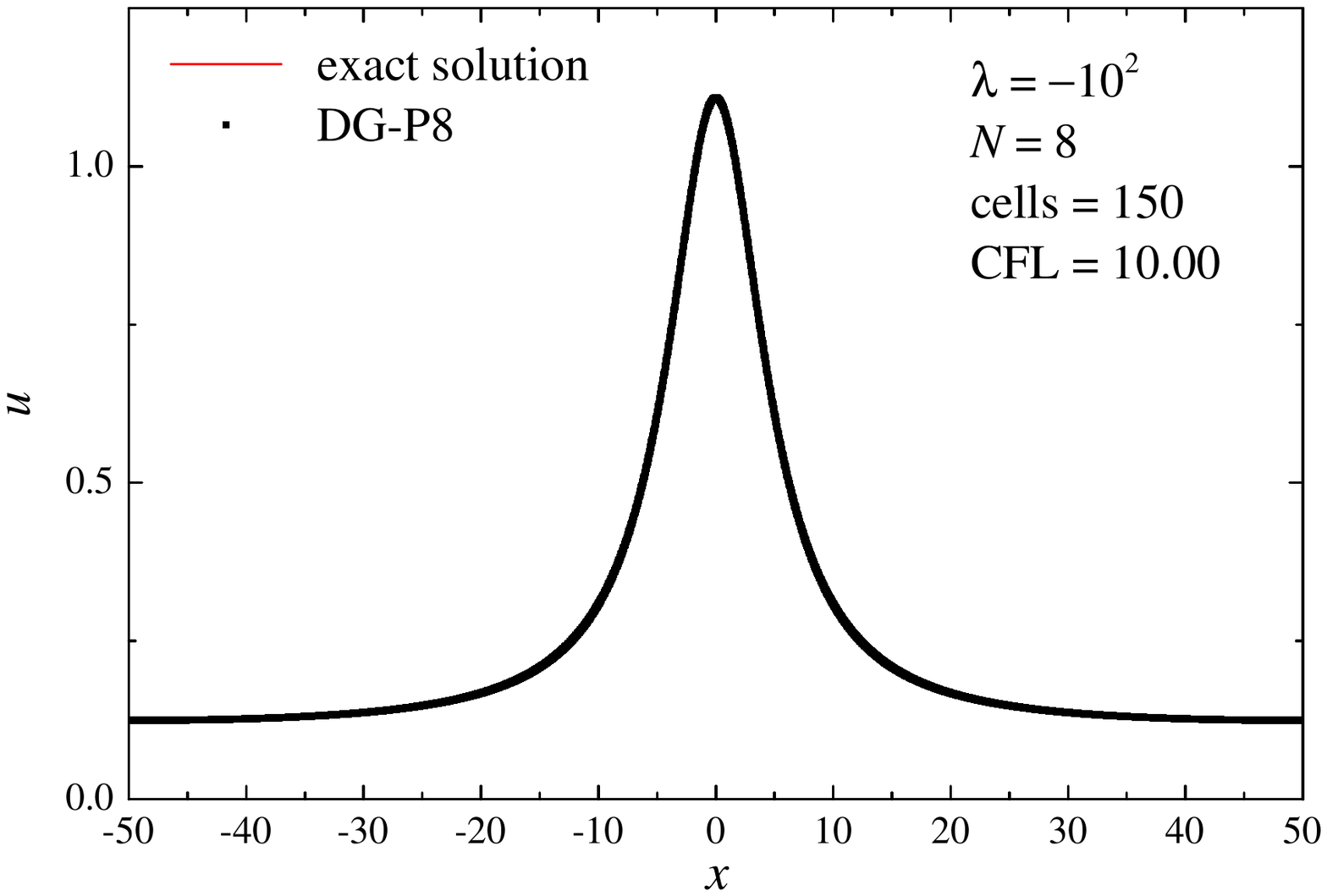}
\vspace{-8mm}\caption{\label{fig:test_adveq_stiff_z100_lrnz_1d_cfl_1000:8}}
\end{subfigure}\\[-2.5mm]
\begin{subfigure}{0.24\textwidth}
\includegraphics[width=\textwidth]{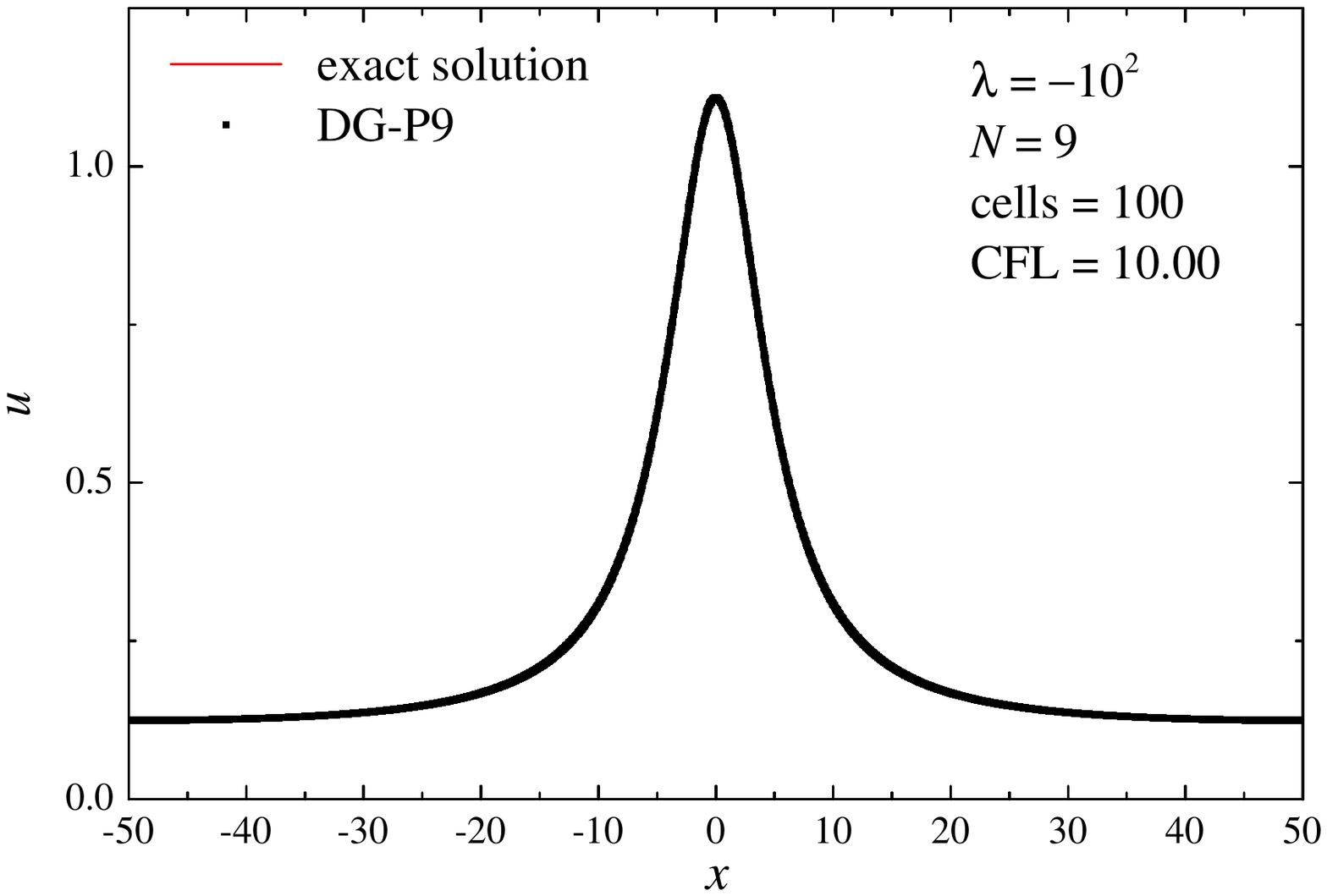}
\vspace{-8mm}\caption{\label{fig:test_adveq_stiff_z100_lrnz_1d_cfl_1000:9}}
\end{subfigure}
\begin{subfigure}{0.24\textwidth}
\includegraphics[width=\textwidth]{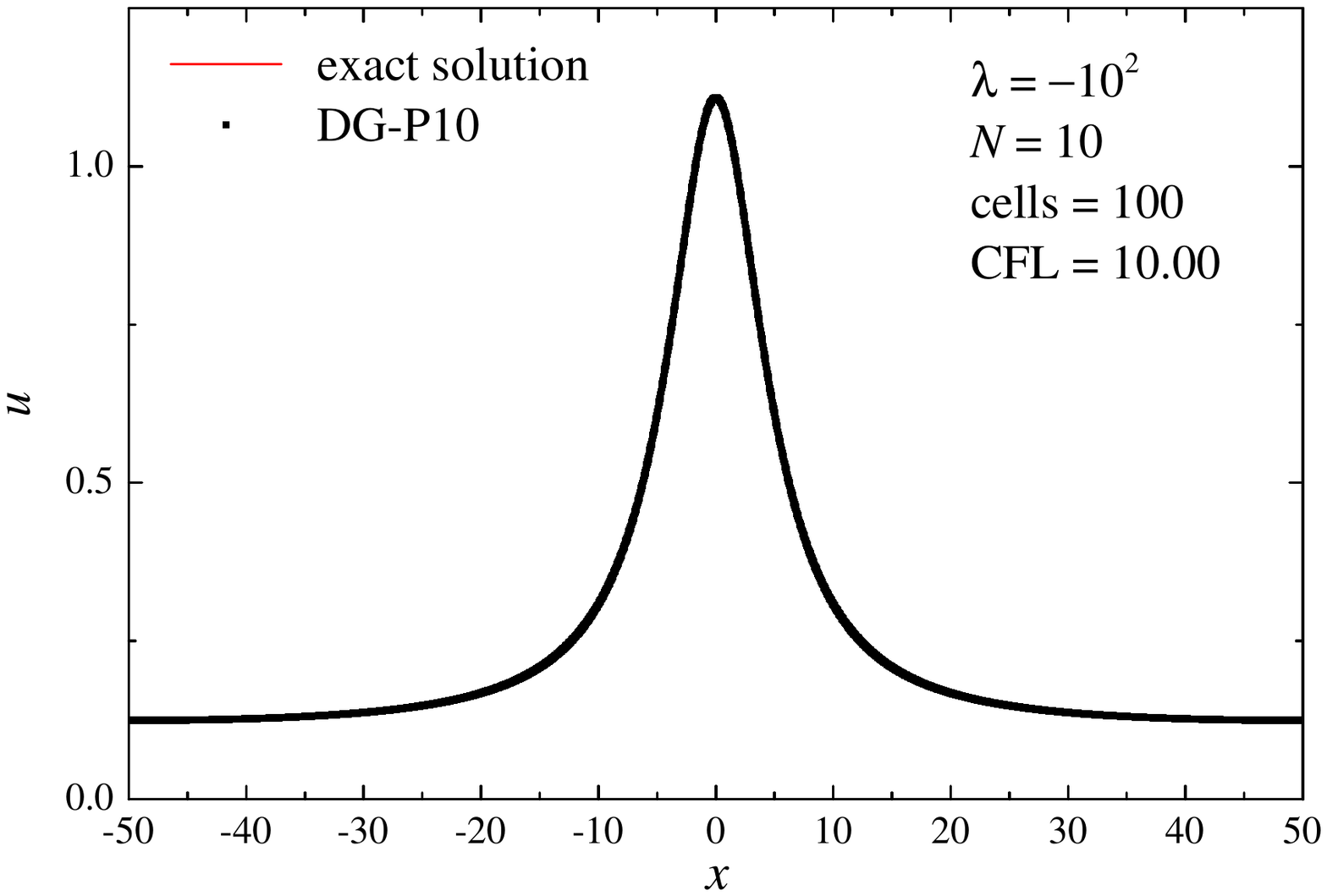}
\vspace{-8mm}\caption{\label{fig:test_adveq_stiff_z100_lrnz_1d_cfl_1000:10}}
\end{subfigure}
\begin{subfigure}{0.24\textwidth}
\includegraphics[width=\textwidth]{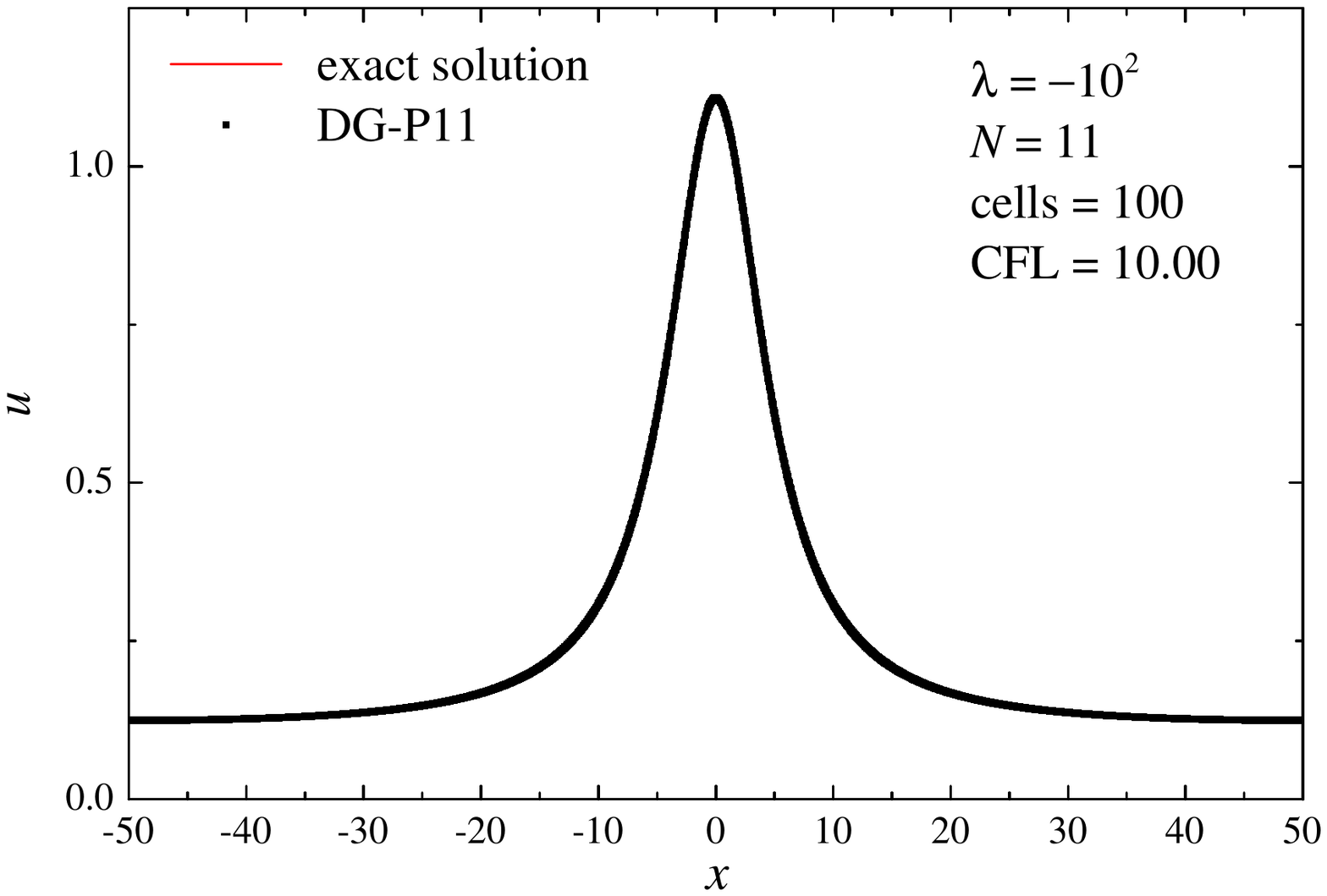}
\vspace{-8mm}\caption{\label{fig:test_adveq_stiff_z100_lrnz_1d_cfl_1000:11}}
\end{subfigure}
\begin{subfigure}{0.24\textwidth}
\includegraphics[width=\textwidth]{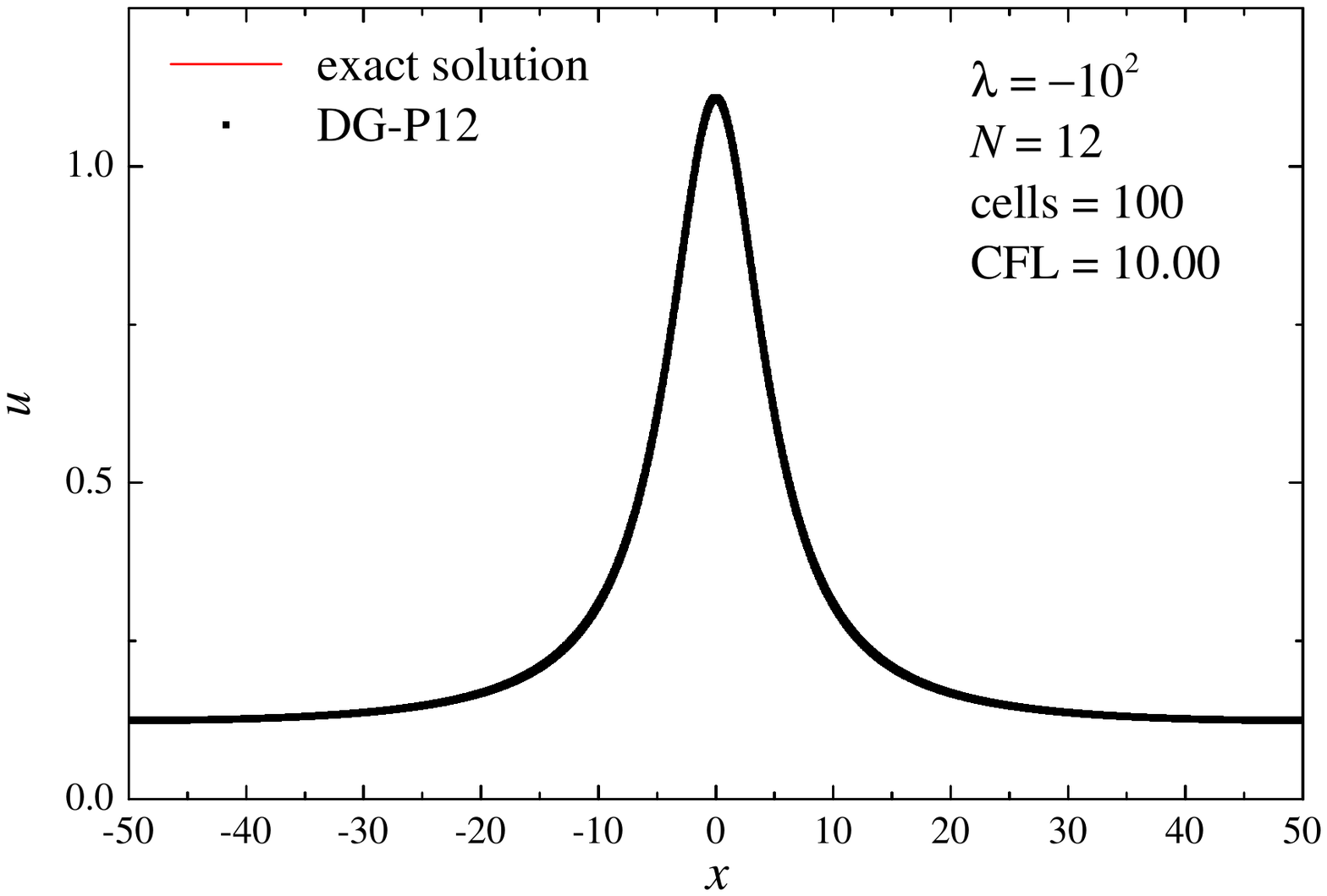}
\vspace{-8mm}\caption{\label{fig:test_adveq_stiff_z100_lrnz_1d_cfl_1000:12}}
\end{subfigure}
\caption{%
Coordinate dependencies of the solution $u(x, t_{f})$ to the linear advection-reaction equation (\ref{eq:adv_eq_src}) at the final time $t_{f}$, obtained by the ADER-DG numerical method with the LST-DG predictor, for the degrees $N = 1, \ldots, 12$ of the basis polynomials for the Courant number $\mathrm{CFL} = 10.0$ and the decay parameter $\nu = -10^{2}$. The red line represents the exact analytical solution.
}
\label{fig:test_adveq_stiff_z100_lrnz_1d_cfl_1000}
\end{figure}
\begin{figure}[h!]
\captionsetup[subfigure]{%
	position=bottom,
	font+=smaller,
	textfont=normalfont,
	singlelinecheck=off,
	justification=raggedright
}
\centering
\begin{subfigure}{0.24\textwidth}
\includegraphics[width=\textwidth]{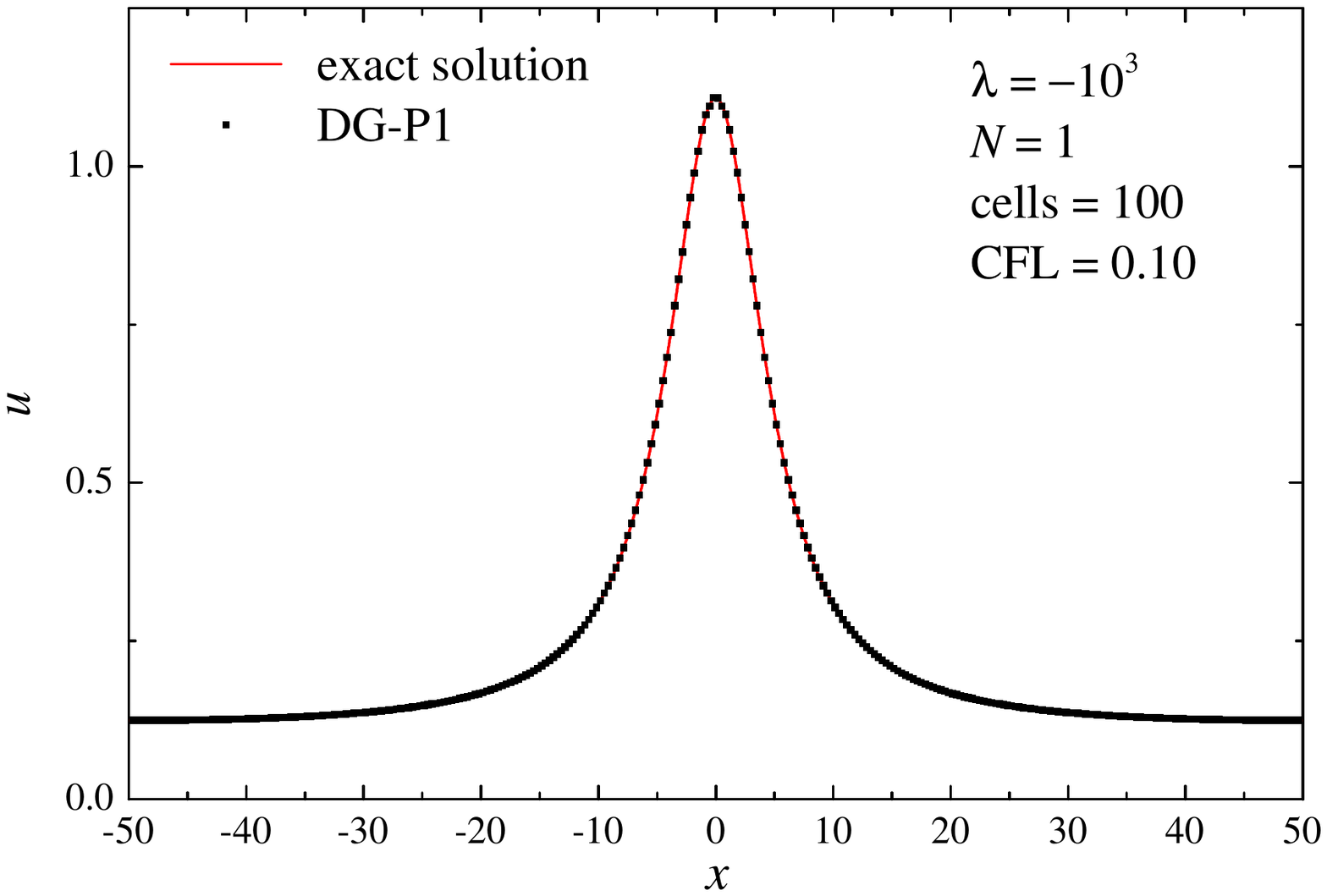}
\vspace{-8mm}\caption{\label{fig:test_adveq_stiff_z1000_lrnz_1d_cfl_010:1}}
\end{subfigure}
\begin{subfigure}{0.24\textwidth}
\includegraphics[width=\textwidth]{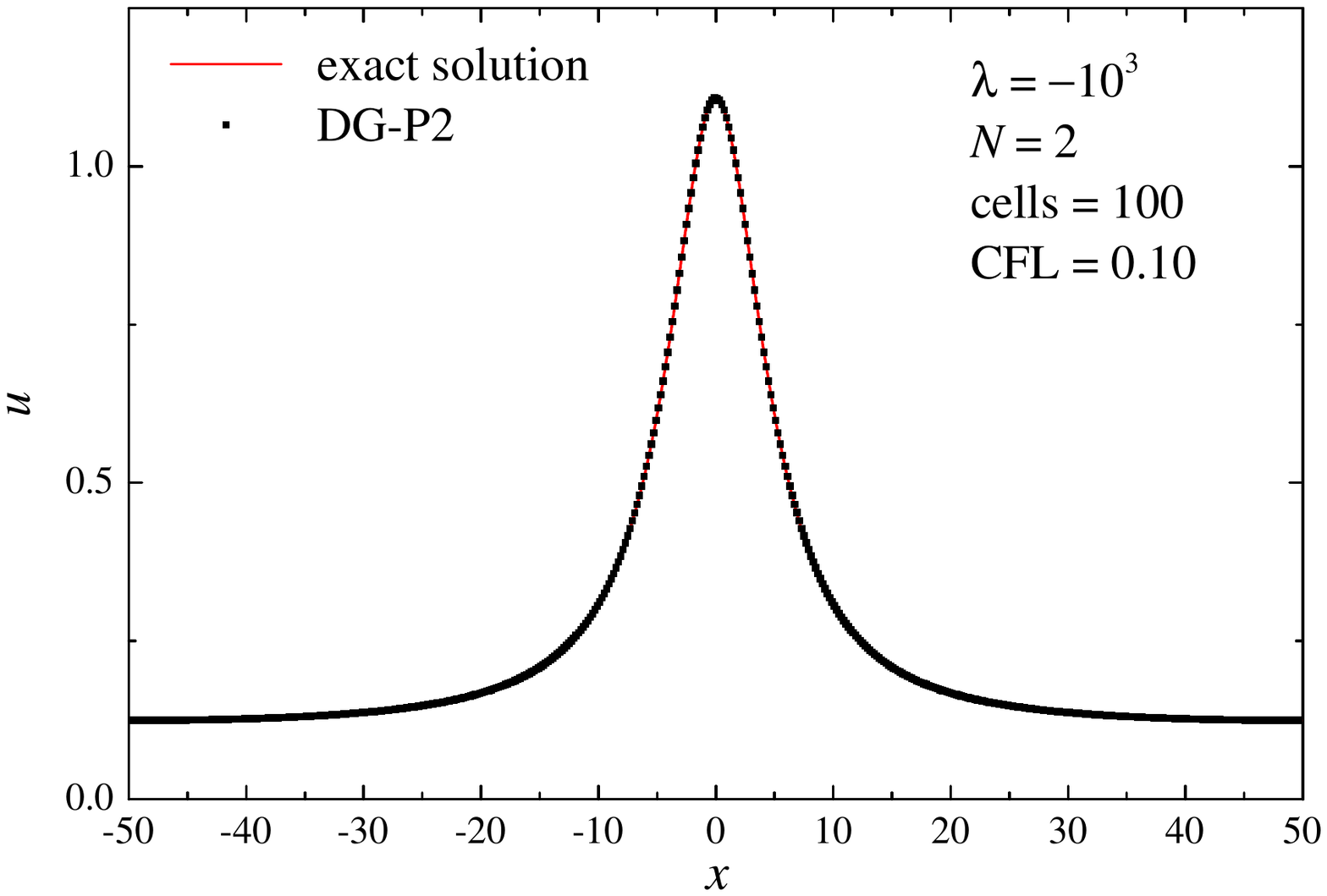}
\vspace{-8mm}\caption{\label{fig:test_adveq_stiff_z1000_lrnz_1d_cfl_010:2}}
\end{subfigure}
\begin{subfigure}{0.24\textwidth}
\includegraphics[width=\textwidth]{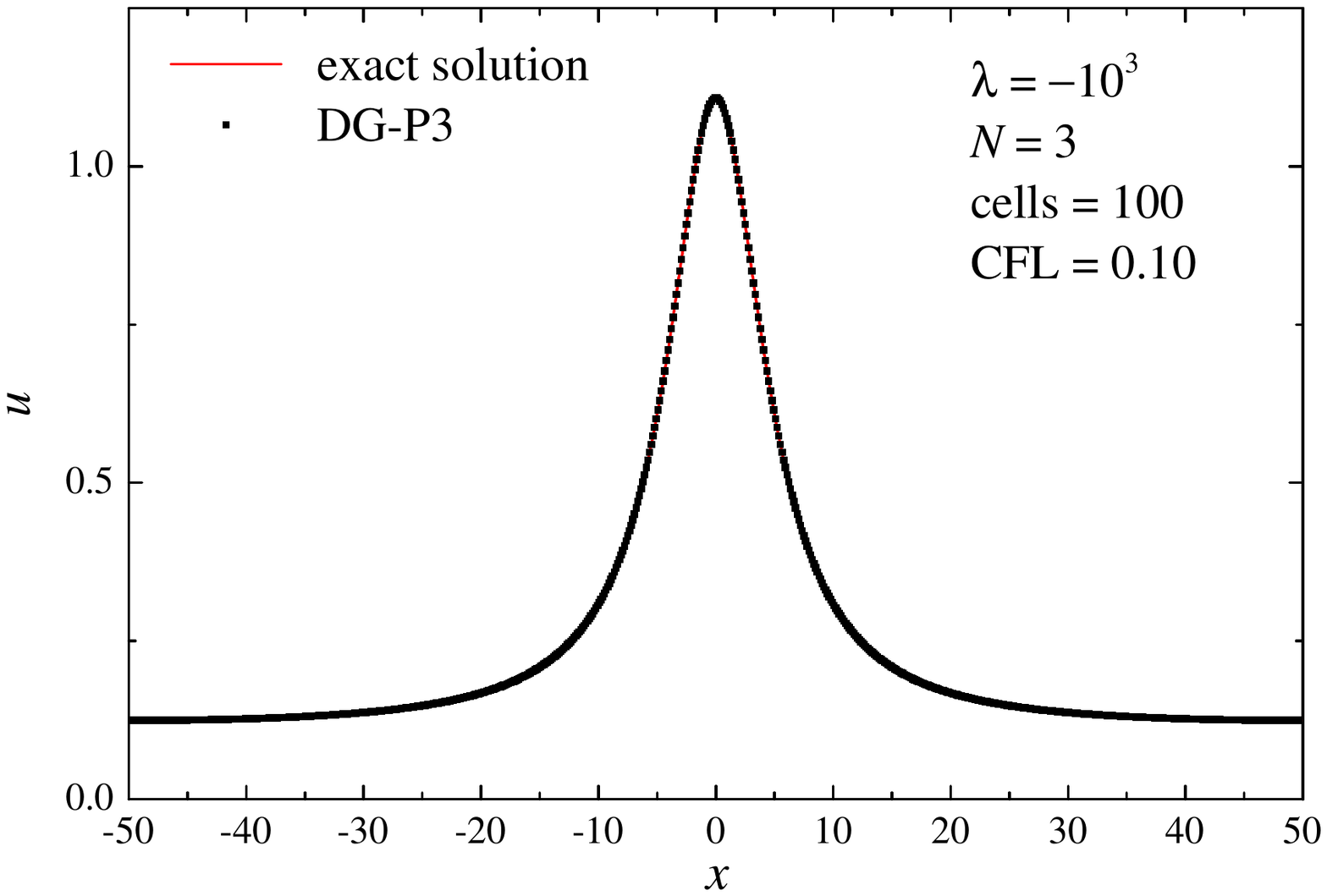}
\vspace{-8mm}\caption{\label{fig:test_adveq_stiff_z1000_lrnz_1d_cfl_010:3}}
\end{subfigure}
\begin{subfigure}{0.24\textwidth}
\includegraphics[width=\textwidth]{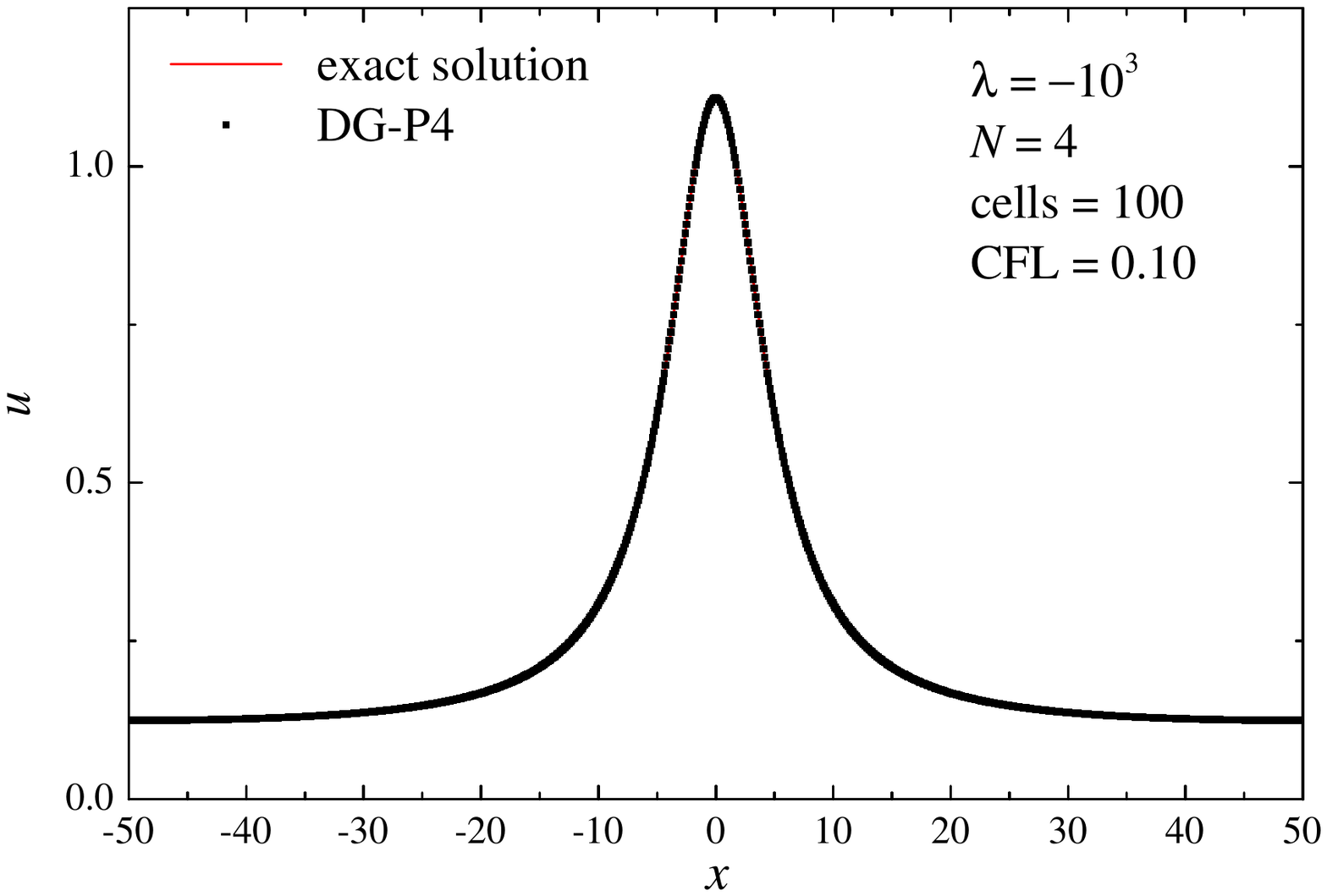}
\vspace{-8mm}\caption{\label{fig:test_adveq_stiff_z1000_lrnz_1d_cfl_010:4}}
\end{subfigure}\\[-2.5mm]
\begin{subfigure}{0.24\textwidth}
\includegraphics[width=\textwidth]{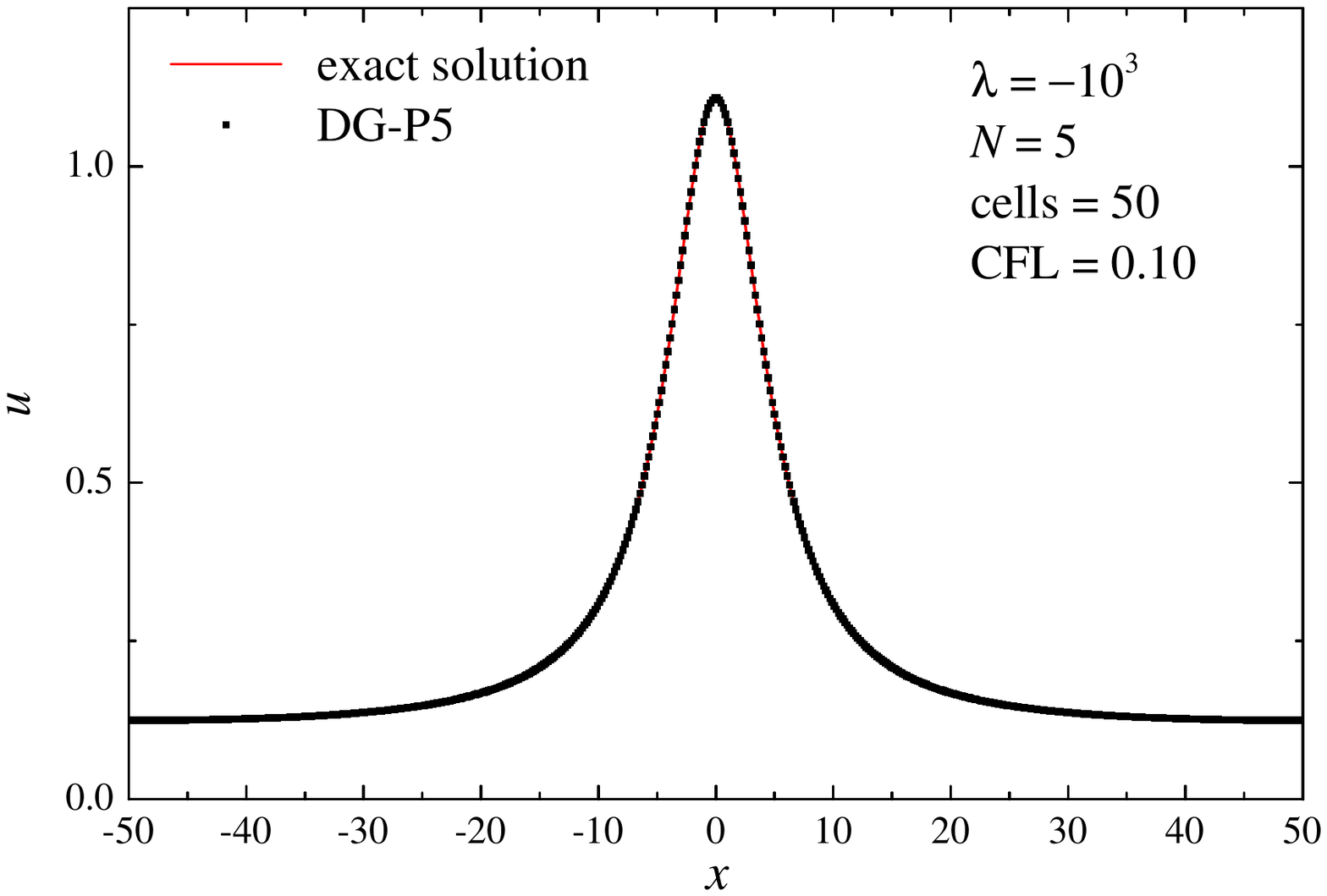}
\vspace{-8mm}\caption{\label{fig:test_adveq_stiff_z1000_lrnz_1d_cfl_010:5}}
\end{subfigure}
\begin{subfigure}{0.24\textwidth}
\includegraphics[width=\textwidth]{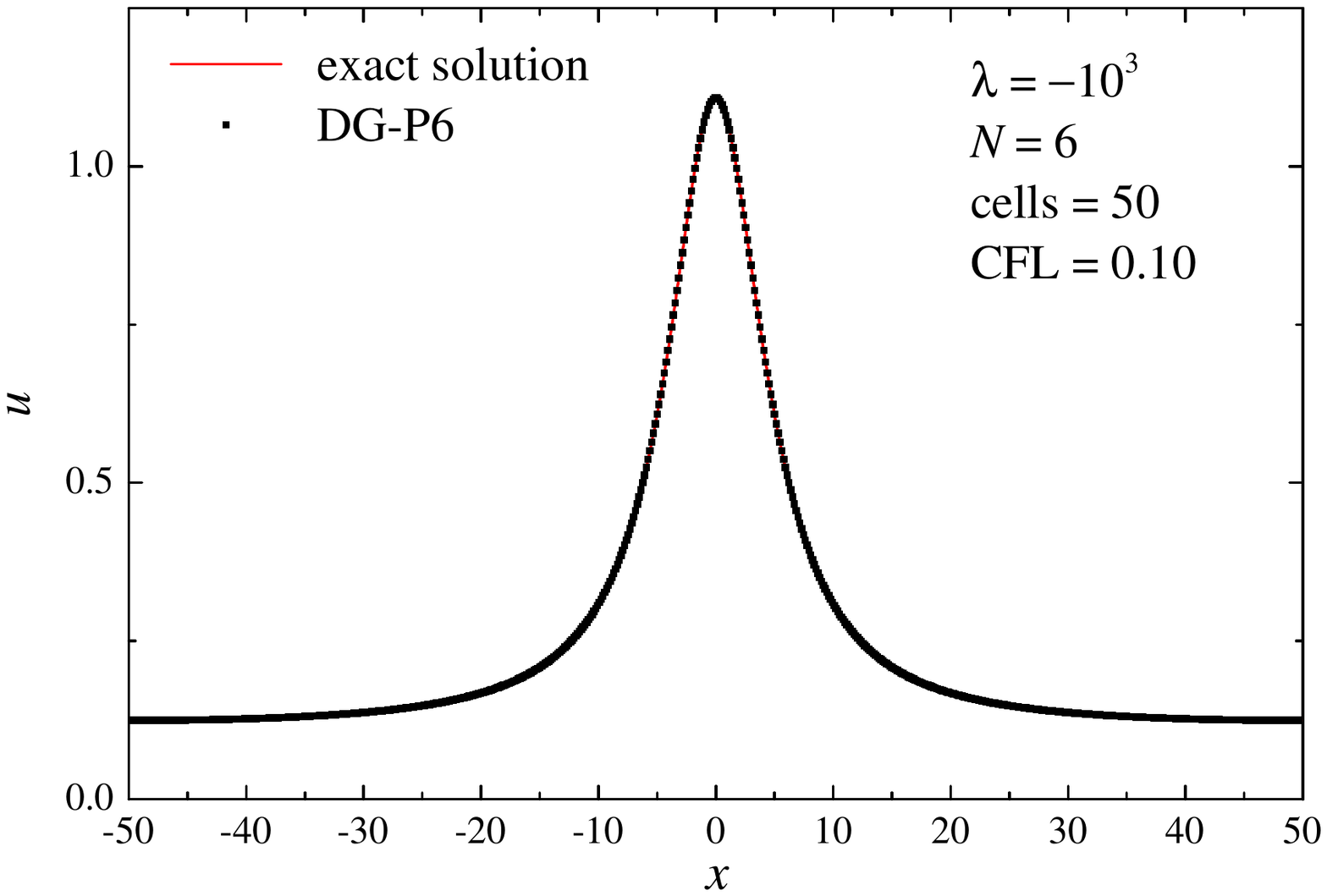}
\vspace{-8mm}\caption{\label{fig:test_adveq_stiff_z1000_lrnz_1d_cfl_010:6}}
\end{subfigure}
\begin{subfigure}{0.24\textwidth}
\includegraphics[width=\textwidth]{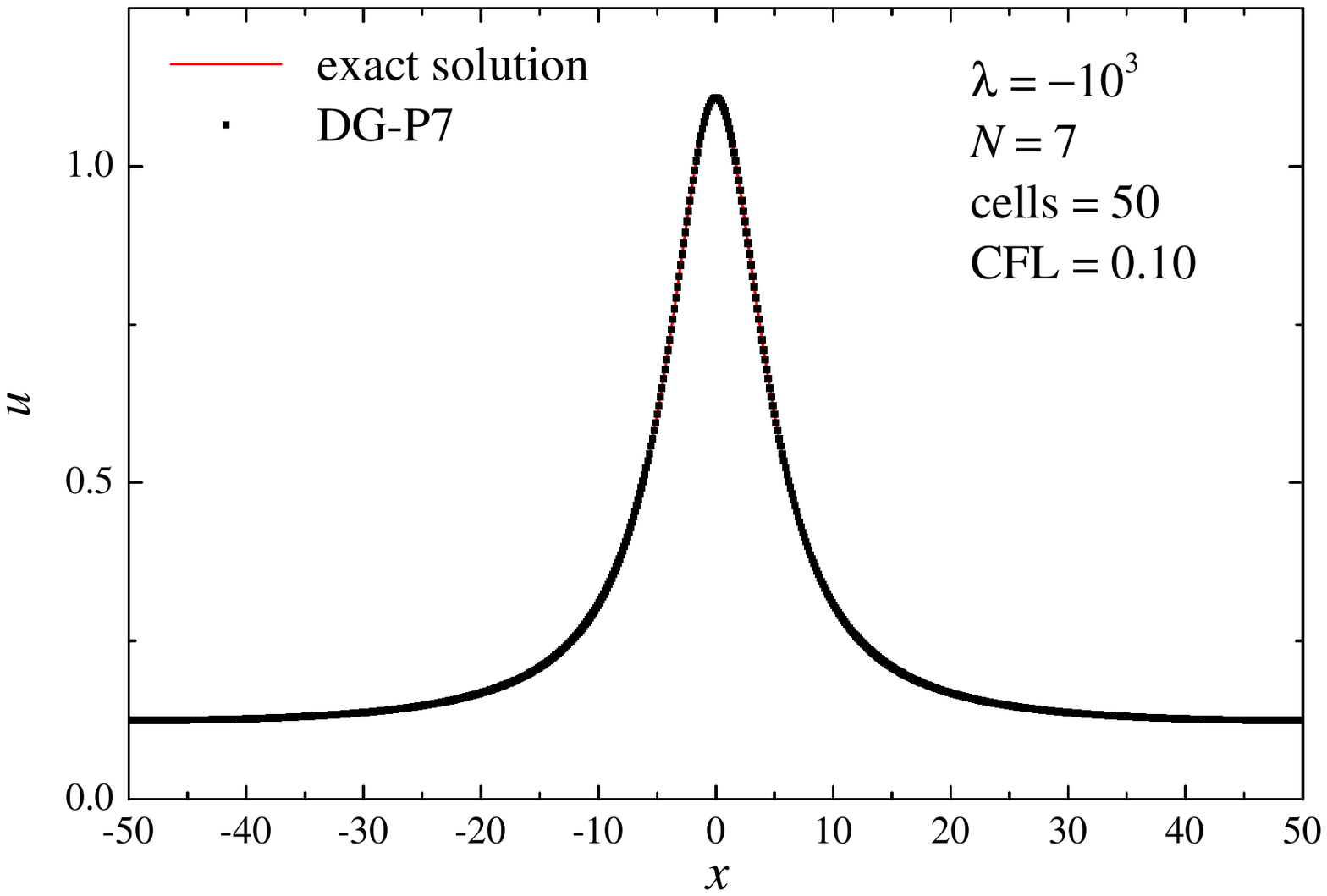}
\vspace{-8mm}\caption{\label{fig:test_adveq_stiff_z1000_lrnz_1d_cfl_010:7}}
\end{subfigure}
\begin{subfigure}{0.24\textwidth}
\includegraphics[width=\textwidth]{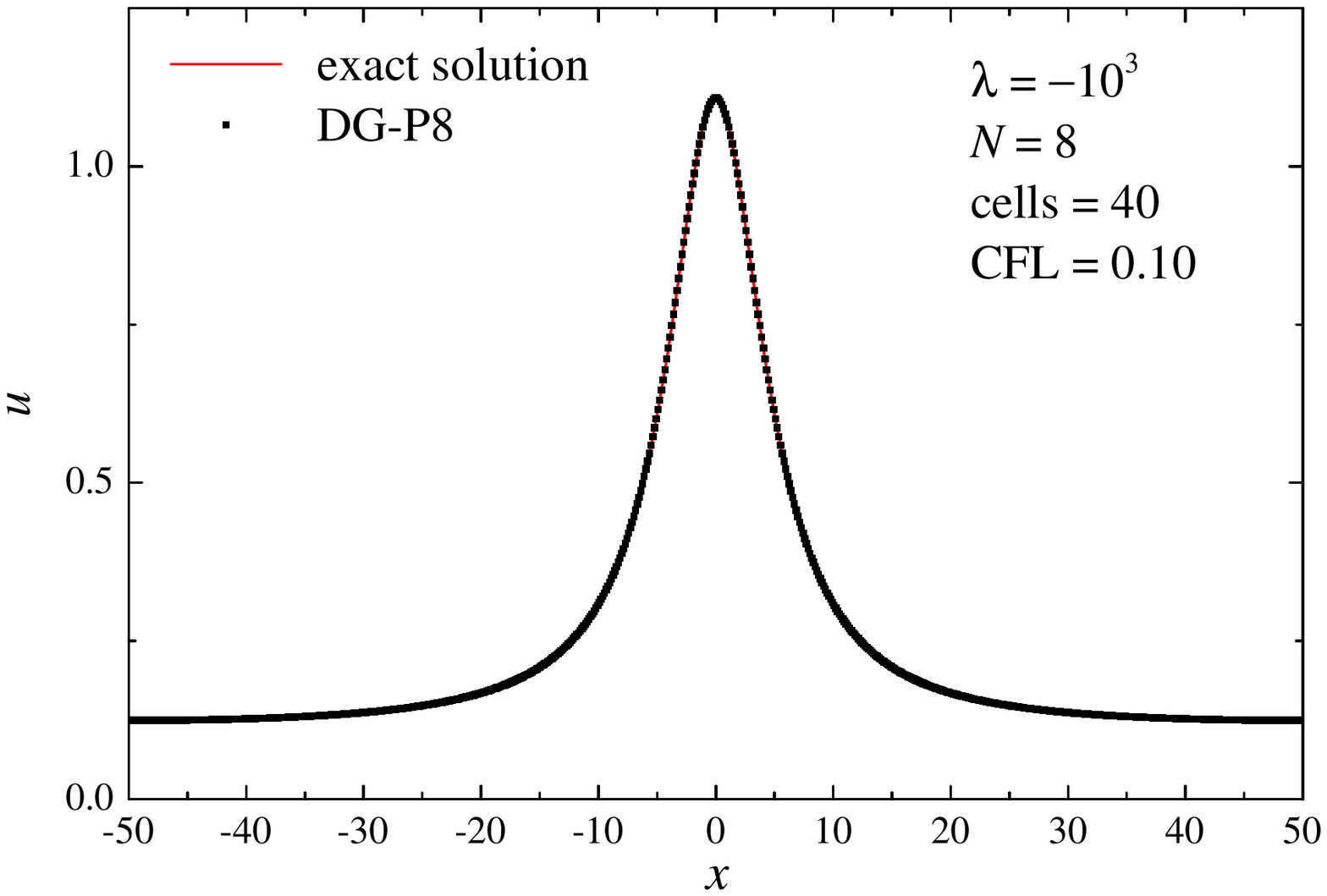}
\vspace{-8mm}\caption{\label{fig:test_adveq_stiff_z1000_lrnz_1d_cfl_010:8}}
\end{subfigure}\\[-2.5mm]
\begin{subfigure}{0.24\textwidth}
\includegraphics[width=\textwidth]{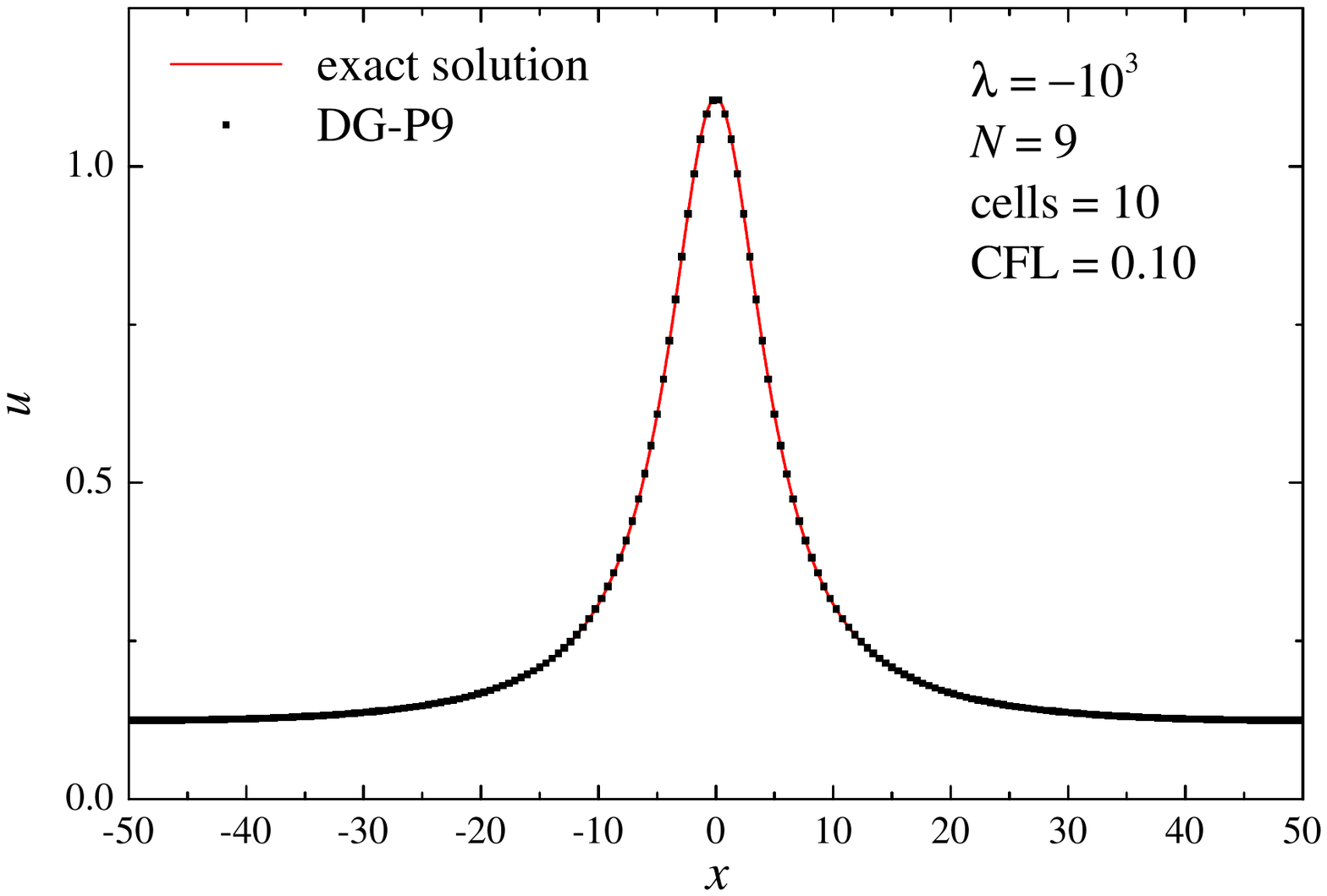}
\vspace{-8mm}\caption{\label{fig:test_adveq_stiff_z1000_lrnz_1d_cfl_010:9}}
\end{subfigure}
\begin{subfigure}{0.24\textwidth}
\includegraphics[width=\textwidth]{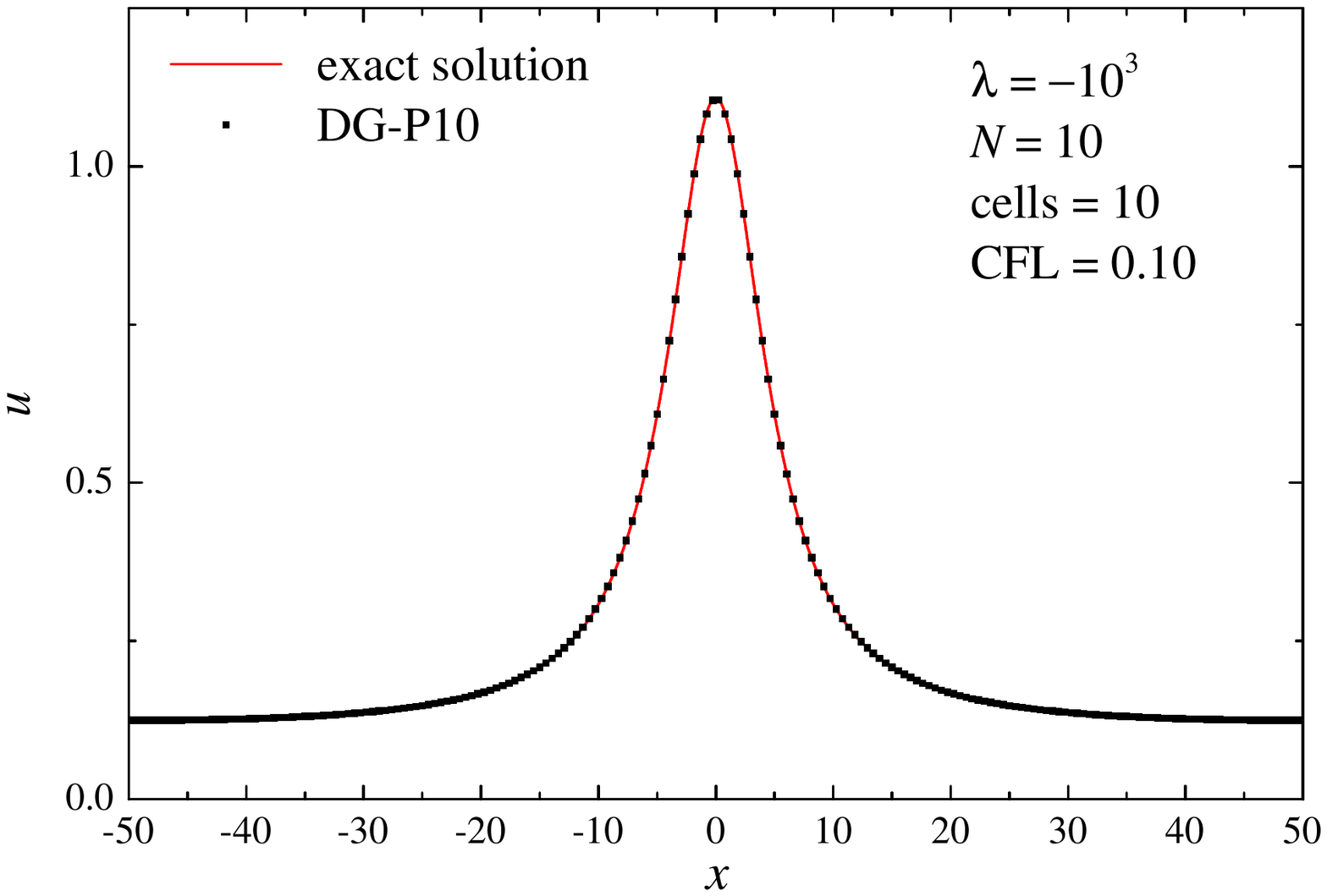}
\vspace{-8mm}\caption{\label{fig:test_adveq_stiff_z1000_lrnz_1d_cfl_010:10}}
\end{subfigure}
\begin{subfigure}{0.24\textwidth}
\includegraphics[width=\textwidth]{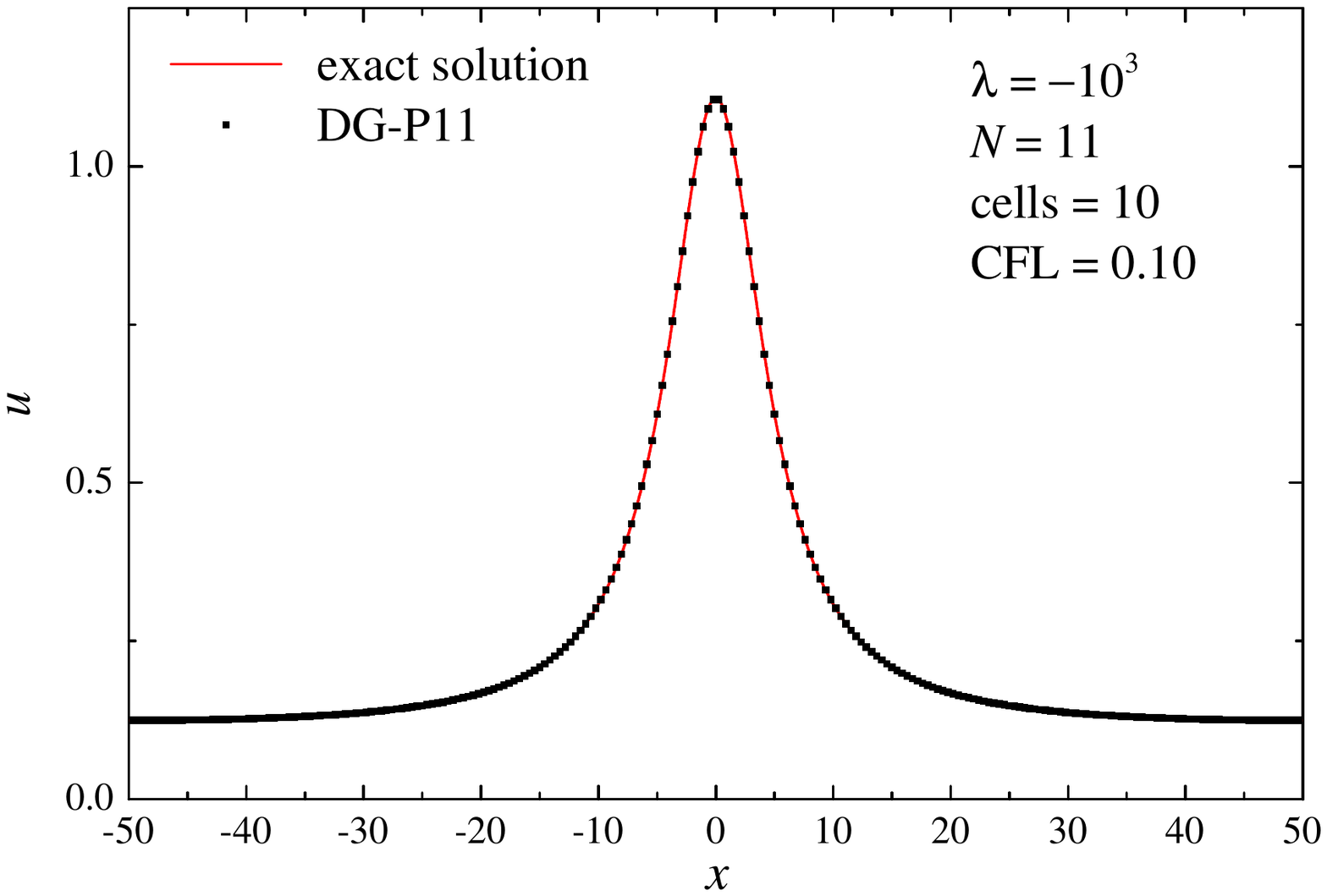}
\vspace{-8mm}\caption{\label{fig:test_adveq_stiff_z1000_lrnz_1d_cfl_010:11}}
\end{subfigure}
\begin{subfigure}{0.24\textwidth}
\includegraphics[width=\textwidth]{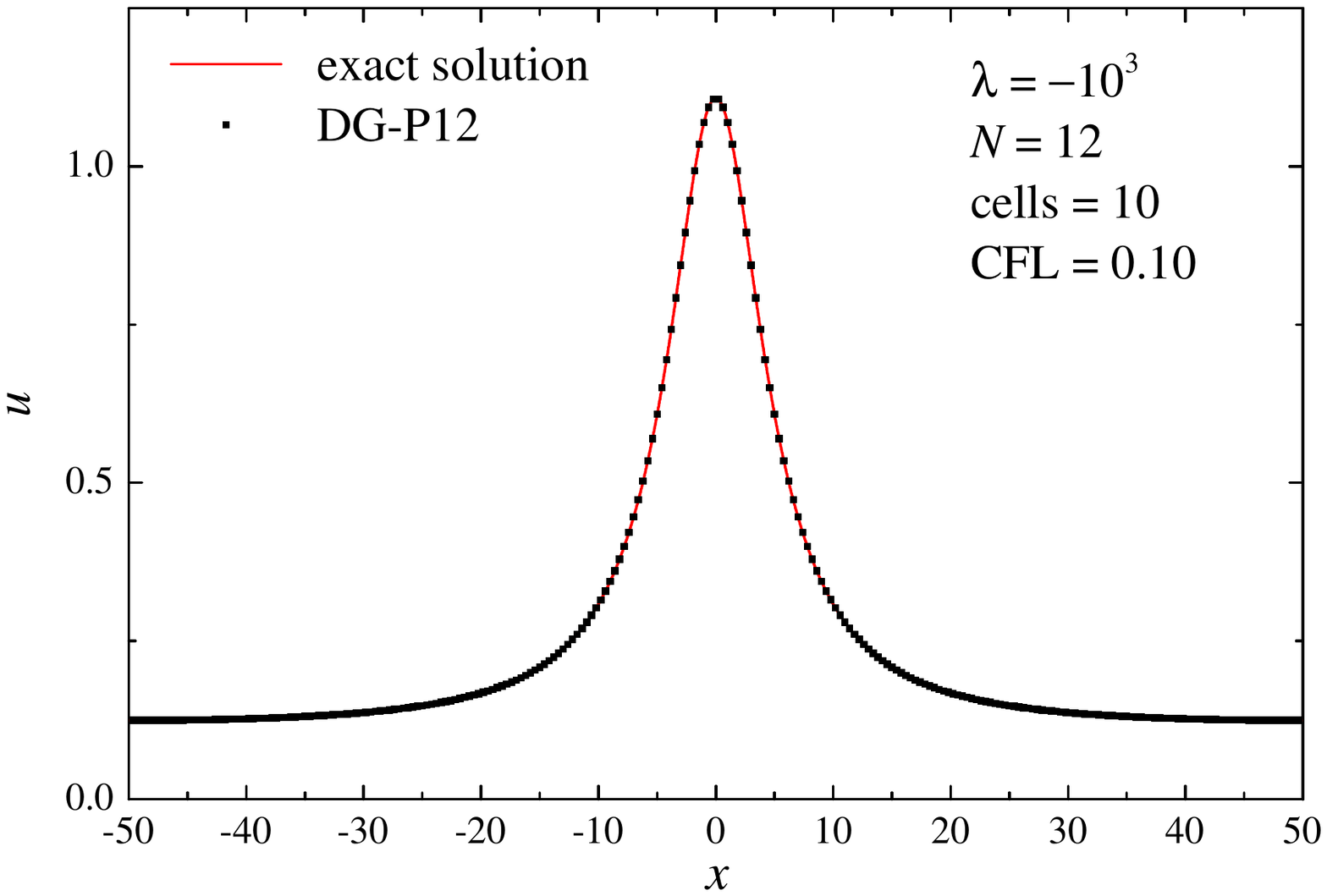}
\vspace{-8mm}\caption{\label{fig:test_adveq_stiff_z1000_lrnz_1d_cfl_010:12}}
\end{subfigure}
\caption{%
Coordinate dependencies of the solution $u(x, t_{f})$ to the linear advection-reaction equation (\ref{eq:adv_eq_src}) at the final time $t_{f}$, obtained by the ADER-DG numerical method with the LST-DG predictor, for the degrees $N = 1, \ldots, 12$ of the basis polynomials for the Courant number $\mathrm{CFL} = 0.1$ and the decay parameter $\nu = -10^{3}$. The red line represents the exact analytical solution.
}
\label{fig:test_adveq_stiff_z1000_lrnz_1d_cfl_010}
\end{figure}
\begin{figure}[h!]
\captionsetup[subfigure]{%
	position=bottom,
	font+=smaller,
	textfont=normalfont,
	singlelinecheck=off,
	justification=raggedright
}
\centering
\begin{subfigure}{0.24\textwidth}
\includegraphics[width=\textwidth]{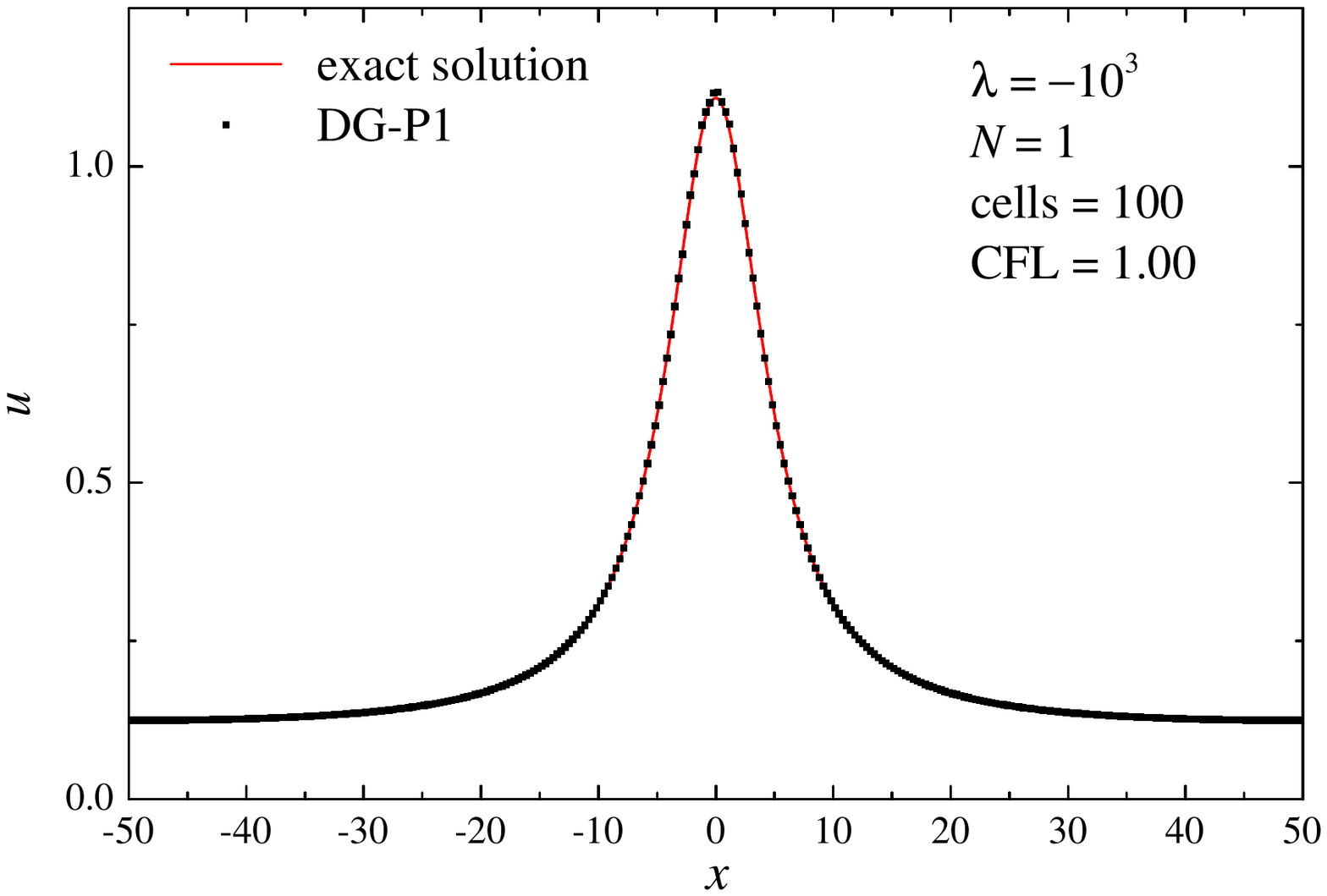}
\vspace{-8mm}\caption{\label{fig:test_adveq_stiff_z1000_lrnz_1d_cfl_100:1}}
\end{subfigure}
\begin{subfigure}{0.24\textwidth}
\includegraphics[width=\textwidth]{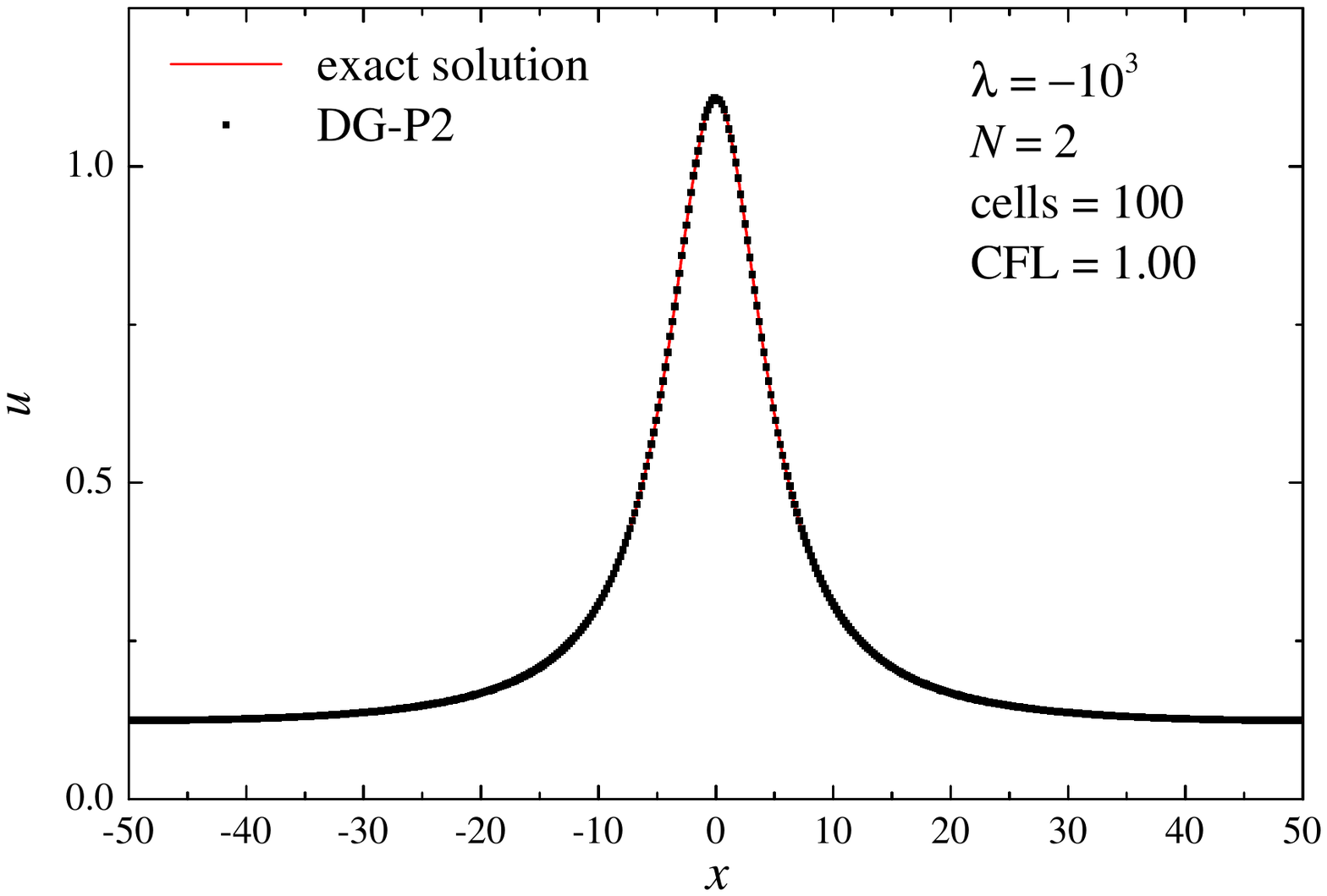}
\vspace{-8mm}\caption{\label{fig:test_adveq_stiff_z1000_lrnz_1d_cfl_100:2}}
\end{subfigure}
\begin{subfigure}{0.24\textwidth}
\includegraphics[width=\textwidth]{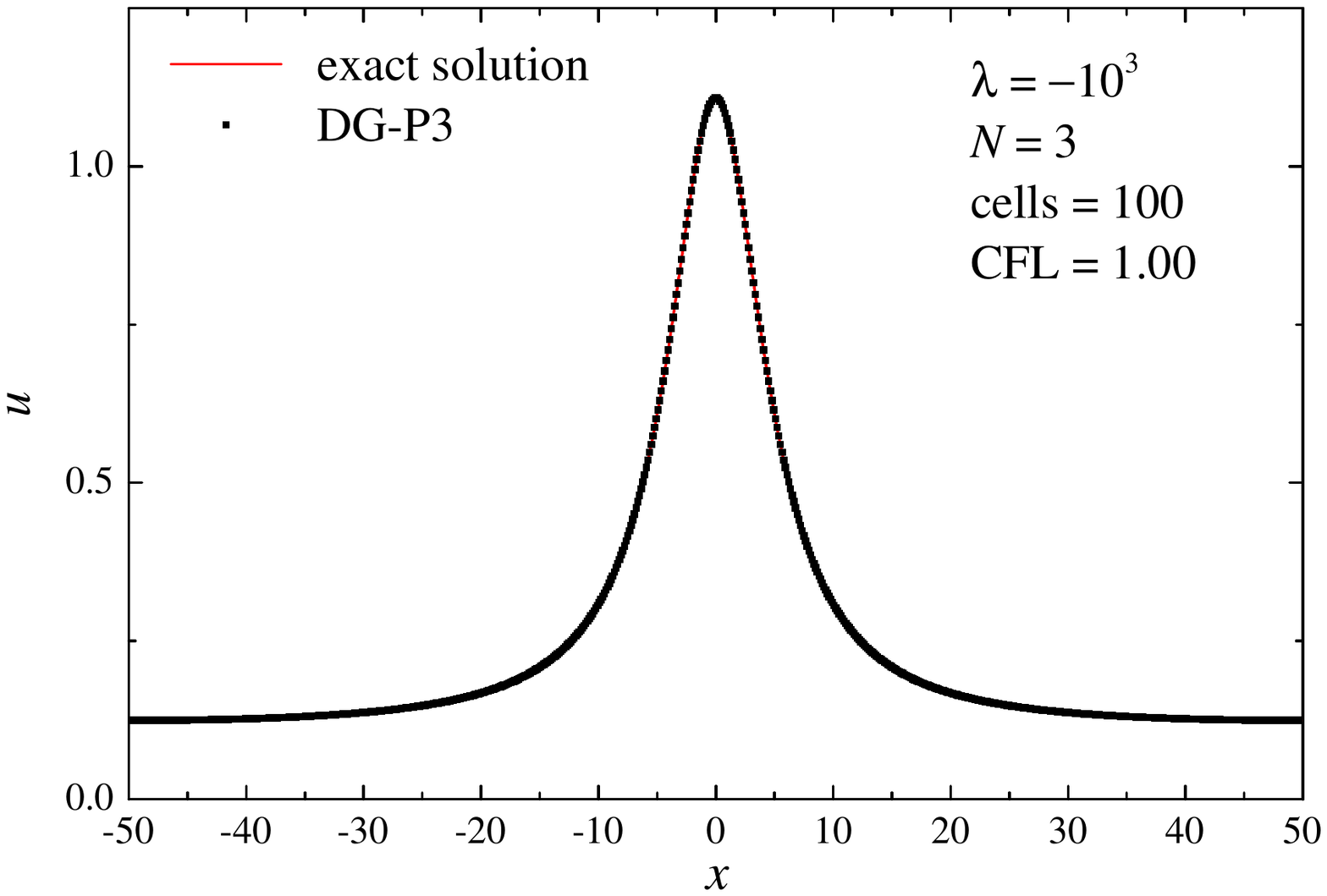}
\vspace{-8mm}\caption{\label{fig:test_adveq_stiff_z1000_lrnz_1d_cfl_100:3}}
\end{subfigure}
\begin{subfigure}{0.24\textwidth}
\includegraphics[width=\textwidth]{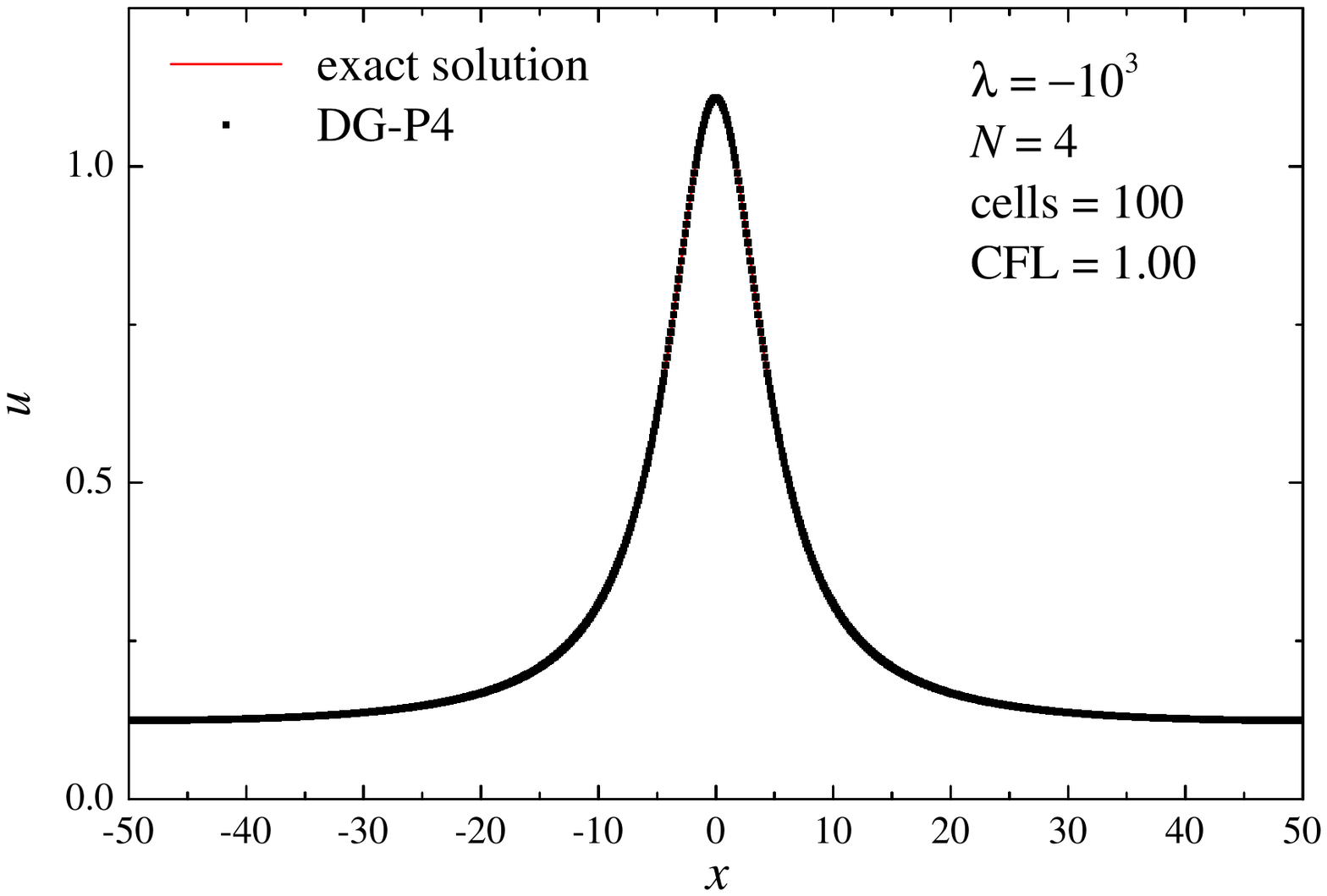}
\vspace{-8mm}\caption{\label{fig:test_adveq_stiff_z1000_lrnz_1d_cfl_100:4}}
\end{subfigure}\\[-2.5mm]
\begin{subfigure}{0.24\textwidth}
\includegraphics[width=\textwidth]{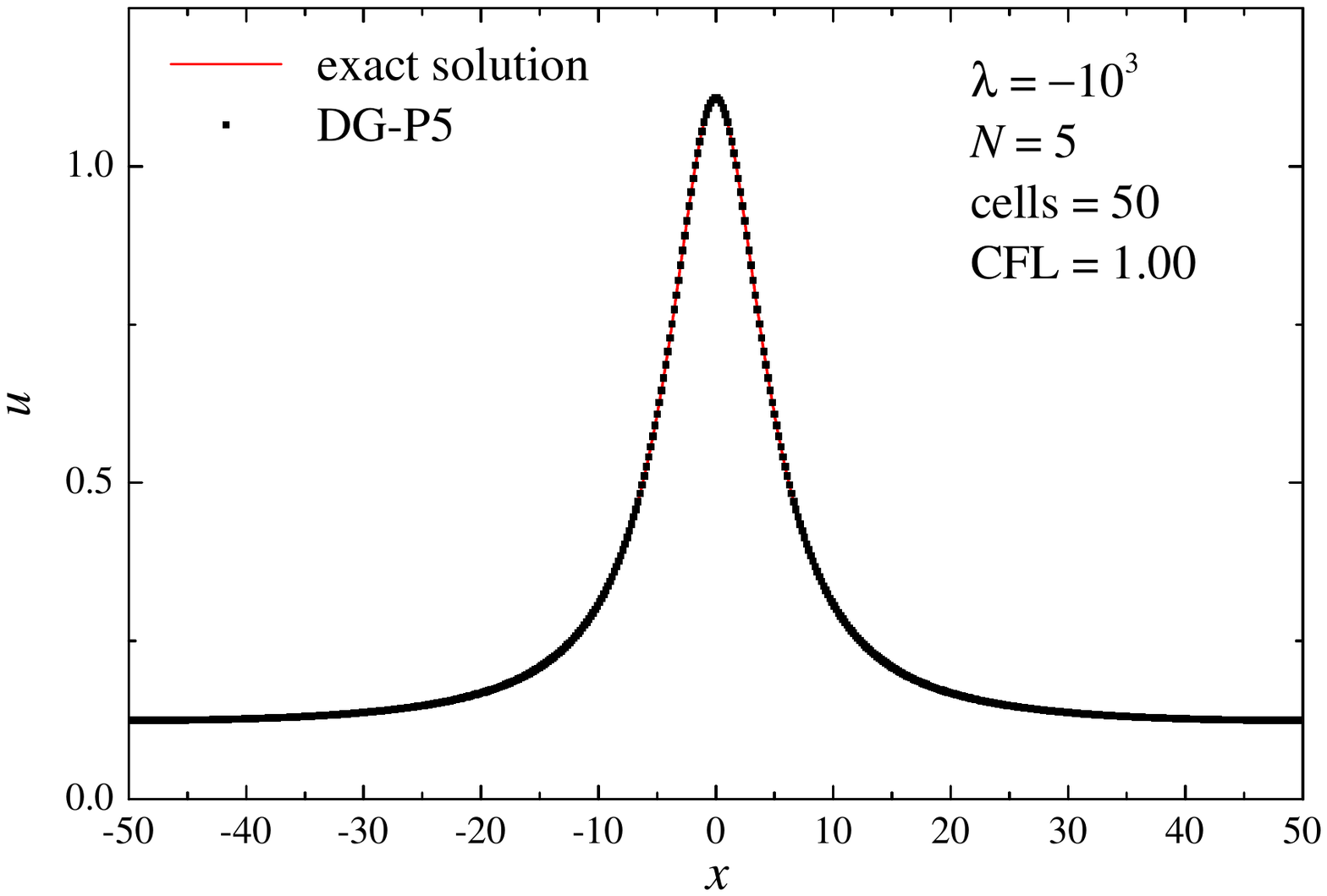}
\vspace{-8mm}\caption{\label{fig:test_adveq_stiff_z1000_lrnz_1d_cfl_100:5}}
\end{subfigure}
\begin{subfigure}{0.24\textwidth}
\includegraphics[width=\textwidth]{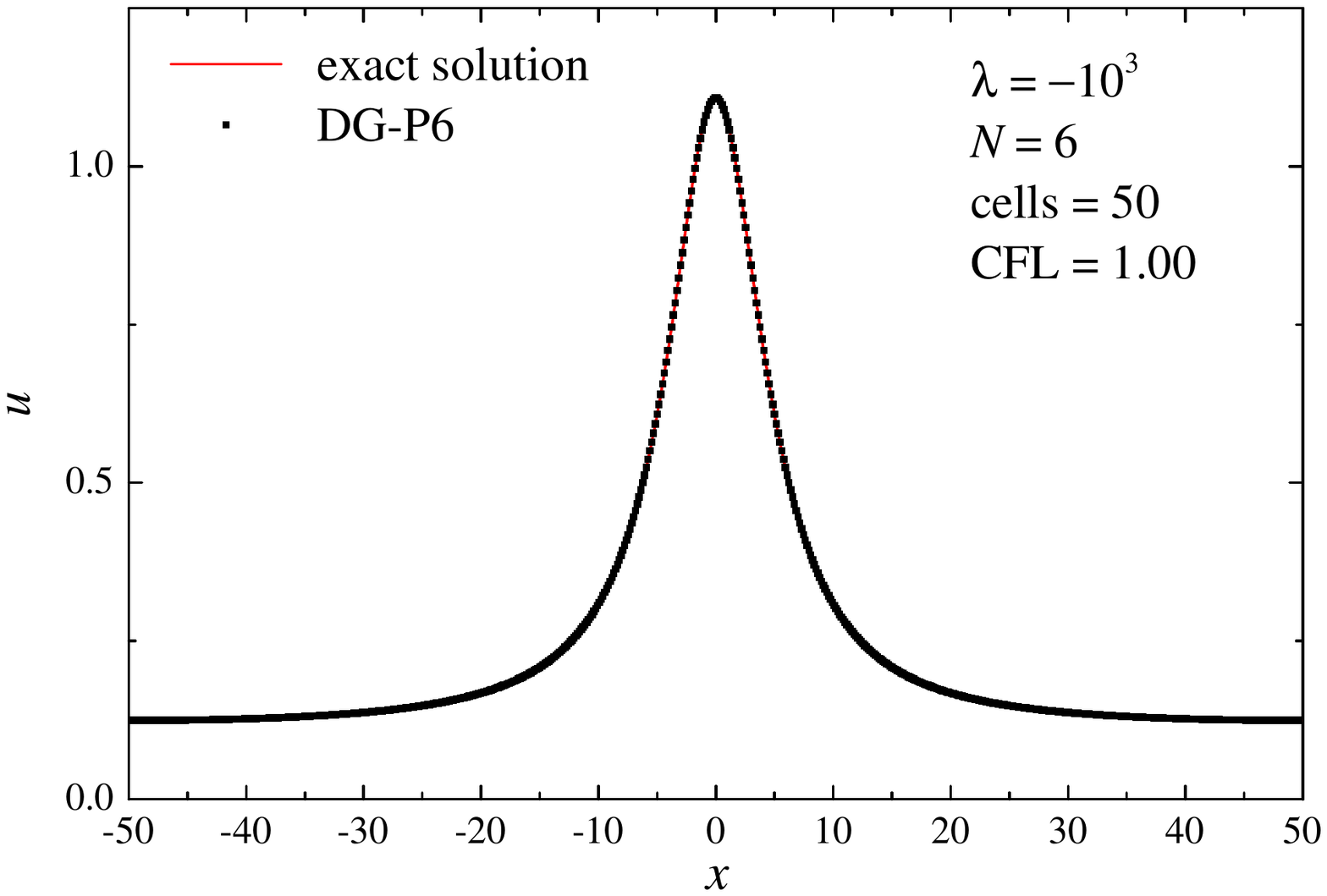}
\vspace{-8mm}\caption{\label{fig:test_adveq_stiff_z1000_lrnz_1d_cfl_100:6}}
\end{subfigure}
\begin{subfigure}{0.24\textwidth}
\includegraphics[width=\textwidth]{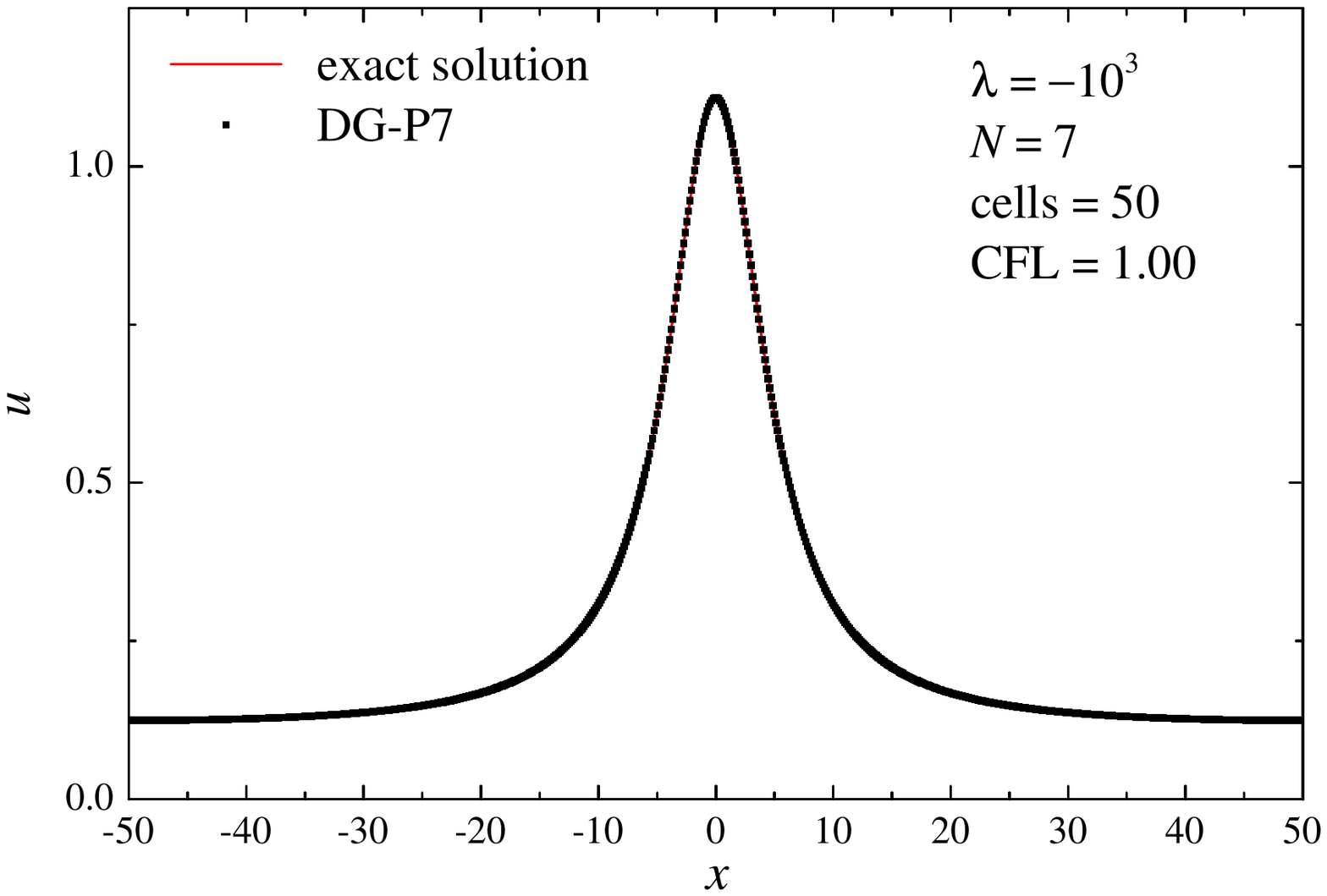}
\vspace{-8mm}\caption{\label{fig:test_adveq_stiff_z1000_lrnz_1d_cfl_100:7}}
\end{subfigure}
\begin{subfigure}{0.24\textwidth}
\includegraphics[width=\textwidth]{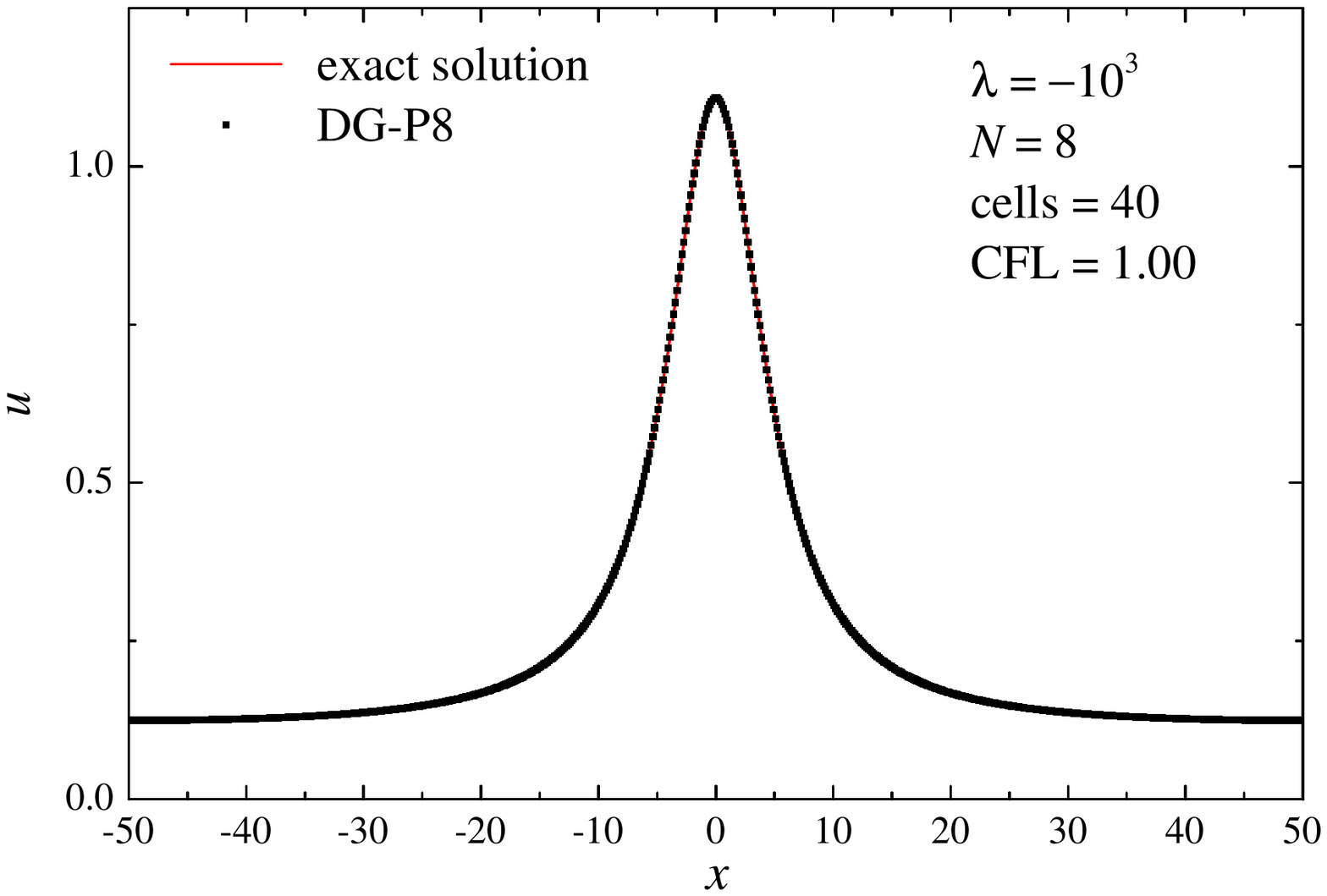}
\vspace{-8mm}\caption{\label{fig:test_adveq_stiff_z1000_lrnz_1d_cfl_100:8}}
\end{subfigure}\\[-2.5mm]
\begin{subfigure}{0.24\textwidth}
\includegraphics[width=\textwidth]{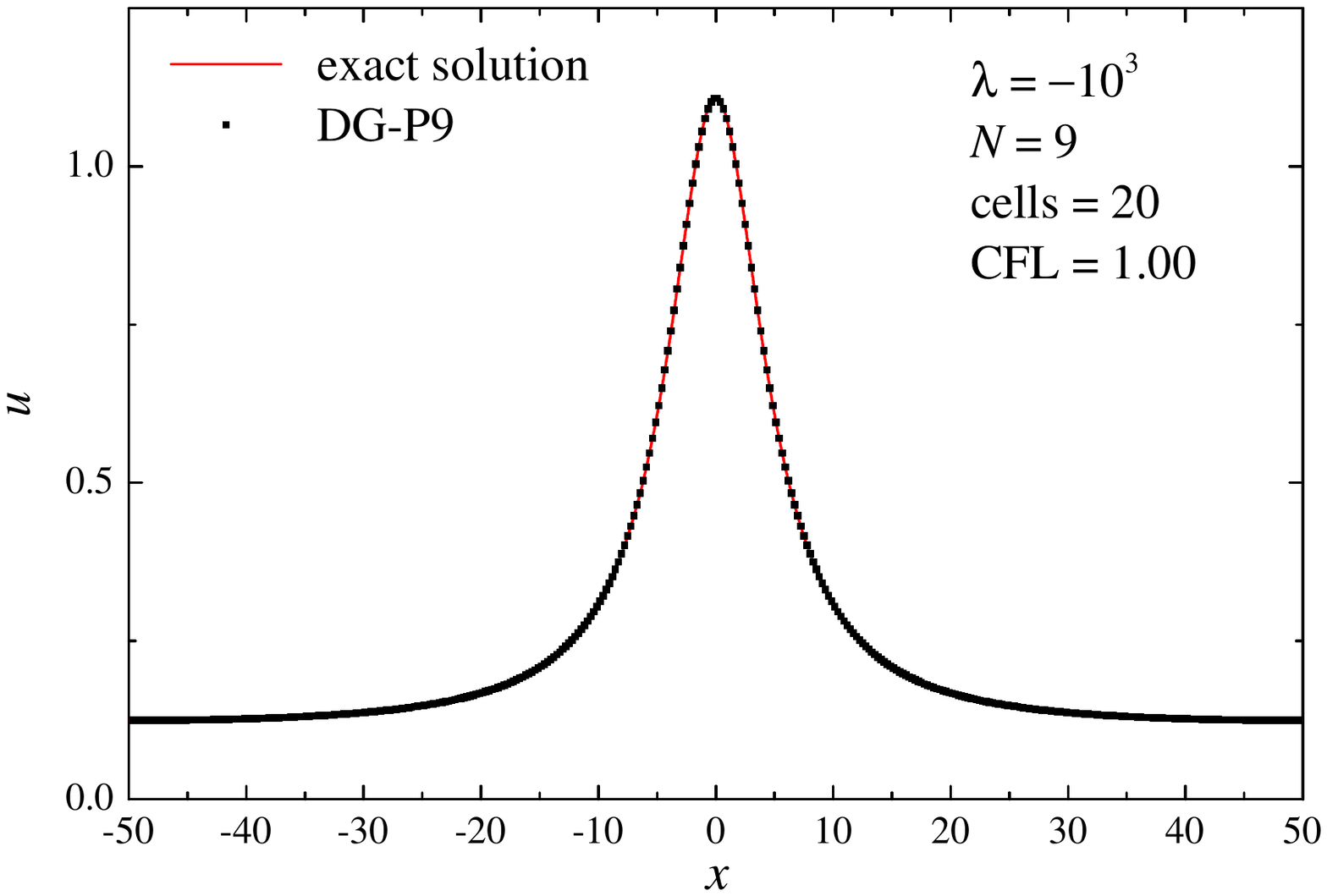}
\vspace{-8mm}\caption{\label{fig:test_adveq_stiff_z1000_lrnz_1d_cfl_100:9}}
\end{subfigure}
\begin{subfigure}{0.24\textwidth}
\includegraphics[width=\textwidth]{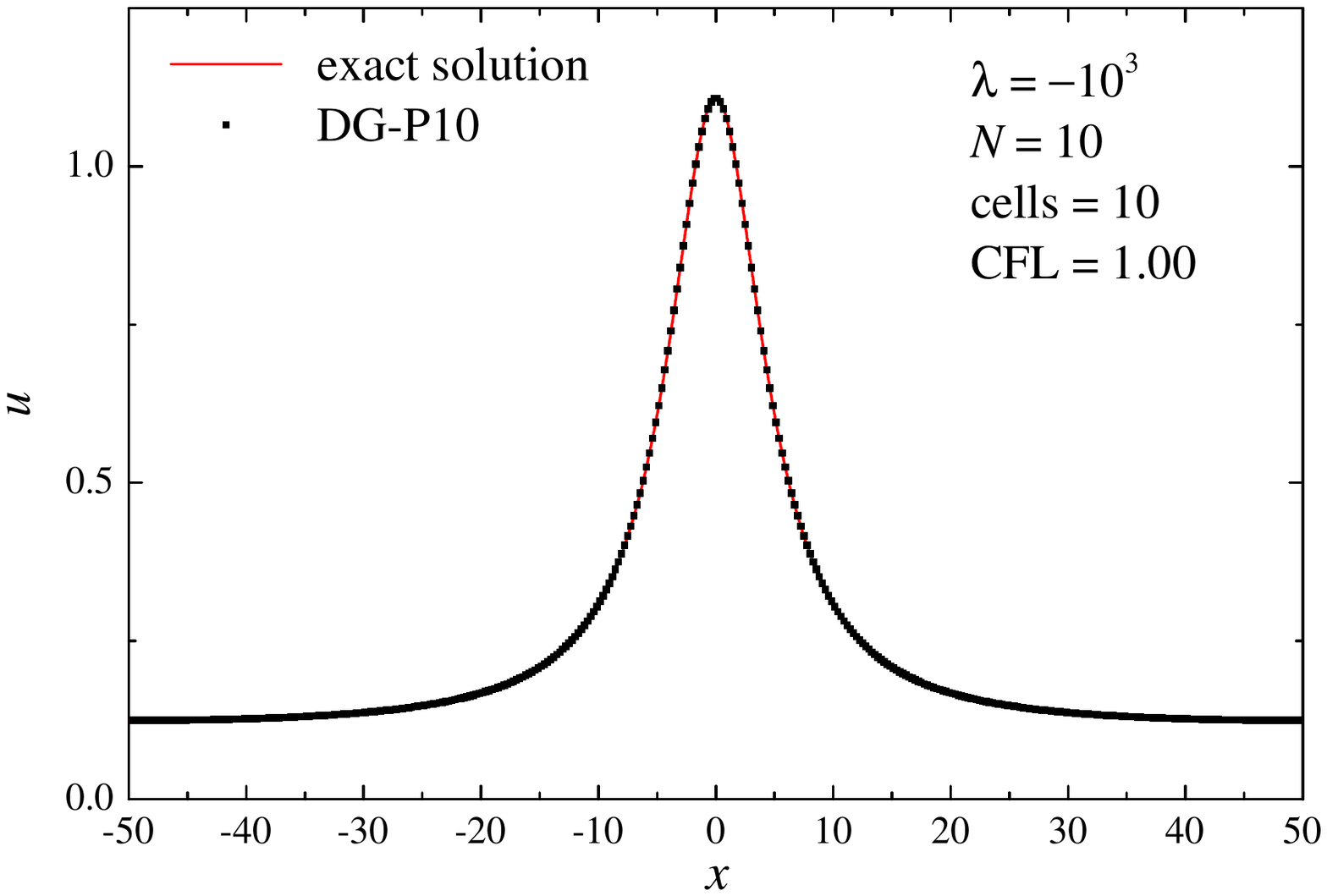}
\vspace{-8mm}\caption{\label{fig:test_adveq_stiff_z1000_lrnz_1d_cfl_100:10}}
\end{subfigure}
\begin{subfigure}{0.24\textwidth}
\includegraphics[width=\textwidth]{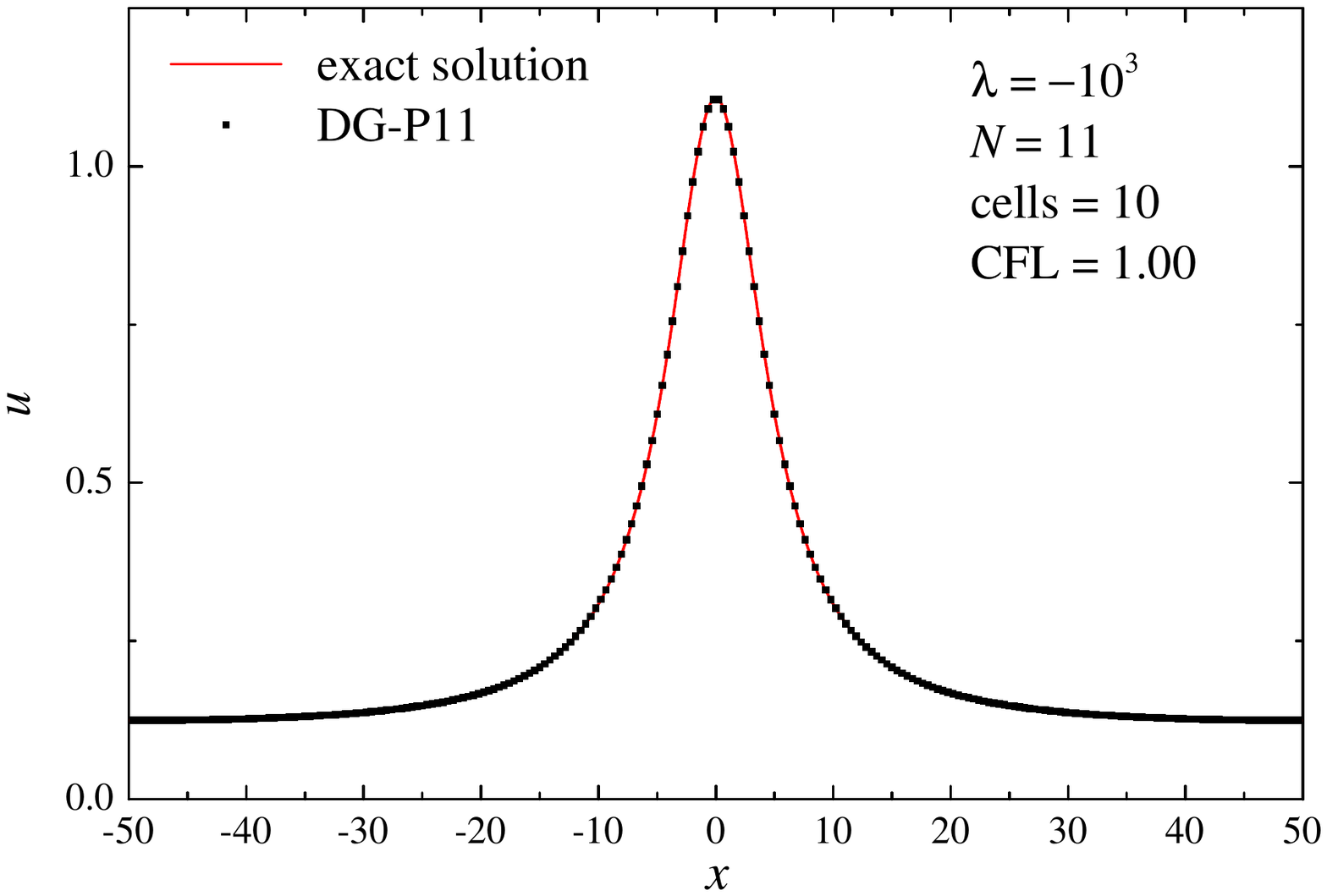}
\vspace{-8mm}\caption{\label{fig:test_adveq_stiff_z1000_lrnz_1d_cfl_100:11}}
\end{subfigure}
\begin{subfigure}{0.24\textwidth}
\includegraphics[width=\textwidth]{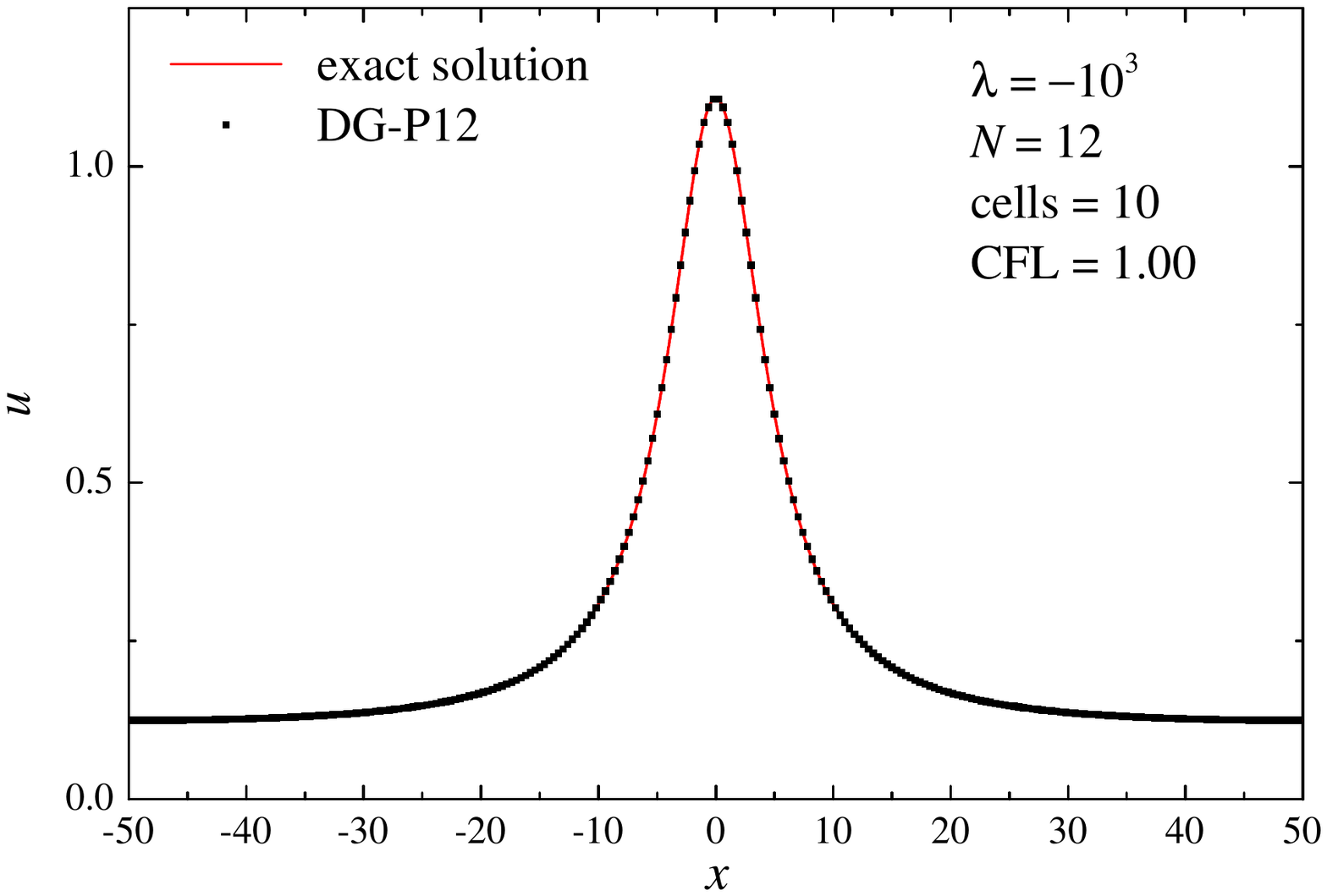}
\vspace{-8mm}\caption{\label{fig:test_adveq_stiff_z1000_lrnz_1d_cfl_100:12}}
\end{subfigure}
\caption{%
Coordinate dependencies of the solution $u(x, t_{f})$ to the linear advection-reaction equation (\ref{eq:adv_eq_src}) at the final time $t_{f}$, obtained by the ADER-DG numerical method with the LST-DG predictor, for the degrees $N = 1, \ldots, 12$ of the basis polynomials for the Courant number $\mathrm{CFL} = 1.0$ and the decay parameter $\nu = -10^{3}$. The red line represents the exact analytical solution.
}
\label{fig:test_adveq_stiff_z1000_lrnz_1d_cfl_100}
\end{figure}
\begin{figure}[h!]
\captionsetup[subfigure]{%
	position=bottom,
	font+=smaller,
	textfont=normalfont,
	singlelinecheck=off,
	justification=raggedright
}
\centering
\begin{subfigure}{0.24\textwidth}
\includegraphics[width=\textwidth]{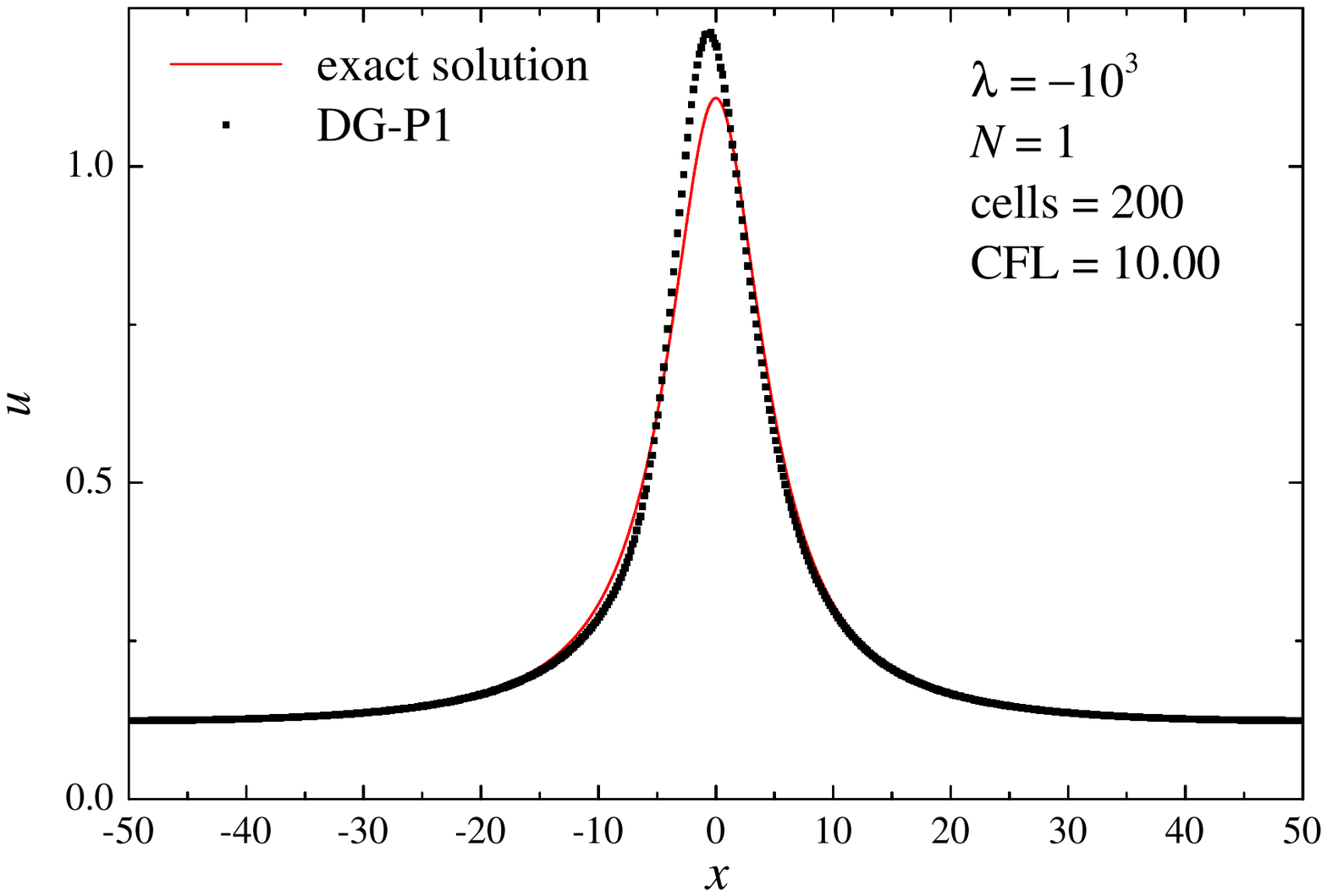}
\vspace{-8mm}\caption{\label{fig:test_adveq_stiff_z1000_lrnz_1d_cfl_1000:1}}
\end{subfigure}
\begin{subfigure}{0.24\textwidth}
\includegraphics[width=\textwidth]{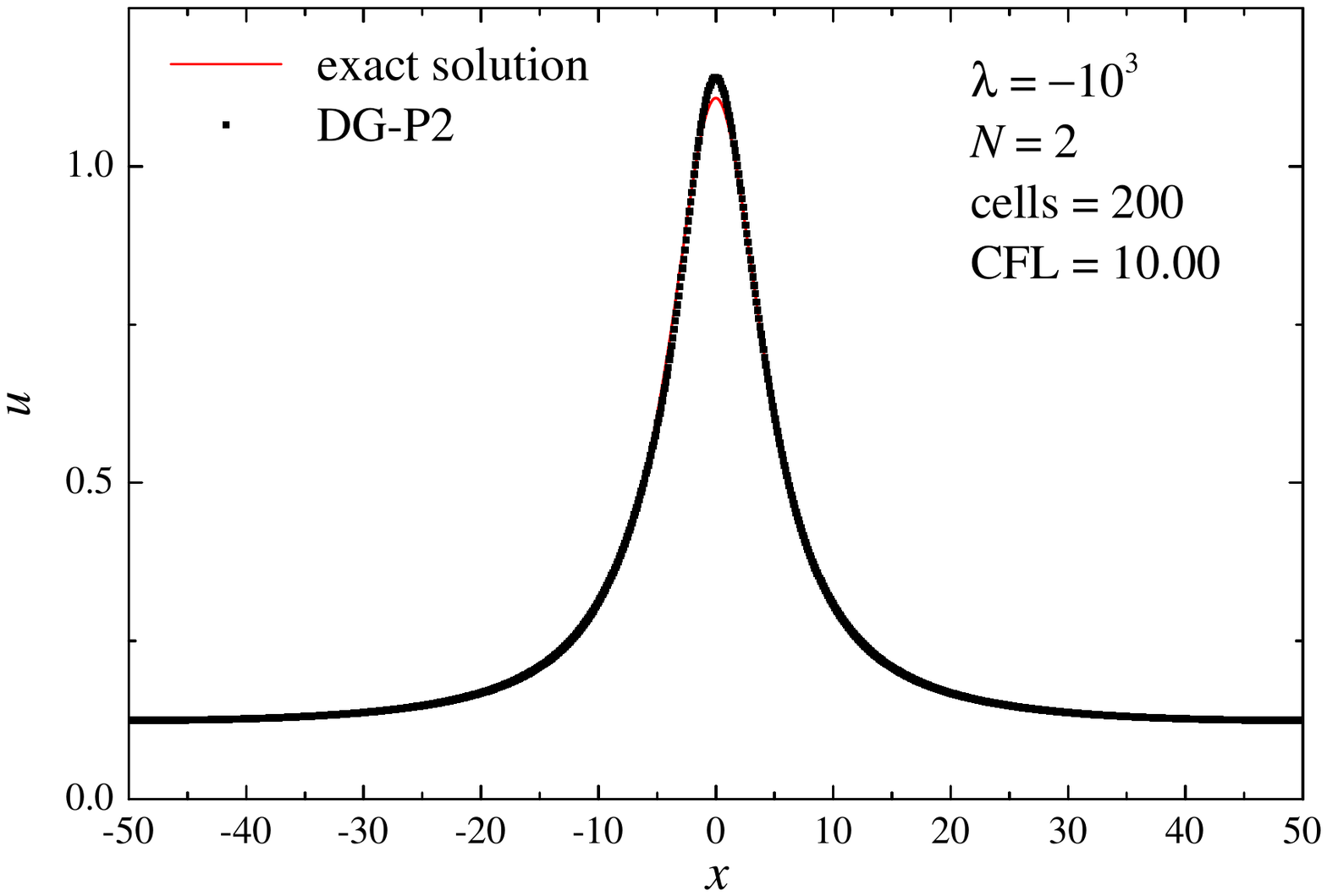}
\vspace{-8mm}\caption{\label{fig:test_adveq_stiff_z1000_lrnz_1d_cfl_1000:2}}
\end{subfigure}
\begin{subfigure}{0.24\textwidth}
\includegraphics[width=\textwidth]{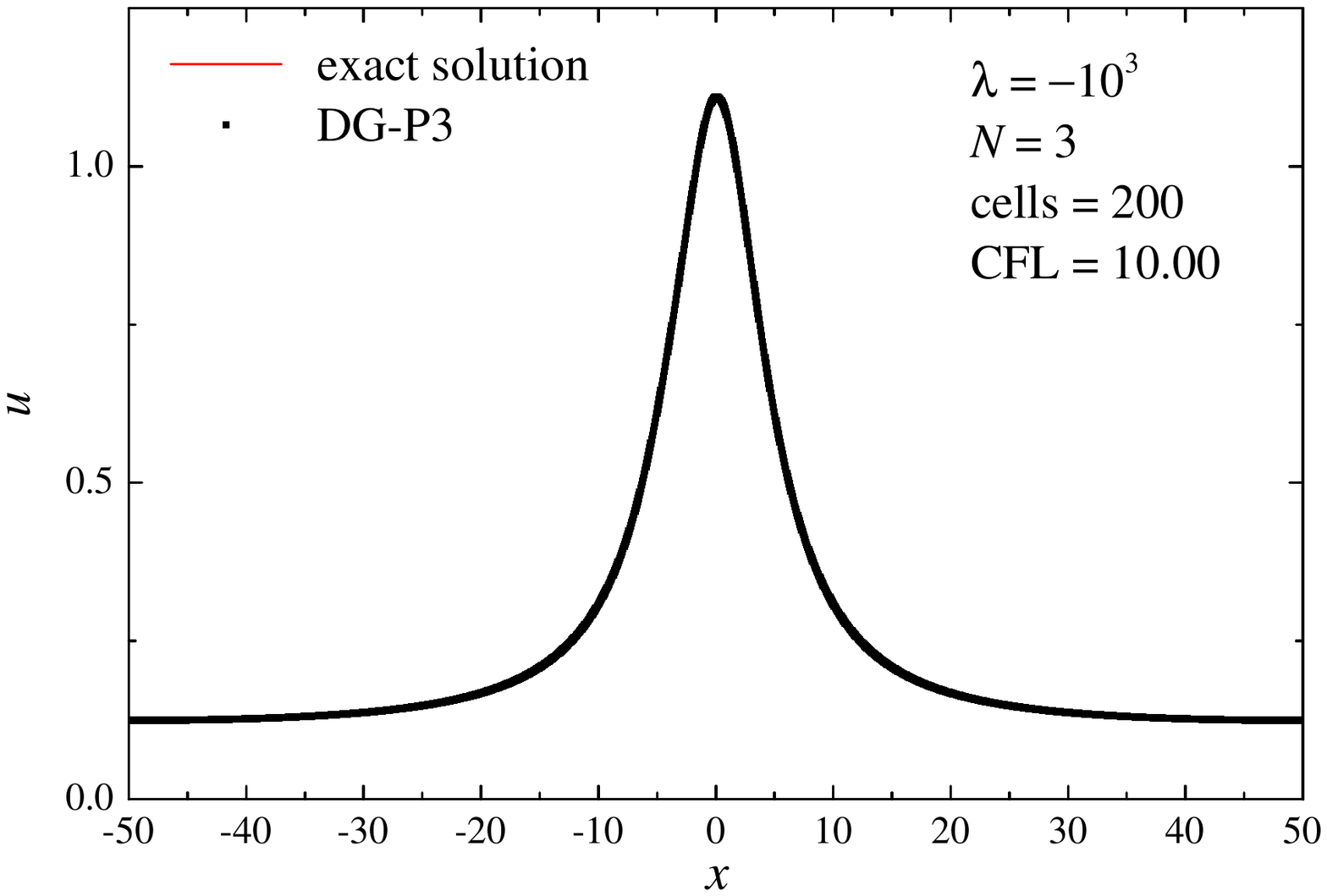}
\vspace{-8mm}\caption{\label{fig:test_adveq_stiff_z1000_lrnz_1d_cfl_1000:3}}
\end{subfigure}
\begin{subfigure}{0.24\textwidth}
\includegraphics[width=\textwidth]{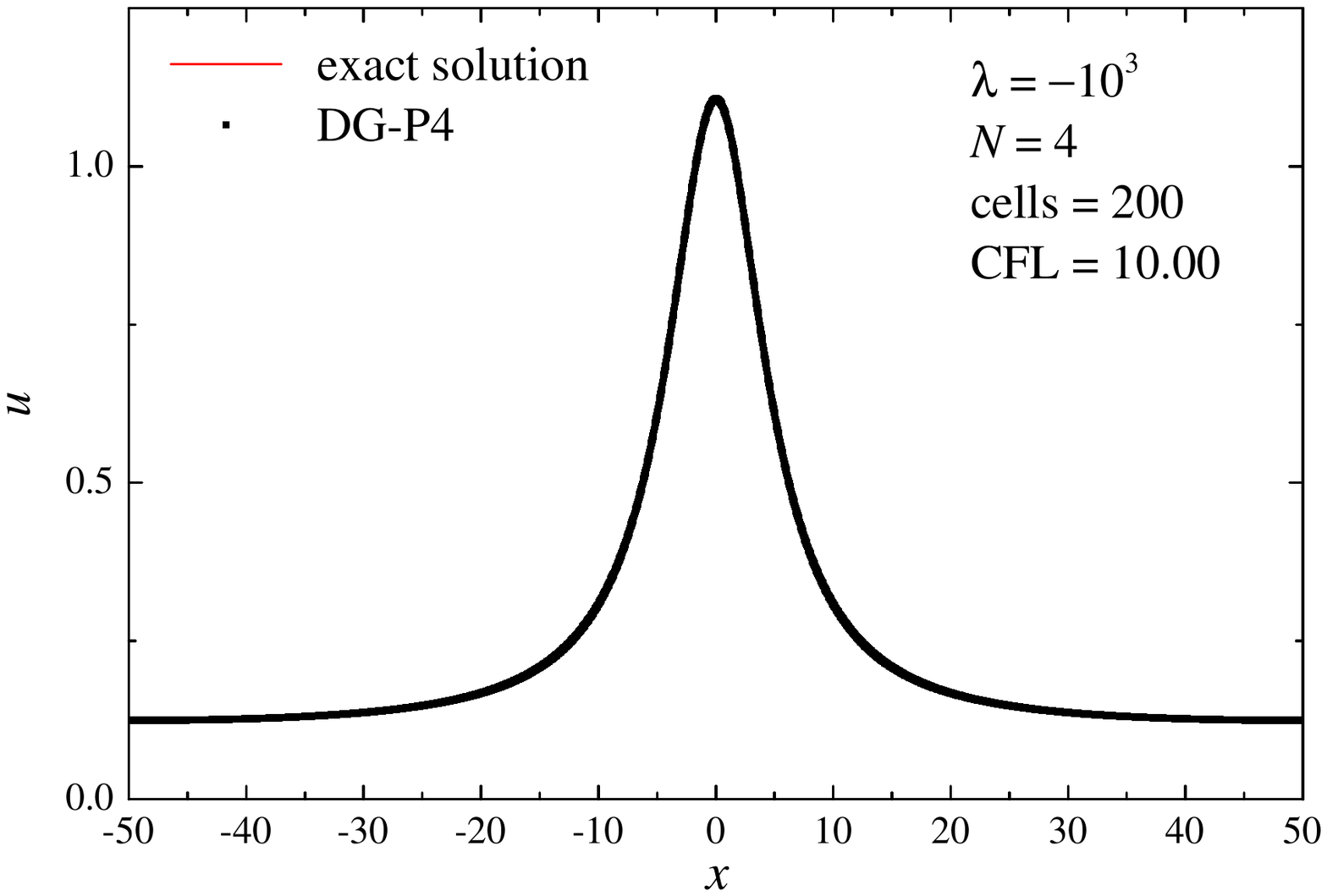}
\vspace{-8mm}\caption{\label{fig:test_adveq_stiff_z1000_lrnz_1d_cfl_1000:4}}
\end{subfigure}\\[-2.5mm]
\begin{subfigure}{0.24\textwidth}
\includegraphics[width=\textwidth]{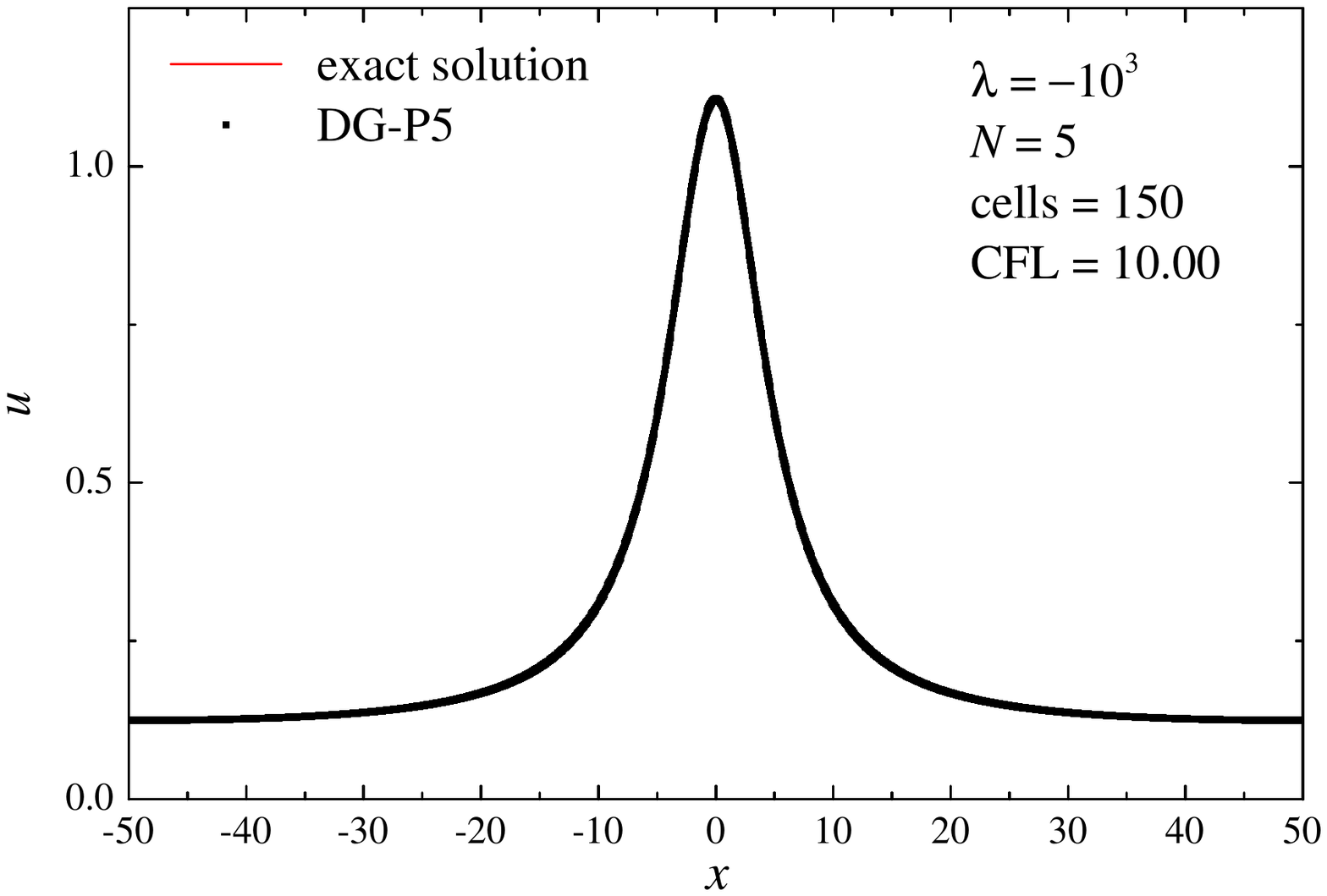}
\vspace{-8mm}\caption{\label{fig:test_adveq_stiff_z1000_lrnz_1d_cfl_1000:5}}
\end{subfigure}
\begin{subfigure}{0.24\textwidth}
\includegraphics[width=\textwidth]{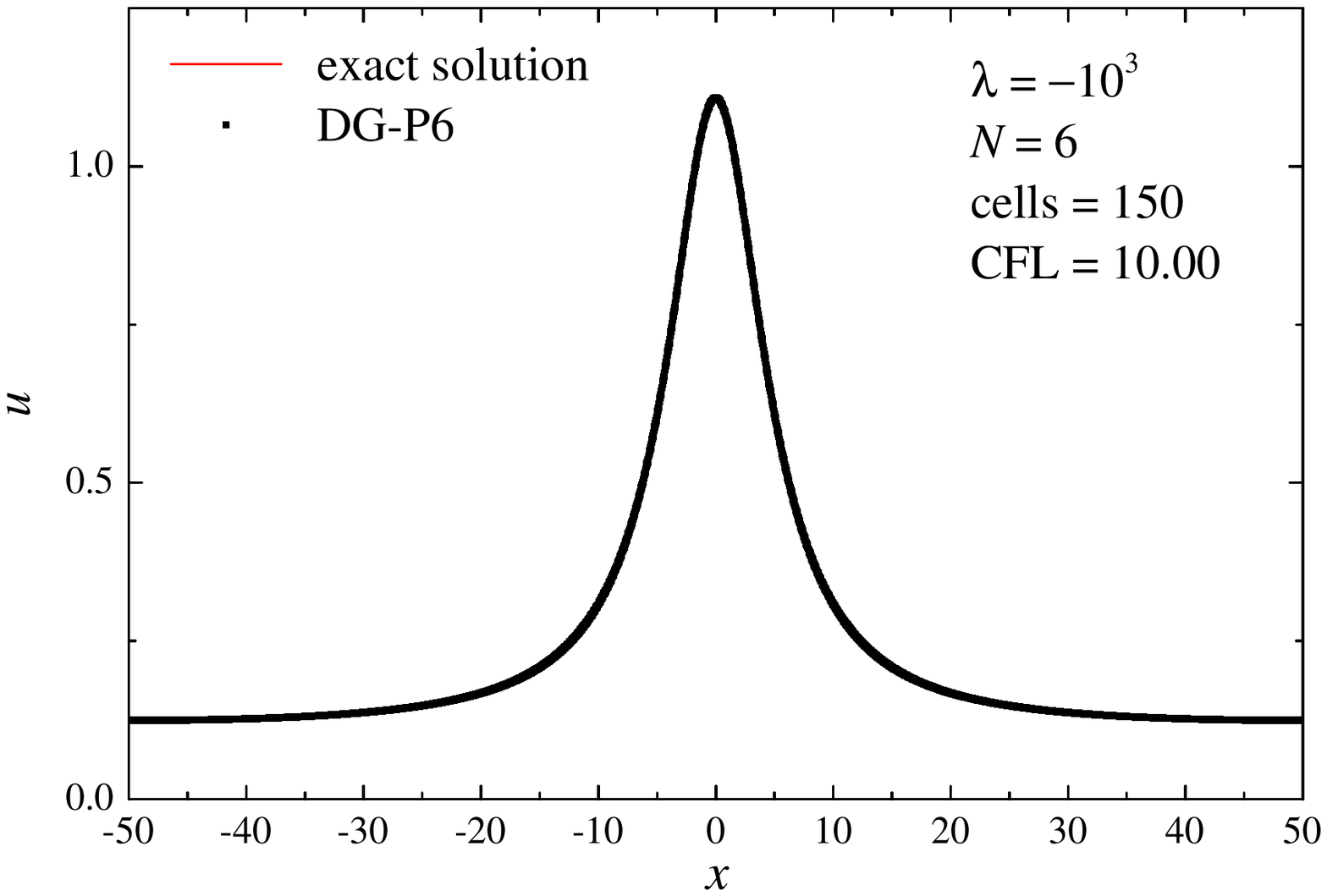}
\vspace{-8mm}\caption{\label{fig:test_adveq_stiff_z1000_lrnz_1d_cfl_1000:6}}
\end{subfigure}
\begin{subfigure}{0.24\textwidth}
\includegraphics[width=\textwidth]{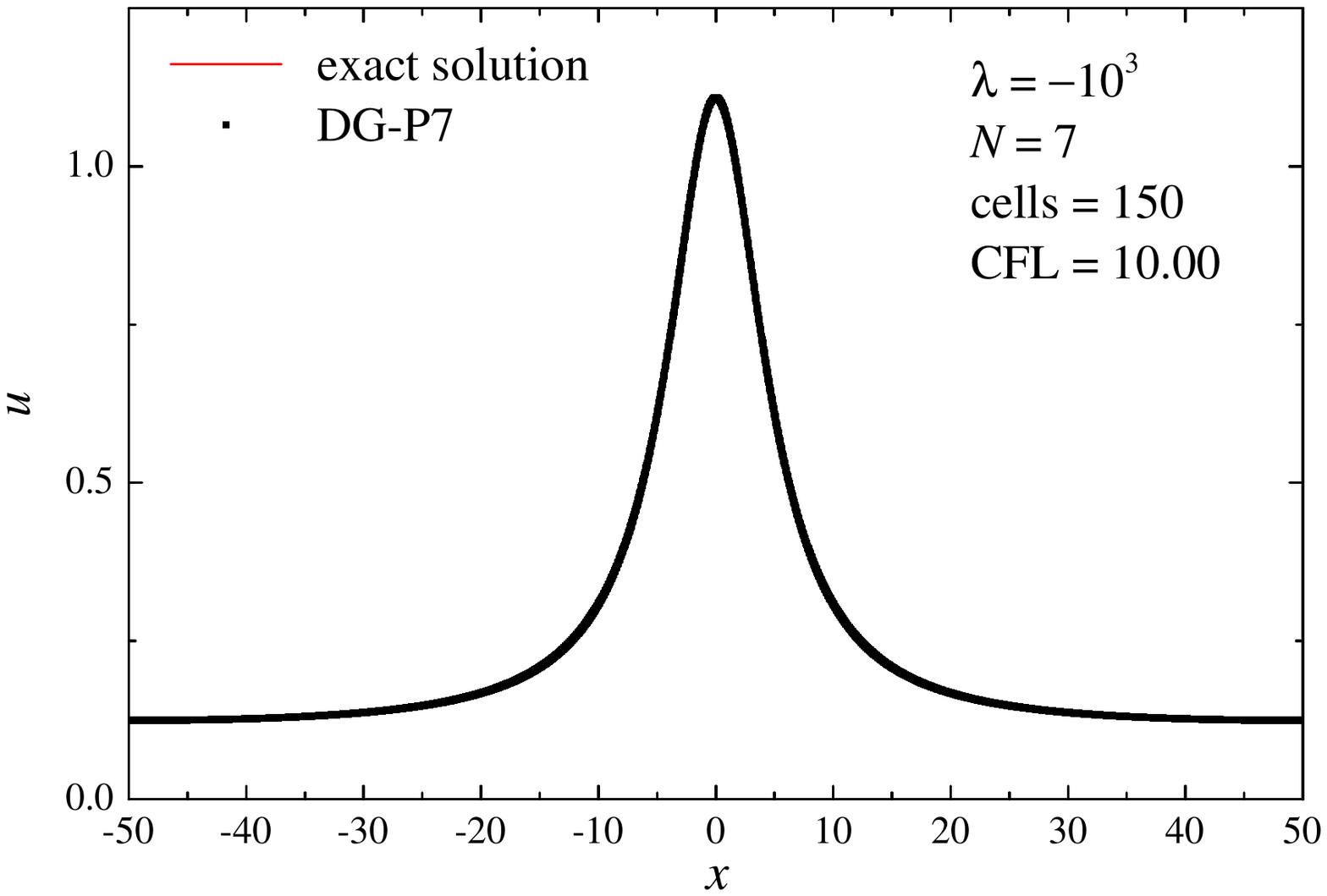}
\vspace{-8mm}\caption{\label{fig:test_adveq_stiff_z1000_lrnz_1d_cfl_1000:7}}
\end{subfigure}
\begin{subfigure}{0.24\textwidth}
\includegraphics[width=\textwidth]{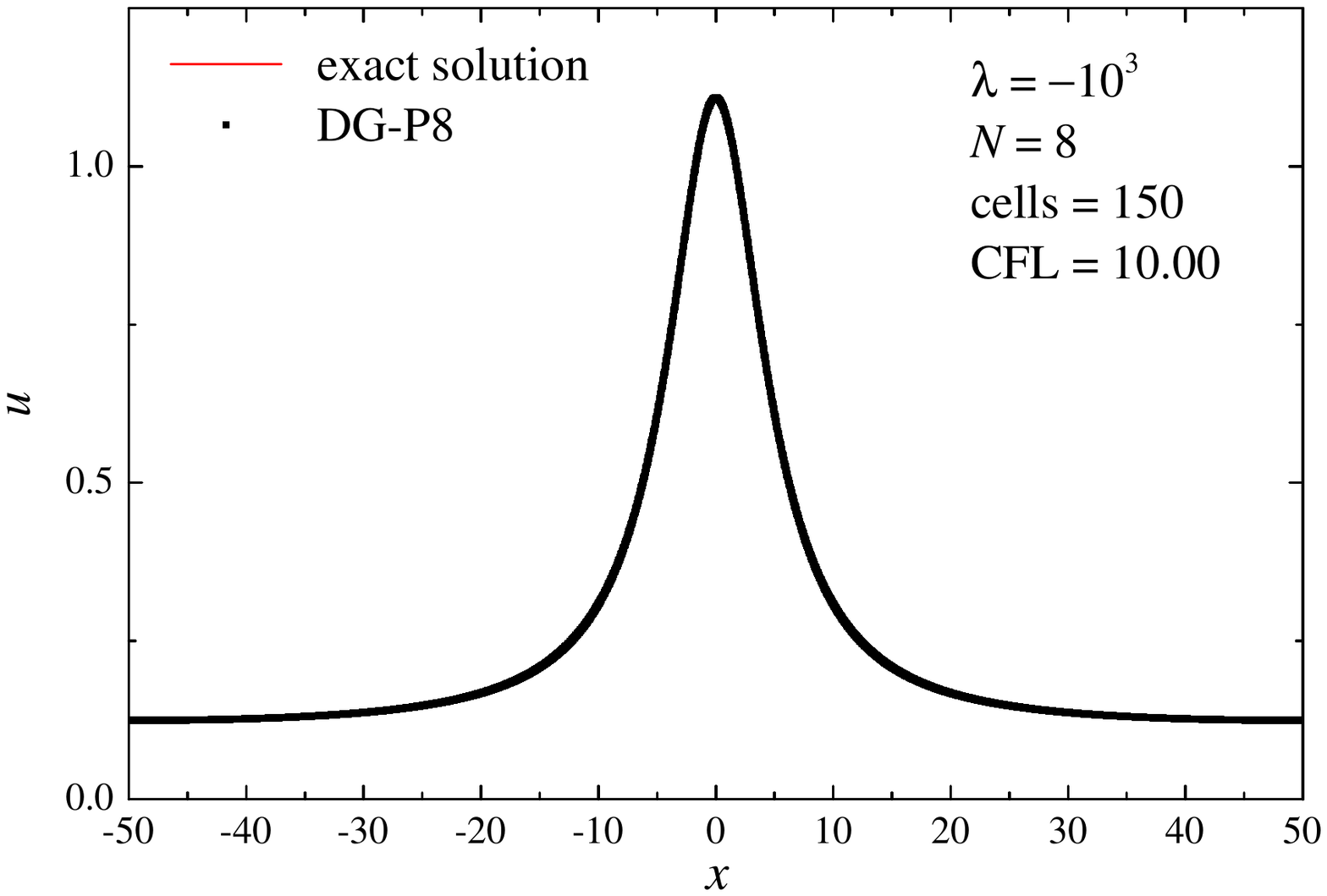}
\vspace{-8mm}\caption{\label{fig:test_adveq_stiff_z1000_lrnz_1d_cfl_1000:8}}
\end{subfigure}\\[-2.5mm]
\begin{subfigure}{0.24\textwidth}
\includegraphics[width=\textwidth]{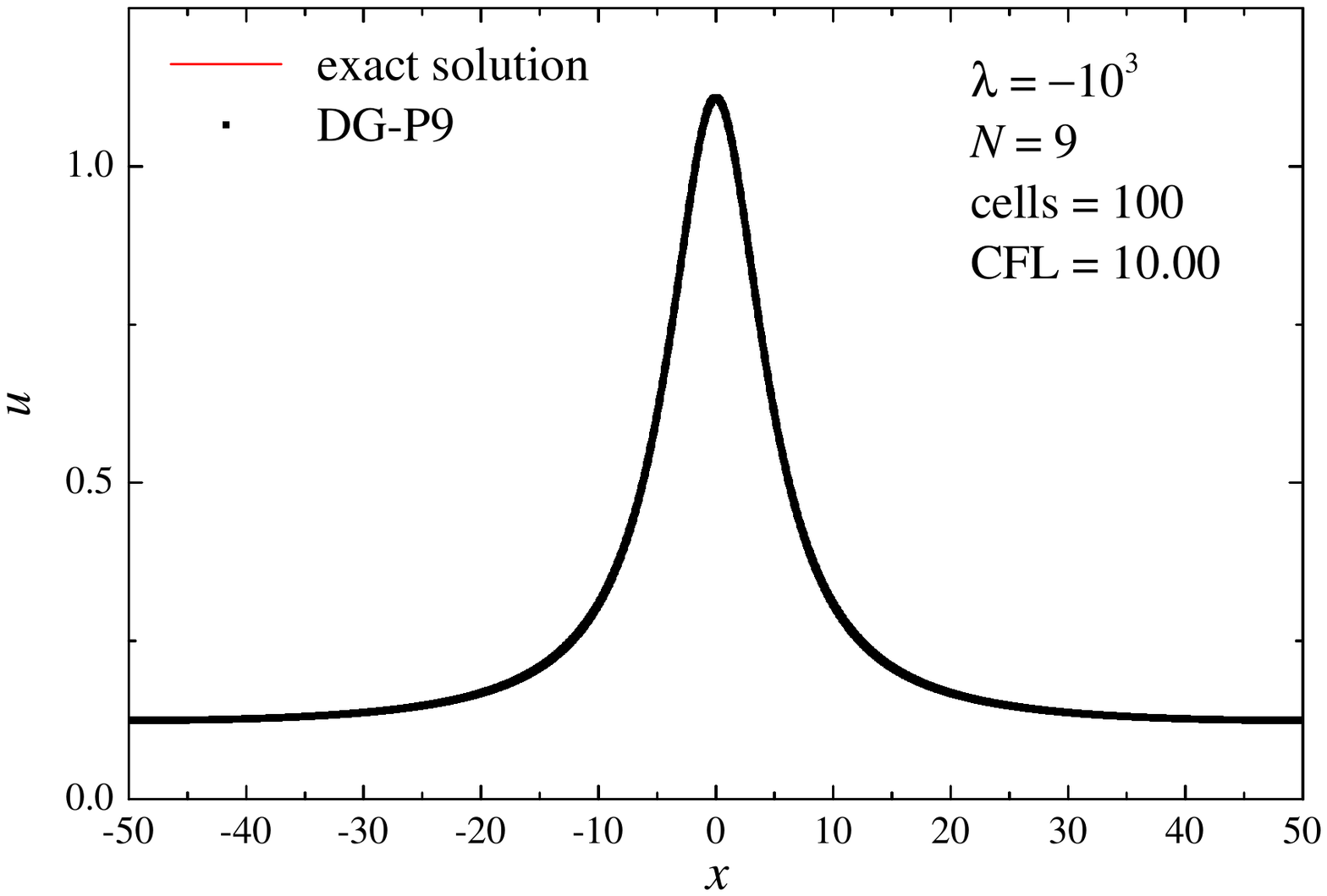}
\vspace{-8mm}\caption{\label{fig:test_adveq_stiff_z1000_lrnz_1d_cfl_1000:9}}
\end{subfigure}
\begin{subfigure}{0.24\textwidth}
\includegraphics[width=\textwidth]{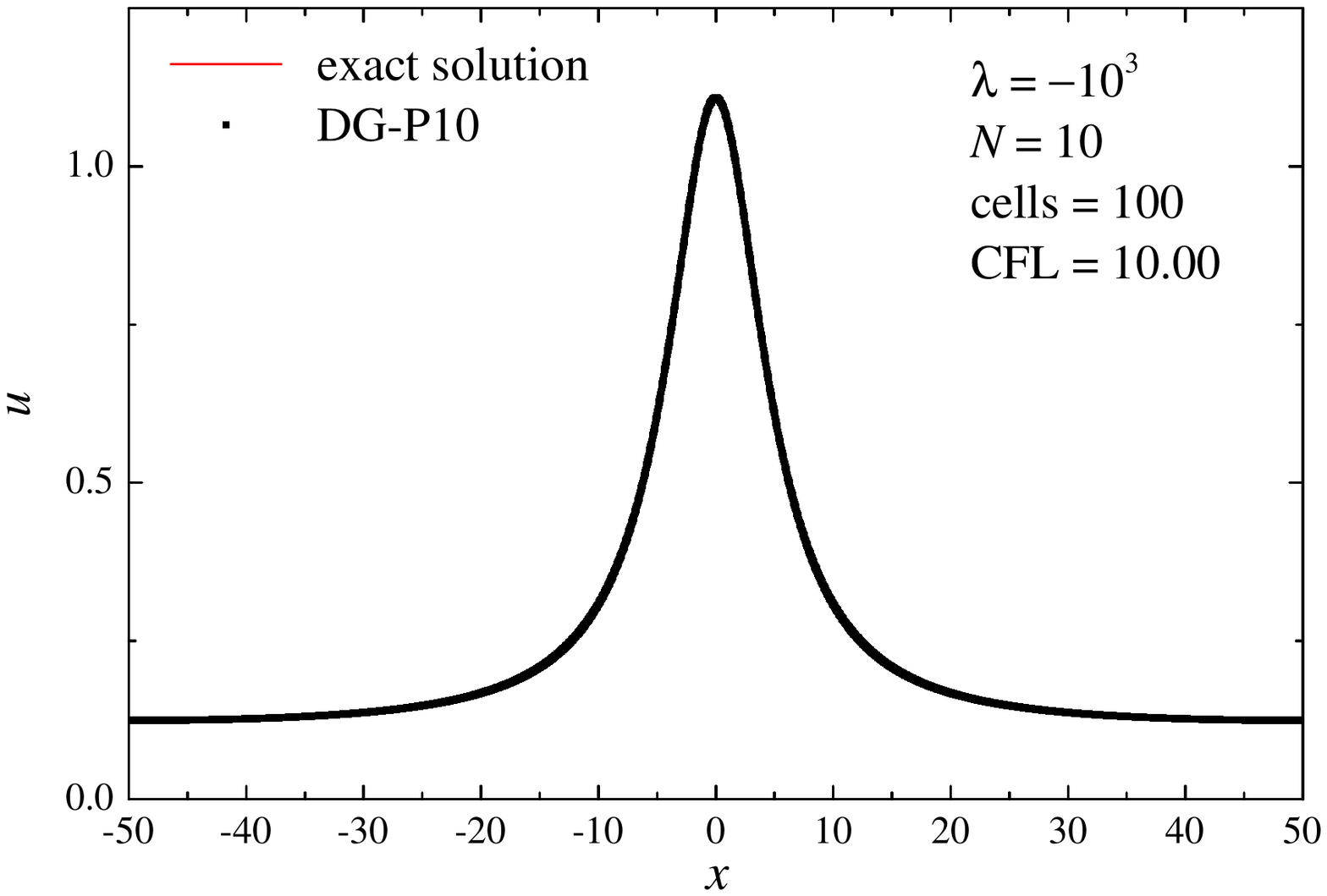}
\vspace{-8mm}\caption{\label{fig:test_adveq_stiff_z1000_lrnz_1d_cfl_1000:10}}
\end{subfigure}
\begin{subfigure}{0.24\textwidth}
\includegraphics[width=\textwidth]{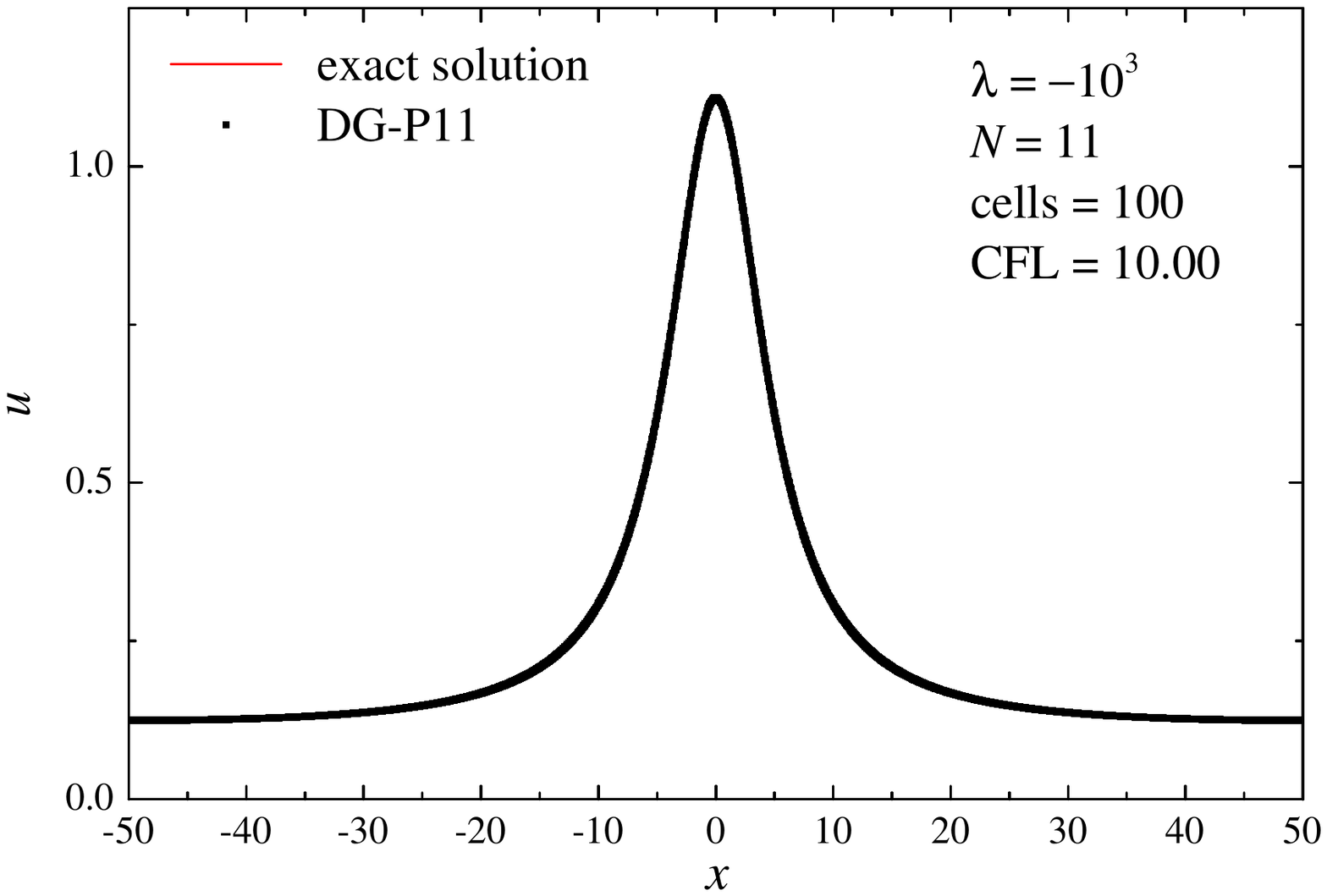}
\vspace{-8mm}\caption{\label{fig:test_adveq_stiff_z1000_lrnz_1d_cfl_1000:11}}
\end{subfigure}
\begin{subfigure}{0.24\textwidth}
\includegraphics[width=\textwidth]{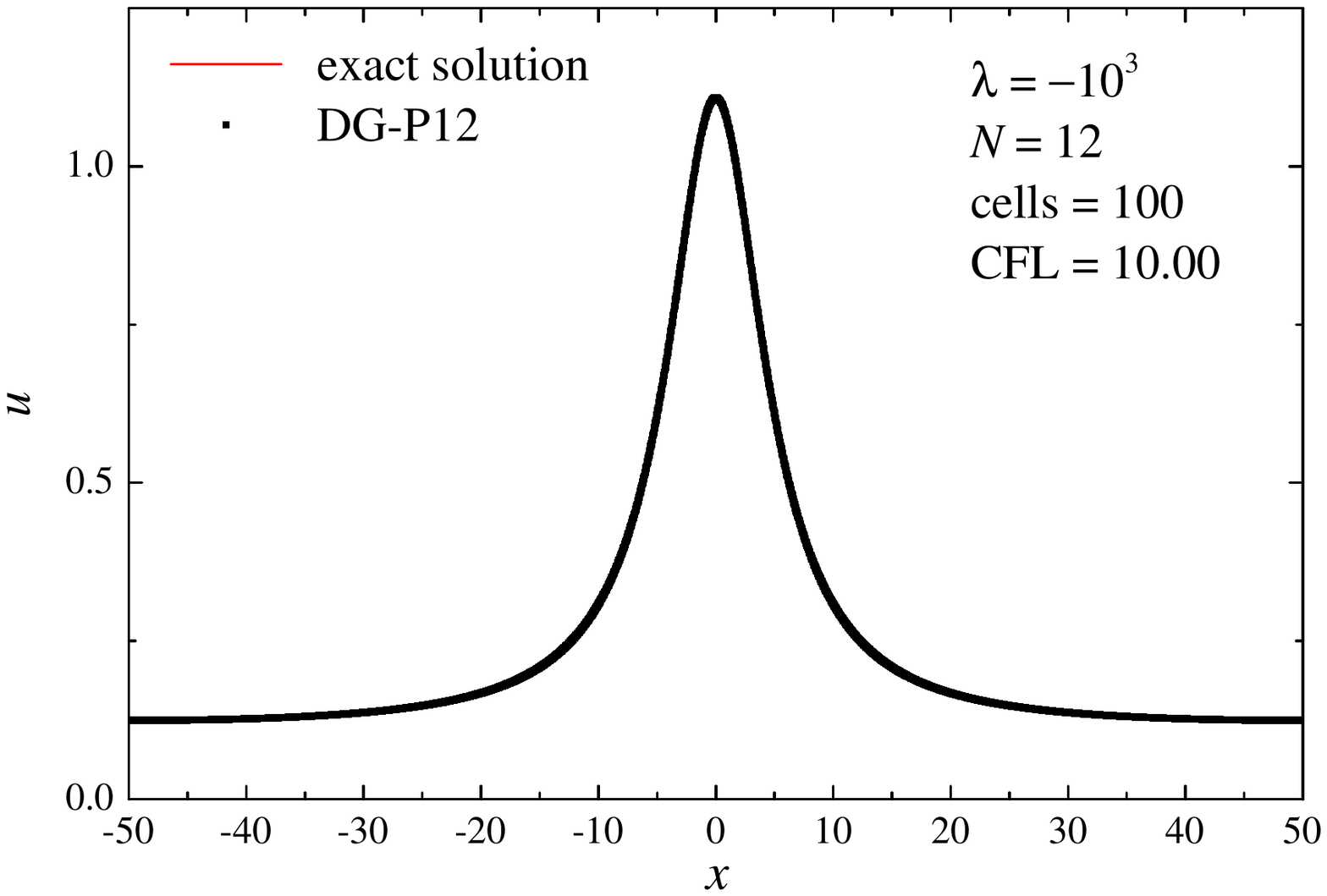}
\vspace{-8mm}\caption{\label{fig:test_adveq_stiff_z1000_lrnz_1d_cfl_1000:12}}
\end{subfigure}
\caption{%
Coordinate dependencies of the solution $u(x, t_{f})$ to the linear advection-reaction equation (\ref{eq:adv_eq_src}) at the final time $t_{f}$, obtained by the ADER-DG numerical method with the LST-DG predictor, for the degrees $N = 1, \ldots, 12$ of the basis polynomials for the Courant number $\mathrm{CFL} = 10.0$ and the decay parameter $\nu = -10^{3}$. The red line represents the exact analytical solution.
}
\label{fig:test_adveq_stiff_z1000_lrnz_1d_cfl_1000}
\end{figure}

\begin{table*}[h!]
\centering
\small
\caption{%
Empirical orders of convergence $p$ of the ADER-DG numerical method with the LST-DG predictor, obtained based on the solution to the profile (\ref{eq:lorentz_bell}) advection-reaction problem for equation (\ref{eq:adv_eq_s_src}), calculated in functional norms $\mathcal{L}_{1}$, $\mathcal{L}_{2}$, $\mathcal{L}_{\infty}$ of errors $\epsilon$ (\ref{eq:norms_of_errors}), for degrees $N = 1, \ldots, 12$ of basis polynomials for $\mathrm{CFL} = 0.1$. Column Th. corresponds to $p = N+1$.
\label{tab:test_adveq_stiff_z100_lrnz_1d_cfl_010}
}
\setlength{\tabcolsep}{3pt}
\begin{adjustbox}{angle=90}
\begin{tabular}{@{}|c|r|ccc|ccc|c||c|r|ccc|ccc|c|@{}}
\toprule
\multicolumn{18}{|c|}{$\lambda = 10^{2}$, $\mathrm{CFL} = 0.1$} \\
\midrule
~$N$~ & cells 
& ~~$\epsilon_{\mathcal{L}_{1}}$~~ & ~~$\epsilon_{\mathcal{L}_{2}}$~~ & ~~$\epsilon_{\mathcal{L}_{\infty}}$~~ 
& ~~$p_{\mathcal{L}_{1}}$~~ & ~$p_{\mathcal{L}_{2}}$~~ & ~~$p_{\mathcal{L}_{\infty}}$~~ & Th. &
~$N$~ & cells 
& ~~$\epsilon_{\mathcal{L}_{1}}$~~ & ~~$\epsilon_{\mathcal{L}_{2}}$~~ & ~~$\epsilon_{\mathcal{L}_{\infty}}$~~
& ~~$p_{\mathcal{L}_{1}}$~~ & ~~$p_{\mathcal{L}_{2}}$~~ & ~~$p_{\mathcal{L}_{\infty}}$~~ & Th. \\
\midrule
1		&	$100$	&	1.70E$-$02	&	5.19E$-$03	&	6.80E$-$03	&	--		&	--		&	--		&	2 & 
7		&	$50$	&	3.71E$-$08	&	2.55E$-$08	&	1.11E$-$07	&	--		&	--		&	--		&	8\\
&	$200$	&	4.18E$-$03	&	1.28E$-$03	&	1.68E$-$03	&	2.02	&	2.02	&	2.02	& & 
&	$100$	&	3.46E$-$10	&	2.47E$-$10	&	1.18E$-$09	&	6.74	&	6.69	&	6.55	&	\\
&	$300$	&	1.85E$-$03	&	5.69E$-$04	&	7.47E$-$04	&	2.01	&	2.01	&	1.99	& & 
&	$150$	&	1.08E$-$11	&	7.15E$-$12	&	3.54E$-$11	&	8.55	&	8.74	&	8.66	&	\\
&	$400$	&	1.04E$-$03	&	3.20E$-$04	&	4.21E$-$04	&	2.00	&	2.00	&	1.99	& & 
&	$200$	&	1.23E$-$12	&	5.79E$-$13	&	2.45E$-$12	&	7.57	&	8.73	&	9.27	&	\\
\midrule
2		&	$100$	&	1.24E$-$03	&	4.31E$-$04	&	4.71E$-$04	&	--		&	--		&	--		&	3 & 
8		&	$20$	&	2.76E$-$05	&	1.24E$-$05	&	3.03E$-$05	&	--		&	--		&	--		&	9\\
&	$200$	&	1.58E$-$04	&	5.57E$-$05	&	6.09E$-$05	&	2.97	&	2.95	&	2.95	& & 
&	$40$	&	4.34E$-$08	&	2.98E$-$08	&	1.22E$-$07	&	9.31	&	8.70	&	7.96	&	\\
&	$300$	&	4.69E$-$05	&	1.66E$-$05	&	1.81E$-$05	&	2.99	&	2.99	&	3.00	& & 
&	$80$	&	1.32E$-$10	&	8.94E$-$11	&	4.09E$-$10	&	8.36	&	8.38	&	8.22	&	\\
&	$400$	&	1.98E$-$05	&	6.98E$-$06	&	7.61E$-$06	&	3.01	&	3.01	&	3.00	& & 
&	$120$	&	4.50E$-$12	&	3.10E$-$12	&	1.80E$-$11	&	8.33	&	8.29	&	7.71	&	\\
\midrule
3		&	$100$	&	3.62E$-$05	&	1.46E$-$05	&	2.55E$-$05	&	--		&	--		&	--		&	4 & 
9		&	$ 5$	&	3.51E$-$01	&	1.11E$-$01	&	2.07E$-$01	&	--		&	--		&	--		&	10\\
&	$200$	&	1.92E$-$06	&	7.58E$-$07	&	1.47E$-$06	&	4.24	&	4.27	&	4.11	& & 
&	$10$	&	9.49E$-$04	&	4.54E$-$04	&	1.36E$-$03	&	8.53	&	7.93	&	7.25	&	\\
&	$300$	&	3.64E$-$07	&	1.43E$-$07	&	2.83E$-$07	&	4.09	&	4.11	&	4.07	& & 
&	$15$	&	1.59E$-$04	&	8.86E$-$05	&	2.84E$-$04	&	4.40	&	4.03	&	3.86	&	\\
&	$400$	&	1.14E$-$07	&	4.48E$-$08	&	8.81E$-$08	&	4.04	&	4.05	&	4.05	& & 
&	$20$	&	3.81E$-$06	&	2.65E$-$06	&	1.09E$-$05	&	12.97	&	12.20	&	11.33	&	\\
\midrule
4		&	$100$	&	3.31E$-$06	&	1.64E$-$06	&	3.54E$-$06	&	--		&	--		&	--		&	5 & 
10		&	$ 5$	&	1.47E$-$01	&	5.15E$-$02	&	1.20E$-$01	&	--		&	--		&	--		&	11\\
&	$200$	&	1.10E$-$07	&	5.52E$-$08	&	1.23E$-$07	&	4.91	&	4.89	&	4.85	& & 
&	$10$	&	3.44E$-$04	&	1.33E$-$04	&	3.81E$-$04	&	8.73	&	8.59	&	8.30	&	\\
&	$300$	&	1.43E$-$08	&	7.19E$-$09	&	1.60E$-$08	&	5.02	&	5.02	&	5.03	& & 
&	$15$	&	3.16E$-$05	&	2.36E$-$05	&	9.83E$-$05	&	5.89	&	4.27	&	3.34	&	\\
&	$400$	&	3.34E$-$09	&	1.67E$-$09	&	3.70E$-$09	&	5.07	&	5.08	&	5.09	& & 
&	$20$	&	1.10E$-$06	&	6.73E$-$07	&	2.78E$-$06	&	11.67	&	12.37	&	12.39	&	\\
\midrule
5		&	$50$	&	5.78E$-$06	&	2.73E$-$06	&	7.17E$-$06	&	--		&	--		&	--		&	6 & 
11		&	$ 5$	&	1.58E$-$01	&	5.37E$-$02	&	1.18E$-$01	&	--		&	--		&	--		&	12\\
&	$100$	&	1.08E$-$07	&	6.04E$-$08	&	1.91E$-$07	&	5.74	&	5.50	&	5.23	& & 
&	$10$	&	1.95E$-$04	&	9.67E$-$05	&	3.37E$-$04	&	9.67	&	9.12	&	8.46	&	\\
&	$150$	&	7.86E$-$09	&	4.01E$-$09	&	1.28E$-$08	&	6.46	&	6.69	&	6.67	& & 
&	$15$	&	1.54E$-$05	&	9.84E$-$06	&	4.00E$-$05	&	6.26	&	5.64	&	5.26	&	\\
&	$200$	&	1.24E$-$09	&	6.02E$-$10	&	1.64E$-$09	&	6.42	&	6.59	&	7.14	& & 
&	$20$	&	3.52E$-$07	&	1.92E$-$07	&	6.48E$-$07	&	13.13	&	13.69	&	14.33	&	\\
\midrule
6		&	$50$	&	8.57E$-$07	&	4.26E$-$07	&	1.09E$-$06	&	--		&	--		&	--		&	7 & 
12		&	$ 5$	&	5.82E$-$02	&	2.13E$-$02	&	5.68E$-$02	&	--		&	--		&	--		&	13\\
&	$100$	&	8.16E$-$09	&	5.21E$-$09	&	2.05E$-$08	&	6.71	&	6.36	&	5.74	& & 
&	$10$	&	6.72E$-$05	&	3.92E$-$05	&	1.54E$-$04	&	9.76	&	9.08	&	8.52	&	\\
&	$150$	&	5.23E$-$10	&	3.41E$-$10	&	1.28E$-$09	&	6.78	&	6.72	&	6.83	& & 
&	$15$	&	3.61E$-$06	&	2.83E$-$06	&	1.36E$-$05	&	7.21	&	6.49	&	5.99	&	\\
&	$200$	&	7.12E$-$11	&	4.65E$-$11	&	1.88E$-$10	&	6.93	&	6.93	&	6.67	& & 
&	$20$	&	6.45E$-$08	&	4.63E$-$08	&	2.39E$-$07	&	13.99	&	14.30	&	14.05	&	\\
\bottomrule
\end{tabular}
\end{adjustbox}
\end{table*}
\begin{table*}[h!]
\centering
\small
\caption{%
Empirical orders of convergence $p$ of the ADER-DG numerical method with the LST-DG predictor, obtained based on the solution to the profile (\ref{eq:lorentz_bell}) advection-reaction problem for equation (\ref{eq:adv_eq_s_src}), calculated in functional norms $\mathcal{L}_{1}$, $\mathcal{L}_{2}$, $\mathcal{L}_{\infty}$ of errors $\epsilon$ (\ref{eq:norms_of_errors}), for degrees $N = 1, \ldots, 12$ of basis polynomials for $\mathrm{CFL} = 1.0$. Column Th. corresponds to $p = N+1$.
\label{tab:test_adveq_stiff_z100_lrnz_1d_cfl_100}
}
\setlength{\tabcolsep}{3pt}
\begin{adjustbox}{angle=90}
\begin{tabular}{@{}|c|r|ccc|ccc|c||c|r|ccc|ccc|c|@{}}
\toprule
\multicolumn{18}{|c|}{$\lambda = 10^{2}$, $\mathrm{CFL} = 1.0$} \\
\midrule
~$N$~ & cells 
& ~~$\epsilon_{\mathcal{L}_{1}}$~~ & ~~$\epsilon_{\mathcal{L}_{2}}$~~ & ~~$\epsilon_{\mathcal{L}_{\infty}}$~~ 
& ~~$p_{\mathcal{L}_{1}}$~~ & ~$p_{\mathcal{L}_{2}}$~~ & ~~$p_{\mathcal{L}_{\infty}}$~~ & Th. &
~$N$~ & cells 
& ~~$\epsilon_{\mathcal{L}_{1}}$~~ & ~~$\epsilon_{\mathcal{L}_{2}}$~~ & ~~$\epsilon_{\mathcal{L}_{\infty}}$~~
& ~~$p_{\mathcal{L}_{1}}$~~ & ~~$p_{\mathcal{L}_{2}}$~~ & ~~$p_{\mathcal{L}_{\infty}}$~~ & Th. \\
\midrule
1		&	$100$	&	4.89E$-$02	&	1.50E$-$02	&	1.65E$-$02	&	--		&	--		&	--		&	2 & 
7		&	$50$	&	1.09E$-$06	&	7.75E$-$07	&	2.81E$-$06	&	--		&	--		&	--		&	8\\
&	$200$	&	1.09E$-$02	&	3.24E$-$03	&	3.55E$-$03	&	2.17	&	2.22	&	2.22	& & 
&	$100$	&	2.79E$-$09	&	2.34E$-$09	&	1.22E$-$08	&	8.61	&	8.37	&	7.85	&	\\
&	$300$	&	4.64E$-$03	&	1.37E$-$03	&	1.54E$-$03	&	2.10	&	2.12	&	2.07	& & 
&	$150$	&	8.56E$-$11	&	6.85E$-$11	&	3.50E$-$10	&	8.59	&	8.71	&	8.76	&	\\
&	$400$	&	2.54E$-$03	&	7.52E$-$04	&	8.61E$-$04	&	2.09	&	2.09	&	2.01	& & 
&	$200$	&	1.07E$-$11	&	5.49E$-$12	&	3.05E$-$11	&	7.23	&	8.78	&	8.49	&	\\
\midrule
2		&	$100$	&	8.79E$-$03	&	3.36E$-$03	&	4.44E$-$03	&	--		&	--		&	--		&	3 & 
8		&	$40$	&	1.97E$-$07	&	1.00E$-$07	&	3.43E$-$07	&	--		&	--		&	--		&	9\\
&	$200$	&	1.12E$-$03	&	4.40E$-$04	&	6.01E$-$04	&	2.97	&	2.93	&	2.88	& & 
&	$80$	&	9.14E$-$10	&	7.19E$-$10	&	3.54E$-$09	&	7.75	&	7.13	&	6.60	&	\\
&	$300$	&	3.32E$-$04	&	1.31E$-$04	&	1.79E$-$04	&	3.00	&	2.99	&	2.99	& & 
&	$120$	&	4.24E$-$11	&	3.00E$-$11	&	1.39E$-$10	&	7.57	&	7.83	&	7.99	&	\\
&	$400$	&	1.40E$-$04	&	5.53E$-$05	&	7.61E$-$05	&	3.01	&	3.00	&	2.98	& & 
&	$160$	&	7.95E$-$12	&	2.81E$-$12	&	1.66E$-$11	&	5.81	&	8.24	&	7.38	&	\\
\midrule
3		&	$100$	&	1.95E$-$04	&	9.77E$-$05	&	2.07E$-$04	&	--		&	--		&	--		&	4 & 
9		&	$20$	&	2.37E$-$05	&	1.20E$-$05	&	3.44E$-$05	&	--		&	--		&	--		&	10\\
&	$200$	&	8.29E$-$06	&	3.64E$-$06	&	7.50E$-$06	&	4.56	&	4.74	&	4.78	& & 
&	$40$	&	4.61E$-$08	&	3.69E$-$08	&	1.67E$-$07	&	9.00	&	8.34	&	7.69	&	\\
&	$300$	&	1.38E$-$06	&	5.65E$-$07	&	9.62E$-$07	&	4.41	&	4.60	&	5.06	& & 
&	$60$	&	2.05E$-$09	&	1.68E$-$09	&	8.21E$-$09	&	7.68	&	7.61	&	7.43	&	\\
&	$400$	&	4.00E$-$07	&	1.60E$-$07	&	2.33E$-$07	&	4.32	&	4.38	&	4.92	& & 
&	$80$	&	1.11E$-$10	&	1.01E$-$10	&	6.01E$-$10	&	10.16	&	9.77	&	9.09	&	\\
\midrule
4		&	$100$	&	2.65E$-$05	&	1.45E$-$05	&	3.91E$-$05	&	--		&	--		&	--		&	5 & 
10		&	$10$	&	1.55E$-$03	&	5.23E$-$04	&	7.23E$-$04	&	--		&	--		&	--		&	11\\
&	$200$	&	8.70E$-$07	&	5.07E$-$07	&	1.46E$-$06	&	4.93	&	4.84	&	4.74	& & 
&	$20$	&	3.24E$-$06	&	1.37E$-$06	&	3.48E$-$06	&	8.90	&	8.58	&	7.70	&	\\
&	$300$	&	1.14E$-$07	&	6.72E$-$08	&	1.96E$-$07	&	5.01	&	4.98	&	4.95	& & 
&	$30$	&	8.26E$-$08	&	4.68E$-$08	&	1.62E$-$07	&	9.05	&	8.33	&	7.57	&	\\
&	$400$	&	2.68E$-$08	&	1.58E$-$08	&	4.62E$-$08	&	5.04	&	5.03	&	5.02	& & 
&	$40$	&	3.77E$-$09	&	2.36E$-$09	&	8.34E$-$09	&	10.73	&	10.38	&	10.30	&	\\
\midrule
5		&	$50$	&	8.08E$-$05	&	5.35E$-$05	&	1.52E$-$04	&	--		&	--		&	--		&	6 & 
11		&	$10$	&	6.94E$-$04	&	2.27E$-$04	&	3.55E$-$04	&	--		&	--		&	--		&	12\\
&	$100$	&	8.12E$-$07	&	5.35E$-$07	&	1.65E$-$06	&	6.64	&	6.64	&	6.52	& & 
&	$20$	&	1.24E$-$06	&	6.91E$-$07	&	2.53E$-$06	&	9.13	&	8.36	&	7.13	&	\\
&	$150$	&	5.13E$-$08	&	3.42E$-$08	&	1.33E$-$07	&	6.81	&	6.78	&	6.22	& & 
&	$30$	&	1.22E$-$08	&	7.56E$-$09	&	2.97E$-$08	&	11.39	&	11.14	&	10.96	&	\\
&	$200$	&	7.45E$-$09	&	4.77E$-$09	&	1.93E$-$08	&	6.70	&	6.85	&	6.70	& & 
&	$40$	&	3.66E$-$10	&	2.51E$-$10	&	1.39E$-$09	&	12.19	&	11.83	&	10.65	&	\\
\midrule
6		&	$50$	&	3.97E$-$06	&	2.29E$-$06	&	6.98E$-$06	&	--		&	--		&	--		&	7 & 
12		&	$10$	&	2.16E$-$04	&	7.72E$-$05	&	5.09E$-$05	&	--		&	--		&	--		&	13\\
&	$100$	&	6.95E$-$08	&	4.73E$-$08	&	1.57E$-$07	&	5.84	&	5.60	&	5.47	& & 
&	$20$	&	1.59E$-$07	&	6.44E$-$08	&	1.61E$-$07	&	10.41	&	10.23	&	8.30	&	\\
&	$150$	&	4.49E$-$09	&	3.26E$-$09	&	1.48E$-$08	&	6.76	&	6.60	&	5.83	& & 
&	$30$	&	1.85E$-$09	&	1.14E$-$09	&	5.08E$-$09	&	10.98	&	9.94	&	8.52	&	\\
&	$200$	&	6.13E$-$10	&	4.54E$-$10	&	2.04E$-$09	&	6.92	&	6.85	&	6.88	& & 
&	$40$	&	8.24E$-$11	&	4.37E$-$11	&	1.93E$-$10	&	10.82	&	11.34	&	11.36	&	\\
\bottomrule
\end{tabular}
\end{adjustbox}
\end{table*}
\begin{table*}[h!]
\centering
\small
\caption{%
Empirical orders of convergence $p$ of the ADER-DG numerical method with the LST-DG predictor, obtained based on the solution to the profile (\ref{eq:lorentz_bell}) advection-reaction problem for equation (\ref{eq:adv_eq_s_src}), calculated in functional norms $\mathcal{L}_{1}$, $\mathcal{L}_{2}$, $\mathcal{L}_{\infty}$ of errors $\epsilon$ (\ref{eq:norms_of_errors}), for degrees $N = 1, \ldots, 12$ of basis polynomials for $\mathrm{CFL} = 10.0$. Column Th. corresponds to $p = N+1$.
\label{tab:test_adveq_stiff_z100_lrnz_1d_cfl_1000}
}
\setlength{\tabcolsep}{3pt}
\begin{adjustbox}{angle=90}
\begin{tabular}{@{}|c|r|ccc|ccc|c||c|r|ccc|ccc|c|@{}}
\toprule
\multicolumn{18}{|c|}{$\lambda = 10^{2}$, $\mathrm{CFL} = 10.0$} \\
\midrule
~$N$~ & cells 
& ~~$\epsilon_{\mathcal{L}_{1}}$~~ & ~~$\epsilon_{\mathcal{L}_{2}}$~~ & ~~$\epsilon_{\mathcal{L}_{\infty}}$~~ 
& ~~$p_{\mathcal{L}_{1}}$~~ & ~$p_{\mathcal{L}_{2}}$~~ & ~~$p_{\mathcal{L}_{\infty}}$~~ & Th. &
~$N$~ & cells 
& ~~$\epsilon_{\mathcal{L}_{1}}$~~ & ~~$\epsilon_{\mathcal{L}_{2}}$~~ & ~~$\epsilon_{\mathcal{L}_{\infty}}$~~
& ~~$p_{\mathcal{L}_{1}}$~~ & ~~$p_{\mathcal{L}_{2}}$~~ & ~~$p_{\mathcal{L}_{\infty}}$~~ & Th. \\
\midrule
1		&	$200$	&	1.02E$+$00	&	2.60E$-$01	&	1.46E$-$01	&	--		&	--		&	--		&	2 & 
7		&	$150$	&	1.16E$-$03	&	4.47E$-$04	&	3.17E$-$04	&	--		&	--		&	--		&	8\\
&	$400$	&	2.65E$-$01	&	6.94E$-$02	&	4.05E$-$02	&	1.95	&	1.91	&	1.84	& & 
&	$300$	&	7.23E$-$06	&	2.99E$-$06	&	2.32E$-$06	&	7.33	&	7.22	&	7.09	&	\\
&	$600$	&	1.18E$-$01	&	3.11E$-$02	&	1.85E$-$02	&	1.99	&	1.98	&	1.93	& & 
&	$450$	&	3.09E$-$07	&	1.30E$-$07	&	1.03E$-$07	&	7.77	&	7.73	&	7.69	&	\\
&	$800$	&	6.65E$-$02	&	1.75E$-$02	&	1.05E$-$02	&	2.00	&	2.00	&	1.96	& & 
&	$600$	&	3.18E$-$08	&	1.35E$-$08	&	1.07E$-$08	&	7.91	&	7.89	&	7.86	&	\\
\midrule
2		&	$200$	&	2.19E$-$01	&	6.59E$-$02	&	3.24E$-$02	&	--		&	--		&	--		&	3 & 
8		&	$150$	&	3.50E$-$04	&	1.39E$-$04	&	9.29E$-$05	&	--		&	--		&	--		&	9\\
&	$400$	&	2.97E$-$02	&	9.20E$-$03	&	4.73E$-$03	&	2.89	&	2.84	&	2.78	& & 
&	$300$	&	1.16E$-$06	&	4.98E$-$07	&	3.57E$-$07	&	8.23	&	8.12	&	8.02	&	\\
&	$600$	&	8.92E$-$03	&	2.78E$-$03	&	1.44E$-$03	&	2.96	&	2.95	&	2.93	& & 
&	$450$	&	3.33E$-$08	&	1.45E$-$08	&	1.06E$-$08	&	8.76	&	8.72	&	8.67	&	\\
&	$800$	&	3.78E$-$03	&	1.18E$-$03	&	6.17E$-$04	&	2.98	&	2.98	&	2.96	& & 
&	$600$	&	2.58E$-$09	&	1.13E$-$09	&	8.29E$-$10	&	8.90	&	8.89	&	8.86	&	\\
\midrule
3		&	$200$	&	5.11E$-$02	&	1.68E$-$02	&	1.07E$-$02	&	--		&	--		&	--		&	4 & 
9		&	$100$	&	2.69E$-$03	&	9.95E$-$04	&	6.62E$-$04	&	--		&	--		&	--		&	10\\
&	$400$	&	3.60E$-$03	&	1.23E$-$03	&	8.22E$-$04	&	3.83	&	3.77	&	3.70	& & 
&	$200$	&	8.32E$-$06	&	3.54E$-$06	&	2.78E$-$06	&	8.34	&	8.14	&	7.90	&	\\
&	$600$	&	7.26E$-$04	&	2.49E$-$04	&	1.69E$-$04	&	3.95	&	3.93	&	3.90	& & 
&	$300$	&	1.86E$-$07	&	8.22E$-$08	&	6.73E$-$08	&	9.37	&	9.28	&	9.17	&	\\
&	$800$	&	2.31E$-$04	&	7.94E$-$05	&	5.40E$-$05	&	3.99	&	3.98	&	3.96	& & 
&	$400$	&	1.15E$-$08	&	5.13E$-$09	&	4.27E$-$09	&	9.68	&	9.64	&	9.59	&	\\
\midrule
4		&	$200$	&	1.19E$-$02	&	4.16E$-$03	&	2.39E$-$03	&	--		&	--		&	--		&	5 & 
10		&	$100$	&	1.11E$-$03	&	4.20E$-$04	&	2.72E$-$04	&	--		&	--		&	--		&	11\\
&	$400$	&	4.34E$-$04	&	1.58E$-$04	&	9.41E$-$05	&	4.77	&	4.72	&	4.66	& & 
&	$200$	&	1.95E$-$06	&	8.50E$-$07	&	6.30E$-$07	&	9.16	&	8.95	&	8.76	&	\\
&	$600$	&	5.87E$-$05	&	2.16E$-$05	&	1.29E$-$05	&	4.93	&	4.91	&	4.89	& & 
&	$300$	&	3.01E$-$08	&	1.36E$-$08	&	1.04E$-$08	&	10.28	&	10.20	&	10.12	&	\\
&	$800$	&	1.40E$-$05	&	5.19E$-$06	&	3.11E$-$06	&	4.97	&	4.96	&	4.95	& & 
&	$400$	&	1.42E$-$09	&	6.43E$-$10	&	5.03E$-$10	&	10.62	&	10.60	&	10.53	&	\\
\midrule
5		&	$150$	&	1.29E$-$02	&	4.58E$-$03	&	3.04E$-$03	&	--		&	--		&	--		&	6 & 
11		&	$100$	&	4.60E$-$04	&	1.77E$-$04	&	1.23E$-$04	&	--		&	--		&	--		&	12\\
&	$300$	&	2.81E$-$04	&	1.07E$-$04	&	7.76E$-$05	&	5.52	&	5.42	&	5.29	& & 
&	$200$	&	4.57E$-$07	&	2.04E$-$07	&	1.67E$-$07	&	9.97	&	9.76	&	9.52	&	\\
&	$450$	&	2.63E$-$05	&	1.02E$-$05	&	7.53E$-$06	&	5.84	&	5.80	&	5.75	& & 
&	$300$	&	4.84E$-$09	&	2.23E$-$09	&	1.91E$-$09	&	11.22	&	11.13	&	11.03	&	\\
&	$600$	&	4.78E$-$06	&	1.86E$-$06	&	1.38E$-$06	&	5.93	&	5.91	&	5.89	& & 
&	$400$	&	1.93E$-$10	&	8.03E$-$11	&	7.24E$-$11	&	11.21	&	11.56	&	11.38	&	\\
\midrule
6		&	$150$	&	3.87E$-$03	&	1.43E$-$03	&	8.89E$-$04	&	--		&	--		&	--		&	7 & 
12		&	$100$	&	1.90E$-$04	&	7.46E$-$05	&	5.05E$-$05	&	--		&	--		&	--		&	13\\
&	$300$	&	4.50E$-$05	&	1.79E$-$05	&	1.19E$-$05	&	6.43	&	6.32	&	6.23	& & 
&	$200$	&	1.07E$-$07	&	4.88E$-$08	&	3.81E$-$08	&	10.79	&	10.58	&	10.37	&	\\
&	$450$	&	2.84E$-$06	&	1.15E$-$06	&	7.72E$-$07	&	6.82	&	6.78	&	6.74	& & 
&	$300$	&	8.03E$-$10	&	3.67E$-$10	&	3.01E$-$10	&	12.07	&	12.06	&	11.94	&	\\
&	$600$	&	3.88E$-$07	&	1.58E$-$07	&	1.07E$-$07	&	6.92	&	6.90	&	6.88	& & 
&	$400$	&	6.31E$-$11	&	1.45E$-$11	&	3.40E$-$11	&	8.84	&	11.23	&	7.57	&	\\
\bottomrule
\end{tabular}
\end{adjustbox}
\end{table*}
\begin{table*}[h!]
\centering
\small
\caption{%
Empirical orders of convergence $p$ of the ADER-DG numerical method with the LST-DG predictor, obtained based on the solution to the profile (\ref{eq:lorentz_bell}) advection-reaction problem for equation (\ref{eq:adv_eq_s_src}), calculated in functional norms $\mathcal{L}_{1}$, $\mathcal{L}_{2}$, $\mathcal{L}_{\infty}$ of errors $\epsilon$ (\ref{eq:norms_of_errors}), for degrees $N = 1, \ldots, 12$ of basis polynomials for $\mathrm{CFL} = 0.1$. Column Th. corresponds to $p = N+1$.
\label{tab:test_adveq_stiff_z1000_lrnz_1d_cfl_010}
}
\setlength{\tabcolsep}{3pt}
\begin{adjustbox}{angle=90}
\begin{tabular}{@{}|c|r|ccc|ccc|c||c|r|ccc|ccc|c|@{}}
\toprule
\multicolumn{18}{|c|}{$\lambda = 10^{3}$, $\mathrm{CFL} = 0.1$} \\
\midrule
~$N$~ & cells 
& ~~$\epsilon_{\mathcal{L}_{1}}$~~ & ~~$\epsilon_{\mathcal{L}_{2}}$~~ & ~~$\epsilon_{\mathcal{L}_{\infty}}$~~ 
& ~~$p_{\mathcal{L}_{1}}$~~ & ~$p_{\mathcal{L}_{2}}$~~ & ~~$p_{\mathcal{L}_{\infty}}$~~ & Th. &
~$N$~ & cells 
& ~~$\epsilon_{\mathcal{L}_{1}}$~~ & ~~$\epsilon_{\mathcal{L}_{2}}$~~ & ~~$\epsilon_{\mathcal{L}_{\infty}}$~~
& ~~$p_{\mathcal{L}_{1}}$~~ & ~~$p_{\mathcal{L}_{2}}$~~ & ~~$p_{\mathcal{L}_{\infty}}$~~ & Th. \\
\midrule
1		&	$100$	&	1.70E$-$02	&	5.20E$-$03	&	6.83E$-$03	&	--		&	--		&	--		&	2 & 
7		&	$50$	&	3.96E$-$08	&	2.78E$-$08	&	1.21E$-$07	&	--		&	--		&	--		&	8\\
&	$200$	&	4.18E$-$03	&	1.28E$-$03	&	1.69E$-$03	&	2.02	&	2.02	&	2.02	& & 
&	$100$	&	3.67E$-$10	&	2.64E$-$10	&	1.29E$-$09	&	6.75	&	6.72	&	6.56	&	\\
&	$300$	&	1.85E$-$03	&	5.70E$-$04	&	7.48E$-$04	&	2.01	&	2.01	&	2.01	& & 
&	$150$	&	1.45E$-$11	&	7.69E$-$12	&	4.07E$-$11	&	7.98	&	8.73	&	8.52	&	\\
&	$400$	&	1.04E$-$03	&	3.20E$-$04	&	4.20E$-$04	&	2.00	&	2.00	&	2.01	& & 
&	$200$	&	4.82E$-$12	&	9.98E$-$13	&	3.73E$-$12	&	3.82	&	7.10	&	8.30	&	\\
\midrule
2		&	$100$	&	1.23E$-$03	&	4.28E$-$04	&	4.71E$-$04	&	--		&	--		&	--		&	3 & 
8		&	$20$	&	2.74E$-$05	&	1.23E$-$05	&	2.99E$-$05	&	--		&	--		&	--		&	9\\
&	$200$	&	1.56E$-$04	&	5.51E$-$05	&	6.11E$-$05	&	2.97	&	2.96	&	2.95	& & 
&	$40$	&	4.39E$-$08	&	3.05E$-$08	&	1.19E$-$07	&	9.28	&	8.65	&	7.97	&	\\
&	$300$	&	4.65E$-$05	&	1.64E$-$05	&	1.82E$-$05	&	2.99	&	2.98	&	2.98	& & 
&	$60$	&	1.65E$-$09	&	9.66E$-$10	&	3.65E$-$09	&	8.09	&	8.51	&	8.60	&	\\
&	$400$	&	1.97E$-$05	&	6.95E$-$06	&	7.73E$-$06	&	2.99	&	2.99	&	2.98	& & 
&	$80$	&	1.41E$-$10	&	9.23E$-$11	&	4.17E$-$10	&	8.56	&	8.16	&	7.54	&	\\
\midrule
3		&	$100$	&	3.67E$-$05	&	1.49E$-$05	&	2.65E$-$05	&	--		&	--		&	--		&	4 & 
9		&	$ 5$	&	3.50E$-$01	&	1.11E$-$01	&	2.06E$-$01	&	--		&	--		&	--		&	10\\
&	$200$	&	1.92E$-$06	&	7.67E$-$07	&	1.53E$-$06	&	4.26	&	4.28	&	4.12	& & 
&	$10$	&	9.66E$-$04	&	4.61E$-$04	&	1.37E$-$03	&	8.50	&	7.91	&	7.23	&	\\
&	$300$	&	3.62E$-$07	&	1.44E$-$07	&	2.89E$-$07	&	4.12	&	4.12	&	4.11	& & 
&	$15$	&	1.58E$-$04	&	8.81E$-$05	&	2.82E$-$04	&	4.46	&	4.08	&	3.90	&	\\
&	$400$	&	1.12E$-$07	&	4.47E$-$08	&	9.01E$-$08	&	4.07	&	4.07	&	4.05	& & 
&	$20$	&	4.09E$-$06	&	2.85E$-$06	&	1.18E$-$05	&	12.71	&	11.93	&	11.05	&	\\
\midrule
4		&	$100$	&	3.30E$-$06	&	1.64E$-$06	&	3.55E$-$06	&	--		&	--		&	--		&	5 & 
10		&	$ 5$	&	1.47E$-$01	&	5.17E$-$02	&	1.21E$-$01	&	--		&	--		&	--		&	11\\
&	$200$	&	1.11E$-$07	&	5.61E$-$08	&	1.26E$-$07	&	4.89	&	4.87	&	4.81	& & 
&	$10$	&	3.41E$-$04	&	1.32E$-$04	&	3.76E$-$04	&	8.75	&	8.61	&	8.33	&	\\
&	$300$	&	1.49E$-$08	&	7.52E$-$09	&	1.70E$-$08	&	4.96	&	4.96	&	4.94	& & 
&	$15$	&	3.22E$-$05	&	2.41E$-$05	&	1.00E$-$04	&	5.82	&	4.20	&	3.26	&	\\
&	$400$	&	3.55E$-$09	&	1.80E$-$09	&	4.11E$-$09	&	4.98	&	4.98	&	4.94	& & 
&	$20$	&	1.11E$-$06	&	6.75E$-$07	&	2.75E$-$06	&	11.71	&	12.43	&	12.50	&	\\
\midrule
5		&	$50$	&	5.98E$-$06	&	2.85E$-$06	&	7.62E$-$06	&	--		&	--		&	--		&	6 & 
11		&	$ 5$	&	1.58E$-$01	&	5.32E$-$02	&	1.17E$-$01	&	--		&	--		&	--		&	12\\
&	$100$	&	1.11E$-$07	&	6.34E$-$08	&	2.06E$-$07	&	5.75	&	5.49	&	5.21	& & 
&	$10$	&	1.97E$-$04	&	9.94E$-$05	&	3.49E$-$04	&	9.65	&	9.06	&	8.38	&	\\
&	$150$	&	8.05E$-$09	&	4.21E$-$09	&	1.47E$-$08	&	6.47	&	6.69	&	6.51	& & 
&	$15$	&	1.51E$-$05	&	9.66E$-$06	&	3.93E$-$05	&	6.33	&	5.75	&	5.39	&	\\
&	$200$	&	1.25E$-$09	&	6.27E$-$10	&	2.08E$-$09	&	6.48	&	6.63	&	6.81	& & 
&	$20$	&	3.60E$-$07	&	1.98E$-$07	&	7.01E$-$07	&	12.98	&	13.52	&	14.00	&	\\
\midrule
6		&	$50$	&	8.56E$-$07	&	4.26E$-$07	&	1.15E$-$06	&	--		&	--		&	--		&	7 & 
12		&	$ 5$	&	5.83E$-$02	&	2.14E$-$02	&	5.72E$-$02	&	--		&	--		&	--		&	13\\
&	$100$	&	8.26E$-$09	&	5.33E$-$09	&	2.12E$-$08	&	6.70	&	6.32	&	5.76	& & 
&	$10$	&	6.80E$-$05	&	3.96E$-$05	&	1.56E$-$04	&	9.75	&	9.07	&	8.52	&	\\
&	$150$	&	5.44E$-$10	&	3.58E$-$10	&	1.34E$-$09	&	6.71	&	6.66	&	6.81	& & 
&	$15$	&	3.70E$-$06	&	2.90E$-$06	&	1.40E$-$05	&	7.18	&	6.45	&	5.95	&	\\
&	$200$	&	7.72E$-$11	&	5.02E$-$11	&	2.06E$-$10	&	6.79	&	6.83	&	6.53	& & 
&	$20$	&	6.77E$-$08	&	4.90E$-$08	&	2.54E$-$07	&	13.90	&	14.19	&	13.92	&	\\
\bottomrule
\end{tabular}
\end{adjustbox}
\end{table*}
\begin{table*}[h!]
\centering
\small
\caption{%
Empirical orders of convergence $p$ of the ADER-DG numerical method with the LST-DG predictor, obtained based on the solution to the profile (\ref{eq:lorentz_bell}) advection-reaction problem for equation (\ref{eq:adv_eq_s_src}), calculated in functional norms $\mathcal{L}_{1}$, $\mathcal{L}_{2}$, $\mathcal{L}_{\infty}$ of errors $\epsilon$ (\ref{eq:norms_of_errors}), for degrees $N = 1, \ldots, 12$ of basis polynomials for $\mathrm{CFL} = 1.0$. Column Th. corresponds to $p = N+1$.
\label{tab:test_adveq_stiff_z1000_lrnz_1d_cfl_100}
}
\setlength{\tabcolsep}{3pt}
\begin{adjustbox}{angle=90}
\begin{tabular}{@{}|c|r|ccc|ccc|c||c|r|ccc|ccc|c|@{}}
\toprule
\multicolumn{18}{|c|}{$\lambda = 10^{3}$, $\mathrm{CFL} = 1.0$} \\
\midrule
~$N$~ & cells 
& ~~$\epsilon_{\mathcal{L}_{1}}$~~ & ~~$\epsilon_{\mathcal{L}_{2}}$~~ & ~~$\epsilon_{\mathcal{L}_{\infty}}$~~ 
& ~~$p_{\mathcal{L}_{1}}$~~ & ~$p_{\mathcal{L}_{2}}$~~ & ~~$p_{\mathcal{L}_{\infty}}$~~ & Th. &
~$N$~ & cells 
& ~~$\epsilon_{\mathcal{L}_{1}}$~~ & ~~$\epsilon_{\mathcal{L}_{2}}$~~ & ~~$\epsilon_{\mathcal{L}_{\infty}}$~~
& ~~$p_{\mathcal{L}_{1}}$~~ & ~~$p_{\mathcal{L}_{2}}$~~ & ~~$p_{\mathcal{L}_{\infty}}$~~ & Th. \\
\midrule
1		&	$100$	&	4.95E$-$02	&	1.52E$-$02	&	1.68E$-$02	&	--		&	--		&	--		&	2 & 
7		&	$50$	&	1.06E$-$06	&	7.57E$-$07	&	2.72E$-$06	&	--		&	--		&	--		&	8\\
&	$200$	&	1.12E$-$02	&	3.32E$-$03	&	3.68E$-$03	&	2.14	&	2.20	&	2.19	& & 
&	$100$	&	2.87E$-$09	&	2.32E$-$09	&	1.12E$-$08	&	8.53	&	8.35	&	7.92	&	\\
&	$300$	&	4.86E$-$03	&	1.43E$-$03	&	1.55E$-$03	&	2.06	&	2.08	&	2.13	& & 
&	$150$	&	1.22E$-$10	&	7.15E$-$11	&	3.67E$-$10	&	7.79	&	8.58	&	8.42	&	\\
&	$400$	&	2.71E$-$03	&	7.91E$-$04	&	8.52E$-$04	&	2.03	&	2.05	&	2.08	& & 
&	$200$	&	5.24E$-$11	&	1.05E$-$11	&	3.24E$-$11	&	2.93	&	6.68	&	8.44	&	\\
\midrule
2		&	$100$	&	8.81E$-$03	&	3.37E$-$03	&	4.45E$-$03	&	--		&	--		&	--		&	3 & 
8		&	$40$	&	1.98E$-$07	&	1.01E$-$07	&	3.29E$-$07	&	--		&	--		&	--		&	9\\
&	$200$	&	1.13E$-$03	&	4.43E$-$04	&	6.05E$-$04	&	2.97	&	2.93	&	2.88	& & 
&	$60$	&	6.76E$-$09	&	4.32E$-$09	&	1.80E$-$08	&	8.32	&	7.76	&	7.17	&	\\
&	$300$	&	3.35E$-$04	&	1.32E$-$04	&	1.81E$-$04	&	2.99	&	2.98	&	2.98	& & 
&	$80$	&	9.54E$-$10	&	6.97E$-$10	&	3.37E$-$09	&	6.81	&	6.34	&	5.82	&	\\
&	$400$	&	1.41E$-$04	&	5.60E$-$05	&	7.72E$-$05	&	3.00	&	2.99	&	2.96	& & 
&	$100$	&	2.32E$-$10	&	1.30E$-$10	&	6.18E$-$10	&	6.34	&	7.52	&	7.60	&	\\
\midrule
3		&	$100$	&	1.98E$-$04	&	9.92E$-$05	&	2.10E$-$04	&	--		&	--		&	--		&	4 & 
9		&	$10$	&	3.28E$-$03	&	1.09E$-$03	&	6.52E$-$04	&	--		&	--		&	--		&	10\\
&	$200$	&	8.57E$-$06	&	3.80E$-$06	&	8.33E$-$06	&	4.53	&	4.70	&	4.66	& & 
&	$20$	&	2.35E$-$05	&	1.18E$-$05	&	3.39E$-$05	&	7.13	&	6.52	&	4.27	&	\\
&	$300$	&	1.46E$-$06	&	5.97E$-$07	&	1.24E$-$06	&	4.36	&	4.57	&	4.69	& & 
&	$30$	&	2.91E$-$07	&	1.29E$-$07	&	3.64E$-$07	&	10.83	&	11.14	&	11.18	&	\\
&	$400$	&	4.28E$-$07	&	1.67E$-$07	&	3.17E$-$07	&	4.27	&	4.42	&	4.74	& & 
&	$40$	&	4.52E$-$08	&	3.60E$-$08	&	1.60E$-$07	&	6.48	&	4.44	&	2.87	&	\\
\midrule
4		&	$100$	&	2.66E$-$05	&	1.46E$-$05	&	3.91E$-$05	&	--		&	--		&	--		&	5 & 
10		&	$10$	&	1.56E$-$03	&	5.26E$-$04	&	7.13E$-$04	&	--		&	--		&	--		&	11\\
&	$200$	&	8.91E$-$07	&	5.18E$-$07	&	1.49E$-$06	&	4.90	&	4.81	&	4.71	& & 
&	$20$	&	3.26E$-$06	&	1.37E$-$06	&	3.38E$-$06	&	8.90	&	8.59	&	7.72	&	\\
&	$300$	&	1.19E$-$07	&	6.99E$-$08	&	2.03E$-$07	&	4.97	&	4.94	&	4.92	& & 
&	$30$	&	8.39E$-$08	&	4.68E$-$08	&	1.61E$-$07	&	9.03	&	8.32	&	7.51	&	\\
&	$400$	&	2.82E$-$08	&	1.67E$-$08	&	4.87E$-$08	&	4.99	&	4.98	&	4.97	& & 
&	$40$	&	3.98E$-$09	&	2.38E$-$09	&	8.10E$-$09	&	10.60	&	10.36	&	10.38	&	\\
\midrule
5		&	$50$	&	8.00E$-$05	&	5.29E$-$05	&	1.51E$-$04	&	--		&	--		&	--		&	6 & 
11		&	$10$	&	6.97E$-$04	&	2.29E$-$04	&	3.53E$-$04	&	--		&	--		&	--		&	12\\
&	$100$	&	8.25E$-$07	&	5.41E$-$07	&	1.67E$-$06	&	6.60	&	6.61	&	6.50	& & 
&	$20$	&	1.23E$-$06	&	6.82E$-$07	&	2.46E$-$06	&	9.15	&	8.39	&	7.17	&	\\
&	$150$	&	5.33E$-$08	&	3.53E$-$08	&	1.38E$-$07	&	6.76	&	6.73	&	6.15	& & 
&	$30$	&	1.22E$-$08	&	7.35E$-$09	&	2.97E$-$08	&	11.38	&	11.17	&	10.89	&	\\
&	$200$	&	7.81E$-$09	&	4.97E$-$09	&	1.93E$-$08	&	6.68	&	6.82	&	6.84	& & 
&	$40$	&	5.86E$-$10	&	2.57E$-$10	&	1.34E$-$09	&	10.54	&	11.66	&	10.78	&	\\
\midrule
6		&	$50$	&	3.93E$-$06	&	2.26E$-$06	&	6.88E$-$06	&	--		&	--		&	--		&	7 & 
12		&	$10$	&	2.19E$-$04	&	7.84E$-$05	&	5.17E$-$05	&	--		&	--		&	--		&	13\\
&	$100$	&	6.96E$-$08	&	4.73E$-$08	&	1.56E$-$07	&	5.82	&	5.58	&	5.46	& & 
&	$20$	&	1.62E$-$07	&	6.56E$-$08	&	1.57E$-$07	&	10.40	&	10.22	&	8.36	&	\\
&	$150$	&	4.62E$-$09	&	3.33E$-$09	&	1.50E$-$08	&	6.69	&	6.55	&	5.78	& & 
&	$30$	&	2.17E$-$09	&	1.14E$-$09	&	4.97E$-$09	&	10.64	&	9.99	&	8.51	&	\\
&	$200$	&	6.56E$-$10	&	4.74E$-$10	&	2.11E$-$09	&	6.78	&	6.78	&	6.82	& & 
&	$40$	&	3.35E$-$10	&	7.25E$-$11	&	2.32E$-$10	&	6.49	&	9.58	&	10.66	&	\\
\bottomrule
\end{tabular}
\end{adjustbox}
\end{table*}
\begin{table*}[h!]
\centering
\small
\caption{%
Empirical orders of convergence $p$ of the ADER-DG numerical method with the LST-DG predictor, obtained based on the solution to the profile (\ref{eq:lorentz_bell}) advection-reaction problem for equation (\ref{eq:adv_eq_s_src}), calculated in functional norms $\mathcal{L}_{1}$, $\mathcal{L}_{2}$, $\mathcal{L}_{\infty}$ of errors $\epsilon$ (\ref{eq:norms_of_errors}), for degrees $N = 1, \ldots, 12$ of basis polynomials for $\mathrm{CFL} = 10.0$. Column Th. corresponds to $p = N+1$.
\label{tab:test_adveq_stiff_z1000_lrnz_1d_cfl_1000}
}
\setlength{\tabcolsep}{3pt}
\begin{adjustbox}{angle=90}
\begin{tabular}{@{}|c|r|ccc|ccc|c||c|r|ccc|ccc|c|@{}}
\toprule
\multicolumn{18}{|c|}{$\lambda = 10^{3}$, $\mathrm{CFL} = 10.0$} \\
\midrule
~$N$~ & cells 
& ~~$\epsilon_{\mathcal{L}_{1}}$~~ & ~~$\epsilon_{\mathcal{L}_{2}}$~~ & ~~$\epsilon_{\mathcal{L}_{\infty}}$~~ 
& ~~$p_{\mathcal{L}_{1}}$~~ & ~$p_{\mathcal{L}_{2}}$~~ & ~~$p_{\mathcal{L}_{\infty}}$~~ & Th. &
~$N$~ & cells 
& ~~$\epsilon_{\mathcal{L}_{1}}$~~ & ~~$\epsilon_{\mathcal{L}_{2}}$~~ & ~~$\epsilon_{\mathcal{L}_{\infty}}$~~
& ~~$p_{\mathcal{L}_{1}}$~~ & ~~$p_{\mathcal{L}_{2}}$~~ & ~~$p_{\mathcal{L}_{\infty}}$~~ & Th. \\
\midrule
1		&	$200$	&	1.02E$+$00	&	2.61E$-$01	&	1.46E$-$01	&	--		&	--		&	--		&	2 & 
7		&	$150$	&	1.17E$-$03	&	4.51E$-$04	&	3.20E$-$04	&	--		&	--		&	--		&	8\\
&	$400$	&	2.67E$-$01	&	6.98E$-$02	&	4.08E$-$02	&	1.94	&	1.90	&	1.84	& & 
&	$300$	&	7.36E$-$06	&	3.05E$-$06	&	2.36E$-$06	&	7.32	&	7.21	&	7.08	&	\\
&	$600$	&	1.19E$-$01	&	3.14E$-$02	&	1.84E$-$02	&	1.98	&	1.97	&	1.96	& & 
&	$450$	&	3.17E$-$07	&	1.33E$-$07	&	1.05E$-$07	&	7.76	&	7.72	&	7.67	&	\\
&	$800$	&	6.73E$-$02	&	1.77E$-$02	&	1.04E$-$02	&	1.99	&	1.99	&	1.98	& & 
&	$600$	&	3.29E$-$08	&	1.39E$-$08	&	1.11E$-$08	&	7.87	&	7.86	&	7.84	&	\\
\midrule
2		&	$200$	&	2.20E$-$01	&	6.61E$-$02	&	3.25E$-$02	&	--		&	--		&	--		&	3 & 
8		&	$150$	&	3.54E$-$04	&	1.40E$-$04	&	9.39E$-$05	&	--		&	--		&	--		&	9\\
&	$400$	&	2.98E$-$02	&	9.23E$-$03	&	4.75E$-$03	&	2.89	&	2.84	&	2.78	& & 
&	$300$	&	1.19E$-$06	&	5.08E$-$07	&	3.65E$-$07	&	8.22	&	8.11	&	8.01	&	\\
&	$600$	&	8.95E$-$03	&	2.80E$-$03	&	1.45E$-$03	&	2.96	&	2.95	&	2.92	& & 
&	$450$	&	3.46E$-$08	&	1.50E$-$08	&	1.10E$-$08	&	8.72	&	8.68	&	8.64	&	\\
&	$800$	&	3.80E$-$03	&	1.19E$-$03	&	6.20E$-$04	&	2.98	&	2.97	&	2.96	& & 
&	$600$	&	2.80E$-$09	&	1.18E$-$09	&	8.84E$-$10	&	8.74	&	8.84	&	8.76	&	\\
\midrule
3		&	$200$	&	5.12E$-$02	&	1.68E$-$02	&	1.07E$-$02	&	--		&	--		&	--		&	4 & 
9		&	$100$	&	2.72E$-$03	&	1.00E$-$03	&	6.68E$-$04	&	--		&	--		&	--		&	10\\
&	$400$	&	3.62E$-$03	&	1.23E$-$03	&	8.27E$-$04	&	3.82	&	3.77	&	3.70	& & 
&	$200$	&	8.47E$-$06	&	3.60E$-$06	&	2.83E$-$06	&	8.33	&	8.12	&	7.88	&	\\
&	$600$	&	7.33E$-$04	&	2.52E$-$04	&	1.70E$-$04	&	3.94	&	3.92	&	3.90	& & 
&	$300$	&	1.92E$-$07	&	8.45E$-$08	&	6.92E$-$08	&	9.34	&	9.25	&	9.15	&	\\
&	$800$	&	2.34E$-$04	&	8.05E$-$05	&	5.47E$-$05	&	3.97	&	3.96	&	3.95	& & 
&	$400$	&	1.21E$-$08	&	5.33E$-$09	&	4.43E$-$09	&	9.62	&	9.61	&	9.56	&	\\
\midrule
4		&	$200$	&	1.19E$-$02	&	4.18E$-$03	&	2.40E$-$03	&	--		&	--		&	--		&	5 & 
10		&	$50$	&	1.21E$-$01	&	3.46E$-$02	&	1.70E$-$02	&	--		&	--		&	--		&	11\\
&	$400$	&	4.38E$-$04	&	1.60E$-$04	&	9.50E$-$05	&	4.77	&	4.71	&	4.66	& & 
&	$100$	&	1.13E$-$03	&	4.25E$-$04	&	2.76E$-$04	&	6.75	&	6.34	&	5.95	&	\\
&	$600$	&	5.95E$-$05	&	2.19E$-$05	&	1.31E$-$05	&	4.92	&	4.90	&	4.88	& & 
&	$150$	&	3.23E$-$05	&	1.35E$-$05	&	9.58E$-$06	&	8.76	&	8.51	&	8.28	&	\\
&	$800$	&	1.43E$-$05	&	5.28E$-$06	&	3.17E$-$06	&	4.96	&	4.95	&	4.94	& & 
&	$200$	&	1.99E$-$06	&	8.70E$-$07	&	6.44E$-$07	&	9.68	&	9.53	&	9.38	&	\\
\midrule
5		&	$150$	&	1.29E$-$02	&	4.61E$-$03	&	3.06E$-$03	&	--		&	--		&	--		&	6 & 
11		&	$50$	&	7.50E$-$02	&	2.17E$-$02	&	1.08E$-$02	&	--		&	--		&	--		&	12\\
&	$300$	&	2.84E$-$04	&	1.08E$-$04	&	7.84E$-$05	&	5.51	&	5.41	&	5.29	& & 
&	$100$	&	4.66E$-$04	&	1.80E$-$04	&	1.25E$-$04	&	7.33	&	6.92	&	6.44	&	\\
&	$450$	&	2.67E$-$05	&	1.03E$-$05	&	7.64E$-$06	&	5.83	&	5.79	&	5.74	& & 
&	$150$	&	9.76E$-$06	&	4.16E$-$06	&	3.25E$-$06	&	9.53	&	9.28	&	9.00	&	\\
&	$600$	&	4.87E$-$06	&	1.89E$-$06	&	1.41E$-$06	&	5.91	&	5.90	&	5.87	& & 
&	$200$	&	4.70E$-$07	&	2.10E$-$07	&	1.72E$-$07	&	10.55	&	10.39	&	10.21	&	\\
\midrule
6		&	$150$	&	3.90E$-$03	&	1.44E$-$03	&	8.95E$-$04	&	--		&	--		&	--		&	7 & 
12		&	$50$	&	4.64E$-$02	&	1.37E$-$02	&	6.97E$-$03	&	--		&	--		&	--		&	13\\
&	$300$	&	4.56E$-$05	&	1.82E$-$05	&	1.20E$-$05	&	6.42	&	6.31	&	6.22	& & 
&	$100$	&	1.93E$-$04	&	7.58E$-$05	&	5.14E$-$05	&	7.91	&	7.49	&	7.08	&	\\
&	$450$	&	2.90E$-$06	&	1.18E$-$06	&	7.88E$-$07	&	6.79	&	6.76	&	6.72	& & 
&	$150$	&	2.95E$-$06	&	1.28E$-$06	&	9.59E$-$07	&	10.31	&	10.06	&	9.82	&	\\
&	$600$	&	3.99E$-$07	&	1.63E$-$07	&	1.10E$-$07	&	6.90	&	6.88	&	6.86	& & 
&	$200$	&	1.11E$-$07	&	5.04E$-$08	&	3.93E$-$08	&	11.41	&	11.26	&	11.11	&	\\
\bottomrule
\end{tabular}
\end{adjustbox}
\end{table*}

\begin{figure}[h!]
\captionsetup[subfigure]{%
	position=bottom,
	font+=smaller,
	textfont=normalfont,
	singlelinecheck=off,
	justification=raggedright
}
\centering
\begin{subfigure}{0.24\textwidth}
\includegraphics[width=\textwidth]{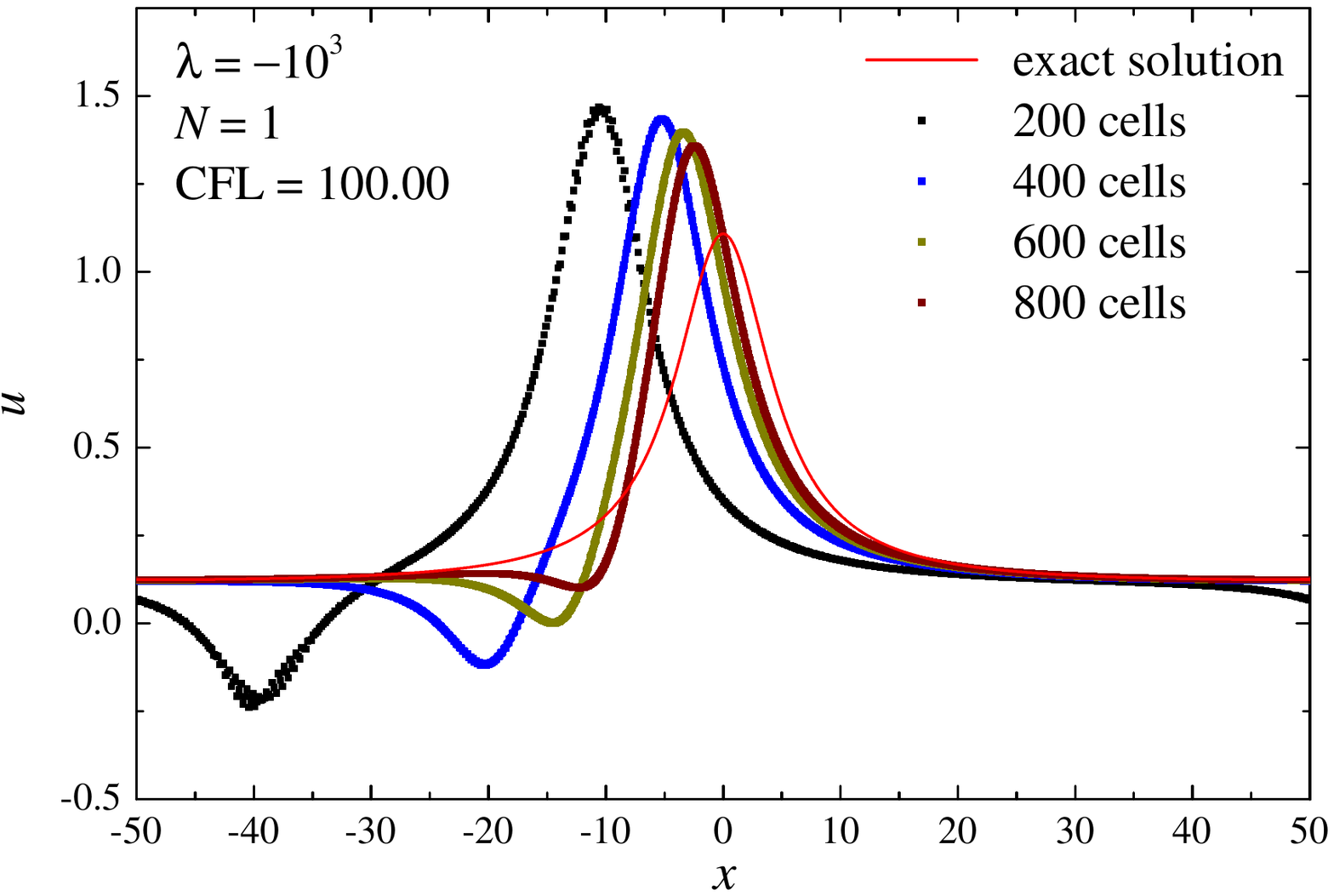}
\vspace{-8mm}\caption{\label{fig:test_adveq_stiff_z1000_lrnz_1d_cfl_10000:1}}
\end{subfigure}
\begin{subfigure}{0.24\textwidth}
\includegraphics[width=\textwidth]{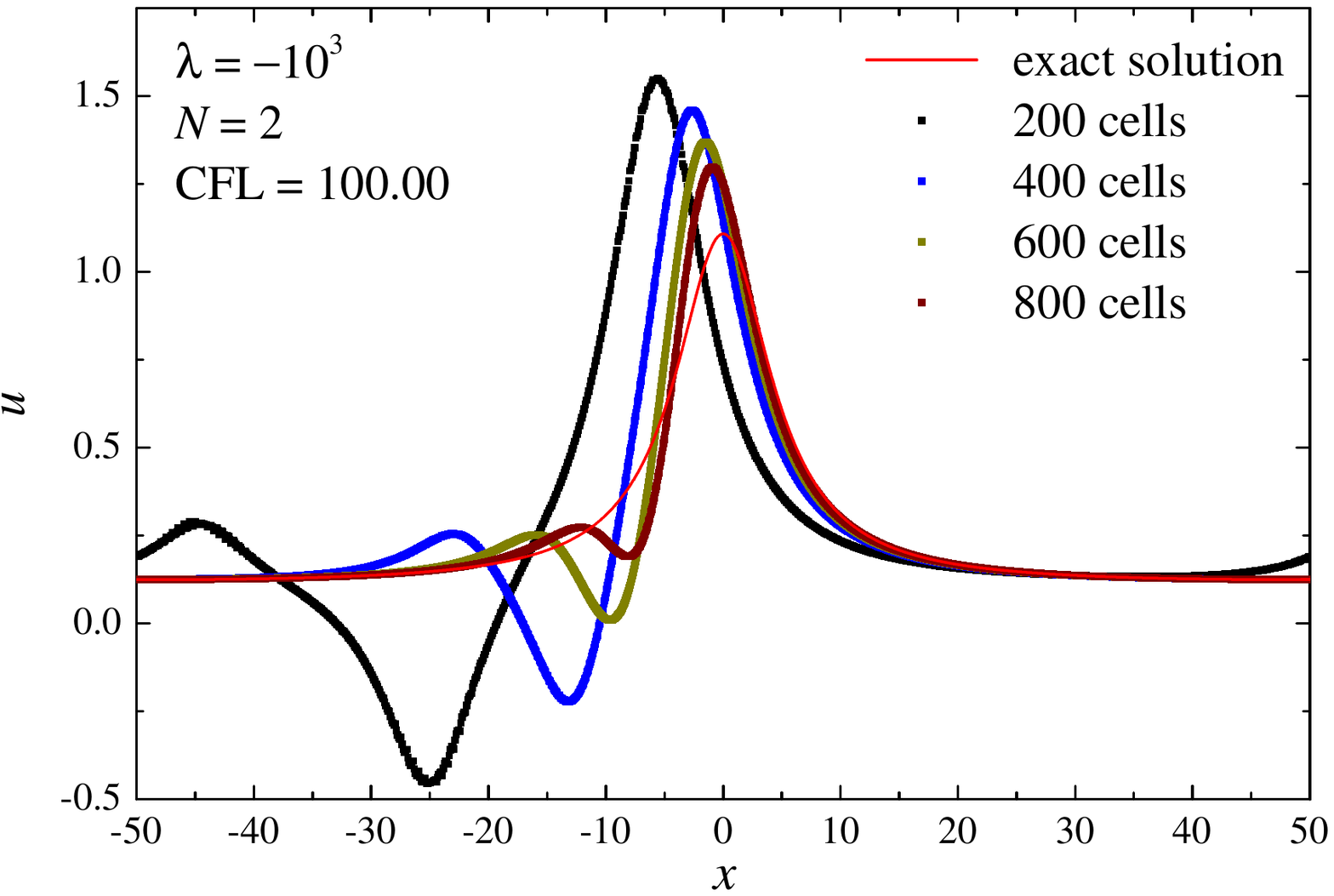}
\vspace{-8mm}\caption{\label{fig:test_adveq_stiff_z1000_lrnz_1d_cfl_10000:2}}
\end{subfigure}
\begin{subfigure}{0.24\textwidth}
\includegraphics[width=\textwidth]{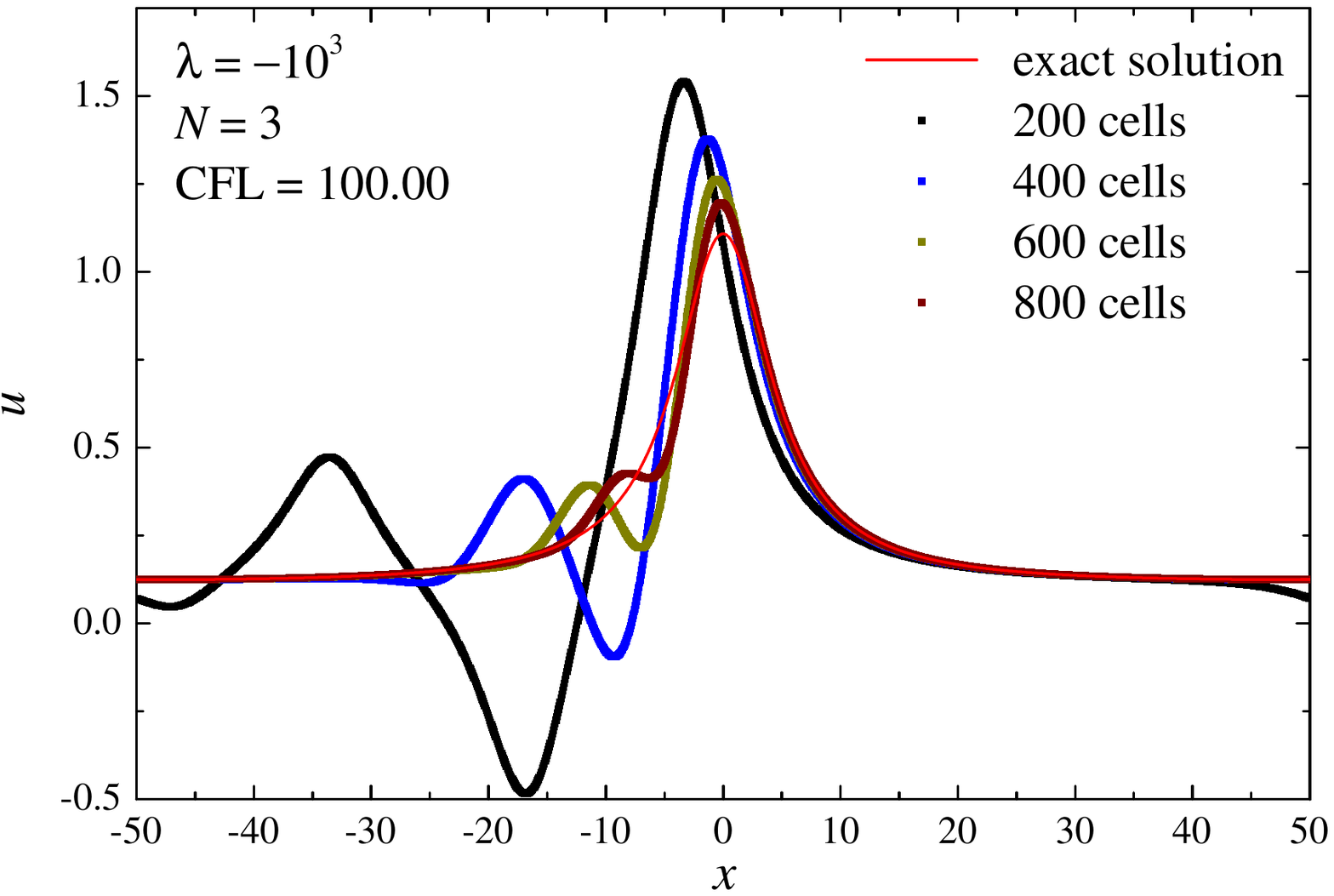}
\vspace{-8mm}\caption{\label{fig:test_adveq_stiff_z1000_lrnz_1d_cfl_10000:3}}
\end{subfigure}
\begin{subfigure}{0.24\textwidth}
\includegraphics[width=\textwidth]{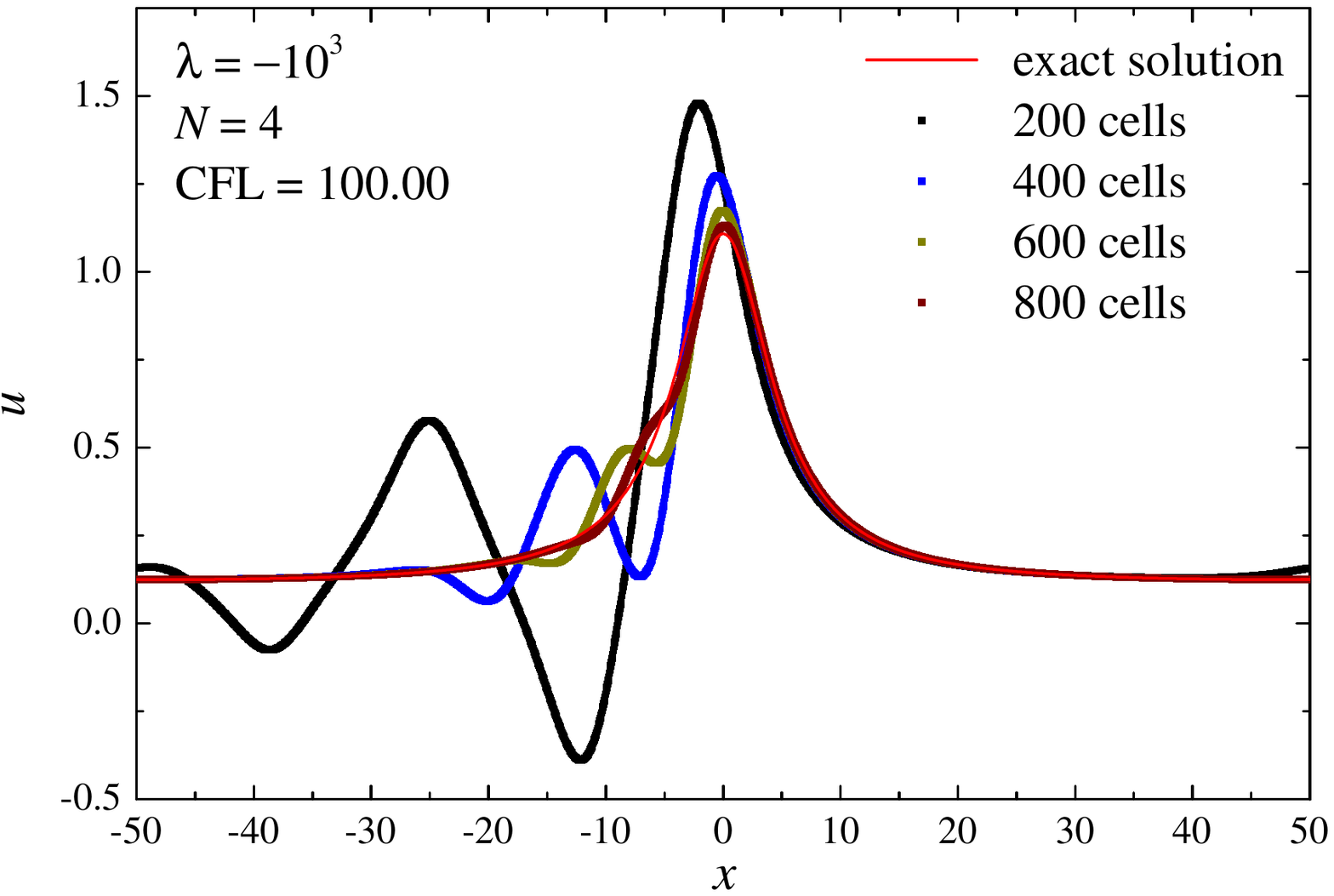}
\vspace{-8mm}\caption{\label{fig:test_adveq_stiff_z1000_lrnz_1d_cfl_10000:4}}
\end{subfigure}\\[-2.5mm]
\begin{subfigure}{0.24\textwidth}
\includegraphics[width=\textwidth]{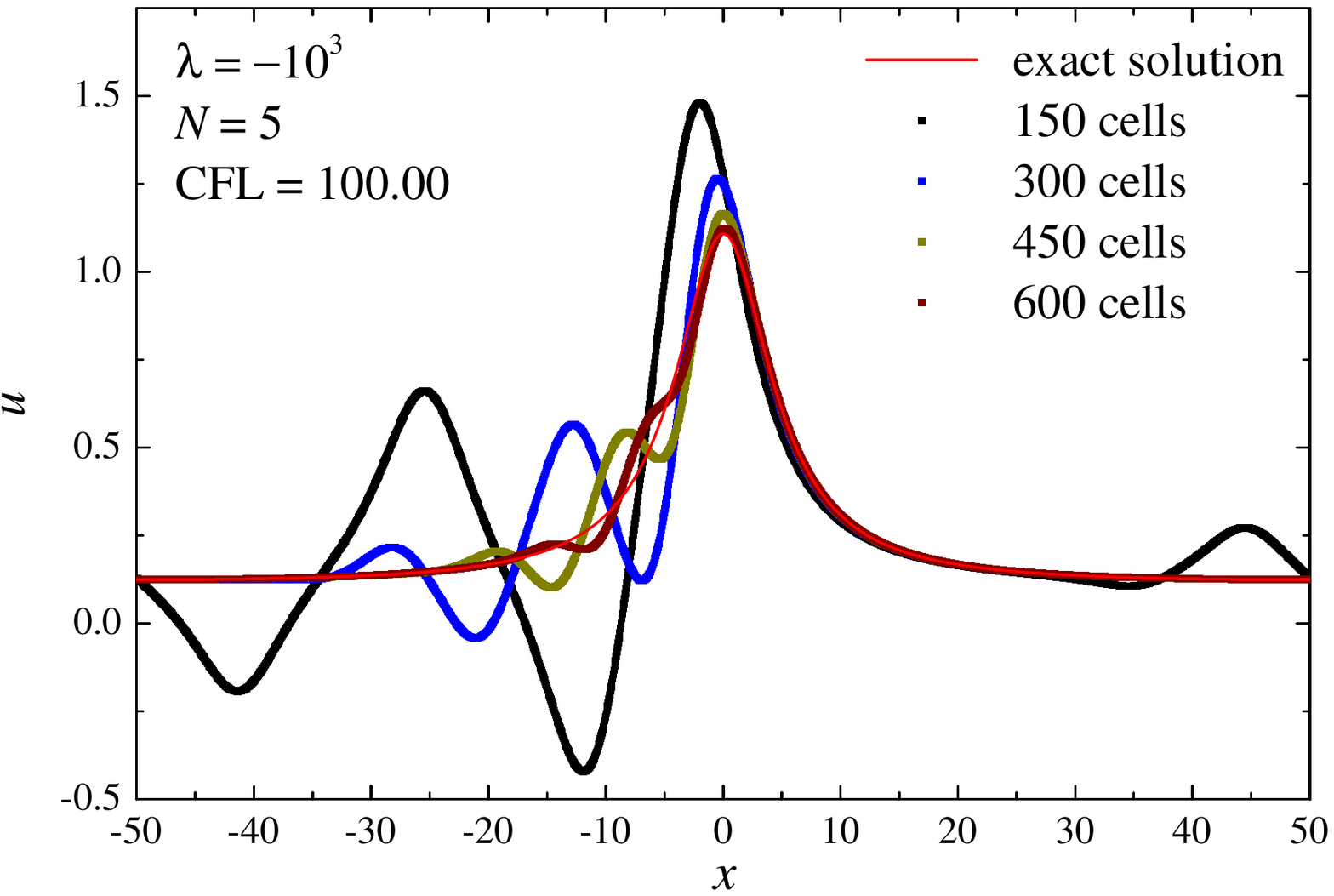}
\vspace{-8mm}\caption{\label{fig:test_adveq_stiff_z1000_lrnz_1d_cfl_10000:5}}
\end{subfigure}
\begin{subfigure}{0.24\textwidth}
\includegraphics[width=\textwidth]{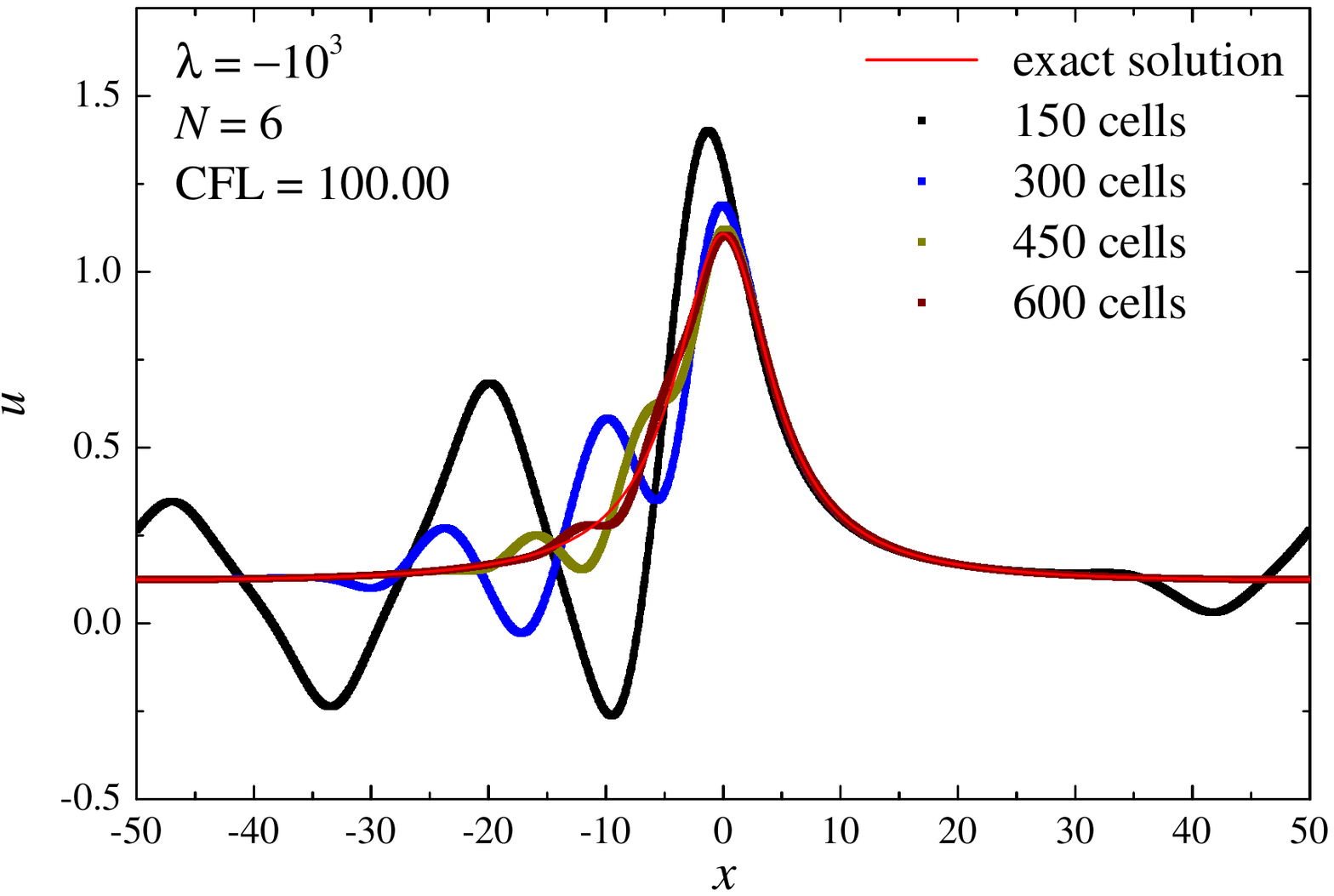}
\vspace{-8mm}\caption{\label{fig:test_adveq_stiff_z1000_lrnz_1d_cfl_10000:6}}
\end{subfigure}
\begin{subfigure}{0.24\textwidth}
\includegraphics[width=\textwidth]{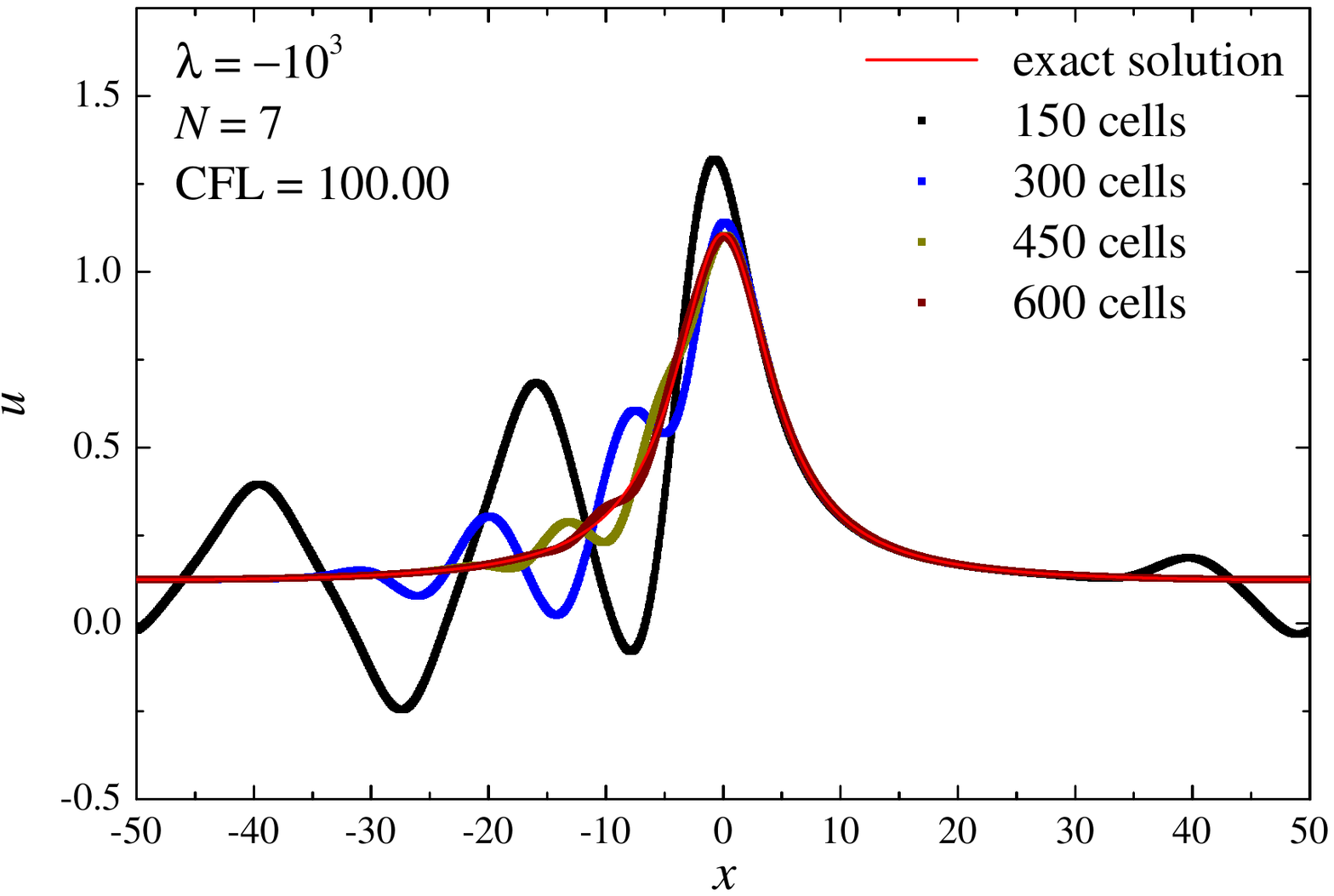}
\vspace{-8mm}\caption{\label{fig:test_adveq_stiff_z1000_lrnz_1d_cfl_10000:7}}
\end{subfigure}
\begin{subfigure}{0.24\textwidth}
\includegraphics[width=\textwidth]{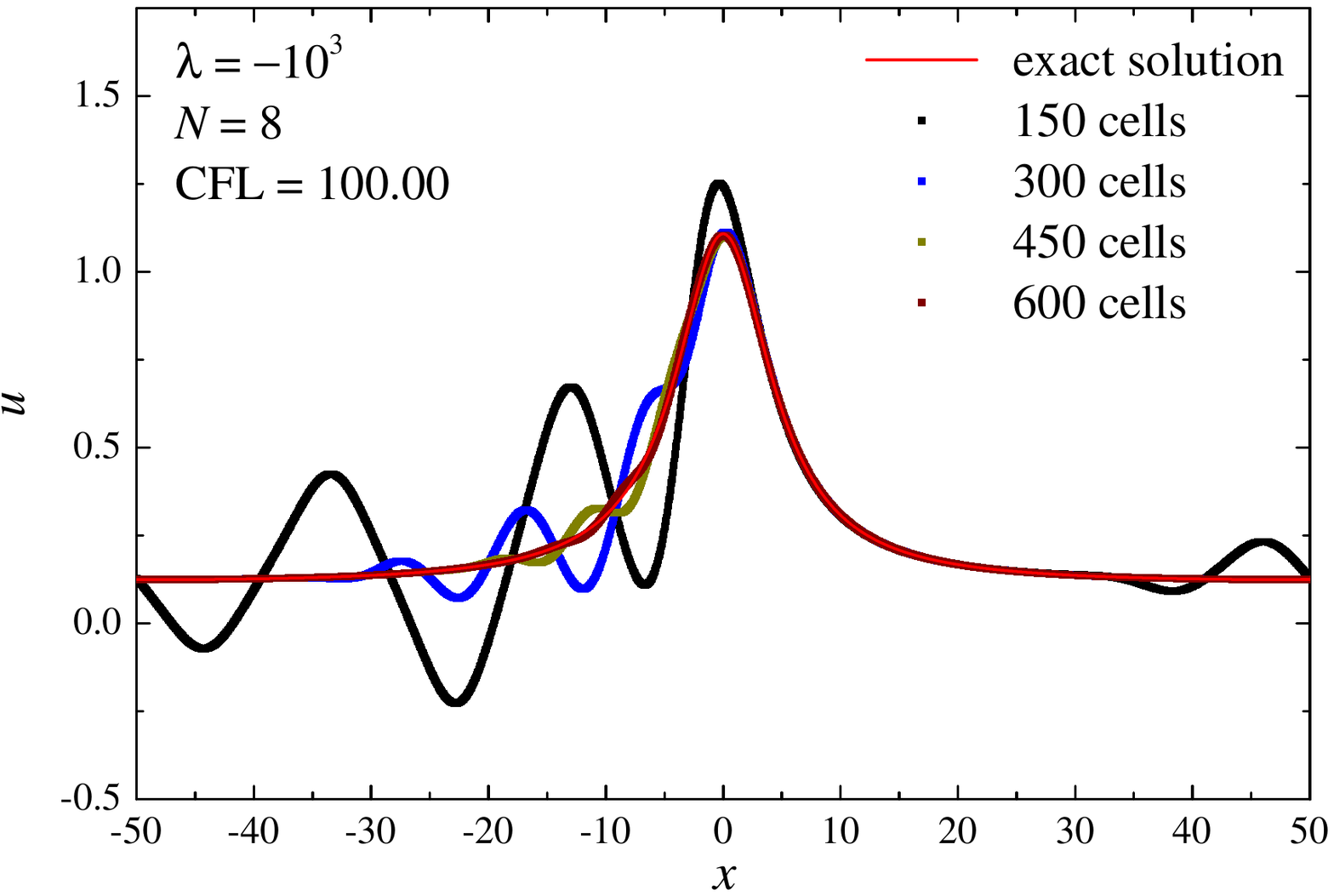}
\vspace{-8mm}\caption{\label{fig:test_adveq_stiff_z1000_lrnz_1d_cfl_10000:8}}
\end{subfigure}\\[-2.5mm]
\begin{subfigure}{0.24\textwidth}
\includegraphics[width=\textwidth]{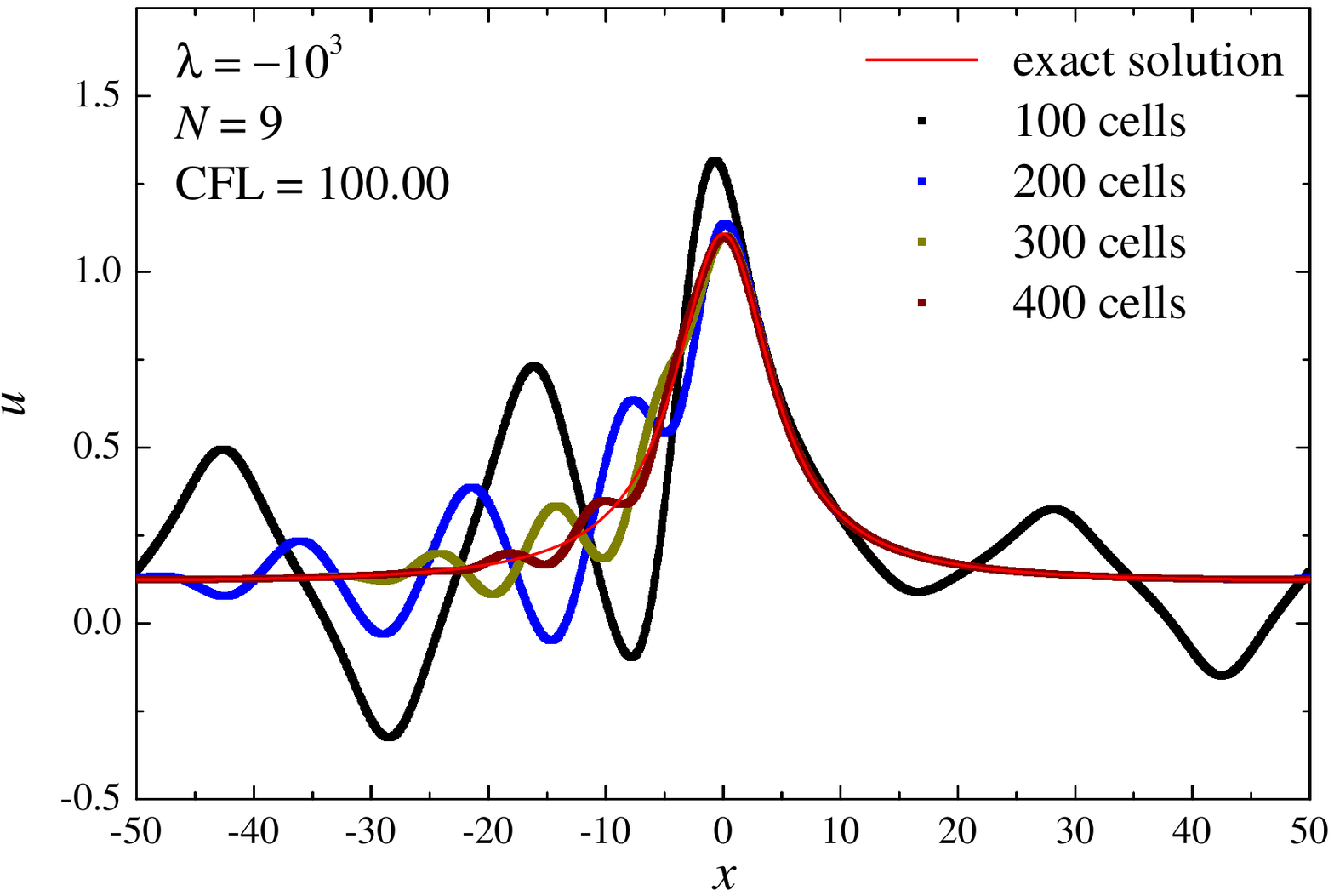}
\vspace{-8mm}\caption{\label{fig:test_adveq_stiff_z1000_lrnz_1d_cfl_10000:9}}
\end{subfigure}
\begin{subfigure}{0.24\textwidth}
\includegraphics[width=\textwidth]{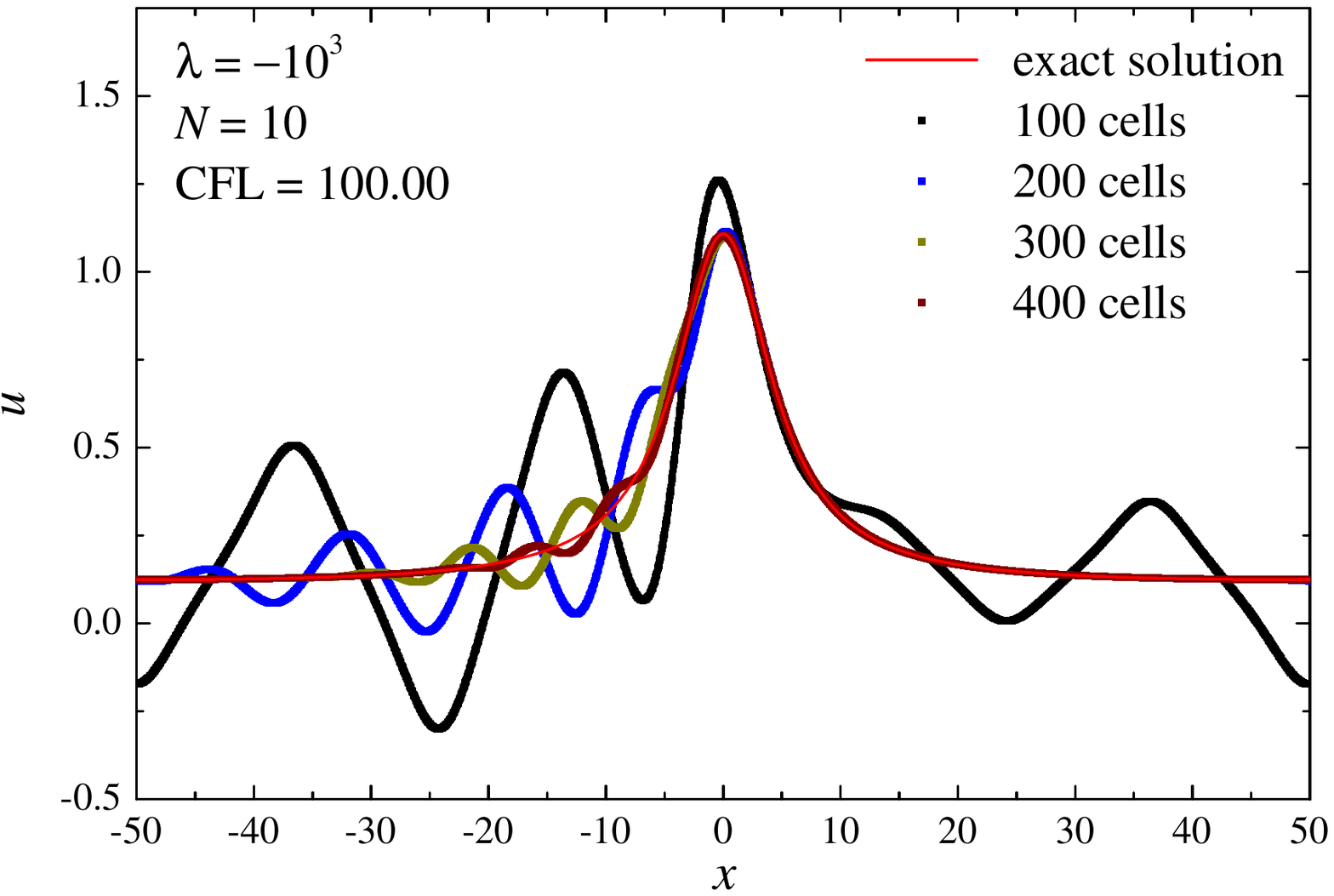}
\vspace{-8mm}\caption{\label{fig:test_adveq_stiff_z1000_lrnz_1d_cfl_10000:10}}
\end{subfigure}
\begin{subfigure}{0.24\textwidth}
\includegraphics[width=\textwidth]{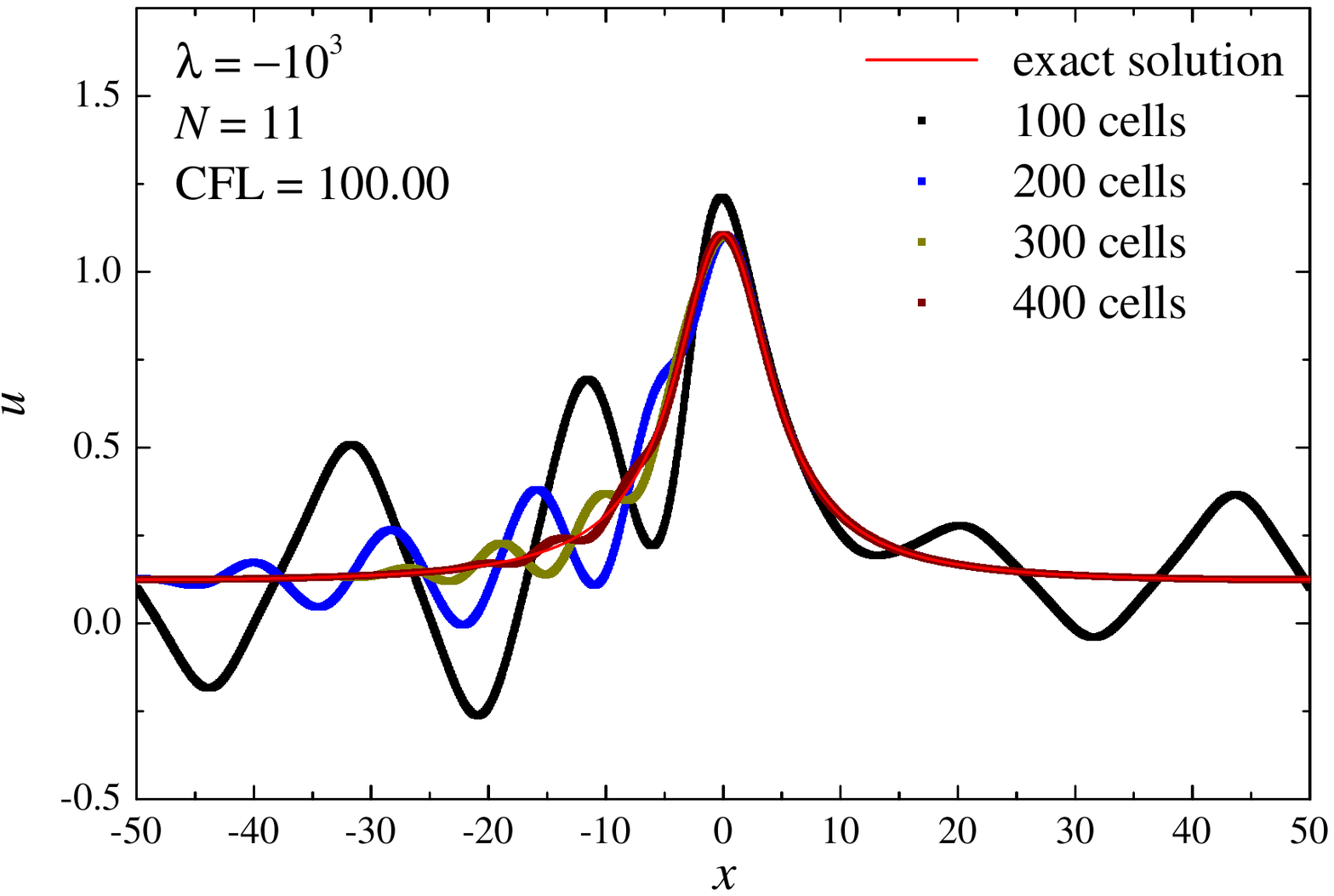}
\vspace{-8mm}\caption{\label{fig:test_adveq_stiff_z1000_lrnz_1d_cfl_10000:11}}
\end{subfigure}
\begin{subfigure}{0.24\textwidth}
\includegraphics[width=\textwidth]{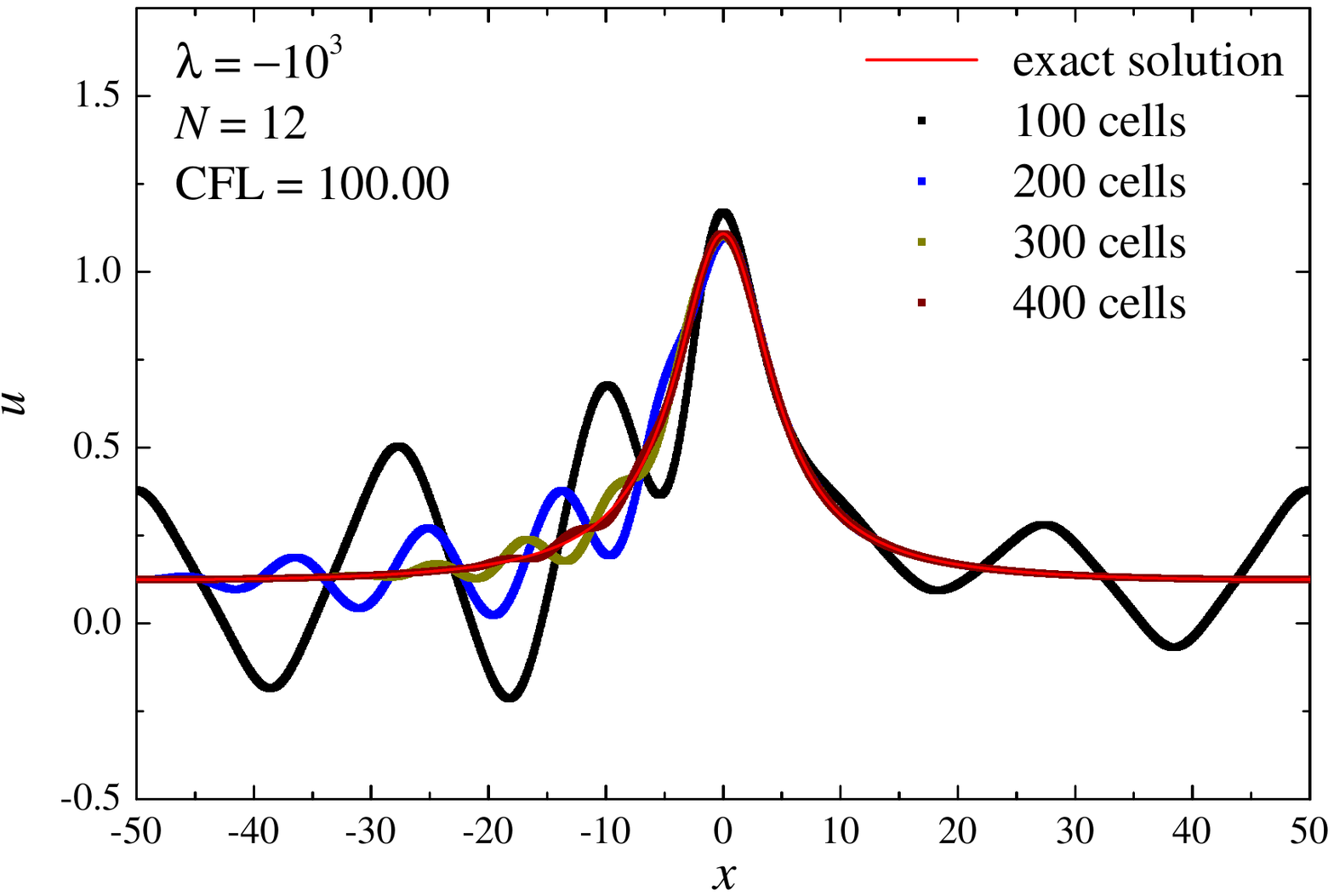}
\vspace{-8mm}\caption{\label{fig:test_adveq_stiff_z1000_lrnz_1d_cfl_10000:12}}
\end{subfigure}
\caption{%
Coordinate dependencies of the solution $u(x, t_{f})$ to the linear advection-reaction equation (\ref{eq:adv_eq_src}) at the final time $t_{f}$, obtained by the ADER-DG numerical method with the LST-DG predictor, for the degrees $N = 1, \ldots, 12$ of the basis polynomials for the Courant number $\mathrm{CFL} = 100.0$ and the decay parameter $\nu = -10^{3}$. The red line represents the exact analytical solution.
}
\label{fig:test_adveq_stiff_z1000_lrnz_1d_cfl_10000}
\end{figure}
\begin{figure}[h!]
\captionsetup[subfigure]{%
	position=bottom,
	font+=smaller,
	textfont=normalfont,
	singlelinecheck=off,
	justification=raggedright
}
\centering
\begin{subfigure}{0.24\textwidth}
\includegraphics[width=\textwidth]{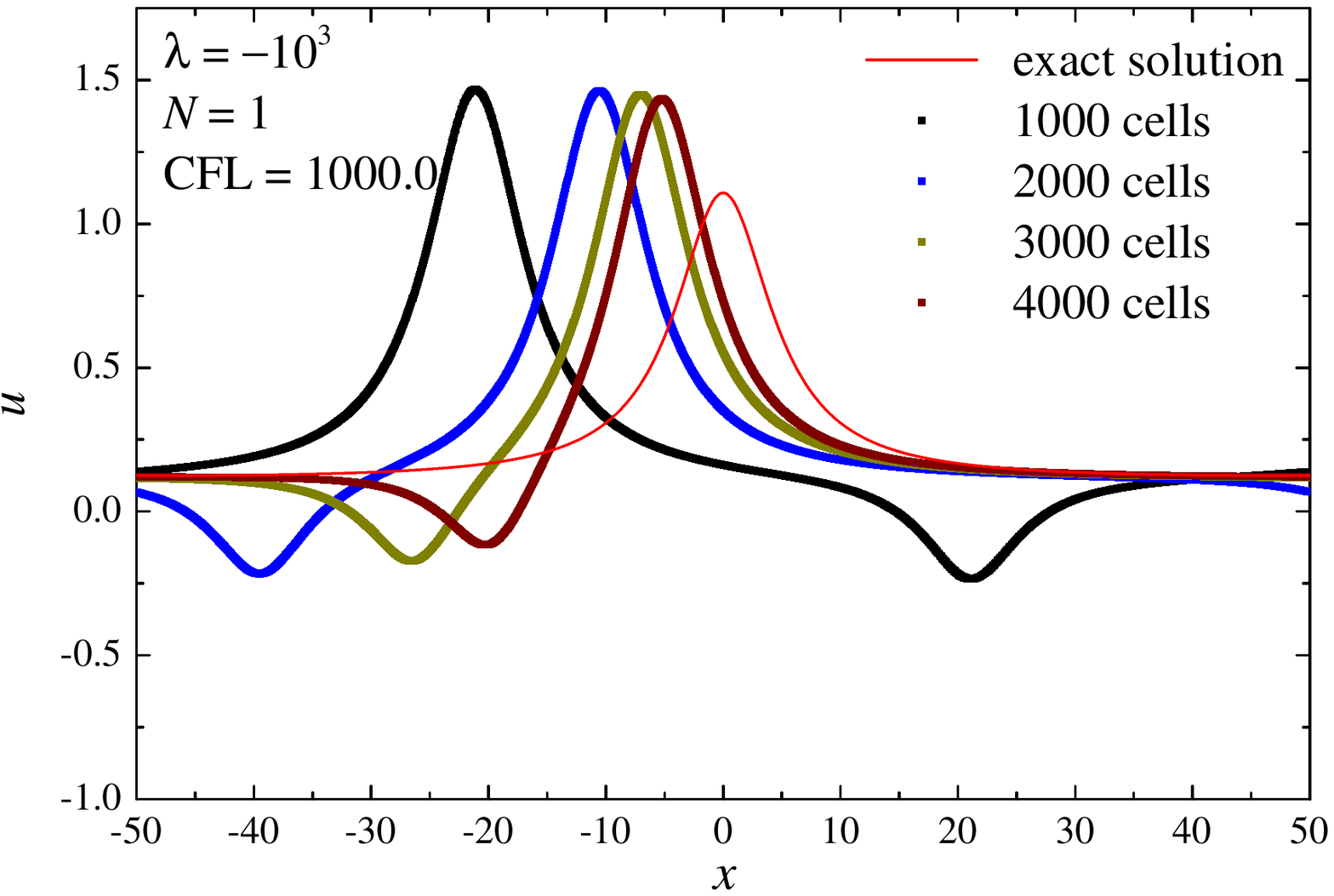}
\vspace{-8mm}\caption{\label{fig:test_adveq_stiff_z1000_lrnz_1d_cfl_100000:1}}
\end{subfigure}
\begin{subfigure}{0.24\textwidth}
\includegraphics[width=\textwidth]{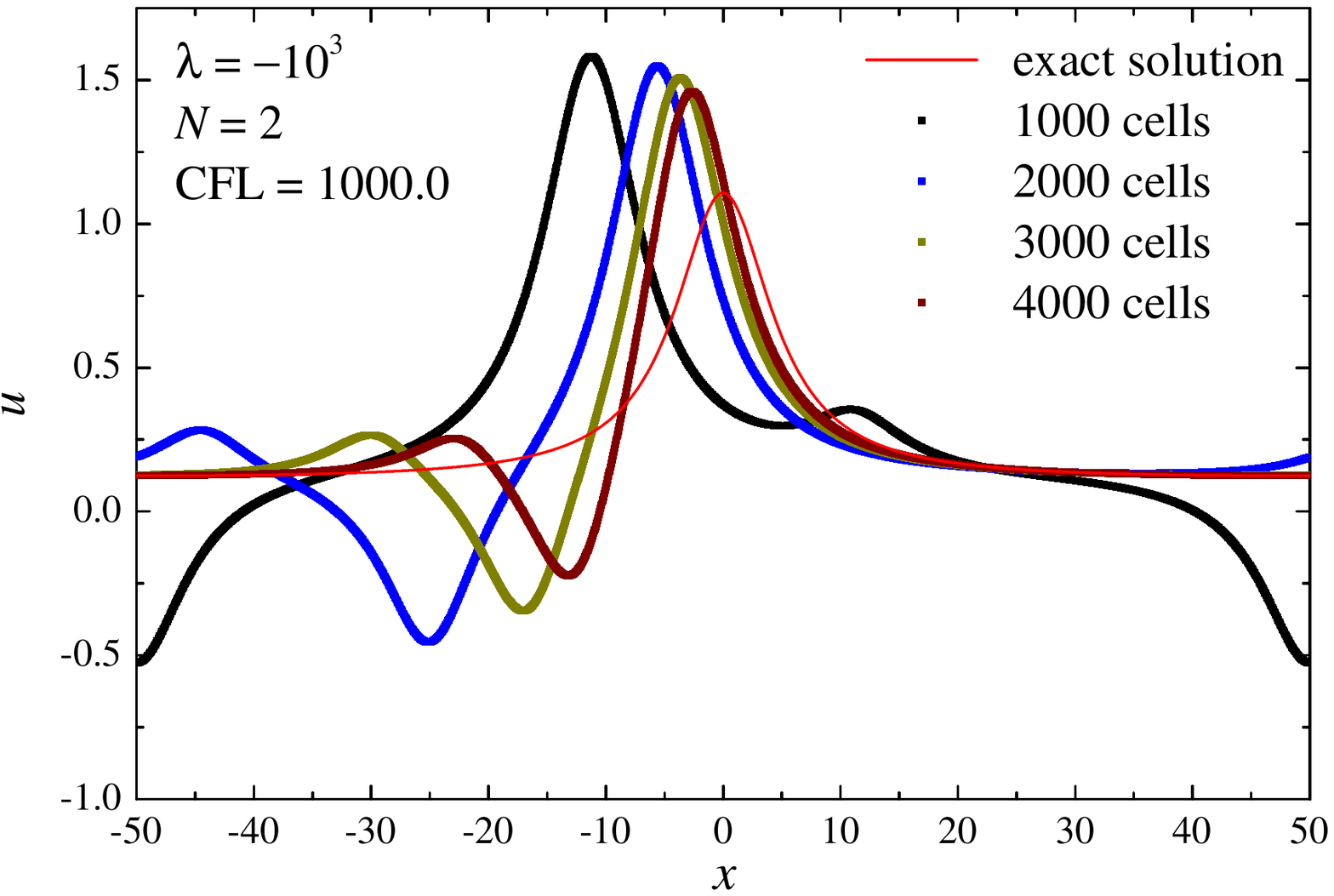}
\vspace{-8mm}\caption{\label{fig:test_adveq_stiff_z1000_lrnz_1d_cfl_100000:2}}
\end{subfigure}
\begin{subfigure}{0.24\textwidth}
\includegraphics[width=\textwidth]{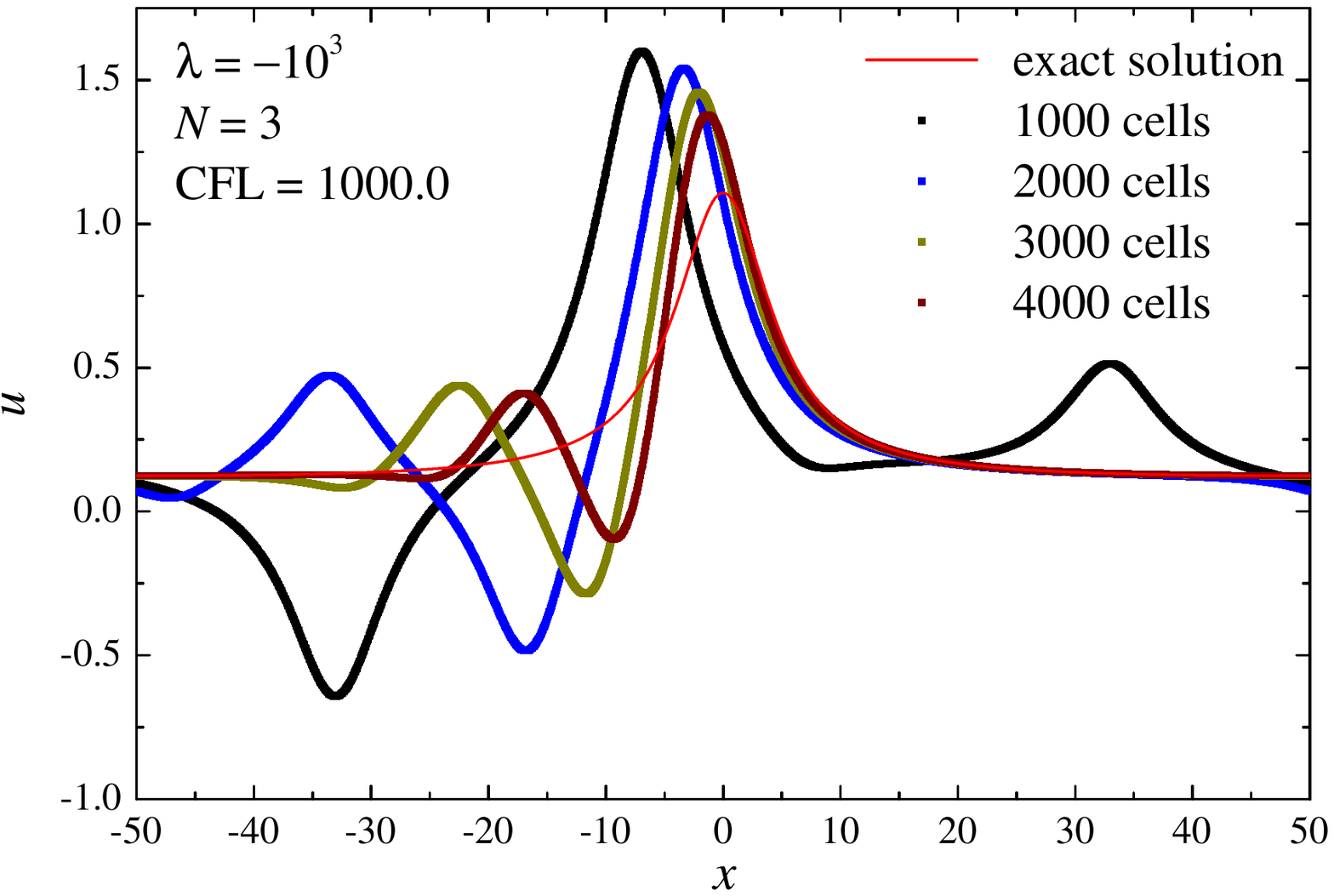}
\vspace{-8mm}\caption{\label{fig:test_adveq_stiff_z1000_lrnz_1d_cfl_100000:3}}
\end{subfigure}
\begin{subfigure}{0.24\textwidth}
\includegraphics[width=\textwidth]{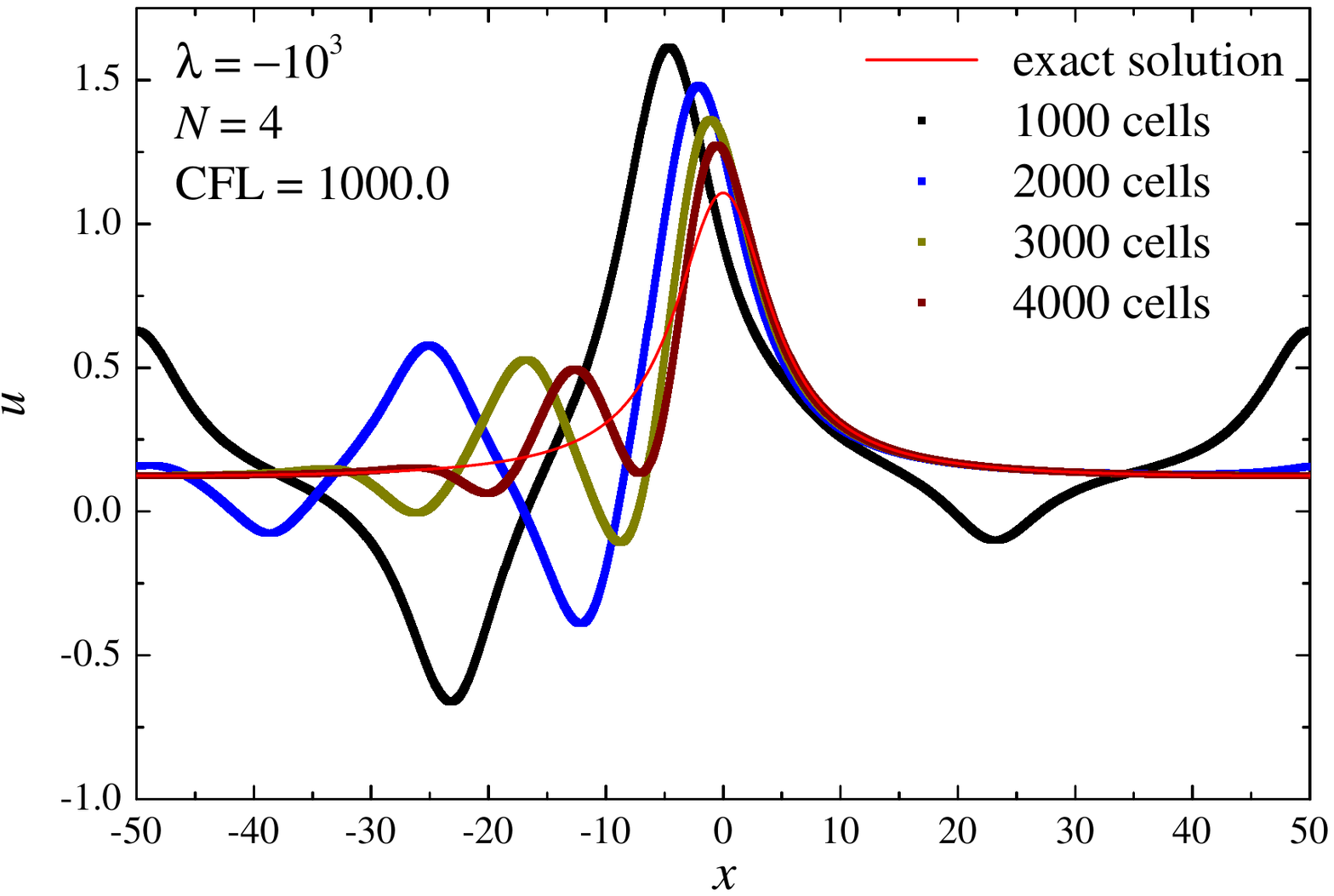}
\vspace{-8mm}\caption{\label{fig:test_adveq_stiff_z1000_lrnz_1d_cfl_100000:4}}
\end{subfigure}\\[-2.5mm]
\begin{subfigure}{0.24\textwidth}
\includegraphics[width=\textwidth]{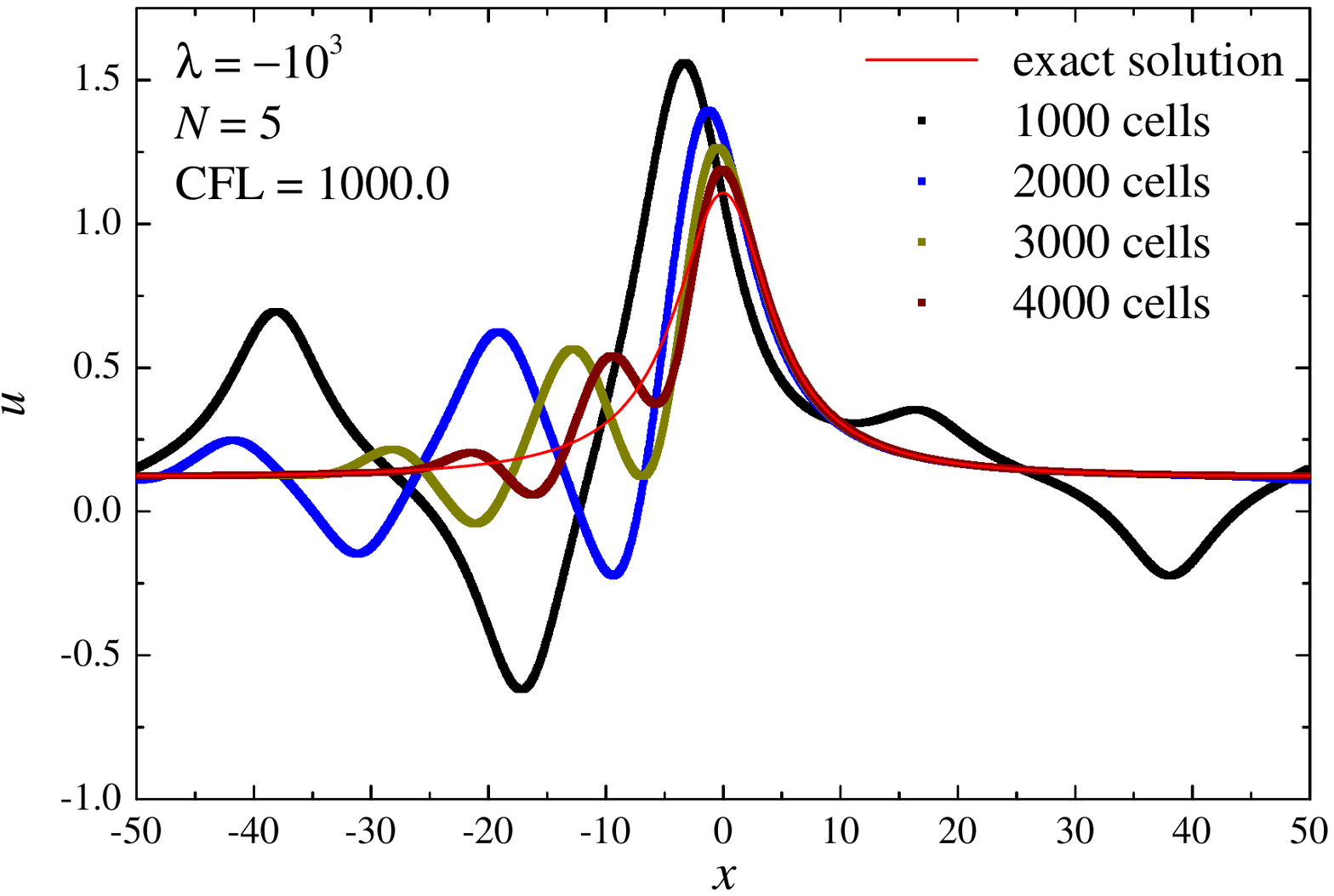}
\vspace{-8mm}\caption{\label{fig:test_adveq_stiff_z1000_lrnz_1d_cfl_100000:5}}
\end{subfigure}
\begin{subfigure}{0.24\textwidth}
\includegraphics[width=\textwidth]{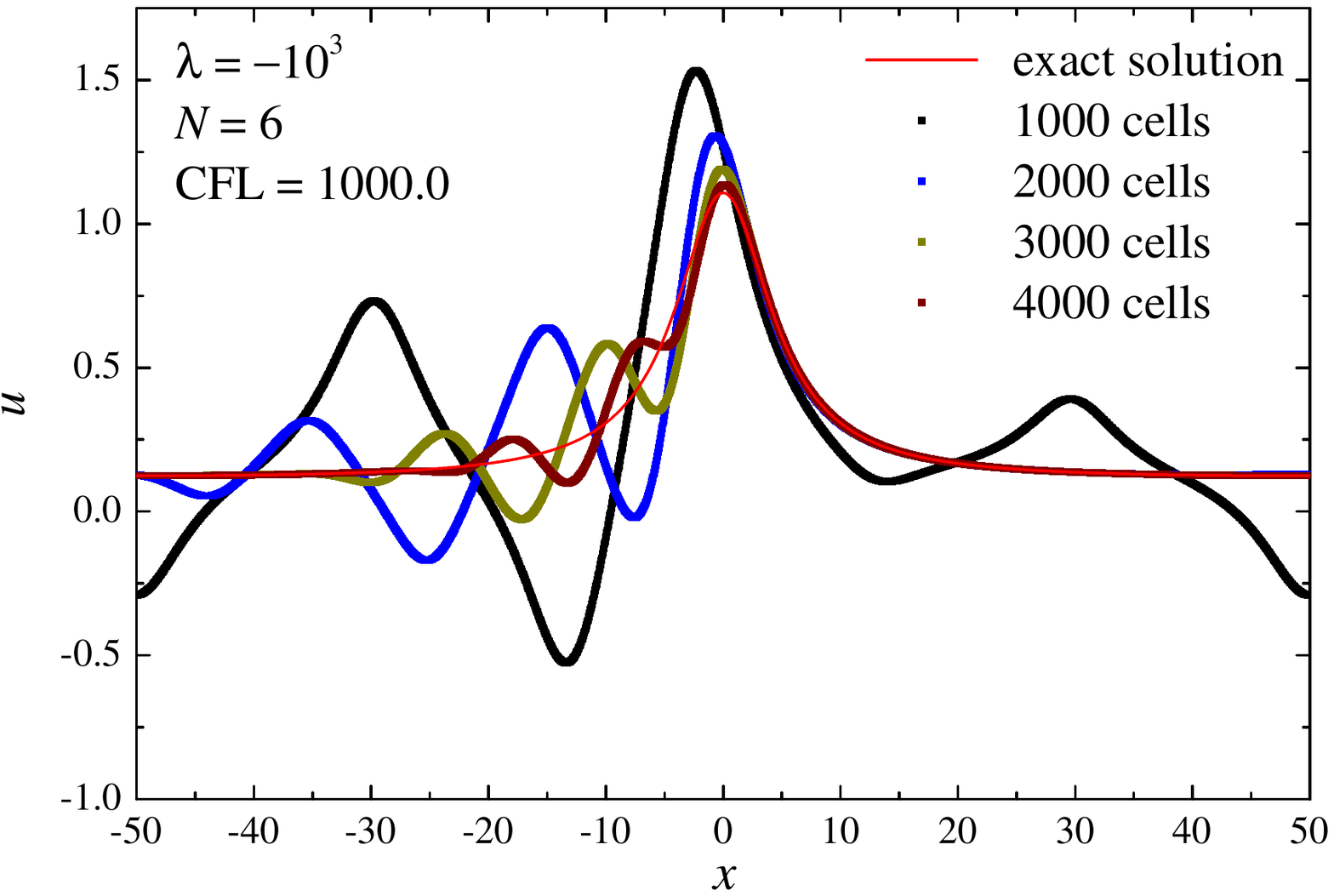}
\vspace{-8mm}\caption{\label{fig:test_adveq_stiff_z1000_lrnz_1d_cfl_100000:6}}
\end{subfigure}
\begin{subfigure}{0.24\textwidth}
\includegraphics[width=\textwidth]{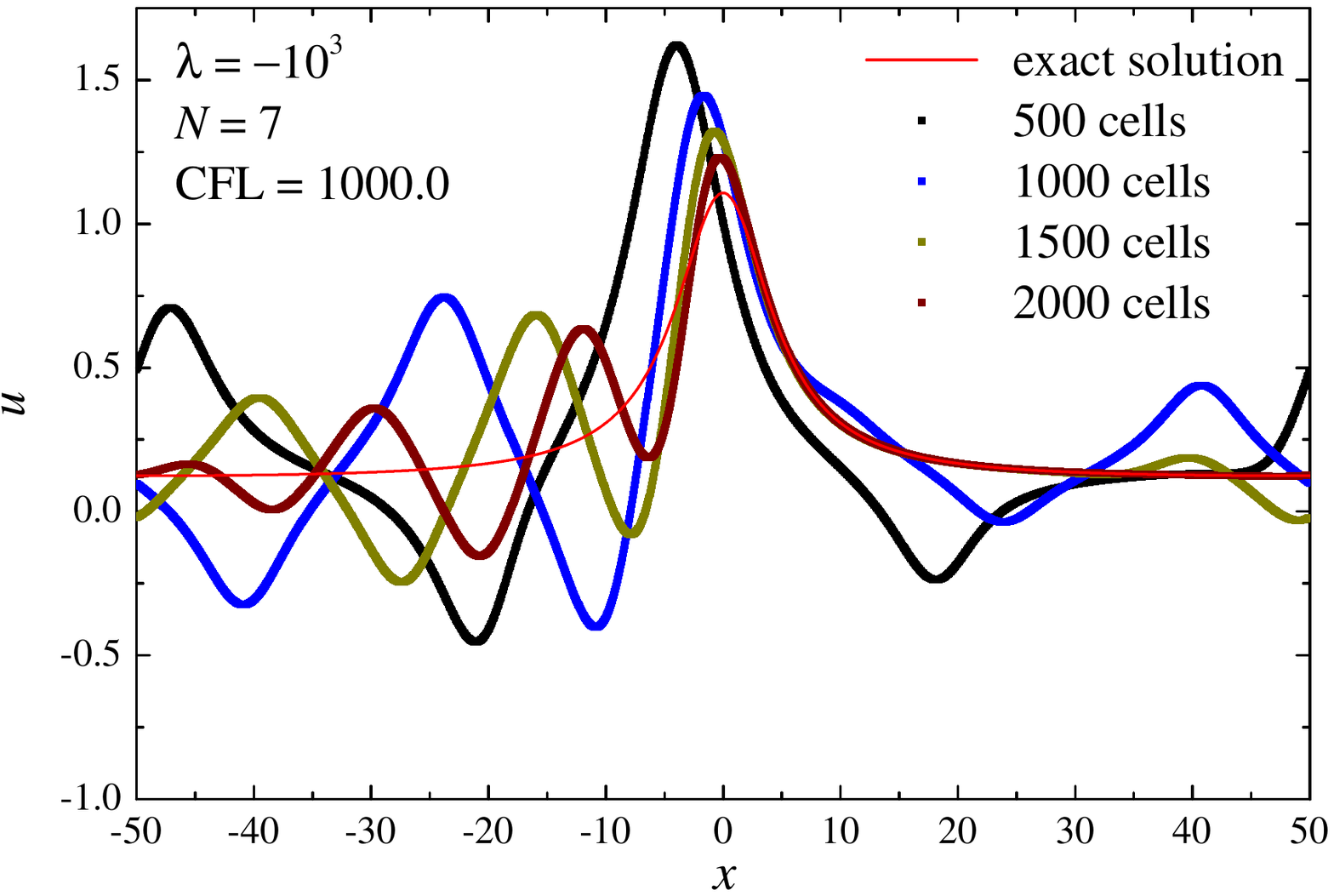}
\vspace{-8mm}\caption{\label{fig:test_adveq_stiff_z1000_lrnz_1d_cfl_100000:7}}
\end{subfigure}
\begin{subfigure}{0.24\textwidth}
\includegraphics[width=\textwidth]{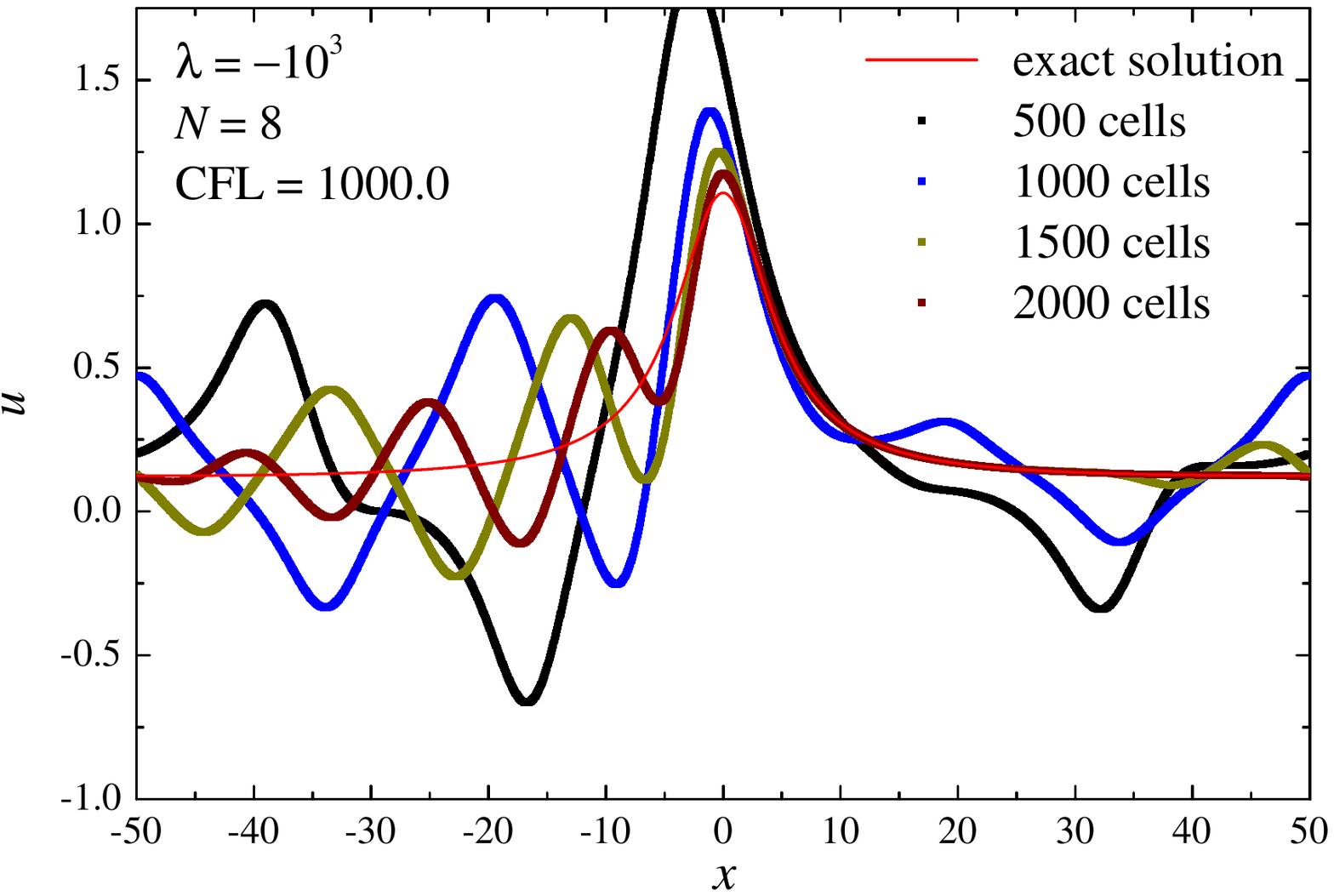}
\vspace{-8mm}\caption{\label{fig:test_adveq_stiff_z1000_lrnz_1d_cfl_100000:8}}
\end{subfigure}\\[-2.5mm]
\begin{subfigure}{0.24\textwidth}
\includegraphics[width=\textwidth]{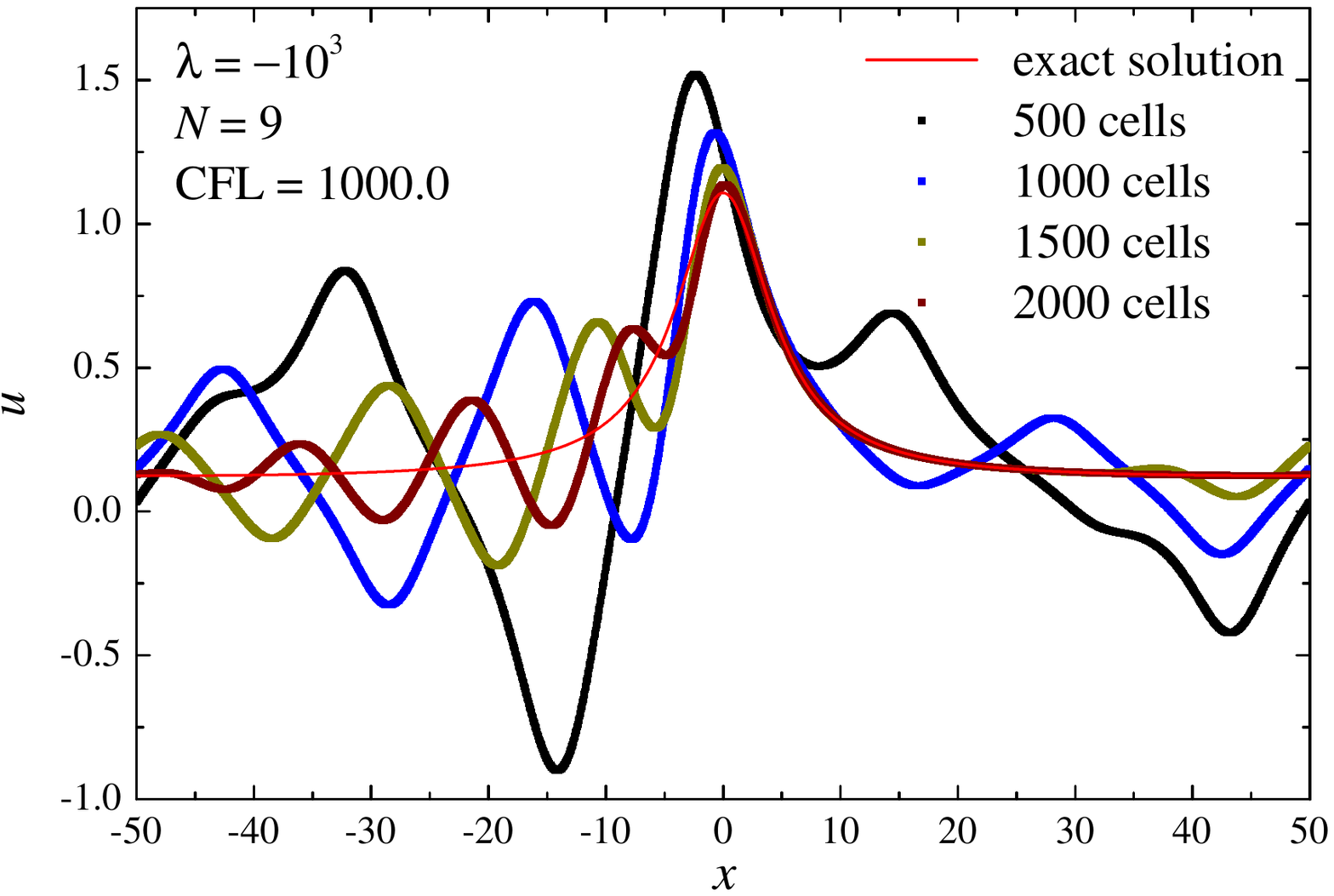}
\vspace{-8mm}\caption{\label{fig:test_adveq_stiff_z1000_lrnz_1d_cfl_100000:9}}
\end{subfigure}
\begin{subfigure}{0.24\textwidth}
\includegraphics[width=\textwidth]{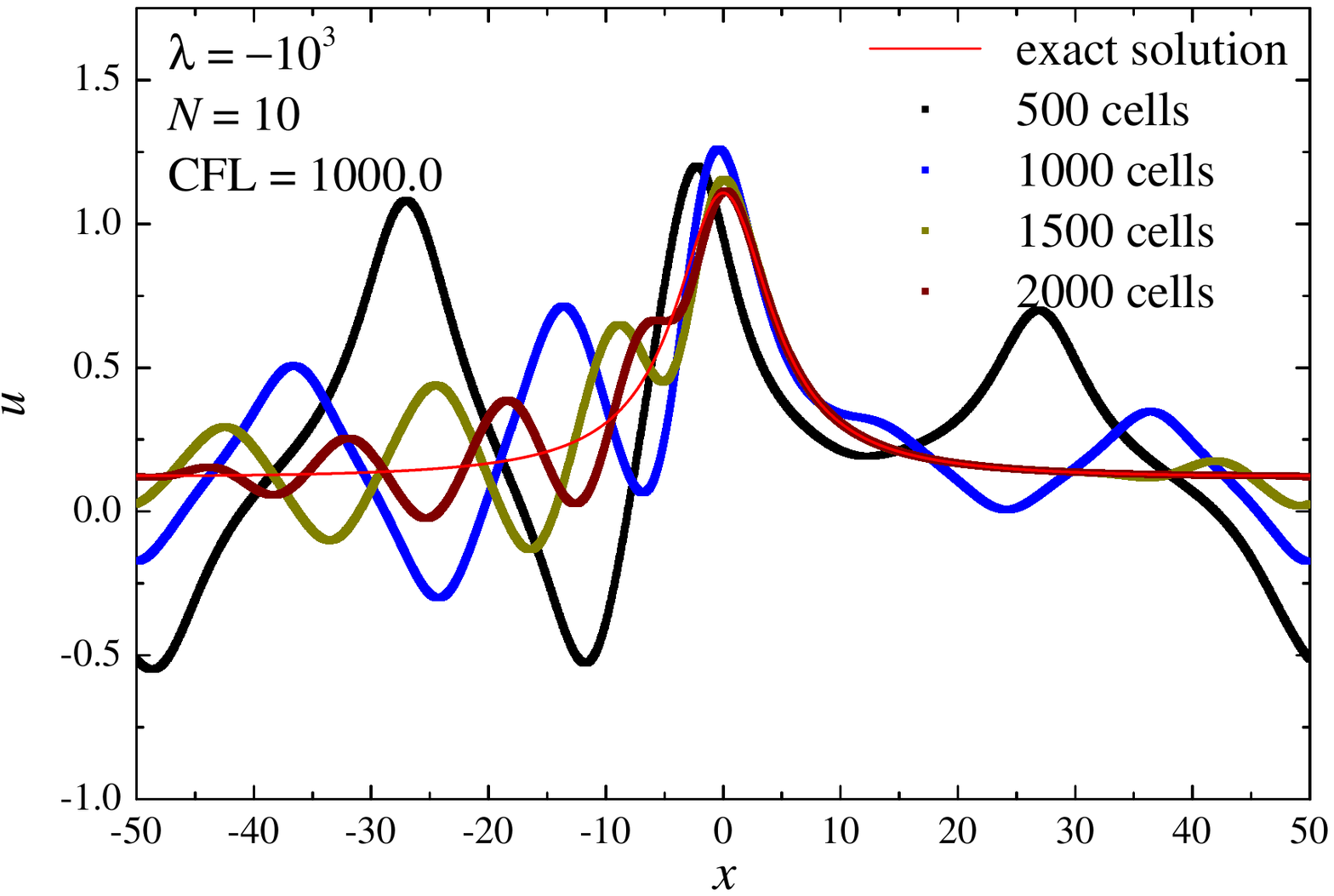}
\vspace{-8mm}\caption{\label{fig:test_adveq_stiff_z1000_lrnz_1d_cfl_100000:10}}
\end{subfigure}
\begin{subfigure}{0.24\textwidth}
\includegraphics[width=\textwidth]{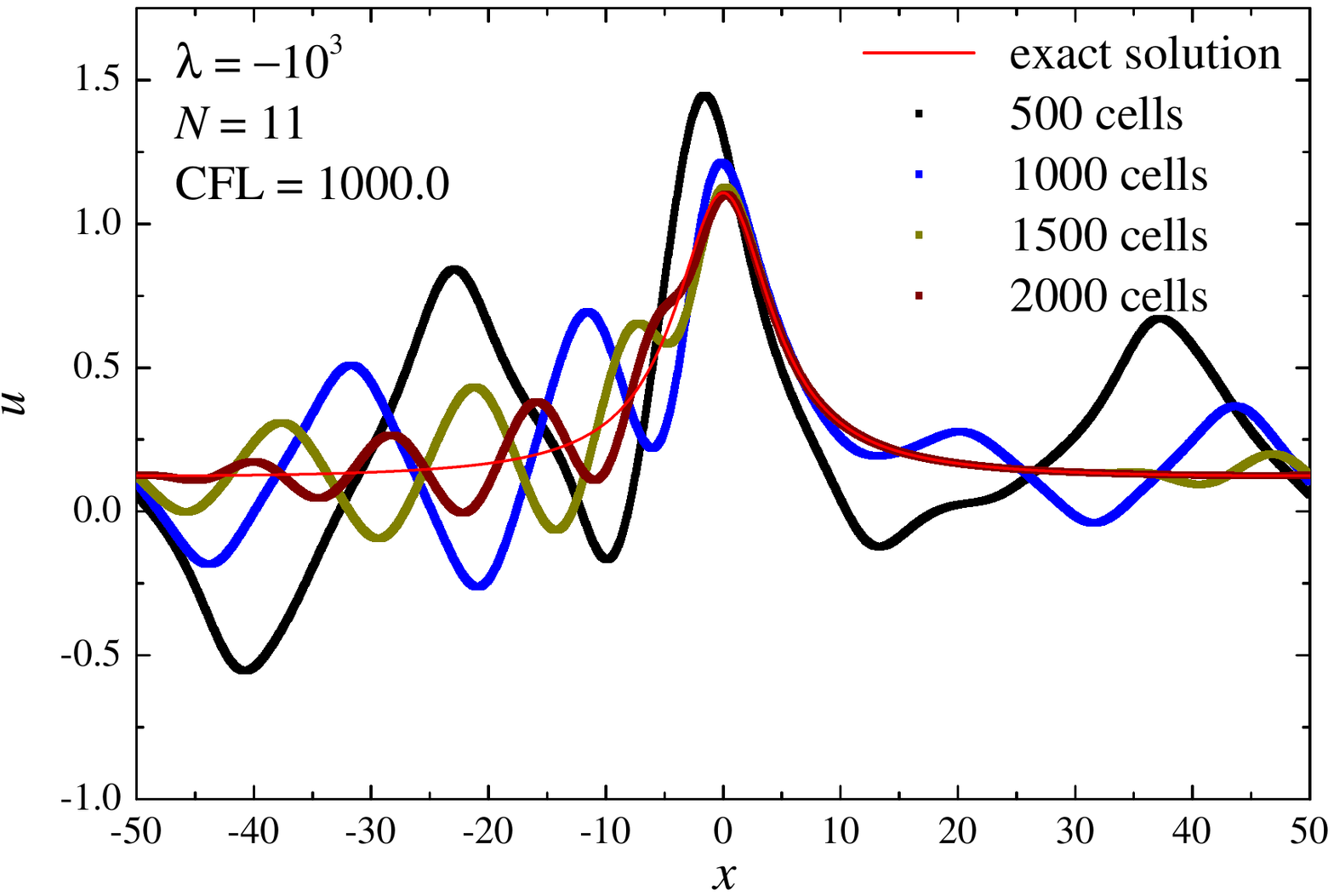}
\vspace{-8mm}\caption{\label{fig:test_adveq_stiff_z1000_lrnz_1d_cfl_100000:11}}
\end{subfigure}
\begin{subfigure}{0.24\textwidth}
\includegraphics[width=\textwidth]{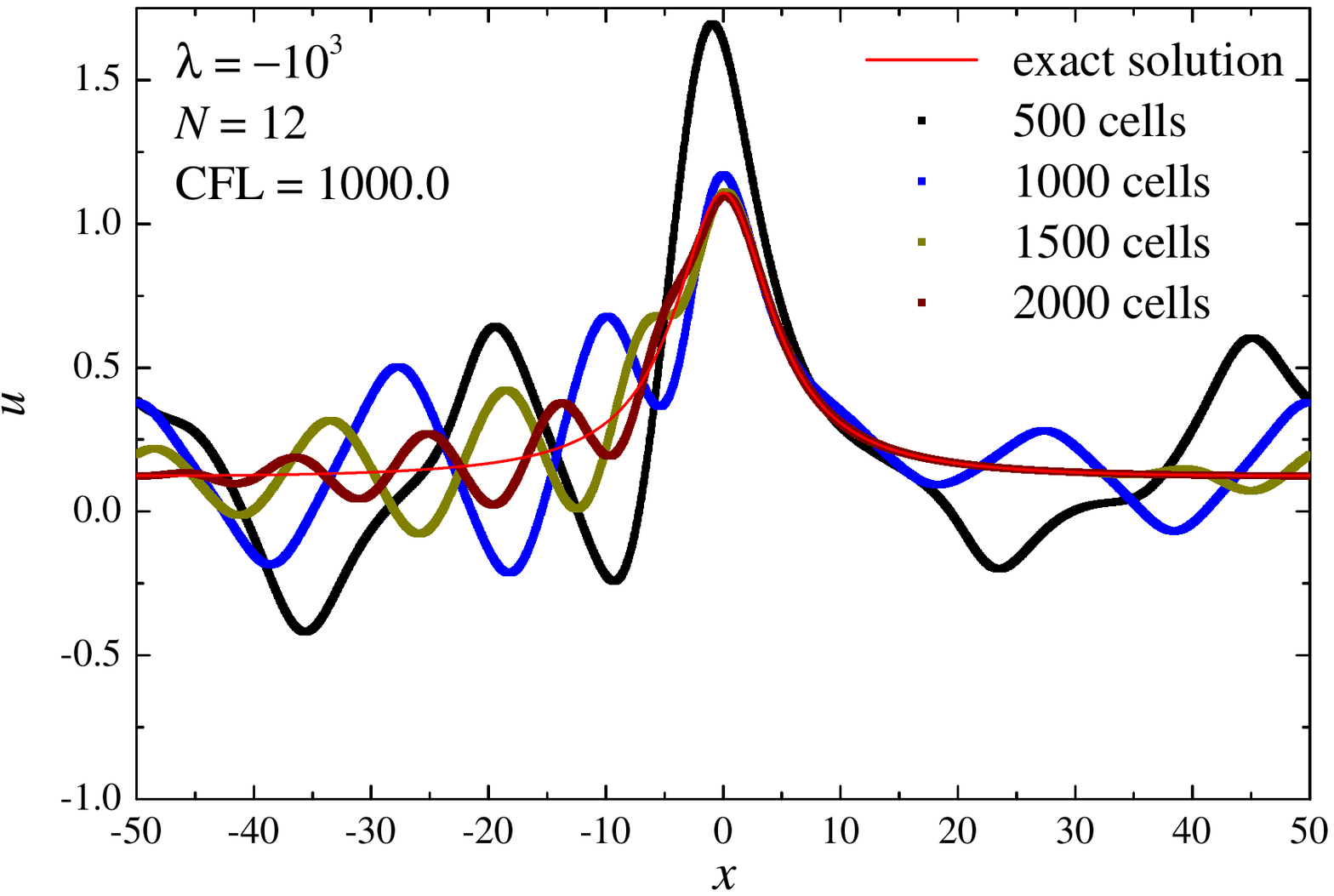}
\vspace{-8mm}\caption{\label{fig:test_adveq_stiff_z1000_lrnz_1d_cfl_100000:12}}
\end{subfigure}
\caption{%
Coordinate dependencies of the solution $u(x, t_{f})$ to the linear advection-reaction equation (\ref{eq:adv_eq_src}) at the final time $t_{f}$, obtained by the ADER-DG numerical method with the LST-DG predictor, for the degrees $N = 1, \ldots, 12$ of the basis polynomials for the Courant number $\mathrm{CFL} = 1000.0$ and the decay parameter $\nu = -10^{3}$. The red line represents the exact analytical solution.
}
\label{fig:test_adveq_stiff_z1000_lrnz_1d_cfl_100000}
\end{figure}

Figures~\ref{fig:test_adveq_stiff_z100_lrnz_1d_cfl_010},~\ref{fig:test_adveq_stiff_z100_lrnz_1d_cfl_100},~\ref{fig:test_adveq_stiff_z100_lrnz_1d_cfl_1000},~\ref{fig:test_adveq_stiff_z1000_lrnz_1d_cfl_010},~\ref{fig:test_adveq_stiff_z1000_lrnz_1d_cfl_100} and~\ref{fig:test_adveq_stiff_z1000_lrnz_1d_cfl_1000} show the representative examples of the obtained coordinate dependencies of the solution $u(x, t_{f})$ to the linear advection-reaction equation (\ref{eq:adv_eq_s_src}) at the final time $t_{f} = 10^{6}$, obtained by the ADER-DG numerical method with the LST-DG predictor, for the degrees $N = 1, \ldots, 12$ of the basis polynomials: for the decay parameter $\nu = -10^{2}$ in Figures~\ref{fig:test_adveq_stiff_z100_lrnz_1d_cfl_010},~\ref{fig:test_adveq_stiff_z100_lrnz_1d_cfl_100},~\ref{fig:test_adveq_stiff_z100_lrnz_1d_cfl_1000} and for the decay parameter $\nu = -10^{3}$ in Figures~\ref{fig:test_adveq_stiff_z1000_lrnz_1d_cfl_010},~\ref{fig:test_adveq_stiff_z1000_lrnz_1d_cfl_100},~\ref{fig:test_adveq_stiff_z1000_lrnz_1d_cfl_1000}; for the Courant number $\mathrm{CFL} = 0.1$ in Figures~\ref{fig:test_adveq_stiff_z100_lrnz_1d_cfl_010},~\ref{fig:test_adveq_stiff_z1000_lrnz_1d_cfl_010}, for the Courant number $\mathrm{CFL} = 1.0$ in Figures~\ref{fig:test_adveq_stiff_z100_lrnz_1d_cfl_100},~\ref{fig:test_adveq_stiff_z1000_lrnz_1d_cfl_100} and for the Courant number $\mathrm{CFL} = 10.0$ in Figures~\ref{fig:test_adveq_stiff_z100_lrnz_1d_cfl_1000},~\ref{fig:test_adveq_stiff_z1000_lrnz_1d_cfl_1000}. To compare the numerical solution with the exact analytical solution and to visually analyze the quality of the numerical solution, the red line in all these Figures represents the coordinate dependence of the exact analytical solution $u_{\rm exact}(x, t_{f})$, which in this case exactly matches the initial condition $u_{\rm init}(x)$. The presented results show that even for such large values of the Courant number $\mathrm{CFL}$, the numerical solution $u(x, t_{f})$ is in excellent qualitative agreement with the exact analytical solution $u_{\rm exact}(x, t_{f})$. No artifacts of the numerical solution $u(x, t_{f})$ were revealed when compared with the exact analytical solution $u_{\rm exact}(x, t_{f})$. For high polynomial degrees $N \geqslant 8$--$9$, good agreement between the numerical solution $u(x, t_{f})$ and the exact analytical solution $u(x, t_{f})$ is observed even for very coarse meshes consisting of only $10$-$20$ coordinate discretization domains $\Omega_{i}$ (coordinate discretization step $\Dx = 5.0$--$10.0$), which corresponds to $z \approx 1.0$--$2.0 \cdot 10^{4}$ for a Courant number of $\mathrm{CFL} = 0.1$, $z \approx 1.0$--$2.0 \cdot 10^{5}$ for a Courant number $\mathrm{CFL} = 1.0$, and $z \approx 1.0$--$2.0 \cdot 10^{6}$ for a Courant number $\mathrm{CFL} = 10.0$. As noted above, $z = |\nu|\Dtn{n}$ can be considered as a certain effective reaction Courant number $\mathrm{CFL}_{\rm react} = \Dtn{n}/\tau_{\rm react} \equiv z$, where $\tau_{\rm react} = |\nu|^{-1}$ is the time during which the amount of the reagent in the monomolecular reaction will decrease by a factor of $e$; therefore, the presented results clearly demonstrate the efficiency of the ADER-DG numerical method with the LST-DG predictor in case $\mathrm{CFL}_{\rm react} \gg \mathrm{CFL} \gg \mathrm{CFL}_{\rm max}(N, 0)$, in this case, the difference can be at least $10^{3}$--$10^{5}$ times.

Empirical convergence orders $p$ are calculated based on the dependence $\epsilon \propto \Dx^{p}$ of the numerical solution error $\epsilon$ on the coordinate grid step $\Dx$. Three classical norms $\mathcal{L}_{1}$, $\mathcal{L}_{2}$, $\mathcal{L}_{\infty}$ of the numerical solution errors $\epsilon$ are used to determine the empirical orders $p$ of convergence:
\begin{equation}\label{eq:norms_of_errors}
\begin{split}
\epsilon_{\mathcal{L}_{1}} = \int\limits_{\Omega}dx\, |u(x, t_{f}) - u_{\rm exact}(x)|,\qquad
\epsilon_{\mathcal{L}_{2}}^{2} = \int\limits_{\Omega}dx\, |u(x, t_{f}) - u_{\rm exact}(x)|^{2},\qquad
\epsilon_{\mathcal{L}_{\infty}} = \operatorname{ess}\sup\limits_{\hspace{-5mm}x\in\Omega} |u(x, t_{f}) - u_{\rm exact}(x)|,\vphantom{\int\limits_{\Omega}}
\end{split}
\end{equation}
where $u(x, t_{f})$ is the numerical solution, $u_{\rm exact}(x)$ is the exact analytical solution to the problem. The integrals in (\ref{eq:norms_of_errors}) are calculated using the GL quadrature formula (\ref{eq:gl_rule}) with very high degree $N = 60$ of polynomial $\tilde{L}_{N}$, and the $\operatorname{ess}\sup$ is calculated as $\max$ on a subgrid containing $1000$ nodes in each discretization domain $\Omega_{i}$ of the coordinate mesh. The empirical convergence orders $p$ are calculated for studied three values of the Courant number $\mathrm{CFL}$: $\mathrm{CFL} = 0.1$, $1.0$, $10.0$. Tables~\ref{tab:test_adveq_stiff_z100_lrnz_1d_cfl_010},~\ref{tab:test_adveq_stiff_z100_lrnz_1d_cfl_100},~\ref{tab:test_adveq_stiff_z100_lrnz_1d_cfl_1000},~\ref{tab:test_adveq_stiff_z1000_lrnz_1d_cfl_010},~\ref{tab:test_adveq_stiff_z1000_lrnz_1d_cfl_100} and~\ref{tab:test_adveq_stiff_z1000_lrnz_1d_cfl_1000} present the obtained empirical convergence orders $p$: for the decay parameter $\nu = -10^{2}$ in Tables~\ref{tab:test_adveq_stiff_z100_lrnz_1d_cfl_010},~\ref{tab:test_adveq_stiff_z100_lrnz_1d_cfl_100},~\ref{tab:test_adveq_stiff_z100_lrnz_1d_cfl_1000} and for the decay parameter $\nu = -10^{3}$ in Tables~\ref{tab:test_adveq_stiff_z1000_lrnz_1d_cfl_010},~\ref{tab:test_adveq_stiff_z1000_lrnz_1d_cfl_100},~\ref{tab:test_adveq_stiff_z1000_lrnz_1d_cfl_1000}; for the Courant number $\mathrm{CFL} = 0.1$ in Tables~\ref{tab:test_adveq_stiff_z100_lrnz_1d_cfl_010},~\ref{tab:test_adveq_stiff_z1000_lrnz_1d_cfl_010}, for the Courant number $\mathrm{CFL} = 1.0$ in Tables~\ref{tab:test_adveq_stiff_z100_lrnz_1d_cfl_100},~\ref{tab:test_adveq_stiff_z1000_lrnz_1d_cfl_100} and for the Courant number $\mathrm{CFL} = 10.0$ in Tables~\ref{tab:test_adveq_stiff_z100_lrnz_1d_cfl_1000},~\ref{tab:test_adveq_stiff_z1000_lrnz_1d_cfl_1000}. It is evident that as the Courant number $\mathrm{CFL}$ changes in the stability region $\mathrm{CFL} \leqslant \mathrm{CFL}_{\rm max}(N)$, the empirical convergence orders $p$ are agree well with the approximation orders $p = N+1$ calculated in Section~\ref{sec:approx_anal} ``Approximation analysis'', which are presented in the Th. column of Tables~\ref{tab:test_adveq_stiff_z100_lrnz_1d_cfl_010},~\ref{tab:test_adveq_stiff_z100_lrnz_1d_cfl_100},~\ref{tab:test_adveq_stiff_z100_lrnz_1d_cfl_1000},~\ref{tab:test_adveq_stiff_z1000_lrnz_1d_cfl_010},~\ref{tab:test_adveq_stiff_z1000_lrnz_1d_cfl_100} and~\ref{tab:test_adveq_stiff_z1000_lrnz_1d_cfl_1000}. A comparison of the empirical convergence orders $p$ with the expected values $p = N+1$ shows that in most cases of the degrees $1 \leqslant N \leqslant 12$ of the basis polynomials, the values of the Courant number $\mathrm{CFL}$ and the decay parameter $\nu$, the empirical values $p$ are in good agreement with the expected values $p = N+1$; although, of course, some observed differences should also be noted. For the Courant number $\mathrm{CFL} = 0.1$, there is a good agreement between the empirical convergence orders and the expected values in all studied cases of the degrees of the polynomials and the decay parameter; however, small deviations, both downwards and upwards (some ``superconvergence''), are expected and usually occur in such empirical quantitative estimates~\cite{ader_stiff_3, ader_stiff_4}. In the case of the Courant number $\mathrm{CFL} = 1.0$, for the decay parameter $\nu = -10^{2}$, a good agreement between the empirical convergence orders and the expected values occurs for polynomial degrees $N \leqslant 8$--$9$, while in the case of very high degrees $N > 9$ of the polynomials, a small decrease in the convergence orders $p$ compared to the expected values $p = N+1$ occurs. This phenomenon, in a test like this, is explained by the simple linear additive accumulation of computational error, which is associated with a very large time $t_{f} - t_{0} = 10^{6}$. Similar behavior has already been observed using the ADER-DG numerical method with the LST-DG predictor in other cases~\cite{ader_stiff_4}. For a Courant number $\mathrm{CFL} = 1.0$, a similar situation occurs for a decay parameter of $\nu = -10^{3}$: good agreement between the empirical convergence orders $p$ and the expected values $p = N+1$ occurs for polynomial degrees $N \leqslant 8$--$9$, while in the case of very high degrees $N > 9$ of the polynomials, a small decrease in convergence orders $p$ compared to the expected values $p = N+1$ occurs, but this decrease is slightly greater than for a decay parameter of $\nu = -10^{2}$. These features are also preserved for a Courant number $\mathrm{CFL} = 10.0$ for decay parameters $\nu = -10^{2}$ and $\nu = -10^{3}$. Therefore, it can be concluded that the ADER-DG numerical method with the LST-DG predictor allows one to effectively obtain a numerical solution for strong stiffness $|\nu| \gg 1$ (and $\mathrm{CFL}_{\rm react} \gg \mathrm{CFL}$) problems even at very high Courant numbers $\mathrm{CFL} \gg \mathrm{CFL}_{\rm max}(N, 0)$, while demonstrating arbitrarily high convergence orders $p$ of the numerical solution.

Figures~\ref{fig:test_adveq_stiff_z1000_lrnz_1d_cfl_10000} and~\ref{fig:test_adveq_stiff_z1000_lrnz_1d_cfl_100000} show the representative examples of the obtained coordinate dependencies of the solution $u(x, t_{f})$ to the linear advection-reaction equation (\ref{eq:adv_eq_s_src}) at the final time $t_{f} = 10^{6}$, obtained by the ADER-DG numerical method with the LST-DG predictor, for the extremely large values of Courant numbers $\mathrm{CFL} = 100.0$ and $1000.0$ for the decay parameter $\nu = -10^{3}$, for the degrees $N = 1, \ldots, 12$ of the basis polynomials. These results are presented only to demonstrate that when using extremely large Courant numbers $\mathrm{CFL} = 100.0$ for a problem with large decay parameter $\nu = -10^{3}$, the solution can exhibit extremely complex artifacts (and difficult to predict in nonlinear cases). The presented coordinate dependencies show that convergence of the numerical solution $u(x, t_{f})$ (at least point-wise) to the exact analytical solution $u_{\rm exact}(x, t_{f})$, although it takes place, is characterized by significant non-uniformity. For a quantitative assessment, it suffices to note that the discretization time step $\Dtn{n} = 100.0$ for the case of $100$ discretization domains $\Omega_{i}$ (in this case, in just a single step $t^{n}$ to $t^{n+1} = t^{n} + \Dtn{n}$, the solution profile $u(x, t^{n})$ traverses the entire definition domain $\Omega$ and must return to its original form), and $10^{4}$ such steps are performed to obtain the final numerical solution $u(x, t_{f})$. With such large time discretization steps $\Dtn{n}$, the dispersion properties of the ADER-DG numerical method with the LST-DG predictor and the deviation of the numerical dispersion law $\omega^{\rm num} = \omega^{\rm num}(k)$ (more precisely, a set of dispersion laws $\{\omega^{\rm num}_{s}(k)\}_{s}$) of the ADER-DG numerical method with the LST-DG predictor from the analytical dispersion law $\omega(k) = ak - i\nu$ of the linear advection-reaction equation (\ref{eq:adv_eq_src}) become significant --- this behavior of the numerical solution can be considered expected.

The presented results clearly demonstrate that the ADER-DG numerical method with the LST-DG predictor has significantly higher boundary values of the Courant number $\mathrm{CFL}$ in the case of a linear advection-reaction equation (\ref{eq:adv_eq_s_src}) with a highly stiff source term. Moreover, with increasing problem stiffness, which is characterized by the parameter $|z| \gg 1$, the boundary values of the Courant number $\mathrm{CFL}_{\rm max}(N, z)$ increase. It is evident that even at very high Courant numbers $\mathrm{CFL}$ --- up to $\mathrm{CFL} = 10.0$, which is almost $100$ times greater than the boundary value of the Courant number $\mathrm{CFL}_{\rm max}(N, 0)$ for a linear advection equation in the case of degree $N = 3$ basis polynomials, and almost $1000$ times greater --- in the case of degree $N = 12$ basis polynomials, the ADER-DG numerical method with the LST-DG predictor maintains convergence with the convergence order $p = N+1$. For even larger Courant numbers $\mathrm{CFL}$, up to $\mathrm{CFL} = 100.0$ and even $\mathrm{CFL} = 1000.0$, the ADER-DG numerical method with the LST-DG predictor allows one to obtain a stable numerical solution; however, in these cases, significant and clearly visible numerical artifacts arise. These results clearly confirm the rigorous and partially empirical theory developed in Section~\ref{sec:approx_anal} ``Approximation analysis'' and in Section~\ref{sec:stab_anal} ``Stability analysis''.

\subsection{ZND-detonation waves}
\label{sec:app:euler}

The results presented in the previous Subsection~\ref{sec:app:lin_adv} ``Linear advection-reaction equation'' pertain only to the accuracy and stability of ADER-DG numerical method with the LST-DG predictor for linear problems (\ref{eq:adv_eq_s_src}). Clearly, the general case of quasilinear systems of equations is of interest for solving real-world problems. The results of the stability analysis of the numerical method for a linear advection equation (equation (\ref{eq:adv_eq_src}) with $\nu = 0$) can be exactly extended to the case of a system of quasilinear hyperbolic equations without source terms by switching from the divergent notation of flux terms to a notation in terms of the Jacobian matrix and the divergence of conservative variables; this analysis and its limitations, related to the dissipative properties of Riemann solvers~\cite{Toro_solvers_2009}, for the ADER-DG numerical method with LST-DG predictor considered in this work was performed in~\cite{ader_dg_stab_rig}, as well as in~\cite{ader_dg_stab} for other selected spatial and space-time bases. However, in the case of a linear advection-reaction equation (\ref{eq:adv_eq_src}) and a system of quasilinear equations with source terms, the study becomes significantly more complex~\cite{GodlewskiRaviart_2021, Dafermos_2016}, primarily because the Jacobian matrices of the flux terms and source terms are usually not diagonalized simultaneously in the same basis. Therefore, if the left-hand side of a system of quasilinear equations with source terms is written in terms of characteristic variables (quasilinear advection equations with Riemann invariants), the right-hand side will not exhibit strict local splitting of the system of equations. In this work, a system of quasilinear hyperbolic equations with source terms is chosen in the following form:
\begin{equation}\label{eq:nonlin_sys_source}
\pderf{\mathbf{U}}{t} + \pderf{\mathbf{F}(\mathbf{U})}{x} = \mathbf{S}(\mathbf{U}; x, t),\quad
(x, t)\in\Omega\times\mathcal{T},\quad
\Omega\subset\mathcal{R},\quad
\mathcal{T} = \left[t_{0}, t_{f}\right]\subset\mathcal{R},
\end{equation}
where classical notations are used: $\mathbf{U}: \Omega\times\mathcal{T}\rightarrow\mathcal{R}^{L}$ is the vector of conservative variables, $\mathbf{F}: \mathcal{R}^{L}\rightarrow\mathcal{R}^{L}$ is the vector of flux terms, $\mathbf{S}: \mathcal{R}^{L}\times\Omega\times\mathcal{T}\rightarrow\mathcal{R}^{L}$ is the vector of source terms $\mathbf{U}$, for which dependence is allowed not only on the vector of conservative variables, but also on the space-time point $(x, t)$, $L$ is the number of conservative variables and the number of equations in the system of quasilinear hyperbolic equations (\ref{eq:nonlin_sys_source}), and the remaining notations are identical to those in the description of the equation (\ref{eq:adv_eq_src}). To determine the relationship between the theory constructed in this work for the linear advection-reaction equation (\ref{eq:adv_eq_src}) and the system of quasilinear equations with source terms (\ref{eq:nonlin_sys_source}), a transformation of the original system of quasilinear hyperbolic equations with source terms (\ref{eq:nonlin_sys_source}) is performed on the left-hand side:
\begin{equation}\label{eq:nonlin_sys_source_prim}
\pderf{\mathbf{U}}{t} + \mathrm{A}(\mathbf{U})\pderf{\mathbf{U}}{x} = \mathbf{S}(\mathbf{U}),\qquad
\mathrm{A}(\mathbf{U}) = \pderf{\mathbf{F}}{\mathbf{U}}(\mathbf{U}),
\end{equation}
where $\mathrm{A}: \mathcal{R}^{L}\rightarrow\mathcal{R}^{L \times L}$ is the Jacobian matrix of the flux terms $\mathbf{F}(\mathbf{U})$, and the explicit dependence of the source terms $\mathbf{S}$ on the space-time point $(x, t)$ is temporarily (before the formulation of the ADER-DG numerical method with LST-DG predictor from expression (\ref{eq:nonlin:eq_mapped})) not considered, so as not to further require too strong conditions from the linearization point $\mathbf{U}_{0}$. To linearize a system of quasilinear hyperbolic equations with source (\ref{eq:nonlin_sys_source_prim}) terms in the neighborhood of a sufficiently smooth solution $\mathbf{U}_{0}$ at the time $t$ (and a small domain $[t, t + \delta t]$ of this time $t$ --- the smoothness of the solution may be violated, including from smooth initial conditions), a small deviation $\delta\mathbf{U} = \mathbf{U} - \mathbf{U}_{0}$ is introduced, for which the system of hyperbolic equations (\ref{eq:nonlin_sys_source_prim}) is linearized in the following form:
\begin{equation}\label{eq:nonlin_sys_source_lin}
\pderf{(\delta\mathbf{U})}{t} + \mathrm{A}_{0}\pderf{(\delta\mathbf{U})}{x} = \mathrm{B}_{0}\delta\mathbf{U},\qquad
\mathrm{B} = \pderf{\mathbf{S}}{\mathbf{U}}(\mathbf{U}),\qquad
\mathrm{B}_{0} = \mathrm{B}(\mathbf{U}_{0}),
\end{equation}
where $\mathrm{B}: \mathcal{R}^{L}\rightarrow\mathcal{R}^{L \times L}$ is the Jacobian matrix of the source terms $\mathbf{S}(\mathbf{U})$, and the notation $\mathrm{A}_{0} = \mathrm{A}(\mathbf{U}_{0})$ is introduced. The hyperbolicity condition of a (non-strict) system of equations allows us to write the matrix in the form $\mathrm{A}_{0} = \mathrm{L}^{-1}\Lambda\mathrm{L}$ (from the form $\Lambda = \mathrm{L}\mathrm{A}_{0}\mathrm{L}^{-1}$), where $\Lambda$ is a diagonal matrix: $\Lambda = \mathrm{diag}(\{\mu_{l}\}_{l})$, with eigenvalues $\mu_{l}\in\sigma(\Lambda)$ on the diagonal, $\mathrm{L}$ is the matrix of left eigenvectors, as a result, the linearized system of equations (\ref{eq:nonlin_sys_source_lin}) takes the following form:
\begin{equation}\label{eq:nonlin_sys_source_lin_char}
\pderf{\mathbf{W}}{t} + \Lambda\pderf{\mathbf{W}}{x} = [\mathrm{L}\mathrm{B}_{0}\mathrm{L}^{-1}]\mathbf{W},
\end{equation}
where $\mathbf{W} = \mathrm{L}\delta\mathbf{U}$ is the vector of characteristic variables of the linearized system of equations (\ref{eq:nonlin_sys_source_lin}). Therefore, the main complication of this problem compared to the case without sources is clearly visible: the matrix $\mathrm{L}\mathrm{B}_{0}\mathrm{L}^{-1}$ is generally not diagonal, meaning that the source ``mixes'' the characteristic modes $\mathbf{W}(x, t)$ (or quasi-characteristic modes). In practice~\cite{OranBoris_2005, Nagnibeda_2009, Lunev_2017, Anderson_Hypersonic_2019}, the source term $\mathbf{S}(\mathbf{U})$ often contains components characterized by high stiffness (particularly in modeling reacting flows~\cite{ader_stiff_3, ader_stiff_4}), which can be characterized by the presence of very large absolute negative eigenvalues $|\nu_{l}| \gg 1$ in the matrix spectrum $\{\nu_{l}\}_{l} = \sigma(\mathrm{L}\mathrm{B}_{0}\mathrm{L}^{-1})$. The eigenvalues $\nu_{l}$ have the inverse time dimension, which, in the case of classical stiffness, characterizes a certain leading local decrement $\nu_{l} = \tau_{\rm react}^{-1}$ of the solution harmonic damping; therefore, an effective Courant number $\mathrm{CFL}_{\rm react} = \Dtn{n}/\tau_{\rm react}$ can be introduced. In this case, the stability conditions for the problem for the linear advection-reaction equation (\ref{eq:adv_eq_src}) and for the system of quasilinear hyperbolic equations (\ref{eq:nonlin_sys_source}) with a source term are qualitatively equivalent: the main influence of the source term $\mathbf{S}(\mathbf{U})$ is associated precisely with a very large eigenvalue $|\nu_{l}| \gg 1$ (or subset of eigenvalues). Therefore, it is expected that in the case of significantly stiff problems, especially those associated with modeling reactive flows containing detonation waves, an expansion of the stability region $[0, \mathrm{CFL}_{\rm max}(N)]$ of the ADER-DG numerical method with the LST-DG predictor should occur; however, of course, it does not necessarily exactly quantitatively correspond to the boundary values of the Courant number $\mathrm{CFL}_{\rm max}(N, z)$ for the linear advection-reaction equation (\ref{eq:adv_eq_src}), which is associated with the existence of other eigenvalues $\{\nu_{l}\}_{l}$ of the matrix $\mathrm{L}\mathrm{B}_{0}\mathrm{L}^{-1}$ of even a linearized system (\ref{eq:nonlin_sys_source_lin_char}) of quasilinear hyperbolic equations (\ref{eq:nonlin_sys_source}) with a source term.

The numerical ADER-DG method with the LST-DG predictor for the case of the system of quasilinear hyperbolic equations (\ref{eq:nonlin_sys_source}) with source terms is constructed similarly to the case of a linear inhomogeneous advection-reaction equation (\ref{eq:adv_eq_s_src}), as presented above in Subsection~\ref{sec:app:lin_adv} ``Linear advection-reaction equation'' of this Section, and to the case of a homogeneous advection-reaction equation (\ref{eq:adv_eq_src}), as presented above in Section~\ref{sec:descr} ``General description of the numerical method''. By mapping (\ref{eq:mapping}) the arguments $(t, x)$ of the coordinate-time discretization domain $\Xi_{i}^{n}$ on the rescaled arguments $(\tau, \xi)$ reference domain $\omega^{2} = [0, 1]^{2}$ the original system of quasilinear hyperbolic equations (\ref{eq:nonlin_sys_source}) with source terms is written in the following form:
\begin{equation}\label{eq:nonlin:eq_mapped}
\pderf{\mathbf{u}}{\tau} + \pderf{\mathbf{f}(\mathbf{u})}{\xi} = \mathbf{s}(\mathbf{u}; \tau, \xi),\quad (\tau, \xi) \in \omega^{2},
\end{equation}
where the rescaled variables $(\mathbf{u},\, \mathbf{f}(\mathbf{u}),\, \mathbf{s}(\mathbf{u}; \tau, \xi))$ are chosen as follows:
\begin{equation}\label{eq:nonlin:vars_mapped}
\mathbf{u} \equiv \mathbf{U},\quad
\mathbf{f}(\mathbf{u}) = \frac{\Dtn{n}}{\Dx}\mathbf{F}(\mathbf{U}),\quad
\mathbf{s}(\mathbf{u}; \tau, \xi) = \Dtn{n}\mathbf{S}(\mathbf{U}; x(\xi), t(\tau)),
\end{equation}
which is similar to expression (\ref{eq:adv_eq_mapped}) for the case of a homogeneous equation (\ref{eq:adv_eq_src}). The ADER-DG numerical method is formulated using the weak form of the system of equations:
\begin{equation}\label{eq:nonlin:eq_weak}
\intrefdom{\tau}\intrefdom{\xi}\,\varphi_{k}(\xi)\left[\pderf{\mathbf{u}}{\tau} + \pderf{\mathbf{f}(\mathbf{u})}{\xi} - \mathbf{s}(\mathbf{u}; \tau, \xi)\right] = 0.
\end{equation}
By introducing the DG representations of the solution at the current $t^{n}$ and next $t^{n+1}$ time steps in the following form (which is similar to expressions (\ref{eq:u_n_def})):
\begin{equation}\label{eq:vec_u_n_def}
\mathbf{u}^{n}(\xi) = \sum\limits_{k = 0}^{N} \hat{\mathbf{u}}^{n}_{k} \varphi_{k}(\xi),\quad
\mathbf{u}^{n+1}(\xi) = \sum\limits_{k = 0}^{N} \hat{\mathbf{u}}^{n+1}_{k} \varphi_{k}(\xi),
\end{equation}
and performing integration by parts in the flux term in the equation (\ref{eq:nonlin:eq_weak}), the following form of the completely single-step discrete finite element scheme of the ADER-DG numerical method is obtained:
\begin{equation}\label{eq:vec_ader_dg_corr}
\begin{split}
w_{k}\left[\hat{\mathbf{u}}^{n+1}_{k} - \hat{\mathbf{u}}^{n}_{k}\right] & -
\intrefdom{\tau}\intrefdom{\xi}\,\varphi_{k}'(\xi) \mathbf{f}(\mathbf{q}(\tau, \xi)) +
\intrefdom{\tau}\,\Big[
  \varphi_{k}(1) \mathbf{f}^{\rm RP}\left(\mathbf{q}(\tau, 1),\, \mathbf{q}^{+}(\tau, 0)\right)\\ & -
  \varphi_{k}(0) \mathbf{f}^{\rm RP}\left(\mathbf{q}^{-}(\tau, 1),\, \mathbf{q}(\tau, 0)\right)
\Big] = \intrefdom{\tau}\intrefdom{\xi}\,\varphi_{k}(\xi) \mathbf{s}(\mathbf{q}(\tau, \xi); \tau, \xi),
\end{split}
\end{equation}
where $\mathbf{f}^{\rm RP}(\mathbf{u}^{L}, \mathbf{u}^{R})$ is the rescaled (similar flux in the expression (\ref{eq:nonlin:vars_mapped})) solution of the classical Riemann problem, and $q: \omega^{2}\rightarrow\mathcal{R}$ is the local discrete space-time solution used in ADER methods to calculate local, flux and source terms, and is determined for each coordinate-time element $\Xi_{i}^{n}$ separately. The HLLE solver~\cite{HLLE_solver_1, HLLE_solver_2, Toro_solvers_2009} to calculate the fluxes, which is known for its high dissipativity and stability is used in this work. It should be noted that similar results are obtained using Rusanov solver~\cite{Rusanov_solver, Toro_solvers_2009}, sometimes referred to as the local Lax-Friedrichs flux~\cite{ader_dg_ideal_flows}. The local discrete space-time solution $q$ is a numerical solution of the equation (\ref{eq:adv_eq_mapped}) in coordinate-time discretization domain $\Xi_{i}^{n}$:
\begin{equation}\label{eq:vec_local_sol_def}
\mathbf{q}(\tau, \xi) = \sum\limits_{\mathrm{p}} \hat{\mathbf{q}}_{\mathrm{p}} \Theta_{\mathrm{p}}(\tau, \xi)
                      = \sum\limits_{p_{0} = 0}^{N}\sum\limits_{p_{1} = 0}^{N} \hat{\mathbf{q}}_{p_{0}p_{1}} \varphi_{p_{0}}(\tau)\varphi_{p_{1}}(\xi).
\end{equation}
The chosen functional basis $\{\varphi_{k}\}$ (\ref{eq:phi_def}) allows for the use of the point-wise evaluation for the expansion coefficients of functions $f: \omega\rightarrow\mathcal{R}$ in this basis:
\begin{equation}
\begin{split}
&\mathbf{f}(\mathbf{q}(\tau, \xi)) \mapsto
\sum\limits_{p_{0} = 0}^{N}\sum\limits_{p_{1} = 0}^{N} \hat{\mathbf{f}}_{p_{0}p_{1}}\varphi_{p_{0}}(\tau)\varphi_{p_{1}}(\xi),\quad
\hat{\mathbf{f}}_{p_{0}p_{1}} = \mathbf{f}(\hat{\mathbf{q}}_{p_{0}p_{1}}),\\
&\mathbf{s}(\mathbf{q}(\tau, \xi); \tau, \xi) \mapsto
\sum\limits_{p_{0} = 0}^{N}\sum\limits_{p_{1} = 0}^{N} \hat{\mathbf{s}}_{p_{0}p_{1}}\varphi_{p_{0}}(\tau)\varphi_{p_{1}}(\xi),\quad
\hat{\mathbf{s}}_{p_{0}p_{1}} = \mathbf{s}(\hat{\mathbf{q}}_{p_{0}p_{1}}; t(\tau_{p_{0}}), x(\xi_{p_{1}})),
\end{split}
\end{equation}
where the expansion coefficients can be conveniently combined into single vector structures: $\hat{\mathbf{f}} = \{\hat{\mathbf{f}}_{\pi_{0}(\mathrm{p})\pi_{1}(\mathrm{p})}\}_{\mathrm{p}}$ and $\hat{\mathbf{s}} = \{\hat{\mathbf{s}}_{\pi_{0}(\mathrm{p})\pi_{1}(\mathrm{p})}\}_{\mathrm{p}}$, where $\mathrm{p}$ is multi-index (\ref{eq:local_sol_idx}). The calculation of flux terms at the interfaces of neighboring discretization domains $\Omega_{i}$ and $\Omega_{i \pm 1}$ is carried out using the GL quadrature rule (\ref{eq:gl_rule}) as follows:
\begin{equation}
\begin{split}
\intrefdom{\tau}\,\mathbf{f}^{\rm RP}\left(\mathbf{q}^{L}(\tau, 1),\, \mathbf{q}^{R}(\tau, 0)\right) \mapsto &
\sum\limits_{p_{0} = 0}^{N} w_{p_{0}} \mathbf{f}^{\rm RP}\left(
  \sum\limits_{q_{0} = 0}^{N}\sum\limits_{q_{1} = 0}^{N} \hat{\mathbf{q}}^{L}_{q_{0}q_{1}} \varphi_{q_{0}}(\tau_{p_{0}})\varphi_{q_{1}}(1),
  \sum\limits_{q_{0} = 0}^{N}\sum\limits_{q_{1} = 0}^{N} \hat{\mathbf{q}}^{R}_{q_{0}q_{1}} \varphi_{q_{0}}(\tau_{p_{0}})\varphi_{q_{1}}(0)
\right)\\ = &
\sum\limits_{p_{0} = 0}^{N} w_{p_{0}} \mathbf{f}^{\rm RP}\left(
  \sum\limits_{q_{0} = 0}^{N}\sum\limits_{q_{1} = 0}^{N} \hat{\mathbf{q}}^{L}_{p_{0}q_{1}} \varphi_{q_{1}}(1),
  \sum\limits_{q_{0} = 0}^{N}\sum\limits_{q_{1} = 0}^{N} \hat{\mathbf{q}}^{R}_{p_{0}q_{1}} \varphi_{q_{1}}(0)
\right).
\end{split}
\end{equation}
The system of equations for obtaining the local discrete space-time solution $q$ is considered in weak form:
\begin{equation}\label{eq:nonlin:eq_weak_lstdg}
\intrefdom{\tau}\intrefdom{\xi}\,\varphi_{k}(\tau)\varphi_{k}(\xi)\left[\pderf{\mathbf{q}}{\tau} + \pderf{\mathbf{f}(\mathbf{q})}{\xi} - \mathbf{s}(\mathbf{q}; \tau, \xi)\right] = 0,\quad
\mathbf{q}(0, \xi) = \mathbf{u}^{n}(\xi),
\end{equation}
where the initial conditions in the problem (\ref{eq:nonlin:eq_weak_lstdg}) are also considered in weak form, so a discontinuity is allowed for the solution $q(\tau, \xi)$ at the initial time $\tau = 0$. Substituting the chosen solution representation (\ref{eq:vec_local_sol_def}) into the equation (\ref{eq:nonlin:eq_weak_lstdg}) leads to the following expression:
\begin{equation}\label{eq:vec_lstdg_pred_eq_tmp_1}
\begin{split}
\sum\limits_{q_{0} = 0}^{N}\sum\limits_{q_{1} = 0}^{N} &\left[\varphi_{p_{0}}(1)\varphi_{q_{0}}(1) - \intrefdom{\tau}\,\varphi_{p_{0}}'(\tau)\varphi_{q_{0}}(\tau)\right]
\intrefdom{\xi}\,\varphi_{p_{1}}(\xi)\varphi_{q_{1}}(\xi)\cdot\hat{\mathbf{q}}_{q_{0}q_{1}}\\
&+\sum\limits_{q_{0} = 0}^{N}\sum\limits_{q_{1} = 0}^{N}
  \intrefdom{\tau}\,\varphi_{p_{0}}(\tau)\varphi_{q_{0}}(\tau)\,\intrefdom{\xi}\,\varphi_{p_{1}}(\xi)\varphi_{q_{1}}'(\xi)\cdot\mathbf{f}(\hat{\mathbf{q}}_{q_{0}q_{1}}) \\
&-\sum\limits_{q_{0} = 0}^{N}\sum\limits_{q_{1} = 0}^{N}
  \intrefdom{\tau}\,\varphi_{p_{0}}(\tau)\varphi_{q_{0}}(\tau)\,\intrefdom{\xi}\,\varphi_{p_{1}}(\xi)\varphi_{q_{1}}(\xi)\cdot
  \mathbf{s}(\hat{\mathbf{q}}_{q_{0}q_{1}}; t(\tau_{q_{0}}), x(\xi_{q_{1}})) \\ & =
\sum\limits_{k = 0}^{N} \varphi_{p_{0}}(0) \intrefdom{\xi}\,\varphi_{p_{1}}(\xi)\varphi_{k}(\xi)\cdot\hat{\mathbf{u}}^{n}_{k},
\end{split}
\end{equation}
where integration by parts over time is performed in the time term to include the initial condition $\mathbf{q}(0, \xi) = \mathbf{u}^{n}(\xi)$ in weak form in the system of equations. Using the previously introduced notations (\ref{eq:matrices_K_def}) for the matrices $\mathrm{K}^{(\tau)}$, $\mathrm{K}^{(\xi)}$, $\mathrm{M}^{(\xi)}$ and $\mathrm{F}^{(0)}$, the expression (\ref{eq:vec_lstdg_pred_eq_tmp_1}) is transformed into a fairly convenient form:
\begin{equation}
\begin{split}
\sum\limits_{q_{0} = 0}^{N}\sum\limits_{q_{1} = 0}^{N}\Bigg[
  \mathrm{K}^{(\tau)}_{p_{0}p_{1},q_{0}q_{1}} \hat{\mathbf{q}}_{q_{0}q_{1}} +
  \mathrm{K}^{(\xi)}_{p_{0}p_{1},q_{0}q_{1}} \mathbf{f}(\hat{\mathbf{q}}_{q_{0}q_{1}}) -
  \mathrm{M}^{(\xi)}_{p_{0}p_{1},q_{0}q_{1}} \mathbf{s}(\hat{\mathbf{q}}_{q_{0}q_{1}}; t(\tau_{q_{0}}), x(\xi_{q_{1}}))
\Bigg] =
\sum\limits_{k = 0}^{N} \mathrm{F}^{(0)}_{p_{0}p_{1},k} \hat{\mathbf{u}}^{n}_{k},
\end{split}
\end{equation}
which, as a result of introducing matrix notations for the rescaled variables $(\mathbf{u},\, \mathbf{f}(\mathbf{u}),\, \mathbf{s}(\mathbf{u}; \tau, \xi))$ (\ref{eq:nonlin:vars_mapped}) (``array of structures'' (AoS) in the terminology~\cite{ader_eff_blas}):
\begin{equation}
\begin{split}
&\mathbf{q} = \Big\{\left[\hat{\mathbf{q}}_{\pi_{0}(\mathrm{p}), \pi_{1}(\mathrm{p})}\right]_{i}\Big\}_{\mathrm{p}, i} \in \mathcal{R}^{(N+1)^{2} \times L},\qquad
 \mathbf{u} = \Big\{\left[\hat{\mathbf{u}}^{n}_{k}\right]_{i}\Big\}_{k, i} \in \mathcal{R}^{(N+1) \times L},\\
&\mathbf{f}(\mathbf{q}) = \Big\{\left[
 \mathbf{f}\left(\hat{\mathbf{q}}_{\pi_{0}(\mathrm{p}), \pi_{1}(\mathrm{p})}\right)
 \right]_{i}\Big\}_{\mathrm{p}, i}: \mathcal{R}^{(N+1)^{2} \times L} \rightarrow \mathcal{R}^{(N+1)^{2} \times L},\\
&\mathbf{s}(\mathbf{q}) = \Big\{\left[
 \mathbf{s}\left(\hat{\mathbf{q}}_{\pi_{0}(\mathrm{p}), \pi_{1}(\mathrm{p})};\, t\left(\tau_{\pi_{0}(\mathrm{p})}\right), 
 x\left(\xi_{\pi_{1}(\mathrm{p})}\right)\right)
 \right]_{i}\Big\}_{\mathrm{p}, i} : \mathcal{R}^{(N+1)^{2} \times L} \rightarrow \mathcal{R}^{(N+1)^{2} \times L},
\end{split}
\end{equation}
is transformed to a compact form:
\begin{equation}
\begin{split}
\mathrm{K}^{(\tau)} \mathbf{q} + \mathrm{K}^{(\xi)} \mathbf{f}(\mathbf{q}) - \mathrm{M}^{(\xi)} \mathbf{s}(\mathbf{q}) = \mathrm{F}^{(0)} \mathbf{u}^{n},
\end{split}
\end{equation}
which is transformed into a form convenient for using the Picard iterative method:
\begin{equation}
\begin{split}
\mathbf{q} + \left\{\left[\mathrm{K}^{(\tau)}\right]^{-1}\mathrm{K}^{(\xi)}\right\}\mathbf{f}(\mathbf{q}) -
\left\{\left[\mathrm{K}^{(\tau)}\right]^{-1}\mathrm{M}^{(\xi)}\right\}\mathbf{s}(\mathbf{q}) =
\left\{\left[\mathrm{K}^{(\tau)}\right]^{-1}\mathrm{F}^{(0)}\right\}\mathbf{u}^{n},
\end{split}
\end{equation}
which, as a result of using the relations (\ref{eq:Kinv_mult_KMF}) for the matrices $\mathrm{K}^{(\tau)}$, $\mathrm{K}^{(\xi)}$, $\mathrm{M}^{(\xi)}$ and $\mathrm{F}^{(0)}$ takes the following form:
\begin{equation}
\begin{split}
\mathbf{q} + \Big[\chi \otimes \mathrm{D}\Big]\mathbf{f}(\mathbf{q}) - \Big[\chi \otimes I\Big]\mathbf{s}(\mathbf{q}) = \mathbf{1} \otimes \mathbf{u}^{n},
\end{split}
\end{equation}
which is friendly for using the matrix-matrix multiplication-addition function \texttt{gemm} of the \texttt{BLAS} interface~\cite{ader_eff_blas}. In the case of source terms $\mathbf{S}(\mathbf{U}; x, t)$ characterized by strong stiffness, Picard iterative method takes the following form:
\begin{equation}\label{eq:nonlin:picard}
\begin{split}
\mathbf{q}^{(l+1)} - \Big[\chi \otimes I\Big]\mathbf{s}\left(\mathbf{q}^{(l+1)}\right) = 
\mathbf{1} \otimes \mathbf{u}^{n} - \Big[\chi \otimes \mathrm{D}\Big]\mathbf{f}\left(\mathbf{q}^{(l)}\right).
\end{split}
\end{equation}
In this case, if the system of nonlinear algebraic equations on the left-hand side of (\ref{eq:nonlin:picard}) is correctly resolved with respect to the desired $\mathbf{q}^{(l+1)}$, the iterative process (\ref{eq:nonlin:picard}) is guaranteed to converge according to the Banach fixed-point theorem~\cite{Zanotti_lects_2016}, due to the nilpotency of the matrix $\mathrm{D}$: $\mathrm{D}^{N+1} = 0$, since all its eigenvalues $\sigma(\mathrm{D})$ are zero. In~\cite{Jackson_2017}, Newton's method is recommended for solving the system of nonlinear equations on the left-hand side (\ref{eq:nonlin:picard}) in the case of strongly stiff source terms $\mathbf{S}(\mathbf{U}; x, t)$. However, in this work, the solution $\mathbf{q}^{(l+1)}$ of the system of nonlinear equations on the left-hand side (\ref{eq:nonlin:picard}) is also calculated using an additional internal iterative Picard method, due to the violation of the smoothness of the functional dependence of the source term function $\mathbf{S}(\mathbf{U}; x, t)$ on conservative variables $\mathbf{U}$.

Unlike the case of linear problems (\ref{eq:adv_eq_src}) and (\ref{eq:adv_eq_s_src}), in the case of a quasilinear hyperbolic system of equations (\ref{eq:nonlin_sys_source}), the violation of the smoothness of the solution $\mathbf{U}(x, t)$ and the formation of discontinuities in the solution can occur even under smooth initial conditions~\cite{Rozhdestvenskii_1983, Kulikovskii_2019, Handbook_NMHP_Basic_2016, Handbook_NMHP_Applied_2017}, which is a well-known feature of these nonlinear problems. Therefore, the ADER-DG numerical method with the LST-DG predictor, which is a fundamentally linear method of arbitrarily high order (and according to the well-known Godunov theorem, there are no linear monotone numerical methods higher than first order for hyperbolic problems), is generally unsuitable for solving a quasilinear hyperbolic system of equations (\ref{eq:nonlin_sys_source}) in the general case of arbitrary initial conditions (there are initial conditions for which the solution retains monotonicity over time $t$, for example, the case of simple linear advection or the advection of an isentropic vortex) without some modifications associated with the correction of the solution when the monotonicity of the numerical solution $\mathbf{u}^{n+1}$ is violated. In the present work, it is used the classical approach~\cite{ader_dg_dev_1, ader_dg_dev_2, ader_dg_ideal_flows, ader_dg_diss_flows, ader_dg_grmhd, ader_dg_gr_prd, ader_dg_gr_z4_2024, ader_dg_PNPM, ader_dg_axioms} based on a posteriori correction of the numerical solution by a robust finite-volume limiter, which is a development of the well-proven MOOD paradigm~\cite{mood_par_1, mood_par_2, mood_par_3, mood_par_4}. A detailed description of the approach used in this work is presented in~\cite{ader_dg_dev_1, ader_dg_dev_2, ader_dg_ideal_flows, ader_dg_diss_flows} and in~\cite{ader_stiff_3, ader_stiff_4, ader_eff_blas}. A brief description of this approach can be formulated as follows: the numerical solution at a new time step $t^{n+1}$, obtained by the ADER-DG numerical method with the LST-DG predictor, is the so-called candidate solution $\mathbf{u}^{*}$, for which either an already existing monotonicity violation may occur, or based on the use of this solution in subsequent time steps this monotonicity violation will occur; therefore, a set of criteria is used for this candidate solution $\mathbf{u}^{*}$ to determine its admissibility; in cells where the candidate solution is not admissibility (so-called troubled cells), a highly accurate correction of the numerical solution is performed using a robust finite-volume constraint, which obtains the solution in subcells of the original cells (this preserves the high-precision subgrid resolution of the ADER-DG numerical method with the LST-DG predictor). Two main criteria are used as admissibility criteria for the candidate numerical solution $\mathbf{u}^{*}$: a physical admissibility detector (PAD) and the numerical admissibility detector (NAD), which is based on the use of the relaxed discrete maximum principle (DMP) in the polynomial sense. To describe the cell state, a troubled cells indicator $\beta$ is used, which is equal to $0$ for a cell with an admissible candidate numerical solution $\mathbf{u}^{*}$ and equal to $1$ for a troubled cell. In the cells where the candidate numerical solution is not admissible according to at least one admissibility criterion, a subgrid is created, consisting of subcells onto which the numerical solution $\mathbf{u}^{n}$ from the previous time step $t^{n}$ is projected (the candidate numerical solution $\mathbf{u}^{*}$ at the new time step $t^{n+1}$ is not further used in these cells) onto the subcells using the suitable piecewise-constant projection operator $\hat{\mathbb{P}}$. Then, in the created subgrid, a finite-volume limiter is called for each subcell. In this work, the second-order finite-volume ADER-WENO scheme with the LST-DG predictor is selected as this limiter~\cite{ader_weno_lstdg_ideal, ader_weno_lstdg_diss}. The numerical solution obtained by the finite-volume limiter at the new time step in the entire subgrid is then projected onto the DG representation of the solution $\mathbf{u}^{n+1}$ at the new time step $t^{n+1}$ using the suitable high order accurate reconstruction operator $\hat{\mathbb{R}}$, for which the relation holds $\hat{\mathbb{R}}\cdot\hat{\mathbb{P}} = 1$. Due to the property $\hat{\mathbb{P}}\cdot\hat{\mathbb{R}} \neq 1$ of the operators $\hat{\mathbb{R}}$ and $\hat{\mathbb{P}}$, in cases where a cell retains the troubled cells indicator $\beta = 1$ over several time steps, the finite-volume solution at the previous time step $t^{n}$, used to calculate the solution by the finite-volume solution limiter at a new time step $t^{n+1}$, is taken to be equal to the finite-volume solution obtained at the previous time step $t^{n}$, and not based on the suitable piecewise-constant projection of the DG representation of the solution $\mathbf{u}^{n}$.  It is usually accepted~\cite{ader_dg_dev_1, ader_dg_dev_2, ader_dg_ideal_flows, ader_dg_diss_flows} to split the cell into subcells with $N_{s} = 2N+1$ subcells for each Cartesian coordinate (in the one-dimensional case, only for one Cartesian coordinate), which is associated with the estimation of the Courant number $\mathrm{CFL}$ using the classical Cockburn-Shu formula (\ref{eq:cfl_cs}) for classical DG methods~\cite{Cockburn_base_1, Cockburn_base_2, Cockburn_base_3, Cockburn_base_4, Cockburn_base_5}. In this work, despite the known overestimation of this estimate specifically for the ADER-DG numerical method with the LST-DG predictor~\cite{ader_dg_stab_rig, ader_dg_stab}, this generally accepted approach is used, as it has demonstrated its applicability in solving problems of modeling reactive flows with detonation waves~\cite{ader_stiff_3, ader_stiff_4}, which are characterized by strong stiffness. It is clear that technically this set of actions is implemented significantly differently, in particular, subgrids are usually already created for all mesh cells; the finite-volume representation in subcells is also calculated for all cells, since it is used to calculate and evaluate expressions for the admissibility criteria (for which, in particular, a certain tolerance window~\cite{ader_dg_dev_1, ader_dg_dev_2} is introduced into the relaxed discrete maximum principle in the polynomial sense); also, to calculate the finite-volume numerical solution in subcells, the finite-volume limiter also requires subcell finite-volume values from the subcells of neighboring cells (so-called halo subcells). It is also necessary to note an important technical aspect: in the case when a cell with the troubled cells indicator $\beta = 0$, in which the numerical solution $\mathbf{u}^{n+1}$ at the new time step $t^{n+1}$ is chosen equal to the candidate numerical solution $\mathbf{u}^{*}$, borders on a troubled cell (with the troubled cells indicator $\beta = 1$), then it is necessary, after calculating the finite-volume solution at the new time step $t^{n+1}$ in the troubled cell, to also recalculate the solution $\mathbf{u}^{n+1}$ in this cell with the troubled cells indicator $\beta = 0$ --- it is necessary to choose the flux at the interface with the troubled cell equal to the flux of the finite-volume method, but with the opposite sign (in the one-dimensional case, this requires the flow only in the subcell closest to the interface), and recalculate the flux term in a completely single-step ADER-DG scheme (\ref{eq:vec_ader_dg_corr}) --- this is necessary to maintain the conservatism of the ADER-DG numerical method with the LST-DG predictor. A more detailed description is presented in~\cite{ader_dg_axioms, exahype, fron_phys, ader_dg_hpc_impl_1, ader_dg_hpc_impl_2, ader_dg_hpc_impl_3, ader_dg_hpc_impl_4, ader_eff_blas}.

As an application of the ADER-DG numerical method with the LST-DG predictor to solving quasilinear hyperbolic systems (\ref{eq:nonlin_sys_source}) with a source term characterized by strong stiffness, the problem of describing the formation and development of a detonation wave in a two-component gas medium with a monomolecular reaction $A \rightarrow B$, where $A$ is the reactant and $B$ is the reaction product, is chosen. This problem, on the one hand, is fairly well studied and classical~\cite{Lee_2008, OranBoris_2005}, and on the other hand, is a very good classical test for studying the applicability of numerical methods of computational fluid dynamics to solving problems of modeling multicomponent reacting flows~\cite{frac_steps_detwave_sim_2000, chem_kin_hrs_rev_1, chem_kin_hrs_rev_2, chem_kin_hrs_weno, RTS_split_2019, correct_det_wave_speed_2017}. The main point is the formation of a non-physical artifact in the form of a weak detonation wave traveling ahead of the main shock wave and ``burning'' the reactant $A$ in front of it~\cite{frac_steps_detwave_sim_2000, correct_det_wave_speed_2017}. This leads to an incorrect description of the detonation front as such --- its integrity is violated, where there should be a clear leading shock front and a subsequent domain of reactionary conversion of reactants $A$ into products $B$ with the release of energy, which should be characterized by a classical chemical peak (the Zeldovich peak), which is one of the main features of such ZND detonation waves~\cite{Lee_2008} (Zel'dovich, von Neumann, and D\"{o}ring). It should also be noted that the use of classical numerical methods of computational fluid dynamics in the case of studying reacting flows with fast reaction processes usually requires the inclusion in the numerical method of some method of splitting into physical processes~\cite{frac_steps_detwave_sim_2000}, similar to the Marchuk-Strang splitting, when hydrodynamic phenomena and the kinetics of reaction processes are calculated at individual steps of the splitting method. This approach is one of the main ones in applied methods for the numerical study of detonation processes. However, it is well known that the use of splitting methods by physical processes leads to a decrease in the order of approximation and the order of convergence of the numerical method, usually to the first order; possible symmetrization of the splitting method to increase the effective order of approximation for use in conjunction with computational fluid dynamics methods is usually too complex and not applied, and can also lead to other, possibly more complex and unexpected, artifacts of the numerical solution~\cite{chem_kin_hrs_weno}. Even in the case of using high-order numerical methods, one often has to resort to the idea of using methods of splitting into physical processes~\cite{chem_kin_hrs_weno}. In this study of the ADER-DG numerical method with the LST-DG predictor, no methods of splitting into physical processes are used.

The quasilinear hyperbolic system of equations (\ref{eq:nonlin_sys_source}) with a source term for describing the flow of the two-component gas medium with a monomolecular reaction is chosen in the following form:
\begin{equation}\label{eq:euler_react_eqs_def}
\pderfoper{t}\left[
\begin{array}{c}
\rho\\ \rho u\\ \varepsilon\\ \rho c_{1}\\ \rho c_{2}
\end{array}
\right]\,+\,\pderfoper{x}\left[
\begin{array}{c}
\rho u\\ \rho u^{2} + p\\ (\varepsilon + p)u\\ \rho c_{1} u\\ \rho c_{2} u
\end{array}
\right] = \left[
\begin{array}{c}
0\\ 0\\ \rho \vartheta q_{0}\\ -\rho\vartheta \\ +\rho \vartheta
\end{array}
\right],
\end{equation}
where $\rho$ is the density, $u$ is the velocity, $p$ is the pressure, $\varepsilon = e + \frac12 \rho u^{2}$ is the energy density, $e$ is the internal energy density, $c_{1}$ and $c_{2}$ are the mass concentrations of the reactant $A$ and product $B$, respectively, which are determined through the partial densities $\rho_{1} = \rho c_{1}$ and $\rho_{2} = \rho c_{2}$ of the reactant and product, respectively, $\vartheta$ is the rate of the monomolecular reaction $A \rightarrow B$ occurring in the two-component reacting medium, and $q_{0}$ is the specific energy yield of the reaction. The caloric equation of state of the medium is chosen in the form of the equation of state for a perfect gas $p = (\gamma - 1)e$, with an adiabatic constant $\gamma = 1.4$. The kinetic function $\vartheta$ of a monomolecular reaction $A \rightarrow B$ occurring in a two-component reacting medium is chosen in the form of the law of mass action $\vartheta = k(T) c_{1}$, where $k(T)$ is the reaction rate constant, $T$ is the temperature, for which a discrete ignition temperature dependence is chosen:
\begin{equation}
k(T) = \left\{
\begin{split}
& \frac{1}{\tau_{0}},\hphantom{\ 0} T \geqslant T_{\rm ign},\\
& \ 0,\hphantom{\frac{1}{\tau_{0}}} T < T_{\rm ign},
\end{split}
\right.
\end{equation}
where $\tau_{0}$ is the time scale of the chemical reaction (some induction time), and $T_{\rm ign}$ is the initiation temperature. Dimensional scales and parameters of medium (such as molar mass $\mu$) are chosen such that $T = p/\rho$. The specific energy yield of the reaction in these units is chosen $q_{0} = 1$. Two main demonstration limiting cases were chosen: ``slow'' reaction kinetics with $\tau_{0} = 10^{-1}$ and ``fast'' reaction kinetics with $\tau_{0} = 4\cdot10^{-3}$, which correspond to weak stiffness and strong stiffness occurring in the system (\ref{eq:euler_react_eqs_def}), respectively. This choice is based on the results presented in the works~\cite{frac_steps_detwave_sim_2000, ader_stiff_3, ader_stiff_4}, and allows us to correctly determine the capabilities of the numerical method under study in a wide range of parameters of the reacting medium. It can be said that in the case of weak stiffness, the use of special methods for directly simulating a detonation wave is not required --- classical numerical methods of computational fluid dynamics, adapted for solving problems of weak stiffness, will allow one to obtain a correct numerical solution. In the case of strong stiffness, it is necessary to be quite strict in the choice of the numerical method. The quasilinear hyperbolic systems (\ref{eq:euler_react_eqs_def}) is a system of Euler equations, an expanded system of quasilinear advection-reaction equations to account for the multicomponent composition and reactive properties of the medium; this model of multicomponent reacting flows is common in practical calculations~\cite{OranBoris_2005, Lunev_2017, Anderson_Hypersonic_2019, Nagnibeda_2009, Fortov_ExtremeStates_2011, Fortov_HEDP_2016}. The last two equations of the Euler system of equations for the reacting medium (\ref{eq:euler_react_eqs_def}) can be represented in the following form:
\begin{equation}
\pderf{(\rho\mathbf{c})}{t} + \pderf{(\rho\mathbf{c} u)}{x} = \rho\boldsymbol{\vartheta},
\end{equation}
where $\mathbf{c} = [c_{1}\, c_{2}]^{T}$ is the vector of reaction components, and $\boldsymbol{\vartheta} = [-\vartheta\, +\!\vartheta]^{T}$ is the vector of reaction kinetic functions; then, taking into account the continuity equation --- the first equation of the system of equations (\ref{eq:euler_react_eqs_def}), the last two equations of the system of equations (\ref{eq:euler_react_eqs_def}) reduces to the form of a quasilinear advection-reaction equation:
\begin{equation}
\pderf{\mathbf{c}}{t} + u\pderf{\mathbf{c}}{x} = \boldsymbol{\vartheta}.
\end{equation}
It should also be noted that the admissibility criteria used for the candidate numerical solution in this case of the system of equations are chosen in the form described in~\cite{ader_stiff_3, ader_stiff_4}, where the physical admissibility detector was also expanded to include a condition on mass concentrations $c_{k}\in[0, 1]$ --- cells with a mass concentration $c_{k} < 0$ or $c_{k} > 1$ were designated as troubled cells by the PAD. It is important to note the following feature of the implementation of the ADER-DG numerical method with the LST-DG predictor and a posteriori correction of the solution by finite-volume limiter used in this work, compared to the implementations used in~\cite{ader_stiff_3, ader_stiff_4}. In~\cite{ader_stiff_3}, this method based on decreasing the global discretization time step $\Dtn{n}$ with increasing stiffness created by the source terms (\ref{eq:nonlin_sys_source}) was used to obtain a numerical solution, which allowed us to obtain a very high-quality numerical solution. In~\cite{ader_stiff_4}, the approach proposed in~\cite{ader_stiff_3} was significantly modified, allowing the use of local discretization time steps $\Dtn{n}$ without requiring a decrease in the global discretization time step $\Dtn{n}$, which is in some sense equivalent to adaptive mesh refinement (AMR), but only for the time, rather than space coordinate. In the case~\cite{ader_stiff_4}, the division of the discretization time $\Dtn{n}$ step was carried out only to obtain a local discrete space-time solution $\mathbf{q}$ (\ref{eq:vec_local_sol_def}) using the LST-DG predictor, and not to calculate the final DG representation of the solution $\mathbf{u}^{n+1}$ (\ref{eq:vec_u_n_def}) (which, in the context of the paradigm of a posteriori correction of the numerical solution, is a candidate numerical solution $\mathbf{u}^{*}$). In this work, the investigation of the capabilities of the ADER-DG numerical method with the LST-DG predictor is carried out rather than the possible related refinement procedures, the influence of which is usually very difficult to strictly isolate. Therefore, in this work, the precisely numerical ADER-DG method with the LST-DG predictor and a posteriori correction of the solution using a finite-volume limiter is used, without using any special procedures for correcting the numerical solution depending on the stiffness created by the source terms (\ref{eq:nonlin_sys_source}). Therefore, the software implementations of the ADER-DG numerical method with the LST-DG predictor and a posteriori correction of the solution by a finite-volume limiter, developed in~\cite{ader_stiff_3, ader_stiff_4, ader_eff_blas}, are used, but the procedures implementing the correction of the discretization time step $\Dtn{n}$ (global or local) are disabled. The following initial and final times are chosen: $t_{0} = 0.0$ and $t_{f} = 0.4$. The boundary conditions are specified in the form of free outflow conditions. It is important to emphasize the following feature: the above-conducted stability study of the ADER-DG numerical method with the LST-DG predictor in Section~\ref{sec:stab_anal} ``Stability analysis'' is based on the case of periodic boundary conditions; however, in this application, the case of free-flow boundary conditions is considered. Clearly, the transition to other boundary conditions may affect the boundary values of the Courant number $\mathrm{CFL}_{\rm max}$; however, typically, all stability studies~\cite{ader_dg_stab_rig, ader_dg_stab, ader_dg_PNPM, PNPM_DG_2008} of numerical methods similar to the ADER-DG numerical method with the LST-DG predictor studied in this work are performed for periodic boundary conditions, and the results are subsequently applied to other types of boundary conditions. Typically, when using coordinate meshes with a large number of cells, the spectrum of the problem with non-periodic boundary conditions can be quite close to the spectrum of the problem with periodic boundary conditions. It is clear that the results of the stability analysis must also be verified using examples of more common problems, which is the problem studied in this Subsection. The initial conditions are chosen in the form of the Chapman-Jouguet (CJ) conditions, which define a stationary detonation wave in the instantaneous detonation approximation, which were calculated in the work~\cite{frac_steps_detwave_sim_2000}:
\begin{equation}\label{eq:cjdw_1d_init}
\begin{split}
&\rho(x, t_{0}) = \left\{
\begin{array}{ll}
1.4, & \mathrm{if}\ x \leqslant 0.5; \\
0.887565, & \mathrm{if}\ x >\, 0.5; \\
\end{array}
\right.\\
&u(x, t_{0}) = \left\{
\begin{array}{ll}
0.0, & \mathrm{if}\ x \leqslant 0.5; \\
-0.577350, & \mathrm{if}\ x >\, 0.5; \\
\end{array}
\right.\\
&p(x, t_{0}) = \left\{
\begin{array}{ll}
1.0, & \mathrm{if}\ x \leqslant 0.5; \\
0.191709, & \mathrm{if}\ x >\, 0.5; \\
\end{array}
\right.
\end{split}\qquad\qquad
\begin{split}
&c_{1}(x, t_{0}) = \left\{
\begin{array}{ll}
1.0, & \mathrm{if}\ x \leqslant 0.5; \\
10^{-14}, & \mathrm{if}\ x >\, 0.5; \\
\end{array}
\right.\\
&c_{2}(x, t_{0}) = \left\{
\begin{array}{ll}
10^{-14}, & \mathrm{if}\ x \leqslant 0.5; \\
1.0, & \mathrm{if}\ x >\, 0.5; \\
\end{array}
\right.\\
&\rho_{1}(x, t_{0}) = \rho(x, t_{0}) c_{1}(x, t_{0}),\\
&\rho_{2}(x, t_{0}) = \rho(x, t_{0}) c_{2}(x, t_{0}),\\
\end{split}
\end{equation}
where it is assumed that unburned gas flows onto the burnt gas, which leads to the formation of a detonation wave propagating through the unburnt gas to the right. A small value $10^{-14}$ of mass concentrations $c_{1}$ and $c_{2}$, instead of strictly $0$, was chosen to prevent the occurrence of negative concentrations immediately at the start of the calculation process, which could lead to a meaninglessly large increase in the number of troubled cells in the solution. From the point of view of energy release and flow energy balance, these values of reagent concentration do not have any significant effect. The approximate speed of the detonation wave $D = 1$ for these initial conditions in the case of a ``fast'' reaction, when the classical ZND detonation structure is formed, taking into account the speed of the hydrodynamic flow~\cite{frac_steps_detwave_sim_2000}; the exact value $D = 1$ occurs in the case of instantaneous detonation. The reference solution in cases of ``slow'' and ``fast'' reactions was obtained using a finite-volume ADER-WENO2 method with LST-DG predictor on a spatial mesh of finite-volume $5000$ cells --- the works~\cite{ader_stiff_2, ader_stiff_4} showed that the finite-volume ADER-WENO method allows one to obtain a correct numerical solution without using additional procedures for recalculating the solution, so this method was chosen as the method for obtaining a reference solution. The cases of ``slow'' and ``fast'' reactions, corresponding to the cases of weak and strong stiffness of the system of equations, are considered. Courant numbers for calculations are selected in the form $\mathrm{CFL} = \mathrm{C}\cdot\mathrm{CFL}_{\rm CS}(N)$ (\ref{eq:cfl_cs}), where $\mathrm{C}$ is not less than $0.9$, which corresponds to the value used in~\cite{ader_stiff_4}, which was applied to solve similar problems of detonation wave formation and propagation. The following values of the effective Courant number $\mathrm{C}$ are selected: $0.90$, $1.00$, $1.05$, $1.10$, $1.15$, $1.20$, $1.25$, $1.30$, $1.35$, the Courant numbers $\mathrm{CFL} = \mathrm{C}\cdot\mathrm{CFL}_{\rm CS}(N)$ for which are greater than the boundary value of the Courant number $\mathrm{CFL}^{\rm NS}_{\rm max}(N) \equiv \mathrm{CFL}_{\rm max}(N, 0)$ for a system of equations without sources in all cases except $N = 1$ and $N = 2$ with effective Courant number $\mathrm{C} = 0.90$.

\begin{figure}[h!]
\captionsetup[subfigure]{%
	position=bottom,
	font+=smaller,
	textfont=normalfont,
	singlelinecheck=off,
	justification=raggedright
}
\centering
\begin{subfigure}{0.24\textwidth}
\includegraphics[width=\textwidth]{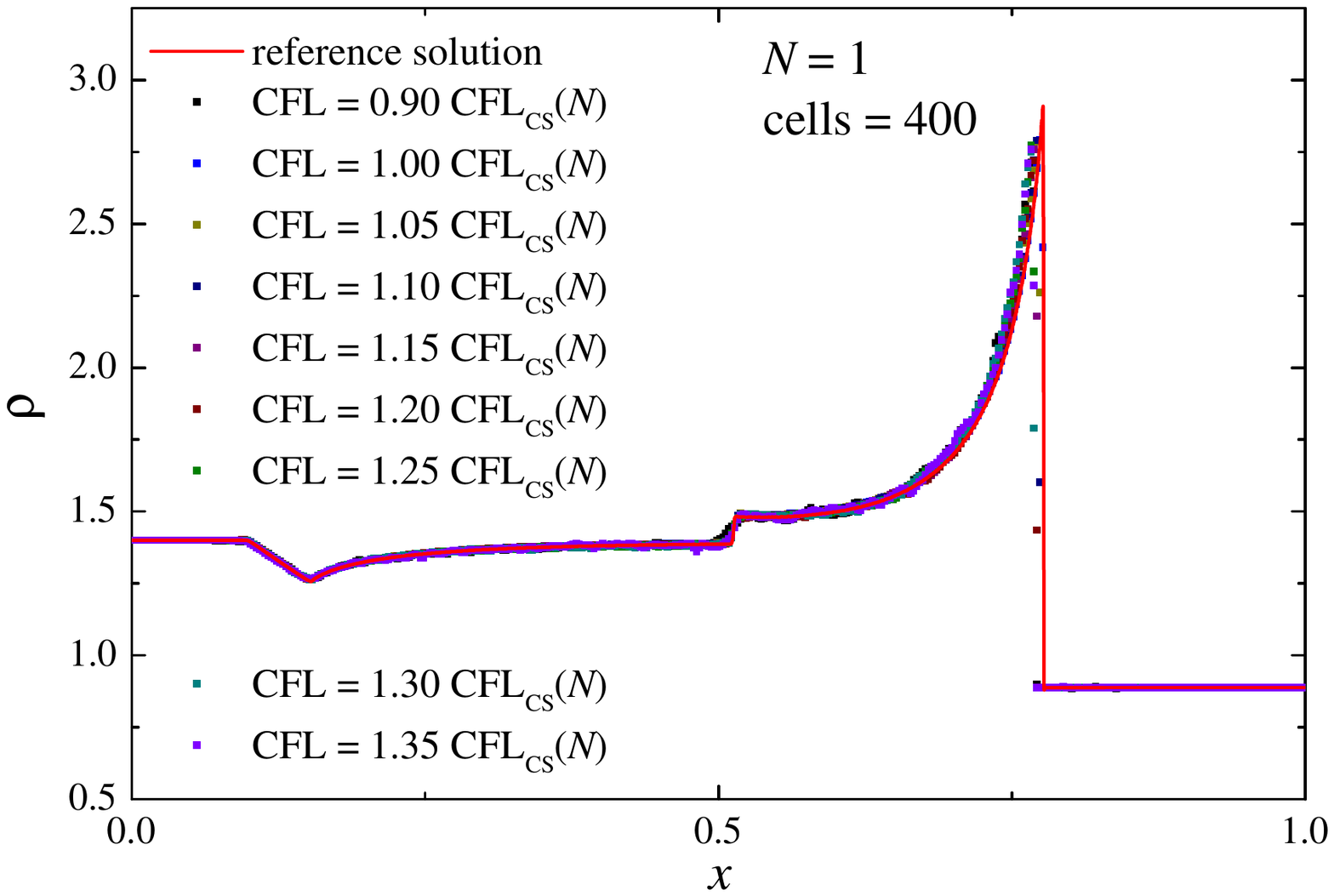}
\vspace{-8mm}\caption{\label{fig:test_znds_1d_degrees_1_3:degree_1_den}}
\end{subfigure}
\begin{subfigure}{0.24\textwidth}
\includegraphics[width=\textwidth]{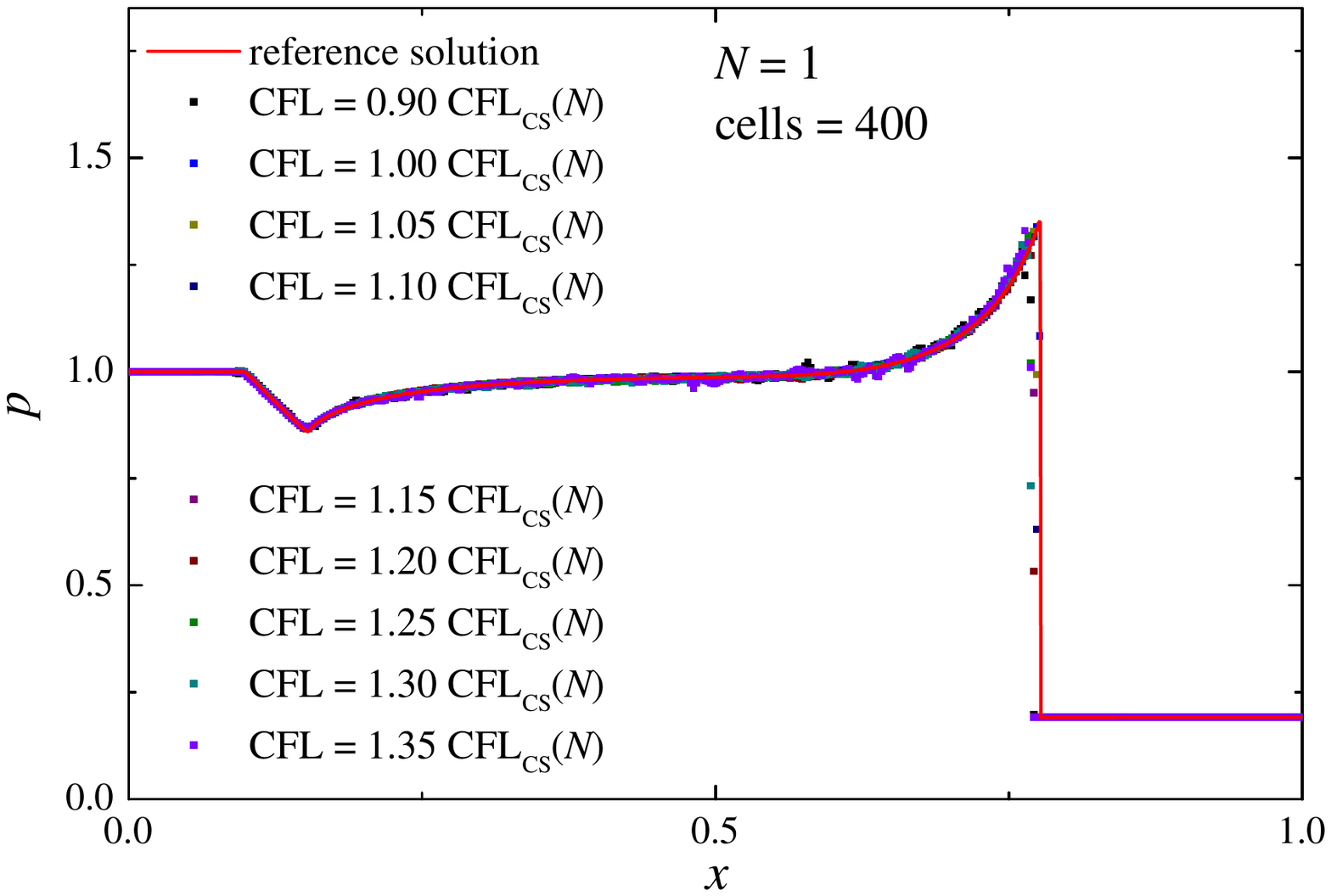}
\vspace{-8mm}\caption{\label{fig:test_znds_1d_degrees_1_3:degree_1_p}}
\end{subfigure}
\begin{subfigure}{0.24\textwidth}
\includegraphics[width=\textwidth]{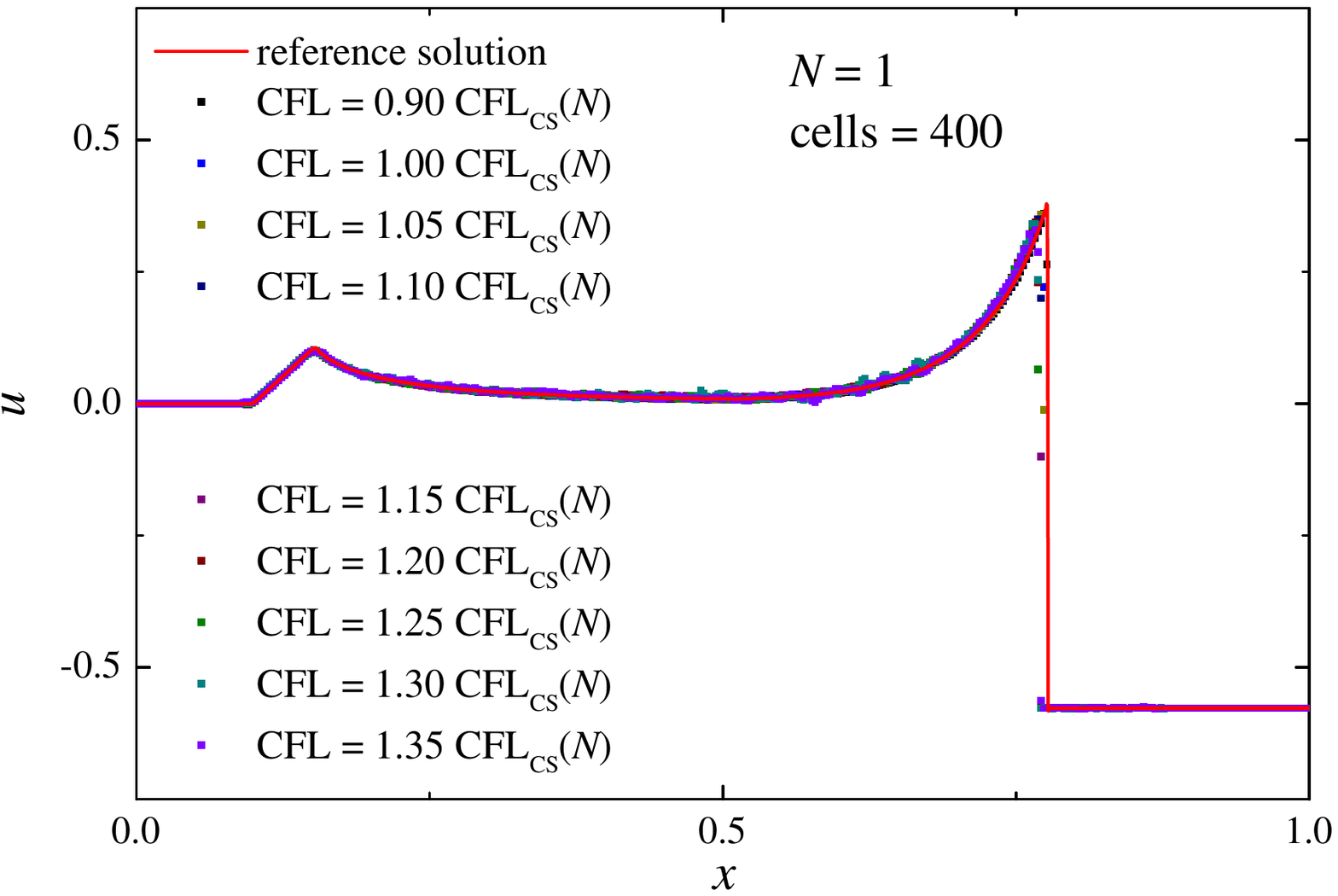}
\vspace{-8mm}\caption{\label{fig:test_znds_1d_degrees_1_3:degree_1_u}}
\end{subfigure}
\begin{subfigure}{0.24\textwidth}
\includegraphics[width=\textwidth]{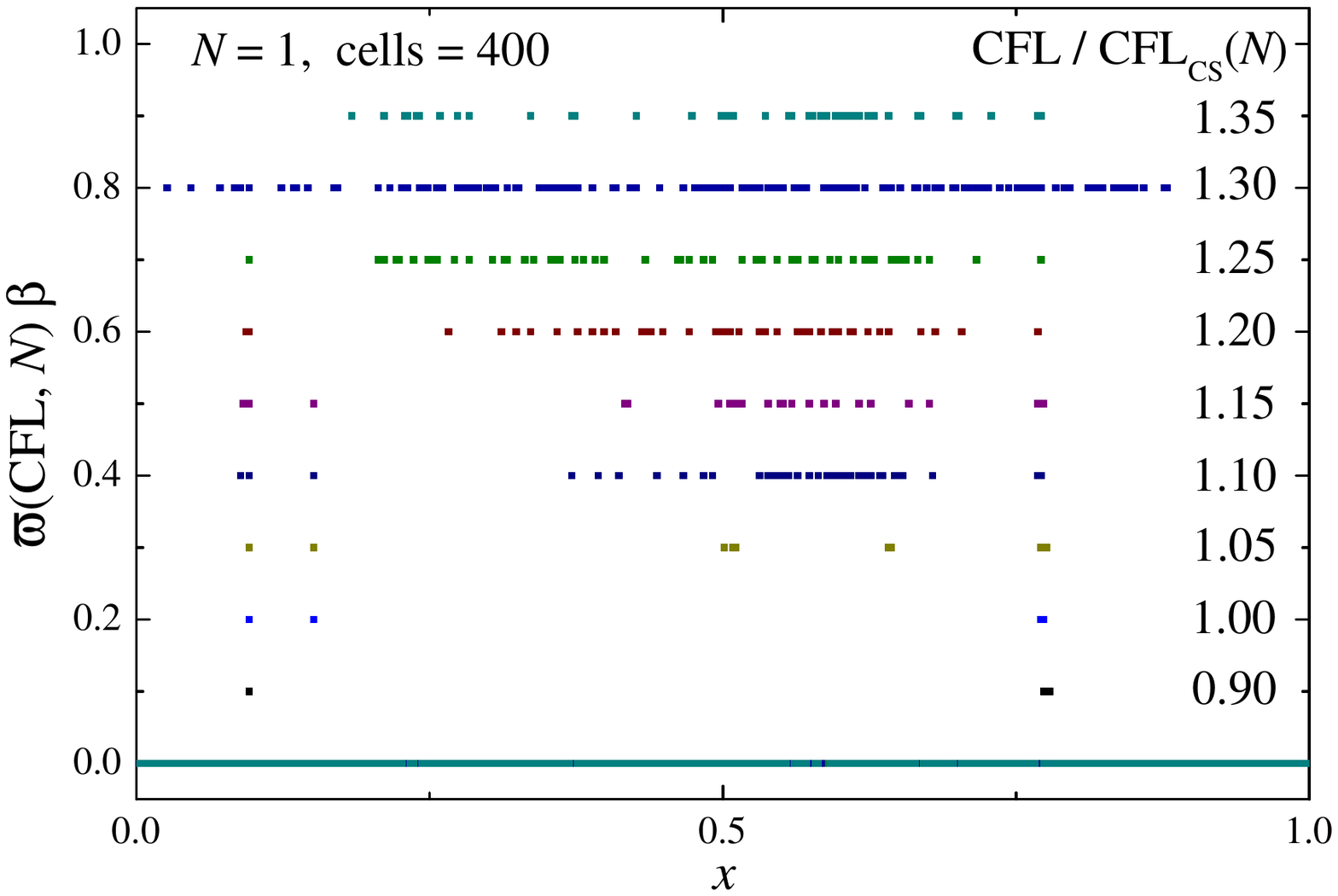}
\vspace{-8mm}\caption{\label{fig:test_znds_1d_degrees_1_3:degree_1_beta}}
\end{subfigure}\\[-2mm]
\begin{subfigure}{0.24\textwidth}
\includegraphics[width=\textwidth]{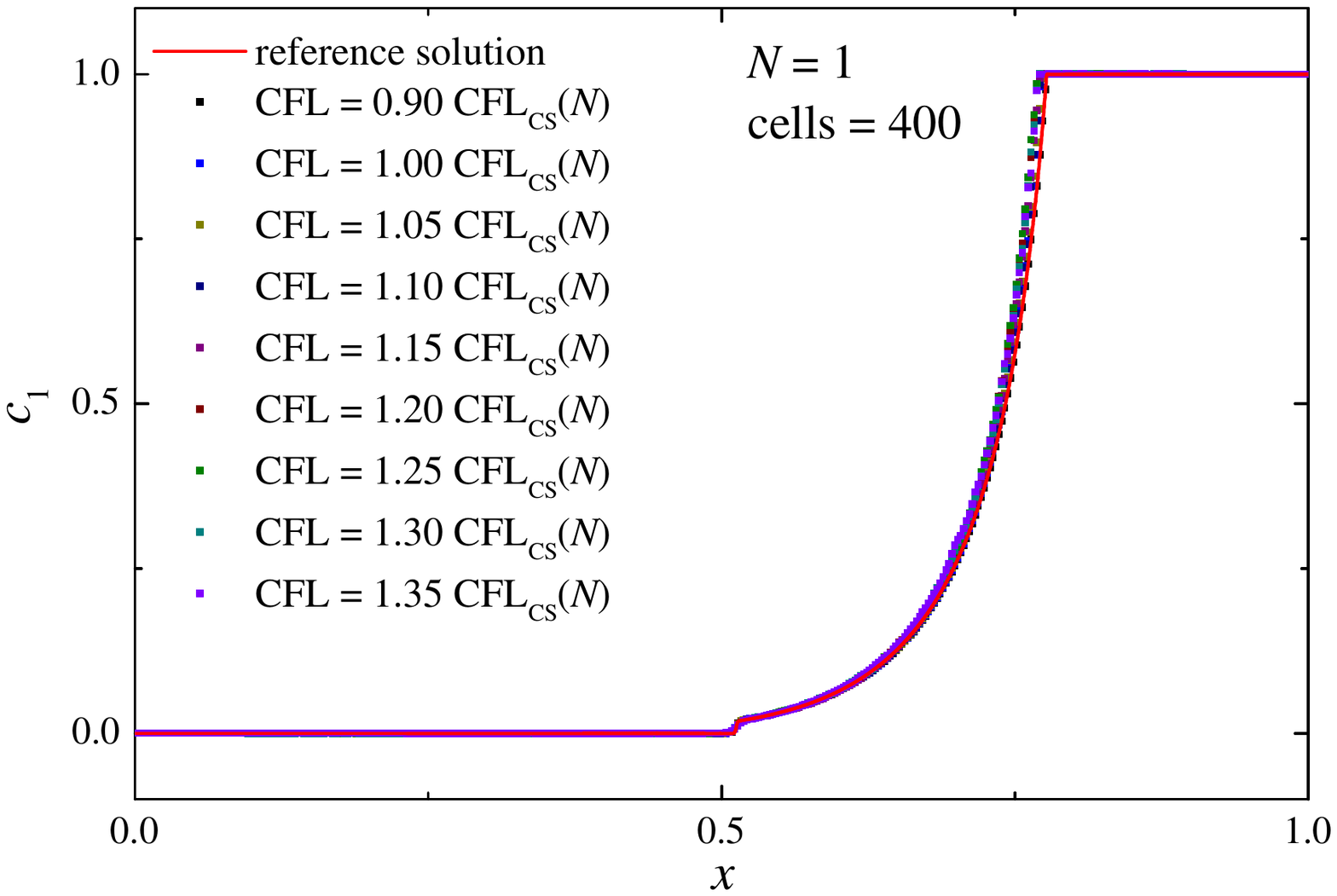}
\vspace{-8mm}\caption{\label{fig:test_znds_1d_degrees_1_3:degree_1_c1}}
\end{subfigure}
\begin{subfigure}{0.24\textwidth}
\includegraphics[width=\textwidth]{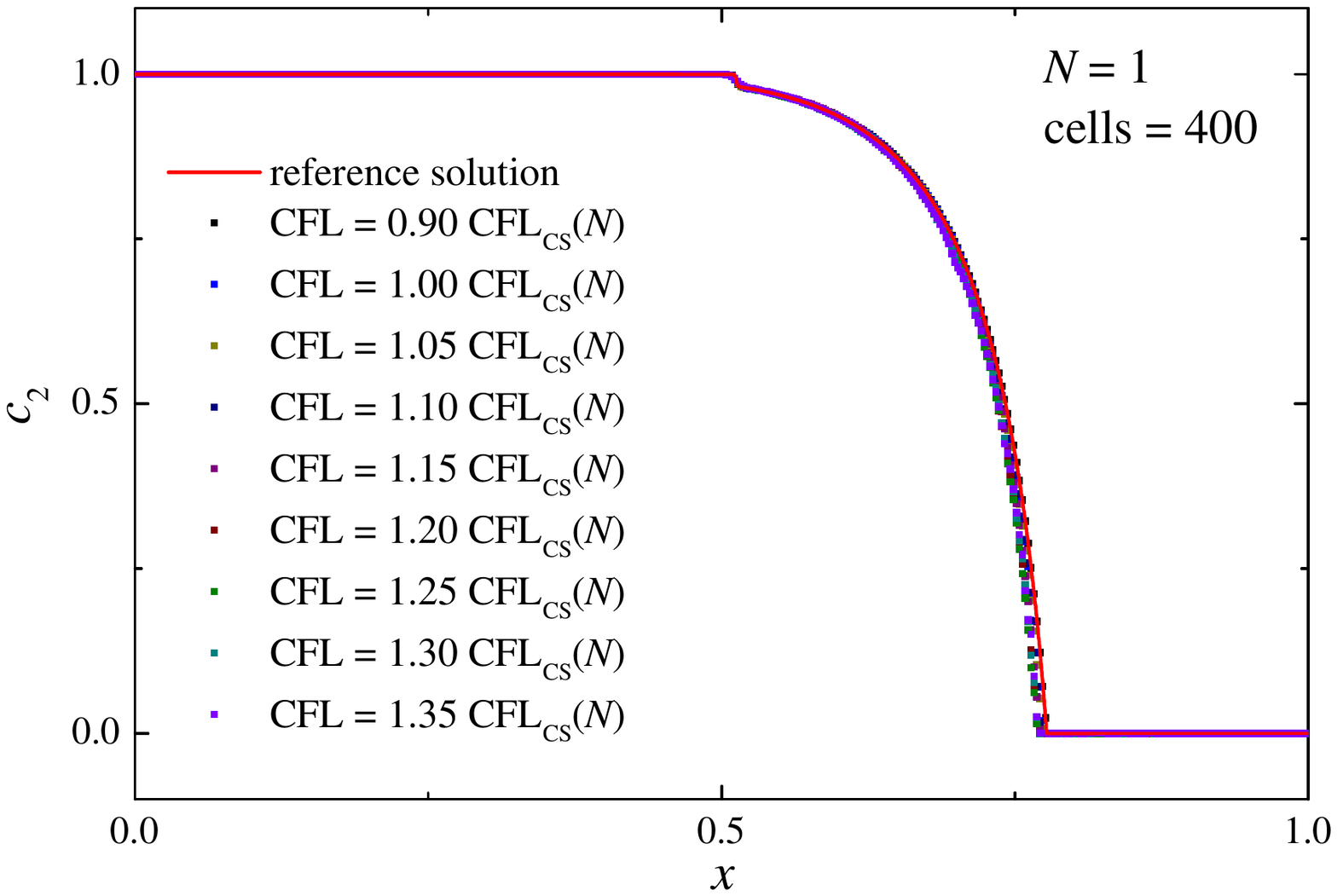}
\vspace{-8mm}\caption{\label{fig:test_znds_1d_degrees_1_3:degree_1_c2}}
\end{subfigure}
\begin{subfigure}{0.24\textwidth}
\includegraphics[width=\textwidth]{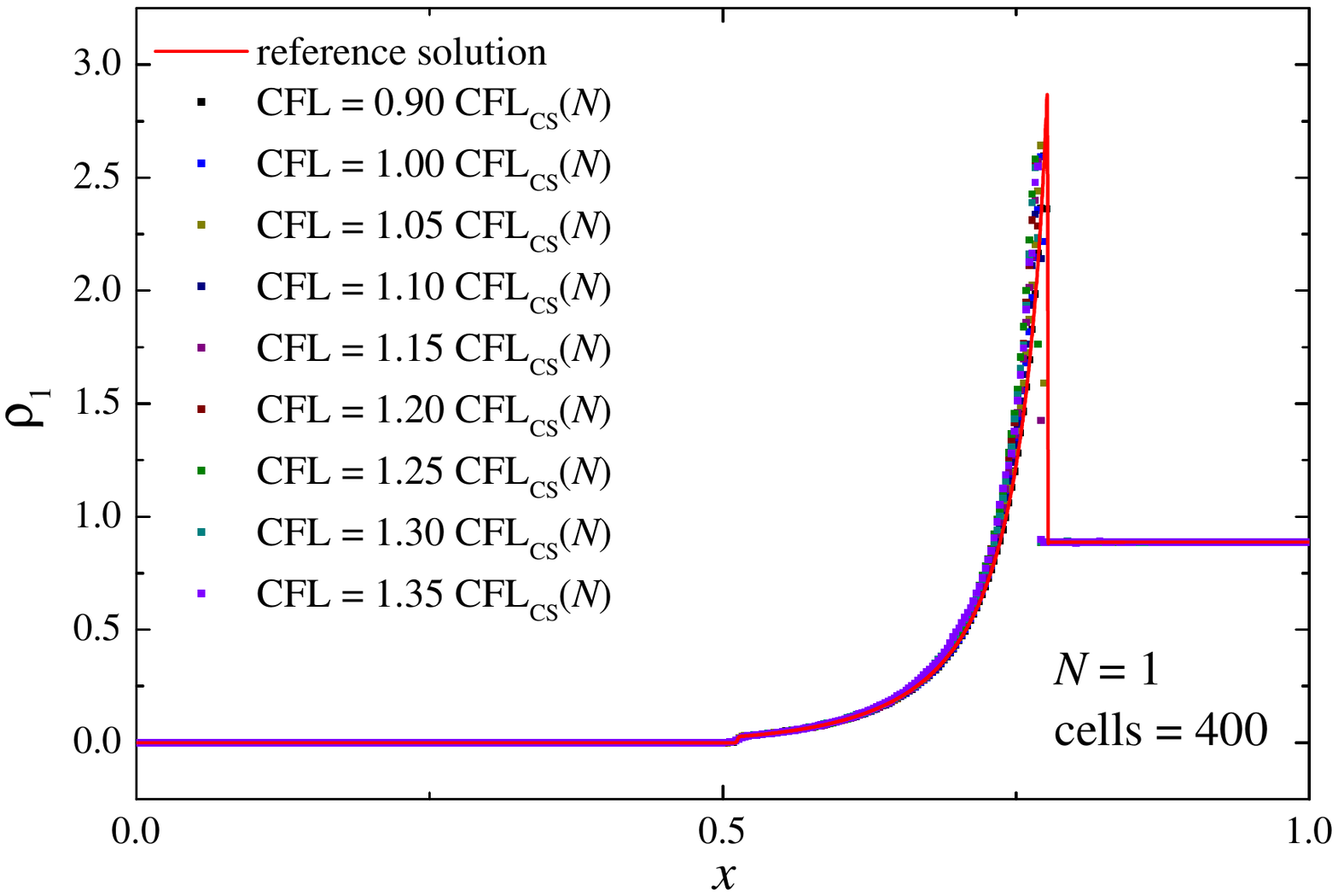}
\vspace{-8mm}\caption{\label{fig:test_znds_1d_degrees_1_3:degree_1_den1}}
\end{subfigure}
\begin{subfigure}{0.24\textwidth}
\includegraphics[width=\textwidth]{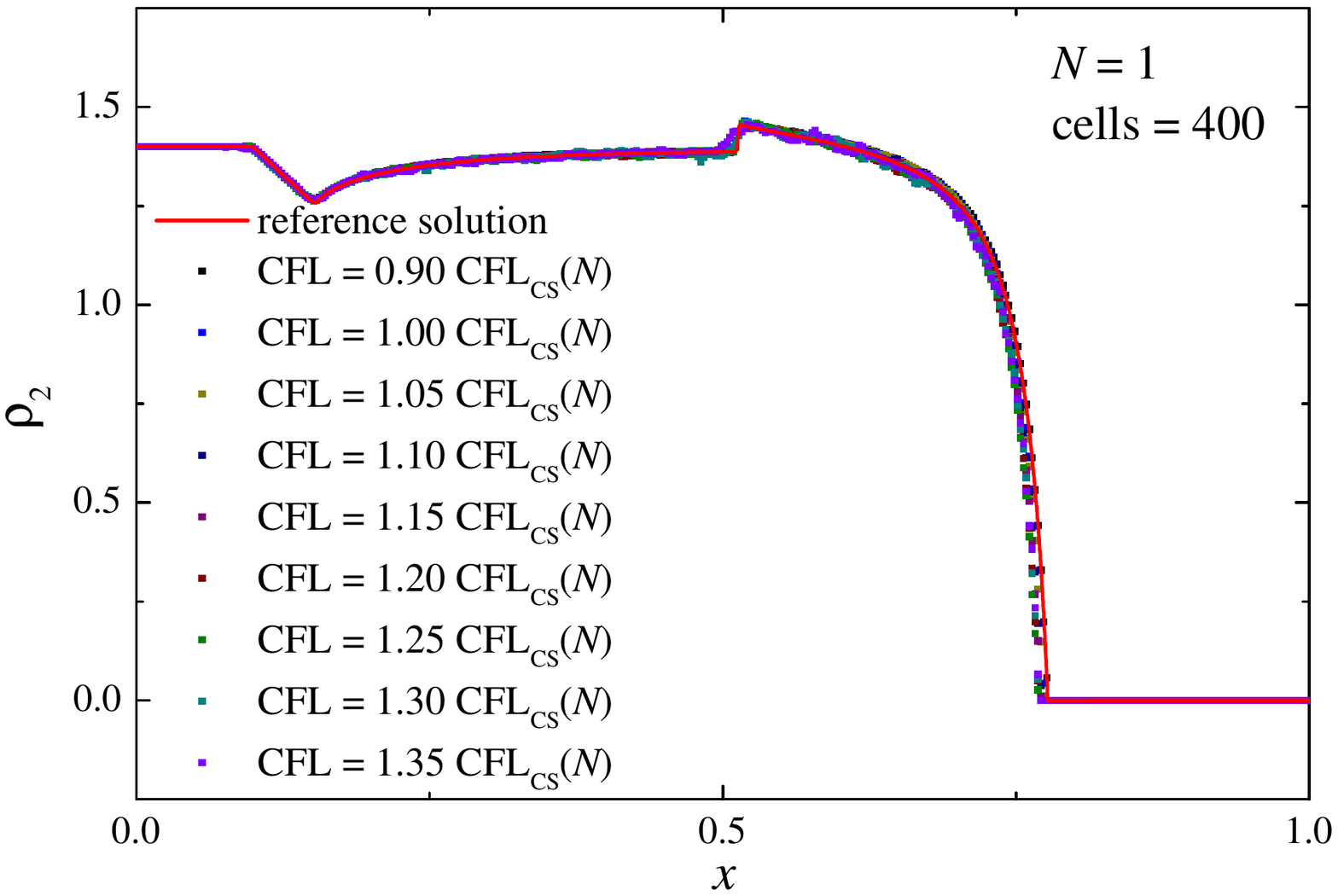}
\vspace{-8mm}\caption{\label{fig:test_znds_1d_degrees_1_3:degree_1_den2}}
\end{subfigure}\\[-2mm]
\begin{subfigure}{0.24\textwidth}
\includegraphics[width=\textwidth]{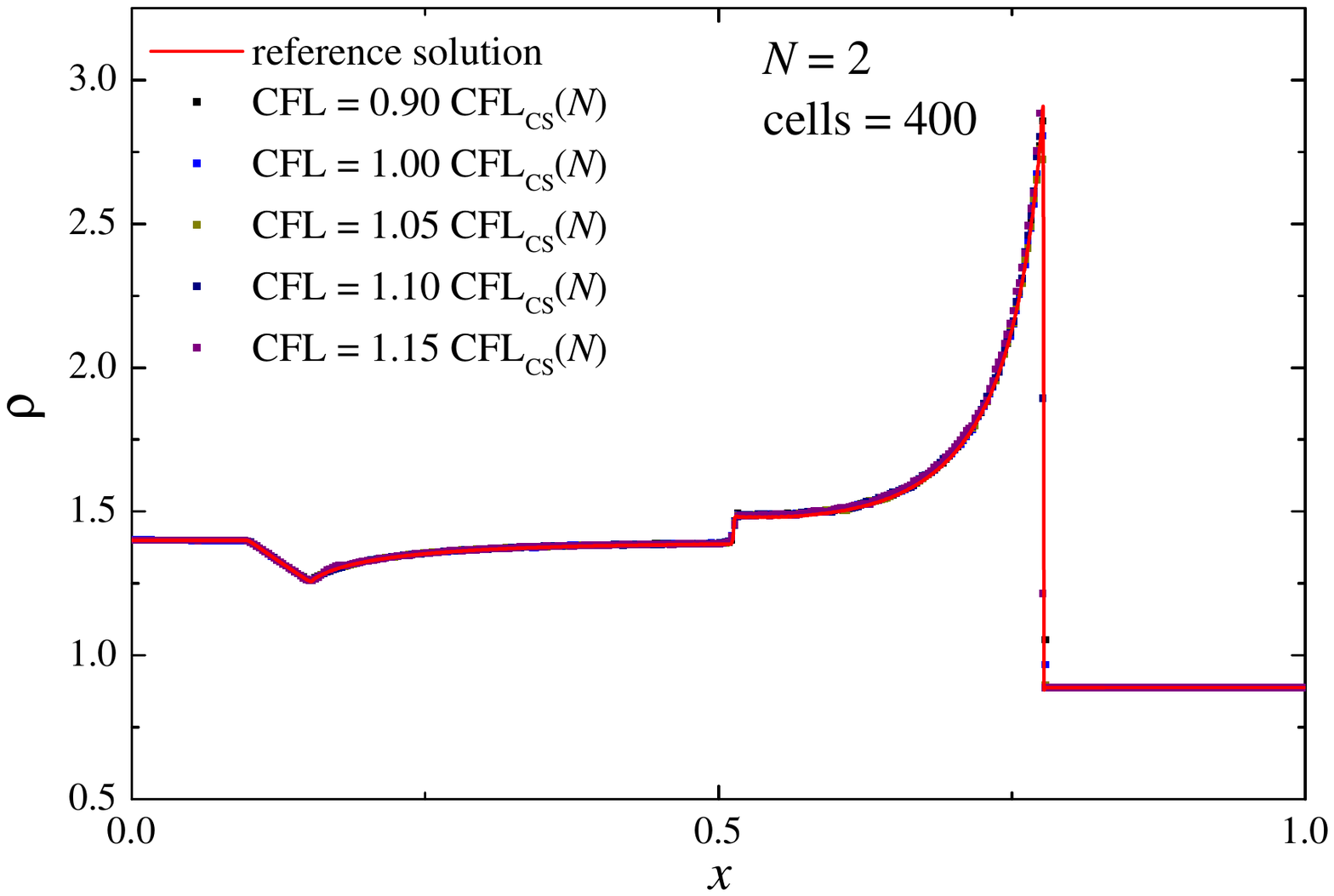}
\vspace{-8mm}\caption{\label{fig:test_znds_1d_degrees_1_3:degree_2_den}}
\end{subfigure}
\begin{subfigure}{0.24\textwidth}
\includegraphics[width=\textwidth]{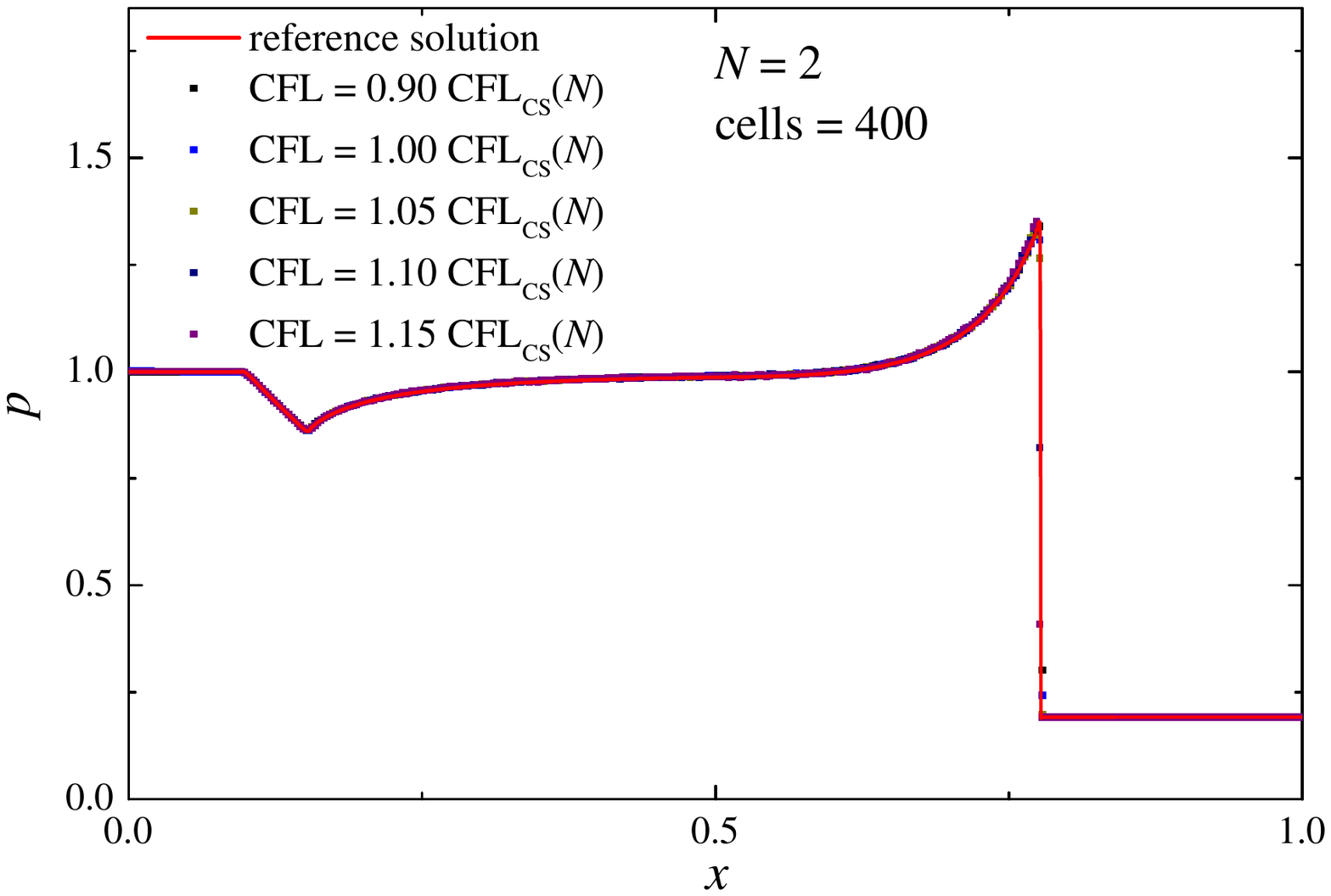}
\vspace{-8mm}\caption{\label{fig:test_znds_1d_degrees_1_3:degree_2_p}}
\end{subfigure}
\begin{subfigure}{0.24\textwidth}
\includegraphics[width=\textwidth]{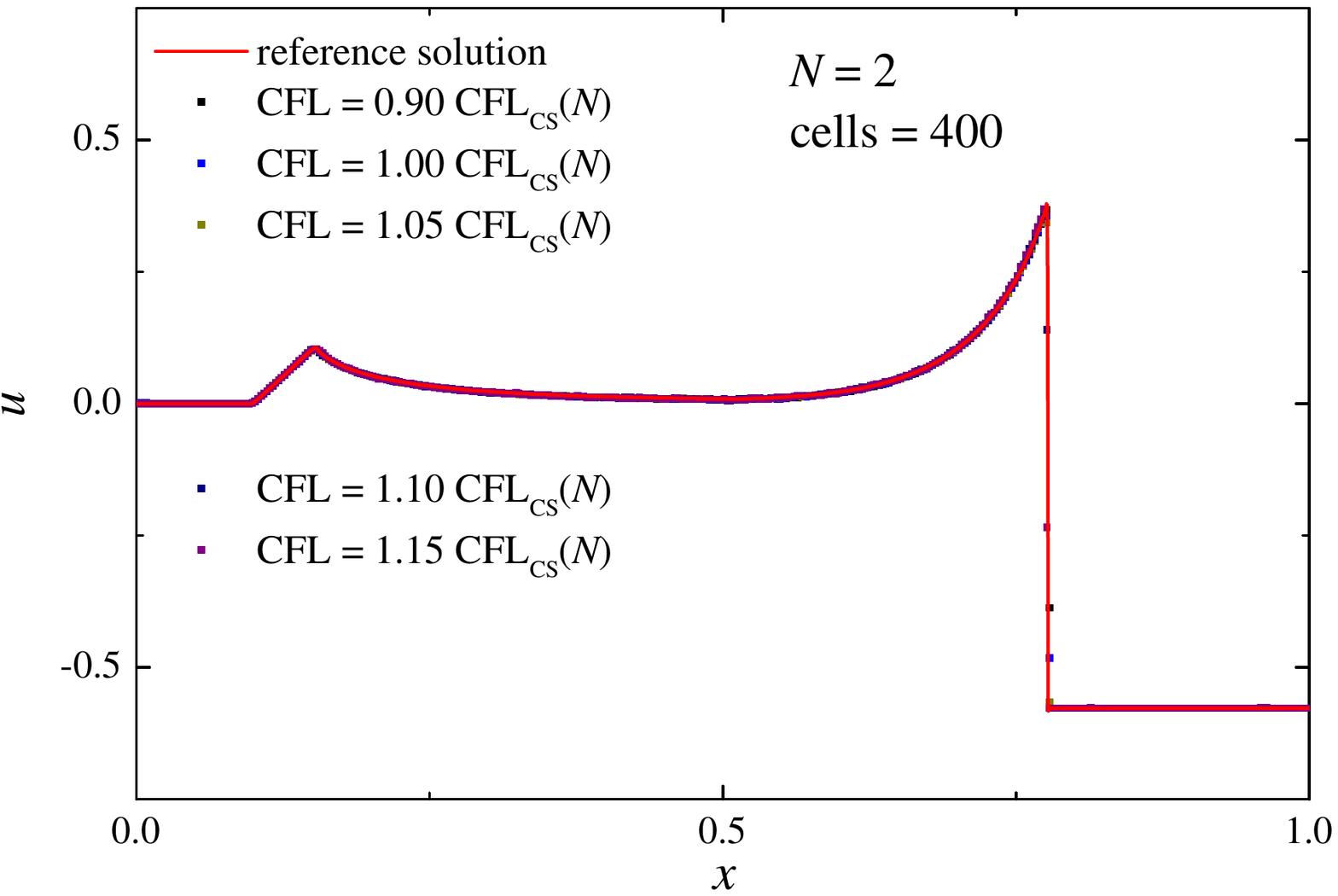}
\vspace{-8mm}\caption{\label{fig:test_znds_1d_degrees_1_3:degree_2_u}}
\end{subfigure}
\begin{subfigure}{0.24\textwidth}
\includegraphics[width=\textwidth]{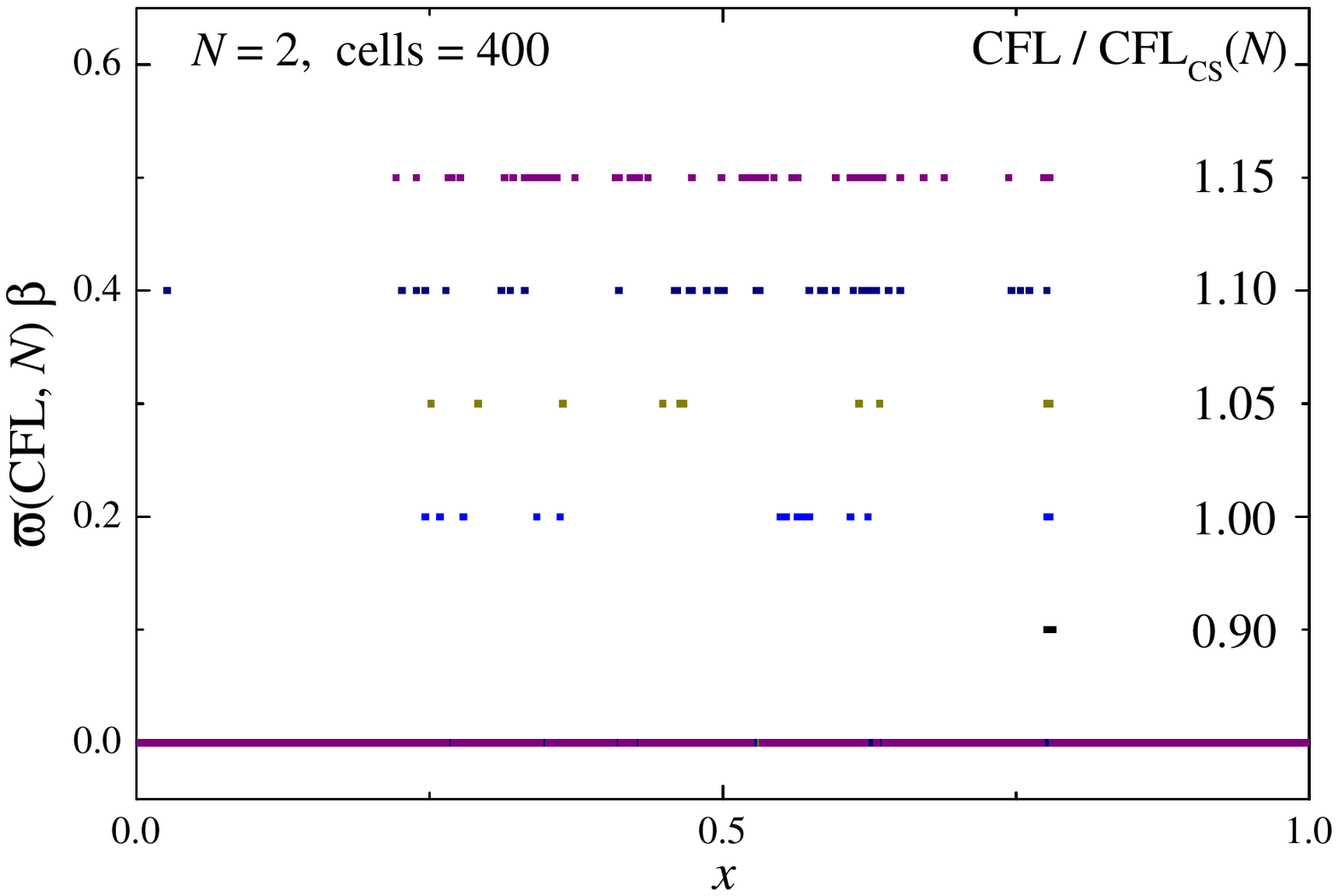}
\vspace{-8mm}\caption{\label{fig:test_znds_1d_degrees_1_3:degree_2_beta}}
\end{subfigure}\\[-2mm]
\begin{subfigure}{0.24\textwidth}
\includegraphics[width=\textwidth]{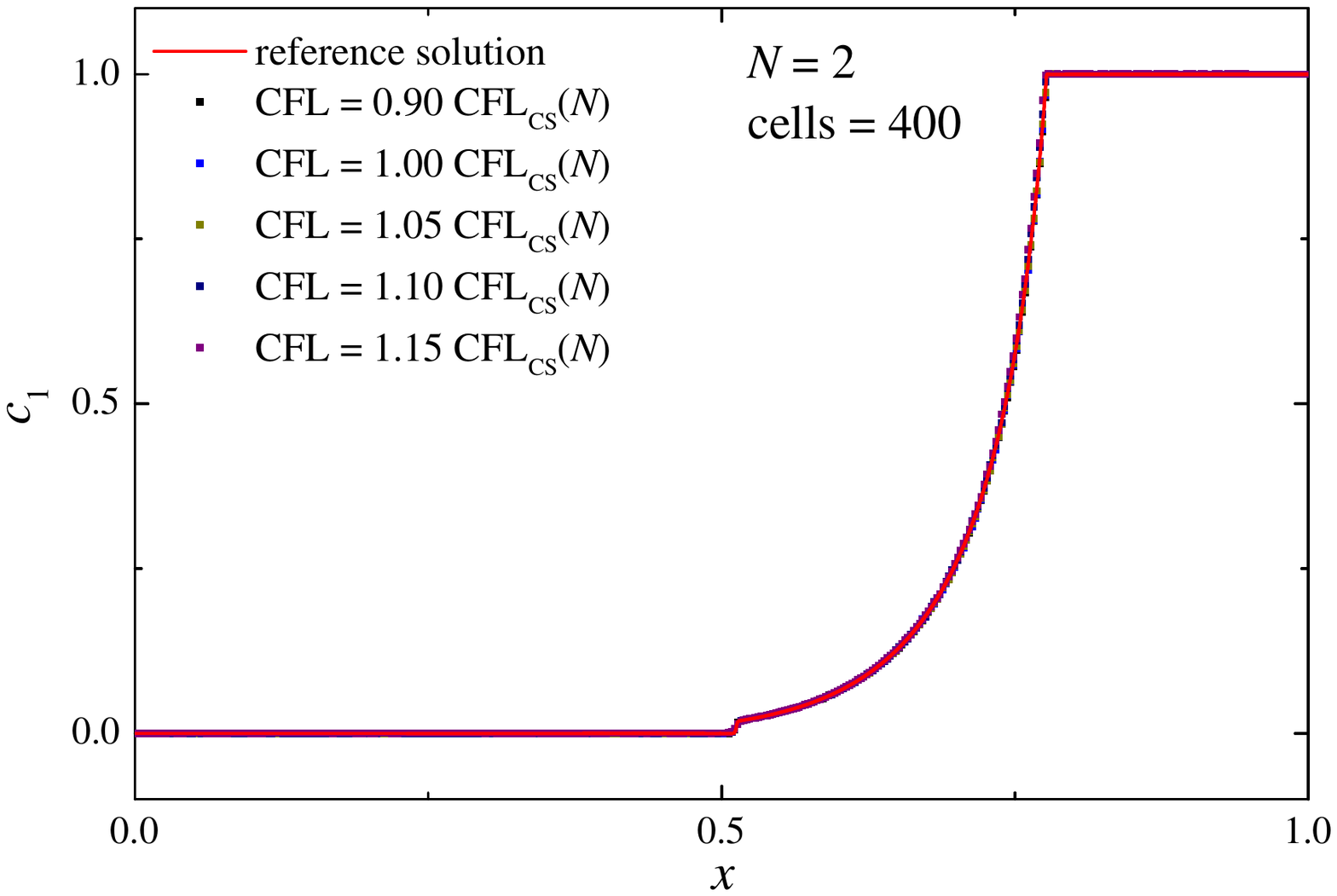}
\vspace{-8mm}\caption{\label{fig:test_znds_1d_degrees_1_3:degree_2_c1}}
\end{subfigure}
\begin{subfigure}{0.24\textwidth}
\includegraphics[width=\textwidth]{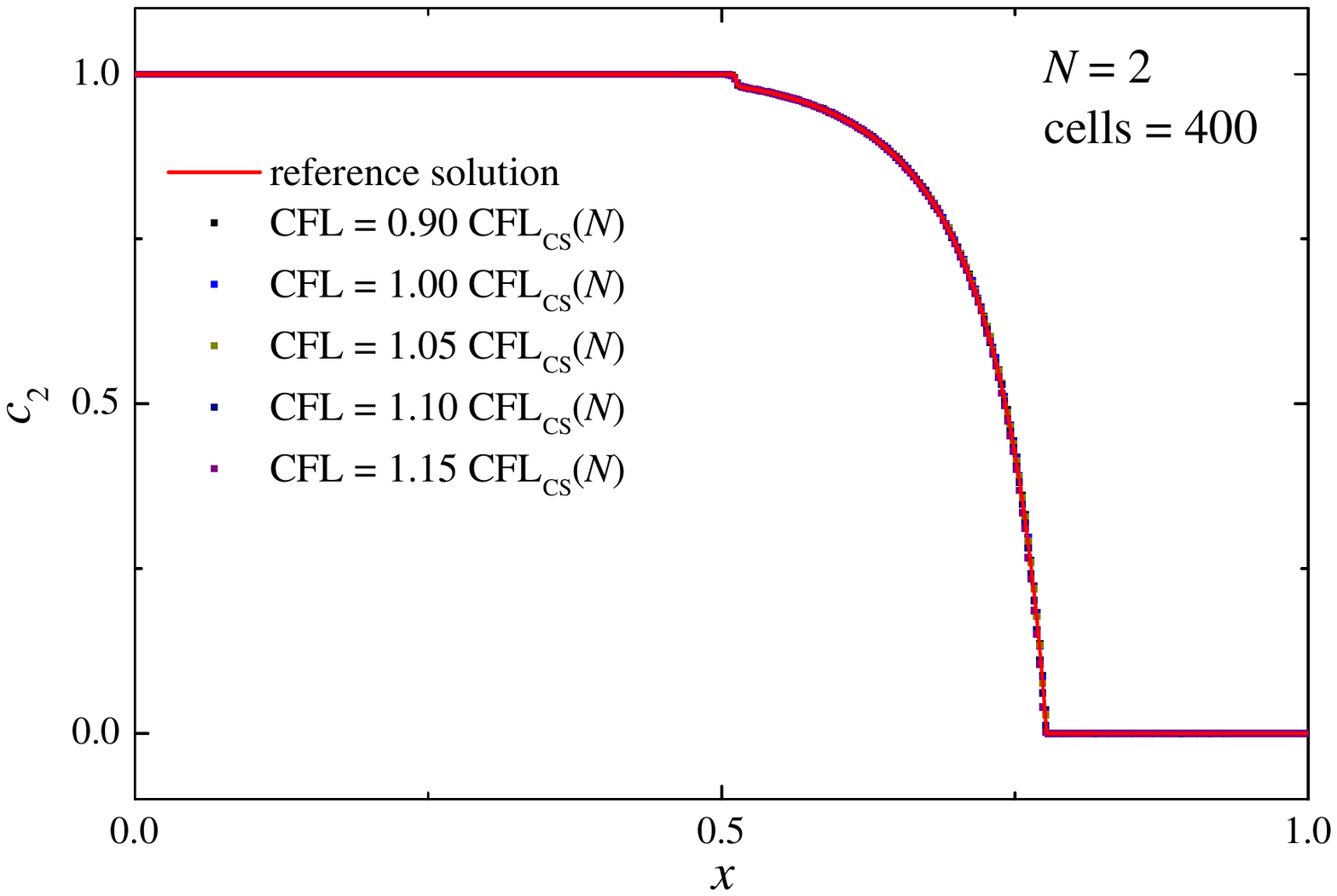}
\vspace{-8mm}\caption{\label{fig:test_znds_1d_degrees_1_3:degree_2_c2}}
\end{subfigure}
\begin{subfigure}{0.24\textwidth}
\includegraphics[width=\textwidth]{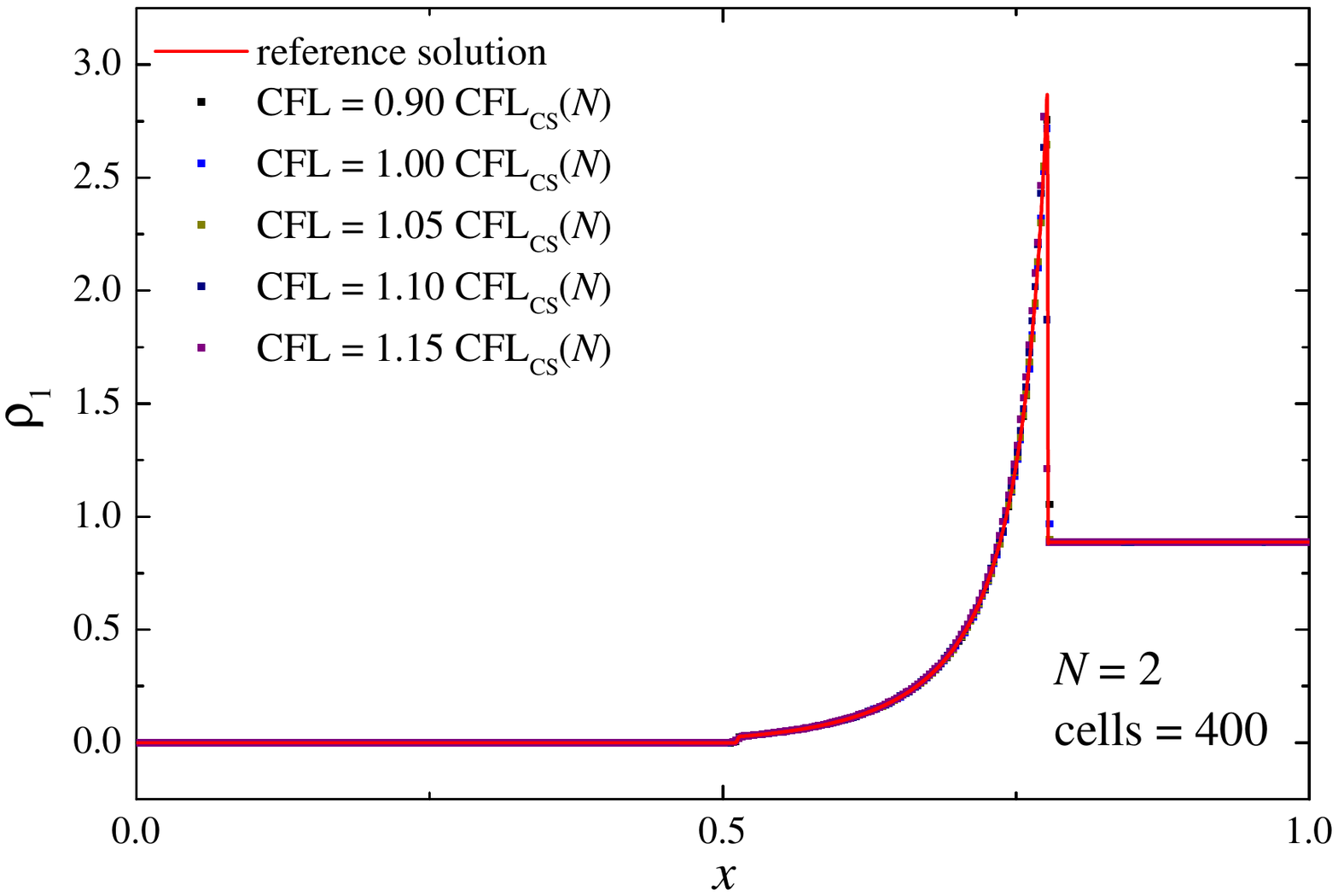}
\vspace{-8mm}\caption{\label{fig:test_znds_1d_degrees_1_3:degree_2_den1}}
\end{subfigure}
\begin{subfigure}{0.24\textwidth}
\includegraphics[width=\textwidth]{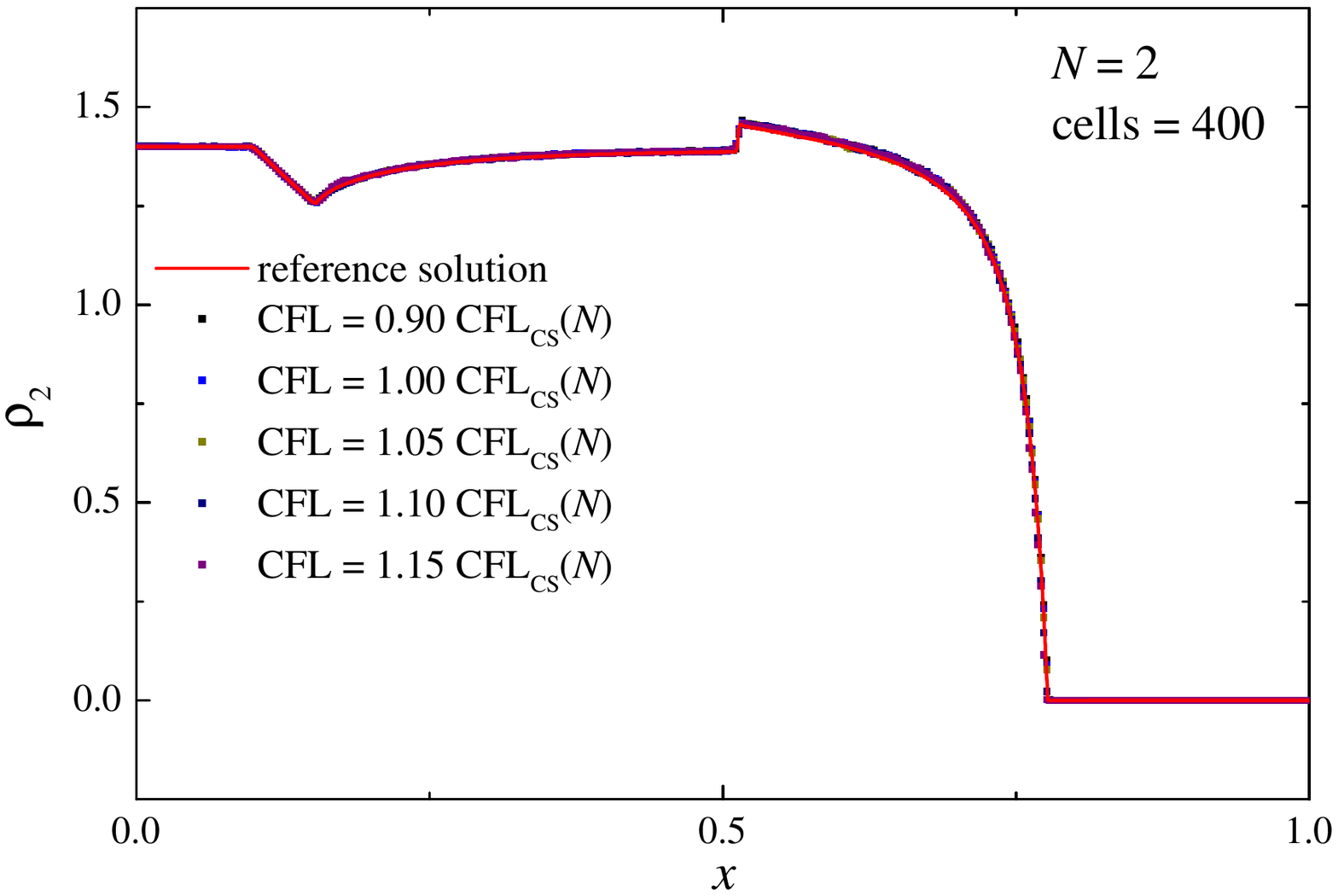}
\vspace{-8mm}\caption{\label{fig:test_znds_1d_degrees_1_3:degree_2_den2}}
\end{subfigure}\\[-2mm]
\begin{subfigure}{0.24\textwidth}
\includegraphics[width=\textwidth]{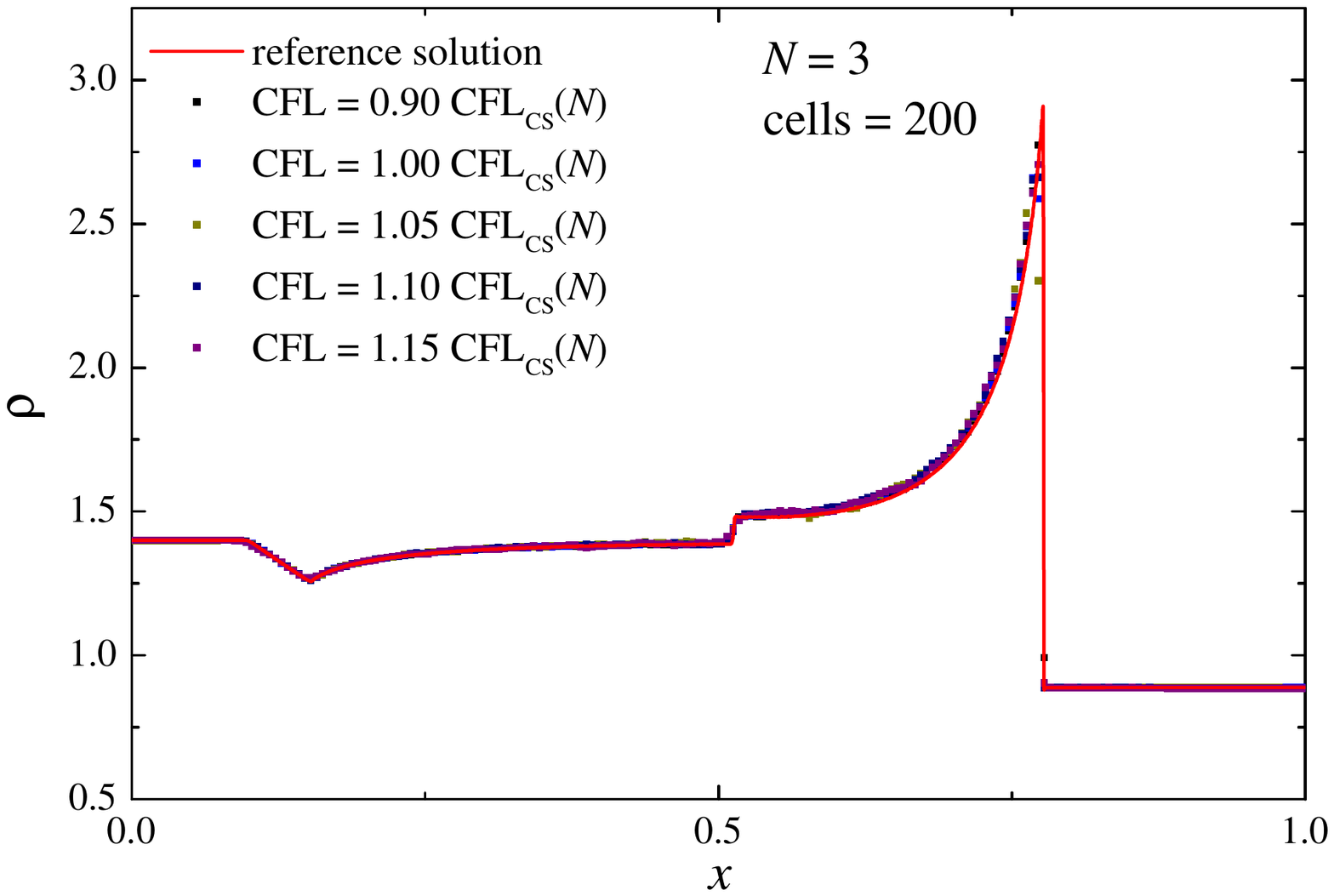}
\vspace{-8mm}\caption{\label{fig:test_znds_1d_degrees_1_3:degree_3_den}}
\end{subfigure}
\begin{subfigure}{0.24\textwidth}
\includegraphics[width=\textwidth]{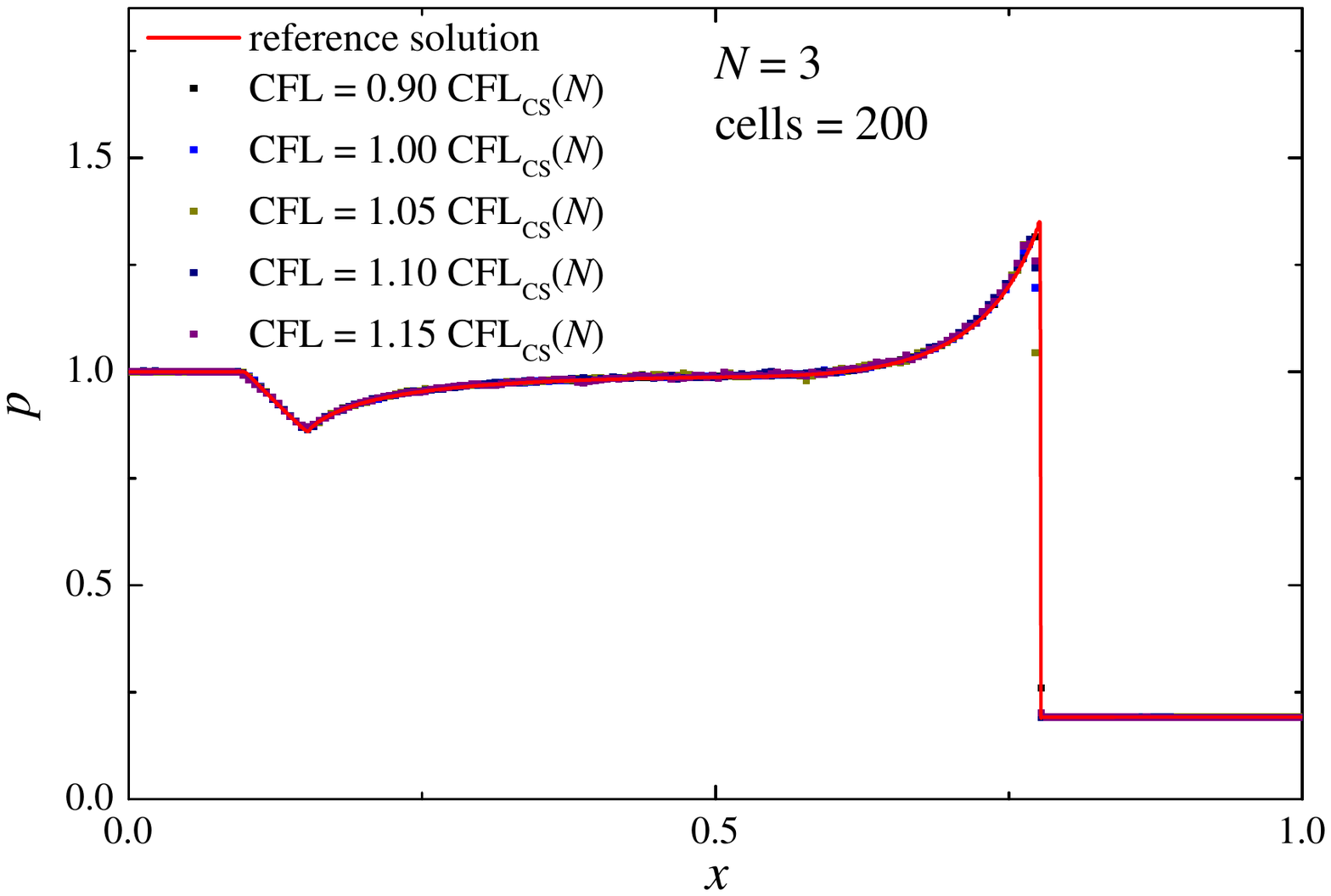}
\vspace{-8mm}\caption{\label{fig:test_znds_1d_degrees_1_3:degree_3_p}}
\end{subfigure}
\begin{subfigure}{0.24\textwidth}
\includegraphics[width=\textwidth]{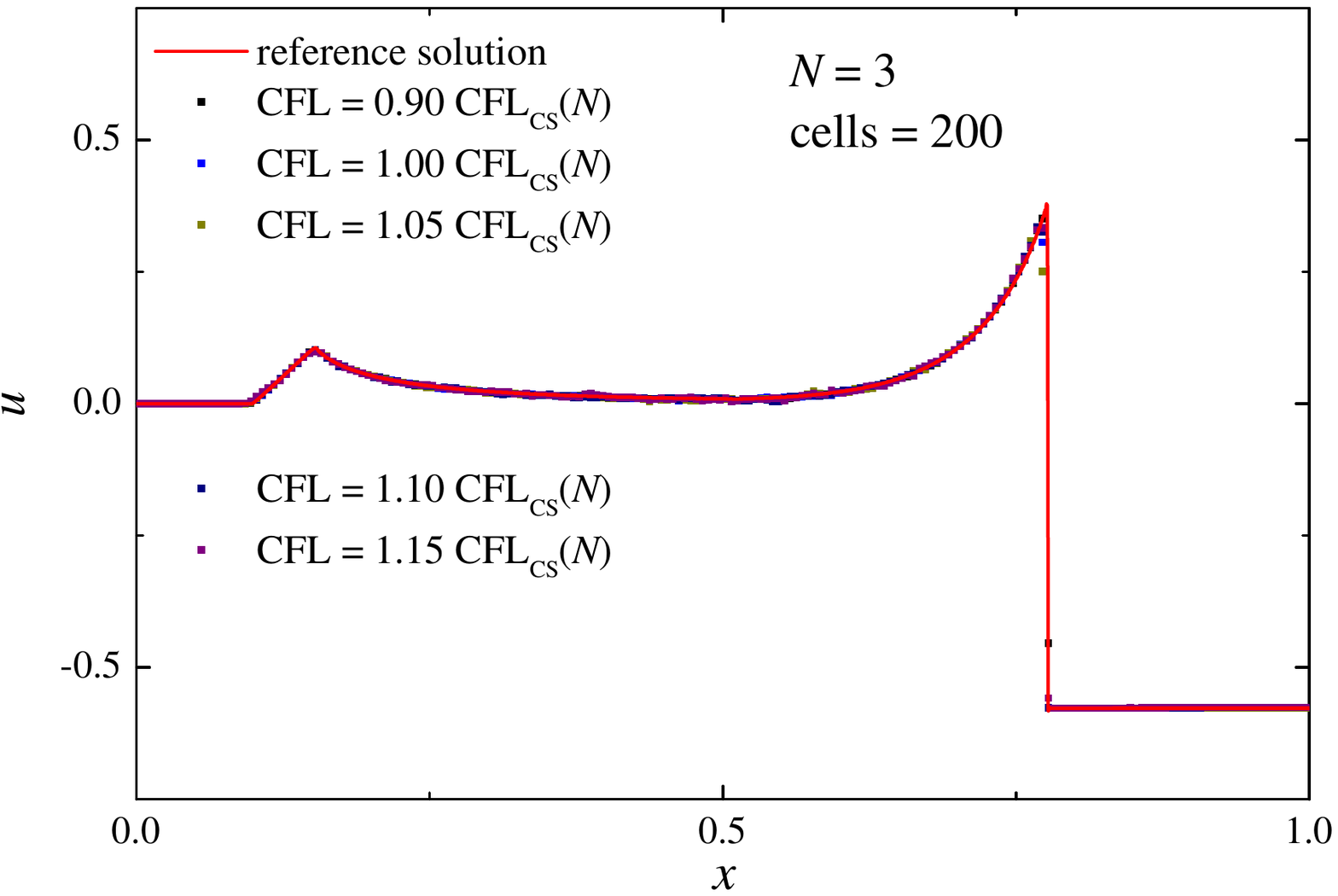}
\vspace{-8mm}\caption{\label{fig:test_znds_1d_degrees_1_3:degree_3_u}}
\end{subfigure}
\begin{subfigure}{0.24\textwidth}
\includegraphics[width=\textwidth]{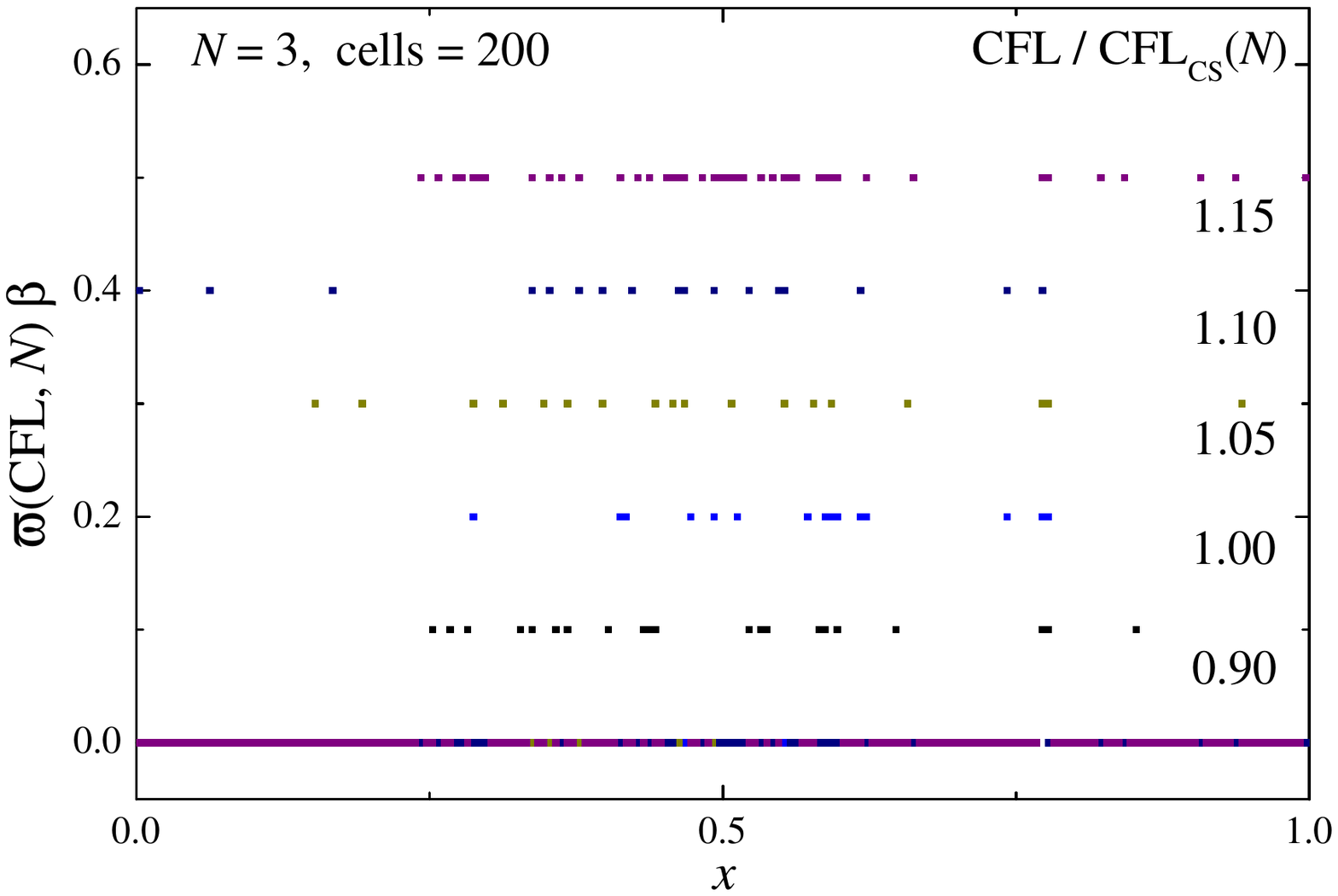}
\vspace{-8mm}\caption{\label{fig:test_znds_1d_degrees_1_3:degree_3_beta}}
\end{subfigure}\\[-2mm]
\begin{subfigure}{0.24\textwidth}
\includegraphics[width=\textwidth]{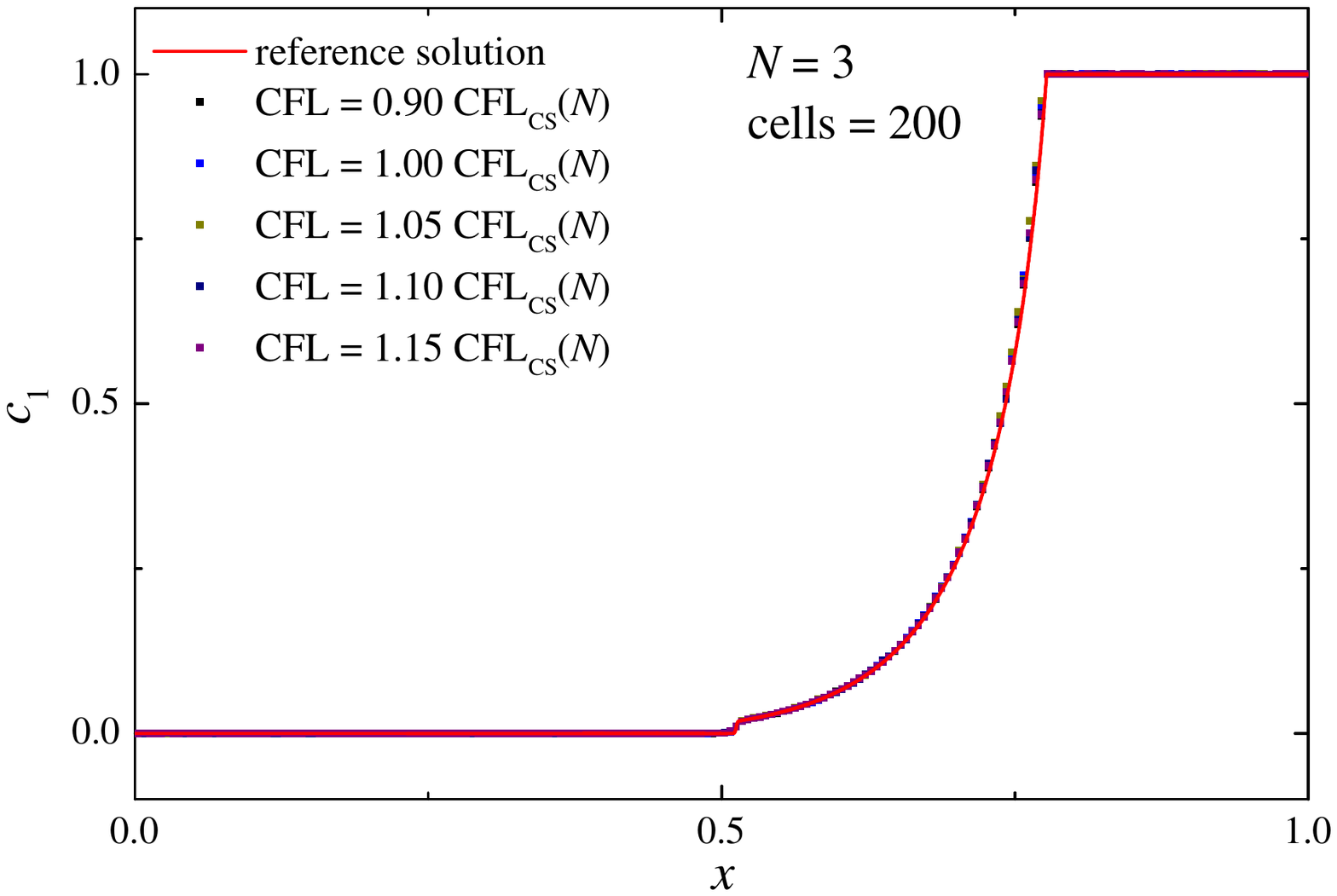}
\vspace{-8mm}\caption{\label{fig:test_znds_1d_degrees_1_3:degree_3_c1}}
\end{subfigure}
\begin{subfigure}{0.24\textwidth}
\includegraphics[width=\textwidth]{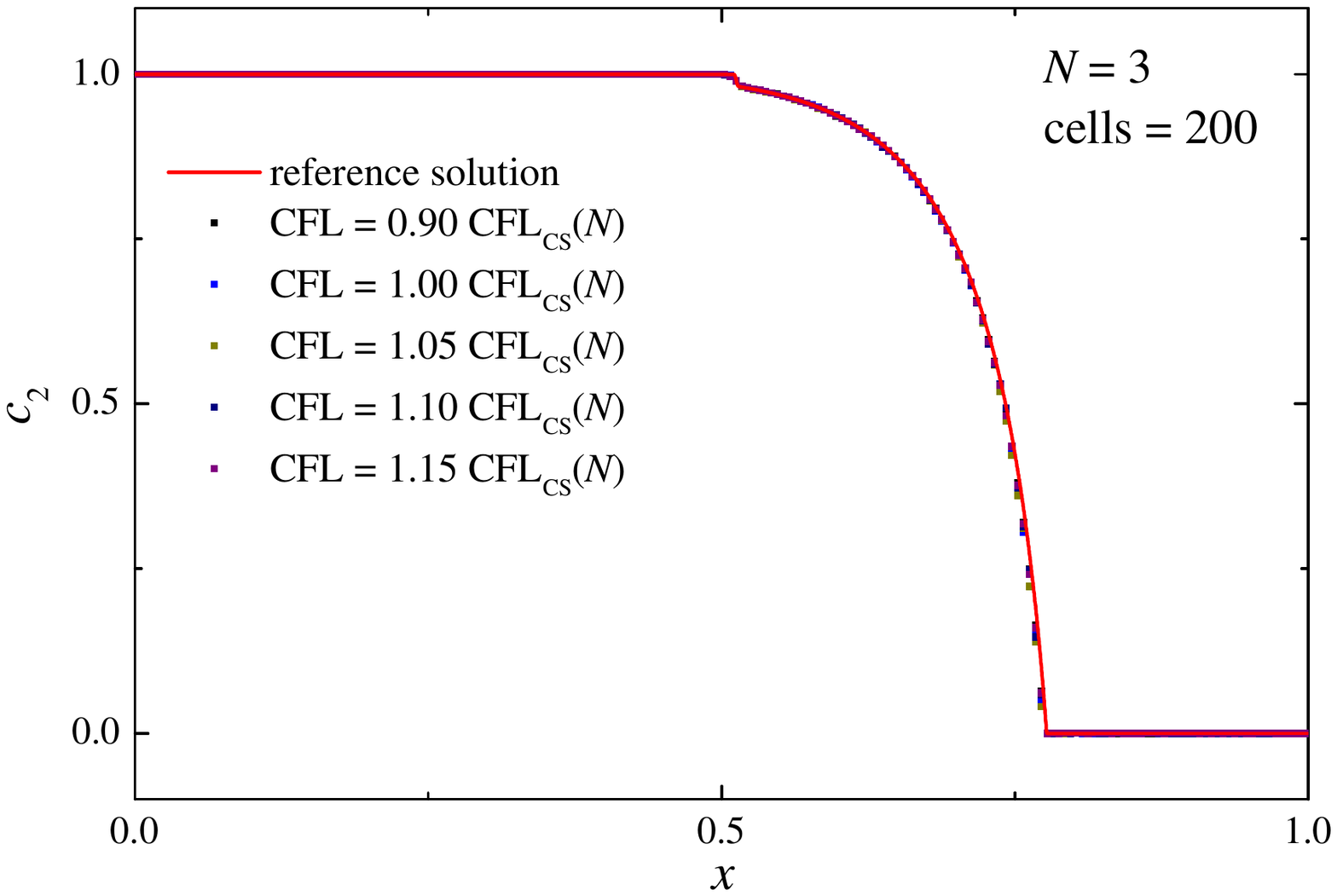}
\vspace{-8mm}\caption{\label{fig:test_znds_1d_degrees_1_3:degree_3_c2}}
\end{subfigure}
\begin{subfigure}{0.24\textwidth}
\includegraphics[width=\textwidth]{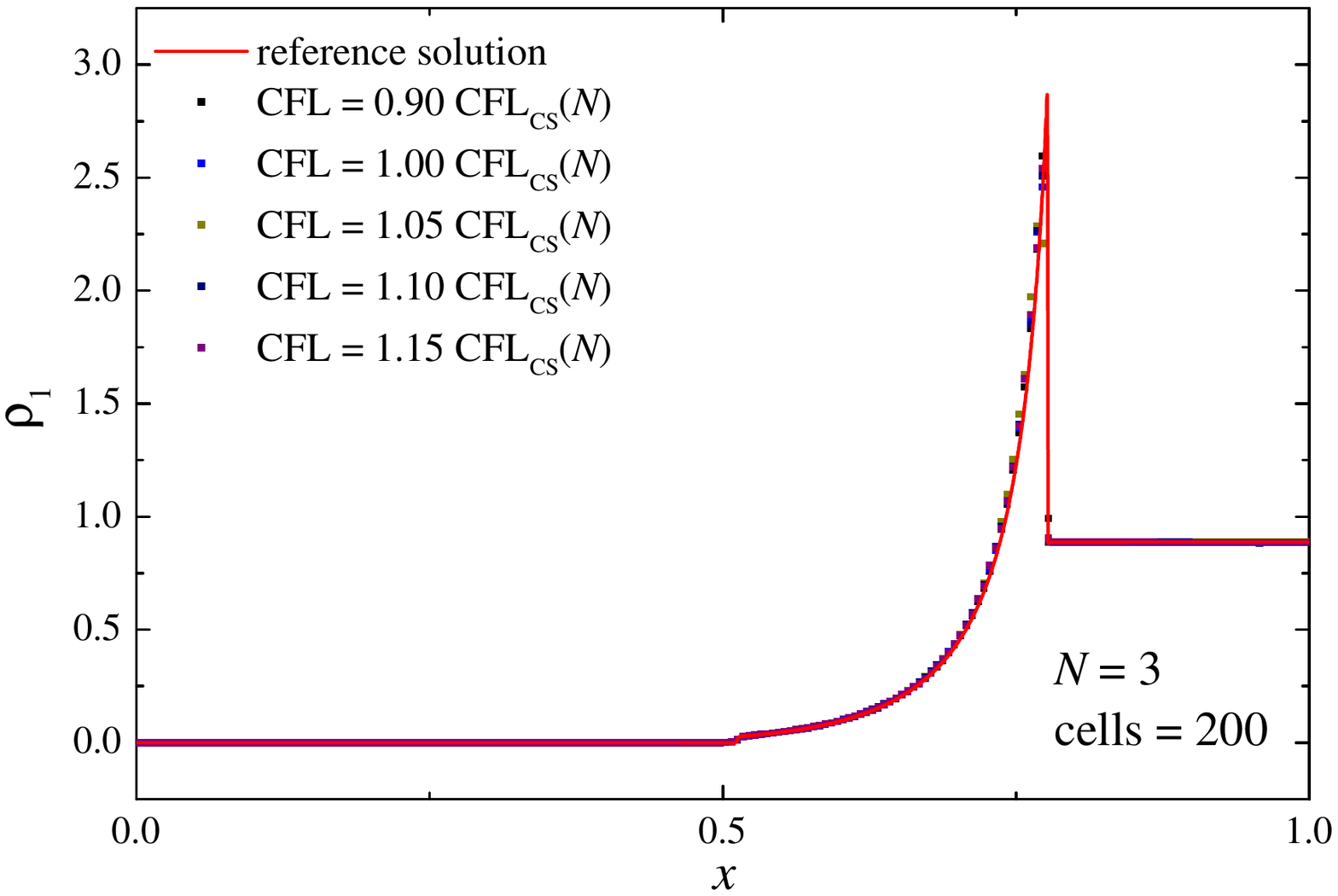}
\vspace{-8mm}\caption{\label{fig:test_znds_1d_degrees_1_3:degree_3_den1}}
\end{subfigure}
\begin{subfigure}{0.24\textwidth}
\includegraphics[width=\textwidth]{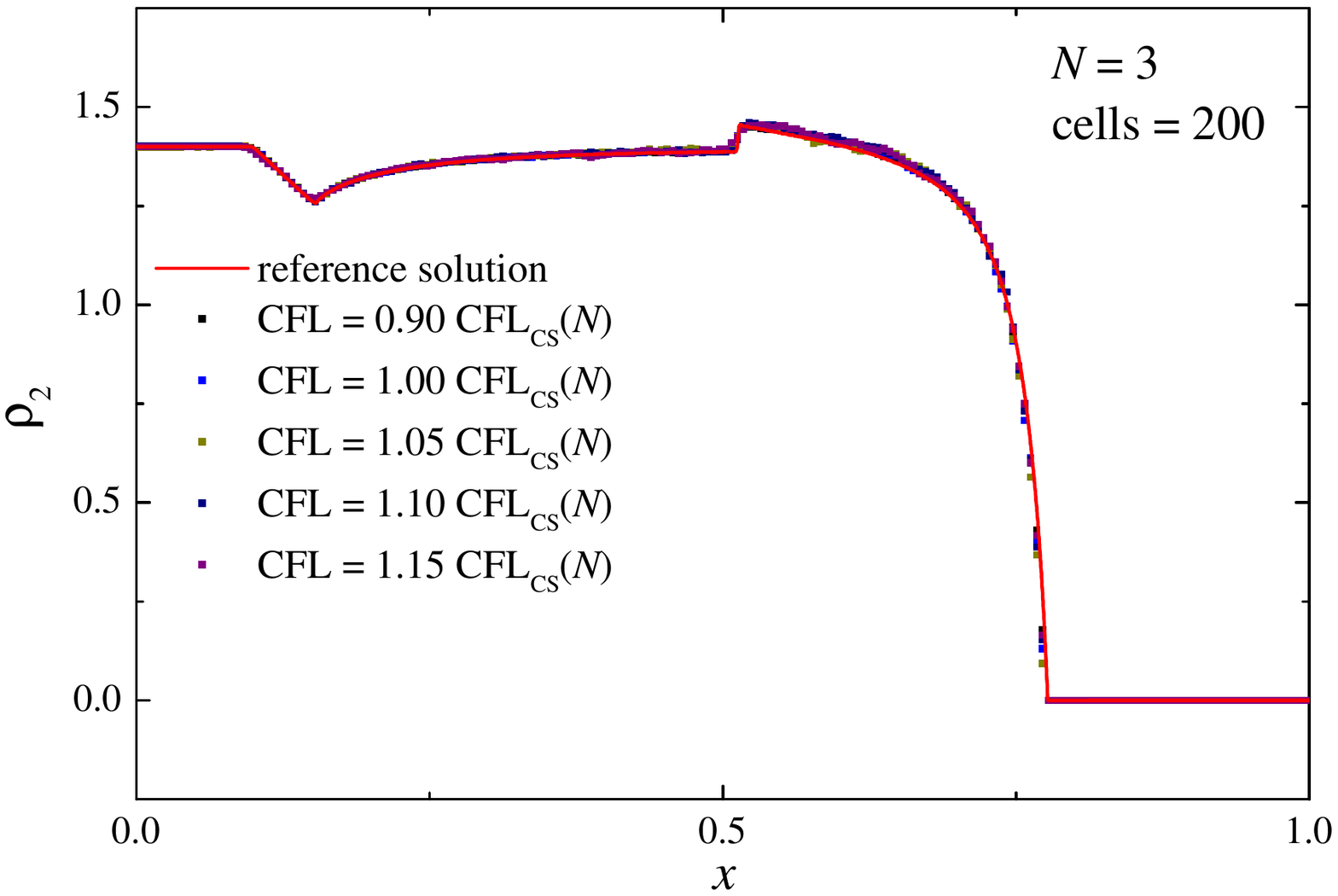}
\vspace{-8mm}\caption{\label{fig:test_znds_1d_degrees_1_3:degree_3_den2}}
\end{subfigure}\\[-2mm]
\caption{%
Numerical solution of the one-dimensional problem of the plane detonation wave formation in a two-component medium with a ``slow'' reaction (weak stiff case, a detailed statement of the problem is presented in the text), obtained using the ADER-DG method for the basis polynomials degree $N = 1$ ((\subref{fig:test_znds_1d_degrees_1_3:degree_1_den}), (\subref{fig:test_znds_1d_degrees_1_3:degree_1_p}), (\subref{fig:test_znds_1d_degrees_1_3:degree_1_u}), (\subref{fig:test_znds_1d_degrees_1_3:degree_1_c1}), (\subref{fig:test_znds_1d_degrees_1_3:degree_1_c2}), (\subref{fig:test_znds_1d_degrees_1_3:degree_1_den1}), (\subref{fig:test_znds_1d_degrees_1_3:degree_1_den2})), $N = 2$ ((\subref{fig:test_znds_1d_degrees_1_3:degree_2_den}), (\subref{fig:test_znds_1d_degrees_1_3:degree_2_p}), (\subref{fig:test_znds_1d_degrees_1_3:degree_2_u}), (\subref{fig:test_znds_1d_degrees_1_3:degree_2_c1}), (\subref{fig:test_znds_1d_degrees_1_3:degree_2_c2}), (\subref{fig:test_znds_1d_degrees_1_3:degree_2_den1}), (\subref{fig:test_znds_1d_degrees_1_3:degree_2_den2})), $N = 3$ ((\subref{fig:test_znds_1d_degrees_1_3:degree_3_den}), (\subref{fig:test_znds_1d_degrees_1_3:degree_3_p}), (\subref{fig:test_znds_1d_degrees_1_3:degree_3_u}), (\subref{fig:test_znds_1d_degrees_1_3:degree_3_c1}), (\subref{fig:test_znds_1d_degrees_1_3:degree_3_c2}), (\subref{fig:test_znds_1d_degrees_1_3:degree_3_den1}), (\subref{fig:test_znds_1d_degrees_1_3:degree_3_den2})) with a posteriori limitation of the solution by an ADER-WENO2 finite-volume limiter, at the final time $t_{f} = 0.4$. The graphs show the coordinate dependence of density $\rho$ ((\subref{fig:test_znds_1d_degrees_1_3:degree_1_den}), (\subref{fig:test_znds_1d_degrees_1_3:degree_2_den}), (\subref{fig:test_znds_1d_degrees_1_3:degree_3_den})), pressure $p$ ((\subref{fig:test_znds_1d_degrees_1_3:degree_1_p}), (\subref{fig:test_znds_1d_degrees_1_3:degree_2_p}), (\subref{fig:test_znds_1d_degrees_1_3:degree_3_p})), flow velocity $u$ ((\subref{fig:test_znds_1d_degrees_1_3:degree_1_u}), (\subref{fig:test_znds_1d_degrees_1_3:degree_2_u}), (\subref{fig:test_znds_1d_degrees_1_3:degree_3_u})), troubled cells indicator $\beta$ multiplied by the separating function $\varpi(\mathrm{CFL}, N)$ ((\subref{fig:test_znds_1d_degrees_1_3:degree_1_beta}), (\subref{fig:test_znds_1d_degrees_1_3:degree_2_beta}), (\subref{fig:test_znds_1d_degrees_1_3:degree_3_beta})), mass concentrations $c_{1}$ ((\subref{fig:test_znds_1d_degrees_1_3:degree_1_c1}), (\subref{fig:test_znds_1d_degrees_1_3:degree_2_c1}), (\subref{fig:test_znds_1d_degrees_1_3:degree_3_c1})) and $c_{2}$ ((\subref{fig:test_znds_1d_degrees_1_3:degree_1_c2}), (\subref{fig:test_znds_1d_degrees_1_3:degree_2_c2}), (\subref{fig:test_znds_1d_degrees_1_3:degree_3_c2})) of the reactant $A$ and product $B$, respectively, and partial densities $\rho_{1}$ ((\subref{fig:test_znds_1d_degrees_1_3:degree_1_den1}), (\subref{fig:test_znds_1d_degrees_1_3:degree_2_den1}), (\subref{fig:test_znds_1d_degrees_1_3:degree_3_den1})) and $\rho_{2}$ ((\subref{fig:test_znds_1d_degrees_1_3:degree_1_den2}), (\subref{fig:test_znds_1d_degrees_1_3:degree_2_den2}), (\subref{fig:test_znds_1d_degrees_1_3:degree_3_den2})) of the reactant $A$ and product $B$, respectively. The square symbols represent the subcells finite-volume representation of the numerical solution; the red solid lines represents the reference solution of the problem.
}
\label{fig:test_znds_1d_degrees_1_3}
\end{figure}
\begin{figure}[h!]
\captionsetup[subfigure]{%
	position=bottom,
	font+=smaller,
	textfont=normalfont,
	singlelinecheck=off,
	justification=raggedright
}
\centering
\begin{subfigure}{0.24\textwidth}
\includegraphics[width=\textwidth]{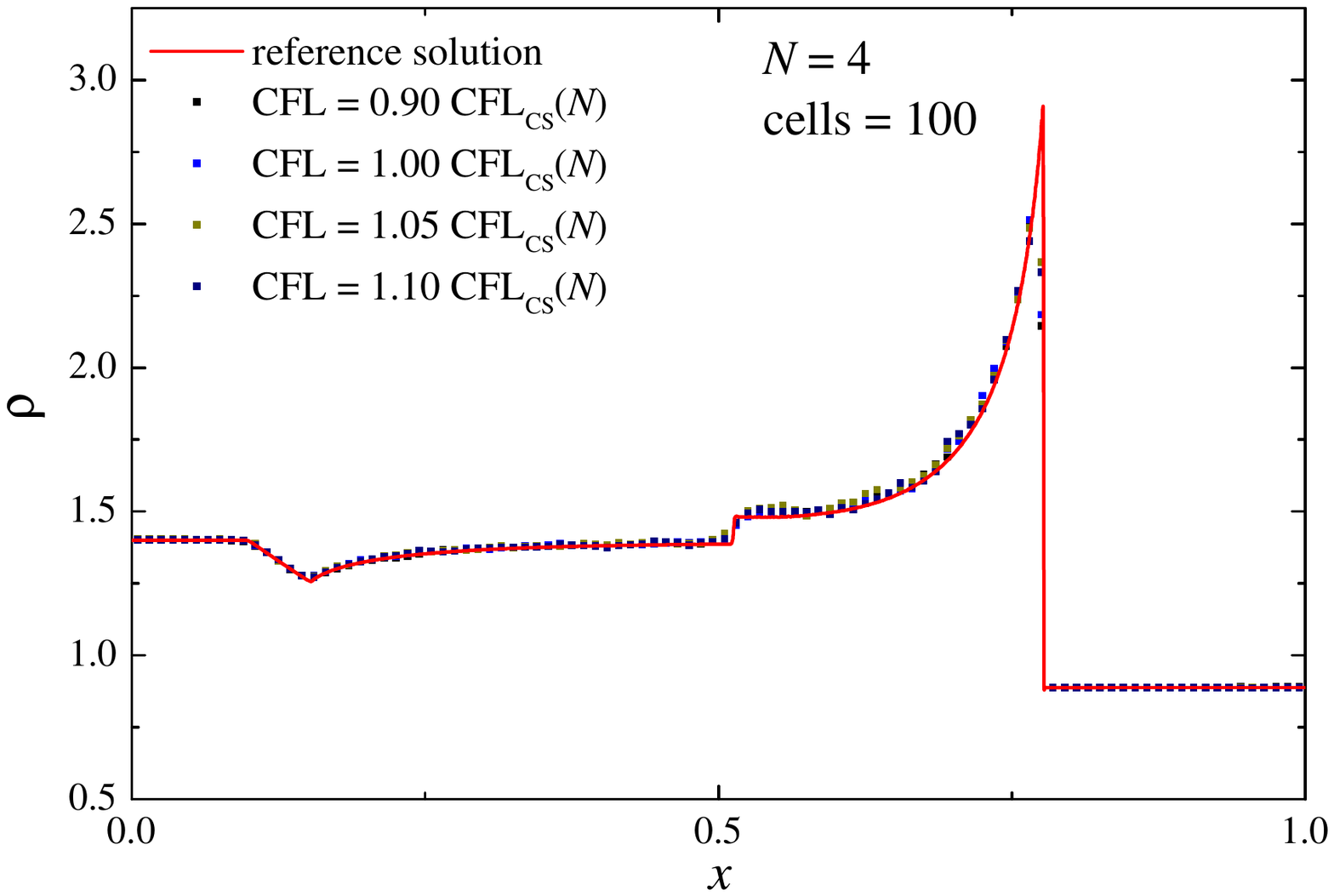}
\vspace{-8mm}\caption{\label{fig:test_znds_1d_degrees_4_6:degree_4_den}}
\end{subfigure}
\begin{subfigure}{0.24\textwidth}
\includegraphics[width=\textwidth]{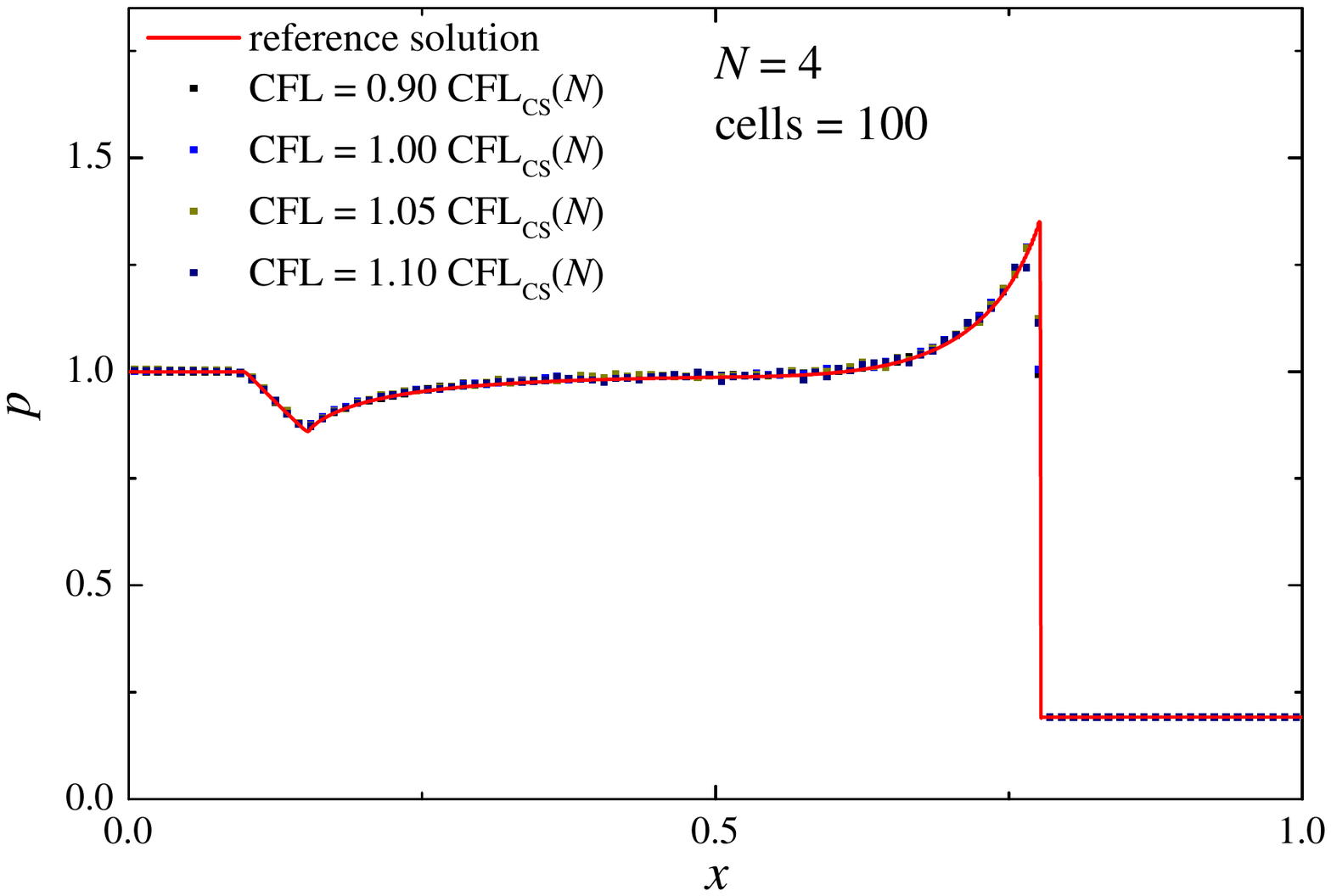}
\vspace{-8mm}\caption{\label{fig:test_znds_1d_degrees_4_6:degree_4_p}}
\end{subfigure}
\begin{subfigure}{0.24\textwidth}
\includegraphics[width=\textwidth]{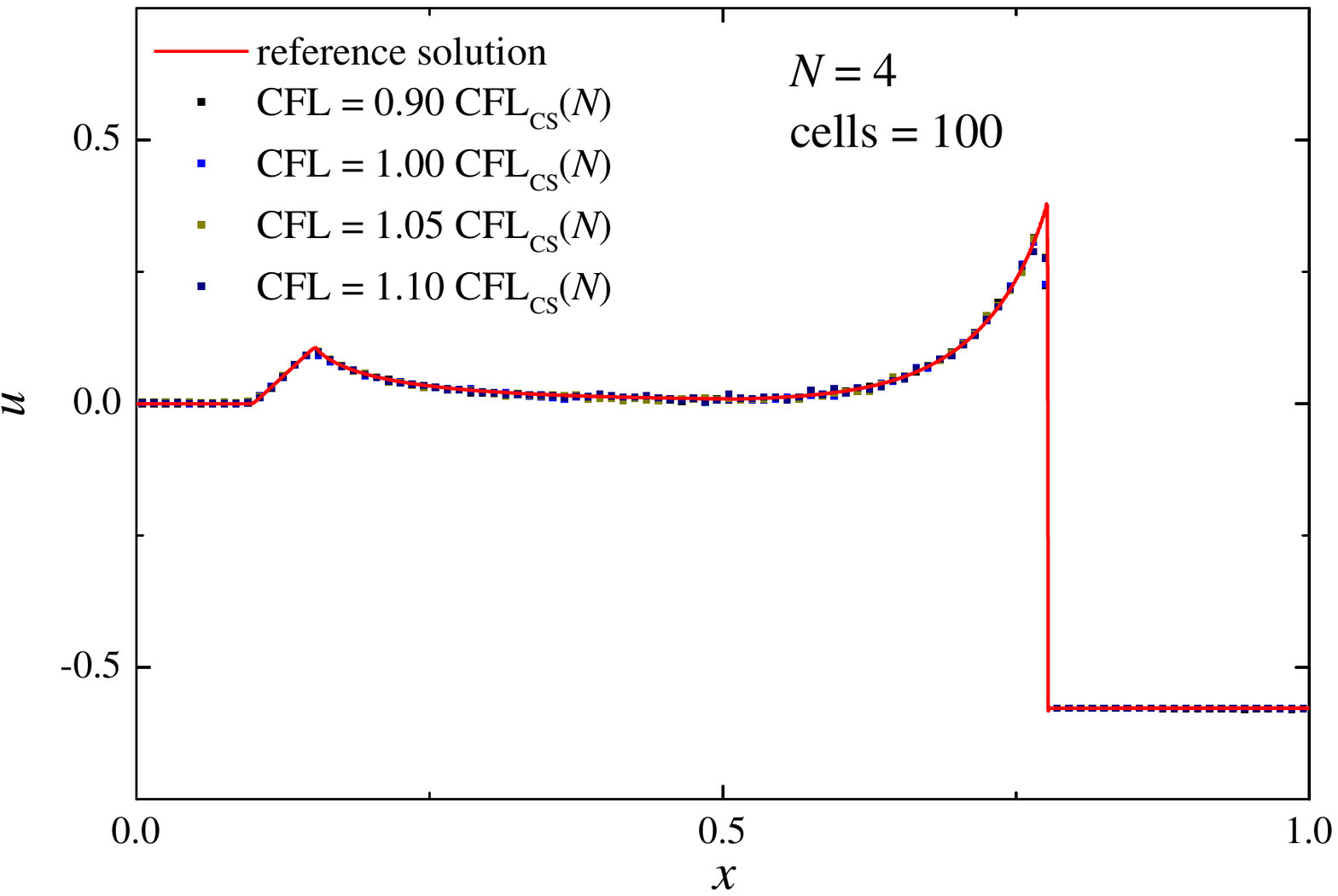}
\vspace{-8mm}\caption{\label{fig:test_znds_1d_degrees_4_6:degree_4_u}}
\end{subfigure}
\begin{subfigure}{0.24\textwidth}
\includegraphics[width=\textwidth]{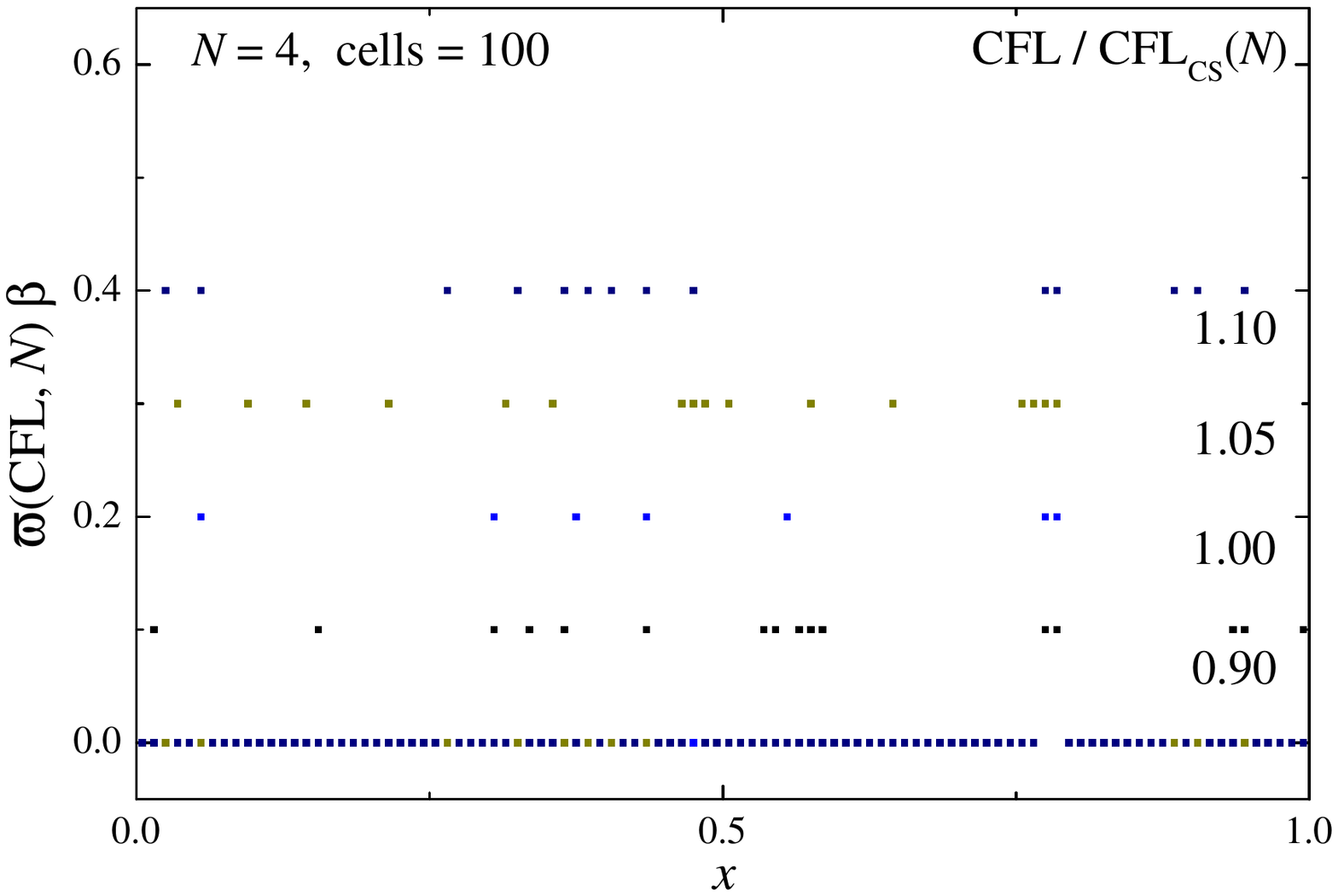}
\vspace{-8mm}\caption{\label{fig:test_znds_1d_degrees_4_6:degree_4_beta}}
\end{subfigure}\\[-2mm]
\begin{subfigure}{0.24\textwidth}
\includegraphics[width=\textwidth]{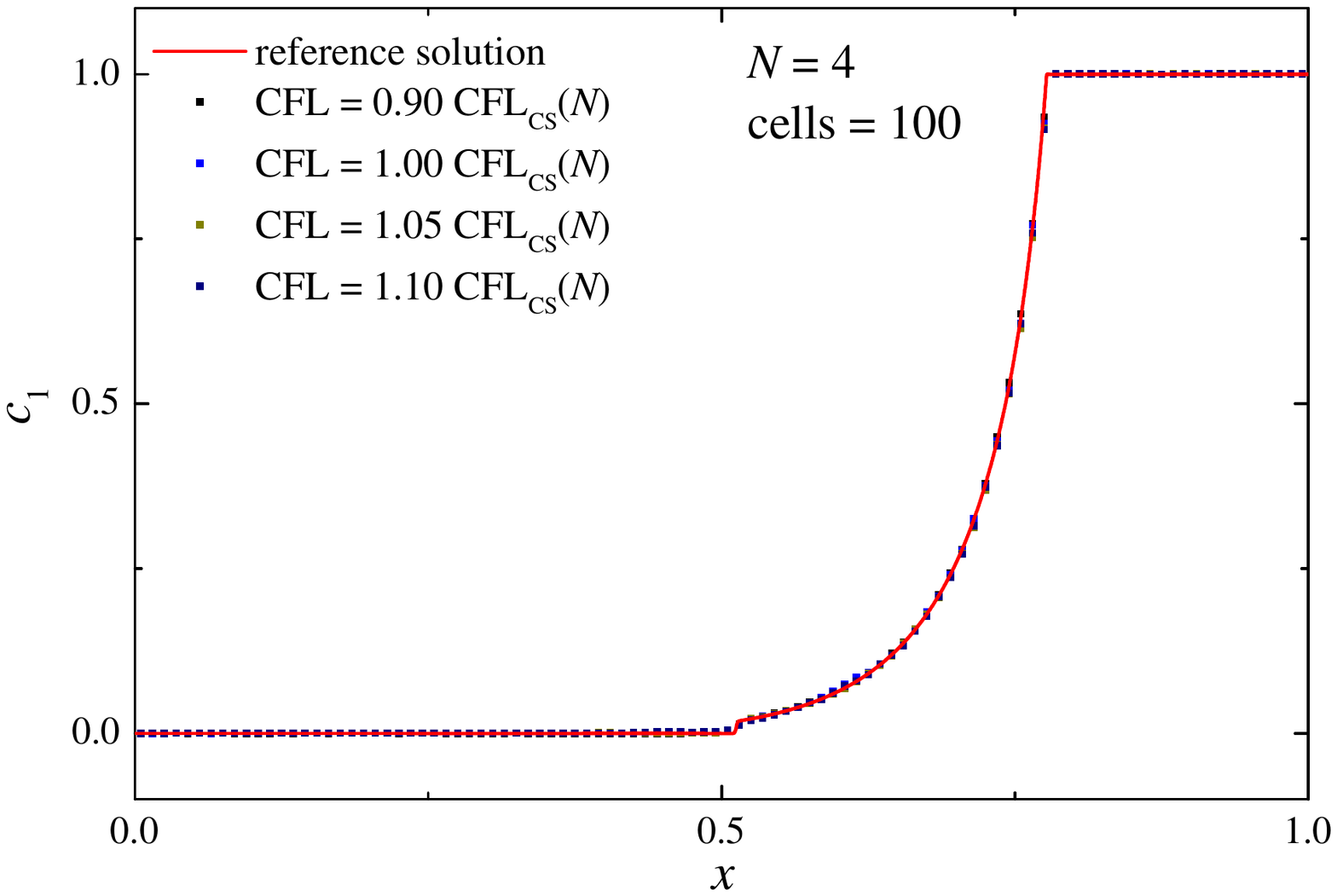}
\vspace{-8mm}\caption{\label{fig:test_znds_1d_degrees_4_6:degree_4_c1}}
\end{subfigure}
\begin{subfigure}{0.24\textwidth}
\includegraphics[width=\textwidth]{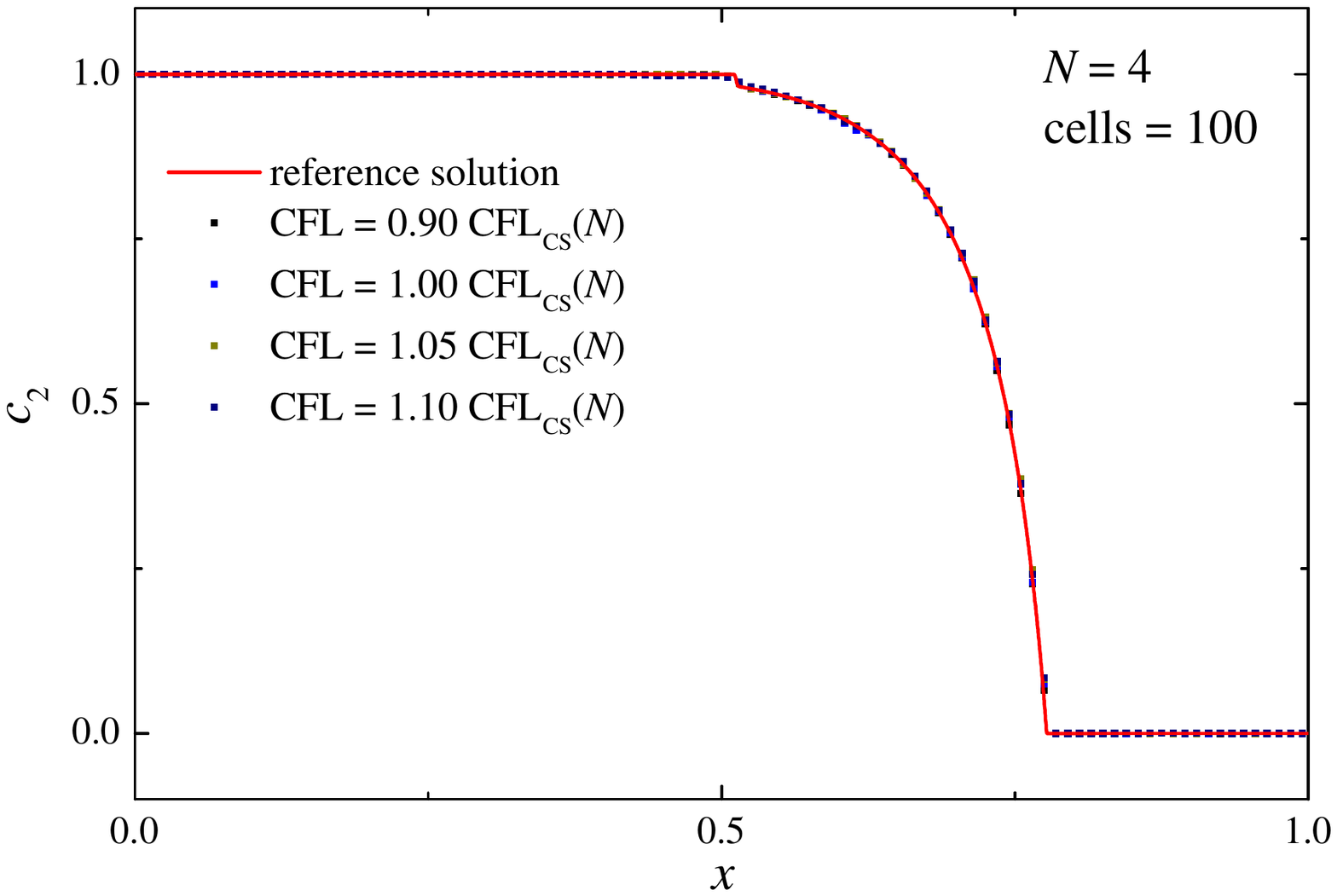}
\vspace{-8mm}\caption{\label{fig:test_znds_1d_degrees_4_6:degree_4_c2}}
\end{subfigure}
\begin{subfigure}{0.24\textwidth}
\includegraphics[width=\textwidth]{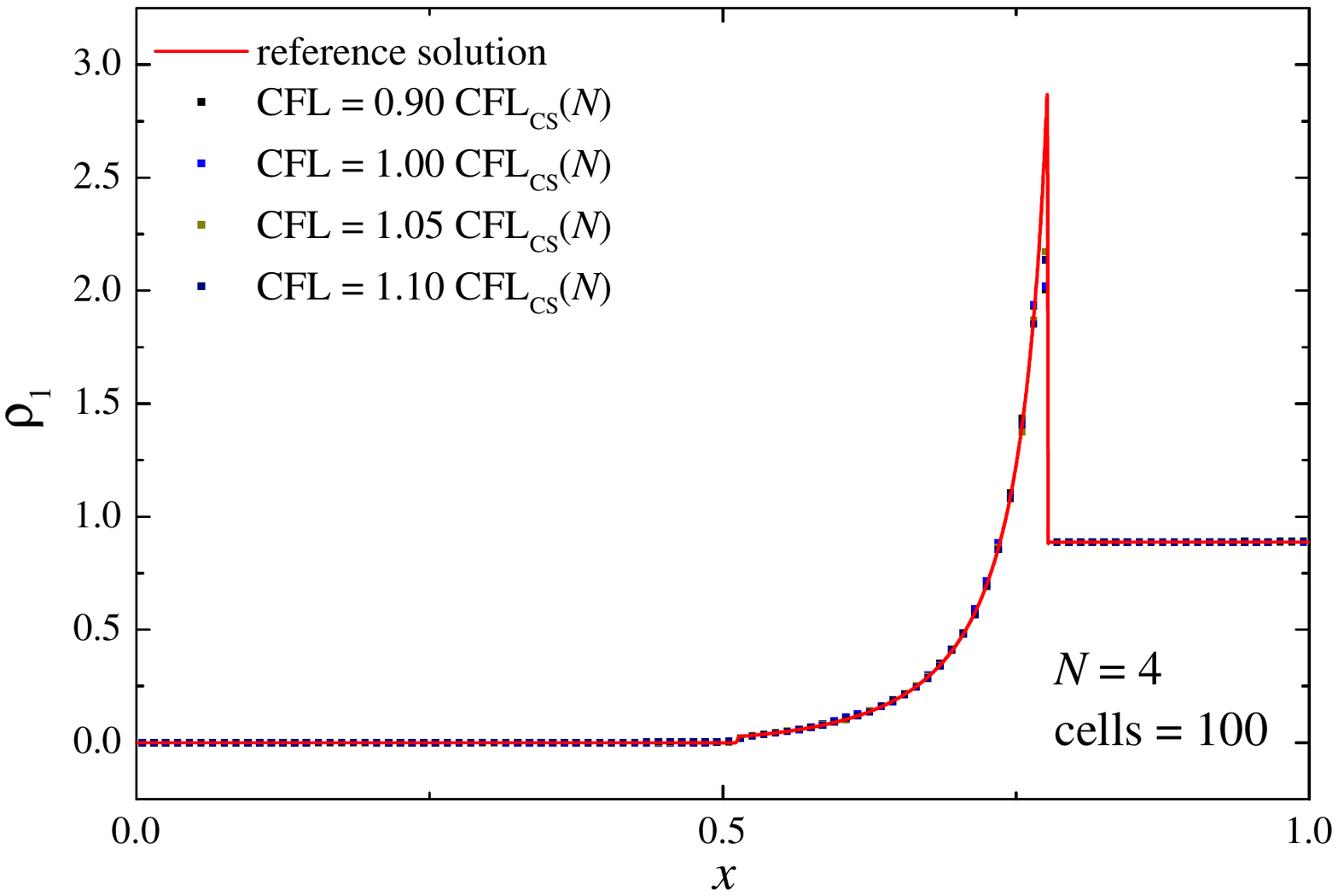}
\vspace{-8mm}\caption{\label{fig:test_znds_1d_degrees_4_6:degree_4_den1}}
\end{subfigure}
\begin{subfigure}{0.24\textwidth}
\includegraphics[width=\textwidth]{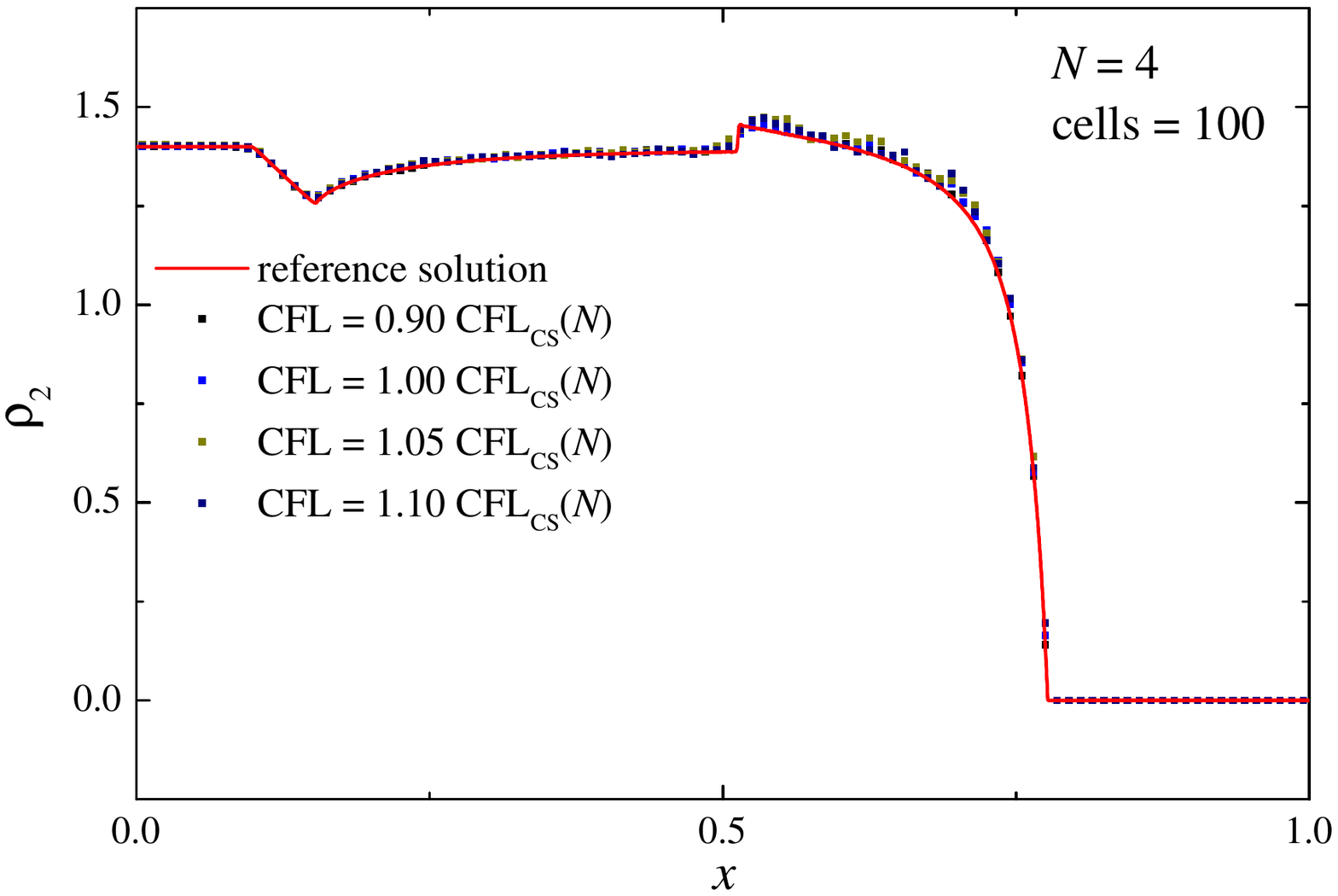}
\vspace{-8mm}\caption{\label{fig:test_znds_1d_degrees_4_6:degree_4_den2}}
\end{subfigure}\\[-2mm]
\begin{subfigure}{0.24\textwidth}
\includegraphics[width=\textwidth]{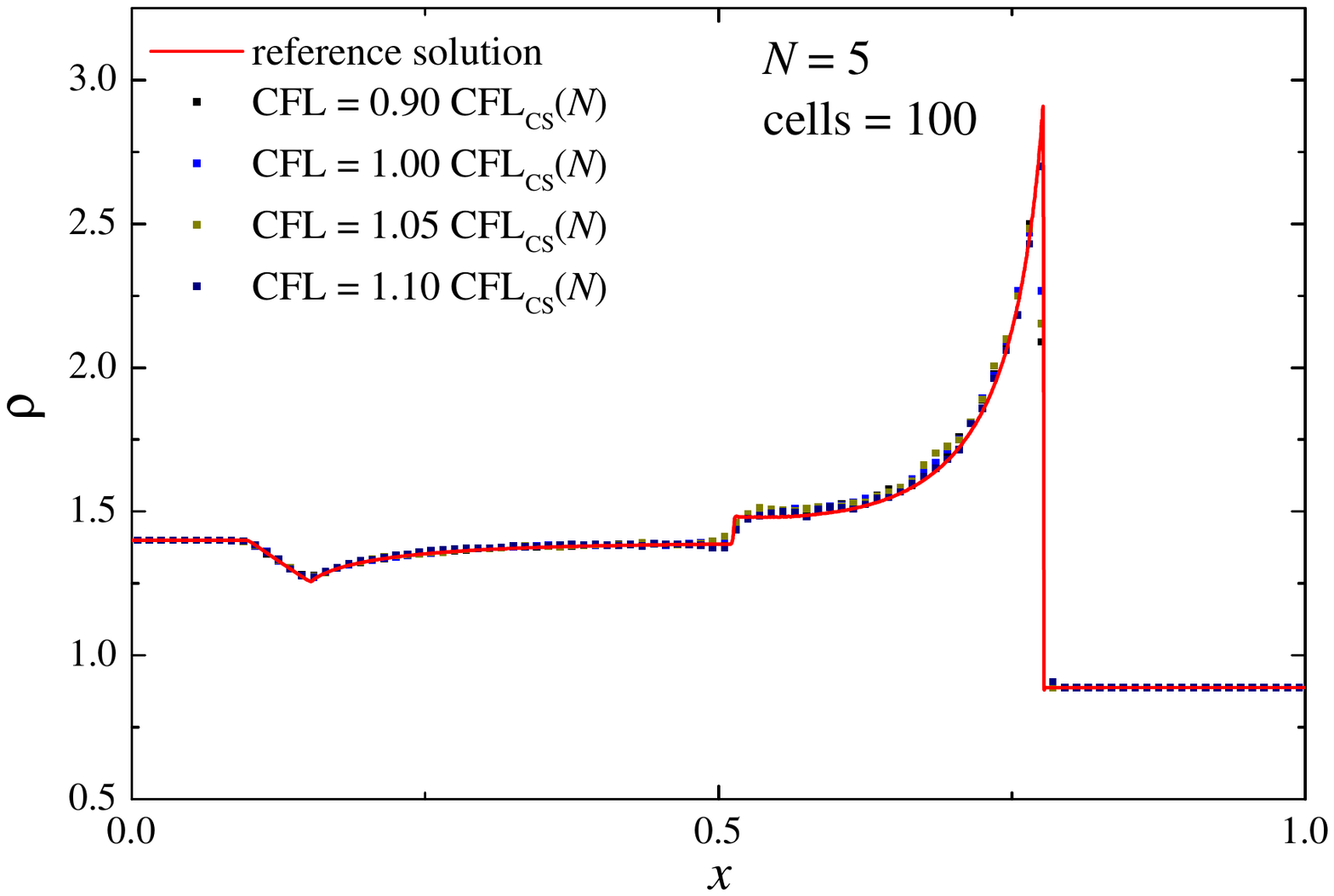}
\vspace{-8mm}\caption{\label{fig:test_znds_1d_degrees_4_6:degree_5_den}}
\end{subfigure}
\begin{subfigure}{0.24\textwidth}
\includegraphics[width=\textwidth]{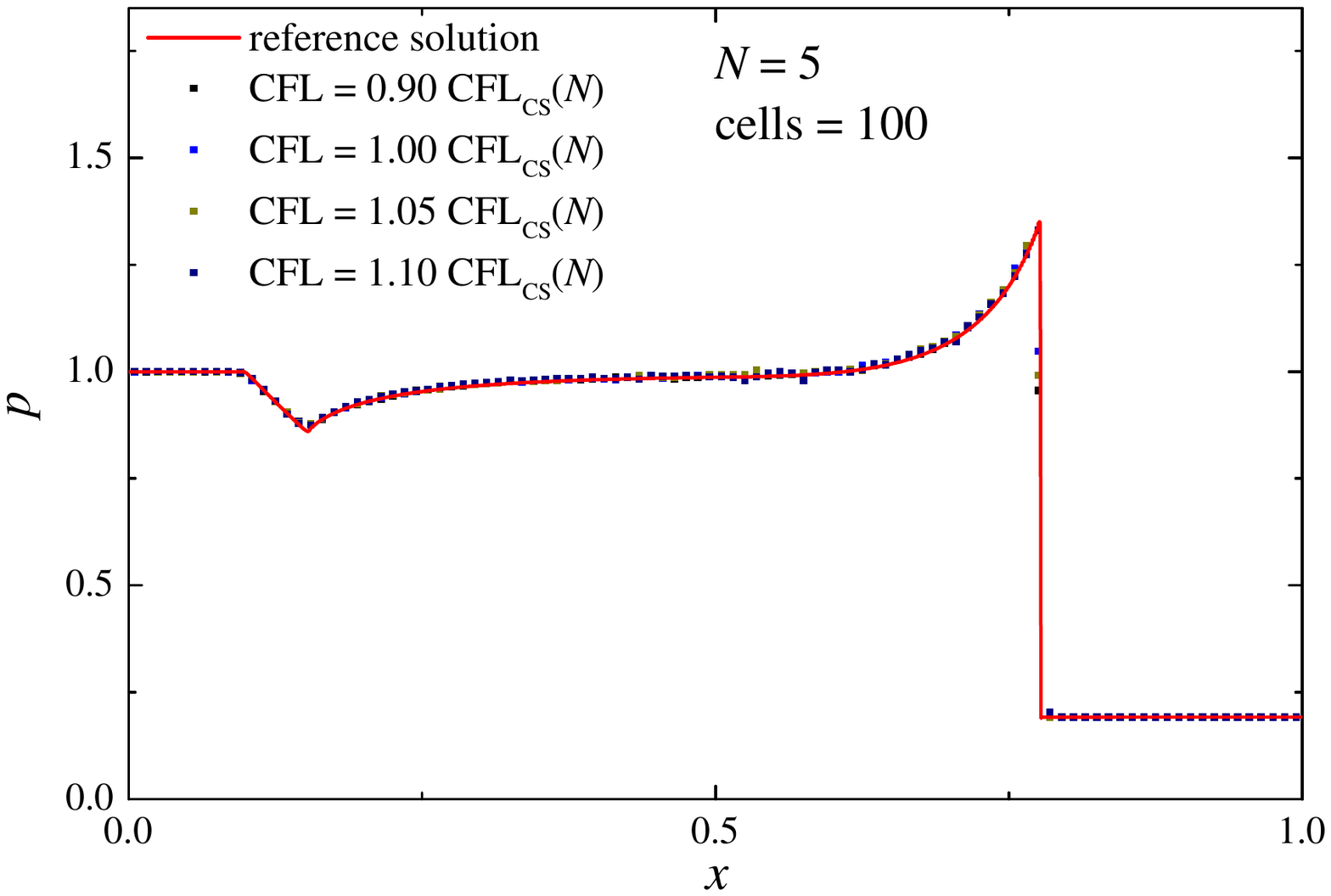}
\vspace{-8mm}\caption{\label{fig:test_znds_1d_degrees_4_6:degree_5_p}}
\end{subfigure}
\begin{subfigure}{0.24\textwidth}
\includegraphics[width=\textwidth]{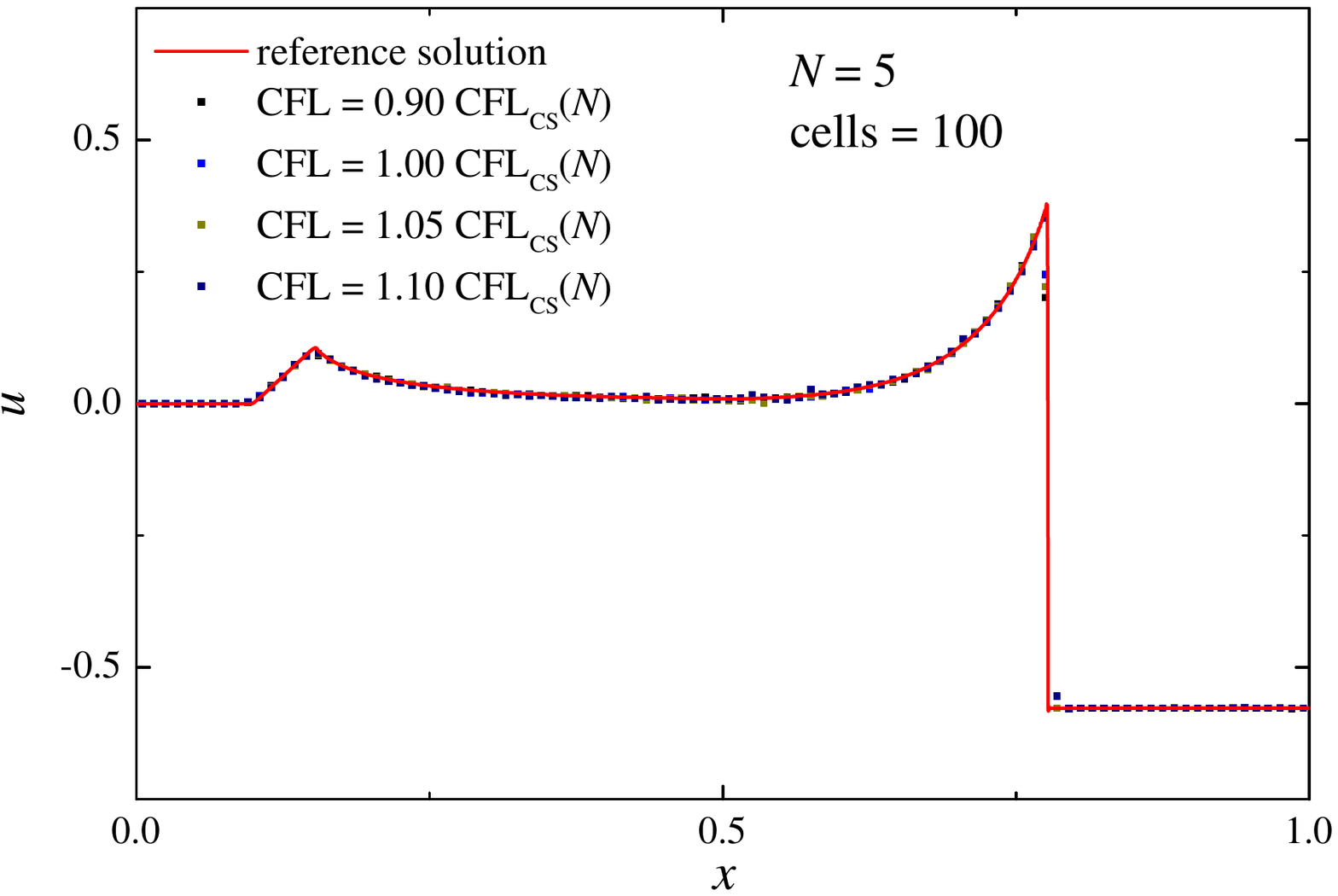}
\vspace{-8mm}\caption{\label{fig:test_znds_1d_degrees_4_6:degree_5_u}}
\end{subfigure}
\begin{subfigure}{0.24\textwidth}
\includegraphics[width=\textwidth]{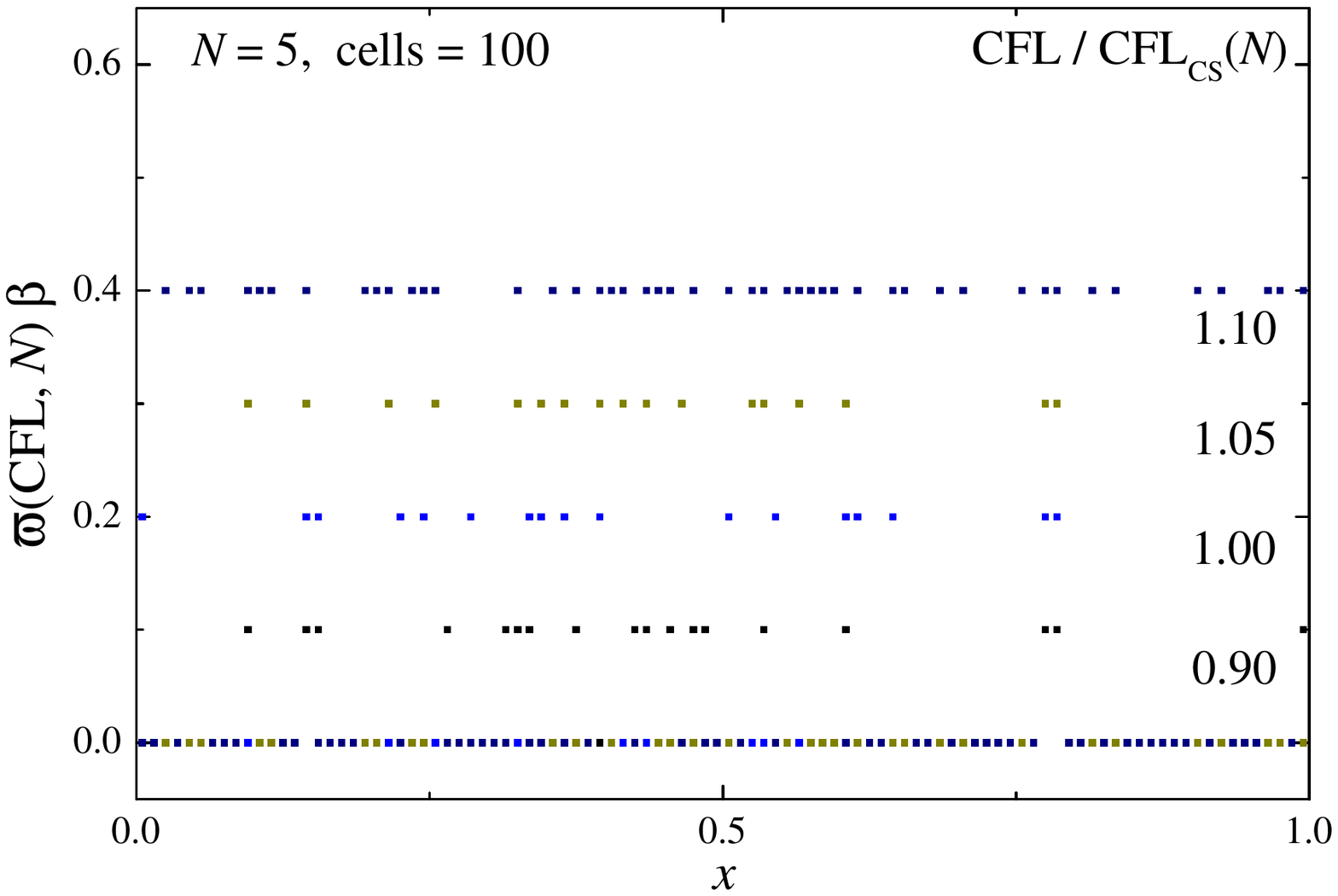}
\vspace{-8mm}\caption{\label{fig:test_znds_1d_degrees_4_6:degree_5_beta}}
\end{subfigure}\\[-2mm]
\begin{subfigure}{0.24\textwidth}
\includegraphics[width=\textwidth]{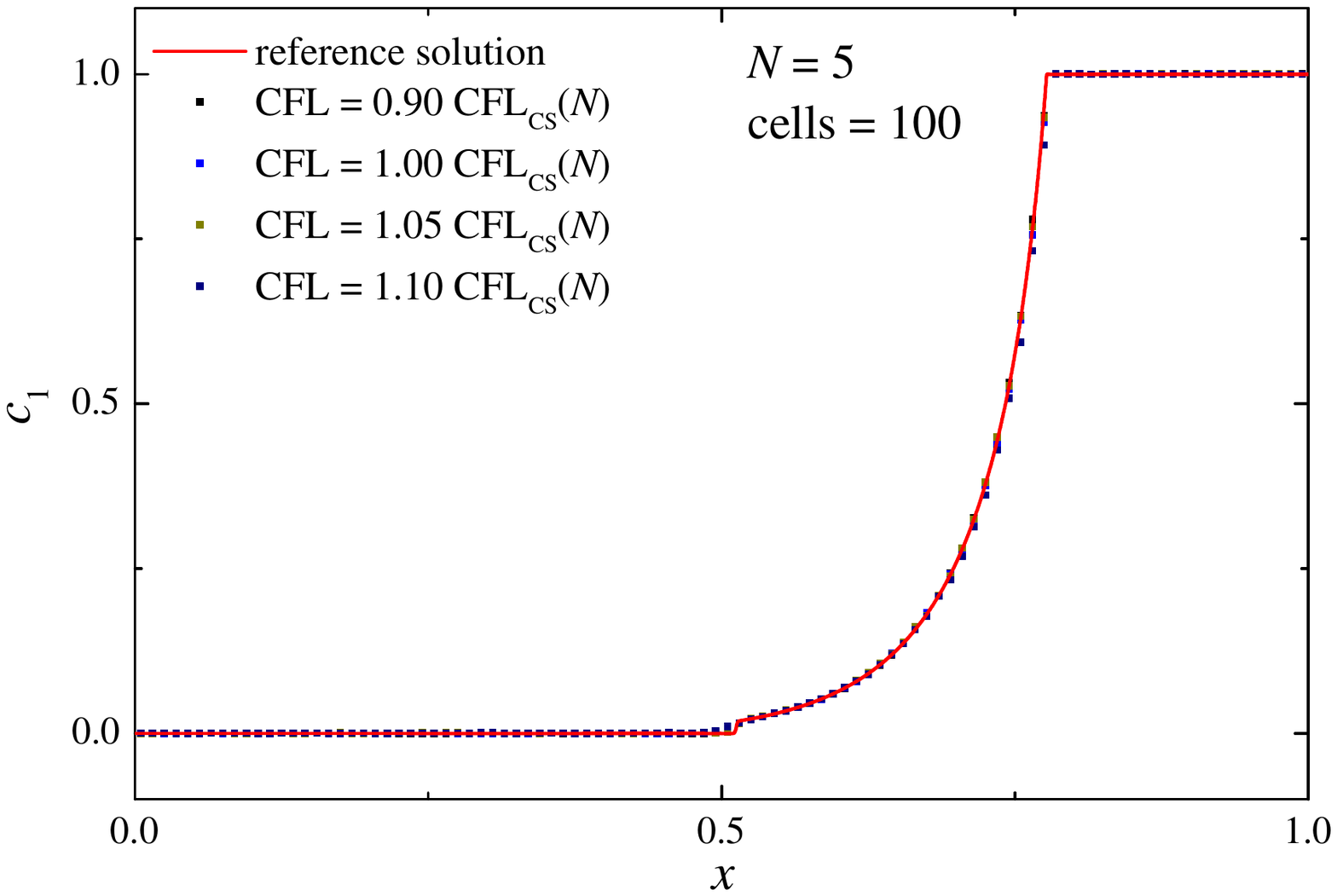}
\vspace{-8mm}\caption{\label{fig:test_znds_1d_degrees_4_6:degree_5_c1}}
\end{subfigure}
\begin{subfigure}{0.24\textwidth}
\includegraphics[width=\textwidth]{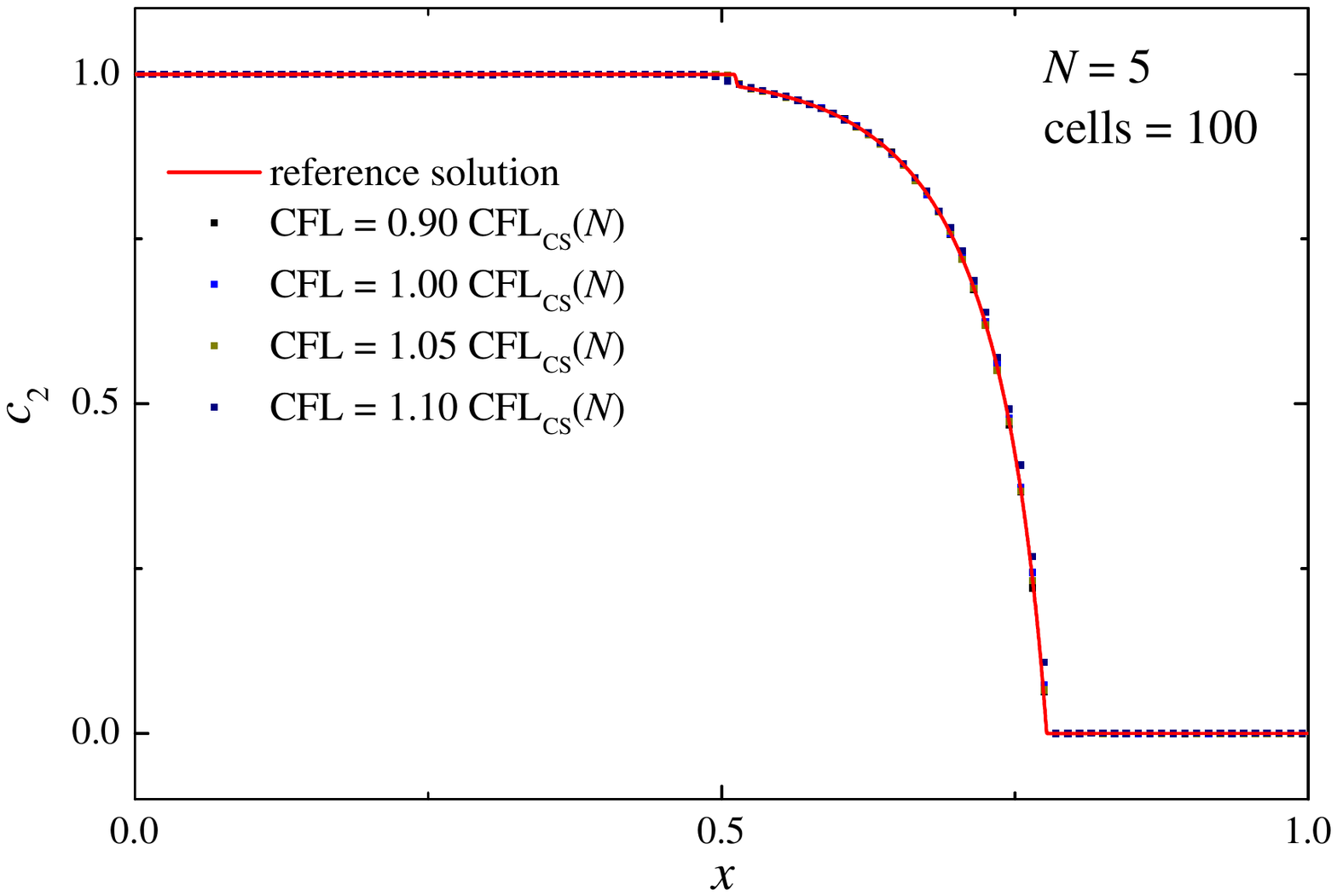}
\vspace{-8mm}\caption{\label{fig:test_znds_1d_degrees_4_6:degree_5_c2}}
\end{subfigure}
\begin{subfigure}{0.24\textwidth}
\includegraphics[width=\textwidth]{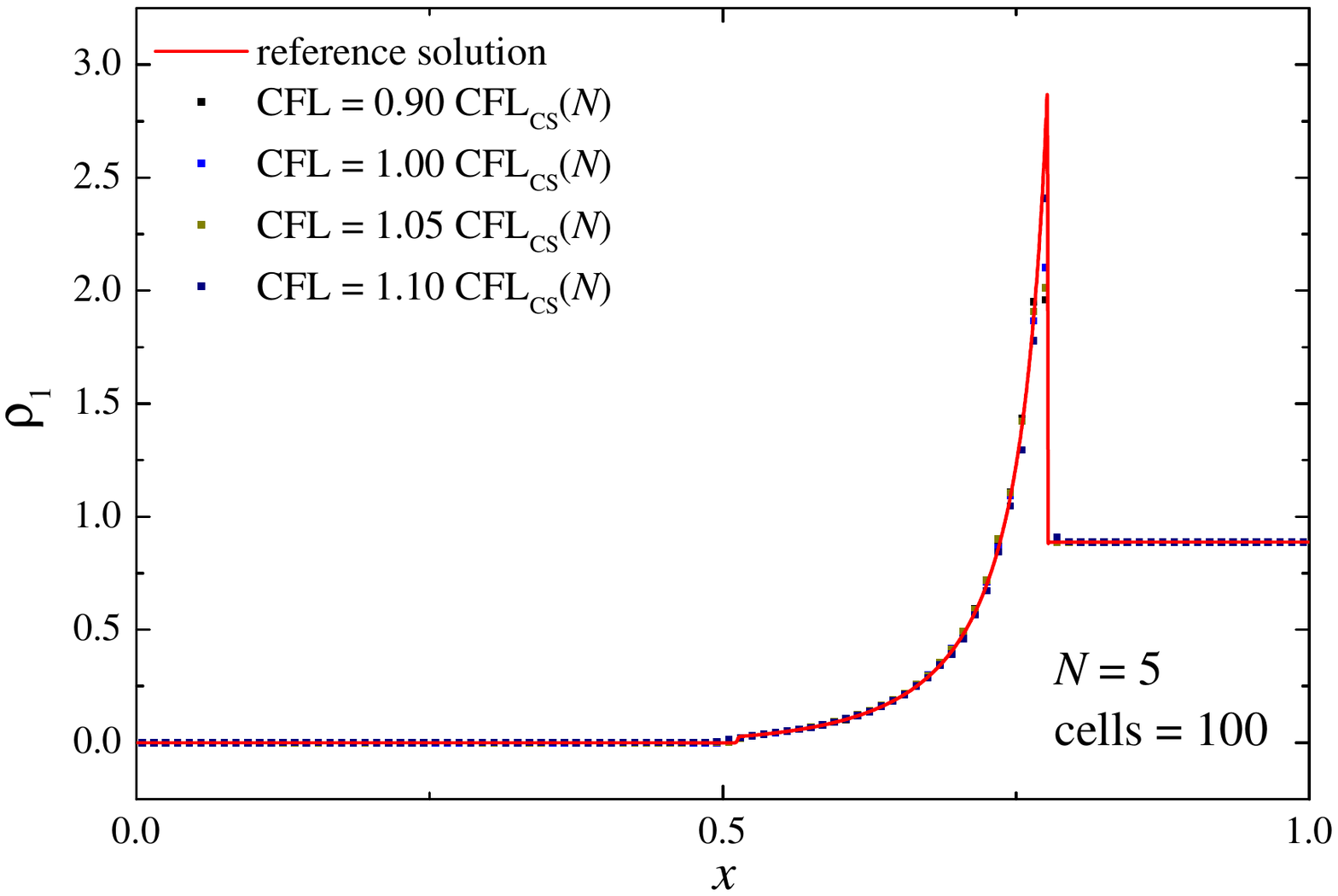}
\vspace{-8mm}\caption{\label{fig:test_znds_1d_degrees_4_6:degree_5_den1}}
\end{subfigure}
\begin{subfigure}{0.24\textwidth}
\includegraphics[width=\textwidth]{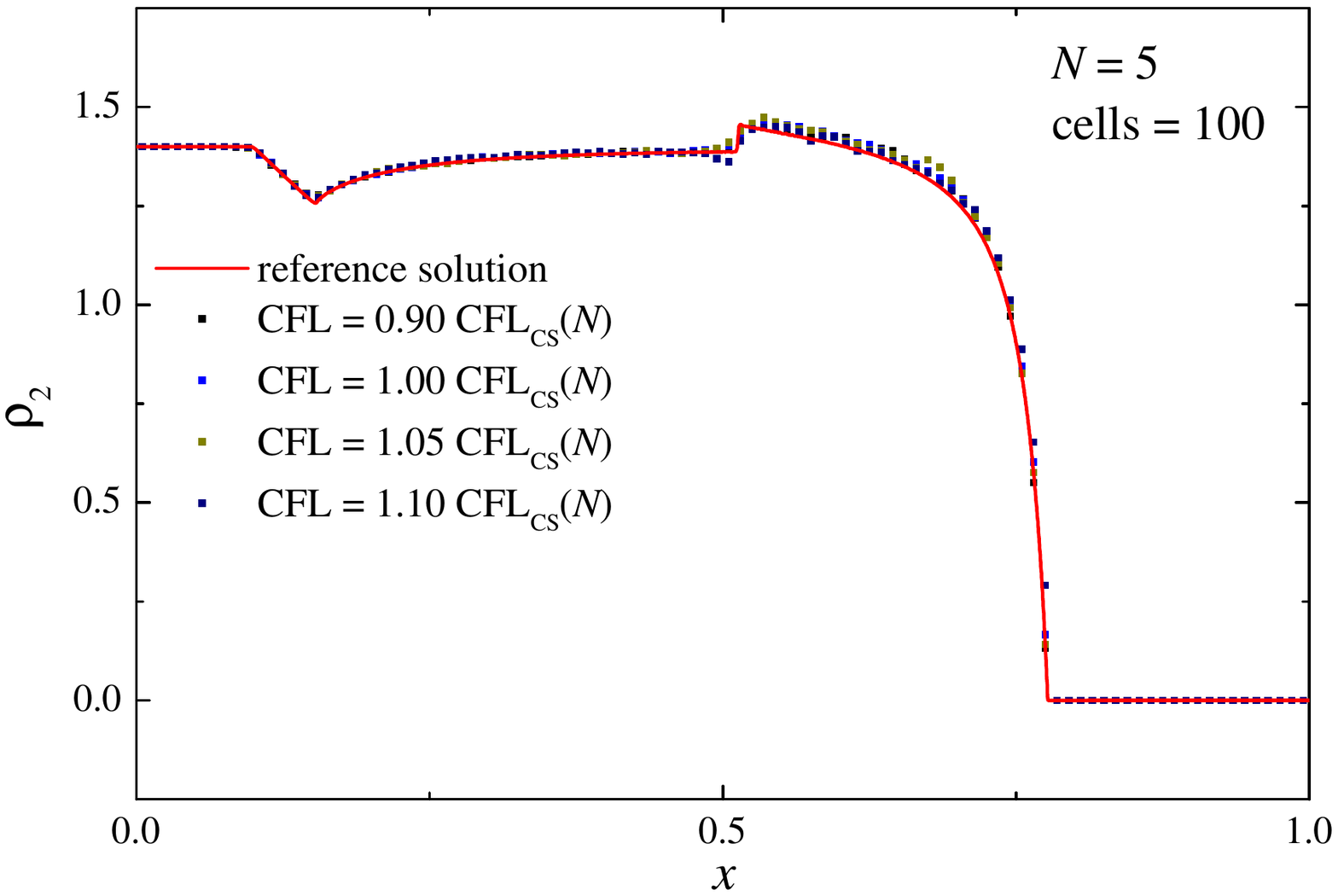}
\vspace{-8mm}\caption{\label{fig:test_znds_1d_degrees_4_6:degree_5_den2}}
\end{subfigure}\\[-2mm]
\begin{subfigure}{0.24\textwidth}
\includegraphics[width=\textwidth]{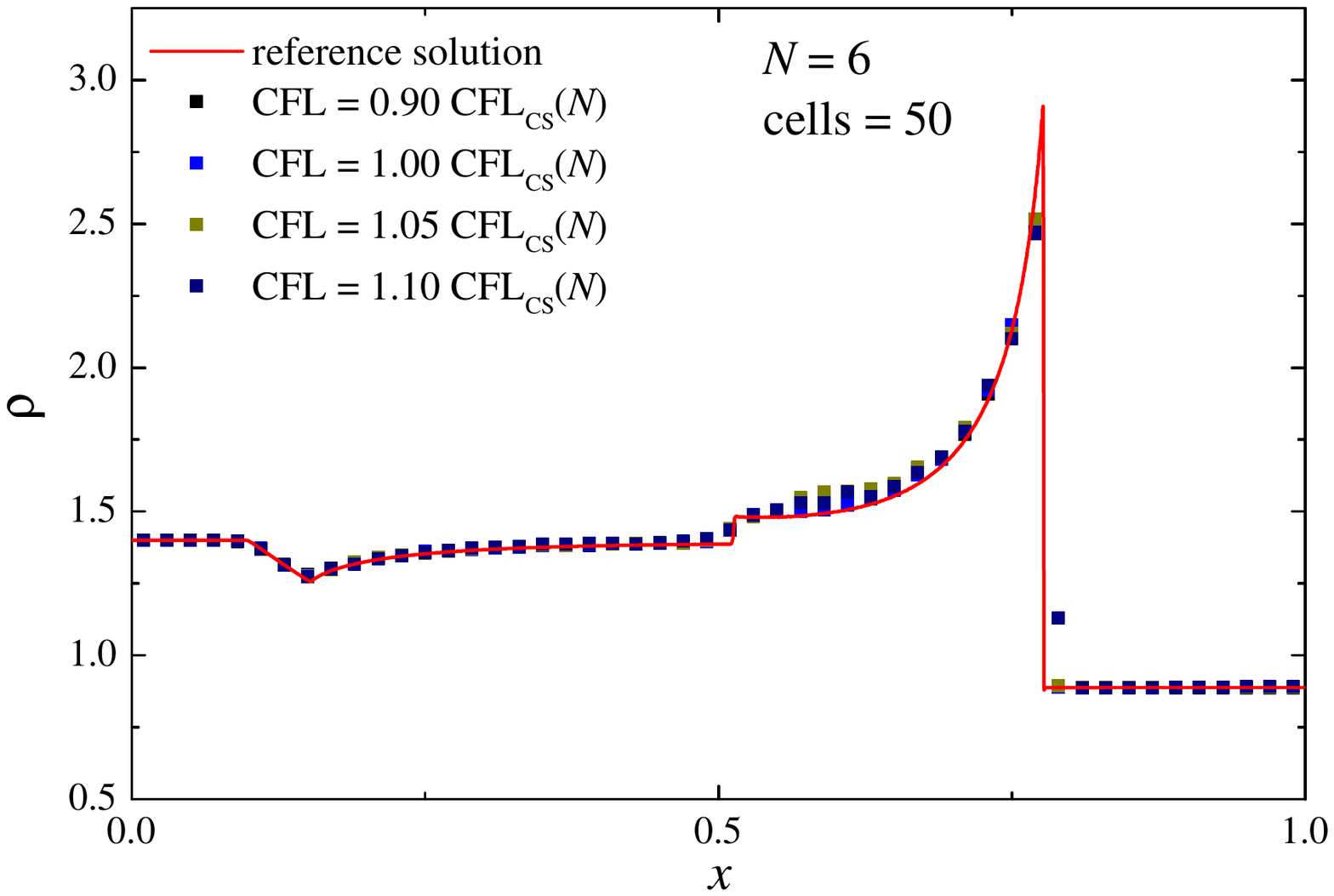}
\vspace{-8mm}\caption{\label{fig:test_znds_1d_degrees_4_6:degree_6_den}}
\end{subfigure}
\begin{subfigure}{0.24\textwidth}
\includegraphics[width=\textwidth]{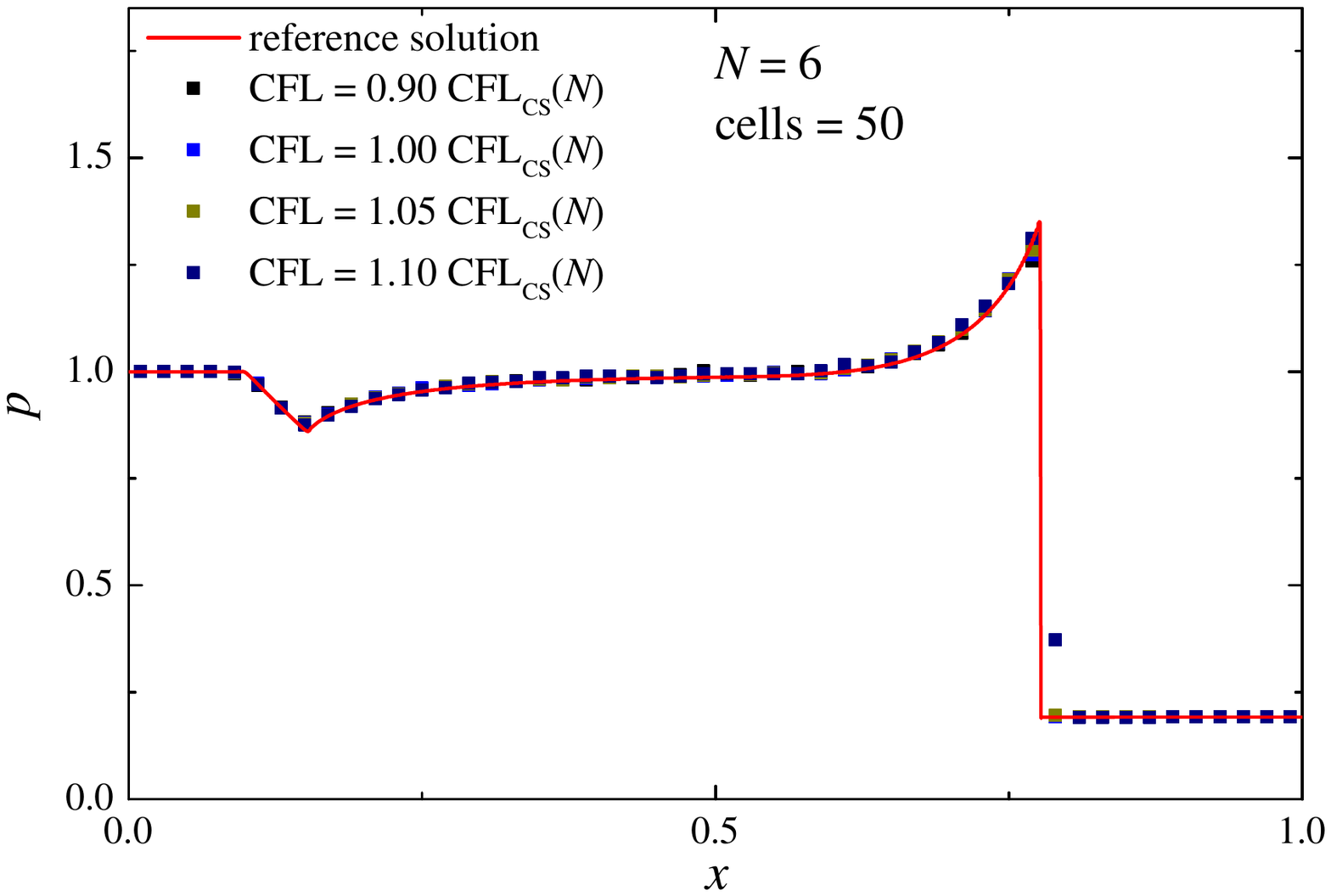}
\vspace{-8mm}\caption{\label{fig:test_znds_1d_degrees_4_6:degree_6_p}}
\end{subfigure}
\begin{subfigure}{0.24\textwidth}
\includegraphics[width=\textwidth]{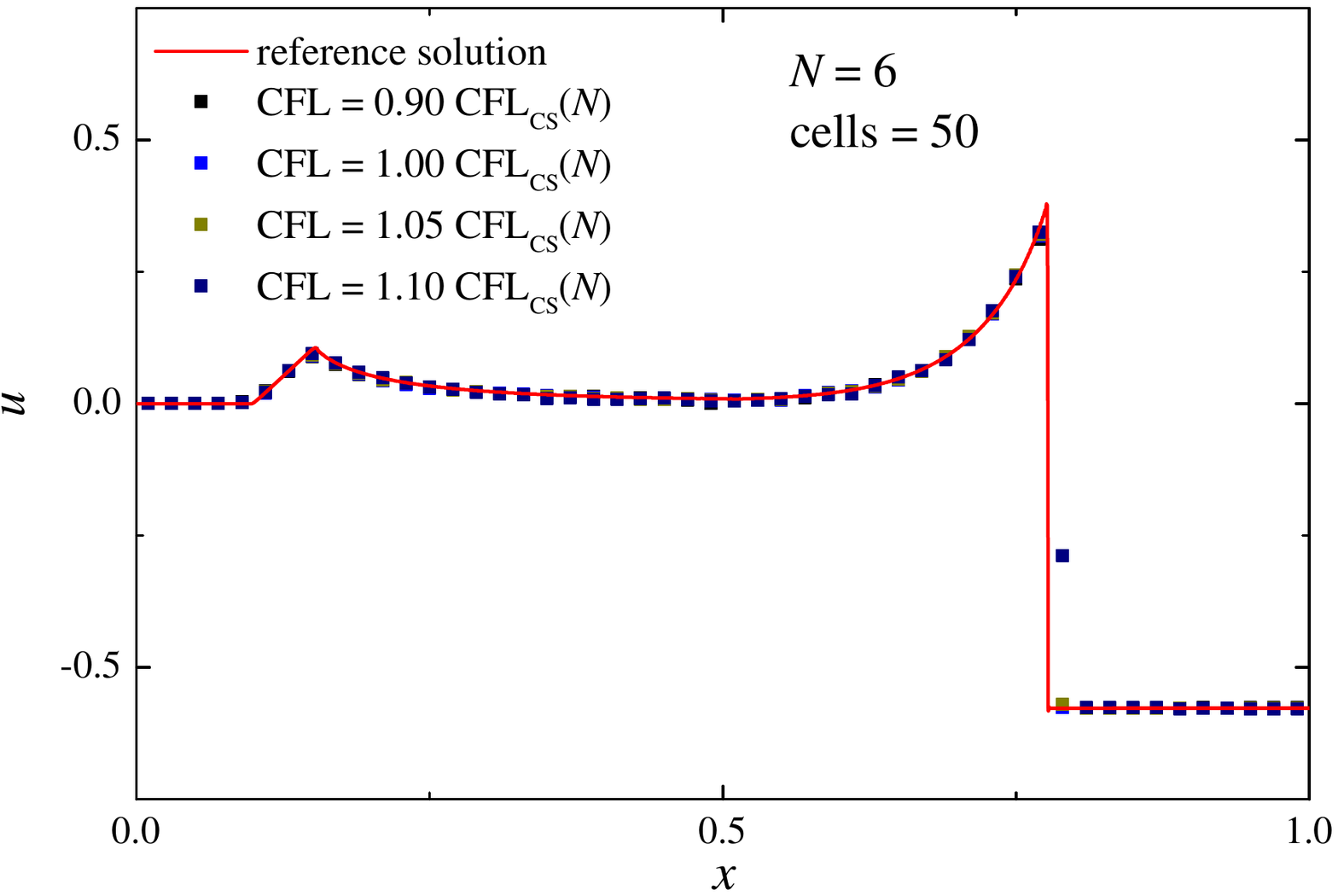}
\vspace{-8mm}\caption{\label{fig:test_znds_1d_degrees_4_6:degree_6_u}}
\end{subfigure}
\begin{subfigure}{0.24\textwidth}
\includegraphics[width=\textwidth]{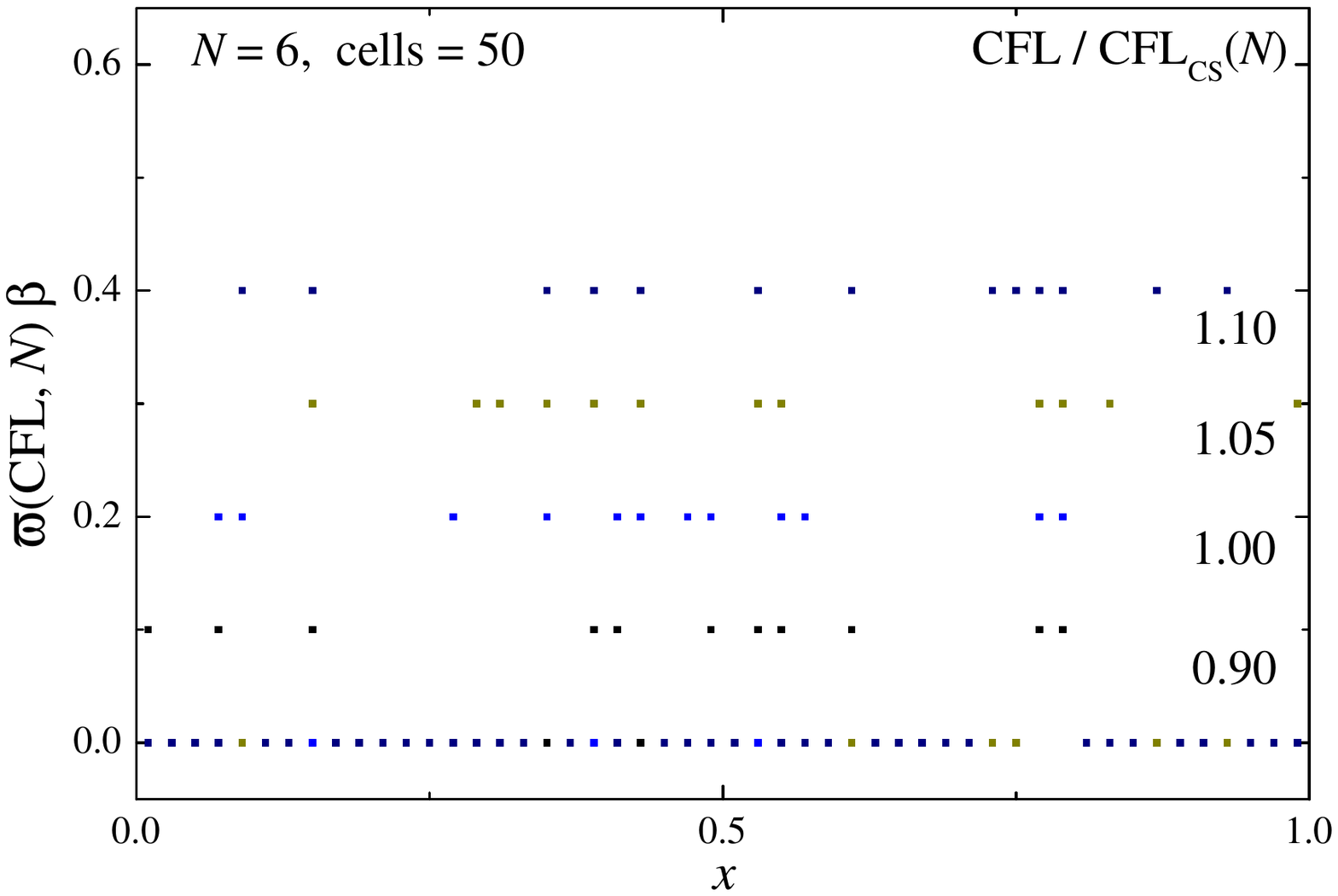}
\vspace{-8mm}\caption{\label{fig:test_znds_1d_degrees_4_6:degree_6_beta}}
\end{subfigure}\\[-2mm]
\begin{subfigure}{0.24\textwidth}
\includegraphics[width=\textwidth]{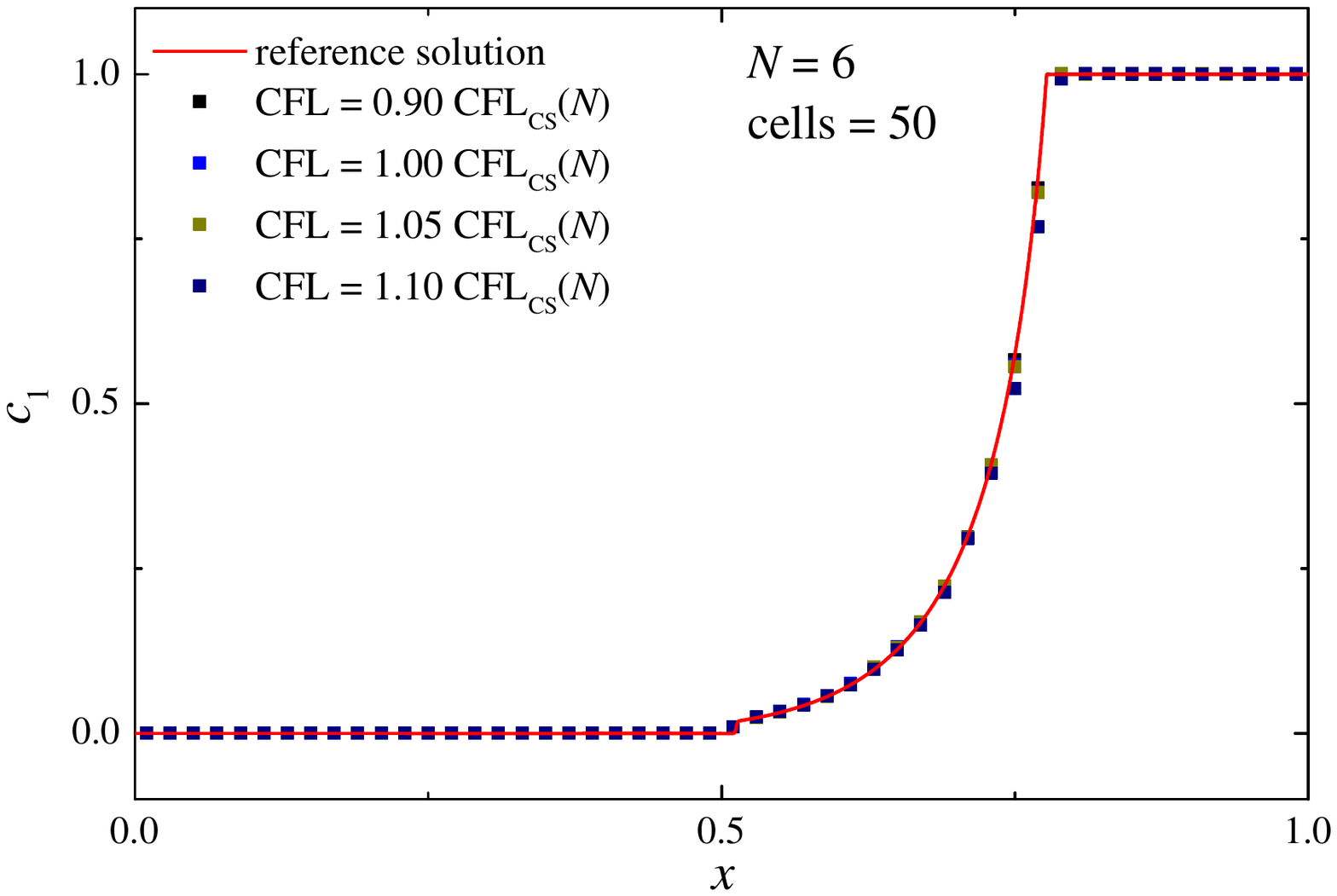}
\vspace{-8mm}\caption{\label{fig:test_znds_1d_degrees_4_6:degree_6_c1}}
\end{subfigure}
\begin{subfigure}{0.24\textwidth}
\includegraphics[width=\textwidth]{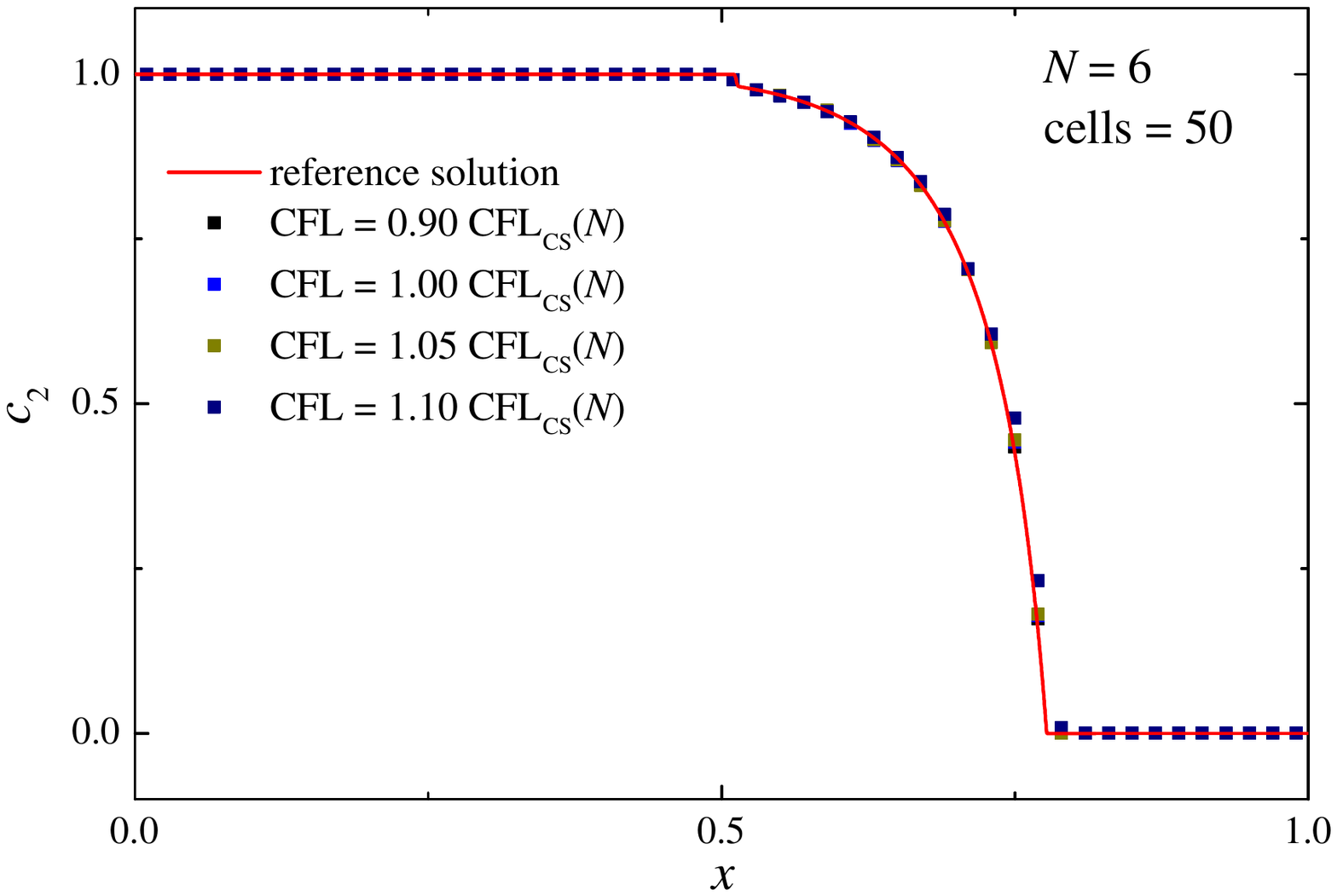}
\vspace{-8mm}\caption{\label{fig:test_znds_1d_degrees_4_6:degree_6_c2}}
\end{subfigure}
\begin{subfigure}{0.24\textwidth}
\includegraphics[width=\textwidth]{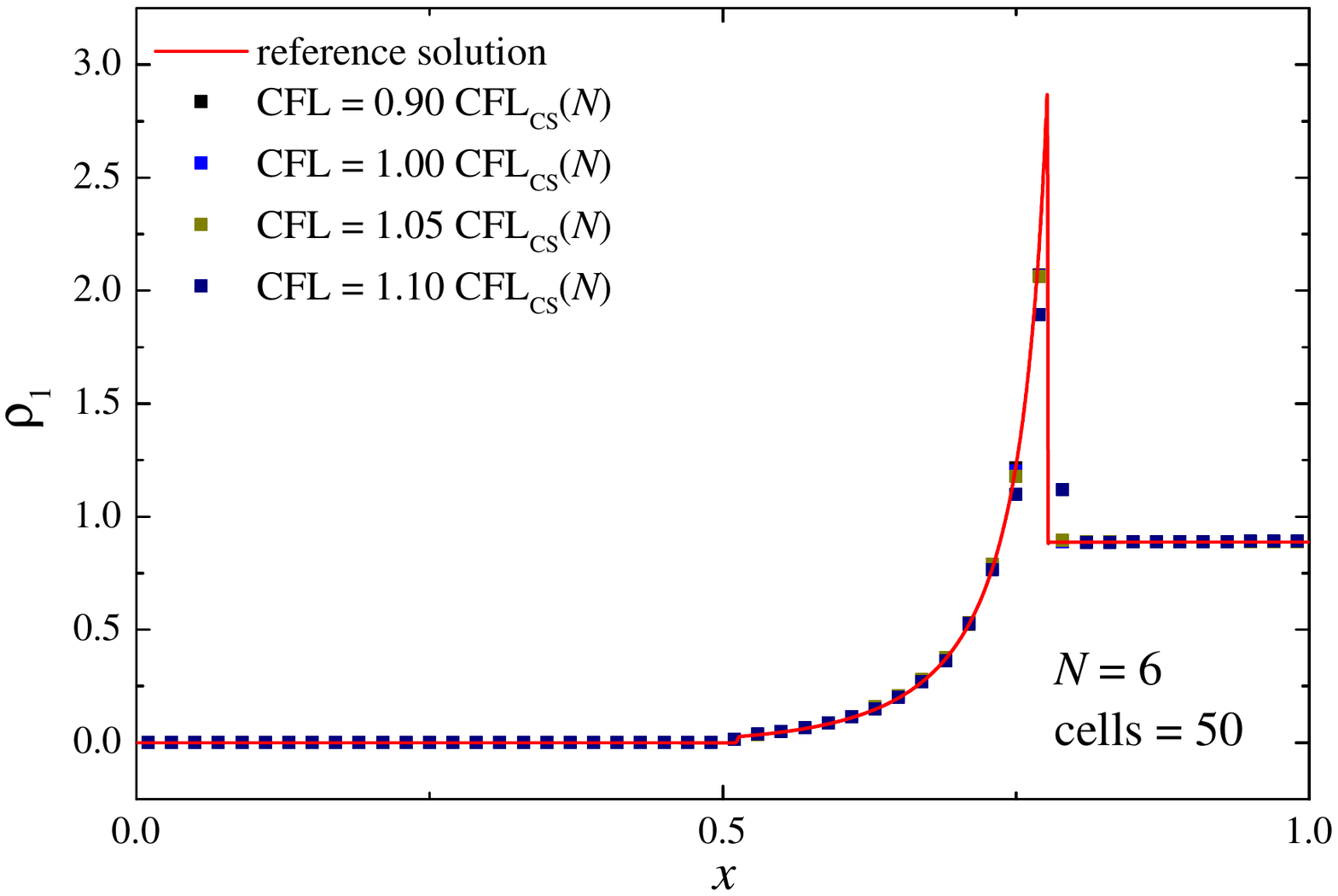}
\vspace{-8mm}\caption{\label{fig:test_znds_1d_degrees_4_6:degree_6_den1}}
\end{subfigure}
\begin{subfigure}{0.24\textwidth}
\includegraphics[width=\textwidth]{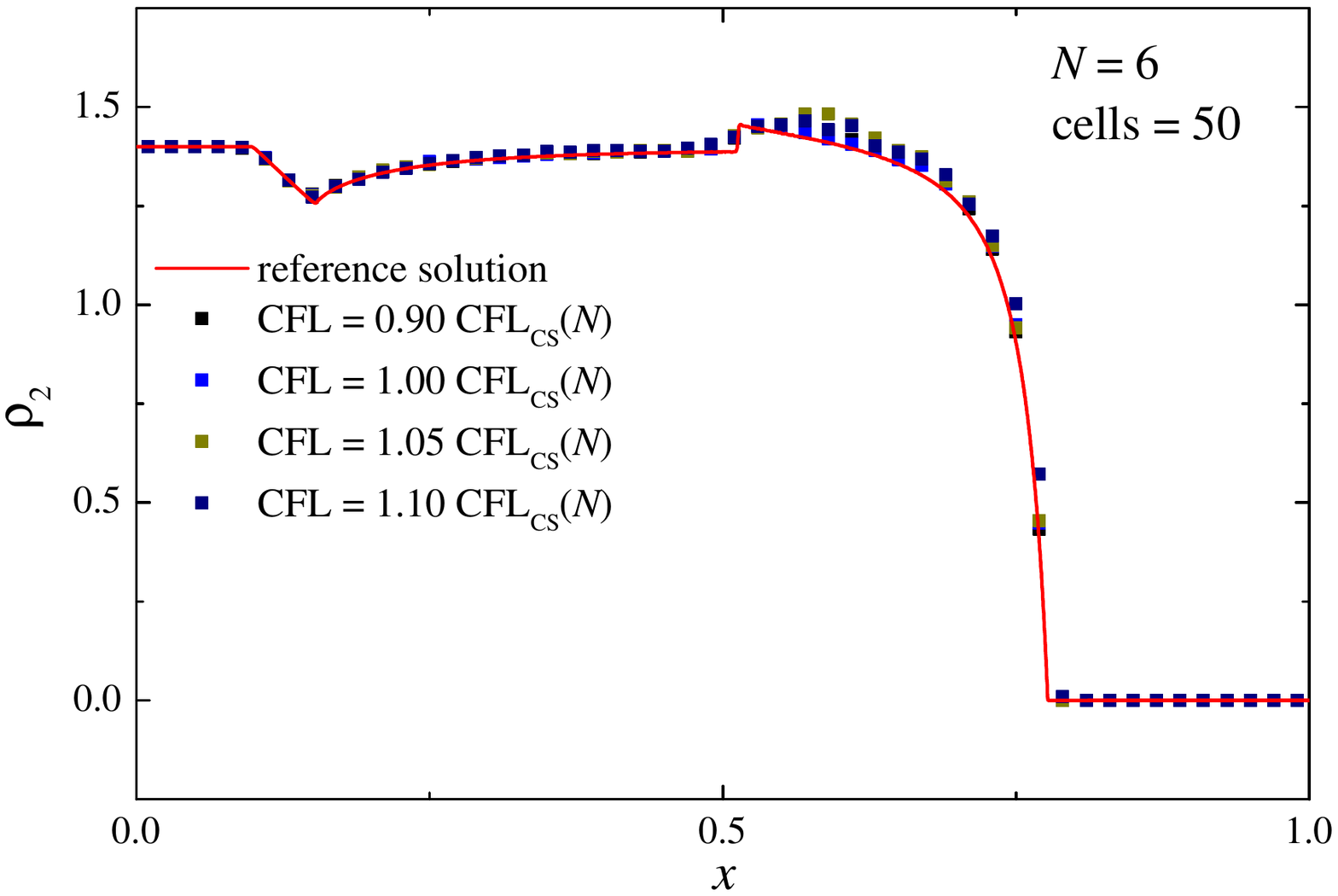}
\vspace{-8mm}\caption{\label{fig:test_znds_1d_degrees_4_6:degree_6_den2}}
\end{subfigure}
\caption{%
Numerical solution of the one-dimensional problem of the plane detonation wave formation in a two-component medium with a ``slow'' reaction (weak stiff case, a detailed statement of the problem is presented in the text), obtained using the ADER-DG method for the basis polynomials degree $N = 4$ ((\subref{fig:test_znds_1d_degrees_4_6:degree_4_den}), (\subref{fig:test_znds_1d_degrees_4_6:degree_4_p}), (\subref{fig:test_znds_1d_degrees_4_6:degree_4_u}), (\subref{fig:test_znds_1d_degrees_4_6:degree_4_c1}), (\subref{fig:test_znds_1d_degrees_4_6:degree_4_c2}), (\subref{fig:test_znds_1d_degrees_4_6:degree_4_den1}), (\subref{fig:test_znds_1d_degrees_4_6:degree_4_den2})), $N = 5$ ((\subref{fig:test_znds_1d_degrees_4_6:degree_5_den}), (\subref{fig:test_znds_1d_degrees_4_6:degree_5_p}), (\subref{fig:test_znds_1d_degrees_4_6:degree_5_u}), (\subref{fig:test_znds_1d_degrees_4_6:degree_5_c1}), (\subref{fig:test_znds_1d_degrees_4_6:degree_5_c2}), (\subref{fig:test_znds_1d_degrees_4_6:degree_5_den1}), (\subref{fig:test_znds_1d_degrees_4_6:degree_5_den2})), $N = 6$ ((\subref{fig:test_znds_1d_degrees_4_6:degree_6_den}), (\subref{fig:test_znds_1d_degrees_4_6:degree_6_p}), (\subref{fig:test_znds_1d_degrees_4_6:degree_6_u}), (\subref{fig:test_znds_1d_degrees_4_6:degree_6_c1}), (\subref{fig:test_znds_1d_degrees_4_6:degree_6_c2}), (\subref{fig:test_znds_1d_degrees_4_6:degree_6_den1}), (\subref{fig:test_znds_1d_degrees_4_6:degree_6_den2})) with a posteriori limitation of the solution by an ADER-WENO2 finite-volume limiter, at the final time $t_{f} = 0.4$. The graphs show the coordinate dependence of density $\rho$ ((\subref{fig:test_znds_1d_degrees_4_6:degree_4_den}), (\subref{fig:test_znds_1d_degrees_4_6:degree_5_den}), (\subref{fig:test_znds_1d_degrees_4_6:degree_6_den})), pressure $p$ ((\subref{fig:test_znds_1d_degrees_4_6:degree_4_p}), (\subref{fig:test_znds_1d_degrees_4_6:degree_5_p}), (\subref{fig:test_znds_1d_degrees_4_6:degree_6_p})), flow velocity $u$ ((\subref{fig:test_znds_1d_degrees_4_6:degree_4_u}), (\subref{fig:test_znds_1d_degrees_4_6:degree_5_u}), (\subref{fig:test_znds_1d_degrees_4_6:degree_6_u})), troubled cells indicator $\beta$ multiplied by the separating function $\varpi(\mathrm{CFL}, N)$ ((\subref{fig:test_znds_1d_degrees_4_6:degree_4_beta}), (\subref{fig:test_znds_1d_degrees_4_6:degree_5_beta}), (\subref{fig:test_znds_1d_degrees_4_6:degree_6_beta})), mass concentrations $c_{1}$ ((\subref{fig:test_znds_1d_degrees_4_6:degree_4_c1}), (\subref{fig:test_znds_1d_degrees_4_6:degree_5_c1}), (\subref{fig:test_znds_1d_degrees_4_6:degree_6_c1})) and $c_{2}$ ((\subref{fig:test_znds_1d_degrees_4_6:degree_4_c2}), (\subref{fig:test_znds_1d_degrees_4_6:degree_5_c2}), (\subref{fig:test_znds_1d_degrees_4_6:degree_6_c2})) of the reactant $A$ and product $B$, respectively, and partial densities $\rho_{1}$ ((\subref{fig:test_znds_1d_degrees_4_6:degree_4_den1}), (\subref{fig:test_znds_1d_degrees_4_6:degree_5_den1}), (\subref{fig:test_znds_1d_degrees_4_6:degree_6_den1})) and $\rho_{2}$ ((\subref{fig:test_znds_1d_degrees_4_6:degree_4_den2}), (\subref{fig:test_znds_1d_degrees_4_6:degree_5_den2}), (\subref{fig:test_znds_1d_degrees_4_6:degree_6_den2})) of the reactant $A$ and product $B$, respectively. The square symbols represent the subcells finite-volume representation of the numerical solution; the red solid lines represents the reference solution of the problem.
}
\label{fig:test_znds_1d_degrees_4_6}
\end{figure}
\begin{figure}[h!]
\captionsetup[subfigure]{%
	position=bottom,
	font+=smaller,
	textfont=normalfont,
	singlelinecheck=off,
	justification=raggedright
}
\centering
\begin{subfigure}{0.24\textwidth}
\includegraphics[width=\textwidth]{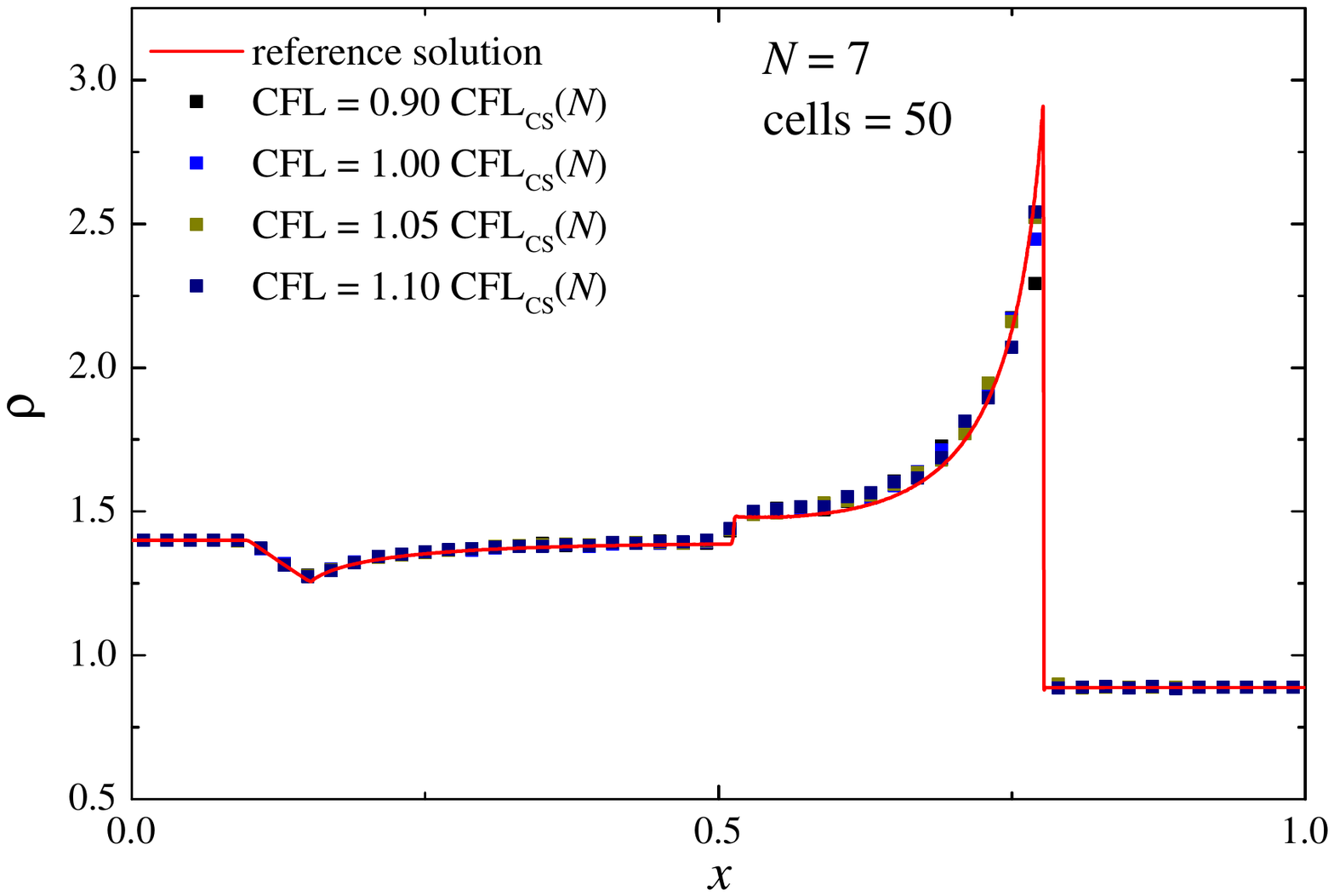}
\vspace{-8mm}\caption{\label{fig:test_znds_1d_degrees_7_9:degree_7_den}}
\end{subfigure}
\begin{subfigure}{0.24\textwidth}
\includegraphics[width=\textwidth]{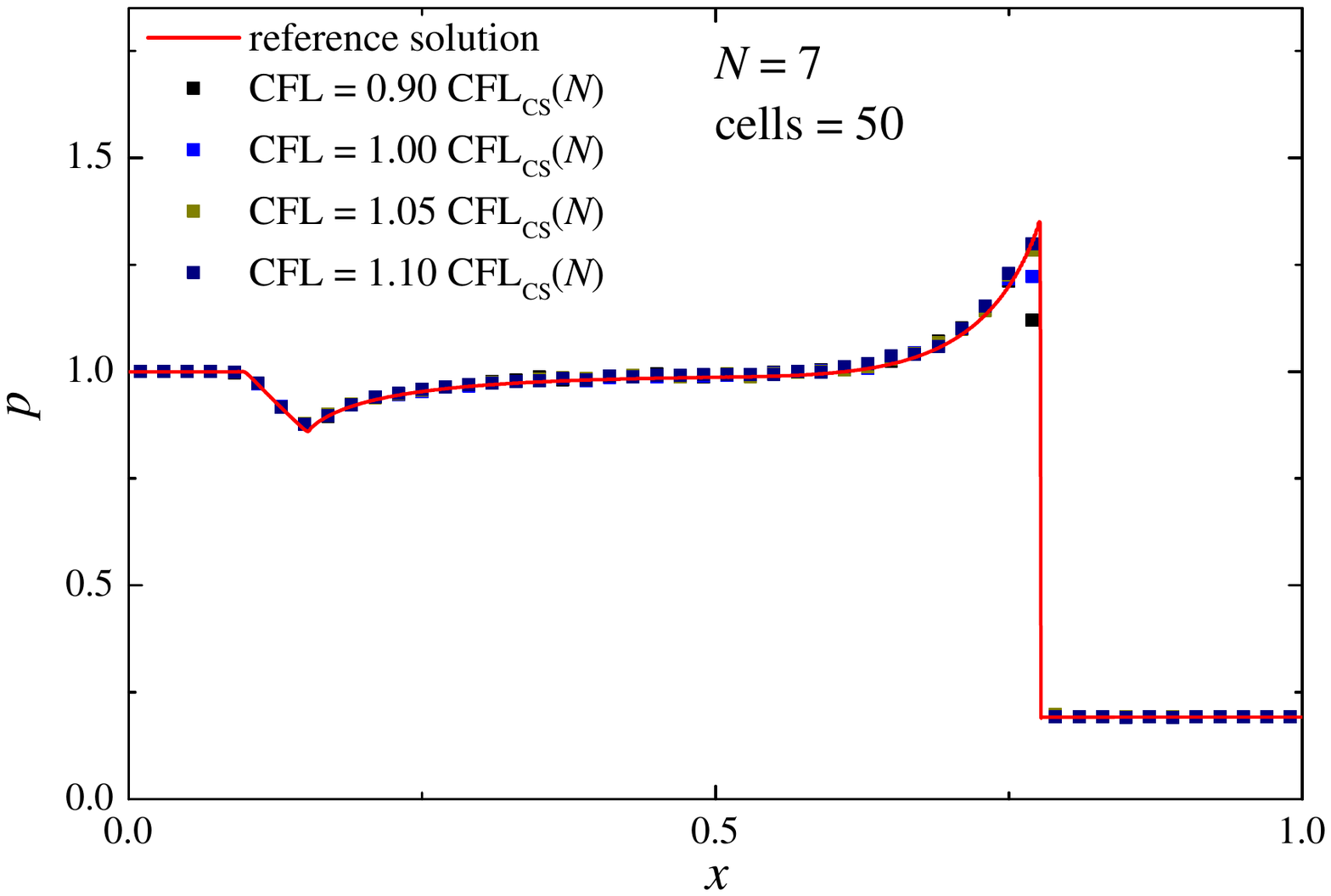}
\vspace{-8mm}\caption{\label{fig:test_znds_1d_degrees_7_9:degree_7_p}}
\end{subfigure}
\begin{subfigure}{0.24\textwidth}
\includegraphics[width=\textwidth]{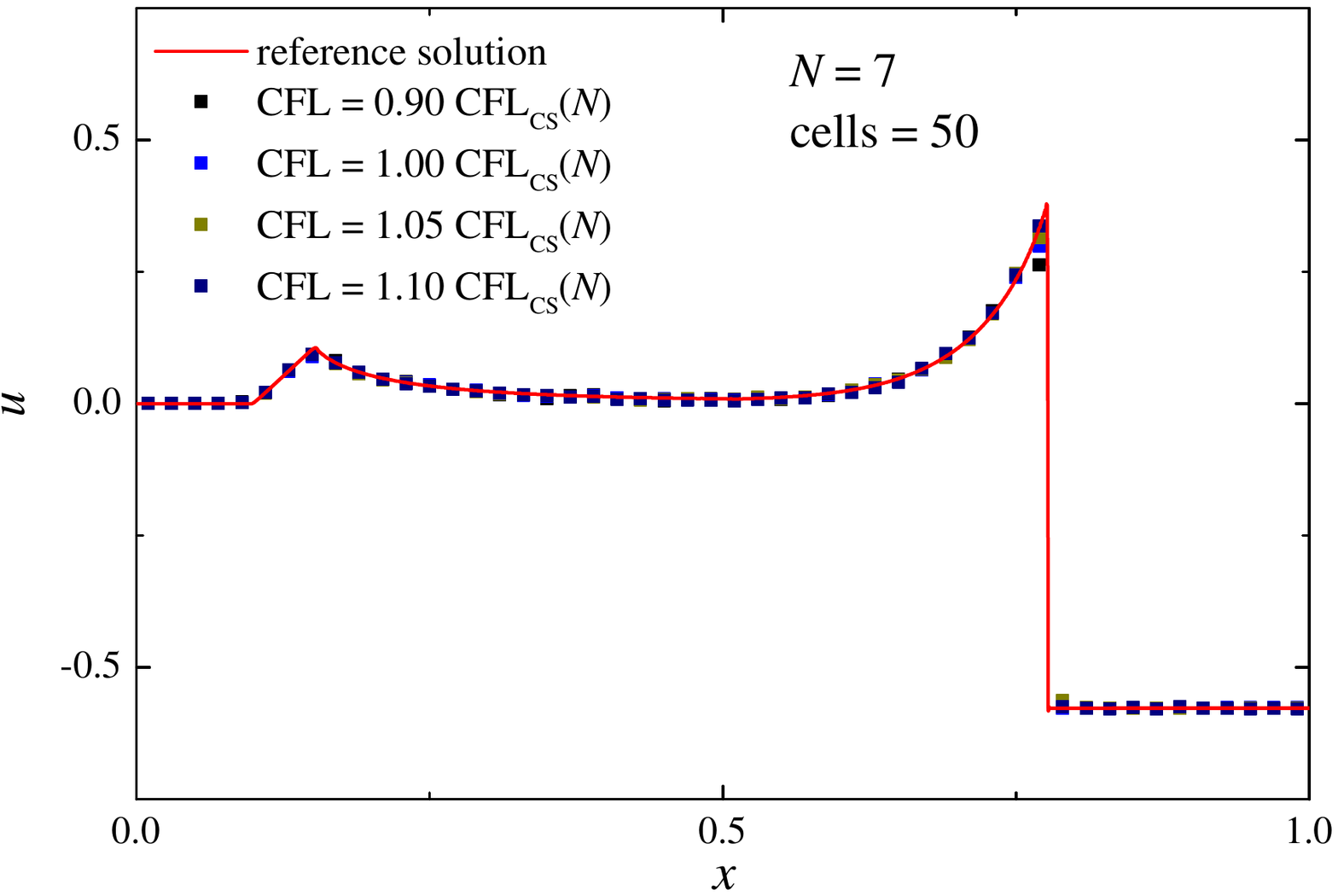}
\vspace{-8mm}\caption{\label{fig:test_znds_1d_degrees_7_9:degree_7_u}}
\end{subfigure}
\begin{subfigure}{0.24\textwidth}
\includegraphics[width=\textwidth]{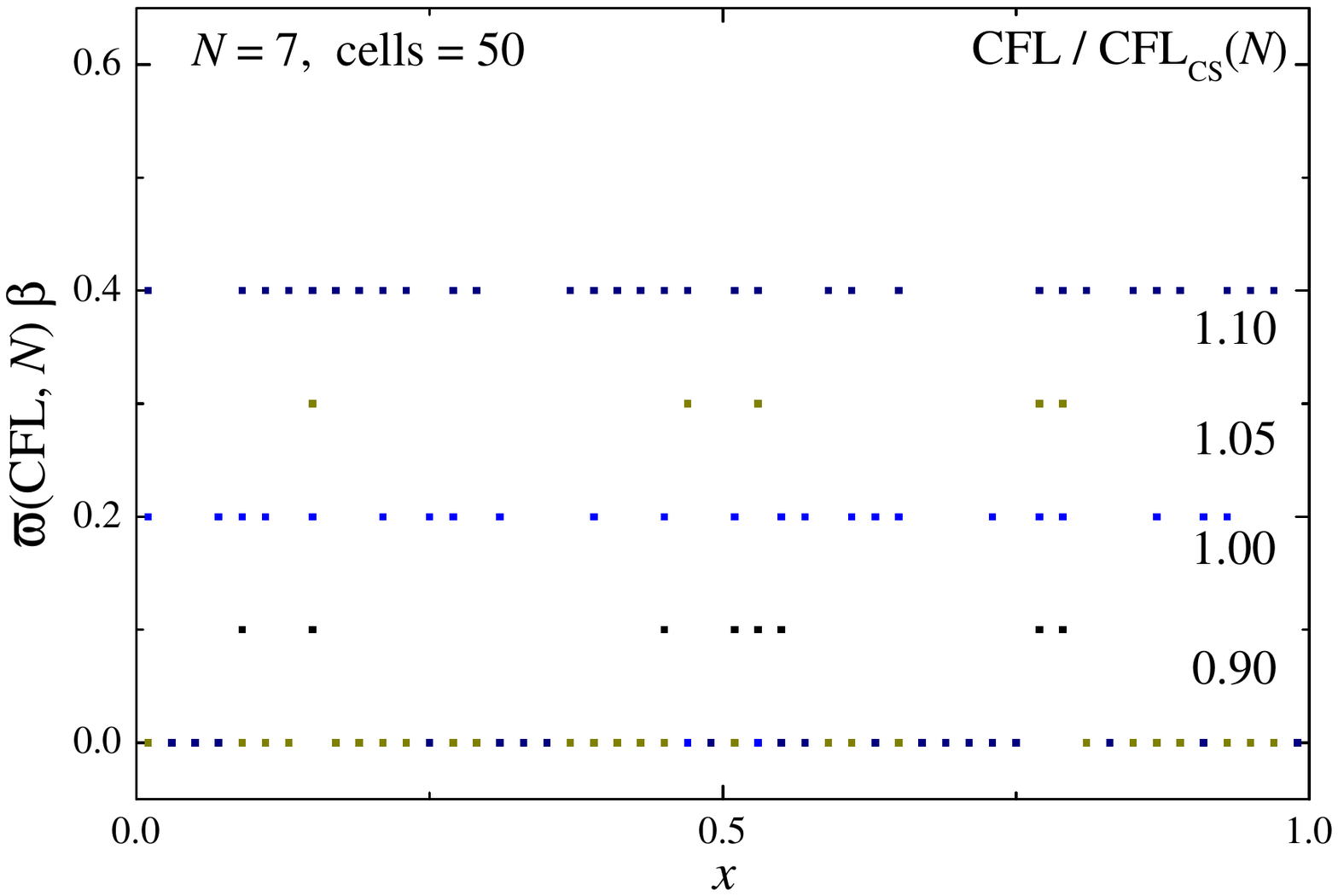}
\vspace{-8mm}\caption{\label{fig:test_znds_1d_degrees_7_9:degree_7_beta}}
\end{subfigure}\\[-2mm]
\begin{subfigure}{0.24\textwidth}
\includegraphics[width=\textwidth]{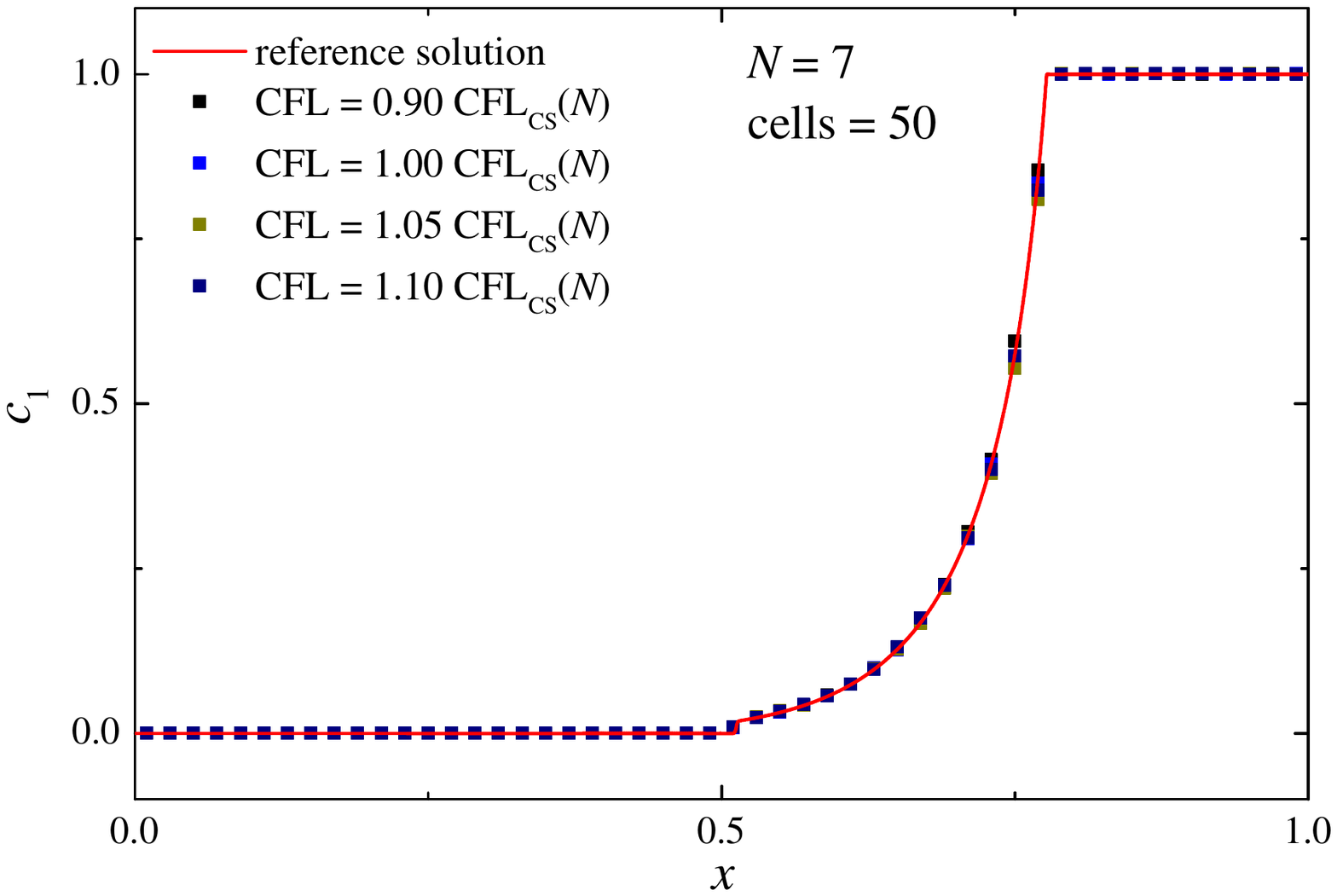}
\vspace{-8mm}\caption{\label{fig:test_znds_1d_degrees_7_9:degree_7_c1}}
\end{subfigure}
\begin{subfigure}{0.24\textwidth}
\includegraphics[width=\textwidth]{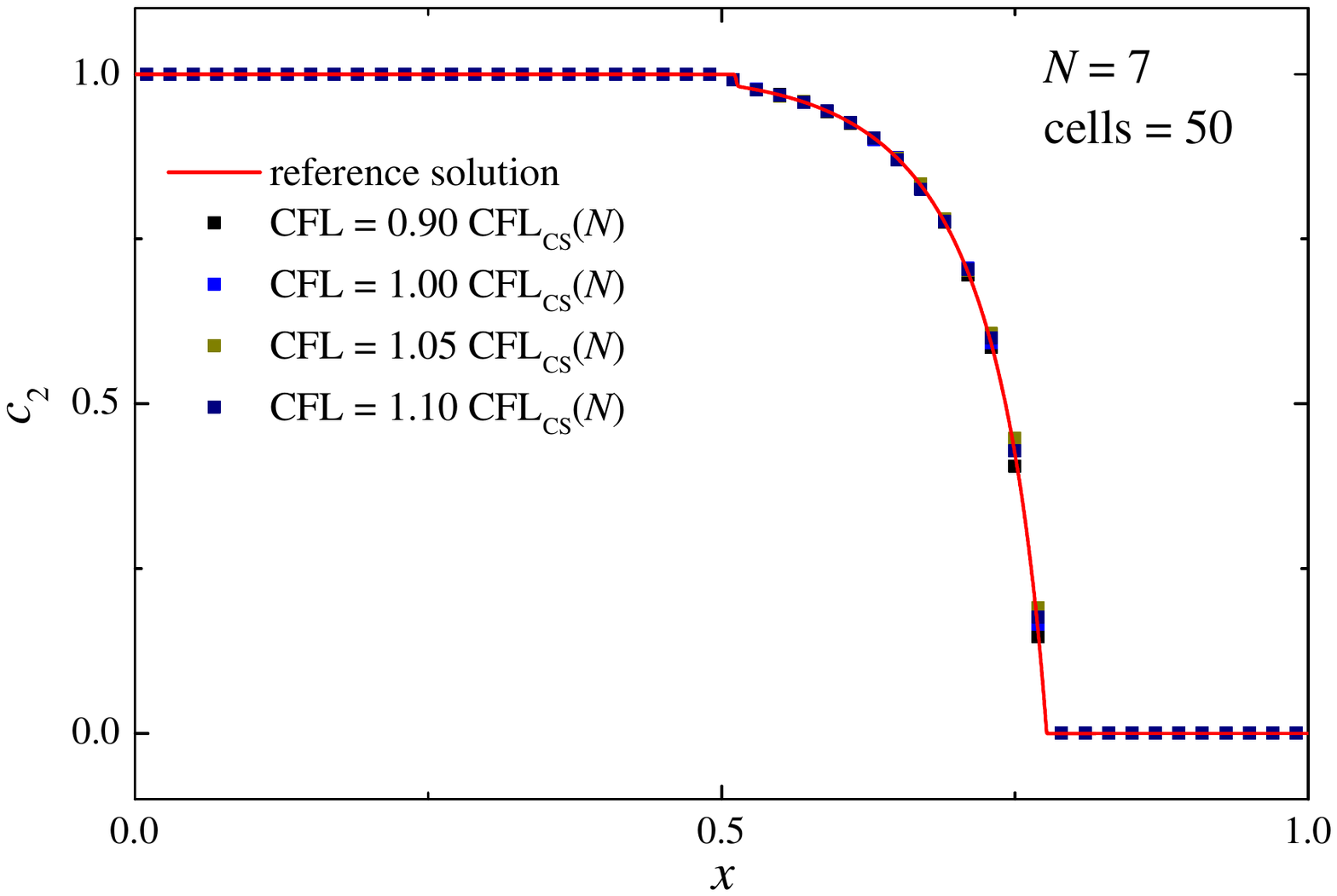}
\vspace{-8mm}\caption{\label{fig:test_znds_1d_degrees_7_9:degree_7_c2}}
\end{subfigure}
\begin{subfigure}{0.24\textwidth}
\includegraphics[width=\textwidth]{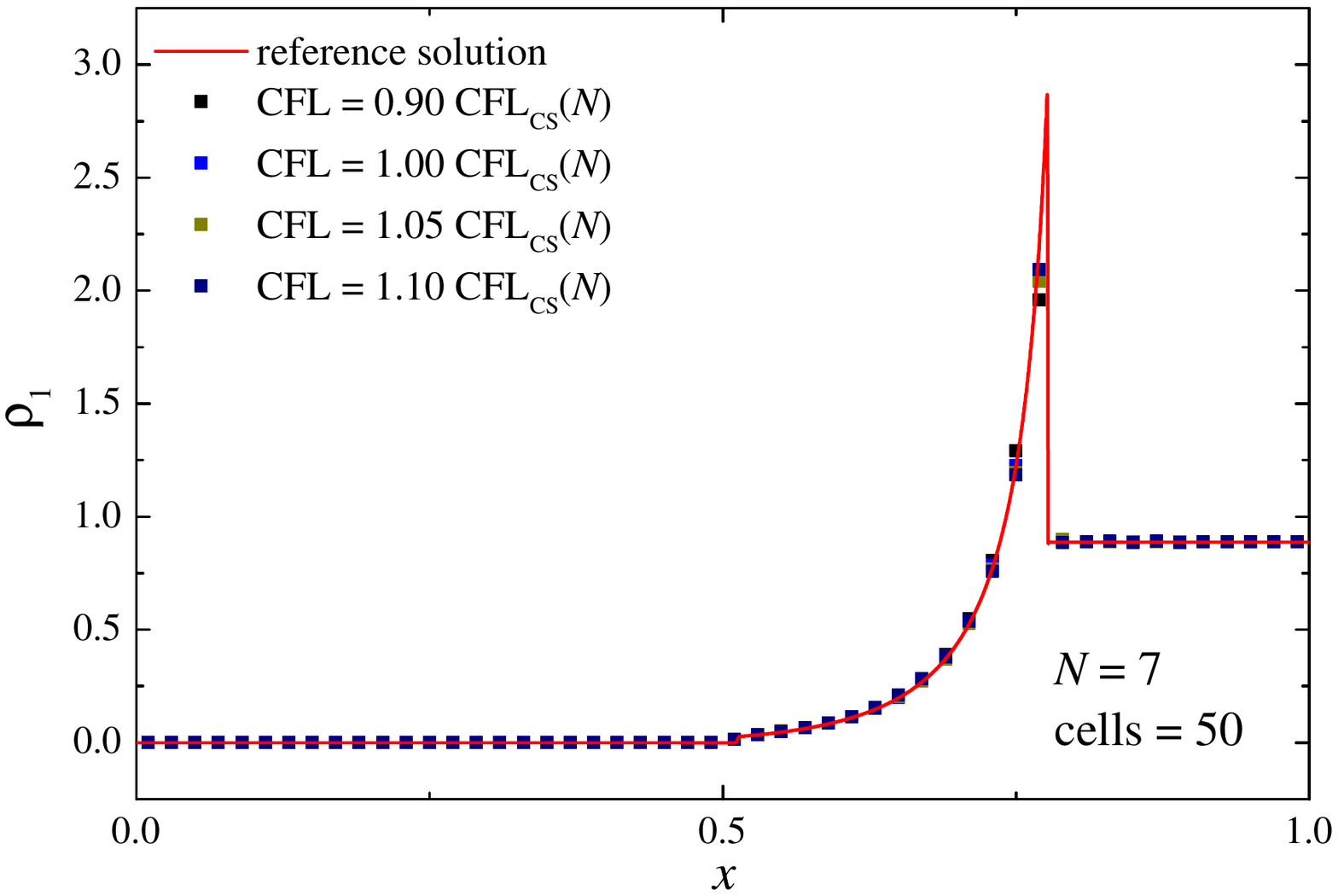}
\vspace{-8mm}\caption{\label{fig:test_znds_1d_degrees_7_9:degree_7_den1}}
\end{subfigure}
\begin{subfigure}{0.24\textwidth}
\includegraphics[width=\textwidth]{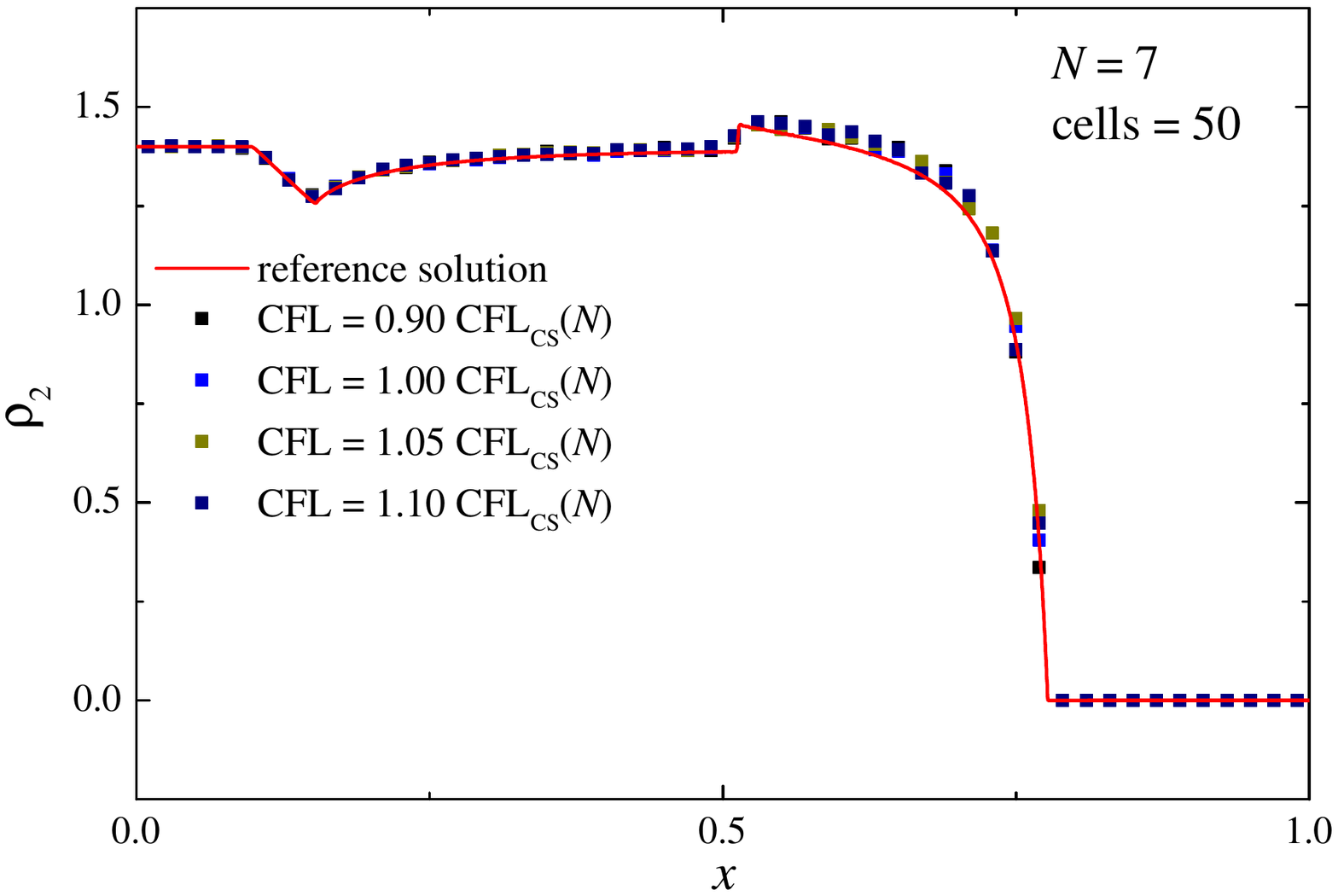}
\vspace{-8mm}\caption{\label{fig:test_znds_1d_degrees_7_9:degree_7_den2}}
\end{subfigure}\\[-2mm]
\begin{subfigure}{0.24\textwidth}
\includegraphics[width=\textwidth]{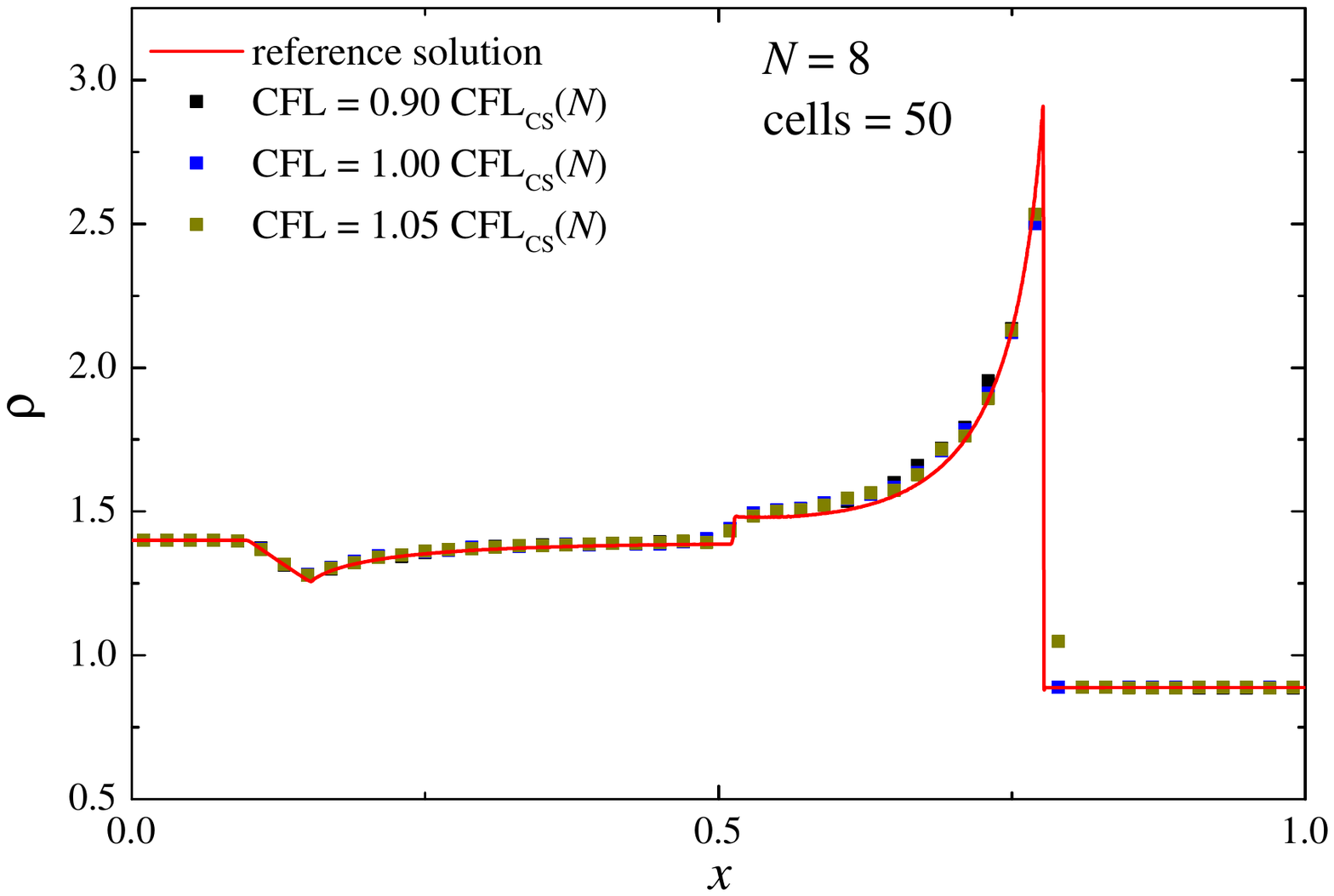}
\vspace{-8mm}\caption{\label{fig:test_znds_1d_degrees_7_9:degree_8_den}}
\end{subfigure}
\begin{subfigure}{0.24\textwidth}
\includegraphics[width=\textwidth]{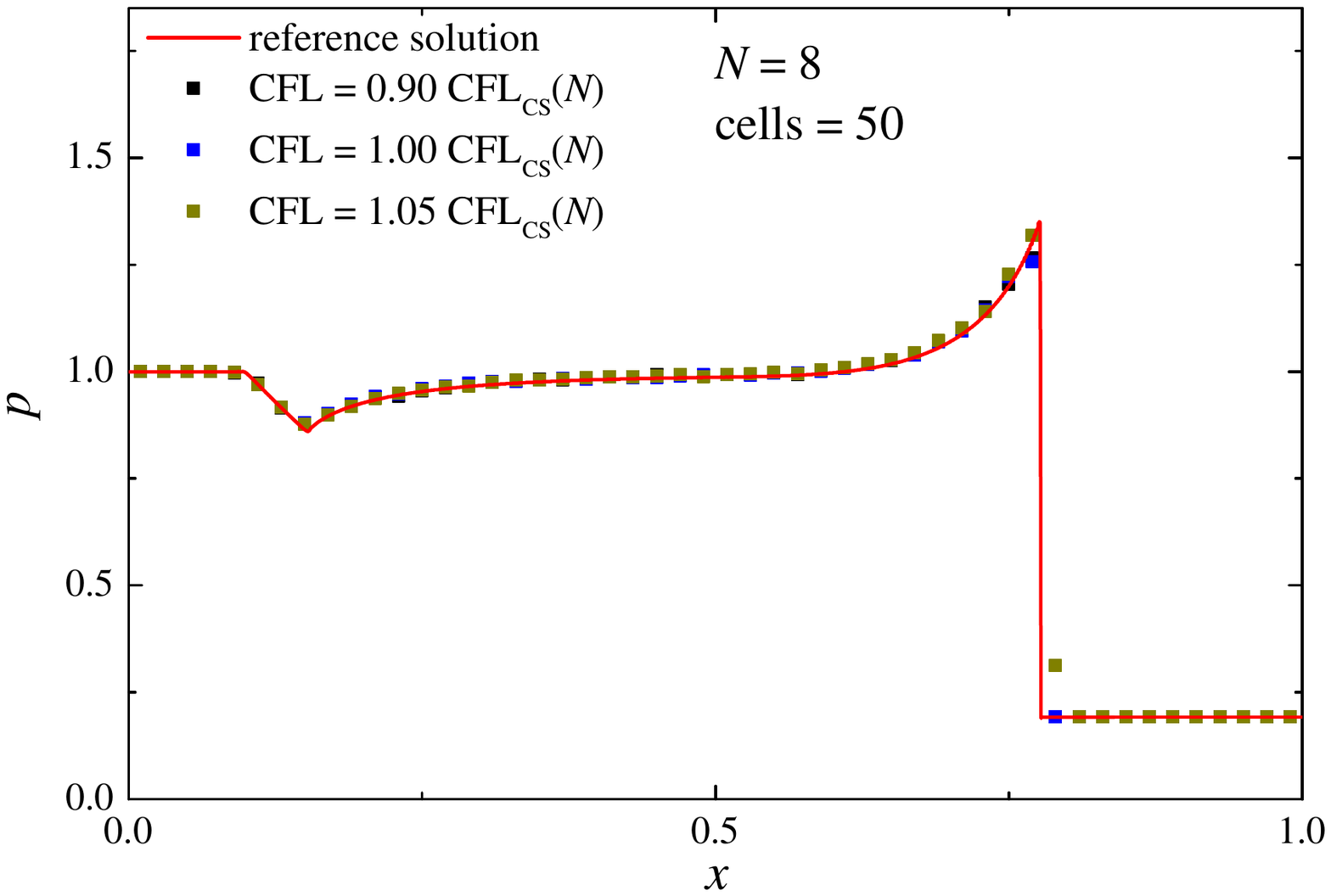}
\vspace{-8mm}\caption{\label{fig:test_znds_1d_degrees_7_9:degree_8_p}}
\end{subfigure}
\begin{subfigure}{0.24\textwidth}
\includegraphics[width=\textwidth]{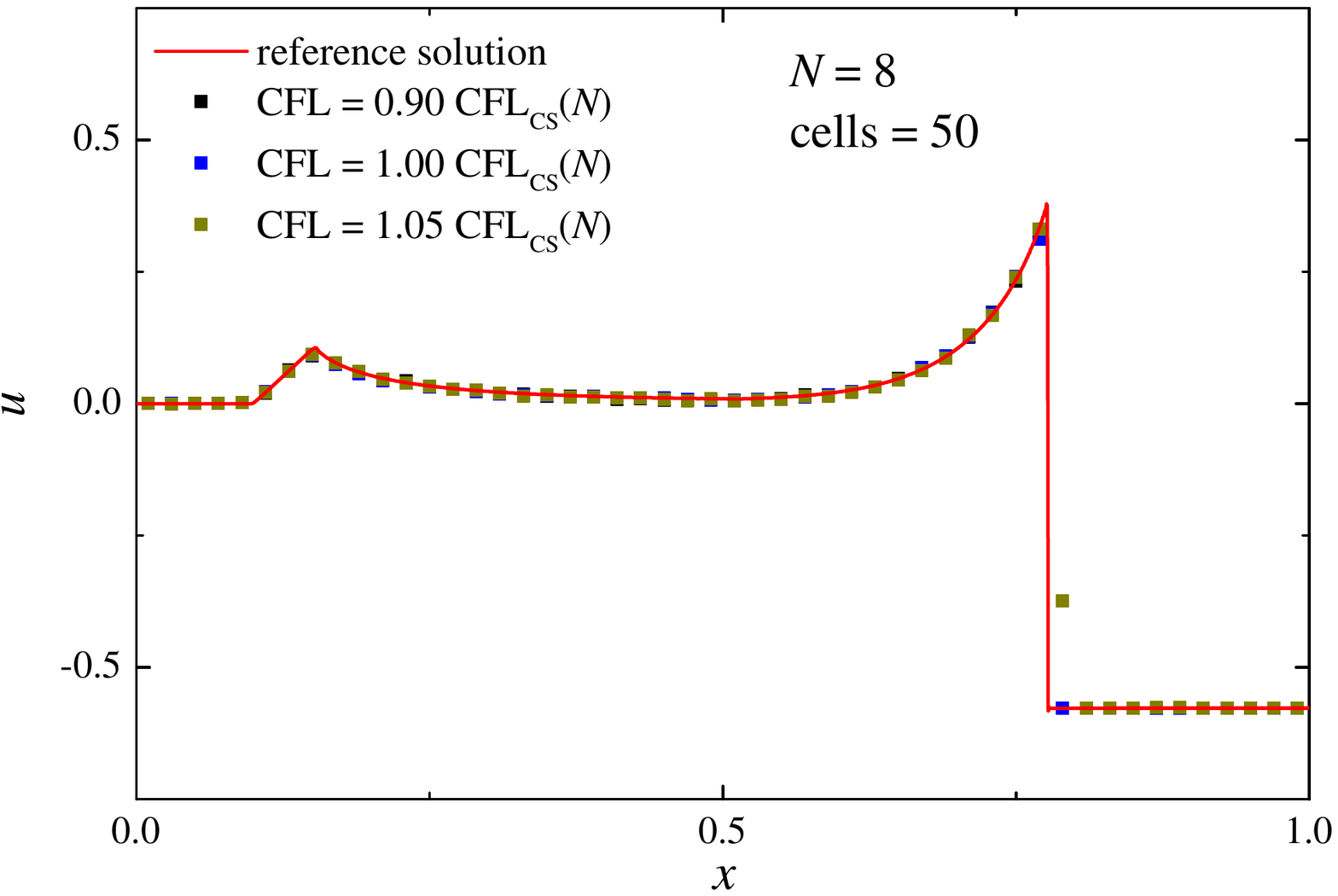}
\vspace{-8mm}\caption{\label{fig:test_znds_1d_degrees_7_9:degree_8_u}}
\end{subfigure}
\begin{subfigure}{0.24\textwidth}
\includegraphics[width=\textwidth]{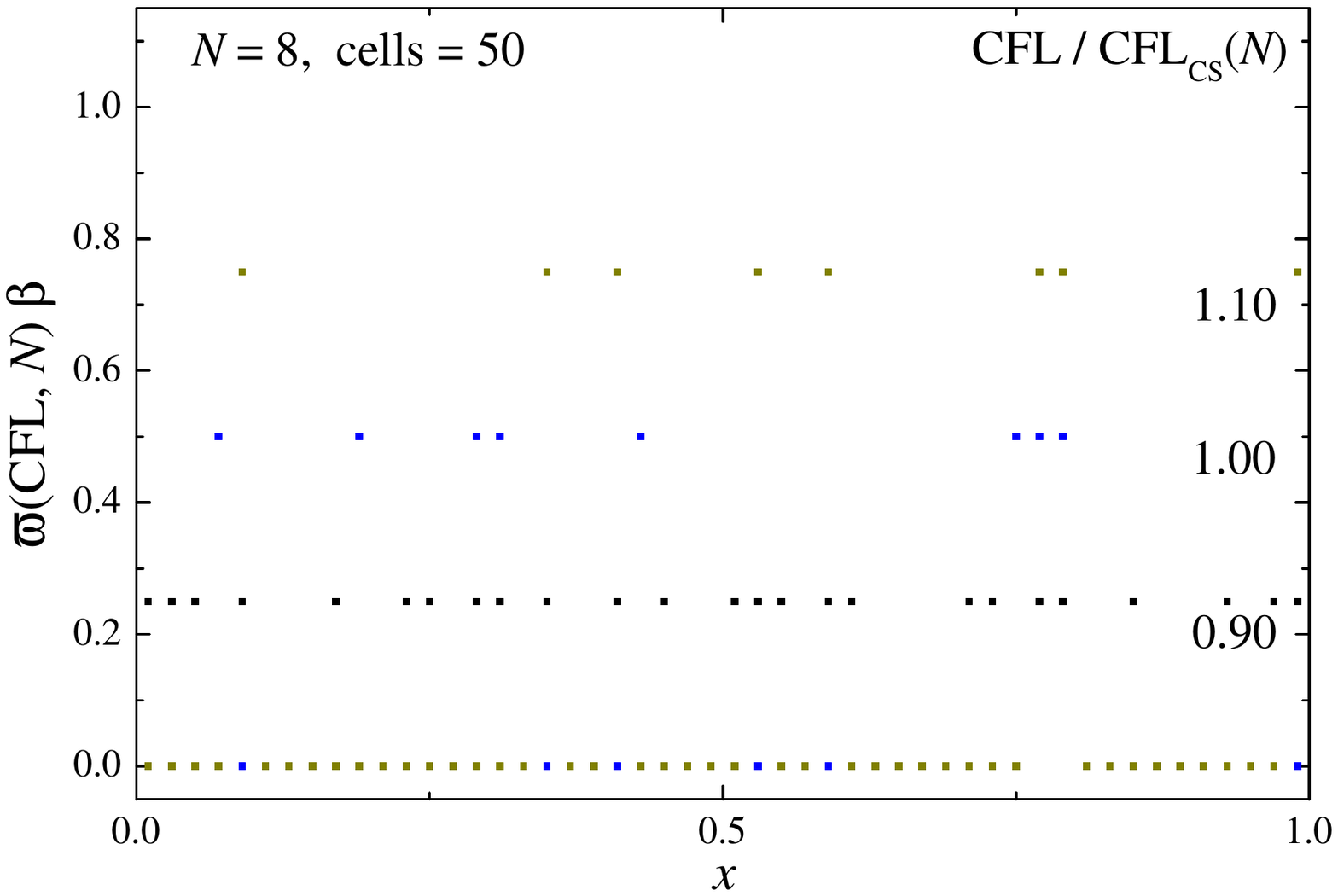}
\vspace{-8mm}\caption{\label{fig:test_znds_1d_degrees_7_9:degree_8_beta}}
\end{subfigure}\\[-2mm]
\begin{subfigure}{0.24\textwidth}
\includegraphics[width=\textwidth]{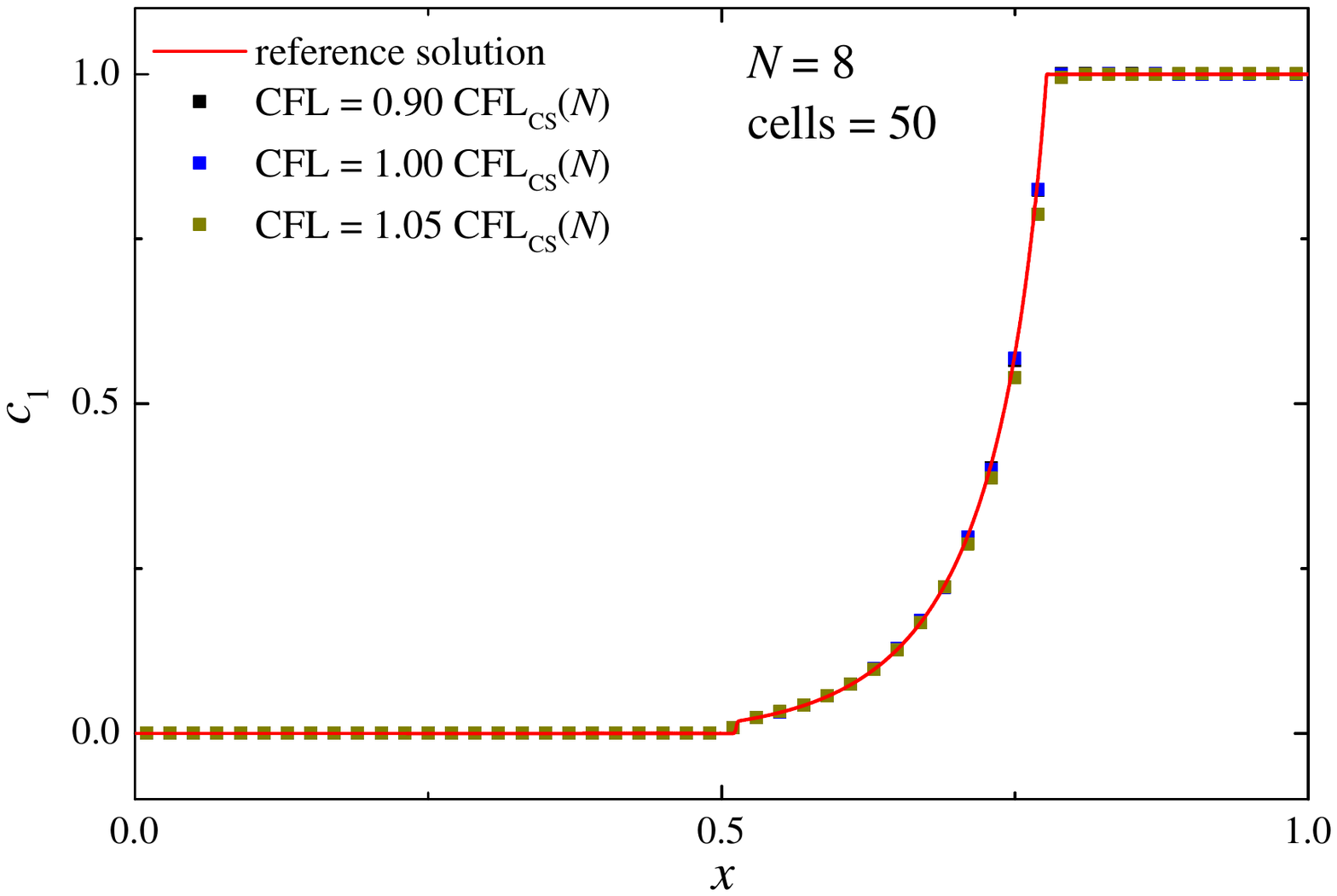}
\vspace{-8mm}\caption{\label{fig:test_znds_1d_degrees_7_9:degree_8_c1}}
\end{subfigure}
\begin{subfigure}{0.24\textwidth}
\includegraphics[width=\textwidth]{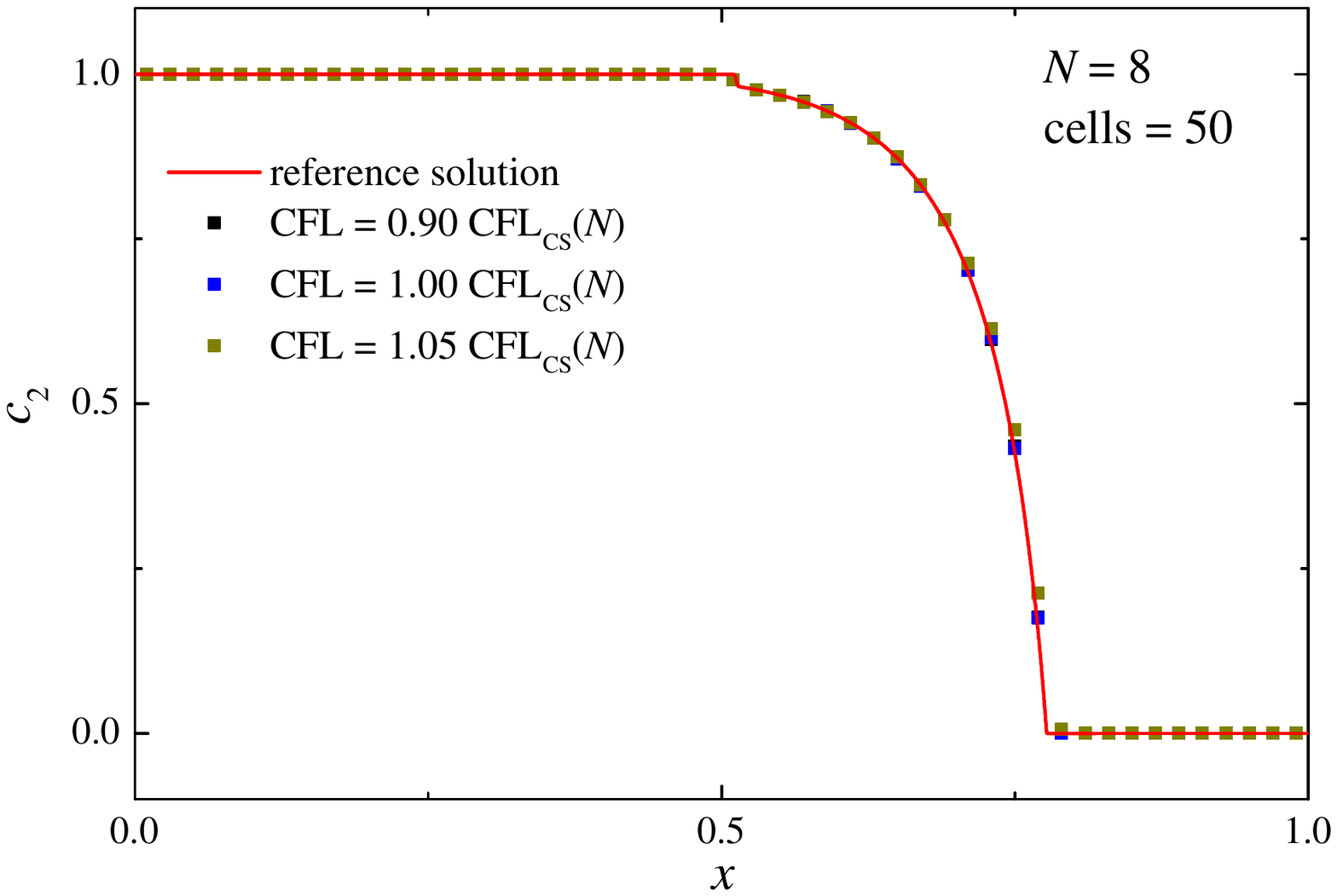}
\vspace{-8mm}\caption{\label{fig:test_znds_1d_degrees_7_9:degree_8_c2}}
\end{subfigure}
\begin{subfigure}{0.24\textwidth}
\includegraphics[width=\textwidth]{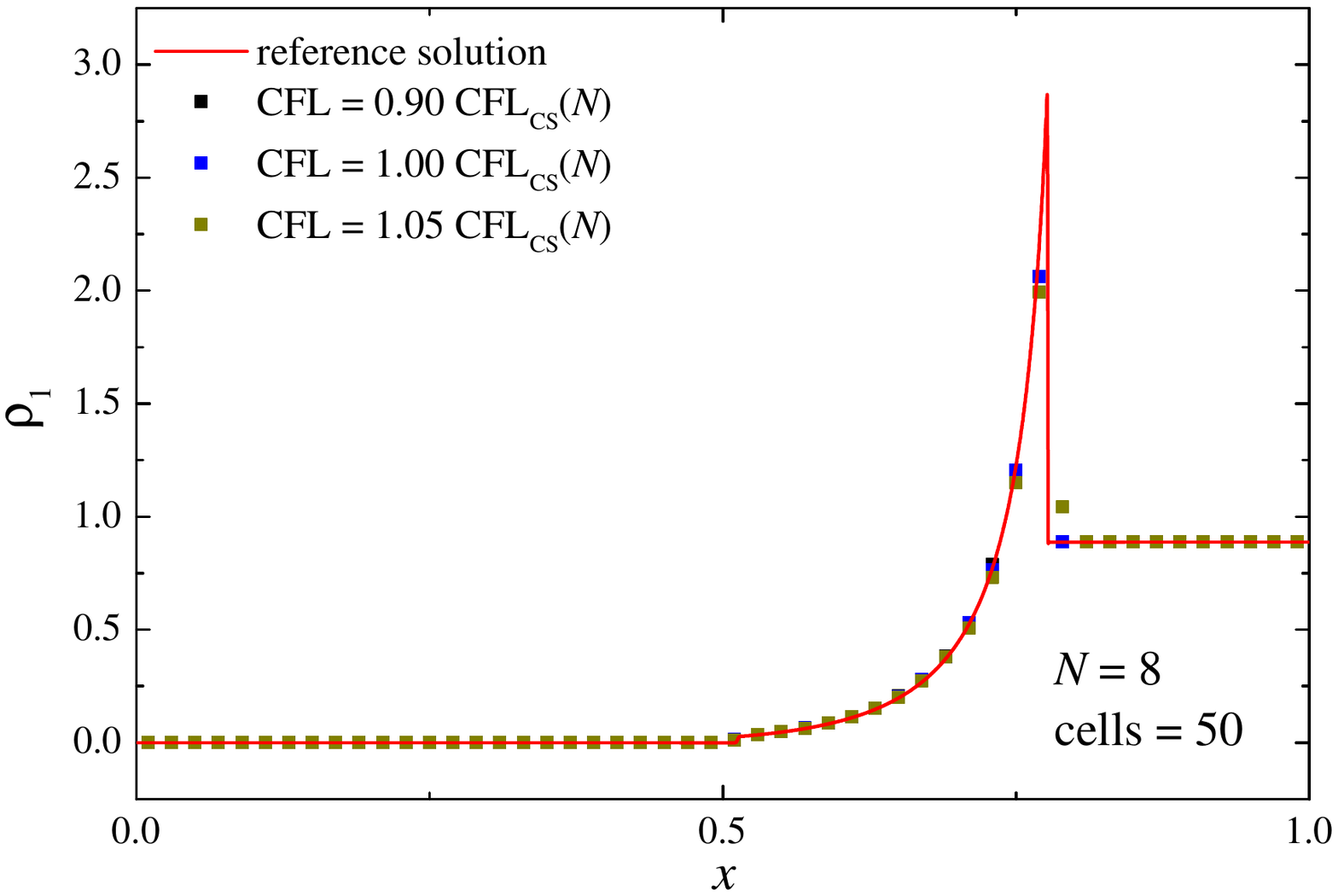}
\vspace{-8mm}\caption{\label{fig:test_znds_1d_degrees_7_9:degree_8_den1}}
\end{subfigure}
\begin{subfigure}{0.24\textwidth}
\includegraphics[width=\textwidth]{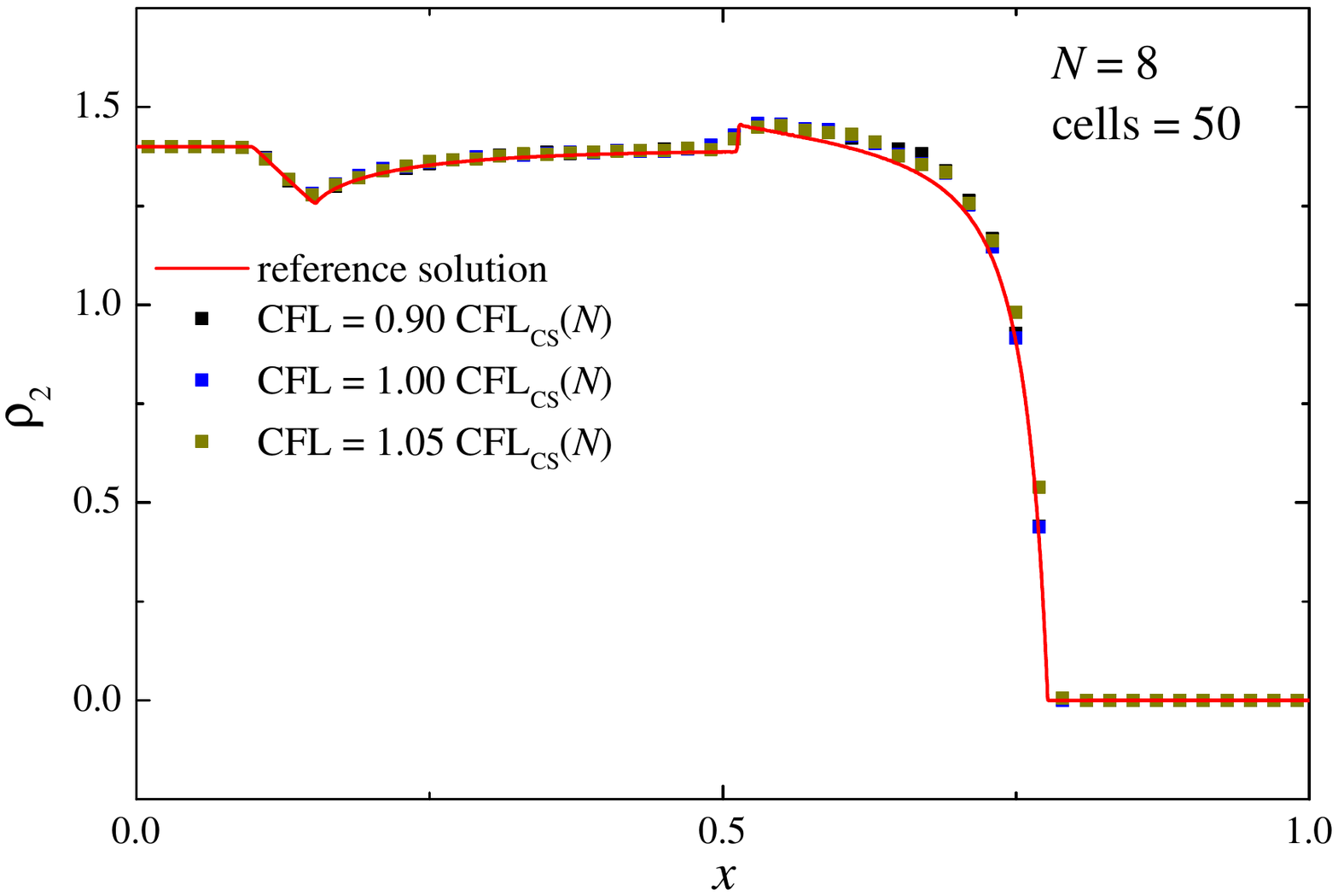}
\vspace{-8mm}\caption{\label{fig:test_znds_1d_degrees_7_9:degree_8_den2}}
\end{subfigure}\\[-2mm]
\begin{subfigure}{0.24\textwidth}
\includegraphics[width=\textwidth]{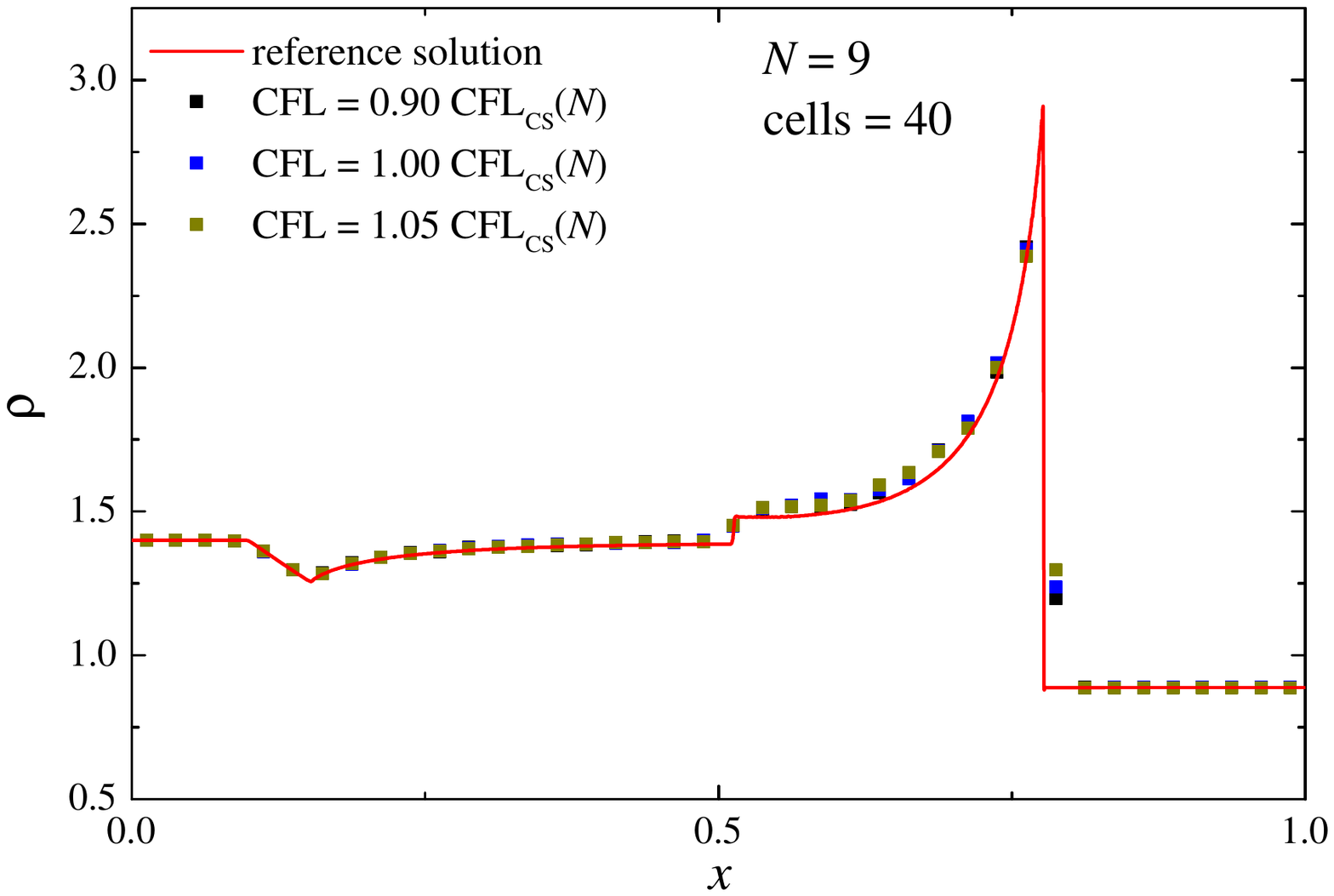}
\vspace{-8mm}\caption{\label{fig:test_znds_1d_degrees_7_9:degree_9_den}}
\end{subfigure}
\begin{subfigure}{0.24\textwidth}
\includegraphics[width=\textwidth]{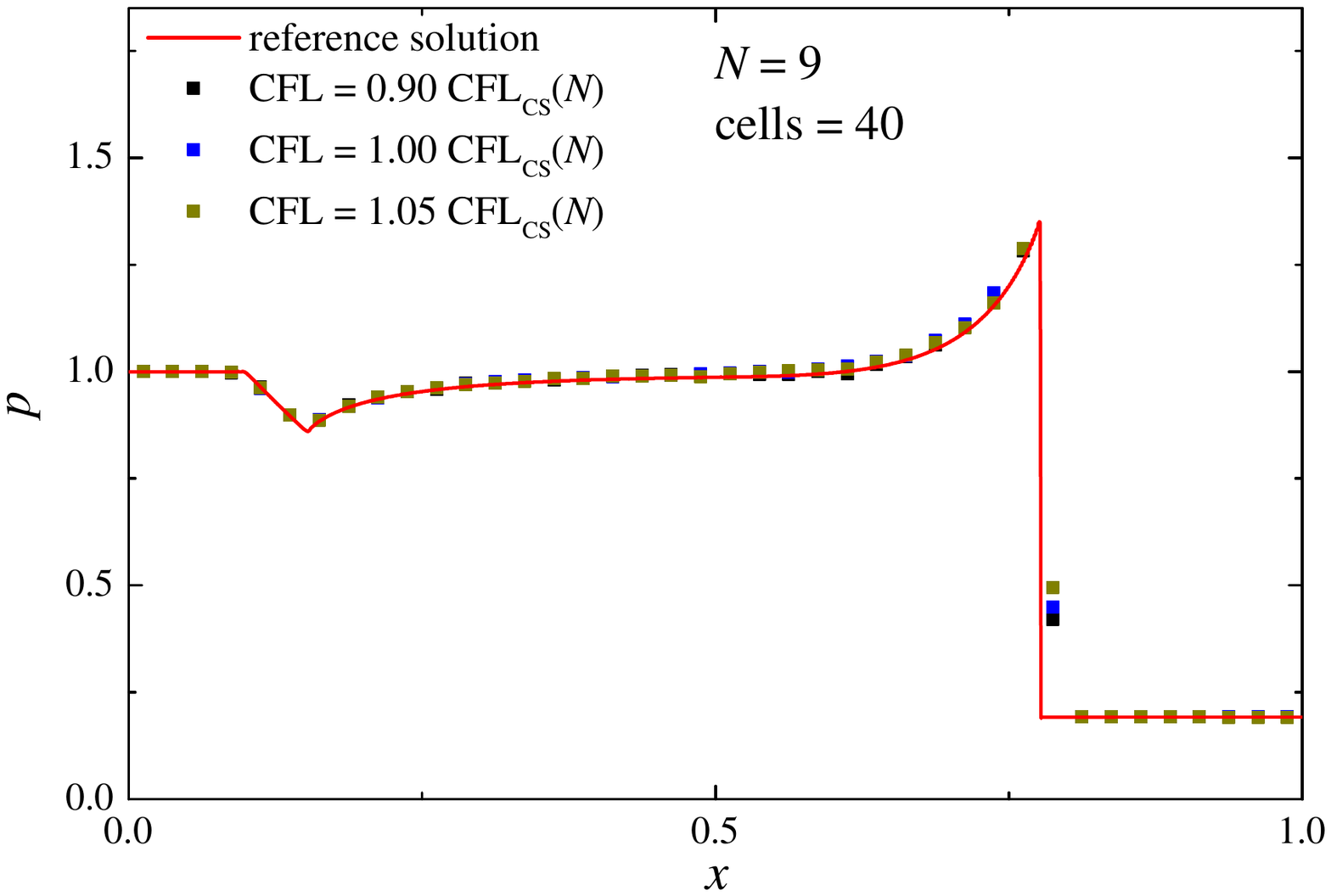}
\vspace{-8mm}\caption{\label{fig:test_znds_1d_degrees_7_9:degree_9_p}}
\end{subfigure}
\begin{subfigure}{0.24\textwidth}
\includegraphics[width=\textwidth]{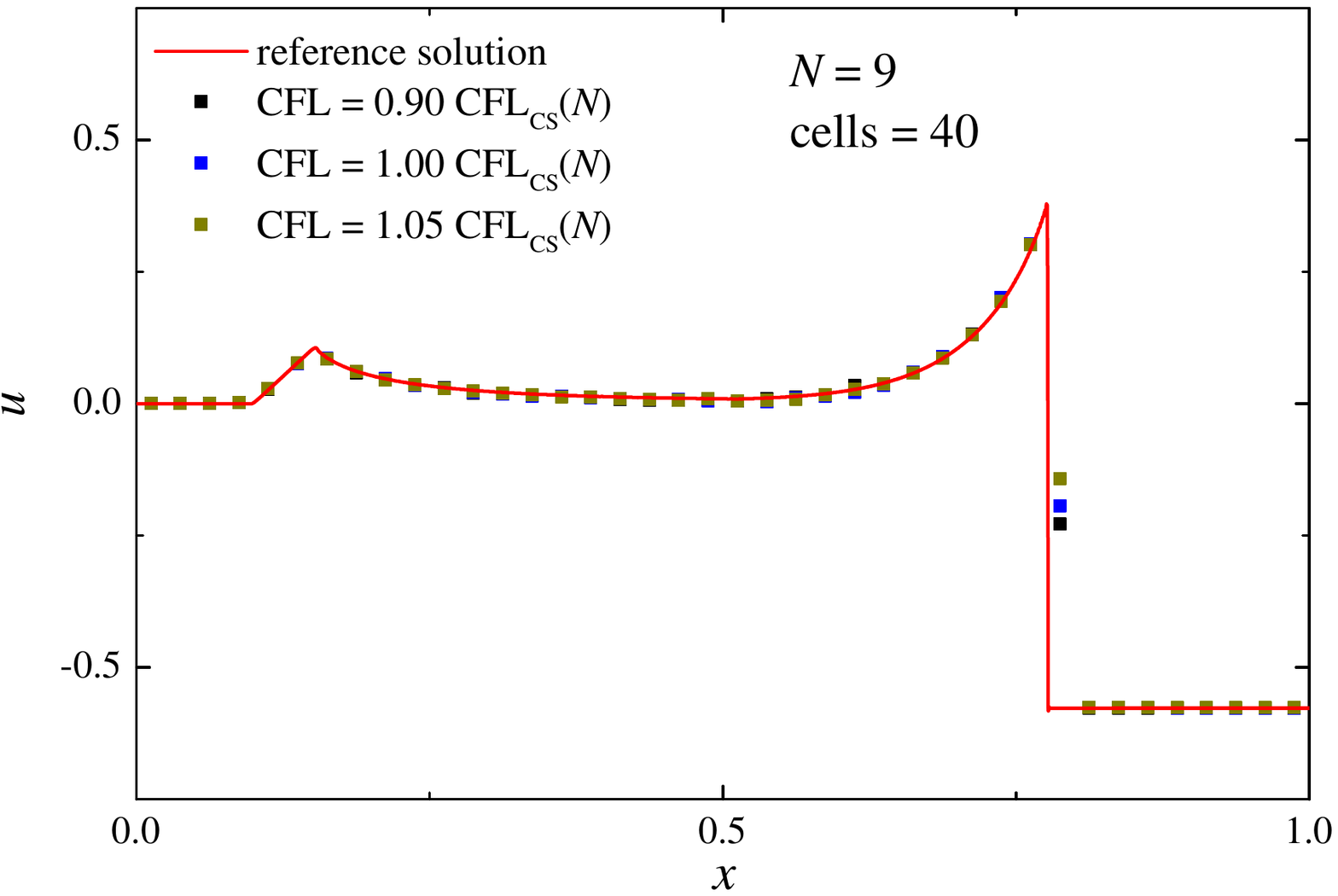}
\vspace{-8mm}\caption{\label{fig:test_znds_1d_degrees_7_9:degree_9_u}}
\end{subfigure}
\begin{subfigure}{0.24\textwidth}
\includegraphics[width=\textwidth]{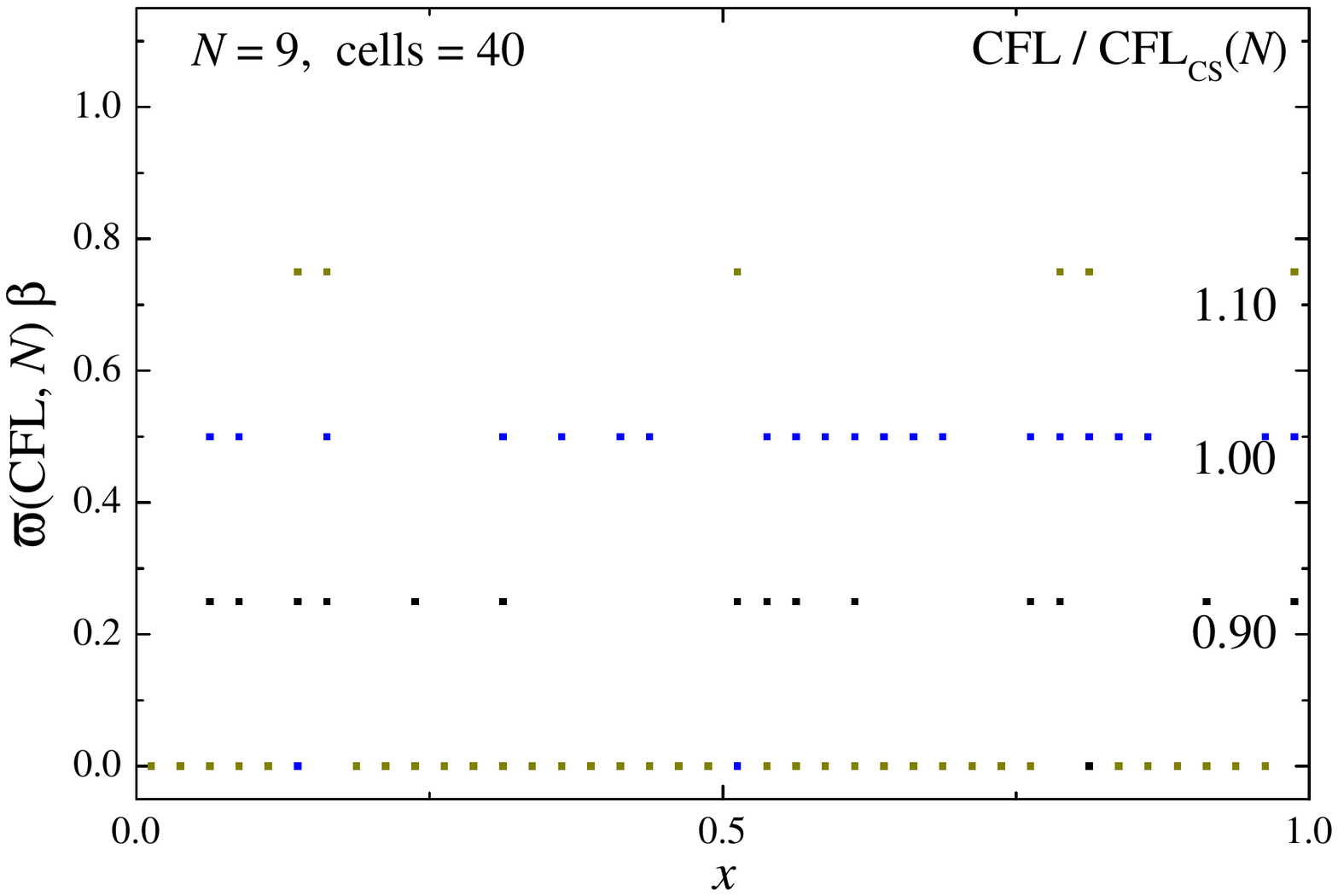}
\vspace{-8mm}\caption{\label{fig:test_znds_1d_degrees_7_9:degree_9_beta}}
\end{subfigure}\\[-2mm]
\begin{subfigure}{0.24\textwidth}
\includegraphics[width=\textwidth]{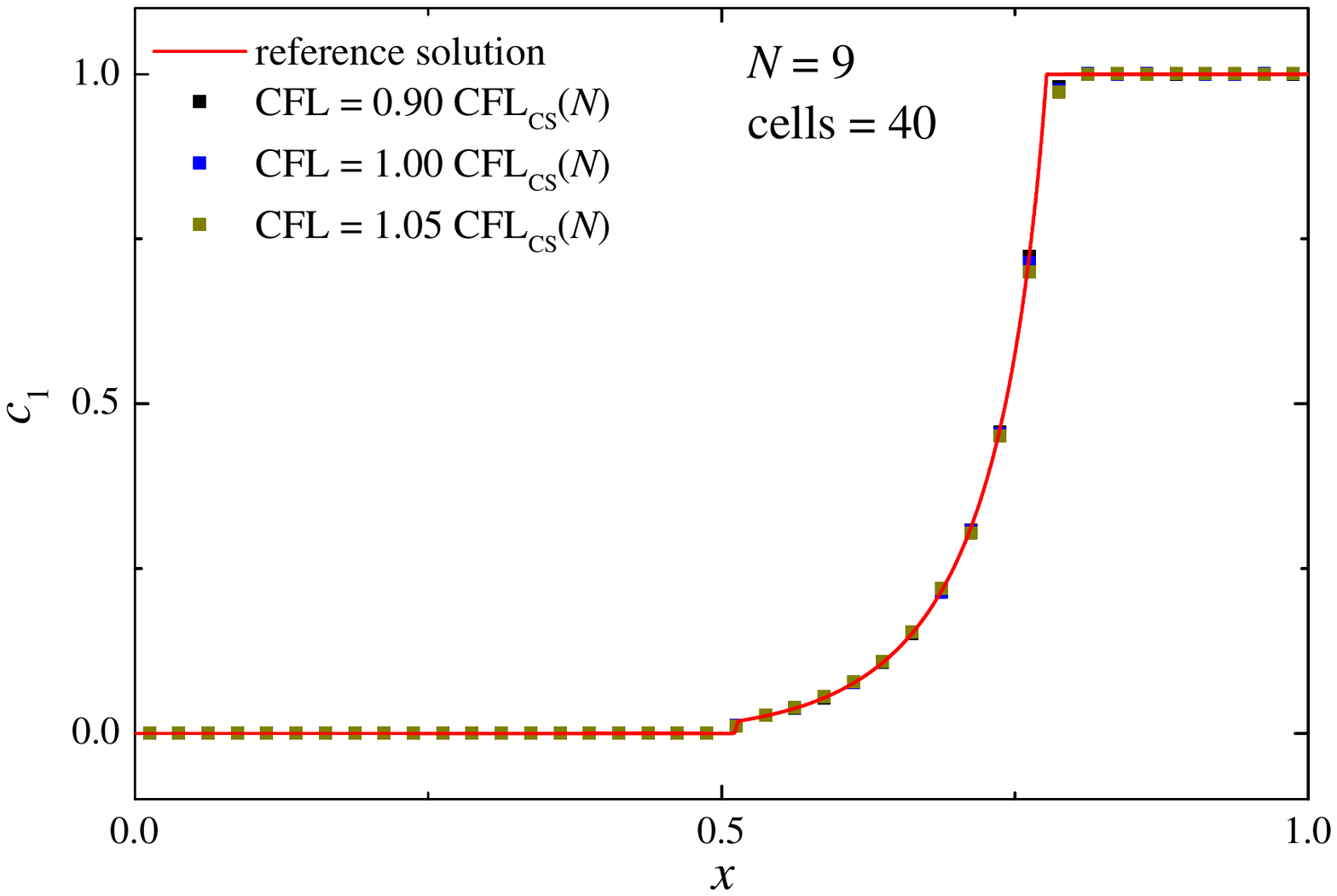}
\vspace{-8mm}\caption{\label{fig:test_znds_1d_degrees_7_9:degree_9_c1}}
\end{subfigure}
\begin{subfigure}{0.24\textwidth}
\includegraphics[width=\textwidth]{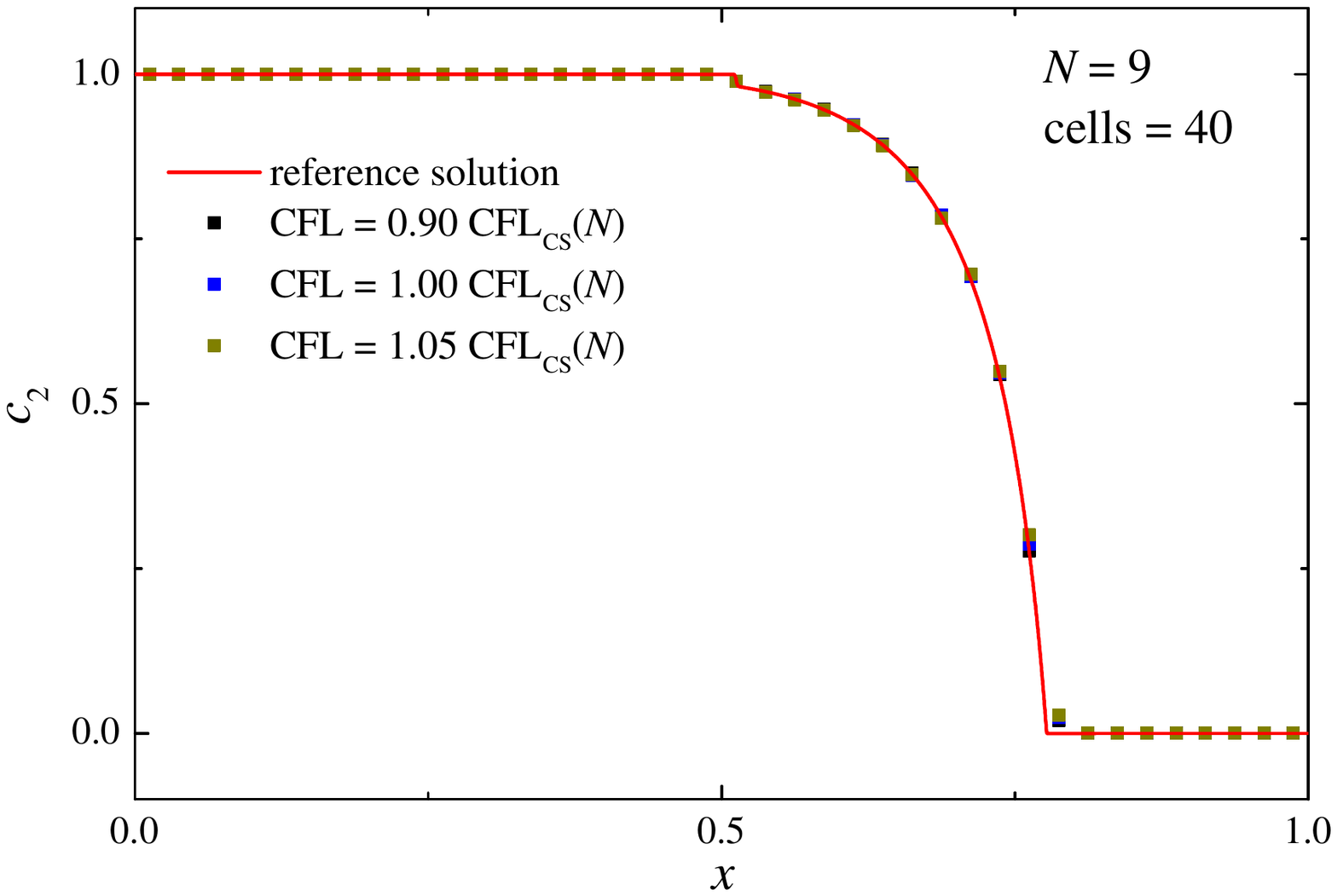}
\vspace{-8mm}\caption{\label{fig:test_znds_1d_degrees_7_9:degree_9_c2}}
\end{subfigure}
\begin{subfigure}{0.24\textwidth}
\includegraphics[width=\textwidth]{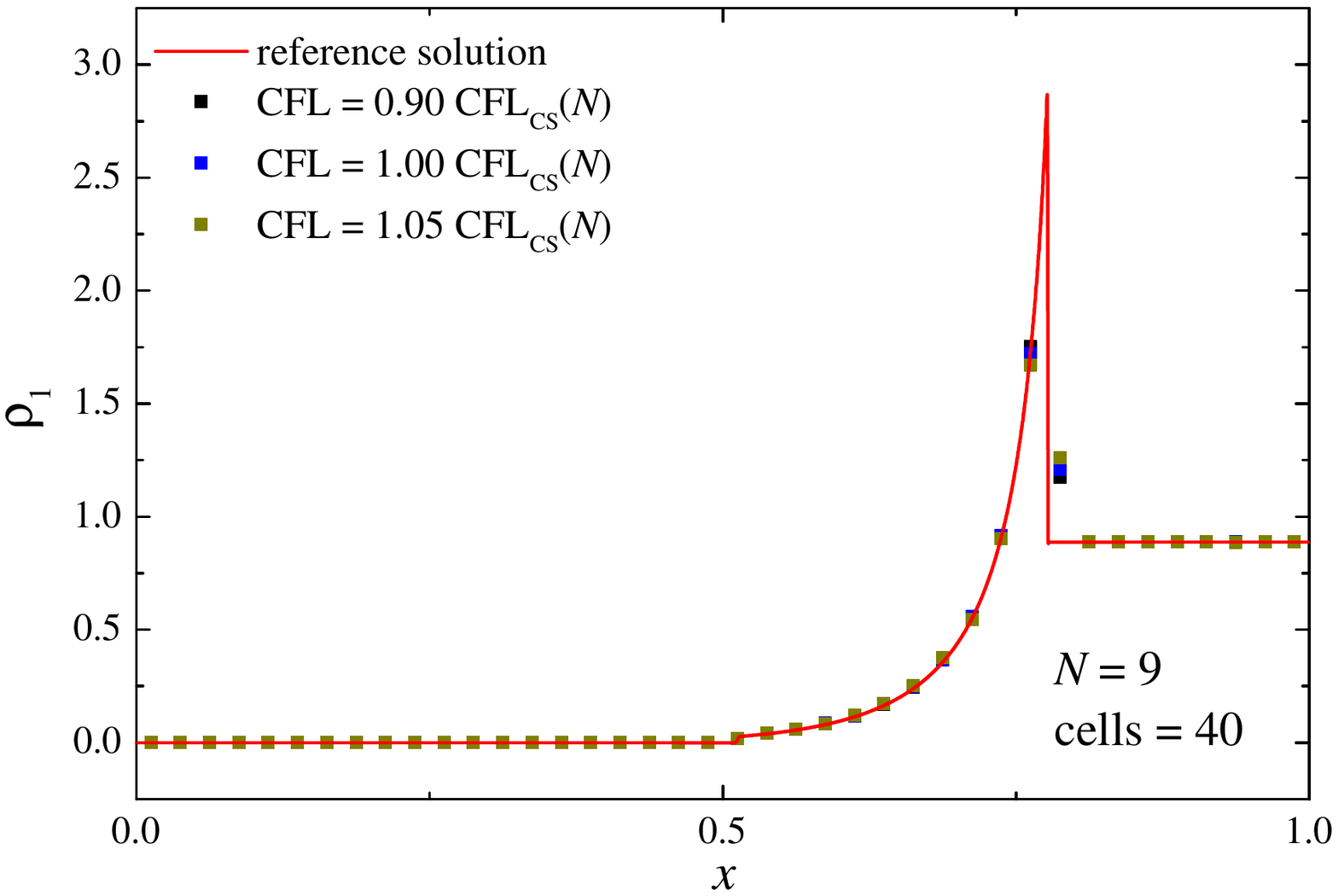}
\vspace{-8mm}\caption{\label{fig:test_znds_1d_degrees_7_9:degree_9_den1}}
\end{subfigure}
\begin{subfigure}{0.24\textwidth}
\includegraphics[width=\textwidth]{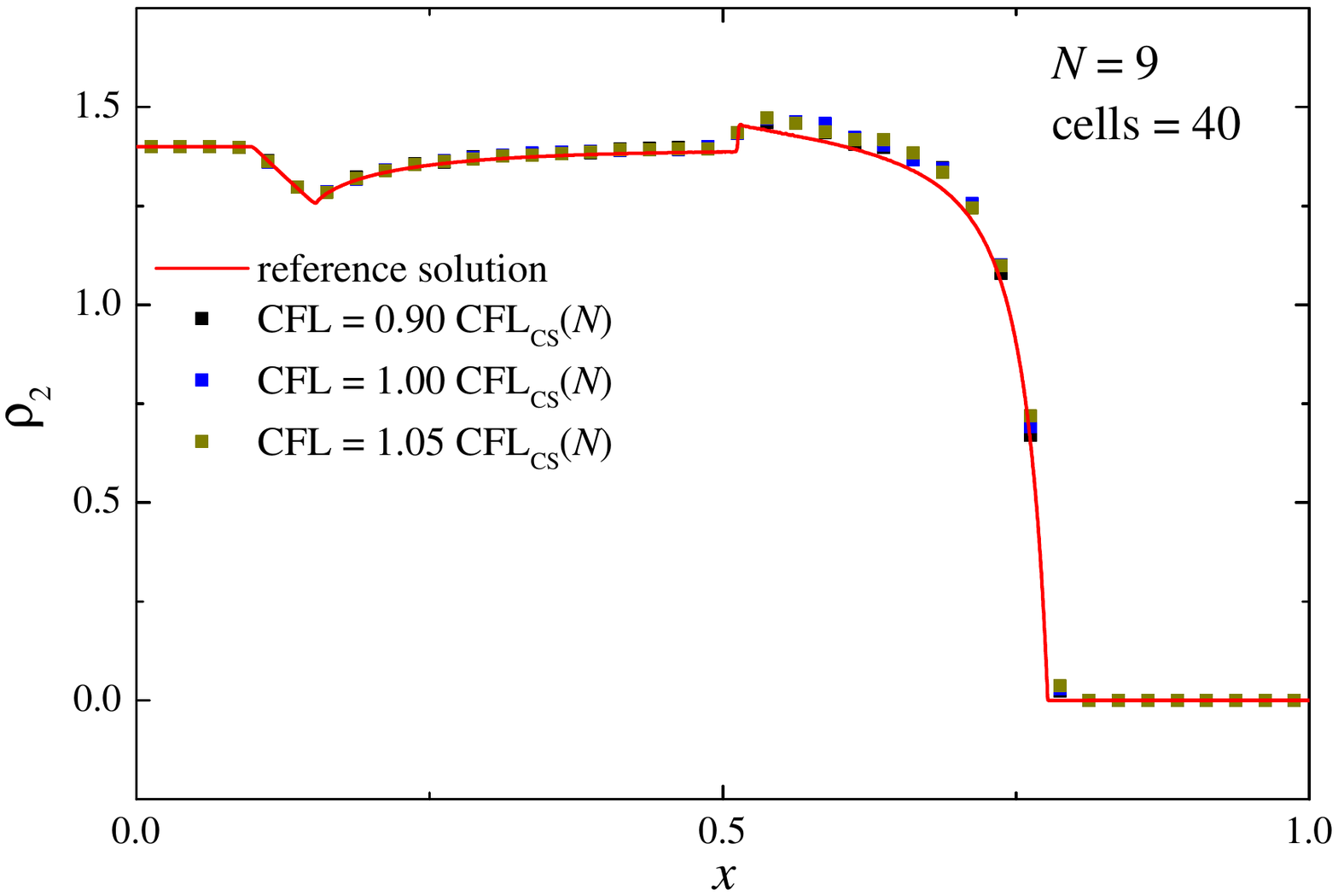}
\vspace{-8mm}\caption{\label{fig:test_znds_1d_degrees_7_9:degree_9_den2}}
\end{subfigure}
\caption{%
Numerical solution of the one-dimensional problem of the plane detonation wave formation in a two-component medium with a ``slow'' reaction (weak stiff case, a detailed statement of the problem is presented in the text), obtained using the ADER-DG method for the basis polynomials degree $N = 7$ ((\subref{fig:test_znds_1d_degrees_7_9:degree_7_den}), (\subref{fig:test_znds_1d_degrees_7_9:degree_7_p}), (\subref{fig:test_znds_1d_degrees_7_9:degree_7_u}), (\subref{fig:test_znds_1d_degrees_7_9:degree_7_c1}), (\subref{fig:test_znds_1d_degrees_7_9:degree_7_c2}), (\subref{fig:test_znds_1d_degrees_7_9:degree_7_den1}), (\subref{fig:test_znds_1d_degrees_7_9:degree_7_den2})), $N = 8$ ((\subref{fig:test_znds_1d_degrees_7_9:degree_8_den}), (\subref{fig:test_znds_1d_degrees_7_9:degree_8_p}), (\subref{fig:test_znds_1d_degrees_7_9:degree_8_u}), (\subref{fig:test_znds_1d_degrees_7_9:degree_8_c1}), (\subref{fig:test_znds_1d_degrees_7_9:degree_8_c2}), (\subref{fig:test_znds_1d_degrees_7_9:degree_8_den1}), (\subref{fig:test_znds_1d_degrees_7_9:degree_8_den2})), $N = 9$ ((\subref{fig:test_znds_1d_degrees_7_9:degree_9_den}), (\subref{fig:test_znds_1d_degrees_7_9:degree_9_p}), (\subref{fig:test_znds_1d_degrees_7_9:degree_9_u}), (\subref{fig:test_znds_1d_degrees_7_9:degree_9_c1}), (\subref{fig:test_znds_1d_degrees_7_9:degree_9_c2}), (\subref{fig:test_znds_1d_degrees_7_9:degree_9_den1}), (\subref{fig:test_znds_1d_degrees_7_9:degree_9_den2})) with a posteriori limitation of the solution by an ADER-WENO2 finite-volume limiter, at the final time $t_{f} = 0.4$. The graphs show the coordinate dependence of density $\rho$ ((\subref{fig:test_znds_1d_degrees_7_9:degree_7_den}), (\subref{fig:test_znds_1d_degrees_7_9:degree_8_den}), (\subref{fig:test_znds_1d_degrees_7_9:degree_9_den})), pressure $p$ ((\subref{fig:test_znds_1d_degrees_7_9:degree_7_p}), (\subref{fig:test_znds_1d_degrees_7_9:degree_8_p}), (\subref{fig:test_znds_1d_degrees_7_9:degree_9_p})), flow velocity $u$ ((\subref{fig:test_znds_1d_degrees_7_9:degree_7_u}), (\subref{fig:test_znds_1d_degrees_7_9:degree_8_u}), (\subref{fig:test_znds_1d_degrees_7_9:degree_9_u})), troubled cells indicator $\beta$ multiplied by the separating function $\varpi(\mathrm{CFL}, N)$ ((\subref{fig:test_znds_1d_degrees_7_9:degree_7_beta}), (\subref{fig:test_znds_1d_degrees_7_9:degree_8_beta}), (\subref{fig:test_znds_1d_degrees_7_9:degree_9_beta})), mass concentrations $c_{1}$ ((\subref{fig:test_znds_1d_degrees_7_9:degree_7_c1}), (\subref{fig:test_znds_1d_degrees_7_9:degree_8_c1}), (\subref{fig:test_znds_1d_degrees_7_9:degree_9_c1})) and $c_{2}$ ((\subref{fig:test_znds_1d_degrees_7_9:degree_7_c2}), (\subref{fig:test_znds_1d_degrees_7_9:degree_8_c2}), (\subref{fig:test_znds_1d_degrees_7_9:degree_9_c2})) of the reactant $A$ and product $B$, respectively, and partial densities $\rho_{1}$ ((\subref{fig:test_znds_1d_degrees_7_9:degree_7_den1}), (\subref{fig:test_znds_1d_degrees_7_9:degree_8_den1}), (\subref{fig:test_znds_1d_degrees_7_9:degree_9_den1})) and $\rho_{2}$ ((\subref{fig:test_znds_1d_degrees_7_9:degree_7_den2}), (\subref{fig:test_znds_1d_degrees_7_9:degree_8_den2}), (\subref{fig:test_znds_1d_degrees_7_9:degree_9_den2})) of the reactant $A$ and product $B$, respectively. The square symbols represent the subcells finite-volume representation of the numerical solution; the red solid lines represents the reference solution of the problem.
}
\label{fig:test_znds_1d_degrees_7_9}
\end{figure}
\begin{figure}[h!]
\captionsetup[subfigure]{%
	position=bottom,
	font+=smaller,
	textfont=normalfont,
	singlelinecheck=off,
	justification=raggedright
}
\centering
\begin{subfigure}{0.24\textwidth}
\includegraphics[width=\textwidth]{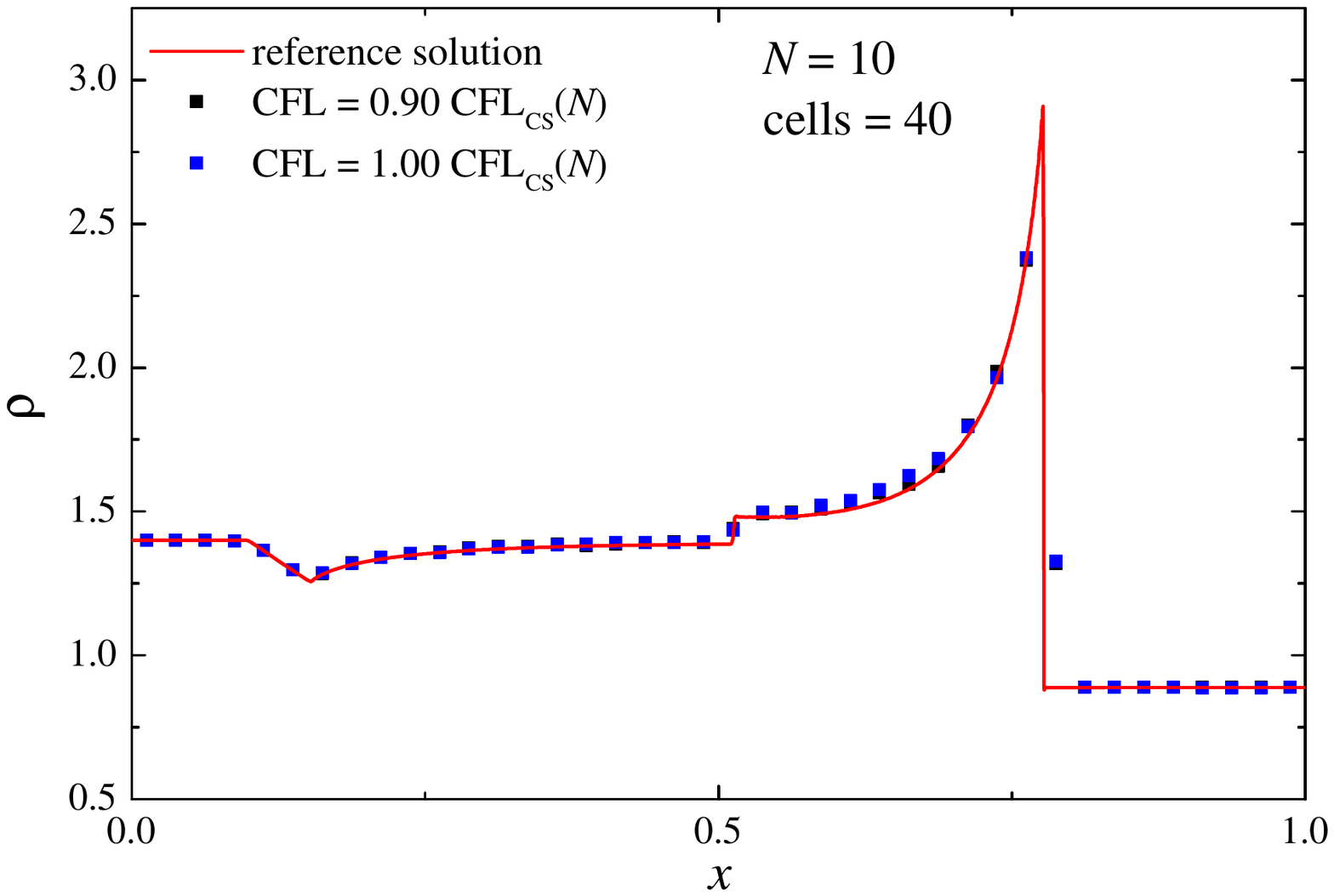}
\vspace{-8mm}\caption{\label{fig:test_znds_1d_degrees_10_12:degree_10_den}}
\end{subfigure}
\begin{subfigure}{0.24\textwidth}
\includegraphics[width=\textwidth]{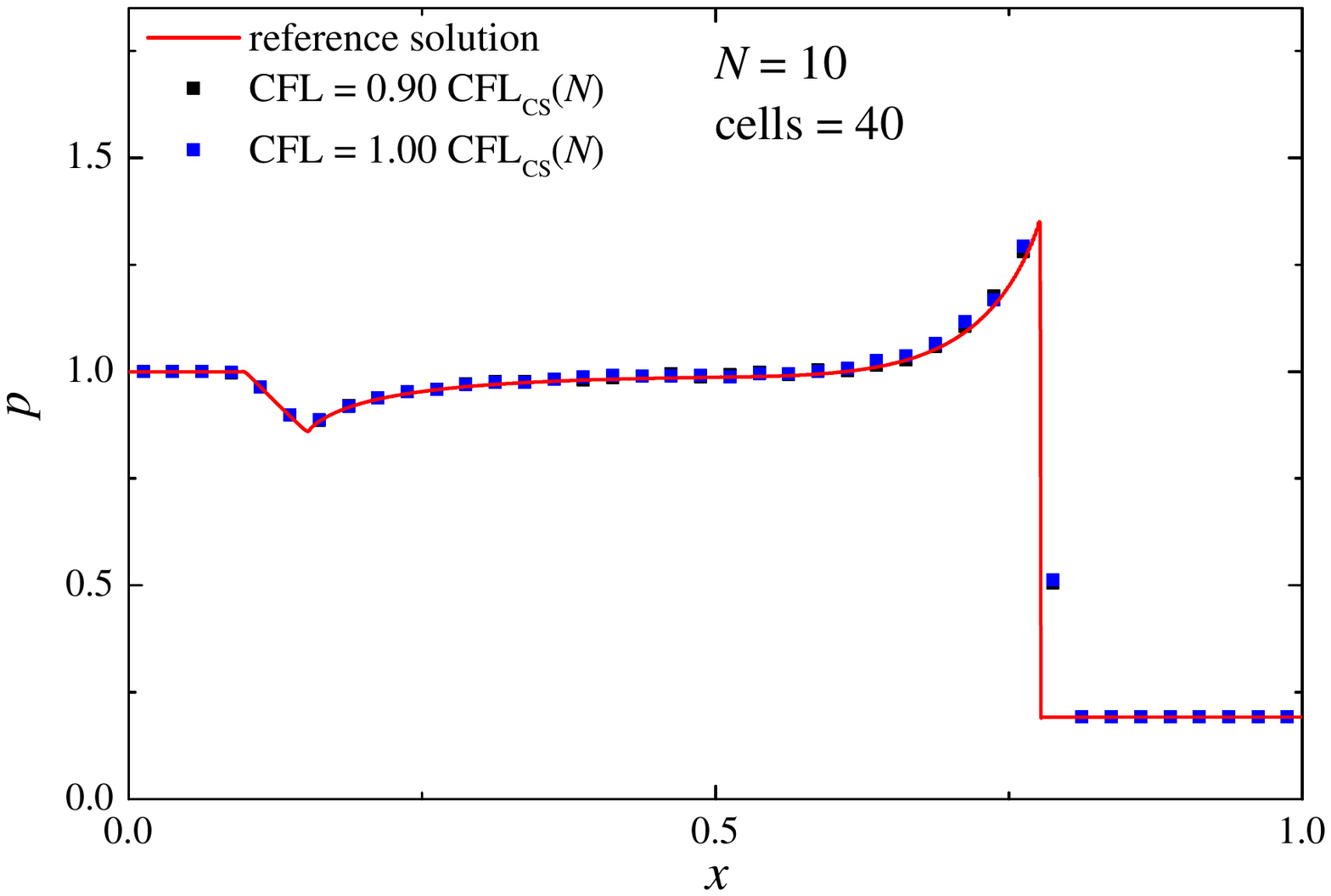}
\vspace{-8mm}\caption{\label{fig:test_znds_1d_degrees_10_12:degree_10_p}}
\end{subfigure}
\begin{subfigure}{0.24\textwidth}
\includegraphics[width=\textwidth]{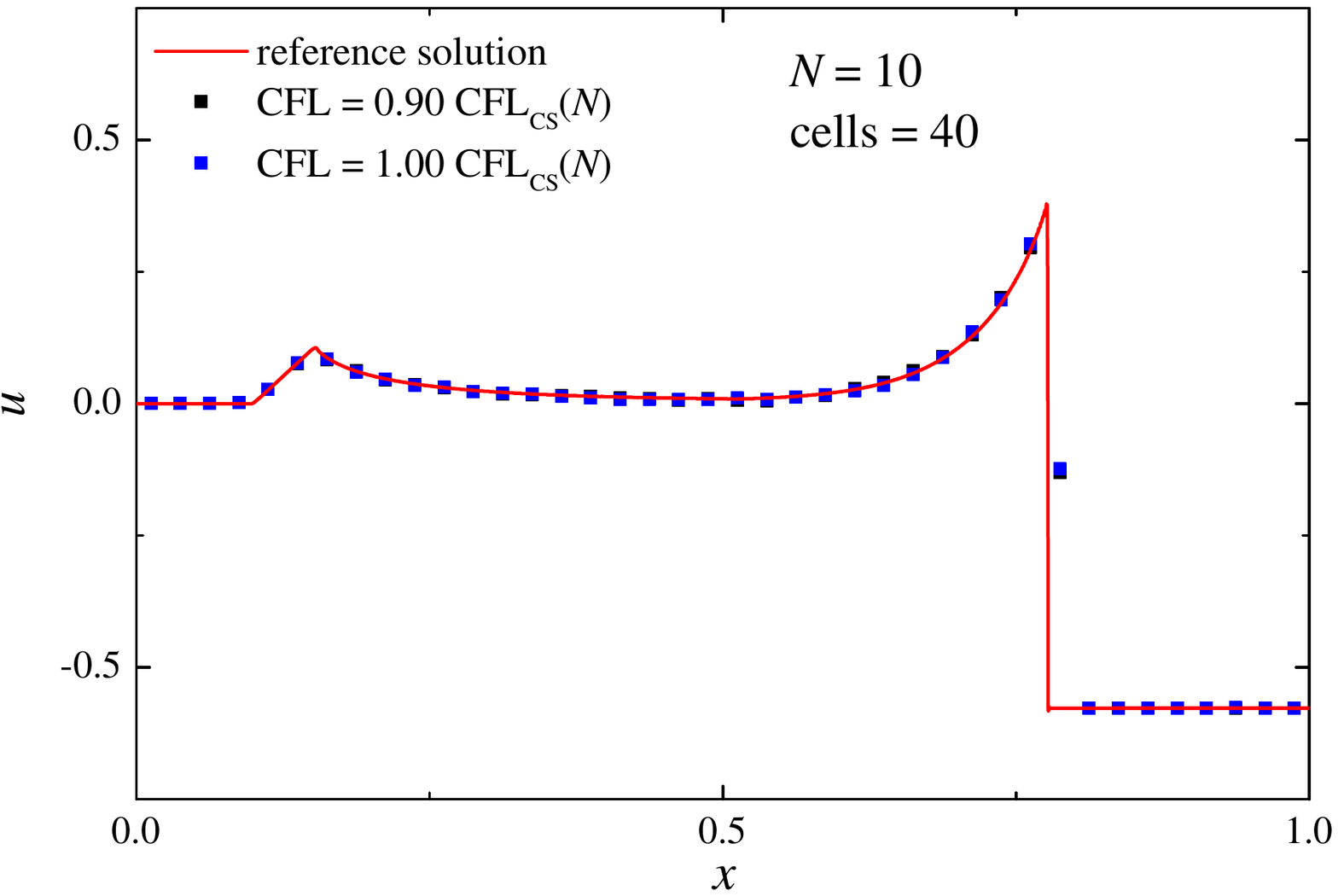}
\vspace{-8mm}\caption{\label{fig:test_znds_1d_degrees_10_12:degree_10_u}}
\end{subfigure}
\begin{subfigure}{0.24\textwidth}
\includegraphics[width=\textwidth]{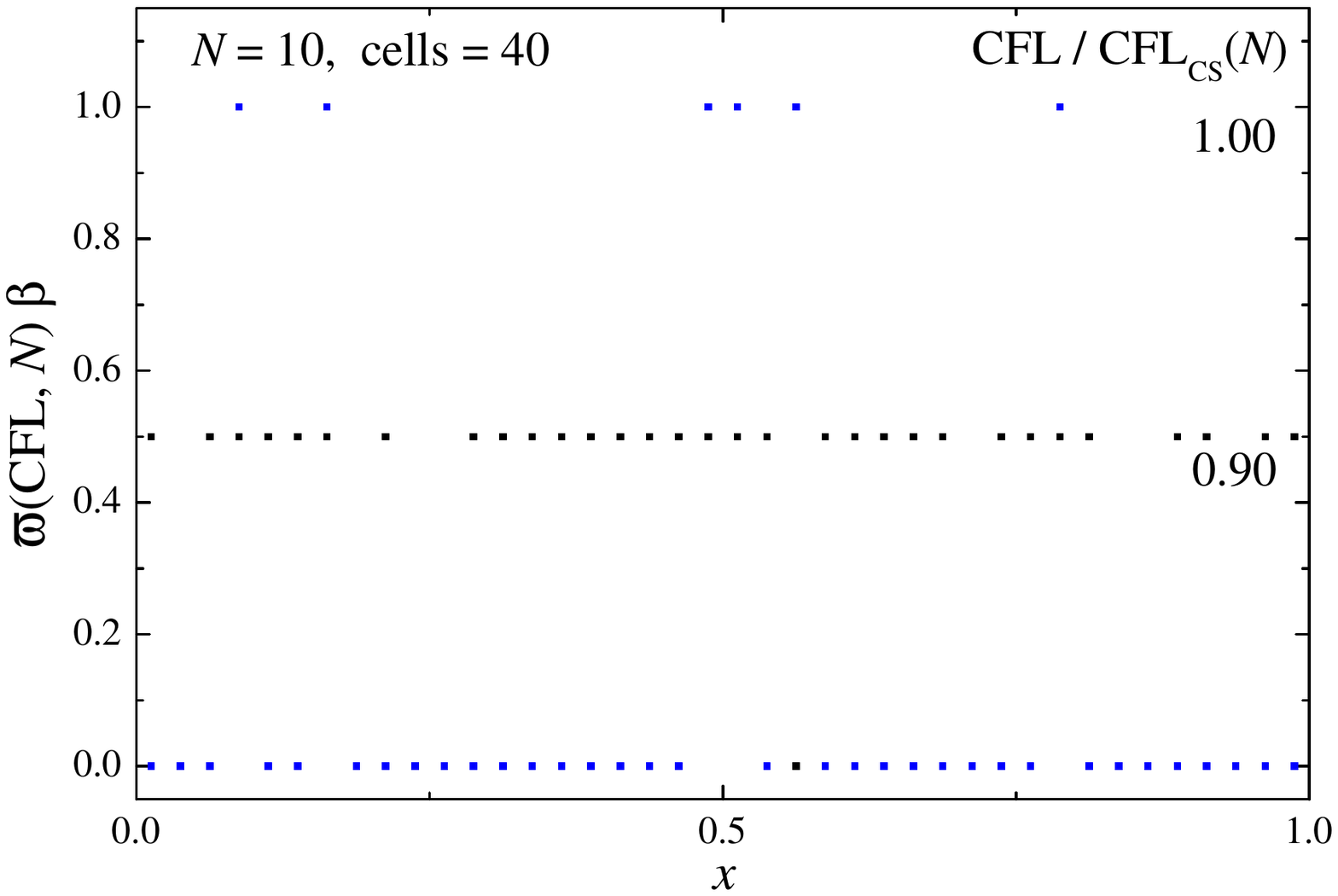}
\vspace{-8mm}\caption{\label{fig:test_znds_1d_degrees_10_12:degree_10_beta}}
\end{subfigure}\\[-2mm]
\begin{subfigure}{0.24\textwidth}
\includegraphics[width=\textwidth]{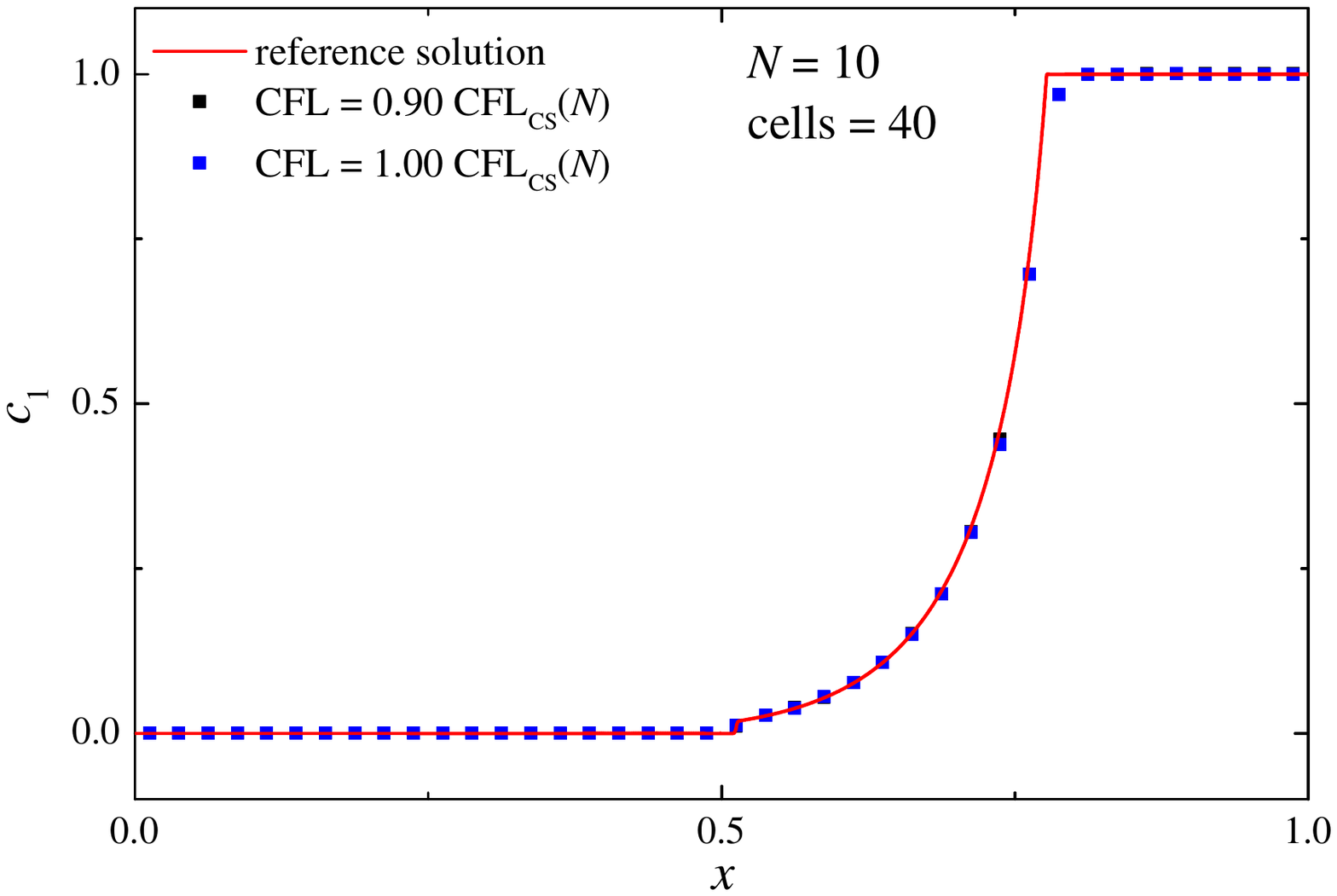}
\vspace{-8mm}\caption{\label{fig:test_znds_1d_degrees_10_12:degree_10_c1}}
\end{subfigure}
\begin{subfigure}{0.24\textwidth}
\includegraphics[width=\textwidth]{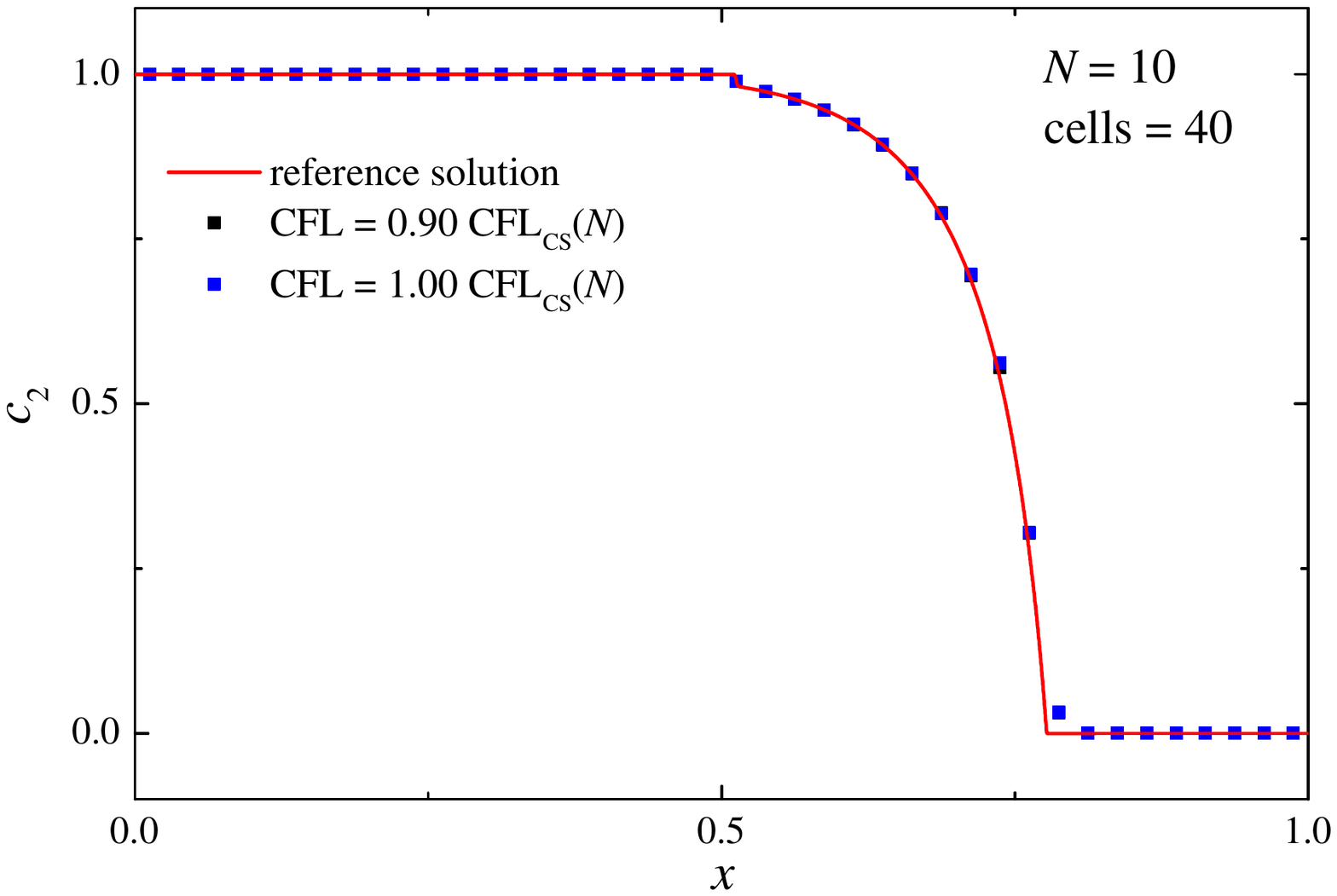}
\vspace{-8mm}\caption{\label{fig:test_znds_1d_degrees_10_12:degree_10_c2}}
\end{subfigure}
\begin{subfigure}{0.24\textwidth}
\includegraphics[width=\textwidth]{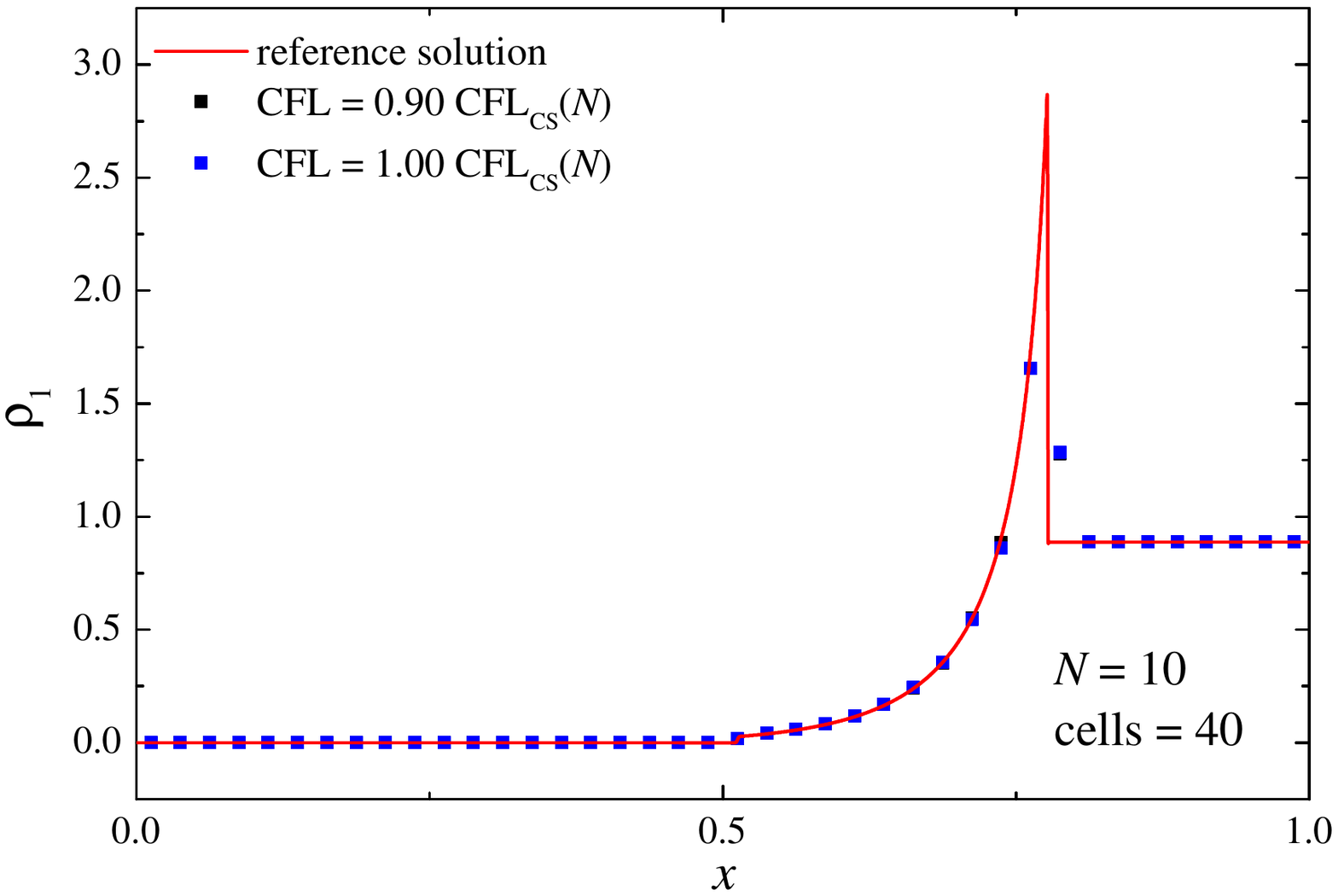}
\vspace{-8mm}\caption{\label{fig:test_znds_1d_degrees_10_12:degree_10_den1}}
\end{subfigure}
\begin{subfigure}{0.24\textwidth}
\includegraphics[width=\textwidth]{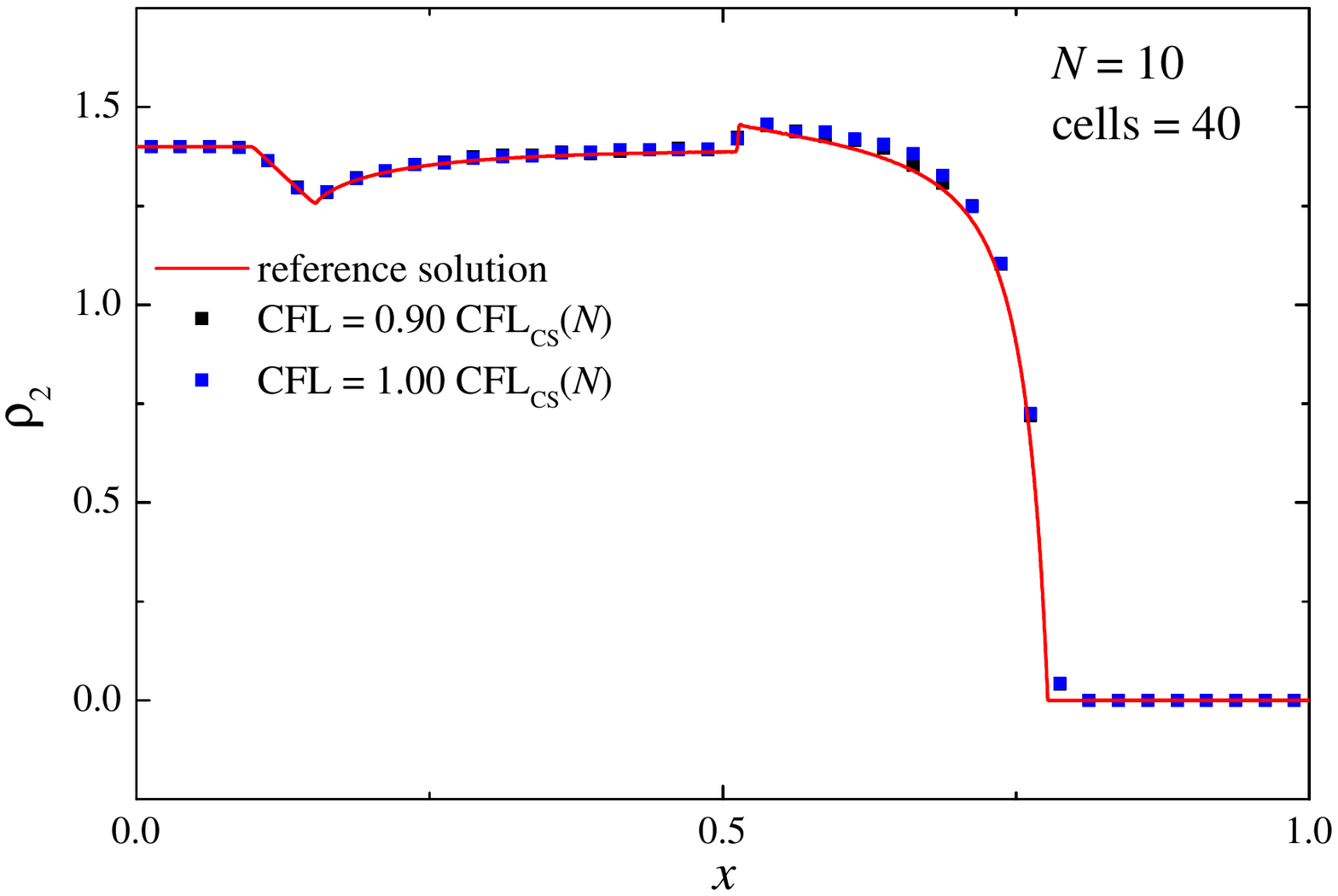}
\vspace{-8mm}\caption{\label{fig:test_znds_1d_degrees_10_12:degree_10_den2}}
\end{subfigure}\\[-2mm]
\begin{subfigure}{0.24\textwidth}
\includegraphics[width=\textwidth]{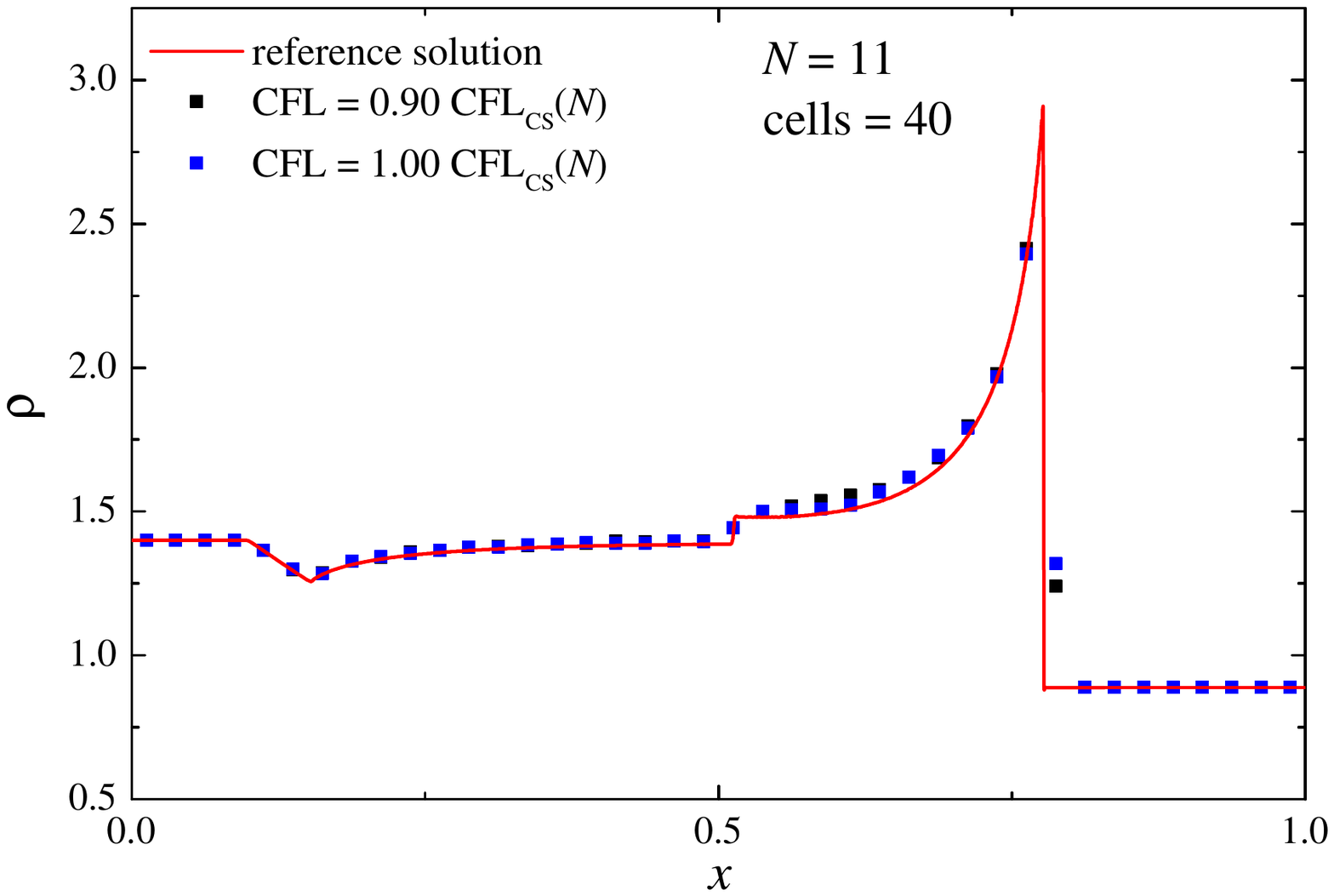}
\vspace{-8mm}\caption{\label{fig:test_znds_1d_degrees_10_12:degree_11_den}}
\end{subfigure}
\begin{subfigure}{0.24\textwidth}
\includegraphics[width=\textwidth]{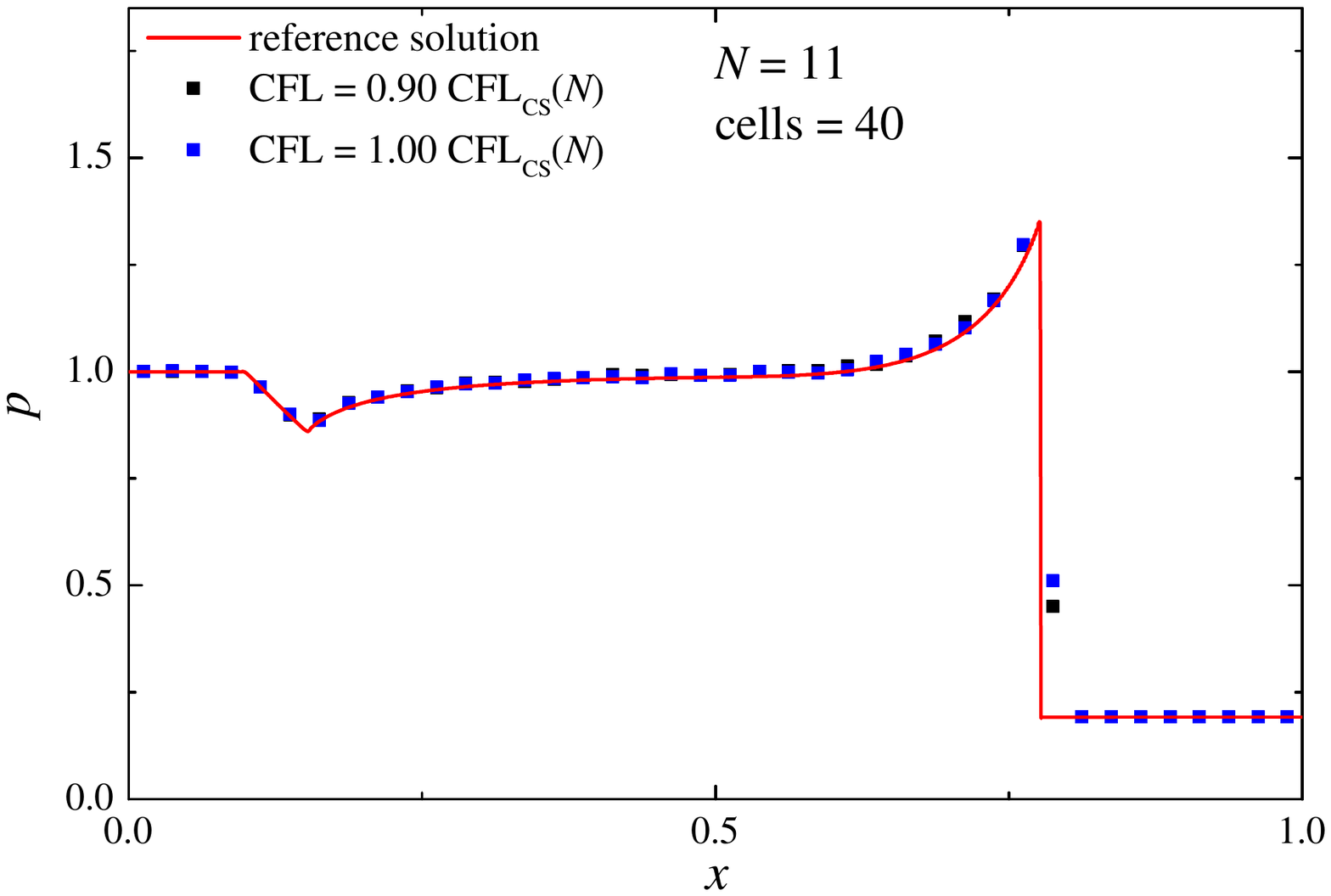}
\vspace{-8mm}\caption{\label{fig:test_znds_1d_degrees_10_12:degree_11_p}}
\end{subfigure}
\begin{subfigure}{0.24\textwidth}
\includegraphics[width=\textwidth]{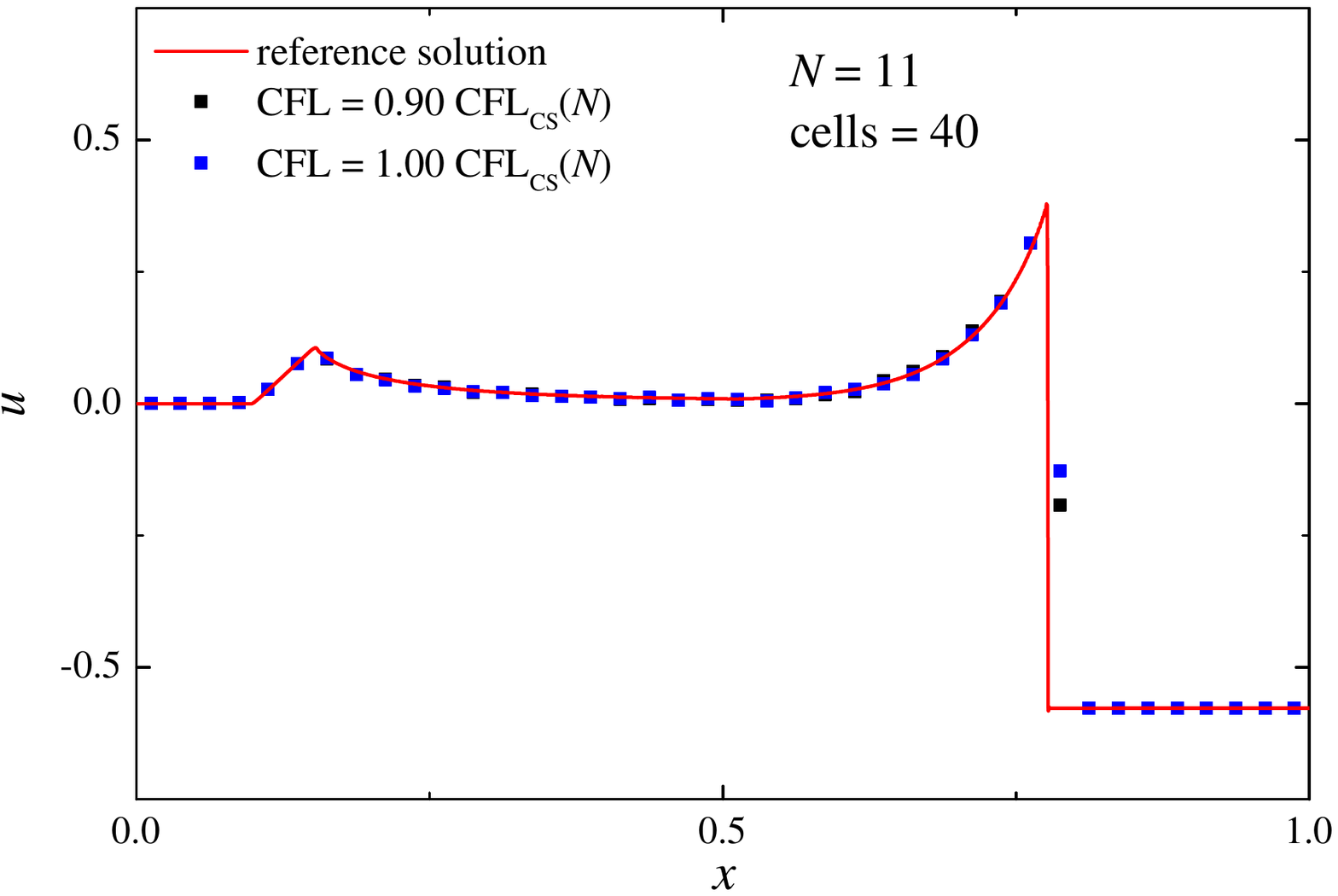}
\vspace{-8mm}\caption{\label{fig:test_znds_1d_degrees_10_12:degree_11_u}}
\end{subfigure}
\begin{subfigure}{0.24\textwidth}
\includegraphics[width=\textwidth]{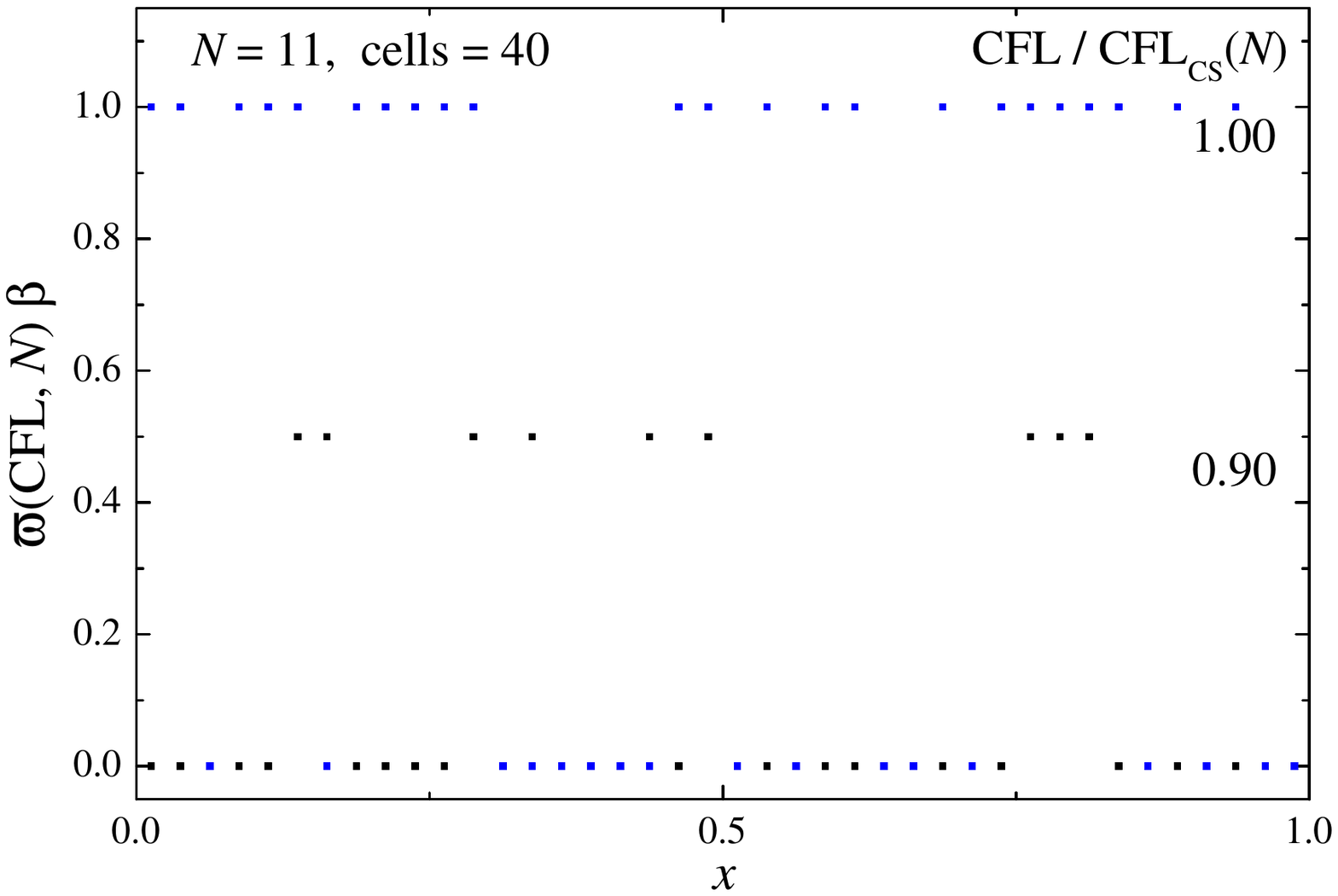}
\vspace{-8mm}\caption{\label{fig:test_znds_1d_degrees_10_12:degree_11_beta}}
\end{subfigure}\\[-2mm]
\begin{subfigure}{0.24\textwidth}
\includegraphics[width=\textwidth]{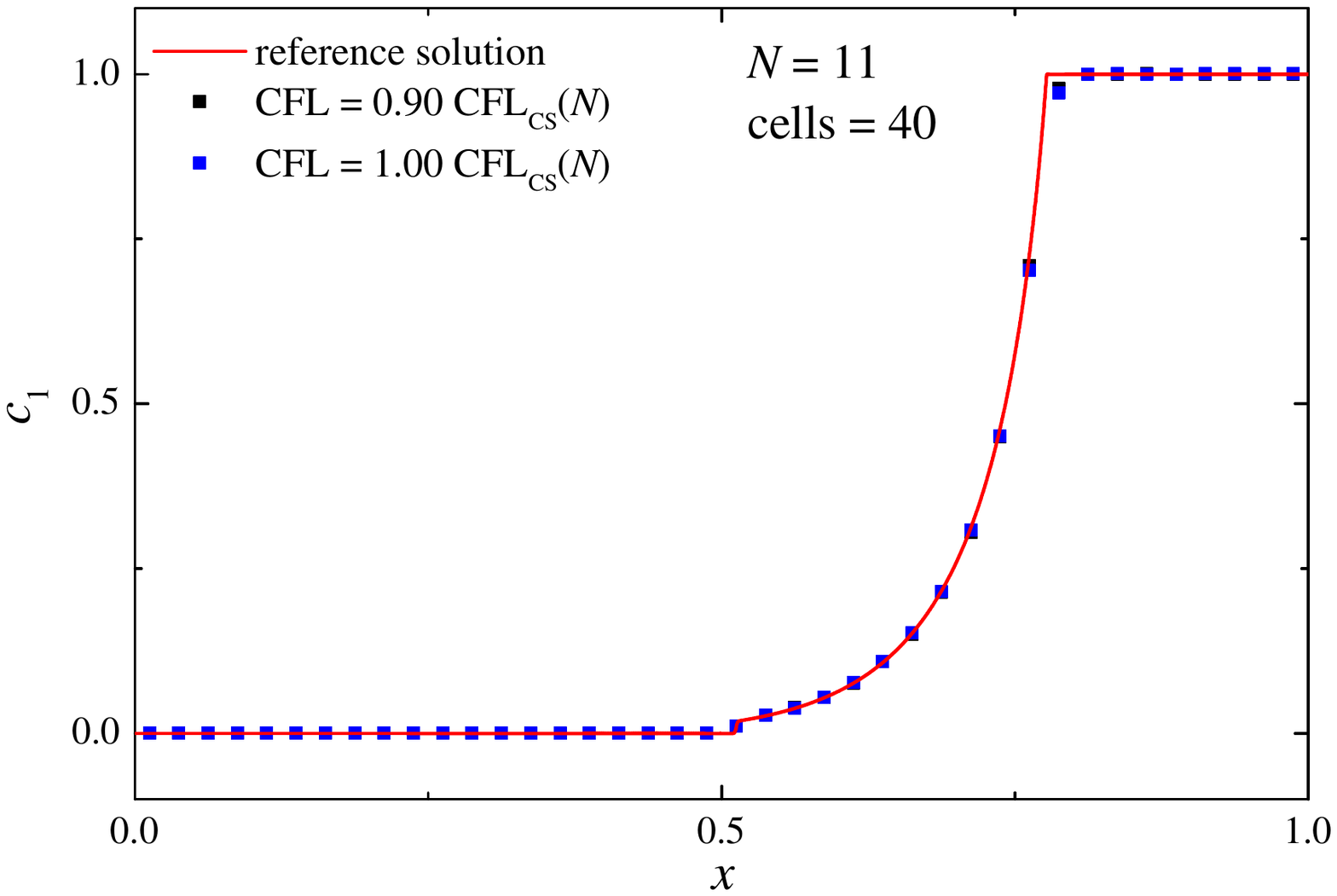}
\vspace{-8mm}\caption{\label{fig:test_znds_1d_degrees_10_12:degree_11_c1}}
\end{subfigure}
\begin{subfigure}{0.24\textwidth}
\includegraphics[width=\textwidth]{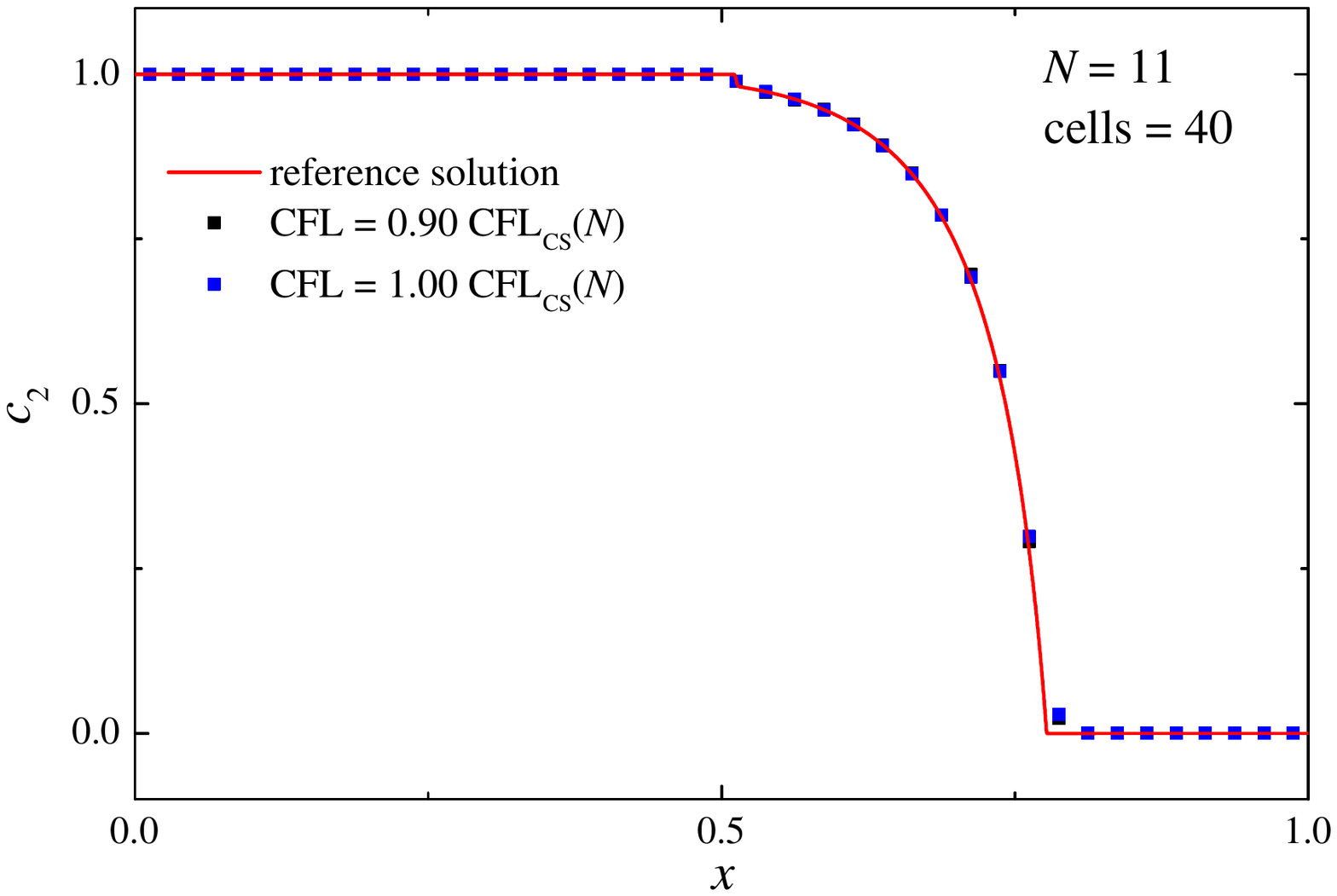}
\vspace{-8mm}\caption{\label{fig:test_znds_1d_degrees_10_12:degree_11_c2}}
\end{subfigure}
\begin{subfigure}{0.24\textwidth}
\includegraphics[width=\textwidth]{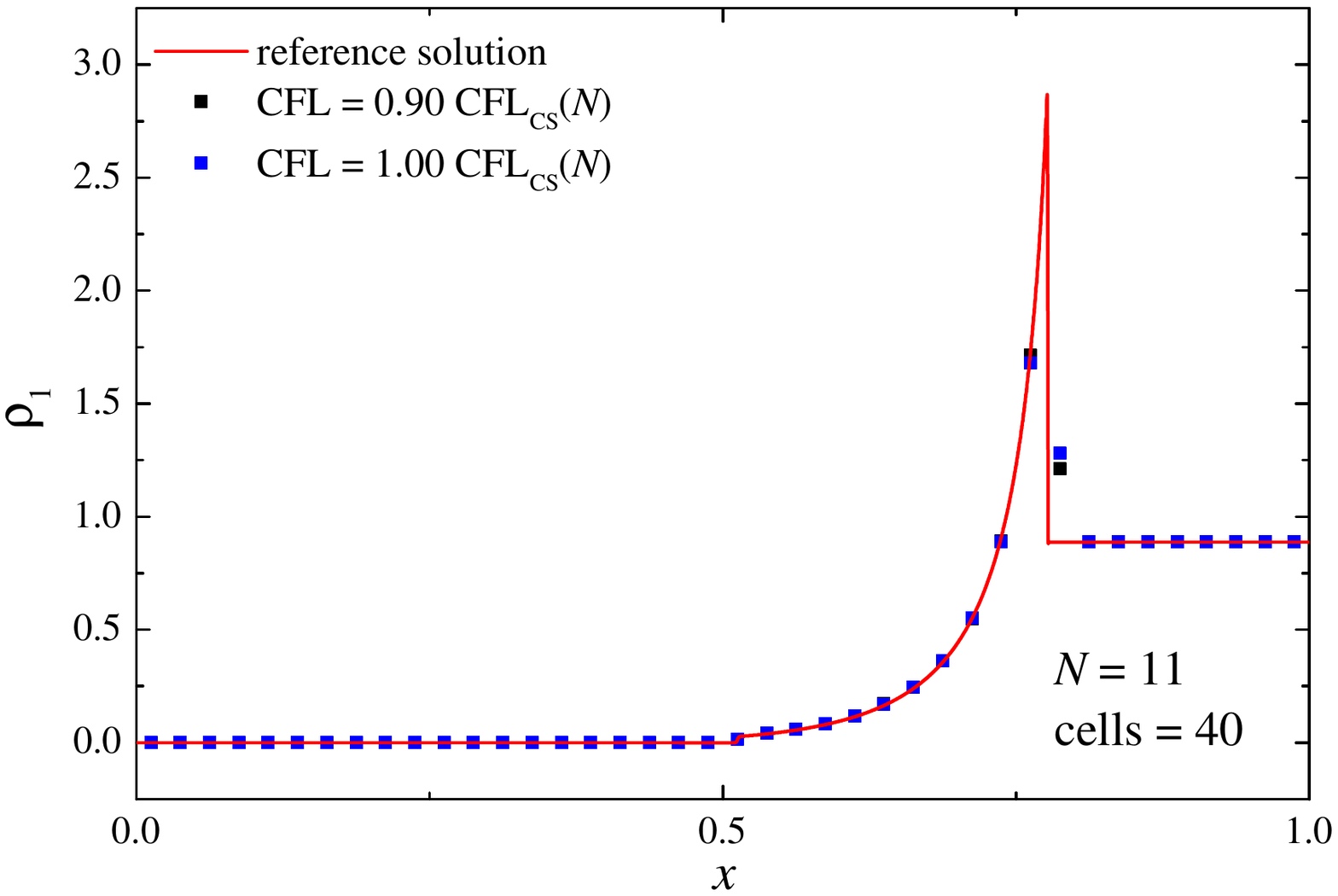}
\vspace{-8mm}\caption{\label{fig:test_znds_1d_degrees_10_12:degree_11_den1}}
\end{subfigure}
\begin{subfigure}{0.24\textwidth}
\includegraphics[width=\textwidth]{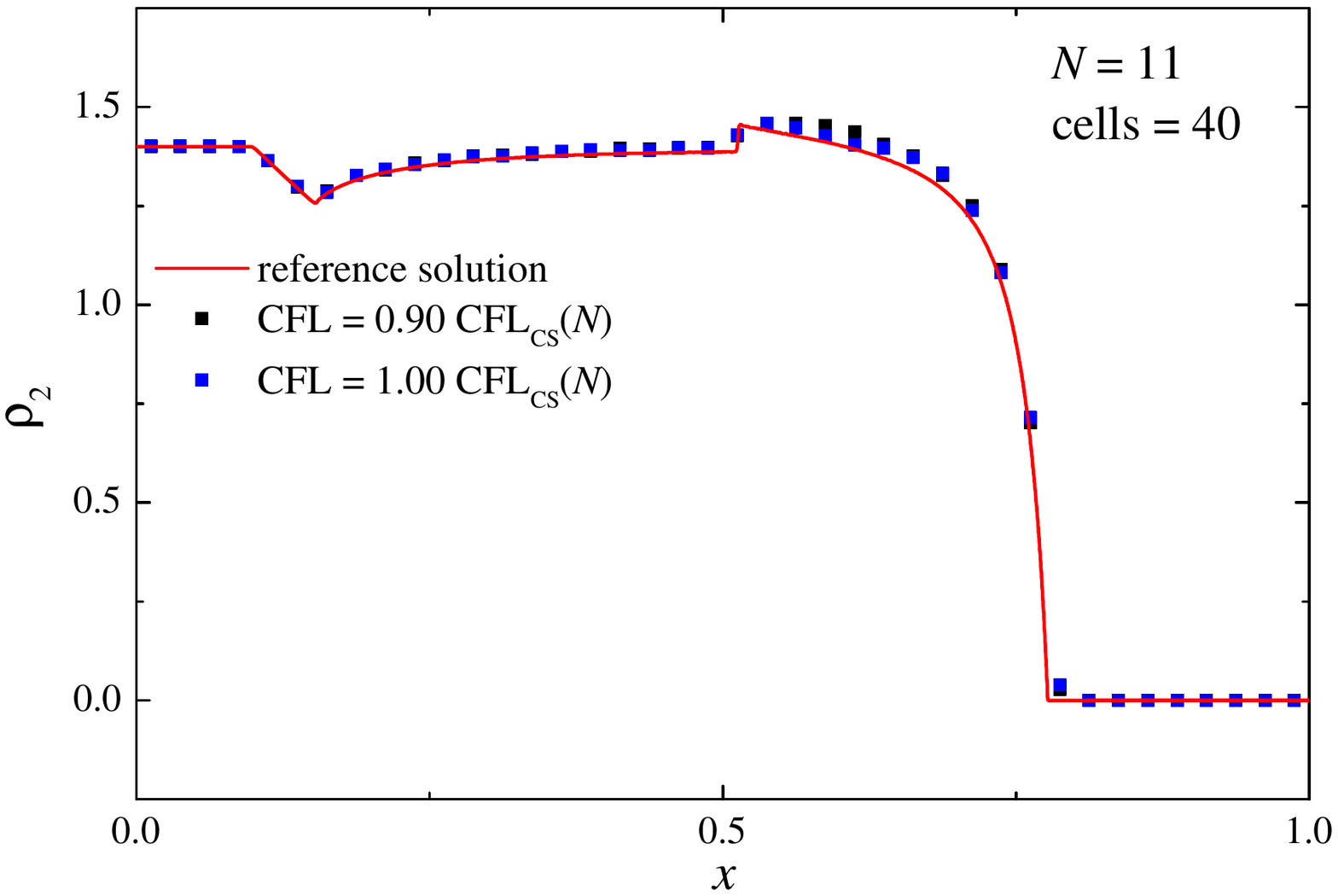}
\vspace{-8mm}\caption{\label{fig:test_znds_1d_degrees_10_12:degree_11_den2}}
\end{subfigure}\\[-2mm]
\begin{subfigure}{0.24\textwidth}
\includegraphics[width=\textwidth]{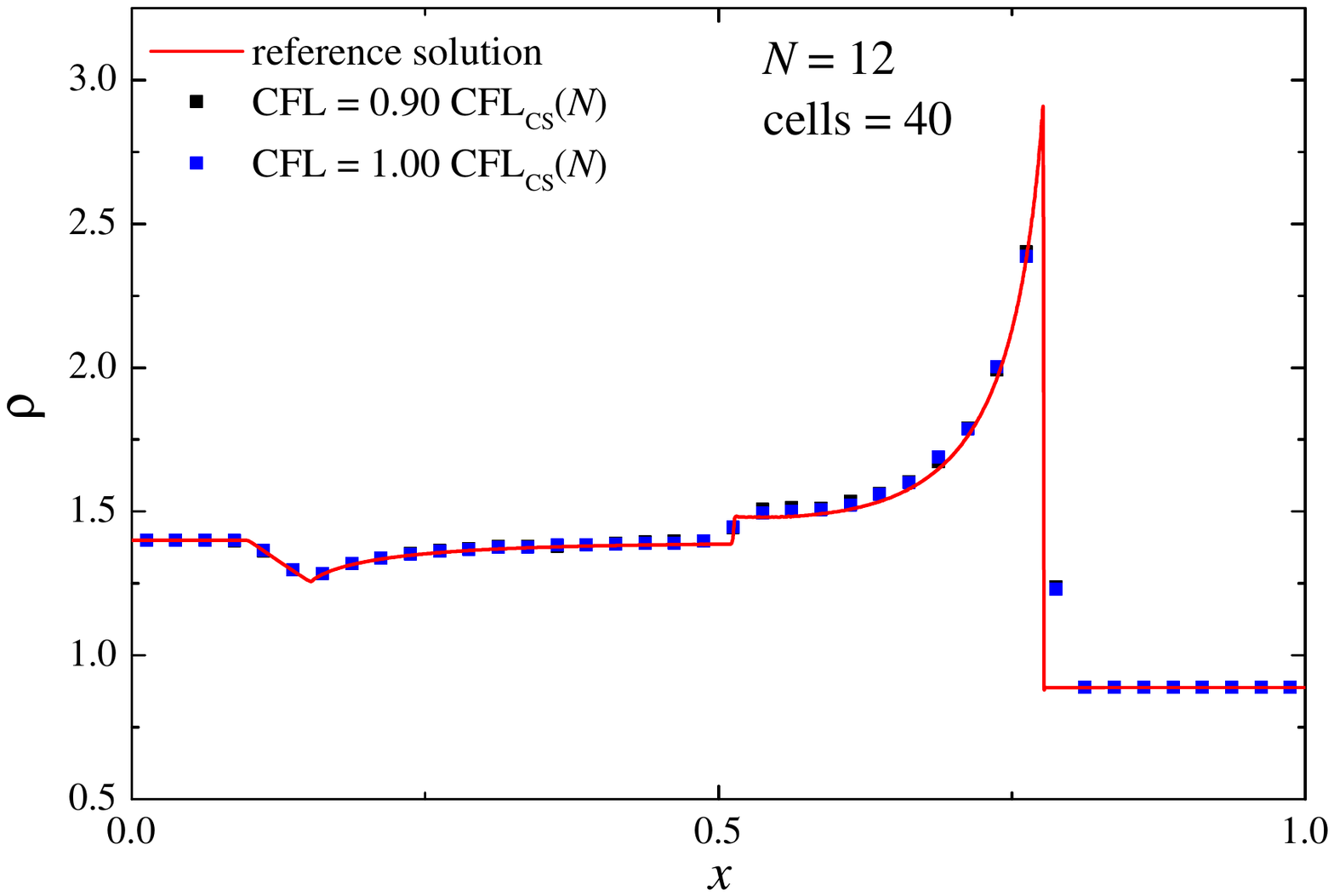}
\vspace{-8mm}\caption{\label{fig:test_znds_1d_degrees_10_12:degree_12_den}}
\end{subfigure}
\begin{subfigure}{0.24\textwidth}
\includegraphics[width=\textwidth]{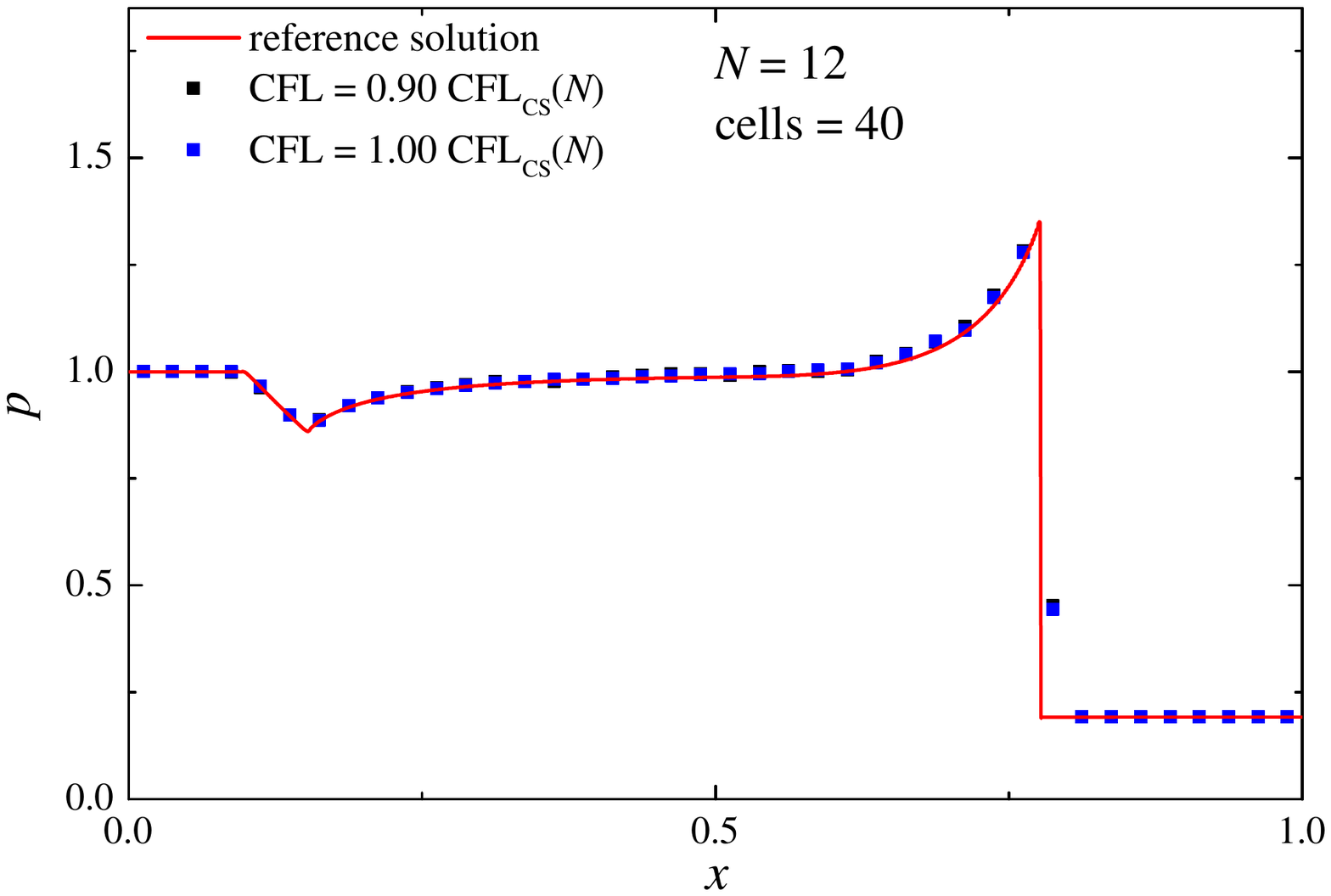}
\vspace{-8mm}\caption{\label{fig:test_znds_1d_degrees_10_12:degree_12_p}}
\end{subfigure}
\begin{subfigure}{0.24\textwidth}
\includegraphics[width=\textwidth]{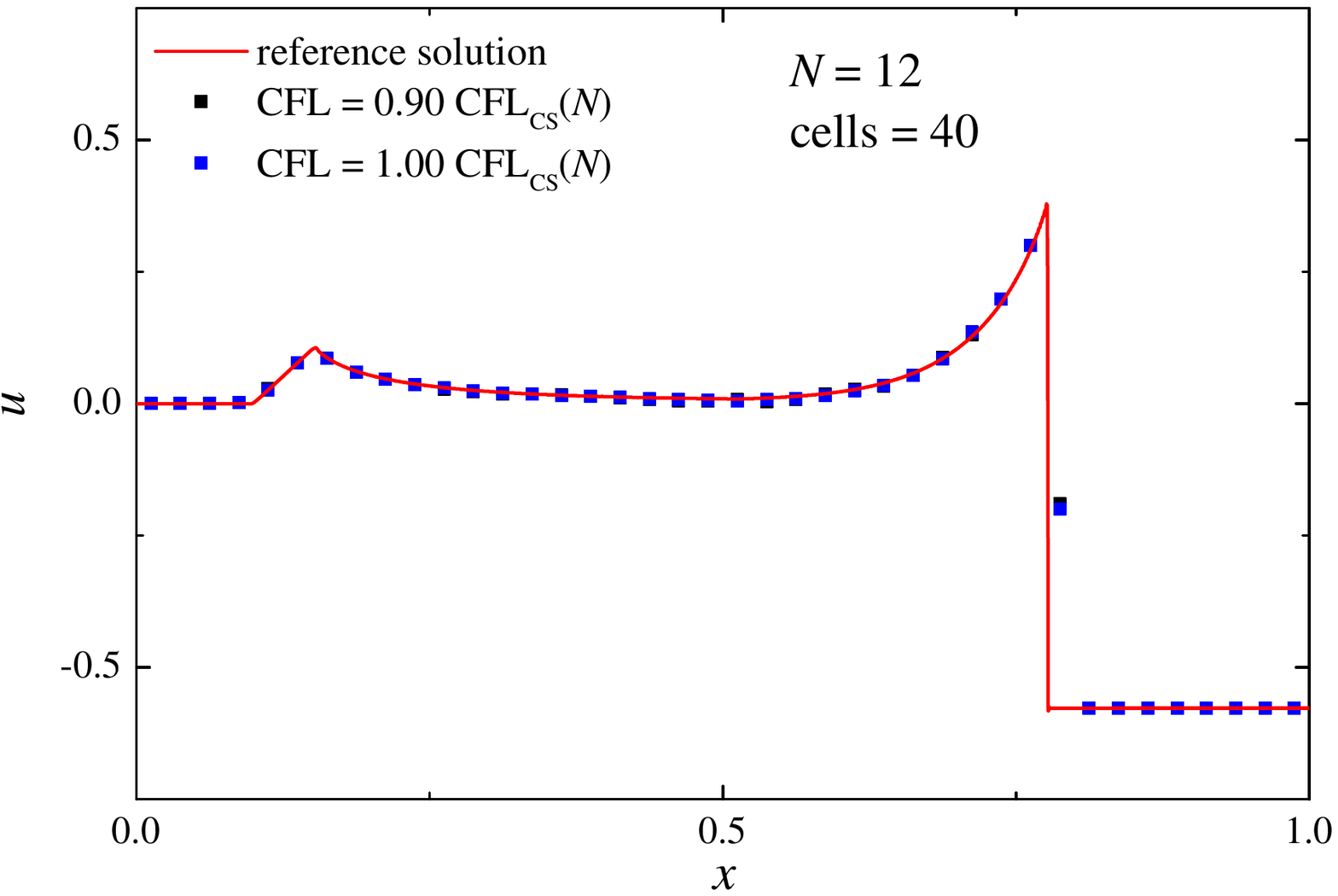}
\vspace{-8mm}\caption{\label{fig:test_znds_1d_degrees_10_12:degree_12_u}}
\end{subfigure}
\begin{subfigure}{0.24\textwidth}
\includegraphics[width=\textwidth]{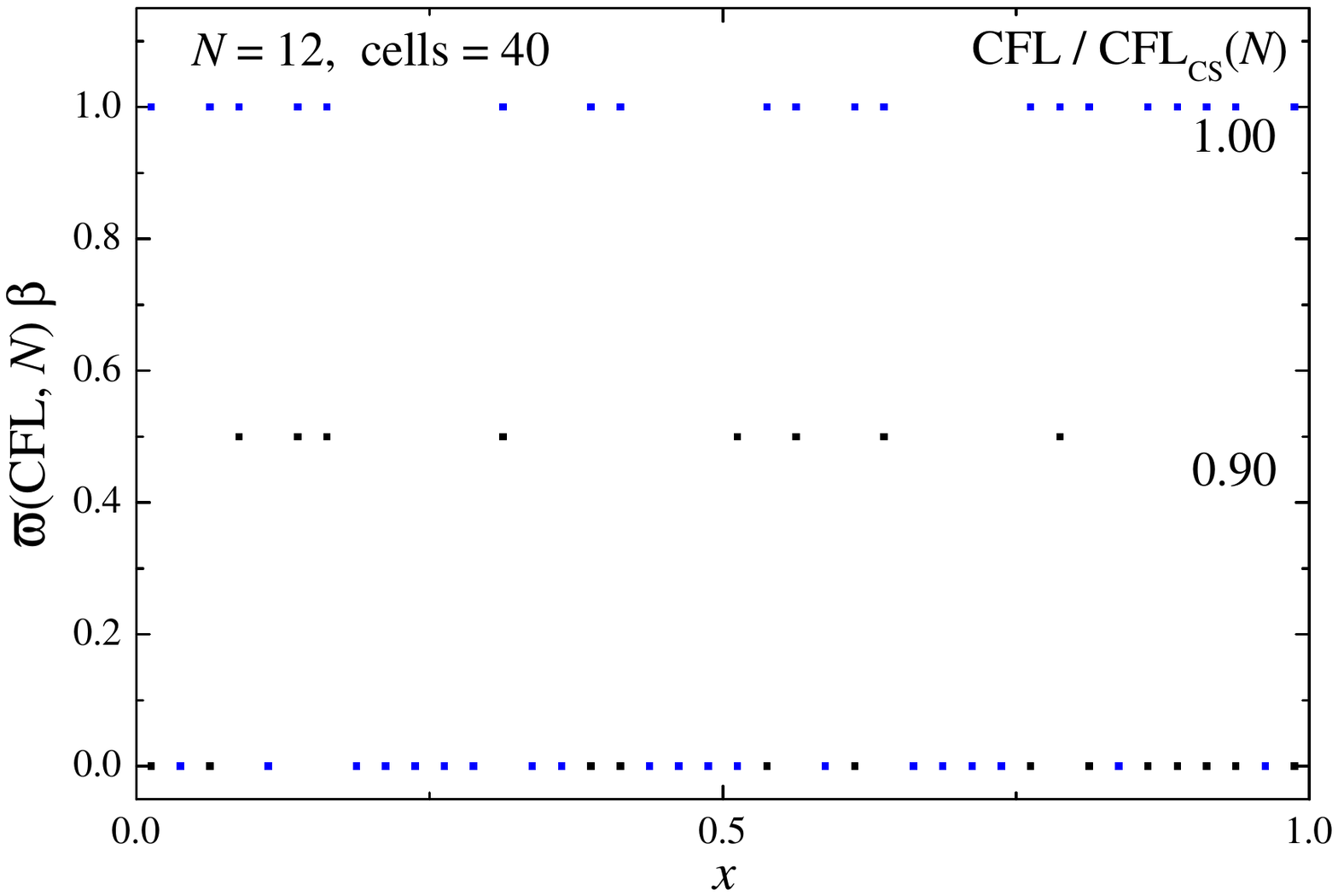}
\vspace{-8mm}\caption{\label{fig:test_znds_1d_degrees_10_12:degree_12_beta}}
\end{subfigure}\\[-2mm]
\begin{subfigure}{0.24\textwidth}
\includegraphics[width=\textwidth]{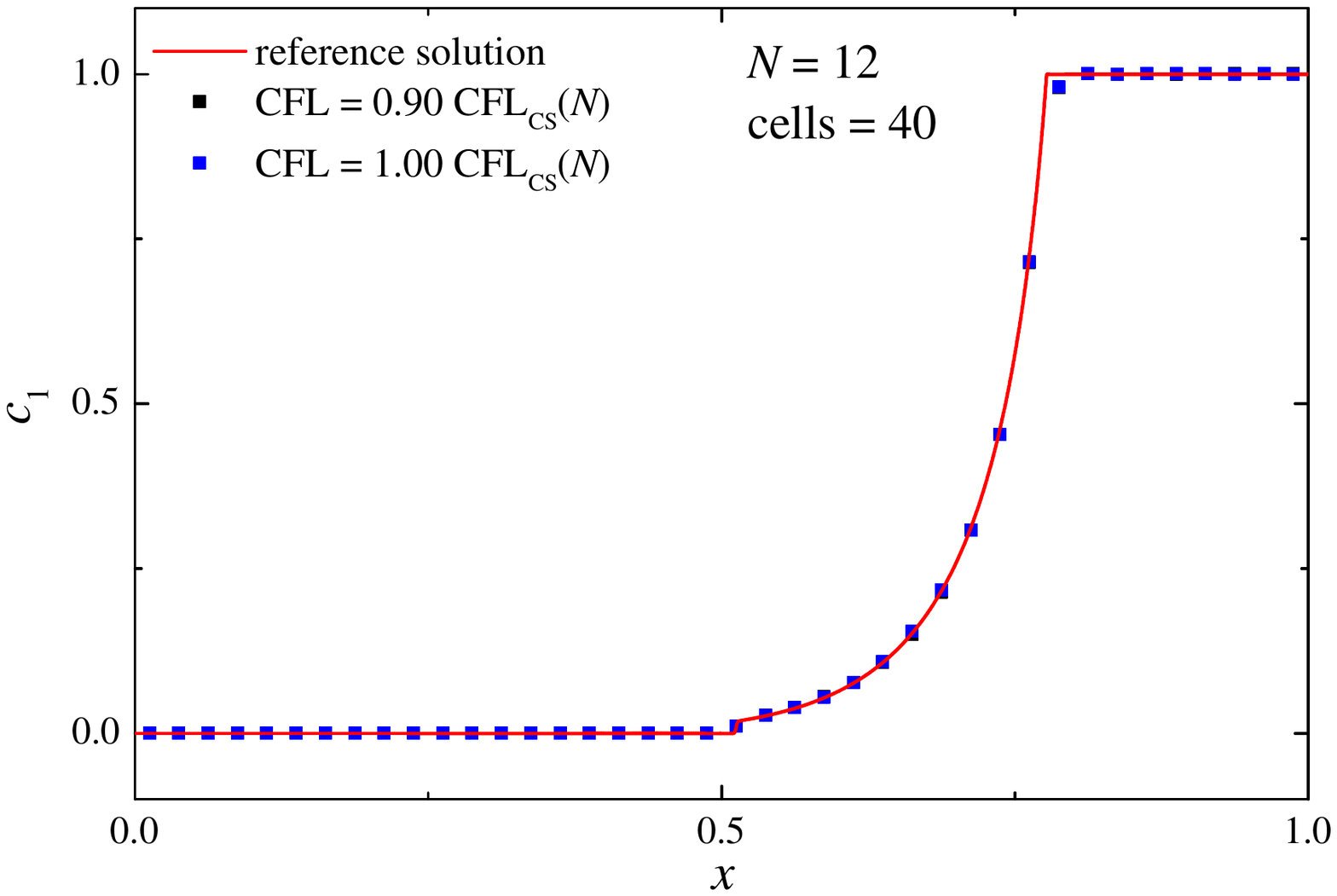}
\vspace{-8mm}\caption{\label{fig:test_znds_1d_degrees_10_12:degree_12_c1}}
\end{subfigure}
\begin{subfigure}{0.24\textwidth}
\includegraphics[width=\textwidth]{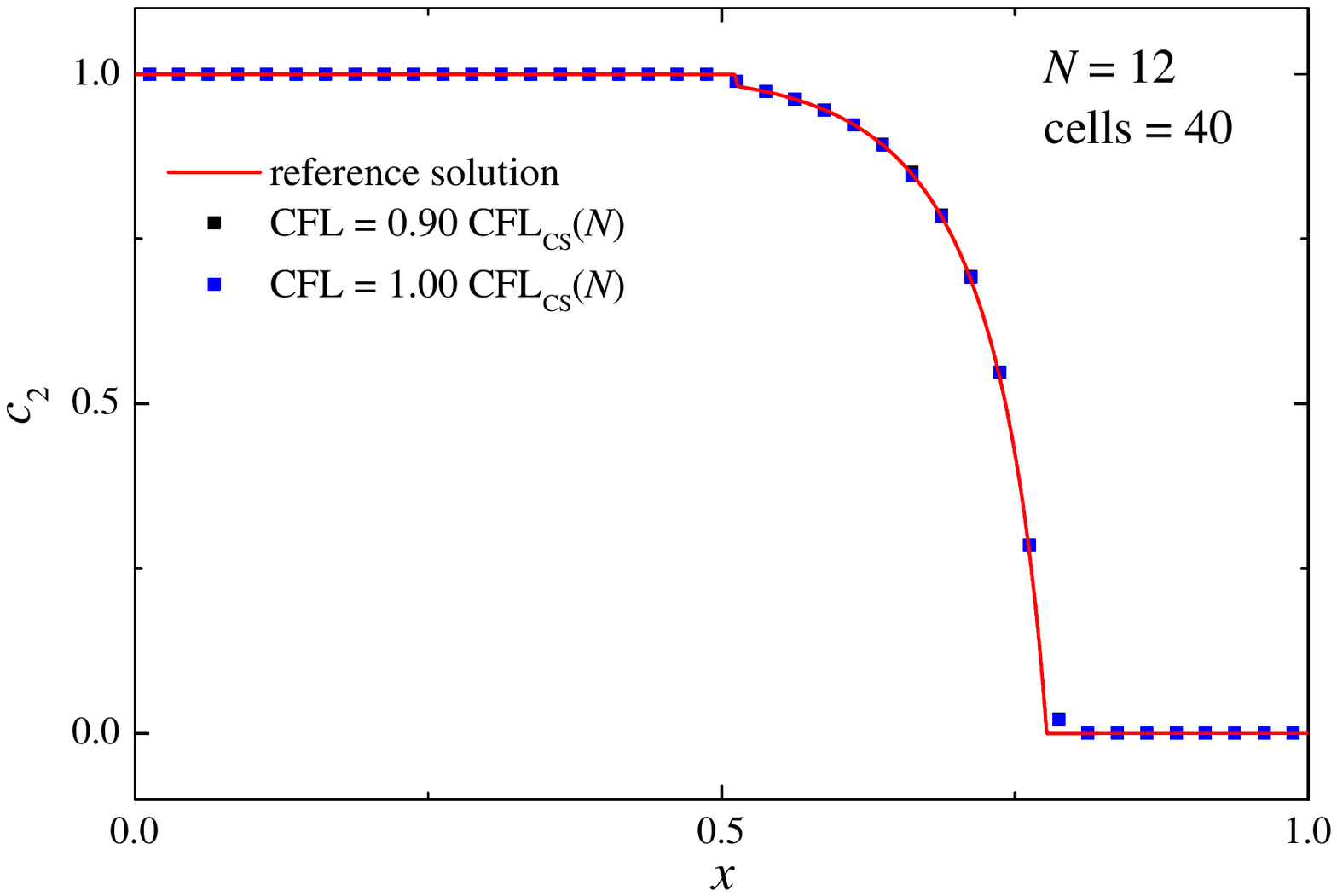}
\vspace{-8mm}\caption{\label{fig:test_znds_1d_degrees_10_12:degree_12_c2}}
\end{subfigure}
\begin{subfigure}{0.24\textwidth}
\includegraphics[width=\textwidth]{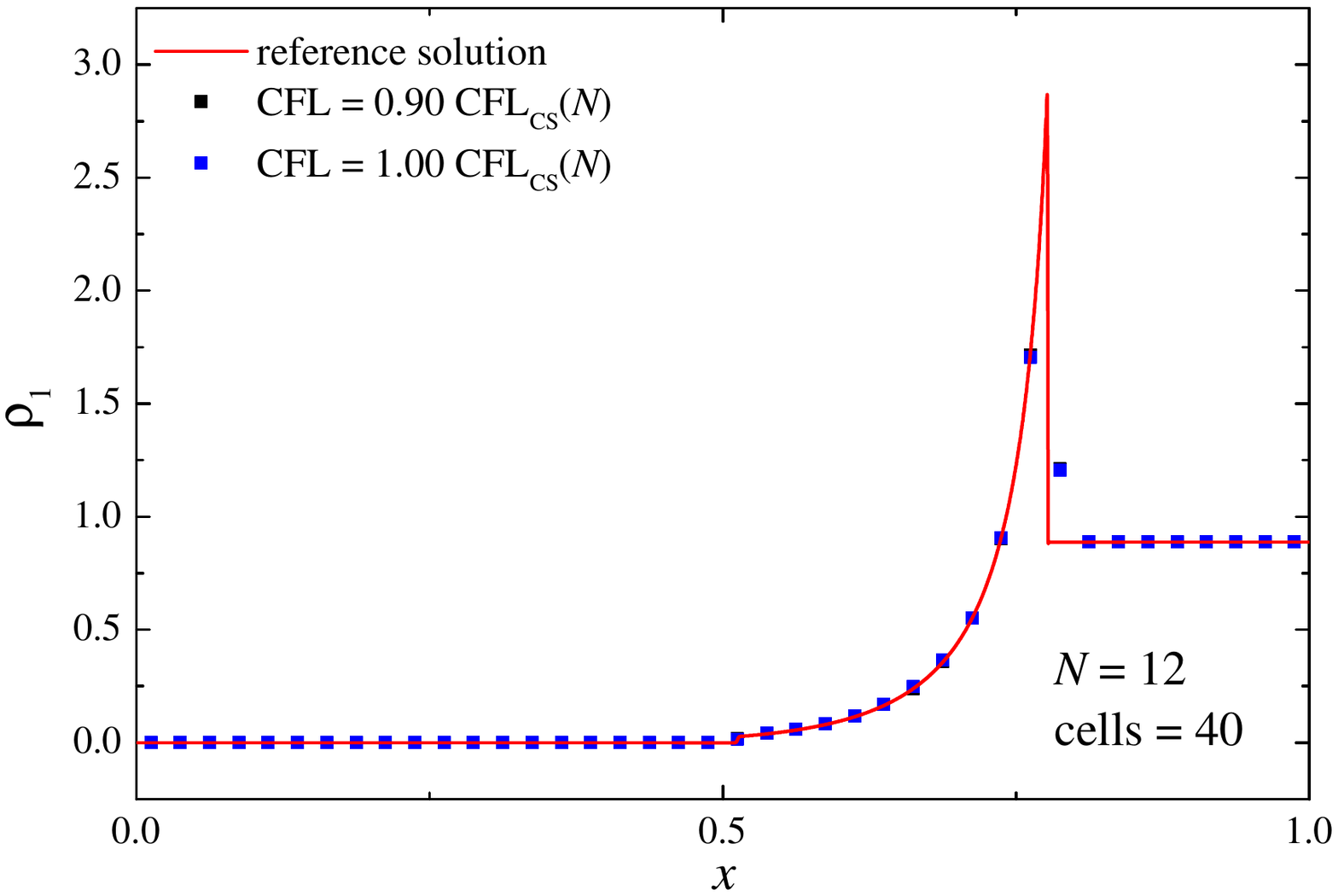}
\vspace{-8mm}\caption{\label{fig:test_znds_1d_degrees_10_12:degree_12_den1}}
\end{subfigure}
\begin{subfigure}{0.24\textwidth}
\includegraphics[width=\textwidth]{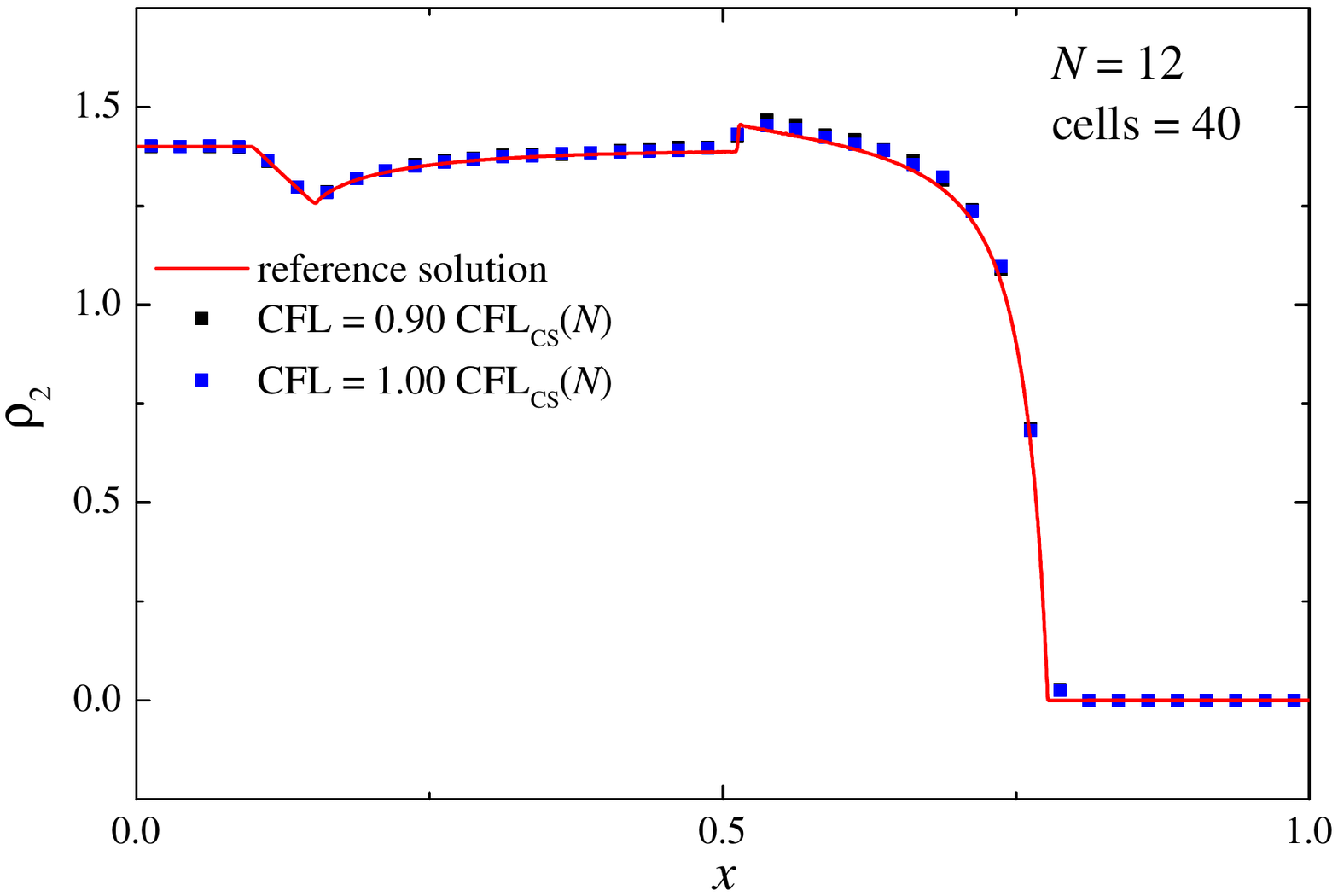}
\vspace{-8mm}\caption{\label{fig:test_znds_1d_degrees_10_12:degree_12_den2}}
\end{subfigure}
\caption{%
Numerical solution of the one-dimensional problem of the plane detonation wave formation in a two-component medium with a ``slow'' reaction (weak stiff case, a detailed statement of the problem is presented in the text), obtained using the ADER-DG method for the basis polynomials degree $N = 10$ ((\subref{fig:test_znds_1d_degrees_10_12:degree_10_den}), (\subref{fig:test_znds_1d_degrees_10_12:degree_10_p}), (\subref{fig:test_znds_1d_degrees_10_12:degree_10_u}), (\subref{fig:test_znds_1d_degrees_10_12:degree_10_c1}), (\subref{fig:test_znds_1d_degrees_10_12:degree_10_c2}), (\subref{fig:test_znds_1d_degrees_10_12:degree_10_den1}), (\subref{fig:test_znds_1d_degrees_10_12:degree_10_den2})), $N = 11$ ((\subref{fig:test_znds_1d_degrees_10_12:degree_11_den}), (\subref{fig:test_znds_1d_degrees_10_12:degree_11_p}), (\subref{fig:test_znds_1d_degrees_10_12:degree_11_u}), (\subref{fig:test_znds_1d_degrees_10_12:degree_11_c1}), (\subref{fig:test_znds_1d_degrees_10_12:degree_11_c2}), (\subref{fig:test_znds_1d_degrees_10_12:degree_11_den1}), (\subref{fig:test_znds_1d_degrees_10_12:degree_11_den2})), $N = 12$ ((\subref{fig:test_znds_1d_degrees_10_12:degree_12_den}), (\subref{fig:test_znds_1d_degrees_10_12:degree_12_p}), (\subref{fig:test_znds_1d_degrees_10_12:degree_12_u}), (\subref{fig:test_znds_1d_degrees_10_12:degree_12_c1}), (\subref{fig:test_znds_1d_degrees_10_12:degree_12_c2}), (\subref{fig:test_znds_1d_degrees_10_12:degree_12_den1}), (\subref{fig:test_znds_1d_degrees_10_12:degree_12_den2})) with a posteriori limitation of the solution by an ADER-WENO2 finite-volume limiter, at the final time $t_{f} = 0.4$. The graphs show the coordinate dependence of density $\rho$ ((\subref{fig:test_znds_1d_degrees_10_12:degree_10_den}), (\subref{fig:test_znds_1d_degrees_10_12:degree_11_den}), (\subref{fig:test_znds_1d_degrees_10_12:degree_12_den})), pressure $p$ ((\subref{fig:test_znds_1d_degrees_10_12:degree_10_p}), (\subref{fig:test_znds_1d_degrees_10_12:degree_11_p}), (\subref{fig:test_znds_1d_degrees_10_12:degree_12_p})), flow velocity $u$ ((\subref{fig:test_znds_1d_degrees_10_12:degree_10_u}), (\subref{fig:test_znds_1d_degrees_10_12:degree_11_u}), (\subref{fig:test_znds_1d_degrees_10_12:degree_12_u})), troubled cells indicator $\beta$ multiplied by the separating function $\varpi(\mathrm{CFL}, N)$ ((\subref{fig:test_znds_1d_degrees_10_12:degree_10_beta}), (\subref{fig:test_znds_1d_degrees_10_12:degree_11_beta}), (\subref{fig:test_znds_1d_degrees_10_12:degree_12_beta})), mass concentrations $c_{1}$ ((\subref{fig:test_znds_1d_degrees_10_12:degree_10_c1}), (\subref{fig:test_znds_1d_degrees_10_12:degree_11_c1}), (\subref{fig:test_znds_1d_degrees_10_12:degree_12_c1})) and $c_{2}$ ((\subref{fig:test_znds_1d_degrees_10_12:degree_10_c2}), (\subref{fig:test_znds_1d_degrees_10_12:degree_11_c2}), (\subref{fig:test_znds_1d_degrees_10_12:degree_12_c2})) of the reactant $A$ and product $B$, respectively, and partial densities $\rho_{1}$ ((\subref{fig:test_znds_1d_degrees_10_12:degree_10_den1}), (\subref{fig:test_znds_1d_degrees_10_12:degree_11_den1}), (\subref{fig:test_znds_1d_degrees_10_12:degree_12_den1})) and $\rho_{2}$ ((\subref{fig:test_znds_1d_degrees_10_12:degree_10_den2}), (\subref{fig:test_znds_1d_degrees_10_12:degree_11_den2}), (\subref{fig:test_znds_1d_degrees_10_12:degree_12_den2})) of the reactant $A$ and product $B$, respectively. The square symbols represent the subcells finite-volume representation of the numerical solution; the red solid lines represents the reference solution of the problem.
}
\label{fig:test_znds_1d_degrees_10_12}
\end{figure}

\begin{figure}[h!]
\captionsetup[subfigure]{%
	position=bottom,
	font+=smaller,
	textfont=normalfont,
	singlelinecheck=off,
	justification=raggedright
}
\centering
\begin{subfigure}{0.24\textwidth}
\includegraphics[width=\textwidth]{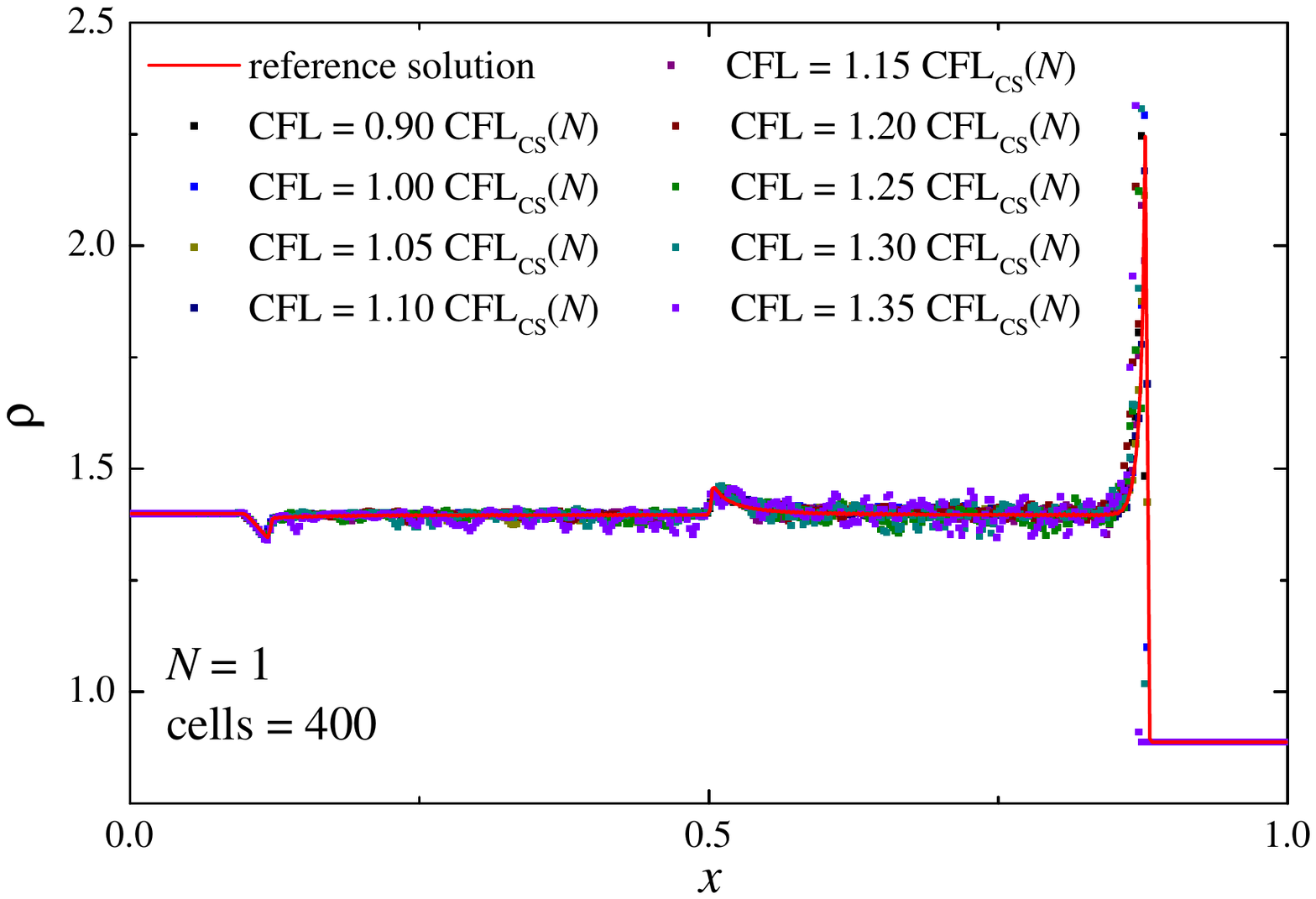}
\vspace{-8mm}\caption{\label{fig:test_zndf_1d_degrees_1_3:degree_1_den}}
\end{subfigure}
\begin{subfigure}{0.24\textwidth}
\includegraphics[width=\textwidth]{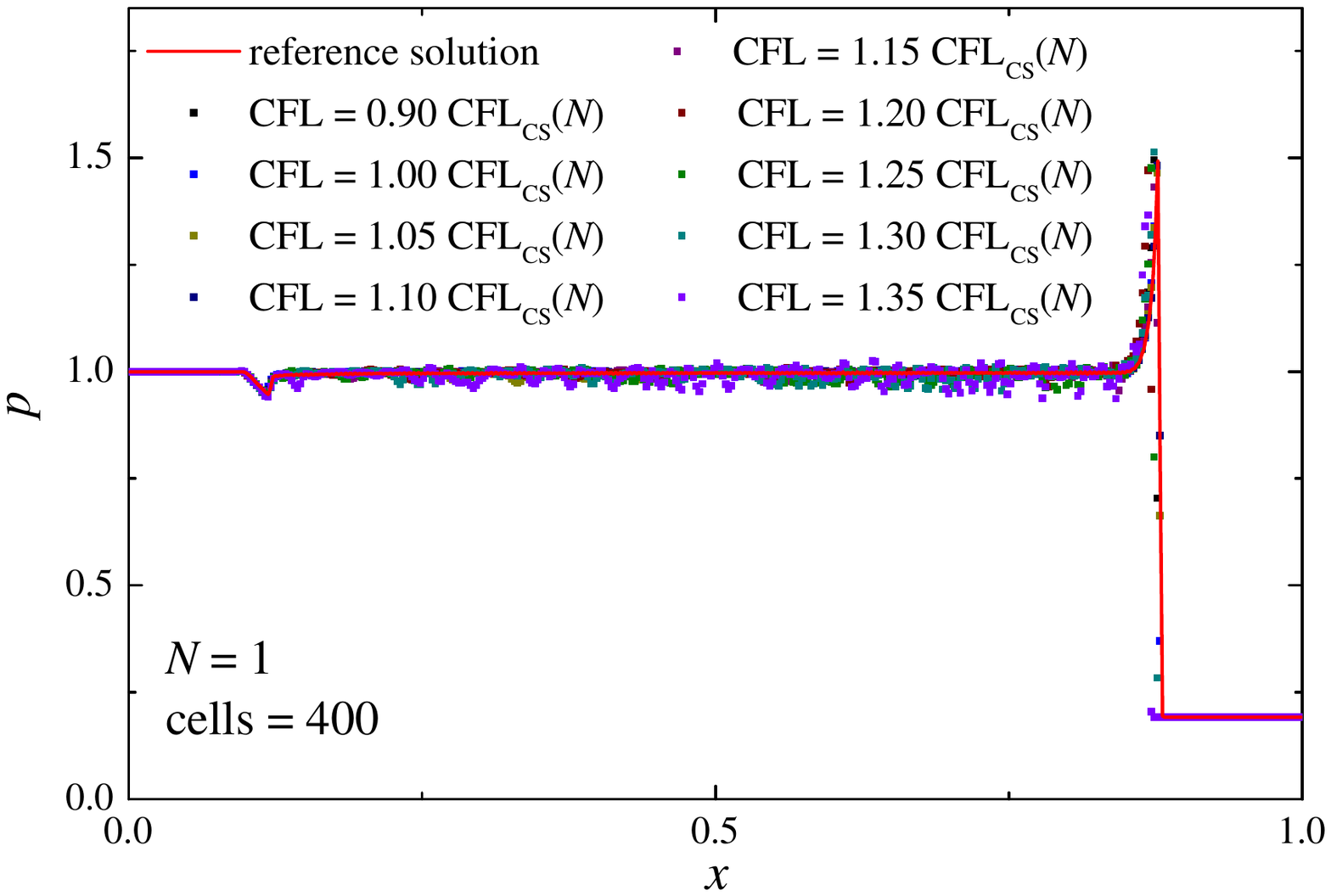}
\vspace{-8mm}\caption{\label{fig:test_zndf_1d_degrees_1_3:degree_1_p}}
\end{subfigure}
\begin{subfigure}{0.24\textwidth}
\includegraphics[width=\textwidth]{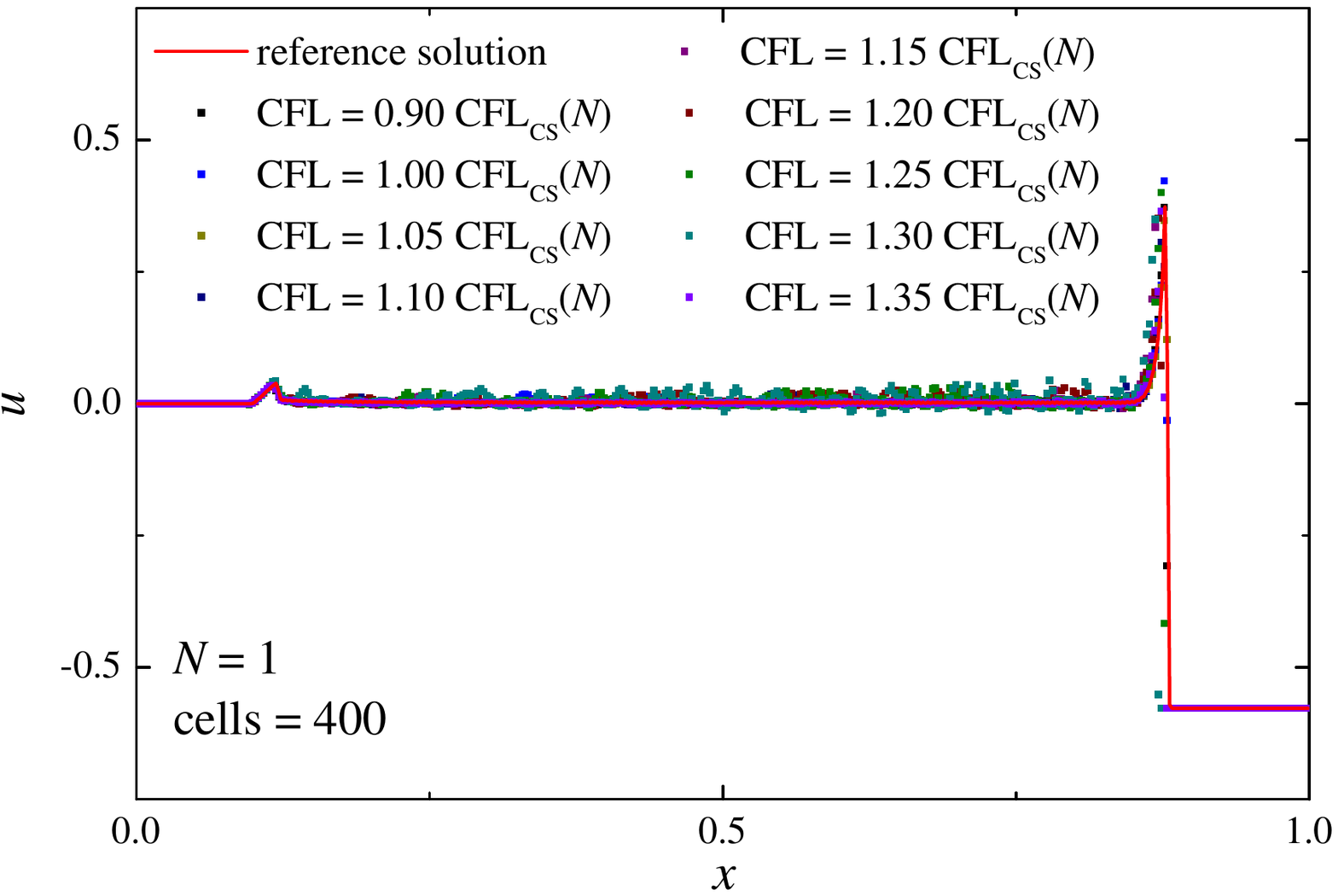}
\vspace{-8mm}\caption{\label{fig:test_zndf_1d_degrees_1_3:degree_1_u}}
\end{subfigure}
\begin{subfigure}{0.24\textwidth}
\includegraphics[width=\textwidth]{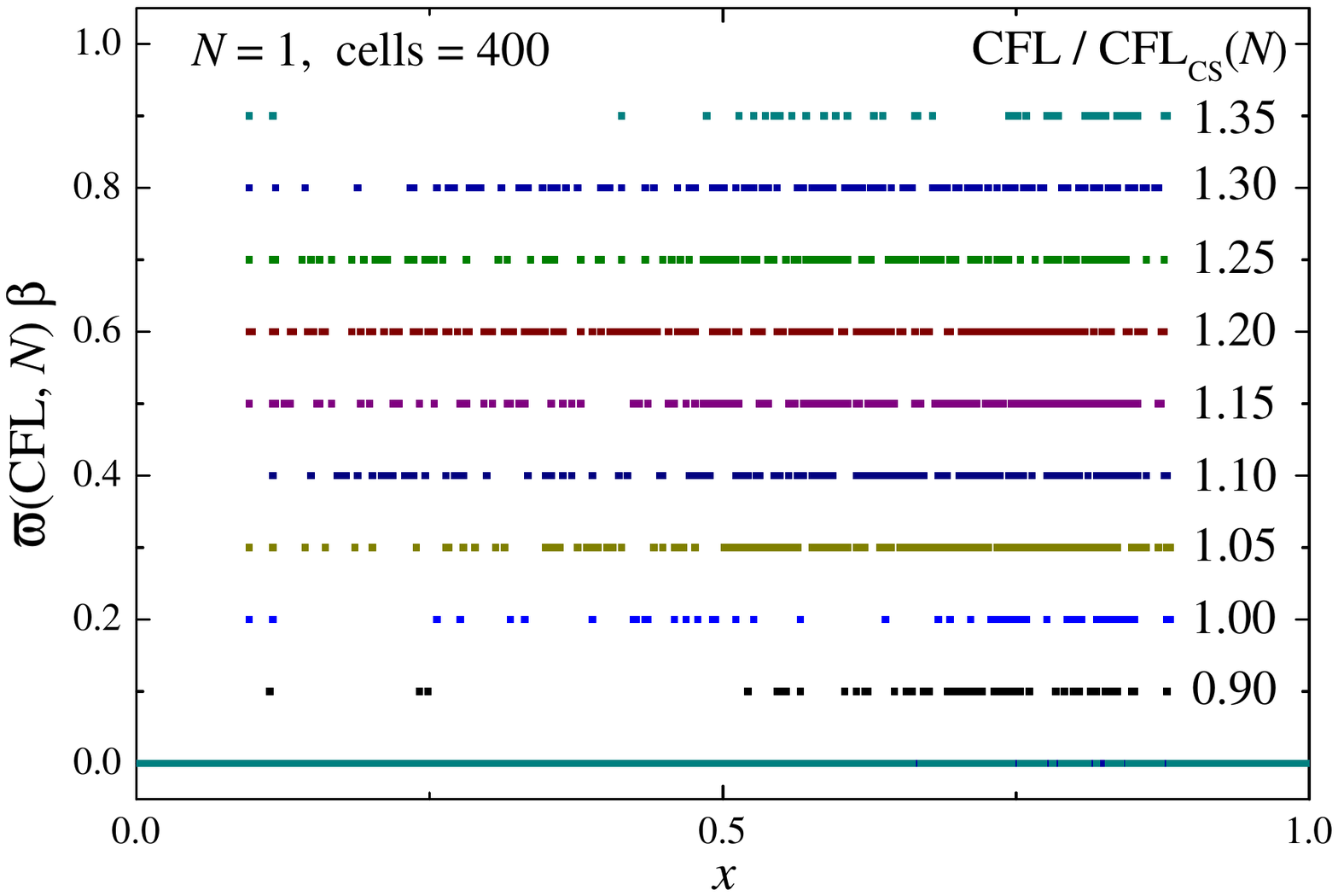}
\vspace{-8mm}\caption{\label{fig:test_zndf_1d_degrees_1_3:degree_1_beta}}
\end{subfigure}\\[-2mm]
\begin{subfigure}{0.24\textwidth}
\includegraphics[width=\textwidth]{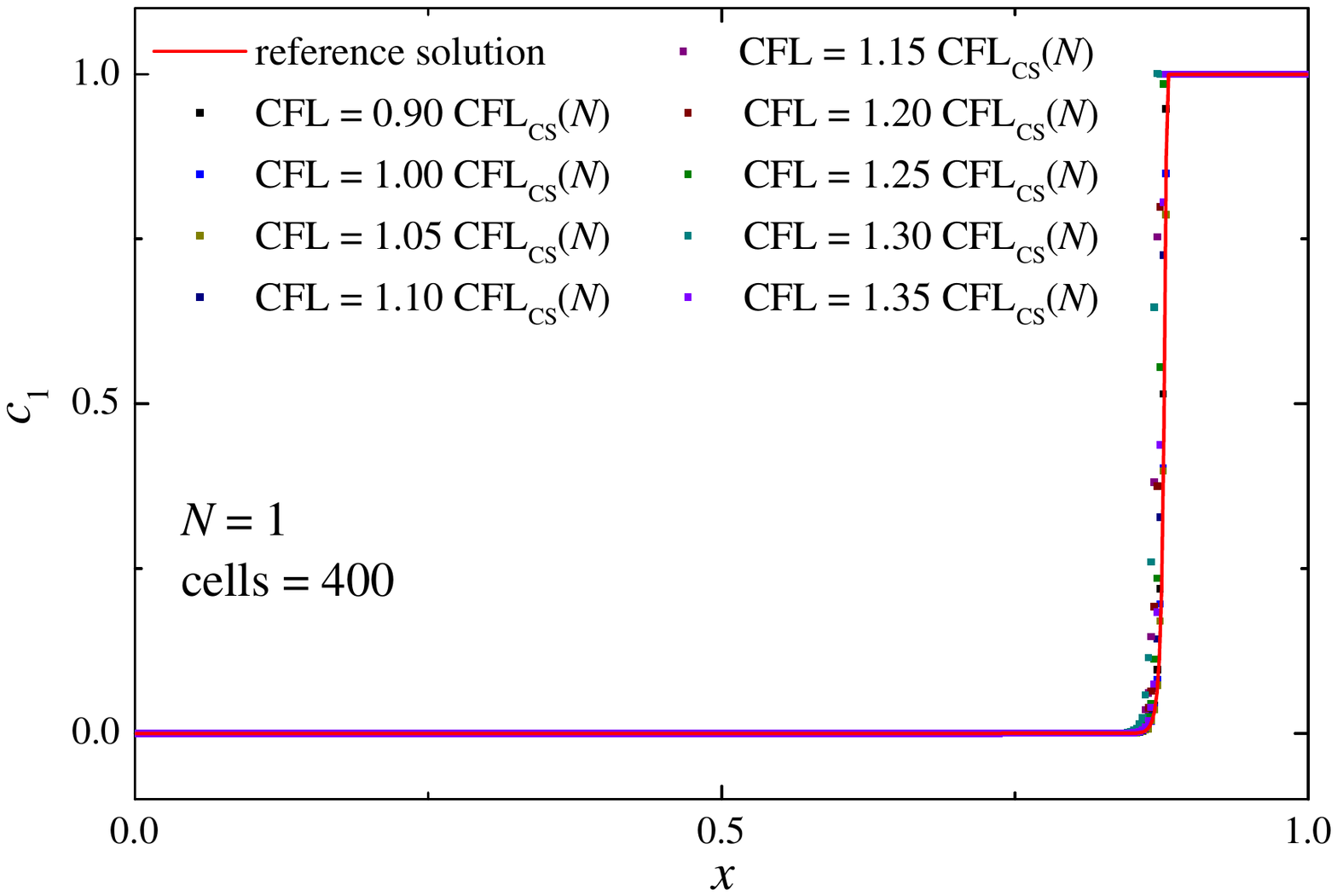}
\vspace{-8mm}\caption{\label{fig:test_zndf_1d_degrees_1_3:degree_1_c1}}
\end{subfigure}
\begin{subfigure}{0.24\textwidth}
\includegraphics[width=\textwidth]{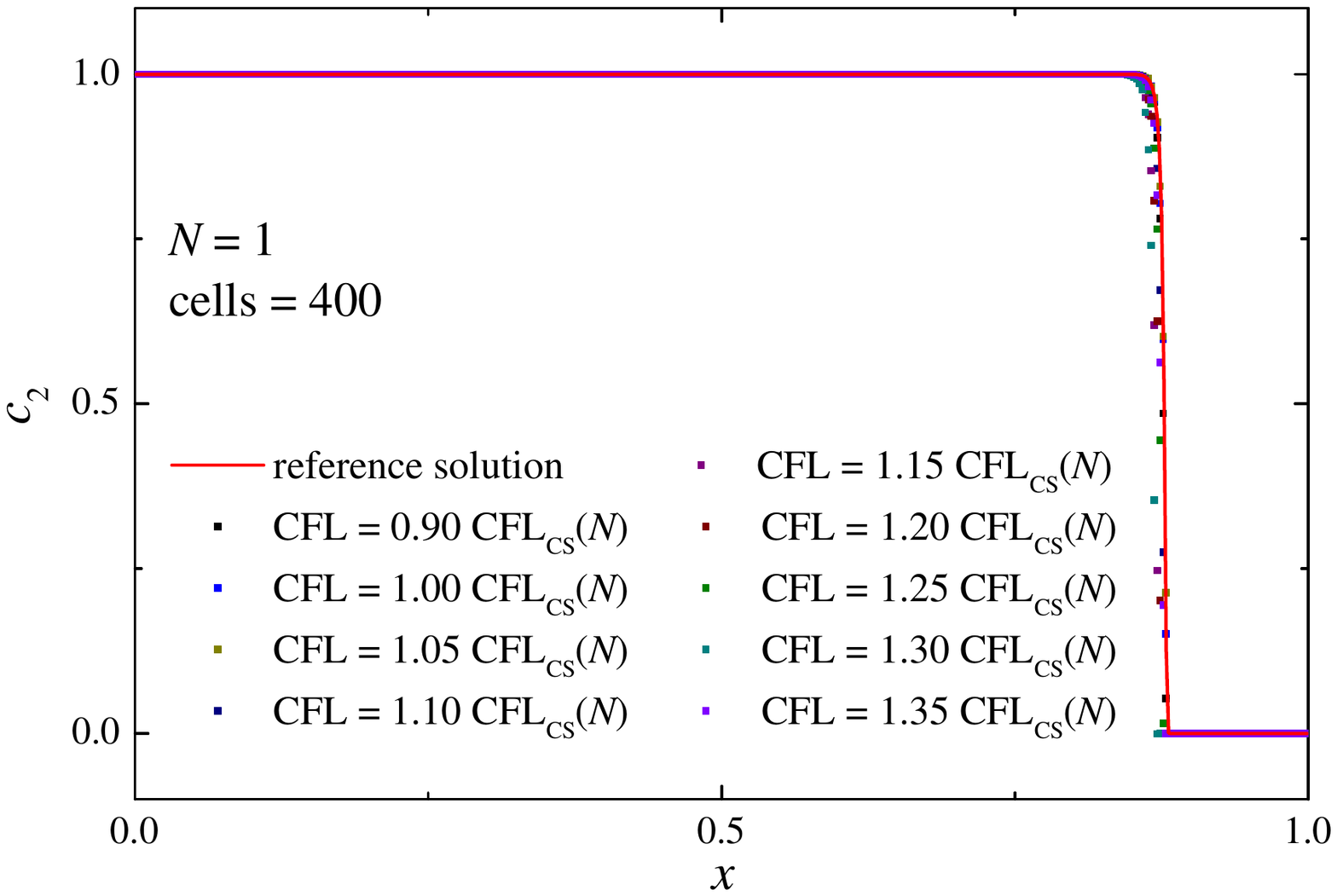}
\vspace{-8mm}\caption{\label{fig:test_zndf_1d_degrees_1_3:degree_1_c2}}
\end{subfigure}
\begin{subfigure}{0.24\textwidth}
\includegraphics[width=\textwidth]{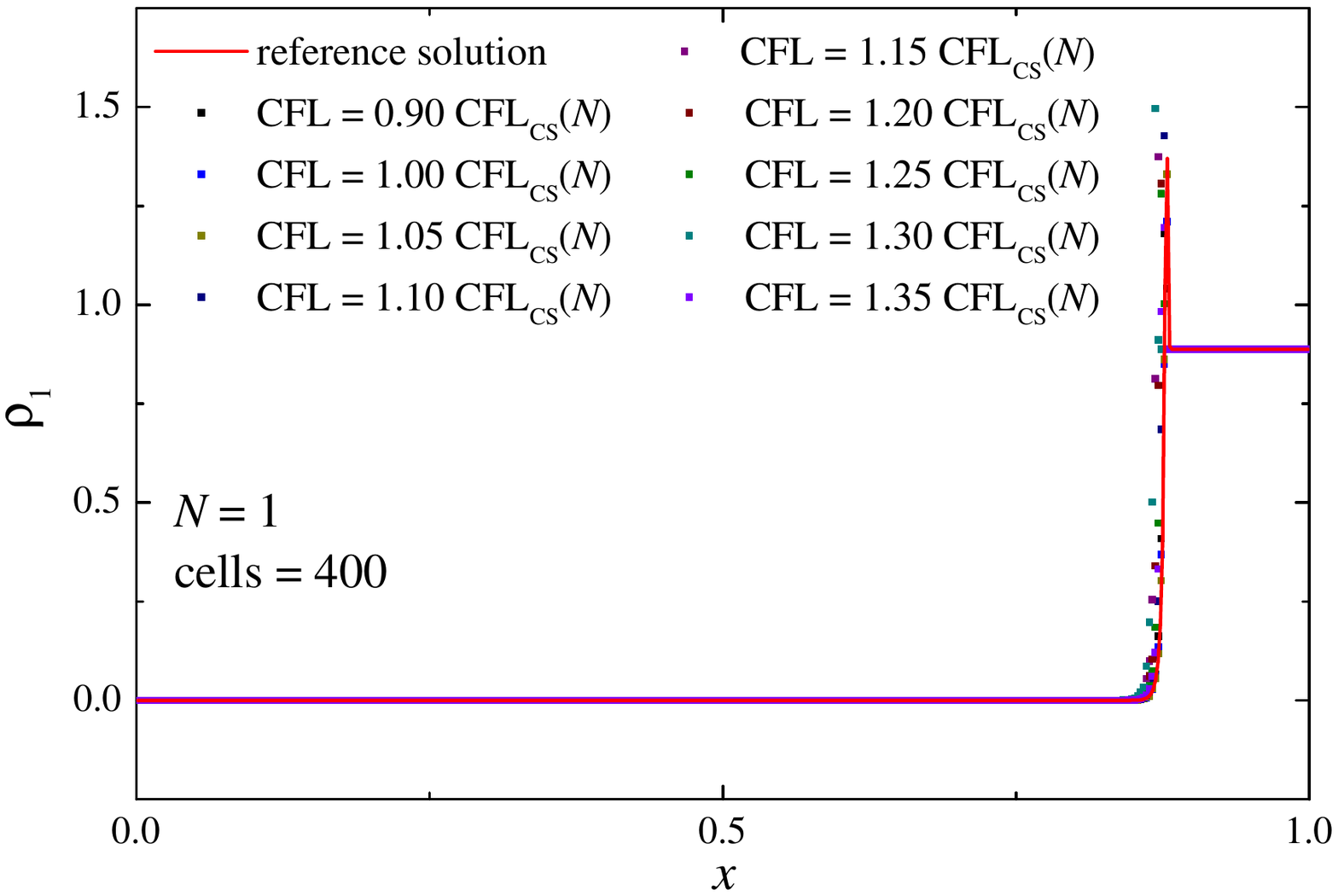}
\vspace{-8mm}\caption{\label{fig:test_zndf_1d_degrees_1_3:degree_1_den1}}
\end{subfigure}
\begin{subfigure}{0.24\textwidth}
\includegraphics[width=\textwidth]{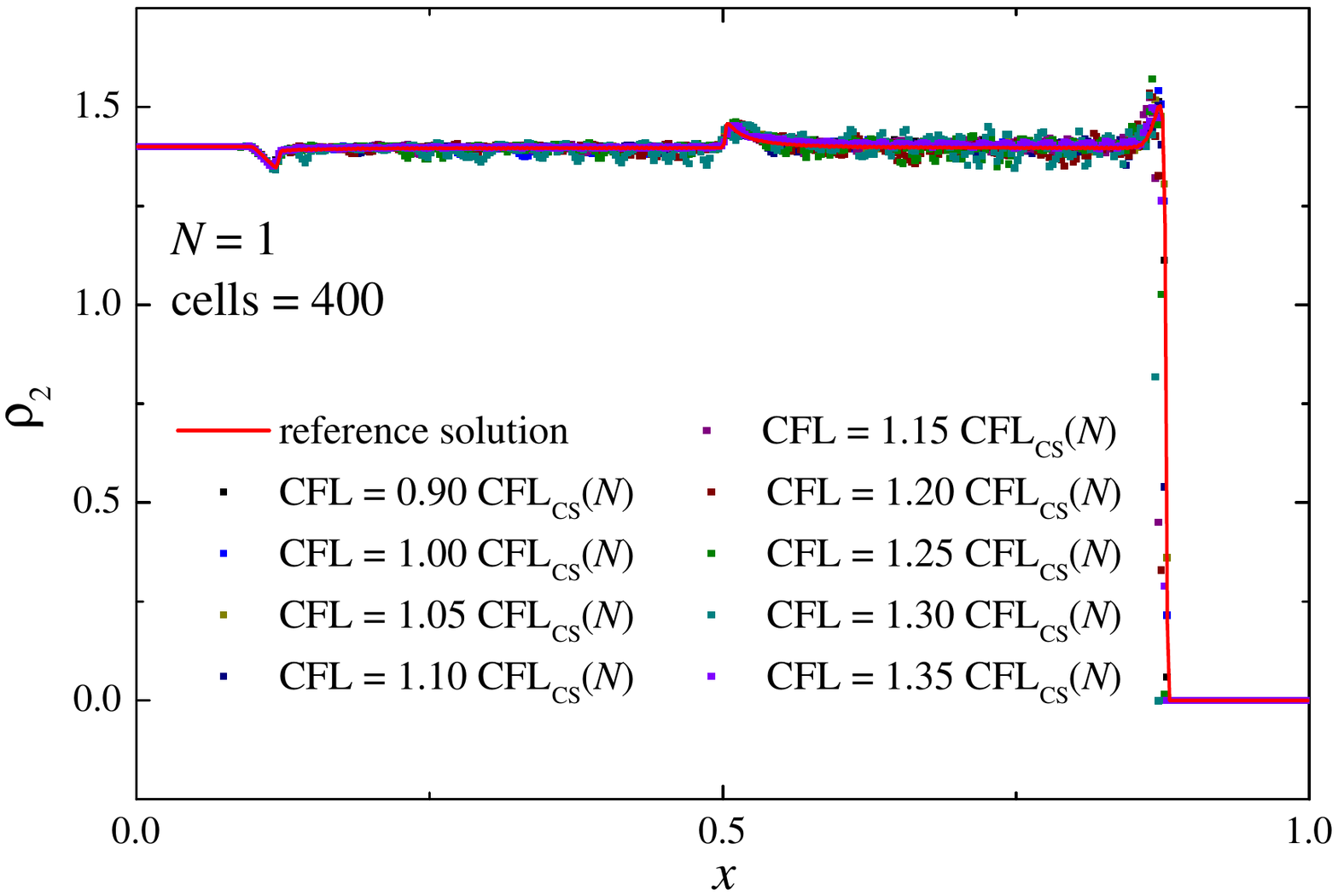}
\vspace{-8mm}\caption{\label{fig:test_zndf_1d_degrees_1_3:degree_1_den2}}
\end{subfigure}\\[-2mm]
\begin{subfigure}{0.24\textwidth}
\includegraphics[width=\textwidth]{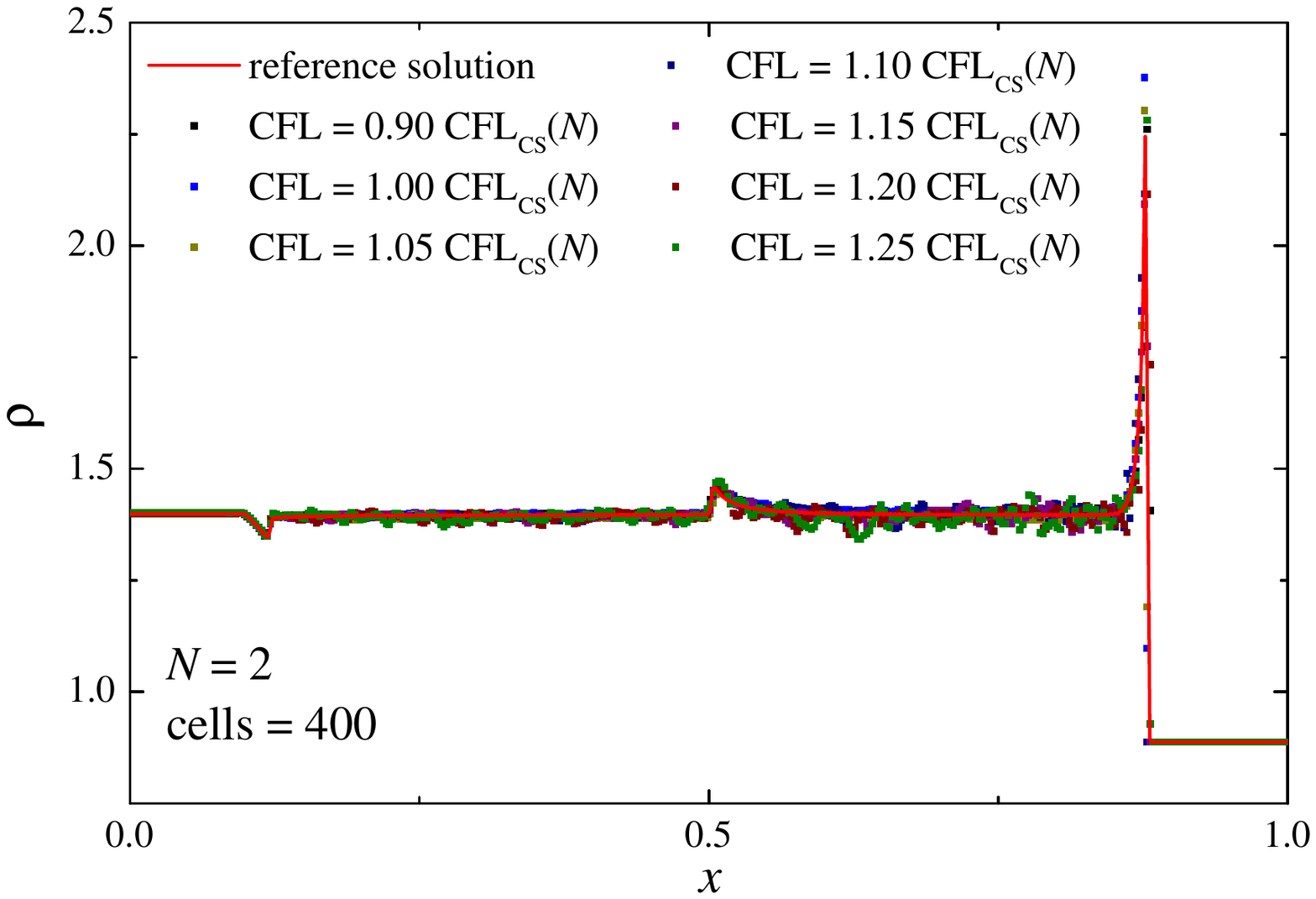}
\vspace{-8mm}\caption{\label{fig:test_zndf_1d_degrees_1_3:degree_2_den}}
\end{subfigure}
\begin{subfigure}{0.24\textwidth}
\includegraphics[width=\textwidth]{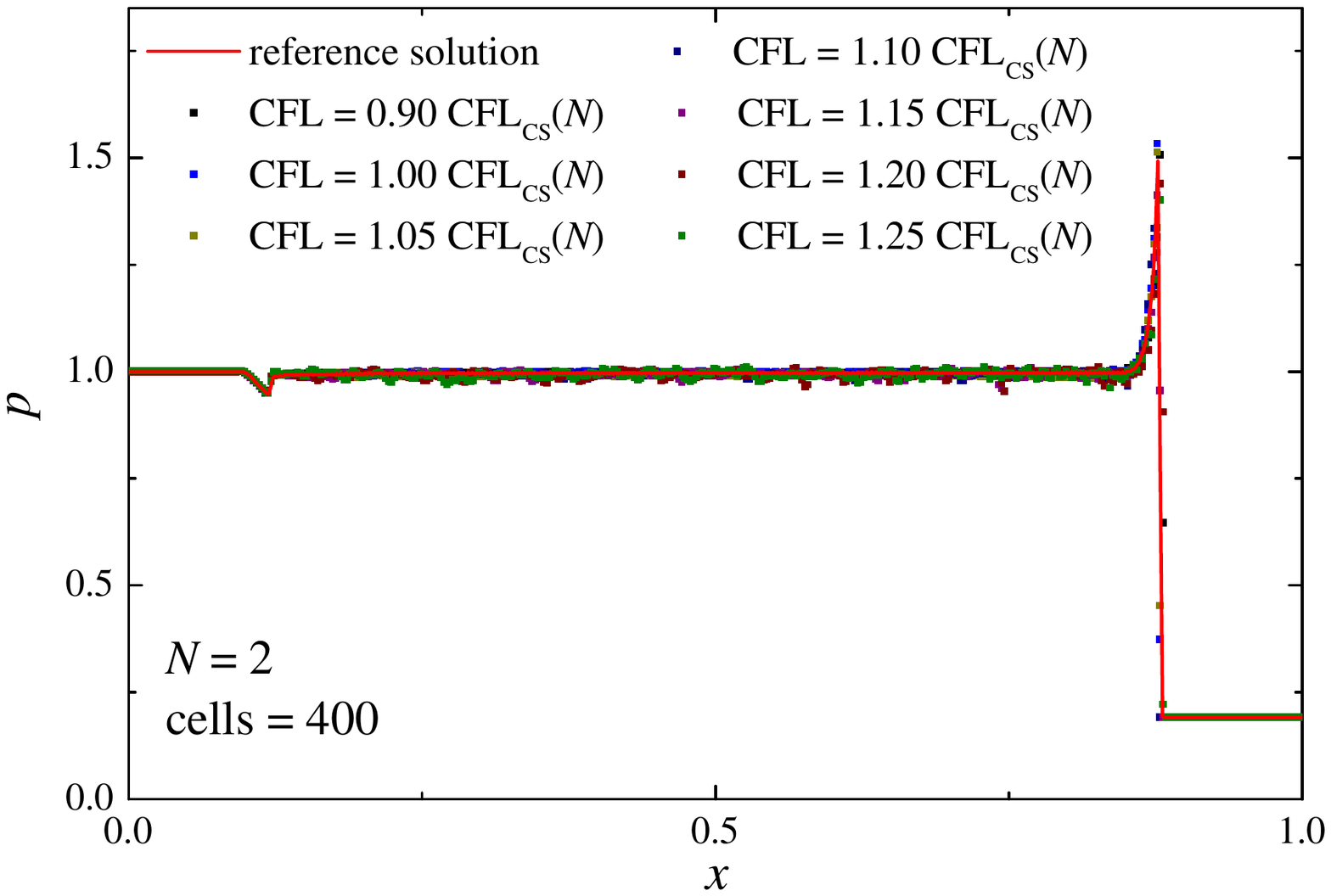}
\vspace{-8mm}\caption{\label{fig:test_zndf_1d_degrees_1_3:degree_2_p}}
\end{subfigure}
\begin{subfigure}{0.24\textwidth}
\includegraphics[width=\textwidth]{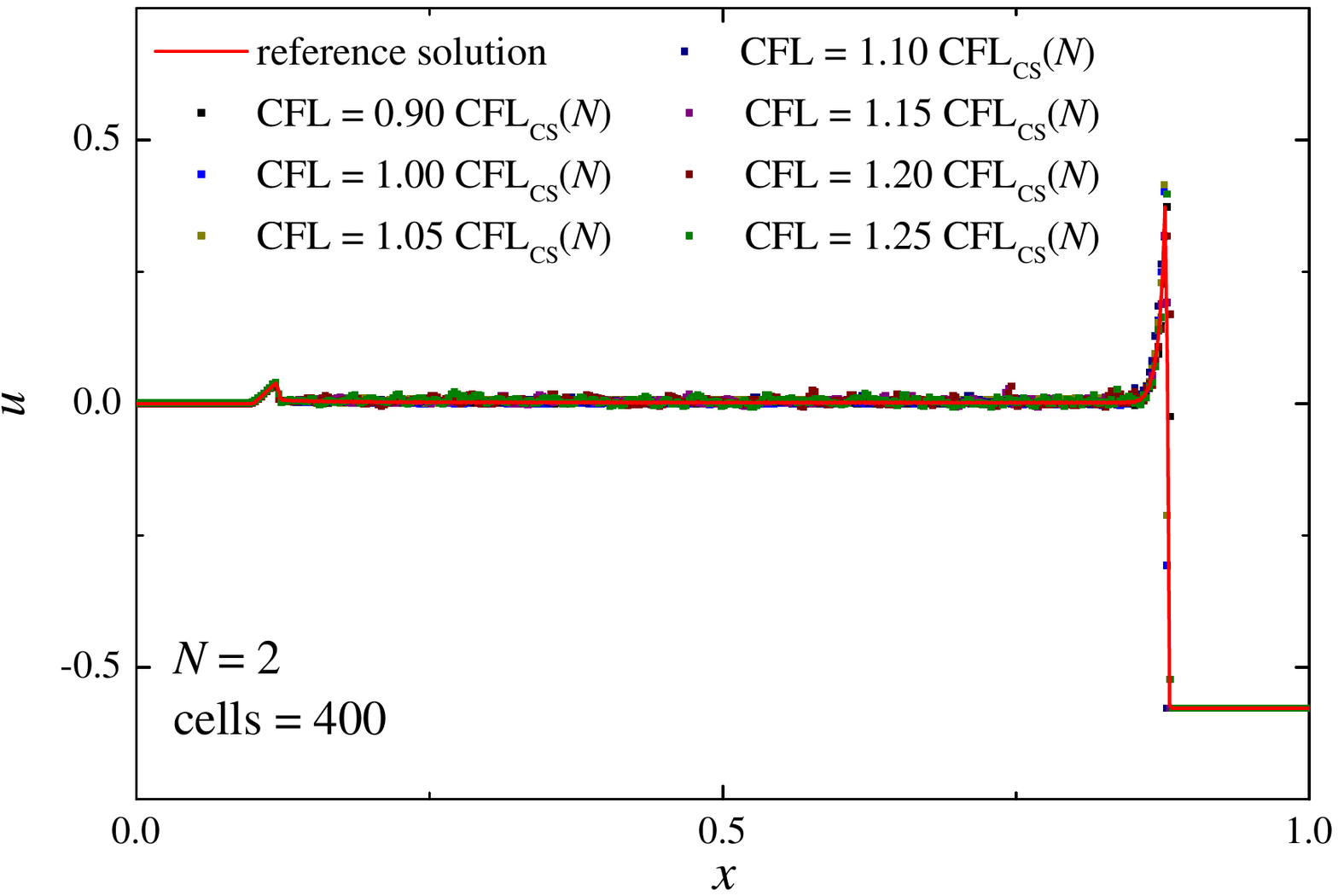}
\vspace{-8mm}\caption{\label{fig:test_zndf_1d_degrees_1_3:degree_2_u}}
\end{subfigure}
\begin{subfigure}{0.24\textwidth}
\includegraphics[width=\textwidth]{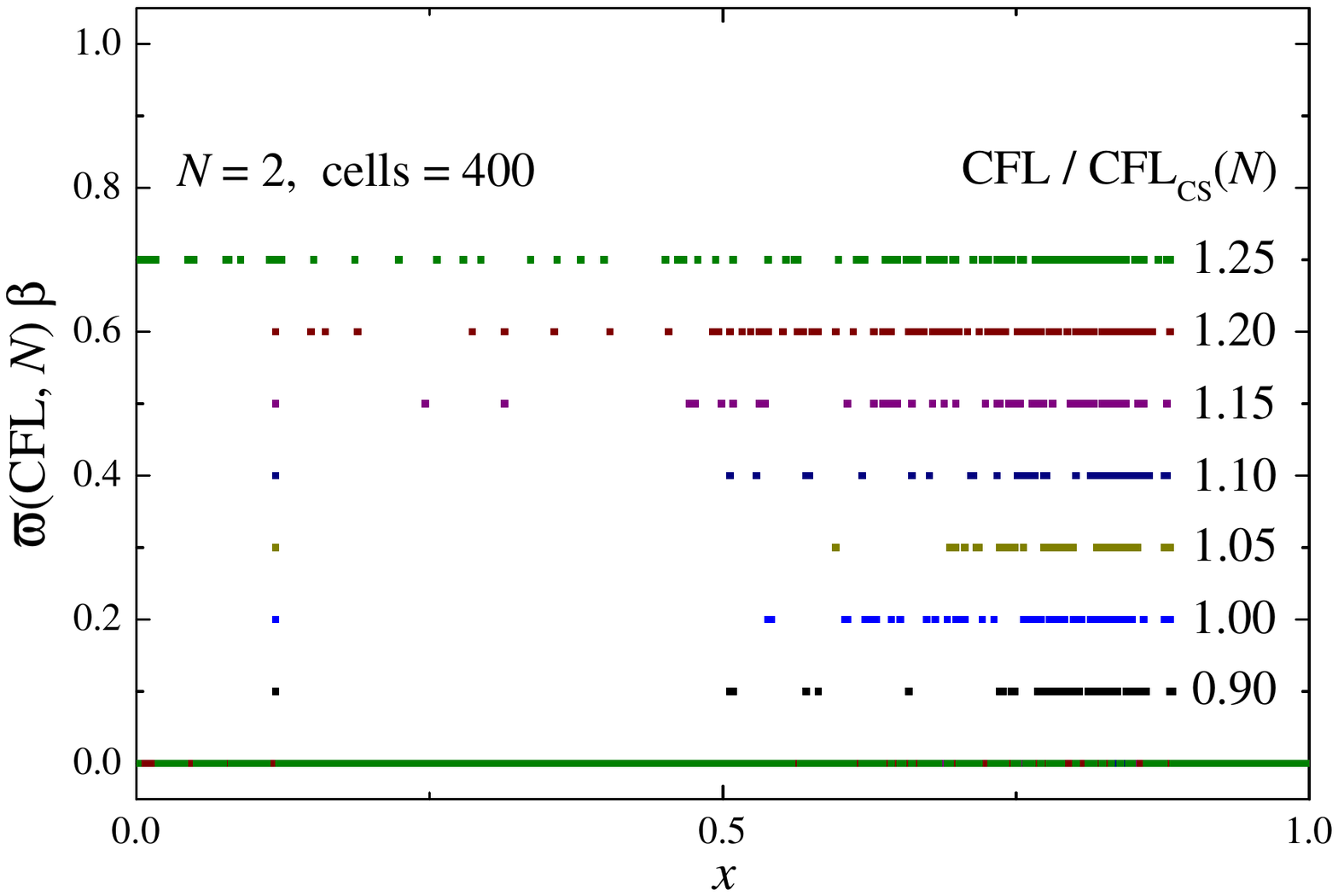}
\vspace{-8mm}\caption{\label{fig:test_zndf_1d_degrees_1_3:degree_2_beta}}
\end{subfigure}\\[-2mm]
\begin{subfigure}{0.24\textwidth}
\includegraphics[width=\textwidth]{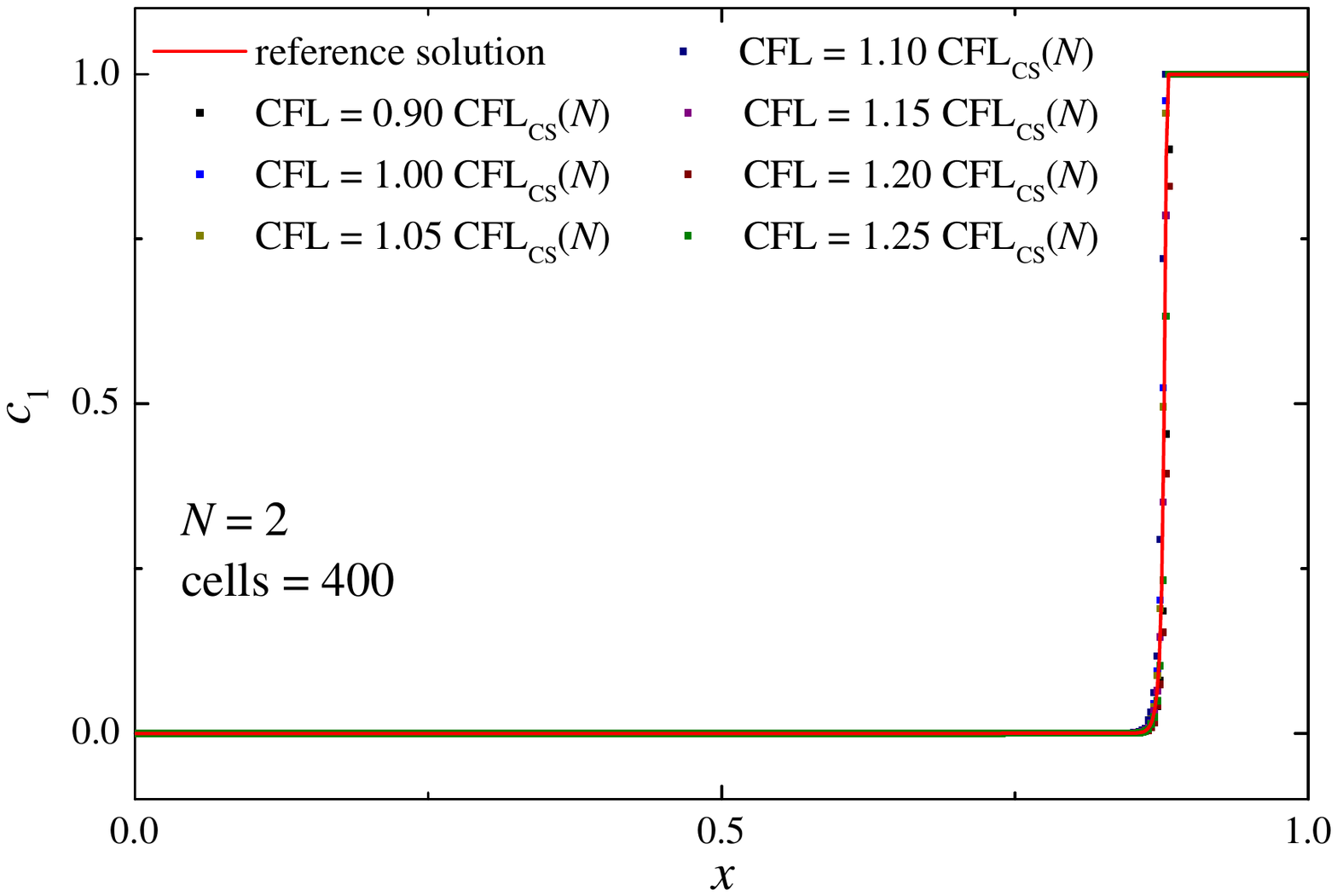}
\vspace{-8mm}\caption{\label{fig:test_zndf_1d_degrees_1_3:degree_2_c1}}
\end{subfigure}
\begin{subfigure}{0.24\textwidth}
\includegraphics[width=\textwidth]{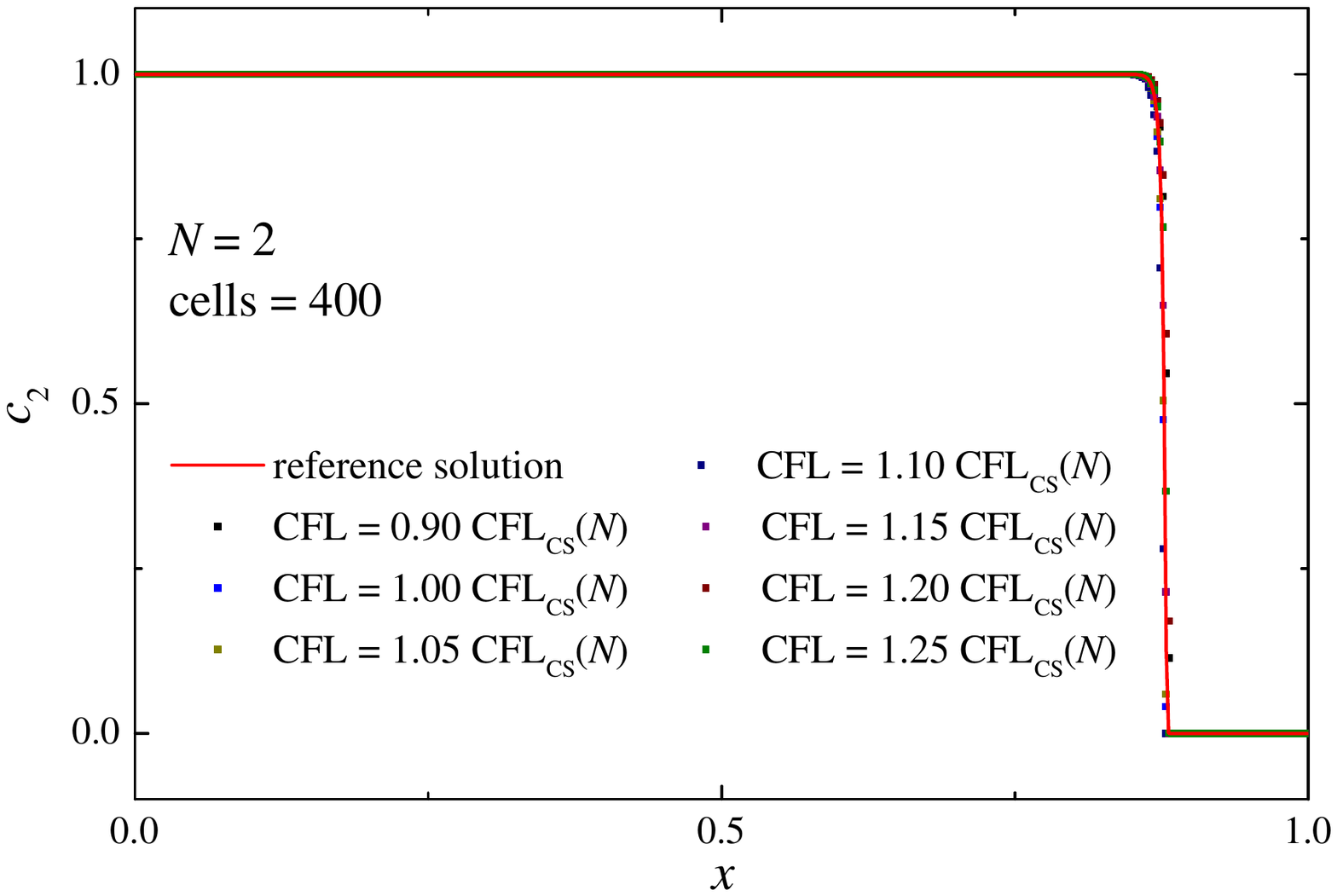}
\vspace{-8mm}\caption{\label{fig:test_zndf_1d_degrees_1_3:degree_2_c2}}
\end{subfigure}
\begin{subfigure}{0.24\textwidth}
\includegraphics[width=\textwidth]{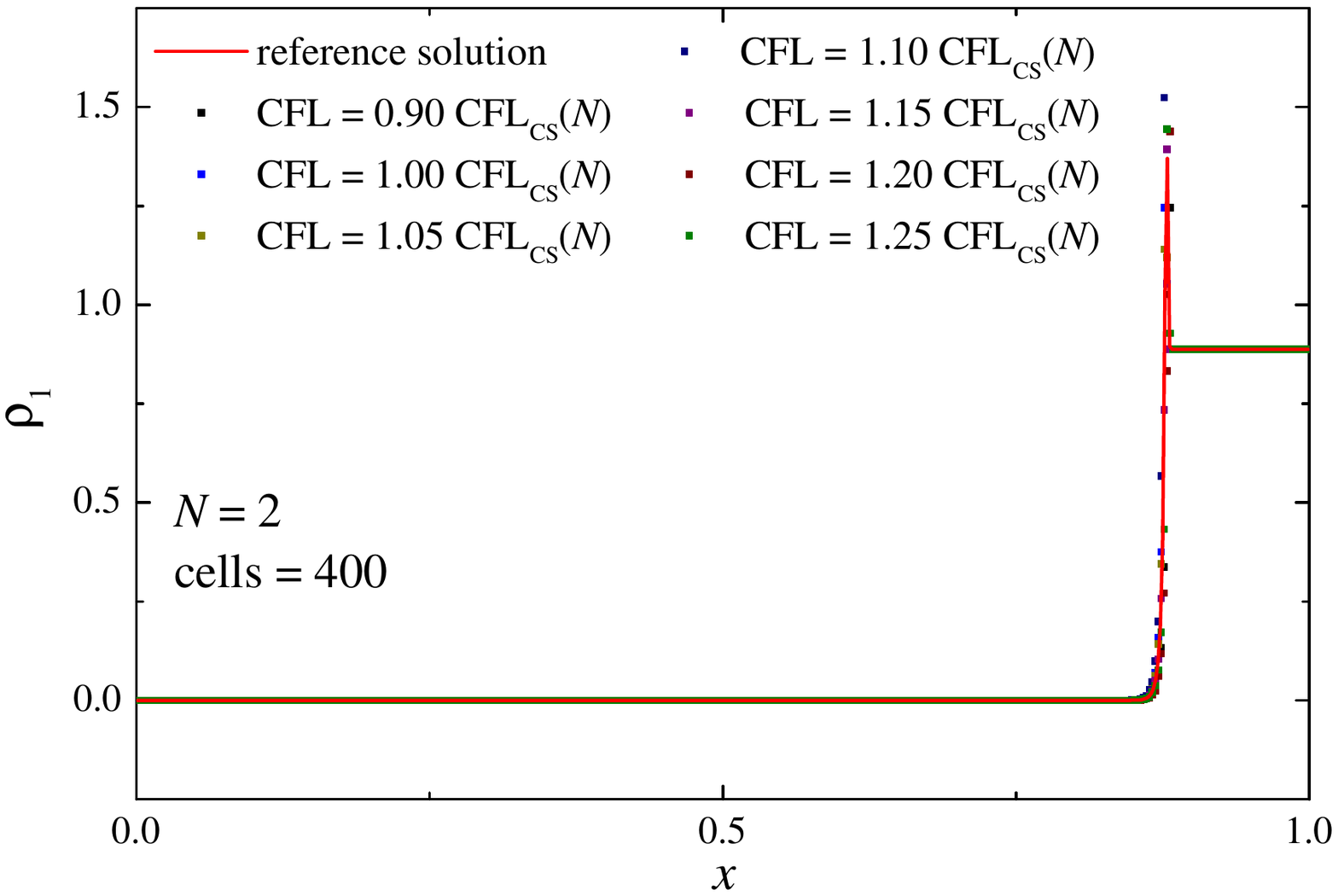}
\vspace{-8mm}\caption{\label{fig:test_zndf_1d_degrees_1_3:degree_2_den1}}
\end{subfigure}
\begin{subfigure}{0.24\textwidth}
\includegraphics[width=\textwidth]{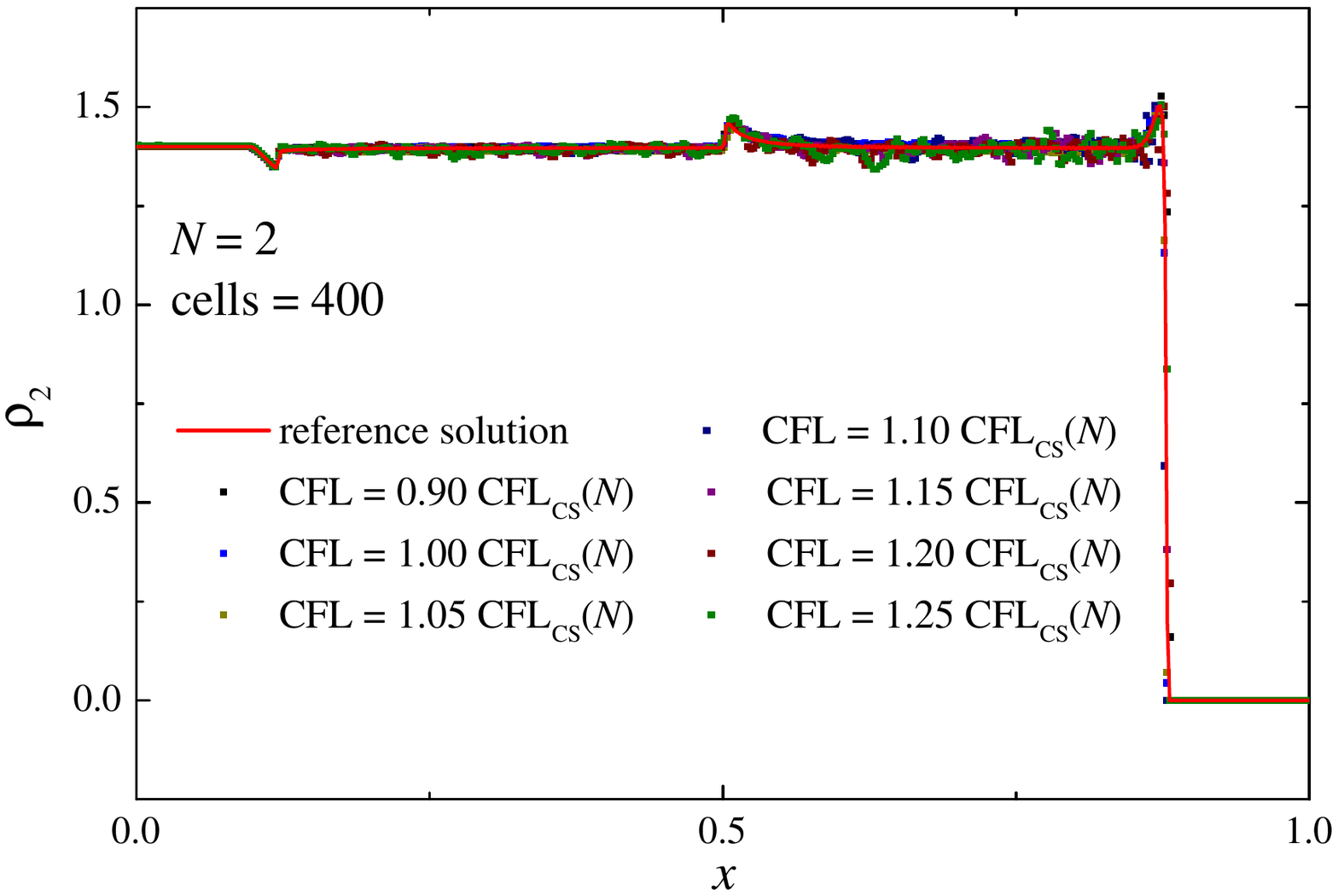}
\vspace{-8mm}\caption{\label{fig:test_zndf_1d_degrees_1_3:degree_2_den2}}
\end{subfigure}\\[-2mm]
\begin{subfigure}{0.24\textwidth}
\includegraphics[width=\textwidth]{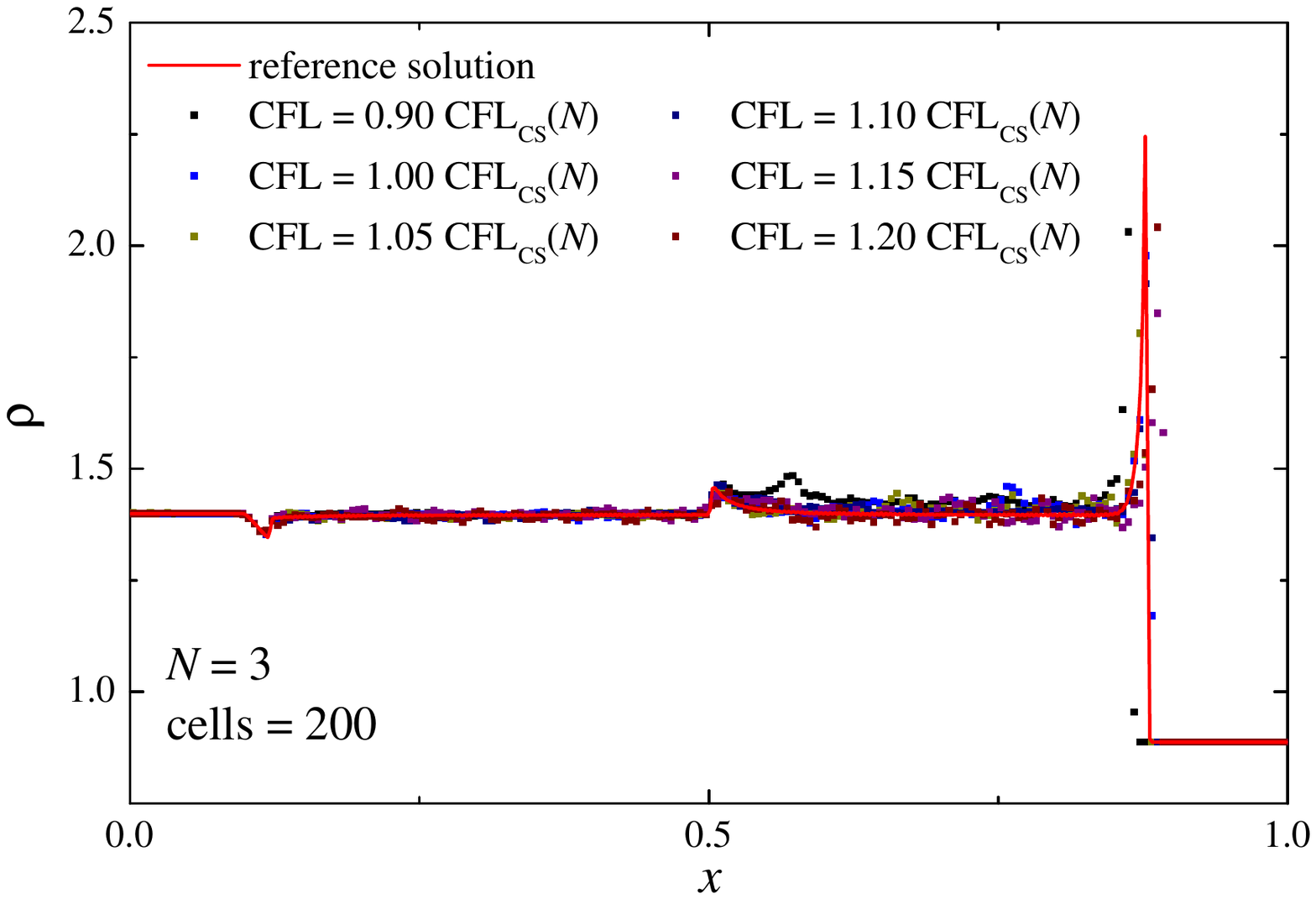}
\vspace{-8mm}\caption{\label{fig:test_zndf_1d_degrees_1_3:degree_3_den}}
\end{subfigure}
\begin{subfigure}{0.24\textwidth}
\includegraphics[width=\textwidth]{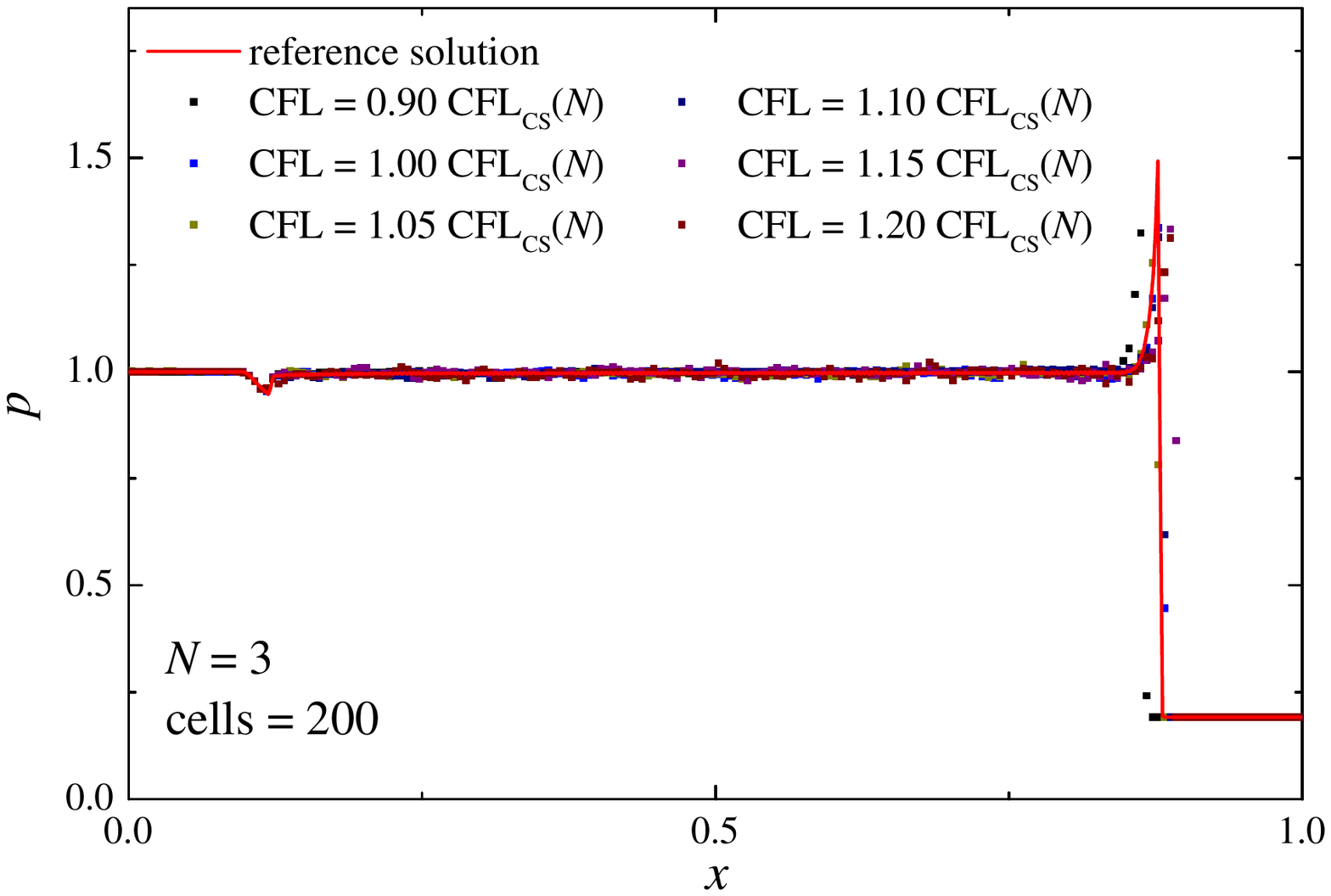}
\vspace{-8mm}\caption{\label{fig:test_zndf_1d_degrees_1_3:degree_3_p}}
\end{subfigure}
\begin{subfigure}{0.24\textwidth}
\includegraphics[width=\textwidth]{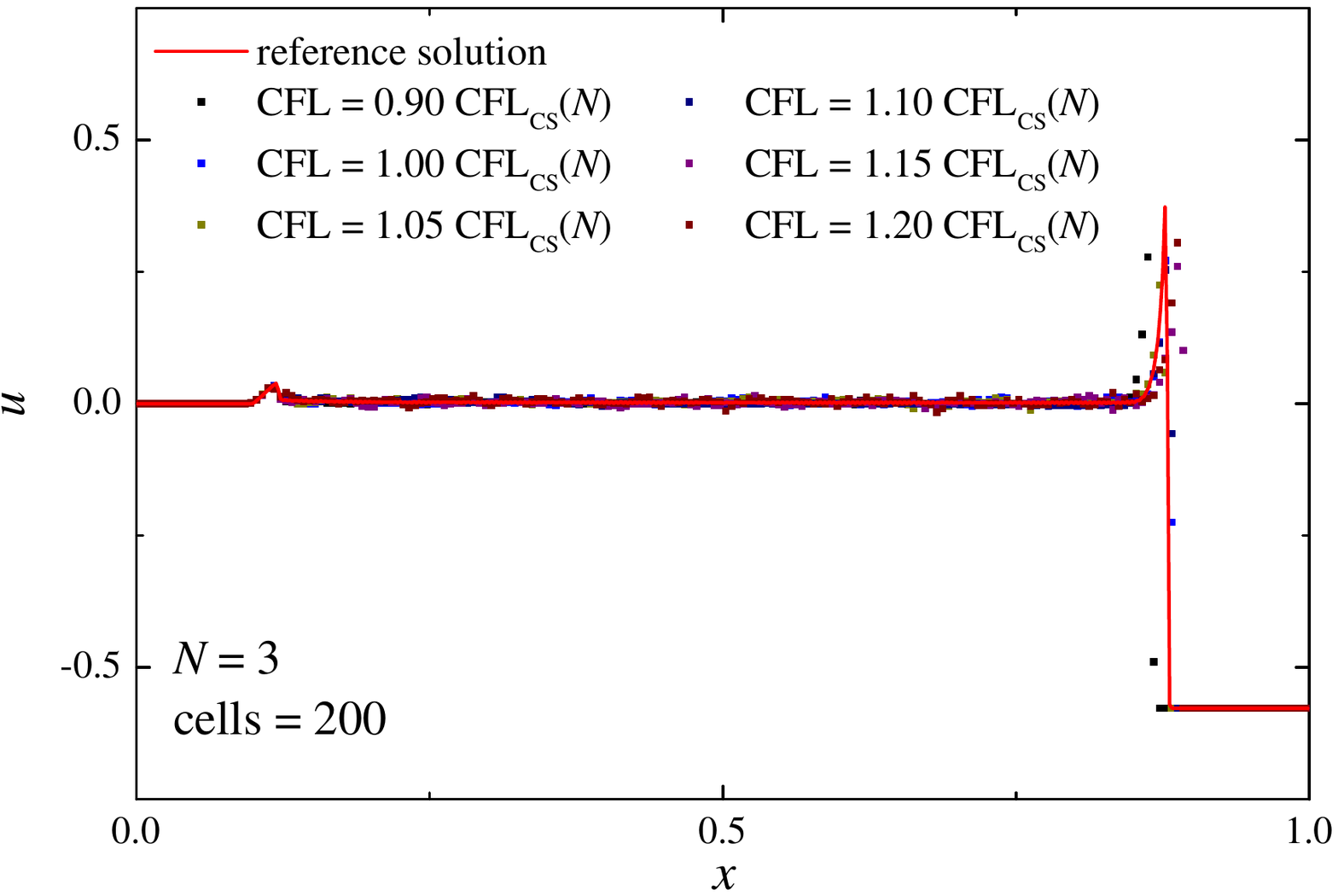}
\vspace{-8mm}\caption{\label{fig:test_zndf_1d_degrees_1_3:degree_3_u}}
\end{subfigure}
\begin{subfigure}{0.24\textwidth}
\includegraphics[width=\textwidth]{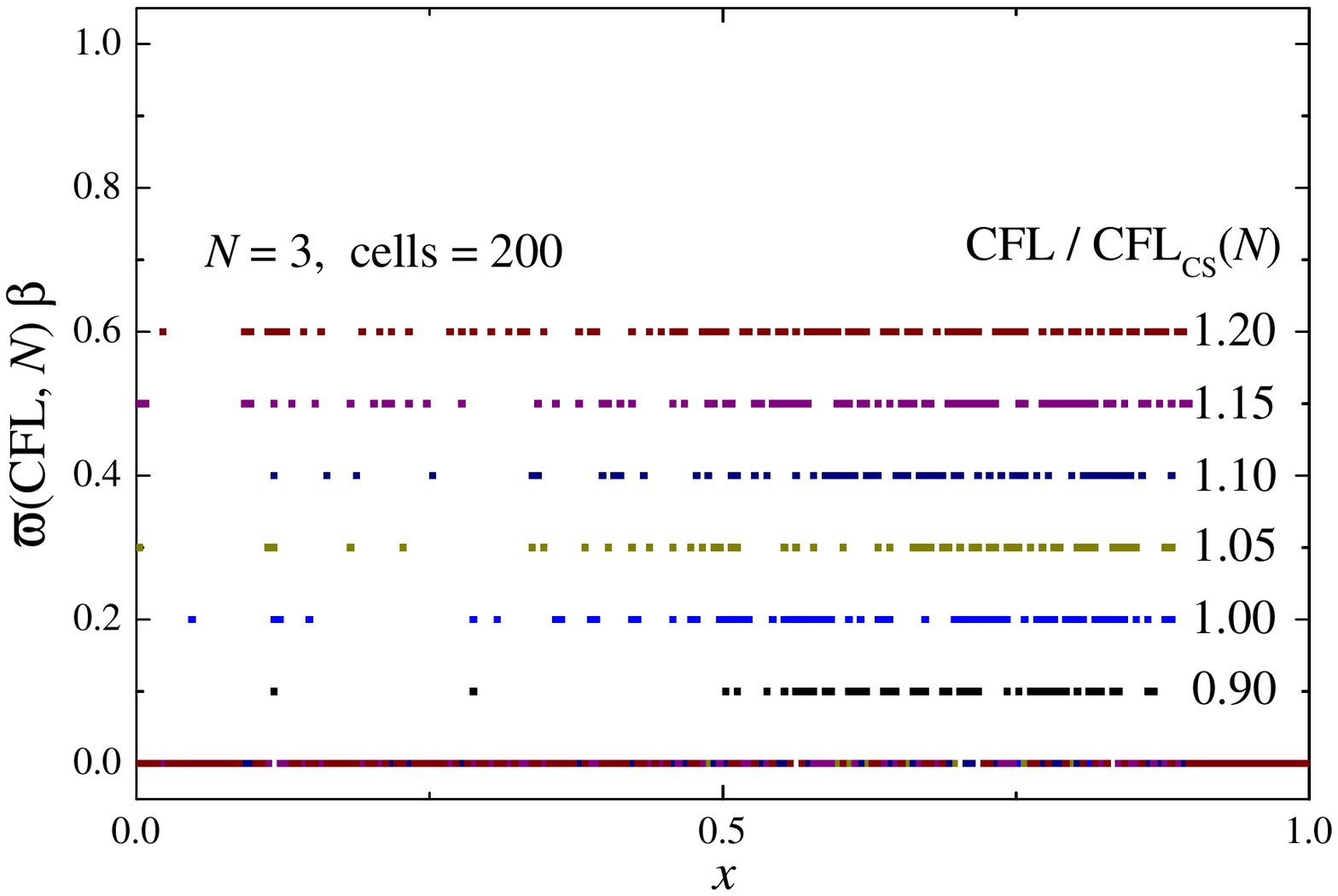}
\vspace{-8mm}\caption{\label{fig:test_zndf_1d_degrees_1_3:degree_3_beta}}
\end{subfigure}\\[-2mm]
\begin{subfigure}{0.24\textwidth}
\includegraphics[width=\textwidth]{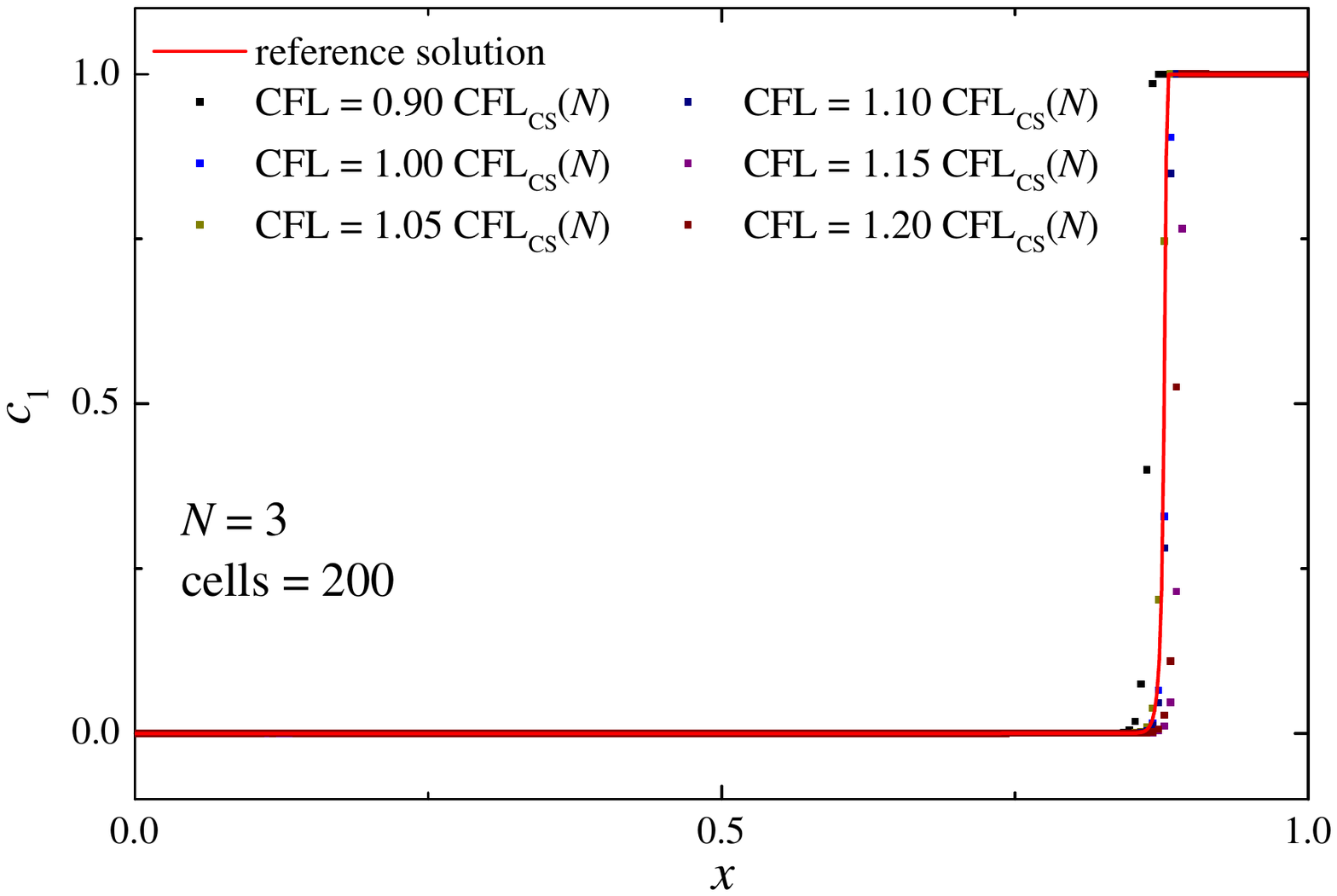}
\vspace{-8mm}\caption{\label{fig:test_zndf_1d_degrees_1_3:degree_3_c1}}
\end{subfigure}
\begin{subfigure}{0.24\textwidth}
\includegraphics[width=\textwidth]{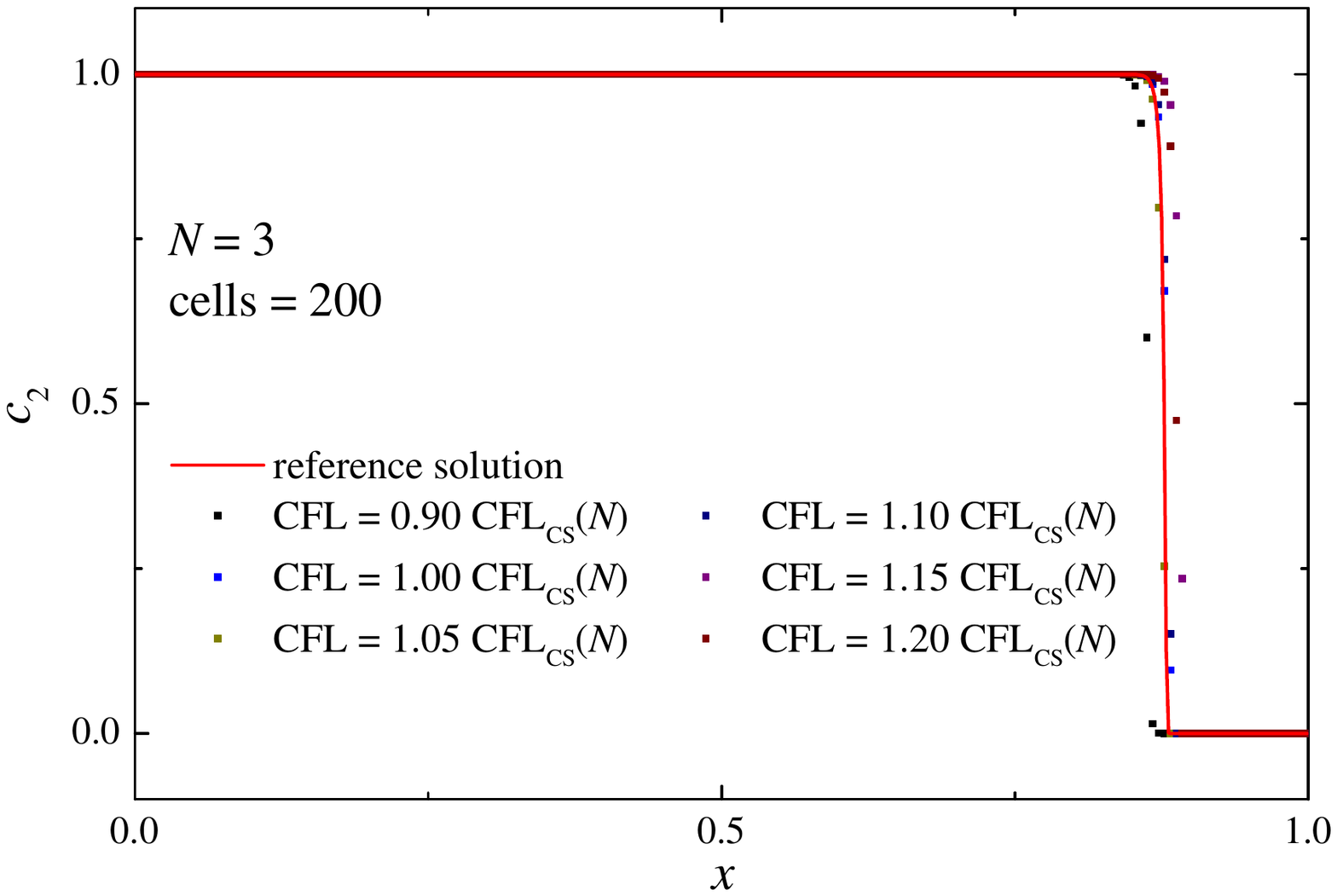}
\vspace{-8mm}\caption{\label{fig:test_zndf_1d_degrees_1_3:degree_3_c2}}
\end{subfigure}
\begin{subfigure}{0.24\textwidth}
\includegraphics[width=\textwidth]{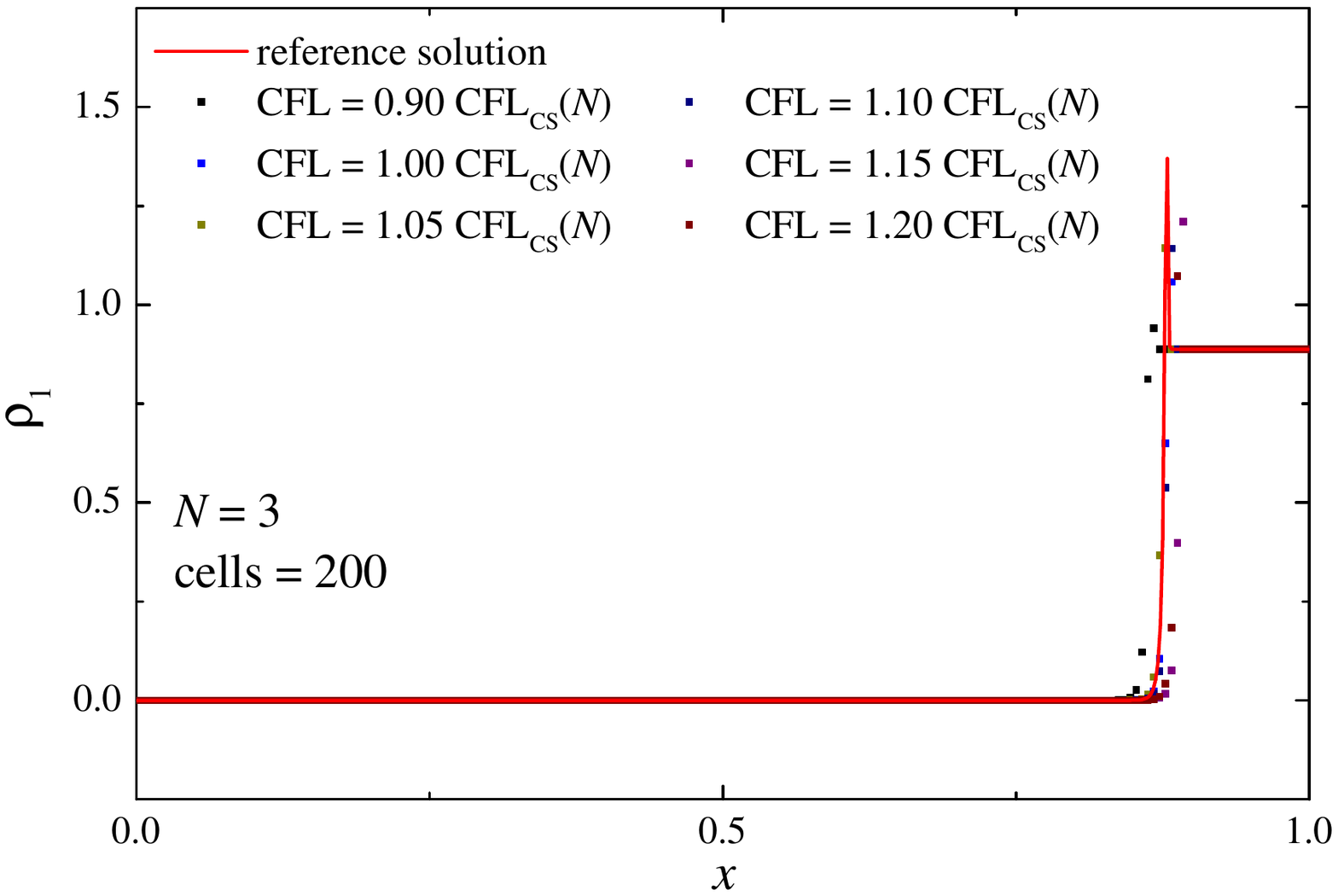}
\vspace{-8mm}\caption{\label{fig:test_zndf_1d_degrees_1_3:degree_3_den1}}
\end{subfigure}
\begin{subfigure}{0.24\textwidth}
\includegraphics[width=\textwidth]{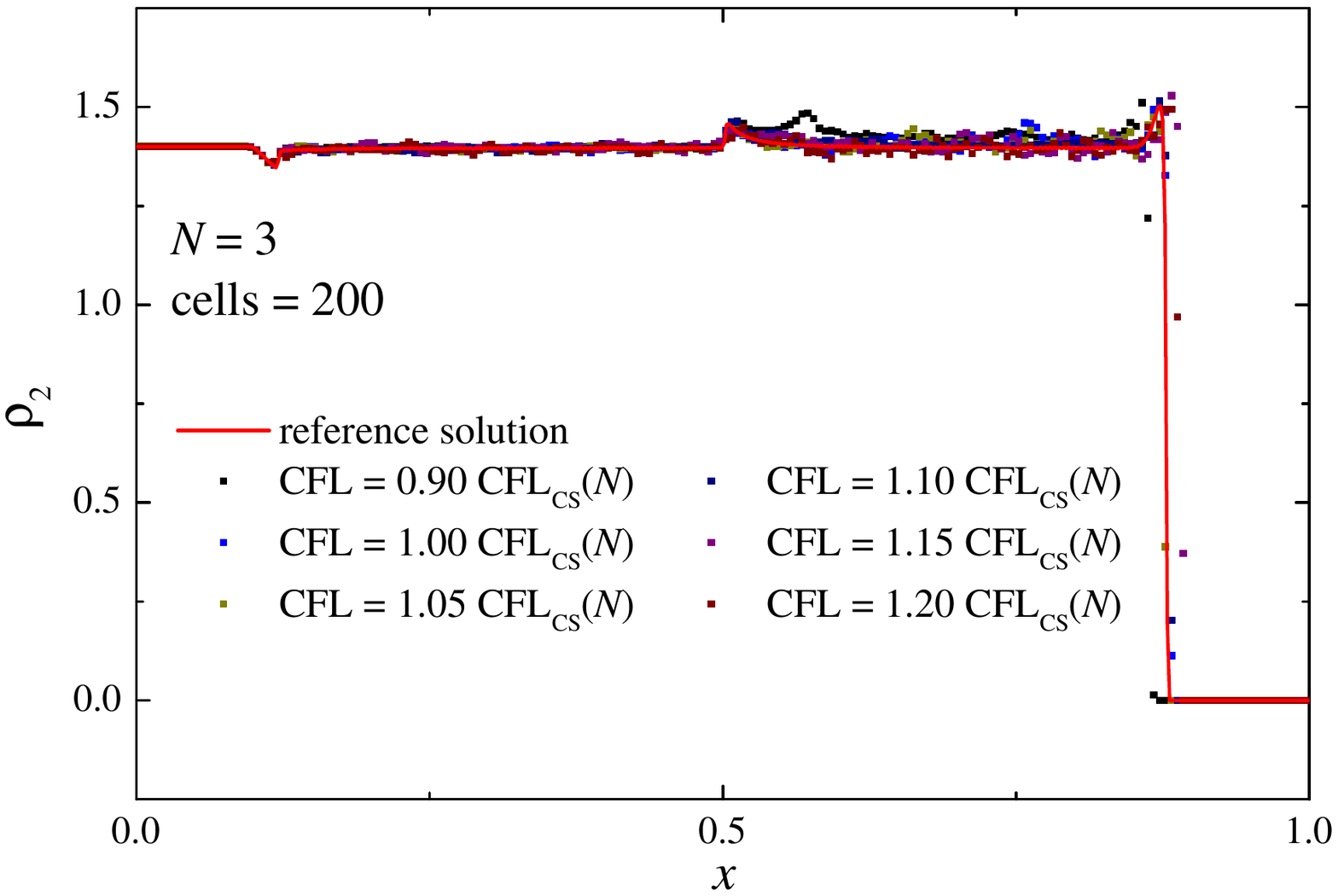}
\vspace{-8mm}\caption{\label{fig:test_zndf_1d_degrees_1_3:degree_3_den2}}
\end{subfigure}\\[-2mm]
\caption{%
Numerical solution of the one-dimensional problem of the plane detonation wave formation in a two-component medium with a ``fast'' reaction (strong stiff case, a detailed statement of the problem is presented in the text), obtained using the ADER-DG method for the basis polynomials degree $N = 1$ ((\subref{fig:test_zndf_1d_degrees_1_3:degree_1_den}), (\subref{fig:test_zndf_1d_degrees_1_3:degree_1_p}), (\subref{fig:test_zndf_1d_degrees_1_3:degree_1_u}), (\subref{fig:test_zndf_1d_degrees_1_3:degree_1_c1}), (\subref{fig:test_zndf_1d_degrees_1_3:degree_1_c2}), (\subref{fig:test_zndf_1d_degrees_1_3:degree_1_den1}), (\subref{fig:test_zndf_1d_degrees_1_3:degree_1_den2})), $N = 2$ ((\subref{fig:test_zndf_1d_degrees_1_3:degree_2_den}), (\subref{fig:test_zndf_1d_degrees_1_3:degree_2_p}), (\subref{fig:test_zndf_1d_degrees_1_3:degree_2_u}), (\subref{fig:test_zndf_1d_degrees_1_3:degree_2_c1}), (\subref{fig:test_zndf_1d_degrees_1_3:degree_2_c2}), (\subref{fig:test_zndf_1d_degrees_1_3:degree_2_den1}), (\subref{fig:test_zndf_1d_degrees_1_3:degree_2_den2})), $N = 3$ ((\subref{fig:test_zndf_1d_degrees_1_3:degree_3_den}), (\subref{fig:test_zndf_1d_degrees_1_3:degree_3_p}), (\subref{fig:test_zndf_1d_degrees_1_3:degree_3_u}), (\subref{fig:test_zndf_1d_degrees_1_3:degree_3_c1}), (\subref{fig:test_zndf_1d_degrees_1_3:degree_3_c2}), (\subref{fig:test_zndf_1d_degrees_1_3:degree_3_den1}), (\subref{fig:test_zndf_1d_degrees_1_3:degree_3_den2})) with a posteriori limitation of the solution by an ADER-WENO2 finite-volume limiter, at the final time $t_{f} = 0.4$. The graphs show the coordinate dependence of density $\rho$ ((\subref{fig:test_zndf_1d_degrees_1_3:degree_1_den}), (\subref{fig:test_zndf_1d_degrees_1_3:degree_2_den}), (\subref{fig:test_zndf_1d_degrees_1_3:degree_3_den})), pressure $p$ ((\subref{fig:test_zndf_1d_degrees_1_3:degree_1_p}), (\subref{fig:test_zndf_1d_degrees_1_3:degree_2_p}), (\subref{fig:test_zndf_1d_degrees_1_3:degree_3_p})), flow velocity $u$ ((\subref{fig:test_zndf_1d_degrees_1_3:degree_1_u}), (\subref{fig:test_zndf_1d_degrees_1_3:degree_2_u}), (\subref{fig:test_zndf_1d_degrees_1_3:degree_3_u})), troubled cells indicator $\beta$ multiplied by the separating function $\varpi(\mathrm{CFL}, N)$ ((\subref{fig:test_zndf_1d_degrees_1_3:degree_1_beta}), (\subref{fig:test_zndf_1d_degrees_1_3:degree_2_beta}), (\subref{fig:test_zndf_1d_degrees_1_3:degree_3_beta})), mass concentrations $c_{1}$ ((\subref{fig:test_zndf_1d_degrees_1_3:degree_1_c1}), (\subref{fig:test_zndf_1d_degrees_1_3:degree_2_c1}), (\subref{fig:test_zndf_1d_degrees_1_3:degree_3_c1})) and $c_{2}$ ((\subref{fig:test_zndf_1d_degrees_1_3:degree_1_c2}), (\subref{fig:test_zndf_1d_degrees_1_3:degree_2_c2}), (\subref{fig:test_zndf_1d_degrees_1_3:degree_3_c2})) of the reactant $A$ and product $B$, respectively, and partial densities $\rho_{1}$ ((\subref{fig:test_zndf_1d_degrees_1_3:degree_1_den1}), (\subref{fig:test_zndf_1d_degrees_1_3:degree_2_den1}), (\subref{fig:test_zndf_1d_degrees_1_3:degree_3_den1})) and $\rho_{2}$ ((\subref{fig:test_zndf_1d_degrees_1_3:degree_1_den2}), (\subref{fig:test_zndf_1d_degrees_1_3:degree_2_den2}), (\subref{fig:test_zndf_1d_degrees_1_3:degree_3_den2})) of the reactant $A$ and product $B$, respectively. The square symbols represent the subcells finite-volume representation of the numerical solution; the red solid lines represents the reference solution of the problem.
}
\label{fig:test_zndf_1d_degrees_1_3}
\end{figure}
\begin{figure}[h!]
\captionsetup[subfigure]{%
	position=bottom,
	font+=smaller,
	textfont=normalfont,
	singlelinecheck=off,
	justification=raggedright
}
\centering
\begin{subfigure}{0.24\textwidth}
\includegraphics[width=\textwidth]{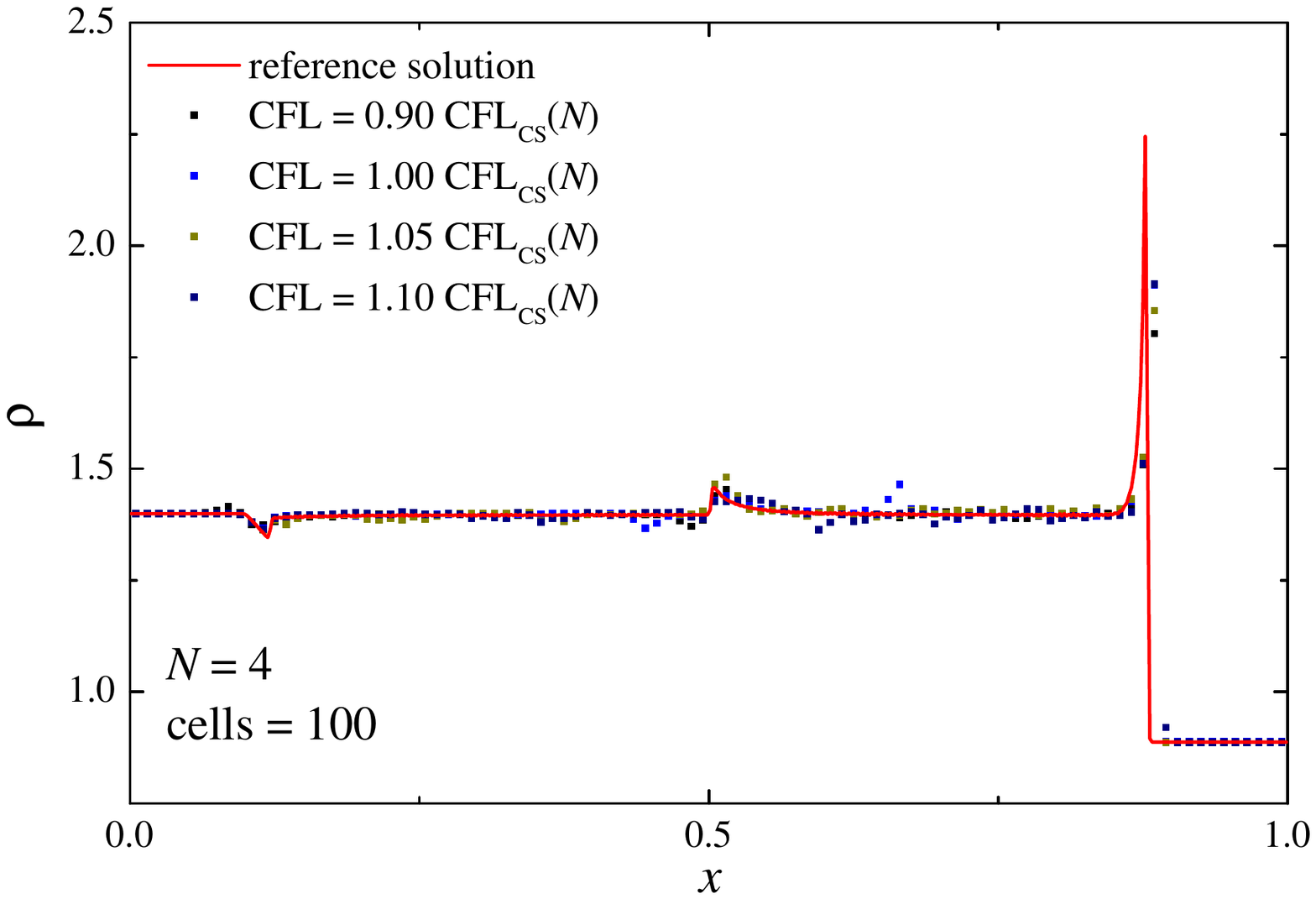}
\vspace{-8mm}\caption{\label{fig:test_zndf_1d_degrees_4_6:degree_4_den}}
\end{subfigure}
\begin{subfigure}{0.24\textwidth}
\includegraphics[width=\textwidth]{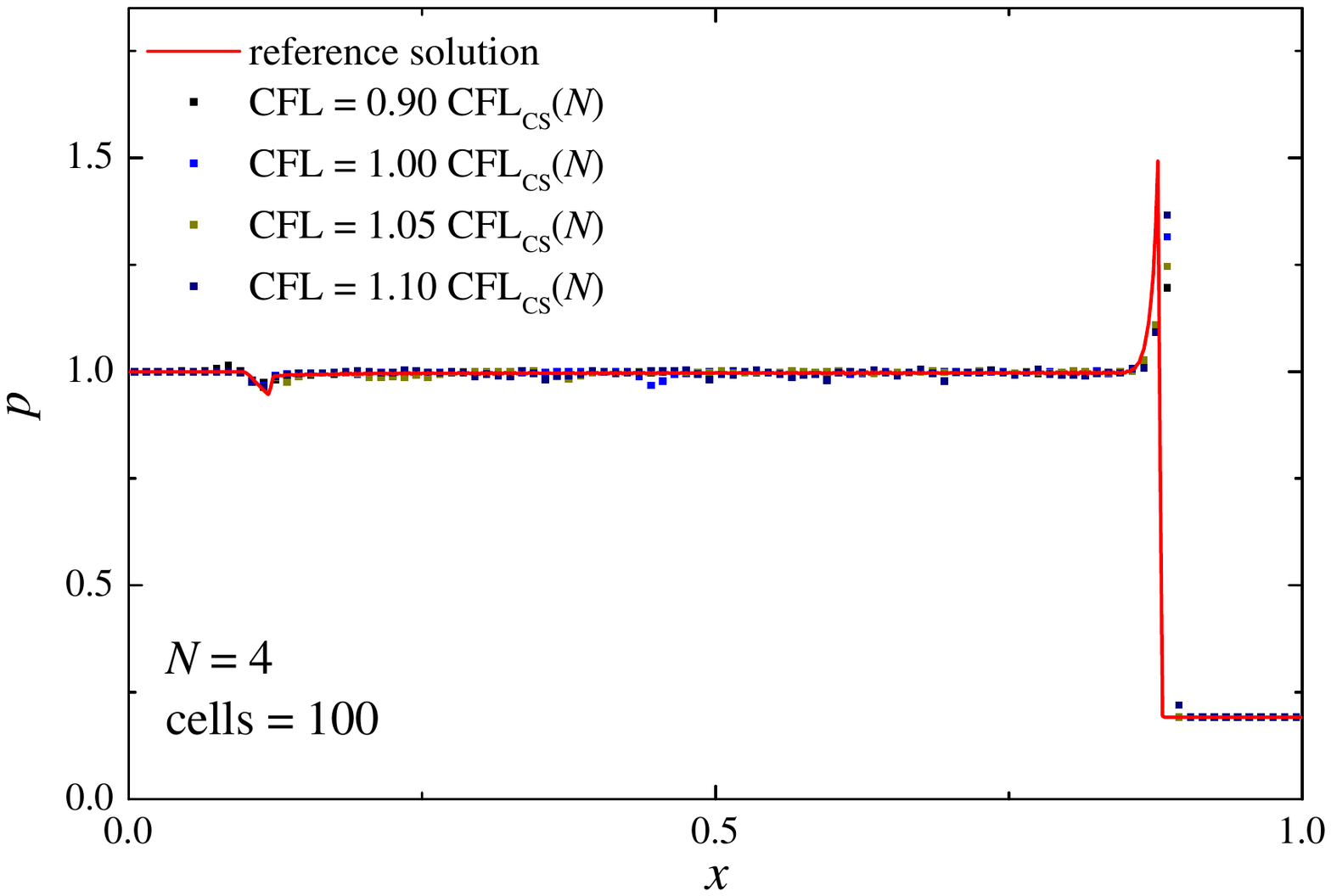}
\vspace{-8mm}\caption{\label{fig:test_zndf_1d_degrees_4_6:degree_4_p}}
\end{subfigure}
\begin{subfigure}{0.24\textwidth}
\includegraphics[width=\textwidth]{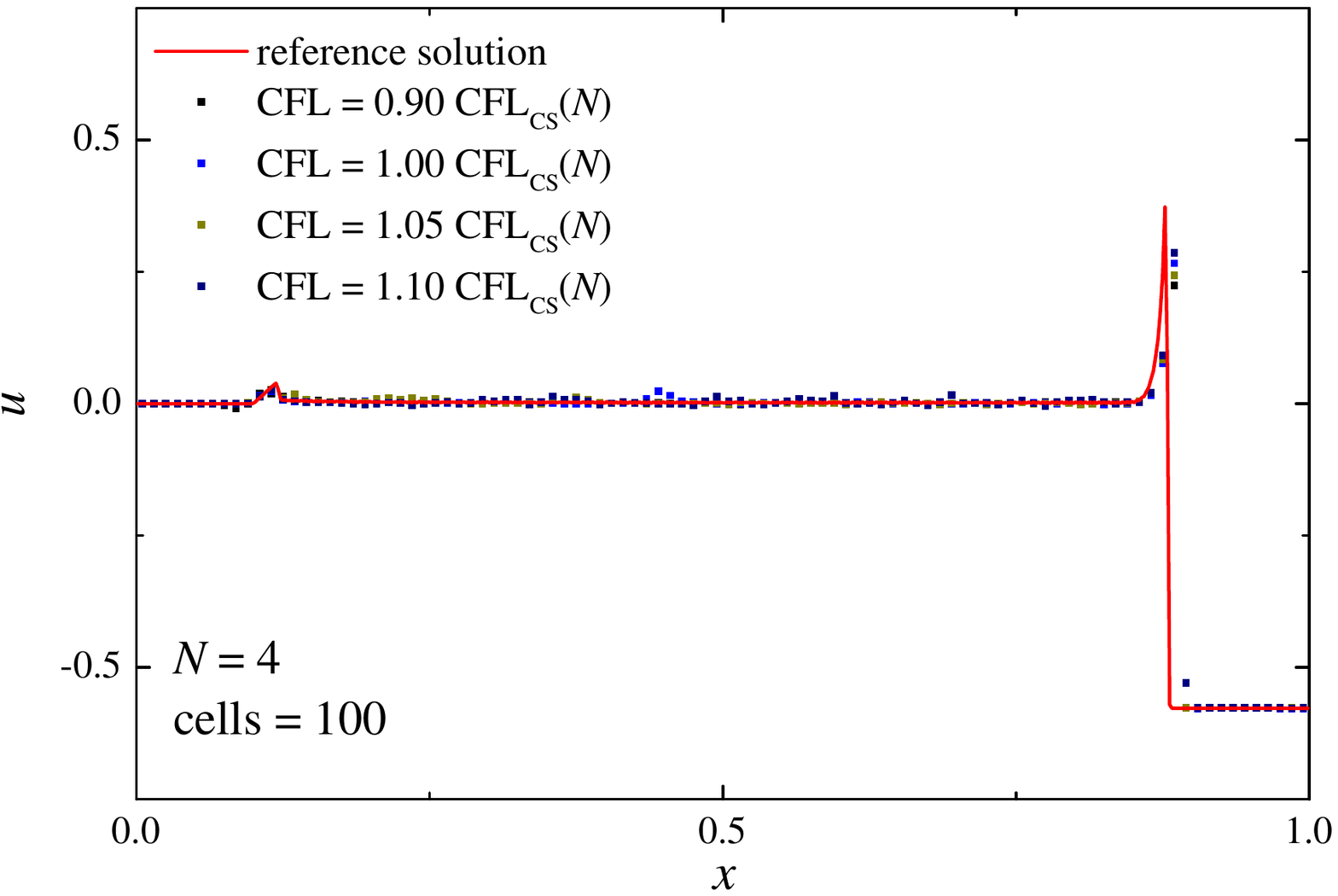}
\vspace{-8mm}\caption{\label{fig:test_zndf_1d_degrees_4_6:degree_4_u}}
\end{subfigure}
\begin{subfigure}{0.24\textwidth}
\includegraphics[width=\textwidth]{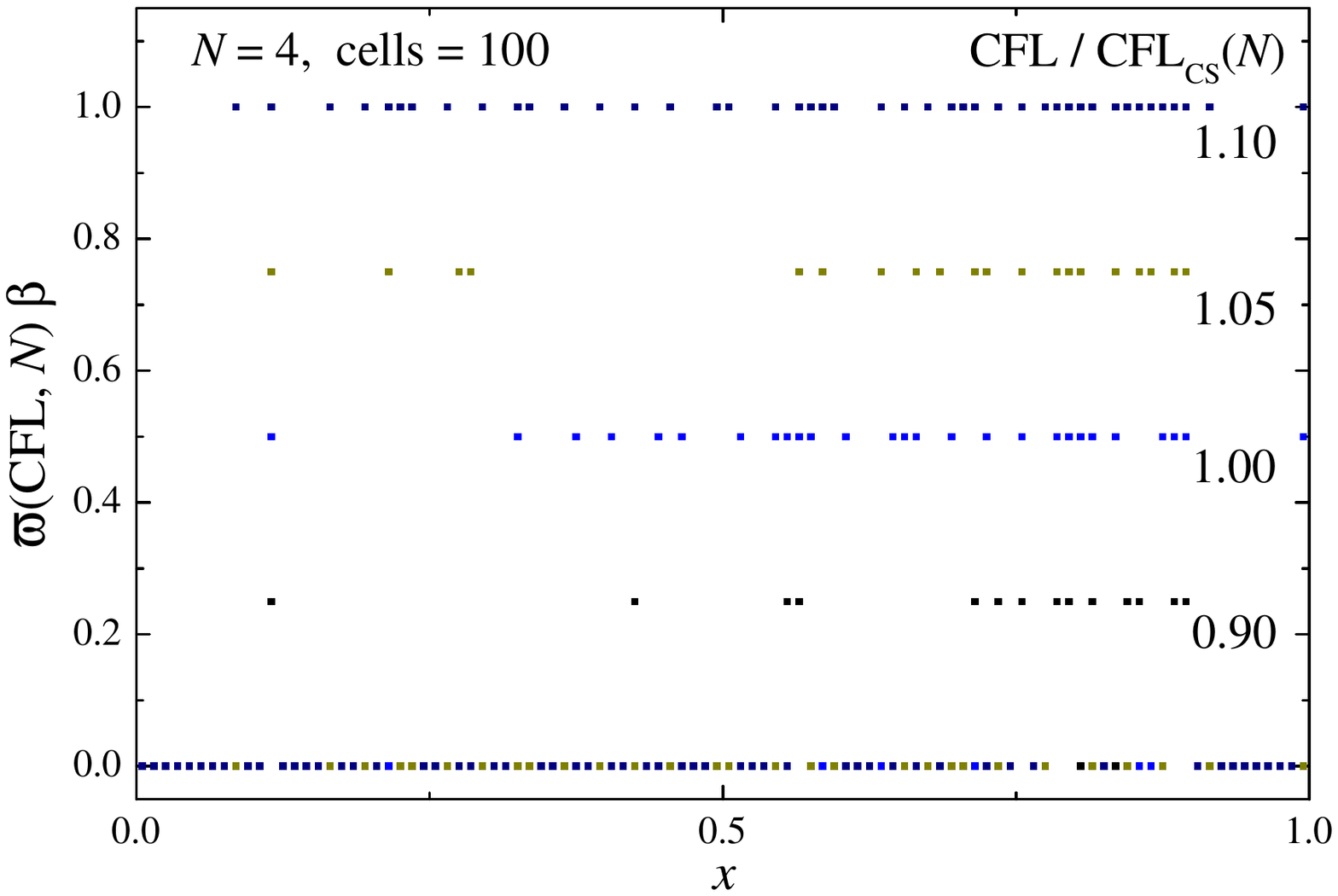}
\vspace{-8mm}\caption{\label{fig:test_zndf_1d_degrees_4_6:degree_4_beta}}
\end{subfigure}\\[-2mm]
\begin{subfigure}{0.24\textwidth}
\includegraphics[width=\textwidth]{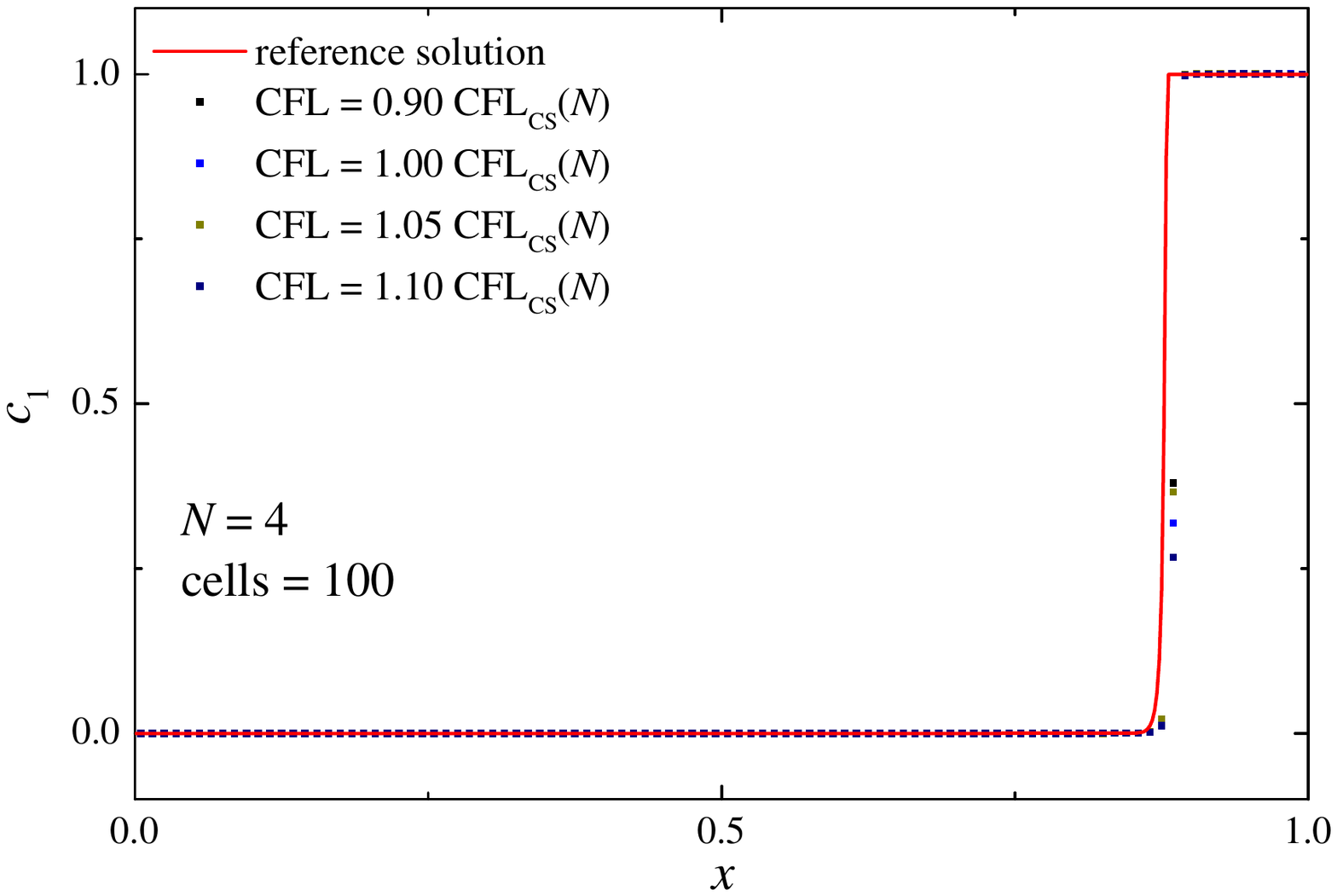}
\vspace{-8mm}\caption{\label{fig:test_zndf_1d_degrees_4_6:degree_4_c1}}
\end{subfigure}
\begin{subfigure}{0.24\textwidth}
\includegraphics[width=\textwidth]{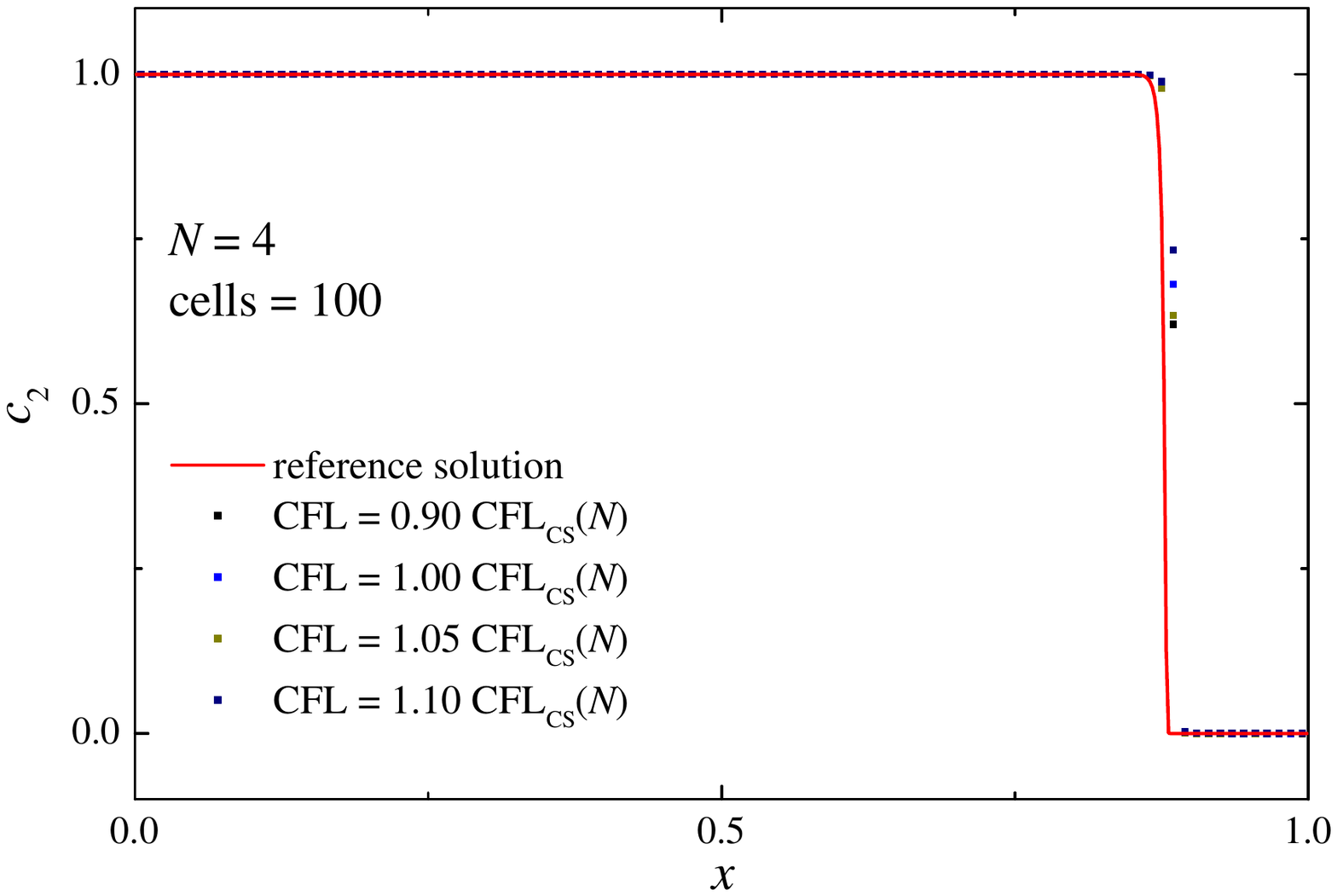}
\vspace{-8mm}\caption{\label{fig:test_zndf_1d_degrees_4_6:degree_4_c2}}
\end{subfigure}
\begin{subfigure}{0.24\textwidth}
\includegraphics[width=\textwidth]{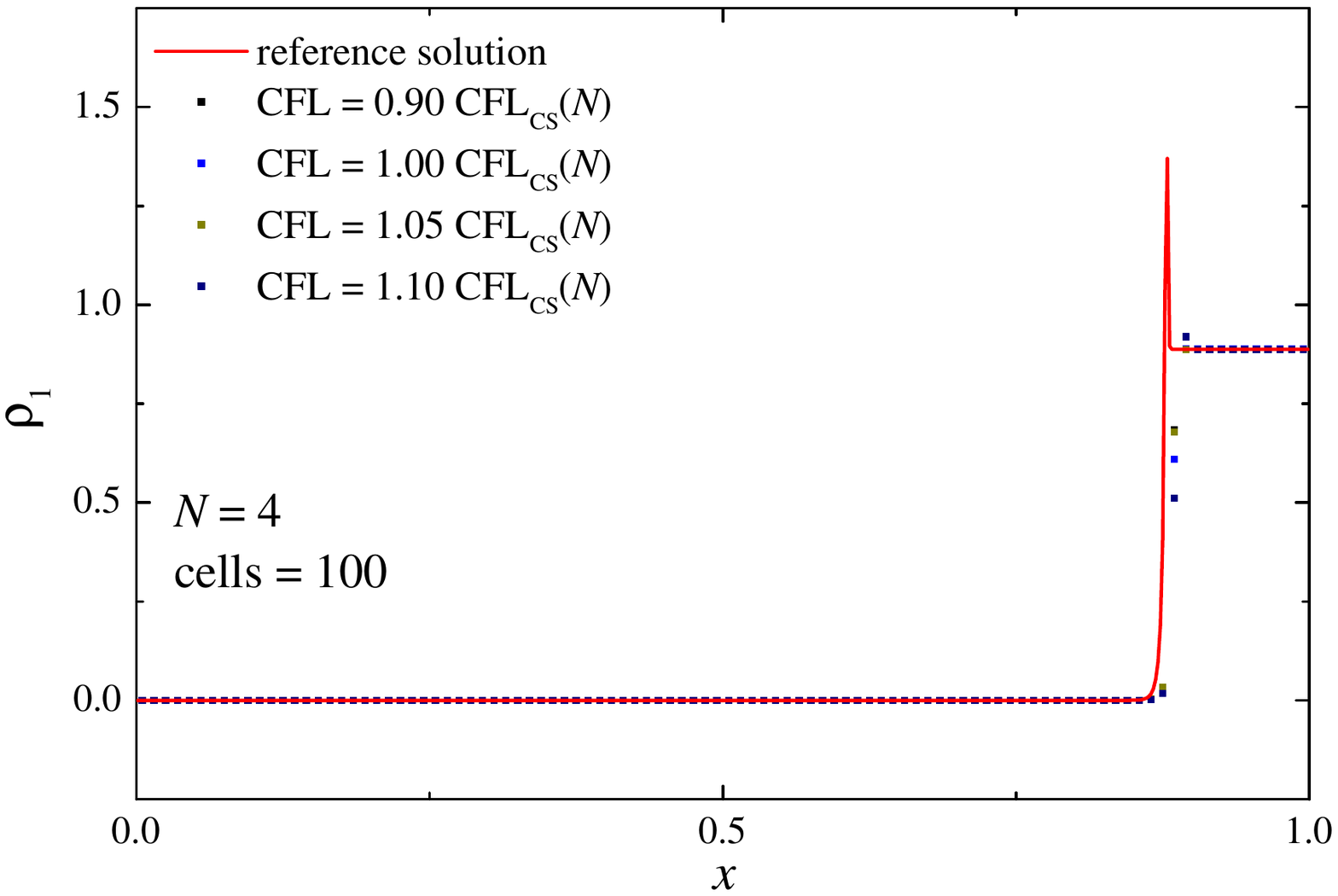}
\vspace{-8mm}\caption{\label{fig:test_zndf_1d_degrees_4_6:degree_4_den1}}
\end{subfigure}
\begin{subfigure}{0.24\textwidth}
\includegraphics[width=\textwidth]{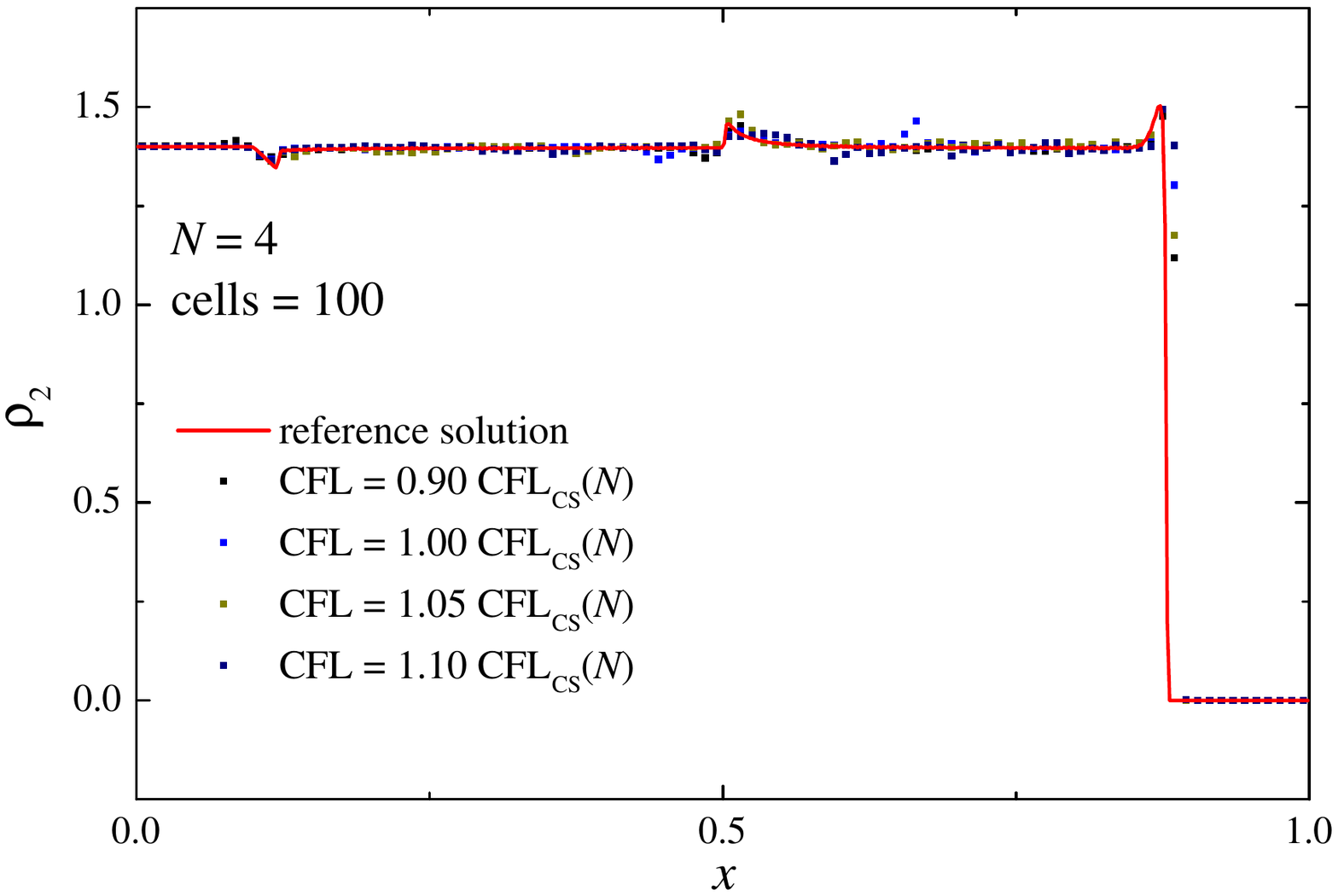}
\vspace{-8mm}\caption{\label{fig:test_zndf_1d_degrees_4_6:degree_4_den2}}
\end{subfigure}\\[-2mm]
\begin{subfigure}{0.24\textwidth}
\includegraphics[width=\textwidth]{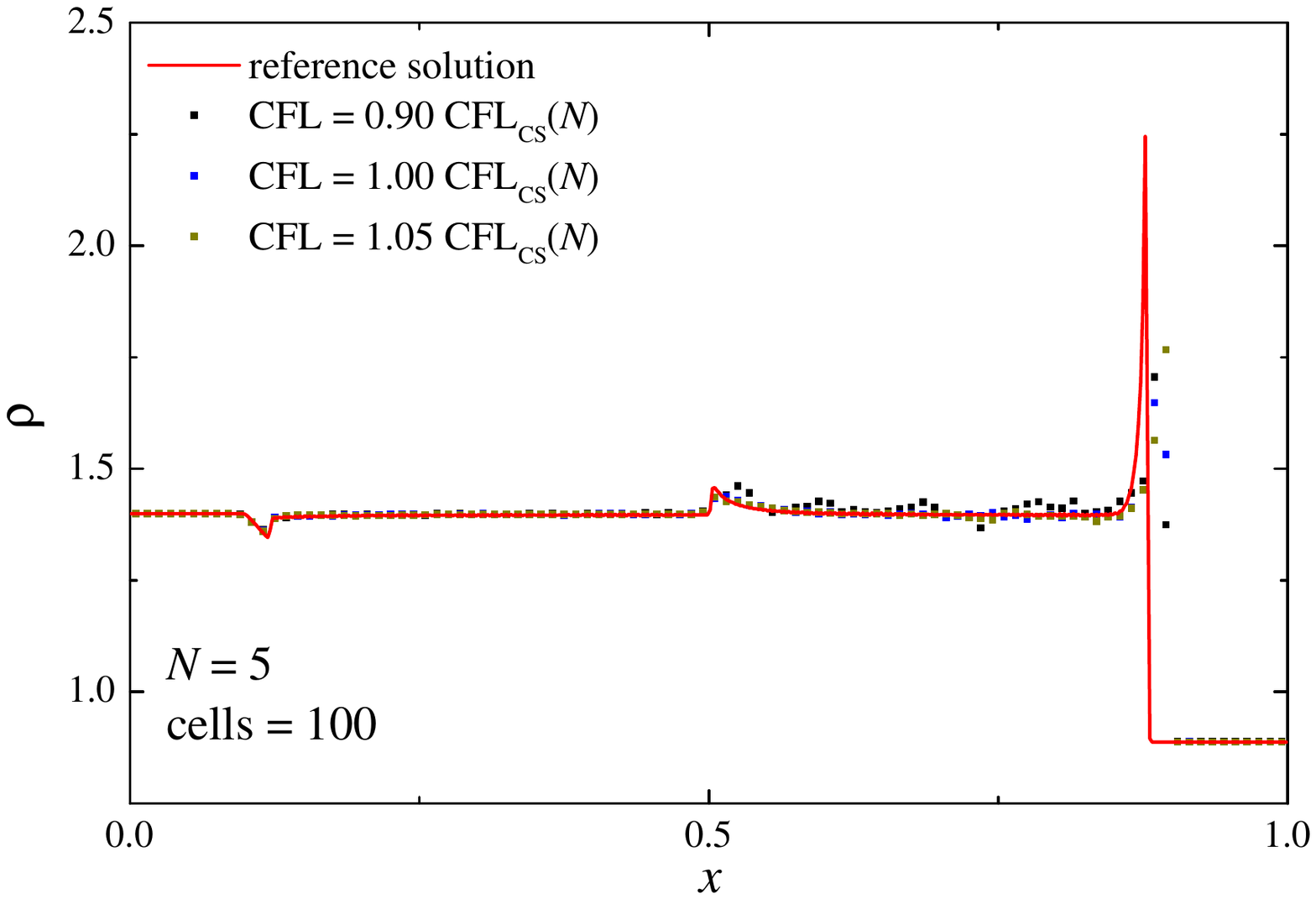}
\vspace{-8mm}\caption{\label{fig:test_zndf_1d_degrees_4_6:degree_5_den}}
\end{subfigure}
\begin{subfigure}{0.24\textwidth}
\includegraphics[width=\textwidth]{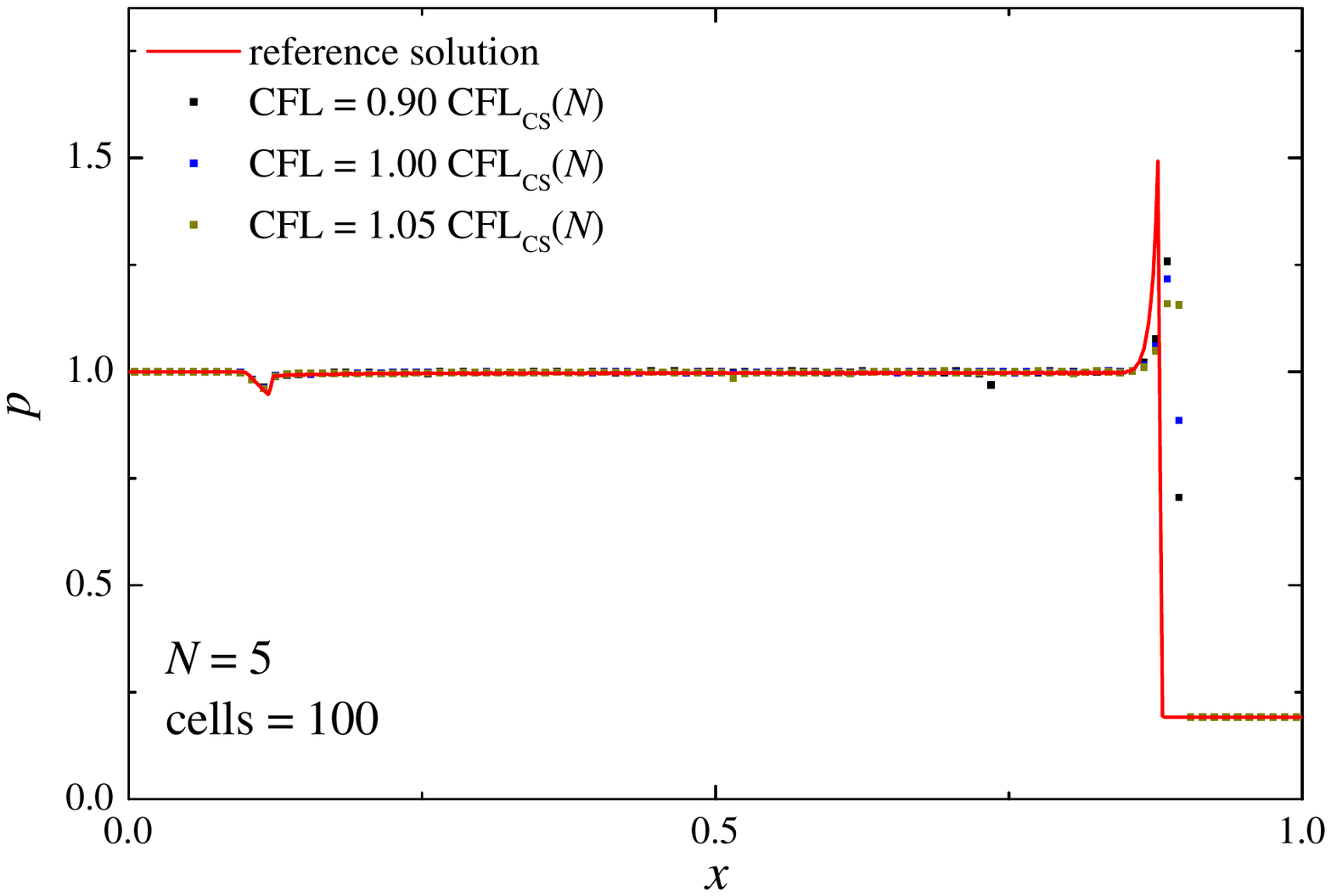}
\vspace{-8mm}\caption{\label{fig:test_zndf_1d_degrees_4_6:degree_5_p}}
\end{subfigure}
\begin{subfigure}{0.24\textwidth}
\includegraphics[width=\textwidth]{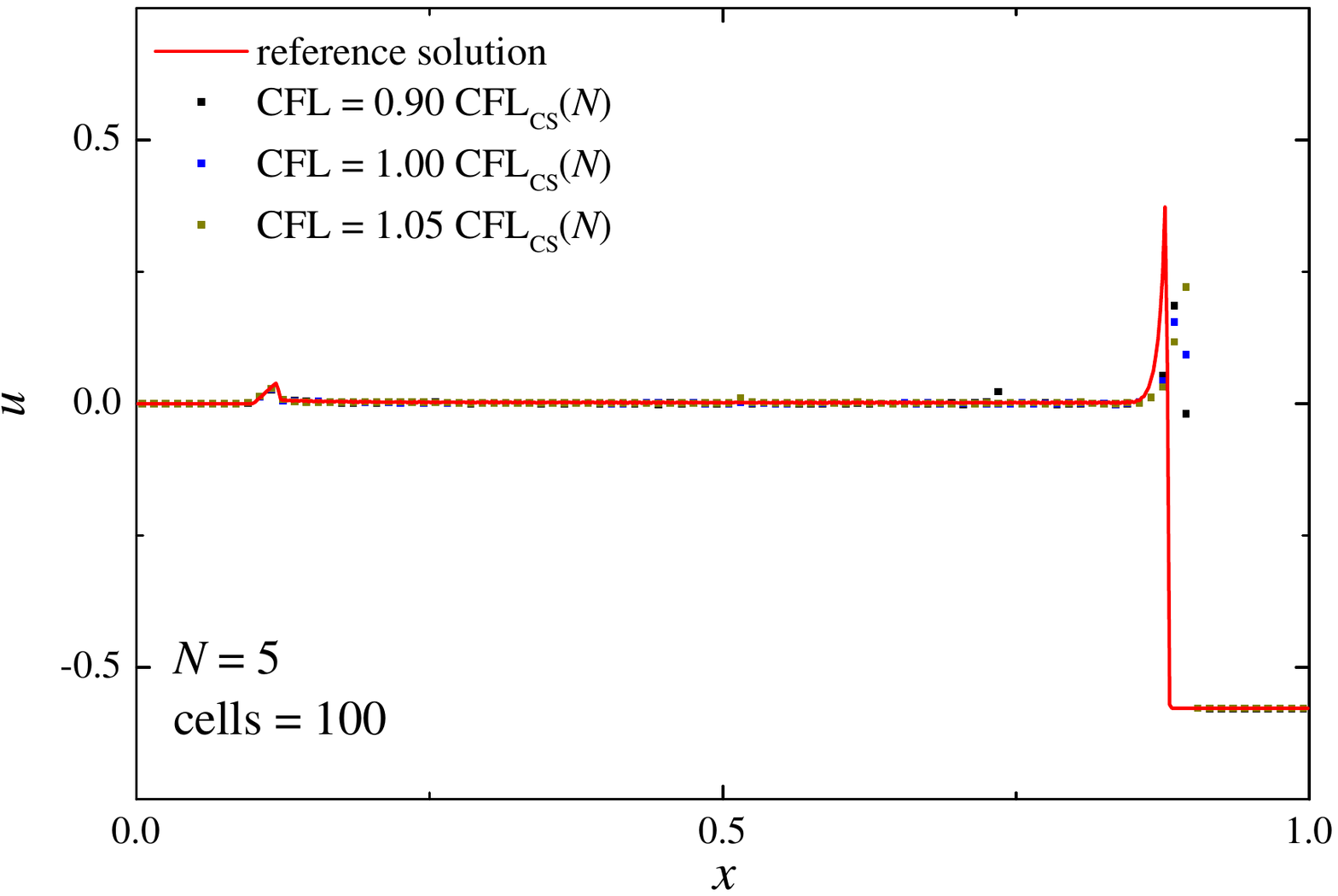}
\vspace{-8mm}\caption{\label{fig:test_zndf_1d_degrees_4_6:degree_5_u}}
\end{subfigure}
\begin{subfigure}{0.24\textwidth}
\includegraphics[width=\textwidth]{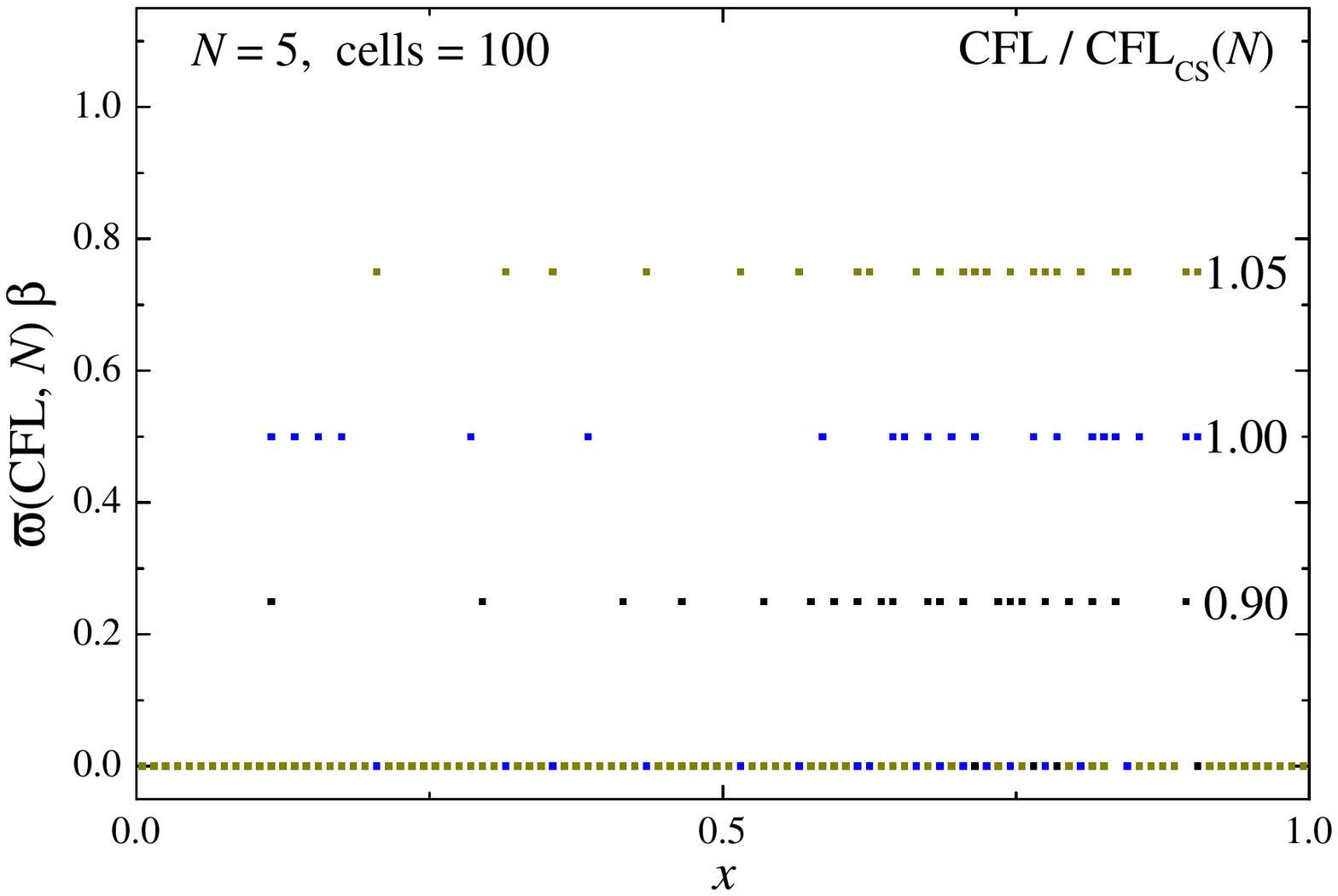}
\vspace{-8mm}\caption{\label{fig:test_zndf_1d_degrees_4_6:degree_5_beta}}
\end{subfigure}\\[-2mm]
\begin{subfigure}{0.24\textwidth}
\includegraphics[width=\textwidth]{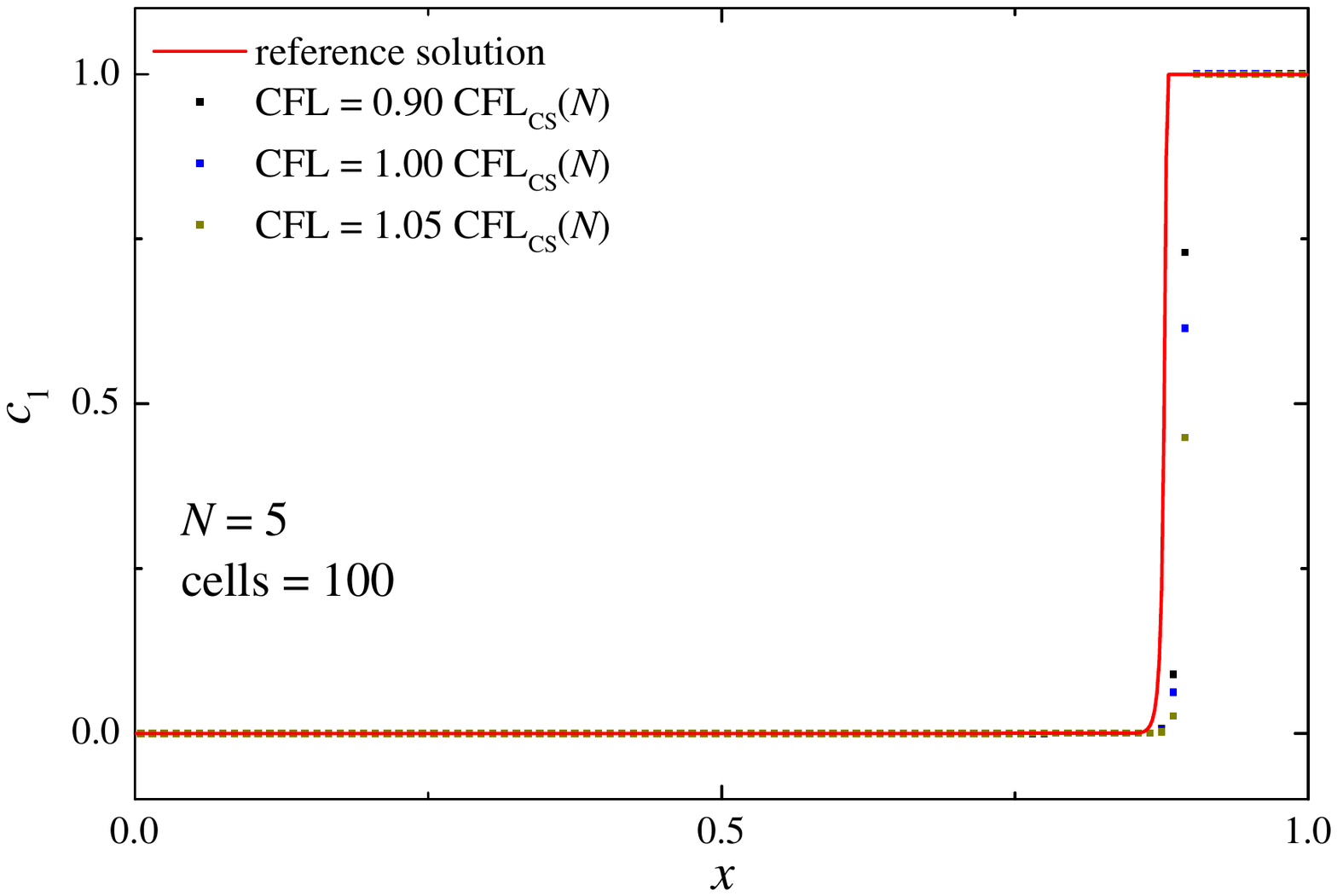}
\vspace{-8mm}\caption{\label{fig:test_zndf_1d_degrees_4_6:degree_5_c1}}
\end{subfigure}
\begin{subfigure}{0.24\textwidth}
\includegraphics[width=\textwidth]{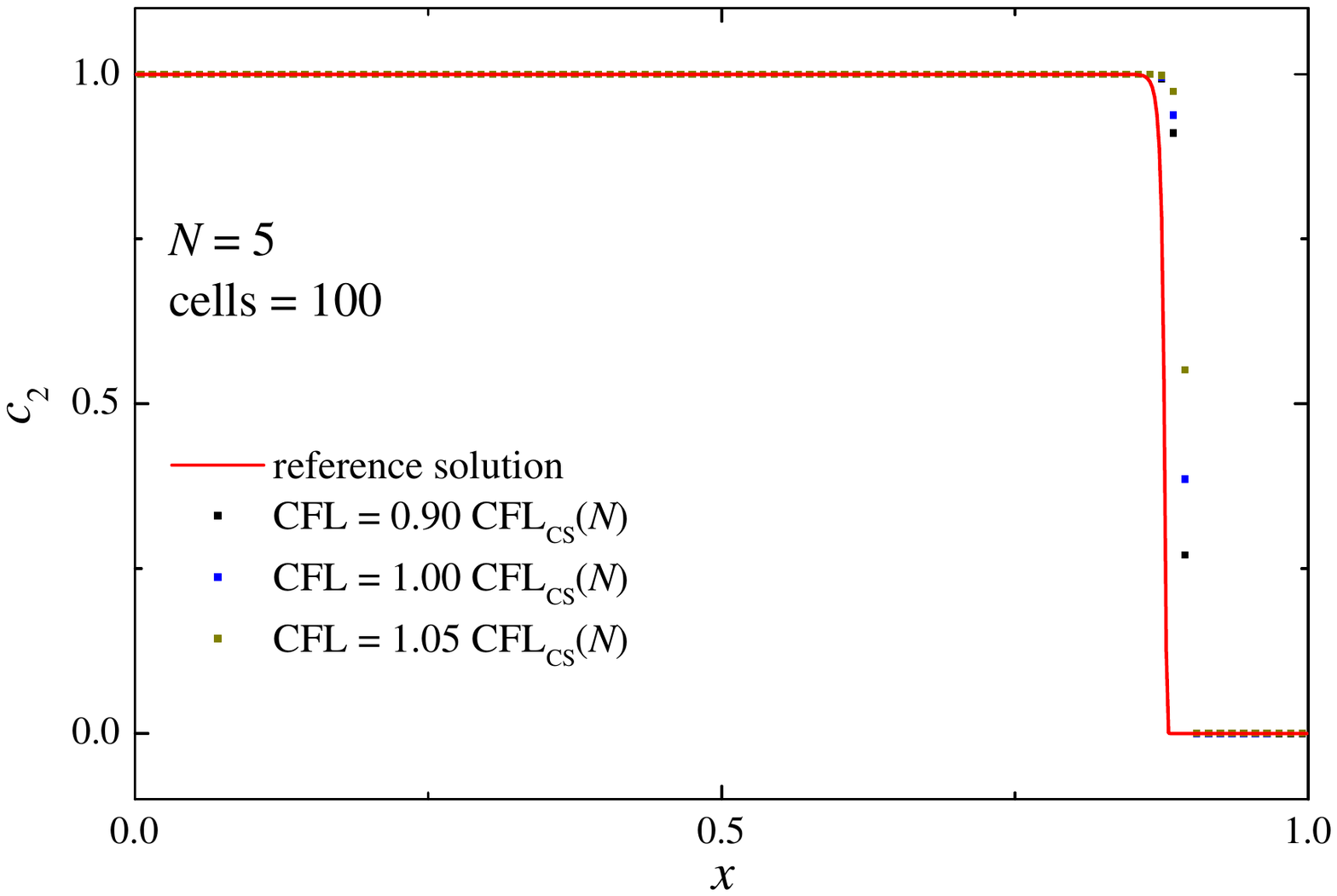}
\vspace{-8mm}\caption{\label{fig:test_zndf_1d_degrees_4_6:degree_5_c2}}
\end{subfigure}
\begin{subfigure}{0.24\textwidth}
\includegraphics[width=\textwidth]{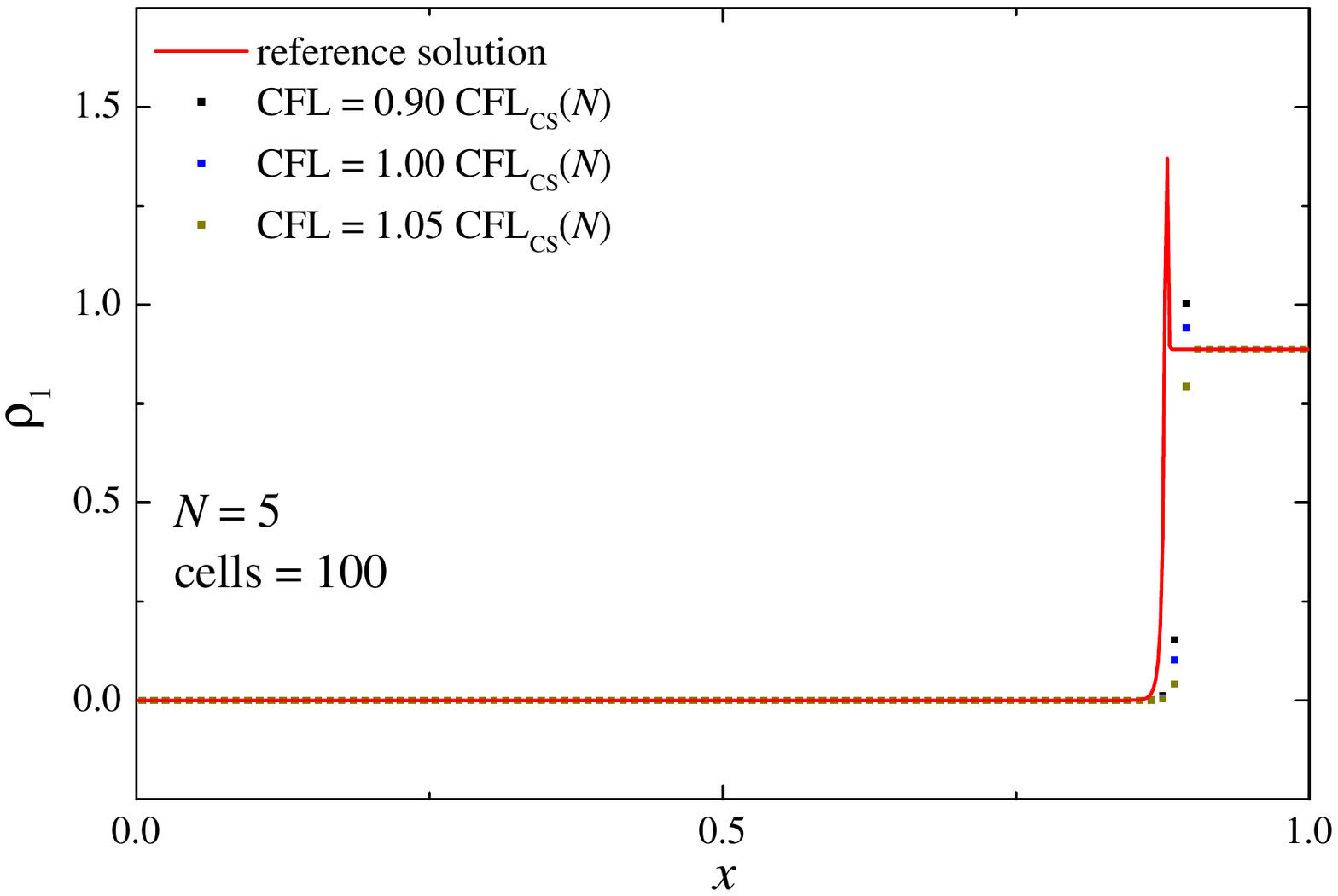}
\vspace{-8mm}\caption{\label{fig:test_zndf_1d_degrees_4_6:degree_5_den1}}
\end{subfigure}
\begin{subfigure}{0.24\textwidth}
\includegraphics[width=\textwidth]{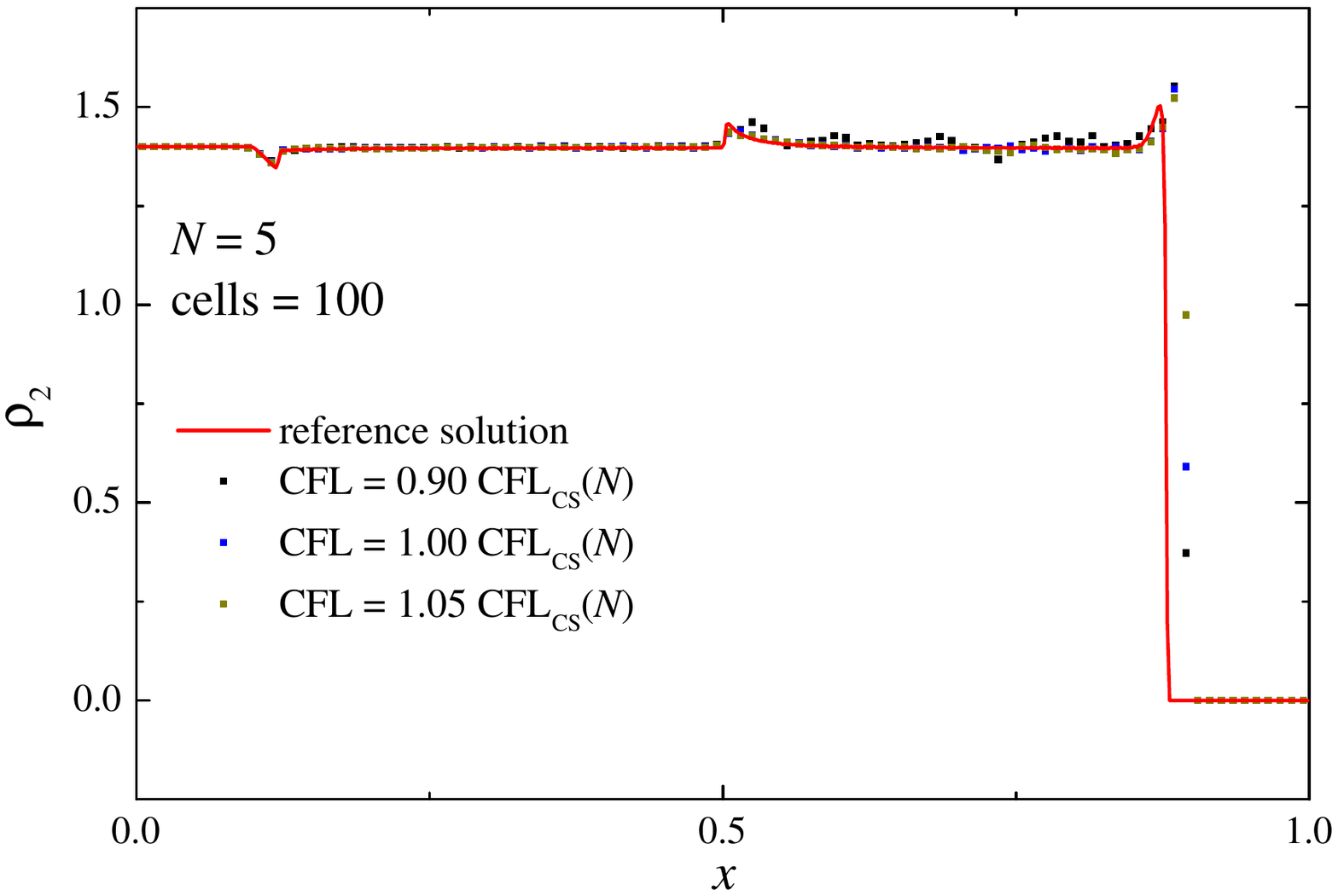}
\vspace{-8mm}\caption{\label{fig:test_zndf_1d_degrees_4_6:degree_5_den2}}
\end{subfigure}\\[-2mm]
\begin{subfigure}{0.24\textwidth}
\includegraphics[width=\textwidth]{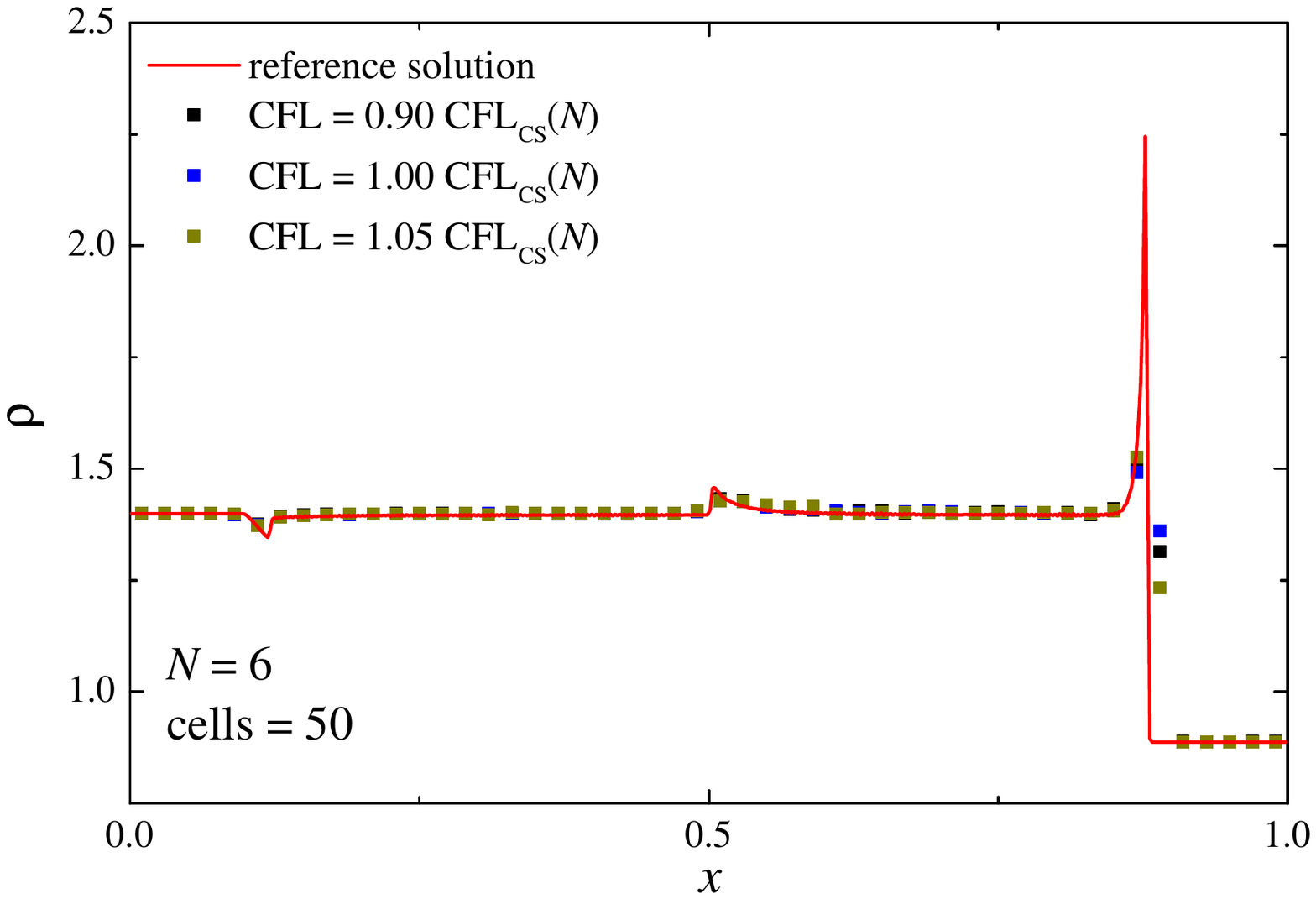}
\vspace{-8mm}\caption{\label{fig:test_zndf_1d_degrees_4_6:degree_6_den}}
\end{subfigure}
\begin{subfigure}{0.24\textwidth}
\includegraphics[width=\textwidth]{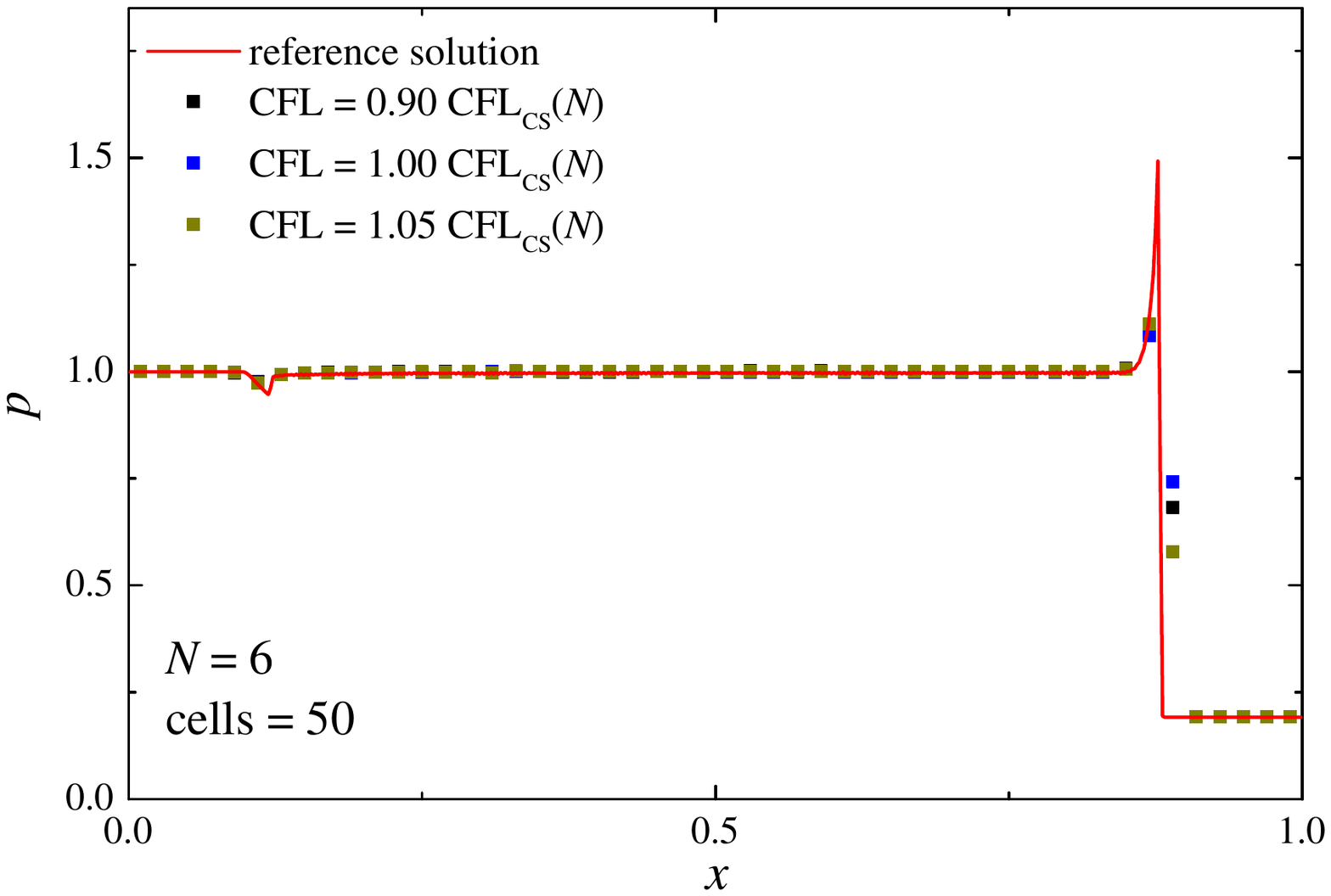}
\vspace{-8mm}\caption{\label{fig:test_zndf_1d_degrees_4_6:degree_6_p}}
\end{subfigure}
\begin{subfigure}{0.24\textwidth}
\includegraphics[width=\textwidth]{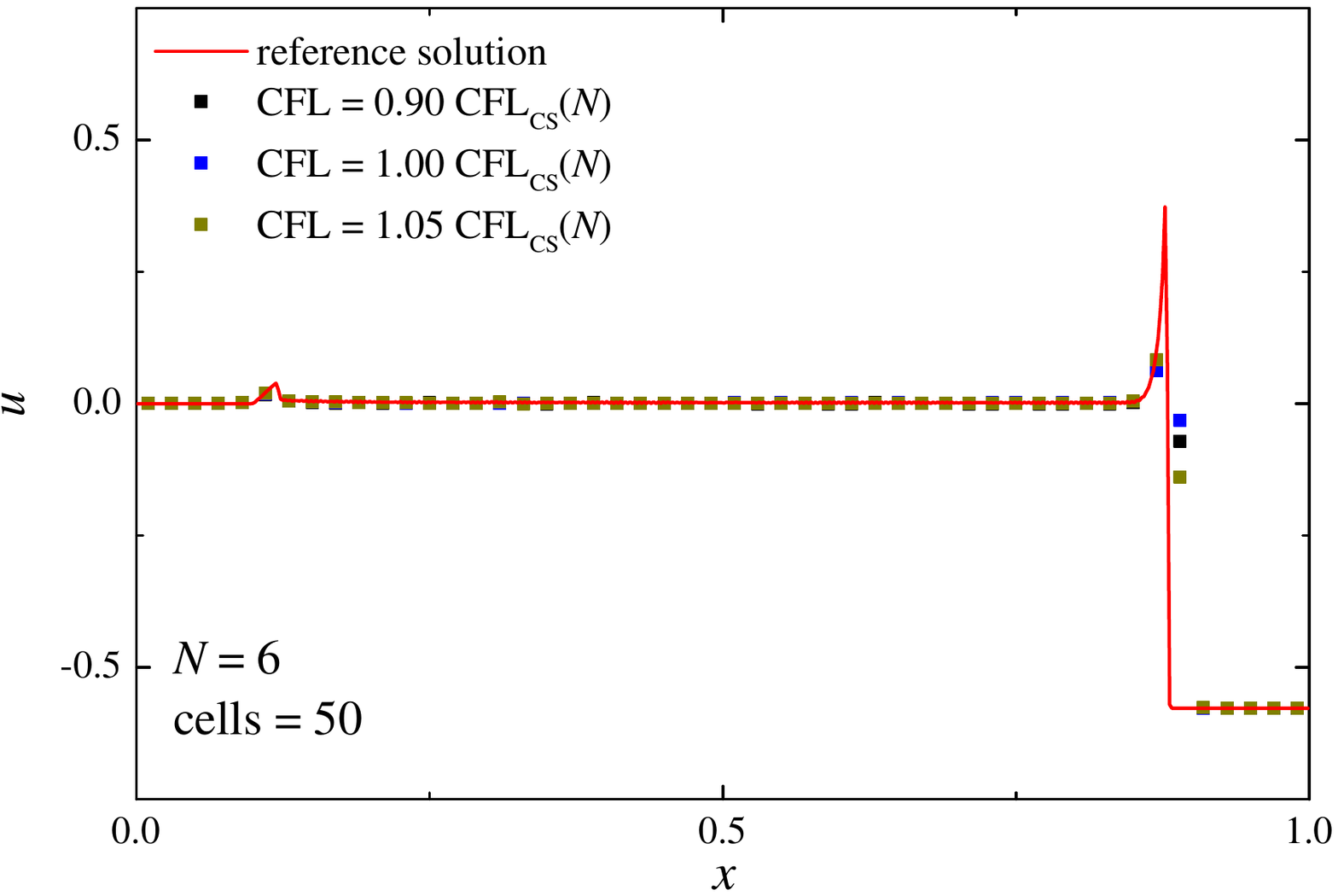}
\vspace{-8mm}\caption{\label{fig:test_zndf_1d_degrees_4_6:degree_6_u}}
\end{subfigure}
\begin{subfigure}{0.24\textwidth}
\includegraphics[width=\textwidth]{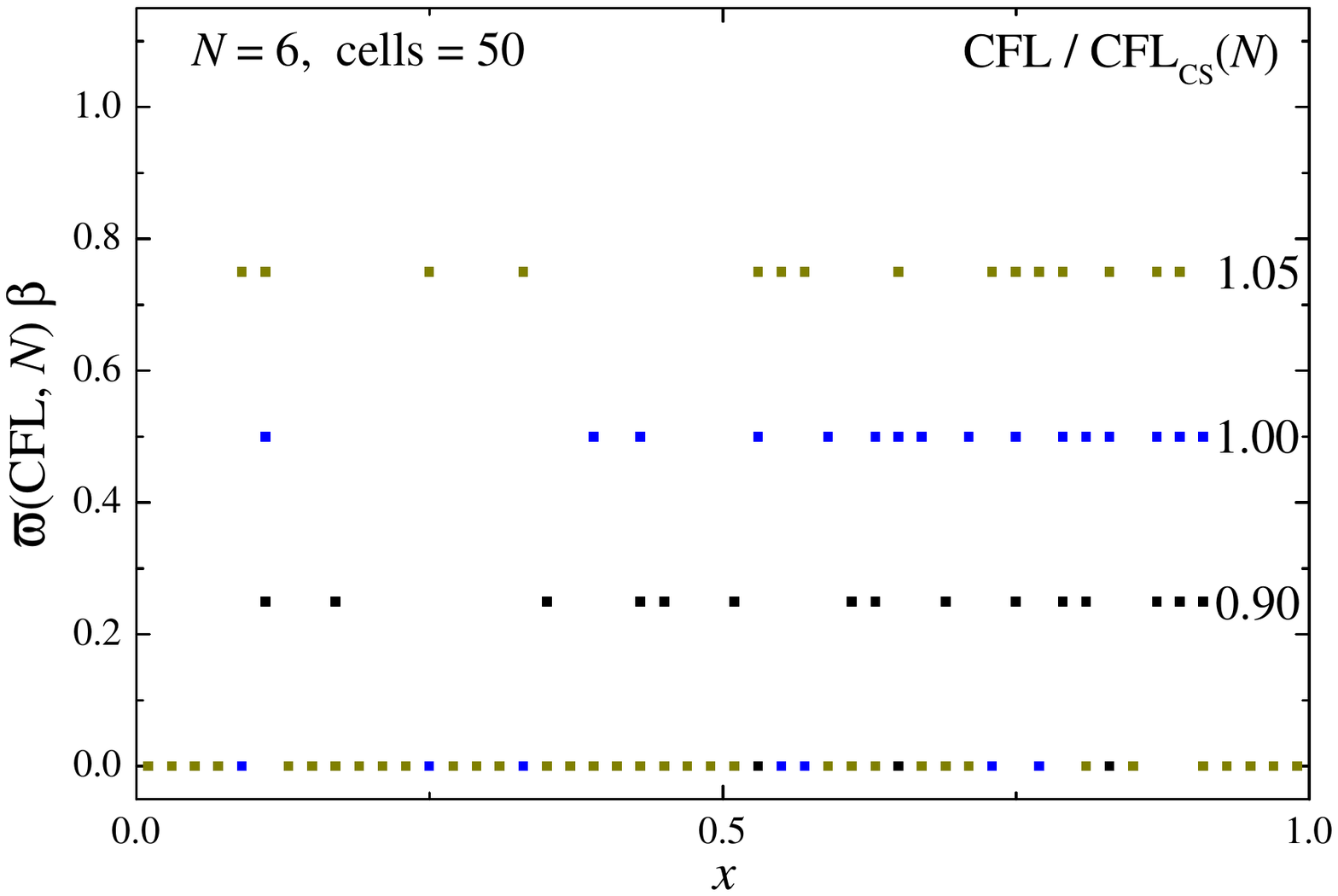}
\vspace{-8mm}\caption{\label{fig:test_zndf_1d_degrees_4_6:degree_6_beta}}
\end{subfigure}\\[-2mm]
\begin{subfigure}{0.24\textwidth}
\includegraphics[width=\textwidth]{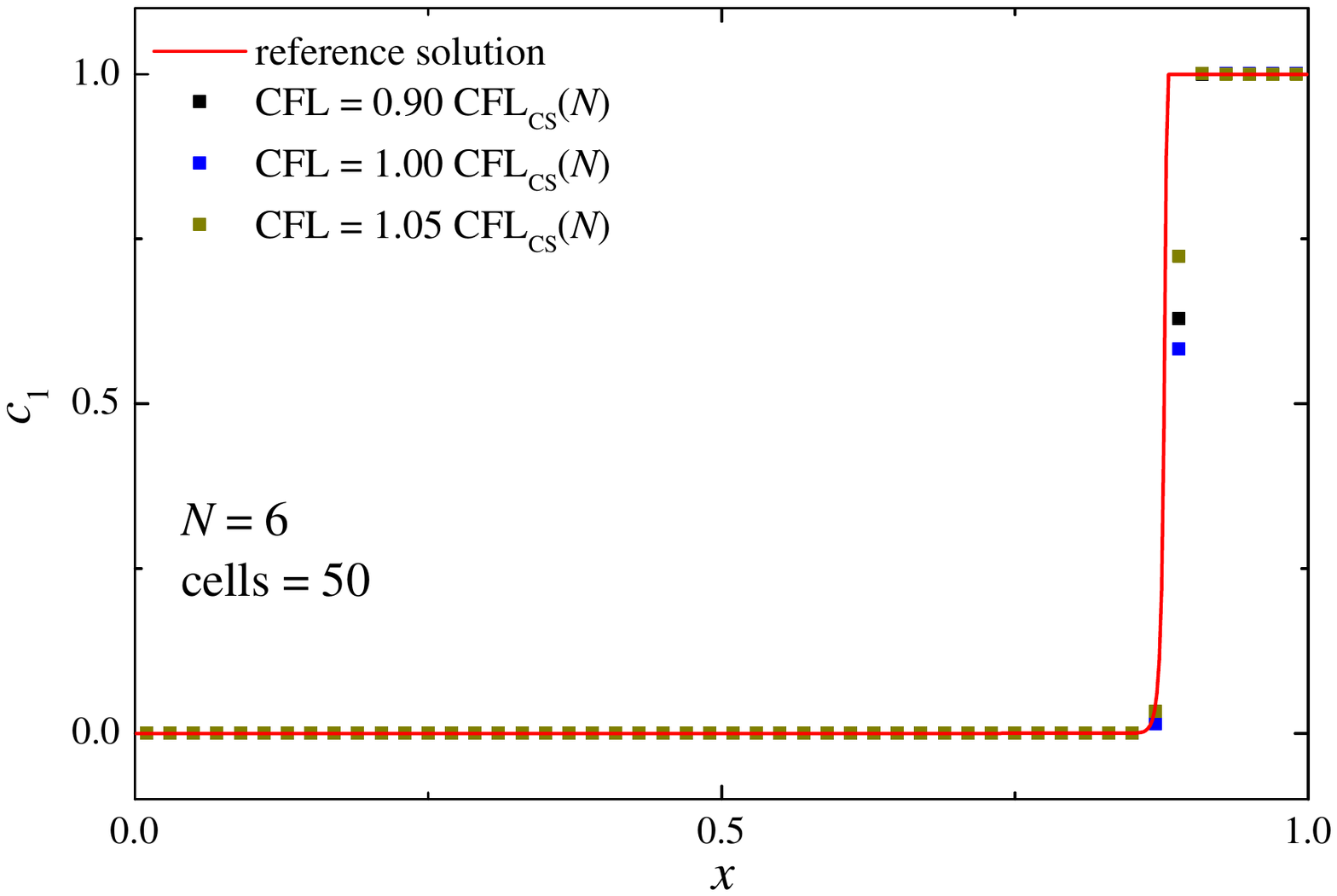}
\vspace{-8mm}\caption{\label{fig:test_zndf_1d_degrees_4_6:degree_6_c1}}
\end{subfigure}
\begin{subfigure}{0.24\textwidth}
\includegraphics[width=\textwidth]{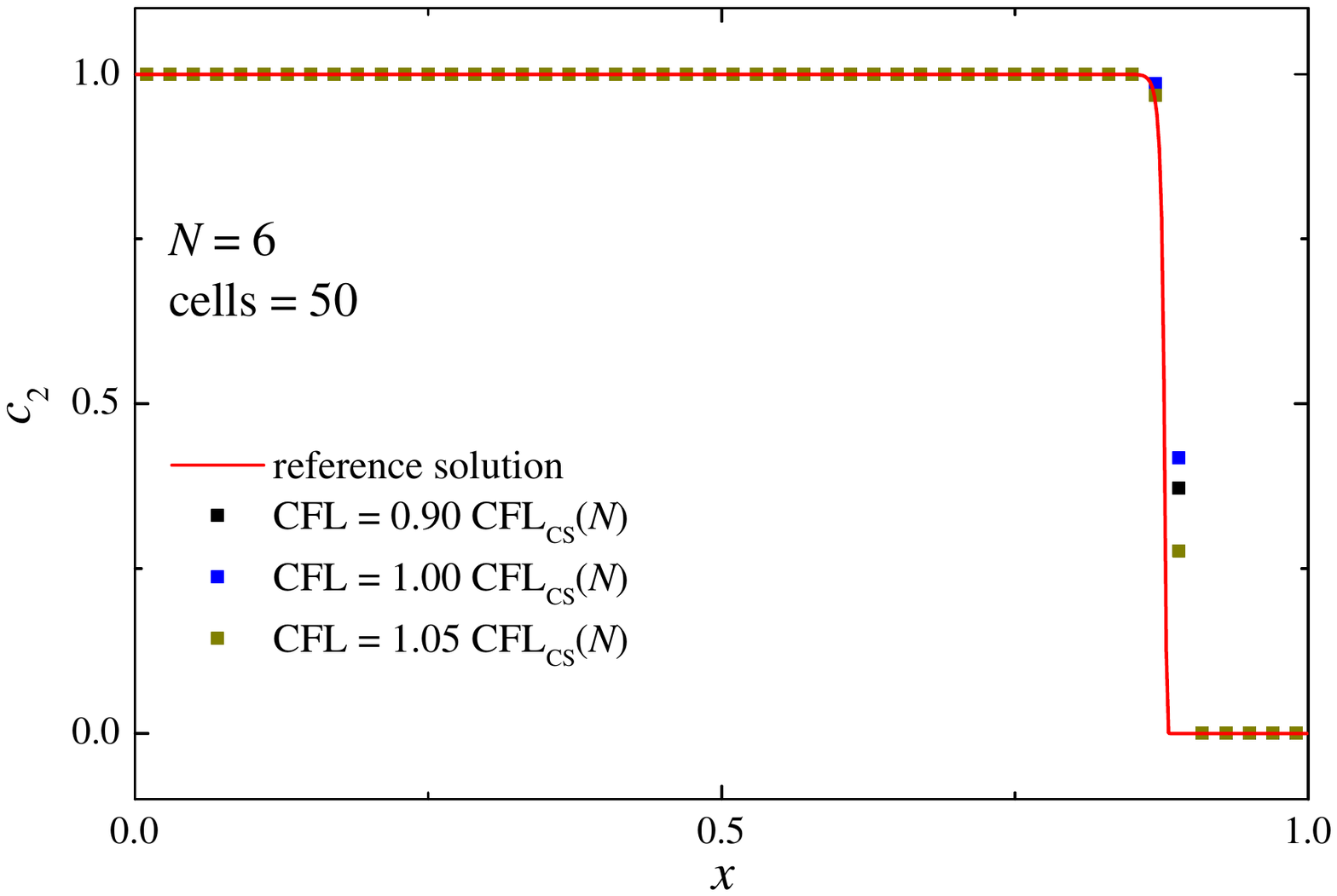}
\vspace{-8mm}\caption{\label{fig:test_zndf_1d_degrees_4_6:degree_6_c2}}
\end{subfigure}
\begin{subfigure}{0.24\textwidth}
\includegraphics[width=\textwidth]{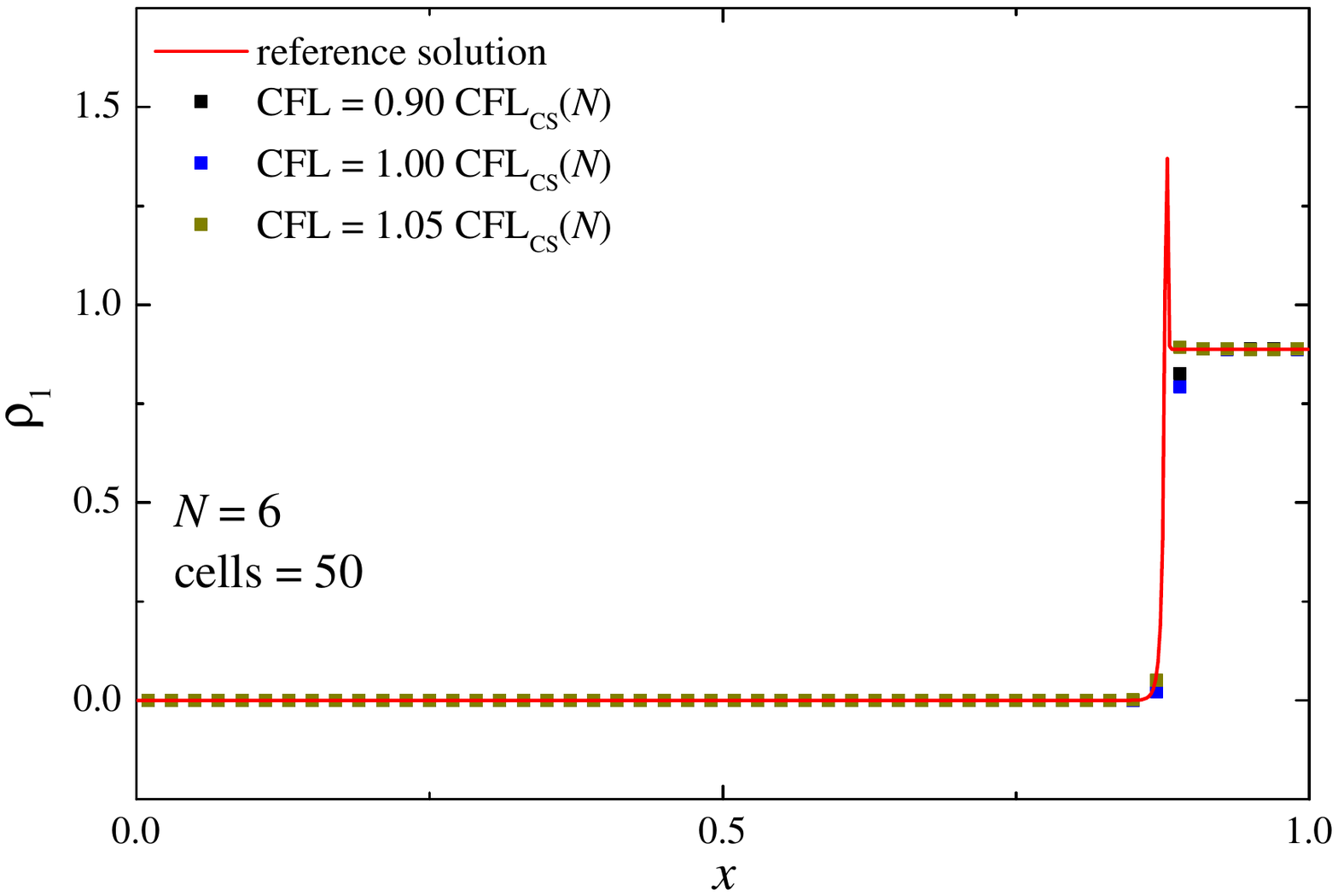}
\vspace{-8mm}\caption{\label{fig:test_zndf_1d_degrees_4_6:degree_6_den1}}
\end{subfigure}
\begin{subfigure}{0.24\textwidth}
\includegraphics[width=\textwidth]{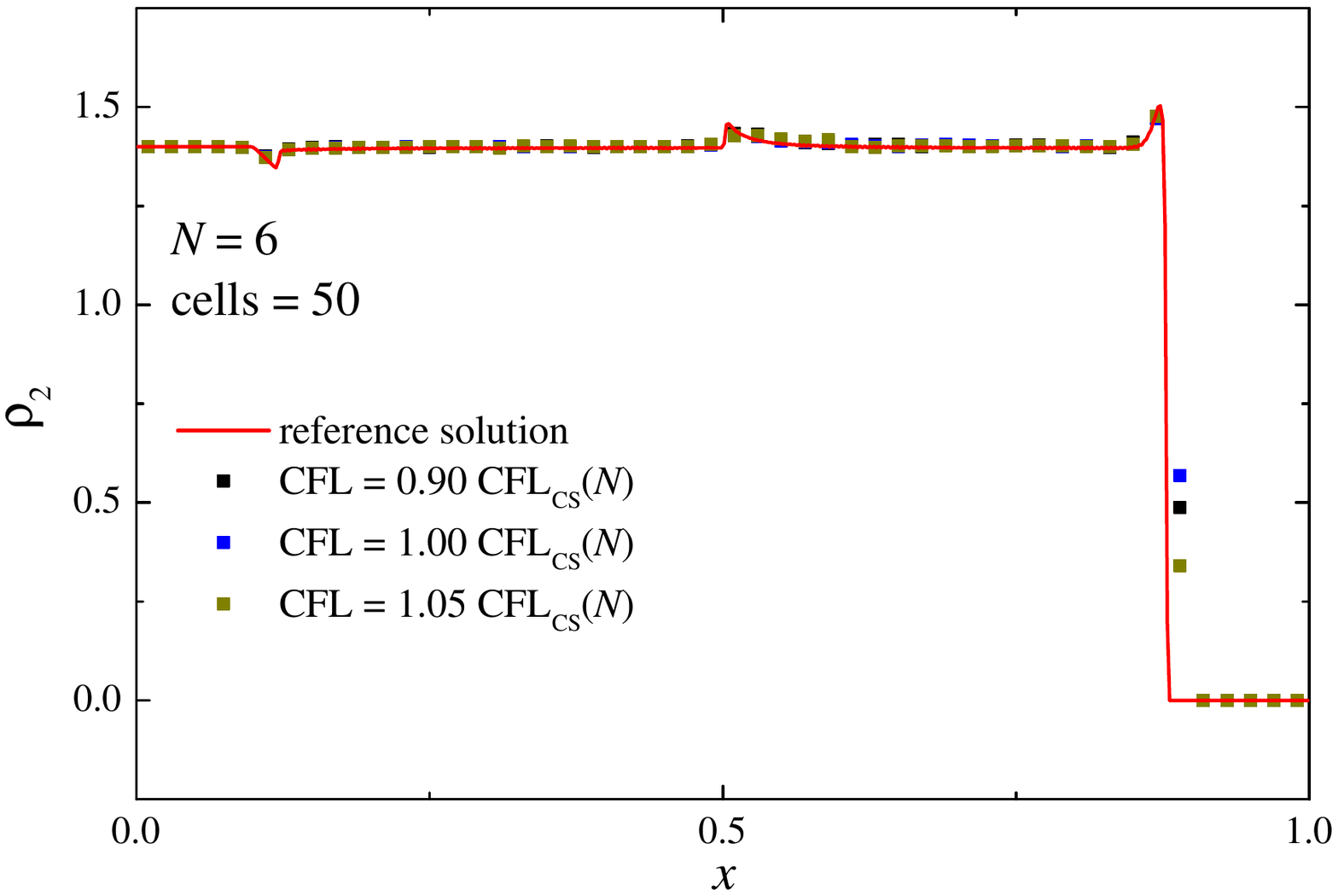}
\vspace{-8mm}\caption{\label{fig:test_zndf_1d_degrees_4_6:degree_6_den2}}
\end{subfigure}
\caption{%
Numerical solution of the one-dimensional problem of the plane detonation wave formation in a two-component medium with a ``fast'' reaction (strong stiff case, a detailed statement of the problem is presented in the text), obtained using the ADER-DG method for the basis polynomials degree $N = 4$ ((\subref{fig:test_zndf_1d_degrees_4_6:degree_4_den}), (\subref{fig:test_zndf_1d_degrees_4_6:degree_4_p}), (\subref{fig:test_zndf_1d_degrees_4_6:degree_4_u}), (\subref{fig:test_zndf_1d_degrees_4_6:degree_4_c1}), (\subref{fig:test_zndf_1d_degrees_4_6:degree_4_c2}), (\subref{fig:test_zndf_1d_degrees_4_6:degree_4_den1}), (\subref{fig:test_zndf_1d_degrees_4_6:degree_4_den2})), $N = 5$ ((\subref{fig:test_zndf_1d_degrees_4_6:degree_5_den}), (\subref{fig:test_zndf_1d_degrees_4_6:degree_5_p}), (\subref{fig:test_zndf_1d_degrees_4_6:degree_5_u}), (\subref{fig:test_zndf_1d_degrees_4_6:degree_5_c1}), (\subref{fig:test_zndf_1d_degrees_4_6:degree_5_c2}), (\subref{fig:test_zndf_1d_degrees_4_6:degree_5_den1}), (\subref{fig:test_zndf_1d_degrees_4_6:degree_5_den2})), $N = 6$ ((\subref{fig:test_zndf_1d_degrees_4_6:degree_6_den}), (\subref{fig:test_zndf_1d_degrees_4_6:degree_6_p}), (\subref{fig:test_zndf_1d_degrees_4_6:degree_6_u}), (\subref{fig:test_zndf_1d_degrees_4_6:degree_6_c1}), (\subref{fig:test_zndf_1d_degrees_4_6:degree_6_c2}), (\subref{fig:test_zndf_1d_degrees_4_6:degree_6_den1}), (\subref{fig:test_zndf_1d_degrees_4_6:degree_6_den2})) with a posteriori limitation of the solution by an ADER-WENO2 finite-volume limiter, at the final time $t_{f} = 0.4$. The graphs show the coordinate dependence of density $\rho$ ((\subref{fig:test_zndf_1d_degrees_4_6:degree_4_den}), (\subref{fig:test_zndf_1d_degrees_4_6:degree_5_den}), (\subref{fig:test_zndf_1d_degrees_4_6:degree_6_den})), pressure $p$ ((\subref{fig:test_zndf_1d_degrees_4_6:degree_4_p}), (\subref{fig:test_zndf_1d_degrees_4_6:degree_5_p}), (\subref{fig:test_zndf_1d_degrees_4_6:degree_6_p})), flow velocity $u$ ((\subref{fig:test_zndf_1d_degrees_4_6:degree_4_u}), (\subref{fig:test_zndf_1d_degrees_4_6:degree_5_u}), (\subref{fig:test_zndf_1d_degrees_4_6:degree_6_u})), troubled cells indicator $\beta$ multiplied by the separating function $\varpi(\mathrm{CFL}, N)$ ((\subref{fig:test_zndf_1d_degrees_4_6:degree_4_beta}), (\subref{fig:test_zndf_1d_degrees_4_6:degree_5_beta}), (\subref{fig:test_zndf_1d_degrees_4_6:degree_6_beta})), mass concentrations $c_{1}$ ((\subref{fig:test_zndf_1d_degrees_4_6:degree_4_c1}), (\subref{fig:test_zndf_1d_degrees_4_6:degree_5_c1}), (\subref{fig:test_zndf_1d_degrees_4_6:degree_6_c1})) and $c_{2}$ ((\subref{fig:test_zndf_1d_degrees_4_6:degree_4_c2}), (\subref{fig:test_zndf_1d_degrees_4_6:degree_5_c2}), (\subref{fig:test_zndf_1d_degrees_4_6:degree_6_c2})) of the reactant $A$ and product $B$, respectively, and partial densities $\rho_{1}$ ((\subref{fig:test_zndf_1d_degrees_4_6:degree_4_den1}), (\subref{fig:test_zndf_1d_degrees_4_6:degree_5_den1}), (\subref{fig:test_zndf_1d_degrees_4_6:degree_6_den1})) and $\rho_{2}$ ((\subref{fig:test_zndf_1d_degrees_4_6:degree_4_den2}), (\subref{fig:test_zndf_1d_degrees_4_6:degree_5_den2}), (\subref{fig:test_zndf_1d_degrees_4_6:degree_6_den2})) of the reactant $A$ and product $B$, respectively. The square symbols represent the subcells finite-volume representation of the numerical solution; the red solid lines represents the reference solution of the problem.
}
\label{fig:test_zndf_1d_degrees_4_6}
\end{figure}
\begin{figure}[h!]
\captionsetup[subfigure]{%
	position=bottom,
	font+=smaller,
	textfont=normalfont,
	singlelinecheck=off,
	justification=raggedright
}
\centering
\begin{subfigure}{0.24\textwidth}
\includegraphics[width=\textwidth]{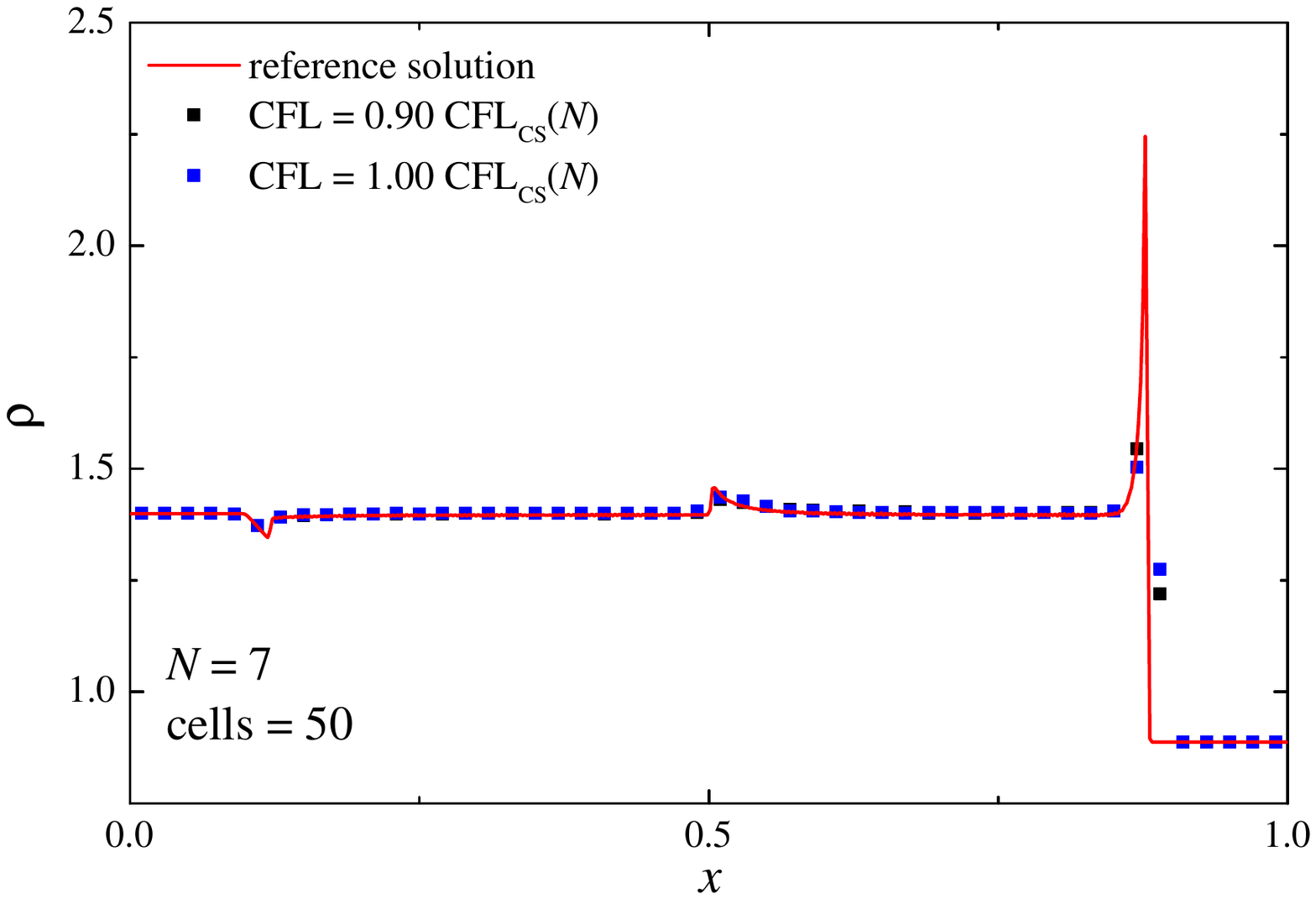}
\vspace{-8mm}\caption{\label{fig:test_zndf_1d_degrees_7_9:degree_7_den}}
\end{subfigure}
\begin{subfigure}{0.24\textwidth}
\includegraphics[width=\textwidth]{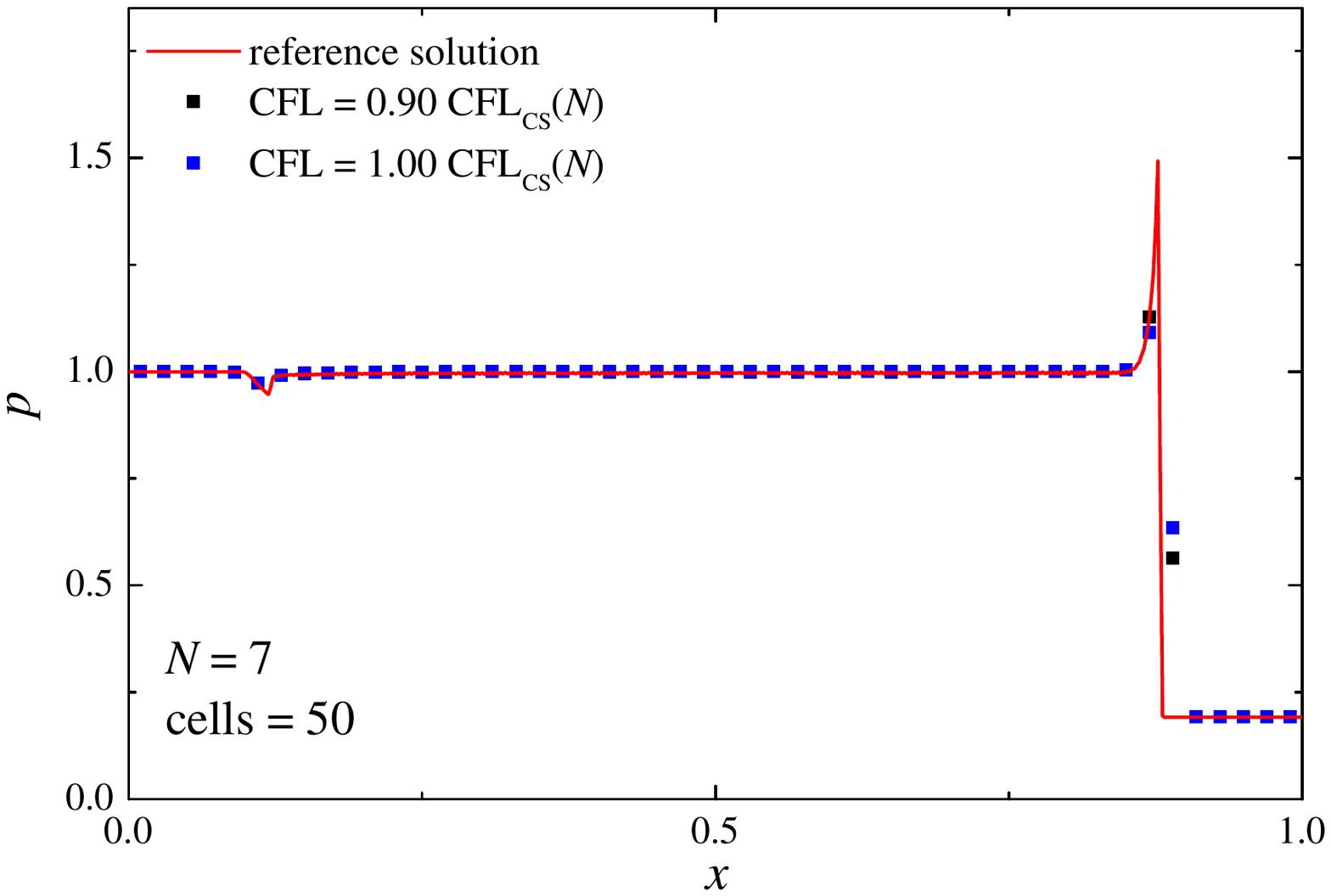}
\vspace{-8mm}\caption{\label{fig:test_zndf_1d_degrees_7_9:degree_7_p}}
\end{subfigure}
\begin{subfigure}{0.24\textwidth}
\includegraphics[width=\textwidth]{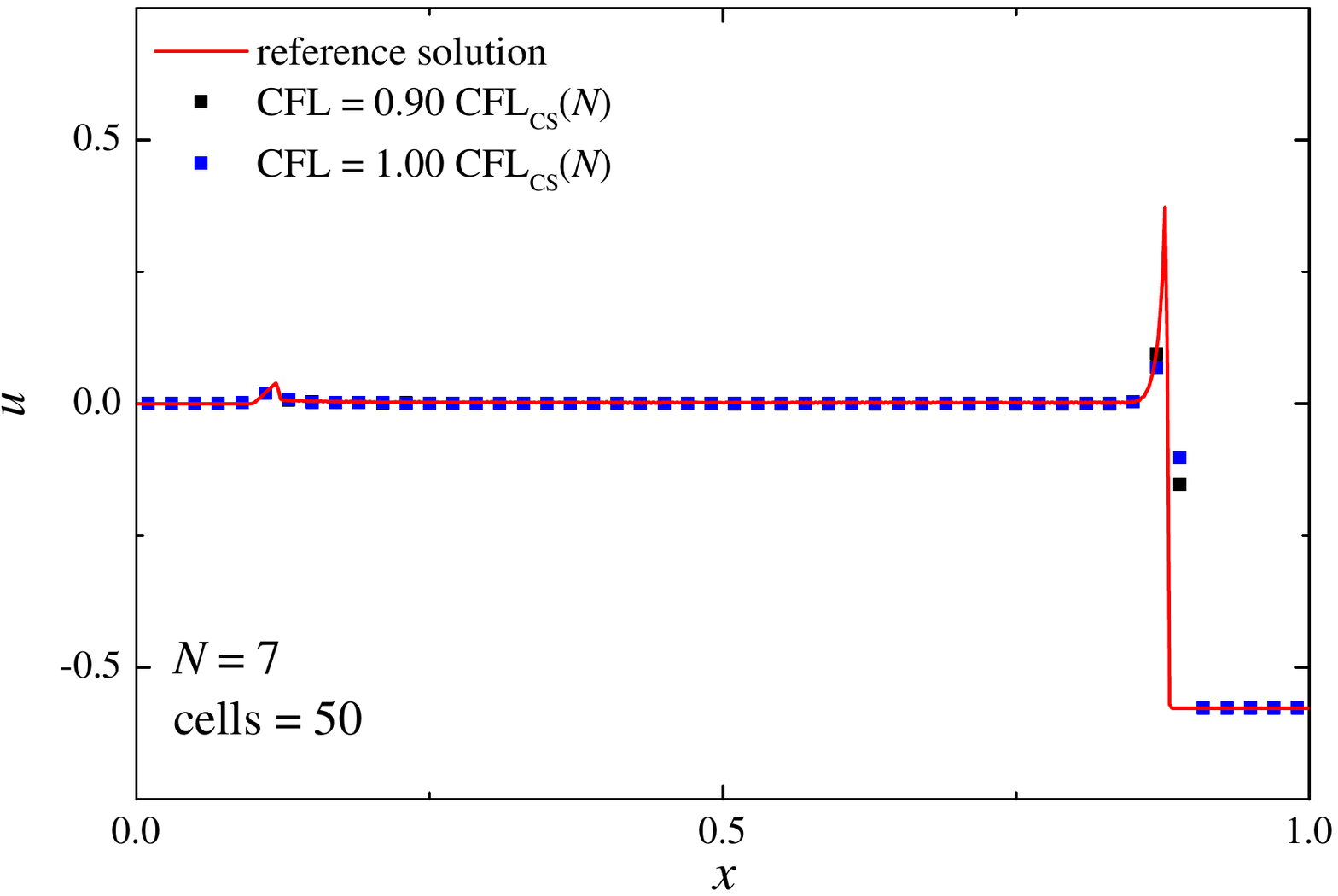}
\vspace{-8mm}\caption{\label{fig:test_zndf_1d_degrees_7_9:degree_7_u}}
\end{subfigure}
\begin{subfigure}{0.24\textwidth}
\includegraphics[width=\textwidth]{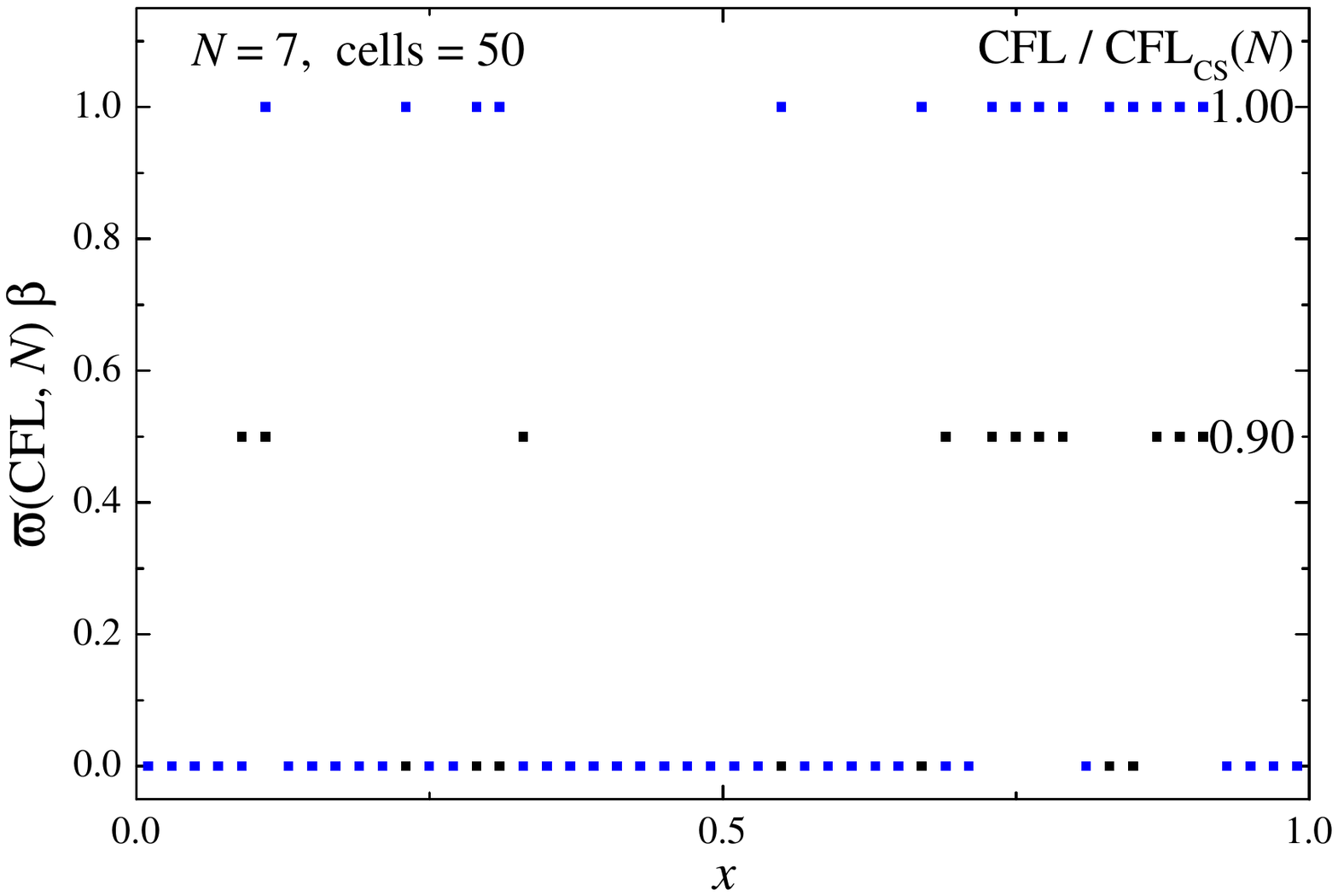}
\vspace{-8mm}\caption{\label{fig:test_zndf_1d_degrees_7_9:degree_7_beta}}
\end{subfigure}\\[-2mm]
\begin{subfigure}{0.24\textwidth}
\includegraphics[width=\textwidth]{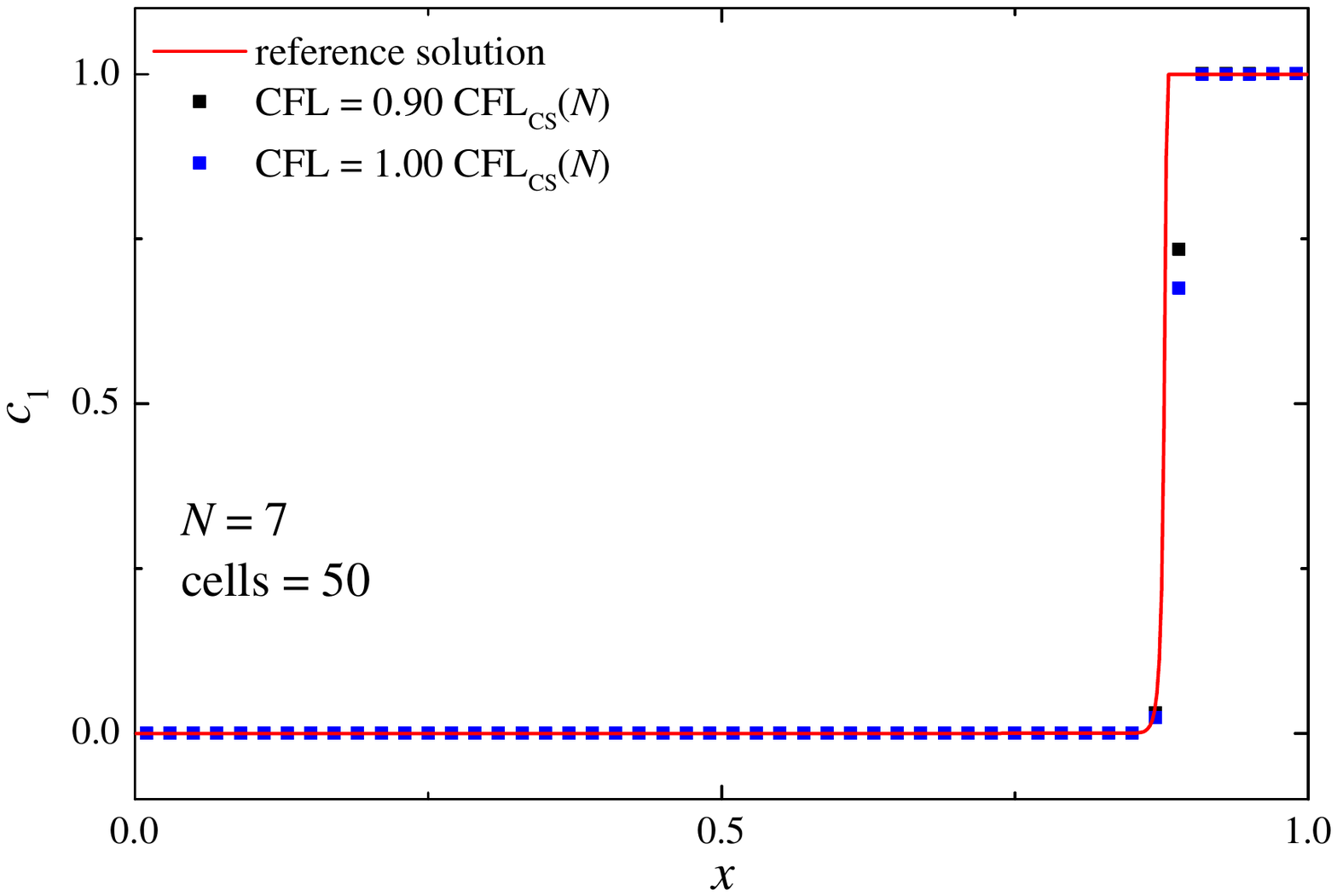}
\vspace{-8mm}\caption{\label{fig:test_zndf_1d_degrees_7_9:degree_7_c1}}
\end{subfigure}
\begin{subfigure}{0.24\textwidth}
\includegraphics[width=\textwidth]{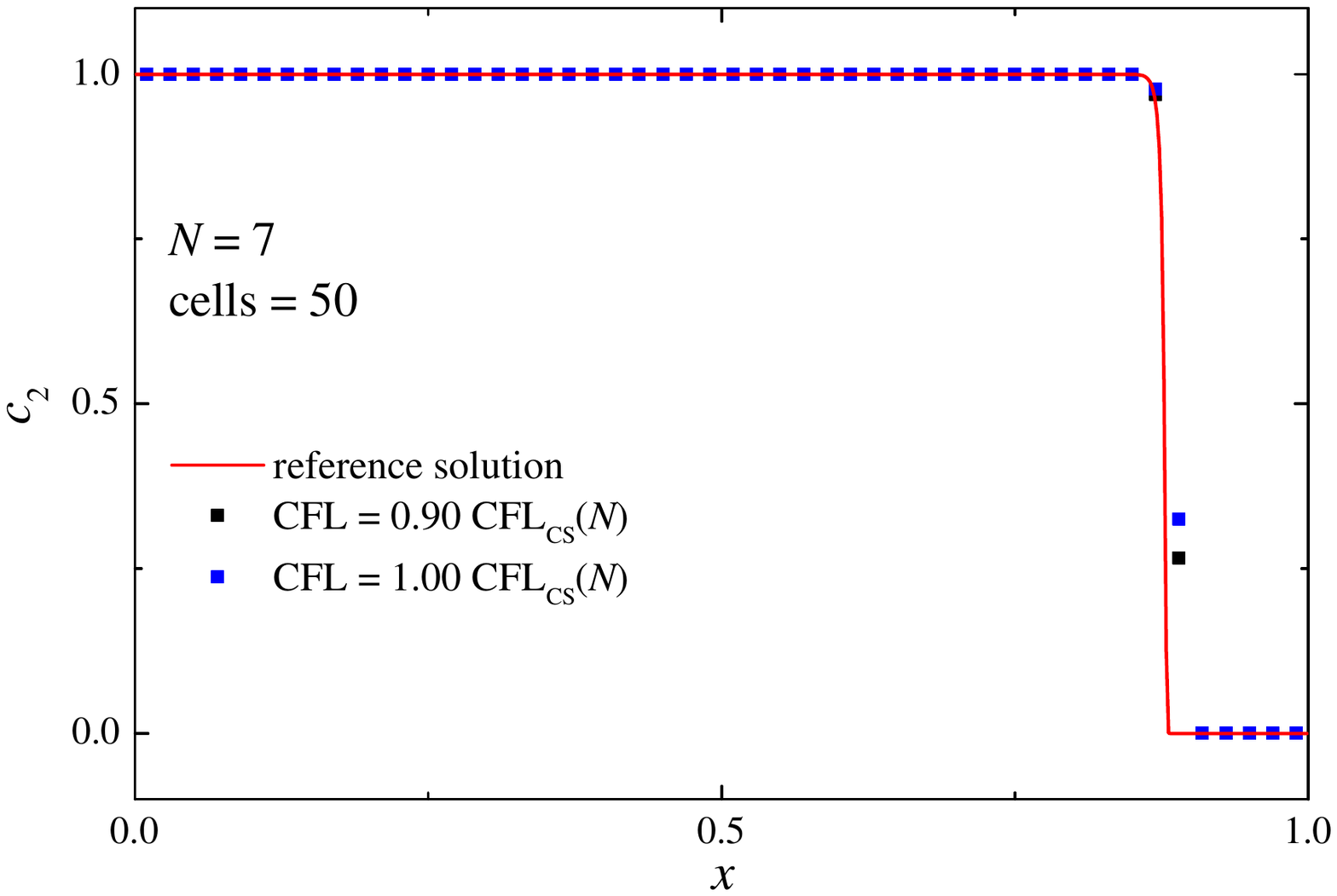}
\vspace{-8mm}\caption{\label{fig:test_zndf_1d_degrees_7_9:degree_7_c2}}
\end{subfigure}
\begin{subfigure}{0.24\textwidth}
\includegraphics[width=\textwidth]{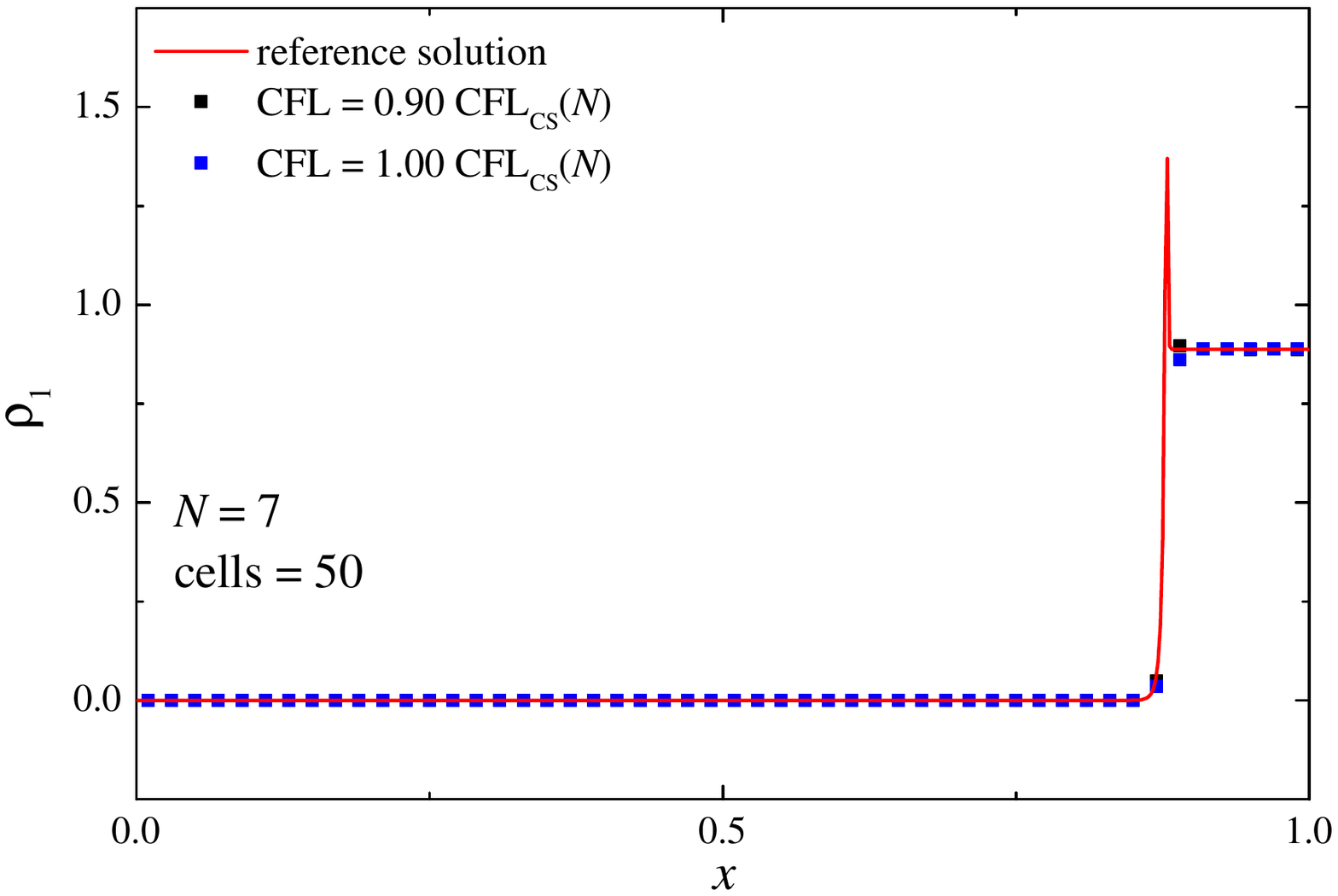}
\vspace{-8mm}\caption{\label{fig:test_zndf_1d_degrees_7_9:degree_7_den1}}
\end{subfigure}
\begin{subfigure}{0.24\textwidth}
\includegraphics[width=\textwidth]{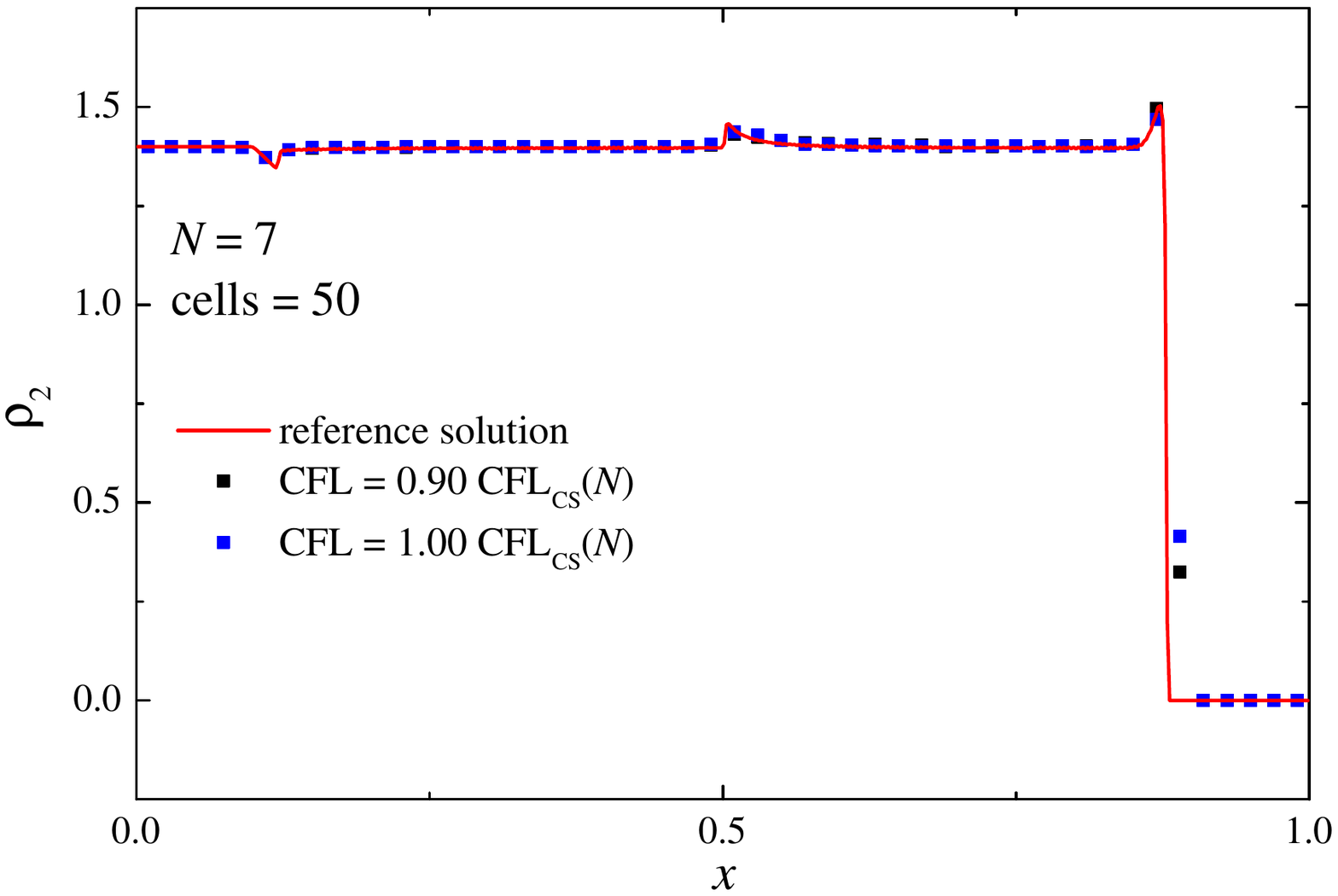}
\vspace{-8mm}\caption{\label{fig:test_zndf_1d_degrees_7_9:degree_7_den2}}
\end{subfigure}\\[-2mm]
\begin{subfigure}{0.24\textwidth}
\includegraphics[width=\textwidth]{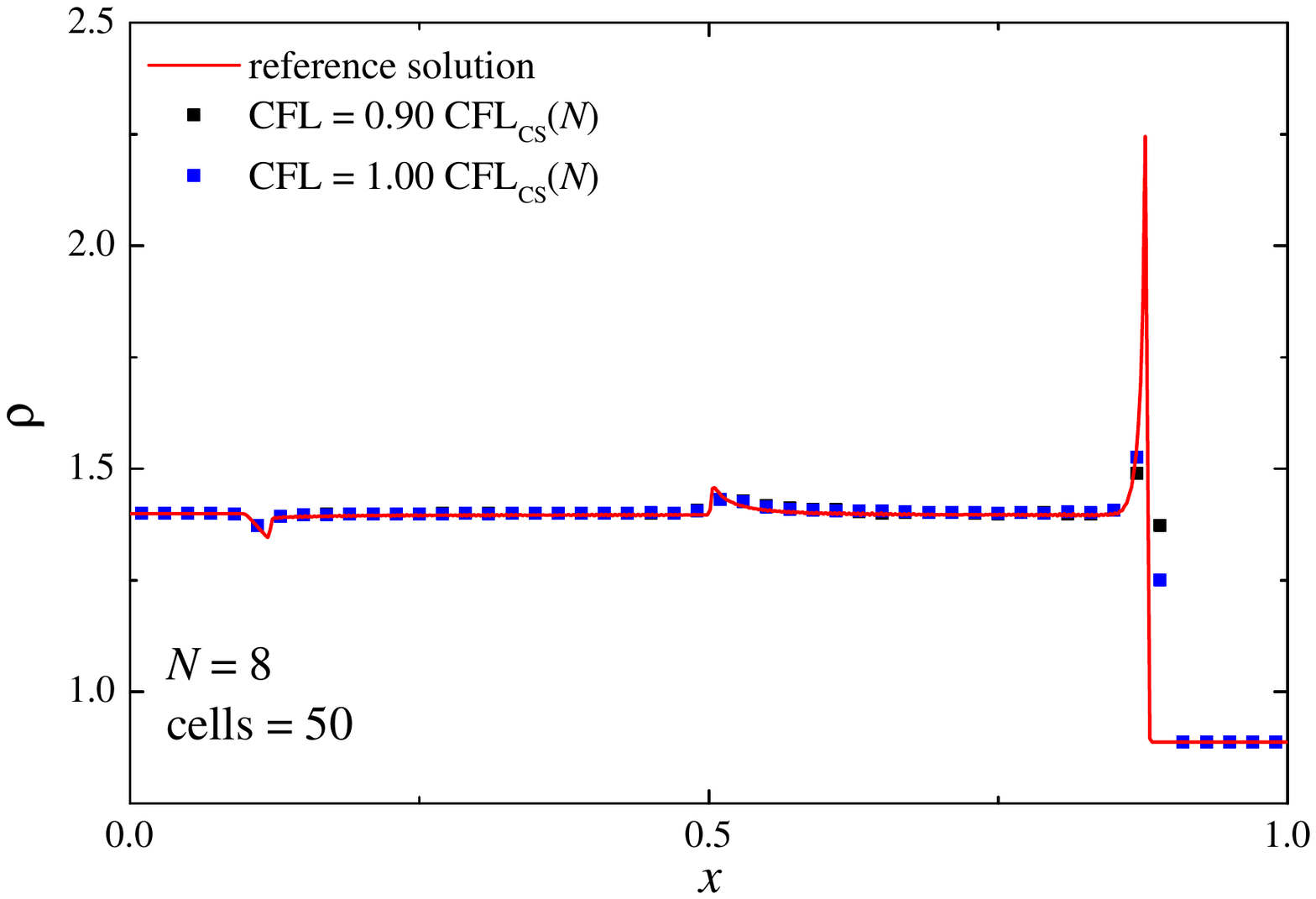}
\vspace{-8mm}\caption{\label{fig:test_zndf_1d_degrees_7_9:degree_8_den}}
\end{subfigure}
\begin{subfigure}{0.24\textwidth}
\includegraphics[width=\textwidth]{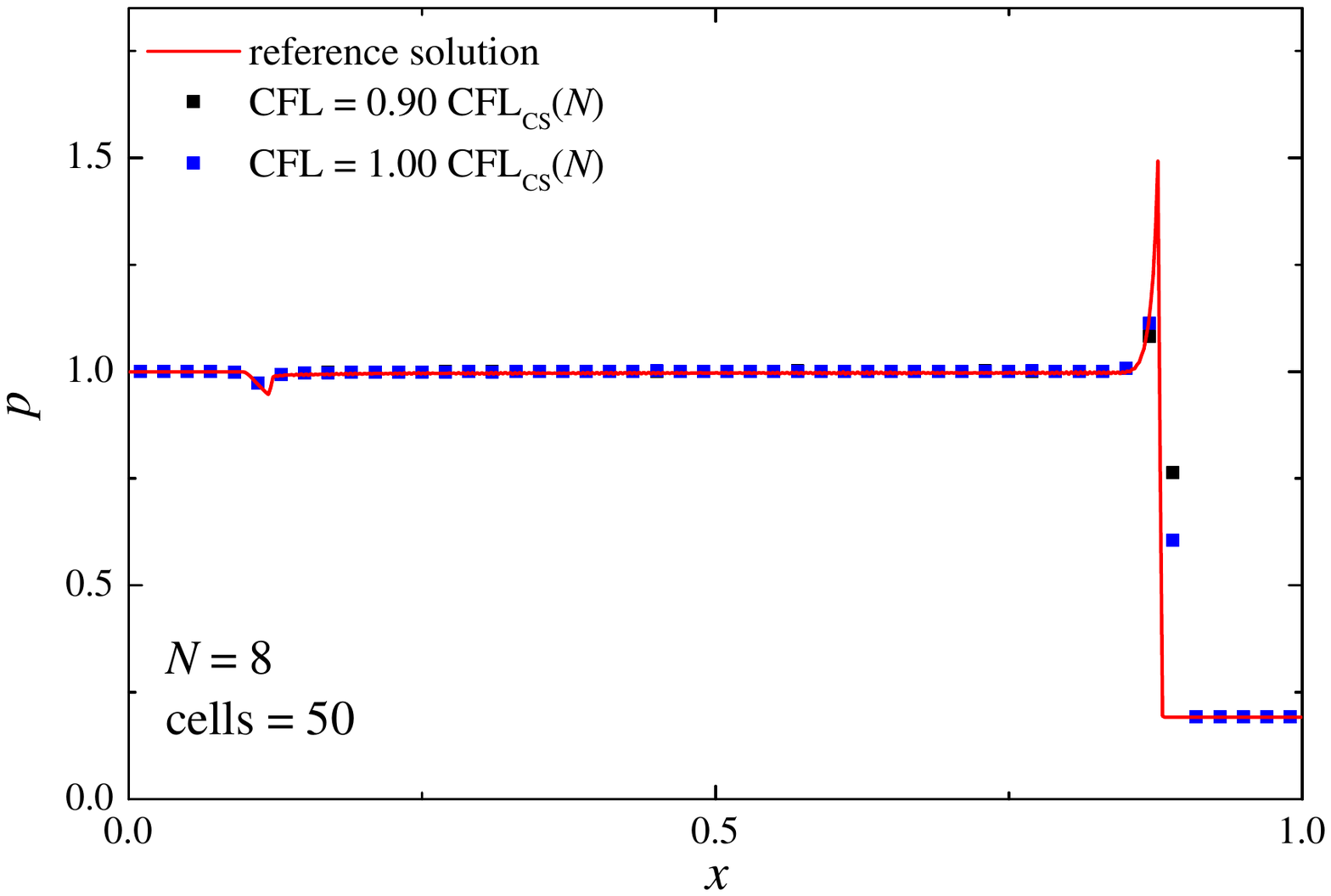}
\vspace{-8mm}\caption{\label{fig:test_zndf_1d_degrees_7_9:degree_8_p}}
\end{subfigure}
\begin{subfigure}{0.24\textwidth}
\includegraphics[width=\textwidth]{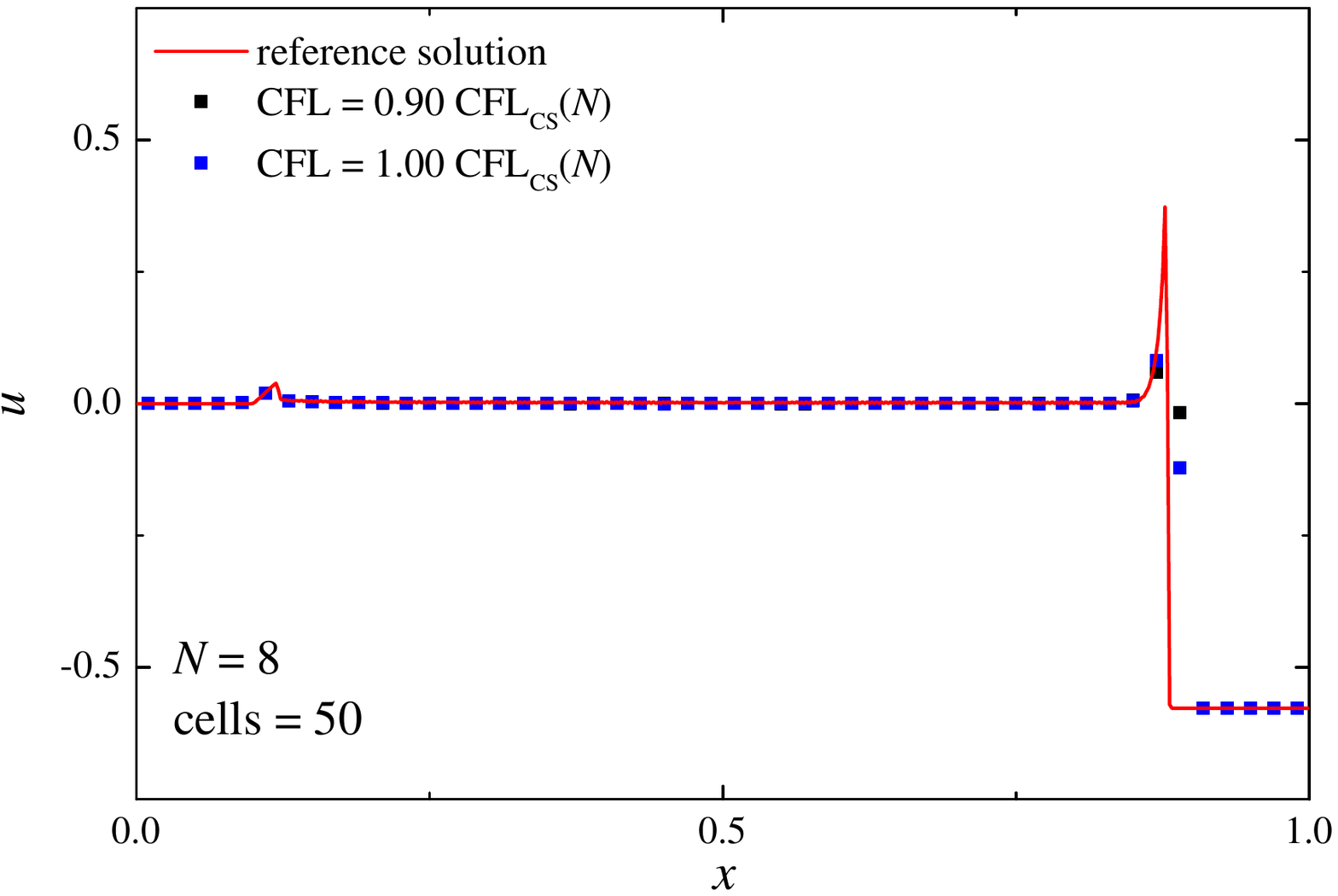}
\vspace{-8mm}\caption{\label{fig:test_zndf_1d_degrees_7_9:degree_8_u}}
\end{subfigure}
\begin{subfigure}{0.24\textwidth}
\includegraphics[width=\textwidth]{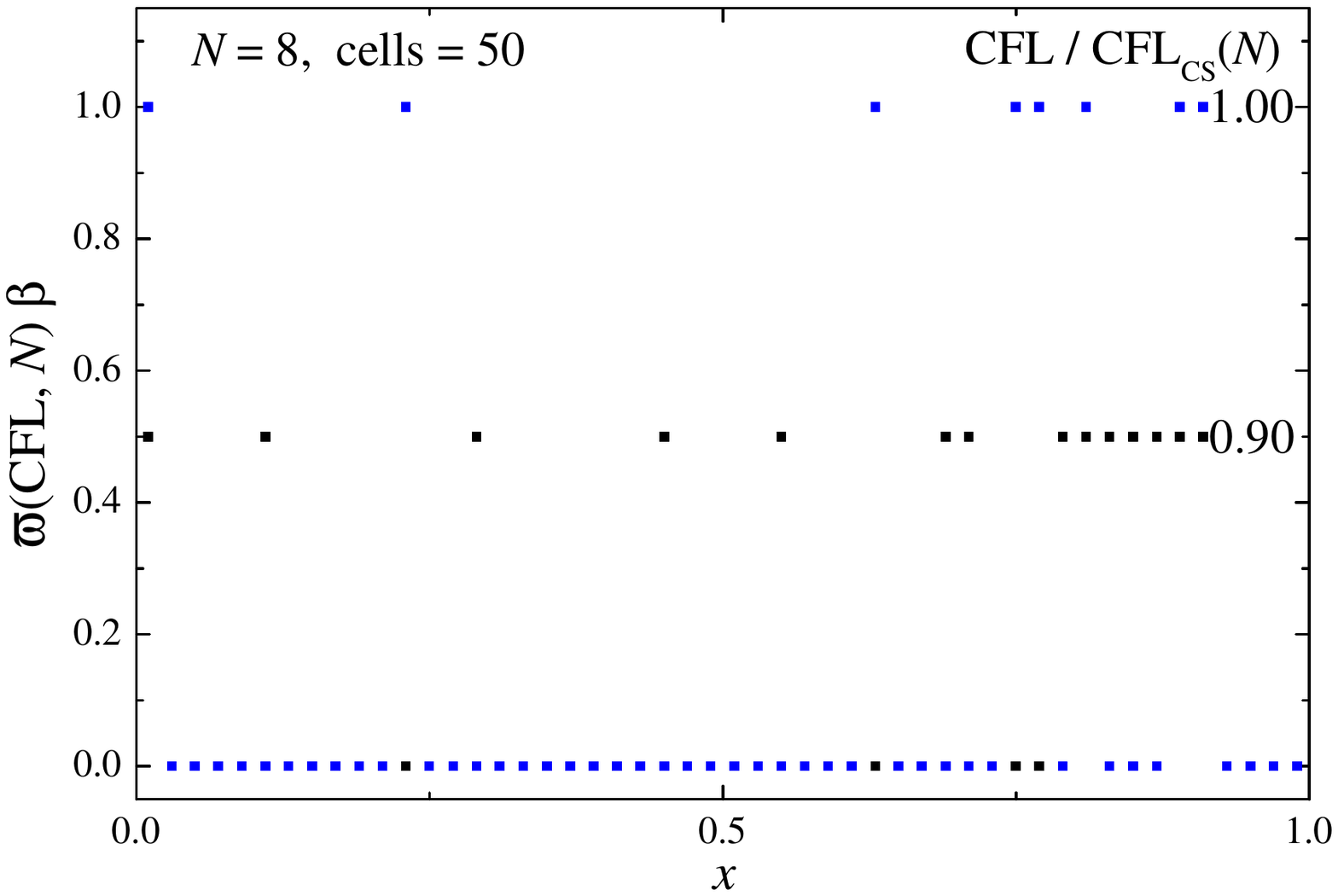}
\vspace{-8mm}\caption{\label{fig:test_zndf_1d_degrees_7_9:degree_8_beta}}
\end{subfigure}\\[-2mm]
\begin{subfigure}{0.24\textwidth}
\includegraphics[width=\textwidth]{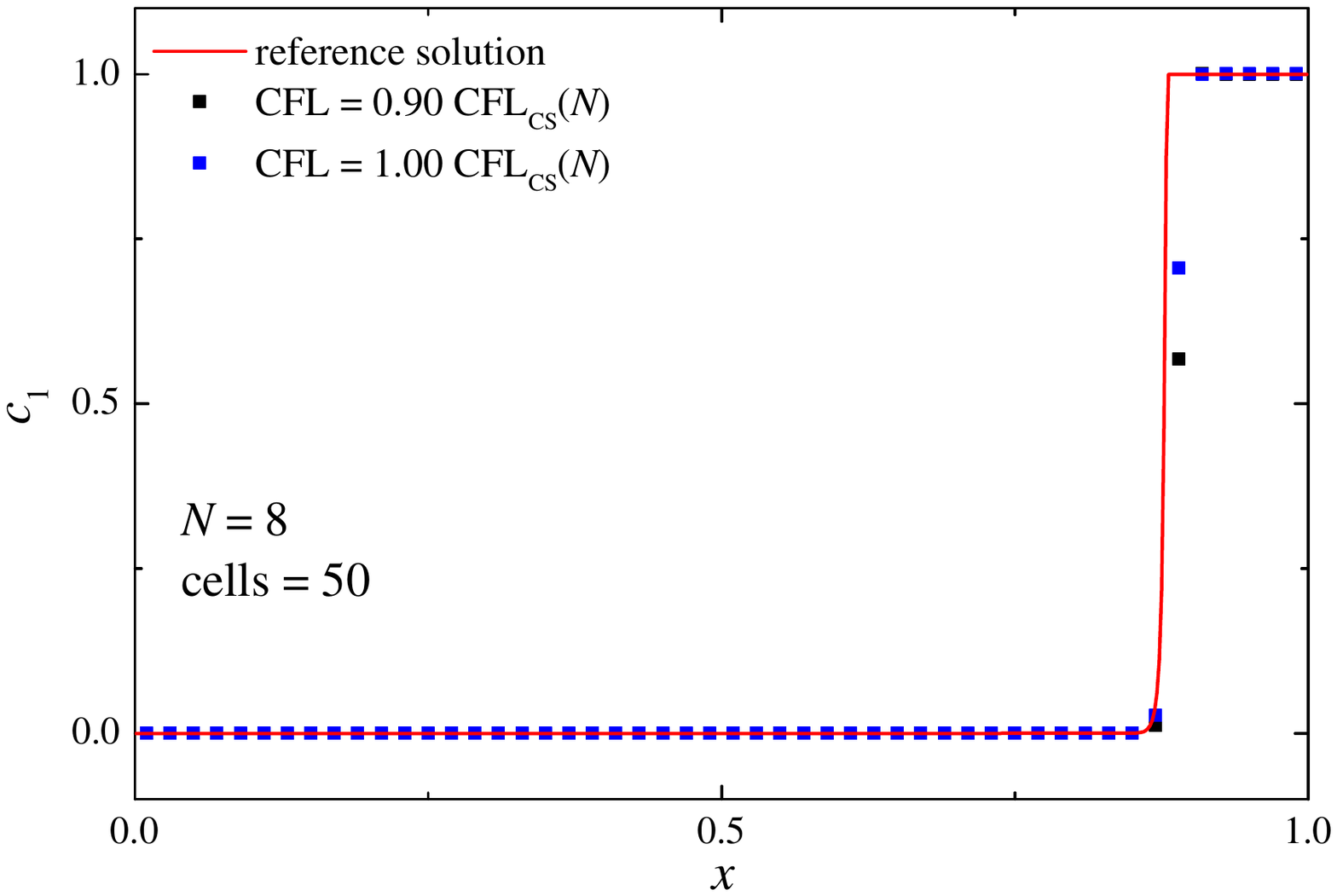}
\vspace{-8mm}\caption{\label{fig:test_zndf_1d_degrees_7_9:degree_8_c1}}
\end{subfigure}
\begin{subfigure}{0.24\textwidth}
\includegraphics[width=\textwidth]{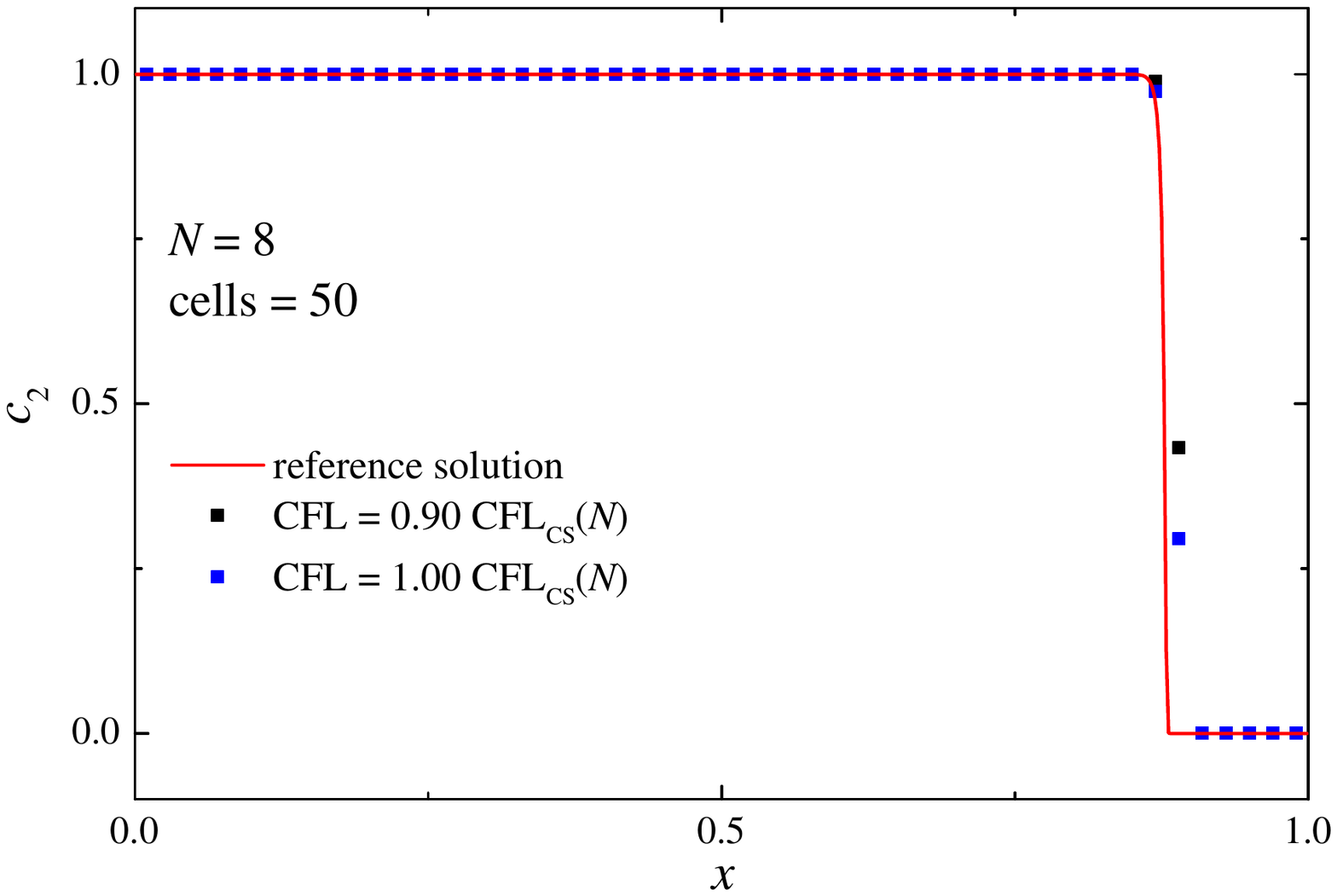}
\vspace{-8mm}\caption{\label{fig:test_zndf_1d_degrees_7_9:degree_8_c2}}
\end{subfigure}
\begin{subfigure}{0.24\textwidth}
\includegraphics[width=\textwidth]{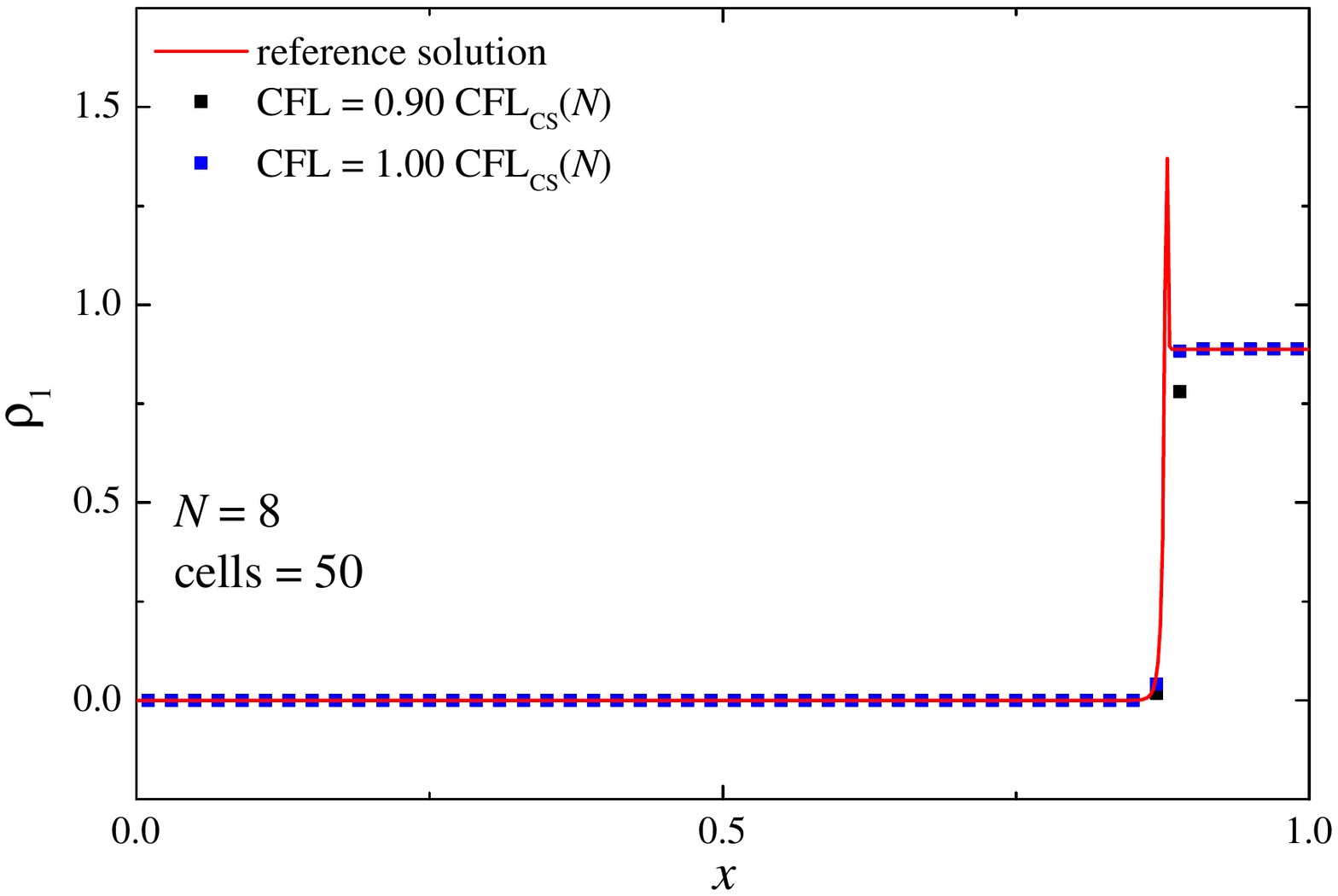}
\vspace{-8mm}\caption{\label{fig:test_zndf_1d_degrees_7_9:degree_8_den1}}
\end{subfigure}
\begin{subfigure}{0.24\textwidth}
\includegraphics[width=\textwidth]{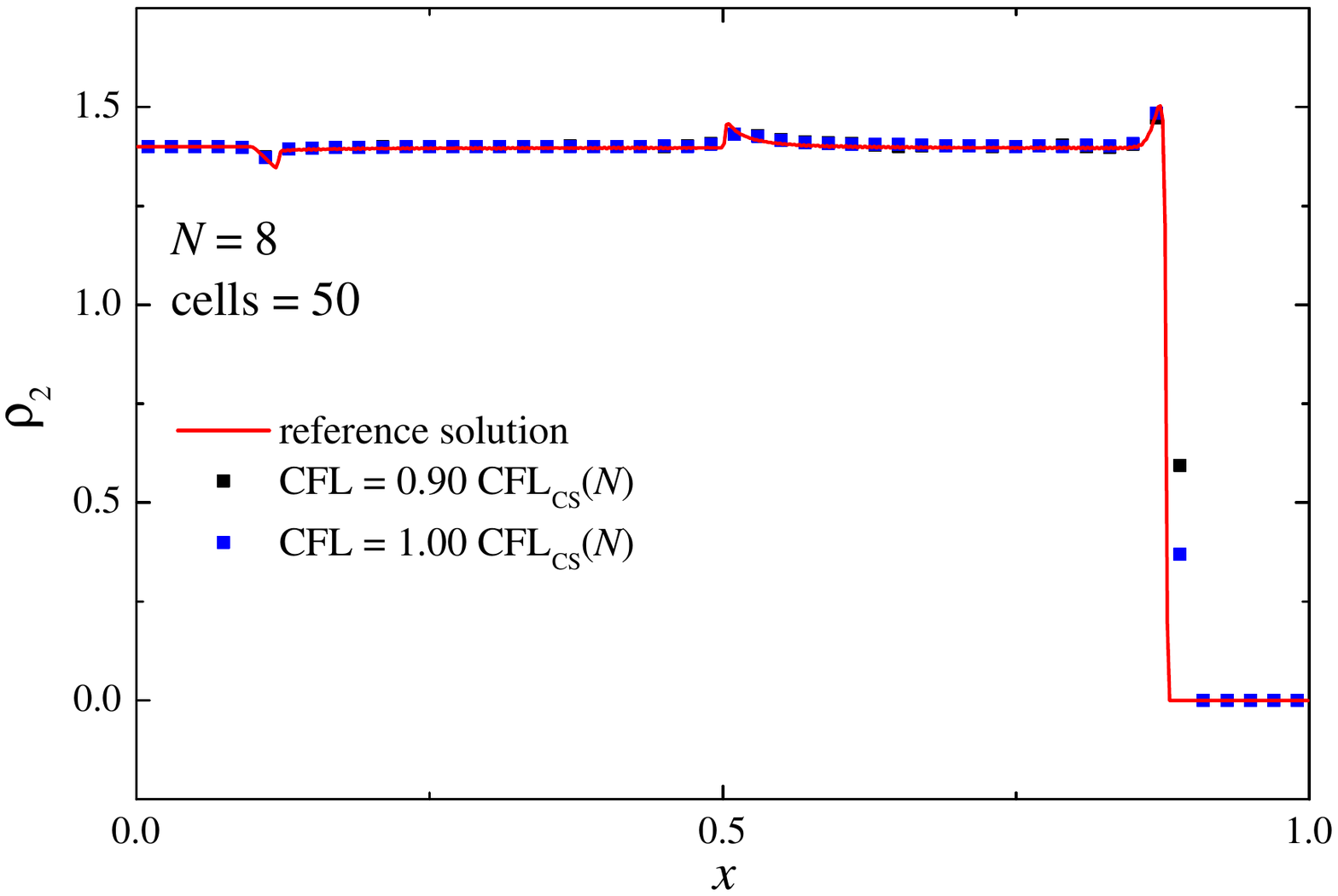}
\vspace{-8mm}\caption{\label{fig:test_zndf_1d_degrees_7_9:degree_8_den2}}
\end{subfigure}\\[-2mm]
\begin{subfigure}{0.24\textwidth}
\includegraphics[width=\textwidth]{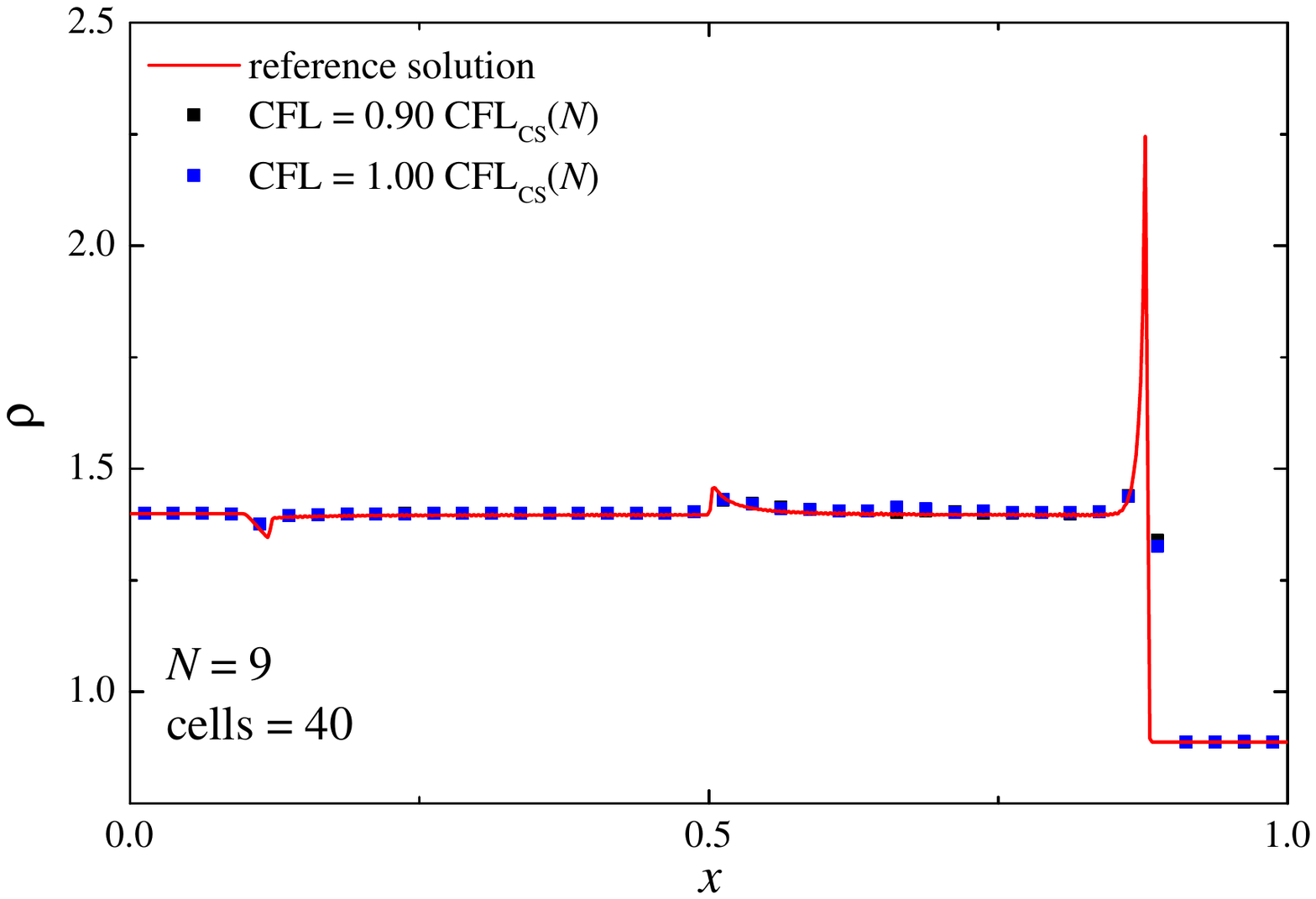}
\vspace{-8mm}\caption{\label{fig:test_zndf_1d_degrees_7_9:degree_9_den}}
\end{subfigure}
\begin{subfigure}{0.24\textwidth}
\includegraphics[width=\textwidth]{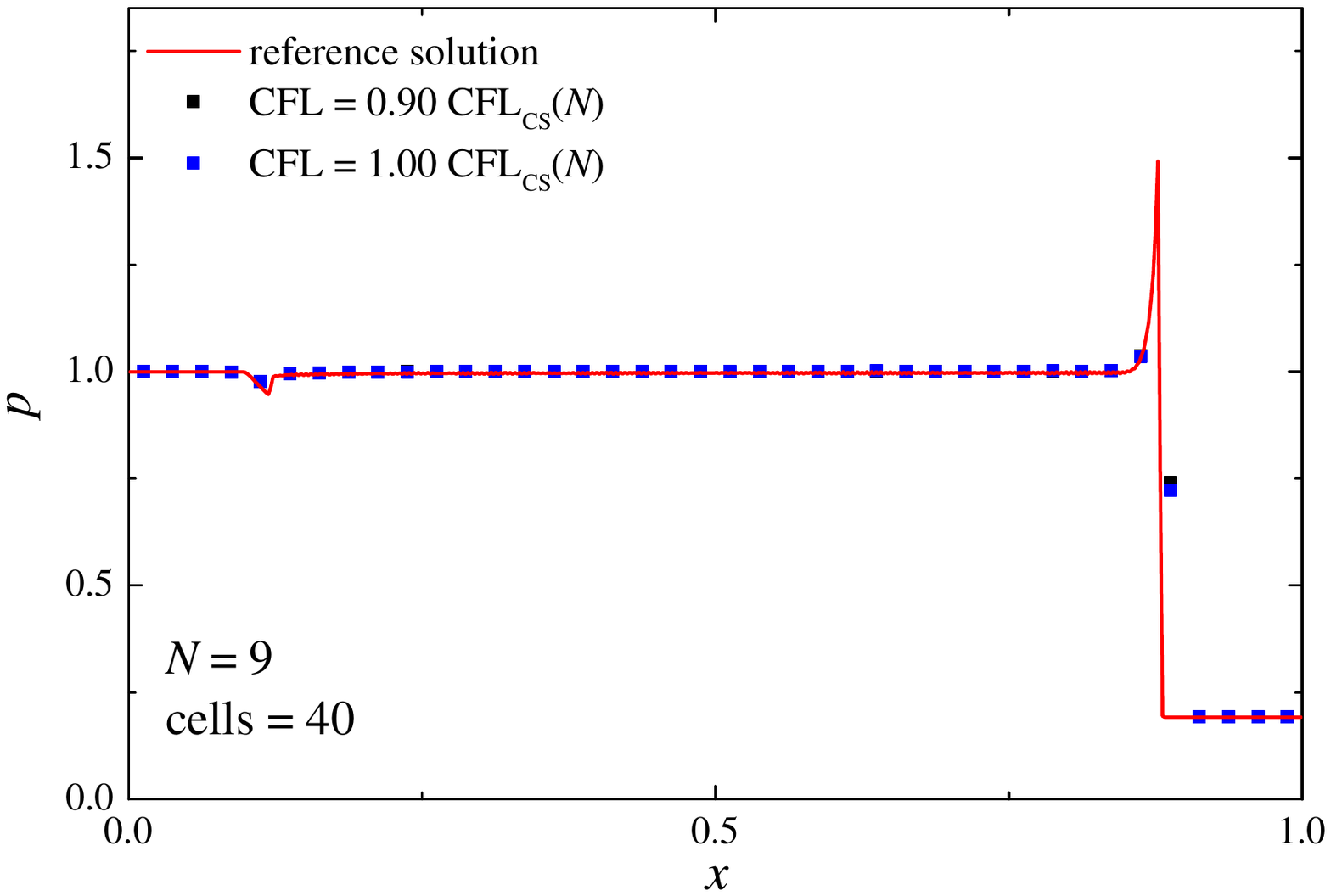}
\vspace{-8mm}\caption{\label{fig:test_zndf_1d_degrees_7_9:degree_9_p}}
\end{subfigure}
\begin{subfigure}{0.24\textwidth}
\includegraphics[width=\textwidth]{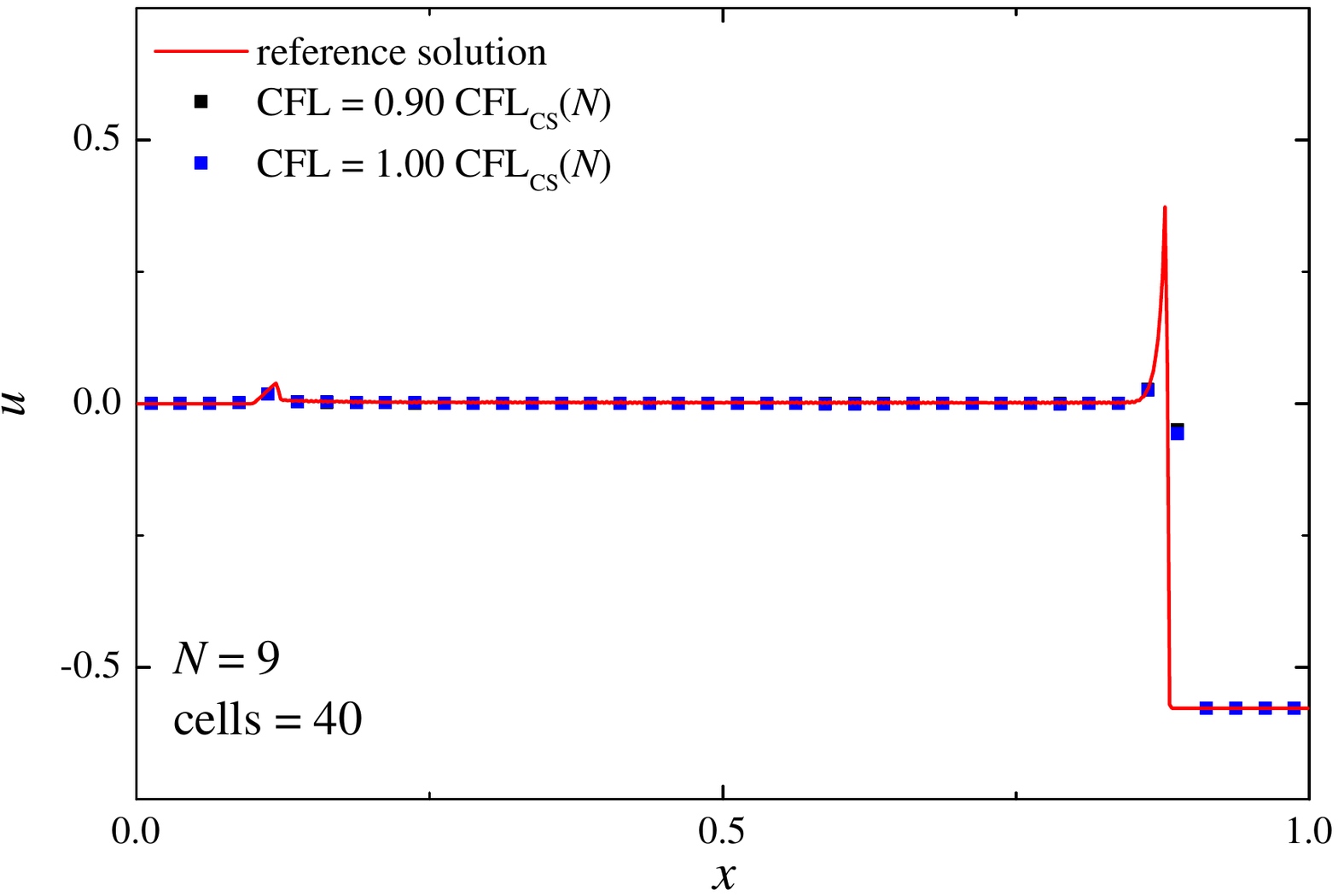}
\vspace{-8mm}\caption{\label{fig:test_zndf_1d_degrees_7_9:degree_9_u}}
\end{subfigure}
\begin{subfigure}{0.24\textwidth}
\includegraphics[width=\textwidth]{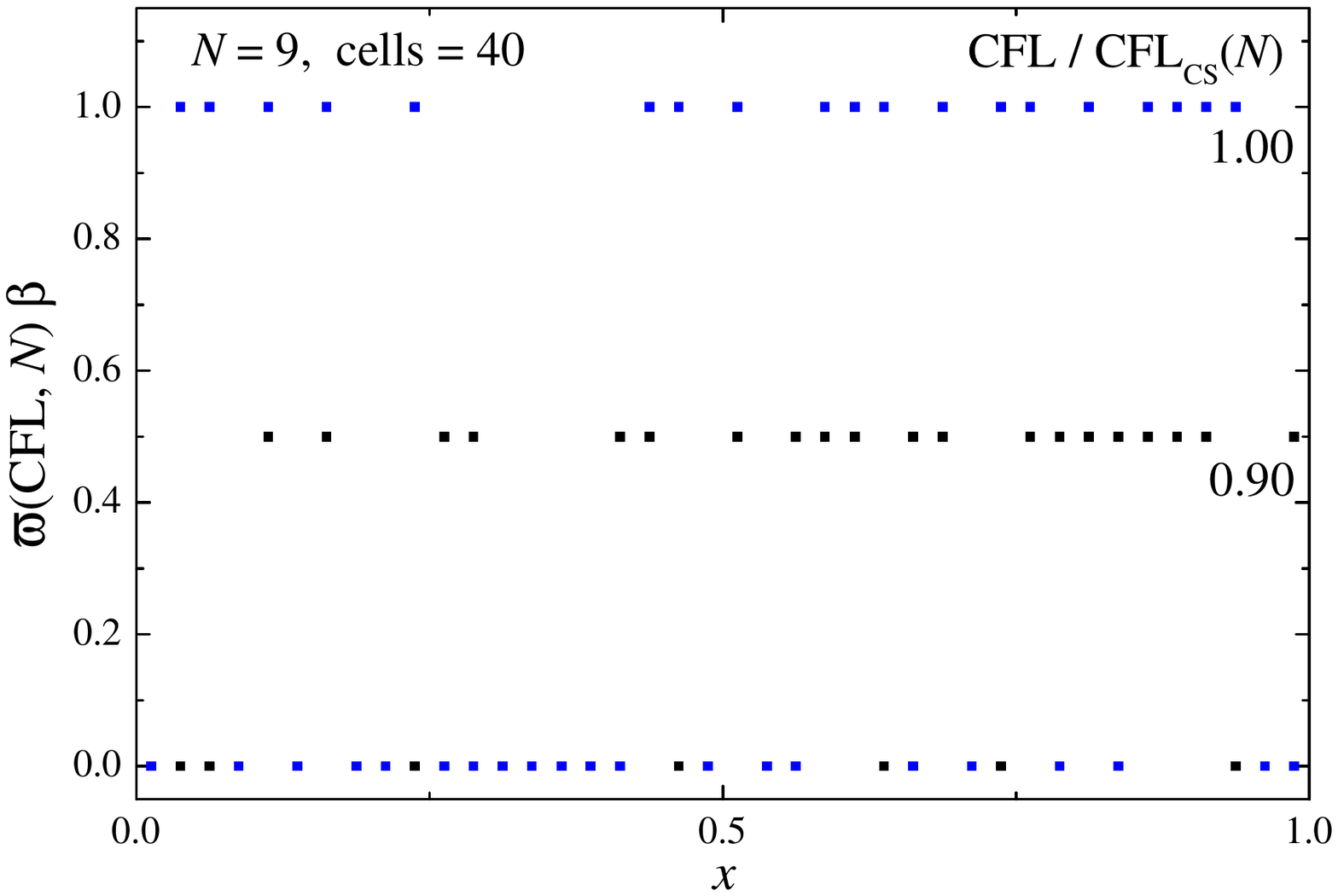}
\vspace{-8mm}\caption{\label{fig:test_zndf_1d_degrees_7_9:degree_9_beta}}
\end{subfigure}\\[-2mm]
\begin{subfigure}{0.24\textwidth}
\includegraphics[width=\textwidth]{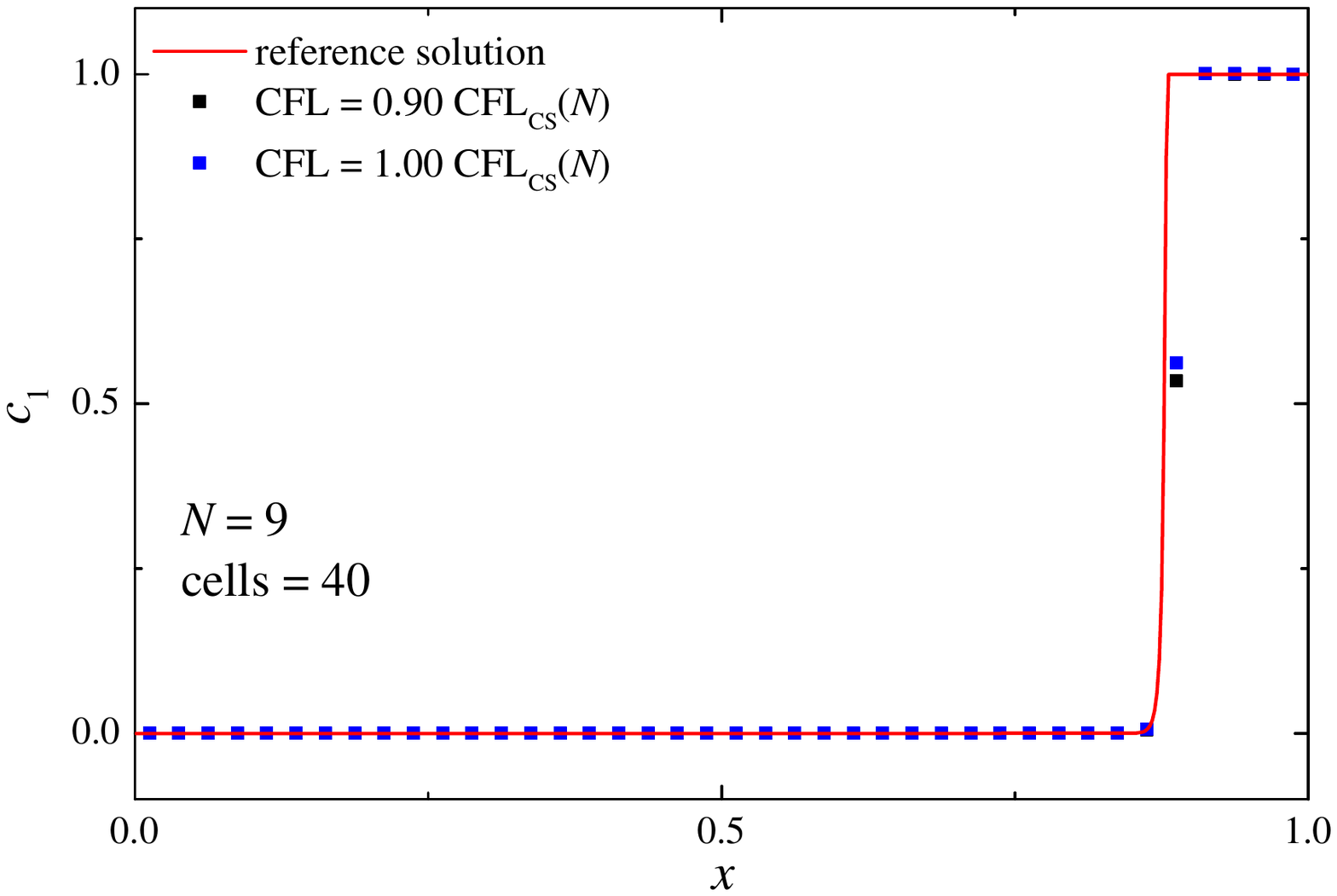}
\vspace{-8mm}\caption{\label{fig:test_zndf_1d_degrees_7_9:degree_9_c1}}
\end{subfigure}
\begin{subfigure}{0.24\textwidth}
\includegraphics[width=\textwidth]{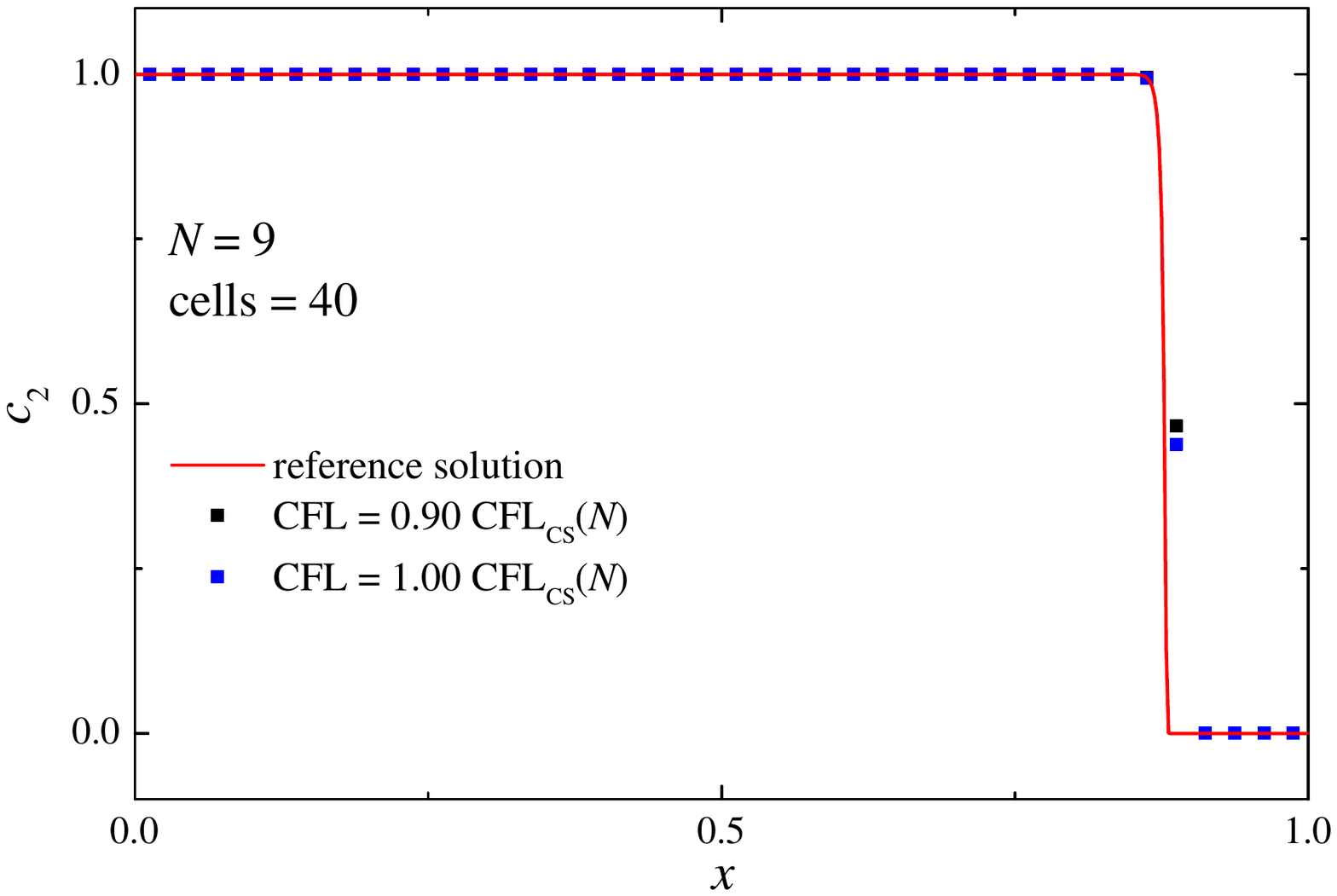}
\vspace{-8mm}\caption{\label{fig:test_zndf_1d_degrees_7_9:degree_9_c2}}
\end{subfigure}
\begin{subfigure}{0.24\textwidth}
\includegraphics[width=\textwidth]{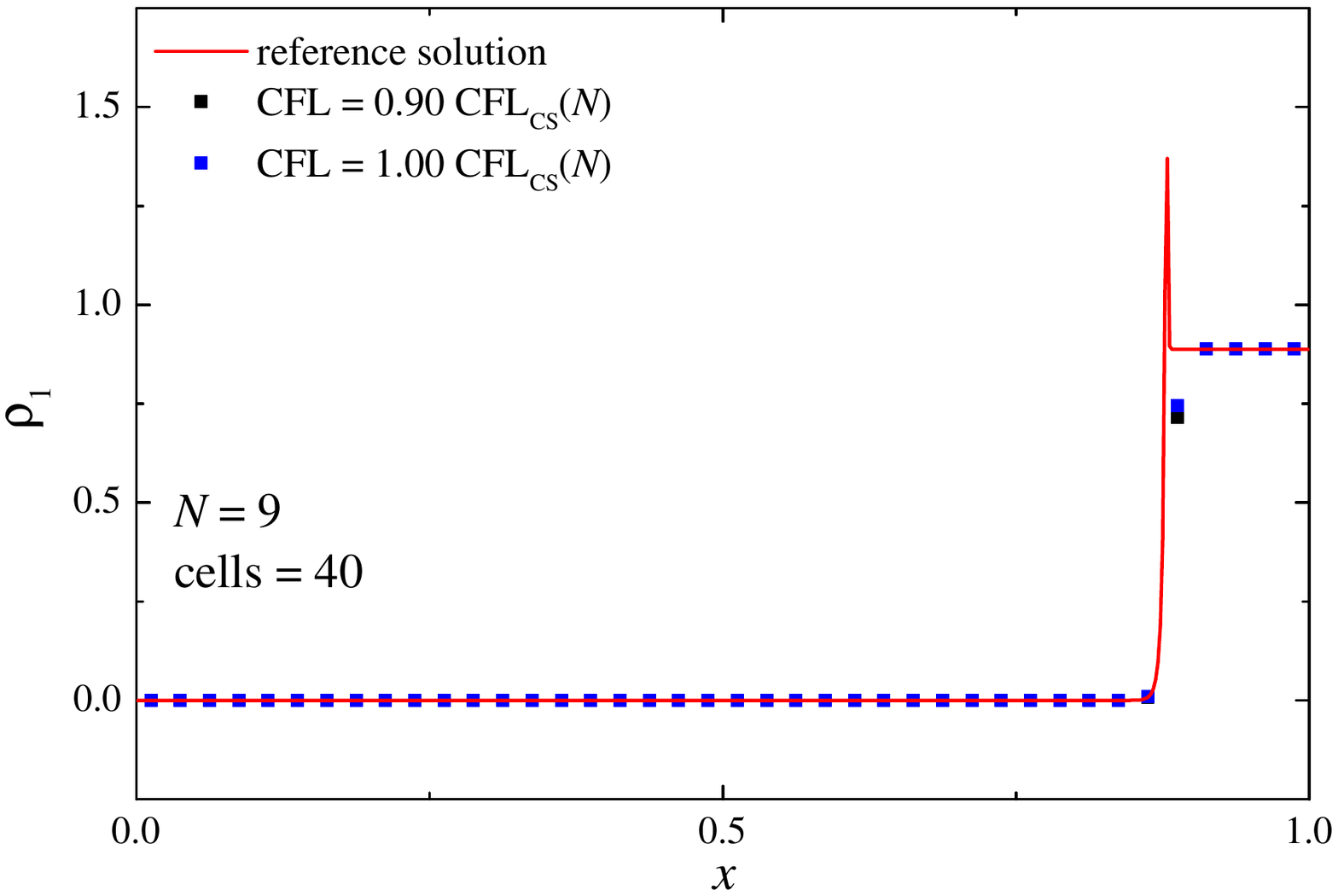}
\vspace{-8mm}\caption{\label{fig:test_zndf_1d_degrees_7_9:degree_9_den1}}
\end{subfigure}
\begin{subfigure}{0.24\textwidth}
\includegraphics[width=\textwidth]{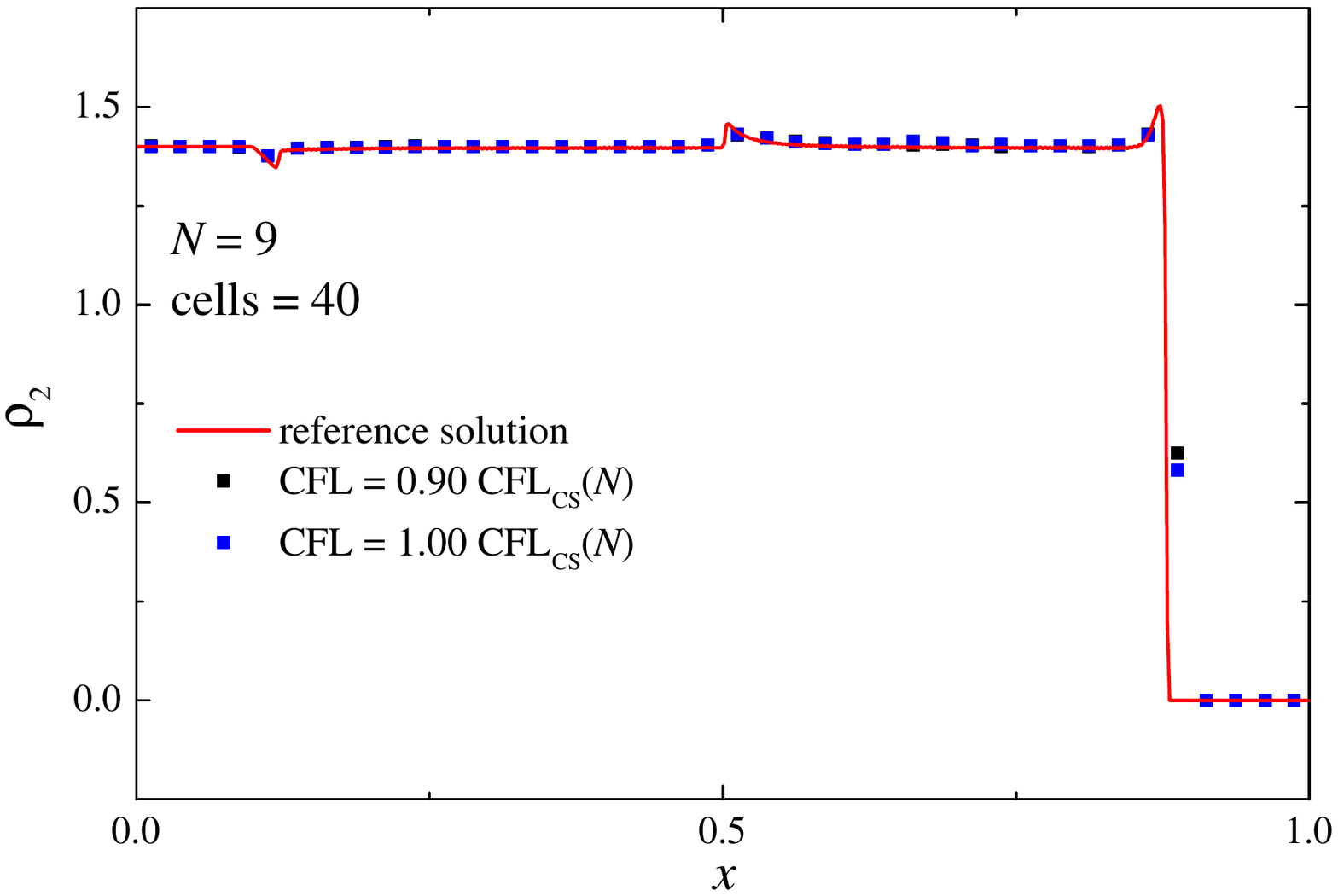}
\vspace{-8mm}\caption{\label{fig:test_zndf_1d_degrees_7_9:degree_9_den2}}
\end{subfigure}
\caption{%
Numerical solution of the one-dimensional problem of the plane detonation wave formation in a two-component medium with a ``fast'' reaction (strong stiff case, a detailed statement of the problem is presented in the text), obtained using the ADER-DG method for the basis polynomials degree $N = 7$ ((\subref{fig:test_zndf_1d_degrees_7_9:degree_7_den}), (\subref{fig:test_zndf_1d_degrees_7_9:degree_7_p}), (\subref{fig:test_zndf_1d_degrees_7_9:degree_7_u}), (\subref{fig:test_zndf_1d_degrees_7_9:degree_7_c1}), (\subref{fig:test_zndf_1d_degrees_7_9:degree_7_c2}), (\subref{fig:test_zndf_1d_degrees_7_9:degree_7_den1}), (\subref{fig:test_zndf_1d_degrees_7_9:degree_7_den2})), $N = 8$ ((\subref{fig:test_zndf_1d_degrees_7_9:degree_8_den}), (\subref{fig:test_zndf_1d_degrees_7_9:degree_8_p}), (\subref{fig:test_zndf_1d_degrees_7_9:degree_8_u}), (\subref{fig:test_zndf_1d_degrees_7_9:degree_8_c1}), (\subref{fig:test_zndf_1d_degrees_7_9:degree_8_c2}), (\subref{fig:test_zndf_1d_degrees_7_9:degree_8_den1}), (\subref{fig:test_zndf_1d_degrees_7_9:degree_8_den2})), $N = 9$ ((\subref{fig:test_zndf_1d_degrees_7_9:degree_9_den}), (\subref{fig:test_zndf_1d_degrees_7_9:degree_9_p}), (\subref{fig:test_zndf_1d_degrees_7_9:degree_9_u}), (\subref{fig:test_zndf_1d_degrees_7_9:degree_9_c1}), (\subref{fig:test_zndf_1d_degrees_7_9:degree_9_c2}), (\subref{fig:test_zndf_1d_degrees_7_9:degree_9_den1}), (\subref{fig:test_zndf_1d_degrees_7_9:degree_9_den2})) with a posteriori limitation of the solution by an ADER-WENO2 finite-volume limiter, at the final time $t_{f} = 0.4$. The graphs show the coordinate dependence of density $\rho$ ((\subref{fig:test_zndf_1d_degrees_7_9:degree_7_den}), (\subref{fig:test_zndf_1d_degrees_7_9:degree_8_den}), (\subref{fig:test_zndf_1d_degrees_7_9:degree_9_den})), pressure $p$ ((\subref{fig:test_zndf_1d_degrees_7_9:degree_7_p}), (\subref{fig:test_zndf_1d_degrees_7_9:degree_8_p}), (\subref{fig:test_zndf_1d_degrees_7_9:degree_9_p})), flow velocity $u$ ((\subref{fig:test_zndf_1d_degrees_7_9:degree_7_u}), (\subref{fig:test_zndf_1d_degrees_7_9:degree_8_u}), (\subref{fig:test_zndf_1d_degrees_7_9:degree_9_u})), troubled cells indicator $\beta$ multiplied by the separating function $\varpi(\mathrm{CFL}, N)$ ((\subref{fig:test_zndf_1d_degrees_7_9:degree_7_beta}), (\subref{fig:test_zndf_1d_degrees_7_9:degree_8_beta}), (\subref{fig:test_zndf_1d_degrees_7_9:degree_9_beta})), mass concentrations $c_{1}$ ((\subref{fig:test_zndf_1d_degrees_7_9:degree_7_c1}), (\subref{fig:test_zndf_1d_degrees_7_9:degree_8_c1}), (\subref{fig:test_zndf_1d_degrees_7_9:degree_9_c1})) and $c_{2}$ ((\subref{fig:test_zndf_1d_degrees_7_9:degree_7_c2}), (\subref{fig:test_zndf_1d_degrees_7_9:degree_8_c2}), (\subref{fig:test_zndf_1d_degrees_7_9:degree_9_c2})) of the reactant $A$ and product $B$, respectively, and partial densities $\rho_{1}$ ((\subref{fig:test_zndf_1d_degrees_7_9:degree_7_den1}), (\subref{fig:test_zndf_1d_degrees_7_9:degree_8_den1}), (\subref{fig:test_zndf_1d_degrees_7_9:degree_9_den1})) and $\rho_{2}$ ((\subref{fig:test_zndf_1d_degrees_7_9:degree_7_den2}), (\subref{fig:test_zndf_1d_degrees_7_9:degree_8_den2}), (\subref{fig:test_zndf_1d_degrees_7_9:degree_9_den2})) of the reactant $A$ and product $B$, respectively. The square symbols represent the subcells finite-volume representation of the numerical solution; the red solid lines represents the reference solution of the problem.
}
\label{fig:test_zndf_1d_degrees_7_9}
\end{figure}
\begin{figure}[h!]
\captionsetup[subfigure]{%
	position=bottom,
	font+=smaller,
	textfont=normalfont,
	singlelinecheck=off,
	justification=raggedright
}
\centering
\begin{subfigure}{0.24\textwidth}
\includegraphics[width=\textwidth]{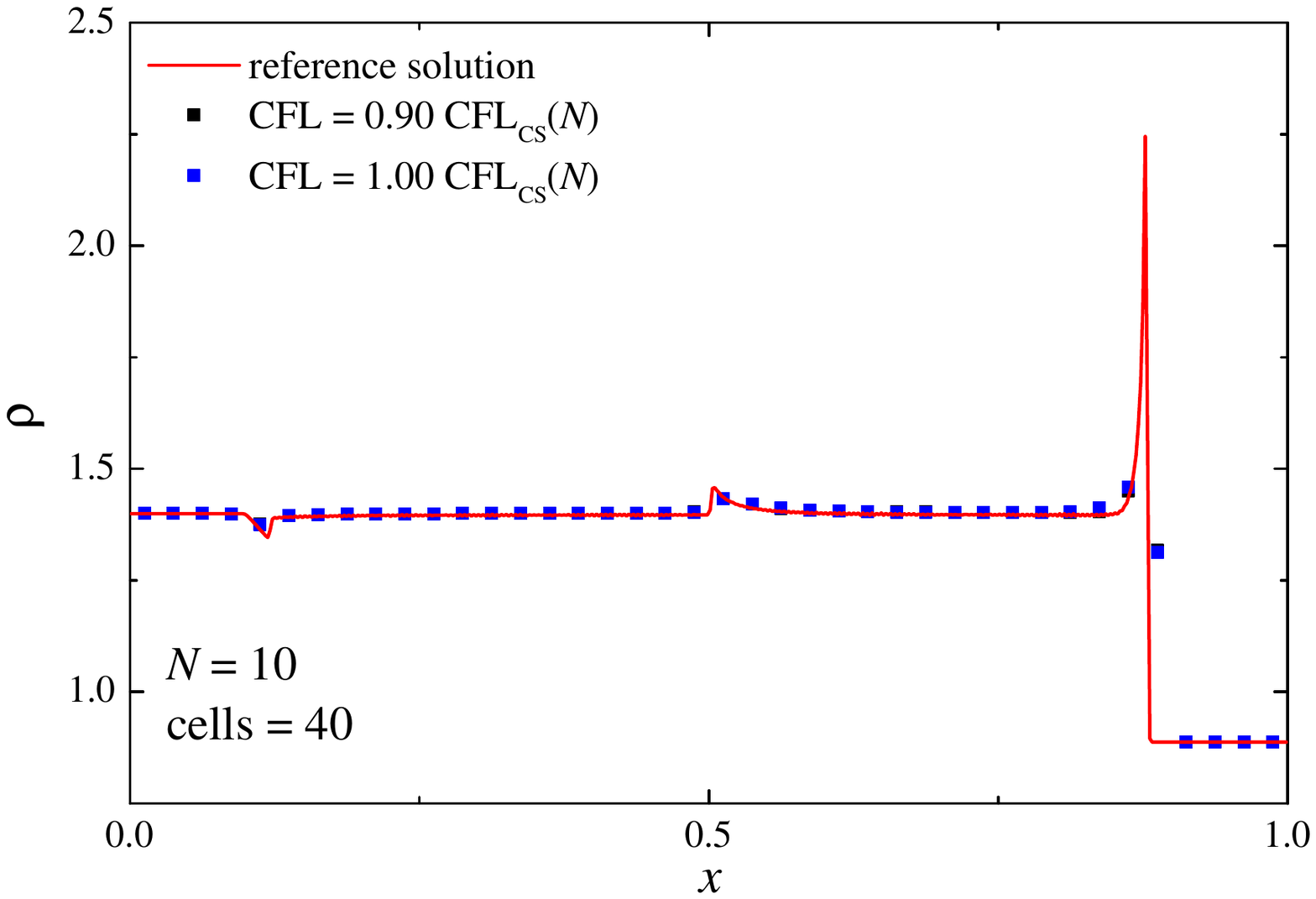}
\vspace{-8mm}\caption{\label{fig:test_zndf_1d_degrees_10_12:degree_10_den}}
\end{subfigure}
\begin{subfigure}{0.24\textwidth}
\includegraphics[width=\textwidth]{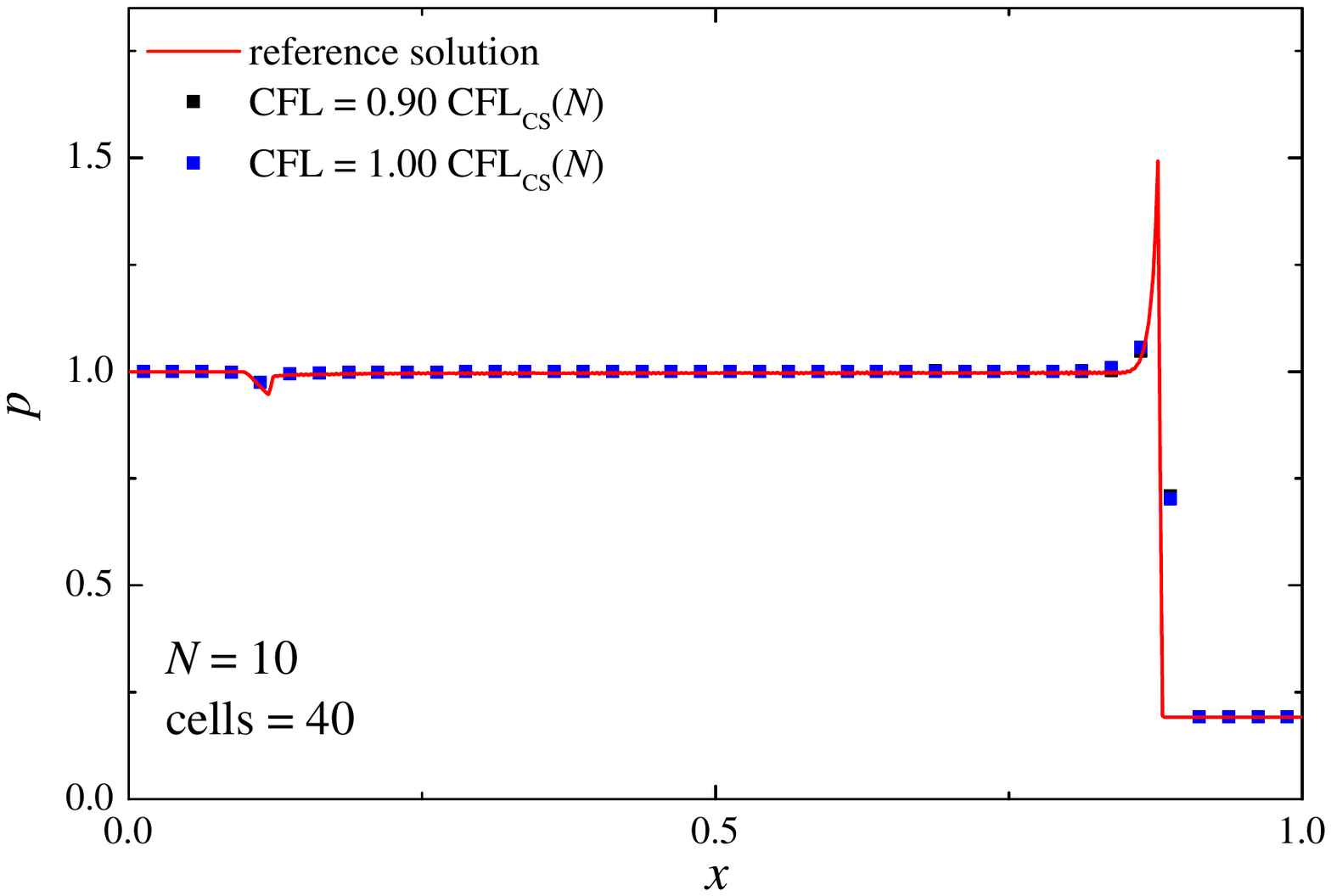}
\vspace{-8mm}\caption{\label{fig:test_zndf_1d_degrees_10_12:degree_10_p}}
\end{subfigure}
\begin{subfigure}{0.24\textwidth}
\includegraphics[width=\textwidth]{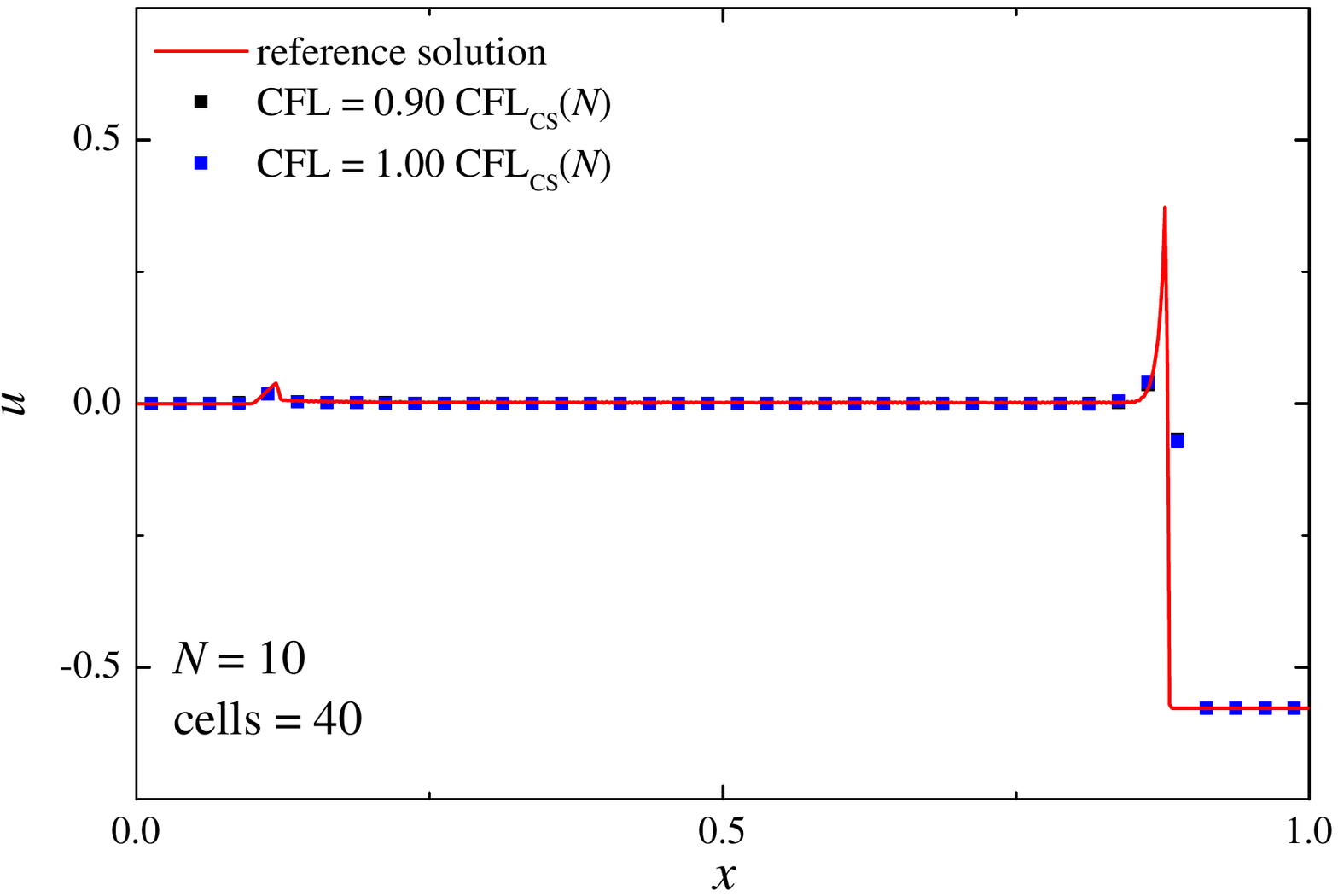}
\vspace{-8mm}\caption{\label{fig:test_zndf_1d_degrees_10_12:degree_10_u}}
\end{subfigure}
\begin{subfigure}{0.24\textwidth}
\includegraphics[width=\textwidth]{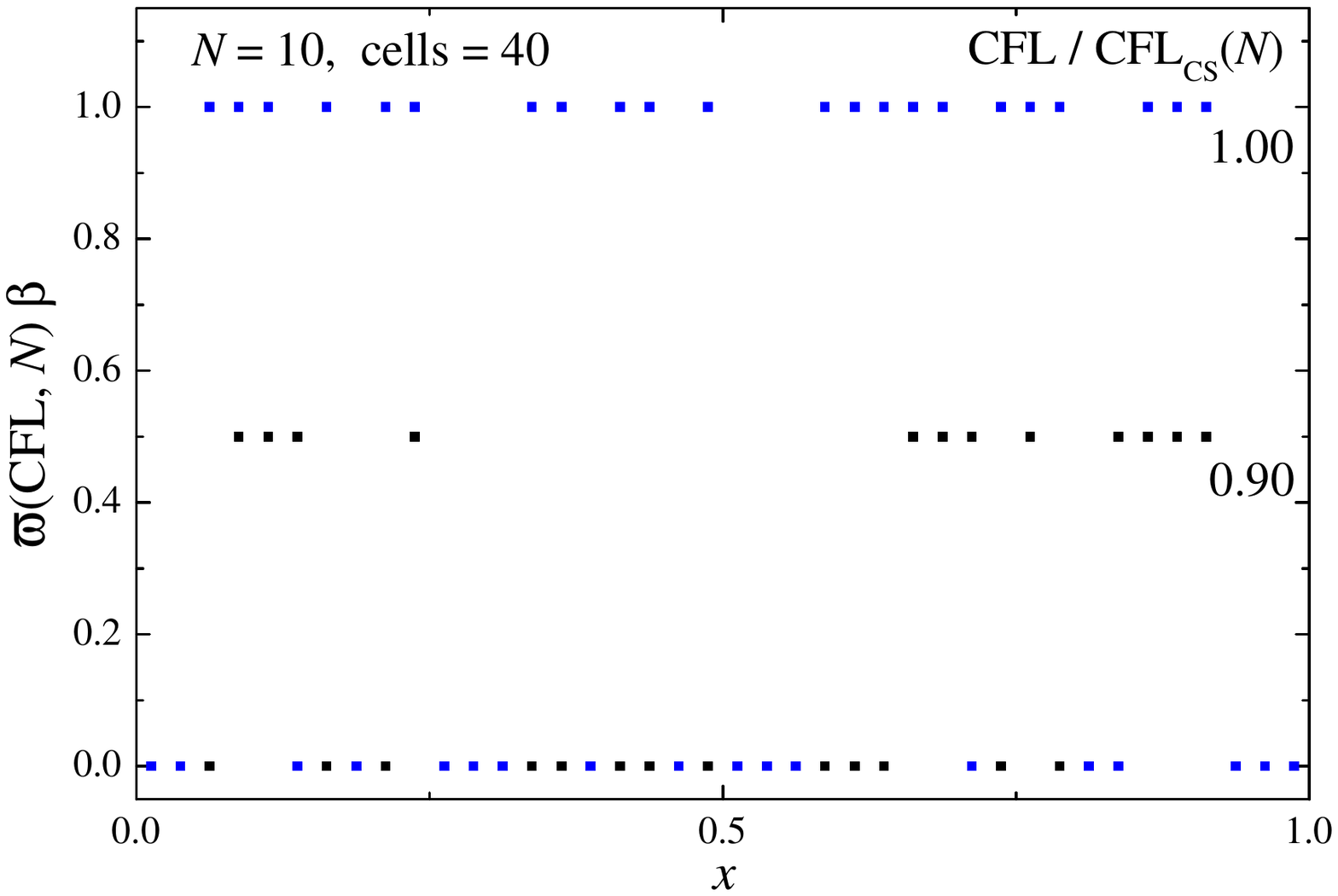}
\vspace{-8mm}\caption{\label{fig:test_zndf_1d_degrees_10_12:degree_10_beta}}
\end{subfigure}\\[-2mm]
\begin{subfigure}{0.24\textwidth}
\includegraphics[width=\textwidth]{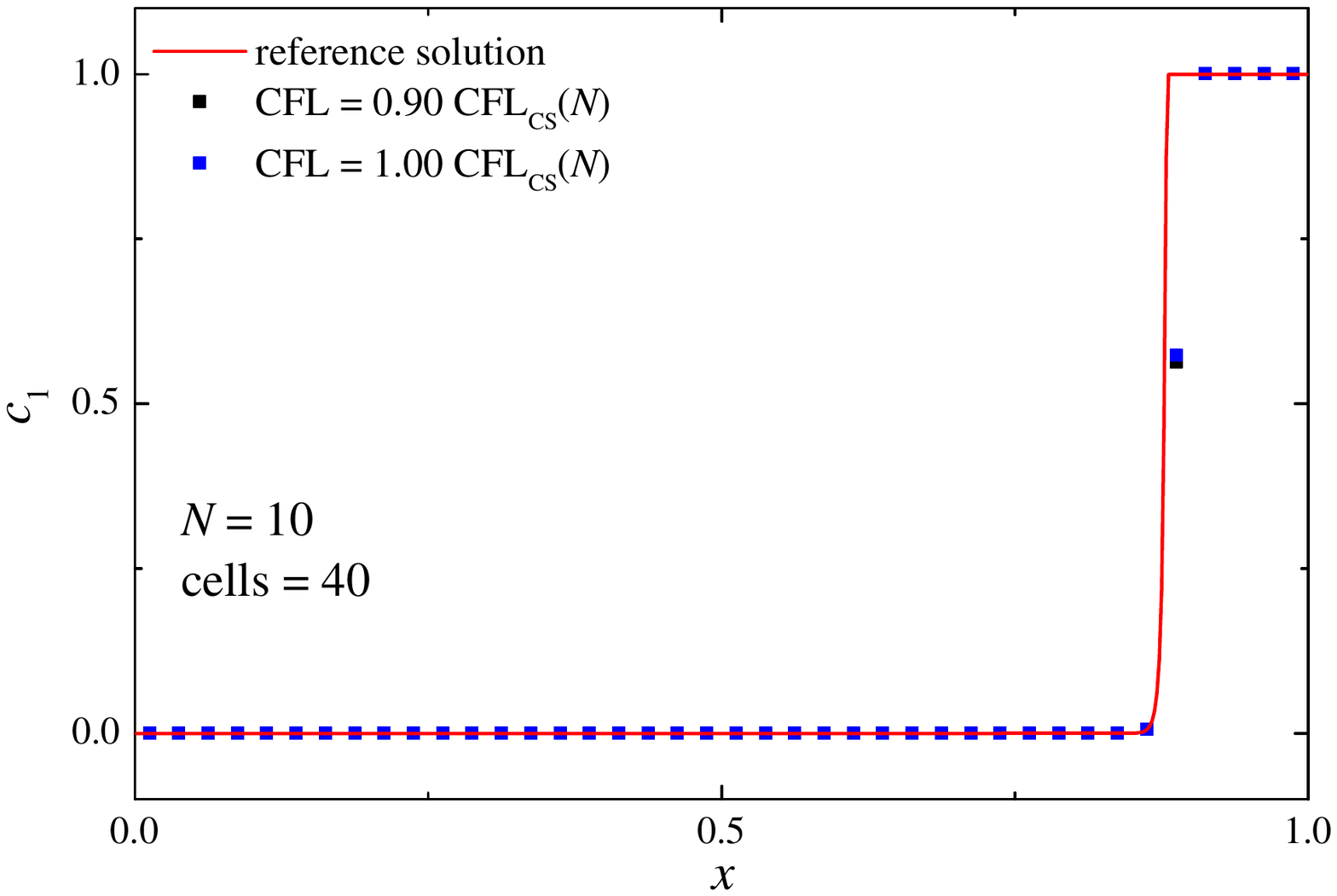}
\vspace{-8mm}\caption{\label{fig:test_zndf_1d_degrees_10_12:degree_10_c1}}
\end{subfigure}
\begin{subfigure}{0.24\textwidth}
\includegraphics[width=\textwidth]{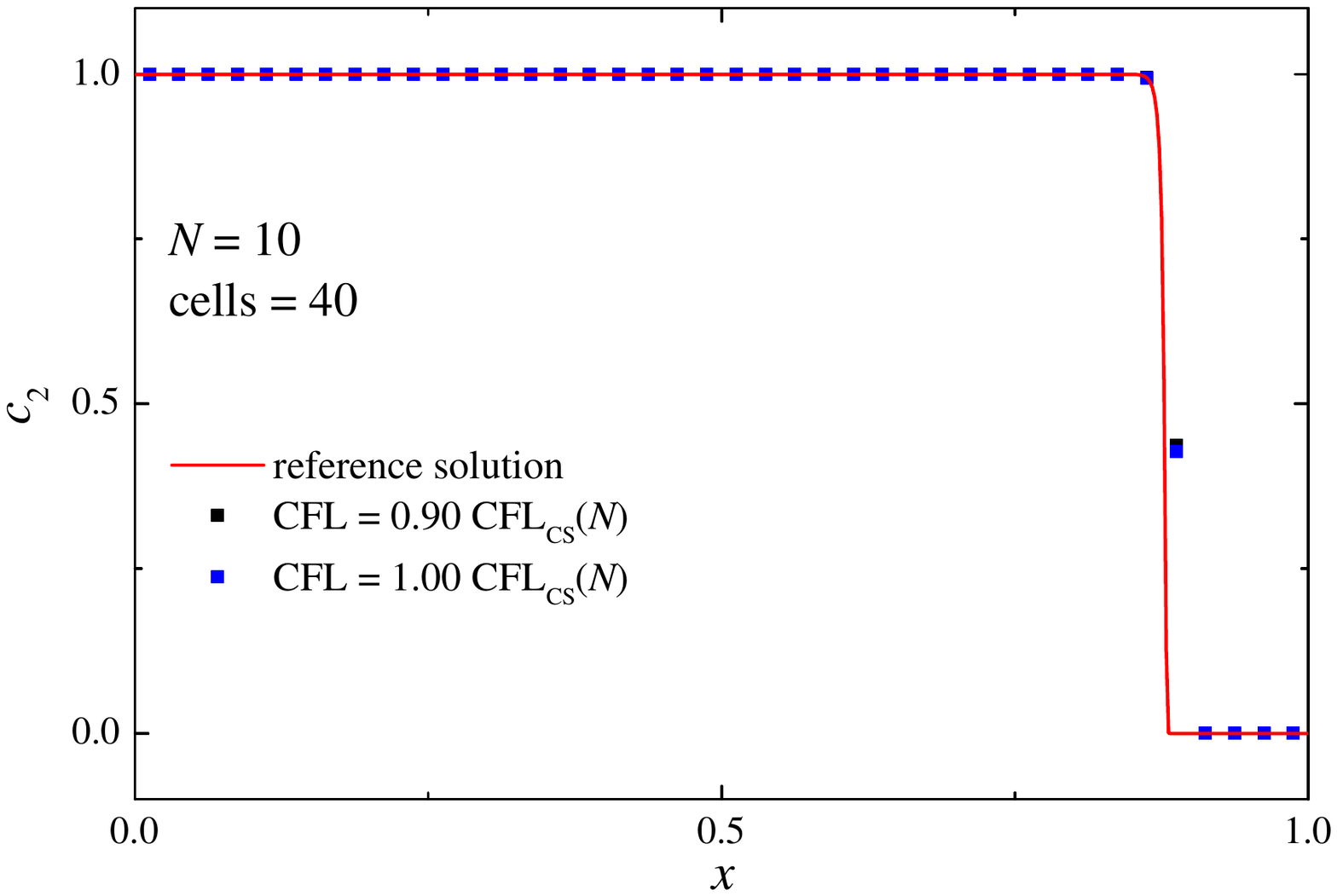}
\vspace{-8mm}\caption{\label{fig:test_zndf_1d_degrees_10_12:degree_10_c2}}
\end{subfigure}
\begin{subfigure}{0.24\textwidth}
\includegraphics[width=\textwidth]{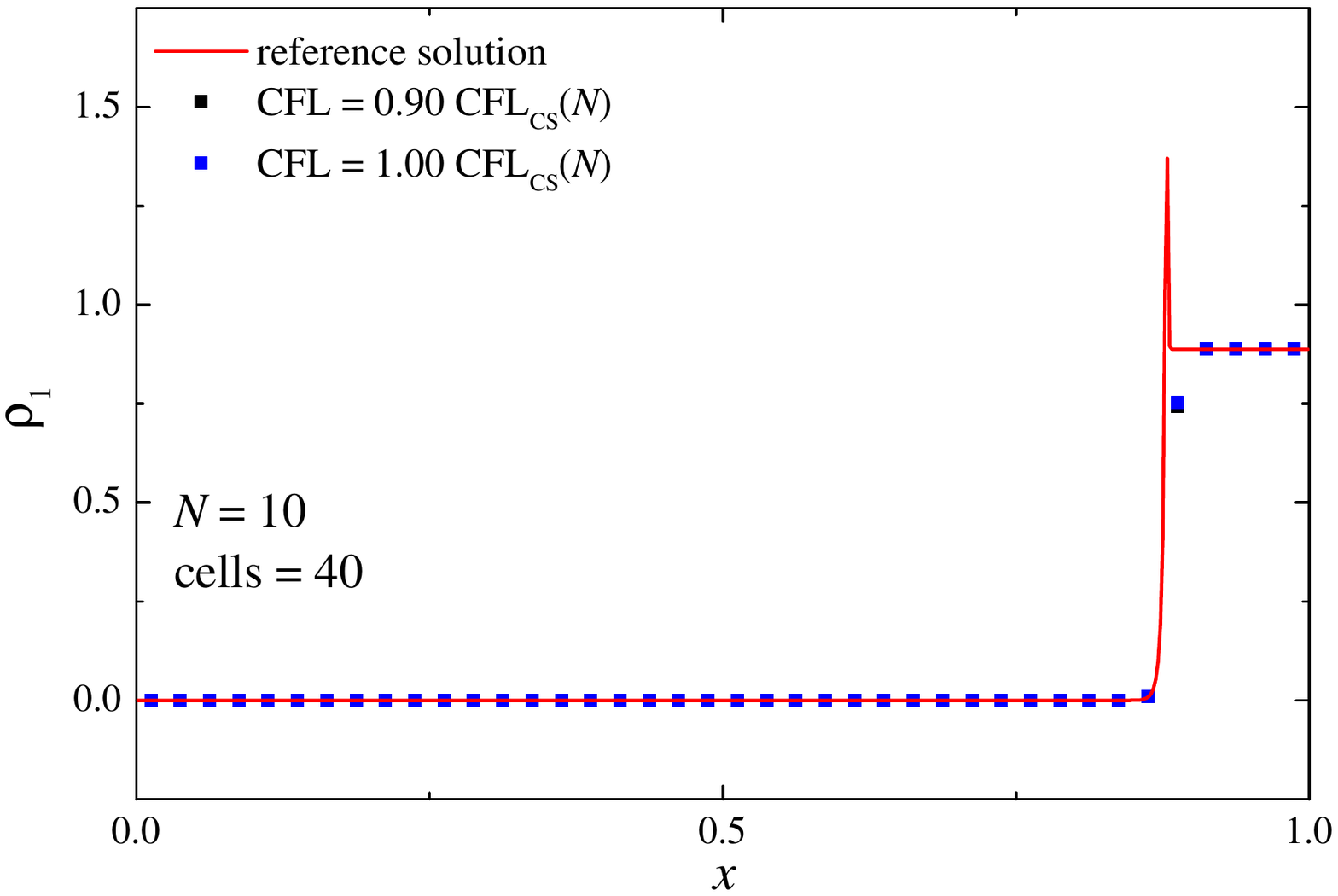}
\vspace{-8mm}\caption{\label{fig:test_zndf_1d_degrees_10_12:degree_10_den1}}
\end{subfigure}
\begin{subfigure}{0.24\textwidth}
\includegraphics[width=\textwidth]{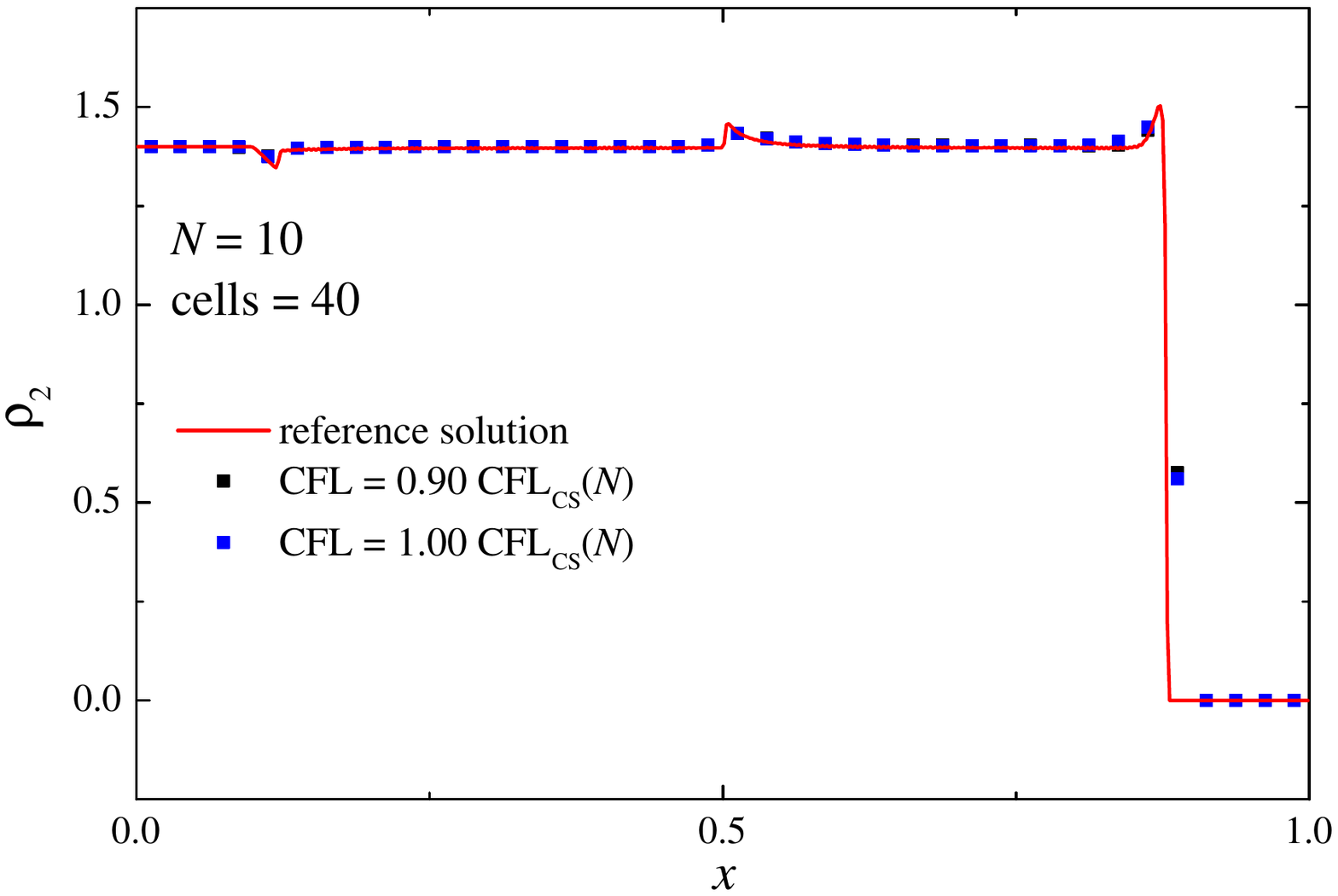}
\vspace{-8mm}\caption{\label{fig:test_zndf_1d_degrees_10_12:degree_10_den2}}
\end{subfigure}\\[-2mm]
\begin{subfigure}{0.24\textwidth}
\includegraphics[width=\textwidth]{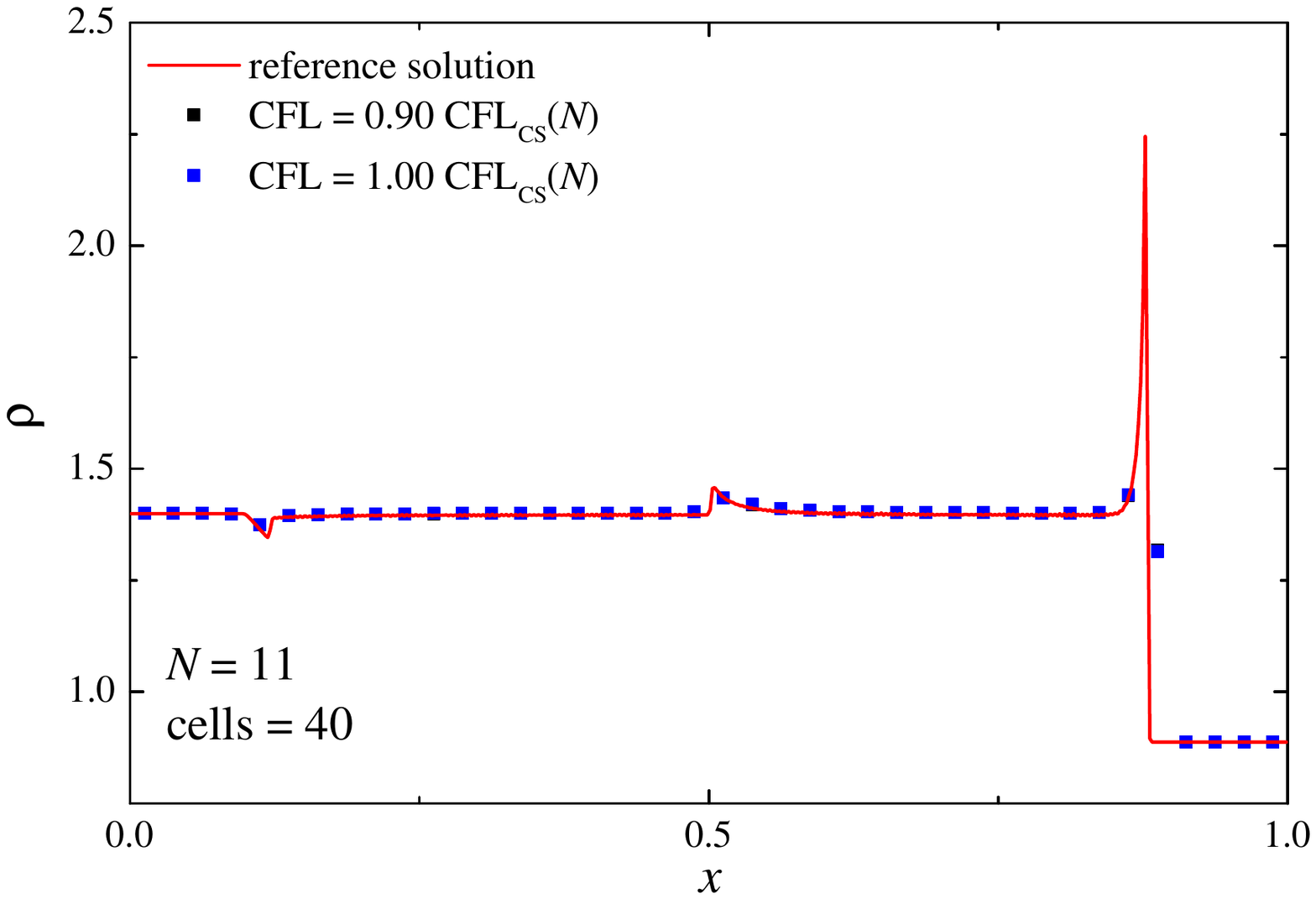}
\vspace{-8mm}\caption{\label{fig:test_zndf_1d_degrees_10_12:degree_11_den}}
\end{subfigure}
\begin{subfigure}{0.24\textwidth}
\includegraphics[width=\textwidth]{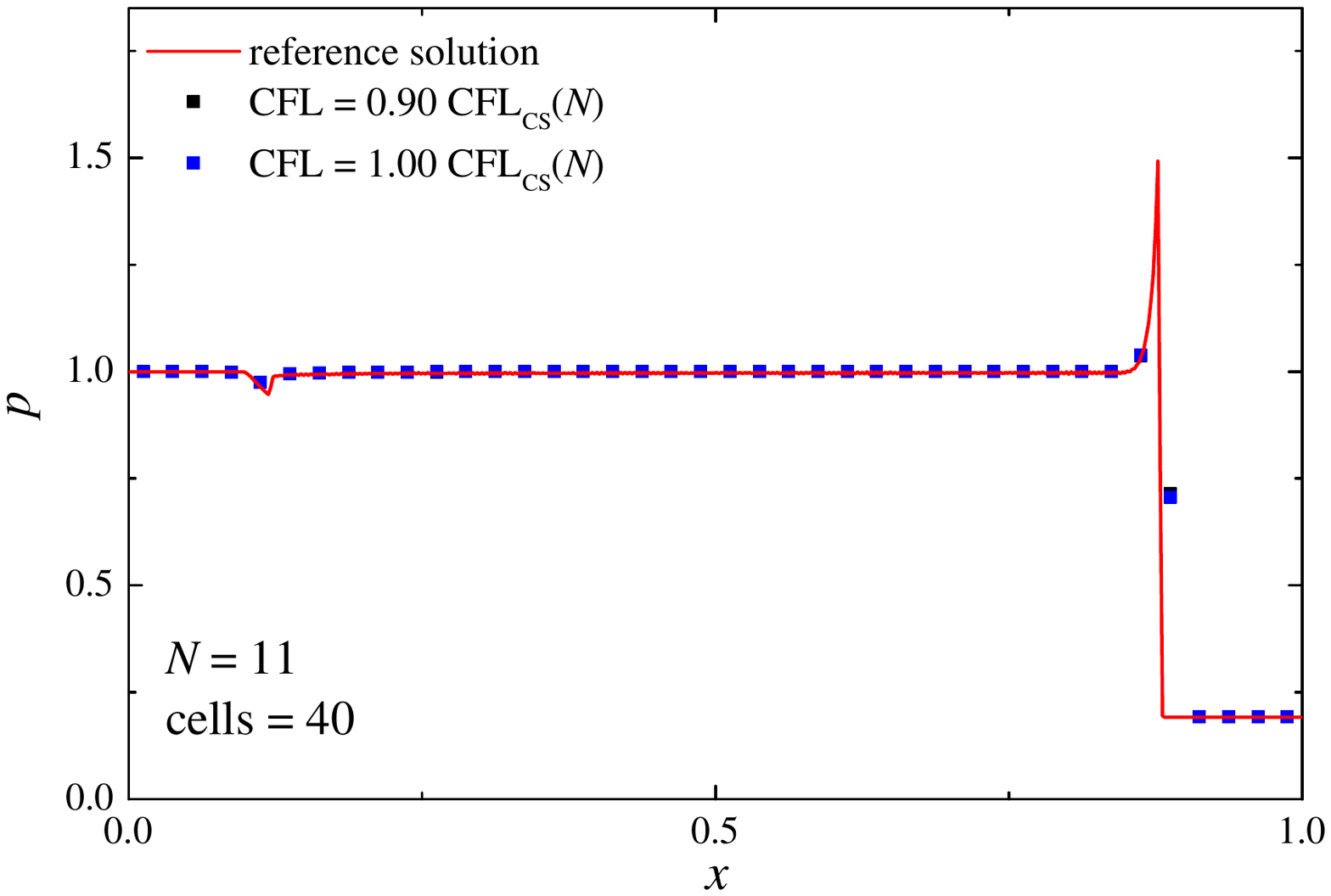}
\vspace{-8mm}\caption{\label{fig:test_zndf_1d_degrees_10_12:degree_11_p}}
\end{subfigure}
\begin{subfigure}{0.24\textwidth}
\includegraphics[width=\textwidth]{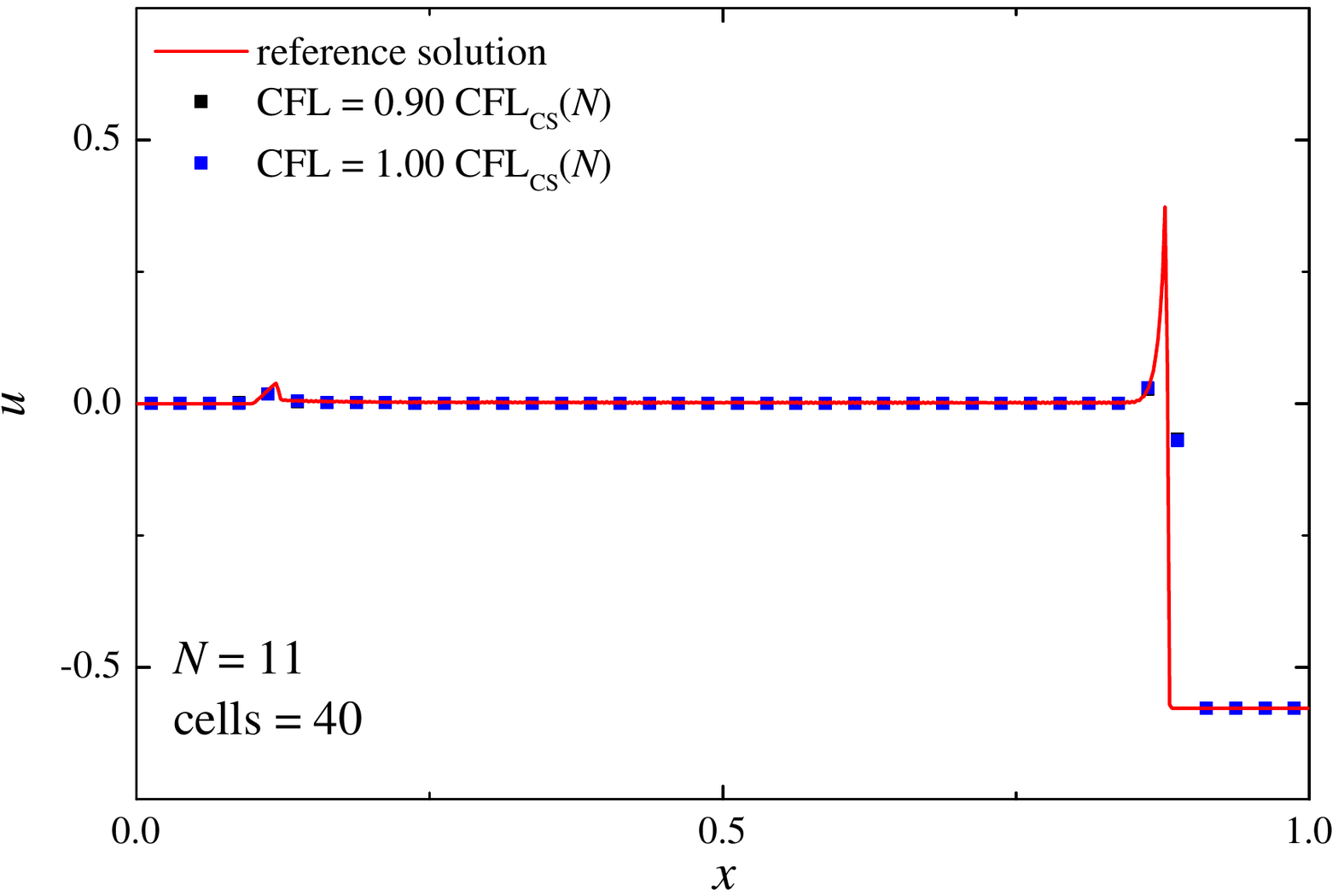}
\vspace{-8mm}\caption{\label{fig:test_zndf_1d_degrees_10_12:degree_11_u}}
\end{subfigure}
\begin{subfigure}{0.24\textwidth}
\includegraphics[width=\textwidth]{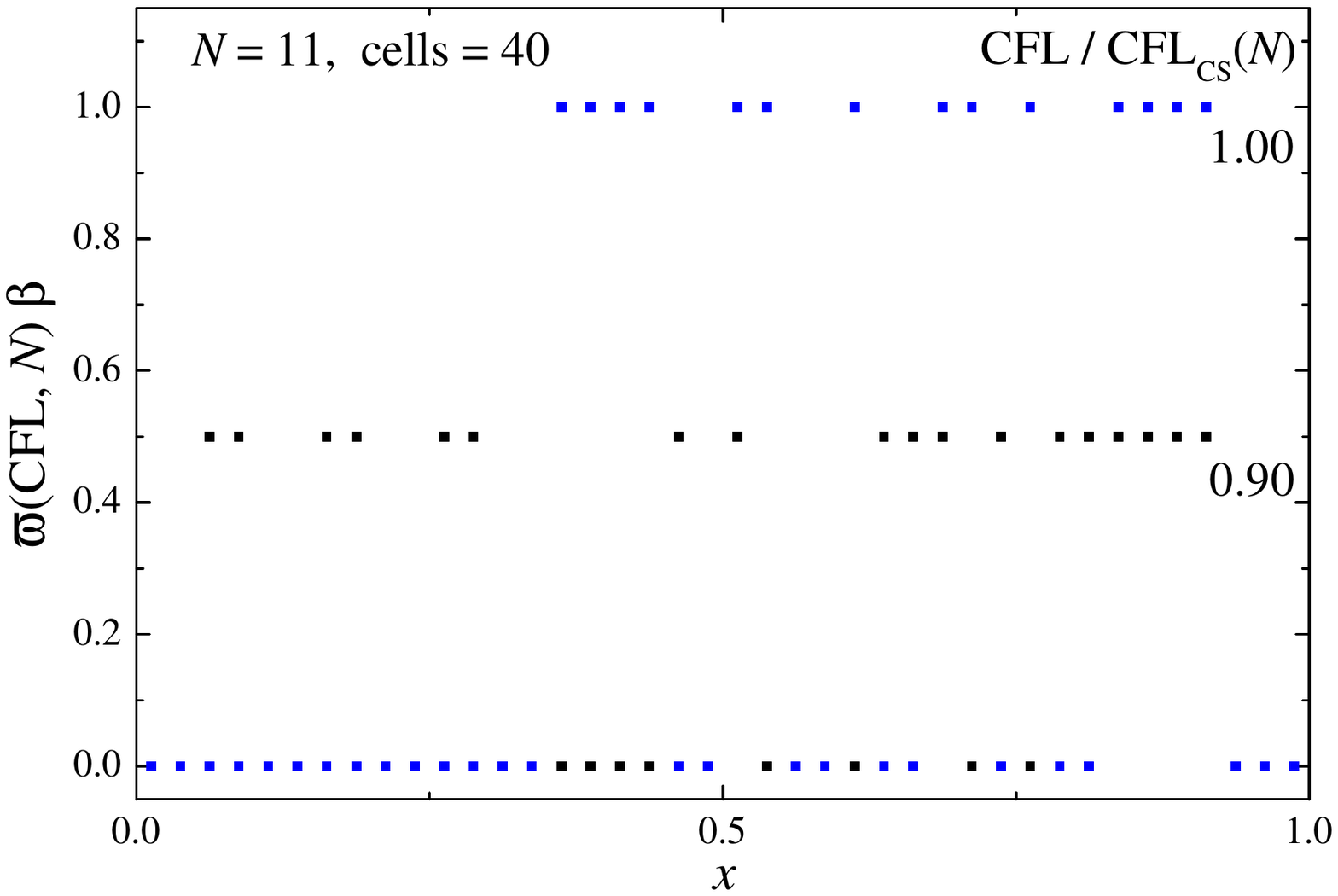}
\vspace{-8mm}\caption{\label{fig:test_zndf_1d_degrees_10_12:degree_11_beta}}
\end{subfigure}\\[-2mm]
\begin{subfigure}{0.24\textwidth}
\includegraphics[width=\textwidth]{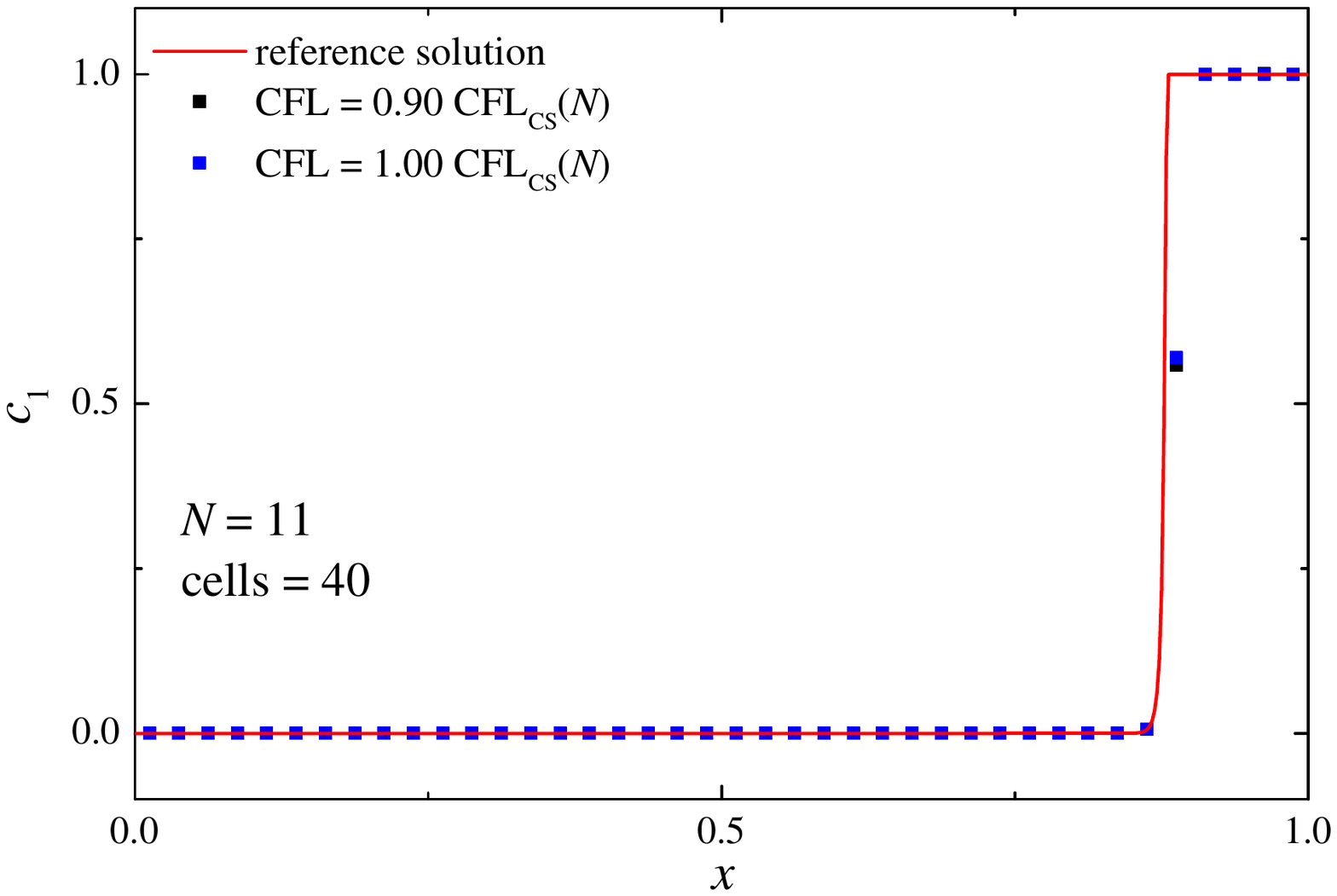}
\vspace{-8mm}\caption{\label{fig:test_zndf_1d_degrees_10_12:degree_11_c1}}
\end{subfigure}
\begin{subfigure}{0.24\textwidth}
\includegraphics[width=\textwidth]{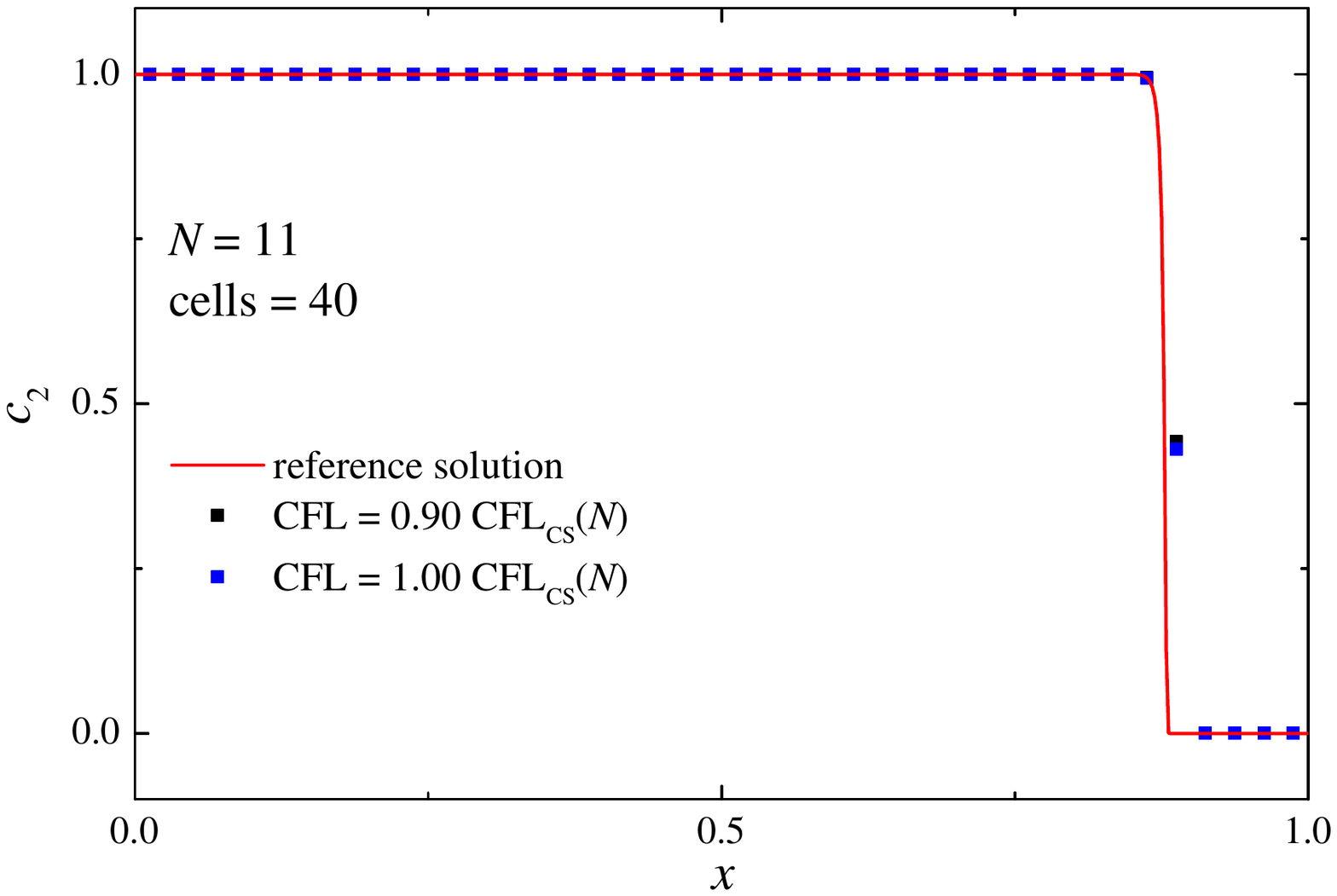}
\vspace{-8mm}\caption{\label{fig:test_zndf_1d_degrees_10_12:degree_11_c2}}
\end{subfigure}
\begin{subfigure}{0.24\textwidth}
\includegraphics[width=\textwidth]{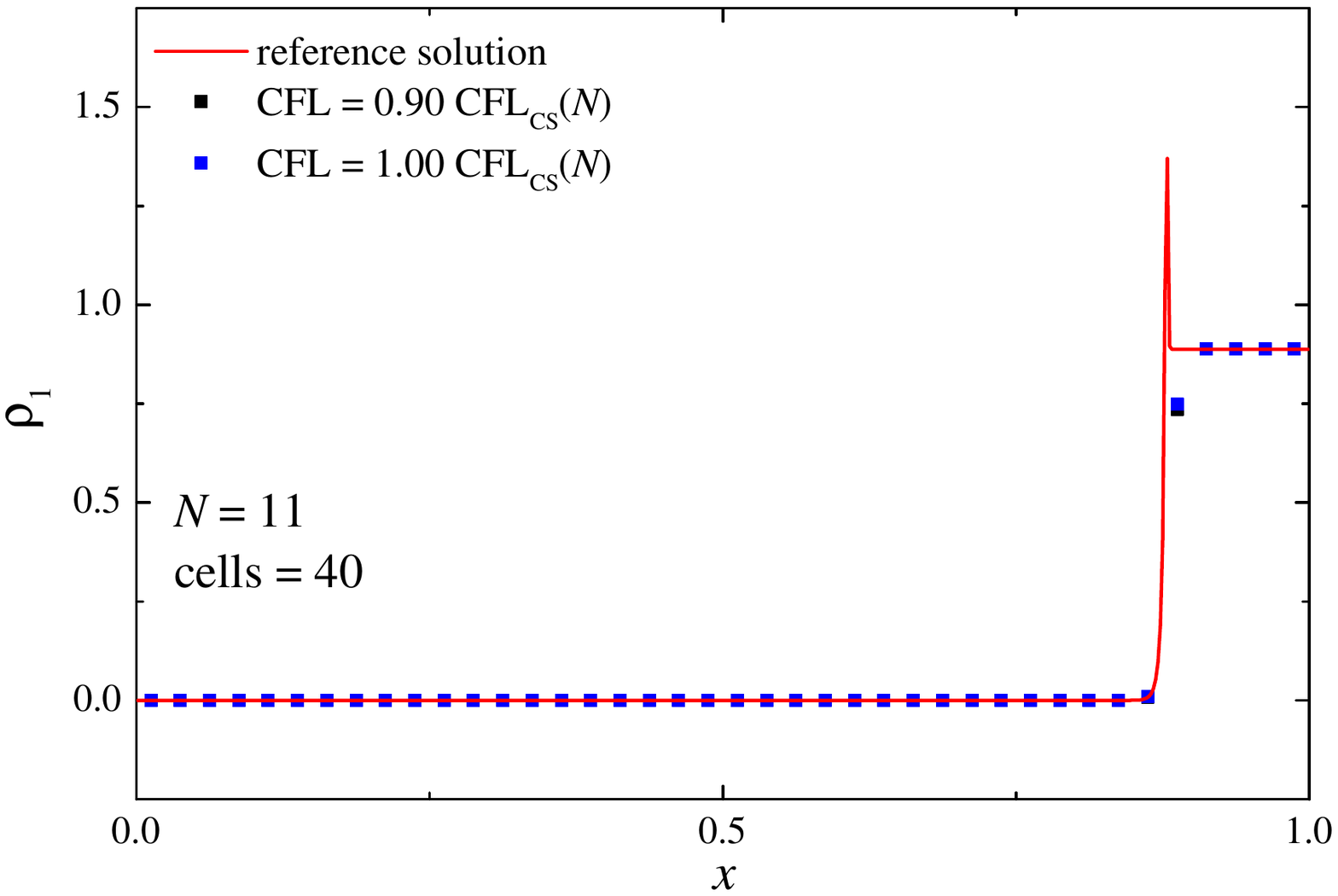}
\vspace{-8mm}\caption{\label{fig:test_zndf_1d_degrees_10_12:degree_11_den1}}
\end{subfigure}
\begin{subfigure}{0.24\textwidth}
\includegraphics[width=\textwidth]{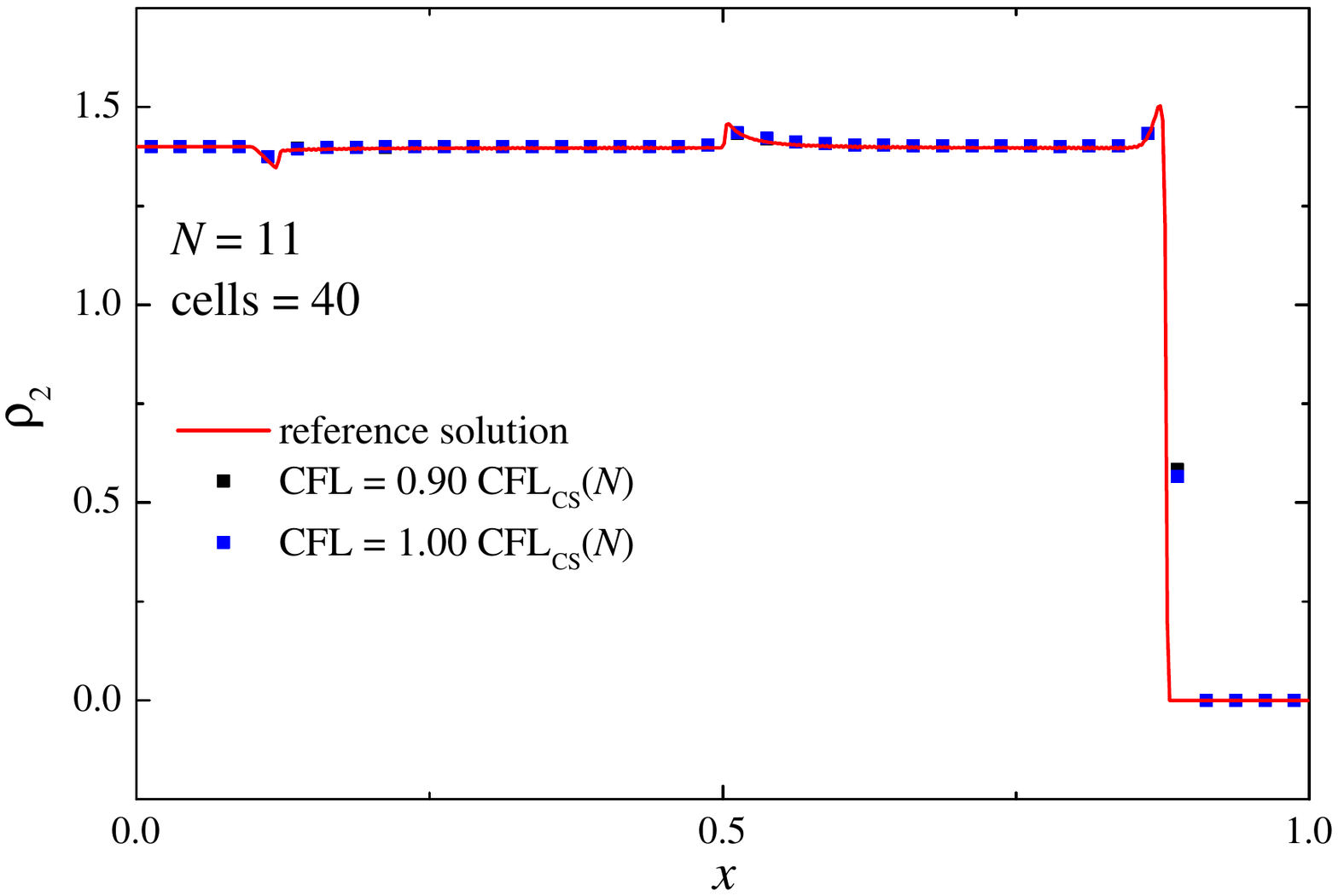}
\vspace{-8mm}\caption{\label{fig:test_zndf_1d_degrees_10_12:degree_11_den2}}
\end{subfigure}\\[-2mm]
\begin{subfigure}{0.24\textwidth}
\includegraphics[width=\textwidth]{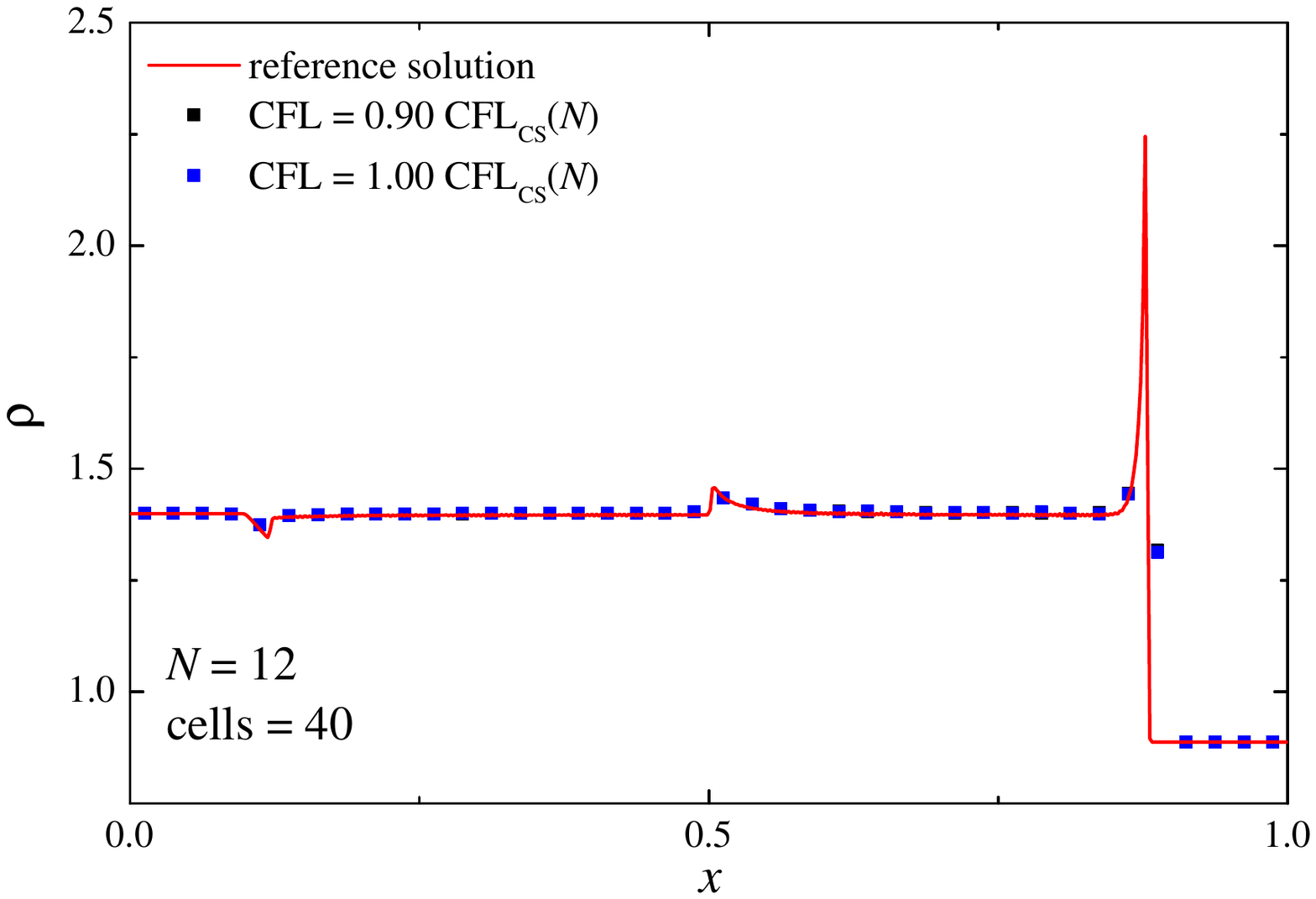}
\vspace{-8mm}\caption{\label{fig:test_zndf_1d_degrees_10_12:degree_12_den}}
\end{subfigure}
\begin{subfigure}{0.24\textwidth}
\includegraphics[width=\textwidth]{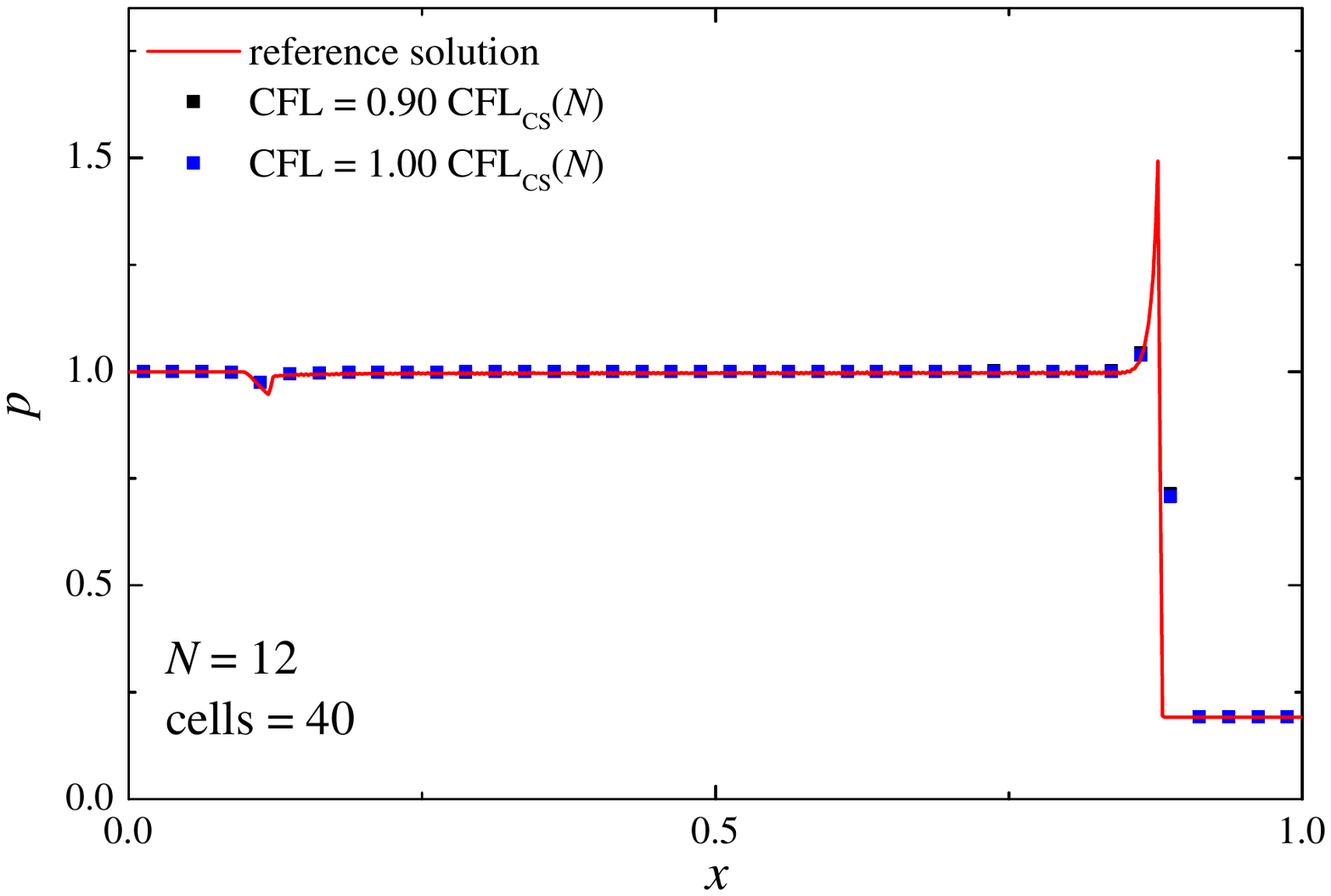}
\vspace{-8mm}\caption{\label{fig:test_zndf_1d_degrees_10_12:degree_12_p}}
\end{subfigure}
\begin{subfigure}{0.24\textwidth}
\includegraphics[width=\textwidth]{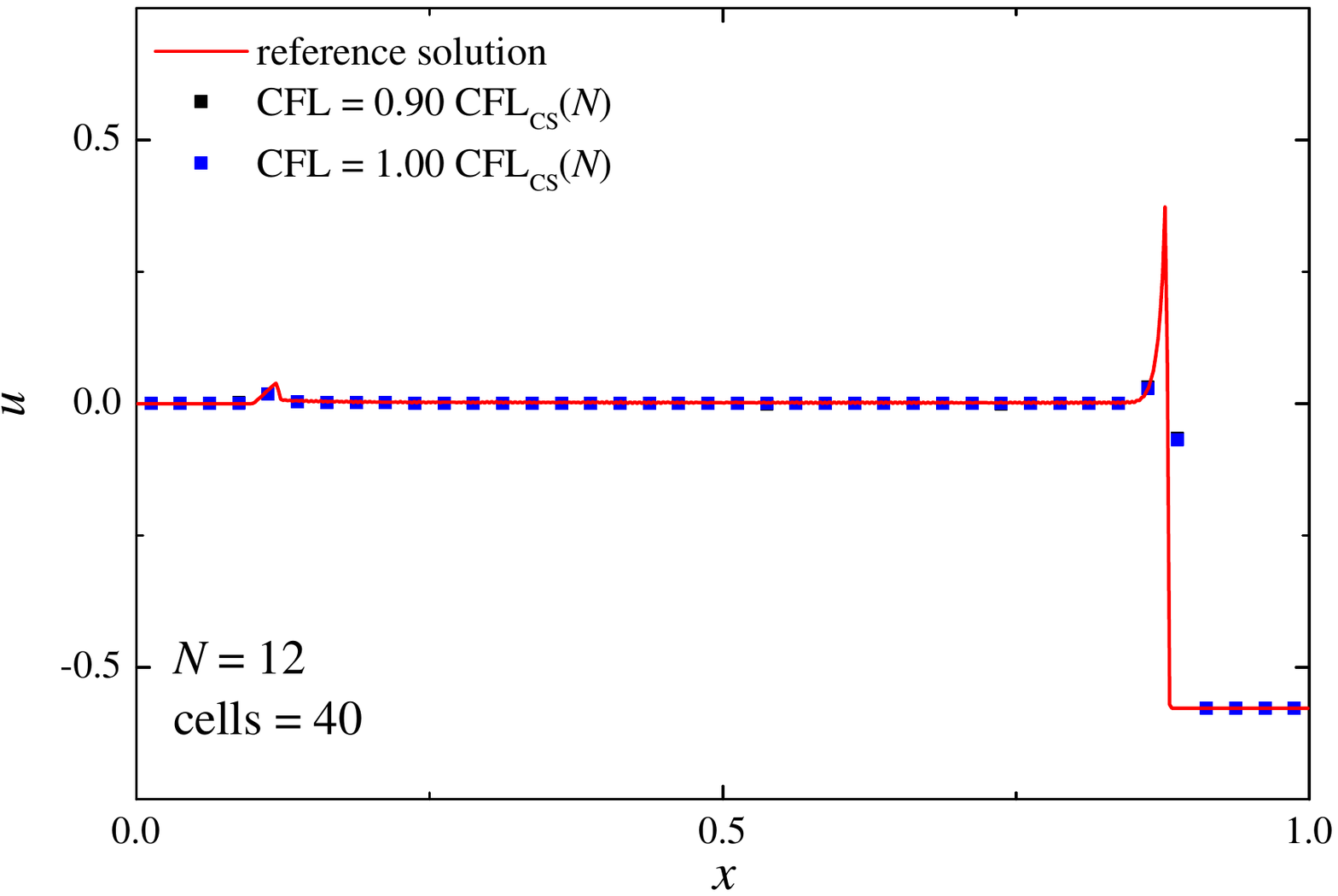}
\vspace{-8mm}\caption{\label{fig:test_zndf_1d_degrees_10_12:degree_12_u}}
\end{subfigure}
\begin{subfigure}{0.24\textwidth}
\includegraphics[width=\textwidth]{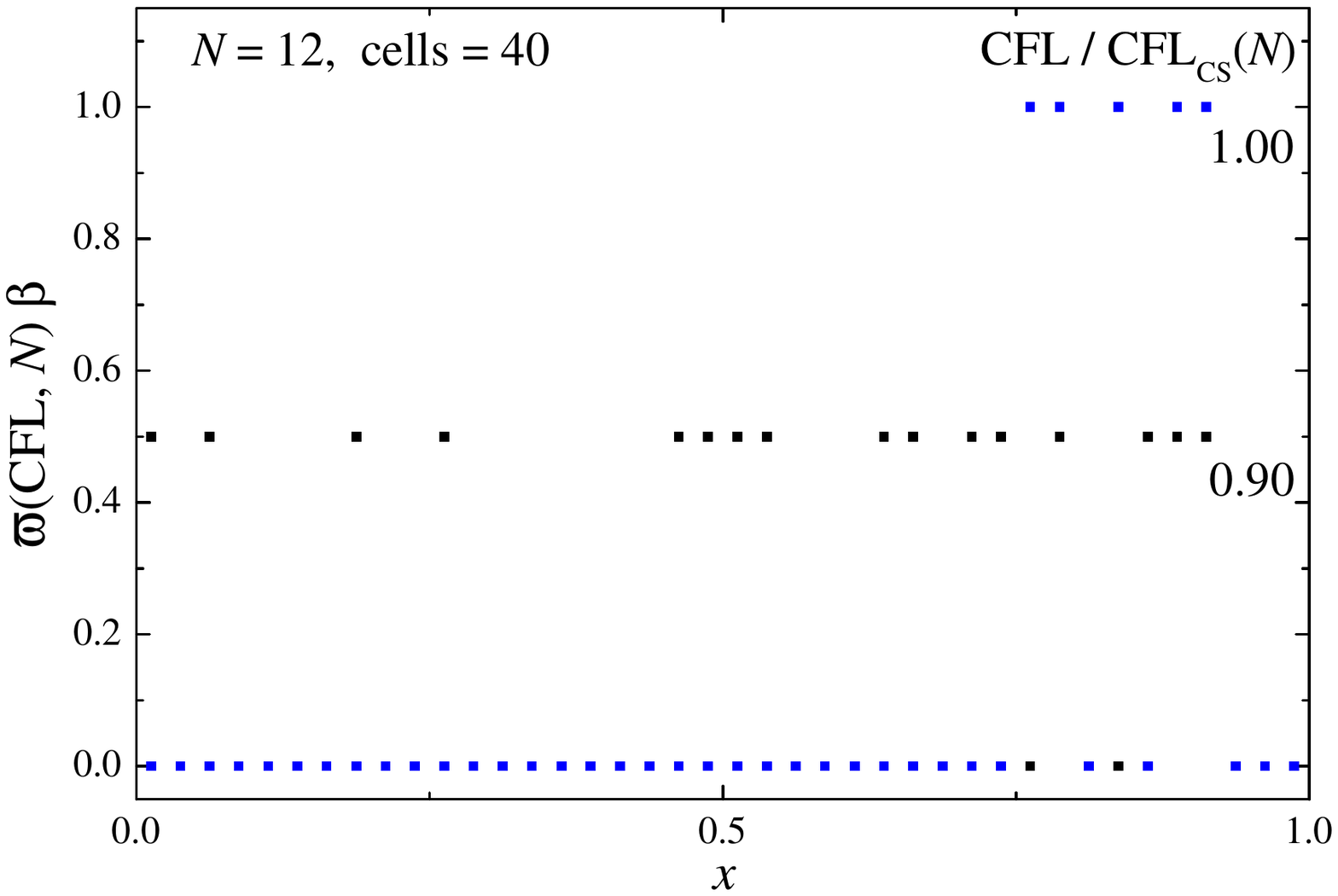}
\vspace{-8mm}\caption{\label{fig:test_zndf_1d_degrees_10_12:degree_12_beta}}
\end{subfigure}\\[-2mm]
\begin{subfigure}{0.24\textwidth}
\includegraphics[width=\textwidth]{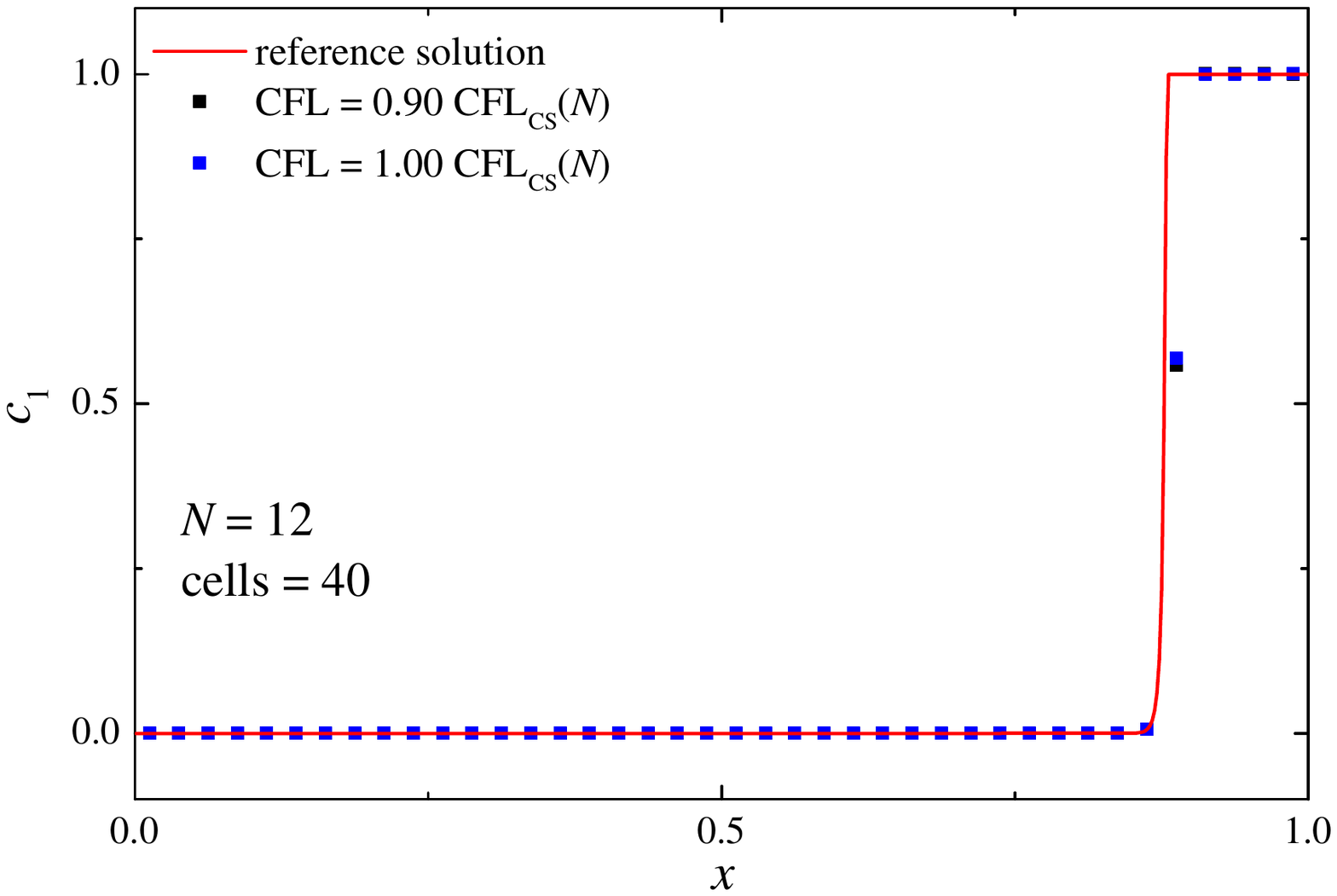}
\vspace{-8mm}\caption{\label{fig:test_zndf_1d_degrees_10_12:degree_12_c1}}
\end{subfigure}
\begin{subfigure}{0.24\textwidth}
\includegraphics[width=\textwidth]{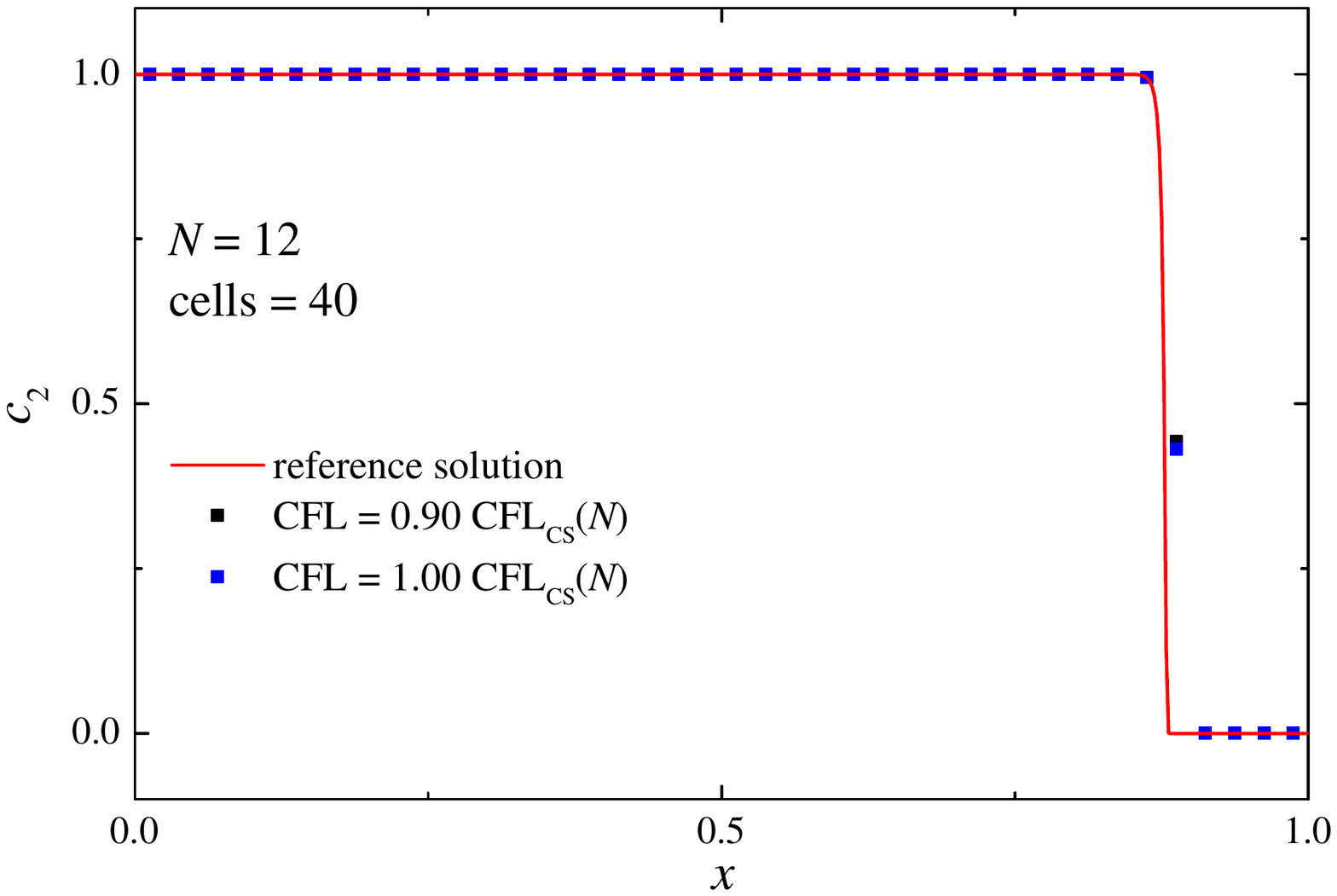}
\vspace{-8mm}\caption{\label{fig:test_zndf_1d_degrees_10_12:degree_12_c2}}
\end{subfigure}
\begin{subfigure}{0.24\textwidth}
\includegraphics[width=\textwidth]{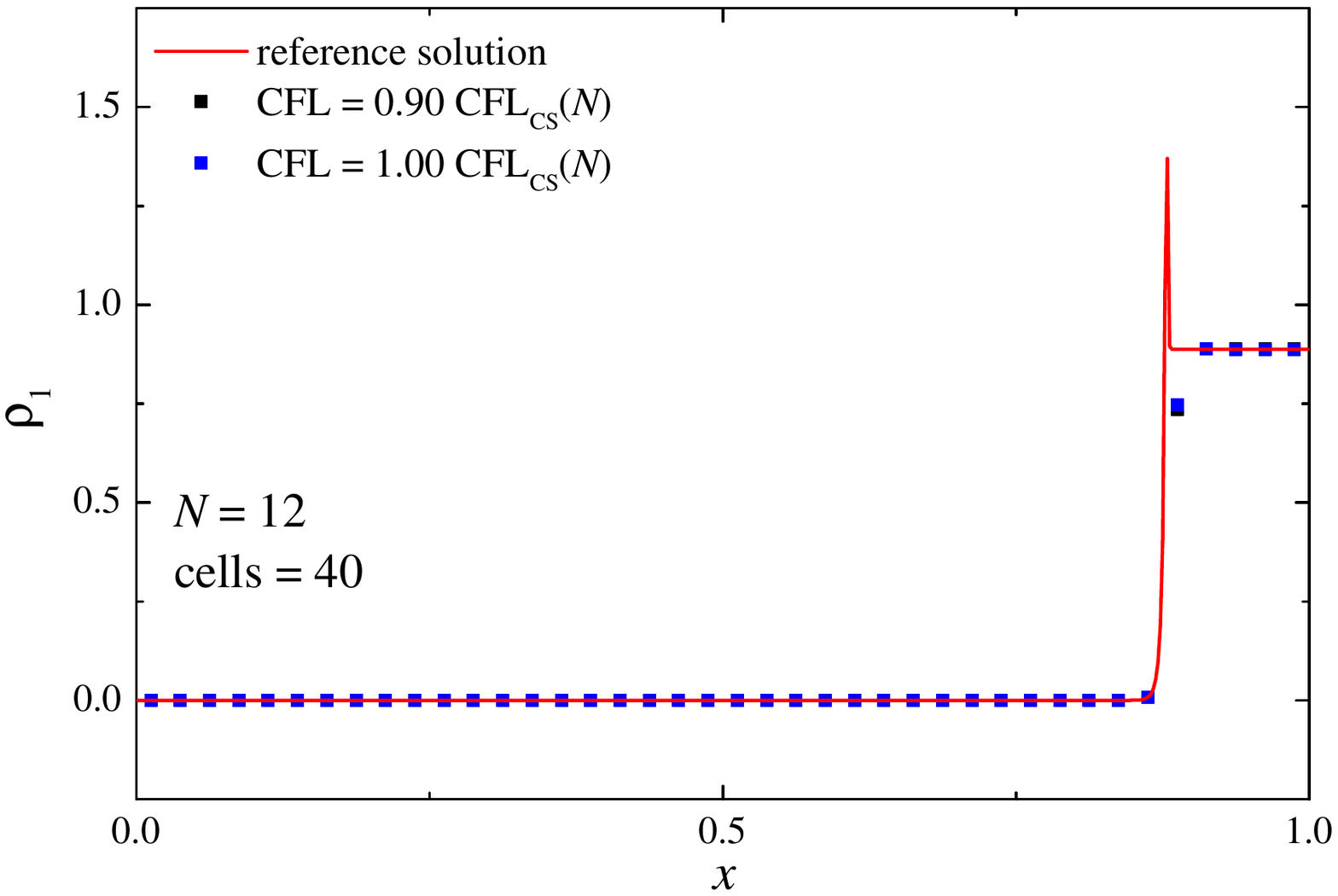}
\vspace{-8mm}\caption{\label{fig:test_zndf_1d_degrees_10_12:degree_12_den1}}
\end{subfigure}
\begin{subfigure}{0.24\textwidth}
\includegraphics[width=\textwidth]{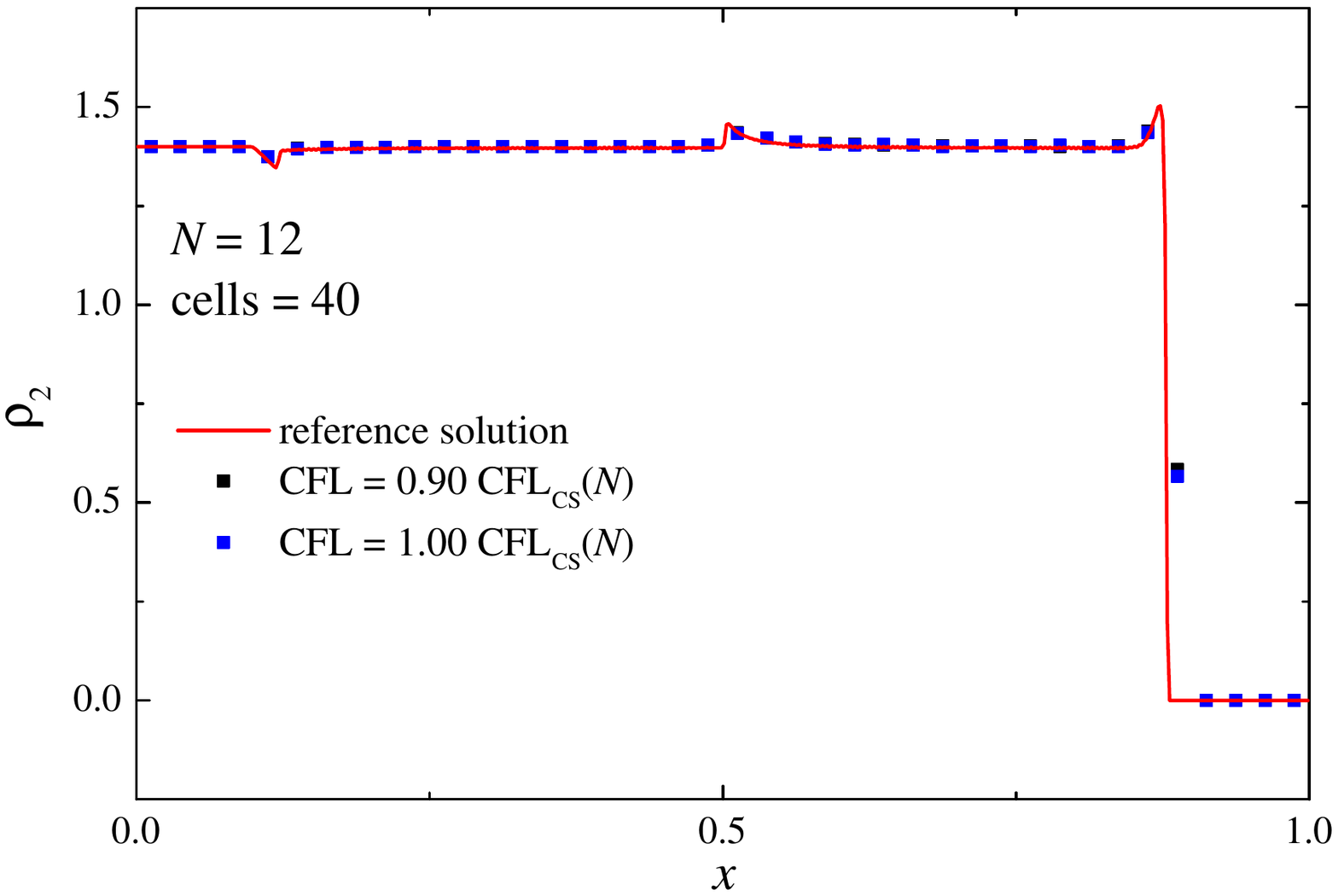}
\vspace{-8mm}\caption{\label{fig:test_zndf_1d_degrees_10_12:degree_12_den2}}
\end{subfigure}
\caption{%
Numerical solution of the one-dimensional problem of the plane detonation wave formation in a two-component medium with a ``fast'' reaction (strong stiff case, a detailed statement of the problem is presented in the text), obtained using the ADER-DG method for the basis polynomials degree $N = 10$ ((\subref{fig:test_zndf_1d_degrees_10_12:degree_10_den}), (\subref{fig:test_zndf_1d_degrees_10_12:degree_10_p}), (\subref{fig:test_zndf_1d_degrees_10_12:degree_10_u}), (\subref{fig:test_zndf_1d_degrees_10_12:degree_10_c1}), (\subref{fig:test_zndf_1d_degrees_10_12:degree_10_c2}), (\subref{fig:test_zndf_1d_degrees_10_12:degree_10_den1}), (\subref{fig:test_zndf_1d_degrees_10_12:degree_10_den2})), $N = 11$ ((\subref{fig:test_zndf_1d_degrees_10_12:degree_11_den}), (\subref{fig:test_zndf_1d_degrees_10_12:degree_11_p}), (\subref{fig:test_zndf_1d_degrees_10_12:degree_11_u}), (\subref{fig:test_zndf_1d_degrees_10_12:degree_11_c1}), (\subref{fig:test_zndf_1d_degrees_10_12:degree_11_c2}), (\subref{fig:test_zndf_1d_degrees_10_12:degree_11_den1}), (\subref{fig:test_zndf_1d_degrees_10_12:degree_11_den2})), $N = 12$ ((\subref{fig:test_zndf_1d_degrees_10_12:degree_12_den}), (\subref{fig:test_zndf_1d_degrees_10_12:degree_12_p}), (\subref{fig:test_zndf_1d_degrees_10_12:degree_12_u}), (\subref{fig:test_zndf_1d_degrees_10_12:degree_12_c1}), (\subref{fig:test_zndf_1d_degrees_10_12:degree_12_c2}), (\subref{fig:test_zndf_1d_degrees_10_12:degree_12_den1}), (\subref{fig:test_zndf_1d_degrees_10_12:degree_12_den2})) with a posteriori limitation of the solution by an ADER-WENO2 finite-volume limiter, at the final time $t_{f} = 0.4$. The graphs show the coordinate dependence of density $\rho$ ((\subref{fig:test_zndf_1d_degrees_10_12:degree_10_den}), (\subref{fig:test_zndf_1d_degrees_10_12:degree_11_den}), (\subref{fig:test_zndf_1d_degrees_10_12:degree_12_den})), pressure $p$ ((\subref{fig:test_zndf_1d_degrees_10_12:degree_10_p}), (\subref{fig:test_zndf_1d_degrees_10_12:degree_11_p}), (\subref{fig:test_zndf_1d_degrees_10_12:degree_12_p})), flow velocity $u$ ((\subref{fig:test_zndf_1d_degrees_10_12:degree_10_u}), (\subref{fig:test_zndf_1d_degrees_10_12:degree_11_u}), (\subref{fig:test_zndf_1d_degrees_10_12:degree_12_u})), troubled cells indicator $\beta$ multiplied by the separating function $\varpi(\mathrm{CFL}, N)$ ((\subref{fig:test_zndf_1d_degrees_10_12:degree_10_beta}), (\subref{fig:test_zndf_1d_degrees_10_12:degree_11_beta}), (\subref{fig:test_zndf_1d_degrees_10_12:degree_12_beta})), mass concentrations $c_{1}$ ((\subref{fig:test_zndf_1d_degrees_10_12:degree_10_c1}), (\subref{fig:test_zndf_1d_degrees_10_12:degree_11_c1}), (\subref{fig:test_zndf_1d_degrees_10_12:degree_12_c1})) and $c_{2}$ ((\subref{fig:test_zndf_1d_degrees_10_12:degree_10_c2}), (\subref{fig:test_zndf_1d_degrees_10_12:degree_11_c2}), (\subref{fig:test_zndf_1d_degrees_10_12:degree_12_c2})) of the reactant $A$ and product $B$, respectively, and partial densities $\rho_{1}$ ((\subref{fig:test_zndf_1d_degrees_10_12:degree_10_den1}), (\subref{fig:test_zndf_1d_degrees_10_12:degree_11_den1}), (\subref{fig:test_zndf_1d_degrees_10_12:degree_12_den1})) and $\rho_{2}$ ((\subref{fig:test_zndf_1d_degrees_10_12:degree_10_den2}), (\subref{fig:test_zndf_1d_degrees_10_12:degree_11_den2}), (\subref{fig:test_zndf_1d_degrees_10_12:degree_12_den2})) of the reactant $A$ and product $B$, respectively. The square symbols represent the subcells finite-volume representation of the numerical solution; the red solid lines represents the reference solution of the problem.
}
\label{fig:test_zndf_1d_degrees_10_12}
\end{figure}

\begin{table*}[h!]
\centering
\small
\caption{%
Effective Courant numbers $\mathrm{C}$ and Courant numbers $\mathrm{C}\cdot\mathrm{CFL}_{\rm CS}(N)$ for which the test of formation and propagation of a ZND detonation wave is correctly passed (\done) or failed (\fail) in the ``slow'' reaction case (the weak stiff case) and the ``fast'' reaction case (the strong stiff case) for the numerical method ADER-DG with the LST-DG predictor for degrees $N = 1$, $2$, $3$ and $4$ of basis polynomials. For comparison, the boundary values of the Courant number $\mathrm{CFL}_{\rm max}^{\rm NS}(N) \equiv \mathrm{CFL}_{\rm max}(N, 0)$ for the numerical method in the case of a problem (\ref{eq:adv_eq_src}) without a source term ($\nu = 0$) and the ratio of the Courant number $\mathrm{C}\cdot\mathrm{CFL}_{\rm CS}(N)$ used in the test to the $\mathrm{CFL}_{\rm max}^{\rm NS}(N)$ are presented; the Courant numbers $\mathrm{CFL}_{\rm CS}(N)$ are defined by Cockburn-Shu formula (\ref{eq:cfl_cs}).
}
\label{tab:cfls_impr_znd_degrees_1_4}
\begin{tabular}{@{}|c|c|c|c|c|c|c|@{}}
\toprule
$N$ & $\mathrm{CFL}_{\rm max}^{\rm NS}(N)$ & ~~~~$\mathrm{C}$~~~~ & $\mathrm{C}\cdot\mathrm{CFL}_{\rm CS}(N)$ & 
$\frac{\displaystyle\mathrm{C}\cdot\mathrm{CFL}_{\rm CS}(N)}{\displaystyle\mathrm{CFL}_{\rm max}^{\rm NS}(N)}$ &
\parbox{1.5cm}{weak\\ stiff case} & \parbox{1.5cm}{strong\\ stiff case} \\
\midrule
$1$ & $0.333333$ & $0.90$ & $0.300000$ & $0.9000$ & \done & \done \\
  &  & $1.00$ & $0.333333$ & $1.0000$ & \done & \done \\
  &  & $1.05$ & $0.350000$ & $1.0500$ & \done & \done \\
  &  & $1.10$ & $0.366667$ & $1.1000$ & \done & \done \\
  &  & $1.15$ & $0.383333$ & $1.1500$ & \done & \done \\
  &  & $1.20$ & $0.400000$ & $1.2000$ & \done & \done \\
  &  & $1.25$ & $0.416667$ & $1.2500$ & \done & \done \\
  &  & $1.30$ & $0.433333$ & $1.3000$ & \done & \done \\
  &  & $1.35$ & $0.450000$ & $1.3500$ & \done & \done \\
\midrule
$2$ & $0.170820$ & $0.90$ & $0.180000$ & $1.0537$ & \done & \done \\
  &  & $1.00$ & $0.200000$ & $1.1708$ & \done & \done \\
  &  & $1.05$ & $0.210000$ & $1.2294$ & \done & \done \\
  &  & $1.10$ & $0.220000$ & $1.2879$ & \done & \done \\
  &  & $1.15$ & $0.230000$ & $1.3464$ & \done & \done \\
  &  & $1.20$ & $0.240000$ & $1.4050$ & \fail & \done \\
  &  & $1.25$ & $0.250000$ & $1.4635$ & \fail & \done \\
  &  & $1.30$ & $0.260000$ & $1.5221$ & \fail & \fail \\
  &  & $1.35$ & $0.270000$ & $1.5806$ & \fail & \fail \\
\midrule
$3$ & $0.103929$ & $0.90$ & $0.128571$ & $1.2371$ & \done & \done \\
  &  & $1.00$ & $0.142857$ & $1.3746$ & \done & \done \\
  &  & $1.05$ & $0.150000$ & $1.4433$ & \done & \done \\
  &  & $1.10$ & $0.157143$ & $1.5120$ & \done & \done \\
  &  & $1.15$ & $0.164286$ & $1.5807$ & \done & \done \\
  &  & $1.20$ & $0.171429$ & $1.6495$ & \fail & \done \\
  &  & $1.25$ & $0.178571$ & $1.7182$ & \fail & \fail \\
  &  & $1.30$ & $0.185714$ & $1.7869$ & \fail & \fail \\
  &  & $1.35$ & $0.192857$ & $1.8557$ & \fail & \fail \\
\midrule
$4$ & $0.069831$ & $0.90$ & $0.100000$ & $1.4320$ & \done & \done \\
  &  & $1.00$ & $0.111111$ & $1.5911$ & \done & \done \\
  &  & $1.05$ & $0.116667$ & $1.6707$ & \done & \done \\
  &  & $1.10$ & $0.122222$ & $1.7503$ & \done & \done \\
  &  & $1.15$ & $0.127778$ & $1.8298$ & \fail & \fail \\
  &  & $1.20$ & $0.133333$ & $1.9094$ & \fail & \fail \\
  &  & $1.25$ & $0.138889$ & $1.9889$ & \fail & \fail \\
  &  & $1.30$ & $0.144444$ & $2.0685$ & \fail & \fail \\
  &  & $1.35$ & $0.150000$ & $2.1480$ & \fail & \fail \\
\bottomrule
\end{tabular}
\end{table*}
\begin{table*}[h!]
\centering
\small
\caption{%
Effective Courant numbers $\mathrm{C}$ and Courant numbers $\mathrm{C}\cdot\mathrm{CFL}_{\rm CS}(N)$ for which the test of formation and propagation of a ZND detonation wave is correctly passed (\done) or failed (\fail) in the ``slow'' reaction case (the weak stiff case) and the ``fast'' reaction case (the strong stiff case) for the numerical method ADER-DG with the LST-DG predictor for degrees $N = 5$, $6$, $7$ and $8$ of basis polynomials. For comparison, the boundary values of the Courant number $\mathrm{CFL}_{\rm max}^{\rm NS}(N) \equiv \mathrm{CFL}_{\rm max}(N, 0)$ for the numerical method in the case of a problem (\ref{eq:adv_eq_src}) without a source term ($\nu = 0$) and the ratio of the Courant number $\mathrm{C}\cdot\mathrm{CFL}_{\rm CS}(N)$ used in the test to the $\mathrm{CFL}_{\rm max}^{\rm NS}(N)$ are presented; the Courant numbers $\mathrm{CFL}_{\rm CS}(N)$ are defined by Cockburn-Shu formula (\ref{eq:cfl_cs}).
}
\label{tab:cfls_impr_znd_degrees_5_8}
\begin{tabular}{@{}|c|c|c|c|c|c|c|@{}}
\toprule
$N$ & $\mathrm{CFL}_{\rm max}^{\rm NS}(N)$ & ~~~~$\mathrm{C}$~~~~ & $\mathrm{C}\cdot\mathrm{CFL}_{\rm CS}(N)$ & 
$\frac{\displaystyle\mathrm{C}\cdot\mathrm{CFL}_{\rm CS}(N)}{\displaystyle\mathrm{CFL}_{\rm max}^{\rm NS}(N)}$ &
\parbox{1.5cm}{weak\\ stiff case} & \parbox{1.5cm}{strong\\ stiff case} \\
\midrule
$5$ & $0.050116$ & $0.90$ & $0.081818$ & $1.6326$ & \done & \done \\
  &  & $1.00$ & $0.090909$ & $1.8140$ & \done & \done \\
  &  & $1.05$ & $0.095455$ & $1.9047$ & \done & \done \\
  &  & $1.10$ & $0.100000$ & $1.9954$ & \done & \fail \\
  &  & $1.15$ & $0.104545$ & $2.0861$ & \fail & \fail \\
  &  & $1.20$ & $0.109091$ & $2.1768$ & \fail & \fail \\
  &  & $1.25$ & $0.113636$ & $2.2675$ & \fail & \fail \\
  &  & $1.30$ & $0.118182$ & $2.3582$ & \fail & \fail \\
  &  & $1.35$ & $0.122727$ & $2.4489$ & \fail & \fail \\
\midrule
$6$ & $0.037701$ & $0.90$ & $0.069231$ & $1.8363$ & \done & \done \\
  &  & $1.00$ & $0.076923$ & $2.0403$ & \done & \done \\
  &  & $1.05$ & $0.080769$ & $2.1424$ & \done & \done \\
  &  & $1.10$ & $0.084615$ & $2.2444$ & \done & \fail \\
  &  & $1.15$ & $0.088462$ & $2.3464$ & \fail & \fail \\
  &  & $1.20$ & $0.092308$ & $2.4484$ & \fail & \fail \\
  &  & $1.25$ & $0.096154$ & $2.5504$ & \fail & \fail \\
  &  & $1.30$ & $0.100000$ & $2.6524$ & \fail & \fail \\
  &  & $1.35$ & $0.103846$ & $2.7545$ & \fail & \fail \\
\midrule
$7$ & $0.029384$ & $0.90$ & $0.060000$ & $2.0419$ & \done & \done \\
  &  & $1.00$ & $0.066667$ & $2.2688$ & \done & \done \\
  &  & $1.05$ & $0.070000$ & $2.3822$ & \done & \fail \\
  &  & $1.10$ & $0.073333$ & $2.4957$ & \done & \fail \\
  &  & $1.15$ & $0.076667$ & $2.6091$ & \fail & \fail \\
  &  & $1.20$ & $0.080000$ & $2.7226$ & \fail & \fail \\
  &  & $1.25$ & $0.083333$ & $2.8360$ & \fail & \fail \\
  &  & $1.30$ & $0.086667$ & $2.9495$ & \fail & \fail \\
  &  & $1.35$ & $0.090000$ & $3.0629$ & \fail & \fail \\
\midrule
$8$ & $0.023542$ & $0.90$ & $0.052941$ & $2.2488$ & \done & \done \\
  &  & $1.00$ & $0.058824$ & $2.4987$ & \done & \done \\
  &  & $1.05$ & $0.061765$ & $2.6236$ & \done & \fail \\
  &  & $1.10$ & $0.064706$ & $2.7485$ & \fail & \fail \\
  &  & $1.15$ & $0.067647$ & $2.8735$ & \fail & \fail \\
  &  & $1.20$ & $0.070588$ & $2.9984$ & \fail & \fail \\
  &  & $1.25$ & $0.073529$ & $3.1233$ & \fail & \fail \\
  &  & $1.30$ & $0.076471$ & $3.2483$ & \fail & \fail \\
  &  & $1.35$ & $0.079412$ & $3.3732$ & \fail & \fail \\
\bottomrule
\end{tabular}
\end{table*}
\begin{table*}[h!]
\centering
\small
\caption{%
Effective Courant numbers $\mathrm{C}$ and Courant numbers $\mathrm{C}\cdot\mathrm{CFL}_{\rm CS}(N)$ for which the test of formation and propagation of a ZND detonation wave is correctly passed (\done) or failed (\fail) in the ``slow'' reaction case (the weak stiff case) and the ``fast'' reaction case (the strong stiff case) for the numerical method ADER-DG with the LST-DG predictor for degrees $N = 9$, $10$, $11$ and $12$ of basis polynomials. For comparison, the boundary values of the Courant number $\mathrm{CFL}_{\rm max}^{\rm NS}(N) \equiv \mathrm{CFL}_{\rm max}(N, 0)$ for the numerical method in the case of a problem (\ref{eq:adv_eq_src}) without a source term ($\nu = 0$) and the ratio of the Courant number $\mathrm{C}\cdot\mathrm{CFL}_{\rm CS}(N)$ used in the test to the $\mathrm{CFL}_{\rm max}^{\rm NS}(N)$ are presented; the Courant numbers $\mathrm{CFL}_{\rm CS}(N)$ are defined by Cockburn-Shu formula (\ref{eq:cfl_cs}).
}
\label{tab:cfls_impr_znd_degrees_9_12}
\begin{tabular}{@{}|c|c|c|c|c|c|c|@{}}
\toprule
$N$ & $\mathrm{CFL}_{\rm max}^{\rm NS}(N)$ & ~~~~$\mathrm{C}$~~~~ & $\mathrm{C}\cdot\mathrm{CFL}_{\rm CS}(N)$ & 
$\frac{\displaystyle\mathrm{C}\cdot\mathrm{CFL}_{\rm CS}(N)}{\displaystyle\mathrm{CFL}_{\rm max}^{\rm NS}(N)}$ &
\parbox{1.5cm}{weak\\ stiff case} & \parbox{1.5cm}{strong\\ stiff case} \\
\midrule
$9$ & $0.019282$ & $0.90$ & $0.047368$ & $2.4566$ & \done & \done \\
  &  & $1.00$ & $0.052632$ & $2.7296$ & \done & \done \\
  &  & $1.05$ & $0.055263$ & $2.8660$ & \done & \fail \\
  &  & $1.10$ & $0.057895$ & $3.0025$ & \fail & \fail \\
  &  & $1.15$ & $0.060526$ & $3.1390$ & \fail & \fail \\
  &  & $1.20$ & $0.063158$ & $3.2755$ & \fail & \fail \\
  &  & $1.25$ & $0.065789$ & $3.4120$ & \fail & \fail \\
  &  & $1.30$ & $0.068421$ & $3.5484$ & \fail & \fail \\
  &  & $1.35$ & $0.071053$ & $3.6849$ & \fail & \fail \\
\midrule
$10$ & $0.016081$ & $0.90$ & $0.042857$ & $2.6651$ & \done & \done \\
  &  & $1.00$ & $0.047619$ & $2.9612$ & \done & \done \\
  &  & $1.05$ & $0.050000$ & $3.1093$ & \fail & \fail \\
  &  & $1.10$ & $0.052381$ & $3.2573$ & \fail & \fail \\
  &  & $1.15$ & $0.054762$ & $3.4054$ & \fail & \fail \\
  &  & $1.20$ & $0.057143$ & $3.5534$ & \fail & \fail \\
  &  & $1.25$ & $0.059524$ & $3.7015$ & \fail & \fail \\
  &  & $1.30$ & $0.061905$ & $3.8496$ & \fail & \fail \\
  &  & $1.35$ & $0.064286$ & $3.9976$ & \fail & \fail \\
\midrule
$11$ & $0.013616$ & $0.90$ & $0.039130$ & $2.8739$ & \done & \done \\
  &  & $1.00$ & $0.043478$ & $3.1932$ & \done & \done \\
  &  & $1.05$ & $0.045652$ & $3.3528$ & \fail & \fail \\
  &  & $1.10$ & $0.047826$ & $3.5125$ & \fail & \fail \\
  &  & $1.15$ & $0.050000$ & $3.6722$ & \fail & \fail \\
  &  & $1.20$ & $0.052174$ & $3.8318$ & \fail & \fail \\
  &  & $1.25$ & $0.054348$ & $3.9915$ & \fail & \fail \\
  &  & $1.30$ & $0.056522$ & $4.1511$ & \fail & \fail \\
  &  & $1.35$ & $0.058696$ & $4.3108$ & \fail & \fail \\
\midrule
$12$ & $0.011676$ & $0.90$ & $0.036000$ & $3.0832$ & \done & \done \\
  &  & $1.00$ & $0.040000$ & $3.4258$ & \done & \done \\
  &  & $1.05$ & $0.042000$ & $3.5971$ & \fail & \fail \\
  &  & $1.10$ & $0.044000$ & $3.7684$ & \fail & \fail \\
  &  & $1.15$ & $0.046000$ & $3.9397$ & \fail & \fail \\
  &  & $1.20$ & $0.048000$ & $4.1110$ & \fail & \fail \\
  &  & $1.25$ & $0.050000$ & $4.2823$ & \fail & \fail \\
  &  & $1.30$ & $0.052000$ & $4.4536$ & \fail & \fail \\
  &  & $1.35$ & $0.054000$ & $4.6249$ & \fail & \fail \\
\bottomrule
\end{tabular}
\end{table*}

The analysis begins with the case of a ``slow'' reaction (the weak stiff case). Figures~\ref{fig:test_znds_1d_degrees_1_3},~\ref{fig:test_znds_1d_degrees_4_6},~\ref{fig:test_znds_1d_degrees_7_9} and~\ref{fig:test_znds_1d_degrees_10_12} show the results of the numerical solution of the problem of formation and development of a ZND detonation wave in the case of ``slow'' reaction (weak stiff case), obtained using the ADER-DG method for the basis polynomials degree $N = 1$, $2$, $3$ (Figure~\ref{fig:test_znds_1d_degrees_1_3}), $N = 4$, $5$, $6$ (Figure~\ref{fig:test_znds_1d_degrees_4_6}), $N = 7$, $8$, $9$ (Figure~\ref{fig:test_znds_1d_degrees_7_9}), $N = 10$, $11$, $12$ (Figure~\ref{fig:test_znds_1d_degrees_10_12}), at the final time $t_{f} = 0.4$. All these Figures show the coordinates dependencies of density $\rho(x, t_{f})$, pressure $p(x, t_{f})$, flow velocity $u(x, t_{f})$, troubled cells indicator $\beta(x, t_{f})$ (multiplied by the separating function $\varpi(\mathrm{CFL}, N)$), mass concentrations $c_{1}(x, t_{f})$ and $c_{2}(x, t_{f})$ of the reactant $A$ and product $B$, respectively, and partial densities $\rho_{1}(x, t_{f})$ and $\rho_{2}(x, t_{f})$ of the reactant $A$ and product $B$, respectively. The troubled cells indicator $\beta$ takes only two values: $\beta = 0$ and $\beta = 1$, with the value $\beta = 1$ of primary interest, which demonstrates a troubled cell, in which the solution is limited by a finite-volume limiter. Therefore, its presentation in the Figures for several values of the Courant number $\mathrm{CFL}$ is very inconvenient for analysis. Therefore, for convenience of presentation, the separation function $\varpi(\mathrm{CFL}, N)$ is introduced, taking unique values for each value of the Courant number $\mathrm{CFL}$, and monotonically increasing with the Courant number $\mathrm{CFL}$. The separation function $\varpi(\mathrm{CFL}, N)$ is multiplied by the troubled cells indicator $\beta$, as a result of which the presented levels $\beta = 1$ for different values of the Courant number $\mathrm{CFL}$ are separated --- the positive values of the separating function $\mathrm{CFL}$ are chosen only for reasons of convenience of troubled cells indicator $\beta$ presentation on the graphs and do not carry any further meaning: the points with $\beta = 0$ is also equivalent to the points with  $\varpi(\mathrm{CFL}, N)\beta = 0$, and the points with $\beta = 1$ in this case is equivalent to the points with $\varpi(\mathrm{CFL}, N)\beta \in (0,\, 1]$. In each Figure presented, coordinate dependencies are presented only for those values of the Courant number $\mathrm{CFL} = \mathrm{C}\cdot\mathrm{CFL}_{\rm CS}(N)$ for which correct results are obtained; in other cases of the Courant numbers $\mathrm{CFL}$, the solution is obtained \texttt{nan} (``Not a number''~\cite{ieee754}). The number of spatial mesh cells is determined by the minimum number for which a correct and adequate numerical solution is obtained. Such detailed specification of the coordinate dependencies of the numerical solution is necessary for an accurate assessment of the quality of the numerical solution in terms of reproducing the structural features of the detonation wave~\cite{Lee_2008, OranBoris_2005}: not only how does the propagation of hydrodynamic disturbances in the medium occur, but also how does the process of reactant burnout and product accumulation occur, and how does the compression of the unburned substance occur during the reaction (which is clearly demonstrated by the graphs of the coordinate dependencies of the partial densities $\rho_{i}$ of the reactants $A$ and products $B$). The case of a ``slow'' reaction (a weak stiff case) is interesting in that the characteristic reaction times, although significantly shorter than the characteristic hydrodynamic times, lead to the formation of a fairly extended burnout region behind the shock front (which is clearly demonstrated by the graphs of the coordinate dependencies of the mass concentrations $c_{i}$ of the reactants $A$ and products $B$). Therefore, the stationary structure of the detonation wave is determined not only by the energy release in a very narrow region near the chemical peak, but also by an extended region behind the shock front. This makes the overall structure of the detonation wave quite complex and requires that the numerical method very accurately reproduce not only the hydrodynamic but also the kinetic (in terms of reaction rates) properties of the system under study. From the presented results, it can be immediately noted that in all studied cases of the basis polynomials degrees $N = 1, \ldots, 12$, the correct numerical solution is obtained for values of the Courant number $\mathrm{CFL} > \mathrm{CFL}^{\rm NS}_{\rm max}(N)$ --- higher than the expected values of the Courant number $\mathrm{CFL}^{\rm NS}_{\rm max}(N)$ for the case of a system of equations without sources (even taking into account the small correction $+5\%$ to the boundary value of the Courant number $\mathrm{CFL}^{\rm NS}_{\rm max}(N)$ for the dissipativity of the HLLE~\cite{Toro_solvers_2009} or Rusanov~\cite{Rusanov_solver} approximate Riemann solvers, indicated in~\cite{ader_dg_stab_rig}), while in all cases the correct solution is also obtained in cases $\mathrm{CFL} \geqslant \mathrm{CFL}_{\rm CS}(N)$ --- not lower than the boundary value of the Courant number $\mathrm{CFL}_{\rm CS}(N)$ (\ref{eq:cfl_cs}) obtained by Cockburn and Shu~\cite{Cockburn_base_1, Cockburn_base_2, Cockburn_base_3, Cockburn_base_4, Cockburn_base_5} for classical DG methods. For basis polynomials degree $N = 1$, the numerical solution is correct and characterized of high quality, despite only $400$ spatial mesh cells $\Omega_{i}$: for all effective Courant numbers $\mathrm{C} = 0.90, \ldots, 1.35$ studied, all features of the reference solution are correctly reproduced in the numerical solution. The ZND-detonation wave front maintains its integrity, and its sharp component is smoothed over $2$--$3$ spatial mesh cells. In this case, the best results in terms of the number of troubled cells are achieved at effective Courant numbers of $\mathrm{C} = 0.90$--$1.05$. With a further increase in the effective Courant number $\mathrm{C}$, the number of troubled cells noticeably increases up to an effective Courant number $\mathrm{C} = 1.30$, after which the monotonous increase in the number of troubled cells is replaced by a decrease in their number. With an increase in the basis polynomials degree $N$, and in the case of degree $N = 2$ of basis polynomials, the quality of the numerical solution significantly improved: for all effective Courant numbers $\mathrm{C} = 0.90, \ldots, 1.15$ for which a correct numerical solution is obtained, all features of the reference solution are correctly reproduced in the numerical solution. The detonation wave front retains its integrity, but the amplitude of the ``small-scale oscillations'' is significantly decreased. This is due to a significant decrease in the number of troubled cells for the effective Courant numbers $\mathrm{C} = 0.90, \ldots, 1.15$ for which a numerical solution is obtained. However, in this case, the maximum effective Courant number $\mathrm{C}_{\rm max} = 1.15$ for which a correct numerical solution is still obtained. The detonation wave front is smoothed by $2$ spatial mesh cells. With a further increase in the basis polynomials degree $N$, for degree $N = 3$ of basis polynomials, a similarly high-quality numerical solution is achieved for a spatial mesh of $200$ cells $\Omega_{i}$; however, the number of troubled cells begins to increase. In this case, the spatial resolution of the detonation wave front is qualitatively improved --- it is smoothed by $1$--$2$ spatial mesh cells. The maximum Courant number $\mathrm{C}_{\rm max} = 1.15$ for which a correct numerical solution is obtained in the case $N = 3$. For degrees $N = 4$ and $5$ of basis polynomials, a high-quality numerical solution is achieved for a spatial mesh of $100$ cells, but the maximum Courant number $\mathrm{C}_{\rm max} = 1.10$ for which a correct numerical solution is obtained. In case $N = 4$, the number of troubled cells exhibits non-monotonic behavior with increasing effective Courant number $\mathrm{C}$, with the minimum number being achieved at the largest available effective Courant number of $\mathrm{C} = 1.10$. In case $N = 5$, the minimum number of troubled cells is achieved at an effective Courant number $\mathrm{C} = 1.05$. With a further increase in the basis polynomials degree $N$, for degree $N = 6$ of basis polynomials, a high-quality numerical solution is achieved for a spatial mesh of $50$ cells $\Omega_{i}$, while the relative number of troubled cells remains approximately the same as for degree $N = 4$ of basis polynomials. The maximum effective Courant number $\mathrm{C}_{\rm max} = 1.10$ for which a correct numerical solution is still obtained. In the case of degree $N = 7$ of basis polynomials, the general behavior of the numerical solution with increasing effective Courant number approximately corresponds to the case of degree $N = 5$. In the case of very high degrees of basis polynomials $N \geqslant 8$, the highest quality and most accurate numerical solution is achieved already with an effective Courant number $\mathrm{C} = 1.10$. In the case $N \geqslant 9$, the quality of the numerical solution does not deteriorate even on very coarse meshes --- $40$ cells of the spatial mesh. In the cases of degrees $N = 10$--$12$ of basis polynomials, the maximum Courant number $\mathrm{C}_{\rm max} = 1.00$ for which a correct numerical solution is still obtained. Moreover, the best results for degree $N = 10$ are achieved with an effective Courant number $\mathrm{C} = 1.00$, while for degrees $N = 11$ and $12$ --- with an effective Courant number of $\mathrm{C} = 0.90$. In the case of very high basis polynomials degrees $N \geqslant 7$, it is interesting to note that, despite such a coarse mesh, smoothing of the detonation wave front occurs per cell of the spatial mesh, and this despite the fact that the actual peaks of the coordinate dependencies of the solution (as evident from the reference solution) are located within the cells of the coordinate mesh.

The analysis continues with the case of a ``fast'' reaction (the strong stiff case). Figures~\ref{fig:test_zndf_1d_degrees_1_3},~\ref{fig:test_zndf_1d_degrees_4_6},~\ref{fig:test_zndf_1d_degrees_7_9} and~\ref{fig:test_zndf_1d_degrees_10_12} show the results of the numerical solution of the problem of formation and development of a ZND detonation wave in the case of ``fast'' reaction (strong stiff case), obtained using the ADER-DG method for the basis polynomials degree $N = 1$, $2$, $3$ (Figure~\ref{fig:test_zndf_1d_degrees_1_3}), $N = 4$, $5$, $6$ (Figure~\ref{fig:test_zndf_1d_degrees_4_6}), $N = 7$, $8$, $9$ (Figure~\ref{fig:test_zndf_1d_degrees_7_9}), $N = 10$, $11$, $12$ (Figure~\ref{fig:test_zndf_1d_degrees_10_12}), at the final time $t_{f} = 0.4$. All these Figures show the coordinates dependencies of density $\rho(x, t_{f})$, pressure $p(x, t_{f})$, flow velocity $u(x, t_{f})$, troubled cells indicator $\beta(x, t_{f})$ (multiplied by the separating function $\varpi(\mathrm{CFL}, N)$), mass concentrations $c_{1}(x, t_{f})$ and $c_{2}(x, t_{f})$ of the reactant $A$ and product $B$, respectively, and partial densities $\rho_{1}(x, t_{f})$ and $\rho_{2}(x, t_{f})$ of the reactant $A$ and product $B$, respectively. The presentation of the troubled cells indicator $\beta$ on graphs using the separating function $\varpi(\mathrm{CFL}, N)$ is done similarly to the case of a ``slow'' reaction (the weak stiff case) in Figures~\ref{fig:test_znds_1d_degrees_1_3},~\ref{fig:test_znds_1d_degrees_4_6},~\ref{fig:test_znds_1d_degrees_7_9} and~\ref{fig:test_znds_1d_degrees_10_12}: the separation function $\varpi(\mathrm{CFL}, N)$ is multiplied by the troubled cells indicator $\beta$, as a result of which the presented levels $\beta = 1$ for different values of the Courant number $\mathrm{CFL}$ are separated --- the positive values of the separating function $\mathrm{CFL}$ are chosen only for reasons of convenience of troubled cells indicator $\beta$ presentation on the graphs and do not carry any further meaning: the points with $\beta = 0$ is also equivalent to the points with  $\varpi(\mathrm{CFL}, N)\beta = 0$, and the points with $\beta = 1$ in this case is equivalent to the points with $\varpi(\mathrm{CFL}, N)\beta \in (0,\, 1]$. In each Figure presented, coordinate dependencies are presented only for those values of the Courant number $\mathrm{CFL} = \mathrm{C}\cdot\mathrm{CFL}_{\rm CS}(N)$ for which correct results are obtained; in other cases of the Courant numbers $\mathrm{CFL}$, the solution is obtained \texttt{nan} (``Not a number''~\cite{ieee754}). The number of spatial mesh cells is determined by the minimum number for which a correct and adequate numerical solution is obtained. The case of ``fast'' reaction (the strong stiff case) is interesting because the characteristic reaction times are significantly shorter than the characteristic hydrodynamic times, and this difference leads to the formation of a very narrow burnout region behind the shock front. Therefore, the stationary structure of the ZND-detonation wave is determined precisely by the correctness and effectiveness of detailed spatial and time resolution of the energy yield process in a very small region immediately behind the passing shock wave, in the vicinity of the chemical peak (as clearly demonstrated by the graphs of the coordinate dependencies of the mass concentrations $c_{i}$ of the reactants $A$ and products $B$). This makes the complete structure of the ZND-detonation wave quite complex and requires the numerical method to very accurately reproduce the extremely fast (compared to hydrodynamic processes alone) kinetic (in terms of reaction rates) processes in the system under study. Therefore, such a detailed specification of the coordinate dependencies of the numerical solution is necessary for a correct assessment of the quality of the numerical solution in terms of reproducing the features of the ZND-detonation wave structure~\cite{frac_steps_detwave_sim_2000}: not only how the process of propagation of hydrodynamic disturbances in the medium occurs, but also how the process of reagent burning and product accumulation occurs, how the compression of the unburned substance occurs during the reaction (which is clearly demonstrated by the graphs of the coordinate dependencies of the partial densities $\rho_{i}$ of the reactants $A$ and products $B$). In this case of ``fast'' reaction (a strong stiff case), if the numerical method is unable to effectively and accurately reproduce (both in time and space) the correct process of product-reactant synthesis, a well-known non-physical artifact is formed in the numerical solution: a fast weak detonation wave that overtakes the shock front~\cite{frac_steps_detwave_sim_2000, OranBoris_2005}. As a result, the reactant ahead of the shock front burns out prematurely, and the stationary detonation wave that should form in this case does not form. A combination of the weak detonation wave located ahead and the shock wave traveling at an increasing distance from it is formed. These non-physical artifacts are well known, especially when using methods of splitting into physical processes~\cite{frac_steps_detwave_sim_2000, chem_kin_hrs_rev_1, chem_kin_hrs_rev_2, chem_kin_hrs_weno, RTS_split_2019}, including high-precision numerical methods~\cite{chem_kin_hrs_weno}. The ``physical'' nature of this non-physical artifacts is presented in~\cite{correct_det_wave_speed_2017}. From the presented results, it can be immediately noted that in all studied cases of the basis polynomials degrees $N = 1, \ldots, 12$, the correct numerical solution is obtained for values of the Courant number $\mathrm{CFL} > \mathrm{CFL}^{\rm NS}_{\rm max}(N)$ --- higher than the expected values of the Courant number $\mathrm{CFL}^{\rm NS}_{\rm max}(N)$ for the case of a system of equations without sources (even taking into account the small correction $+5\%$ to the boundary value of the Courant number $\mathrm{CFL}^{\rm NS}_{\rm max}(N)$ for the dissipativity of the HLLE~\cite{Toro_solvers_2009} or Rusanov~\cite{Rusanov_solver} approximate Riemann solvers, indicated in~\cite{ader_dg_stab_rig}), while in all cases the correct solution is also obtained in cases $\mathrm{CFL} \geqslant \mathrm{CFL}_{\rm CS}(N)$ --- not lower than the boundary value of the Courant number $\mathrm{CFL}_{\rm CS}(N)$ (\ref{eq:cfl_cs}) obtained by Cockburn and Shu~\cite{Cockburn_base_1, Cockburn_base_2, Cockburn_base_3, Cockburn_base_4, Cockburn_base_5} for classical DG methods. In the case of the lowest of the considered degrees $N$ of the basis polynomials --- $N = 1$, the solution is correct for all studied values of the effective Courant number $\mathrm{C}$, and the correct numerical solution is achieved already for a spatial mesh of $400$ cells. However, the best quality solution is obtained for the effective Courant number of 0.90, and with its further increase, the number of troubled cells becomes very large, reaching $50$--$60\%$ in the cases of the effective Courant numbers $\mathrm{C} = 1.15$--$1.30$, slightly decreasing at $\mathrm{C} = 1.35$. In the case of $\mathrm{C} = 0.90$, the detonation front is smoothed by $1$--$2$ cells of the spatial mesh, and with the increase of the effective Courant number $\mathrm{C}$, the smoothing of the front increases to $2$--$3$ cells of the spatial mesh. An interesting feature of the numerical solution should be noted for the effective Courant numbers $\mathrm{C} > 1.25$ --- the detonation wave front begins to lag by one cell behind the detonation front in the reference solution; This is, in general, a typical effect for low-order numerical methods on fairly coarse (for this method order) spatial meshes. In the case of degree $N = 2$ of basis polynomials, the quality of the numerical solution increases significantly compared to case $N = 1$, and also for a spatial mesh of $400$ cells. However, the maximum value of the effective Courant number $\mathrm{C} = 1.25$, for which a correct solution is obtained. Moreover, the highest quality numerical solution is obtained at $\mathrm{C} = 1.05$, where there is the smallest number of troubled cells. In the case of $\mathrm{C} = 1.05$, the detonation front is smoothed by $1$--$2$ cells of the spatial mesh, and for other effective Courant numbers $\mathrm{C}$, the smoothing of the front increases to $3$ cells of the spatial mesh. In the case of degree $N = 3$ of basis polynomials, the correct numerical solution is achieved already for a spatial mesh of $200$ cells, but the maximum value of the effective Courant number $\mathrm{C} = 1.20$, for which a correct solution is obtained. In terms of the qualitative properties of the numerical solution, it is evident that it is even somewhat worse than in the case of degree 2 basis polynomials and a spatial mesh of $400$ cells, due to the increase in the number of troubled cells. It should be noted that in the case of an effective Courant number of $\mathrm{C} = 0.90$, a certain artifact of the numerical solution is also observed in the form of a non-physical contact discontinuity located slightly ahead of the main small contact discontinuity in the vicinity of the initial discontinuity point $x = 0.5$. For basis polynomials of degrees $N = 4$ and $5$, a correct numerical solution is achieved for a spatial mesh of $100$ cells, but the maximum effective Courant number is $\mathrm{C} = 1.10$ for case 4 and $\mathrm{C} = 1.05$ for case $N = 5$, for which a correct solution is obtained. The quality of the numerical solution is quite high, but in all cases, a slight gap of $1$--$2$ spatial mesh cells is observed between the detonation front in the numerical solution and the detonation front in the reference solution. The detonation front itself is smoothed by 1-2 spatial mesh cells. In the case of degree $N = 6$ of basis polynomials, the correct numerical solution is achieved for a spatial mesh of $50$ cells, and the maximum effective Courant number $\mathrm{C} = 1.05$, for which a correct solution is obtained, remains the same as in the case $N = 5$. In the case $N = 6$, the detonation front in the numerical solution also stays behind the detonation front in the reference solution by one spatial mesh cell, while the detonation front itself is smoothed by one spatial mesh cell. In the case of high degrees of basis polynomials $N \geqslant 7$, the qualitative properties of the numerical solution improve significantly, which is expected for the ADER-DG numerical methods with the LST-DG predictor. However, in all these cases, the maximum effective Courant number $\mathrm{C} = 1.00$, for which a correct solution is obtained. For degrees $N = 7$ and $8$ of basis polynomials, a correct numerical solution is achieved for a spatial mesh of $50$ cells, and for $N \geqslant 9$ --- for a spatial mesh of $40$ cells, which is an extremely coarse mesh for a numerical solution of this case of a ``fast'' reaction (the strong stiff case). The numerical solution correctly and accurately reproduces all the features of the reference solution, not only the detonation front itself and the chemical spike, but also the contact discontinuity and the small rarefaction wave. The detonation front itself is smoothed by no more than one spatial mesh cell.

The obtained results demonstrate that the numerical ADER-DG method with the LST-DG predictor allows solving fairly complex systems of hyperbolic equations with stiff source terms. Moreover, the boundary value of the Courant number, even in the case of quasilinear systems of hyperbolic equations (rather than a single linear equation), is higher than in the case of a system of equations without a stiff source term and is not even lower than estimates $\mathrm{CFL}_{\rm CS}(N)$ from the Cockburn-Shu formula (\ref{eq:cfl_cs}). However, the presented results are not very convenient for comparison with the boundary values of the Courant number $\mathrm{CFL}_{\rm max}^{\rm NS}(N) \equiv \mathrm{CFL}_{\rm max}(N, 0)$~\cite{ader_dg_stab_rig} in the case of quasilinear systems of hyperbolic equations without a stiff source term. Tables~\ref{tab:cfls_impr_znd_degrees_1_4},~\ref{tab:cfls_impr_znd_degrees_5_8} and~\ref{tab:cfls_impr_znd_degrees_9_12} show the effective Courant numbers $\mathrm{C}$ and Courant numbers $\mathrm{C}\cdot\mathrm{CFL}_{\rm CS}(N)$ for which the test of formation and propagation of a ZND detonation wave is correctly passed or failed in the ``slow'' reaction case (the weak stiff case) and the ``fast'' reaction case (the strong stiff case) for the numerical method ADER-DG with the LST-DG predictor for degrees $N = 1, \ldots, 12$ of basis polynomials: $N = 1$, $2$, $3$, $4$ (Table~\ref{tab:cfls_impr_znd_degrees_1_4}), $N = 5$, $6$, $7$, $8$ (Table~\ref{tab:cfls_impr_znd_degrees_5_8}), $N = 9$, $10$, $11$, $12$ (Table~\ref{tab:cfls_impr_znd_degrees_9_12}). For comparison, the boundary values of the Courant number $\mathrm{CFL}_{\rm max}^{\rm NS}(N) \equiv \mathrm{CFL}_{\rm max}(N, 0)$ for the numerical method in the case of a problem (\ref{eq:adv_eq_src}) without a source term ($\nu = 0$) and the ratio of the Courant number $\mathrm{C}\cdot\mathrm{CFL}_{\rm CS}(N)$ used in the test to the $\mathrm{CFL}_{\rm max}^{\rm NS}(N)$ are presented. From the results presented in these Tables, it is evident that in the case of degree $N = 1$ of basis polynomials, the correct numerical solution is obtained at least up to $\mathrm{CFL}_{\rm max} \approx 1.35\,\mathrm{CFL}_{\rm max}^{\rm NS}(1)$ in both cases of weak and strong stiffness of the source term. In the case of degree $N = 2$ of basis polynomials, in the case of weak stiffness of the source term, the correct numerical solution is obtained up to $\mathrm{CFL}_{\rm max} \approx 1.35\,\mathrm{CFL}_{\rm max}^{\rm NS}(2)$, and in the case of strong stiffness of the source term --- up to $\mathrm{CFL}_{\rm max} \approx 1.46\,\mathrm{CFL}_{\rm max}^{\rm NS}(2)$. In case degree $N = 3$ of basis polynomials, in the case of weak stiffness of the source term, the correct numerical solution is obtained up to $\mathrm{CFL}_{\rm max} \approx 1.58\,\mathrm{CFL}_{\rm max}^{\rm NS}(3)$, and in the case of strong stiffness of the source term --- up to $\mathrm{CFL}_{\rm max} \approx 1.65\,\mathrm{CFL}_{\rm max}^{\rm NS}(3)$. In case degree $N = 4$ of basis polynomials, in both cases of weak and strong stiffness of the source term, the correct numerical solution is obtained up to $\mathrm{CFL}_{\rm max} \approx 1.75\,\mathrm{CFL}_{\rm max}^{\rm NS}(4)$. It is evident that up to degree $N = 4$, the boundary value of the Courant number $\mathrm{CFL}_{\rm max}$ in the case of strong stiffness of the source term was not less than in the case of weak stiffness. In case degree $N = 5$, this tendency changes its character: in the case of weak stiffness of the source term, the correct numerical solution is obtained up to $\mathrm{CFL}_{\rm max} \approx 2.00\,\mathrm{CFL}_{\rm max}^{\rm NS}(5)$, and in the case of strong stiffness of the source term --- up to $\mathrm{CFL}_{\rm max} \approx 1.90\,\mathrm{CFL}_{\rm max}^{\rm NS}(5)$. In case degree $N = 6$, the correct numerical solution is obtained up to $\mathrm{CFL}_{\rm max} \approx 2.24\,\mathrm{CFL}_{\rm max}^{\rm NS}(6)$ in the case of weak stiffness of the source term, and up to $\mathrm{CFL}_{\rm max} \approx 2.14\,\mathrm{CFL}_{\rm max}^{\rm NS}(6)$ in the case of strong stiffness of the source term. In case degree $N = 7$ of basis polynomials, in the case of weak stiffness of the source term, the correct numerical solution is obtained up to $\mathrm{CFL}_{\rm max} \approx 2.50\,\mathrm{CFL}_{\rm max}^{\rm NS}(7)$, and in the case of strong stiffness of the source term --- up to $\mathrm{CFL}_{\rm max} \approx 2.27\,\mathrm{CFL}_{\rm max}^{\rm NS}(7)$. In case degree $N = 8$ of basis polynomials, in the case of weak stiffness of the source term, the correct numerical solution is obtained up to $\mathrm{CFL}_{\rm max} \approx 2.62\,\mathrm{CFL}_{\rm max}^{\rm NS}(8)$, and in the case of strong stiffness of the source term --- up to $\mathrm{CFL}_{\rm max} \approx 2.50\,\mathrm{CFL}_{\rm max}^{\rm NS}(8)$. In case degree $N = 9$ of basis polynomials, which among the studied degrees $N$ of polynomials closes the sequence of degrees of polynomials $N = 1, \ldots, 12$ for which the opposite tendency took place, the following results take place: in the case of weak stiffness of the source term, the correct numerical solution is obtained up to $\mathrm{CFL}_{\rm max} \approx 2.87\,\mathrm{CFL}_{\rm max}^{\rm NS}(9)$, and in the case of strong stiffness of the source term --- up to $\mathrm{CFL}_{\rm max} \approx 2.73\,\mathrm{CFL}_{\rm max}^{\rm NS}(9)$. In case degree $N = 10$ of basis polynomials, in both cases of weak and strong stiffness of the source term, the correct numerical solution is obtained up to $\mathrm{CFL}_{\rm max} \approx 2.96\,\mathrm{CFL}_{\rm max}^{\rm NS}(10)$, and in cases degrees $N = 11$ and $12$ of basis polynomials, in both cases of weak and strong stiffness of the source term --- $\mathrm{CFL}_{\rm max} \approx 3.19\,\mathrm{CFL}_{\rm max}^{\rm NS}(11)$ and $\approx 3.43\,\mathrm{CFL}_{\rm max}^{\rm NS}(12)$, respectively. A clear trend is observed: as the degree of the basis polynomials increases, a boundary value of the Courant number $\mathrm{CFL}_{\rm max}$ occurs on the scale of the Courant number $\mathrm{CFL}_{\rm max}^{\rm NS}$ of the problem without stiff source terms. For very high degrees $N \geqslant 12$ of basis polynomials, the boundary value of the Courant number $\mathrm{CFL}_{\rm max}$ of the problem with stiff source terms is $3.0$ or more times greater than the boundary value of the Courant number $\mathrm{CFL}_{\rm max}^{\rm NS}$ of the problem without stiff source terms. Clearly, this is not tens or hundreds of times higher, as in the case of the linear advection-reaction equation (\ref{eq:adv_eq_s_src}) presented in Subsection~\ref{sec:app:lin_adv} ``Linear advection-reaction equation'', but it is still an impressive result. It should be emphasized that for very high polynomial degrees $N \geqslant 10$, the boundary value of the Courant number $\mathrm{CFL}_{\rm max}$ of the problem with stiff source terms is well described by the Cockburn-Shu formula (\ref{eq:cfl_cs}).

The presented results clearly demonstrate that the ADER-DG numerical method with the LST-DG predictor exhibits significantly higher boundary values of the Courant number $\mathrm{CFL}_{\rm max}$ for a system of quasilinear hyperbolic equations (\ref{eq:nonlin_sys_source}) with a stiff source term. However, unlike the linear advection-reaction equation (\ref{eq:adv_eq_src}), in this case the increase in the boundary value of the Courant number is not as significant, due to the completely different stiffness structure in systems of quasilinear equations (\ref{eq:nonlin_sys_source}) with source terms. At the same time, these results explain well how it was possible to use such large effective Courant number $\mathrm{C} = 0.90$ when modeling ZND detonation waves using the ADER-DG numerical method with the LST-DG predictor~\cite{ader_stiff_4}, and that the ADER-DG numerical method with the LST-DG predictor is, in general, well suited for solving such problems with strong stiffness of the source terms.

\section*{Conclusion}
\addcontentsline{toc}{section}{Conclusion}
\label{concl}

In conclusion, it should be noted that this work constructs a fairly general stability theory for the ADER-DG numerical method with the LST-DG predictor for hyperbolic problems with stiff source terms. This theory is based on the analysis of the von Neumann stability criterion (spectral stability criterion) for a linear advection-reaction equation (\ref{eq:adv_eq_src}) with a decay term. It is shown that with increasing source term stiffness, the boundary value of the Courant number $\mathrm{CFL}_{\rm max}$, which determines the stability boundary $\mathrm{CFL}\in[0, \mathrm{CFL}_{\rm max}]$ of the numerical method, increases significantly, almost linearly with the parameter $|z| = \nu\Dtn{n}$, which represents the effective ``reaction Courant number'' $\mathrm{CFL}_{\rm react} \equiv |z|$ and characterizes the relative scale of the characteristic reaction time $\sim |\nu|^{-1}$ in the system. The resulting boundary values of the Courant number $\mathrm{CFL}_{\rm max}$, which characterizes the relative hydrodynamic times, are related to the parameter $|z|$ as $\mathrm{CFL}_{\rm max} \ll |z|$ in region $|z| \gg 1$. A preliminary detailed qualitative explanation of such a non-intuitive behavior of the numerical method in the case of the existence of strong stiff terms in the system is presented, based on an analogy with the classical IMEX finite-difference scheme for the linear advection-reaction equation, and this analogy received a fairly accurate quantitative confirmation in the present work.

The developed theory was based on the empirical argument that loss of stability occurs when one of the eigenvalues $\{\lambda_{k}\}$ of the matrix $\mathrm{R}(\mathrm{CFL}, z, \theta)$ (\ref{eq:r_matrix_in_matrix_form}) of the evolution operator $R$ for a single time step $\Dtn{n}$ is reached. This empirical argument was confirmed by a set of high-precision calculations. In this case, unlike the case of a linear advection equation, where this argument was confirmed only with a small accuracy (see the detailed discussion in~\cite{ader_dg_stab_rig} and~\cite{ader_dg_stab}), in the case of a linear advection-reaction equation, small unstable eigenvalues with phase $\arg\lambda_{k} \in [-\pi/4, +\pi/4]$ did not arise even for values of the parameter $|z| > 10^{-5}$--$10^{-6}$ (which are of no practical interest), and especially in the region $|z| \gg 1$. This result guarantees that in the case of applying the ADER-DG numerical method with the LST-DG predictor to solving stiffness problems, anomalously small boundary values of the Courant number $\mathrm{CFL}_{\rm max}$ (calculated in~\cite{ader_dg_stab} and discussed in~\cite{ader_dg_stab_rig}) do not arise. Calculations have shown that with an increase in the parameter $|z|$, a significant compression of the spectrum of the matrix $\mathrm{R}(\mathrm{CFL}, z, \theta)$ (\ref{eq:r_matrix_in_matrix_form}) of the evolution operator $R$ for a single time step $\Dtn{n}$ occurs, which has a behavior similar to the behavior of the stability function of the classical IMEX finite-difference scheme for a linear advection-reaction equation. Additionally, a function $F(\varphi; \mathrm{CFL}, z)$ was derived, in whose expression the phase $\theta$ was completely eliminated, and whose root $F(\varphi; \mathrm{CFL}, z) = 0$ at a certain phase $\phi = \arg\lambda_{k}$ and Courant number $\mathrm{CFL}$ should indicate the stability boundary. Calculations showed that the roots of equation $F(\varphi; \mathrm{CFL}, z) = 0$ occurs precisely for phase $\varphi = \pi$, further confirming empirical considerations regarding the loss of stability precisely when one of the eigenvalues $\{\lambda_{k}\}$ of matrix $\mathrm{R}(\mathrm{CFL}, z, \theta)$ (\ref{eq:r_matrix_in_matrix_form}) of the evolution operator $R$ for a single time step $\Dtn{n}$ reaches $\lambda = -1$.

The developed theory allows one to calculate the boundary value of the Courant number $\mathrm{CFL}_{\rm max}(N, z)$ for any degree $N$ of the basis polynomials and any value of the parameter $z$. The calculation requires solving two polynomial equations of degree $N+1$ for the boundary value of the Courant number $\mathrm{CFL}_{\rm max}(N, z)$: $\Phi_{-}(\mathrm{CFL}; N, z) = 0$ and $\Phi_{+}(\mathrm{CFL}; N, z) = 0$, and selecting the smallest positive real root $\mathrm{CFL}\in\mathcal{R}_{+}$ among all the roots. The coefficients of these polynomial equations are polynomials of the parameter $z$. This problem of finding the roots of a polynomial equations is computationally very simple, especially compared to the problem of very precisely determining the lower bound of the solution set of the von Neumann stability criterion inequality $\rho[\mathrm{R}(\mathrm{CFL}, z, \theta)] \leqslant 1$ for the spectral radius $\rho$ of the matrix $\mathrm{R}(\mathrm{CFL}, z, \theta)$ (\ref{eq:r_matrix_in_matrix_form}) of the evolution operator $R$ for a single time step $\Dtn{n}$, especially in the case $|z| \gg 1$, when calculating the eigenvalues can be computationally unstable, and a very high level of detailing of the dependence on phase $\theta$ might be required, for example, in the form of discretization of the phase definition domain $[0, 2\pi]$ with a very fine or adaptive grid. A rigorous direct proof of the approximation of the ADER-DG numerical method with the LST-DG predictor is presented. Approximation orders $p = N+1$ for arbitrary degrees $N$ of basis polynomials are rigorously derived. It was also rigorously proven that in the case of a constant initial condition, the ADER-DG numerical method with the LST-DG predictor strictly demonstrates a ``decay property'' similar in essence to the ``constant-preserving property'' in the case of a linear advection equation, while the solution characterizing this ``decay property'' has an approximation order $p = 2N+1$, which is consistent with the theoretical results for the application of the ADER-DG numerical method with the local time-DG predictor to solving the initial value problem for an ODE system~\cite{ader_dg_ivp_ode, ader_dg_ivp_dae, ader_dg_ivp_ode_sinum}.

The dependencies of the boundary value of the Courant number $\mathrm{CFL}_{\rm max}(N, z)$ on the parameter $z$ were calculated over a wide range of parameter $-z\in[10^{-6},\, 10^{3}]$ values for degrees $N = 1, \ldots, 12$ of basis polynomials. It was found that in the region of small parameter $|z| \ll 1$, a plateau occurs with a Courant number $\mathrm{CFL}_{\rm max}(N, z)$ corresponding to the case $\mathrm{CFL}_{\rm max}(N, 0)$ of the advection equation without a source term ($z = 0$). In the region $|z| \gg 1$, a practically strict linear increase in the boundary value of the Courant number $\mathrm{CFL}_{\rm max}(N, z)$ with $z$ occurs. The slope coefficients $r(N)$ of this linear growth law were calculated for degrees $N = 1, \ldots, 12$ of the basis polynomials. It was shown that with increasing degree $N$ of the basis polynomials, this slope coefficient $r(N)$ decreases. However, as a result of normalizing the slope coefficient $r(A)$ to the boundary value of the Courant number $\mathrm{CFL}_{\rm max}(N, 0)$ for the linear advection equation (or to the minimum value of the Courant number $\mathrm{CFL}_{\rm max}(N, z_{\rm min})$), which follows from a qualitative consideration of the analogy between the ADER-DG numerical method with the LST-DG predictor and the classical IMEX scheme for the linear advection-reaction equation, it was revealed that this decrease of $r(N)$ is associated precisely with a decrease in the Courant number $\mathrm{CFL}_{\rm max}(N, 0)$ for the linear advection equation with increasing degree $N$ of basis polynomials, and this also additionally confirmed the correctness of the chosen qualitative analogy and consideration. The obtained boundary values of the Courant number $\mathrm{CFL}_{\rm max}(N, z)$ clearly indicate that the following relationship of characteristic scales holds: $\mathrm{CFL}_{\rm react} \gg \mathrm{CFL}_{\rm max}(N, z) \gg \mathrm{CFL}_{\rm max}(N, 0)$, which clearly demonstrates the good applicability of the ADER-DG numerical method with the LST-DG predictor for solving problems with stiff source terms --- local reaction processes (or other local processes corresponding to stiff source terms) are dynamically resolved on time scales $\Dtn{n}$ significantly longer than the characteristic reaction time $\sim |\nu|^{-1}$: $\Dtn{n} \gg |\nu|^{-1}$. Between the two characteristic regimes $|z| \ll 1$ and $|z| \gg 1$, which correspond to the case of a very small source term and a strong stiff source term, a rather narrow region of decreasing boundary value of the Courant number $\mathrm{CFL}_{\rm max}(N, z)$ with respect to the parameter $z$ was discovered in region $|z| \sim 1$ (at point $z_{\rm min} < 1$). However, although this decrease of $\mathrm{CFL}_{\rm max}(N, z)$ increased with increasing degree $N$ of the basis polynomials, it did not exceed $\Delta \mathrm{CFL}_{\rm max}(N, z) \approx 5$--$6\%$ even for degree $N = 12$ of the basis polynomials and asymptotically tended to it (this is covered by the dissipativity of Riemann solvers in applied calculations~\cite{ader_dg_stab_rig}). This phenomenon cannot be explained within the framework of the used qualitative consideration with an analogy with the IMEX finite-difference scheme, and this can be attributed to the ``subtle'' native properties of the ADER-DG numerical method with the LST-DG predictor.

The set of numerical experiments was carried out using the ADER-DG numerical method with the LST-DG predictor to solve the linear advection-reaction equation (\ref{eq:adv_eq_s_src}) to verify the developed theory. Large values of the decay parameter $\nu$ were chosen: $\nu = -10^{2}$ and $\nu = -10^{3}$, the wide coordinate definition domain $\Omega = [-50.0,\, +50.0]$, coarse spatial meshes (up to 10 discretization domains $\Omega_{i}$, with coordinate discretization step $\Dx = 10.0$), the large time range $t_{f} - t_{0} = 1.0 \cdot 10^{6}$ of integration, and the Courant numbers $\mathrm{CFL} = 0.1$, $1.0$ and $10.0$ were chosen. The obtained results showed that the ADER-DG numerical method with the LST-DG predictor allows one to obtain a highly accurate correct numerical solution even under such strong conditions. The calculated empirical convergence orders in various classical grid norms showed good agreement with the expected value of $p = N+1$, within the limits expected for these numerical methods~\cite{ader_stiff_3, ader_stiff_4}. Calculations were also carried out for the Courant numbers $\mathrm{CFL} = 100.0$ and $1000.0$, the results of which showed that in such cases of extremely large time steps $\Dtn{n}$ (for which the solution profile must traverses the entire coordinate definition domain $\Omega$ in one time step $\Dtn{n}$), the dispersion effects of the ADER-DG numerical method with the LST-DG predictor have a significant impact and non-physical artifacts of the solution arise, however, point-wise convergence is preserved. These results clearly confirm the rigorous and partially empirical developed theory.

To test the applicability of the obtained results to solving ``real-world'' problems, a set of numerical experiments was carried out related to solving the system of Euler equations for a two-component reacting medium with a monomolecular reaction with energy release and with divergent initiation (\ref{eq:euler_react_eqs_def}). Two cases of initiation time were considered: the case of a ``slow'' reaction (the weak stiff case) and the case of a ``fast'' reaction (the strong stiff case). The latter case corresponds to the classical situation of formation and propagation of a ZND detonation wave. The obtained results showed that the ADER-DG numerical method with the LST-DG predictor allows one to obtain a highly accurate numerical solution of these problems even on fairly coarse grids. The numerical solution correctly and accurately reproduces all the features of the reference solution, not only the detonation front itself and the chemical spike, but also the contact discontinuity and the small rarefaction wave. It was shown that the correct numerical solution is obtained even when using Courant numbers $\mathrm{CFL} > 3\,\mathrm{CFL}_{\rm max}(N, 0)$, including and in the case of the Courant number from the Cockburn-Shu estimate $\mathrm{CFL}_{\rm CS}(N)$ (\ref{eq:cfl_cs})~\cite{Cockburn_base_1, Cockburn_base_2, Cockburn_base_3, Cockburn_base_4, Cockburn_base_5} for classical DG methods. These results demonstrate that the obtained theoretical results are also correct in the case of quasilinear systems of equations with stiff source terms, and that the ADER-DG numerical method with the LST-DG predictor is well applicable to solving such problems.

\bmhead{Acknowledgments}
The reported study was supported by the Russian Science Foundation grant No. 26-71-10011 \url{https://rscf.ru/en/project/26-71-10011/}.

\subsection*{Data Availability Statement}
\addcontentsline{toc}{subsection}{Data Availability Statement}
Data will be made available on reasonable request.

\section*{Declarations}
\addcontentsline{toc}{section}{Declarations}

\subsection*{Conflict of interest}
\addcontentsline{toc}{subsection}{Conflict of interest}
The authors declare that they have no conflict of interest.

\addcontentsline{toc}{section}{References}


\end{document}